\documentclass[rmp,reprint,aps,floatfix]{revtex4-2}
\usepackage{amssymb,amsmath}
\usepackage{graphicx,bm}
\usepackage[colorlinks,citecolor=blue]{hyperref}
\begin{document}

\title{Origin of the heaviest elements: The rapid neutron-capture process}
  
\author{John J.~Cowan}
\email{jjcowan1@ou.edu}
\affiliation{\mbox{HLD
  Department of Physics and Astronomy, University of Oklahoma},
  \mbox{440 W. Brooks St., Norman, OK 73019, USA}}

\author{Christopher~Sneden}
\email{chris@astro.as.utexas.edu}
\affiliation{\mbox{Department of Astronomy, University of Texas},
  \mbox{2515 Speedway,
  Austin, TX 78712-1205, USA}}

\author{James E.~Lawler}
\email{jelawler@wisc.edu}
\affiliation{\mbox{Physics Department, University of
    Wisconsin-Madison},
  \mbox{1150 University Avenue,
  Madison, WI 53706-1390, USA}}

\author{Ani~Aprahamian}
\email{aapraham@nd.edu}
\affiliation{\mbox{Department of Physics and Joint Institute for Nuclear
  Astrophysics, University of Notre Dame},
 \mbox{225 Nieuwland Science Hall,
  Notre Dame, IN 46556, USA}}

\author{Michael~Wiescher}
\email{Michael.C.Wiescher.1@nd.edu}
\affiliation{\mbox{Department of Physics and Joint Institute for Nuclear
  Astrophysics, University of Notre Dame},
 \mbox{225 Nieuwland Science Hall,
  Notre Dame, IN 46556, USA}}

\author{Karlheinz~Langanke}
\email{k.langanke@gsi.de}
\affiliation{GSI Helmholtzzentrum f\"ur Schwerionenforschung,
  \mbox{Planckstra{\ss}e~1, 64291 Darmstadt, Germany}}
\affiliation{\mbox{and Institut f{\"u}r Kernphysik (Theoriezentrum),
    Fachbereich Physik}, Technische Universit{\"a}t
    Darmstadt, \mbox{Schlossgartenstra{\ss}e 2, 64298 Darmstadt, Germany}}

\author{Gabriel~Mart\'{\i}nez-Pinedo}
\email{g.martinez@gsi.de}
\affiliation{GSI Helmholtzzentrum f\"ur Schwerionenforschung,
  \mbox{Planckstra{\ss}e~1, 64291 Darmstadt, Germany;}}
\affiliation{\mbox{Institut f{\"u}r Kernphysik (Theoriezentrum),
    Fachbereich Physik}, Technische Universit{\"a}t
    Darmstadt, \mbox{Schlossgartenstra{\ss}e 2, 64298 Darmstadt, Germany;}}
\affiliation{and Helmholtz Forschungsakademie Hessen f\"ur FAIR, GSI
    Helmholtzzentrum f\"ur Schwerionenforschung,
  \mbox{Planckstra{\ss}e~1,
    64291 Darmstadt, Germany}} 

\author{Friedrich-Karl~Thielemann}
\email{f-k.thielemann@unibas.ch}
\affiliation{\mbox{Department of Physics, University of Basel},
  \mbox{Klingelbergstrasse~82,
  4056 Basel, Switzerland}}
\affiliation{and GSI
  Helmholtzzentrum f\"ur Schwerionenforschung,
  \mbox{Planckstra{\ss}e~1,
  64291 Darmstadt, Germany}}

\date{\today}

\begin{abstract}
  The production of about half of the heavy elements found in nature is
  assigned to a specific astrophysical nucleosynthesis process: the
  rapid neutron capture process (r-process). Although this idea has
  been postulated more than six decades ago, the full understanding
  faces two types of uncertainties/open questions: (a) The
  nucleosynthesis path in the nuclear chart runs close to the
  neutron-drip line, where presently only limited experimental
  information is available, and one has to rely strongly on
  theoretical predictions for nuclear properties.  (b) While for many
  years the occurrence of the r-process has been associated with
  supernovae, where the innermost ejecta close to the central neutron
  star were supposed to be neutron-rich, more recent studies have cast
  substantial doubts on this environment. Possibly only a weak
  r-process, with no or negligible production of the third r-process
  peak, can be accounted for, while much more neutron-rich conditions,
  including an r-process path with fission-cycling, are likely
  responsible for the majority of the heavy r-process elements. Such
  conditions could result during the ejection of initially highly
  neutron-rich matter, as found in neutron stars, or during the fast
  ejection of matter which has prior experienced strong
  electron-captures at high densities. Possible scenarios are the
  mergers of neutron stars, neutron-star black hole mergers, but
  include also rare classes of supernovae as well as
  hypernovae/collapsars with polar jet ejecta and possibly also
  accretion disk outflows related to the collapse of fast rotating
  massive stars.  The composition of the ejecta from each event
  determines the temporal evolution of the r-process abundances during
  the ``chemical'' evolution of the Galaxy. Stellar r-process
  abundance observations, have provided insights into, and constraints
  on the frequency of and conditions in the responsible stellar
  production sites.  One of them, neutron star mergers, was just
  identified thanks to the observation of the r-process kilonova
  electromagnetic transient following the Gravitational Wave event
  GW170817. These observations, increasingly more precise due to
  improved experimental atomic data and high resolution observations,
  have been particularly important in defining the heavy element
  abundance patterns of the old halo stars, and thus determining the
  extent, and nature, of the earliest nucleosynthesis in our
  Galaxy. Combining new results and important breakthroughs in the
  related nuclear, atomic and astronomical fields of science, this
  review attempts to provide an answer to the question ``How Were the
  Elements from Iron to Uranium Made?''
\end{abstract}

\maketitle

\tableofcontents

\section{Introduction and historical reviews}
\label{sec:intr-hist-revi}

At present we know of 118 elements from charge number $Z=1$ (H) to
$Z=118$ (Og). Eighty of them have at least one stable isotope (up to
$Z=82$, Pb) with $Z=43$ (Tc) and $Z=61$ (Pm) being unstable. Another
11 elements up to $Z=94$ (Pu) [with the exception of $Z=93$, Np] are
naturally occurring on earth with sufficiently long half-lives, while
the remaining ones with short half-lives have only been either
produced in laboratory or possibly also astrophysical
environments. The question of how this took place in the Universe is a
long-standing one. Presently we know that of the natural
elements/isotopes only $^{1,2}$H, $^{3,4}$He and $^{7}$Li originate in
the Big Bang, with problems remaining in understanding the abundance
of $^7$Li \citep{Cyburt.Olive.ea:2016,Pitrou.Coc.ea:2018}.  All other
elements were synthesized in stars, the first ones forming a few
hundred million years after the Big Bang.  The majority of stars,
which have long evolutionary phases, are powered by fusion
reactions. Major concepts for stellar burning were laid out in the
1950s \citep{Burbidge.Burbidge.ea:1957,Cameron:1957}, including the
then called x-process which today is understood via spallation of
nuclei by cosmic rays \citep[e.g.][]{Prantzos:2012}.  During their
evolution, and in explosive end phases, massive stars can synthesize
elements from C through Ti, the iron-peak elements (e.g.,
$21 \leq Z \leq 30$ from Sc to Zn) and beyond, \citep[as outlined over
many years,
e.g.][]{Howard.Arnett.ea:1972,Woosley.Heger:2007,Wanajo.Mueller.ea:2018,Curtis.ea:2019}.
The major result is, however, that the production of heavier nuclei up
to Pb, Bi, and the actinides requires free neutrons, as
charged-particle reactions in stellar evolution and explosions lead
typically to full chemical or quasi equilibria which favor the
abundance of nuclei with the highest nuclear binding energies,
occurring in the Fe-peak~\citep{Hix.ea:2007}.

A (very) small number of these heavy isotopes can be produced as a
result of charged-particle and photon-induced reactions in explosive
nucleosynthesis, the so called (proton-rich) p-process
\citep[e.g.][and references
therein]{Arnould.Goriely:2003,Travaglio.Rauscher.ea:2018,Nishimura.Rauscher.ea:2018},
and possibly a further contribution resulting from interactions with
neutrinos in such environments, including the $\nu$
process~\citep{Woosley.Hartmann.ea:1990,Heger.Kolbe.ea:2005,Suzuki.Kajino:2013,Sieverding.Martinez-Pinedo.ea:2019}
and $\nu$p
process~\citep{Froehlich.Martinez-Pinedo.ea:2006,Pruet.Hoffman.ea:2006,Wanajo:2006}.

\begin{figure}[htb]
  \includegraphics[width=\linewidth]{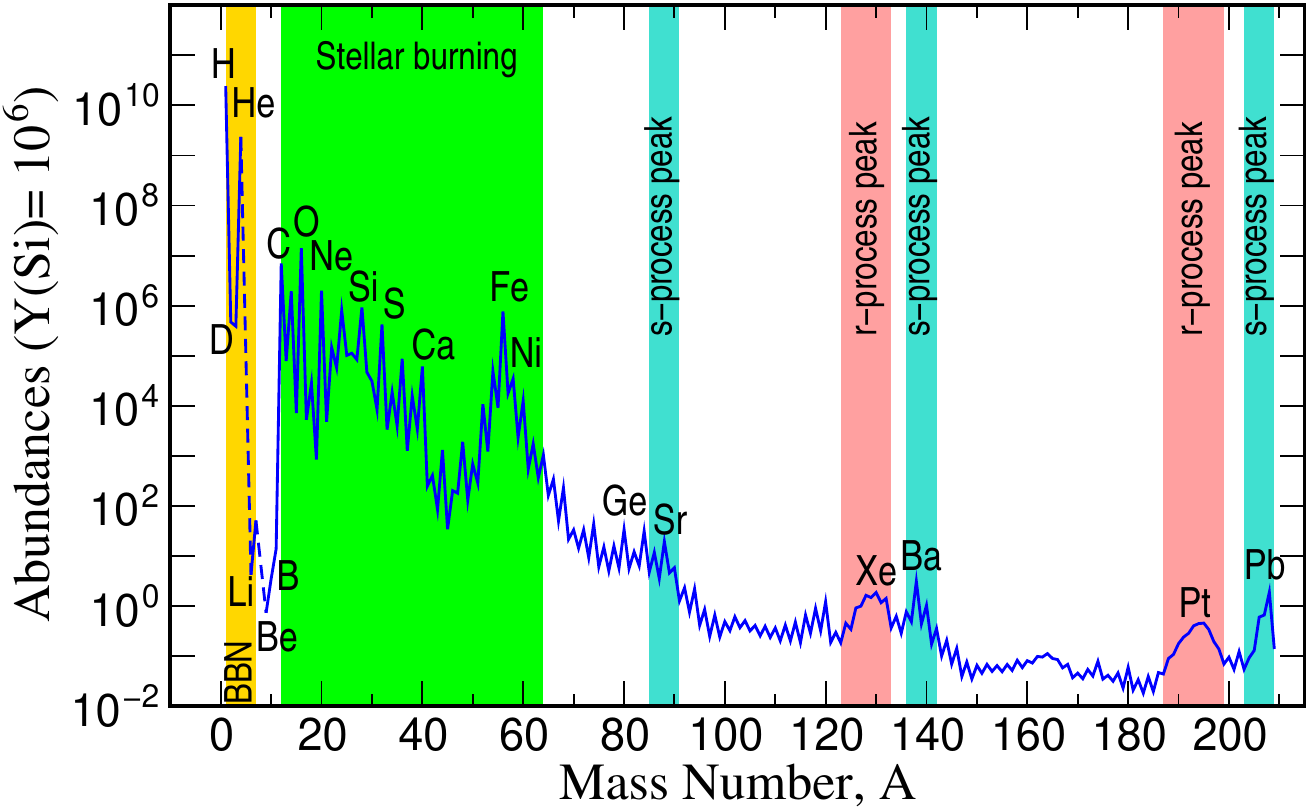}
  \caption{Abundances, $Y_i$, of elements and their isotopes in the
    solar system as a function of mass number $A_i=Z_i+N_i$.
    $A_i Y_i $ is equal to the mass fraction of isotope $i$, the sum
    of mass fractions amounts to 1, $\sum_iA_i Y_i=1$. The present
    figure utilizes a scaling, leading (for historical reasons) to an
    abundance of $10^6$ for the element Si. Element ratios are
    obtained from solar spectra, the isotopic ratios from primitive
    meteorites and terrestrial values
    \citep{Asplund.Grevesse.ea:2009,lodders09}. These values represent
    a snapshot in time of the abundances within the gas that formed
    the solar system. \label{fig:solarabund}}
\end{figure}

The two main processes involving the capture of free neutrons are the
slow (s)-process and the rapid (r)-process (due to low or high
densities of neutrons available and the resulting reaction timescales
of neutron captures). In the s-process, taking place during stellar
evolution and passing through nuclei near stability, there is
sufficient time for beta-decay between two neutron captures.  The
process timescale ranges from hundreds to thousands of years. For many
of these nuclei experimental data are available \citep[see
e.g.][]{Kaeppeler.Gallino.ea:2011,karakas14,Reifarth.Lederer.Kaeppeler:2014}.
In order to allow for the production of the heaviest nuclei over a
timescale of seconds, the r-process operates far from stability, which
requires high neutron densities. This involves highly unstable nuclei,
for which still little experimental data are available. In addition,
the quest for the stellar origin of the required conditions involved a
large number of speculations for many decades~\citep[e.g.][]
{Cowan.Thielemann.Truran:1991,arnould07}. [There are also
observational indications of intermediate neutron capture processes
between the s and the r-process, e.g., the i-process \citep{cowan77},
possibly occurring in super-AGB stars \citep{Jones.Ritter.ea:2016}.]
Fig.~\ref{fig:solarabund} gives an overview of the major contributions
to the solar system abundances.  It includes the Big Bang (light
elements H, He, Li and their isotopes $^{1,2}$H, $^{3,4}$He and
$^7$Li, given in yellow), plus stellar sources, contributing via winds
and explosions to the interstellar medium until the formation of the
solar system. These stellar burning abundances result from
charged-particle reactions up to the Fe-group in stellar evolution and
explosions (green), and neutron capture processes. The latter are a
superposition of (understood) slow neutron captures (s-process) in
helium burning of stars (with abundance maxima at closed neutron
shells for stable nuclei, turquoise), and a rapid neutron capture
process (r-process, pink) leading to abundance maxima shifted to
lighter nuclei in comparison to the s-process.  We note, however, that
the contributions of the i, p, $\nu$, and $\nu$p-processes are minor
and thus are not readily apparent in this figure.  The focus of this
review will be on the r-process and the understanding of how the
corresponding isotopes were synthesized in nature.

Over the years there have been a number of comprehensive reviews on
this topic \citep[for a selected list see e.g.][and references
therein]{Hillebrandt:1978,Cowan.Thielemann.Truran:1991,Qian.Wasserburg:2007,arnould07,sneden08,Thielemann.Arcones.ea:2011,
  thielemann17a,Thielemann.Eichler.ea:2017}, as well as recent parallel efforts \citep{Horowitz.ea:2019,Kajino.Aoki.ea:2019}.  In
order to get clues on the r-process origin, a wide range of subtopics
need to be addressed: (1) nuclear physics input to understand the
nucleosynthesis path far from stability, (2) nucleosynthesis modeling
to find out conditions for neutron densities and temperatures which
can reproduce the r-process abundances found in nature, (3)
determining whether proposed astrophysical sites can match such
conditions, (4) observations of stellar abundances throughout galactic
history in order to find out which of these sites can contribute
during which period of galactic evolution, (5) in order to do so with
good precision a detailed study of the atomic physics is required for
identifying the strengths of absorption lines needed to determine
abundances, and (6) detections of long-lived radioactive species that
can hint towards understanding the frequencies of r-process events in
the Galaxy.  Thus, a number of connected fields, including atomic
physics, nuclear physics, stellar spectroscopy, stellar (explosion)
modeling, and galactic chemical evolution are involved in attempting
to answer the long-standing problem of ``How Were the Elements from
Iron to Uranium Made?'', one of the \textsl{Eleven Science Questions
  for the New Century} addressed by the National Academy of Sciences
in 2003~\cite{NAP10079}. Detailed discussions will follow in later
sections, here we list a number of considered scenarios.

While there have been many parametric studies in the early days,
assuming a set of neutron densities and temperatures
\citep[e.g.][]{seeger.fowler.clayton:1965,Kodama.Takahashi:1975,Kratz.Gabelmann.ea:1986,Kratz.Thielemann.ea:1988,Kratz.Bitouzet.ea:1993,Freiburghaus.Rembges.ea:1999,Pfeiffer.Kratz.ea:2001},
the long-standing question is, where an r-process with neutron
densities of $10^{26}$~cm$^{-3}$ and higher, producing highly unstable
neutron-rich isotopes of all heavy elements and permitting a fast
build-up of the heaviest elements up to the actinides, can take place.

As will be discussed later with respect to observations, there are
indications that a ``weak'' and a ``strong'' r-process occur
in nature, and the ``strong'' component is probably the dominant one,
accounting essentially for solar-system r-process abundances. But some
old stars, although displaying abundances of r-process elements,
including Eu, show a strongly declining trend towards heavy elements,
and it is not clear whether the third r-process peak with $A=195$ or
even the actinides are present. In our review we will focus mostly on
the strong r-process, but discuss observations and possible sites of
the weak r-process as well. There have been many suggestions relating
the site of the strong r-process to
\begin{enumerate}
\item the innermost ejecta of regular core-collapse supernovae CCSNe
  \citep[e.g.][]{Schramm:1973,Sato:1974,Hillebrandt.Kodama.Takahashi:1976,Hillebrandt:1978,Woosley.Wilson.ea:1994,Takahashi.Witti.Janka:1994,Witti.Janka.Takahashi:1994,Qian.Woosley:1996,hoffman.woosley.qian:1997,Thompson.Burrows.Meyer:2001,Wanajo.Kajino.ea:2001,Terasawa.Sumiyoshi.ea:2001b,Qian.Wasserburg:2007,Farouqi.Kratz.ea:2010,Roberts.Woosley.Hoffman:2010,Roberts.Reddy.Shen:2012,Martinez-Pinedo.Fischer.ea:2012,Arcones.Thielemann:2013,Mirizzi:2015}. However,
  despite all remaining uncertainties in the explosion mechanism,
  recent conclusions are that at most a weak r-process can occur under
  these conditions
  \cite{Wanajo.Janka.Mueller:2011,Martinez-Pinedo.Fischer.ea:2012,Roberts.Reddy.Shen:2012,Curtis.ea:2019},
  because weak interactions with electron neutrinos and anti-neutrinos
  from the newly formed hot proto-neutron star will either make
  initially neutron-rich matter less neutron-rich or even proton-rich
  or, in case of slightly neutron-rich matter, sufficiently high
  entropies are not attained. Another option for a weak r-process
  exists in so-called quark deconfinement supernovae, where after the
  collapse of a massive star, leading to a proto-neutron star, a
  quark-hadron phase transition sets in which causes the subsequent
  supernova explosion
  \citep{Fischer.Bastian.ea:2018,Fischer.Wu.ea:2020}.
\item Outer layers of supernova explosions, e.g.\ the helium layer
  where neutrons are created by $(\alpha,n)$-reactions, were also
  suggested
  \cite{Truran.Cowan.Cameron:1978,Thielemann.Arnould.Hillebrandt:1979,Cowan.Cameron.Truran:1980,Hillebrandt.ea:1981,Klapdor.ea:1981,Cameron.Cowan.Truran:1983,Cowan.Cameron.Truran:1983,Thielemann.Metzinger.Klapdor:1983,Cowan.Cameron.Truran:1985},
  later also the collapsing ONeMg core of massive stars
  \cite{Wheeler.Cowan.Hillebrandt:1998}. The emergence of realistic
  pre-explosion stellar models made this site less likely. Further
  options include --- for low abundances of heavy elements in the
  early Galaxy --- sufficient amounts of neutrons in the He-shell,
  provided via neutrino
  interactions~\citep{Epstein.Colgate.Haxton:1988,Nadyozhin.Panov:2007}. But
  this scenario, with low neutron number densities, would not be able
  to produce the solar r-process pattern with its correct peak
  locations~\citep{Banerjee.Haxton.Qian:2011,Qian:2014,Banerjee.Qian.ea:2016}.
\item Special classes of core-collapse events of massive stars with
  fast rotation and high magnetic fields. They can either lead to
  highly magnetized neutron stars (magnetars) and neutron-rich jet
  ejecta (MHD-jet supernovae) along the polar
  axis~\citep{symbalisty85,cameron03,nishimura06,Winteler.Kaeppeli.ea:2012,Moesta.Richers.ea:2014,Nishimura.Takiwaki.Thielemann:2015,Moesta.ea:2015,Nishimura.Sawai.ea:2017,moesta18,halevi18,Obergaulinger.Just.Aloy:2018}
  or to black holes, polar jets, and black hole accretion disk
  outflows (hypernovae/collapsars).  The latter have been attributed
  to neutron-rich jet
  ejecta~\citep[e.g.][]{Fujimoto.Nishimura.Hashimoto:2008,Ono.Hashimoto.ea:2012}
  and/or the creation of r-process elements in black hole accretion
  disks~\citep[e.g.][]{Pruet.Woosley.Hoffman:2003,Pruet.Thompson.Hoffman:2004,Siegel.Barnes.Metzger:2019}. The
  first type of events showed quite some promise for producing
  r-process ejecta, but the necessity that very high pre-collapse
  magnetic fields exist puts constraints on this scenario. The second
  option (collapsars) stands for a high-angular momentum subset of
  rotating stars which form black holes in combination with
  long-duration gamma-ray bursts (GRB). A variant of this, based on
  the spiraling in of a neutron star via merging with a giant in a
  binary system (leading eventually to accretion, black hole
  formation, and a black hole accretion disk) has been suggested
  by~\citet{Grichener.Soker:2019}.
\item Ejecta from binary neutron star (or BH-neutron star) mergers
  have been studied for many years before the first detection of such
  an
  event~\citep[e.g.][]{Lattimer.Schramm:1974,symbalisty82,Eichler.Livio.ea:1989,Freiburghaus.Rosswog.Thielemann:1999,Rosswog.Davies.ea:2000,Rosswog.Korobkin.ea:2014,Wanajo.Sekiguchi.ea:2014,Goriely.Bauswein.Janka:2011,Just.Bauswein.ea:2015,Eichler.Arcones.ea:2015,Goriely.Bauswein.ea:2015,Ramirez-Ruiz.Trenti.ea:2015,Mendoza-Temis.Wu.ea:2015,Shibagaki.Kajino.ea:2016,Wu.Fernandez.Martinez.ea:2016,Lippuner.Fernandez.Roberts.ea:2017,Thielemann.Eichler.ea:2017}.
  After the gravitational wave detection GW170817 of a neutron star
  merger with a combined total mass of about 2.74~M$_\odot$,
  \citep{Abbott.Abbott.ea:2017,Abbott.Abbott.ea:2018b}, accompanied by
  a kilonova observation supporting the production of heavy
  elements~\citep[see
  e.g.][]{Metzger:2017,Tanaka.Utsumi.ea:2017,Villar.ea:2017}, this
  type of event has attracted special attention~\citep[see reviews
  e.g.\ by][and references
  therein]{rosswog18,Horowitz.ea:2019,Shibata.Hotokezaka:2019}.  More
  recent gravitational wave observations point to further neutron star
  mergers~\citep[e.g. GW190425 with a combined total mass of $\sim$
  3.4~M$_\odot$,][]{Abbott.ea:2020}, or even neutron star-black hole
  mergers e.g.~GW190426 and GW190814 with a combined total mass in
  excess of 7~M$_\odot$ and
  25M$_\odot$~\citep{Lattimer:2019b,Abbott.ea:2020b,Abbott.Abbott.ea:2020}. The three last
  mentioned events had no observed electromagnetic counterpart, due to
  either non-existence or non-detection, the latter related to a large
  distance and/or missing precise
  directions~\citep{Foley.Coulter.ea:2020,Kyutoku.Fujibayashi.ea:2020,Barbieri.Salafia.ea:2020,Ackley.Others:2020}. Whether
  the smaller 2.6 M$_\odot$ binary member in GW190814 is actually a
  very massive neutron star or a very small black hole is still
  debated \citep{Godzieba.Radice.Bernuzzi:2020}.
\end{enumerate}

Most of the astrophysical sites mentioned above involve ejection of
material from high densities and involve a neutron star or black hole
produced during core-collapse or a compact binary merger. Hence the
high density equation of state that ultimately determines the
transition from a neutron star to a black hole plays an important role
in the modeling of these objects. We will not discuss this topic
further but refer the interested reader to recent reviews on the
nuclear equation of
state~\citep{Lattimer:2012,Hebeler.Holt.ea:2015,Oezel.Freire:2016,Oertel.Hempel.ea:2017,Tews.Margueron.Reddy:2019,Bauswein.Stergioulas:2019}.

Before discussing the r-process astrophysical sources in detail, a lot
of groundwork has to be laid out.
Section~\ref{sec:observations-jjc-cs} provides an overview of
observations (including the atomic physics for their correct
interpretation), section~\ref{sec:basic-working} the basic working of
an r-process and which conditions are needed for its successful
operation, sections~\ref{sec:exper-devel-r}
and~\ref{sec:nuclear-modeling-r} discuss the impact played by nuclear
physics (with experimental and theoretical investigations), and
section~\ref{sec:astr-sites-their} passes through the astrophysical
sites which can fulfill the required
conditions. Section~\ref{sec:chemev} combines these astrophysical
sites and how their role in galactic evolution connects to
section~\ref{sec:observations-jjc-cs}. Finally in the summary
(section~\ref{sec:conclusions}), after having presented all possible
connections, we discuss remaining issues and open questions, i.e.\
whether a single r-process site has been identified by now, or whether
we still might need several sources to explain observations throughout
galactic evolution.

\section{Observations}
\label{sec:observations-jjc-cs}

\subsection{Stellar Abundances of Neutron-Capture Elements in
  Metal-Poor Stars}
\label{sec:stellar-abunds}
Stellar abundance observations over decades have provided fresh
evidence about the nature and extent of heavy element nucleosynthesis.
In the case of the s-process there is direct observational evidence of
in situ stellar nucleosynthesis with the observation of the
radioactive element Tc, discovered first by \citet{merrill52}.
Additional stellar abundance studies have strongly linked this type of
nucleosynthesis to very evolved He shell-burning asymptotic giant
branch stars
\citep[e.g.][]{busso99,Kaeppeler.Gallino.ea:2011,karakas14}.  There is
no similar example for the r-process, related to nucleosynthesis
during stellar evolution, as it requires rather extensive neutron
fluxes only obtainable in explosive events. Some elements are only
formed exclusively or almost so in the r-process, such as Eu, Os, Ir,
Pt, Th and U.  Their presence in old galactic very metal-poor (VMP)
halo stars is a clear indication that this process occurred in violent
astrophysical sites early in the history of the Galaxy \citep[see
e.g.][and references therein]{sneden08,Thielemann.Eichler.ea:2017}.

\begin{figure}[htb]
  \includegraphics[width=\linewidth]{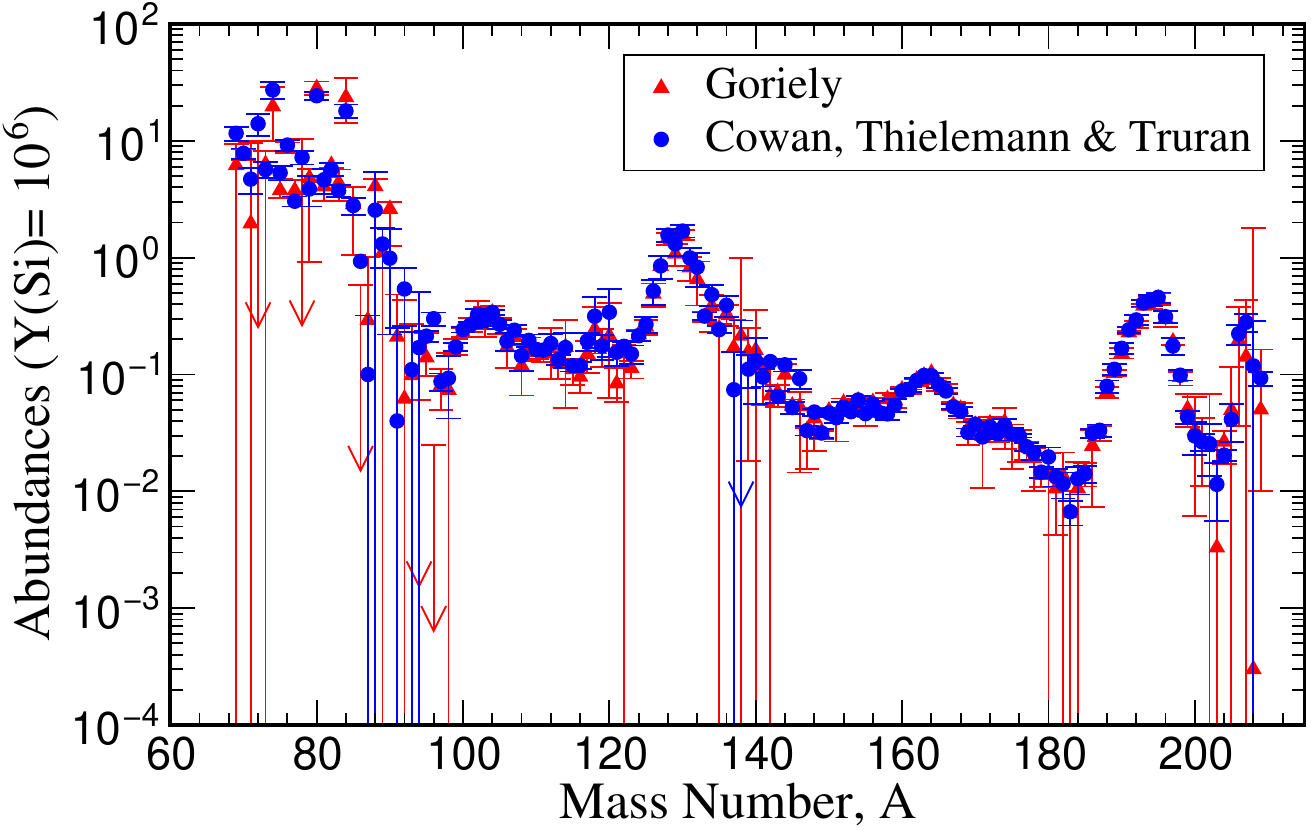}
  \caption{Solar r-process abundances as determined
    by~\citet{Cowan.Thielemann.Truran:1991}
    and~\citet{Goriely:1999}. The largest uncertainties are clearly
    visible for $A\lesssim 100$ (weak s process region) and around
    lead.\label{fig:solarr}}
\end{figure}

Identification of r-process-rich stars began with the discovery of
overabundances of neutron-capture elements in the field red giant
HD~115444 \citep{griffin82}.  This was followed by the identification
of an r-process pattern in the well known bright giant HD~122563, even
though its overall neutron-capture element level is depressed relative
to Fe (\citealt{sneden83}, see also the more extensive analysis of
\citealt{honda06}).  An initial abundance survey in metal-poor (MP)
stars \citep{gilroy88} considered 20 red giants, finding a common and
easily spotted pattern of increasing overabundances from Ba~($Z=56$)
to Eu ($Z=63$) among the rare-earth elements.  With better echelle
spectrographic data came discoveries of many more r-process-rich
stars, leading \citet{beers05} to sub-classify them as ``r-I'' with
$0.3 \leq \text{[Eu/Fe]} \leq +1.0$ and $\text{[Ba/Eu]} < 0$, and as
``r-II'' with $\text{[Eu/Fe]}>+1.0$ and $\text{[Ba/Eu]}< 0$.

The most detailed deconvolution of abundances into nucleosynthetic
contributions exists for the solar system, as we have accurate
abundances down to the isotopic level as a result of meteoritic and
solar atmospheric measurements \citep[e.g.][see
Fig.~\ref{fig:solarabund}]{cameron59,Asplund.Grevesse.ea:2009,lodders09}.
Identifying the r-process contributions to the solar system
neutron-capture abundances is usually accomplished by first
determining the s-process fractions,
\citep[e.g.][]{Kaeppeler:1999,arlandini99,Burris.Pilachowski.ea:2000,Kaeppeler.Gallino.ea:2011}.
The remaining (residual) amount of the total elemental abundance is
assumed to be the solar r-process contribution (see
Figures~\ref{fig:solarabund} and~\ref{fig:solarr}).  Aside from the
so-called
p-process~\citep{Arnould.Goriely:2003,rauscher13,Nishimura.Rauscher.ea:2018}
that accounts for the minor heavy element isotopes on the proton-rich
side of the valley of instability, as well as the
$\nu$-process~\citep{Woosley.Hartmann.ea:1990} and the $\nu$p-process
\citep{Froehlich.Martinez-Pinedo.ea:2006}, only the s and r-processes
are needed to explain nearly all of the solar heavy element
abundances.

\begin{figure}[htb]
  \includegraphics[width=\linewidth]{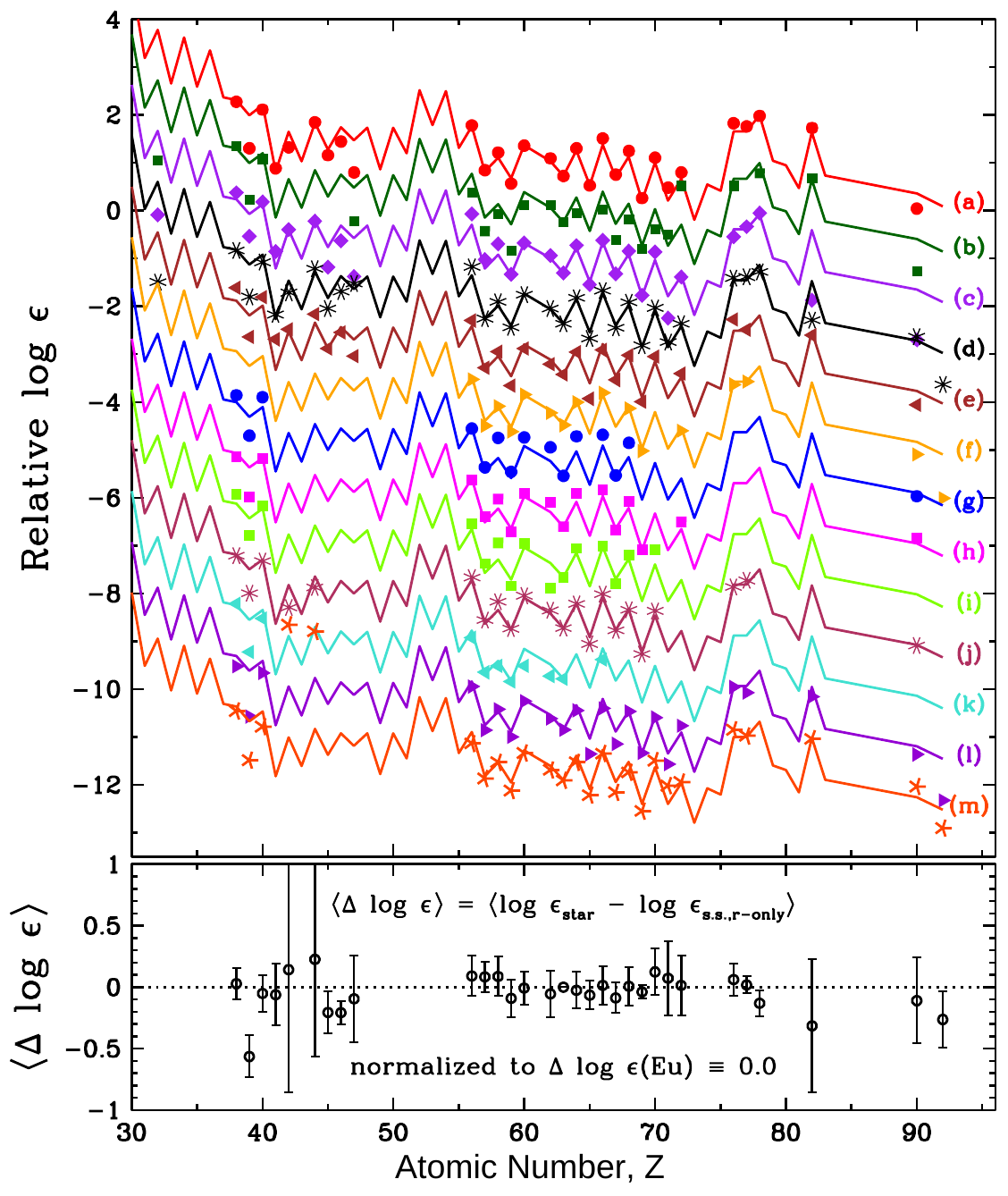}
  \caption{Top panel:  neutron-capture abundances in 13 r-II stars
    (points) and the scaled solar-system r-process-only abundances
    of~\cite{siqueira13}, mostly adopted from
    \cite{Simmerer.Sneden.ea:2004}. 
    The stellar and solar system distributions have been normalized
    to agree for element Eu (Z~=~63), and than vertical shifts
    have been applied in each case for plotting clarity.
    The stellar abundance sets are:
    (a) CS~22892-052, \cite{Sneden.Cowan:2003}; 
    (b) HD~115444, \cite{Westin.Sneden.ea:2000};
    (c) BD+17 3248, \cite{Cowan.Sneden.ea:2002};
    (d) CS~31082-001, \cite{siqueira13};
    (e) HD~221170, \cite{Ivans.Simmerer.ea:2006};
    (f) HD~1523+0157, \cite{frebel07};
    (g) CS~29491-069, \cite{Hayek.Wiesendahl.ea:2009};
    (h) HD~1219-0312, \cite{Hayek.Wiesendahl.ea:2009};
    (i) CS~22953-003, \cite{Francois.Depagne.ea:2007};
    (j) HD~2252-4225, \cite{Mashonkina:2014};
    (k) LAMOST J110901.22+075441.8, \cite{Li.Aoki.ea:2015};
    (l) RAVE J203843.2-002333, \cite{Placco.Holmbeck.ea:2017};
    (m) 2MASS J09544277+5246414, \citep{Holmbeck.Beers.ea:2018}.
    Bottom panel: mean abundance differences for the 13 stars
    with respect to the solar system r-process values.
    \label{rstars}}
\end{figure}

Early observations of CS~22892-052 \citep{sneden94,sneden03} and later
CS~31082-001 \citep{hill02,siqueira13} and references therein),
indicated a ``purely'' or ``complete'' solar system r-process
abundance pattern (see Figure~\ref{rstars}). The total abundances of
these, mostly rare-earth, elements in the stars were smaller than in
the Sun but with the same relative proportions, i.e., scaled.  This
indicated that these stars, that likely formed early in the history of
the Galaxy, experienced already a pollution by a robust r-process.

However, the growing literature on abundance analyses of VMP stars has
added to our knowledge of the average r-process pattern, and has
served to highlight departures from that pattern.  Additions to the
observational results since the review of \citet{sneden08} include
\citet{Roederer.Sneden.ea:2010,roederer14,Li.Aoki.ea:2015,roederer16,Roederer:2017,Aoki.Ishimura.ea:2017,Yong.Morris.ea:2017,Hansen.Holmbeck.ea:2018,Sakari.Placco.ea:2018,Roederer.Hattori.Valluri:2018,Ezzeddine_2020}.
These additional observations have shown that there is a complex
relationship between light and heavy neutron-capture elements:
\citet{Travaglio.Gallino.ea:2004,Cowan.Sneden.ea:2005,Hansen.Primas:2011,Hansen.Primas.ea:2012,Aoki.Beers.ea:2013,Ural.Cescutti.ea:2015,Wu.Fernandez.Martinez.ea:2016}.
In particular it has been found in some stars that there is
significant observed star-to-star abundance scatter of lighter
neutron-capture elements (Z~$\leq$~50), opposite to the heavier ones
(Z~$\geq$~56), as shown in Fig.~\ref{rstars}.  For heavy
neutron-capture elements, particularly among the well-studied rare
earths, an r-process origin does not always mean perfect agreement
with the solar r-process pattern.  So-called ``truncated'' (or
incomplete or limited) r-process stars have been identified with sharp
abundance falloffs toward the heavy end of the rare earths
\cite{honda06,Honda.Aoki.ea:2007,Roederer.Cowan.ea:2010,Boyd.Famiano.ea:2012}.
These observed abundance patterns can be described as having a range
of r-process ``completeness'' with some stars showing only a partial
agreement. The differences in these abundance patterns have led to a
flurry of stellar models and calculations to identify a site or sites
for the r-process, and to determine why stars show differences in
these heavy element patterns.  In addition to the suggested
operation of a ``weak'' r-process, two further processes have
gained currency: the so-called Lighter Element Primary Process
\citep[LEPP of still unknown origin;][]{Travaglio.Gallino.ea:2004},
and the i~process (\citealt{cowan77}; see also \citealt{denissenkov17}
and references therein).  (While the LEPP and the i~process may
explain certain individual stellar abundances, their contributions to
the total solar system (SS) abundances appear to be very small.)

\begin{figure}[htb]
  \includegraphics[width=\linewidth]{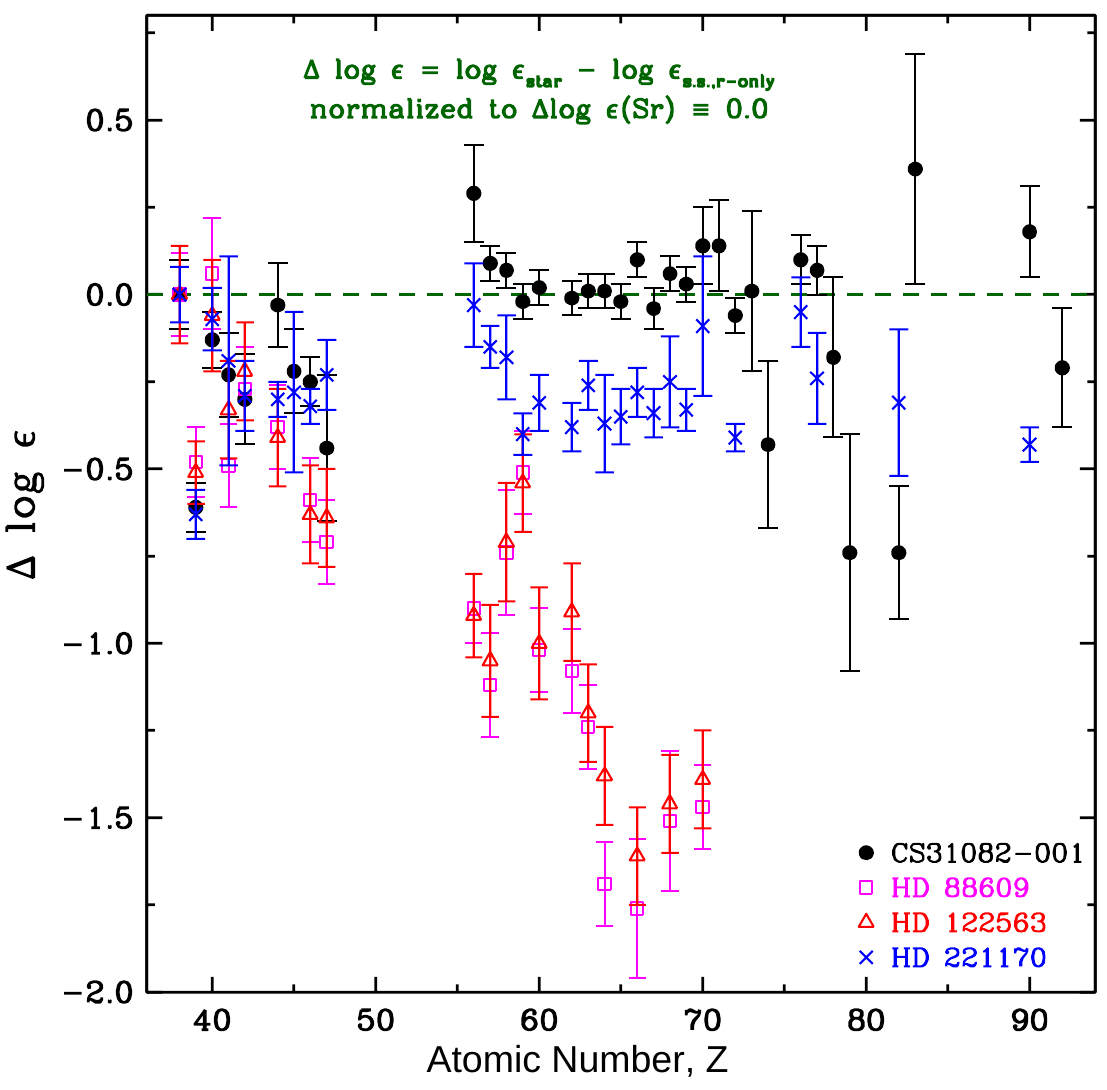}
  \caption{Differences between stellar and r-process-only solar system 
    (s.s.) abundances for four very MP stars with 
    r-process abundance mixes, after Figure~5 of \citet{Honda.Aoki.ea:2007}                   and Fig.~11 of \citet{Roederer.Sneden.ea:2010}.  
    The ``s.s.,r-only'' abundances are those of 
    \citet{siqueira13}, mostly from \citet{Simmerer.Sneden.ea:2004}.  
    The stellar abundance sets are: CS31082-001, \cite{siqueira13}; 
    HD~88609, \cite{Honda.Aoki.ea:2007}; HD~122563, \cite{honda06}; 
    and HD~221170, \cite{Ivans.Simmerer.ea:2006}.
    \label{ssrdiffs}}
\end{figure}

An r-process pattern (defined here as $\text{[Eu/Ba]}>~+0.3$) can be
seen even in MP stars with bulk deficiencies in neutron-capture
elements: In Fig.~\ref{ssrdiffs} we show differences in abundances
between stellar observations and those of the solar system attributed
only to the r-process.  Fig.~\ref{ssrdiffs} is similar in structure to
those of~\citet{Honda.Aoki.ea:2007} and
\citet{Roederer.Cowan.ea:2010}.  As defined in the figure, if
$\Delta\log \epsilon=0$, then the stellar neutron-capture abundance
set is identical to the solar-system r-process-only distribution.
This is clearly the case for elements in the atomic numbers range
$Z=57$--78, e.g.\ La--Pt in CS31082-001 \citep{siqueira13}.  All
extremely r-process-rich stars (classified as ``r-II'':
$\text{[Eu/Fe]} >+1$) have similar abundance runs in the heavy
neutron-capture elements, as discussed above.  However, many MP stars
with a clear dominance of the r-process, as defined by
$\text{[Eu/Ba]}>+0.3$, have abrupt drop-offs in abundances through the
rare-earth domain.  The most dramatic examples are the truncated
r-process stars shown in Fig.~\ref{ssrdiffs}: HD~122563
\citep{honda06} and HD~88609 \citep{Honda.Aoki.ea:2007}.  Intermediate
cases are abound, as shown in \citet{Roederer.Sneden.ea:2010}.

To understand the types and nature of the nucleosynthesis, along with
identifying the stellar sites and the identities of the first stars in
our Galaxy, demands highly precise stellar abundance
observations. Those require both high-resolution spectrographic
measurements and accurate atomic data.  Thus, the discovery of MP
stars renewed efforts to improve atomic data for many heavy (beyond
the Fe-group) neutron-capture elements \citep[see e.g.,][]{sneden09},
as discussed below in section~\ref{sec:atomic-elements-stars}.

\subsection{Atomic Data for the Analysis of neutron-capture Elements
  in Metal-Poor Stars}
\label{sec:atomic-elements-stars}

Despite a great need for improved transition probabilities, the
identification of lines from neutron-capture elements in stellar
spectra has been possible for most elements using readily available
laboratory data from about the middle of the 20th Century.
Wavelengths of spectral lines of such elements were measured during
the first half of the 20th Century using large grating spectrographs
such as 10~m Rowland circle instruments.  These early wavelength
measurements often achieved 1~part per million (ppm) accuracy and were
compiled in the well known Atomic Energy Level series by
\citet{moore71} and for the Rare-Earth Elements by \citet{martin71}.
The latter of these two works includes more data from Fourier
transform spectrometers (FTSs) and thus achieved $\simeq 0.01$~ppm or
10~ppb accuracy in many cases.  All of these spectroscopic data are
now available online\footnote{\url{http://physics.nist.gov/asd}}.
Although modern optical frequency comb lasers could add many
additional digits to energy levels, this technology has not yet been
widely applied because of the difficulty in simultaneously using it on
large numbers of spectral lines.

The situation with respect to transition probabilities changed with
the development of tunable dye lasers originally by \citet{sorokin66}
in the US and \citet{schafer66} in Germany.  Although it took some
time to thoroughly control dye laser performance, many research groups
had organic dye lasers with broad tunability, narrow bandwidths
(comparable to or less than Doppler widths), short (few nsec) pulse
durations, and repetition rates in the 10s of Hz.  Non-linear
techniques, using crystals and/or gas cells, are needed to access IR
and UV wavelengths, and those were also increasingly available.  The
remaining challenge is to make free atoms and ions of various elements
in the periodic table in an optically thin sample with a low collision
rate.  There are several methods, including sputtering metal cathodes,
in a low pressure gas cell \citep{hannaford81}, laser driven plasma
sources \citep[e.g.][]{svanberg94}), and the hollow cathode atom/ion
beam source~\citep{duquette81,salih83}.  The broadly tunable organic
dye lasers, in combination with a technique to make low pressure
samples of metal atoms and ions, opened the possibility of using
time-resolved laser-induced-fluorescence (TRLIF) to measure accurate
and precise (about a few \%) radiative lifetimes of upper levels on
interest in atoms and ions.  These lifetimes provide an accurate and
precise total decay rate for transition probabilities from the
selected upper level.

Emission branching fractions (BFs) in rich spectra still represented a
challenge.  The same visible and UV capable FTS instruments
\citep[e.g.][]{brault76}), used to improve energy levels, became the
``work horse'' of efforts on BFs in complex spectra.  Reference
Ar~\textsc{i} and \textsc{ii} lines became internal standards for many
laboratory spectra from hollow cathode lamps recorded using FTS
instruments \citep[and references therein]{whaling93}.  The advantages
of interferometric instruments such as the 1~m FTS of the National
Solar Observatory on Kitt Peak, AZ, were critical for BF measurements
in complex spectra.  This instrument has a large etendue common to all
interferometric spectrometers, wavenumber accuracy to 1 part in
10$^8$, a limit of resolution as small as 0.01~cm$^{-1}$, broad
spectral coverage from the UV to IR, and the capability of recording a
million point spectrum in minutes \citep{brault76}.  Hollow cathode
lamps which yield emission spectra for neutral and singly ionized
atoms are available for essentially the entire periodic table.

Interest in rare-earth elements is a natural part of studies of
neutron-capture elements in MP stars.  Atoms and ions with open
f-shells have a great many transitions in the optical (see
  section~\ref{sec:kilon-an-electr} for their relevance in kilonova
  light curves).  Rare-earths have important applications in general
lighting and in optoelectronics because of their rich visible spectra.
Rare-earth elements in MP stars are convenient for spectroscopic
studies in the optical region accessible to ground based telescopes.
Europium is a nearly pure r-process element and lanthanum is a nearly
pure s-process element in solar system material.  Although none of the
r-process peaks are in the rare-earth row, the accessibility from the
ground is a major advantage for rare earths.

\begin{figure}[htb]
  \includegraphics[width=\linewidth]{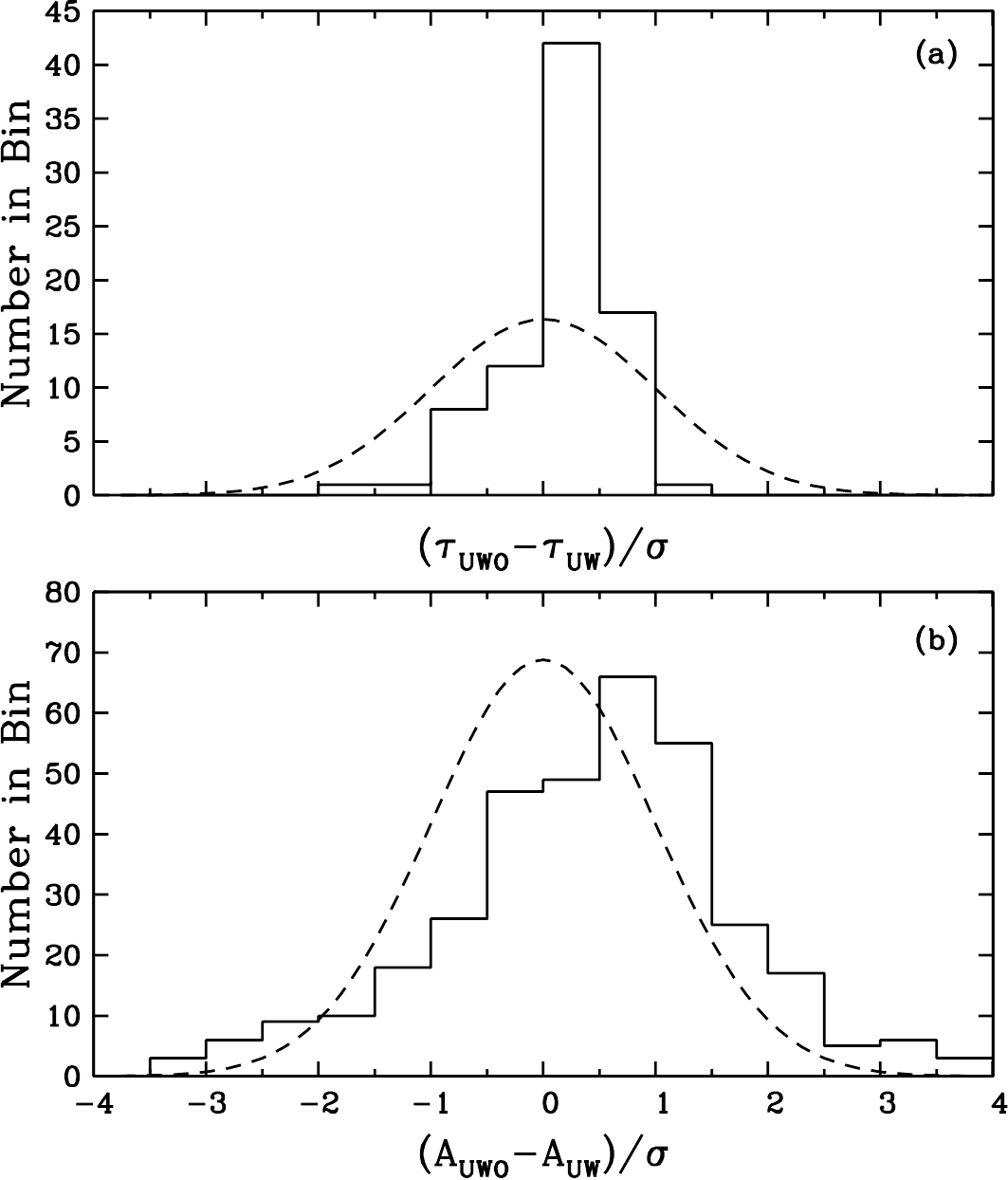}
  \caption{Comparisons of laboratory data on Sm~\textsc{ii} from University of 
    Western Ontario (UWO) and University of Wisconsin (UW) groups,
    adapted from Figs.~4 and 5 of \citet{lawler08}.
    In panel (a) a histogram of differences in lifetimes ($\tau$ divided 
    by their uncertainties added in quadrature) is shown, along with
    a dashed line representing a one standard deviation Gaussian.
    In panel (b) we show a similar histogram and Gaussian representation
    for transition probabilities (A-values).
    \label{labhisto}}
\end{figure}

Rare-earth elements tend to be singly ionized in the photospheres of
F, G, and K stars of interest for many elemental abundance studies.
The spectrum of singly ionized samarium (Sm~\textsc{ii}) received
special attention~\citep{lawler06,rehse06}, \citet{lawler08} completed
comparisons from the two sets of measurements.  {Fig.~\ref{labhisto}
  shows a histogram of lifetime measurement differences between the
  two studies with a one standard deviation Gaussian superposed, and a
  similar histogram comparison for Einstein A coefficients which
  include BFs.}  It is clear from these histograms that radiative
lifetime uncertainties are overly conservative and BFs uncertainties
are satisfactory but perhaps slightly too optimistic in at least one
of two sets of measurements.

Uncertainties in radiative lifetimes from TRLIF experiments have
proven to be easier to minimize than uncertainties in emission BFs.
Various techniques can conveniently be used to check for optical depth
(vary the atom/ion beam intensity), to check for collisional effects
(throttle a vacuum pump), and to eliminate errors from Zeeman quantum
beats (zero the magnetic field in the experimental region for short
lifetimes and introduce a high, 30~Gauss, magnetic field for long
lifetimes).  Most importantly benchmark lifetimes in simple spectra
such as He~\textsc{i}, Be~\textsc{i}, Be~\textsc{ii}, Mg~\textsc{ii},
etc., which are well known from accurate theory, can be periodically
re-measured as an end-to-end test of the TRLIF experiment
\cite{denhartog02}.  There are multiple challenges in BF measurements.
It is essential to have a reliable relative radiometric calibration,
and a source that is optically thin for strong lines of interest.  One
must resolve lines of interest from nearby blending partners and line
identifications must be correct.  These latter two constraints are
most easily achieved using FTS instruments due to their exceptional
resolving power and absolute wave number accuracy and precision.  Weak
lines from an upper level of interest are clearly most vulnerable to
blending, poor signal-to-noise ratios (S/N), and other problems.
Uncertainty migrates to weak lines because BFs from an upper level of
interest sum to unity by definition.

Elements with wide hyperfine structure and/or a wide range of isotopes
require some additional effort, but in most cases the needed hyperfine
splitting (hfs) data can be extracted from FTS spectra.  The existence
of even a few hfs data from single frequency laser measurements is
helpful since such data can serve to constrain nonlinear least square
fitting of partially resolved hfs patterns in FTS data.  Laboratory
transition probability measurements on rare-earth ions were summarized
during a study of Ce~\textsc{ii} by \citet{lawler09} and were applied
to five r-process rich very MP stars in a companion paper by
\citet{sneden09}.  The most striking conclusion from the decade long
rare-earth study is that the relative r-process abundance pattern is
stable over time and space.
Third r-process peak elements, including Os, Ir, and Pt were observed
in MP stars by \citet{Cowan.Sneden.ea:2005}.  Some useful lines of
Os~\textsc{i} and Ir~\textsc{i} are accessible to ground based
studies.  Unfortunately lines suitable for abundance studies of many
lighter neutron-capture elements are not accessible to ground based
observations.  Elements near the first r-process peak such as a As and
Se have their valence electrons in nearly closed p-shells.  The huge
gap between the ground and first resonance levels exists in both the
neutral and ion energy level structure, although the neutral atom
population is dominant in most stars of interest for both of these
elements.  A similar problem arises for Te at the second r-process
peak with only deep UV lines.  Fortunately HST time was allocated for
a study of Te~\textsc{i} lines in multiple MP
stars~\citep{roederer12b}.  The success of the Te study inspired a
careful search through the HST archives for one or more stars with
sufficiently deep UV spectral coverage for observations on all three
r-process peaks~\citep{roederer12a}.  Unfortunately the star HD~160617
is likely the only such star with sufficient deep UV spectral
coverage.  Laboratory data sets for many of the lighter r-process
elements are included.  Laboratory data sets for many of the lighter
r-process elements could be improved, but a successor telescope to HST
with a high resolution spectrograph and UV capability will be needed
to exploit improvements in the laboratory data.

The discovery of a single line of U~II in an MP star (complicated by
being located on the shoulder of a much stronger Fe~\textsc{i} line)
by~\citep{cayrel01} was a milestone in stellar spectroscopy.
Despite this complication, there is some confidence in its
identification.  Thorium is also an element of choice for stellar
chronometry~\citep[e.g.][]{sneden03}.

\subsection{Abundance Trends in Galactic and Extragalactic Stars}
\label{sec:gce}

As already discussed in section~\ref{sec:stellar-abunds}, the galactic
MP stars show indications of neutron-capture abundances, in fact, it
appears as if ALL such stars (to an observational limit) exhibit some
level of neutron-capture abundances.  In addition, observations have
indicated the presence of elements such as Ba in nearby dwarf
spheroidal galaxies
\citep[e.g.][]{Shetrone.Bolte.Stetson:1998,Shetrone.Venn.ea:2003,Venn.Tolstoy.ea:2003,Skuladottir.Hansen.ea:2019}.
Recently there has been evidence of these elements in ultra-faint
dwarf (UFD) galaxies, structures of only about $10^4$ M$_\odot$ and
possibly being also the building blocks and substructures of the early
Galaxy \citep{Brauer.Ji.ea:2019}. By now more than 10 UFDs are
discovered around our Galaxy, being very metal-poor with metallicities
of $\text{[Fe/H]}\approx -3$
\citep{Kirby.ea:2013,Frebel.Norris:2015,Ji.Simon.ea:2019,Simon:2019},
and most of them show very low r-process enhancements. However, one of
them (Reticulum II) shows highly r-process enhanced stars comparable
to galactic r-process rich stars such as CS 22892-052
\citep{Roederer:2013,ji16,Roederer:2017,Ji.Frebel:2018} which seems to
go back to one very early r-process event.  In addition to Reticulum
II, a further dwarf galaxy, Tucana III, has recently been observed and
also shows r-process features
\citep{Hansen.Simon.ea:2017,Marshall.ea:2019}.

\begin{figure}[htb]
  \includegraphics[width=\linewidth]{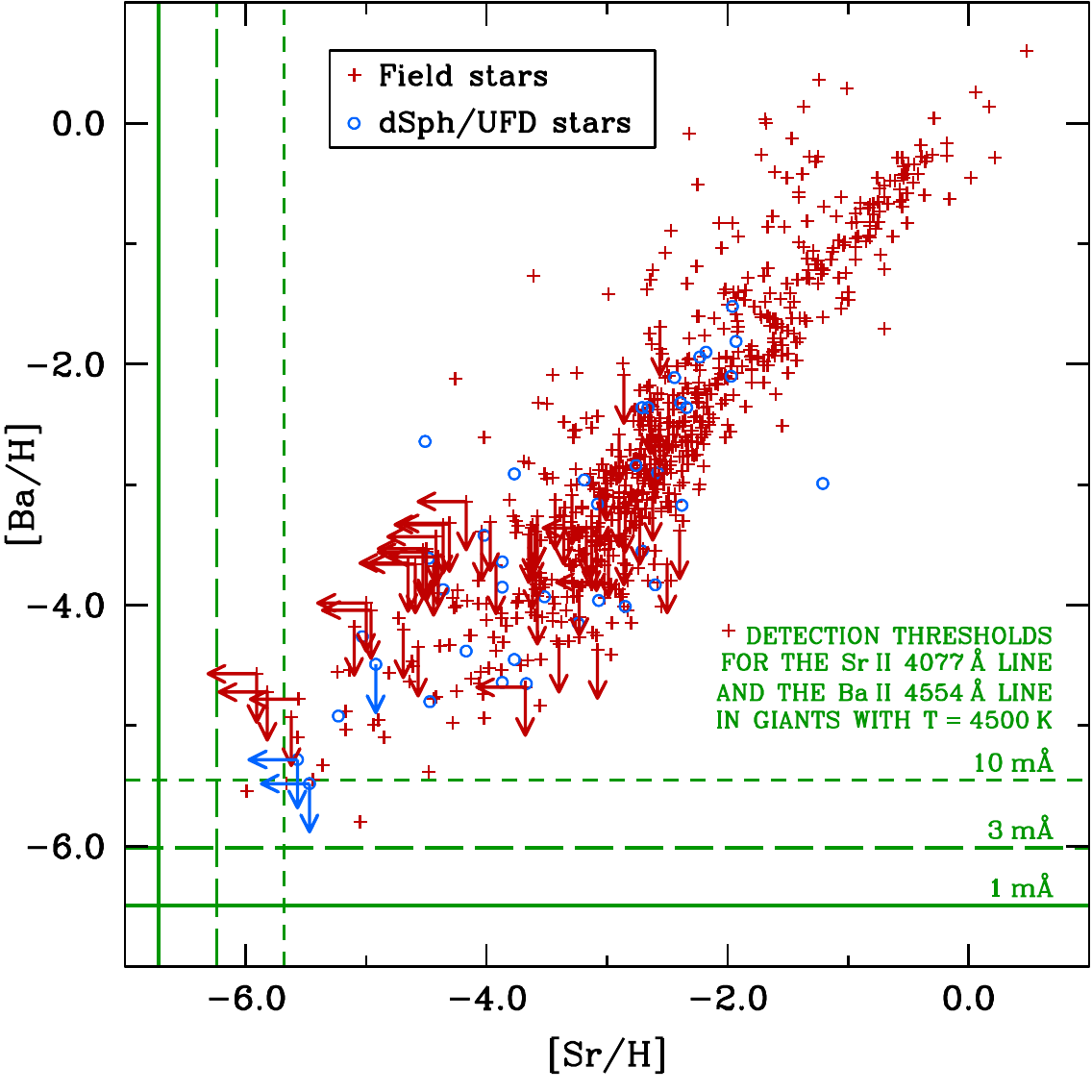}
  \caption{Abundances of [Sr/Fe] vs. [Ba/Fe] in a large number of galactic 
    and extragalactic stars from \citet{Roederer:2013} and references therein (reproduced by permission of the AAS).
    \label{roederer13f1}}
\end{figure}

We show in Fig.~\ref{roederer13f1}, \citep[taken from][and references
therein]{Roederer:2013}, a compilation of abundances in both galactic
and extragalactic stars. In these observations the Sr abundance acts
as a surrogate for the overall metallicity of these stars and Ba
indicates the enrichment of neutron-capture elements.  The figure
illustrates that stars down to the lowest metallicities contain Sr
and/or Ba. In a solar mix these are predominantly s-process elements,
i.e.\ their s-process isotopes dominate in present solar
abundances. If massive stars with fast rotation rates contributed
already some s-process in early galactic evolution
\cite{frischknecht16}, this could be due to such s-process
sources. However, global trends, where observed elemental or isotopic
ratios can be deconvolved into s- and r-process contributions, show an
s-process appearance only in later periods of galactic
evolution. Thus, this compilation strongly suggests that all of these
stars have been enriched in r-process material, which also has
implications for early nucleosynthesis in galaxies.

\begin{figure}[htb]
  \includegraphics[width=\linewidth]{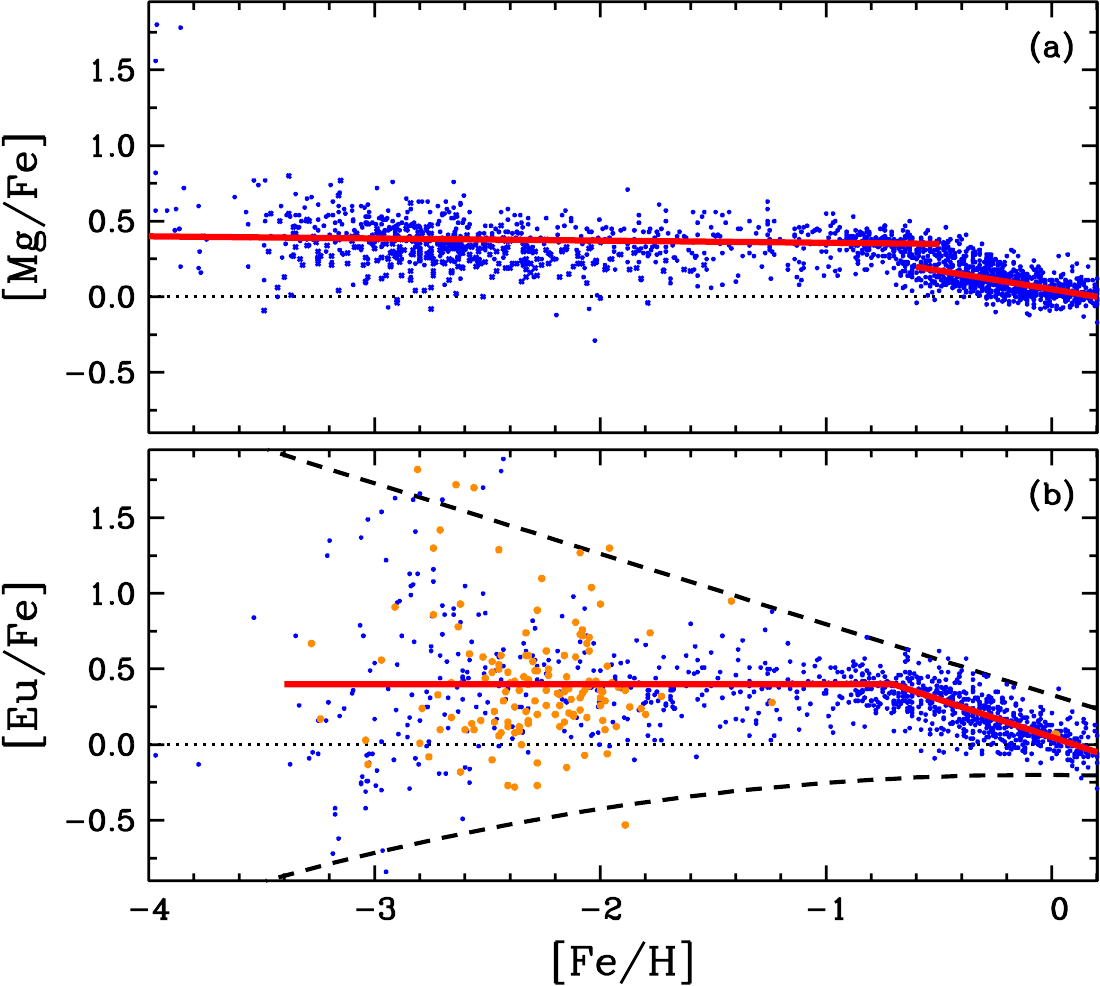}
  \caption{Abundances as a function of metallicity for [Mg/Fe] (panel a) 
    and [Eu/Fe] (panel b). 
    This is an update of Fig.~14 in \cite{sneden08}.
    Red solid lines are approximate fits to the averages of halo,
    thick disk, and thin disk stars. 
    Black dashed lines in panel (b) highlight the growing star-to-star
    scatter in [Eu/Fe] with decreasing metallicity.
    Individual data points are taken from \citet{fulbright00, 
      hill02,reddy03,cayrel04b,Simmerer.Sneden.ea:2004,
      cohen04,Barklem.Christlieb.ea:2005,reddy06,Francois.Depagne.ea:2007,
      bensby14,roederer14,battistini16}.
    \label{mgeufe}}
\end{figure}

Clues about early galactic nucleosynthesis are also found in
comparison of elements with different nucleosynthesic origin.  We show
one such comparison in Figure~\ref{mgeufe}, observed in halo stars,
i.e.\ containing elements synthesized prior to the formation of these
stars.  It is evident that alpha elements (such as Mg) appear early in
galactic evolution at low metallicities, originating from fast
evolving massive stars and core-collapse supernovae as their final
endpoints. Such events occur with a high frequency during galactic
evolution and show little scatter. Common r-process elements, like Eu,
display, however, an extensive scatter. These observations, combined
with those from ultrafaint dwarf galaxies, indicate that the heavy
r-process elements are made in rare events which contribute
significant amounts of material, when they occur (see
Fig.~\ref{mgeufe}).  Such abundance comparisons can be used to put
constraints on the site (or sites) for the r-process in terms of (a)
ejecta compositions, (b) amounts of r-process ejecta, and (c) their
mixing with the extended interstellar medium in order to understand
the history of element formation in the Galaxy (i.e., Galactic
Chemical Evolution, GCE). With respect to (a)
\citet{Ji.Drout.Hansen:2019} have analyzed extended sets of low
metallicity observations with $\text{[Fe/H]}<-2.5$ (attempting to
select stars being polluted only by single events) and
$\text{[Eu/Ba]}>0.4$ (to ensure a pure r-process origin, avoiding
s-process contributions) with the aim to find the typical lanthanide
(plus actinide) fraction $X_{\text{La}}$ among the global r-process
element distribution. This permits on the one hand to look for
variations among old stars, indicating apparently a different result
for the bulk of low metallicity stars with
$\log X_{\text{La}}\approx -1.8$, while the most r-process enriched
stars with $[\text{Eu/Fe}]>0.7$ have $\log X_{\text{La}}>-1.5$. This
measure will also permit comparisons to future kilonova events, if
observations allow to determine this quantity (see
section~\ref{sec:kilon-an-electr}).  With respect to (b) and (c) of
the list above we will return to galactic evolution issues later in
section~\ref{sec:chemev}, after having presented the nucleosynthesis
yields of different astrophysical sites.

The eventual demise of the Hubble Space Telescope, able to obtain
high-quality UV observations, will hamper future progress in the
observation of heavy elements in low-metallicity stars. The James Webb
Space Telescope (JWST), the scientific ``successor'' of HST, will have
no UV capability but an IR capability. Identification of
neutron-capture element lines in the IR region could provide new
avenues for understanding the operation and nature of the r-process
(see subsection~\ref{sec:atomic-elements-stars}).

\subsection{The role of long-lived radioactive species }
\label{sec:radioactive}

\begin{figure}[htbp]
  \includegraphics[width=1.0\linewidth]{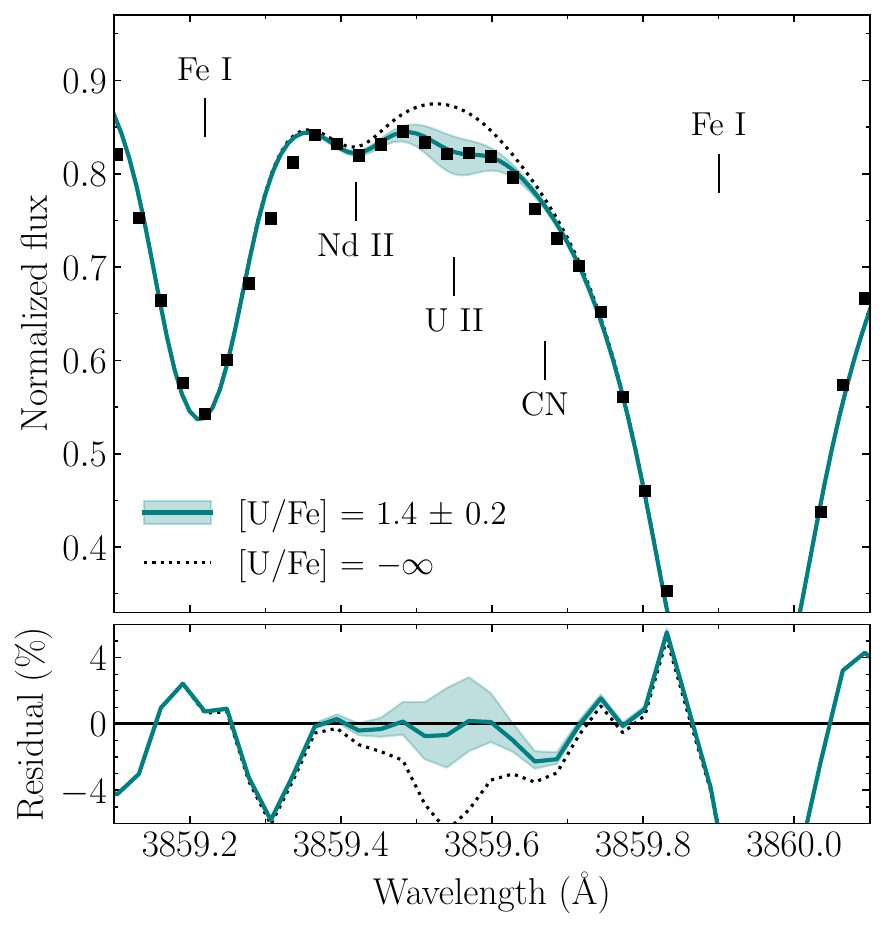}
  \caption{Synthesis and derived abundance for U in the star 
    2MASS J09544277+5246414 from Figure 2 of
    \citet{Holmbeck.Beers.ea:2018}, reproduced by permission of the
    AAS.   \label{Usyn1}}
\end{figure}

Identification and detailed spectroscopic analysis of a handful of
r-II stars, e.g., CS~22892-052~\citep{sneden94,sneden03}
CS~31082-001~\citep[and references therein]{hill02,siqueira13}, and
HE~1523-0901~\citep{frebel07} brought forth detections of the
long-lived very heavy neutron-capture radioactive elements Th
($t_{1/2} =13.0$~Gyr) and U~($t_{1/2} =4.6$~Gyr), which can only be
made in the r-process, and in addition neutron-capture element
abundances ranging from $Z\approx30$ to 92, indicating also an
r-process pattern. This makes detailed comparisons possible between
observations and r-process theory. More Th detections have been made
since then, and more recently U has also been detected in some halo
stars.  Due to its shorter half-life, its abundance is inherently
smaller and detections are difficult. Shown in Fig.~\ref{Usyn1} from
\cite{Holmbeck.Beers.ea:2018} is a uranium detection in 2MASS
J09544277+5246414, the most actinide-enhanced r-II star known.

\begin{figure}[htbp]
  \includegraphics[width=1.0\linewidth]{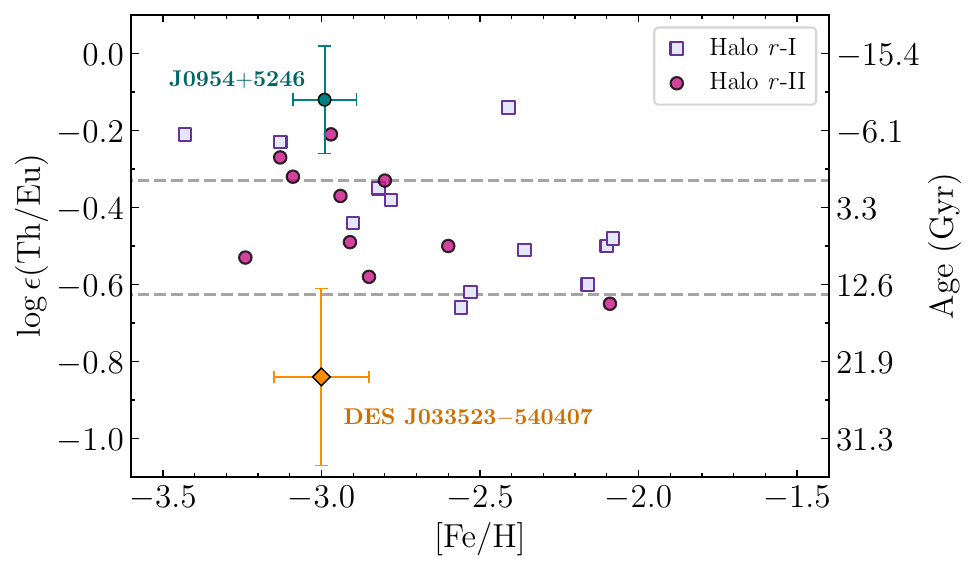}
  \caption{Th/Eu ratios for stars with detected thorium abundances
    from \citet{Holmbeck.Beers.ea:2018}, reproduced by permission of
    the AAS. One can see that at low metallicities around
    $\text{[Fe/H]}\approx-3$ quite a number of so-called actinide boost stars
    can be found. If utilizing initial r-process production ratios
    which fit solar r-abundances \cite{Schatz.ea:2002}, unreasonable,
    and even negative, ages of these stars are obtained, not at all
    consistent with their metallicity, which points to the formation
    of these stars in the very early Galaxy. 
    \label{boost}}
\end{figure}

These Th and U discoveries led to cosmochronology estimates,
independent of a cosmological model, based solely on decay half-lives
of involved isotopes. This method requires, however, Th/U ratios from
theoretical r-process predictions (geared to fit a solar r-process
pattern) plus the observed abundance ratios. It enabled estimates on
the decay-time since the birth of a star (when the addition of new
material from nucleosynthesis sites stopped) and promising results
were
obtained~\cite{Cowan.Thielemann.Truran:1991,Cowan.Pfeiffer.ea:1999,Kratz.Moeller.ea:2000,Schatz.ea:2002,Hill.Christlieb.ea:2017}.
The same can in principle also be done utilizing the Th/Eu ratio for
some stars, yielding values in concordance with cosmological age
estimates (see above).  The fact that some stars seem to have
experienced an ``actinide boost'', i.e.\ an enhanced amount of Th and
U in comparison to lighter r-process elements, could point back to a
non-universal r-process production pattern and possibly varying
r-process compositions from different productions sites. This made the
[Th/Eu] chronology uncertain or non-reliable for such stars
\citep[e.g.][]{cayrel01,honda04}, having experienced a non-solar
r-process contribution, while the [U/Th] did not show these anomalies
\cite{Mashonkina:2014}. Such an actinide boost is mostly found in
stars with metallicities $\text{[Fe/H]}\approx-3$.  This indicates
that (a) an r-process was already contributing in very early galactic
evolution, but also (b) with possibly varying conditions for producing
the heaviest elements, dependent on the r-process site.  Unfortunately
it has proved difficult to obtain U detections in many stars and it
remains to be seen how such actinide boosts are distributed as a
function of metallicity \citep[see Fig.~\ref{boost}
from][]{Holmbeck.Beers.ea:2018}.

In addition to observations of long-lived radioactive species seen via
the spectra of stars throughout galactic evolution, there have also
been detections in deep-sea sediments, indicating more recent
additions of these elements to the earth. While the discussion
in~\ref{sec:gce} points to rare strong r-process events in the early
Galaxy, the latter detections, suggest the same in recent
history. Long-lived radioactive species can act as witness of recent
additions to the solar system, dependent on their half-lives. For a
review on the signature of radioactive isotopes alive in the early
solar system see e.g. \citet{davis14}. Two specific isotopes have been
utilized in recent years to measure such activities in deep sea
sediments.  One of them, $^{60}$Fe, has a half-life of
$2.6\times 10^6$~yr and can indicate recent additions from events
occurring up to several million years ago. $^{60}$Fe is produced
during the evolution and explosion of massive stars, leading to
supernovae \citep{thielemann11,wanajo13,limongi18,thielemann18}.  It
is found in deep-sea sediments which incorporated stellar debris from
a nearby explosion about two million years ago
\citep{knie04,ludwig16,Wallner:2016,Soerensen.Svensmark.Grae:2017}. Such
a contribution is consistent with a supernova origin and related
occurrence frequencies, witnessing the last nearby event. Another
isotope utilized, $^{244}$Pu, has a half-life of $8.1\times 10^7$~yr
and would contain a collection from quite a number of contributing
events. If the strong r-process would take place in every
core-collapse supernova from massive stars, about
10$^{-4}$--10$^{-5}$~M$_\odot$ of r-process matter would need to be
ejected per event in order to explain the present day solar abundances
(see Fig.~\ref{rosswogr}). The recent $^{244}$Pu
detection~\citep{Wallner.Faestermann.ea:2015} is lower than expected
from such predictions by two orders of magnitude, suggesting that
considerable actinide nucleosynthesis is very rare (permitting
substantial decay since the last nearby event). This indicates that
(regular) core-collapse supernovae did not contribute significantly to
the strong r-process in the solar neighborhood for the past few
hundred million years, but does not exclude a weak r-process
contribution with very minor Eu
production~\citep{Fields.ea:2019,Wallner.Others:2019}.  Thus, in
addition to the inherent problems of (regular) core-collapse supernova
models (to be discussed in later sections) to provide conditions
required for a strong r-process --- also producing the actinides in
solar r-process proportions --- these observational constraints from
nearby events also challenge them as source of main r-process
contributions. A recent careful study of the origin of the strong
r-process with continuous accretion of interstellar dust grains into
the inner solar system \cite{hotokezaka15} concluded that the
experimental findings \cite{Wallner.Faestermann.ea:2015} are in
agreement with an r-process origin from a rare event. This can explain
the $^{244}$Pu existing initially in the very early solar system as
well as the low level of more recent additions witnessed in deep-sea
sediments over the past few hundred million years.

\subsection{Kilonova observations}
\label{sec:nsm}

\begin{figure*}[htb]
  \includegraphics[width=0.60\linewidth]{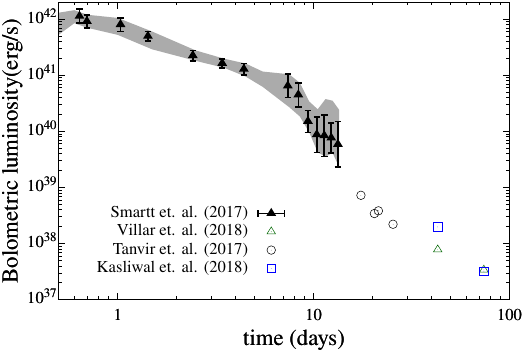}%
  \includegraphics[width=0.39\linewidth]{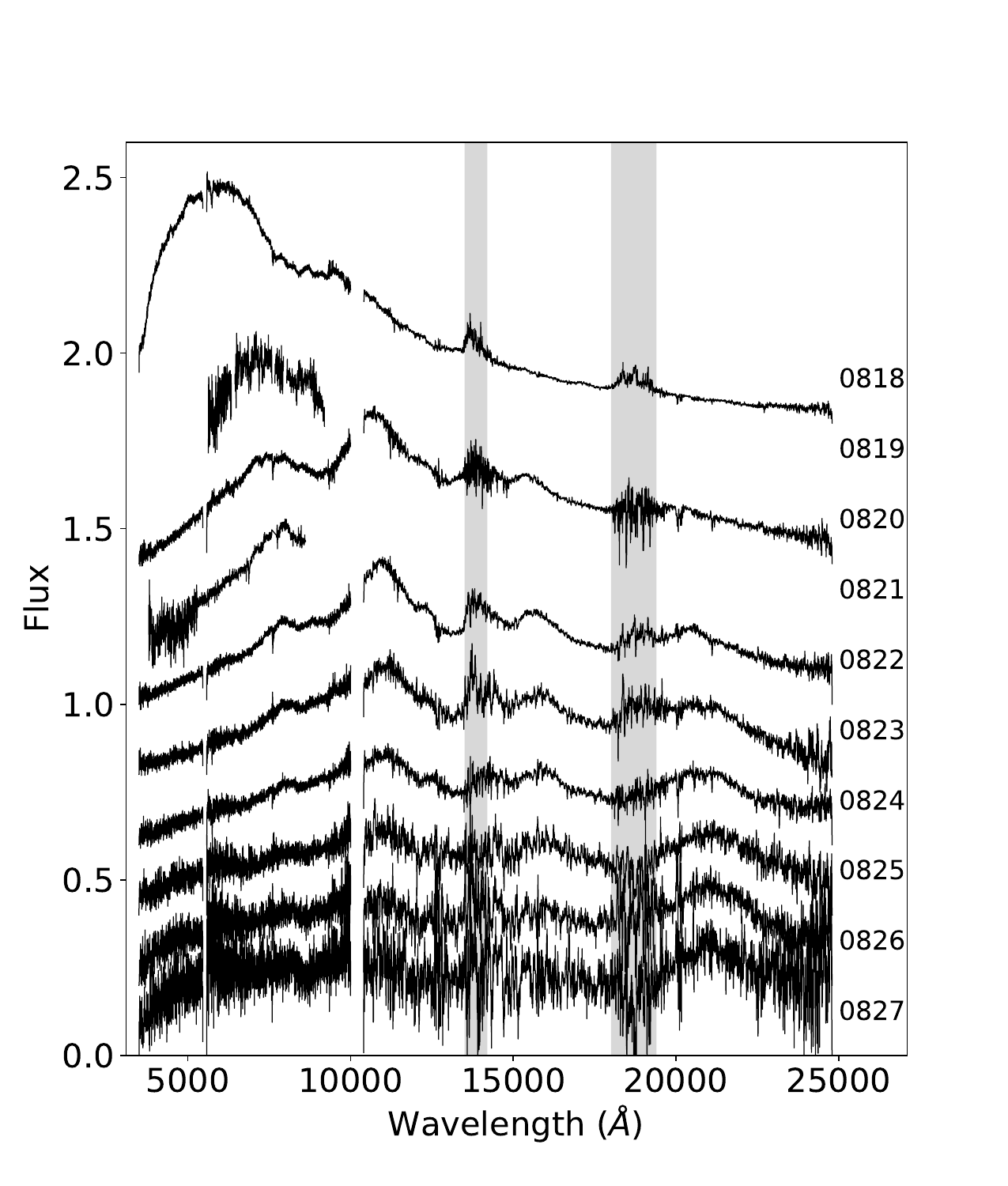}
  \caption{(left panel) Bolometric light curve of AT 2017gfo, the
    kilonova associated with GW170817. The filled black triangles are
    from~\citet{smartt17}. Uncertainties derived from the range of
    values given in the
    literature~\citep{Waxman.Ofek.ea:2017,cowperthwaite17,smartt17}
    are shown as a grey band. Also shown are lower limits on the
    late-time luminosity as inferred from the Ks band with
    VLT/HAWK-I~\citep{Tanvir.Others:2017} (empty circles) and the
    4.5~$\mu$m detections by the \emph{Spitzer Space Telescope}
    from~\citet{Villar.Others:2018} (empty triangles)
    and~\citet{Kasliwal.Kasen.ea:2019} (empty squares)~\citep[adapted
    from][]{Wu.Barnes.ea:2019}. (right panel) Evolution of the
    kilonova flux spectrum during the first 10 days. Each spectrum is
    labelled by the observation epoch. The shaded areas mark the
    wavelength ranges with very low atmospheric
    transmission~\citep[reprinted by permission from Springer
    Nature,][]{Pian.Others:2017}. \label{fig:kiloobs}}
\end{figure*}

For many years a connection between observations of short-duration
gamma-ray bursts (sGRBs), supernova-like electromagnetic transients
(macronovae/kilonovae), and compact binary mergers has been
postulated~\citep[see e.g.][]{Piran:2004}.  The first observational
evidence of an excess of near infrared emission over the standard GRB
afterglow came in 2013 with the observation of GRB 130603B
by~\citet{tanvir13}\footnote{see \url{https://kilonova.space} for an
  up to date catalog of kilonova observations} and suggested a thermal
component consistent with kilonova emissions. Further, evidence has
been obtained from a reanalysis of the GRB
060614~\citep{Yang.Jin.ea:2015}, GRB
050709~\citep{Jin.Hotokezaka.ea:2016}, and GRB
070809~\citep{Jin.Covino.ea:2020} afterglow data including a first
estimate of the kilonova emission
temperature~\citep{Jin.Li.ea:2015}. See~\citet{Gompertz.Others:2018}
for a review of kilonova candidates associated to short GRB
observations.

Following the seminal work of~\citet{lipacz97}, first predictions of
light curves powered by radioactive decay were done
by~\citet{Metzger.Martinez.ea:2010,Roberts.Kasen.ea:2011,Goriely.Bauswein.Janka:2011}. These
initial studies used grey opacities appropriate to the Fe-rich ejecta
in type Ia SNe and predicted peak luminosities at timescales of a day
in the blue. However, the opacity of heavy r-process elements is
substantially higher due to the high density of line transitions
associated with the complex atomic structure of lanthanides and
actinides. This lead to a light curve peak at timescales of a week in
the
red/near-infrared~\cite{kasen13,Barnes.Kasen:2013,Tanaka.Hotokezaka:2013}.
\citet{Metzger.Bauswein.ea:2015} speculated on the possibility that in
fast expanding ejecta unburned neutrons are left and lead via their
decay to a ultraviolet/blue precursor event. Early blue emission has
also been suggested to originate from the hot cocoon that surrounds
the GRB jet as it crosses the
ejecta~\citep{Gottlieb.Nakar.Piran:2018}.

On August 17, 2017 the gravitational wave event GW170817 was
observed~\cite{Abbott.Abbott.ea:2017} and identified as merger of two
neutron stars. With the combination of gravitational wave signals and
electromagnetic observations, its location was
identified~\citep{Abbott:2017}, a (weak) sGRB
detected~\citep{Abbott17c}~\citep[weak probably due to an off-axis
observation,][]{Wu.Macfadyen:2018,Mooley.Nakar.ea:2018}, accompanied
by secondary X-ray and radio signals.

Within eleven hours of the merger the electromagnetic transient, named
AT~2017gfo, was observed in the ultraviolet, optical and near infrared
wavelength bands in the galaxy
NGC~4993~\citep{arcavi17,chornock17,coulter17,cowperthwaite17,drout17,evans17,kasliwal17,nicholl17,Pian.Others:2017,smartt17,soaressantos17,Tanvir.Others:2017}.
The left panel of Fig~\ref{fig:kiloobs} shows the bolometric light
curve for the two-week-long epoch of detailed observations adapted
from~\citet{Wu.Barnes.ea:2019}. The figure also includes late-time
observations from the Ks band with the
VLT/HAWK-I~\citep{Tanvir.Others:2017} and the 4.5 $\mu$m detections by
the \emph{Spitzer Space
  Telescope}~\citep{Villar.Others:2018,Kasliwal.Kasen.ea:2019}. The
right panel shows the evolution of the kilonova flux spectra from the
X-shooter VLT spectrograph during the first 10 days
from~\citet{Pian.Others:2017}. Further analysis of these spectra even
led to the first identification of an element,
Sr~\citep{Watson.Hansen.ea:2019}.

The luminosity and its evolution agreed with predictions for the light
powered by the radioactive decay of heavy nuclei synthesized via the
r-process in the neutron-rich merger
ejecta~\citep{lipacz97,Metzger.Martinez.ea:2010,Roberts.Kasen.ea:2011,Barnes.Kasen:2013,rosswog18}
(see section~\ref{sec:kilon-an-electr}). Additional evidence is
provided by the spectral/color evolution, that requires models with at
least two-components. Simulations suggest that at least three
components are necessary to account for the ejecta of neutron star
mergers: dynamic, winds, and secular outflows from the
disk~\citep{Perego.Radice.Bernuzzi:2017}. Combining all
observations~\citet{Villar.ea:2017} find a best-fit kilonova model
consisting of three-components: a ``blue'' lanthanide-poor component
(opacity $\kappa = 0.5$~cm$^{2}$~g$^{-1}$) with
${M}_{\text{ej}}\approx 0.020$~M$_\odot$, moving with a velocity of
approximately $0.27\ c$, an intermediate opacity ``purple'' component
($\kappa = 3$~cm$^{2}$~g$^{-1}$) with
${M}_{\text{ej}}\approx 0.047$~M$_\odot$ at $0.15\ c$, and a ``red''
lanthanide-rich component ($\kappa = 10$~cm$^{2}$~g$^{-1}$) with
${M}_{ej}\approx 0.011$ M$_\odot$ at $0.14\ c$. The three-component
model is compatible with a two-component model containing only blue
and red components. The blue component is expected to contain light
r-process elements with a negligible mass fraction of
lanthanides/actinides
$X_{\text{lan}} \lesssim 10^{-4}$~\citep{Kasen.Metzger.ea:2017}. The
mass fraction of lanthanides/actinides necessary to account for the
reddening of the spectra has been inferred to be
$X_{\text{lan}}\sim
10^{-3}$--$10^{-2}$~\citep{Kasen.Metzger.ea:2017,Tanaka.Utsumi.ea:2017,Waxman.Ofek.ea:2017}
and hence contains both light and heavy r-process material assuming
solar proportions. The purple component corresponds to ejecta with a
small, but non-negligible, lanthanide fraction. The early blue
emission has been interpreted to originate from the fastest outer
layers of the ejecta originating from material ejected in the polar
direction and containing exclusively light r-process
nuclei~\citep{metzger14,nicholl17,drout17}~\citep[see however,][for
alternative
explanations]{Waxman.Ofek.ea:2017,Kawaguchi.Shibata.Tanaka:2018}. The
later transition of the emission colors to the near infrared suggest
ejecta containing high r-process elements originating from the
post-merger accretion disk ejecta given their smaller velocities and
larger
masses~\citep{Siegel.Metzger:2017,Siegel.Metzger:2018,Kasen.Metzger.ea:2017,Perego.Radice.Bernuzzi:2017,Fernandez.Tchekhovskoy.ea:2019,Siegel:2019}~(see
section~\ref{sec:neutron-star-neutron}). The total amount of ejecta
has been estimated to be
$M_{\text{ej}} \approx
0.03$--0.08~M$_\odot$~\citep{Kasen.Metzger.ea:2017,kasliwal17,cowperthwaite17,Perego.Radice.Bernuzzi:2017,Villar.ea:2017,Waxman.Ofek.ea:2017,Kawaguchi.Shibata.Tanaka:2018}. However,
those estimates are based on spherically symmetric models while more
realistic geometries may lead to different
predictions~\citep{Kawaguchi.Shibata.Tanaka:2020,Korobkin.Wollaeger.ea:2020}.
This milestone observation provided the first direct indication that
r-process elements are produced in neutron-star mergers including
estimates of the amount of ejecta, composition and
morphology. Additional information about kilonova modeling and the
connection of these observations with models of compact binary mergers
can be found in section~\ref{sec:kilon-an-electr}.

\section{Basic Working of the r-process and necessary environment
  conditions}
\label{sec:basic-working}

\subsection{Modeling Composition Changes in Astrophysical Plasmas}
\label{sec:model-comp-chang}

Before discussing the working of the r-process in detail, a short
introduction into the methods should be given, how the build-up of
elements in astrophysical plasmas can be described and determined.
The mechanism to model composition changes is based on nuclear
reactions, occurring in environments with a given temperature and
density.  Integrating the reaction cross section $\sigma(E)$ over the
energy distribution of reacting partners at a given $T$, abbreviated
as $\langle\sigma v\rangle(T)$, determines the probability for
reactions to happen.  For most conditions in stellar evolution and
explosions a Maxwell-Boltzmann distribution is attained~\citep[e.g.][]
{Clayton:1968,rolfs.rodney:1988,Iliadis:2007,Lippuner.Roberts:2017}.
Nuclear decays can be expressed via the decay constant $\lambda$,
related to the half-life of a nucleus $t_{1/2}$ via
$\lambda=\ln 2 /t_{1/2}$.  Interactions with photons
(photodisintegrations) are described by the integration of the
relevant cross section over the energies of the photon Planck
distribution for the local temperature. This results in an effective
(temperature-dependent) ``decay constant'' $\lambda(T)$. Reactions
with electrons (electron captures on
nuclei)~\citep[e.g.][]{Fuller.Fowler.Newman:1980,Langanke.Martinez-Pinedo:2001,Langanke.Martinez-Pinedo:2003,Juodagalvis.Langanke.ea:2010}
or
neutrinos~\citep[e.g.][]{Langanke.Kolbe:2001,Langanke.Kolbe:2002,Kolbe.Langanke.ea:2003}
can be treated in a similar way, also resulting in effective decay
constants $\lambda$, which can depend on temperature $T$ and density
$\rho$ (determining for electrons whether degenerate or non-degenerate
Fermi distributions are in place). The $\lambda$'s for neutrinos
require their energy distributions~\citep{Tamborra.Mueller.ea:2012}
from detailed radiation transport, not necessarily reflecting the
local conditions~\citep[see e.g.][]{Liebendoerfer.Rampp.ea:2005,%
  Liebendoerfer.Whitehouse.ea:2009,Richers.Nagakura.ea:2017,%
  Janka:2016,Burrows.Vartanyan.ea:2018,Pan.Mattes.ea:2019}.

All these reactions contribute to changes of the abundances $Y_i$,
related to number densities $n_i=\rho Y_i/m_u$ and mass fractions of
the corresponding nuclei via $X_i=A_i Y_i$, where $A_i$ is the mass
number of nucleus $i$, $\sum_i X_i = 1$, $\rho$ denotes the density of
the medium, and $m_u$ the atomic mass unit.  The reaction network
equations for the time derivatives of the abundances $Y_i$ include
three types of terms~\citep[e.g.,][]{Hix.Thielemann:1999b}

\begin{eqnarray}
  \frac{d Y_i}{dt} & = & \sum_j P^i _j\ \lambda_j Y_j
                         + \sum_{j,k} P^i _{j,k}\ \frac{\rho}{m_u} \langle j,k\rangle Y_j
                         Y_k \label{4:network}\\ 
                   & & + \sum_{j,k,l} P^i _{j,k,l}\ \frac{\rho^2}{m_u^2}
                       \langle j,k,l\rangle Y_j Y_k  Y_l. \nonumber,
\end{eqnarray}
summing over all reaction partners related to the different summation
indices. The P's include an integer (positive or negative) factor
$N^i$ (appearing with one, two or three lower indices for one-body,
two-body, or three-body reactions), describing whether (and how often)
nucleus $i$ is created or destroyed in this reaction. Additional
correction factors $1/m!$ are applied for two-body and three-body
reactions in case two or even three identical partners are involved.
This leads to $P_j^i=N_j^i$, $P_{j,k}^i=N_{j,k}^i/m(j,k)!$, or
$P_{j,k,l}^i=N_{j,k,l}^i/m(j,k,l)!$. $m(i,j)$ is equal to 1 for
$i\ne j$ and 2 for $i=j$, $m(i,j,k)$ can have the values 1 (for
non-identical reaction partners), 2 for two identical partners, and 3
for the identical partners. Thus, this (additional) correction factor
is 1 for non-identical reaction partners, 1/2=1/2! for two identical
partners or even 1/6=1/3! for three identical partners. The
$\lambda$'s stand for decay rates (including decays,
photodisintegrations, electron captures and neutrino-induced
reactions), $\langle j,k\rangle$ for $\langle\sigma v\rangle$ of
reactions between nuclei $j$ and $k$.  Although in astrophysical
environments true three-body reactions are negligible, a sequence of
two two-body reactions --- with an intermediate extremely short-lived
nucleus --- is typically written as a three-body reaction term,
resulting in the expression
$\langle
j,k,l\rangle$~\citep{nomoto85,Goerres.Wiescher.Thielemann:1995}. The
nuclei involved in the first reaction, including the highly unstable
intermediate nucleus, are typically in chemical equilibrium (see
below).  A survey of computational methods to solve nuclear networks
is given in
\citet{Hix.Thielemann:1999b,timmes99,Hix.Meyer:2006,Lippuner.Roberts:2017}.
The solution of the above set of differential equations provides the
changes of individual nuclear abundances for any burning process in
astrophysical environments, requiring the inclusion of all possible
reactions and the relevant nuclear physics input\footnote{for data
  repositories see e.g. \url{https://jinaweb.org/reaclib/db},
  \url{https://nucastro.org/reaclib.html},
  \url{http://www.kadonis.org}, and
  \url{http://www.astro.ulb.ac.be/pmwiki/Brusslib/HomePage}.}.  In
astrophysical applications the composition changes determined by
Eq.~\eqref{4:network} cause related energy generation which couples to
the thermodynamics and hydrodynamics of the event
\citep{Mueller:1986}. For large reaction networks that can be
computationally rather expensive due to two effects: (a) nuclear
reaction time scales vary by orders of magnitude and the resulting
reaction networks represent so-called stiff systems of differential
equations which can only be solved with implicit computational
methods, requiring huge systems of non-linear equations with several
Newton-Raphson iterations. (b) the size of time steps needed to follow
nuclear composition changes can be much smaller than those relevant
for hydrodynamic changes. For these reasons in most cases the problem
is split in a hydrodynamics/thermodynamics part with a limited
reaction network, sufficient for the correct energy generation, and
postprocessing of the obtained thermodynamic conditions with a
detailed nucleosynthesis network~\citep[see e.g.][and references
therein]{Ebinger.Curtis.ea:2019,Curtis.ea:2019}.

If matter experiences explosive burning at high temperatures and
densities, the reaction rates for fusion reactions and the
photodisintegration rates (due to a Planck photon distribution
extending to high energies) are large. This will lead to chemical
equilibria, i.e.\ balancing of forward and backward flows in
reactions, in particular also for proton or neutron capture reactions
$p+(Z,A)\rightleftarrows (Z+1,A+1) +\gamma$ and
$n+(Z,A)\rightleftarrows (Z,A+1) +\gamma$, corresponding to a relation
between the chemical potentials $\mu _p + \mu (Z,A) = \mu (Z+1,A+1)$
and $\mu_n + \mu(Z,A) = \mu(Z,A+1)$, as the chemical potential of
photons vanishes.  If this is not only the case for a particular
reaction, but across the whole nuclear chart, the complete reaction
sequence is in chemical equilibrium, i.e.\
$Z \mu_p + N \mu_n = \mu(Z,A)$, termed complete chemical or also
nuclear statistical equilibrium
(NSE)~\citep[e.g.][]{Clayton:1968,Hix.Thielemann:1999a}.
For Boltzmann distributions (which apply in general in astrophysical
plasmas, with the exception of highly degenerate conditions, where
Fermi distributions have to be utilized for the chemical potentials)
\citep[see
e.g.][]{Thielemann.Truran:1986,Bravo.Garcia:1999,Yakovlev.ea:2006,Haensel.Potekhin.Yakovlev:2007},
the abundances of nuclei can be expressed by nuclear properties like
the binding energies $B(Z,A)$, the abundances of free neutrons and
protons, and environment conditions like temperatures $T$ and
densities $\rho$, leading to the abundance of nucleus $i$ \citep[with
$Z_i$ protons and $N_i$ neutrons or $A_i = Z_i + N_i$
nucleons,][]{Clayton:1968}

\begin{eqnarray}
  Y_i&=&Y_n^{N_i} Y_p^{Z_i}\frac{G_i(T) A^{3/2}_i}{2^{A_i}}
         \left(\frac{\rho}{m_u}\right)^{A_i-1} \label{NSE1} \\ 
     & & \times \left(\frac{2\pi\hbar^2}{m_ukT}\right)^{3(A_i-1)/2}
         \exp\left(\frac{B_i}{kT}\right),\nonumber  
\end{eqnarray}
where $B_i$ is the nuclear binding energy of the nucleus. $G_i$
corresponds to the partition function of nucleus $i$, as the ground
and excited state population is in thermal equilibrium.  Reactions
moderated by the weak interaction, i.e.  $\beta$-decays, electron
captures, and charged-current neutrino interactions, change the
overall proton to nucleon ratio $Y_e=\sum Z_i Y_i$ and occur on longer
time scales than particle captures and photodisintegrations. They are
not necessarily in equilibrium and have to be followed
explicitly. Thus, as a function of time the NSE abundances depend on
density $\rho(t)$, temperature $T(t)$, and proton-to-nucleon ratio
$Y_e(t)$, which is determined by weak interactions that act on longer
timescales and have not necessarily reached an equilibrium. Mass and
charge conservation lead to these two equations that determine the
values of $Y_n$ and $Y_p$:

\begin{eqnarray}
  \sum_i A_i Y_i&=&Y_n+Y_p + \nonumber \\ 
                & & \sum_{i,(A_i>1)}
                    (Z_i+N_i)Y_i(\rho,T,Y_n,Y_p)=1 \label{NSE2} \\
  \sum_i Z_i Y_i&=&Y_p+\sum_{i,(Z_i>1)} Z_i~Y_i(\rho,T,Y_n,Y_p) \ =Y_e.\nonumber
\end{eqnarray}

In general, very high densities favor heavy nuclei, due to the high
power of $\rho^{A_i-1}$, and very high temperatures favor light
nuclei, due to $(kT)^{-3(A_i-1)/2}$ in Eq.~\eqref{NSE1}. In the
intermediate regime $\exp (B_i/kT)$ favors tightly bound nuclei with
the highest binding energies in the mass range $A=50$--60 of the
Fe-group, but depending on the given $Y_e$. The width of the
composition distribution is determined by the
temperature~\citep[already derived in][]{Clayton:1968}.

Under certain conditions, i.e.\ not sufficiently high temperatures
when not all reactions are fast enough, especially due to small
reaction rates caused by too small $Q$-values, i.e.\ proton or neutron
binding energies across magic proton or neutron numbers (closed
shells), not a full NSE emerges but only certain areas of the nuclear
chart are in equilibrium, called quasi-equilibrium groups (or
QSE). This happens e.g.\ during early or late phases of explosive
burning, before or after conditions for a full NSE have been fulfilled
(In the latter case this is referred to as ``freeze-out'').  A typical
situation is a break-up in three groups, the Fe-group above Ca
($N=Z=20$), the Si-group between Ne ($N=Z=10$) and Ca, the light group
from neutrons and protons up to He, and nuclei not in equilibrium from
there up to Ne, as discussed in great detail
in~\cite{Hix.Thielemann:1999a}.

\begin{figure}[htbp]
  \includegraphics[width=1.0\linewidth]{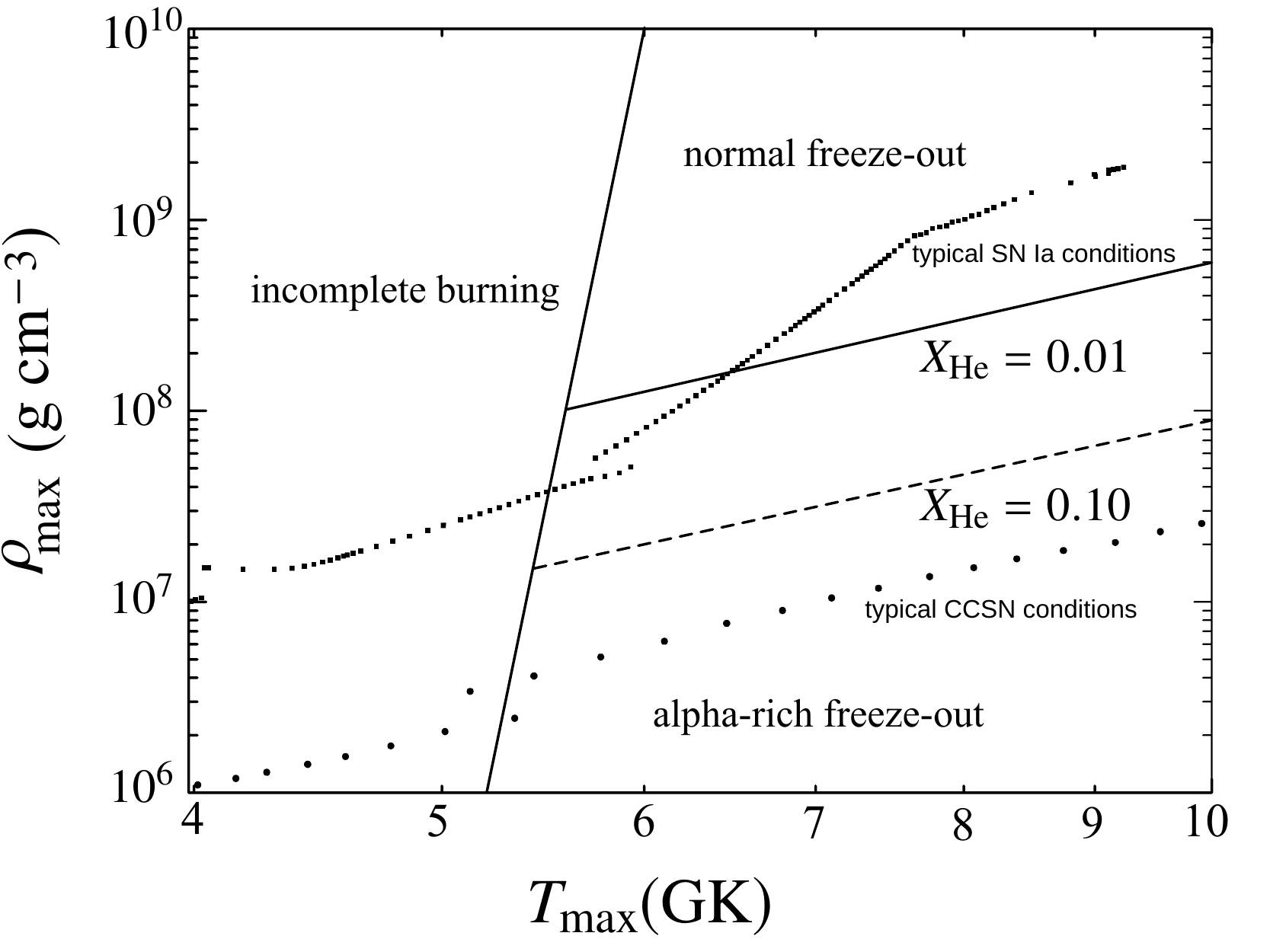}
  \caption{Plane of maximum temperatures and
    densities~\citep{Woosley.Arnett.Clayton:1973}, indicating for
    explosive Si-burning the boundaries of conditions after freeze-out
    of charged-particle reactions with an adiabatic expansion for an
    electron fraction $Y_e=0.498$. High densities, permitting
    ``three-body'' reactions to build-up C (and heavier nuclei), lead
    a normal freeze-out NSE composition. For lower densities unburned
    $\alpha$-particles remain, i.e.\ the final outcome is a so-called
    $\alpha$-rich freeze-out (see the lines of remaining He mass
    fractions $X_{\text{He}}$). The figure also displays typical
    conditions experienced in Si-burning mass zones of type Ia and
    core-collapse supernovae (CCSN), determining the nucleosynthesis
    outcome of such explosions (figure adapted from
    \citet{Thielemann.Nomoto.Yokoi:1986,Thielemann.Hashimoto.ea:1990,Thielemann.Nomoto.Hashimoto:1996}).
    \label{explosi}}
\end{figure}

A so-called $\alpha$-rich charged-particle freeze-out is a special
case of such QSE conditions, when the build-up of nuclei beyond $^4$He
is hampered by the need of ``three-body'' reaction sequences,
involving highly unstable $^8$Be (e.g.
$\alpha+\alpha+\alpha \rightarrow{} ^{12}\text{C} + \gamma$ or
$\alpha+\alpha+n \rightleftarrows{} ^9$Be) which are strongly
dependent on the density of matter (see Fig.~\ref{explosi}). The first
part of these reaction sequences involves a chemical equilibrium for
$\alpha+\alpha \rightleftarrows{} ^{8}$Be which is strongly shifted to
the left side of the reaction equation, due to the half-life of
$^{8}$Be ($t_{1/2}=6.7 \times 10^{-17}$s). Reasonable amounts of
$^{8}$Be, which permit the second stage of these reaction sequences
via an alpha or neutron-capture, can only be built-up for high
densities. The reaction rates for the combined (three-body) reactions
have a quadratic dependence on density in comparison to a linear
density dependence in two-body fusion reactions. Therefore, for low
densities the NSE cannot be kept and an overabundance of alpha
particles ($^4$He) remains, permitting only a (much) smaller fraction
of heavier elements to be formed than in an NSE (determined by binding
energies of nuclei). This $\alpha$-rich freeze-out leads to two
features: (a) the abundance of nuclei heavier than $^4$He is
(strongly) reduced in comparison to their NSE abundances, and (b) the
abundance maximum of the (fewer) heavy nuclei is shifted (via final
alpha captures) to heavier nuclei in comparison to an NSE\@.  While
this maximum would normally be around Fe and Ni (the highest binding
energies) with $A=50$--60, it can be shifted up to $A$ about 90 for
$\alpha$ mass fractions $X_{He}$ after freeze-out in excess of 40\%
\citep[see Figs.\ref{explosi}, \ref{fig_far1} and][]
{hoffman.woosley.qian:1997,Freiburghaus.Rembges.ea:1999}.

Other quasi-equilibrium conditions are encountered in proton or
neutron-rich environments. The first case permits proton captures and
reverse photodisintegrations, causing QSE-clusters along isotonic
lines in the nuclear chart, connected via $\beta^+$-decays and/or
$(\alpha,p)$-reactions on longer timescales~\citep[see
e.g.][]{rembges97}.  For the second case
\citet{Burbidge.Burbidge.ea:1957} \citep[followed up later by][]
{seeger.fowler.clayton:1965} postulated already in their 1957 review
that isotopic lines in the nuclear chart are in quasi-equilibrium for
neutron-rich r-process conditions (i.e.\ via neutron captures and
their reverse photodisintegrations), connected via $\beta^-$-decays on
longer timescales. The latter will be discussed in more detail with
respect to the r-process. Quasiequilibrium along isotonic and isotopic
lines typically acts close to the proton or neutron drip lines,
respectively. Thus, small reaction $Q$-values are involved for proton
or neutron captures, and also only small photon energies for the
inverse reactions are needed to establish such an equilibrium. This
changes temperature requirements somewhat.  While a full NSE, close to
stability with Q-values of the order 8--10~MeV, is only established
for temperatures around 4--5~GK \citep[as a rule of thumb temperatures
$kT$ need to exceed $Q/30$,][]{thielemann18}, for Q-values of the
order 1--2~MeV close to the drip-lines such equilibria can still be
established at temperatures exceeding about 1--1.5~GK.

\subsection{Special features of the r-process and the role of neutron
  densities and temperatures}
\label{sec:special-features-r}

The previous subsection explained how a complete NSE or QSE-subgroups
can be established. Here we want to discuss the special case of
QSE-subgroups along isotopic chains, before entering a description of
the possible sites which permit such quasi-equilibria. When
charged-particle reactions are frozen, the only connection between
isotopic chains is given by weak processes, i.e.\ $\beta$-decay, or
for large mass numbers fission and alpha decay (producing lighter
nuclei). High neutron densities make the timescales for neutron
capture much faster than those for $\beta$-decay and can produce
nuclei with neutron separation energies $S_n\sim 2$~MeV and less
(close to the neutron-drip line, where $S_n$ goes down to~0).  This is
the energy gained ($Q$-value) when capturing a neutron on nucleus
$A-1$ or the photon energy required to release a neutron from nucleus
$A$ via photo-disintegration.
For temperatures around 1~GK, $(\gamma,n)$ photodisintegrations in a
thermal plasma can still be very active for such small reaction
$S_n$-values (see above).  With both reaction directions being faster
than astrophysical (and $\beta$-decay) timescales a chemical
equilibrium between neutron captures and photodisintegrations is
attained. This establishes (quasi-)equilibrium clusters along isotopic
chains of heavy nuclei. The abundance distribution in each isotopic
chain follows the ratio of two neighboring isotopes

\begin{eqnarray}
  \label{Sn}
  \frac{Y(Z,A+1)}{Y(Z,A)} &=&n_n \frac{G(Z,A+1)}{2G(Z,A)}
                              \left[\frac{A+1}{A}\right]^{3/2} \\
                          & &\times \left[\frac{2\pi
                              \hbar^2}{m_ukT}\right]^{3/2}
                              \exp\left(\frac{S_n(A+1)}{kT}\right),
                              \nonumber
\end{eqnarray}
with partition functions $G$, the nuclear-mass unit $m_u$, and the
neutron-separation (or binding) energy of nucleus $(Z,A+1)$,
$S_n(A+1)$.  This relation for a chemical equilibrium of neutron
captures and photo-disintegrations in an isotopic chain follows from
utilizing the appropriate chemical potentials (see the previous
subsection) or equivalently due to the fact that the cross sections
for these reactions and their reverses are linked via detailed balance
between individual states in the initial and the final nucleus of each
capture reaction. The abundance ratios are dependent only on
$n_n=\rho Y_n/m_u$, $T$, and $S_n$. $S_n$ introduces the dependence on
nuclear masses, i.e.\ a nuclear-mass model for these very neutron-rich
unstable nuclei. Under the assumption of an
$(n,\gamma)\rightleftarrows (\gamma,n)$ equilibrium, no detailed
knowledge of neutron-capture cross sections is needed.

\begin{figure}[htb]
  \includegraphics[width=\linewidth]{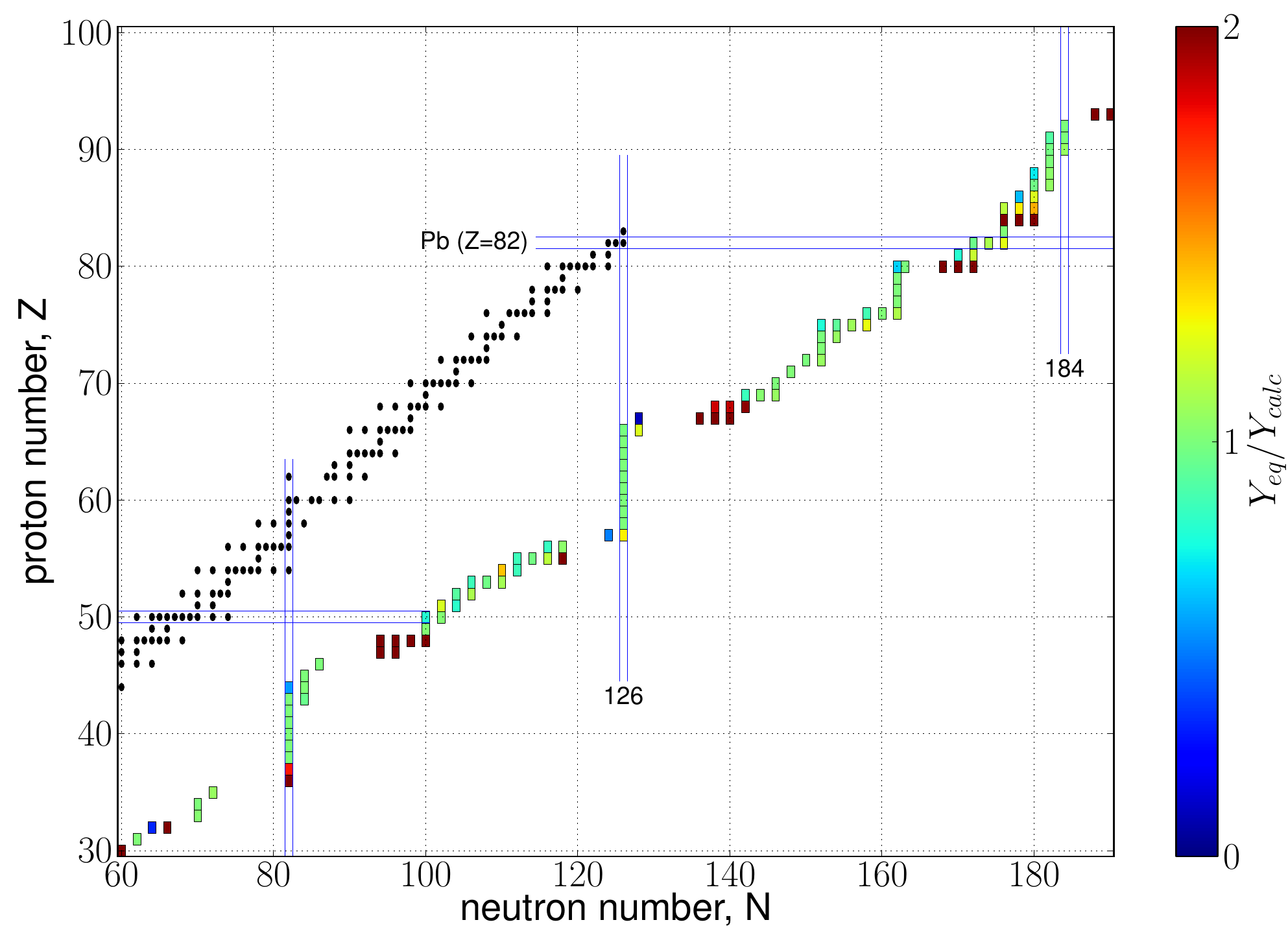}
  \caption{Shown is the line of stability (black solid circles), the
    proton and neutron magic numbers (blue horizontal and vertical
    lines), and an r-process path. The special conditions here are
    taken from a neutron star merger environment which will be
    discussed further below \citep[][reproduced by permission of the
    AAS]{Eichler.Arcones.ea:2015}. The position of the path follows
    from a chemical equilibrium between neutron captures and
    photo-disintegrations in each isotopic chain
    ($(n,\gamma)\rightleftarrows (\gamma,n)$ equilibrium). However,
    the calculation was performed with a complete nuclear network,
    containing more than 3000 nuclei. The colors along the path
    indicate how well the full network calculation follows such an
    $(n,\gamma) \rightleftarrows (\gamma,n)$ equilibrium. It can be
    seen that such full calculations agree with this equilibrium
    approach within a factor of 2 along the r-process path, which
    continues to the heaviest nuclei. }
  \label{fig_path} 
\end{figure}

Given that $Y(A+1)/Y(A)$ first rises with increasing neutron excess
before it decreases further out [caused by the last two factors in
Eq.(\ref{Sn})], leads to abundance maxima in each isotopic chain which
are only determined by the neutron number density $n_n$ and the
temperature $T$.  Approximating $Y(Z,A+1)/Y(Z,A) \simeq 1$ at the
maximum, as well as $G(Z,A+1) \approx G(Z,A)$, the neutron-separation
energy $S_n$ has to be the same for the abundance maxima in all
isotopic chains, defining the so-called r-process path.

Fig.~\ref{fig_path} shows such an r-process path for
$S_n \simeq 2$~MeV, when utilizing masses based on the Finite Range
Droplet Model FRDM \citep{moeller95}; for a detailed discussion of
nuclear properties far from stability see
Secs.~\ref{sec:exper-devel-r} and~\ref{sec:nuclear-modeling-r}. In
environments with sufficiently high neutron densities, the r-process
continues to extremely heavy nuclei and finally encounters the neutron
shell closure $N=184$, where fission plays a dominant role.
Fig.~\ref{fig_path} displays also the line of stability.  As the speed
along the r-process path is determined by $\beta$-decays, being
longest closer to stability, abundance maxima will occur at the top
end of the kinks in the r-process path at neutron shell closures
$N=50, 82, 126$. After decay to stability at the end of the process,
these maxima appear at the corresponding mass numbers $A$. These $A$'s
are smaller than those of stable nuclei for the same neutron shell
closures. The latter experience the smallest neutron capture cross
sections and cause the s-process maxima.

\begin{figure}[htbp]
  \centering
  \includegraphics[width=0.91\linewidth]{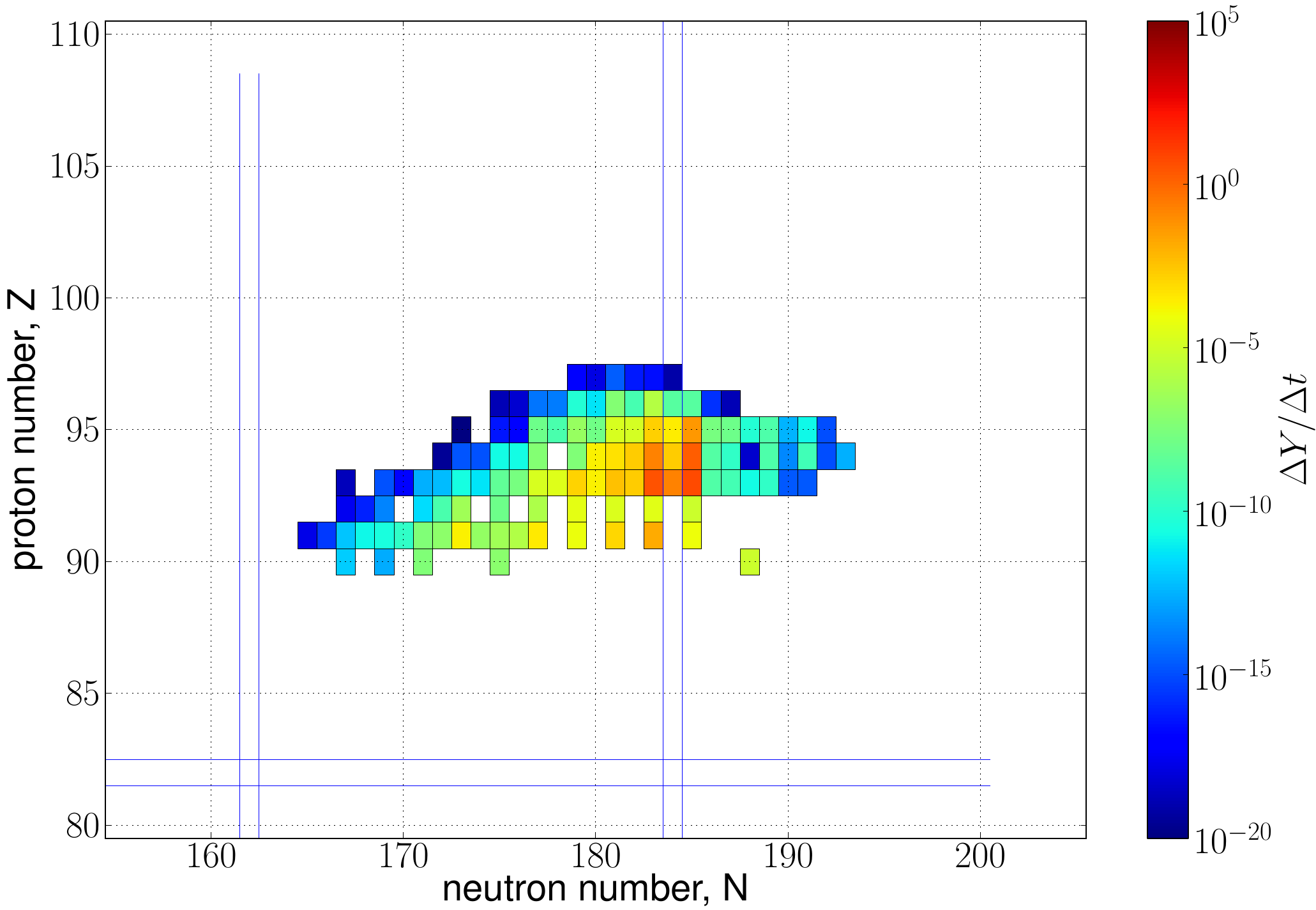}\\
  \includegraphics[width=0.91\linewidth]{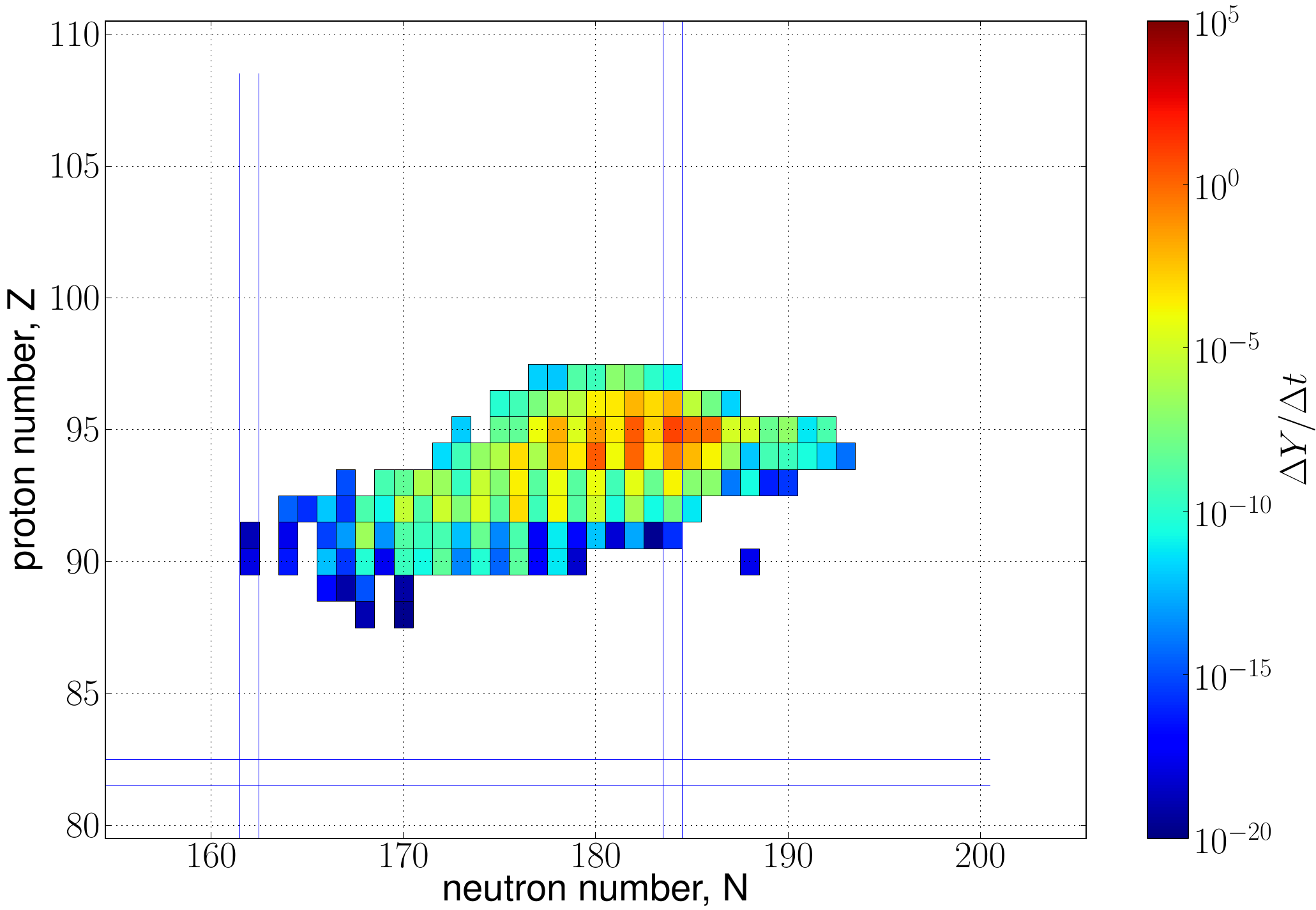}\\
  \includegraphics[width=0.91\linewidth]{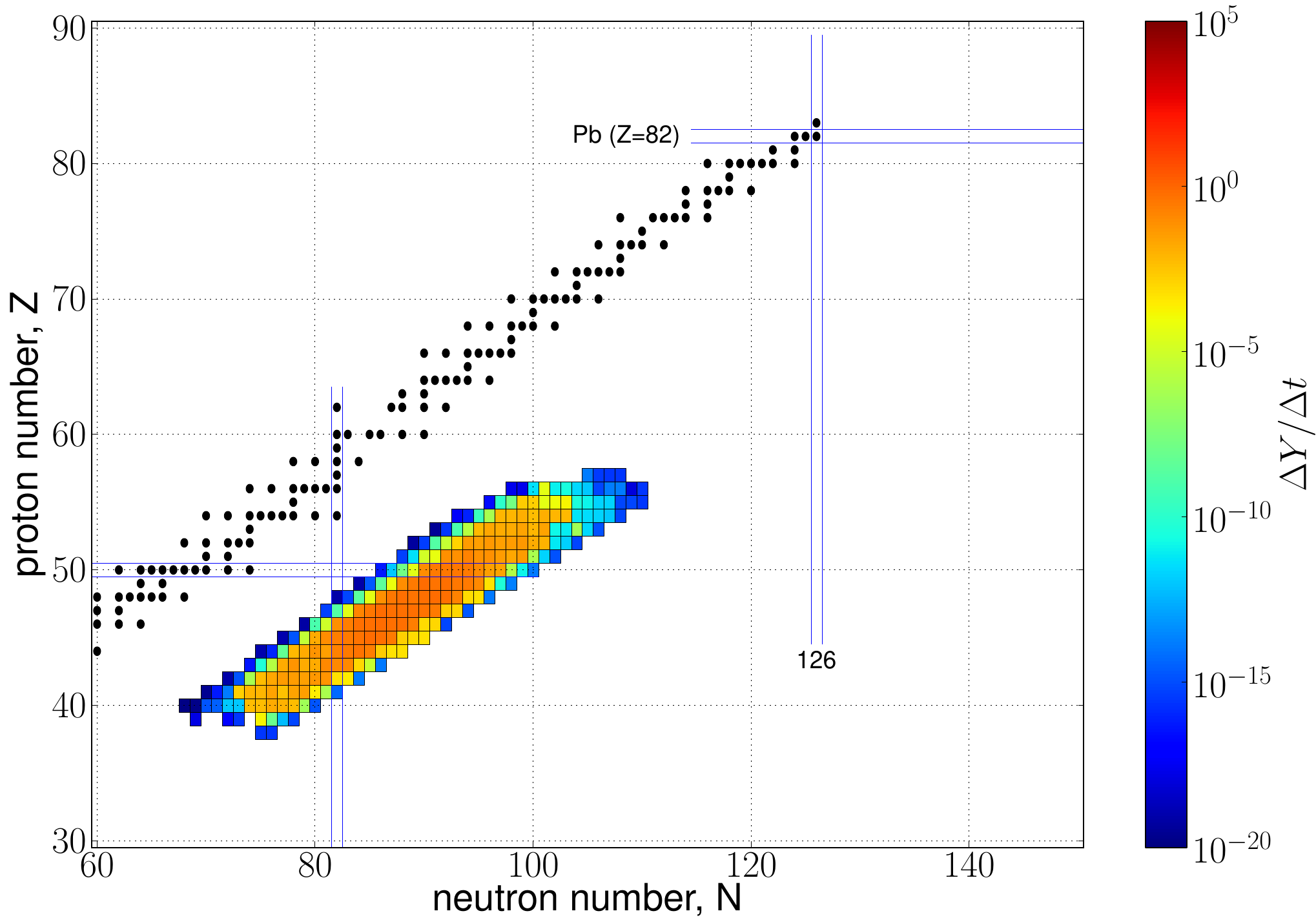}
  \caption{Color-coded time derivatives of nuclear abundances $Y$
    during an r-process simulation \citep[][reproduced by permission
    of the AAS]{Eichler.Arcones.ea:2015}, 
    describing the destruction via neutron-induced fission (top) and
    $\beta$-delayed fission (center)~\citep{Panov.Korneev.ea:2010} and
    (bottom) the production of fission fragments~\citep{kelic08}.  The
    largest destruction rates occur at and close to the neutron
    closure $N=184$, due to the smallest fission barriers encountered
    at these locations. Fission fragments are produced in a broad
    distribution, ranging in mass numbers $A$ from 115 to 155.  }
  \label{fig_fission} 
\end{figure}

Fig.~\ref{fig_fission} shows the regions of the nuclear chart where
fission dominates and the location of fission fragments for various
mass models and fission barriers.  Nuclear properties like mass
models, fission, and weak interactions will be discussed in extended
detail in the following two sections.
Early r-process calculations always made use of an
$(n,\gamma)\rightleftarrows(\gamma,n)$ equilibrium,
but had to assume neutron densities, temperatures, and a specific
duration time (before the final decay to stability, via $\beta$-decay and
$\beta$-delayed neutron emission)
\citep{Burbidge.Burbidge.ea:1957,seeger.fowler.clayton:1965,Kodama.Takahashi:1975}. They
realized that with such calculations not a unique set of conditions
could reproduce solar r-process abundances. Within this approach, and
with increasing knowledge of nuclear properties, Kratz and
collaborators provided a large series of parameter studies \citep[see
e.g.][]{Kratz.Bitouzet.ea:1993,Pfeiffer.Kratz.ea:2001}. An optimal
  fit for the three r-process peaks and the amount of matter
  in the actinides required a superposition of four components.

\begin{figure}[htb]
  \centering
  \includegraphics[width=\linewidth]{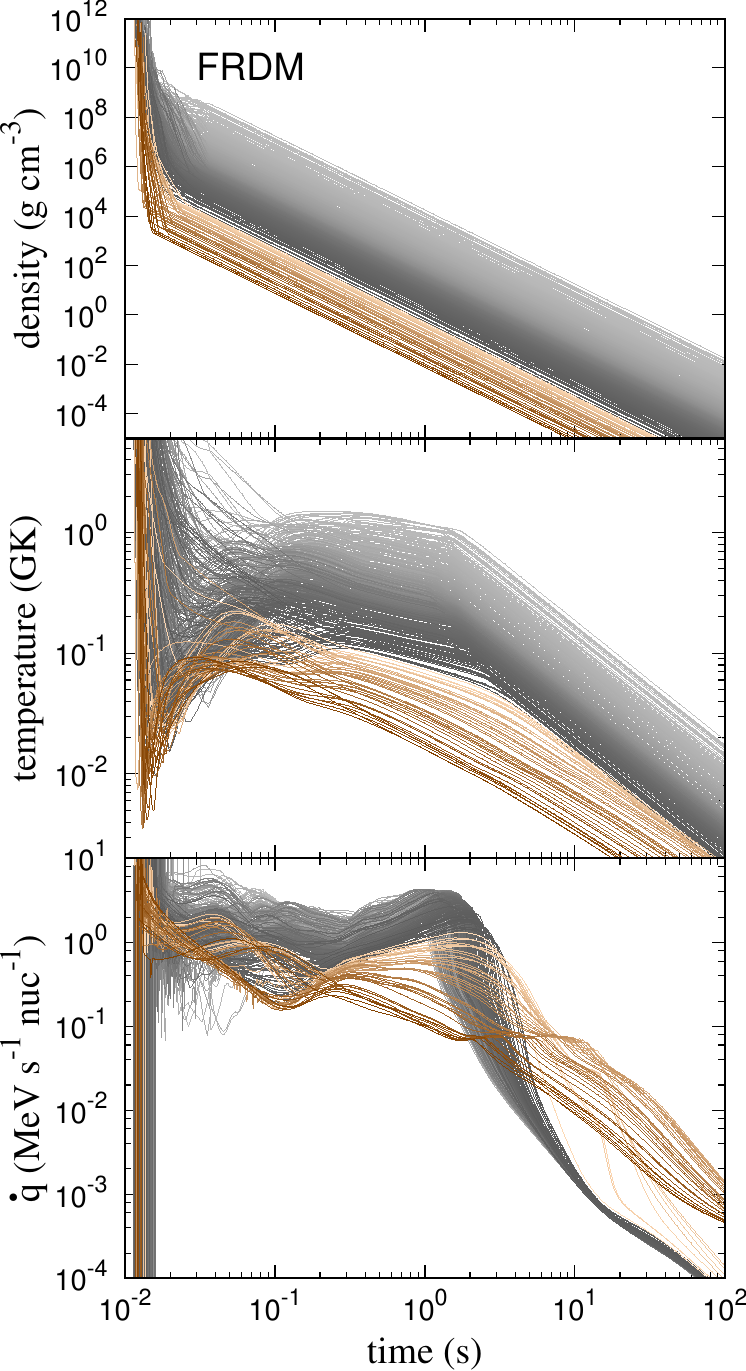}
  \caption{Evolution of temperature, density, and nuclear energy
    generation for different trajectories corresponding to material
    ejected dynamically in neutron star mergers. Gray darker (brown
    lighter) lines correspond to the ``slow (fast) ejecta'' discussed
    in section~\ref{sec:dynamic-ejecta}. For both gray and brown
    lines, light (dark) colors correspond to hotter (colder)
    conditions~\citep[adapted
    from][]{Mendoza-Temis.Wu.ea:2015}\label{fig:revol}}.
\end{figure}

Dynamical calculations with varying $n_n(t)$, and $T(t)$, and
discarding the $(n,\gamma)\rightleftarrows(\gamma,n)$ equilibrium,
follow the abundance changes in detail
\citep[e.g.][]{Blake.Schramm:1976,Truran.Cowan.Cameron:1978,Cowan.Cameron.Truran:1980,Cameron.Cowan.Truran:1983,Cowan.Cameron.Truran:1985}.
These calculations showed that the r-process can operate under two
different regimes with very different nuclear physics
demands~\citep{Wanajo:2007,Arcones.Martinez-Pinedo:2011}: a ``hot''
r-process in which the temperatures are large enough to reach
$(n,\gamma)\rightleftarrows(\gamma,n)$ equilibrium and a ``cold''
r-process in which the temperatures are so low that photodissociation
reactions are irrelevant~\citep{Blake.Schramm:1976}. Notice that the
differentiation between ``hot'' and ``cold'' refers to the temperature
conditions during the neutron capture phase and not during the earlier
phase when the seeds are formed; see
Sec.~\ref{sec:how-obtain-required}. Material could be initially very
cold and later reheated by nuclear-processes, resulting in a hot
r-process or initially very hot and during the expansion cool to low
temperatures producing a cold r-process. In general, astrophysical
environments produce a broad range of conditions in which both high
and low temperatures are reached, this is illustrated in
Fig.~\ref{fig:revol}. In some cases the material can reach such low
densities that free neutrons remain after the r-process (brown lines
in the figure) with potentially important observational
consequences~\citep{Metzger.Bauswein.ea:2015}. The figure also shows
the nuclear energy generation during the r-process (lower panel) that
is particularly relevant for very neutron-rich conditions, as expected
in dynamic ejecta from neutron star mergers (see
section~\ref{sec:dynamic-ejecta}), and for simulations of kilonova
r-process electromagnetic transients~(see
section~\ref{sec:kilon-an-electr}).

\begin{figure}[htb]
  \centering
  \includegraphics[width=\linewidth]{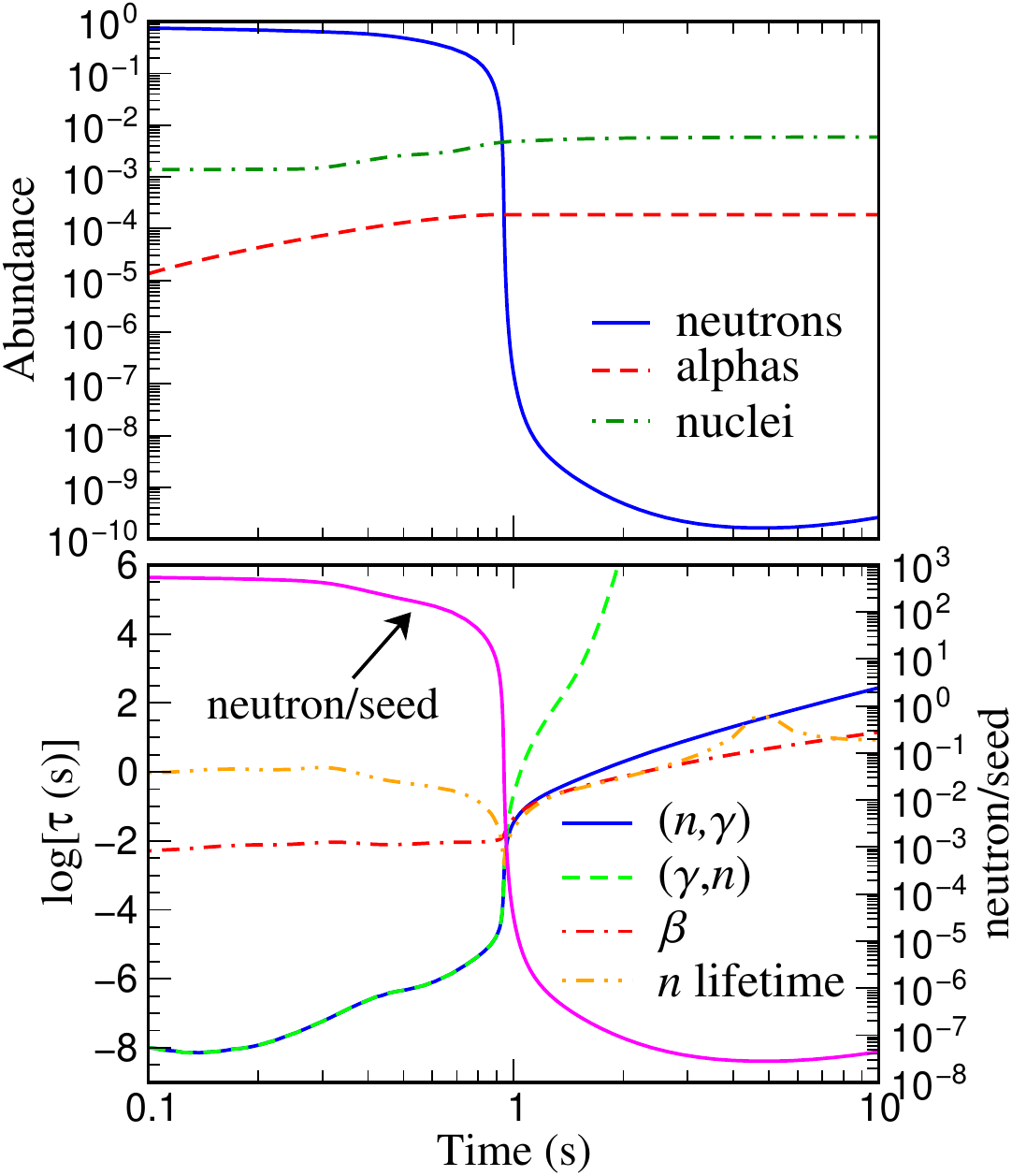}
  \caption{(Upper panel) Evolution of the abundances of neutrons,
    alphas and heavy nuclei during the r-process. (Bottom panel)
    Evolution of the neutron-to-seed ratio and average timescales for
    neutron captures, photodissociation and $\beta$-decays. The
    absolute value of the neutron lifetime, defined in the text, is
    also shown. Notice that for $t\lesssim 1$~s neutron capture and
    photodissociation timescales coincide (courtesy of
    M.-R. Wu).\label{fig:atevol}} 
\end{figure}

Following the freeze-out of charged-particle reactions,
the r-process consists typically of two phases: an initial phase
dominated by neutron captures and, depending on temperature,
photodissociations and a later phase in which neutron captures and
$\beta$-decays operate on very similar time scales during the decay to
stability in what is typically known as r-process freeze-out. The
transition between both phases occurs when the neutron-to-seed ratio
(i.e.\ the ratio of free neutrons to heavy nuclei) reaches values
close to one. This is illustrated in Fig.~\ref{fig:atevol}. The upper
panel shows the evolution of the abundances of neutrons, alphas and
heavy nuclei for a typical trajectory from those shown in
Fig.~\ref{fig:revol}. The lower panel shows the effective neutron
lifetime, $\tau_n$, the average radiative neutron capture timescale
per nucleus, $\tau_{(n,\gamma)}$, the average photodissociation
timescale per nucleus, $\tau_{(\gamma,n)}$, and the average
$\beta$-decay timescale per nucleus, $\tau_\beta$, defined as the
inverse of their average destruction rates per nucleus for the
respective processes:

\begin{subequations}
  \label{eq:timescales}
  \begin{equation}
    \label{eq:taun}
    \frac{1}{\tau_n} = \left|\frac{1}{Y_n}\frac{d Y_n}{dt}\right|
  \end{equation}
  \begin{equation}
    \label{eq:taung}
    \frac{1}{\tau_{(n,\gamma)}} = \frac{\sum_{Z,A} Y(Z,A)
      n_n\langle\sigma v\rangle_{A,Z}}{\sum_{Z,A} Y(Z,A)} 
  \end{equation}
  \begin{equation}
    \label{eq:taugn}
    \frac{1}{\tau_{(\gamma,n)}} = \frac{\sum_{Z,A} Y(Z,A)
      \lambda_{\gamma}(Z,A)}{\sum_{Z,A} Y(Z,A)} 
  \end{equation}
  \begin{equation}
    \label{eq:taubeta}
    \frac{1}{\tau_{\beta}} = \frac{\sum_{Z,A} Y(Z,A)
      \lambda_{\beta}(Z,A)}{\sum_{Z,A} Y(Z,A)} 
  \end{equation}
\end{subequations}
Thus, the latter three equations provide the neutron capture rate on
an average seed nucleus (averaged over all nuclei with their
abundances $Y(Z,A)$),
the photo-disintegration $(\gamma,n)$ rate and the $\beta$-decay rate,
respectively, being the inverse to the corresponding average reaction
time scales.

Fig.~\ref{fig:atevol} shows results of calculations with an initial
neutron-to-seed (nucleus) ratio $n_s\sim 600$, allowing for several
fission cycles before the end of the r-process, i.e.\ subsequent
sequences of fission, leading to (lighter) fission fragments which can
again capture neutrons until heavier nuclei are produced, encountering
fission again and the production of fission fragments. The impact of
fission cycling, doubling the number of heavy nuclei with each cycle
can be seen in the upper panel of the figure by the increase in the
abundance of heavy nuclei. A similar increase is also seen in the
abundances of alpha particles mainly due to the $\alpha$-decay of
translead nuclei. At early times ($<1$~s) the neutron abundance is
large and changes slowly with time. This is a consequence of the
almost identical $(n,\gamma)$ and $(\gamma,n)$ timescales as the
temperatures are large enough to maintain
$(n,\gamma)\rightleftarrows(\gamma,n)$ equilibrium. Neglecting the
production of neutrons by $\beta$-decay and fission one finds the
following relation for $1/\tau_n$, corresponding to
Eq.(\ref{eq:taun}), which is equal to the difference between average
neutron destructions via neutron capture and productions via
photo-disintegrations per nucleus, divided by the neutron-to-seed
(nucleus) ratio $n_s$

\begin{figure}[htb]
  \centering
  \includegraphics[width=\linewidth]{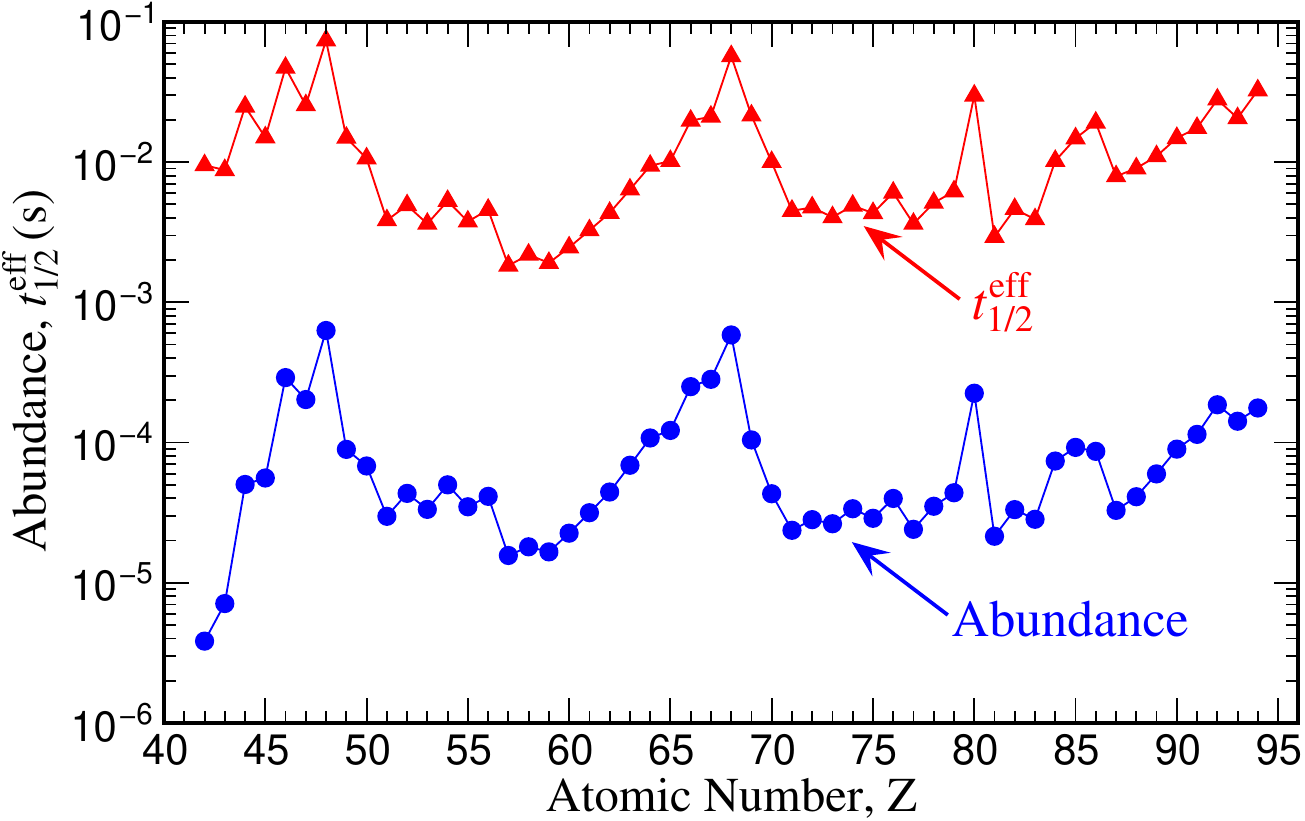}
  \caption{r-process elemental abundances at freeze-out compared with
    the effective $\beta$-decay half-live of an isotopic
    chain.\label{fig:beta-vs-abund}}
\end{figure}

\begin{equation}
  \label{eq:taunvstaung}
  \frac{1}{\tau_n} = \frac{1}{n_s}
  \left(\frac{1}{\tau_{(n,\gamma)}}-\frac{1}{\tau_{(\gamma,n)}}\right) 
\end{equation}
illustrating the important role played by the neutron-to-seed
ratio. Whenever $n_s > 1$, the effective neutron lifetime is large and
there is enough time for the r-process to pass via successive
$\beta$-decays through many isotopic chains and to reach a
$\beta$-flow
equilibrium~\citep{Kratz.Bitouzet.ea:1993,Freiburghaus.Rembges.ea:1999}. The
time derivative $\dot Y(Z)$ of the abundance in a whole isotopic chain
$Y(Z)=\sum_N Y(Z,N)$, due to $\beta$-decays, is given by

\begin{eqnarray}
  \label{eq:betaflow}
\sum_N \lambda_\beta(Z,N) Y(Z,N) & = & Y(Z) \sum_N \lambda_\beta(Z,N) \frac{Y(Z,N)}{Y(Z)} \nonumber \\
                                                   & = & Y(Z)
                                                         \lambda_\beta^{\text{eff}}(Z)
                                                         = \text{const},
\end{eqnarray}
with $\lambda_\beta^{\text{eff}}(Z)$ being the effective decay rate of
the whole chain. In a $\beta$-flow equilibrium this flux is constant
through all affected $Z$'s. As decay rates are related to half-lives
via $\lambda=\ln 2/t_{1/2}$, the abundance of a complete isotopic
chain is proportional to its effective $\beta$-decay half-life
$Y(Z) \propto t^{\text{eff}}_{1/2}(Z)$ (see
Fig.~\ref{fig:beta-vs-abund}).

During this early phase, the r-process path is mainly determined by
the two-neutron separation energies, $S_{2n}$ as only even neutron
number nuclei are present, due to the pairing effect on binding
energies (see Fig.~\ref{fig_path}). Typically, $S_{2n}$ values
decrease smoothly with neutron excess with a sudden decrease at magic
neutron numbers. However, for several mass models the $S_{2n}$ are
either constant or show a saddle point behavior in regions where there
is a transition from deformed to spherical nuclei (or vice-versa) just
before or after magic shell closures. This leads also to saddle points
in contour lines of constant $S_n$ in the nuclear chart and translates
to the appearance of gaps in the r-process path (see
Fig.~\ref{fig_path}) producing troughs in the abundance
distribution~\citep[][]{Thielemann.Kratz.ea:1994,Arcones.Martinez-Pinedo:2011}
before the onset of the freeze-out of neutron captures. These troughs
have been extensively discussed in the literature~\citep[see e.g.][and
references
therein]{Chen.Dobaczewski.ea:1995,Pfeiffer.Kratz.Thielemann:1997} as a
signature of quenching of the $N=82$ shell gap.
However, whether the related behavior of neutron separation
  energies is due to quenching of shell
effects far from stability or insufficiency in the challenging
treatment of nuclei around the transition from well deformed to
spherical is still
debated~\citep[see e.g.][]{Grawe.Langanke.Martinez-Pinedo:2007}.

Before freeze-out the nuclei with the strongest impact in the
r-process dynamics are those with the longest $\beta$-decay
half-lives. These are the nuclei closest to the stability at or just
after the magic shell closures. Uncertainties in the nuclear physics
properties of those nuclei may have a strong impact on the final
abundances. This is particularly the case for nuclei located after the
$N=82$ shell closure. This is confirmed by sensitivity
studies~\citep[see e.g.][]{Mumpower.Surman.ea:2016} that explore the
impact on r-process abundances due to variations of nuclear
properties.

Once the r-process reaches $n_s\approx 1$, there is an important
change in the dynamics. Nuclei start to compete for the few available
neutrons and the effective neutron lifetime decreases dramatically,
see Eq.~\eqref{eq:taunvstaung} and Fig.(\ref{fig:atevol}). The neutron lifetime increases again
once the $\beta$-decay timescale becomes shorter than the $(n,\gamma)$
timescale, resulting in a more gradual decline of $n_s$ at later
times. The evolution after $n_s \lesssim 1$ is known as r-process
freeze-out. During this phase, the timescales of neutron captures and
$\beta$-decays become similar. It is precisely the competition between
neutron captures and $\beta$-decays (often followed by neutron
emission) during the decay to stability that is responsible of
smoothing the r-process abundances. Just before the freeze-out the
abundances exhibit strong oscillations versus mass number. However,
after freeze-out they are rather smooth in agreement with the solar
system r-process abundances. This is a characteristic feature of the
r-process when compared to the s-process. In the latter case, there is (almost) 
never a competition between $\beta$-decays and neutron captures and
hence the abundances show a strong sensitivity on $A$.

Any process that produces neutrons during freeze-out can affect the
final abundances. This includes $\beta$-delayed neutron emission and
fission with the first one dominating for $n_s \lesssim 150$. One
should keep in mind that the impact of neutron production is
non-local, in the sense that neutrons can be produced in one region of
the nuclear chart and captured in another. The freeze-out is
responsible of shaping the final abundances. The rare-earth peak is
known to be formed during the r-process freeze-out. At low $n_s$, this
is due to a competition between neutron captures and
$\beta$-decays~\citep{Surman.Engel.ea:1997}, at high $n_s$, when
fission is important, the fission yields play also an important
role~\citep{Steinberg.Wilkins:1978,panov08,Goriely.Sida.ea:2013,Eichler.Arcones.ea:2015}. The
freeze-out has also a strong impact on the abundances at the r-process
peaks at $A\sim 130$ and 195~\citep[see
e.g.][]{Mendoza-Temis.Wu.ea:2015}.  Due to the large $n_s$, material
accumulates at the $N=184$ shell closure with $A\sim 280$. During the
decay to $\beta$-stability the material fissions, producing nuclei
with $A\lesssim 140$ and neutrons. Depending on the fission rates and
yields used it may result in very different final
abundances~\citep{Eichler.Arcones.ea:2015,Goriely.Martinez-Pinedo:2015,Vassh.Mumpower.ea:2019,Giuliani.Martinez-Pinedo.ea:2019}. Neutrons
emitted by fission have a strong impact on the abundances of the 3rd
r-process peak. Depending on the masses of nuclei around
$N=130$~\citep{Mendoza-Temis.Wu.ea:2015} and the $\beta$-decay rates
of nuclei with $Z\gtrsim 80$~\citep{Eichler.Arcones.ea:2015} the peak
could be shifted to higher mass numbers when compared with solar
system abundances.

After having discussed here the general working of and the nuclear
physics input for an r-process, the following subsection
discusses how to obtain the required neutron-to-seed rations, before
discussing the astrophysical sites in
section~\ref{sec:astr-sites-their}. However, independently, the
influence of nuclear uncertainties should be analyzed, which will be
done in sections~\ref{sec:exper-devel-r}
and~\ref{sec:nuclear-modeling-r}.  They can also affect the validity
of suitable astrophysical environments. Recent studies of the impact
of mass models, $\beta$-decay half-lives, and fission rates and
fragment distributions have been (among others) performed by
\citet{Arcones.Martinez-Pinedo:2011,Eichler.Arcones.ea:2015,Goriely:2015,Mendoza-Temis.Wu.ea:2015,Mumpower.Surman.ea:2016,Marketin.Huther.ea:2016,panov16,Mumpower.Kawano.ea:2018,Vassh.Vogt.ea:2019,Giuliani.Martinez-Pinedo.ea:2019,Vassh.Mumpower.ea:2019}.

\subsection{How to obtain the required neutron-to-seed ratios}
\label{sec:how-obtain-required}

Explosive environments with high temperatures exceeding about 5~GK,
lead to a nuclear statistical equilibrium NSE, consisting of neutrons,
protons, and $\alpha$-particles, as discussed
subsection~\ref{sec:model-comp-chang}. The $Y_e$, which affects the NSE
composition, is given by the initial abundances and the weak
interactions, which determine the overall neutron to proton (free and
in nuclei) ratio.  Essentially all sites of interest for the r-process, 
whether starting out with hot conditions or emerging from
cold neutron star material, which heats up during the build-up of
heavier nuclei, pass through such a phase. Thus, both cases will lead
to similar compositions of light particles and nuclei before the
subsequent cooling and expansion of matter, still being governed
initially by the trend of keeping matter in NSE before the charged-particle
freeze-out and the onset of neutron captures in the r-process.

\begin{figure}[htb]
  \includegraphics[width=\linewidth]{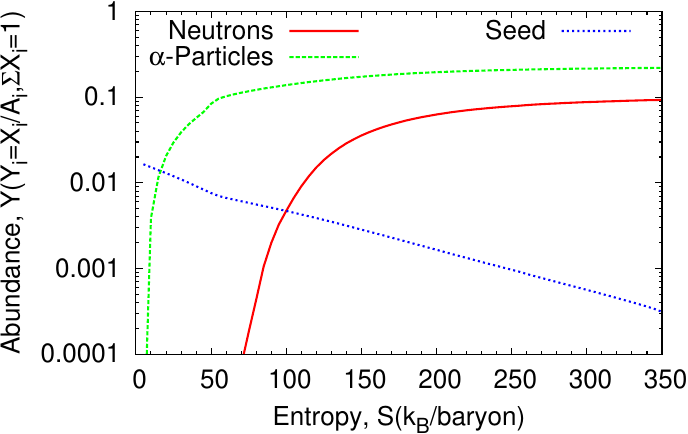}
  \caption{Abundances of neutrons $Y_n$, $^4$He ($\alpha$-particles)
    $Y_\alpha$, and so-called seed nuclei $Y_{\text{seed}}$ (in the mass
    range $50\leq A\leq100$), resulting after the charged-particle
    freeze-out of explosive burning (for a $Y_e=0.45$), as a function
    of entropy in the explosively expanding plasma \citep[][reproduced by permission of the AAS]{Farouqi.Kratz.ea:2010}. It can be realized that the
    ratio of neutrons to seed nuclei ($n_s=Y_n/Y_{\text{seed}}$)
    increases with entropy. The number of neutrons per seed nucleus
    determines whether the heaviest elements (actinides) can be
    produced in a strong r-process, requiring
    $A_{\text{seed}}+n_s \gtrsim 230$.  \label{fig_far1} }
\end{figure}

In hot environments the total entropy is dominated by the black-body
photon gas (radiation). In such radiation-dominated plasmas the
entropy is proportional to
$T^3/\rho$~\citep{Hartmann.Woosley.El:1985,woosley92,Meyer:1993,Witti.Janka.Takahashi:1994},
i.e.\ the combination of high temperatures and low densities leads to
high entropies. Thus, high entropies lead to an $\alpha$-rich
freeze-out (see Fig.~\ref{explosi}), and --- dependent on the entropy
--- only small amounts of Fe-group (and heavier) elements are
produced, essentially the matter which passed the three-body bottle
neck reactions (triple-alpha or $\alpha\alpha$n) transforming He to Be
and/or C. This can also be realized when examining
Fig.~\ref{fig_far1}, obtained from detailed nucleosynthesis
calculations, not assuming any equilibrium conditions.  Initially NSE
has been obtained. However, dependent on the entropy, different types
of charged-particle freeze-out occur, paving the way to the subsequent
evolution.

The calculation for Fig.~\ref{fig_far1}, starts out with matter in an
NSE composition for $Y_e=0.45$, at $T_0=8$~GK and a density $\rho_0$
corresponding to the given entropy.  The expansion from those
conditions follows on a so-called free-fall timescale
$t_{ff}=(3\pi/(32G\rho_0))^{1/2}$ (the timescale on which a
homogeneous gas cloud of initial density $\rho_0$ would
contract). This timescale is comparable to the expansion caused by an
explosion.  Fig.~\ref{fig_far1} shows how --- with increasing
entropies --- the alpha mass-fraction ($X_\alpha = 4 Y_\alpha$) is
approaching a constant value and the amount of heavier elements (which
would provide the seed nuclei for a later r-process) is going to
zero. This is similar to the big bang, where extremely high entropies
permit essentially only the production of elements up to He, and tiny
amounts of Li.  Opposite to the Big Bang, experiencing very
proton-rich conditions, the $Y_e=0.45$ chosen here is slightly
neutron-rich, leading at high entropies predominantly to He and free
neutrons. The small amount of heavier nuclei after this
charged-particle freeze-out (in the mass range of $A=50$--100),
depending on the entropy or $\alpha$-richness of the freeze-out, can
then act as seed nuclei for capture of the free neutrons. Once the
charged-particle freeze-out has occurred, resulting in a high
neutron-to-seed ratio $n_s$, the actual r-process --- powered by the
rapid capture of neutrons --- can start, at temperatures below 3~GK\@.
Whether this r-process is a ``hot'' or ``cold'' one, as discussed in
subsection~\ref{sec:special-features-r}, depends on the resulting
neutron densities and temperatures.

\begin{figure}[htb]
  \includegraphics[width=\linewidth]{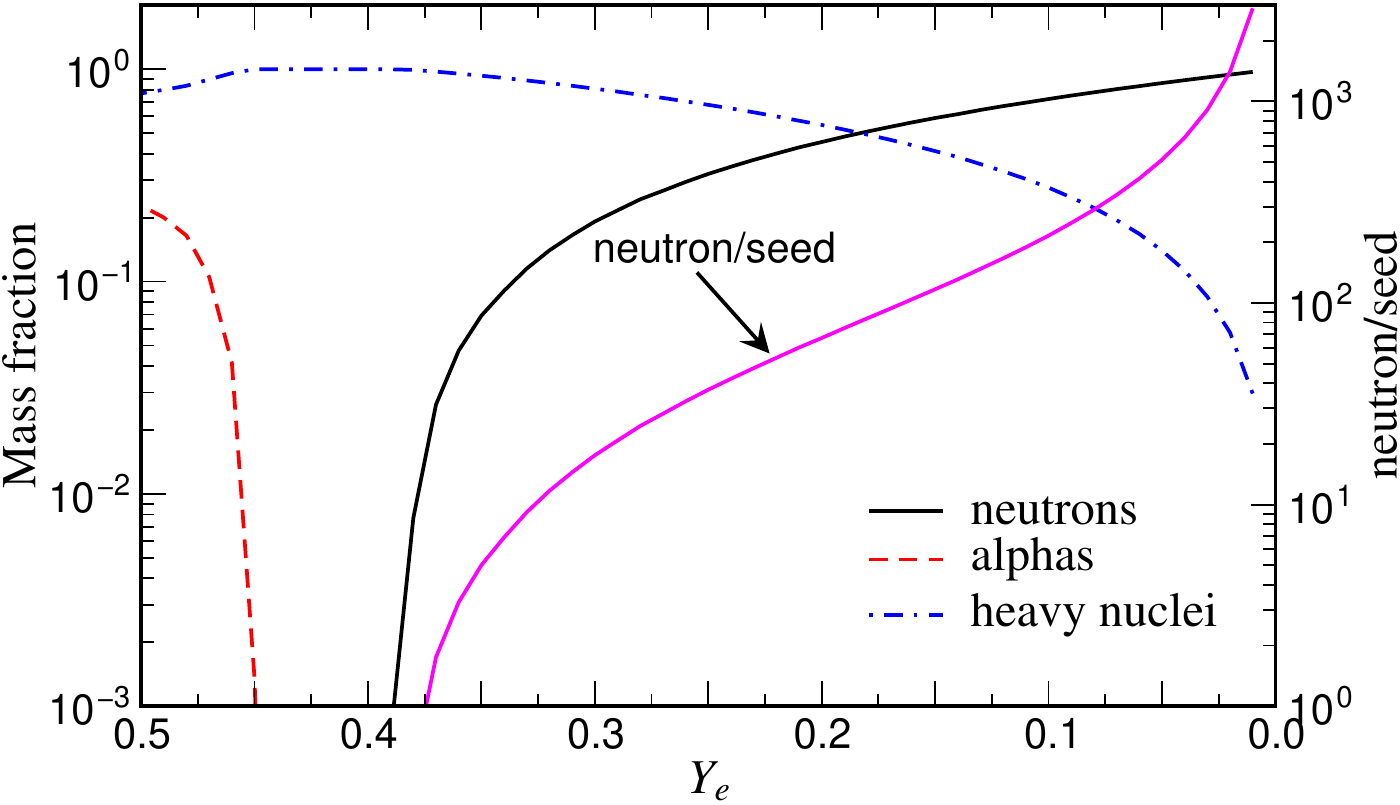}
  \caption{Evolution of the mass fractions of neutrons,
    alphas and heavy nuclei versus $Y_e$ (left y-axis scale) and
    neutron-to-seed (right y-axis scale) at a temperature of 3~GK for
    an adiabatic expansion with an entropy per nucleon $s=20\
    k_B$. \label{fig-gabr}}
\end{figure}

For both cases, the neutron-to-seed ratio $n_s$ determines the mass
range of nuclei to be produced. Starting with $A=50$--100 nuclei, the
production of lanthanides requires $n_s \sim 50$ while for actinides
an $n_s \sim 150$ is needed. 

When considering the result of these investigations, there remain two
options for a strong r-process in matter which was heated sufficiently
to pass through NSE: (a) for moderately neutron-rich conditions with
$Y_e$ not much smaller than 0.5, only very high entropies can provide
the necessary environment \citep[see Fig.\ref{fig_far1} and][]{Freiburghaus.Rembges.ea:1999}.  (b) For very low entropies, when after charged-particle
freeze-out only (NSE-)seed nuclei and free neutrons remain ($Y_n+A_{\text{seed}}Y _{\text{seed}}=1$ with
$Y_e=Z_{\text{seed}}Y _{\text{seed}}$) the $n_s$
ratio $Y_n/Y_{\text{seed}}$ becomes essentially entropy-independent,
$n_s \approx A_{\text{seed}} [Z_{\text{seed}}/(A_{\text{seed}}
Y_e)-1]$, such that only very neutron-rich matter
($Y_e \lesssim 0.15$) can support a strong
r-process. Fig.~\ref{fig-gabr} shows such a case of low entropies per
nucleon using $s=20\ k_B$, typical for matter ejected in neutron star
mergers. It shows several quantities as a function of $Y_e$. Comparing
with Fig.~\ref{fig_far1} at $S=20$ ($s=20\ k_B$) and for $Y=0.45$ one finds
consistent results, i.e.\ essentially only alphas and heavy nuclei (no
free neutrons) with typical charges $Z_{\text{seed}} \approx 28$ and
$A_{\text{seed}} \approx 63$. Only for $Y_e \lesssim 0.38$ free
neutrons start to appear and lanthanides are produced for
$Y_e \lesssim 0.25$~\citep[see][for a systematic study of the
astrophysical conditions necessary to produce
lanthanides]{Lippuner.Roberts:2015}.

\section{Experimental developments for r-process studies}
\label{sec:exper-devel-r}

The r-process path runs through nuclei with extreme neutron excess.
Most of these nuclei have yet not been produced in the laboratory and
their properties are experimentally unknown.  Hence the major nuclear
physics input required for r-process simulation must largely be
modelled, although recent measurements have led to a significant
decrease in uncertainties.  Intermediate-mass r-process nuclei could
be produced at existing radioactive ion-beam facilities like
CERN/Isolde, GSI and, and more recently in particular,
RIKEN\@. Significant advance, however, is expected in the future when
key r-process nuclei, including those around the third r-process peak,
become accessible at the next-generation RIB facilities like FAIR and
FRIB\@. This will be discussed in more detail in the following
sections. Data taken at these facilities will not only directly
substitute theory predictions, but will also serve as stringent and
valuable constraints to advance model predictions for even then not
accessible nuclei.

The next two sections deal with the nuclear ingredients needed for
r-process simulations. At first we will discuss the various
experimental approaches to produce and study neutron-rich nuclei and
summarize the experimental data relevant for r-process nucleosynthesis
which have been achieved recently. In the next section we present the
nuclear models applied to interpret the experimental results and
derive the vast nuclear data sets needed for large scale
simulations. We will focus on theoretical advances achieved by
improved models and experimental constraints and guidance, and finally
discuss the impact on the improved nuclear data on our understanding
of r-process nucleosynthesis.

\begin{figure}[htb]
 \includegraphics[width=\linewidth]{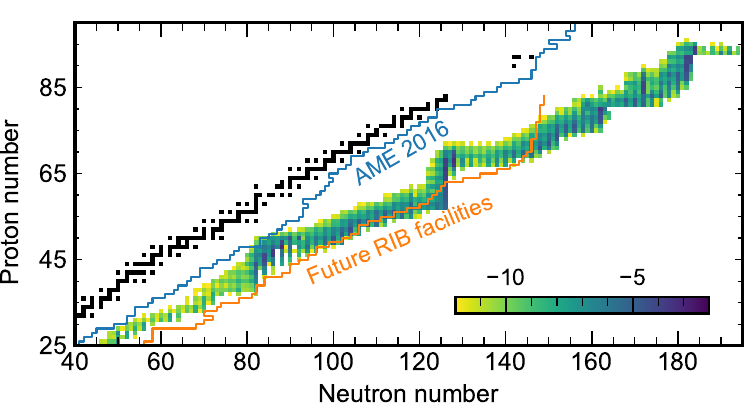}
  \caption{The nuclear chart with stable nuclei indicated by black squares, the
  limit of known masses from the Atomic Mass Evaluation 2016 (blue line), the
  future reach of Radioactive Ion Beam (RIB) facilities (like FRIB
  or FAIR), and abundance results (color coded) at neutron freeze-out
  from an r-process calculation taken
  from~\citep{Giuliani.Martinez-Pinedo.ea:2019}, utilizing the FRDM
  mass model, combined with Thomas-Fermi fission barriers. (courtesy
  Samuel A. Giuliani)\label{fig:ribfac}}
\end{figure}

There has been considerable experimental effort over the last forty
years to explore the nuclear physics of the r-process and the
structure and properties of r-process nuclei along the projected
reaction path; a goal that is nearly equivalent with exploring the
evolution of nuclear structure towards the limits of stability.  This
was one of the strong motivations towards the development of
facilities capable of producing radioactive ion beams. The experiments
have concentrated on measurement of masses, $\beta$~decay and
$\beta$-delayed neutron emission probabilities of neutron-rich nuclei
towards and even at the anticipated r-process path.  More recently new
methods are developed for the study of neutron capture reactions on
nuclei near or at the r-process path. A multitude of experimental
probes have been used to facilitate the production and separation of
very neutron-rich short-lived nuclei and to measure their specific
properties.  The traditional tools in the past ranged from extracting
fission products from reactors to the use of spontaneous fission
sources, to the analysis of short-lived reaction products at
spallation and fragmentation facilities.  The enormous progress
in producing neutron-rich isotopes with increasing intensity and
resolution was enabled by the simultaneous progress in the development
of new experimental techniques and detectors.  The traditional
approach of tape-collection and decay analysis leading to half-life
and $\beta$-endpoint determination for single separation products was
replaced by large scale ring experiments for measuring hundreds of
masses at once, or complementary to that by sophisticated trapping
experiments for determining the masses and decay properties of
individual neutron-rich nuclei with unprecedented accuracy.  The
utilization of these facilities and techniques produced and will
produce a wealth of data, which primarily address the needs for
knowledge about the properties of neutron-rich nuclei near or at the
r-process path (see Fig.\ref{fig:ribfac}).

These studies provided important information and input for r-process
simulations based on the $(n,\gamma)\rightleftarrows(\gamma,n)$
equilibrium assumptions. Specific challenges remained, such as the
$(n,\gamma)$ nuclear cross section reaction data for simulating the
r-process nucleosynthesis after freeze-out. The associated reaction
cross sections and reaction rates that are now used in dynamic
r-process simulations rely entirely on statistical model calculations,
utilizing Hauser-Feshbach codes like SMOKER,
NON-SMOKER\footnote{\url{https://nucastro.org/reaclib.html}}, and
TALYS\footnote{\url{http://www.talys.eu/more-about-talys}}. It is not
clear how reliable these predictions for the $(n,\gamma)$ reaction
rates are and how valid they are for reactions on neutron-rich
closed-shell nuclei that are characterized by low $Q$-values (neutron
binding energies)~\citep{Rauscher.Thielemann.Kratz:1997}. In fact, it
is expected that far from stability the direct capture component
dominates, permitting possibly an
$(n,\gamma)\rightleftarrows(\gamma,n)$ equilibrium down to low
temperatures \cite{Mathews1983}, but how reliable are the predictions
for the strength of such direct capture components
\cite{Goriely:1998,arnould07}?  A direct measurement of neutron
capture reactions on short-lived neutron-rich nuclei is challenging
and certainly not feasible within the near future.  The experimental
developments have focused on two approaches that combine theory and
experiment in order to get at the neutron capture cross-sections, the
$\beta$-Oslo method \cite{Spyrou2014, Spyrou2017, Tornyi2014,
  Guttormsen1987} and surrogate reactions
\cite{Escher2012,Kozub2012,Manning.Arbanas.ea:2019,Tang.Key.ea:2020},
mostly $(d,p)$ to get access to $(n,\gamma)$ rates. New initiatives
have also recently been proposed for direct neutron capture studies at
ring experiments \cite{Reifarth2017}. This proposal is particularly
challenging since the idea is to combine, for the first time, a method
that couples radioactive beam with radioactive target experiments.

Below, we discuss in detail the facilities and approaches presently
used for the production of neutron-rich nuclei on or near the
r-process path, along with some of the noteworthy experimental
developments in tools and techniques that allow measurements of
nuclear masses, $\beta$-decay rates, $\beta$-delayed neutron emission
probabilities, and neutron capture rates of nuclei required as inputs
for reliable simulations of the r-process.


\subsection{Production of neutron-rich isotopes}
The biggest challenge in experimentally studying isotopes near or at
the r-process path is the production of these isotopes in sufficient
abundances to explore their properties. This is closely correlated
with the selectivity of the separators necessary to select the
isotopes in question and the sensitivity of the detectors for
measuring the respective properties. The overall production of rare
isotopes has not significantly improved over the last decades due to
the cross section limitations in the production reactions, or the
energetics of the facilities to produce beams. However, substantial
improvements have been made in the selection process due to innovative
techniques in the isotope separation through electromagnetic systems
and the increasing utilization of laser based separation
techniques. Also, enormous progress has been made on the detection
side, and the development of ion trapping techniques has overcome many
of the statistical limitations in the more traditional measurements of
lifetimes, masses, and direct measurements of $\beta$-delayed neutron
emission probabilities of the very neutron-rich nuclei. This section
addresses the production of neutron-rich nuclei in reactors, in
spontaneous fission sources, in fission products in accelerators, at
spallation sources or ISOL facilities, and at fragmentation
facilities.

\subsubsection{Nuclear reactors and fission product sources}

One of the traditional methods for the production of neutron-rich
nuclei is the extraction and separation of neutron-rich fission
products from high flux nuclear reactors. Pioneering work has been
done at the TRISTAN separator \cite{TristanI, TristanII} at the Ames
Laboratory Research Reactor in Iowa which was moved in the mid 1970's
to the 60~MW High Flux Beam Reactor (HFBR) at Brookhaven National
Laboratory \cite{Crease2000}. The fission products were ionized,
extracted from a $^{235}$U target placed in an ion source located in a
beam line close to the reactor core, and separated.  A similar
separator was installed at the High Flux Reactor of the Institute Laue
Langevin (ILL) in Grenoble, where the LOHENGRIN fission fragment
separator is used to extract and analyze fission products in order to
study their decay properties \cite{Lohengrin}. This facility was
complemented by the installation of thermoionization separators, OSTIS
I \& II\@. The OSTIS separator concept \cite{OstisI, OstisII} was based
on the use of an external neutron guide line bombarding an external
$^{235}$U source. This approach allowed the measurement of
shorter-lived fission products since it reduced the transport times to
the ion-source.  Studies of neutron-rich nuclei were also performed at
smaller reactors, even at the Californium fission source
(CARIBU)~\citep{Pardo.Savard.Janssens:2016} at Argonne National
Laboratory, as long as separators were available to select the desired
fission product.  Measurements of masses, decay half-lives,
$\beta$-delayed neutron emission probabilities, and $\gamma$-ray decay
properties were performed with the best techniques available,
utilizing moving tape systems.  The measurement of neutron-rich
isotopes reached close to some of the r-process trajectories, in
particular for the alkali isotopes. The measurements on the
neutron-rich Rubidium isotopes made at that time~\citep{Kratz:1984}
are only rivaled now some thirty years
later~\citep{Lorusso.Nishimura.ea:2015}. The main handicaps, for
reaching the r-process path and mapping the very neutron-rich nuclei,
were the fission product distribution and long extraction times for
the fission products. All of these measurements had limitations that
were overcome with advances and technical developments in
detectors, including neutron detection technologies based on
$^6$Li-glass, $^3$He tubes, and $^{3}$He spectrometer systems
\cite{Ohm,Yeh.Clark.ea:1983}.

\begin{figure*}[htb]
  \includegraphics[width=\linewidth]{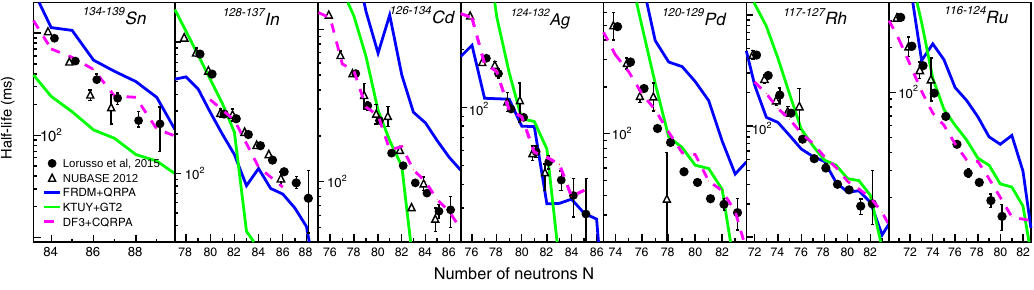}
  \caption{$\beta$-decay half-lives measured
    by~\citet{Lorusso.Nishimura.ea:2015} (solid circles) for a number
    of isotopic chains as a function of neutron number, compared with
    the 2012 NUBASE evaluation~\citep{Audi.Kondev.ea:2012} (open
    triangles) and the predictions of the models: Finite Range Droplet
    Model with Quasi Random Phase Approximation FRDM-QRPA (dark
    blue)~\citep{Moeller.Pfeiffer.Kratz:2003}, Gross Theory KTUY-GT2
    (light
    green)~\citep{Tachibana.Yamada.Yoshida:1990,Koura.Tachibana.ea:2005},
    Fayans Density Functional with Continuum Quasi Random Phase
    Approximation DF3-CQRPA (magenta
    dashed)~\citep{Borzov.Cuenca-Garcia.ea:2008} when
    available.~\citep[figure
    from][]{Lorusso.Nishimura.ea:2015}\label{fig:lorusso}}
\end{figure*}

\subsubsection{Spallation sources and ISOL techniques}

The on-line separators for fission products were complemented by
ISOLDE (Isotope Separator On-Line DEtector), designed in the
mid-sixties for separating spallation products produced by impinging
600~MeV protons from the synchro-cyclotron at CERN on a stationary
target.  The spallation of the heavy target nuclei produced a
distribution of target fragments, which were extracted and filtered to
separate the desired isotope. The time required for extraction placed
a lower limit on the half-life of isotopes which could be produced by
this method. Once extracted, the isotopes were directed to one of
several detector stations for measuring the decay properties. The
ISOLDE separator was moved in 1990 to the CERN PS booster to increase
the yield of the spallation products \cite{Isolde}. In a two-step
process, the 1 GeV proton beam from the PS-Booster, impinging on a Ta
or W rod positioned in close proximity to the uranium-carbide target,
produced the fast spallation neutrons to induce fission. This method
was essential for suppressing the proton-rich isobaric spallation
products that dominated the spallation yield.  The new ISOLDE system
was one of the most successful sources for neutron-rich isotopes and
dominated the production of neutron-rich isotopes for nearly two
decades. The implementation of laser ion-source techniques for
improving the $Z$-selectivity was a significant improvement to
studies of neutron-rich nuclei. The laser ionization of the fission
products led to a significant reduction in isobar background. This was
further improved by using the hyperfine splitting to select or
separate specific isomers in neutron-rich isotopes. These gradual
improvements of ion source and separator techniques finally led to the
detailed measurement of the r-process waiting point nucleus
$^{130}$Cd, the first milestone in reaching and mapping the r-process
path \citep{Kratz.Pfeiffer.ea:2000,Dillmann2003}, as well as the
recent mass measurements of $^{129-131}$Cd nuclei \cite{Atanasov2015}
near the doubly closed magic shell nucleus of $^{132}$Sn, a capability
that is unmatched to date.

\subsubsection{Fragmentation sources}

Another successful technique for the production of neutron-rich
isotopes is the use of fragmentation for the production of
neutron-rich, or fusion-evaporation for the production of
neutron-deficient, isotopes in heavy ion reactions. The GSI Online
Mass Separator was one of the first instruments to utilize
fusion-evaporation for studying isotopes far from stability using
heavy-ion beams from the UNILAC accelerator. The reaction products
were stopped in a catcher inside an ion source, from where they were
extracted as singly charged (atomic or molecular) ions and
re-accelerated to 60 keV. These beams were implanted, yielding sources
for $\beta$ or particle decay spectroscopy \cite{Roeckl}. This method,
however, was more suitable for the study of neutron-deficient isotopes
and remained not competitive with fission product or spallation based
production of neutron-rich isotopes. This instrument was gradually
replaced by the Projectile Fragment Separator FRS to focus on the
neutron-rich side of the line of stability
\cite{Geissel}. Fragmentation was based on smashing a high energy
heavy ion beam on light target material and collecting the fragments
through electromagnetic separator systems for subsequent on-line
analysis. The great advantage of this technique over the traditional
ISOL approach was that even short-lived isotopes could be studied if
properly separated. Fragmentation played an increasingly important
role for $\beta$-decay studies of r-process isotopes both at the
fragment separators at GSI \cite{Kurcewicz}, at NSCL/MSU in the US
\cite{Quinn}, and RIKEN in Japan \cite{Lorusso.Nishimura.ea:2015}. The
measurement of a very neutron-rich, doubly closed shell nucleus,
$^{78}$Ni, at the on-set of the r-process, presented a particularly
impressive example on the new relevance of fragment separators for the
study of r-process nuclei \cite{Hosmer.Schatz.ea:2005}. The
simultaneous measurement of the half-lives of 110 neutron-rich nuclei
near the $N=82$ closed shell at RIKEN \cite{Lorusso.Nishimura.ea:2015}
proved a substantial step forward for studies of r-process nuclei (see
Fig.~\ref{fig:lorusso}). Another example is the systematic study of
$\beta$-decay half-lives and $\beta$-delayed neutron emission
processes using the BELEN $^3$He detector array for 20 heavier
isotopes of Au, Hg, Tl, Pb, and Bi in the neutron-rich mass region
above the neutron shell closure $N=126$~\citep{Caballero2016} to probe
the feeding pattern of the third r-process peak at $A\approx195$ in
astrophysical
studies~\citep{Caballero.Arcones.ea:2014,Eichler.Arcones.ea:2015}.

\subsection{Experimental Achievements in Measuring Nuclear Properties}

In the following section, we want to discuss in more detail the
experimental progress in measuring the different nuclear parameters
that have been achieved over the four decades of studying nuclei far
off the neutron-rich side of stability.

The production of neutron-rich isotopes at ISOL-based systems,
both at reactors as well as at spallation facilities, was mainly
limited by the chemistry and extraction time from the ion
source. Large effort went in the development of suitable target
materials and ion source techniques \cite{Ravn}. The choice of
isotopes for the study of masses, half-lives and other decay
properties was often dictated by the availability of isotope products
rather than by physics priorities \cite{Kratz}.  

However, during the last decade fragmentation techniques improved
enormously. They allowed the measurement of much shorter lived
neutron-rich radioisotopes, since they were not handicapped by
chemical delay processes that were typical for ISOL target
systems. With the right target and projectile combination they were
able to reach far beyond the range accessible by ISOL facilities. Yet
in other cases, such as noble gas and alkali elements, the chemistry
conditions are advantageous for ISOL production techniques yielding
superior beam intensities. Recent measurements of neutron-rich Rb
isotopes~\citep{Lorusso.Nishimura.ea:2015} are still at the the limits
reached at ISOLDE twenty years
earlier~\citep{Kratz:1984,Lhersonneau.Gabelmann.ea:1995,Lhersonneau.Pfeiffer.ea:1995}.

Based on the availability of these complementary isotope production
modes, during the last decades several new technical developments led
to an enormous improvement in the study of r-process masses and decay
properties. These were partly driven by the development of larger
high efficiency detection devices, but also by new techniques using
storage ring technology to determine masses of multiple isotopes at
once, instead of painstakingly extracting and probing one isotope
after the other. Other advances were based on the development of laser
traps designed to trap only a few of the selected and collected
neutron-rich isotopes and determine their masses and decay
characteristics with unprecedented accuracy.  The most significant
developments will be discussed in the following sections.

\subsubsection{The experimental study of nuclear masses}
Mass measurements of selected isotopes near stability were
traditionally performed using mass spectrometers based on magnetic and
electric sector fields for separating single isotopes. Modern
experimental methods of mass measurement of rare isotopes are
generally based on three experimental techniques. Time-of-flight mass
spectrometry (TOF-MS)~\citep{Meisel.George:2013} is based on the
velocity measurement of short-lived isotopes produced in fragment
processes that are analyzed in single-pass spectrometers. The other
techniques are frequency-based spectrometry of isotopes in storage
rings using Schottky pick-up signals of rapidly circulating
particles~\cite{Litvinov2004} and Penning traps capable of making
measurements with even single trapped particles
\cite{Blaum:2006,Blaum2013}.  The TOF and frequency-based methods are
often mentioned as direct mass measurement methods
because unknown masses (in fact, mass-to-charge ratios) are directly
determined by calibration with well-known masses.

Indirect methods usually rely on the measurement of the energy balance
in reaction or decay processes of the isotopes in question. The
unknown mass is calculated from known ones in the reaction or decays,
plus the determined Q values. This classical approach requires a
substantial production of the radioactive isotopes in question to
ensure sufficient statistical reliability of the data.

\subsubsection*{Mass measurements in storage rings}

Spectrometry of masses at storage rings allows the simultaneous
measurements of many nuclei. The ions produced at fragmentation
facilities are then stored in storage rings where the
relative frequencies of ion revolutions or relative revolution times
of the stored ions are related to their relative mass to charge ratios
and velocities \cite{Yan2016}. In order to measure the masses of the
ions in storage rings, the ions are cooled in order to
minimize the velocity spread. The cooling process requires time, and
therefore limits the half-lives that can be measured. This was the
principle of the Schottky Mass Measurement Method
\cite{Radon2000,Litvinov2004}. Another approach, named the Isochronous
Mass 
Spectrometry \cite{Hausmann2000, Hausmann2001, Sun2008} removed the
limitation on half-lives since it does not depend on cooling. This ISM
approach resulted in a reduction of the velocity spread by injection
of the ions into the isochronous ion optical mode of the ring.  That
is, the fast and slow ions of the same species are deliberately placed
in the longer and shorter orbital paths, respectively, of the ring in
order to yield essentially the same revolution frequency and therefore
a reduced velocity spread. Two facilities use these methods for mass
measurements at storage rings, the GSI Helmholtz Center in Germany and
the Institute of Modern Physics in Lanzhou, China.  While these two
methods have been used mainly for neutron-deficient nuclei, the masses
of $^{129,130,131}$Cd have been recently measured at
GSI~\citep{Knoebel.Diwisch.ea:2016}. There are ongoing plans to
implement the same approaches at the Radioactive Ion Beam Factory in
RIKEN and the future FAIR facility at GSI.

\subsubsection*{Mass measurements in traps}

Measurements of masses in traps have yielded the most precise and
accurate mass measurements to date, and present a significant advance
over any other methods, including the storage rings and the
traditional $\beta$-endpoint measurements. There are basically two
types of traps.  Paul traps are based on radio-frequency confinement
of ions and Penning traps use electromagnetic fields to trap ions
(magnetic fields for radial confinement and electrostatic ones for
axial trapping). Coupling of traps to fragmentation or spallation
facilities or coupling to spontaneous fissioning sources has
tremendously extended the reach of high precision and high accuracy
measurements, setting new worldwide standards for studies of this very
fundamental property of the nucleus and its impact on simulations of
the r-process \citep{Blaum:2006,Blaum2013}. There are now numerous
facilities worldwide, including the Canadian Penning Trap at Argonne
National Laboratory; LEBIT at the National Superconducting Cyclotron
Laboratory at MSU in the USA; TITAN at TRIUMF in Canada; JYFLTRAP in
Jyvaskyla, Finland; SHIPTRAP at GSI Darmstadt and MAFFTRAP in Munich,
Germany; ISOLTRAP at CERN; and RIKEN trap at the SLOWRI facility in
Japan. Many of the facilities have implemented or intend to implement
MRTOF devices (multi-reflection time-of-flight spectrographs) to
increase the purity of the ions as well as the range of short-lived
exotic nuclei that can be measured.  ISOLTRAP at CERN was a pioneer in
the field of traps, reaching an uncertainty of a few parts in $10^8$
with a resolving power of up to $10^6$ with nuclear half-lives in the
order of seconds \cite{Eliseev2013}.  Exotic neutron-rich nuclei with
shorter half-lives required much greater resolving power and led to
the introduction of the phase-imaging ion-cyclotron resonance
technique \citep{Eliseev2013}. This new technique is based on
determining the frequency of the ion by the projection of the ion
motion in the trap onto a high resolution position sensitive
micro-channel plate detector.  The method has been shown to increase
the resolving power forty-fold, and at the same time, to tremendously
increase the speed with which measurements can be made. The higher
precision of the measurements has a strong impact on attempts to
distinguish sites of the
r-process~\cite{Mumpower2015,Mumpower.Surman.ea:2016,Mumpower2017}. Recent
highlights include precision mass measurement of neutron-rich
Neodymium and Samarium Isotopes at the CARIBU
facility~\citep{Orford.Vassh.ea:2018}, of neutron-rich rare-earth
isotopes at JYFLTRAP~\citep{Vilen.Kelly.ea:2018}, and neutron-rich
Gallium isotopes at TITAN~\citep{Reiter.Ayet.ea:2020}.

\subsubsection{Beta-Decay Studies}

Beta-decay measurements are critical for the determination of the
half-lives of nuclei along the r-process path and for investigating
the decay patterns that form the final r-process abundance
distribution along the line of stability. Beta-decay measurements are
typically challenged by the detection efficiency of electrons and
neutron-rich ions. This not only requires high production rates at
radioactive beam facilities but also sophisticated detector
arrangements. New pioneering results for half-lives along the
r-process path have been measured at RIKEN using stacking of eight
silicon double-sided strip detectors such as WAS3ABi (Wide- range
Active Silicon-Strip Stopper Array for Beta and ion
detection)~\cite{Lorusso.Nishimura.ea:2015,Wu.Nishimura.ea:2017}
surrounded by an array of 84 High Purity Germanium detectors (HPGe) of
the EURICA array~\cite{Soderstrom2013}. The results are shown in
Fig.~\ref{fig:lorusso}, with a substantially lower experimental
uncertainty than previous results, but also indicating substantial
disagreements with the theoretical half-life predictions (see also
section~\ref{sec:nuclear-modeling-r}).  While silicon has been an
excellent choice as detector material, other materials such as Ge with
higher Z have recently been commissioned at the NSCL fragmentation
facility for use with $\beta$-decay experiments. The GeDSSD array
(Germanium Double-Sided Silicon Detector) shows 50\% electron
efficiency and greater mechanical stability in allowing the
manufacture of thicker detectors \cite{Larson2013}.  TRIUMF in Canada
has also developed the Scintillating Electron-Positron Tagging Array
(SCEPTAR) comprised of 20 thin plastic scintillator beta detectors
that surround the implantation point of radioactive ion beams inside a
central vacuum chamber surrounded by 16 Clover type, large volume
Germanium detectors. SCEPTAR has been shown to have an efficiency of
$\sim$80\% for electrons emitted from radioactive decays, and also
provides information on their directions of emission in order to veto
background in the surrounding GRIFFIN HPGe detectors from the
bremsstrahlung radiation produced by the stopping of the energetic
beta particles.  Neutron-rich nuclei are transported to the center of
GRIFFIN by a moving tape collector (MTC) system. The efficiency of
  this approach has been beautifully demonstrated with the
  measurements of the $\beta$-decay half-lives of ground state and two
  isomeric states in $^{131}$In and the subsequent $\gamma$-decay
  patterns of $^{131}$Sn~\cite{Dunlop.Svensson.ea:2019}. The
  sensitivity of this experiment allowed the first detection of
  $\gamma$-rays following $\beta$-delayed neutron decay for $^{131}$In
  $\rightarrow$ $^{130}$Sn, an important decay branch for many
  r-process nuclei.

\subsubsection{Beta-delayed neutron emission probability measurements}

Beta-delayed neutron emission changes the availability of neutrons and is particularly important since the
delayed neutrons can significantly change the abundances of
neutron-rich nuclei during freeze-out. While the probabilities of
$\beta$-delayed neutron emission ($P_n$) are of great impact for
r-process simulations, as well as nuclear power reactor designs
\citep[for early work see e.g.][]{Kratz.Herrmann:1973,Kratz.Schroeder.Ohm.ea:1982}, the
experimental situation is quite poor since very few of the
$P_n$-values have been measured. Facilities capable of producing
neutron-rich nuclei by fragmentation, spallation, or fission sources
have invested in a variety of approaches to measure theses
$P_n$-values.

A number of neutron detection techniques were applied, but multiple
counter systems, consisting of a number of $^3$He counters embedded in
a paraffin matrix to thermalize the neutrons for better efficiency,
emerged as a standard approach for these kind of studies. One more
recent example was the neutron counter NERO \cite{Nero}, developed at
the NSCL/MSU to measure the $P_n$-values of neutron-rich isotopes in
the lower mass range that was accessible using fragment production and
separation at the A1900 separator. NERO consists of sixty $^3$He
counters embedded in a polyethylene matrix surrounding the collection
station to maximize counting efficiency. The efficiency was tested
using a $^{252}$Cf spontaneous fissioning source and the energy
detection range of the detectors was expanded using ($\alpha, n$)
reactions on various target materials. NERO was utilized primarily for
the study of medium mass nuclei in the Co to Cu region \cite{CoCu} and
in the range of of very neutron-rich Y, Mo, and Zr isotopes
\cite{Pereira}, pushing the experiments to nuclei in the N=82 closed
neutron shell region \cite{Montes}.

The BELEN (BEta-deLayEd Neutron) detector array is another development
of a neutron counter that follows the same concept as NERO\@. BELEN
was conceived as a modular detector that has been developed in
preparation for experiments at FAIR\@. Specifically, the DEcay
SPECtroscopy (DESPEC) experiment at FAIR is planned for the
measurement of $\beta$-decays in an array of Double Sided Silicon
Detectors (DSSD) called AIDA, in coincidence with the $^3$He neutron
detectors of BELEN, in order to measure $P_n$-values of exotic
nuclei. BELEN-20 consists of two concentric rings of $^3$He counters
(8 and 12 counters, respectively), arranged inside a polyethylene
neutron moderator. Early measurements included testing the system by
transporting a beam of ions to the center of the neutron detector in
front of a Si detector to measure the
$\beta$-decay~\citep{Gomez-Hornillos2011}. The most current version of
BELEN includes 48 $^3$He tubes~\citep{Calvino2014,Agramunt2016} and
was recently tested, similarly to BELEN-20, at Jyvaskyla, with fission
products produced from the proton-induced fission of Thorium.  Fission
products were swept away by a Helium gas jet system into a double
Penning trap system that acts as a mass separator, resulting in a
relatively pure beam of $\beta$-decaying products. The transport
system takes on the order of a few hundred milliseconds and imposes a
limitation on the lifetimes that can be studied. The BELEN detector is
soon to become a part of the largest ever neutron detector of its kind
as a part of a 160 $^3$He counter arrangement being built by the
BRIKEN collaboration for measurements of exotic nuclei at RIKEN\@. The
challenges that remain are the trade-offs between the highest
efficiencies and the best energy resolutions for the detection of
neutrons. First measurements at the GSI fragment separator focused on
the study of nuclei around the $N=126$ closed shell in the Au to Rn
range, in order to determine the half-lives of these isotopes and the
$P_n$-values, to compare with theoretical model predictions~\citep{Caballero2016}. This work demonstrated that the FRDM+QRPA
(Finite-Range Droplet Model coupled with the Quasiparticle Random
Phase Approximation) predictions differ sometimes by up to an order of
magnitude from the experimental values. These discrepancies between
theoretical predictions and experimental result underline the
importance of such studies for exploring the evolution of nuclear
structure towards the r-process path and beyond.

Recently, a new technique was demonstrated in which the challenges of
neutron detection are circumvented by measuring the nuclear recoil
\cite{Yee2013}, instead of the neutron energy, using traps. The traps
can confine a radioactive ion and basically $\beta$-decay at rest. The
emitted radiation emerges with minimal scattering, allowing the
measurement of the ion recoil. The $\beta$ is measured in coincidence
with the ion, recoiling due to neutron emission, resulting in a
time of flight spectrum. The proof of principle was demonstrated with a 
$^{252}$Cf fission source, where fission fragments are
thermalized in a large volume gas catcher, extracted, bunched, trapped
and mass separated in a Penning trap, then delivered into a
$\beta$-decay Paul Trap (BPT). The $\beta$-particles are detected in a
$\Delta$E-E plastic scintillator while the recoil ions are detected in
a microchannel plate detector. The technique allows the measurements
of exotic isotopes with half-lives as short as 50 ms while avoiding
some of the complications of neutron measurements \citep{Munson.Siegl.ea:2018,Siegl.Scielzo.ea:2018}.

\begin{figure}[htb]
  \includegraphics[width=\linewidth]{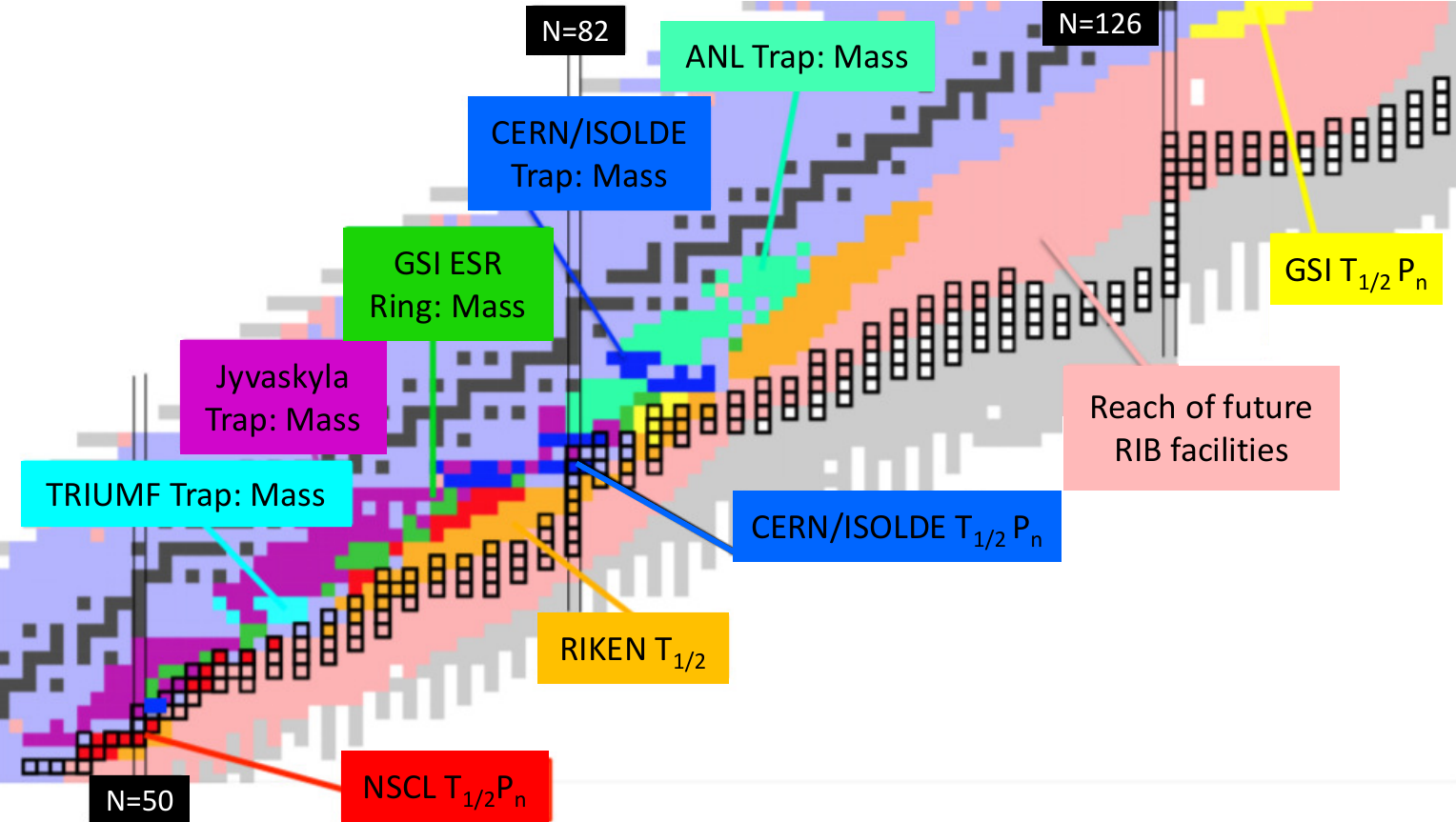}
  \caption{A summary of recent efforts undertaken at experimental
    facilities world-wide \citep[adapted from][]{Horowitz.ea:2019} in
    order to attain precise (a) 
    nuclear masses and (b) $\beta$-decay properties like half-lives
    $T_{1/2}$ and delayed neutron probabilities $P_n$. The individual
    results at TRIUMF, Jyvaskyla, GSI, CERN, ANL, NSCL, and RIKEN
    (discussed earlier in this section) are indicated via
    color-coding. Shown is also the region in the nuclear chart in
    reach of the RIB facilities. \label{fig:ani1}}
\end{figure}

Fig.~\ref{fig:ani1} provides a summary of the recent experimental
efforts and achievements discussed in this and the previous subsection

\subsection{Experiments towards Neutron Capture Rates}

For a long time, the determination of neutron capture rates on
neutron-rich nuclei has been considered of secondary relevance for the
simulation of r-process nucleosynthesis and scenarios. This is due to
the fact that the r-process is governed by an
$(n,\gamma)\rightleftarrows(\gamma,n)$ equilibrium, where the actual
reaction rates cancel out as described earlier. However, after
freeze-out the equilibrium is no longer maintained and neutron capture
reactions on the neutron-rich reaction products may well shift the
abundance distribution towards heavier nuclei. Sensitivity studies
with variations of the neutron capture rates by factors of ten can
result in significant variations in the resulting abundances of the
heavy elements~\citep{Mumpower2015}. However, the experimental
measurements of neutron capture on exotic beams pose significant
challenges, both in the production of the exotic nuclei as well as the
neutrons and in turn in the measurements of the reaction rates.

While the direct measurement of neutron capture reactions on stable
and even long-lived radioactive for the s-process has been very
successful \cite{Guerrero2017}, a similar approach to study neutron
capture on short-lived neutron-rich isotopes provides considerable
challenges. Most of the r-process neutron capture rates rely on
theoretical predictions based on the Hauser-Feshbach statistical model
formalism \cite{Rauscher.Thielemann.Kratz:1997,Goriely:1998}. To test
and verify these predictions a number of indirect methods have been
developed over the last decade. This includes the so-called Oslo
method \cite{Guttormsen1987} as well as the surrogate reaction
technique
\cite{Escher2012,Kozub2012,Manning.Arbanas.ea:2019,Tang.Key.ea:2020},
while new methods are being envisioned towards a direct experimental
approach.

\subsubsection{Neutron Capture on neutron-rich nuclei: $\beta$-Oslo
  method}
\label{sec:neutr-capt-neutr}

The Oslo method involves the extraction of level densities and
$\gamma$-ray strength functions by the measurements of the total
de-excitation of a nucleus as a function of energy. The different
excitation ranges to be studied are populated by different nuclear
reaction modes that can range from light ion transfer reactions to
inelastic scattering techniques. This approach requires high intensity
beams and the direct measurements of cross sections
\cite{Guttormsen1987} to obtain the level density and strength
function data with sufficient statistics for extracting neutron
capture cross sections. The recent adaptation of the Oslo method has
been demonstrated in the $\beta$-Oslo method in which the
$\beta$-decay of a neutron-rich nucleus populates the levels at high
excitation range and the subsequent $\gamma$-decay is measured using
total absorption spectroscopy~\cite{Spyrou2017}. A first version of
this approach was developed on the basis of $\beta$-decay data
obtained at the ILL Grenoble and at ISOLDE at
CERN~\citep{Kratz1983,Leist1985}. A benchmark test for quantifying the
method was the successful comparison between the level density
analysis from the study of $^{87}$Br$(\beta^-n){}^{86}$Kr through
neutron unbound states in $^{87}$Kr and the direct
$^{86}$Kr$(n,\gamma){}^{87}$Kr resonant neutron capture
data~\citep{Raman1983}. An important aspect in this work is the fact
that the extracted level density is based on the analysis of the
neutron decay data, selecting configurations prone to neutron capture
(and not solely on the $\gamma$-decay analysis), which contain all
possible excitation modes.

The present $\beta$-Oslo method, however,
rests mostly on the analysis of $\gamma$-decay of highly excited
states. Neutron unbound states, populated by the $\beta$-decay are
less likely to be observed because they primarily decay into the
particle rather than the $\gamma$ channel as observed in the early
studies \cite{Raman1983}.  Nevertheless, the study of the
$\beta$-delayed $\gamma$-decay is a useful tool for determining level
densities up to the threshold. The new approach relies on the use of a
4$\pi$ summing detector device instead of a single Ge detector to
analyze the $\gamma$-decay pattern. The spectra are then unfolded as a
function of excitation energy in order to determine the nuclear level
density and the $\gamma$-strength function. The neutron capture cross
section is derived by folding the level density and $\gamma$-ray
strength function with a nucleon-nucleus optical model potential,
adopting statistical assumptions for the neutron transmission
channels. The analysis depends critically on a number of assumptions
with respect to level density normalization and the optical potential,
which possibly introduces systematic uncertainties. However, the
largest uncertainty is in the assumption of the density of neutron
unbound states above the threshold and the associated neutron strength
distribution. This is typically determined from systematics and
statistical model simulations. It works well near the stability where
the level density above the neutron threshold is high.  It becomes
more questionable when the method is applied to nuclei at the
r-process path, where the neutron thresholds and therefore the level
density are much lower.  A number of measurements have been performed
and the extracted results agree well with the predictions of
Hauser-Feshbach simulations~\citep{Spyrou2014} and the uncertainty
range in the prediction is claimed to be significantly
reduced~\citep{Liddick2016}. 

The approach suggests a certain
redundancy since the experimental data do not consider the neutron
strength function above the threshold, but adopt the one predicted by
the same statistical model against which the predicted reaction rates
are being tested. A study of the systematic uncertainties
by~\citet{Spyrou2017} suggests that the overall uncertainty in the rates
obtained by the $\beta$-Oslo method is within a factor of $\sim 3$
which is comparable to the uncertainty range of case-optimized Hauser
Feshbach calculations \citep{Beard2014}.

\subsubsection{Neutron capture by $(d,p)$ surrogate reactions}

Single particle transfer reactions such as $(d,p)$ have emerged as a
powerful tool for probing the single particle structure of
neutron-rich nuclei near the r-process path. First $(d,p)$ transfer
measurements, using radioactive $^{130,132}$Sn beams on CD2
(deuterated polyethylene) targets at the Holifield Radioactive Ion
Beam Facility (HRIBF) of the Oak Ridge National Laboratory, led to a
better understanding of the single particle structure of bound states
in $^{131,133}$Sn \cite{Kozub2012,%
  nature09048}. The extracted single particle spectroscopic factors
allowed calculating the direct reaction components for neutron capture
reactions. Higher energy unbound states were not observed. The
observation of such states is critical for extracting reliably the
single resonant or statistical resonant contributions expected for
high level density compound nuclei in $(n,\gamma)$ reactions. More
recently, a similar study has been performed at ISOLDE aiming at the
determination of the neutron shell structure below lead and beyond
$N=126$ by probing the neutron excitations in $^{207}$Hg in the
reaction $^{206}$Hg$(d,p)^{207}$Hg in inverse
kinematics~\citep{Tang.Key.ea:2020}.

The study of the unbound regions of neutron-rich compound nuclei in
$(n,\gamma)$ reactions near the r-process path is the primary goal of
the surrogate reaction approach where single particle transfer
reactions are utilized to bypass the challenges of measuring neutron
capture cross sections on short lived nuclei. Neutron transfer
reactions such as $(d,p)$ or $(d,p\gamma)$ are frequently highlighted as
surrogates for direct neutron capture studies \cite{Escher2012}. In
surrogate reactions the neutron is carried within a ``Trojan''
projectile and brought to react with the target. The neutron-capture
cross sections on the target nucleus can be extracted by measuring the
proton in the final stage \cite{Escher2006,Forssen2007}. First
benchmark experiments have been performed at the 88-inch cyclotron at
Lawrence Berkeley National Laboratory probing the
$^{171,173}$Yb$(n,\gamma)$ cross section via the surrogate reaction
$^{171,173}$Yb$(d,p\gamma)$, using a high intensity deuterium beam
\cite{Hatarik2010}. The extracted neutron capture cross sections
agreed within 15\% with direct measurements \cite{Wisshak2000} at
energies above 90 keV, while at lower energies 
considerably larger discrepancies are observed.

In the case of neutron capture on short-lived nuclei, inverse
kinematics techniques will be necessary with short-lived radioactive
beams interacting with a deuterium target. Neutron-transfer
measurement on a radioactive r-process nucleus needs large area
silicon detector arrays at backward angles in coincidence with an
ionization counter at forward angles to detect the beam-like recoils
to reduce the beam induced background. Such a system was developed as
the Oak Ridge-Rutgers University Barrel Array (ORRUBA)
\cite{Pain2007}. The ORRUBA detector has been used in the center of
the Gammasphere Ge-array in a combination called Gammasphere Orruba
Dual DEtectors for $(d,p\gamma)$ studies using stable $^{95}$Mo beams
\cite{Cizewski2018} but no conclusive results have been
presented. Extracting the neutron capture cross section out of the
surrogate reaction measurements offers its own challenges since it
requires proper treatment of nuclear model parameters. Deviations
between the results of direct measurements and surrogate reaction
studies may reflect insufficient treatment and separation between
different reaction mechanisms, such as direct transfer and break-up
components \cite{Avrigeanu2016}. While promising as a method, a deeper
understanding of the reaction mechanism seems necessary
\cite{Potel2015}. This is also indicated in a recent paper on neutron
capture reactions on neutron-rich Sn nuclei using the surrogate
method~\cite{Manning.Arbanas.ea:2019}. The goal was not to
determine the resonant contributions, but the direct capture
components to low excited states. The spectroscopy method indicated a
few single states rather than a broad level distribution. The results
deviate with theoretical predictions at a fixed neutron energy of
30~keV, but a broader analysis, including resonances over a wider
range of methods, would be necessary for a detailed evaluation.

\subsubsection{Neutron-capture in ring experiments}
Recently a new method has been proposed for the direct study of
neutron capture on short-lived nuclei, using high intensity
radioactive beams in a storage ring on a thermalized neutron target
gas produced on-line by proton-induced spallation
reactions. \cite{Reifarth2017}. The cross section of the neutron
capture reactions would be measured in inverse kinematics, detecting
the heavy ion recoils in the ring, using for example the Schottky
method developed at the GSI storage ring facilities
\cite{Nolden2011}. This concept is an expansion of earlier work that
proposed the use of the high neutron flux in a reactor core as
possible target environment with the radioactive beam passing through
the reactor core in a storage ring \cite{Reifarth2014}. A number of
simulations demonstrate that both methods seems feasible albeit
technically challenging, since it requires the combination of a
storage ring facility with either a spallation or fission neutron
source. There is a half-life limit that is mostly determined by the
production rate at the radioactive ion facility or the beam intensity
and the beam losses due to interactions with the rest gas in the
ring. Yet, such a facility would allow for the first time to address
the challenges of neutron capture reaction measurements on neutron
rich radioactive isotopes with half-lives less than a minute in the
decay products of r-process neutron-rich nuclei.



\section{Nuclear modeling of r-process input}
\label{sec:nuclear-modeling-r}

\subsection{Nuclear masses}
\label{sec:nuclear-masses}

\begin{figure}[htbp]
  \centering
  \includegraphics[width=\linewidth]{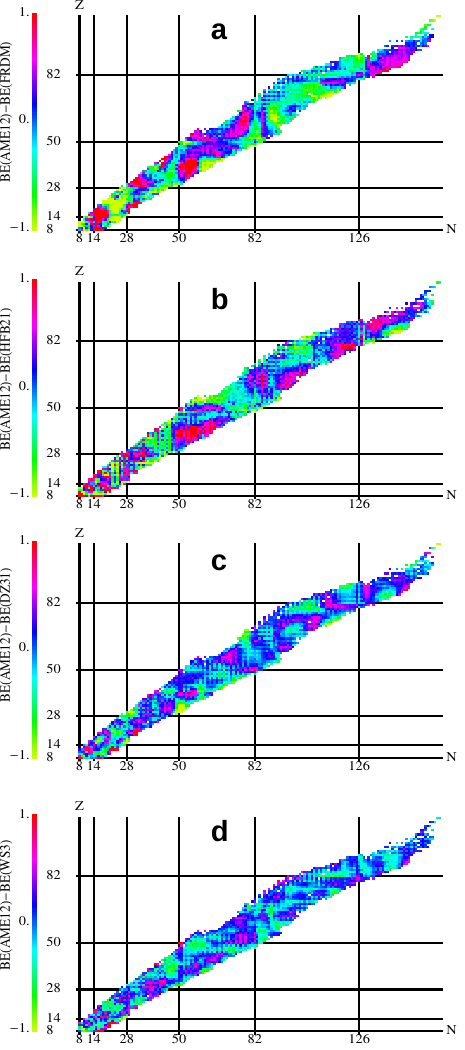}
  \caption{Differences, in MeV, between experimental, taken from the
    2012 version of the atomic mass evaluation
    AME12~\citep{Wang.Audi.ea:2012}, and theoretical binding
    energies. The following mass models are shown:
    \citep[FRDM-1992][a]{moeller95}, \citep[HFB-21][b]{Goriely.Chamel.Pearson:2010}, \citep[DZ31][c]{Duflo.Zuker:1995},
    \citep[WS3][d]{Liu.Wang.ea:2011}. \citep[Figure
    from][]{Mendoza-Temis:2014}.\label{fig:mass-deviations}}
\end{figure}

The most basic nuclear property for any r-process calculation is the
mass of the nuclei involved. It determines the threshold energy for
the main reactions during the r-process: $\beta$~decay, neutron capture
and photo-dissociation. Neutron separation energies, $S_n$, are
particularly important if the r-process proceeds in
$(n,\gamma) \rightleftarrows (\gamma,n)$ equilibrium, as the reaction
path is then fixed at a constant value of $S_n$, for given values of
neutron density and temperature of the astrophysical environment. The
most commonly used mass tabulations can be grouped in three different
approaches: a) microscopic-macroscopic models like the finite-range
droplet model (FRDM) approach
\cite{moeller95,moeller12,Moeller.Sierk.ea:2012,Moeller.Sierk.ea:2015,moeller15}, the
Extended Thomas-Fermi model with Strutinski Integral (ETFSI) approach
\cite{Aboussir.Pearson.ea:1995}, the extended Bethe-Weizs\"acker
formula \cite{Kirson:2008} and the Weizs\"acker-Skyrme mass models
\cite{Wang.Liu.Wu:2010,Liu.Wang.ea:2011}; b) a microscopically
inspired parametrization based on the averaged mean field extracted
from the shell model and extended by Coulomb, pairing and symmetry
energies \cite{Duflo.Zuker:1995}; and c) microscopic models based on
the non-relativistic \cite{Goriely.Chamel.Pearson:2016} or
relativistic \cite{Sun.Meng:2008} mean-field models.

\begin{table}[htb]
  \begin{ruledtabular}
    \caption{Comparison of the root mean square deviation, in keV,
      between mass models and experiment; mass models:
      FRDM-1992~\citep{moeller95},
      HFB-21~\citep{Goriely.Chamel.Pearson:2010}, DZ10,
      DZ31~\citep{Duflo.Zuker:1995}, and WS3 \citep{Liu.Wang.ea:2011},
      experimental values taken from the
      2003~\citep{Audi.Wapstra.Thibault:2003} and 2012
      evaluations~\citep{Wang.Audi.ea:2012}. The columns labeled
      ``full'' consider all masses present in each evaluation while
      the column labeled ``new'' includes only masses found in
      AME-2012 but not in AME-2003. \label{tab:mass-deviations}}
    \begin{tabular}{lccc}
      Model & AME-2003 & AME-2012 & AME-2012 \\
            &  (full)  & (new) &  (full) \\ \hline
      FRDM-1992 &  655 &  765 &  666 \\
      HFB-21 & 576 &  646 &  584 \\
      WS3  &  336 &  424 &  345 \\
      DZ10 &  551 &  880 &  588 \\
      DZ31 &  363 &  665 &  400 \\
    \end{tabular}
  \end{ruledtabular}
\end{table}

All mass models have in common that, by fitting a certain set of
parameters to known experimental data, they are then being used to
predict the properties of all nuclei in the nuclear landscape.  The
models reproduce the experimentally known masses quite well, with mean
deviations between 350 keV and 600 keV (see
Table~\ref{tab:mass-deviations}).  It is quite satisfying to see that,
when in 2012 a new atomic mass evaluation (AME)
\cite{Wang.Audi.ea:2012}, including 219 new experimental masses,
became available the agreement with data worsened only slightly
compared to the comparison with the previous AME\@. However, when
considering only the new experimental masses found in AME-2012 the
agreement deteriorates. As the new masses typically involve more
exotic nuclei than those found in a previous evaluation, they provide
a measure of the capabilities of each model to extrapolate to regions
far from stability. This is in general one of the most challenging
aspects to determine when using a given mass model in r-process
calculations. \citet{Neufcourt.Cao.ea:2018} has recently applied
Bayesian machine-learning techniques to assess the predictive power of
global mass models towards more unstable neutron-rich nuclei and
provide uncertainty quantification of predictions. Nevertheless,
deviations between model and data for neutron-rich nuclei are
typically related to bulk properties that may not dramatically affect
the abundance predictions, e.g.\ the symmetry energy whose value is
known with an uncertainty 3.8~MeV to be in the range
29.7--33.5~MeV~\citep{Hebeler.Lattimer.ea:2013}.

Fig.~\ref{fig:mass-deviations} provides a closer comparison between
models and data. One notices systematic deviations, e.g.\ for neutron
numbers around $N \sim 90$ and 130 just above the neutron shell
closures at $N=82$ and 126 (Fig.~\ref{fig:mass-deviations}).  These
mass regions are known as `transitional regions' where nuclear shapes
change from spherical to deformed configurations, accompanied by a
sudden drop in neutron separation energies. The description of these
shape changes is very sensitive to correlations which are not fully
accounted for in the current mass models.  Noticeable differences
among the various mass models, and the data, are also observed in the
differences of neutron separation energies for odd-A and odd-odd
nuclei~\cite{Arzhanov:2017}, likely pointing to the need for an
improved description of neutron-proton correlations.  A better
description, in particular of the transitional region, requires
beyond-mean-field techniques. A first attempt has been presented
in~\citet{Rodriguez.Arzhanov.Martinez-Pinedo:2015}, based on the
Generator Coordinator Method which considers superpositions of
different shapes and restores the breaking of particle number and
angular momentum as inherent in the Hartree-Fock-Bogoliubov (HFB)
approach. However, first calculations of nuclear masses show only
rather slight effects for nuclei in the $N \sim 90$ range.

\begin{figure}[htbp]
  \centering
  \includegraphics[width=\linewidth]{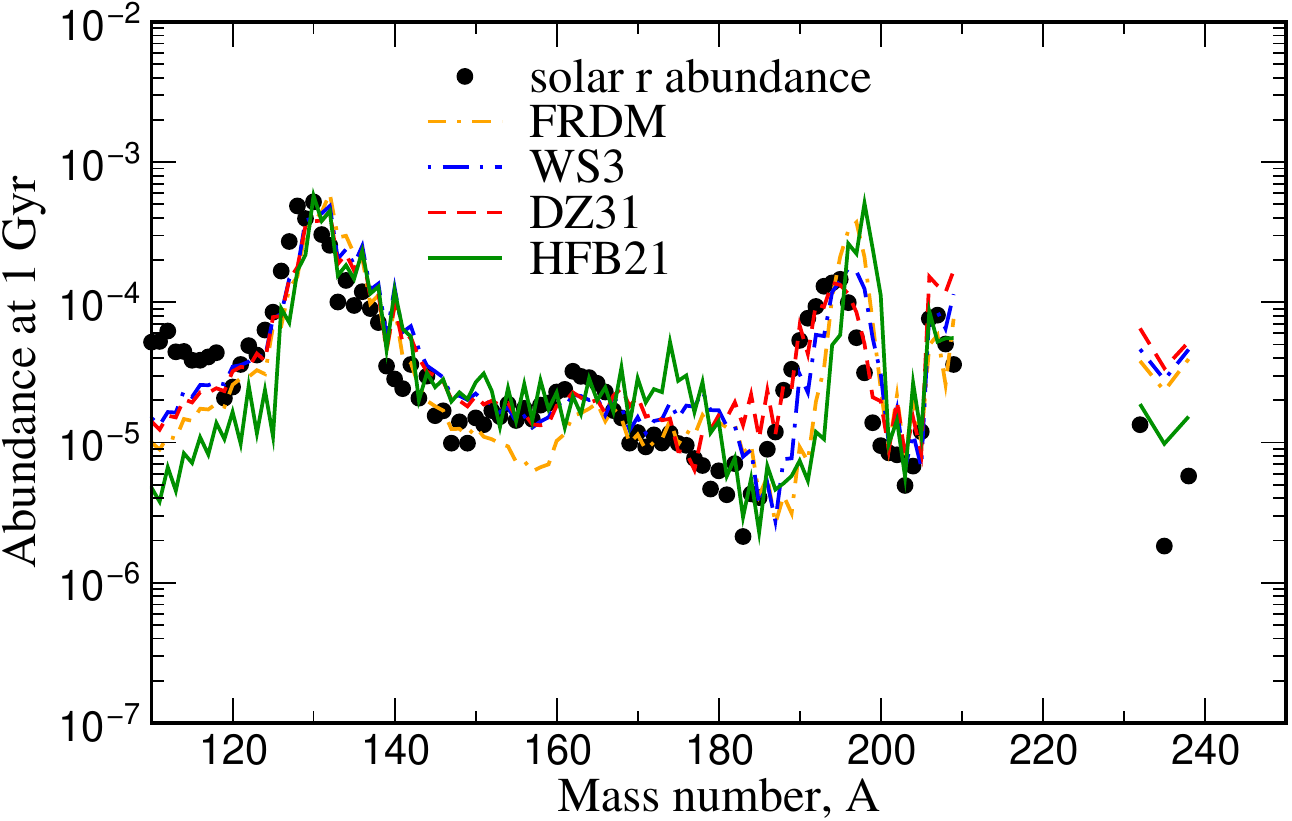}
  \caption{Final mass-integrated r-process abundances obtained in a
    neutron-star merger simulation using four different mass models
    \citep[adopted
    from][]{Mendoza-Temis.Wu.ea:2015}\label{fig:Joel-final}}
\end{figure}

Although the differences between the various mass models might be
rather minute in the transitional regions at $N\sim90$ and 130, they
can have noticeable impact in r-process simulations.  The FRDM
\citep{moeller95} and version 21 of the Brussels
Hartree-Fock-Bogoliubov mass model
\citep[HFB21][]{Goriely.Chamel.Pearson:2010} predict noticeably
smaller neutron separation energies than the
Duflo-Zuker~\citep{Duflo.Zuker:1995} or the Weizs\"acker-Skyrme
\citep[WS3][]{Wang.Liu.Wu:2010,Liu.Wang.ea:2011} models in the
$N \sim 130$ mass range. As a consequence, for the former two mass
tabulations, these nuclei act as obstacles in the r-process mass flow
and produce a third r-process abundance peak which is narrower in
width, overestimated in height, slightly shifted to larger mass
numbers and followed by an abundance trough just above the peak if
compared to simulations using the Duflo-Zuker and WS3 masses and to
observational data (see Fig.~\ref{fig:Joel-final}). At $N \sim 90$ the
FRDM predicts very low neutron separation energies, in contrast to the
other mass models~\cite{Arcones.Martinez-Pinedo:2011}. As is discussed
in \citet{Mendoza-Temis.Wu.ea:2015} these low $S_n$ values have
consequences for the matter flow between the second and third
r-process peaks and result in a narrow peak around $A\sim136$ in the
r-process abundances at freeze-out, which is, however, washed out at
later times due to continuous productions of material in this region
by fission. Similar effects have also been observed
in~\citet{Martin.Arcones.ea:2016} using masses derived from Skyrme
energy density functionals based on different optimization
protocols. This allows for systematic studies of uncertainty bands
under the same underlying physical model for the description of
nuclear masses.

\subsection{Beta-decay half-lives}

Nuclear beta-decays, which change a neutron into a proton, are
responsible for the mass flow to elements with increasingly heavier
$Z$-number.  As the r-process occurs in a dynamical environment, the
time, needed for the succession of beta-decays to produce
thorium and uranium from the seed nuclei available after freeze-out of
charged-particle fusion reactions, is competing with the dynamical
timescale of the explosion, during which matter is transported to larger radii
and lower densities. The latter suppresses the neutron number density
required for the mass flow to heavier nuclei by neutron
captures. Particularly important are beta-decays of nuclei with magic
neutron numbers $N_{\rm mag}$, as the matter flow is hindered by the
reduced neutron separation energies of the nuclei with
$N_{\rm mag} +1$.  Furthermore, due to the extra binding of the magic
nuclei, the $Q$-value of their beta-decays is relatively reduced,
resulting in longer lifetimes.

Calculations of beta-decays require two ingredients: the relative
energy scale between parent and daughter nuclei (Q value) and the
transition strength distribution in the daughter nucleus. We note that
the $Q$ values are large for r-process nuclei due to the extreme
neutron excess.  As a consequence uncertainties in this quantity
(usually of order 0.5--1~MeV) have a mild effect on the half lives,
despite the strong energy dependence of the involved phase space
($E^5$ for allowed Gamow-Teller transition, and even higher powers for
forbidden transitions). However, this strong energy dependence makes
the half-life sensitive to the detailed low-lying strength
distribution, which is also crucial to determine whether the beta-decay is
accompanied by the emission of neutrons, i.e.\ whether the
transition proceeds to states in the daughter nucleus above or below
the neutron threshold (which is only 2--3 MeV in r-process
nuclei). This so-called $\beta$-delayed neutron emission is a source
of free neutrons and plays an important role in determining the final
r-process abundances during the freeze-out of neutron captures
\cite{Arcones.Martinez-Pinedo:2011}.

\begin{figure}[htbp]
  \centering
  \includegraphics[width=\linewidth]{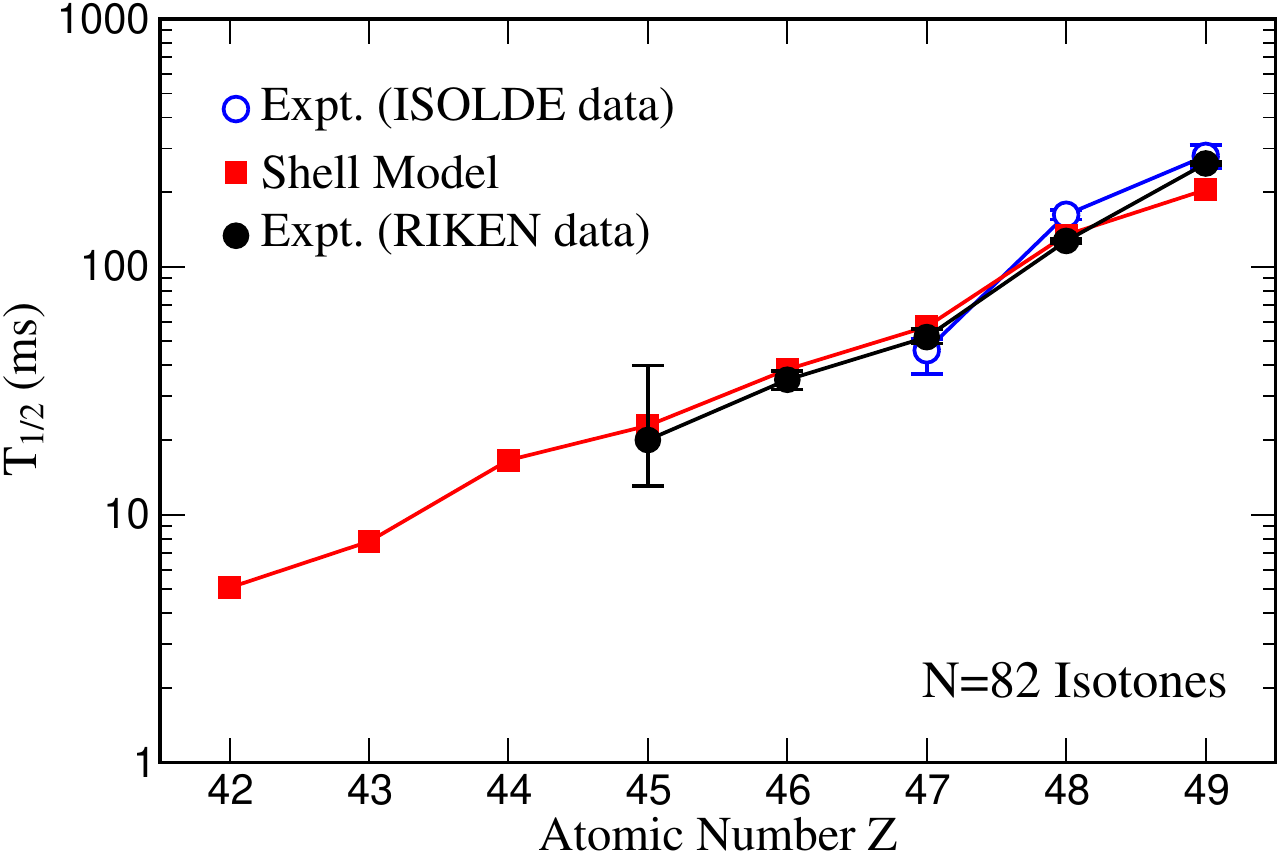}
  \caption{\label{fig:n82halflives}Comparison of
    experimental~\citep{Pfeiffer.Kratz.ea:2001,Dillmann2003,Fogelberg.Gausemel.ea:2004,Lorusso.Nishimura.ea:2015}
    and shell-model half-lives~\citep{Zhi.Caurier.ea:2013} for $N=82$
    r-process nuclei}
\end{figure}

Nucleon-nucleon correlations are responsible for the strong
fragmentation of the transition strengths and for its suppression
compared to the Independent Particle Model. These correlations are
accounted for in the interacting shell model
\cite{Caurier.Martinez-Pinedo.ea:2005}, and in fact large-scale shell
model calculations have been proven as the appropriate tool to
describe nuclear Gamow-Teller
distributions~\cite{Caurier.Langanke.ea:1999,Cole.Anderson.ea:2012}
for stellar weak interaction
processes~\cite{Langanke.Martinez-Pinedo:2000,
  Langanke.Martinez-Pinedo:2003}. Shell-model calculations proved also
very valuable for the calculation of the half-lives of r-process
key-nuclei with magic neutron numbers. For $^{78}$Ni the shell model
predicted a half-life of 127~ms~\citep{Langanke.Martinez-Pinedo:2003},
which was significantly shorter than the value estimated by global
models at the time, and was subsequently experimentally verified
\citep[$110\pm40$ ms,][]{Hosmer.Schatz.ea:2005}.  As is shown in
Fig.~\ref{fig:n82halflives}, the half-lives for the $N=82$ r-process
nuclei recently measured at RIKEN \cite{Lorusso.Nishimura.ea:2015}
agree very well with the earlier shell model values
\cite{Zhi.Caurier.ea:2013}, once the quenching of GT transitions is
adjusted to the new $^{130}$Cd half-life. The shell model calculations
imply that the half-lives for the $N=50$ and 82 r-process nuclei are
dominated by Gamow-Teller transitions and forbidden strengths
contribute only on the few-percent level. This is different for the
$N=126$ r-process waiting points. Here two independent large-scale
shell model
calculations~\cite{Suzuki.Yoshida.ea:2012,Zhi.Caurier.ea:2013} give
evidence that, due to the presence of intruder states with different
parity, forbidden transitions contribute significantly and make the
half lives about a factor of 2 shorter than estimated for pure allowed
transitions. In turn, the shorter half-lives allow for a faster mass
flow through the $N=126$ waiting points. We note that the relevant
forbidden transitions are at low excitation energies, where due to
their enhanced phase space energy dependence they can compete with
allowed transitions, and hence they have a strong impact on the
$\beta$-delayed neutron emission probability.

The shell model is the method-of-choice for $\beta$-decay
calculations. However, due to the model spaces involved, calculations
are only possible for r-process nuclei near closed neutron
shells. Thus, the global beta-decay rates for r-process simulations
have to be modelled by less sophisticated many-body
models. Traditionally these studies were performed by calculation of
the Gamow-Teller strength distributions within the Quasiparticle
Random Phase Approximation on the basis of the Finite Range Droplet
Model \cite{Moeller.Nix.Kratz:1997} or the ETFSI approach
\citep[Extended Thomas Fermi Model with Strutinsky
Integral,][]{Borzov.Goriely:2000}. Experimental data for half-lives of
r-process nuclei around $N=50$ and 82
\cite{Pfeiffer.Kratz.ea:2001,Lorusso.Nishimura.ea:2015} showed that
these estimates were systematically too long. The FRDM+QRPA model was
subsequently extended to include forbidden transitions within the
phenomenological ``gross theory'' \cite{Moeller.Pfeiffer.Kratz:2003}.
A new promising road towards globally calculating half-lives for
r-process nuclei has recently been developed by performing QRPA
studies on top of the self-consistent Hartree-Fock-Bogoliubov
\citep[HFB+QRPA,][]{Engel.Bender.ea:1999} method or density
functionals, either non-relativistic~\citep{Borzov:2003} or
relativistic \citep{Marketin.Vretenar.Ring:2007}.  Recent covariant
density functional
theory~\citep[D3C$^{*}$+QRPA,][]{Marketin.Huther.Martinez-Pinedo:2016}
and Skyrme
finite-amplitude~\cite{Mustonen.Engel:2016,Shafer.Engel.ea:2016,Ney.Engel.Schunck:2020}
studies, which accounted for allowed and forbidden transitions,
yielded noticeably shorter half-lives for medium and heavy nuclei than
obtained by the FRDM+QRPA approach.

Shorter half-lives for r-process nuclei with $Z>80$ have a strong
impact on the position of the third r-process
peak~\cite{Eichler.Arcones.ea:2015} and enhance the mass flow through
the $N=126$ waiting points \cite{Mendoza-Temis.Wu.ea:2015}. The latter
implies more material available for fission, thus affecting the
abundances of the second r-process peak, and the late-time $\alpha$
decays from the decaying r-process matter in a neutron star merger
event~\citep{Wu.Barnes.ea:2019}. Studies of the influence on beta-decays on 
the r-process abundances for different astrophysical sites
have been reported in~\citep{Mumpower.Surman.ea:2016,Shafer.Engel.ea:2016,Kajino.Mathews:2017}.

In principle, the transformation of neutrons into protons can also be
achieved by charged-current ($\nu_e,e^-$) reactions. In fact, there
have been various suggestions how neutrino-induced reactions on nuclei
might affect r-process
nucleosynthesis~\citep[e.g.][]{Qian.Haxton.ea:1997,Haxton.Langanke.ea:1997,Meyer.Mclaughlin.Fuller:1998,Otsuki.Tagoshi.ea:2000,Terasawa.Langanke.ea:2004}. All
these studies were based on the assumption that the r-process operates
in the neutrino-driven wind scenario in the presence of strong
neutrino fluxes. These assumptions are not supported by modern
supernova simulations. In the neutron star merger scenario neutrino
fluxes once the r-process operates are too low to substantially
influence the abundances by charged-current
reactions~\citep{roberts17}.  However, the initial proton-to-neutron
ratio of the matter ejected in neutron star mergers and its spatial
and time dependence is set by weak reactions on free nucleons~(see
section~\ref{sec:astr-sites-their}) 

\subsection{Neutron captures}

During the r-process phase, in which the temperature is large enough
($T \gtrsim 1$~GK), neutron captures and their inverse reactions,
photo-dissociations, are in equilibrium. The rates become, however,
relevant once the nucleosynthesis process drops out of this
equilibrium.  During this period of decreasing temperatures, it is
mainly neutron capture that matters.

The neutron capture and photo-dissociation rates for r-process nuclei
(the latter can be derived by detailed balance from the former) are
traditionally determined within the statistical model.  This assumes a
sufficiently high density of states in the daughter nucleus at the
relevant capture energies just above the neutron threshold, which is
not given for the most neutron-rich nuclei close to the neutron
dripline. A systematic estimate about the range of nuclei for which
the statistical model is applicable to calculate neutron capture rates
is given in \cite{Rauscher.Thielemann.Kratz:1997}. It has been
proposed that for the most neutron-rich nuclei the capture rates
should be calculated within a direct-capture approach based on a
potential \cite{Mathews1983,Rauscher.Bieber.ea:1998,Otsuki.Typel.Martinez-Pinedo:2010,Xu.Goriely:2012,Xu.Goriely.ea:2014}. In such an approach the rate is
often determined by a single resonance in the Gamow window~\citep{Loens.Langanke.ea:2012}. This makes rate predictions quite
uncertain, as nuclear models are not capable to predict the resonance
energies with sufficient accuracy. It has therefore been suggested to
describe the final states by a level density rather than by discrete
levels \cite{Goriely:1997,Ejnisman.Goldman.ea:1998}.  Calculations of
neutron capture rates, which include a statistical component and a
direct contribution, are reported in \citet{Mocelj.Rauscher.ea:2007}.

The main ingredients of statistical model calculations within the
Hauser-Feshbach approach are the nuclear level density, the
$\gamma$-strength function for the decay of the compound state, and
various light-particle potentials. The $\gamma$ transition can occur
with different multipolarities, requiring either different ($E1$) or
equal parities ($M1$, $E2$) between the involved states. To also
fulfil angular momentum selection rules, requires the knowledge of
parity- and angular-momentum-dependent level densities.

There has been significant progress in modelling nuclear level
densities in recent years. With the Shell Model Monte Carlo approach
(SMMC)~\citep{Johnson.Koonin.ea:1992,Koonin.Dean.Langanke:1997} a tool
became available which allowed to determine level densities in
unprecedentedly large model spaces. The method to derive level densities
within the SMMC was presented in \citet{Ormand:1997},
\citet{Nakada.Alhassid:1997}, and \citet{Langanke:1998} and then
systematically extended to explore the parity-dependence
\cite{Alhassid.Liu.Nakada:1999} and angular-momentum-dependence
\cite{Alhassid.Liu.Nakada:2007}. In \citet{Ozen.Alhassid.Nakada:2015}
the collective vibrational and rotational enhancement factors have
been explored, finding that the decay of these enhancement factors is
correlated with the pairing and shape phase transitions. The vanishing
of pairing and its effect on the level density has been studied
in~\citet{Langanke:2006}. In the Bethe Fermi Gas (BFG) level density
formula this vanishing has been described by a temperature-dependent
pairing
parameter~\citep{Grossjean.Feldmeier:1985,Mustafa.Blann.ea:1992,Junghans.deJong.ea:1998}
for which \citet{Langanke:2006} gives a parametrization on the basis
of the SMMC calculations.  SMMC calculations have been performed for
many mid-mass and heavy nuclei.  These include even-even, odd-A and
odd-odd nuclei, allowing to microscopically test the standard
prescription in the BFG level density to describe the systematic
differences in these nuclei due to the pairing effect by a pairing
shift
parameter~\citep[e.g.][]{Cowan.Thielemann.Truran:1991,Rauscher.Thielemann:2000}.

These calculations have initiated and guided
attempts to extend a microscopically derived parity dependence into
phenomenological level density formulae like the BFG approach. This is
achieved by deriving the excitation-energy dependent parity ratio in
the level density by the assumption of Poisson distributed independent
quasi-particles combined occupation numbers obtained from the BCS
(Bardeen-Cooper-Schrieffer) model, in this way including pairing
\citep{Alhassid.Bertsch.ea:2000}.  In \citet{Mocelj.Rauscher.ea:2007}
this approach has been applied to the large set of r-process nuclei
(incorporating also a temperature-dependent pairing parameter
suggested from SMMC studies) and its effects on astrophysically
relevant reaction rates was studied in \citet{Loens.Langanke.ea:2008}.
This improved level density description is part of the statistical
model packages NON-SMOKER and SMARAGD developed by Rauscher
\cite{Rauscher.Thielemann:2001, Rauscher:2011}.

A different path to derive parity- and angular-momentum-dependent
level densities has been followed by Goriely and co-workers, based on
a combinatorial approach within HFB calculations.  Also this approach
has been incorporated into a statistical model package and applied to
the calculation of neutron capture rates for r-process nuclei within
the Brussels Nuclear Library for Astrophysics Applications (BRUSLIB)\footnote{\url{http://www.astro.ulb.ac.be/bruslib}}
data compilation
\cite{Koning.Hilaire.Goriely:2008,Goriely.Hilaire.Koning:2008,Hilaire.Goriely.ea:2010,Goriely.Hilaire.Girod:2012}.

Traditionally the different $\gamma$-strength functions have been
described by global parametrizations
\cite{Cowan.Thielemann.Truran:1991} which were adjusted to
photo-dissociation for $E1$ transitions or electron scattering data for
$M1$ transitions \citep[e.g.][] {Cowan.Thielemann.Truran:1991}.
Recently $E1$ strength functions became available which were
microscopically calculated for individual nuclei within the framework
of the HFB model \cite{Goriely.Khan:2002,Goriely.Khan.Samyn:2004} or
based on the relativistic mean field model
\cite{Litvinova.Loens.ea:2009}.  These calculations support the
presence of enhanced dipole strength at energies just above the
neutron threshold \citep[see also][]{Rauscher:2008}.  Experimentally such enhanced strength is observed
as `pygmy dipole strength' in nuclei with large neutron excess, like
those involved in r-process nucleosynthesis
\cite{Adrich.Klimkiewicz.ea:2005}.  As shown in \citet{Goriely:1998}
this enhanced dipole strength can have significant impact on neutron
capture cross sections.

Dipole $\gamma$-strength functions determined from particle-$\gamma$
coincidence data in neutron pick-up and inelastic scattering data for
several mid-mass nuclei exhibit a remarkable upbend of the strength
towards $E_\gamma =0$
\citep[e.g.][]{Guttormsen.Chankova.ea:2005,Larsen.Chankova.ea:2006,Larsen.Guttormsen.ea:2007}. The
data also allow for the derivation of the nuclear level density,
making a few assumptions \citep[the Oslo
method,][]{Schiller.Bergholt.ea:2000} (see
section~\ref{sec:neutr-capt-neutr}). The impact of this upbend on
neutron capture rates for r-process nuclei has been studied 
\cite{Larsen.Goriely:2010,Larsen.Goriely.ea:2015,Larsen.Spyrou.ea:2019} and a potential
increase of the capture rate by up to two orders of magnitude has been
calculated.  The origin of the low-energy upbend has yet not been
completely identified.  Coherent adding of magnetic moments of
high-$j$ orbitals has been suggested as a possible mechanism for
low-energy $M1$ enhancement
\cite{Schwengner.Frauendorf.Larsen:2013,Brown.Larsen:2014,Schwengner.Frauendorf.Brown:2017},
while a low-energy upbend in the $E1$ strength was obtained within
finite-temperature relativistic QRPA calculations~\cite{Litvinova.Belov:2013}. An upbend in the $M1$ strength function has
also been found in large-scale shell model calculations for selected
$pf$-shell nuclei \cite{Sieja:2017} and
$A\gtrsim 100$~\cite{Sieja:2018}. \citet{Goriely.Hilaire.ea:2018} have
performed large scale calculations of $E1$ and $M1$ strength functions
by a combination of shell-model and Gogny-HFB+QRPA calculations. 

Several general questions regarding basic assumptions made in
statistical model evaluations of capture rates have been addressed in
large-scale shell model calculations of the $M1$ strength functions for
several mid-mass nuclei \citep[similar studies for $E1$ transitions are
yet prohibited by computing limitations as they require the inclusion
of two major shells,][]{Loens.Langanke.ea:2012}. The results are
briefly summarized as: a) The shell-model $M1$ strength functions turned
out to give smaller cross sections than the usually adopted
parametrizations; b) the scissors mode, a fundamental orbital $M1$
excitation observed in deformed nuclei at low energies
\cite{Bohle.Richter.ea:1984}, might lead to a noticeable enhancement
of the capture rates; c) the assumption of the Brink hypothesis, i.e.\
the strength function is the same for all nuclear states
\cite{Brink:1955,Brink:1957} is only valid with moderate accuracy
d) the cross section calculated microscopically by a state-by-state
approach had the largest contribution from a single state with $M1$
excitations just in the Gamow window.  Such a nuclear structure effect
cannot be caught by any global parametrization.  The potential impact
of the $M1$ scissors mode on r-process neutron capture cross sections
has subsequently been revisited by \citet{Mumpower.Kawano.ea:2017}.

The transmission coefficients required in statistical model
calculations of astrophysical rates
\cite{Cowan.Thielemann.Truran:1991,Rauscher.Thielemann:2000} are
calculated on the basis of global optical potentials. For the proton
and neutron potentials several rather reliable potentials exist
\citep[e.g.,][]{Jeukenne.Lejeune.Mahaux:1977,Bauge.Delaroche.Girod:2001,
  Koning.Delaroche:2003,Goriely.Delaroche:2007}.  The situation is
different for the $\alpha$-optical potential. Although several global
potentials exist
\citep[e.g.,][]{McFadden.Satchler:1966,Demetriou.Grama.Goriely:2002,
  Demetriou.Grama.Goriely:2003,Kiss.Mohr.ea:2009,Mohr.Kiss.ea:2013},
none of them is able to consistently describe the existing data at low
energies in statistical model approaches.  Using $^{64}$Zn as an
example, \citet{Mohr.Gyurku.Fulop:2017} explored the sensitivity of
the $\alpha$-induced reaction cross section to the variation of
different alpha optical potential (and other parameters in the
statistical model).  Attempts have been made to cure the
problem. Rauscher suggested that the consideration of Coulomb
excitation leads to a better agreement with data
\cite{Rauscher:2013}. In \citet{Demetriou.Grama.Goriely:2002}
and~\citet{Szuecs.Mohr.ea:2020,Mohr.Fueloep.ea:2020} it was shown that
a modified imaginary part of the optical potential, in particular at
large radii, can improve the reproduction of experimental reaction
data at low energies.  Based on a large set of $\alpha$-induced
reaction data at sub-Coulomb energies,
\citet{Avrigeanu.Avrigeanu.Manailescu:2014,Avrigeanu.Avrigeanu:2015}
have presented a global $\alpha$-optical potential for nuclei in the
mass range $45 \leq A \leq 209$.

\subsection{Fission}
\label{sec:fission}

Fission plays an important role in the r-process, in particular within
the NS-NS merger scenario. Fission determines the region of the
nuclear chart at which the flow of neutron captures and beta-decays
stops~\cite{Thielemann.Metzinger.Klapdor:1983,Petermann.Langanke.ea:2012,Giuliani.Martinez-Pinedo.Robledo:2018,Mumpower.Kawano.ea:2018,Vassh.Vogt.ea:2019,Giuliani.Martinez-Pinedo.ea:2019}. In
the particular case of dynamic cold ejecta from mergers, several
fission cycles are expected to operate before all neutrons are used
\cite{Korobkin.Rosswog.ea:2012,Goriely:2015,Goriely.Martinez-Pinedo:2015}.
Fission has been suggested to be responsible for producing a robust
r-process pattern
\cite{Korobkin.Rosswog.ea:2012,Rosswog.Korobkin.ea:2014,Goriely:2015},
in which the abundances of nuclei with $A \lesssim 140$ are determined
during the r-process freeze-out from the fission yields of nuclei with
$A\lesssim 280$~\cite[see][]{Mendoza-Temis.Wu.ea:2015}. 

The description of fission for r-process nuclei is very challenging as
it sensitively depends on the knowledge of the fission barriers for a
broad range of very neutron-rich nuclei.  In addition, the evolution
of the shell structure as function of neutron excess is very
uncertain. Several competing reaction channels need to be modelled,
including neutron capture, neutron-induced fission, beta-decay,
$\beta$-delayed fission, spontaneous fission, alpha decay and
gamma-induced fission. Hence, parallel to the calculation of fission
barriers one has to develop models for all these different reaction
channels.  Several studies have computed barriers for r-process nuclei
\cite{Howard.Moeller:1980,Myers.Swiatecki:1999,Mamdouh.Pearson.ea:2001,
  Goriely.Hilaire.ea:2009,Erler.Langanke.ea:2012,Moeller.Sierk.ea:2015,Giuliani.Martinez-Pinedo.Robledo:2018}. It
has been shown that the dominating fission channel during r-process
nucleosynthesis is neutron-induced fission~\citep{Panov.Kolbe.ea:2005,Martinez-Pinedo.Mocelj.ea:2007,Petermann.Langanke.ea:2012}.
However, the necessary reaction rates have been computed for a
rather limited set of barriers~\citep{Thielemann.Cameron.Cowan:1989,Panov.Kolbe.ea:2005,Goriely.Hilaire.ea:2009,Panov.Korneev.ea:2010,Giuliani.Martinez-Pinedo.Robledo:2018,Giuliani.Martinez-Pinedo.ea:2019}.
This hinders studies of the sensitivity of the r-process abundances to
the fission barriers. Due to the dominance of neutron-induced fission,
the fission barrier itself is the most important quantity for the
determination of reliable fission rates, as the fission process occurs
at energies just above the fission barrier. In this case, the inertial
mass parameter plays a minor role as tunneling through the barrier has
only a negligible contribution. This fact, however, simplifies
calculations considerably as the calculation of the inertial mass
parameter is rather challenging~\cite{Sadhukhan.Mazurek.ea:2013,Giuliani.Robledo.Rodriguez-Guzman:2014,Giuliani.Martinez-Pinedo.Robledo:2018}.

In addition to the description of the different fission reaction
channels, also the corresponding fission yields, which depend on the
excitation energies of the compound nucleus
\cite{kelic08,Kelic.Ricciardi.Schmidt:2009,Schmidt.Jurado.ea:2016,Sadhukhan.Nazarewicz.Schunck:2016,Zhang.Schuetrumpf.Nazarewicz:2016,Schmitt.Schmidt.Jurado:2018,Schmidt.Jurado:2018,Mumpower.Jaffke.ea:2019,Vassh.Mumpower.ea:2019,Sadhukhan.Giuliani.ea:2020},
have to be known for r-process simulations. As discussed above, the
fission yields determine the abundance of r-process elements in the
second r-process peak and above and can play an important role for
abundance distribution of rare-earth elements~\cite{Bengtsson.Howard:1975,Steinberg.Wilkins:1978,panov08,Goriely.Sida.ea:2013,Eichler.Arcones.ea:2015,Vassh.Vogt.ea:2019}.

It should be emphasized that during the last phase of the r-process
alpha decays compete with fission. This competition determines the
final abundances of Pb, U and Th and of long-lived
actinides. Consequently, an improved description of transuranic
nuclides is necessary for the determination of the r-process
abundances produced in neutron star mergers,
with important consequences for the kilonova
lightcurves~\citep{Hotokezaka.Wanajo.ea:2016,barnes16,Rosswog.Feindt.ea:2017,Zhu.Wollaeger.ea:2018,Wanajo:2018,Wu.Barnes.ea:2019}.

\section{Astrophysical Sites and their ejecta composition}
\label{sec:astr-sites-their}

Section~\ref{sec:basic-working} has discussed conditions that any
astrophysical site should attain in order to produce r-process
nuclei. They reduce to particular combinations of entropy, expansion
time scale, and $Y_e$ in the ejecta. As a minimum requirement the
ejecta should be characterized by a high neutron-to-seed nuclei
ratio. This is certainly the case for neutron-rich matter, pointing
naturally to neutron stars as an important reservoir of
neutrons. However, ejecting material from the deep gravitational field
of a neutron star requires a cataclysmic event. This could be either
associated to the birth of a neutron star in a supernova explosion or
to ejecta from a compact binary merger involving a neutron star,
leading logically to the most promising sites for a strong r-process:
(i) the innermost ejecta of regular core-collapse supernovae, (ii) a
special class of core collapse supernovae (magneto-rotational MHD-jet
supernovae or collapsars), with fast rotation and high magnetic fields
responsible for their explosion mechanism, which can produce
neutron-rich jet ejecta along the poles or from accretion disk
outflows, and (iii) ejecta from binary neutron star mergers or neutron
star black hole systems which are naturally neutron-rich and have been
considered already extensively before the observation of GW170817.

A common feature of these scenarios, which will be discussed in detail
below, is that matter reaches such high temperatures that nuclei are
dissociated into free nucleons, and neutrinos become the main cooling
mechanism. Those neutrinos and in particular electron flavor
(anti)neutrinos can interact with the ejecta and reset the composition
that is commonly determined by a balance between the following
reactions:

\begin{eqnarray}
  \nu_e + n & \rightleftarrows &p+ e^- \\
  \bar\nu_e + p& \rightleftarrows &n+ e^+.
\end{eqnarray}

In the case of neutrino-driven winds and potentially also for neutron
star merger ejecta, the material is subject long enough to these
processes to reach an equilibrium between neutrino and
antineutrino
captures~\citep{Qian.Woosley:1996,Thompson.Burrows.Meyer:2001,Martinez-Pinedo.Fischer.ea:2016},
resulting in

\begin{equation}
  \label{eq:yeeq}
  Y_e=  Y_{e,\text{eq}} = \left[1+\frac{L_{\bar{\nu}_e}
      W_{\bar{\nu}_e}}{L_{\nu_e}W_{\nu_e}} \frac{\varepsilon_{\bar{\nu}_e} - 2 
      \Delta + \Delta^2/\langle 
      E_{\bar{\nu}_e}\rangle}{\varepsilon_{\nu_e} + 2 \Delta + \Delta^2/\langle
      E_{\nu_e}\rangle}\right]^{-1},
\end{equation}
with $L_{\nu_e}$ and $L_{\bar{\nu}_e}$ being the neutrino and
antineutrino luminosities,
$\varepsilon_{\nu} = \langle E_\nu^2\rangle/\langle E_\nu\rangle$ the
ratio between the second moment of the neutrino spectrum and the
average neutrino energy (similarly for antineutrinos),
$\Delta = 1.2933$~MeV the neutron-proton mass difference, and
$W_{\nu} \approx 1+1.01 \langle E_\nu \rangle/(m_u c^2)$,
$W_{\bar{\nu}} \approx 1-7.22 \langle E_{\bar{\nu}} \rangle/(m_u c^2)$
the weak-magnetism correction to the cross sections for neutrino and
antineutrino absorption~\citep{Horowitz:2002} with $m_u$ being the
nucleon mass.

If matter is exposed long enough to neutrinos to reach an
equilibrium, these reactions turn matter neutron-rich, provided the
following condition is fulfilled:

\begin{equation}
  \label{eq:nrichcond}
  \varepsilon_{\bar{\nu}_e} - \varepsilon_{\nu_e} > 4\Delta -
  \left[\frac{L_{\bar{\nu}_e} W_{\bar{\nu}_e}}{L_{\nu_e}W_{\nu_e}}-1\right]
  (\varepsilon_{\bar{\nu}_e}-2\Delta).
\end{equation}

\begin{figure*}[htb]
  \centering
  \includegraphics[width=0.48\linewidth]{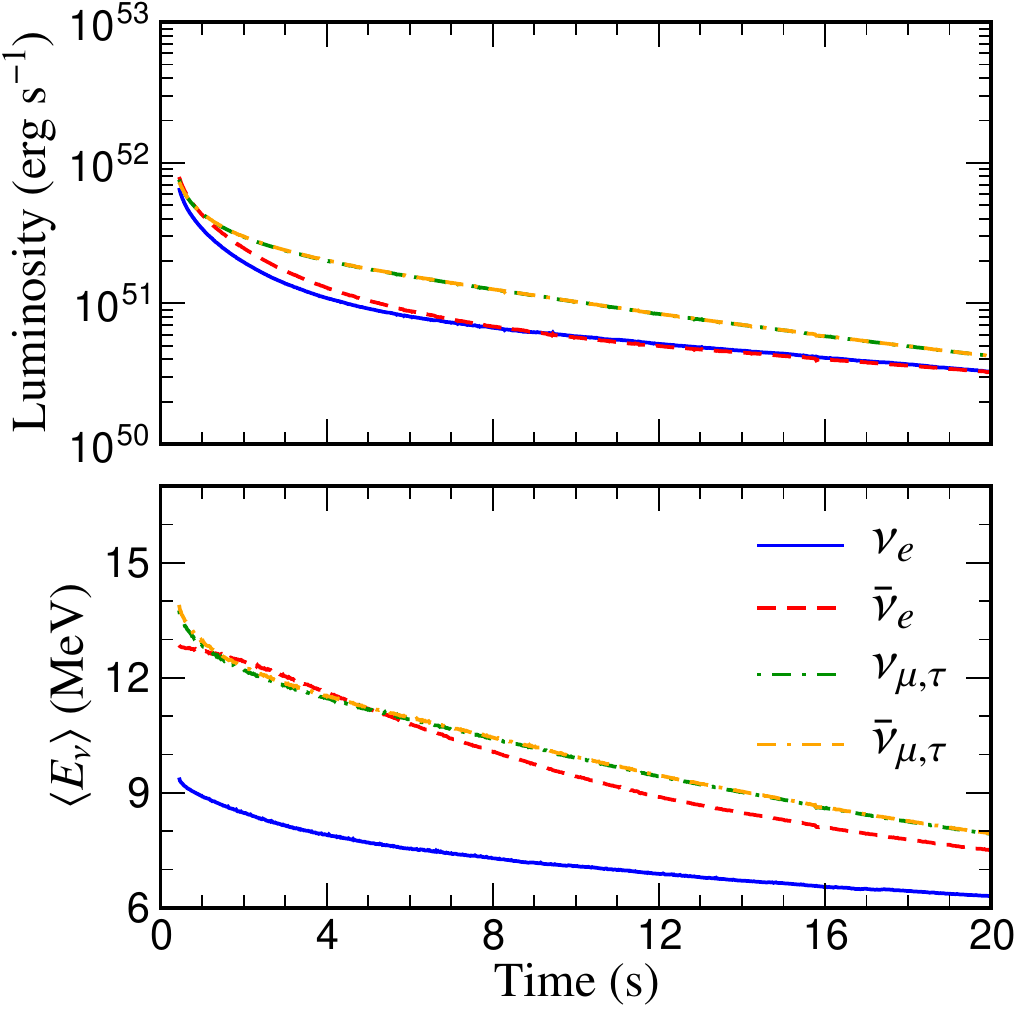}%
  \hspace{0.02\linewidth}%
  \includegraphics[width=0.48\linewidth]{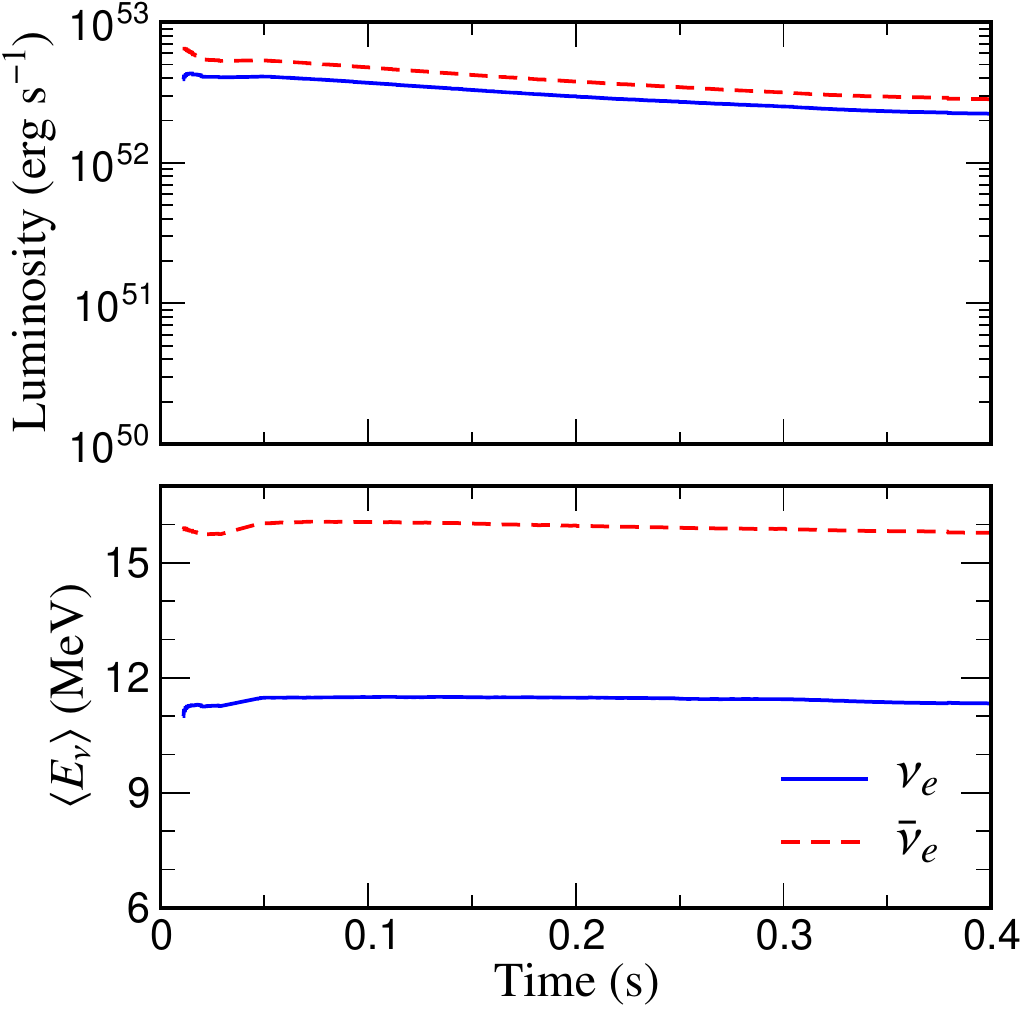}
  \caption{(left panel) Evolution of the luminosities and average
    energies of neutrinos emitted during the protoneutron star cooling
    phase following a core-collapse supernova explosion \citep[adapted
    from][courtesy of Tobias
    Fischer]{Martinez-Pinedo.Fischer.Huther:2014}. (right panel)
    Luminosities and average energies of neutrinos emitted after a
    NS-NS merger that forms a hypermassive neutron star surrounded by
    an accretion disk. \citep[based on the simulations from][courtesy
    of Albino Perego]{Perego.Yasin.Arcones:2017}. \label{fig:lumi-e}}
\end{figure*}

One should keep in mind an important difference between neutrino
emission from protoneutron stars formed in core-collapse supernovae
and the emission from a neutron-star merger remnant (see
Fig.~\ref{fig:lumi-e}). In the supernova case, we deal with the
deleptonization of a hot neutron star and consequently we expect
slightly higher fluxes for $\nu_e$'s than $\bar\nu_e$'s. However, due
to the fact that the $\bar\nu_e$ spectrum is slightly hotter than the
$\nu_e$ spectrum, the luminosities of both flavors are rather
similar. According to Eq.~\eqref{eq:nrichcond} this implies that the
average energies between $\bar\nu_e$ and $\nu_e$ should differ by at
least $4\Delta \approx 5.2$~MeV to produce neutron rich ejecta. Such
large differences are not reached in any modern neutrino-wind
simulation~\cite{huedepohl10,fischer10,Martinez-Pinedo.Fischer.ea:2012,Roberts.Reddy.Shen:2012,Martinez-Pinedo.Fischer.Huther:2014,mirizzi15,Fischer.Guo.ea:2020}.

In
the case of a neutron star merger the initial configuration
corresponds to a cold very neutron-rich neutron star. Due to the
merger dynamics the final merger remnant and accretion disk is heated
to large temperatures. The large temperatures favor the production of
electron-positron pairs and the material tends to protonize towards
the new equilibrium $Y_e$ on timescales of hundreds of ms as
determined by the weak interaction timescale in matter affected by
neutrino interactions
\citep{Beloborodov:2003,Arcones.Martinez-Pinedo.ea:2010}. During this
phase the luminosities and average energies of $\bar\nu_e$ are much
larger than those of $\nu_e$ (see right panels of
Fig.~\ref{fig:lumi-e}), reducing the required energy difference of
Eq.~(\ref{eq:nrichcond}). Hence, even if the impact of neutrino
reactions in mergers is expected to be
substantial~\cite{Wanajo.Sekiguchi.ea:2014,perego14,sekiguchi15,martin15,sekiguchi16,Foucart.Oconnor.ea:2016,Martin.Perego.ea:2018}
the (late) ejecta affected by neutrino interactions are expected to be
still neutron-rich enough to produce a (weak) r-process, while early
dynamic ejecta, emerging from spiral arms after the collision, stay in
any case very neutron-rich and lead to a strong r-process.

There is also an important difference between the nucleosynthesis
operating in neutrino heated ejecta for supernova and mergers. In the
supernova case, due to the high entropies and moderate electron
fractions, the material suffers an $\alpha$-rich freeze-out (see
Fig.~\ref{fig_far1}). Under these conditions, if the material is
subject to strong neutrino fluxes during the phase of alpha formation,
the so-called $\alpha$-effect~\cite{Meyer.Mclaughlin.Fuller:1998} will
drive the composition to $Y_e \approx 0.5$, hindering the occurrence
of an r-process. In the case of merger ejecta, due to the more moderate
entropies, no alpha formation takes place for $Y_e \lesssim 0.45$ (see
Fig.~\ref{fig-gabr}) and hence the $\alpha$-effect plays no role.

The discussion above neglects neutrino flavor transformations and
their impact on the $Y_e$ of the ejected material. In the supernova
case, the existence of very similar spectra for all neutrino flavors hinders the
impact of neutrino active-active flavor transformations~\citep[see
e.g.][]{Wu.Qian.ea:2015}. Active-sterile transformations, involving
sterile neutrinos on the eV mass scale, as suggested by the
reactor~\cite{Mention.Fechner.ea:2011} and
Gallium~\cite{Giunti.Laveder.ea:2012} anomalies, tend to drive the
composition more
neutron-rich~\citep{Nunokawa.Peltoniemi.ea:1997,Mclaughlin.Fetter.ea:1999,Wu.Fischer.ea:2014,Pllumbi.Tamborra.ea:2015}. As
discussed above, in the case of mergers the $\bar\nu_e$ fluxes dominate
over those of $\nu_e$. Hence, the neutrino self-interaction potential
has a different sign than the neutrino matter potential in the
Hamiltonian that describes flavor transformations. This induces
conversions via matter-neutrino
resonances~\cite{malkus12,foucart15,malkus16,zhu16,frensel17} and fast
pairwise conversions~\cite{Wu.Tamborra:2017,Wu.Tamborra.ea:2017}. The
existing investigations clearly point to a potential impact on $Y_e$
and thus on the resulting nucleosynthesis.

After this general outline, discussing in detail how weak interactions
are setting the stage for the resulting $Y_e$ (and entropy), being the
dominant criteria for the operation of an r-process, we will discuss
in the following potential environments/sites related either to
massive stars or compact objects in binary systems. This leaves out
sites of neutron-rich ejecta from core-collapse supernovae 
\citep[see e.g.][]{Hillebrandt:1978}, ruled out since the
neutrino-powered explosion mechanism has been established \citep{Bethe:1990}, as well as an r-process in He-layers
due to the $^{13}$C($\alpha,n)^{16}$O
reaction, ruled out since realistic models of massive stars are
existent \citep{Woosley.Heger.Weaver:2002}. 

\subsection{Possible r-process sites related to massive stars}
\label{sec:massive-stars}

\subsubsection{Neutrino winds from core-collapse supernovae}
\label{sec:neutrino-winds-from}

\begin{figure}[htb]
  \includegraphics[width=\linewidth]{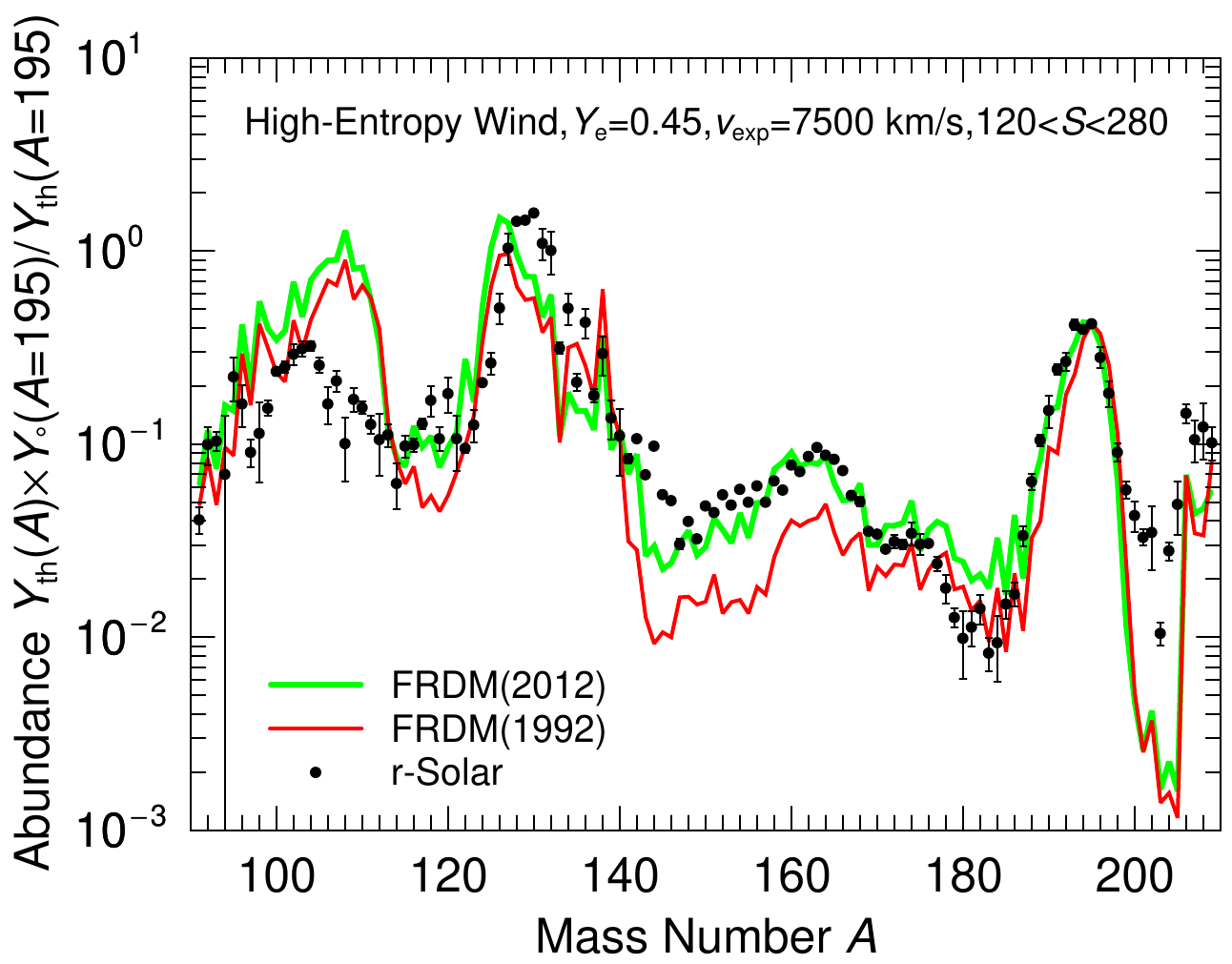}
  \caption{Results of an r-process calculation \citep[][reproduced by
    permission of the AAS]{kratz14}, 
    assuming an initial $Y_e=0.45$, adiabatic expansion of matter in a
    so-called neutrino wind with a given expansion speed $v_{\text{exp}}$ of
    ejected mass shells, and a superposition of entropies $S$ between
    120 and 280 $k_B$/baryon with equal amounts of matter ejected per
    entropy interval.  The plot indicates the changes due to utilizing
    an improved nuclear mass model
    \citep{moeller12,moeller15}.  \label{fig_kl2}}
\end{figure}

Supernovae have been thought to be the origin of the strong r-process
for many years~\citep[see e.g.\ the reviews
by][]{Cowan.Thielemann.Truran:1991,sumiyoshi01,arnould07}. While the
prompt explosion mechanism has been shown to fail \citep{Bethe:1990},
the development of multidimensional neutrino radiation transport
simulations has shown that the neutrino delayed explosion mechanism
remains the most promising scenario to explain the
observations~\citep[see][for reviews]{Kotake.Sumiyoshi.ea:2012,Burrows:2013,Foglizzo.Kazeroni.ea:2015,Janka.Melson.Summa:2016,Mueller:2016,Hix.Lentz.ea:2016,Janka:2017a,Burrows.Vartanyan.ea:2018,Cabezon.Pan.ea:2018,Burrows.Radice.ea:2020}. These
simulations predict that after the onset of the supernova explosion
the hot proto-neutron star enters the so-called Kelvin-Helmholtz
cooling phase. During this phase that lasts around 10~s, the
proto-neutron star deleptonizes, emitting neutrinos of all
flavors. Those neutrinos are responsible of producing an outflow of
matter known as neutrino-driven
wind~\citep{Duncan.Shapiro.Wasserman:1986} that is expected to operate
in each supernova explosion that produces a neutron star. The basic
properties of the wind are well understood, based on semianalytical
models~\citep{Duncan.Shapiro.Wasserman:1986,Qian.Woosley:1996,hoffman.woosley.qian:1997,Thompson.Burrows.Meyer:2001,Otsuki.Tagoshi.ea:2000,Arcones.Thielemann:2013}. These
models relate the nucleosynthesis relevant conditions (see
sect.~\ref{sec:how-obtain-required}) to fundamental properties
including neutrino luminosities, average energy, and mass and radius
of the proto-neutron star. Early simulations and parametric
models~\cite{woosley92,meyer92,Woosley.Wilson.ea:1994,Witti.Janka.Takahashi:1994,Takahashi.Witti.Janka:1994,Freiburghaus.Rembges.ea:1999,Farouqi.Kratz.ea:2010,Arcones.Martinez-Pinedo:2011,kratz14}
led to impressive results. However, large uncertainties remained,
particularly in the determination of entropy and
$Y_e$. Fig.~\ref{fig_kl2}, taken from~\citet{kratz14}, shows that the
solar r-process abundances can be reproduced, especially when
utilizing modern input from nuclear mass models, but requiring a
superposition of entropies of up to 280~$k_B$ per baryon (and a
$Y_e<0.5$).

The development of hydrodynamic
simulations~\citep{Arcones.Janka.Scheck:2007,Arcones.Janka:2011}
showed that such high entropies were out of reach. Nevertheless, they
still allowed for the occurrence of a weak
r-process~\citep{Roberts.Woosley.Hoffman:2010,Arcones.Montes:2011,Akram.Farouqi.ea:2020}. Further
progress, including the development of neutrino radiation
hydrodynamics simulations that follow the whole cooling
phase~\citep{huedepohl10,fischer10,Roberts:2012}, improvements in the
treatment of neutrino opacities in the decoupling
region~\cite{Martinez-Pinedo.Fischer.ea:2012,Roberts.Reddy.Shen:2012,Horowitz.Shen.ea:2012,Rrapaj.Holt.ea:2015,Janka:2016,Roberts.Reddy:2017,Martinez-Pinedo.Fischer.Huther:2014,Bollig.Janka.ea:2017,Fischer.Guo.ea:2020},
and the treatment of convection in the proto-neutron
star~\cite{Roberts.Shen.ea:2012,mirizzi15b} have shown that most or
all of the ejecta are proton-rich. Under these conditions the
nucleosynthesis proceeds via the $\nu p$
process~\cite{Froehlich.Martinez-Pinedo.ea:2006,Pruet.Hoffman.ea:2006,Wanajo:2006},
producing neutron deficient isotopes~\citep{Froehlich.Hauser.ea:2006},
including light p-process nuclei like $^{92}$Mo~\citep[see][]{Martinez-Pinedo.Fischer.Huther:2014,Pllumbi.Tamborra.ea:2015,Wanajo.Mueller.ea:2018,Eichler.Nakamura.ea:2018}.

The above result can be understood by considering that in
neutrino-driven winds matter is ejected by neutrino energy deposition
and is subject to neutrino reactions for sufficiently long time to
permit $Y_e$ to attain the equilibrium value given in
Eq.~\eqref{eq:yeeq}. Modern simulations predict very similar
  spectra of $\nu_e$ and $\bar\nu_e$ leading to proton-rich ejecta (see
Fig.~\ref{fig:lumi-e}). These results are robust against the inclusion
of neutrino flavor transformations between active
flavors~\cite{Wu.Qian.ea:2015,Pllumbi.Tamborra.ea:2015} but may be
affected by the active-sterile flavor
transformations~\cite{Wu.Fischer.ea:2014,Pllumbi.Tamborra.ea:2015}.

\subsubsection{Electron-capture supernovae}

A way out of the problem that neutrino irradiation is turning matter
proton-rich is by considering matter that is ejected promptly with
little exposure to neutrinos. This occurs in the so-called
electron-capture supernovae in the stellar mass range 8-10 M$_\odot$
\citep{jones14}, which could lead to a weak r
process~\citep{Kitaura.Janka.Hillebrandt:2006,Janka.Mueller.ea:2008,Wanajo.Nomoto.ea:2009,Wanajo.Janka.Mueller:2011},
possibly producing nuclei up to Eu, but not up to and beyond the third
r-process peak~\citep[for more details see][]{mirizzi15b}. However,
there are also strong indications, based on multidimensional
hydrodynamic simulations of the oxygen
deflagration~\cite{Jones.Roepke.ea:2016} and nuclear physics data on
the electron capture rate on
$^{20}$Ne~\citep{Martinez-Pinedo.Lam.ea:2014,Kirsebom.Jones.ea:2019,Kirsebom.Hukkanen.ea:2019},
that intermediate mass stars may end their lives as a thermonuclear
supernova triggered by electron captures on $^{20}$Ne~\citep[see][for
a recent review]{Nomoto.Leung:2017b}.

\subsubsection{Neutrino-induced r-process in the He-shell}
One of the major requirements for an r-process to take place is to
attain a sufficiently high neutron-to-seed ratio. As already discussed
above for the high entropy wind, this can also be achieved via a
(very) low seed abundance.  \citet{Banerjee.Haxton.Qian:2011} and
\citet{Banerjee.Qian.ea:2016}, following on an idea
by~\citet{Epstein.Colgate.Haxton:1988}, could show that for
core-collapse supernovae with metallicities as low as
$\text{[Fe/H]}\leq 3$, i.e.\ indicating a very low seed abundance, the
neutrons released in the He-shell by $^4$He$(\bar\nu_e, e^+n)^3$H can
be captured to produce nuclei with mass numbers up to $A=200$ in the
stellar mass range of 11--15~M$_\odot$, which are subsequently ejected
during the supernova explosion. The caveat of this environment is,
that while a sufficiently high neutron-to-seed ratio permits the
production of heavy nuclei via neutron captures, the relatively low
neutron density $n_n$ leads to an abundance pattern between the
r-process and an s-process with peaks shifted to higher masses numbers
than found for the solar r-abundances. Thus, such a process cannot be
an explanation for solar r-process abundances and abundance
patterns observed in low-metallicity stars.

\subsubsection{Quark deconfinement supernovae}

This scenario considers objects which undergo core
collapse and form a central compact
proto-neutron star, but the neutrino emission from the hot
proto-neutron star and accreted matter is not sufficient to prevent a
further collapse with ongoing mass accretion. The question is whether
this second collapse leads directly to black hole formation or can
come to a halt \citep{Fischer.Bastian.ea:2018}.  A specific equation
of state effect was initially introduced by \citet{sagert09} and
\citet{Fischer.Sagert.ea:2011}, with a quark-hadron phase transition
taking place just at the appropriate density/temperature conditions.
When adjusting the equation of state properties to presently observed
maximum neutron star masses, \citet{Fischer.Wu.ea:2020} could show
that in such supernovae explosions, expected for a certain stellar
mass range, an r-process can take place in the innermost ejecta.
When examining their results, they show that abundance up to the third
r-process peak can be obtained, however, the relative abundances
beyond the second r-process peak are strongly suppressed with respect
to solar.

Summarizing the discussion in the previous subsections, there remains
a possibility that core-collapse supernovae can produce r-process
elements but they probably do not support a solar-type r-process up to
the third r-process peak.

\subsubsection{Magneto-rotational supernovae with jets}
\label{sec:magn-rotat-supern}

\begin{figure}[htb]
  \centering
  \includegraphics[width=\linewidth]{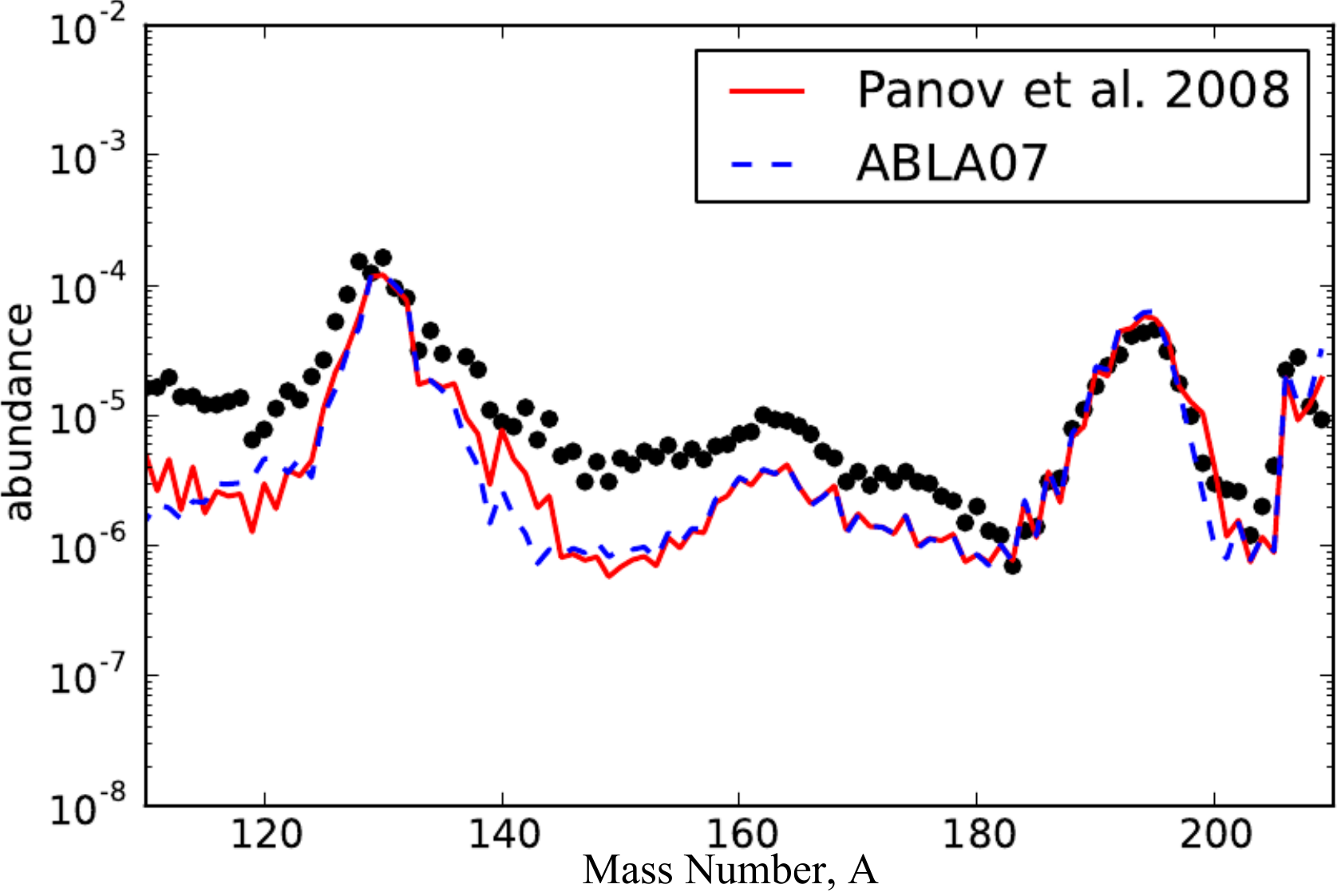}
  \caption{In an MHD-jet supernova the winding up of magnetic field
    lines causes the ``squeezing-out'' of polar jets, along the
    rotation axis \citep{Winteler.Kaeppeli.ea:2012}. This
    environment leads to quite low entropies, much lower than those
    discussed in Fig.~\ref{fig_far1}. But opposite to the $Y_e$-values
    utilized for Fig.~\ref{fig_far1}, the collapse to high densities
    resulted in large amounts of electron captures and $Y_e$-values
    close to 0.1-0.15 (see top part of this figure as well as
    Fig.~\ref{fig-gabr}). Such low $Y_e$'s, similar to neutron star
    merger conditions (where even values as low as 0.03-0.05 can be
    attained, see next subsection) lead to a strong r-process and
    abundance predictions displayed in the bottom part \citep[shown
    for two fission fragment distributions
    utilized][]{kelic08,panov08}. As $Y_e$ is only moderately low, the
    effect of late neutron-capture by fission neutrons is also
    moderate, avoiding a final shift of the third r-process peak as
    indicated in Figs.~\ref{fig_6}
    and~\ref{fig:fast}. \citep[Reprinted by permission from Springer
    Nature,][]{thielemann17a}}
  \label{fig_mhd} 
\end{figure}

Core-collapse with fast rotation and strong magnetic fields is
considered to lead to neutron stars with extremely high magnetic
fields of the order $10^{15}$~G \citep[magnetars, see e.g.][]{Duncan.Thompson:1992,Kramer:2009,Kaspi.Belobodorov:2017,Beniamini.Hotokezaka.Horst.ea:2019},
connected to a special class of supernovae
\citep{Kasen.Bildsten:2010,greiner15,Nicholl.ea:2017}.  Such
supernovae, induced by strong magnetic fields and/or fast rotation of
the stellar core, i.e., magneto-rotational supernovae (MHD-SNe), could
provide an alternative and robust astronomical source
for the r-process \citep{symbalisty85}. Nucleosynthetic studies were
carried out by \citet{nishimura06}, based on MHD simulations
which exhibited a successful r-process in jet-like explosions.
One important question is whether these earlier results, assuming axis
symmetry, also hold in full three-dimensional (3D) simulations, i.e.,
lead to the ejection of jets along the polar axis. Newtonian 3D MHD
simulations with an effective GR gravitational
potential~\citep{Marek.ea:2006} and improved treatment of neutrino
physics were performed by \citet{Winteler.Kaeppeli.ea:2012} for a
15~M$_\odot$ progenitor, utilizing an initial dipole magnetic field of
$5\times 10^{12}$~G and a ratio of magnetic to gravitational binding
energy, $E_{\text{mag}} / W = 2.63 \times 10^{-8}$.  These
calculations supported and confirmed the ejection of polar jets in 3D,
attaining magnetic fields of the order $5\times 10^{15}$~G and
$E_{\text{mag}}/ W = 3.02\times 10^{-4}$ at core-bounce, with a
successful r-process up to and beyond the third r-process peak at
$A=195$~(see e.g. Fig.~\ref{fig_mhd}).  Subsequent general
relativistic simulations in 3D-MHD~\citep{Moesta.Richers.ea:2014},
involving a 25~M$_\odot$ progenitor with an initial magnetic field of
$10^{12}$~G, led in the early phase to jet formation, but experienced
afterwards a kink instability, deforming the jet-like
feature. Probably this marks a transition between jet-like explosions
and deformed explosions, depending on critical limits in stellar mass,
initial rotation, and magnetic fields.

Further high resolution investigations \citep[resolving the
magneto-rotational instability MRI,][]{Moesta.ea:2015} have shown that
this mechanism can produce magnetar-strength magnetic fields and lead
to magneto-rotationally powered explosions, even for smaller initial
magnetic fields, probably causing the majority of
magnetars~\citep{Beniamini.Hotokezaka.Horst.ea:2019}. However, these
are not prompt jet-like explosions on timescales of 10th of ms, but
rather deformed (dual-lobe) explosions on timescales of 100th of ms,
which experienced the above-mentioned kink
instability~\citep{moesta18}.  This underlines that only for very high
initial magnetic fields such kink instabilities and long exposures to
neutrinos (increasing $Y_e$) can be avoided, ensuring a strong
r-process.  Studies by~\citet{halevi18} also analyzed the dependence
on the alignment between rotation axis and magnetic fields, where the
most aligned cases result in the strongest r-process. Recent studies
of this phenomenon have been undertaken~\citep{Obergaulinger.Just.Aloy:2018,Obergaulinger.Aloy:2020,Bugli.ea:2020,Reichert.Obergaulinger.ea:2020}.
The major constraint is the prerequisite of high initial magnetic
fields combined with high rotation rates in order to lead to an early
(prompt) polar jet-like ejection of neutron-rich matter. In delayed
ejections matter experiences interactions with neutrinos, which
enhance $Y_e$ and weaken the strength of the r-process, like in the
supernova neutrino wind (see the above discussion on that topic).

\begin{figure}[tbh]
  \centering
  \includegraphics[width=\linewidth]{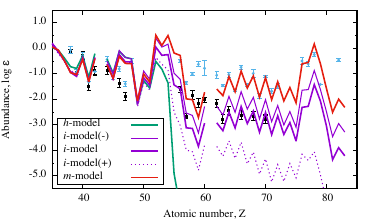}
  \caption{Abundances from nucleosynthesis calculations with varying
    ratios of magnetic field strength vs.\ the neutrino heating of
    regular core-collapse SNe, increasing for the models h, i-, i, i+,
    and m \citep[for details see][reproduced by permission of the
    AAS]{Nishimura.Sawai.ea:2017}. For comparison also abundances from
    MP stars with a weak r-process are shown, i.e. HD122563
    \citep[black squares,][]{honda06}, and solar-type r-process
    observations from CS22892-052 \citep[blue circles,][]{sneden96}.
    Abundances are normalized for Z = 40 of HD122563. Observations of
    low metallicity stars with strong r-process contributions vary for
    abundances below Z=50 \citep{sneden08}.}
  \label{fig_no2}
\end{figure}

\begin{figure}[tbh]
  \centering
  \includegraphics[width=\linewidth]{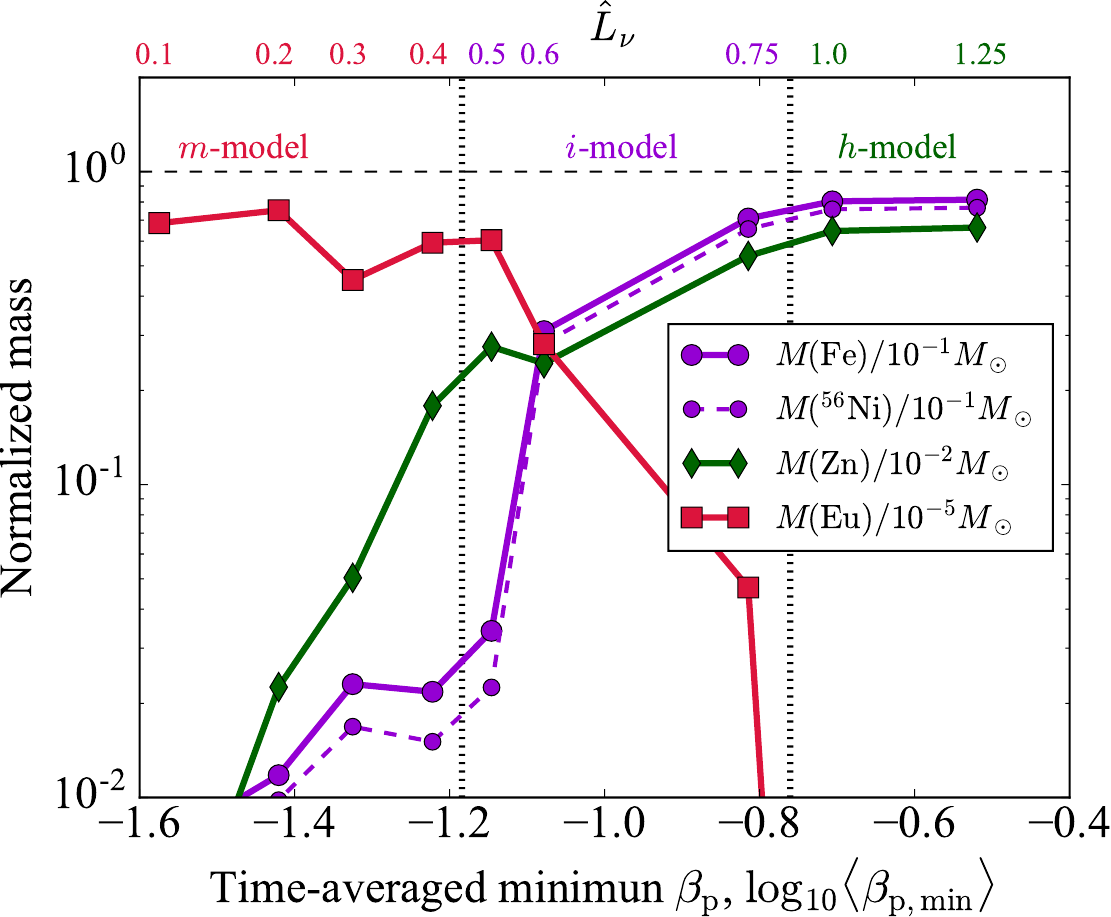}
  \caption{Nucleosynthesis features of rotating core-collapse SN
    models (h, i-, i, i+, m) with varying ratios of neutrino
    luminosity and magnetic field strengths as in
    Fig.~\ref{fig_no2}. Model m represents a strong MHD-jet
    supernova. One can see the transition from a regular core-collapse
    SN pattern, dominated by $^{56}$Ni, total Fe (after decay), and
    Zn, to a strong r-process pattern with a high Eu abundance
    \citep[for details see][reproduced by permission of the AAS]{Nishimura.Sawai.ea:2017}.}
  \label{fig_no3}
\end{figure}

A number of 2D axisymmetric simulations tested nucleosythesis features
\citep[e.g.][]{Nishimura.Takiwaki.Thielemann:2015,Shibagaki.Kajino.ea:2016,Nishimura.Sawai.ea:2017,Reichert.Obergaulinger.ea:2020},
depending on a variety of conditions in terms of rotation rates,
initial magnetic fields, and ratios of neutrino luminosities vs.\
magnetic field strengths.  \citet{Nishimura.Sawai.ea:2017} performed a
series of long-term explosion simulations, based on special
relativistic MHD \citep{takiwaki09,takiwaki11} with outcomes from
prompt magnetic jet over delayed magnetic explosions up to dominantly
neutrino-powered explosions, determined by the ratio of magnetic field
strengths in comparison to neutrino heating. This causes also a
variation of r-process nucleosynthesis results (see
Fig.~\ref{fig_no2}), from full-blown strong r-process environments,
over a weak r-process, not producing nuclei of the third r-process
peak, down to no r-process at all.  Thus, the production for heavy
neutron capture elements varies strongly, being either Fe and Zn
dominated like in regular core-collapse SNe or Eu dominated,
indicating a strong r-process. This is shown in Fig.~\ref{fig_no3}.

The relative fraction which such MHD-jet supernovae contribute to all
core-collapse supernovae depends on the distribution of pre-collapse
magnetic field strength and rotation among progenitor stars, probably
being metallicity dependent. Higher metallicities lead to stronger
stellar wind loss, accompanied by a loss of angular momentum and thus
reducing the fast rotation necessary for this type of SN
explosions. These events would eject only small amounts of Fe-group
nuclei in case of strong
r-processing~\citep{Nishimura.Takiwaki.Thielemann:2015,Nishimura.Sawai.ea:2017}.
Fig.~\ref{fig_no3} shows how the Ni/Eu-ratio (and similarly the
Fe/Eu-ratio) varies strongly as a function of neutrino heating vs.\
magnetic field effects. Thus, these types of supernovae alone would be
able to provide a large spread in Eu/Fe and might even explain the
variations in actinides vs. Eu, seen in a number of cases at low
metallicities~\citep[see e.g.][]{wehmeyer15,thielemann17a}. The
influence of the explosion mechanism of this type of rare supernovae
on their lightcurves and spectra is discussed
in~\citet{Siegel.Barnes.Metzger:2019}.

\subsubsection{Collapsars, Hypernovae, long-duration Gamma-Ray Bursts}

One of the most interesting developments in the study of supernovae
(SNe) is the discovery of some very energetic supernovae~\citep[for a review see e.g.][]{Nomoto.Tominaga.ea:2006}, 
dubbed hypernovae, whose kinetic
energy \citep[in spherically symmetric analysis, see
also][]{Piran:2004} exceeds $10^{52}$ erg.  The most luminous and
powerful of these objects, the Type Ic supernova (SN~Ic) 1998bw \citep[][]{Galama.ea:1998,Patat.Piemonte.ea:1998}, was
probably linked to the gamma-ray burst GRB 980425, thus establishing
for the first time a connection between (long-duration) gamma-ray
bursts (lGRBs) and the well-studied phenomenon of core-collapse SNe.
However, SN~1998bw was exceptional, indicating that it synthesized
$\sim 0.5$~M$_\odot$ of $^{56}$Ni with an estimated explosion energy
of $E \sim 3 \times 10^{52}$~erg \citep[][]{Iwamoto.ea:1998,Woosley.Eastman.ea:1999}.

The question is where these events should be placed in the stellar
mass range and which other features should be related.  For
non-rotating massive stars only the (regular) supernova ``branch''
(with neutron stars as final outcome) can be attained, followed
towards increasing stellar mass by a faint or failed supernova branch
(leading eventually to black holes, but not to gamma-ray bursts and
high ejecta masses).  Thus, massive stars, which fail to explode as
supernovae via neutrino-powered explosions, will eventually experience
the formation of a central black hole \citep[BH, see
e.g.][]{Pan.ea:2018,Kuroda.ea:2018}. However, rotating BHs and the
formation of accretion disks with accretion rates of about
$\approx 0.1$ M$_\odot$~s$^{-1}$ can lead --- for certain conditions
(strong magnetic fields) --- to long duration gamma-ray bursts (lGRBs)
or hypernovae, also dubbed collapsars.  The collapsar model was
proposed by Woosley, MacFadyen and others~\citep[see also][]{Woosley:1993,macfadyen99,macfadyen01,nagataki07,sekiguchi11,nagataki11},
including neutrino heating from the accretion disk and the winding of
strong magnetic fields, causing MHD jets~\citep[e.g.][]{fujimoto06,Ono.Hashimoto.ea:2012,mckinney13,Janiuk.Sukova.Palit:2018}. Early
hydrodynamic simulations (injecting explosion energies artificially)
were performed either by introducing high explosion energies (up to
$10^{52}$ erg) in a spherically symmetric way or aspherically in order
to understand jet-like explosions
\citep{macfadyen99,Nakamura.Umeda.ea:2001,Nomoto.Tominaga.ea:2006,nomoto13,nomoto17}.

The basic (consensus) picture has been the following: explosion
energies can be found up to $5\times 10^{52}$ erg, $^{56}$Ni ejecta up
to 0.5 M$_\odot$, and there exist relativistic jets responsible for
lGRBs.  There exists uncertainty in predicting $Y_e$, due to weak
interactions and especially neutrino transport. The observational
constraint of high $^{56}$Ni ejecta argues for a dominant $Y_e$ in
matter of the order 0.5. High explosion energies also lead to high
entropies and a strong $\alpha$-rich freeze-out, including large
amounts $^{45}$Sc (that is difficult to produce in other
environments), $^{64}$Zn (from $^{64}$Ge-decay), and also other
Fe-group elements.  \citet{Nakamura.Umeda.ea:2001} and
\citet{nomoto17} concluded that larger abundance ratios for (Zn, Co,
V, Ti)/Fe and smaller (Mn, Cr)/Fe ratios are expected than for normal
SNe, a feature which seems to be consistent with observations in
extremely metal-poor (EMP) stars.

Self-consistent modeling of the complete event, from collapse, black
hole formation, accretion disk modeling, jet ejection, and GRB
occurrence is a formidable challenge. Specific investigations, with
respect to the role of weak interactions, magnetic fields, and
resulting nucleosynthesis in the accretion disk and corresponding
outflows, have been undertaken, related either to individual magnetic
bubbles~\citep [e.g.][]
{Pruet.Woosley.Hoffman:2003,Pruet.Thompson.Hoffman:2004} or to the
  main wind outflows~\citep[see
e.g.][]{Beloborodov:2003,Mclaughlin.Surman:2005,surman06,
  Janiuk:2014,Siegel.Metzger:2017,Janiuk:2017,Janiuk.Sapountzis:2018,Siegel.Barnes.Metzger:2019}.
\citet{Beloborodov:2003} found conditions for the minimum accretion
rate required, leading to neutron-rich environments with low $Y_e$'s
at a given radius:

\begin{equation}
  \dot{M_n}= 3.821 \times 10^{-3} \left(\frac{r}{3r_g}\right)^{1/2} \left(\frac{\alpha}{0.1}\right)
  \left(\frac{M_{\text{BH}}}{M_\odot}\right)^2\ \text{M}_\odot\ \text{s}^{-1}. 
\end{equation}
Here $r_g$ is the gravitational radius, $\alpha$ the
disk viscosity, $M_{\text{BH}}$ the mass of the central black
hole. For typical accretion rates of 0.1~M$_\odot$~s$^{-1}$ this can
lead to a low $Y_e$ at small radii in the disk. Larger accretion rates
favor smaller $Y_e$ out to larger radii.

\begin{figure}[tbh]
  \centering
  \includegraphics[width=\linewidth]{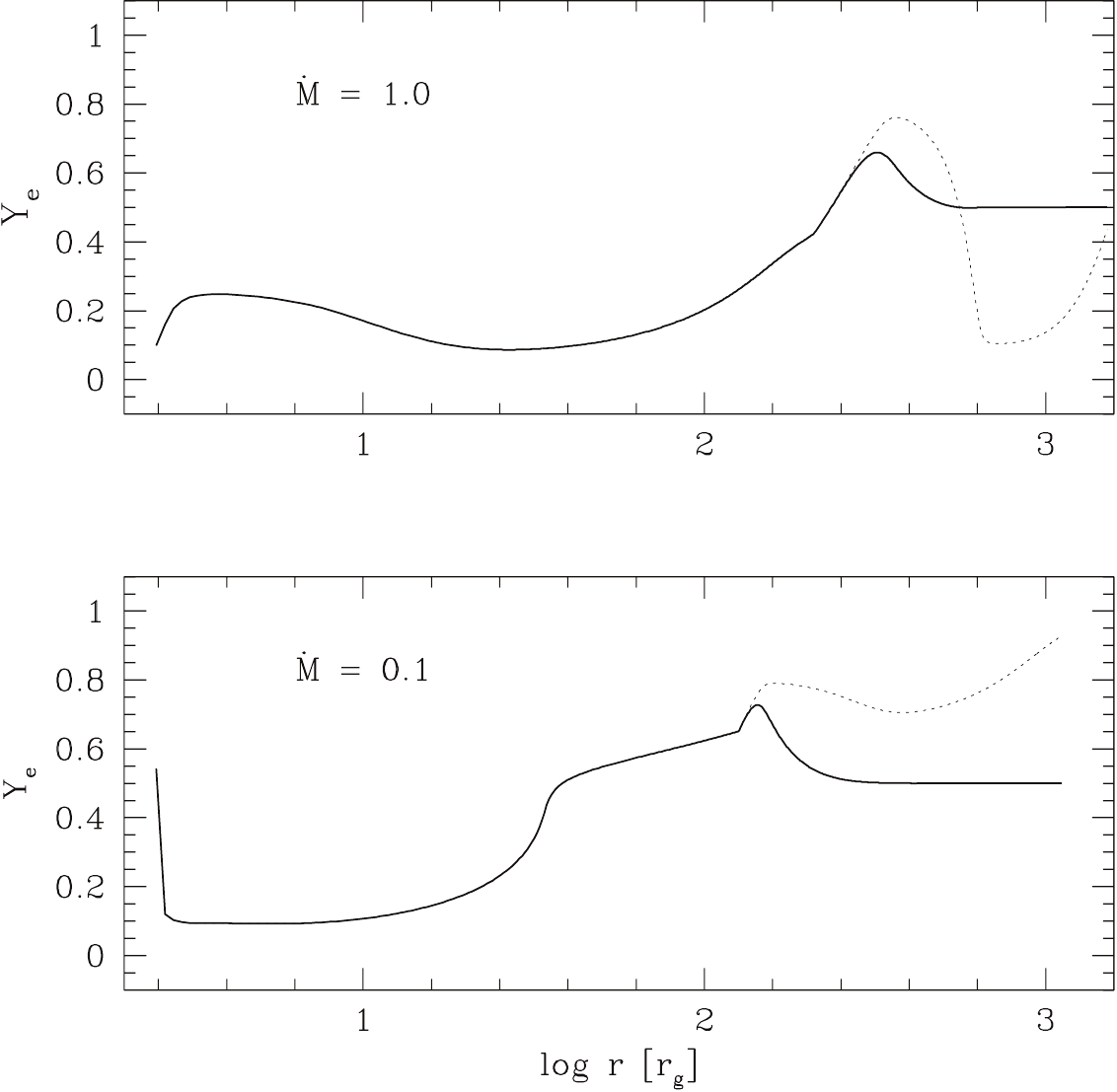}
  \caption{Radial distribution of $Y_e$ (thick solid line) and proton
    fraction (dashed line) in a disk, with $\alpha=0.1$ and
    $M_{\text{BH}}=3$~M$_\odot$, for different accretion rates. $r_g$,
    the gravitational radius, stands for the Schwarzschild radius of
    the black hole. $Y_e$ indicates very neutron-rich conditions deep
    inside the disk but develops aymptotically to values of 0.5 in the
    outer layers above 100~$r_g$ from where the nucleosynthesis
    outflow will occur \citep[from][reproduced by permission from
    A\&A, ESO] {Janiuk:2014}}
  \label{janiuk1}
\end{figure}

Fig.~\ref{janiuk1} shows the $Y_e$ distribution obtained by
\citet{Janiuk:2014} as a function of the radius. While the central
parts of the disk experience a low $Y_e$, its value reaches
$Y_e\approx 0.5$ in the outermost regions and even exceeds beyond 0.5
in intermediate regions. If the disk outflow occurs from the outer
regions, this is consistent with the large $^{56}$Ni ejecta (observed
and) found e.g.\ by
\citet{Pruet.Thompson.Hoffman:2004,surman06,Janiuk:2014,Janiuk:2017,Janiuk.Sapountzis:2018}.
However, \citet{Pruet.Thompson.Hoffman:2004} also speculated that in
case of strong magnetic fields low $Y_e$ matter can be flung out from
more central regions of the disk along magnetic field lines, possibly
causing r-process production.  Additionally, MHD-driven collapsar
models, involving black hole accretion disk systems
\citep{nagataki07,Fujimoto.Nishimura.Hashimoto:2008,harikae09}, have
argued that the jets produced by the central engine of long duration
gamma-ray-bursts can produce heavy r-process nuclei
\citep{fujimoto07,Fujimoto.Nishimura.Hashimoto:2008,Ono.Hashimoto.ea:2012,Nakamura.Kajino.ea:2015}.
However, it should be mentioned that early studies assumed a quite
simplified treatment of the black hole and the required microphysics.
\citet{Siegel.Metzger:2017,Siegel.Barnes.Metzger:2019,Janiuk:2019},
having performed multi-D MHD simulations for accretion disk outflows,
argue that large amounts ($>0.1$~M$_\odot$) of r-process
material can be ejected.  If this scenario materializes, it would be
sufficient to have about one such event per 1\,000--10\,000 core-collapse
supernovae to explain the solar r-process abundances.

The open question is whether both, large amounts of $^{56}$Ni expected
for hypernovae, as well as r-process ejecta, can be produced in the same
event.  \citet{Siegel.Barnes.Metzger:2019} argue that the $^{56}$Ni
would have to come from a preceding supernova explosion phase, leaving
an intermittent neutron star before further accretion causes black
hole formation and a black hole accretion disk.  This brings up the
following questions: (a) At which stellar progenitor masses do we have
a transition from the formation of neutron stars to the formation of
black holes after collapse?  (b) In which transition region are
initially neutron stars formed, causing a regular supernova explosion,
but ongoing accretion leads to a black hole?  (c) For which
progenitor masses are black holes formed directly during collapse and
how can this be observed?  (d) What is the role of rotation and
magnetic fields to cause lGRBs and can we give reliable
nucleosynthesis yields for such events?  (e) Is there a separation in
different types of events, depending on the parameters in (d), leading
either to hypernovae and strong $^{56}$Ni ejecta or systems with a
large outflow of r-process elements?  (f) Are jets and lGRBs occurring
in both types of events?

The scenario suggested by \citet{Siegel.Barnes.Metzger:2019} would
relate to questions (b) and case (c). The main question is whether a
strong supernova/hypernova explosion with large Ni production can take
place before the accretion disk outflows eject r-process material in
the same event or could r-process outflows occur without causing a
hypernova? Further observations will have to constrain such events.

\subsection{Neutron-star and neutron-star / black hole mergers}
\label{sec:neutron-star-neutron}

\begin{figure*}[htb]
  \centering
  \includegraphics[width=\linewidth]{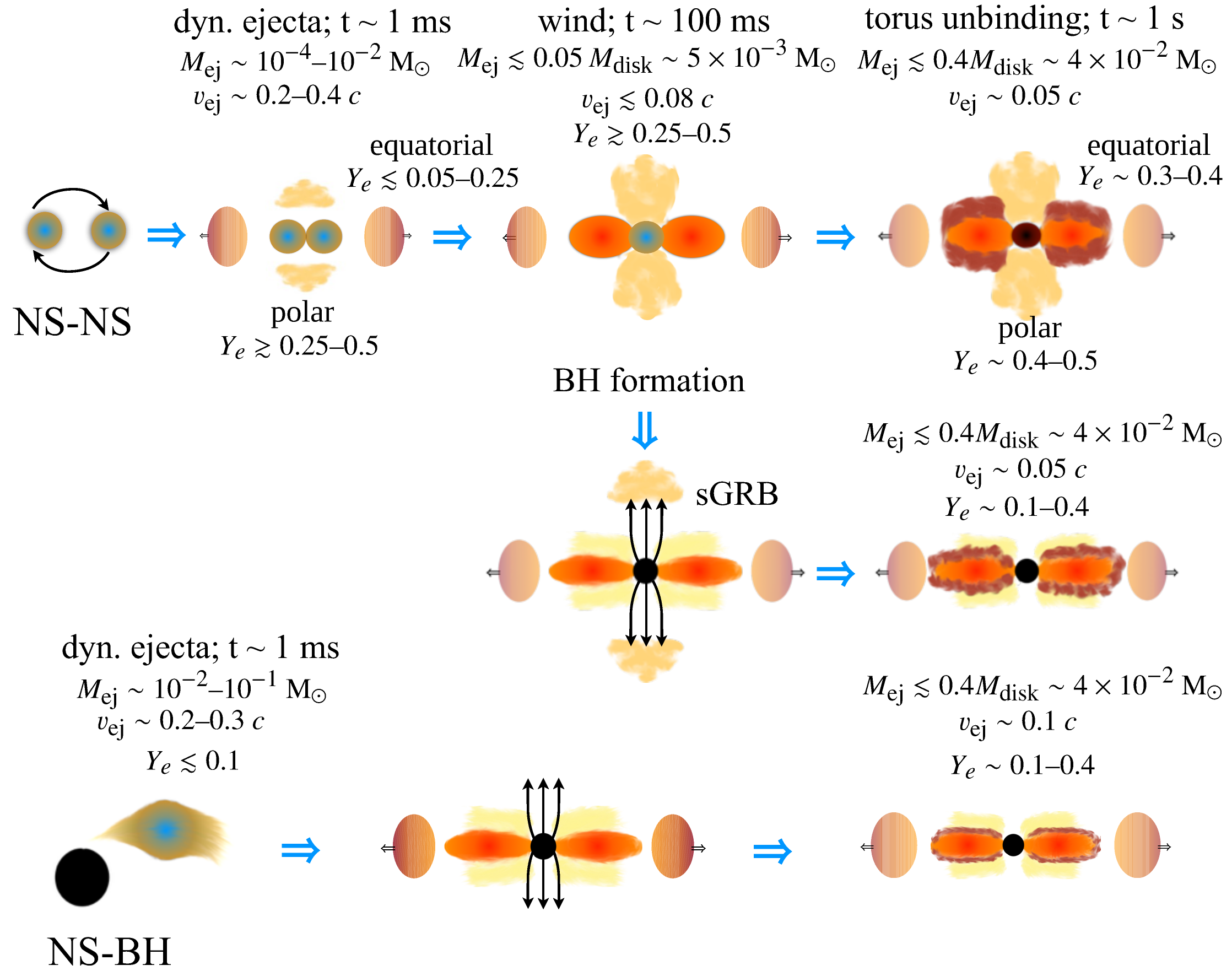}
  \caption{Ejection channels in compact binary mergers including
    estimates based on simulations of the ejecta mass, $Y_e$, and
    velocity during the different ejection phases. The NS-NS merger
    system is shown in the upper part including the two possible
    outcomes: a log lived massive neutron star and a hypermassive
    neutron star that collapses to a black hole on a timescale shorter
    than the disk lifetime. The BH-NS merger is shown in the lower
    part.~\citep[Adapted
    from][]{Rosswog.Feindt.ea:2017}\label{fig:cbmmassloss}}
\end{figure*}

Neutron stars, \citep[for historical references see
e.g.][]{Landau:1932,baadezwicky34,hewish65}, when being part of a
compact binary system, lose energy by emission of gravitational waves
as predicted by General Relativity and are expected to
merge~\citep[e.g.][]{hulsetaylor75}. Observed systems suggest
timescales of $\sim 10^8$ years for this
inspiral~\citep{Weisberg.Huang:2016}, but a larger range of timescales
is expected, dependent on initial separations and excentricities of
the orbit. Simultaneously to the discovery of binary pulsars, it was
suggested that neutron star or neutron star-black hole mergers would
eject r-process
nuclei~\cite{Lattimer.Schramm:1974,lattimer76,symbalisty82}, followed
up by a first detailed analysis of possible abundance
distributions~\cite{meyer88}. Later predictions showed that such
mergers would be accompanied by neutrino and gamma-ray
bursts~\cite{Eichler.Livio.ea:1989}. The first predictions of mass
ejection from neutron star mergers in Newtonian approximation were
given in~\citet{davies94}, \citet{ruffert96}, and
\citet{rosswog99,Rosswog.Davies.ea:2000}. The first detailed
nucleosynthesis prediction was provided
by~\citet{Freiburghaus.Rosswog.Thielemann:1999}.

Thereafter, extensive investigations have been undertaken with
respect to nucleosynthesis predictions
\citep{panov04,panov08,Goriely.Bauswein.Janka:2011,Korobkin.Rosswog.ea:2012,panov13,Bauswein.Goriely.Janka:2013,Goriely.Sida.ea:2013,hotokezaka13,Rosswog.Korobkin.ea:2014,Wanajo.Sekiguchi.ea:2014,Just.Bauswein.ea:2015,Goriely.Bauswein.ea:2015,perego14,Eichler.Arcones.ea:2015,martin15,Mendoza-Temis.Wu.ea:2015,Ramirez-Ruiz.Trenti.ea:2015,hotokezaka15,Shibagaki.Kajino.ea:2016,just16,Wu.Fernandez.Martinez.ea:2016,Radice.Galeazzi.Lippuner.ea:2016,roberts17,Martin.Perego.ea:2018,Wojczuk.Janiuk:2018,Papenfort.Gold.Rezzolla:2018,Radice.Perego.ea:2018a,Holmbeck19a}.
Initial Newtonian approaches~\citep[e.g.][]{Ruffert.Janka:2001} have been replaced with conformally flat
and fully relativistic treatments
\citep[e.g.][]{Shibata.Uryu:2000,oechslin02,oechslin04,Shibata.Uryu:2006,oechslin07,Shibata.Taniguchi:2011,Bauswein.Janka:2012,Bauswein.Goriely.Janka:2013,hotokezaka13,Wanajo.Sekiguchi.ea:2014,sekiguchi15,sekiguchi16,
  Radice.Galeazzi.Lippuner.ea:2016,Baiotti.Rezzolla:2017,Bovard.Martin.ea:2017,Papenfort.Gold.Rezzolla:2018},
and further followed by the inclusion of magnetic
fields~\cite{Price.Rosswog:2006,Anderson.Hirschmann.ea:2008,Liu.Shapiro.ea:2008,Giacomazzo.Rezzolla.Baiotti:2009,Obergaulinger.Aloy.Mueller:2010,Zrake.Macfadyen:2013,kiuchi15,Giacomazzo.Zrake.ea:2015}
as well as their interplay with
neutrinos~\citep{Palenzuela.Liebling.ea:2015,Guilet.Bauswein.ea:2017}

In parallel to neutron star (NS-NS) mergers also neutron star-black
hole (NS-BH) mergers have been
investigated~\citep[e.g.][]{rosswog05,Shibata.Uryu:2006,Chawla.Anderson.ea:2010,Shibata.Taniguchi:2011,Korobkin.Rosswog.ea:2012,Wanajo.Janka:2012,kyutoku13,Foucart.Deaton.ea:2014,mennekens14,Rosswog.Feindt.ea:2017,Brege.Duez.ea:2018}. A
common outcome of a NS-NS merger and in some NS-BH mergers is the
formation of an accretion disk surrounding a central
remnant~\citep[][see below]{Ruffert.Janka.ea:1997,Shibata.Taniguchi:2006}.

From the point of view of r-process nucleosynthesis, simulations
should predict the amount of ejecta, their properties (particularly
$Y_e$), spatial distribution and temporal evolution. In the following,
we discuss the major phases of ejection and the general dependencies
on the merging system. The discussion is mostly based on a
presentation of~\citet{Shibata:2018}, see also
\citet{Shibata.Hotokezaka:2019} and
\citet{Radice.Bernuzzi.Perego:2020}. Fig.~\ref{fig:cbmmassloss}
summarizes the main ejection channels in compact binary mergers and
provides estimates of ejecta mass, $Y_e$, and velocity. See also
Fig.~1 in~\citet{Bartos.Brady.Marka:2013} for estimated behaviors
dependent on the mass of the binary components involved.

Due to the emission of gravitational waves, that reduces the
eccentricity of the orbit, at times close to coalescence NS-NS systems
are expected to have almost circular orbits and spins much smaller
than the orbital frequency~\citep{Rosswog:2015}. During the
coalescence phase matter is ejected dynamically due to angular
momentum conservation on timescales of milliseconds with mildly
relativistic speed
$v \sim
0.2\text{--}0.4$~c~\citep{rosswog99,Rosswog.Davies.ea:2000,Bauswein.Goriely.Janka:2013,Hotokezaka.Kiuchi.ea:2013a,Palenzuela.Liebling.ea:2015,sekiguchi15,sekiguchi16,Radice.Galeazzi.Lippuner.ea:2016,Foucart.Oconnor.ea:2016}.

The amount of dynamic ejecta and their properties depend on the
compactness of the neutron stars and their mass
ratio~\cite{Bauswein.Goriely.Janka:2013,Hotokezaka.Kiuchi.ea:2013a,Radice.Perego.ea:2018a}. Two
components can be distinguished: cold tidal ejecta in the equatorial
plane and shock-heated ejecta originating from the contact interface
with a more isotropic distribution.  Systems with small mass ratios
tend to eject larger amounts of material mainly in the equatorial
region, while for similar masses the shock heated component
dominates~\citep{Bauswein.Goriely.Janka:2013,Hotokezaka.Kiuchi.ea:2013a,Palenzuela.Liebling.ea:2015,Lehner.Liebling.ea:2016}. While
the cold tidal ejecta maintain the original low $Y_e$ of the outer
regions of the neutron star from which they are ejected, the shock
component is heated to very large temperatures. This drives electron
and positron captures which increase $Y_e$ from a very low initial
value. As the material moves away, $Y_e$ is further increased by
$\nu_e$ and $\bar{\nu}_e$
absorption~\citep{Wanajo.Sekiguchi.ea:2014,Goriely.Bauswein.ea:2015,martin15,sekiguchi16,Radice.Galeazzi.Lippuner.ea:2016,Martin.Perego.ea:2018}. The
impact of neutrino absorption is sensitive to the evolution of the
central remnant~\citep{sekiguchi15}.  The total amount of dynamic
ejecta are in the range
$10^{-4}$--$10^{-2}$~M$_\odot$~\cite{Bauswein.Goriely.Janka:2013,Hotokezaka.Kiuchi.ea:2013a,sekiguchi15,sekiguchi16,Lehner.Liebling.ea:2016,Radice.Perego.ea:2018a}
with an angular mass distribution well approximated by
$F(\theta) = \sin^2\theta$~\citep{Perego.Radice.Bernuzzi:2017} and a
$Y_e$ distribution that can reach up to $Y_e\sim 0.5$ in the polar
region~\citep{Shibata.Fujibayashi.ea:2017,Radice.Perego.ea:2018a,Shibata.Hotokezaka:2019}. Magneto-hydrodynamic
instabilities, operating during the merger, can produce a third
component denoted as ``viscous-dynamical''
ejecta~\citep{Radice.Perego.ea:2018b} with asymptotic velocities
extending up to $\sim 0.8\ c$.
The analysis of the tidal deformability from the Gravitational Wave
observations of
GW170817~\citep{Most.Weih.ea:2018,De.Finstad.ea:2018,Abbott.Abbott.ea:2018a,Capano.Tews.ea:2020},
the observation of an electromagnetic transient which disfavors a
prompt collapse to a black
hole~\citep{Margalit.Metzger:2017,Bauswein.Just.ea:2017,Shibata.Fujibayashi.ea:2017,Coughlin.Dietrich.ea:2019,Margalit.Metzger:2019},
together with nuclear physics
constraints~\citep{Fattoyev.Piekarewicz.Horowitz:2018,Annala.Gorda.ea:2018,Tews.Margueron.Reddy:2018,Tews.Margueron.Reddy:2019,Capano.Tews.ea:2020}
favor moderately compact neutron stars with a radius in the range
8.9--13.2~km. In this case, the major source of ejecta is the contact
interface between the neutron
stars~\citep{Bauswein.Goriely.Janka:2013,sekiguchi15,Radice.Perego.ea:2018a}. The
maximum mass of a neutron star has been constrained to
$M_{\text{max}} \lesssim
2.17$~M$_\odot$~\citep{Margalit.Metzger:2017,Shibata.Fujibayashi.ea:2017,Rezzolla.Most.Weih:2018,Ruiz.Shapiro.Tsokaros:2018}
following the observation of GW170817.  Recently another NS-NS merger
GW190425 with a combined total mass of $\sim$ 3.4 M$_\odot$ has been
observed~\citep{Abbott.ea:2020}. The high total mass, together with
the absence of an electromagnetic signal, suggest a prompt collapse to
a black hole~\citep{Foley.Coulter.ea:2020}.  In the case of NS-BH
systems it is necessary that the NS is tidally disrupted by the BH, in
order to eject material. Tidal disruption means that the BH tidal
force is larger than the self-gravity of the NS\@. The amount of
ejected mass depends on the relative competition between the orbital
separation at which tidal disruption occurs and the radius of the
innermost stable circular orbit of the BH\@. The larger this ratio,
the larger is the amount of ejecta. This requires a large NS radius, a
small BH mass or small BH/NS mass ratio, or a high spin for the
BH~\cite{Kyutoku.Ioka.ea:2015,Capano.Tews.ea:2020}. We notice that
mass ejection may occur even if the neutron star is disrupted inside
the innermost stable circular
orbit~\cite{Faber.Baumgarte.ea:2006}. Population synthesis studies
favor a BH/NS mass ratio $\sim
7$~\citep{Belczynski.Dominik.ea:2010}. This, together with the NS
radius constraints mentioned above, suggests that mass ejection will
only take place for a BH with a spin parameter
$\chi = c J/(G M^2) \gtrsim
0.5$~\citep{Foucart:2012,Foucart.Deaton.ea:2013,Foucart.Deaton.ea:2014,Kyutoku.Ioka.ea:2015,Kawaguchi.Kyutoku.ea:2016,Brege.Duez.ea:2018,Kyutoku.Kiuchi.ea:2018}.
For possible $\chi$-values of the recent BH-NS candidate GW190426 see
Fig.~4 of~\citet{Lattimer:2019b} and~\citet{Abbott.Abbott.ea:2020}. The
tidal dynamic ejecta are much more anisotropic than those of NS-NS
mergers. They are mainly concentrated around the orbital plane and
often sweep out only half of the plane. The ejected mass can reach
$\sim 0.1$~M$_\odot$ with asymptotic velocities of 0.2--0.3~$c$. The
material is very neutron-rich $Y_e \lesssim 0.1$ and not affected by
neutrino
irradiation~\citep{Foucart.Deaton.ea:2014,Kyutoku.Kiuchi.ea:2018}.

An equally common outcome of compact binary mergers is the production
of a rotating torus surrounding the newly-formed central object with a
typical mass of
0.1~M$_\odot$~\citep{Ruffert.Janka.ea:1997,Shibata.Taniguchi:2006,Radice.Perego.ea:2018a}. In
the case of BH-NS mergers the central remnant is a BH and we deal with
a neutrino cooled disk that evolves on viscous timescales of
seconds. The study of such systems has evolved from the use of
$\alpha$-viscosity prescriptions to parametrize
dissipation~\cite{Fernandez.Metzger:2013,metzger14,Just.Bauswein.ea:2015,fernandez15,fernandez16,just16,Wu.Fernandez.Martinez.ea:2016,Wojczuk.Janiuk:2018,Fujibayashi.Kiuchi.ea:2018}
to three-dimensional General-Relativistic Magneto-hydrodynamic
simulations~\cite{Siegel.Metzger:2017,Siegel.Metzger:2018,Fernandez.Tchekhovskoy.ea:2019,JaniuksGRB:2019}
in which dissipation emerges naturally via the magneto-rotational
  instability. These works find that up to 40\% of the disk mass,
depending on the BH spin, is unbound in a quasi-spherical fashion. The
electron fraction in the outflow is in the range $Y_e \sim 0.1$--0.4
with velocities
$v \approx
0.1$~c~\citep{Siegel.Metzger:2018,Fernandez.Tchekhovskoy.ea:2019,Christie.Lalakos.ea:2019,Fujibayashi.Shibata.ea:2020}
depending on the efficiency of dissipation in the disk. For the
case of NS-NS mergers, the possibilities for the central object are a
stable NS, a long-lived massive neutron star (MNS, i.e.\ a NS with a
mass above the maximum mass for a non-spinning NS and below the one
for a uniformly rotating NS), a hypermassive neutron star~\citep[HMNS,
i.e.\ a NS with a mass above the maximum mass for a uniformly rotating
NS,][]{Baumgarte.Shapiro.Shibata:2000} or a black hole (BH) depending
primarily on the total mass of the binary,
$M_t$~\citep{Hotokezaka.Kiuchi.ea:2013b,Shibata:2018}. If $M_t$
exceeds a critical value $M_c$, the central object produced by the
merger collapses promptly to a BH on the dynamical time scale of a few
ms~\citep{Sekiguchi.Kiuchi.ea:2011}. On the other hand, if $M_t<M_c$
the resulting HMNS is at least temporarily supported against
gravitational collapse by differential rotation and thermal
pressure~\citep{Hotokezaka.Kiuchi.ea:2013b,Kaplan.Ott.ea:2014}. The
value of $M_c$ depends on the uncertain EoS of nuclear matter,
particularly its stiffness, mainly related to the symmetry
energy~\citep{Baldo.Burgio:2016,Oertel.Hempel.ea:2017}.  The discovery
of massive $\sim 2$~M$_\odot$ neutron
stars~\citep{Demorest.Pennucci.ea:2010,Antoniadis.Freire.ea:2013},
places a lower limit to
$M_c\gtrsim
2.6-2.8$~M$_\odot$~\citep{Hotokezaka.Kiuchi.ea:2013b}. Hydrodynamical
simulations of neutron star mergers for a large sample of
temperature-dependent equations of state, show that the ratio between
critical mass and the maximum mass of a non-rotating NS are tightly
correlated with the compactness of the non-rotating
NS~\citep{Bauswein.Baumgarte.Janka:2013}. This allows to derive
semi-analytical expressions for the critical
mass~\citep{Bauswein.Stergioulas:2017,bauswein16,Bauswein.Stergioulas:2019}
that, combined with the GW170817 constraints on the maximum mass and
radius of NS, give $M_c \approx 2.8$~M$_\odot$. It thus appears likely
that the canonical 1.35 + 1.35~M$_\odot$, including GW170817, binary
merger goes through a HMNS phase. The duration of this phase depends
on angular momentum transport processes (gravitational wave
  emission, magnetic fields, \ldots) and the EoS as it determines the
value of
$M_c$~\cite{Shibata.Taniguchi.Uryu:2005,Shibata.Taniguchi:2006,Kiuchi.Sekiguchi.ea:2009,Hotokezaka.Kyutoku.ea:2011,Kastaun.Ciolfi.Giacomazzo:2016}:
For a soft EoS that results in rather compact initial neutron
stars before the merger, the HMNS collapses to a black hole on
timescales of several 10s of milliseconds, while for a stiff EoS the
HMNS is long-lived with a lifetime longer than the timescales relevant
for matter ejection. The NS radius constraints mentioned above favor
the first case. For the case of prompt collapse to a BH, the BH-torus
system evolves similarly to the BH-NS merger case considered
before. However, systems with large $M_t$ are expected to eject little
mass dynamically and produce a low mass accretion disk. In these
cases, the total amount of ejecta, dynamical plus accretion disk, is
$\sim 10^{-3}$~M$_\odot$ of neutron-rich material,
$Y_e \lesssim 0.1$~\citep{Shibata:2018}.

The HMNS-torus is characterized by a more important role of neutrino
heating that increases the amount of ejecta and raises their $Y_e$ to
values that depend on the lifetime of HMNS
remnant~\citep{metzger14,perego14,Kaplan.Ott.ea:2014,martin15,Lippuner.Fernandez.Roberts.ea:2017,Fujibayashi.Kiuchi.ea:2018}. The
ejecta consist of two components being either neutrino-driven or
viscous-driven (also known as secular). The neutrino-driven component
is ejected mainly in the polar direction with velocities
$v\lesssim 0.08$~c and $Y_e \gtrsim 0.25$ and containing around 5\% of
the disk mass~\citep{martin15,Perego.Radice.Bernuzzi:2017}. The
viscous-driven component occurs mainly in the equatorial direction
with a velocity $v \sim 0.05$~c and contains around 40\% of the disk
mass~\cite{metzger14,Lippuner.Fernandez.Roberts.ea:2017,Fujibayashi.Kiuchi.ea:2018}. The
$Y_e$ distribution depends on the lifetime of the
HMNS~\citep{Fujibayashi.Kiuchi.ea:2018}. If the HMNS survives at least
for the timescale of neutrino cooling of the disk ($\sim 10$~s),
neutrino heating drives $Y_e$ to values above 0.25.  If the HMNS
collapses to a black hole on a timescale shorter than the disk
lifetime, the $Y_e$ distribution is in the range 0.1--0.4, similar to
the BH-torus case.

There exists extensive literature relating these events to short
duration Gamma-Ray Bursts (sGRBs) and kilonovae as electromagnetic
counterparts \citep[see
e.g.,][]{lipacz97,Nakar:2007,Metzger.Berger:2012,tanvir13,Piran.Nakar.Rosswog:2013,Tanaka.Hotokezaka:2013,kasen13,grossman14,Rosswog.Korobkin.ea:2014,metzger14,wanderman15,Rosswog:2015,fryer15,Hotokezaka.Wanajo.ea:2016,barnes16,fernandez16,Rosswog.Feindt.ea:2017,metzger17,Ascenzi.Coughlin.ea:2019}.
Although these objects are also of major importance as strong sources
for gravitational wave
emission~\citep{Shibata.Taniguchi:2011,Baiotti.Rezzolla:2017},
especially after GW170817~\citep{Abbott:2017}, underpinning the
importance of multi-messenger observations, we will 
focus here on the ejected nucleosynthesis composition. In the
following subsections we will concentrate on (i) the dynamic ejecta,
(ii) the post-merger neutrino wind ejecta, and (iii) the late time
viscous or secular outflow from the accretion disk.

\subsubsection{Dynamic ejecta}
\label{sec:dynamic-ejecta}

The dynamic ejecta consist of two components: a cold component
consisting of very neutron-rich matter originating from the outer
regions of the neutron star that is ``thrown out'' via tidal
interaction in the equatorial plane, and a hotter component
originating from the contact interface. The first component is the
  only one present in NS-BH mergers and the second one may constitute
  most of the unbound material in NS-NS mergers with similar masses.
The tidal component was originally found in Newtonian simulations
(first investigations by \citealt{davies94,rosswog99} and more
detailed discussions by \citealt{Korobkin.Rosswog.ea:2012}), while the
contact interface component was found in relativistic simulations,
first within the conformal flatness
approximation~\citep{oechslin07,Goriely.Bauswein.Janka:2011,Bauswein.Goriely.Janka:2013}
and then in fully relativistic
simulations~\citep{Hotokezaka.Kiuchi.ea:2013a}. The latter simulations
neglected the impact of weak processes in the ejecta and hence the
ejected material kept the very neutron-rich conditions corresponding
to $\beta$-equilibrium in the cold neutron star, $Y_e \lesssim 0.01$.

\begin{figure}[htb]
  \includegraphics[width=\linewidth]{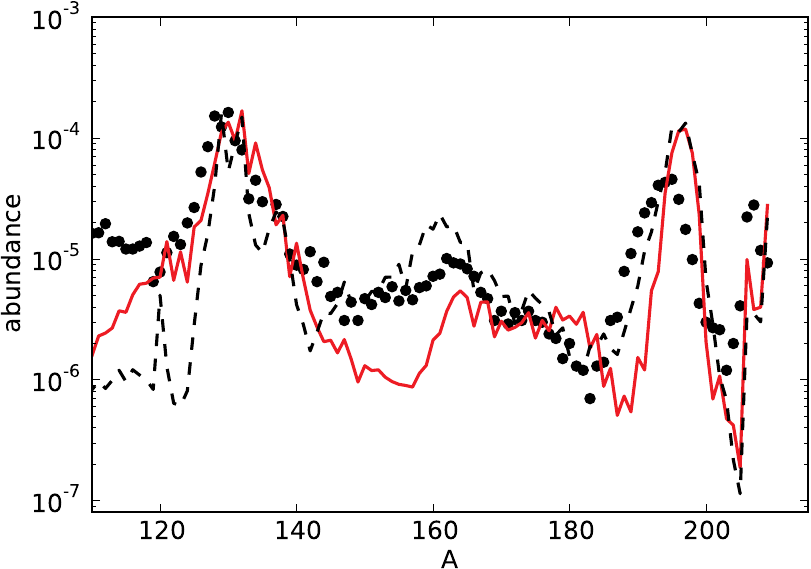}
  \caption{Resulting r-process abundances for dynamic tidal ejecta (in
    comparison to solar values-black dots) from neutron star merger
    simulations \citep{Eichler.Arcones.ea:2015}
    \citep[figure from][reprinted by permission from Springer]{thielemann17a}, making use of
    $\beta$-decay half-lives from \citet{Moeller.Pfeiffer.Kratz:2003}
    (red line) and recent $\beta$-decay half-life predictions
    \citep[][black line]{Marketin.Huther.Martinez-Pinedo:2016}
    together with the fragment distributions from fissioning nuclei of
    \citet{kelic08}.\label{fig_6}}
\end{figure}

The nucleosynthesis in low $Y_e$ ejecta, as found in BH-NS mergers and
the tidal component of NS-NS mergers, has been extensively
studied~\cite{Freiburghaus.Rosswog.Thielemann:1999,Korobkin.Rosswog.ea:2012,Bauswein.Goriely.Janka:2013,Rosswog.Korobkin.ea:2014,Mendoza-Temis.Wu.ea:2015,Eichler.Arcones.ea:2015,Martin.Arcones.ea:2016,Mumpower.Surman.ea:2016,Bovard.Martin.ea:2017}
and found to be independent of the astrophysical
conditions~\citep{Korobkin.Rosswog.ea:2012} but rather sensitive to
the nuclear physics
input~\citep{panov04,panov08,Bauswein.Goriely.Janka:2013,Eichler.Arcones.ea:2015,Mendoza-Temis.Wu.ea:2015,Goriely:2015,Goriely.Martinez-Pinedo:2015,Martin.Arcones.ea:2016,Mumpower.Surman.ea:2016,Shibagaki.Kajino.ea:2016,Thielemann.Eichler.ea:2017,Vassh.Vogt.ea:2019}. For
very neutron-rich ejecta neutron-to-seed ratios can even reach several
1000 and the associated nucleosynthesis becomes insensitive to
the initial composition. The temperature
evolution is characterized by having a high temperature plateau,
$T_{\text{max}}$, (see Fig.~\ref{fig:revol}) whose value is determined
by a competition between the r-process energy generation rate,
$\dot{Q}$, and the expansion dynamical
timescale~\citep{Mendoza-Temis.Wu.ea:2015}:

\begin{equation}
  \label{eq:tmax}
  T_{\text{max}} \approx 0.8\ \text{GK}\left[ \biggl(\frac{\rho}{10^5\ \text{g
        cm}^{-3}}\biggr) \biggl(\frac{\dot{Q}}{4\ \text{MeV
        s}^{-1}}\biggr) \biggl(\frac{\tau_{\text{dyn}}}{10\
      \text{ms}}\biggr)\right]^{1/4},
\end{equation}

Independent of the initial conditions, during the phase of neutron
captures one can have a hot or cold r-process, see
section~\ref{sec:basic-working} and Fig.~\ref{fig:revol} where dark
gray and brown lines correspond to cold r-process conditions and light
gray lines to hot r-process conditions. Typically the expansion of the
material is ``slow'' enough to allow for all neutrons to be
captured. This leads to the occurrence of several fission cycles with
large amounts of very heavy nuclei prone to fission, mainly around
$A\sim 280$, remaining at freeze-out, see
Fig.~\ref{fig_fission}. During the final freeze-out phase the fission
yields of the heaviest nuclei determine the final abundances of nuclei
with $A\lesssim 140$~\citep{Goriely.Martinez-Pinedo:2015}. Fission
also produces large amounts of neutrons that tend to be captured on
the third r-process peak material. Depending on the amount of neutrons
produced and the speed at which they are released, the third r-process
peak can be shifted to higher mass numbers when compared with solar
abundances \citep[see Fig.~\ref{fig_6} from][and for more details on
the effects of fission,
subsection~\ref{sec:fission}]{Eichler.Arcones.ea:2015}. This depends
on the mass model, see Fig.~\ref{fig:Joel-final}, and $\beta$-decay
half-lives~\citep{Marketin.Huther.Martinez-Pinedo:2016,panov16}. In
particular, shorter $\beta$-decay half-lives for very heavy nuclei
result in smaller abundances in the fissioning region and hence less
and faster release of neutrons during the
freeze-out~\citep{Eichler.Arcones.ea:2015}.


Fission rates and yields for neutron-rich heavy and superheavy nuclei are then
fundamental for the determination of the r-process
abundances~\citep{Goriely:2015}. This requires not only the
determination of the region of the nuclear chart where fission
occurs~\citep{Thielemann.Metzinger.Klapdor:1983,Petermann.Langanke.ea:2012,Giuliani.Martinez-Pinedo.Robledo:2018,Giuliani.Matheson.ea:2019},
but also the modeling of all relevant fission channels, including
neutron-induced fission, $\beta$-delayed fission, and spontaneous
fission~\citep{Thielemann.Metzinger.Klapdor:1983,Panov.Kolbe.ea:2005,panov04,Goriely.Hilaire.ea:2009,Panov.Korneev.ea:2010,Mumpower.Kawano.ea:2018,Vassh.Vogt.ea:2019}
and corresponding
yields~\citep{Kelic.Ricciardi.Schmidt:2009,Goriely.Sida.ea:2013,Schmidt.Jurado.ea:2016,Schmitt.Schmidt.Jurado:2018,Schmidt.Jurado:2018,Vassh.Vogt.ea:2019}. Low
$Y_e$ ejecta produce a final abundance distribution that
follows the solar r-process abundance distribution for $A>140$
independently of the fission yields
used~\citep{Goriely.Martinez-Pinedo:2015}. The production of lighter
nuclei is rather sensitive to the fission rates and yields used and
typically no nuclei below $A\sim 110$ are produced in substantial
amounts~\citep{panov08,Goriely.Sida.ea:2013,Mendoza-Temis.Wu.ea:2015,Eichler.Arcones.ea:2015,Vassh.Vogt.ea:2019}. Fission
is also relevant for the production of actinides with important
consequences for late time kilonova light
curves~\citep{barnes16,Rosswog.Feindt.ea:2017,Wanajo:2018,Wu.Barnes.ea:2019,Zhu.Wollaeger.ea:2018,Holmbeck19a}
and U/Th cosmochronometry (see section~\ref{sec:r-proc-cosm}).

\begin{figure}[htb]
  \centering
  \includegraphics[width=\linewidth]{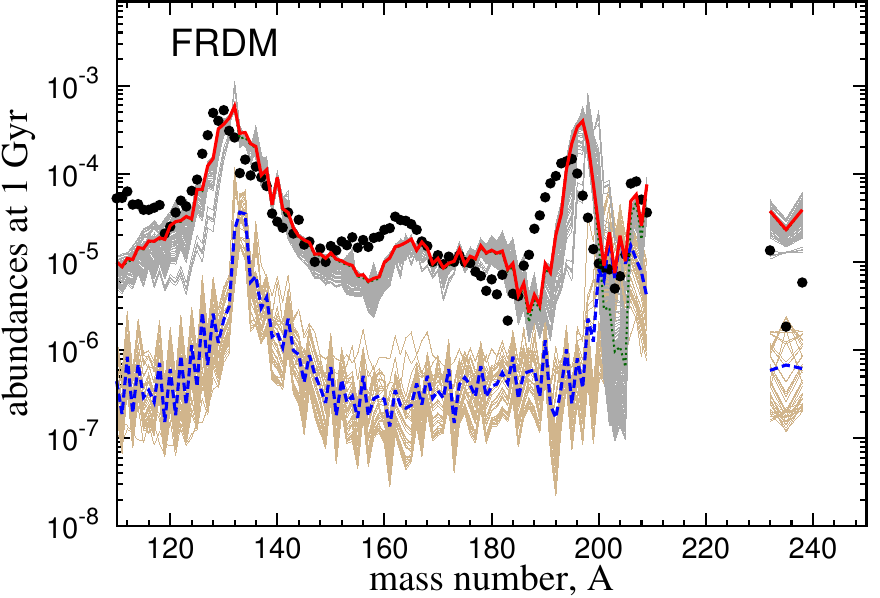}
  \caption{R-process abundances after a decay time of 1~Gyr for all
    trajectories shown in Fig.~\ref{fig:revol}. The dark grey (light
    brown) curves correspond to the abundances of the trajectories of
    the slow (fast) ejecta. The mass-averaged abundances for all
    trajectories (red solid curves), the slow ejecta (green dotted
    curves), and the fast ejecta (blue dashed curves) are also
    shown. The abundances for the slow and fast trajectories and their
    averages have been scaled by the value of their fractional
    contribution to the total ejecta. \citep[Figure adapted
    from][]{Mendoza-Temis.Wu.ea:2015}.\label{fig:fast}}
\end{figure}

Several
studies~\cite{Goriely.Bauswein.ea:2014,Metzger.Bauswein.ea:2015,Mendoza-Temis.Wu.ea:2015,Radice.Perego.ea:2018b,Fernandez.Tchekhovskoy.ea:2019,Ishii.Shigeyama.Tanaka:2018}
have shown that part of the material, up to 10\% in mass, is ejected
very fast and reaches such low densities that the timescale for
neutron captures becomes much longer than the expansion
timescale~\cite{Mendoza-Temis.Wu.ea:2015} (see brown lines on
Fig.~\ref{fig:revol}). Under such conditions most of the neutrons are
not captured, despite of having a large neutron-to-seed ratio. The
final abundances of this ``frustrated'' r-process does not correspond
to solar abundances~(see Fig.~\ref{fig:fast}) and hence it cannot
constitute a major component of the total ejected mass, assuming
mergers are a major r-process site. However,
it can significantly contribute for nuclei around $A\sim 200$ (see
difference between green and red lines in Fig.~\ref{fig:fast}) and can
drive an early (timescales of hours) electromagnetic emission that is
powered by the radioactive decay of the free neutrons left after
completion of the r-process~\citep{Metzger.Bauswein.ea:2015}.

\citet{Wanajo.Sekiguchi.ea:2014} showed that weak processes operating
on the shock heated ejecta of NS-NS mergers can increase the
$Y_e$. They are particular efficient in the polar region, where the
large neutrino fluxes from the HMNS increase substantially the $Y_e$
of the ejecta, provided the HMNS does not collapse promptly to a
BH\@. Depending on the neutrino luminosities, $Y_e$ could be increased
to values between 0.25 and 0.4. While it is currently accepted that
weak processes increase the $Y_e$ of the ejecta, an aspect confirmed
by the kilonova observations discussed in
section~\ref{sec:kilon-an-electr}, there is still a relatively large
spread between the predictions of different
groups~\citep{Shibata.Fujibayashi.ea:2017,sekiguchi16,sekiguchi15,Radice.Galeazzi.Lippuner.ea:2016,Radice.Perego.ea:2018a,Foucart.Oconnor.ea:2016,Bovard.Martin.ea:2017,Foucart.Duez.ea:2018}
related to the different approximations in the treatment of neutrino
radiation transport and/or to differences in the thermodynamical
conditions of matter reached after the
merger~\citep{Perego.Bernuzzi.Radice:2019} as they determine the
magnitude of electron and positron capture processes. Dynamic ejecta
from NS-NS mergers are expected to contribute to the synthesis of a
broad range of r-process nuclei, including both light and heavy, once
weak processes are
considered~\citep{Wanajo.Sekiguchi.ea:2014,Goriely.Bauswein.ea:2015,Martin.Perego.ea:2018}. However,
it should be kept in mind that the predicted amount of high $Y_e$
matter is typically much smaller than found in accretion disk
outflows.

\subsubsection{Neutrino Winds and the Effect of Neutrinos}
\label{sec:neutr-winds-effect}

\begin{figure}[htb]
  \includegraphics[width=\linewidth]{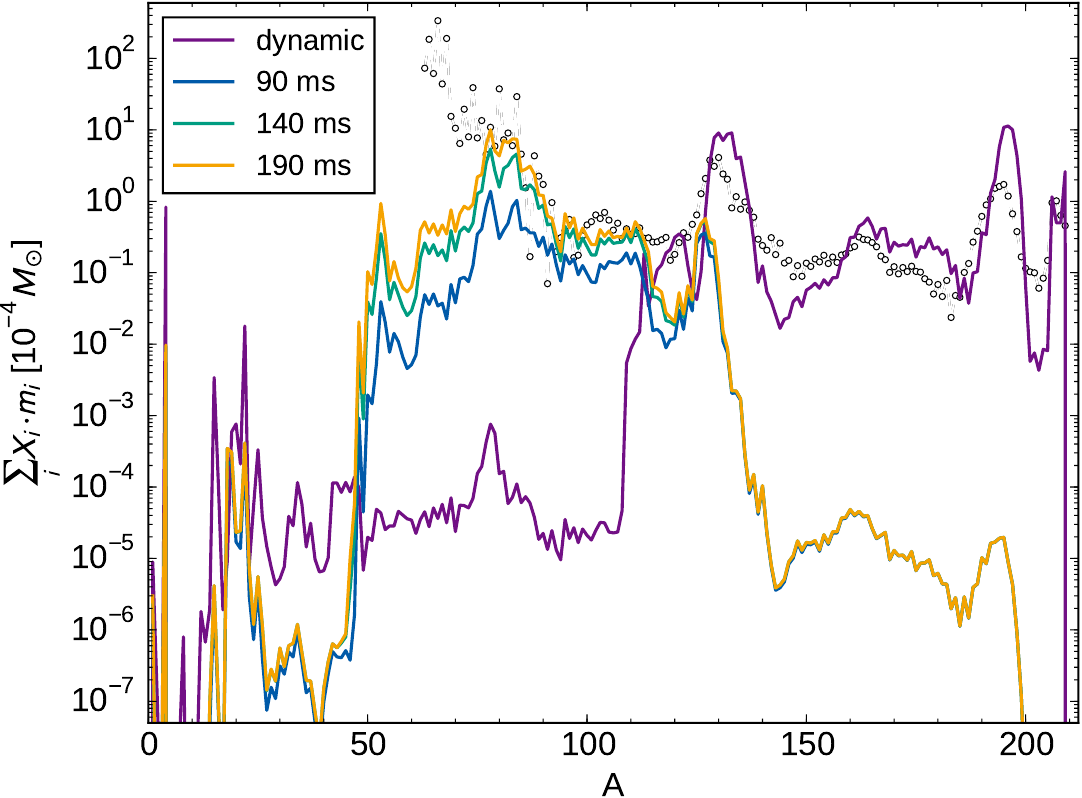}
  \caption{Neutrino wind contribution to neutron star merger ejecta,
    dependent on the delay time between the merger and BH formation
    \cite[][reproduced by permission of the AAS]{martin15}.  In comparison also the dynamic, tidal ejecta of
    \cite{Korobkin.Rosswog.ea:2012} are shown. The neutrino wind,
    ejected dominantly in polar regions, contributes nuclei with
    $A<130$, due to the effect of the neutrinos on
    $Y_e$. \label{fig_7}} 
\end{figure}

In addition to the dynamic ejecta, related directly to the
merging/collision, post merger ejecta will emerge as well. One
component is a ``neutrino-wind'' as found in core-collapse
supernovae. For the typical merging system, the hot central NS
remnant, supported by high temperatures and differential rotation,
will not collapse to a black hole immediately (provided that the
combined total mass of the system $M_t$ is smaller than $M_c$, see the
introductory part of this subsection~\ref{sec:neutron-star-neutron}),
and will be surrounded by a hot and dense torus. Hence, the structure
of the wind is quite different from the isolated NSs usually found in
core-collapse supernovae. The wind outflow occurs mainly in the polar
direction~\citep{Rosswog.Korobkin.ea:2014,perego14,metzger14,martin15}. Matter
is exposed to neutrinos long enough for the material to reach an
equilibrium between electron neutrino and antineutrino absorption,
changing $Y_e$, see Eq.~\eqref{eq:yeeq}, from the initial
(neutron-rich) conditions towards higher values that can even be above
$Y_e=0.5$. Due to the much larger $\bar{\nu}_e$ luminosities and
energy differences between $\bar{\nu}_e$ and $\nu_e$, found during the
post-merger evolution when compared with core-collapse supernova (see
Fig.~\ref{fig:lumi-e}), the peak of the $Y_e$ distribution is expected
to be neutron-rich with
$Y_e \gtrsim
0.25$~\citep{martin15,Lippuner.Fernandez.Roberts.ea:2017,Fujibayashi.Sekiguchi.ea:2017,Fujibayashi.Kiuchi.ea:2018}. This
leads to a weak r-process and produces mainly matter below the second
r-process peak, i.e.\ no lanthanides are produced.  Fig.~\ref{fig_7}
displays the results of~\citet{martin15} for the neutrino wind
component as a function of the delay time until black hole formation.
It can be seen that predominantly nuclei below $A=130$ are produced,
complementing nicely the abundance features originating from dynamic
low $Y_e$ ejecta, which are also displayed and result here from
Newtonian simulations~\citep{Korobkin.Rosswog.ea:2012}.

Similar to the situation in core-collapse supernovae, the properties
of neutrino-wind ejecta, and particularly $Y_e$, are expected to be
sensitive to the spectral differences between $\nu_e$ and
$\bar{\nu}_e$. This requires an accurate prediction of neutrino
luminosities and spectra. Supernova neutrino-wind transport
simulations are nowadays based on accurate numerical solutions of the
Boltzmann transport
equation~\citep[e.g.][]{huedepohl10,fischer10,Roberts:2012} exploiting
the spherically symmetric nature of the problem. In the case of
mergers, which require multidimensional treatments, simulations so far
are based on neutrino leakage
schemes~\citep{perego14,metzger14,Radice.Galeazzi.Lippuner.ea:2016,Ardevol-Pulpillo.Janka.ea:2018}
and $M1$
schemes~\citep{Just.Bauswein.ea:2015,Just.Obergaulinger.Janka:2015,foucart15,Fujibayashi.Sekiguchi.ea:2017,Fujibayashi.Kiuchi.ea:2018}. There
are indications that they may not properly capture the energy
densities and fluxes of neutrinos in the polar
regions~\citep{Just.Bauswein.ea:2015}, hence, affecting the $Y_e$
estimates in the polar
region~\citep{Foucart.Oconnor.ea:2016,Foucart.Duez.ea:2018}. Additional
opacity reactions, like neutrino-pair annihilation, are so far not
considered, but may also play an important role in determining the
properties of the
ejecta~\citep{just16,Fujibayashi.Sekiguchi.ea:2017,Perego.Yasin.Arcones:2017,Fujibayashi.Kiuchi.ea:2018,Foucart.Duez.ea:2018}.

$Y_e$ can also be affected by modifications of neutrino and
antineutrino spectra due to neutrino flavor conversion. There have
been a number of tests to verify such neutrino conversions via
matter-neutrino
resonances~\cite{malkus12,foucart15,malkus16,zhu16,frensel17} and fast
pairwise flavor
conversions~\citep{Wu.Tamborra:2017,Wu.Tamborra.ea:2017}. Due to the
more complicated geometry of a disk environment in comparison to
core-collapse supernovae, most of the calculations are based on
single-angle approximations. Spherically symmetric test calculations
show that the matter-neutrino resonance still occurs in multi-angle
models~\citep{Vlasenko.Mclaughlin:2018}, but with reduced
efficiency. Nevertheless, the existing investigations clearly point to
a potential effect on $Y_e$, and thus the resulting nucleosynthesis
can be affected.

A further wind component, not addressed here, relates to magnetically
driven winds from the central
remnant~\citep[][]{Kiuchi.Kyutoku.ea:2012,Siegel.Ciolfi.ea:2012,Ciolfi.Kastaun.ea:2017,Metzger.Thompson.Quataert:2018}.
However, their nucleosynthesis yields and interaction with neutrino
driven winds have not been explored yet.

\subsubsection{Accretion Disks outflows}
\label{sec:accr-discs-outfl}

The long-term evolution, $t\sim 1$--10~s, of the accretion disc
produces outflows of material powered by viscous heating and nuclear
recombination~\citep{Lee.Ramirez-Ruiz:2007,Beloborodov:2008,Metzger.Piro.Quataert:2009,Fernandez.Metzger:2013}. Those
outflows can contain up to 40\% of the disk mass. The amount of
ejected mass increases with the lifetime of the MNS formed in the
merger, but most importantly for nucleosynthesis the $Y_e$
distribution is dramatically affected by the lifetime of the
MNS~\cite{metzger14}. For a long lived MNS, $t \gtrsim 1$~s, neutrino
irradiation from the MNS results in ejecta with
$Y_e >
0.3$~\citep{metzger14,Lippuner.Fernandez.Roberts.ea:2017,Fujibayashi.Kiuchi.ea:2018}. The
nucleosynthesis in these ejecta is similar to the neutrino-wind ejecta
discussed in the previous subsubsection.

\begin{figure}[htb]
  \includegraphics[width=\linewidth]{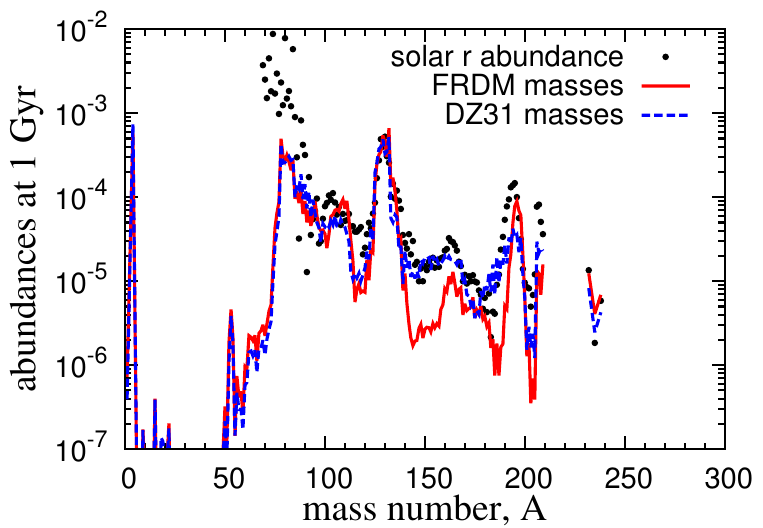}
  \caption{Resulting r-process abundances (in comparison to solar
    values --- black dots) from black hole accretion disk simulations
    \citep[reproduced with permission of Oxford University Press][]{Wu.Fernandez.Martinez.ea:2016}, making use of a black hole
    mass of 3~M$_\odot$, a disk mass of 0.03~M$_\odot$, an initial
    $Y_e$ of 0.1, entropy per baryon of 8~$k_b$, an alpha parameter of
    the viscous disk of 0.03, and a vanishing black hole spin. The
    figure illustrates the impact of two different mass models, FRDM
    and DZ31, in the final abundances.
    \label{fig_8}} 
\end{figure}

For a short-lived MNS, $t \lesssim 1$~s, the impact of
neutrino-irradiation is small and, from the point of view of
nucleosynthesis, outflows from accretion disks formed in NS-NS and
BH-NS mergers give similar results. Early nucleosynthesis studies were
mainly parametric and considered mainly the ``neutrino-driven'' wind
outflow from the surface of the
disk~\citep{Pruet.Woosley.Hoffman:2003,Fujimoto.Hashimoto.ea:2003,Fujimoto.Hashimoto.ea:2004,Pruet.Thompson.Hoffman:2004,Mclaughlin.Surman:2005,surman06,Metzger.Thompson.Quataert:2008,surman08,Dessart.Ott.ea:2009,Kizivat.Martinez-Pinedo.ea:2010,Wanajo.Janka:2012,surman14}.
Detailed simulations, based initially on $\alpha$-viscosity
prescriptions, and more recently on three-dimensional
General-Relativistic magneto-hydrodynamics (for references see the
introductory part of this section~\ref{sec:neutron-star-neutron}),
show that neutrino winds from the accretion disk eject little mass and
that most of the material is ejected by viscous
heating~\citep[e.g.][]{Just.Bauswein.ea:2015}.  The results for disk
outflows by~\citet{Wu.Fernandez.Martinez.ea:2016} are displayed in
Fig.~\ref{fig_8}, which shows the integrated abundance pattern of all
tracer particles.  This underlines that, in principle, disk outflows
alone can produce the whole range of r-process nuclei, with a
significant production of $A \lesssim 130$ nuclei, also reaching the
third peak at $A = 195$ in most of the simulations. The detailed
results depend on the disk viscosity, the initial mass or entropy of
the torus, the black hole spin, and (of course) the nuclear physics
input. The latter is illustrated in Fig.~\ref{fig_8} that compares the
nucleosynthesis results for two different mass models,
FRDM~\cite{moeller95} and Duflo-Zuker~\citep{Duflo.Zuker:1995}. The
production of heavy ($A\gtrsim 195$) nuclei is also affected by
uncertainties of the disk properties discussed
above~\citep{Just.Bauswein.ea:2015,Wu.Fernandez.Martinez.ea:2016,Fernandez.Tchekhovskoy.ea:2019,Christie.Lalakos.ea:2019,Fujibayashi.Shibata.ea:2020}. Recent
$\alpha$-viscous simulations~\citep{Fujibayashi.Shibata.ea:2020} using
torus masses compatible with GW170817, $M\sim 0.1$--0.02~M$_\odot$,
predict relatively large $Y_e$ ejecta mainly due to the disk reaching
dynamical beta-equilibrium~\citep{Arcones.Martinez-Pinedo.ea:2010}
between electron and positron captures. The associated nucleosynthesis
is strongly suppressed in $A>130$ nuclei when compared with
solar. However, such a possible deficit can be counterbalanced by the
dynamic ejecta, as the total nucleosynthesis of the merger includes
the components of the dynamic ejecta, the neutrino wind, and the
accretion disk.

Nucleosynthesis studies in mergers are commonly based on simulation
data that follow the evolution of the ejecta for timescales shorter,
$\sim$~ms, than the r-process nucleosynthesis timescale,
$\sim$~s. This makes it necessary to extrapolate the time evolution of
thermodynamic properties like temperature and density in order to
follow the nucleosynthesis to completion. It is commonly assumed that
the expansion is homologous, $\rho \sim t^{-3}$, with the temperature
evolution determined by the nuclear energy production of the r-process. 
It originates mainly from $\beta$-decays and is in the range
$\dot{Q} \approx 1-4$~MeV~s$^{-1}$~nuc$^{-1}$ (see lower panel of
Fig.~\ref{fig:revol}). \citet{Rosswog.Korobkin.ea:2014} have performed
long-term simulations and found that the r-process energy release does
not qualitatively alter the properties of dynamic
ejecta. \citet{Wu.Fernandez.Martinez.ea:2016} found that r-process
heating can increase the amount of ejecta up to a factor 2 in viscous
outflows from accretion disks and remove an anomalously high abundance
of $A=132$ nuclei \citep[see Fig.~\ref{fig_8}
and][]{Lippuner.Fernandez.Roberts.ea:2017,Siegel.Metzger:2018}. R-process
heating can critically shape the
dynamics of marginally bound ejecta responsible for fall-back
accretion on timescales of seconds to
minutes~\citep{Metzger.Arcones.ea:2010,Desai.Metzger.Foucart:2019}. Late-time
fall-back accretion has been suggested as possible mechanism to
explain the extended X-ray emission observed in some short
GRBs~\citep{Rosswog:2007}. R-process heating on timescales of days to
weeks after the merger has been found responsible for powering the
``kilonova'' electromagnetic
emission~\citep{lipacz97,Metzger.Martinez.ea:2010} as will be
discussed in the next section.

\section{Electromagnetic signatures of r-process nucleosynthesis}
\label{sec:kilon-an-electr}

While we have evidence for the existence of some of the events listed
in the previous section~\ref{sec:astr-sites-their}, i.e.\ among the
sites of subsection~\ref{sec:massive-stars} possibly for
electron-capture supernovae
\citep[e.g.][]{Wanajo.Nomoto.ea:2009,Moriya.Tominaga.ea:2014}, for
supernovae resulting in magnetars
\citep[e.g.][]{Vink:2008,greiner15,Zhou.Vink.ea:2019,Beniamini.Hotokezaka.Horst.ea:2019},
and clearly for hypernovae and lGRB's
\citep[e.g.][]{Nomoto.Tanaka.ea:2010}, there exists no observational
evidence, yet, for their production of heavy r-process elements. This
is different for compact binary mergers
(subsection~\ref{sec:neutron-star-neutron}) since GW170817, a neutron
star merger with the combined mass $M_t$ of about 2.74~M$_\odot$
\citep{Abbott.Others:2017,Abbott.Abbott.ea:2017,Abbott.Abbott.ea:2018b}. The
observation of an electromagnetic counterpart delivered clear
indications for the existence of heavy r-process elements in the
ejecta~\citep{Metzger:2017,Tanaka.Utsumi.ea:2017,Villar.ea:2017}, and
even identified one element, Sr~\citep{Watson.Hansen.ea:2019}. This
will be discussed in detail below. By now, additional gravitational
wave observations point to further neutron star
mergers~\citep[e.g. GW190425 with
$M_t \sim 3.4$~M$_\odot$,][]{Abbott.ea:2020}, or even neutron
star-black hole merger candidates~\citep[e.g. GW190426 with $M_t$ in
excess of 7~M$_\odot$,][]{Lattimer:2019b,Abbott.Abbott.ea:2020}.  However, the latter two
events had no observed accompanying electromagnetic
counterpart~\citep{Hosseinzadeh.Cowperthwaite.ea:2019,Foley.Coulter.ea:2020,Ackley.Others:2020},
due to either non-existence or non-detection, related to a larger
distance and/or missing precise directions. This will hopefully change
with future GW events.

\begin{figure}[htb]
  \centering
  \includegraphics[width=\linewidth]{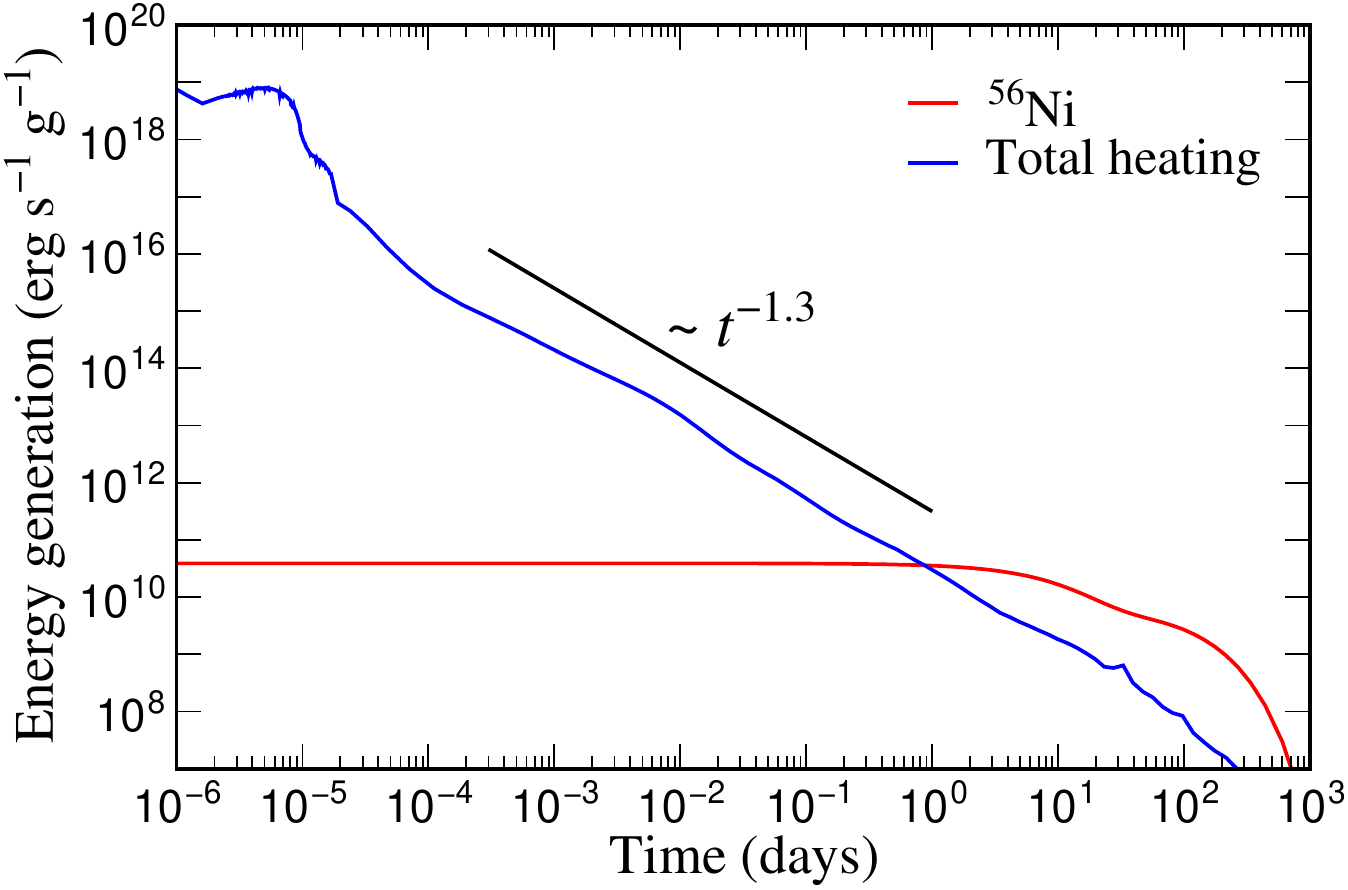}
  \caption{Specific energy generation rate, $\dot{Q}$, in r-process
    ejecta (black line). For comparison the energy production from the
    decay chain
    $^{56}\text{Ni} \rightarrow{}^{56}\text{Co}
    \rightarrow{}^{56}\text{Fe}$ (red) and the analytical estimate
    $\dot{Q} \sim t^{-1.3}$ are also shown~\citep[adapted
    from][]{Metzger.Martinez.ea:2010}. \label{fig:heating}}
\end{figure}

The r-process produces very neutron-rich unstable nuclei on timescales
of a few seconds that decay to stability by a combination of $\beta$,
$\alpha$ and fission decays. These decays produce large amounts of
energy and can lead to an observable electromagnetic emission. The
first suggestion of such an electromagnetic emission was due
to~\citet{Burbidge.Hoyle.ea:1956} who attributed type Ia supernova
light curves to the decay of $^{254}$Cf, produced by the
r-process. Today, we know that both type Ia and type II supernova
light curves are due mainly to the decay of $^{56}$Ni. The study of
light curves and spectra does not only constrain the nucleosynthesis
yields~\citep{Diehl.Timmes:1998,Seitenzahl.Timmes.Magkotsios:2014},
but also provides information about the physical parameters of the
progenitor system and the explosion
itself~\citep{Bersten.Mazzali:2017,Zampieri:2017}. This illustrates
the physics potential of an electromagnetic transient observation
associated with r-process ejecta. It can identify a site where the
r-process occurs~\citep{Metzger.Martinez.ea:2010}, serve as
electromagnetic counterpart to the gravitational wave detection
following a neutron star merger~\cite{Metzger.Berger:2012}, and the
nature of the merging system, NS-NS vs NS-BH, and
remnant~\citep{Margalit.Metzger:2019,Barbieri.Salafia.ea:2019,Kawaguchi.Shibata.Tanaka:2020,Zhu.Yang.ea:2020}. All
these aspects were confirmed by the electromagnetic transient
AT~2017gfo, following the gravitational wave event
GW170817~\citep{Abbott:2017}.

\begin{figure}[htb]
  \includegraphics[width=\linewidth]{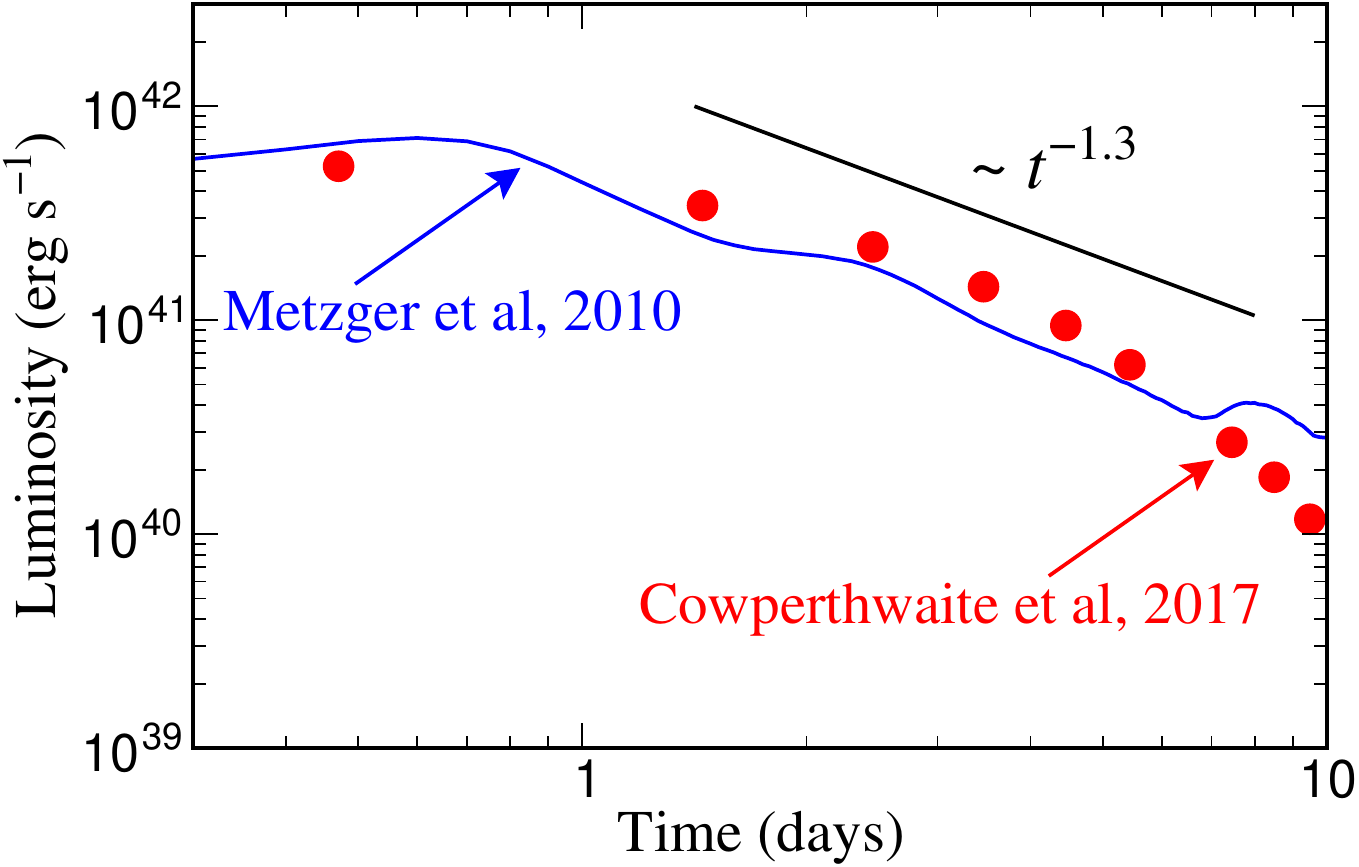}
  \caption{Bolometric light curve of the optical/infrared counterpart,
    AT 2017gfo, of GW170817 (red circles) from multi-band
    photometry~\citep{cowperthwaite17}, compared with the fiducial
    model of~\citet{Metzger.Martinez.ea:2010}. For comparison a line
    with the approximate power-law decay $\dot{Q} \sim t^{-1.3}$ for
    r-process heating (see
    Fig.~\ref{fig:heating}).\label{fig:lumi-vs-2010}}
\end{figure}

\citet{lipacz97} were the first to propose that radioactive ejecta
from a NS-NS merger could power a supernova-like transient. However,
they did not possess a physical model to describe the origin of the
radioactive heating, $\dot{Q}$, and considered two possible limiting
cases: an exponential-law decay and a power-law $\dot{Q} \sim
t^{-1}$. In both cases the normalization was left as a free
parameter. Hence, even if the model predicted the right timescale for
the peak luminosity, it could not determine the absolute luminosity,
spectral peak frequency and time-evolution of the luminosity. Indeed,
their fiducial model reached extremely high values of the luminosity
$\sim 10^{44}$~erg~s$^{-1}$ with a spectral peak in the
ultra-violet. \citet{Kulkarni:2005} considered two possible origins
for the heating: neutron and $^{56}$Ni decay; and named such events
``macronova''. \citet{Metzger.Martinez.ea:2010} were the first to
relate the late time radioactive heating to the decay of freshly
produced r-process nuclei. Based on heating rates derived
self-consistently from a nuclear reaction network, they showed that
the heating rate follows a power law at time scales of a day with a
steeper dependence ($\dot{Q} \sim t^{-1.3}$) than the one assumed
by~\citet{lipacz97}. As shown in Fig.~\ref{fig:heating}, the heating
evolves very differently for r-process material than for
supernova-like ejecta dominated by $^{56}$Ni. A power law dependence
is expected whenever the heating is dominated by a broad distribution
of nuclei all of them decaying exponentially. It can be understood
from basic physics of $\beta$-decay and the properties of neutron-rich
nuclei~\citep{Metzger.Martinez.ea:2010,Hotokezaka.Sari.Piran:2017}. A
similar dependence is found for the decay rate of terrestrial
radioactive waste~\citep{Way.Wigner:1948}.

The work of~\citet{Metzger.Martinez.ea:2010} predicted peak
luminosities $\sim 3\times 10^{41}$~erg~s$^{-1}$ for 0.01~M$_\odot$ of
ejecta, expanding at $v\sim 0.1c$, and a spectral peak at visual
magnitude. As such value corresponds to 1000 times the luminosity of
classical novae they named these events ``kilonova''.
Fig.~\ref{fig:lumi-vs-2010} compares their prediction with the
observation of AT~2017gfo~\citep{cowperthwaite17}. Similar results
were also found by~\citet{Roberts.Kasen.ea:2011} and
\citet{Goriely.Bauswein.Janka:2011}.

The physical processes determining the kilonova light curve
are~\citep[see][for reviews]{fernandez16,metzger17,Tanaka:2016}:

\paragraph{Radioactive heating.} The radioactive heating of r-process
products is expected to follow a power law whenever a large
statistical ensemble of nuclei is produced. This is the case for
ejecta with $Y_e \lesssim 0.2$. For higher $Y_e$ ejecta, the heating
rate has `bumps' as a function of time caused by being dominated by a
few
nuclei~\citep{grossman14,martin15,Lippuner.Roberts:2015,rosswog18,Wanajo:2018}.
However, when averaged over $Y_e$ distributions, as predicted by
simulations, the heating rate at timescales of days to a week (of
greatest relevance to determine the peak luminosity) varies only
slightly within a factor of a few for
$Y_e \lesssim 0.4$~\citep{Lippuner.Roberts:2015,Wu.Barnes.ea:2019}.
R-process nuclei decay in a variety of channels including
$\beta$-decay, $\alpha$-decay and fission. The energy production in
each individual channel is important as the absorption of the energy
depends on the decay products being electrons, photons, alphas and
fission products. For high $Y_e$ ejecta, heating is dominated by
$\beta$-decay and only electrons and photons are relevant with
neutrinos being just an energy loss.  For low $Y_e$ ejecta, actinides
are produced and alpha decay and fission can be substantial for the
energy production and sensitive to the underlying mass
model~\citep{barnes16,Rosswog.Feindt.ea:2017,Wu.Barnes.ea:2019}. The
exact form of the heating depends also on the time after the
freeze-out of neutron captures. At early times of a few hours the
heating may be dominated by neutron decay, assuming a substantial
amount of free neutrons in the outermost layers of the ejecta,
producing an ultraviolet/blue precursor to the kilonova
emission~\citep{Metzger.Bauswein.ea:2015,metzger17}. Early blue
emission can also originate from the hot cocoon that surrounds the GRB
jet as it crosses the ejecta~\citep{Gottlieb.Nakar.Piran:2018}. At
intermediate times up to $\sim 10$~d, $\beta$-decays dominate the
energy production~\citep{barnes16}. The heating rate follows a power
law ($\dot{Q}\sim t^{-1.3}$), provided that it is determined by the decay
of a statistical ensemble of nuclei. As the material expands and the
thermalization of the electrons becomes inefficient, the effective
heating rate, including thermalization effects (discussed later),
follows a power law $\sim t^{-n}$ with a power $n$ that depends on the
time evolution of the thermalization efficiency and the confinement of
electrons in the plasma due to magnetic
fields~\citep{Waxman.Ofek.Kushnir:2019,Kasen.Barnes:2019,Hotokezaka.Nakar:2020}.
At late times of up to 100~d, the heating is dominated by a few
decays~\citep[see][for a complete listing]{Wu.Barnes.ea:2019}, due to
the scarcity of nuclei with the appropriate half-life and hence the
heating can substantially differ from a power law dependence.

\paragraph{Thermalization efficiency.} At early times the ejected
material is very dense and the energy produced by radioactive decay,
except for neutrinos, is completely reabsorbed.  However, with
decreasing density an increasing fraction of the energy is lost, which
is incorporated normally via a time dependent thermalization
efficiency of the energy produced by radioactive
processes~\citep{barnes16}. This efficiency depends on bulk properties
of the ejecta like mass and velocity as they determine the evolution
of the density. It also depends on the presence of magnetic fields and
their geometry. Furthermore, it varies with the decay product and time
evolution of the heating for each particular decay
channel~\citep{Kasen.Barnes:2019}, i.e.\ whether we have a statistical
distribution of decaying nuclei or a heating dominated by a few
isotopes, which is probably more appropriate for late times. Earlier
works considered the thermalization of
$\gamma$-rays~\citep{Hotokezaka.Wanajo.ea:2016} and were later
extended to consider charged particles~\citep{barnes16}. This has been
recently extended to the case of a few decays dominating the
heating~\citep{Kasen.Barnes:2019,Wu.Barnes.ea:2019}. Qualitatively one
finds that the thermalization efficiency for $\gamma$-rays decreases
very rapidly and becomes negligible on timescales of a few 10's of
days. The thermalization efficiency for charged particles, and
particularly alphas and fission products, remains substantial at late
times. This makes kilonova light curves rather sensitive to the
heating contribution of alpha decays and
fission~\citep{barnes16,Rosswog.Feindt.ea:2017,Wu.Barnes.ea:2019,Zhu.Wollaeger.ea:2018,Vassh.Vogt.ea:2019,Giuliani.Martinez-Pinedo.ea:2019}.

\paragraph{Atomic opacities.} A significant electromagnetic luminosity
is only possible once the density decreases sufficiently that photons
can escape the ejecta on the expansion
time-scale~\citep{Arnett:1980,Arnett:1982}. Assuming a homogeneous
spherical distribution of ejecta with mass $M$, expanding homologously
with velocity $v$ and radius $R=v t$, the diffusion time scale of the
ejecta can be approximated as
$t_{\text{diff}} \approx \rho\kappa R^2/(3 c)$, with
$\rho=3 M/(4\pi R^3)$ the density and $\kappa$ the opacity of the
ejecta. Once the ejecta expand enough to become transparent, they
release line radiation. This occurs when the diffusion timescale,
$t_{\text{diff}}$, becomes comparable to the dynamical timescale,
$t=R/v$, and defines the time at which the maximum of the luminosity
is reached~\citep[see e.g.][]{Metzger.Martinez.ea:2010,fernandez16}:

\begin{eqnarray}
  \label{eq:tpeak}
  t_{\text{peak}} & \approx & \left(\frac{\kappa
                              M}{4\pi cv}\right)^{1/2} \\
                  & \approx & 1.5\ \text{days} \left(\frac{M}{0.01
                              M_\odot}\right)^{1/2} \left(\frac{v}{0.1
                              c}\right)^{-1/2} \left(\frac{\kappa}{\text{cm}^2
                              \text{g}^{-1}}\right)^{1/2} \nonumber
\end{eqnarray}

At timescales beyond the peak time the luminosity can be approximated
using Arnett's Law~\citep{Arnett:1980,Arnett:1982}:
$L(t) = M \dot{Q}_{\text{dep}}(t)$.  $\dot{Q}_{\text{dep}}$ is the
energy deposition rate, corrected by the thermalization efficiency, and
can be approximated as
$\dot{Q}_{\text{dep}} \approx \varepsilon\, \times 10^{10}\,
(t/\text{day})^{-\alpha}\ \text{erg}\ \text{s}^{-1} \text{g}^{-1}$,
with $\varepsilon < 1$ the thermalization efficiency. At peak time the
kilonova luminosity is given by:

\begin{eqnarray}
  \label{eq:lpeak}
  L_{\text{peak}} & \approx & 1.1\, \varepsilon \times 10^{41}\ \text{erg}\ \text{s}^{-1} 
  \\
                  & & \left(\frac{M}{0.01 M_\odot}\right)^{1-\alpha/2} \left(\frac{v}{0.1
                      c}\right)^{\alpha/2} \left(\frac{\kappa}{\text{cm}^2
                      \text{g}^{-1}}\right)^{-\alpha/2} \nonumber
\end{eqnarray}
The effective emission temperature can be obtained from the luminosity
using the Stefan-Boltzmann law that together with the Wien
displacement law gives the characteristic wavelength of the emission:

\begin{eqnarray}
  \label{eq:wavepeak}
  \lambda_{\text{peak}} & \approx & 514\ \text{nm} \\
                        & & \left(\frac{M}{0.01
                            M_\odot}\right)^{\alpha/8} \left(\frac{v}{0.1
                            c}\right)^{(2-\alpha)/8} \left(\frac{\kappa}{\text{cm}^2
                            \text{g}^{-1}}\right)^{(2+\alpha)/8} \nonumber
\end{eqnarray}

The above formulas illustrate several characteristic features of
kilonova light curves. Even if the emission mechanism is similar to
supernovae the typical ejecta mass is much smaller and the velocity
larger. The equations illustrate the important role played by the
opacity that is dominated by Doppler-broadened atomic line bound-bound
transitions~\citep{kasen13,Fontes.Fryer.ea:2015,Tanaka.Kato.ea:2018}. Ejecta
containing light r-process elements ($A\lesssim 140$) with $d$-shell
valence electrons possess an opacity
$\kappa \lesssim 1$~cm$^2$~g$^{-1}$. In this case the emission peaks
in the blue after about a day. This was, indeed, the case for
AT~2017gfo~\citep{nicholl17}. If the ejecta contain lanthanide or
actinide nuclei ($A\gtrsim 140$), then the optical opacity is very
high, $\kappa \gtrsim 10\text{--}100$~cm$^2$~g$^{-1}$, due to the
complex structure of $f$-shell valence electrons for these elements,
resulting in a dense forest of lines, and the emission shifts to the
red/infrared~\citep{kasen13,Barnes.Kasen:2013,Tanaka.Hotokezaka:2013,Fontes.Fryer.ea:2017,Fontes.Fryer.ea:2020}. Having
a kilonova observation, like in the case of AT~2017gfo, it is possible
to adjust the multi-wavelength evolution of the light curve using a
variation of the model described above and to determine the amount of
ejecta, velocity and opacity that is a proxy for the composition. As
discussed in section~\ref{sec:nsm}, to reproduce the AT~2017gfo
observations requires at least two different ejecta components, with
three-component models being slightly favored~\citep[see
e.g.][]{Villar.ea:2017}.  This result is consistent with the existence
of several ejecta components in mergers giving rise to different
nucleosynthesis products (see
section~\ref{sec:neutron-star-neutron}). The analysis of sGRB
observations~\citep{Wu.Macfadyen:2018} and the kilonova
transient~\citep{Perego.Radice.Bernuzzi:2017} favors an off-axis
viewing angle of $\sim 30^\circ \deg$. Hence, the early blue phase of
the kilonova light curve has been suggested to originate from
lanthanide-poor polar ejecta~\citep[see e.g.][]{Kasen.Metzger.ea:2017}
\citep[however, see][for an alternative
explanation]{Kawaguchi.Shibata.Tanaka:2018}. This result is consistent
with simulations that predict that weak processes, including
  electron (anti)neutrino absorption, drive the composition to
$Y_e \gtrsim 0.25$. It provides observational evidence of the
important role of neutrinos in determining the composition of the
ejecta. However, there is a tension between the velocity of the ejecta,
$v \approx 0.27\ c$, that is consistent with simulations of dynamical
ejecta, and the large ejecta mass,
${M}_{\text{ej}}\approx 0.020$~M$_\odot$, that is not.
Additional lanthanide-poor material is expected to originate from
the post-merger neutrino-wind ejecta. However, its velocity is
expected to be smaller, unless the wind is magnetically accelerated by
the strongly magnetized HMNS
remnant~\citep{Metzger.Thompson.Quataert:2018}. The amount of material
and velocity of material involved in the ``purple'' and ``red''
components suggest that they originate from post-merger outflows from
the accretion disk~\citep[see
e.g.][]{Kasen.Metzger.ea:2017}. Simulations predict that the ejecta
contain a broad distribution of $Y_e$ and are able to produce both
light and heavy r-process material including the lanthanides/actinides
necessary to account for the high
opacity~\citep{Just.Bauswein.ea:2015,Wu.Fernandez.Martinez.ea:2016}.

It is, indeed, the observation of the lanthanide-rich ``red'' emission
that provided the first observational evidence that neutron-star
mergers produce r-process nuclei. The only element identified in the
spectra is Sr~\citep{Watson.Hansen.ea:2019}, providing further
evidence that weak processes (enhancing $Y_e$) operate in the
ejecta and demonstrating that first r-process-peak elements are
produced in mergers. This is consistent with the inferred lanthanide
mass fraction
$X_{\text{lan}}\sim
10^{-3}$--$10^{-2}$~\citep{Kasen.Metzger.ea:2017,Tanaka.Utsumi.ea:2017,Waxman.Ofek.ea:2017}
that together with the assumption that the GW170817 yield follows
solar proportions requires the production of all r-process nuclei with
additional contributions of trans-iron
nuclei~\citep{Wu.Barnes.ea:2019}.  However, if GW170817 represents a
typical r-process yield from NS mergers, it suggest that an
alternative r-process site may be responsible for the r-process
abundances observed in r-enhanced metal poor stars~\citep[][see also
section~\ref{sec:supernova-vs-r}]{Ji.Drout.Hansen:2019}.

No direct spectroscopic evidence has been obtained pointing to the
production of heavy r-process elements. The high density of lines for
lanthanides/actinides together with the large velocities of the ejecta
produces line blending and smoothens the
spectra~\citep{chornock17}. This aspect has been used to determine the
velocity of the ejecta from spectroscopic information~\citep[see
e.g.][]{chornock17}. Nevertheless, the spectra present peaks that
may probe the abundance of further individual elements beyond
  Sr~\citep[see figure~4 of][]{Kasen.Metzger.ea:2017}. However,
uncertainties in current atomic data hinder a detailed spectral
analysis. The lanthanide/actinide opacities are uncertain because the
atomic states and line strengths of these elements are
not measured experimentally. Theoretically, such high-Z atoms
represent a challenging problem in many-body quantum mechanics, and
hence are based on statistical models that must be calibrated to
experimental
data~\citep{kasen13,Fontes.Fryer.ea:2015,Tanaka.Kato.ea:2018,Radziute.Gaigalas.ea:2020,Tanaka.Kato.ea:2020}. Beyond
identifying the line transitions themselves, there is considerable
uncertainty in how to translate these data into an effective
opacity. The commonly employed ``line expansion opacity''
formalism~\citep{Pinto.Eastman:2000a,Pinto.Eastman:2000b,Li:2019},
based on the Sobolev approximation and applied to kilonovae
by~\citet{Barnes.Kasen:2013} and~\citet{Tanaka.Hotokezaka:2013}, may
break down if the line density is sufficiently high that the
wavelength spacing of strong lines becomes comparable to the intrinsic
thermal width of the
lines~\citep{kasen13,Fontes.Fryer.ea:2015,Fontes.Fryer.ea:2017,Fontes.Fryer.ea:2020}.

Lacking a direct spectroscopic identification of the abundance of
individual elements, recent work has focused in identifying
fingerprints of heavy elements in kilonova light curves. Particularly
promising are late-time observations as the decay heating can be
dominated by a few
nuclei~\citep{Wu.Barnes.ea:2019}. \citet{Kasliwal.Kasen.ea:2019}
suggest heavy isotopes (e.g. $^{140}$Ba, $^{143}$Pr, $^{147}$Nd,
$^{156}$Eu, $^{191}$Os, $^{223}$Ra, $^{225}$Ra, $^{233}$Pa,
$^{234}$Th) with $\beta$-decay half-lives around
14~days. \citet{Wu.Barnes.ea:2019} have shown that at times of weeks to
months, the decay energy input may be dominated by a discrete number
of $\alpha$-decays, $^{223}$Ra (half-life $t_{1/2} = 11.43$~d),
$^{225}$Ac ($t_{1/2} = 10.0$~d, following the $\beta$-decay of
$^{225}$Ra with $t_{1/2} =14.9$~d), and the fissioning isotope
$^{254}$Cf ($t_{1/2} = 60.5$~d)~\citep[see
also][]{Zhu.Wollaeger.ea:2018}, which liberate more energy per decay
and thermalize with greater efficiency than $\beta$-decay products.
Late-time nebular observations of kilonovae, which constrain the
radioactive power, provide the potential to identify signatures of
these individual isotopes, thus confirming the production of heavy
nuclei. In order to constrain the bolometric light to the required
accuracy, multi-epoch and wide-band observations are required with
sensitive instruments like the James Webb Space Telescope.

An alternative mechanism to probe the in-situ production of r-process
nuclei is the identification of X-ray or $\gamma$-ray lines from their
decay similar to the observations of $^{44}$Ti $\gamma$-rays in Cas
A~\citep{Vink.Laming.ea:2001,Renaud.Vink.ea:2006} and
SN~1987A~\citep{Grebenev.Lutovinov.ea:2012}
remnants. \citet{Qian.Vogel.Wasserburg:1998,Qian.Vogel.Wasserburg:1999},
\citet{Wu.Banerjee.ea:2019}, and~\citet{Korobkin.Hungerford.ea:2020}
have provided estimates of $\gamma$-ray fluxes for several r-process
nuclei and~\citet{Ripley.Metzger.ea:2014} have extended those
estimates to X-ray lines. The predicted fluxes are too low to be
detected by current missions, however improvements in detection
techniques may allow for the first detection of a merger remnant in
our Galaxy~\citep[see][for a search strategy]{Wu.Banerjee.ea:2019}.

\section{Abundance evolution in the Galaxy and Origin of the
  r-process}
\label{sec:chemev}

In section~\ref{sec:astr-sites-their} we presented possible
astrophysical sites and the related abundance predictions.  This
section addresses some of the additional features like their
occurrence frequency and its time evolution throughout galactic
history, with the aim to provide an understanding of the impact of
these individual sites on the evolution of the Galaxy.  

\subsection{Supernova vs.\ r-process imprints in early galactic
  evolution}
\label{sec:supernova-vs-r}

Based on the nucleosynthesis predictions for (regular) core-collapse
and for type Ia supernovae, plus their occurrence rates, one finds
that the early phase of the evolution of galaxies is dominated by the
ejecta of (fast evolving) massive stars, i.e.\ those leading to
core-collapse supernovae. While there exist variations for the ejecta
composition of different progenitor masses, average abundance ratios
in the interstellar gas will be found after some time delay when many
such explosions and the mixing of their ejecta with the interstellar
medium have taken place.  These averaged abundance ratios reflect
ejecta yields integrated over the distribution of initial stellar
  masses (initial mass function, IMF).  Type Ia supernovae originate
from exploding white dwarfs in binary systems, i.e. (a) from
slowly evolving stars with initially less than 8~M$_\odot$ in
order to become a white dwarf and (b) requiring time delaying
mass transfer in a binary system before the type Ia supernova
explosion (unless they are produced by very rare collisions of
  white dwarfs).  Thus, such events are delayed in comparison to the
explosion of massive single stars.  Type Ia supernovae, which are only
important at later phases in galactic evolution, dominate the overall
production of Fe and Ni (typically 0.5--0.6~M$_\odot$ per event), but
are only minor contributors to intermediate mass elements
$Z=8$--22. As core-collapse supernovae produce larger amounts of O,
Ne, Mg, Si, S, Ar, Ca, Ti (so-called $\alpha$-elements) than Fe-group
nuclei like Fe and Ni (only of the order 0.1~M$_\odot$), their average
ratio of $\alpha$/Fe is larger than the corresponding solar ratio.

\begin{figure}[htb]
  \centering
  \includegraphics[width=\linewidth]{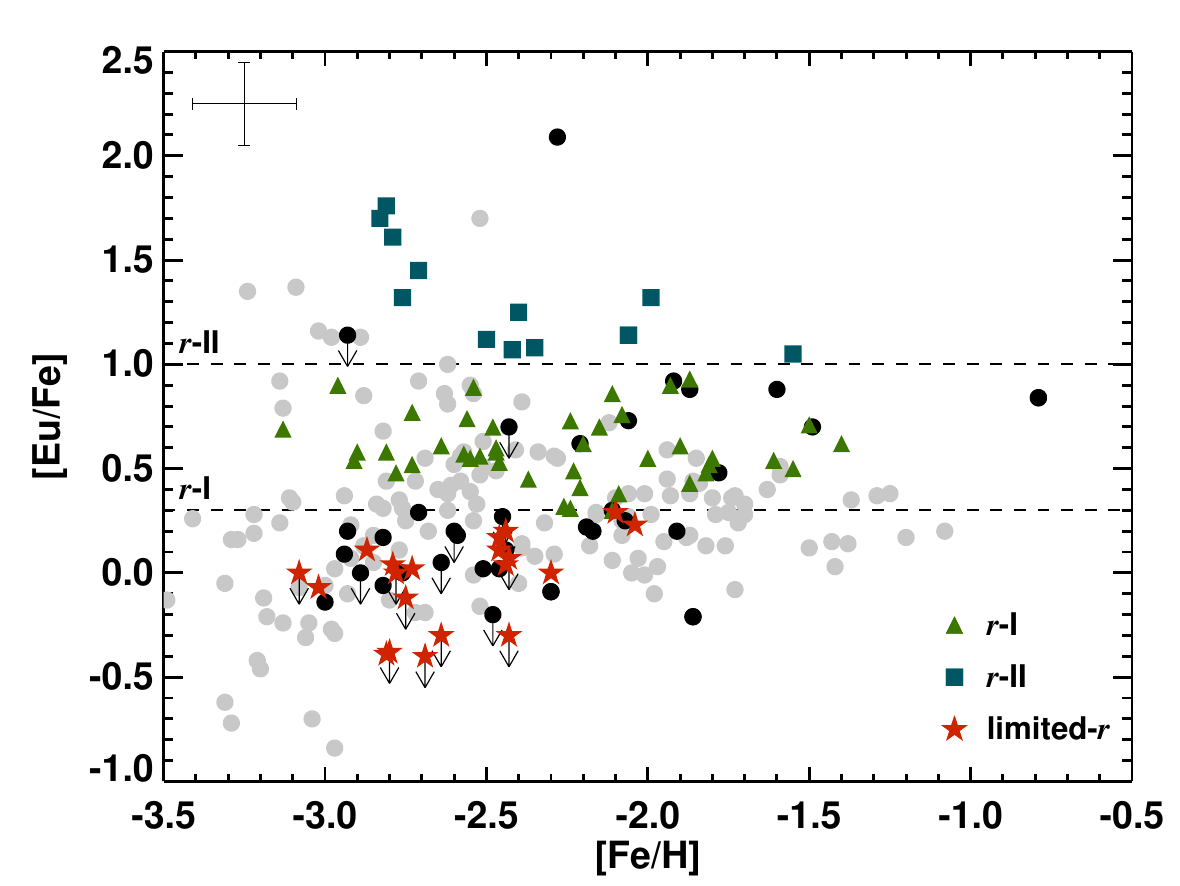}
  \caption{Derived [Eu/Fe] abundances as a function of
    metallicity~\citep[][reproduced by permission of the
    AAS]{Hansen.Holmbeck.ea:2018}: r-I stars (green triangles), r-II
    stars (blue squares), limited-r (red stars), and non
    r-process-enhanced stars (black dots), see classifications defined
    in section~\ref{sec:observations-jjc-cs}; upper limits are shown
    with black arrows. Grey light dots refer to an earlier
    overview~\citep{Roederer.Preston.ea:2014}.}
  \label{HansenEu}
\end{figure}

For most stars, with the exception of evolved stars which blew off
part of their envelope by stellar winds or stars in binary systems
with mass exchange, their surface abundances represent the composition
of the interstellar gas out of which they formed.  Thus, we can look
back into the early history of the Galaxy via the surface abundances
of unevolved low-mass stars, witnessing the composition of the
interstellar medium at the time of their birth.  In Fig.~\ref{mgeufe}
of section~\ref{sec:stellar-abunds} some of these aspects were
displayed, with [Mg/Fe] plotted as a function of metallicity [Fe/H]
for stars in our Galaxy. For Mg (a typical $\alpha$-element) one
sees --- with a relatively small scatter --- a flat value of [Mg/Fe]
between 0.3 and 0.5 up to $\text{[Fe/H]} \le -1$, which decreases down
to solar values at $\text{[Fe/H]}=0$. This can be explained by the
early appearance of core-collapse supernovae from fast evolving
massive, single stars, producing on average $\text{[Mg/Fe]} = 0.4$
\citep[see
e.g.][]{Woosley.Heger.Weaver:2002,Woosley.Heger:2007,limongi18} before
type Ia supernovae set in. The properties of the latter have been
reviewed in
\citet{hillebrandt13,maoz14,Goldstein.Kasen:2018,Livio.Mazzali:2018}
as well as their nucleosynthesis properties
\citep{seitenzahl17,Nomoto.Leung:2017a,Seitenzahl.Chavamian.ea:2019}. These
basic features of galactic evolution have been understood reasonably
well for the majority of
elements~\citep{matteucci86,Timmes.Woosley.Weaver:1995,nomoto13},
while open questions remain in stellar evolution and supernova
explosion mechanisms. This includes the question of the role of more
massive stars, probably ending as black holes
\citep{Heger.Fryer.ea:2003,Ertl.Janka.ea:2016,Sukhbold.Ertl.ea:2016,thielemann18,Ebinger.Curtis.ea:2019,Ebinger.Curtis.ea:2020,Ertl.Woosley.ea:2020},
and for sufficiently high angular momentum related to so-called
hypernovae/long-duration gamma-ray bursts (see
subsection~\ref{sec:massive-stars}) or even more massive pair
instability supernovae.

\begin{figure*}[htb]
  \centering
  \includegraphics[width=0.9\linewidth]{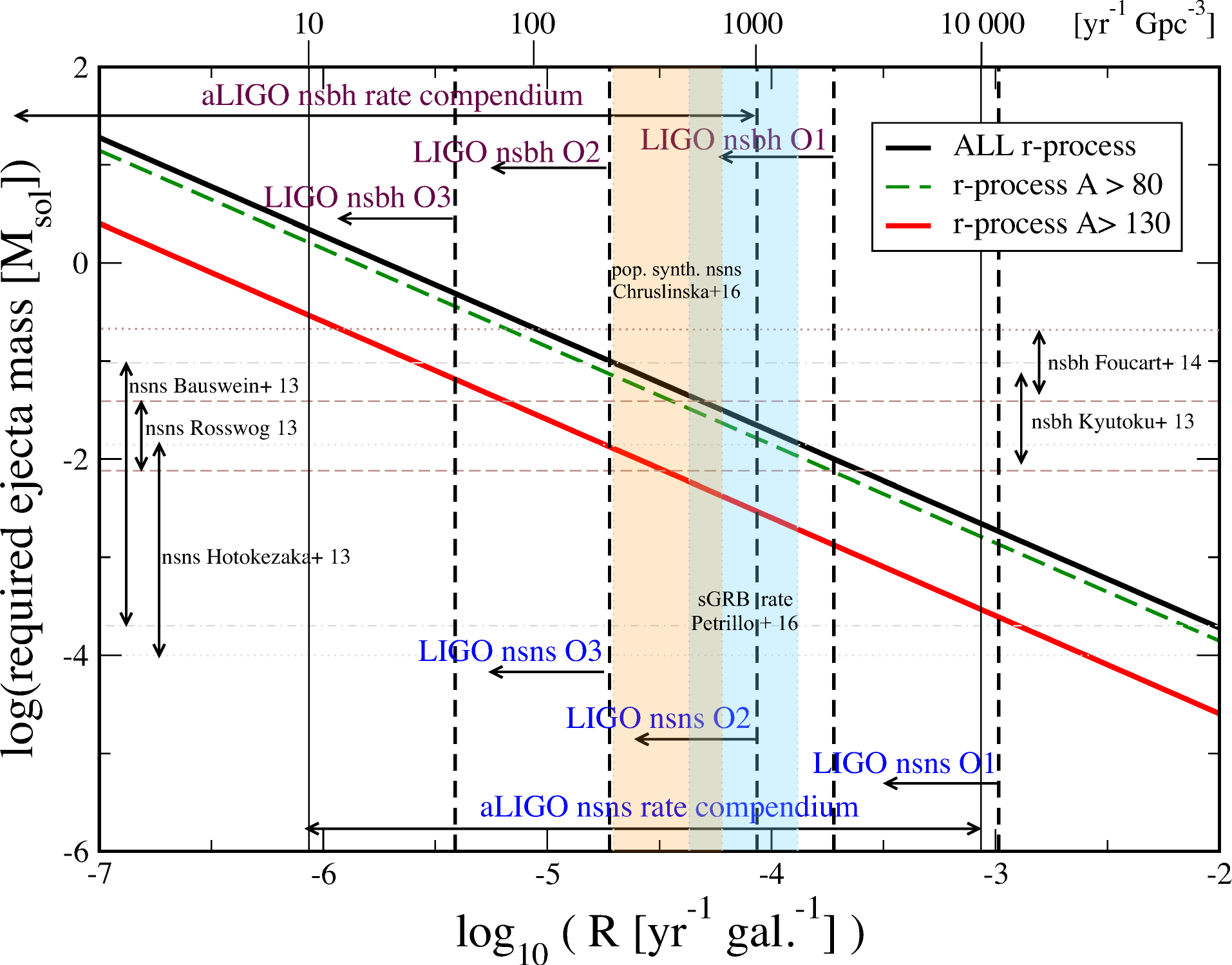}
  \caption{Figure taken from \citet{Rosswog.Feindt.ea:2017}
    (reproduced with permission of IOP Publishing), indicating the
    required r-process ejecta masses as a function of the occurrence
    frequency of the production site (for references we refer to the
    original paper). The figure shows that on a typical SN frequency
    of $10^{-2}$~yr$^{-1}$ about $10^{-4}$ to $10^{-5}$~M$_\odot$ of r-process
    matter would need to be produced, for binary merger ejecta with
    about $10^{-2}$~M$_\odot$ the frequency must be rarer by a factor
    of 100 to 1000, and if 1~M$_\odot$ of r-process matter would be
    ejected in specific events, the frequency must be again lower by
    another factor of 100.  \label{rosswogr}}
\end{figure*}

The solar abundance of Eu is more than 90\% dominated by those
isotopes which are produced in the r-process
\citep{bisterzo15,bisterzo17}. Therefore it is considered as a major
r-process indicator.  The ratio Eu/Fe in the Galaxy, already displayed
in Fig.~\ref{mgeufe} of section~\ref{sec:stellar-abunds} and its
recent update in Fig.~\ref{HansenEu} (above), shows a huge scatter by
more than two orders of magnitude at low metallicities, corresponding
to very early galactic evolution. While the evolution of the average
ratio resembles that of the alpha elements (see Fig.~\ref{mgeufe}),
being of a core-collapse supernova origin and also experiencing a
decline to solar ratios for $\text{[Fe/H]}\ge -1$~\citep[for similar
trends in Mo and Ru see recent observations
by][]{Mishenina.Pignatari.ea:2019}, it is far more complex to
understand Eu and other r-process dominated elements than
$\alpha$-elements like Mg. This is also true for elements whose solar
abundances are not dominated by the r-process, but which show a large
scatter at low metallicities as well, probably also related to
r-process contributions~\citep[e.g.~for Sr and Ba, see][and references
therein]{Mishenina.Pignatari.ea:2019a,Hill.Skuladottir.ea:2019}.  In
this section, we will discuss the suggested origins for the r-process
and the possibility of their discrimination.  A large scatter seems to
indicate a not yet well mixed or averaged interstellar medium,
permitting us to actually see the abundance patterns of individual
events.  The approach to an average [Eu/Fe] value with a small scatter
is only observed in the interval $-2\le\text{[Fe/H]}\le -1$. For
[Mg/Fe] (but also other alpha elements and e.g. Zn and Ge), produced
by supernovae, the approach to average values occurs already at about
$\text{[Fe/H]}=-3$ (see Figs.~\ref{mgeufe} and~\ref{HansenEu}). An
obvious conclusion from this would be that r-process events occur at a
much lower rate than supernovae.  In order to be consistent with total
solar abundances this would need to be compensated by larger amounts
of their ejecta (see Fig.~\ref{rosswogr}).  If the observed abundance
ratio of an r-process element over Fe ([r/Fe]), for example [Eu/Fe],
scatters at low metallicities due to individual events, this could
have two origins: (a) the pollution varies dependent on the birth
location of the observed star with respect to the r-process event
and/or (b) the strength of individual r-process events varies. There
exists also the option that a high-frequency weak r-process site,
related to supernovae, is responsible for the ``limited-r'' sample of
Fig,~\ref{HansenEu}, which only shows a small scatter at low
metallicities.

A further interesting aspect of this analysis is related to the
question whether r-process elements are correlated or not correlated
with other nucleosynthesis products, in order to determine whether
they were co-produced in the same nucleosynthesis site or require a
different origin.  \citet{Cowan.Sneden.ea:2005} compared the
abundances of Fe, Ge, Zr, and r-process Eu in low metallicity
stars. They found a strong correlation of Ge with Fe, indicating the
same nucleosynthesis origin (core-collapse supernovae), a weak
correlation of Zr with Fe, indicating that other sites than
core-collapse supernovae (without or low Fe-ejection) contribute as
well, and no correlation between Eu and Fe, pointing essentially to a
pure r-process origin with negligible Fe-ejection. More recent data
from the SAGA and JINA database~\citep{Sagadatabase,JINAbase:2018} permit a weak correlation
for $\text{[Eu/Fe]}<0.3$, i.e.\ for stars with lower than average
r-process enrichment. Interpreted in a straight-forward way this would
point to a negligible Fe/Eu ratio (in comparison to solar ratios) in
the major r-process sources, while a noticeable co-production of Fe
with Eu is possible in less strong r-process sources, e.g.\ possibly
with a weak r-process. Such cases could again be identified with the
stars labeled ``limited-r'' in Fig.~\ref{HansenEu}.  Not focusing on
Eu as a single r-process indicator, \citet{Ji.Drout.Hansen:2019}
looked for low metallicity stars ($\text{[Fe/H]}<-2.5$, with
compositions indicating an r-process origin, $\text{[Eu/Ba]}>0.4$), at
the typical lanthanide (plus actinide) fraction $X_{\text{La}}$ among
the global r-process element distribution. They found for the bulk of
low metallicity stars $\log X_{\text{La}}\approx -1.8$ and for the
most r-process enriched stars $\log X_{\text{La}}>-1.5$. This might
hint at different sources.

Without pursuing this aspect further at the moment, we list here
preliminary conclusions for the sites discussed in
section~\ref{sec:astr-sites-their} (see references therein):

\begin{enumerate}
  \renewcommand{\theenumi}{\alph{enumi}}

\item Electron-capture supernovae can possibly produce a weak
  r-process, not a strong one. If their existence is not ruled out by
  recent investigations (see section~\ref{sec:observations-jjc-cs})
  and they take place for stars in the interval of 8 to 10~M$_\odot$
  of the initial mass function, they are probably not rare. This
  contradicts a large scatter in [Eu/Fe], but they could be candidates
  for ``limited-r'' observations. \label{item:ecsn}
  
\item The neutrino-induced processes in He-shells of low-metallicity
  massive stars would be frequent events at low metallicities, and
  thus could not lead to a large scatter of e.g. [Eu/Fe]. In addition,
  the related abundance peaks would not be consistent with a strong
  r-process.

\item The regular neutrino-driven core-collapse SNe produce Fe, but at
  most a weak r-process (for an extended set of references see
  section~\ref{sec:neutrino-winds-from}). They are excluded as site of
  a strong r-process, because they do not produce the correct
  abundance pattern and would also be too frequent, not permitting a
  large scatter in [Eu/Fe] at low metallicities. However, they could
  be candidates for ``limited-r'' observations.

\item It is not known if quark deconfinement supernovae exist. While
  present model predictions do not yield a full strong r-process, the
  production of elements up to the actinides is possible, however,
  with the heaviest elements strongly reduced.

\item Magneto-rotational supernovae, starting from a variety of
  initial magnetic fields, possibly enhanced via magneto-rotational
  (MRI) instabilities, can produce magnetars. While more than 10\% of
  neutron stars seem to be born as
  magnetars~\citep{Beniamini.Hotokezaka.Horst.ea:2019}, only
  progenitors with pre-collapse magnetic fields of the order
  $10^{12}$--$10^{13}$~G and fast rotation can lead to fast jet
  ejection with a strong r-process composition. Smaller fields result
  in a weak or no r-process, i.e.\ following a transition to regular
  CCSNe. To be consistent with solar r-process abundances a
  core-collapse supernova with these extreme initial conditions should
  represent 1/100 to 1/1000 of regular core-collapse supernovae (see
  Fig.~\ref{rosswogr}), and produce also a large [Eu/Fe] scatter.
  Such events still require observational
  confirmation.  \label{item:magneto}

\item Collapsars, (i.e.\ high mass stars leading to central black
  holes, long-duration gamma-ray bursts, hypernovae, and accrection
  disk outflows) could also be consistent with the overall r-process
  production in the Galaxy, if they would occur rarer than
  core-collapse supernovae by more than a factor of
  1000~\citep{Siegel:2019}.  In such a scenario they would coproduce
  Fe ($\sim 0.5$~M$_\odot$) and r-process matter ($> 0.1$~M$_\odot$),
  but with negligible ratios in comparison to solar (i.e.\ a ratio of
  $<$5 in mass and about less than 2 in abundances vs.\ about 1000 in a
  solar composition), and thus not lead to a visible correlation,
  consistent
  with~\citet{Cowan.Sneden.ea:2005}. \label{item:collapsars}

\item Compact binary mergers lead to r-process ejecta masses of the
  order $10^{-2}~$M$_\odot$ and small occurrence frequencies (less
  than 1 in 100 CCSNe), similar to the ones required for
  case~(\ref{item:magneto}). Their existence and their contribution to
  heavy elements is observationally proven via gravitational wave,
  (short) GRB, as well as macronova/kilonova observations.  Whether
  the ejected composition is on average consistent with overall solar
  r-abundances will have to be seen in the future. \label{item:cbm}
\end{enumerate}

Summarizing the properties of the events listed above,
the following sites remain as strong r-process candidates:
(\ref{item:magneto}) magneto-rotational jet supernovae,
(\ref{item:collapsars}) collapsars, and (\ref{item:cbm}) compact
binary mergers. Of these (\ref{item:magneto}) and
(\ref{item:collapsars}) would belong to massive stars, i.e.\ occurring
during the earliest instances of galactic evolution. (\ref{item:cbm})
is related to the coalescence of compact objects resulting from the
prior collapse of massive stars and would therefore experience a
delay in their appearance. 

\citet{Macias.Ramirez:2019} suggested a further test to be fulfilled
by the site: the maximum pollution a star of the next generation would
experience, if it is born from a remnant of such an r-process event. A
Sedov-Taylor blast wave of an explosion with $10^{51}$~erg results in
mixing with about $5\times 10^4$~M$_{\odot}$ of interstellar medium
and about $5\times 10^5$~M$_{\odot}$ for the more energetic explosions
of collapsars.  \citet{Montes.Ramirez.ea:2016} come for compact binary
mergers to a similar conclusion as for supernovae. If applying this to
the cases~\ref{item:magneto}, \ref{item:collapsars},
and~\ref{item:cbm}, one would find maximum values of
$\text{[Eu/Fe]}>3$ for collapsars and MHD jet supernovae and
$\text{[Eu/Fe]}\sim 2.3$ for compact binary mergers, appearing at
$\text{[Fe/H]}\approx -3.4$, $-3.9$, and $-2.6$, if the explosions
occurred in a pristine, previously unpolluted ISM (except for the Fe
from two prior CCSNe in case of compact binary mergers). When
comparing this to Fig.~\ref{HansenEu} it might argue against the first
two sites. The question is, however, whether Fe-production by earlier
CCSNe could have reduced [Eu/Fe]. In a similar way uncertainties in
the Fe contribution of the CCSN progenitors of case~\ref{item:cbm} or
further mixing processes could reduce [Fe/H].

The following subsection will address the question how rare and
frequent events can be modeled consistently in galactic chemical
evolution.  Another aspect is how actually early galactic evolution
takes place. There exist indications that (ultra-faint) dwarf galaxies
are the earliest building blocks of galactic evolution and their
merger will finally lead to the evolution of the early Galaxy as a
whole. Due to different gas densities they might experience different
star formation efficiencies and due to a low gravitational pull they might
lose explosive ejecta more easily. This can have an effect on the
point in time (and metallicity) when the first imprints of explosive
ejecta can be observed. These features will be addressed as well.

\subsection{Galactic Chemical Evolution Modelling}

\subsubsection{Homogeneous evolution models}

In chemical evolution models of galaxies it is still common to use the
instantaneous mixing approximation (IMA), i.e.\ assuming that ejecta
compositions were instantaneously and completely mixed throughout the
Galaxy.  Neglecting this complete mixing can explain radial gradients,
but would still assume mixing within large and extended volumes (e.g.\
radial shells). Further developments included infall of primordial
matter into and outflow of enriched material out of the
Galaxy~\citep[for a review of early investigations see
e.g.][]{Audouze:1976,Tinsley:1980}. When taking into account that
(explosive) stellar ejecta enter the interstellar medium (ISM) delayed
with respect to the birth of a star by the duration of its stellar
evolution, detailed predictions for the time evolution of element
abundances can be made. Based on nucleosynthesis predictions for
stellar deaths, a number of detailed analyses have been performed,
from light elements up to the Fe-group
\citep[e.g.][]{matteucci86,Wheeler.Sneden.Truran:1989,Timmes.Woosley.Weaver:1995,Matteucci.Chiappini:2001,Kobayashi.Umeda.Nomoto.ea:2006,Pagel:2009,Matteucci:2012,nomoto13}. Such
approaches have also been applied to understand the enrichment of
heavy elements in the Galaxy (including r-process contributions) as a
function of time or metallicity [Fe/H] \citep[see
e.g.][]{Travaglio.Galli.Gallino.ea:1999,Ishimaru.Wanajo:1999,Dedonder.Vanbeveren:2004,Wanajo.Ishimaru:2006,Matteucci.Romano.Arcones.ea:2014,ishimaru15,Vangioni.ea:2016,Cote.Belczynski.ea:2017,hotokezaka18,Cote.Fryer.ea:2018,Cote.Eichler.ea:2019,Schonrich:2019,Siegel:2019,Grisoni.Cescutti.ea:2020}.

The IMA simplifies a chemical evolution model in terms of mass
movement. In detail, all event outputs are expected to cool down and
mix with the surrounding ISM instantaneously. Thus, a problem of this
simplification is: All stars born at a given time inherit the same
averaged abundance patterns of elements and therefore it is impossible
to reproduce a scatter in the galactic abundances, which is a crucial
aspect, especially at low metallicities. As a consequence a unique
relation between galactic evolution time and metallicity [Fe/H]
emerges, and for each [Fe/H] only a mean value of [X/Fe] (X being the
element of interest to follow in chemical evolution) is obtained.  The
IMA approach can be used to get a quick overview of the trends in
chemical evolution with a considerably lower computational effort for
such a model, and is probably approximately valid, also in case of
rare r-process events, for $\text{[Fe/H]}>-2$. For a detailed study of
especially early chemical evolution, including the reproduction of
spreads in abundance ratios due to local inhomogeneities, however,
this approach is not sufficient.

\subsubsection{Inhomogeneous galactic chemical evolution}

Local inhomogeneities can be produced if only limited amounts of ISM
are polluted by and mixed with the ejecta of an event. The latter
effect is of essential importance, especially at low metallicities,
where portions of the ISM are already polluted by stellar winds and
supernovae, and others are not. Inhomogeneous mixing could produce
large element ratios in strongly polluted areas by only one or few
events. This means that the scatter in [X/Fe] at low metallicities can
be a helpful asset in hinting at the origin of element
X. Inhomogeneous mixing can experience similar [Fe/H] values in
different locations of the Galaxy at different times or different
[Fe/H] values at the same time. In addition, different portions of the
ISM are polluted by different types of events, leading to a scatter at
the same metallicity, which can in fact be utilized as a constraint
for these different stellar ejecta.  This is especially the case in
the very early galactic evolution ($\text{[Fe/H]}\le-2.5$), when
locally (out of a whole initial mass function, IMF) only a few stars
with varying masses might have exploded and imprinted their stellar
neighborhood with their ejecta.  Thus, rare events, which produce
large amounts of element X, would cause a large scatter, being helpful
to identify the production site.  Therefore, more advanced chemical
evolution studies revoked the IMA
\citep[e.g.][]{Chiappini:2001a,Recchi:2001,Argast.Samland.ea:2004,Spitoni:2009,Recchi:2009,vandervoort15,cescutti15,wehmeyer15,shen15,hirai15,hirai17,Haynes.Kobayashi:2019,Wehmeyer.ea:2019,VandeVoort.ea:2019}.
For the reasons summarized above, especially for the abundance
evolution of r-process elements like Eu, such inhomogeneous chemical
evolution models are well suited.

While some of the models mentioned here are of a more stochastic
nature, \citet{Minchev.ea:2014} started with truly chemo ``dynamical''
galactic evolution models. These, as well as
e.g. \citet{vandervoort15,hirai15,shen15,Kobayashi:2016,hirai17,Haynes.Kobayashi:2019},
are based on smoothed-particle hydrodynamics (SPH) simulations. They
can model in a self-consistent way massive mergers of galactic
subsystems (treated as infall in simpler models), energy feedback from
stellar explosions causing outflows (and introduced as such in simpler
models), radial migrations in disk galaxies, mixing and diffusion of
matter in the ISM, and the initiation of a star formation dependence
with progenitor mass. Thus, these global SPH approaches are on the one
hand most suited to model such environments. However, the mass and
smoothing length utilized for the SPH particles will also determine
the resolution, as all matter within one particle is treated in a
homogeneous fashion. This acts like an artificial mixing on such
scales which are not necessarily related to the real mixing
mechanisms.  All events within the total mass of one SPH particle are
treated within a homogeneous reservoir in this approach. Such effects
go into the direction of an IMA on related scales, if they are larger
than a Sedov blast wave which mixes only with a limited amount of ISM
(see subsection~\ref{sec:supernova-vs-r}).
The
chemodynamic code AREPO is probably the most advanced one with
the highest resolution, including also MHD and large-scale mixing
effects self-consistently~\citep{VandeVoort.ea:2019}, however, still
not resolving Sedov-Taylor blast wave scales. This is done in simpler
stochastic approaches \citep[e.g.][]{wehmeyer15,Wehmeyer.ea:2019}
which, however, lack most of the advances included in chemodynamic codes. 

Within all these approaches a challenge remains: How to model
substructures of only about $10^4$~M$_\odot$, observed as ultra-faint
dwarf (UFD) galaxies, possibly being building blocks of the early
Galaxy. In a superposition of IMA treatments \citet{ojima18},
utilizing a variation of sizes of such galactic substructures, made
use of related different star formation rates and different outflows
according to their gravity, and added stochastically neutron star
mergers in these substructures within the range of possible
coalescence delay times. The merging of these substructures is
expected to represent eventually the early Galaxy as a whole.
\citet{Tsujimoto.Shigeyama:2014} discuss how Eu of neutron star merger
ejecta is dispersed in UFDs, where --- due to ejection velocities of
0.1--0.2$c$ --- such heavy elements can experience cosmic-ray type
propagation rather than follow a hydrodynamical treatment and mix
throughout the whole UFD\@. They applied this effect within a
chemodynamical model of hierarchical galaxy formation/evolution
in~\citet{Komiya.Shigeyama:2016}.

The treatment of compact binary mergers needs to connect the early
supernova events that produce the neutron stars and Fe ejecta with the
delayed merger event that produces the r-process ejecta.  Special
binary evolution aspects might apply for such close binary systems and
the resulting supernovae~\citep{Mueller18,MuellerB19}, not necessarily
accompanied by the same amount of Fe/Ni ejecta as for regular CCSNe.
Since explosive events give rise to nucleosynthesis inside a supernova
remnant bubble (given by a Sedov-Taylor blastwave), the abundances of
metals are higher than outside such a remnant. A star which is born
later \emph{inside} such a remnant will inherit more metals than a
star born \emph{outside}. Thus, it is of high importance, \emph{where}
a star is born during galactic evolution, especially in the early
phases. If the later merger, producing large amounts of r-process
matter, is occurring within the supernova remnant bubble, the [Fe/H]
has been set already by the earlier supernova explosions and the
related [Eu/Fe] ratios will appear at the appropriate [Fe/H]. The main
challenge is to have a large [Eu/Fe] ratio at the lowest
metallicities, which could be achieved in the following ways: (a)
neutron star kicks during the supernova explosions act in such a way
that the actual neutron star merger takes place outside the initial Fe
pollution by the preceding supernovae, (b) neutron star-black hole
mergers would have only experienced the Fe-ejecta of one supernova,
(c) large-scale turbulent mixing could lead to the dilution of Fe on
timescales shorter than the coalescence delay time of the mergers.
Such effects are not (yet) necessarily treated correctly by present
models.

\subsection{Connecting observational constraints on r-process
  abundances with different astrophysical sites}
  
In subsection~\ref{sec:supernova-vs-r}, we listed possible production
sites for a strong r-process. They need to fulfill the observational
constraints: (i) in order to lead to a large scatter of [r/Fe] at low
metallicities these events must be rare in comparison to regular
core-collapse supernovae (this does not exclude the latter from being
the site of a weak r-process), and (ii) if they should be the dominant
site responsible for the solar r-abundances, the combination of their
ejecta mass and occurrence frequency must be able to match this
requirement. Three of the listed possible sites may fulfill both
criteria: (\ref{item:magneto}) magneto-rotational jet supernovae,
(\ref{item:collapsars}) collapsars, and (\ref{item:cbm}) compact
binary mergers. Site (\ref{item:cbm}) is a rare, but observed event
with ejecta amounts consistent with solar abundances. Sites
(\ref{item:magneto}) and (\ref{item:collapsars}) are potential
r-process scenarios but still lacking observational confirmation. If
the required high rotation rates and extreme magnetic fields for
magneto-rotational jet supernovae can exist, they would eject similar
amounts of r-process matter as binary mergers. These requirements
would also make them rare events.  (\ref{item:collapsars}) Collapsars,
also known as hypernovae or observed as lGRBs, have been related to
high $^{56}$Ni ejecta, but recently also postulated to eject more than
0.1~M$_\odot$ of r-process
matter~\citep{Siegel.Metzger:2017,Siegel.Barnes.Metzger:2019,Janiuk:2019}. In
such a case, these events should be very rare, even a factor of 10 or
more rarer than compact binary mergers.

A further requirement, in addition to the two discussed above (rarity
and reproducing the total amount of solar r-abundances), is that
galactic evolution modeling should reproduce the observed metallicity
(or time) evolution. As discussed in the previous subsections,
homogeneous approaches with IMA are justified if applied for
metallicities $\text{[Fe/H]}>-2$. This has been especially utilized
for testing the distribution of delay times for neutron star mergers
after the formation of a binary neutron star system. Early
investigations utilized coalescence delay times with a narrow
spread. Population synthesis studies, consistent with the occurrence
of short-duration gamma-ray bursts (sGRBs, related to compact binary
mergers) indicate that the possible delay times follow a distribution
with a large spread, ranging over orders of magnitude with a $t^{-1}$
behavior. Based on such a behavior, studies with the IMA modeling of
chemical
evolution~\citep{Cote.Belczynski.ea:2017,hotokezaka18,Cote.Fryer.ea:2018,Cote.Eichler.ea:2019,Siegel:2019}
come to the conclusion that mergers would not be able to reproduce the
galactic evolution for metallicities $\text{[Fe/H]}>-2$, including the
decline of [Eu/Fe] at $\text{[Fe/H]}=-1$. This would require either a
different delay time
distribution~\citep{Vigna-Gomez.ea:2018,Beniamini.Piran:2019,Simonetti.Matteucci.ea:2019}
or an additional source for the main, strong, r-process.
\citet{Schonrich:2019} suggested another solution: star formation
takes only place in cooled regions of the ISM, i.e.\ not all recently
ejected matter can already be incorporated and stars contain lower
metallicities [Fe/H] than the overall ISM at the time of their
birth. This shifts e.g. [Eu/Fe] ratios to lower [Fe/H] and has a
similar effect as a steeper delay-time distribution.

In order to address the challenges of explaining the [r/Fe] scatter at
metallicities $\text{[Fe/H]}<-2$, inhomogeneous chemical evolution
studies are needed and have been
implemented~\citep{Argast.Samland.ea:2004,vandervoort15,shen15,wehmeyer15,hirai15,mennekens16,Komiya.Shigeyama:2016,hirai17,Wehmeyer.ea:2019,Haynes.Kobayashi:2019,VandeVoort.ea:2019}. At
these low metallicities, there is an important difference between the
scenarios (\ref{item:magneto}), (\ref{item:collapsars}) and
(\ref{item:cbm}): MHD supernova and collapsars result from the final
phases of a single massive star, while in the merger scenario two
supernova explosions are required before the merger happens after a
delay. Scenarios (\ref{item:magneto}) and (\ref{item:collapsars}) can
act at earliest times in galactic evolution, while we have to examine
which effect the delay time in scenario (\ref{item:cbm}) plays.

\begin{figure}[htb]
  \includegraphics[width=\linewidth]{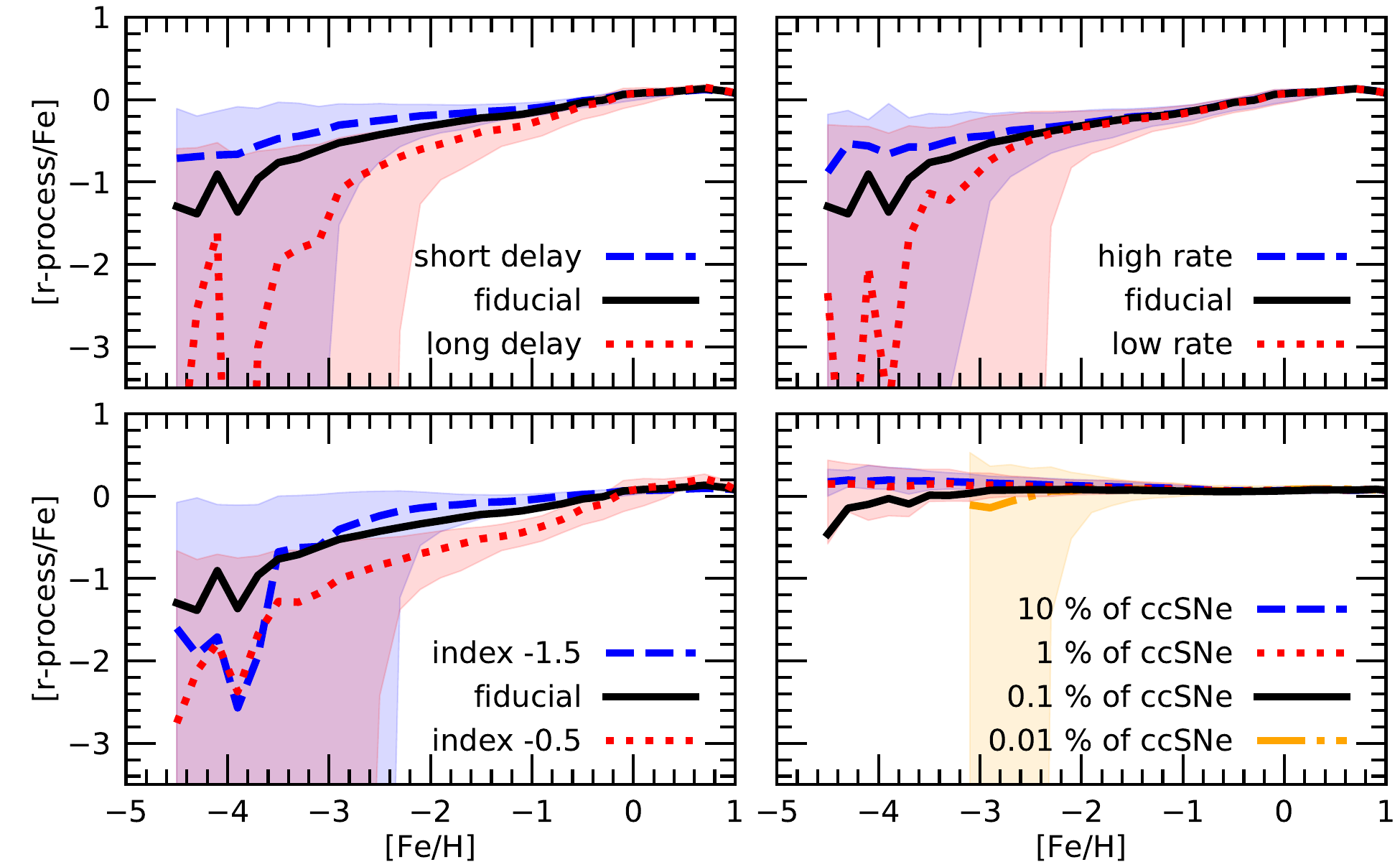}
  \caption{Results from chemodynamical inhomogeneous evolution models
    of \citet{VandeVoort.ea:2019} (reproduced with permission of Oxford
    University Press), predicting median values (lines) and
    distributions of [r/Fe] ratios in newly born stars of the Milky
    Way for three different neutron star merger rates, different
    indices of the coalescence delay time distributions
    $t^{\text{index}}$, and an additional admixture of r-process
    events occurring with rates proportional to CCSNe. The combination
    of additional r-process events proportional to the CCSN rate,
    being of strong importance at low metallicities early in the
    evolution of the Galaxy, and neutron star mergers permits an
    [r/Fe] dependence on [Fe/H] which does not decline towards low
    metallicities.
    \label{Freeke}}
\end{figure}

\begin{figure}
  \includegraphics[width=\linewidth]{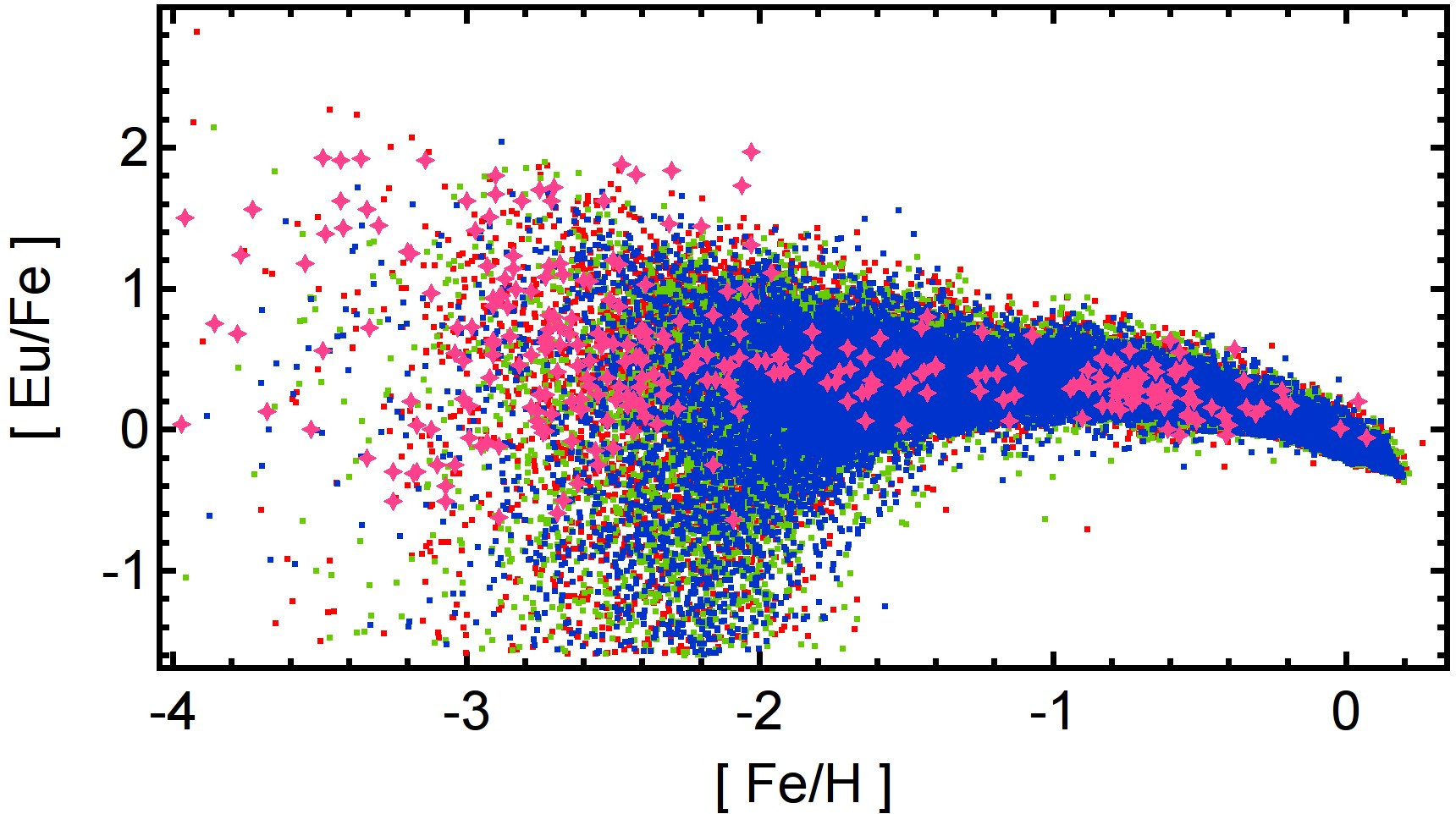}
  \caption {Evolution of [Eu/Fe] in a stochastic inhomogeneous
    galactic chemical evolution model, including both neutron star and
    neutron star-black hole mergers as r-process sites, under the
    assumption that all NS-BH mergers eject r-process matter
    \citep[for details see][reproduced with permission of Oxford
    University Press]{Wehmeyer.ea:2019}.  Magenta crosses represent
    observations, whereas different choices for black hole formation
    are utilized at low metallicities: Red (green, blue) squares
    represent models where all stars $\geq 20$~M$_{\odot}$
    ($\geq 25$~M$_{\odot}$, $\geq 30$~M$_{\odot}$) at metallicites
    $\text{Z}\leq 10^{-2} \text{Z}_{\odot}$ lead to failed SNe and
    black holes at the end of their life. The combination of these two
    r-process sources permits a good fit with observations also at low
    metallicities.\label{weh2}}
\end{figure}

Therefore the question arises, whether in addition to the scatter of
r-process elements like Eu compared to Fe, [Eu/Fe], covering more than
two orders of magnitude, see Figs.~\ref{mgeufe} and~\ref{HansenEu}),
especially also the early appearance of high [Eu/Fe] values, can be
consistent with a ``delayed'' process like compact binary mergers.
\citet{Ramirez-Ruiz.Trenti.ea:2015} suggested much shorter delays,
because collisions of neutron stars that were dynamically assembled in
the first nuclear star clusters could take place shortly after the
supernovae occurrences that produced them, opposite to mergers
resulting from binary evolution. However, the "early" appearance at
very low [Fe/H]-values is not only related to "timing". Also in such a
case the mergers only take place after the progenitor supernovae
already produced Fe. This is similar to compact binary mergers with a
longer delay, as also in that case the supernovae responsible for
producing at least one neutron star produce Fe, which shifts the
appearance of a typical r-process element like Eu to higher
metallicities [Fe/H].  This effect has been discussed in inhomogeneous
galactic evolution models utilizing neutron star mergers, i.e.\ events
with two prior supernovae and their Fe-ejecta
\citep{Argast.Samland.ea:2004,cescutti15,wehmeyer15,Haynes.Kobayashi:2019,VandeVoort.ea:2019}. These
authors come to the conclusion that neutron star mergers have problems
to explain [Eu/Fe] at lowest metallicities, while other/earlier
inhomogenous models came to the conclusion that they can do so
\citep{vandervoort15,shen15,hirai15,Komiya.Shigeyama:2016,hirai17}. The
difference is related to resolution and mixing issues discussed in the
previous subsubsection, but it should be noted that the high
resolution run of \citet{vandervoort15} as well as recent further
investigations~\citep[][see Fig.~\ref{Freeke}]{VandeVoort.ea:2019}
indicate the rise of [Eu/Fe] to occur at too high metallicities.
Whether and how much the use of NS-BH mergers, which explode in an
environment polluted only with Fe by one prior supernova, improve this
situation needs to be seen~\citep[][see
Fig.\ref{weh2}]{Wehmeyer.ea:2019}.

Thus, problems would remain to explain the strong r-process by
NS-mergers alone.  But the path to a binary merger is a complex
one. The mass, ejecta, and explosion energy of the second supernova in
a binary system have to be addressed for a full understanding
\citep{Mueller18,MuellerB19}. This includes the fact that neutron star
kicks from supernova explosions could move the neutron star binary far
out of the reach of the initial supernova remnant which polluted the
local ISM with Ni and Fe
\citep[e.g.][]{fryer.kalogera97,Kalogera.Fryer:1999,Abbott17d,Safarzadeh.Ramirez-Ruiz.ea:2019}.
If in such a way the merger event can be displaced from the original
supernovae, \citet{Wehmeyer.Others:2018} finds that mergers could be
made barely consistent with the [Eu/Fe] observations, if such a
displacement is taken into account and very short coalescence
timescales of $10^6$~yr are used. In addition, triple and multiple
stellar systems can cause different delay time distributions for
neutron star
mergers~\citep{Bonetti.Perego.ea:2018,Hamers.Thompson:2019}. And, as
mentioned above, compact binary mergers include also neutron
star-black hole mergers, which experience only the pollution by one
prior supernova~\citep{Wehmeyer.ea:2019}.

A different issue is the formation of the Galaxy from small
substructures like (ultra-faint) dwarf galaxies (UFDs), which, due to
different gas densities, experience different star formation
efficiencies and, due to small gravity, the loss of metals from
explosive events \citep{Simon:2019}, both shifting the occurrence of
abundance features to lower metallicities. The baryonic mass of these
UFDs, as small as $10^4$~M$_\odot$, is too small to be followed by the
global simulations discussed above, while local simulations have been
performed~\citep[see
e.g.][]{Emerick.Bryan.ea:2018,Corlies.Johnston.ea:2018,Tarumi.Yoshida.ea:2020}.
The IMA, combined with outflow, can probably be utilized as a first
order approximation locally in UFDs. Observations indicate that star
formation continues only for about a few $10^8$~yr, permitting still
to observe features from type Ia supernovae contributions, leading to
the [$\alpha$/Fe] downturn at
$\text{[Fe/H]}<-2$~\citep{Pakhomov.Mashonkina.ea:2019}, which takes
place in the Milky Way only at $-1$. As strong r-process sites are
rare events (by a factor of 100 to 1000 less frequent than both types
of supernovae) only a few UFDs show noticeable r-process
contributions, as observed in Reticulum II and Tucana
III~\citep[e.g.][]{Beniamini.Hotokezaka.Piran:2016,ji16,Ji.Frebel:2018,Marshall.ea:2019},
while displaying underabundances in most
cases~\citep{Simon:2019,Ji.Simon.ea:2019}. Only about 10\% of UFDs
experience a strong early r-process
contribution~\citep{Brauer.Ji.ea:2019}.  An early simpler IMA approach
by \citet{ishimaru15}, recently extended by a stochastic inclusion of
neutron star mergers, which are permitted to vary statistically with
respect to coalescence time scales \citep{ojima18}, indicates a
possible solution to the question whether neutron star mergers alone
could be responsible for the appearance of r-process products at
lowest metallicities.  This relates to the question whether the
apparent uniform r-process abundances observed in stars of UFDs can be
consistent with NS merger scenarios~\citep{Tsujimoto.Shigeyama:2014,Komiya.Shigeyama:2016,Bonetti.Perego.Dotti.ea:2019,Tarumi.Yoshida.ea:2020}.

The discussion in this subsection, focusing on the question whether
galactic chemical evolution studies can determine which of the three
sites --- (\ref{item:magneto}) magneto-rotational jet supernovae,
(\ref{item:collapsars}) collapsars/hypernovae, or (\ref{item:cbm})
compact binary mergers --- are consistent with or required from
observations, came to a still somewhat open result whether neutron
star mergers alone can provide the explanation for a strong r-process
throughout and also in the early Galaxy. Ongoing and future
observations utilizing dynamical tagging of groups in space with
orbital energy as well as angular momentum, combined with their
abundance patterns~\citep{Yuan.Myeong.ea:2020,Gudin.Shank.ea:2020}, might provide further
clues. The behavior of [Eu/Fe] around $\text{[Fe/H]}=-1$ puts
challenges on their delay time distribution, the early rise of [Eu/Fe]
at lowest metallicities can be hardly achieved within global galactic
evolution studies, not resolving scales of ultra-faint dwarf
galaxies. These studies argue for a contribution of
(\ref{item:magneto}) and/or (\ref{item:collapsars}) at early times in
galactic evolution. If UFDs are considered, there seems to be a
possible option out of this conclusion. However, there are independent
observational indications, combining results from the Milky Way and
its dwarf galaxy satellites Sagittarius, Fornax, and Sculptor, that
there exist two distinct r-process contributions from an early quick
source and a delayed source
\citep{Skuladottir.Hansen.ea:2019,Skuladottir.Salvadori:2020}.  Thus,
the answer is still somewhat open, depending also on the question
whether on average mergers alone can reproduce the lanthanide fraction
$X_{\text{La}}$ observed in low metallicity
stars~\citep{Ji.Simon.ea:2019}. It also needs to be shown whether the
turbulent diffusion coefficients, deduced
by~\citet{Beniamini.Hotokezaka:2020} from occurrence frequencies and
production yields of different r-process sites, in order to reproduce
the abundance scatter at low metallicities, agrees with those from
self-consistent simulations \citep{VandeVoort.ea:2019}.

\subsection{Long-lived Radioactivities: r-process cosmochronometers
  and actinide boost stars} 
\label{sec:r-proc-cosm}

A complete list of isotopes with half-lives in the range
$10^7$--$10^{11}$ yr is given in Table~\ref{tab1}. They cover a time span
from a lower limit in excess of the evolution time of massive stars up
to (and beyond) the age of the Universe. Such nuclei can be utilized
as ``chronometers'' for nucleosynthesis processes in galactic evolution
and also serve as a measure for the age of the Galaxy (and thus as a
lower limit for the age of the Universe, see the earlier discussion
in~\ref{sec:radioactive}).  The list is not long. Two of the nuclei
require predictions for the production of the ground and isomeric
states ($^{92}$Nb, $^{176}$Lu).  With the exception of $^{40}$K,
all of the remaining nuclei are heavier than the ``Fe-group'' and can
only be made via neutron capture.

\begin{table}[htb]
  \begin{center}
    \caption {Isotopes with half-lives in the range $10^7$--$10^{11}$
      yr}
    \label{tab1}
    \begin{ruledtabular}
      \begin{tabular}{cccc}
        {Isotope} & {Half-Life} & {Isotope} & {Half-Life}\\
        \hline
        $^{40}$K & $1.3\times 10^9$ yr & $^{205}$Pb & $1.5\times 10^7$ yr\\
        $^{87}$Rb & $4.8\times 10^{10}$ yr & $^{232}$Th & $1.4\times 10^{10}$ yr\\
        $^{92}$Nb & $3.5\times 10^7$ yr & $^{235}$U & $7.0\times 10^8$ yr\\
        $^{129}$I & $1.6\times 10^7$ yr & $^{236}$U & $2.3\times 10^7$ yr\\
        $^{147}$Sm & $1.1\times 10^{11}$ yr & $^{238}$U & $4.5\times 10^9$ yr\\
        $^{176}$Lu & $3.7\times 10^{10}$ yr & $^{244}$Pu & $8.0\times 10^7$ yr\\
        $^{187}$Re & $4.4\times 10^{10}$ yr & $^{247}$Cm & $1.6\times 10^7$ yr
      \end{tabular}
    \end{ruledtabular}
  \end{center}
\end{table}

The nuclei with half-lives comparable to the age of the
Galaxy/Universe, $^{232}$Th and $^{238}$U, as well as all other
actinide isotopes, are products of a single nucleosynthesis process,
the r-process. The possible astrophysical settings have been discussed
in the previous sections. The question is how to predict reliable
production ratios for these long-lived isotopes, if (a) not even the
site is completely clear, and (b) even for a given site nuclear
uncertainties enter.

Nevertheless, for many years such chronometers have been utilized to
attempt predictions for the age of the Galaxy~\citep[see
e.g.][]{Fowler.Hoyle:1960,Schramm.Wasserburg:1970,Cowan.Thielemann.Truran:1991,Panov2017}.
This has been performed initially with simplified chemical evolution
models via (i) the prediction of $^{232}$Th/$^{238}$U and
$^{235}$U/$^{238}$U ratios in r-process calculations, (ii) applying
them in galactic evolution models, which include assumptions about the
histories of star formation rates and r-process production, and
finally (iii) comparing these ratios with meteoritic data, which
provide the $^{232}$Th/$^{238}$U and $^{235}$U/$^{238}$U ratios at the
formation of the solar system.  An advantage (somewhat decreasing the
nuclear uncertainties involved) is that $^{232}$Th and $^{235,238}$U
are populated by $\alpha$-decay chains, summing up the contributions
of a number of nuclei, and therefore uncertainties in the predictions
of the individual isotopes involved average out to some
extent~\cite{Thielemann.Metzinger.Klapdor:1983,Cowan.Thielemann.Truran:1991,Goriely.Clerbaux:1999}.

Observations of elemental Th and U with respect to the r-process
reference element Eu in individual old stars can also be used to
estimate the age. However, in general this requires a chemical
evolution model except for the case of low-metallicity stars.  The
metallicities of the halo stars for which neutron-capture element data
have become available range from
$-3 \lesssim \text{[Fe/H]} \lesssim -2$. Typical, still simple,
galactic chemical evolution calculations suggest roughly the
metallicity-age relation $\text{[Fe/H]} = -1$ at $10^9$~yr,
$\text{[Fe/H]} = -2$ at $10^8$~yr, and $\text{[Fe/H]} = -3$ at
$10^7$~yr \citep[if the IMA could be applied at such low
metallicities, see,
e.g.][]{Tsujimoto.Yoshii.ea:1997,Chiappini:2000}. Even if these
estimates are uncertain by factors of 2--3, very low metallicity stars
most certainly were born when the Galaxy was only $10^7$--$10^8$~yr
old, a tiny fraction of its present age. Thus, the neutron-capture
elements observed in very low metallicity stars were generated in only
one or at most a few prior nucleosynthesis episodes \citep[see
also][]{Beniamini.Hotokezaka:2020}. If several events contributed, the
time interval between these events had to be very short in comparison
to Th decay ages. Thus, it is justified to treat the sum as a single
r-process abundance distribution which undergoes decay from the time
of its incorporation into a (low metallicity) star until its detection
in present observations.

Such considerations can also be employed for the ratio of Th and U to
stable Pb, which has in addition to the s-process contribution (i) a
direct r-process contribution to the $^{206,207,208}$Pb isotopes, (ii)
a contribution due to fast $\alpha$- and $\beta$-decay chains from
unstable nuclei produced in the r-process beyond Pb (decaying within
less than $10^6$~yr), and finally (iii) a contribution from the
long-lived decay chains originating at $^{232}$Th and
$^{235,238}$U~\citep{Roederer.Kratz.ea:2009,Frebel.Kratz:2009}.

The prediction of required isotopic/elemental production ratios,
lacking (until recently) site-specific detailed information, has been
based on parametrized, so-called site-independent fits, utilizing a
superposition of neutron densities which reproduce all solar r-process
abundances from $A = 130$ through the actinides~\citep[see
e.g.][]{Freiburghaus.Rembges.ea:1999,Cowan.Pfeiffer.ea:1999,Goriely.Arnould:2001,Schatz.ea:2002,Kratz2004,Roederer.Kratz.ea:2009}. Alternatively
neutrino wind models were employed with a superposition of
contributing entropy
components~\citep{Freiburghaus.Rembges.ea:1999,Farouqi.Kratz.ea:2010,kratz14,Hill.Christlieb.ea:2017}. \citet{Goriely.Janka:2016}
used steady-state neutrino-driven wind models with adiabatic expansion
and the superposition of many contributing components.

Making use of such so-called site-independent predictions for standard
r-process production ratios, combined with observed abundance ratios
found in low-metallicity stars, gives an indication for the decay time
of radioactive isotopes since the star was born, polluted by an
original r-process pattern.  Typical results for ages of most
low-metallicity r-process enhanced stars are in the range of
12--14~Gyr
\citep{Cowan.Pfeiffer.ea:1999,Schatz.ea:2002,Kratz2004,Roederer.Kratz.ea:2009,Hill.Christlieb.ea:2017}. This
approach assumes that the production ratios of the site(s) responsible
for these observations are consistent with those reproducing a solar
r-process. It should be kept in mind that, independent of the
uncertainties in abundance predictions discussed above, such age
determinations are clearly also affected by observational
uncertainties. \citet{Ludwig.Caffau.ea:2010} provide a detailed
analysis of the affects to be expected, which also apply for the
discussion in the upcoming paragraphs.

Among the stars with observed Th and U, there exist a number of
so-called actinide-boost stars with an enhanced ratio of Th/Eu and
U/Eu in comparison to all other r-process enhanced stars \citep[see
e.g.][]{Roederer.Kratz.ea:2009,Holmbeck.Beers.ea:2018}, observed
especially at low metallicities around $\text{[Fe/H]}\approx -3$.
When utilizing as initial abundance patterns the parametrized fits
discussed above (reproducing solar r-abundances), the age estimates
for those stars are unrealistically low or even negative~\citep[see
right scale of Fig.~\ref{boost} in section~\ref{sec:observations-jjc-cs}, from][]{Holmbeck.Beers.ea:2018}.
Although it appears that most of the elemental abundances in
actinide-boost stars, up to the third r-process peak, are close to a
solar r-process pattern, one should investigate further possible
correlations between the actinide boost and other abundance features.
The question is whether these features point either to a different
site than the dominant one responsible for the solar r-process
abundances or to variations of conditions in the same type of events,
dependent on still unknown aspects \citep{Holmbeck19b,Holmbeck19a}.

The actinide to Eu ratio is related to the path of the r-process and
the timing (a) when the actinides are reached via the r-process flow
and (b) when fission plays a role during the further flow onto heavier
nuclei.  Due to this, the r-process results are dependent on the
proton/nucleon ratio $Y_e$ in the expanding matter, determining the
neutron/seed ratio. Intuitively one could expect that the lowest (most
neutron-rich) $Y_e$'s would lead to the highest actinide production.
\citet{Holmbeck19a} showed, with their nuclear physics input, that
the ratio is highest for a $Y_e$ in the range 0.1--0.15 (see their
Figs. 16 and 17), with the highest values found around
$Y_e=0.125$. Higher $Y_e$-values (i.e.\ less neutron-rich conditions)
lead to a smaller actinide production, because of a less strong
r-process. Lower $Y_e$-values (i.e.\ more or very neutron-rich
conditions) lead also to smaller ratios. This is due to the fact that
an initially higher actinide production is reduced later by fission
cycling, which can be very effective in destroying the actinides. The
details depend on mass models and related fission barriers.

\begin{figure}[htb]
    \centering
    \includegraphics[angle=270,width=1.0\linewidth]{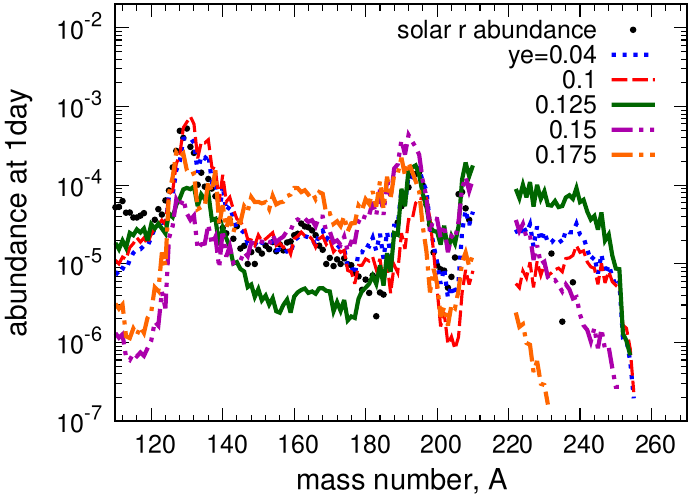}
    \caption{From \citet{Wu:2017}: Utilizing the DZ mass model
      \citep{Duflo.Zuker:1995} and trajectories from \citet{barnes16}
      permits 
    large variations in actinide production, even at low
    $Y_e$. The highest actinide production is found at $Y_e=0.125$}
    \label{Wuye}
\end{figure}

\citet{Wu:2017} presented a similar behavior, as indicated in
Fig.~\ref{Wuye}, using trajectories adopted from~\citet{barnes16},
finding also an actinide boost for $Y_e$-conditions close to 0.125.
\citet{Eichler.Sayar.ea:2019} did an independent study, testing in
detail the influence of nuclear physics uncertainties. They found
slightly higher $Y_e$-values of 0.15 for the maximum actinide
production, but similar conclusions, examining also the actinide
decline for lower $Y_e$'s as a function of the number of resulting
fission cycles.  In all these cases the Th/U ratio, involving two
actinide nuclei close in mass numbers, is not strongly affected by a
variation in $Y_e$.
  
\citet{Holmbeck19b} argue, that a variety of neutron star merger
characteristics (possible due to e.g.\ different binary masses and/or
mass ratios, affecting the total amount of dynamic ejecta - including
tidal tails, neutrino wind, and black hole accretion disk outflows),
can be responsible for varying outcomes, ranging from solar-type
r-process patterns to actinide boosts.  \citet{Wu.Barnes.ea:2019}
discuss how variations in the produced abundance patterns can affect
kilonova lightcurves and spectra, with the aim to identify the exact
pattern for individual observed events.  \citet{Ji.Drout.Hansen:2019}
give hints that the observed merger GW170817 is not as lanthanide- and
actinide-rich as required for the dominant solar r-process site.

Thus, the question remains, based on low-metallicity observations, why
most events lead apparently to a solar r-process pattern and some
others cause an actinide boost~\citep[keep in mind also the discussion
above on observational uncertainties][]{Ludwig.Caffau.ea:2010}. And,
it remains to be seen whether different sites are the reason for these
two features or whether one site, i.e.\ neutron star mergers, can lead
to this variety. Which $Y_e$-interval is resulting from specific
events, stopping above 0.125, including 0.125, or also continuing to
values below 0.125? What are the dominant conditions in MHD jet
supernovae, what are the properties of accretion disk outflows in
collapsars, what is the role of the individual components in compact
binary mergers --- dynamic ejecta (including prompt ejecta and tidal
tails), neutrino winds, accretion disk outflows, and do all of these
subcomponents exist if one of the compact objects is a black hole or
the combined mass is sufficiently high, preventing the intermediate
existence of a hypermassive neutron star?

Conclusions on this are still speculative, but variations among the
different sites should be investigated further \citep[see
e.g.][]{Nishimura.Sawai.ea:2017,Wu.Barnes.ea:2019,Eichler.Sayar.ea:2019,Holmbeck19b,Siegel.Barnes.Metzger:2019,Siegel:2019}.
Improved predictions for all the most probable main r-process sites,
discussed in the previous section (plus possibly exotic scenarios,
\citealp{Fuller.Kusenko.Takhistov:2017}, but see also
\citealp{Camelio18}) can hopefully lead to a one-to-one connection
between responsible production sites and observations (although still
affected both by astrophysical site as well as nuclear physics
uncertainties).

Independent of these considerations, concerning actinide boosts, it
should be kept in mind that chemical evolution
findings~\citep{wehmeyer15,cescutti15,Cote.Belczynski.ea:2017,Cote.Fryer.ea:2018,hotokezaka18,Cote.Eichler.ea:2019,Haynes.Kobayashi:2019,VandeVoort.ea:2019}
and observations
\citep{Skuladottir.Hansen.ea:2019,Skuladottir.Salvadori:2020} seem to
indicate that there exist two distinct r-process contributions from an
early quick source and a delayed source.

In addition to the identification and possible explanation of the
abundance pattern in actinide boost stars, related to long-lived
unstable Th and U isotopes, shorter-lived radioactive isotopes have
been addressed by~\citet{Lugaro.Ott.ea:2018}, \citet{Vescavi.ea:2018},
and \citet{Cote.Yague.ea:2019}. Nuclei with half-lives of a few $10^6$
to $10^7$~yr permit observers to probe recent nucleosynthesis events
in the vicinity of the presolar nebula. In the present context only
nuclei of an r-process origin are of interest here. Of
these~\citet{Cote.Yague.ea:2019} point out $^{129}$I and $^{247}$Cm
with identical half-lives, and~\citet{Cote:2020} utilize them to
measure the strength of the last r-process event affecting the
pre-solar nebula, as indicated from meteoritic data. In this respect
it is also of interest what composition even later occurring events
contribute, affecting only the delivery of $^{244}$Pu onto earth and
the deposition in deep-sea sediments over the past few hundred million
years (see section~\ref{sec:radioactive}). Very recent investigations
by~\citet{Wallner.Others:2019}, possibly indicating that $^{60}$Fe
from the last CCSNe might have been accompanied by a (very) minor
$^{244}$Pu contribution, would possibly permit a frequent weak
r-process, producing very small, but not negligible amounts of
actinides.

\section{Final remarks and conclusions}
\label{sec:conclusions}

In this review we have reported on recent developments and new data
from nuclear and atomic physics experiments and constraints from
astronomical observations. We also discussed their impact, combined
with advances in astrophysical modelling, on our understanding of the
astrophysical r-process. This includes the operation of the r-process,
its potential astrophysical sites and its contribution to the chemical
evolution of our Galaxy. Despite the tremendous progress achieved
since the r-process was proposed by~\citet{Burbidge.Burbidge.ea:1957}
and by~\citet{Cameron:1957}, several open questions still remain. In
these final remarks we specify these challenges, but also the future
opportunities how to overcome them.

With respect to the \textsc{nuclear physics}, which enters decisively
in producing the abundance pattern of the r-process, major
achievements have been accomplished. This is related to experimental
progress in accessing unstable nuclei far from stability, combined
with a growing theoretical understanding of their properties.

\begin{enumerate}

\item Novel detection technologies, employed at operational RIB
  facilities, have allowed to determine \emph{nuclear masses} for
  nuclei far from stability with improved precision. This in turn
  served as stringent constraints to improve (empirical and
  microscopic) global mass models which in r-process simulations
  determine the location of the r-process path in the nuclear
  chart. Further improvements are required to remove uncertainties
  which have decisive consequences for the r-process mass flow across
  neutron-shell closures shaping the final r-process abundance
  distribution.

\item The measurement of \emph{$\beta$-decay half-lives} for
  medium-mass neutron-rich nuclei at and near the r-process path at
  RIKEN has been a major recent achievement. The data are of crucial
  importance for the speed with which the r-process moves matter to
  heavier nuclei and (in combination with the location of the
  r-process path) for the height of peaks and the overall final
  abundance distribution. Progress has also been reached to measure
  $\beta$-delayed neutron emissions (important in the late phases
  during decay back to stability), in particular for nuclei close to
  the $N=126$ shell closure. However, no $\beta$-decay data exist yet
  for $N=126$ (or heavier) nuclei on the r-process path. Such
  measurements, which can be expected from the next-generation RIB
  facilities, will be of crucial relevance to determine the amount of
  matter which is transported beyond the third r-process peak into the
  fission region.

\item \emph{Fission} plays a crucial role in current r-process models,
  in particular related to high neutron density environments, as
  e.g.\ in neutron star mergers. Here fission terminates the flow to
  heavier nuclei beyond the actinides, causing fission-cycling.  This
  returns matter to lighter nuclei and is also a source of neutrons
  which can shape the final abundance pattern. In these models fission
  yields contribute strongly to the second r-process peak, which needs
  confirmation in future work. Fission also affects the heaviest
  long-lived nuclei that are produced by the r-process. Heavy
  neutron-rich nuclei, in particular those at the $N=184$ shell
  closure, are still experimentally out-of-reach. But experimental
  programs are envisioned to push the measurement of fission rates and
  yields to more neutron-rich nuclei than currently accessible. Such
  improvements are also required to address the question whether
  superheavy elements can be produced by the r-process.

\item Simulations identify \emph{$\alpha$-decays}, especially the
  decay-chains originating from actinide nuclei, as important
  contributors to the kilonova light light curve and to determine the
  r-process Pb abundance. Many of these decays are experimentally
  studied. It is an open question whether $\alpha$-decays can compete
  with fission for heavy neutron-rich nuclei.

\item \emph{Neutron captures} (and their inverse photo-disintegration)
  affect the final abundance distribution during the r-process
  freeze-out period. (Fortunately for a large variety of conditions
  during the r-process build-up a chemical equilibrium between these
  two reactions can be maintained and the r-process path is determined
  solely by nuclear masses.) The nuclei involved can have very low
  neutron separation energies (with a low density of states) so that
  direct neutron captures might be favored over compound nucleus
  reactions. Direct measurements of neutron capture rates for
  r-process nuclei are experimentally out-of-reach. Advances have been
  made to develop surrogate techniques to constrain the rates
  indirectly or to reduce nuclear uncertainties entering statistical
  model capture rate evaluations. If the direct capture is dominated
  by individual resonances, the rate can be constrained by indirect
  determination of the resonance parameters.

\end{enumerate}

In the investigation of all these aspects much progress has occurred,
but major uncertainties are remaining, as experiments have touched
nuclei in the r-process path only at a limited number of locations in
the nuclear chart. Besides these nuclear aspects, there exist also
other challenges in \textsc{modelling astrophysical sites} of the
r-process:

\begin{enumerate}

\item Most environments expected to be sites of a strong r-process
  involve objects at high densities and temperatures, making it
  necessary to determine the \emph{nuclear equation of state} at these
  extreme conditions. Constraints for the EoS have been obtained in
  relativistic heavy-ion collisions and by astronomical
  observations. Decisive progress is expected from upcoming nuclear
  experiments at heavy-ion facilities and by dedicated experiments
  probing the nuclear symmetry energy as well as from astronomical
  observations exploiting upgraded and novel detectors. Improved
  knowledge of the nuclear EoS is also important to answer the
  question whether in compact binary mergers a hypermassive neutron
  star exists temporarily or even remains as a final outcome.

\item The modelling of r-process sites requires multi-scale general
  relativistic, multidimensional radiation magneto-hydrodynamics
  simulations. Such calculations are computationally very demanding
  and involve approximations and numerical methods whose reliability
  need to be critically assessed.
  
\item Many of the discussed effects involve the modeling of
  \emph{magnetic fields}, possibly as a major ingredient to predict
  jet ejection. A decisive aspect is whether and how magnetic fields
  can be enhanced during these events, where the magneto-rotational
  instability MRI plays a major role. High resolution
  magneto-hydrodynamics modeling is a field only at the brink of
  getting reliable results for the modelling of complete astrophysical
  sites.

\item As reactions mediated by the weak interaction are not in
  equilibrium, processes like electron/positron captures and
  \emph{neutrino interactions with matter} have to be explicitly
  modelled. Neutrino-flavor transformations, especially via
  matter-neutrino resonances and fast pairwise flavor conversion, have
  been identified to play an essential role in compact binary
  mergers. Weak-interaction reactions are also crucial to determine
  the proton-to-nucleon ratio $Y_e$, which is a key ingredient for
  r-process nucleosynthesis. The adequate treatment of the neutrino
  processes requires multi-D transport simulations. Such studies for
  the complicated geometries involved in neutrino-flavor
  transformations are just in their infancy.

\item In addition to neutrino transport, general \emph{radiation
    transport via photons} is important to predict the electromagnetic
  aftermath of explosions in order to make a connection to
  observational features like lightcurves and spectra. Fundamental
  ingredients for these predictions are the total energy released by
  radioactive matter, its thermalization and the yet unknown atomic
  opacities of (especially multiply-ionized) heavy elements.  Progress
  in this field will permit to test whether the lanthanide fraction
  $X_{\text{La}}$ in observed events is consistent with solar
  r-process abundances

\end{enumerate}

Based on the presently available input for nuclear properties and the
present status with respect to modelling \textsc{possible r-process
  sites}, three major options for sites of a strong r-process have
emerged. For one of these sites (NS mergers) observational evidence
exists, while for the other two observational proofs of r-process
ejecta are still missing.

\begin{enumerate}

\item Models of \emph{compact binary mergers} indicate that they are
  prolific sites of r-process nucleosynthesis, with up to
  10$^{-2}$~M$_\odot$ of r-process matter in the dynamic ejecta and a
  few times 10$^{-2}$~M$_\odot$ from accretion disk outflows. When
  including all components --- dynamic ejecta, neutrino winds, and
  viscous/secular accretions disk outflows --- they produce not only
  the heaviest r-process nuclei but also significant amounts of the
  standard solar r-process abundances for mass numbers with
  $A<130$. The first observation of a neutron star merger (GW170817),
  accompanied by the AT~2017gfo macronova/kilonova thermal afterglow,
  makes this the first proven and confirmed production site of heavy
  r-process elements; variations, depending on the mass of the merged
  object, as well as neutron star --- black hole mergers need further
  observational confirmation and theoretical modeling.

\item There exist observational indications of neutron
stars with surface magnetic fields exceeding 10$^{15}$ Gauss
(magnetars) which could be produced by a rare class of
  \emph{magneto-rotational core-collapse supernovae}. Dependent on
  high initial magnetic field strengths and rotation rates before
  collapse, they might eject r-process matter in polar jets. However,
  better predictions of these initial parameters from stellar
  evolution are needed in order to understand whether fast ejection
  via jets takes place or whether a magneto-rotational instability
  (MRI) will only eventually cause an explosion, ejecting less
  neutronized matter. Investigating the role of the MRI during the
  collapse/explosion phase is impossible without high resolution
  simulations.

\item A very recent multi-D MHD simulation for \emph{accretion disk
    outflows from collapsars}, i.e.\ objects which result from the
  final collapse of massive stars and end in the formation of a black
  hole, suggest that large amounts of r-process material
  ($> 0.1$~M$_\odot$) can be ejected. Further simulations are needed,
  in particular to understand the relation of collapsars to hypernovae
  and long-duration GRBs and their potential for the galactic
  inventory of r-process material.

\end{enumerate}

While the three above mentioned sites are candidates for a strong
r-process, producing heavy elements up to the actinides, there exist
further options to produce a so-called \emph{weak or limited
  r-process}, probably also synthesizing elements up to Eu (and
beyond?), but with a steeper decline as a function of nuclear mass
number than found in the solar system r-abundance pattern; such
possible sites include electron capture supernovae, regular
core-collapse supernovae, and also quark deconfinement supernovae.

All of these models have to be confronted with and scrutinized by
\textsc{astronomical observations}. An interesting aspect is that one
of the listed and confirmed sites is related to binary systems, while
the others are resulting from the evolution of massive single
stars. Observations supporting and constraining r-process sites exist
in a number of ways:

\begin{enumerate}

\item The requirement to reproduce the total amount of r-process
  matter in the Galaxy, puts very stringent constraints on the
  occurrence and frequency of r-process events. For compact binary
  mergers as well as magneto-rotational jet supernovae this frequency
  should be between 1 event per 100 to 1000 regular supernovae. For
  mergers this would be consistent with population synthesis studies,
  while collapsars should be less frequent by a factor of about
  10.

\item The \emph{radioactive r-process tracer} $^{244}$Pu is found in
  \emph{deep sea sediments}, however the observed amount rules out a
  quasi-continuous production of r-process elements as expected for
  sites with occurrence frequencies like supernovae and points to much
  rarer events. There exist indications from deep sea sediments that
  point to events where $^{60}$Fe and $^{244}$Pu are co-produced,
  however with relative amounts of $^{244}$Pu strongly reduced in
  comparison to the solar value. Such events could be related to a
  weak r-process (see above).

\item \emph{Observations indicate the presence of r-process elements
    in halo stars at lowest metallicities}, showing either a complete
  or only a partial (or incomplete) r-process abundance pattern, in
  the latter case possibly pointing to a weak r-process origin. These
  abundance detections act as proof for nucleosynthesis early in the
  history of the Galaxy and provide important clues about the nature
  of the earliest stars and their r-process sites.

\item Observations of lowest metallicity stars in our Galaxy and
  (ultra-faint) dwarf galaxies show substantial variations in
  r-process abundances, indicating a production site with a
  low event rate and consistent high amounts of r-process ejecta in
  order to explain solar abundances; this is also underlined by the
  \emph{large scatter of Eu/Fe} (Eu being an r-process element and Fe
  stemming from core-collapse supernovae at these low metallicities)
  \emph{seen in the earliest stars of the Galaxy}; this is explained
  by a not yet well mixed interstellar medium with respect to
  contributions of products from regular core-collapse supernovae and
  the rare r-process events.

\item Due to the availability of experimental atomic data and high
  resolution, precision observations, r-process abundance
  determinations have improved much over time; this permitted also to
  detect \emph{the presence of long-lived radioactive nuclei like Th
    and U} in the same star, making even a \emph{``dating'' of old
    stars} possible; the observed variations in the actinide to
  intermediate mass r-process elements like Eu, leading to so-called
  \emph{actinide-boost stars}, would even give clues about different
  r-process sites.

\end{enumerate}

The above observations indicate rare events for the strong r-process,
a requirement which is matched by all three candidate sites.
Moreover, the observations related to the overall evolution of heavy
r-process elements in comparison to Fe, and especially its large
scatter at low metallicities, require inhomogeneous \textsc{galactic
  evolution} simulations, which can reproduce this behavior and might
actually point to favored sites:

\begin{enumerate}

\item A major open question is: \emph{can products of the neutron star
    merger r-process alone explain the observations} of a large
  scatter of Eu/Fe and other r-process elements seen already at
  metallicities of $\text{[Fe/H]}\le -3$? As the supernovae which
  produce the neutron stars of a merger already lead to a substantial
  floor of Fe, they enhance [Fe/H]. Thus, the high [Eu/Fe] due to the
  new ejecta would then be seen first at the metallicity [Fe/H]
  inherited from the prior supernovae, if the merger ejecta are mixed
  with the same interstellar medium as the prior supernova ejecta. For
  a typical explosion energy of $10^{51}$~erg and typical densities of
  the interstellar medium, this mixing would occur via a Sedov-Taylor
  blast wave in a range of about $5\times 10^{4}$~M$_\odot$. As matter
  is ejected with similar kinetic energies in neutron star mergers and
  in supernovae, recent investigations indicate that it would mix with
  a similar amount of interstellar medium.  Stochastic inhomogeneous
  chemical evolution calculations, utilizing this effect alone, show
  the appearance of Eu only in the metallicity range
  $-3<\text{[Fe/H]}<-2$ for neutron star mergers. First investigations
  have been performed how this would be affected in the case of
  neutron star-black hole mergers, as they would only lead to
  Fe-ejecta by one prior supernova.

\item Like other hydrodynamic calculations, \emph{large-scale SPH
    simulations} can suffer from resolution problems, which
  overestimate the material mixing.  This mixes Fe with larger amounts
  of interstellar medium and thus causes a decrease in the metallicity
  at which r-process nucleosynthesis sets in.  Global simulations can
  handle turbulent mixing of interstellar medium matter in the early
  Galaxy, however, it is questionable whether they can resolve the
  relevant length scales. Some of these simulations seem to be able to
  reproduce the r-process behavior of low-metallicity observations
  with compact binary mergers, but the most recent ones also favor an
  additional site or source at lowest metallicities.

\item \emph{Neutron star kicks}, resulting from a supernova explosion,
  could have the binary neutron star system move out of its supernova
  remnants (polluted with Fe and Ni), and the merger event could take
  place in galactic regions unpolluted by Fe from earlier supernovae.
  This could permit the ejection of r-process matter in environments
  with a lower [Fe/H], also in the case of neutron star
  mergers. Preliminary simulations with stochastic inhomogeneous
  models (see above) are able to reproduce observations, if
  coalescence delay times are as short as 1 Myr.

\item Another option is that early on, in \emph{galactic substructures
    of the size of dwarf galaxies}, different star formation rates can
  exist, combined with a loss of nucleosynthesis ejecta out of these
  galaxies due to smaller gravity. This can shift the behavior of the
  [Eu/Fe] ratio as a function of metallicity [Fe/H] to lower
  metallicities. When also considering a statistical distribution of
  (down to small) coalescence timescales in the individual
  substructures, the low-metallicity observations could possibly be
  matched, while the merging of these substructures within the early
  Galaxy at later times can be made consistent with the [Eu/Fe]
  decline (similar to alpha elements) at $\text{[Fe/H]}=-1$. However,
  it should be noted that also in dwarf galaxies indications for two
  different sources, an early quick and a delayed r-process
  contribution exist.

\item In somewhat simpler galactic evolution models, employing the
  instantaneous mixing approximation (IMA) and \emph{coalescence delay
    time distributions following a $t^{-1}$ power law}, as expected
  from population synthesis studies and statistics of short duration
  gamma-ray bursts, apparently no model can reproduce the metallicity
  dependence of the r-process/Fe abundance ratios with neutron star
  mergers alone. But further constraints on the delay time
  distribution, and considering hot and cold phases of the
  interstellar medium might help.

\end{enumerate}

The discussion above underlines that it is still inconclusive whether
binary compact mergers alone can explain low metallicity
observations. Although mergers could be responsible for the dominant
amount of r-process products in the solar system and present Galaxy,
an additional component which acts at lowest metallicities may still
be required. The detection of actinide boost stars, found in
particular at metallicities as low as $[\text{Fe/H}]\approx -3$, could
be a further argument for such an additional component.

This review of all aspects of the astrophysical r-process, from
nuclear physics via stellar (explosive) modeling, astronomical
observations, as well as galactic evolution, has shown that
substantial progress has been made since the r-process was
postulated in the 1950s by \citet{Burbidge.Burbidge.ea:1957} and
\citet{Cameron:1957}. But it also shows that, despite the very first
observation of an r-process production site (GW170817) in 2017,
confirming neutron star mergers as probably the most important site,
many open questions remain and further progress on all fronts is
required in a truly interdisciplinary effort, in order to answer them.

Existing and upcoming nuclear facilities world-wide (FAIR, FRIB, HIAF,
RAON, RIKEN, SPIRAL) will allow to produce neutron-rich nuclei along
the r-process path and to determine their properties, including, for
the first time, nuclei of the third r-process peak.  Relativistic
heavy-ion collision experiments envisioned for FAIR, NICA and RHIC
will generate and investigate nuclear matter at the temperatures and
densities as they exist in neutron-star mergers (and core-collapse
supernovae). These exciting experimental prospectives will constrain
and guide advances in global nuclear models, and together they will
decisively reduce the nuclear uncertainties currently hampering
r-process studies.

Observational programs targeting abundances in low-metallicity stars,
like RAVE, Gaia, and APOGEE will be incorporated in future surveys
such as SDSS V, 4MOST, and WEAVE\@. This will provide highest quality
information about the chemical structure of the galactic disc, the
halo, and the bulge.  Multi-messenger astronomy with present and
future gravitational wave detectors like LIGO, VIRGO, KAGRA, and
IndIGO is expected to detect up to 10 compact binary mergers per year,
which can be followed up with observations of related gamma-ray bursts
and the electromagnetic afterglow. This permits to analyse the outcome
of many sources with different viewing angles and will help our
further understanding of this site and possible variations,
producing the heaviest elements in the Universe.

The upcoming experimental facilities and observational tools have to
be supplemented by theory advances to lead us to a deeper and detailed
understanding of the origin of the heavy elements from Fe to U,
produced by the r-process. These theory efforts have to include
improved models for the nuclear equation of state and for neutron-rich
nuclei far from stability, but also for stellar atmospheres, stellar
evolution and explosions, and finally for the chemo-dynamical history of
the Galaxy.

In summary, we are living in exciting times for unravelling the
mysteries of r-process nucleosynthesis.

\begin{acknowledgments}
  We thank a very large number of colleagues and friends for helpful
  collaborations and discussions, but especially Al Cameron and Willy
  Fowler should be mentioned here who motivated many of us to explore
  the topic of this review.  This research was supported in part by:
  NASA grant HST-GO-14232 (JJC), NSF grant AST1616040 (CS), NASA grant
  NNX16AE96G and NSF grant AST-1516182 (JEL), NSF grant Phys-0758100
  and the Joint Institute for Nuclear Astrophysics through NSF Grant
  Phys-0822648 (AA and MW), NSF Grant PHY-1430152 (JINA Center for the
  Evolution of the Elements AA, JJC, and MW), NSF Grant
  No. PHY-1927130 [AccelNet-WOU: International Research Network for
  Nuclear Astrophysics (IReNA)] (A. A., J. J. C., K. L., G. M.-P.,
  F.-K. T., and M. W.), the Extreme Matter Institute EMMI (KL and
  GMP), the Deutsche Forschungsgemeinschaft (DFG, German Research
  Foundation) -- Project-ID 279384907 -- SFB 1245 (GMP), the Swiss SNF
  through grant 200020-132816/1, ERC Advanced Grant FISH 321263 (FKT),
  ERC Advanced Grant KILONOVA 885281 (GMP), and COST Actions
  NewCompstar MP1304 (FKT) and ChETEC CA16117 (FKT and GMP).
\end{acknowledgments}

\bibliography{rprocess}

\begin{thebibliography}{971}%
\makeatletter
\providecommand \@ifxundefined [1]{%
 \@ifx{#1\undefined}
}%
\providecommand \@ifnum [1]{%
 \ifnum #1\expandafter \@firstoftwo
 \else \expandafter \@secondoftwo
 \fi
}%
\providecommand \@ifx [1]{%
 \ifx #1\expandafter \@firstoftwo
 \else \expandafter \@secondoftwo
 \fi
}%
\providecommand \natexlab [1]{#1}%
\providecommand \enquote  [1]{``#1''}%
\providecommand \bibnamefont  [1]{#1}%
\providecommand \bibfnamefont [1]{#1}%
\providecommand \citenamefont [1]{#1}%
\providecommand \href@noop [0]{\@secondoftwo}%
\providecommand \href [0]{\begingroup \@sanitize@url \@href}%
\providecommand \@href[1]{\@@startlink{#1}\@@href}%
\providecommand \@@href[1]{\endgroup#1\@@endlink}%
\providecommand \@sanitize@url [0]{\catcode `\\12\catcode `\$12\catcode
  `\&12\catcode `\#12\catcode `\^12\catcode `\_12\catcode `\%12\relax}%
\providecommand \@@startlink[1]{}%
\providecommand \@@endlink[0]{}%
\providecommand \url  [0]{\begingroup\@sanitize@url \@url }%
\providecommand \@url [1]{\endgroup\@href {#1}{\urlprefix }}%
\providecommand \urlprefix  [0]{URL }%
\providecommand \Eprint [0]{\href }%
\providecommand \doibase [0]{https://doi.org/}%
\providecommand \selectlanguage [0]{\@gobble}%
\providecommand \bibinfo  [0]{\@secondoftwo}%
\providecommand \bibfield  [0]{\@secondoftwo}%
\providecommand \translation [1]{[#1]}%
\providecommand \BibitemOpen [0]{}%
\providecommand \bibitemStop [0]{}%
\providecommand \bibitemNoStop [0]{.\EOS\space}%
\providecommand \EOS [0]{\spacefactor3000\relax}%
\providecommand \BibitemShut  [1]{\csname bibitem#1\endcsname}%
\let\auto@bib@innerbib\@empty
\bibitem [{\citenamefont {Abbott}\ \emph
  {et~al.}(2017{\natexlab{a}})\citenamefont {Abbott} \emph
  {et~al.}}]{Abbott.Others:2017}%
  \BibitemOpen
  \bibfield  {author} {\bibinfo {author} {\bibnamefont {Abbott}, \bibfnamefont
  {B~P}},  \emph {et~al.} (\bibinfo {collaboration} {LIGO Scientific
  Collaboration and Virgo Collaboration})} (\bibinfo {year}
  {2017}{\natexlab{a}}),\ \bibfield  {title} {\enquote {\bibinfo {title}
  {{Estimating the Contribution of Dynamical Ejecta in the Kilonova Associated
  with GW170817}},}\ }\href {https://doi.org/10.3847/2041-8213/aa9478}
  {\bibfield  {journal} {\bibinfo  {journal} {Astrophys. J. Lett.}\ }\textbf
  {\bibinfo {volume} {850}},\ \bibinfo {eid} {L39}}\BibitemShut {NoStop}%
\bibitem [{\citenamefont {Abbott}\ \emph
  {et~al.}(2017{\natexlab{b}})\citenamefont {Abbott} \emph
  {et~al.}}]{Abbott17c}%
  \BibitemOpen
  \bibfield  {author} {\bibinfo {author} {\bibnamefont {Abbott}, \bibfnamefont
  {B~P}},  \emph {et~al.} (\bibinfo {collaboration} {LIGO Scientific
  Collaboration and Virgo Collaboration, Fermi Gamma-ray Burst Monitor,
  INTEGRAL})} (\bibinfo {year} {2017}{\natexlab{b}}),\ \bibfield  {title}
  {\enquote {\bibinfo {title} {{Gravitational Waves and Gamma-Rays from a
  Binary Neutron Star Merger: GW170817 and GRB 170817A}},}\ }\href
  {https://doi.org/10.3847/2041-8213/aa920c} {\bibfield  {journal} {\bibinfo
  {journal} {Astrophys. J. Lett.}\ }\textbf {\bibinfo {volume} {848}},\
  \bibinfo {eid} {L13}}\BibitemShut {NoStop}%
\bibitem [{\citenamefont {Abbott}\ \emph
  {et~al.}(2017{\natexlab{c}})\citenamefont {Abbott} \emph
  {et~al.}}]{Abbott.Abbott.ea:2017}%
  \BibitemOpen
  \bibfield  {author} {\bibinfo {author} {\bibnamefont {Abbott}, \bibfnamefont
  {B~P}},  \emph {et~al.} (\bibinfo {collaboration} {LIGO Scientific
  Collaboration and Virgo Collaboration})} (\bibinfo {year}
  {2017}{\natexlab{c}}),\ \bibfield  {title} {\enquote {\bibinfo {title}
  {{GW170817: Observation of Gravitational Waves from a Binary Neutron Star
  Inspiral}},}\ }\href {https://doi.org/10.1103/PhysRevLett.119.161101}
  {\bibfield  {journal} {\bibinfo  {journal} {Phys. Rev. Lett.}\ }\textbf
  {\bibinfo {volume} {119}},\ \bibinfo {pages} {161101}}\BibitemShut {NoStop}%
\bibitem [{\citenamefont {Abbott}\ \emph
  {et~al.}(2017{\natexlab{d}})\citenamefont {Abbott} \emph
  {et~al.}}]{Abbott:2017}%
  \BibitemOpen
  \bibfield  {author} {\bibinfo {author} {\bibnamefont {Abbott}, \bibfnamefont
  {B~P}},  \emph {et~al.}} (\bibinfo {year} {2017}{\natexlab{d}}),\ \bibfield
  {title} {\enquote {\bibinfo {title} {{Multi-messenger Observations of a
  Binary Neutron Star Merger}},}\ }\href
  {https://doi.org/10.3847/2041-8213/aa91c9} {\bibfield  {journal} {\bibinfo
  {journal} {Astrophys. J. Lett.}\ }\textbf {\bibinfo {volume} {848}},\
  \bibinfo {eid} {L12}}\BibitemShut {NoStop}%
\bibitem [{\citenamefont {Abbott}\ \emph
  {et~al.}(2017{\natexlab{e}})\citenamefont {Abbott} \emph
  {et~al.}}]{Abbott17d}%
  \BibitemOpen
  \bibfield  {author} {\bibinfo {author} {\bibnamefont {Abbott}, \bibfnamefont
  {B~P}},  \emph {et~al.} (\bibinfo {collaboration} {LIGO Scientific
  Collaboration and Virgo Collaboration})} (\bibinfo {year}
  {2017}{\natexlab{e}}),\ \bibfield  {title} {\enquote {\bibinfo {title} {{On
  the Progenitor of Binary Neutron Star Merger GW170817}},}\ }\href
  {https://doi.org/10.3847/2041-8213/aa93fc} {\bibfield  {journal} {\bibinfo
  {journal} {Astrophys. J. Lett.}\ }\textbf {\bibinfo {volume} {850}},\
  \bibinfo {eid} {L40}}\BibitemShut {NoStop}%
\bibitem [{\citenamefont {Abbott}\ \emph {et~al.}(2018)\citenamefont {Abbott}
  \emph {et~al.}}]{Abbott.Abbott.ea:2018a}%
  \BibitemOpen
  \bibfield  {author} {\bibinfo {author} {\bibnamefont {Abbott}, \bibfnamefont
  {B~P}},  \emph {et~al.} (\bibinfo {collaboration} {The LIGO Scientific
  Collaboration and the Virgo Collaboration})} (\bibinfo {year} {2018}),\
  \bibfield  {title} {\enquote {\bibinfo {title} {{GW170817: Measurements of
  Neutron Star Radii and Equation of State}},}\ }\href
  {https://doi.org/10.1103/PhysRevLett.121.161101} {\bibfield  {journal}
  {\bibinfo  {journal} {Phys. Rev. Lett.}\ }\textbf {\bibinfo {volume} {121}},\
  \bibinfo {pages} {161101}}\BibitemShut {NoStop}%
\bibitem [{\citenamefont {Abbott}\ \emph {et~al.}(2019)\citenamefont {Abbott}
  \emph {et~al.}}]{Abbott.Abbott.ea:2018b}%
  \BibitemOpen
  \bibfield  {author} {\bibinfo {author} {\bibnamefont {Abbott}, \bibfnamefont
  {B~P}},  \emph {et~al.} (\bibinfo {collaboration} {LIGO Scientific
  Collaboration and Virgo Collaboration})} (\bibinfo {year} {2019}),\ \bibfield
   {title} {\enquote {\bibinfo {title} {{Properties of the Binary Neutron Star
  Merger GW170817}},}\ }\href {https://doi.org/10.1103/PhysRevX.9.011001}
  {\bibfield  {journal} {\bibinfo  {journal} {Phys. Rev. X}\ }\textbf {\bibinfo
  {volume} {9}},\ \bibinfo {pages} {011001}}\BibitemShut {NoStop}%
\bibitem [{\citenamefont {{Abbott}}\ \emph
  {et~al.}(2020{\natexlab{a}})\citenamefont {{Abbott}} \emph
  {et~al.}}]{Abbott.ea:2020}%
  \BibitemOpen
  \bibfield  {author} {\bibinfo {author} {\bibnamefont {{Abbott}},
  \bibfnamefont {B~P}},  \emph {et~al.}} (\bibinfo {year}
  {2020}{\natexlab{a}}),\ \bibfield  {title} {\enquote {\bibinfo {title}
  {{GW190425: Observation of a Compact Binary Coalescence with Total Mass $\sim
  3.4 M_{\odot}$}},}\ }\href {https://doi.org/10.3847/2041-8213/ab75f5}
  {\bibfield  {journal} {\bibinfo  {journal} {Astrophys. J. Lett.}\ }\textbf
  {\bibinfo {volume} {892}}~(\bibinfo {number} {1}),\ \bibinfo {eid}
  {L3}}\BibitemShut {NoStop}%
\bibitem [{\citenamefont {{Abbott}}\ \emph
  {et~al.}(2020{\natexlab{b}})\citenamefont {{Abbott}} \emph
  {et~al.}}]{Abbott.ea:2020b}%
  \BibitemOpen
  \bibfield  {author} {\bibinfo {author} {\bibnamefont {{Abbott}},
  \bibfnamefont {R}},  \emph {et~al.}} (\bibinfo {year} {2020}{\natexlab{b}}),\
  \bibfield  {title} {\enquote {\bibinfo {title} {{GW190814: Gravitational
  Waves from the Coalescence of a 23 Solar Mass Black Hole with a 2.6 Solar
  Mass Compact Object}},}\ }\href {https://doi.org/10.3847/2041-8213/ab960f}
  {\bibfield  {journal} {\bibinfo  {journal} {Astrophys. J. Lett.}\ }\textbf
  {\bibinfo {volume} {896}}~(\bibinfo {number} {2}),\ \bibinfo {eid}
  {L44}}\BibitemShut {NoStop}%
\bibitem [{\citenamefont {{Abbott}}\ \emph
  {et~al.}(2020{\natexlab{c}})\citenamefont {{Abbott}} \emph
  {et~al.}}]{Abbott.Abbott.ea:2020}%
  \BibitemOpen
  \bibfield  {author} {\bibinfo {author} {\bibnamefont {{Abbott}},
  \bibfnamefont {R}},  \emph {et~al.}} (\bibinfo {year} {2020}{\natexlab{c}}),\
  \bibfield  {title} {\enquote {\bibinfo {title} {{GWTC-2: Compact Binary
  Coalescences Observed by LIGO and Virgo During the First Half of the Third
  Observing Run}},}\ }\href@noop {} {\bibfield  {journal} {\bibinfo  {journal}
  {arXiv e-prints}\ }}\Eprint {https://arxiv.org/abs/2010.14527}
  {arXiv:2010.14527 [gr-qc]} \BibitemShut {NoStop}%
\bibitem [{\citenamefont {{Abohalima}}\ and\ \citenamefont
  {{Frebel}}(2018)}]{JINAbase:2018}%
  \BibitemOpen
  \bibfield  {author} {\bibinfo {author} {\bibnamefont {{Abohalima}},
  \bibfnamefont {Abdu}}, and\ \bibinfo {author} {\bibfnamefont {Anna}\
  \bibnamefont {{Frebel}}}} (\bibinfo {year} {2018}),\ \bibfield  {title}
  {\enquote {\bibinfo {title} {{JINAbase---A Database for Chemical Abundances
  of Metal-poor Stars}},}\ }\href {https://doi.org/10.3847/1538-4365/aadfe9}
  {\bibfield  {journal} {\bibinfo  {journal} {Astrophys. J. Suppl.}\ }\textbf
  {\bibinfo {volume} {238}}~(\bibinfo {number} {2}),\ \bibinfo {eid}
  {36}}\BibitemShut {NoStop}%
\bibitem [{\citenamefont {Aboussir}\ \emph {et~al.}(1995)\citenamefont
  {Aboussir}, \citenamefont {Pearson}, \citenamefont {Dutta},\ and\
  \citenamefont {Tondeur}}]{Aboussir.Pearson.ea:1995}%
  \BibitemOpen
  \bibfield  {author} {\bibinfo {author} {\bibnamefont {Aboussir},
  \bibfnamefont {Y}}, \bibinfo {author} {\bibfnamefont {J.~M.}\ \bibnamefont
  {Pearson}}, \bibinfo {author} {\bibfnamefont {A.~K.}\ \bibnamefont {Dutta}},
  and\ \bibinfo {author} {\bibfnamefont {F.}~\bibnamefont {Tondeur}}} (\bibinfo
  {year} {1995}),\ \bibfield  {title} {\enquote {\bibinfo {title} {{Nuclear
  mass formula via an approximation to the Hartree-Fock method}},}\ }\href
  {https://doi.org/10.1016/S0092-640X(95)90014-4} {\bibfield  {journal}
  {\bibinfo  {journal} {At. Data Nucl. Data Tables}\ }\textbf {\bibinfo
  {volume} {61}},\ \bibinfo {pages} {127--176}}\BibitemShut {NoStop}%
\bibitem [{\citenamefont {{Ackley}}\ \emph {et~al.}(2020)\citenamefont
  {{Ackley}} \emph {et~al.}}]{Ackley.Others:2020}%
  \BibitemOpen
  \bibfield  {author} {\bibinfo {author} {\bibnamefont {{Ackley}},
  \bibfnamefont {K}},  \emph {et~al.}} (\bibinfo {year} {2020}),\ \bibfield
  {title} {\enquote {\bibinfo {title} {{Observational constraints on the
  optical and near-infrared emission from the neutron star-black hole binary
  merger S190814bv}},}\ }\href@noop {} {\bibfield  {journal} {\bibinfo
  {journal} {arXiv e-prints}\ }}\Eprint {https://arxiv.org/abs/2002.01950}
  {arXiv:2002.01950 [astro-ph.SR]} \BibitemShut {NoStop}%
\bibitem [{\citenamefont {Adrich}\ \emph {et~al.}(2005)\citenamefont {Adrich}
  \emph {et~al.}}]{Adrich.Klimkiewicz.ea:2005}%
  \BibitemOpen
  \bibfield  {author} {\bibinfo {author} {\bibnamefont {Adrich}, \bibfnamefont
  {P}},  \emph {et~al.} (\bibinfo {collaboration} {LAND-FRS Collaboration})}
  (\bibinfo {year} {2005}),\ \bibfield  {title} {\enquote {\bibinfo {title}
  {{Evidence for Pygmy and Giant Dipole Resonances in $^{130}$Sn and
  $^{132}$Sn}},}\ }\href {https://doi.org/10.1103/PhysRevLett.95.132501}
  {\bibfield  {journal} {\bibinfo  {journal} {Phys. Rev. Lett.}\ }\textbf
  {\bibinfo {volume} {95}},\ \bibinfo {eid} {132501}}\BibitemShut {NoStop}%
\bibitem [{\citenamefont {{Agramunt}}\ \emph {et~al.}(2016)\citenamefont
  {{Agramunt}} \emph {et~al.}}]{Agramunt2016}%
  \BibitemOpen
  \bibfield  {author} {\bibinfo {author} {\bibnamefont {{Agramunt}},
  \bibfnamefont {J}},  \emph {et~al.}} (\bibinfo {year} {2016}),\ \bibfield
  {title} {\enquote {\bibinfo {title} {{Characterization of a neutron-beta
  counting system with beta-delayed neutron emitters}},}\ }\href
  {https://doi.org/10.1016/j.nima.2015.10.082} {\bibfield  {journal} {\bibinfo
  {journal} {Nucl. Instruments Methods Phys. Res. Sect. A Accel. Spectrometers,
  Detect. Assoc. Equip.}\ }\textbf {\bibinfo {volume} {807}},\ \bibinfo {pages}
  {69--78}}\BibitemShut {NoStop}%
\bibitem [{\citenamefont {{Akram}}\ \emph {et~al.}(2020)\citenamefont
  {{Akram}}, \citenamefont {{Farouqi}}, \citenamefont {{Hallmann}},\ and\
  \citenamefont {{Kratz}}}]{Akram.Farouqi.ea:2020}%
  \BibitemOpen
  \bibfield  {author} {\bibinfo {author} {\bibnamefont {{Akram}}, \bibfnamefont
  {Waheed}}, \bibinfo {author} {\bibfnamefont {Khalil}\ \bibnamefont
  {{Farouqi}}}, \bibinfo {author} {\bibfnamefont {Oliver}\ \bibnamefont
  {{Hallmann}}}, and\ \bibinfo {author} {\bibfnamefont {Karl-Ludwig}\
  \bibnamefont {{Kratz}}}} (\bibinfo {year} {2020}),\ \bibfield  {title}
  {\enquote {\bibinfo {title} {{Nucleosynthesis of light trans-Fe isotopes in
  ccSNe: Implications from presolar SiC-X grains}},}\ }in\ \href
  {https://doi.org/10.1051/epjconf/202022701009} {\emph {\bibinfo {booktitle}
  {European Physical Journal Web of Conferences}}},\ \bibinfo {series}
  {European Physical Journal Web of Conferences}, Vol.\ \bibinfo {volume}
  {227},\ p.\ \bibinfo {pages} {01009}\BibitemShut {NoStop}%
\bibitem [{\citenamefont {Alhassid}\ \emph {et~al.}(2000)\citenamefont
  {Alhassid}, \citenamefont {Bertsch}, \citenamefont {Liu},\ and\ \citenamefont
  {Nakada}}]{Alhassid.Bertsch.ea:2000}%
  \BibitemOpen
  \bibfield  {author} {\bibinfo {author} {\bibnamefont {Alhassid},
  \bibfnamefont {Y}}, \bibinfo {author} {\bibfnamefont {G.~F.}\ \bibnamefont
  {Bertsch}}, \bibinfo {author} {\bibfnamefont {S.}~\bibnamefont {Liu}}, and\
  \bibinfo {author} {\bibfnamefont {H.}~\bibnamefont {Nakada}}} (\bibinfo
  {year} {2000}),\ \bibfield  {title} {\enquote {\bibinfo {title} {{Parity
  Dependence of Nuclear Level Densities}},}\ }\href
  {https://doi.org/10.1103/PhysRevLett.84.4313} {\bibfield  {journal} {\bibinfo
   {journal} {Phys. Rev. Lett.}\ }\textbf {\bibinfo {volume} {84}},\ \bibinfo
  {pages} {4313}}\BibitemShut {NoStop}%
\bibitem [{\citenamefont {{Alhassid}}\ \emph {et~al.}(1999)\citenamefont
  {{Alhassid}}, \citenamefont {{Liu}},\ and\ \citenamefont
  {{Nakada}}}]{Alhassid.Liu.Nakada:1999}%
  \BibitemOpen
  \bibfield  {author} {\bibinfo {author} {\bibnamefont {{Alhassid}},
  \bibfnamefont {Y}}, \bibinfo {author} {\bibfnamefont {S.}~\bibnamefont
  {{Liu}}}, and\ \bibinfo {author} {\bibfnamefont {H.}~\bibnamefont
  {{Nakada}}}} (\bibinfo {year} {1999}),\ \bibfield  {title} {\enquote
  {\bibinfo {title} {{Particle-Number Reprojection in the Shell Model Monte
  Carlo Method: Application to Nuclear Level Densities}},}\ }\href
  {https://doi.org/10.1103/PhysRevLett.83.4265} {\bibfield  {journal} {\bibinfo
   {journal} {Phys. Rev. Lett.}\ }\textbf {\bibinfo {volume} {83}},\ \bibinfo
  {pages} {4265--4268}}\BibitemShut {NoStop}%
\bibitem [{\citenamefont {{Alhassid}}\ \emph {et~al.}(2007)\citenamefont
  {{Alhassid}}, \citenamefont {{Liu}},\ and\ \citenamefont
  {{Nakada}}}]{Alhassid.Liu.Nakada:2007}%
  \BibitemOpen
  \bibfield  {author} {\bibinfo {author} {\bibnamefont {{Alhassid}},
  \bibfnamefont {Y}}, \bibinfo {author} {\bibfnamefont {S.}~\bibnamefont
  {{Liu}}}, and\ \bibinfo {author} {\bibfnamefont {H.}~\bibnamefont
  {{Nakada}}}} (\bibinfo {year} {2007}),\ \bibfield  {title} {\enquote
  {\bibinfo {title} {{Spin Projection in the Shell Model MonteCarlo Method and
  the Spin Distribution of Nuclear Level Densities}},}\ }\href
  {https://doi.org/10.1103/PhysRevLett.99.162504} {\bibfield  {journal}
  {\bibinfo  {journal} {Phys. Rev. Lett.}\ }\textbf {\bibinfo {volume} {99}},\
  \bibinfo {eid} {162504}}\BibitemShut {NoStop}%
\bibitem [{\citenamefont {Anderson}\ \emph {et~al.}(2008)\citenamefont
  {Anderson}, \citenamefont {Hirschmann}, \citenamefont {Lehner}, \citenamefont
  {Liebling}, \citenamefont {Motl}, \citenamefont {Neilsen}, \citenamefont
  {Palenzuela},\ and\ \citenamefont {Tohline}}]{Anderson.Hirschmann.ea:2008}%
  \BibitemOpen
  \bibfield  {author} {\bibinfo {author} {\bibnamefont {Anderson},
  \bibfnamefont {Matthew}}, \bibinfo {author} {\bibfnamefont {Eric~W.}\
  \bibnamefont {Hirschmann}}, \bibinfo {author} {\bibfnamefont {Luis}\
  \bibnamefont {Lehner}}, \bibinfo {author} {\bibfnamefont {Steven~L.}\
  \bibnamefont {Liebling}}, \bibinfo {author} {\bibfnamefont {Patrick~M.}\
  \bibnamefont {Motl}}, \bibinfo {author} {\bibfnamefont {David}\ \bibnamefont
  {Neilsen}}, \bibinfo {author} {\bibfnamefont {Carlos}\ \bibnamefont
  {Palenzuela}}, and\ \bibinfo {author} {\bibfnamefont {Joel~E.}\ \bibnamefont
  {Tohline}}} (\bibinfo {year} {2008}),\ \bibfield  {title} {\enquote {\bibinfo
  {title} {{Magnetized Neutron-Star Mergers and Gravitational-Wave Signals}},}\
  }\href {https://doi.org/10.1103/PhysRevLett.100.191101} {\bibfield  {journal}
  {\bibinfo  {journal} {Phys. Rev. Lett.}\ }\textbf {\bibinfo {volume} {100}},\
  \bibinfo {pages} {191101}}\BibitemShut {NoStop}%
\bibitem [{\citenamefont {Annala}\ \emph {et~al.}(2018)\citenamefont {Annala},
  \citenamefont {Gorda}, \citenamefont {Kurkela},\ and\ \citenamefont
  {Vuorinen}}]{Annala.Gorda.ea:2018}%
  \BibitemOpen
  \bibfield  {author} {\bibinfo {author} {\bibnamefont {Annala}, \bibfnamefont
  {Eemeli}}, \bibinfo {author} {\bibfnamefont {Tyler}\ \bibnamefont {Gorda}},
  \bibinfo {author} {\bibfnamefont {Aleksi}\ \bibnamefont {Kurkela}}, and\
  \bibinfo {author} {\bibfnamefont {Aleksi}\ \bibnamefont {Vuorinen}}}
  (\bibinfo {year} {2018}),\ \bibfield  {title} {\enquote {\bibinfo {title}
  {Gravitational-wave constraints on the neutron-star-matter equation of
  state},}\ }\href {https://doi.org/10.1103/PhysRevLett.120.172703} {\bibfield
  {journal} {\bibinfo  {journal} {Phys. Rev. Lett.}\ }\textbf {\bibinfo
  {volume} {120}},\ \bibinfo {pages} {172703}}\BibitemShut {NoStop}%
\bibitem [{\citenamefont {Antoniadis}\ \emph {et~al.}(2013)\citenamefont
  {Antoniadis} \emph {et~al.}}]{Antoniadis.Freire.ea:2013}%
  \BibitemOpen
  \bibfield  {author} {\bibinfo {author} {\bibnamefont {Antoniadis},
  \bibfnamefont {John}},  \emph {et~al.}} (\bibinfo {year} {2013}),\ \bibfield
  {title} {\enquote {\bibinfo {title} {{A Massive Pulsar in a Compact
  Relativistic Binary}},}\ }\href {https://doi.org/10.1126/science.1233232}
  {\bibfield  {journal} {\bibinfo  {journal} {Science}\ }\textbf {\bibinfo
  {volume} {340}},\ \bibinfo {pages} {1233232}}\BibitemShut {NoStop}%
\bibitem [{\citenamefont {{Aoki}}\ \emph {et~al.}(2017)\citenamefont {{Aoki}},
  \citenamefont {{Ishimaru}}, \citenamefont {{Aoki}},\ and\ \citenamefont
  {{Wanajo}}}]{Aoki.Ishimura.ea:2017}%
  \BibitemOpen
  \bibfield  {author} {\bibinfo {author} {\bibnamefont {{Aoki}}, \bibfnamefont
  {M}}, \bibinfo {author} {\bibfnamefont {Y.}~\bibnamefont {{Ishimaru}}},
  \bibinfo {author} {\bibfnamefont {W.}~\bibnamefont {{Aoki}}}, and\ \bibinfo
  {author} {\bibfnamefont {S.}~\bibnamefont {{Wanajo}}}} (\bibinfo {year}
  {2017}),\ \bibfield  {title} {\enquote {\bibinfo {title} {{Diversity of
  Abundance Patterns of Light Neutron-capture Elements in Very-metal-poor
  Stars}},}\ }\href {https://doi.org/10.3847/1538-4357/aa5d08} {\bibfield
  {journal} {\bibinfo  {journal} {Astrophys. J.}\ }\textbf {\bibinfo {volume}
  {837}},\ \bibinfo {eid} {8}}\BibitemShut {NoStop}%
\bibitem [{\citenamefont {{Aoki}}\ \emph {et~al.}(2013)\citenamefont {{Aoki}},
  \citenamefont {{Beers}}, \citenamefont {{Lee}}, \citenamefont {{Honda}},
  \citenamefont {{Ito}}, \citenamefont {{Takada-Hidai}}, \citenamefont
  {{Frebel}}, \citenamefont {{Suda}}, \citenamefont {{Fujimoto}}, \citenamefont
  {{Carollo}},\ and\ \citenamefont {{Sivarani}}}]{Aoki.Beers.ea:2013}%
  \BibitemOpen
  \bibfield  {author} {\bibinfo {author} {\bibnamefont {{Aoki}}, \bibfnamefont
  {W}}, \bibinfo {author} {\bibfnamefont {T.~C.}\ \bibnamefont {{Beers}}},
  \bibinfo {author} {\bibfnamefont {Y.~S.}\ \bibnamefont {{Lee}}}, \bibinfo
  {author} {\bibfnamefont {S.}~\bibnamefont {{Honda}}}, \bibinfo {author}
  {\bibfnamefont {H.}~\bibnamefont {{Ito}}}, \bibinfo {author} {\bibfnamefont
  {M.}~\bibnamefont {{Takada-Hidai}}}, \bibinfo {author} {\bibfnamefont
  {A.}~\bibnamefont {{Frebel}}}, \bibinfo {author} {\bibfnamefont
  {T.}~\bibnamefont {{Suda}}}, \bibinfo {author} {\bibfnamefont {M.~Y.}\
  \bibnamefont {{Fujimoto}}}, \bibinfo {author} {\bibfnamefont
  {D.}~\bibnamefont {{Carollo}}}, and\ \bibinfo {author} {\bibfnamefont
  {T.}~\bibnamefont {{Sivarani}}}} (\bibinfo {year} {2013}),\ \bibfield
  {title} {\enquote {\bibinfo {title} {{High-resolution Spectroscopy of
  Extremely Metal-poor Stars from SDSS/SEGUE. I. Atmospheric Parameters and
  Chemical Compositions}},}\ }\href
  {https://doi.org/10.1088/0004-6256/145/1/13} {\bibfield  {journal} {\bibinfo
  {journal} {Astron. J.}\ }\textbf {\bibinfo {volume} {145}},\ \bibinfo {eid}
  {13}}\BibitemShut {NoStop}%
\bibitem [{\citenamefont {{Arcavi}}\ \emph {et~al.}(2017)\citenamefont
  {{Arcavi}}, \citenamefont {{Hosseinzadeh}}, \citenamefont {{Howell}},
  \citenamefont {{McCully}}, \citenamefont {{Poznanski}}, \citenamefont
  {{Kasen}}, \citenamefont {{Barnes}}, \citenamefont {{Zaltzman}},
  \citenamefont {{Vasylyev}}, \citenamefont {{Maoz}},\ and\ \citenamefont
  {{Valenti}}}]{arcavi17}%
  \BibitemOpen
  \bibfield  {author} {\bibinfo {author} {\bibnamefont {{Arcavi}},
  \bibfnamefont {I}}, \bibinfo {author} {\bibfnamefont {G.}~\bibnamefont
  {{Hosseinzadeh}}}, \bibinfo {author} {\bibfnamefont {D.~A.}\ \bibnamefont
  {{Howell}}}, \bibinfo {author} {\bibfnamefont {C.}~\bibnamefont {{McCully}}},
  \bibinfo {author} {\bibfnamefont {D.}~\bibnamefont {{Poznanski}}}, \bibinfo
  {author} {\bibfnamefont {D.}~\bibnamefont {{Kasen}}}, \bibinfo {author}
  {\bibfnamefont {J.}~\bibnamefont {{Barnes}}}, \bibinfo {author}
  {\bibfnamefont {M.}~\bibnamefont {{Zaltzman}}}, \bibinfo {author}
  {\bibfnamefont {S.}~\bibnamefont {{Vasylyev}}}, \bibinfo {author}
  {\bibfnamefont {D.}~\bibnamefont {{Maoz}}}, and\ \bibinfo {author}
  {\bibfnamefont {S.}~\bibnamefont {{Valenti}}}} (\bibinfo {year} {2017}),\
  \bibfield  {title} {\enquote {\bibinfo {title} {{Optical emission from a
  kilonova following a gravitational-wave-detected neutron-star merger}},}\
  }\href {https://doi.org/10.1038/nature24291} {\bibfield  {journal} {\bibinfo
  {journal} {Nature}\ }\textbf {\bibinfo {volume} {551}},\ \bibinfo {pages}
  {64--66}}\BibitemShut {NoStop}%
\bibitem [{\citenamefont {{Arcones}}\ and\ \citenamefont
  {{Janka}}(2011)}]{Arcones.Janka:2011}%
  \BibitemOpen
  \bibfield  {author} {\bibinfo {author} {\bibnamefont {{Arcones}},
  \bibfnamefont {A}}, and\ \bibinfo {author} {\bibfnamefont {H.-T.}\
  \bibnamefont {{Janka}}}} (\bibinfo {year} {2011}),\ \bibfield  {title}
  {\enquote {\bibinfo {title} {{Nucleosynthesis-relevant conditions in
  neutrino-driven supernova outflows. II. The reverse shock in two-dimensional
  simulations}},}\ }\href {https://doi.org/10.1051/0004-6361/201015530}
  {\bibfield  {journal} {\bibinfo  {journal} {Astron. \& Astrophys.}\ }\textbf
  {\bibinfo {volume} {526}},\ \bibinfo {eid} {A160}}\BibitemShut {NoStop}%
\bibitem [{\citenamefont {{Arcones}}\ \emph {et~al.}(2007)\citenamefont
  {{Arcones}}, \citenamefont {{Janka}},\ and\ \citenamefont
  {{Scheck}}}]{Arcones.Janka.Scheck:2007}%
  \BibitemOpen
  \bibfield  {author} {\bibinfo {author} {\bibnamefont {{Arcones}},
  \bibfnamefont {A}}, \bibinfo {author} {\bibfnamefont {H.-T.}\ \bibnamefont
  {{Janka}}}, and\ \bibinfo {author} {\bibfnamefont {L.}~\bibnamefont
  {{Scheck}}}} (\bibinfo {year} {2007}),\ \bibfield  {title} {\enquote
  {\bibinfo {title} {{Nucleosynthesis-relevant conditions in neutrino-driven
  supernova outflows. I. Spherically symmetric hydrodynamic simulations}},}\
  }\href {https://doi.org/10.1051/0004-6361:20066983} {\bibfield  {journal}
  {\bibinfo  {journal} {Astron. \& Astrophys.}\ }\textbf {\bibinfo {volume}
  {467}},\ \bibinfo {pages} {1227--1248}}\BibitemShut {NoStop}%
\bibitem [{\citenamefont {Arcones}\ and\ \citenamefont
  {Mart{\'\i}nez-Pinedo}(2011)}]{Arcones.Martinez-Pinedo:2011}%
  \BibitemOpen
  \bibfield  {author} {\bibinfo {author} {\bibnamefont {Arcones}, \bibfnamefont
  {A}}, and\ \bibinfo {author} {\bibfnamefont {G.}~\bibnamefont
  {Mart{\'\i}nez-Pinedo}}} (\bibinfo {year} {2011}),\ \bibfield  {title}
  {\enquote {\bibinfo {title} {Dynamical $r$-process studies within the
  neutrino-driven wind scenario and its sensitivity to the nuclear physics
  input},}\ }\href {https://doi.org/10.1103/PhysRevC.83.045809} {\bibfield
  {journal} {\bibinfo  {journal} {Phys. Rev. C}\ }\textbf {\bibinfo {volume}
  {83}},\ \bibinfo {pages} {045809}}\BibitemShut {NoStop}%
\bibitem [{\citenamefont {{Arcones}}\ \emph {et~al.}(2010)\citenamefont
  {{Arcones}}, \citenamefont {{Mart{\'{\i}}nez-Pinedo}}, \citenamefont
  {{Roberts}},\ and\ \citenamefont
  {{Woosley}}}]{Arcones.Martinez-Pinedo.ea:2010}%
  \BibitemOpen
  \bibfield  {author} {\bibinfo {author} {\bibnamefont {{Arcones}},
  \bibfnamefont {A}}, \bibinfo {author} {\bibfnamefont {G.}~\bibnamefont
  {{Mart{\'{\i}}nez-Pinedo}}}, \bibinfo {author} {\bibfnamefont {L.~F.}\
  \bibnamefont {{Roberts}}}, and\ \bibinfo {author} {\bibfnamefont {S.~E.}\
  \bibnamefont {{Woosley}}}} (\bibinfo {year} {2010}),\ \bibfield  {title}
  {\enquote {\bibinfo {title} {Electron fraction constraints based on nuclear
  statistical equilibrium with beta equilibrium},}\ }\href
  {https://doi.org/10.1051/0004-6361/201014276} {\bibfield  {journal} {\bibinfo
   {journal} {Astron. \& Astrophys.}\ }\textbf {\bibinfo {volume} {522}},\
  \bibinfo {eid} {A25}}\BibitemShut {NoStop}%
\bibitem [{\citenamefont {{Arcones}}\ and\ \citenamefont
  {{Montes}}(2011)}]{Arcones.Montes:2011}%
  \BibitemOpen
  \bibfield  {author} {\bibinfo {author} {\bibnamefont {{Arcones}},
  \bibfnamefont {A}}, and\ \bibinfo {author} {\bibfnamefont {F.}~\bibnamefont
  {{Montes}}}} (\bibinfo {year} {2011}),\ \bibfield  {title} {\enquote
  {\bibinfo {title} {{Production of Light-element Primary Process Nuclei in
  Neutrino-driven Winds}},}\ }\href {https://doi.org/10.1088/0004-637X/731/1/5}
  {\bibfield  {journal} {\bibinfo  {journal} {Astrophys. J.}\ }\textbf
  {\bibinfo {volume} {731}},\ \bibinfo {eid} {5}}\BibitemShut {NoStop}%
\bibitem [{\citenamefont {Arcones}\ and\ \citenamefont
  {Thielemann}(2013)}]{Arcones.Thielemann:2013}%
  \BibitemOpen
  \bibfield  {author} {\bibinfo {author} {\bibnamefont {Arcones}, \bibfnamefont
  {A}}, and\ \bibinfo {author} {\bibfnamefont {F-K}\ \bibnamefont
  {Thielemann}}} (\bibinfo {year} {2013}),\ \bibfield  {title} {\enquote
  {\bibinfo {title} {Neutrino-driven wind simulations and nucleosynthesis of
  heavy elements},}\ }\href {https://doi.org/10.1088/0954-3899/40/1/013201}
  {\bibfield  {journal} {\bibinfo  {journal} {J. Phys. G: Nucl. Part. Phys.}\
  }\textbf {\bibinfo {volume} {40}},\ \bibinfo {pages} {013201}}\BibitemShut
  {NoStop}%
\bibitem [{\citenamefont {{Ardevol-Pulpillo}}\ \emph
  {et~al.}(2019)\citenamefont {{Ardevol-Pulpillo}}, \citenamefont {{Janka}},
  \citenamefont {{Just}},\ and\ \citenamefont
  {{Bauswein}}}]{Ardevol-Pulpillo.Janka.ea:2018}%
  \BibitemOpen
  \bibfield  {author} {\bibinfo {author} {\bibnamefont {{Ardevol-Pulpillo}},
  \bibfnamefont {Ricard}}, \bibinfo {author} {\bibfnamefont {H.~Thomas}\
  \bibnamefont {{Janka}}}, \bibinfo {author} {\bibfnamefont {Oliver}\
  \bibnamefont {{Just}}}, and\ \bibinfo {author} {\bibfnamefont {Andreas}\
  \bibnamefont {{Bauswein}}}} (\bibinfo {year} {2019}),\ \bibfield  {title}
  {\enquote {\bibinfo {title} {{Improved Leakage-Equilibration-Absorption
  Scheme (ILEAS) for Neutrino Physics in Compact Object Mergers}},}\ }\href
  {https://doi.org/10.1093/mnras/stz613} {\bibfield  {journal} {\bibinfo
  {journal} {Mon. Not. Roy. Astron. Soc.}\ }\textbf {\bibinfo {volume} {485}},\
  \bibinfo {pages} {4754--4789}}\BibitemShut {NoStop}%
\bibitem [{\citenamefont {{Argast}}\ \emph {et~al.}(2004)\citenamefont
  {{Argast}}, \citenamefont {{Samland}}, \citenamefont {{Thielemann}},\ and\
  \citenamefont {{Qian}}}]{Argast.Samland.ea:2004}%
  \BibitemOpen
  \bibfield  {author} {\bibinfo {author} {\bibnamefont {{Argast}},
  \bibfnamefont {D}}, \bibinfo {author} {\bibfnamefont {M.}~\bibnamefont
  {{Samland}}}, \bibinfo {author} {\bibfnamefont {F.-K.}\ \bibnamefont
  {{Thielemann}}}, and\ \bibinfo {author} {\bibfnamefont {Y.-Z.}\ \bibnamefont
  {{Qian}}}} (\bibinfo {year} {2004}),\ \bibfield  {title} {\enquote {\bibinfo
  {title} {{Neutron star mergers versus core-collapse supernovae as dominant
  r-process sites in the early Galaxy}},}\ }\href
  {https://doi.org/10.1051/0004-6361:20034265} {\bibfield  {journal} {\bibinfo
  {journal} {Astron. \& Astrophys.}\ }\textbf {\bibinfo {volume} {416}},\
  \bibinfo {pages} {997--1011}}\BibitemShut {NoStop}%
\bibitem [{\citenamefont {{Arlandini}}\ \emph {et~al.}(1999)\citenamefont
  {{Arlandini}}, \citenamefont {{K{\"a}ppeler}}, \citenamefont {{Wisshak}},
  \citenamefont {{Gallino}}, \citenamefont {{Lugaro}}, \citenamefont
  {{Busso}},\ and\ \citenamefont {{Straniero}}}]{arlandini99}%
  \BibitemOpen
  \bibfield  {author} {\bibinfo {author} {\bibnamefont {{Arlandini}},
  \bibfnamefont {C}}, \bibinfo {author} {\bibfnamefont {F.}~\bibnamefont
  {{K{\"a}ppeler}}}, \bibinfo {author} {\bibfnamefont {K.}~\bibnamefont
  {{Wisshak}}}, \bibinfo {author} {\bibfnamefont {R.}~\bibnamefont
  {{Gallino}}}, \bibinfo {author} {\bibfnamefont {M.}~\bibnamefont {{Lugaro}}},
  \bibinfo {author} {\bibfnamefont {M.}~\bibnamefont {{Busso}}}, and\ \bibinfo
  {author} {\bibfnamefont {O.}~\bibnamefont {{Straniero}}}} (\bibinfo {year}
  {1999}),\ \bibfield  {title} {\enquote {\bibinfo {title} {{Neutron Capture in
  Low-Mass Asymptotic Giant Branch Stars: Cross Sections and Abundance
  Signatures}},}\ }\href {https://doi.org/10.1086/307938} {\bibfield  {journal}
  {\bibinfo  {journal} {Astrophys. J.}\ }\textbf {\bibinfo {volume} {525}},\
  \bibinfo {pages} {886--900}}\BibitemShut {NoStop}%
\bibitem [{\citenamefont {{Armbruster}}\ \emph {et~al.}(1976)\citenamefont
  {{Armbruster}}, \citenamefont {{Asghar}}, \citenamefont {{Bocquet}},
  \citenamefont {{Decker}}, \citenamefont {{Ewald}}, \citenamefont {{Greif}},
  \citenamefont {{Moll}}, \citenamefont {{Pfeiffer}}, \citenamefont
  {{Schrader}}, \citenamefont {{Schussler}}, \citenamefont {{Siegert}},\ and\
  \citenamefont {{Wollnik}}}]{Lohengrin}%
  \BibitemOpen
  \bibfield  {author} {\bibinfo {author} {\bibnamefont {{Armbruster}},
  \bibfnamefont {P}}, \bibinfo {author} {\bibfnamefont {M.}~\bibnamefont
  {{Asghar}}}, \bibinfo {author} {\bibfnamefont {J.~P.}\ \bibnamefont
  {{Bocquet}}}, \bibinfo {author} {\bibfnamefont {R.}~\bibnamefont {{Decker}}},
  \bibinfo {author} {\bibfnamefont {H.}~\bibnamefont {{Ewald}}}, \bibinfo
  {author} {\bibfnamefont {J.}~\bibnamefont {{Greif}}}, \bibinfo {author}
  {\bibfnamefont {E.}~\bibnamefont {{Moll}}}, \bibinfo {author} {\bibfnamefont
  {B.}~\bibnamefont {{Pfeiffer}}}, \bibinfo {author} {\bibfnamefont
  {H.}~\bibnamefont {{Schrader}}}, \bibinfo {author} {\bibfnamefont
  {F.}~\bibnamefont {{Schussler}}}, \bibinfo {author} {\bibfnamefont
  {G.}~\bibnamefont {{Siegert}}}, and\ \bibinfo {author} {\bibfnamefont
  {H.}~\bibnamefont {{Wollnik}}}} (\bibinfo {year} {1976}),\ \bibfield  {title}
  {\enquote {\bibinfo {title} {{The recoil separator Lohengrin: Performance and
  special features for experiments}},}\ }\href
  {https://doi.org/10.1016/0029-554X(76)90677-7} {\bibfield  {journal}
  {\bibinfo  {journal} {Nucl. Instruments Methods}\ }\textbf {\bibinfo {volume}
  {139}},\ \bibinfo {pages} {213--222}}\BibitemShut {NoStop}%
\bibitem [{\citenamefont {{Arnett}}(1980)}]{Arnett:1980}%
  \BibitemOpen
  \bibfield  {author} {\bibinfo {author} {\bibnamefont {{Arnett}},
  \bibfnamefont {W~D}}} (\bibinfo {year} {1980}),\ \bibfield  {title} {\enquote
  {\bibinfo {title} {{Analytic solutions for light curves of supernovae of Type
  II}},}\ }\href {https://doi.org/10.1086/157898} {\bibfield  {journal}
  {\bibinfo  {journal} {Astrophys. J.}\ }\textbf {\bibinfo {volume} {237}},\
  \bibinfo {pages} {541--549}}\BibitemShut {NoStop}%
\bibitem [{\citenamefont {{Arnett}}(1982)}]{Arnett:1982}%
  \BibitemOpen
  \bibfield  {author} {\bibinfo {author} {\bibnamefont {{Arnett}},
  \bibfnamefont {W~D}}} (\bibinfo {year} {1982}),\ \bibfield  {title} {\enquote
  {\bibinfo {title} {{Type I supernovae. I - Analytic solutions for the early
  part of the light curve}},}\ }\href {https://doi.org/10.1086/159681}
  {\bibfield  {journal} {\bibinfo  {journal} {Astrophys. J.}\ }\textbf
  {\bibinfo {volume} {253}},\ \bibinfo {pages} {785--797}}\BibitemShut
  {NoStop}%
\bibitem [{\citenamefont {{Arnould}}\ and\ \citenamefont
  {{Goriely}}(2003)}]{Arnould.Goriely:2003}%
  \BibitemOpen
  \bibfield  {author} {\bibinfo {author} {\bibnamefont {{Arnould}},
  \bibfnamefont {M}}, and\ \bibinfo {author} {\bibfnamefont {S.}~\bibnamefont
  {{Goriely}}}} (\bibinfo {year} {2003}),\ \bibfield  {title} {\enquote
  {\bibinfo {title} {The p-process of stellar nucleosynthesis: astrophysics and
  nuclear physics status},}\ }\href
  {https://doi.org/10.1016/S0370-1573(03)00242-4} {\bibfield  {journal}
  {\bibinfo  {journal} {Phys. Rep.}\ }\textbf {\bibinfo {volume} {384}},\
  \bibinfo {pages} {1--84}}\BibitemShut {NoStop}%
\bibitem [{\citenamefont {{Arnould}}\ \emph {et~al.}(2007)\citenamefont
  {{Arnould}}, \citenamefont {{Goriely}},\ and\ \citenamefont
  {{Takahashi}}}]{arnould07}%
  \BibitemOpen
  \bibfield  {author} {\bibinfo {author} {\bibnamefont {{Arnould}},
  \bibfnamefont {M}}, \bibinfo {author} {\bibfnamefont {S.}~\bibnamefont
  {{Goriely}}}, and\ \bibinfo {author} {\bibfnamefont {K.}~\bibnamefont
  {{Takahashi}}}} (\bibinfo {year} {2007}),\ \bibfield  {title} {\enquote
  {\bibinfo {title} {{The r-process of stellar nucleosynthesis: Astrophysics
  and nuclear physics achievements and mysteries}},}\ }\href
  {https://doi.org/10.1016/j.physrep.2007.06.002} {\bibfield  {journal}
  {\bibinfo  {journal} {Phys. Rep.}\ }\textbf {\bibinfo {volume} {450}},\
  \bibinfo {pages} {97--213}}\BibitemShut {NoStop}%
\bibitem [{\citenamefont {Arzhanov}(2017)}]{Arzhanov:2017}%
  \BibitemOpen
  \bibfield  {author} {\bibinfo {author} {\bibnamefont {Arzhanov},
  \bibfnamefont {A}}} (\bibinfo {year} {2017}),\ \emph {\bibinfo {title}
  {Microscopic nuclear mass model for r-process nucleosynthesis}},\ \href@noop
  {} {Ph.D. thesis}\ (\bibinfo  {school} {Technische Universit{\"a}t
  Darmstadt})\BibitemShut {NoStop}%
\bibitem [{\citenamefont {{Ascenzi}}\ \emph {et~al.}(2019)\citenamefont
  {{Ascenzi}}, \citenamefont {{Coughlin}}, \citenamefont {{Dietrich}},
  \citenamefont {{Foley}}, \citenamefont {{Ramirez-Ruiz}}, \citenamefont
  {{Piranomonte}}, \citenamefont {{Mockler}}, \citenamefont
  {{Murguia-Berthier}}, \citenamefont {{Fryer}}, \citenamefont
  {{Lloyd-Ronning}},\ and\ \citenamefont
  {{Rosswog}}}]{Ascenzi.Coughlin.ea:2019}%
  \BibitemOpen
  \bibfield  {author} {\bibinfo {author} {\bibnamefont {{Ascenzi}},
  \bibfnamefont {Stefano}}, \bibinfo {author} {\bibfnamefont {Michael~W.}\
  \bibnamefont {{Coughlin}}}, \bibinfo {author} {\bibfnamefont {Tim}\
  \bibnamefont {{Dietrich}}}, \bibinfo {author} {\bibfnamefont {Ryan~J.}\
  \bibnamefont {{Foley}}}, \bibinfo {author} {\bibfnamefont {Enrico}\
  \bibnamefont {{Ramirez-Ruiz}}}, \bibinfo {author} {\bibfnamefont {Silvia}\
  \bibnamefont {{Piranomonte}}}, \bibinfo {author} {\bibfnamefont {Brenna}\
  \bibnamefont {{Mockler}}}, \bibinfo {author} {\bibfnamefont {Ariadna}\
  \bibnamefont {{Murguia-Berthier}}}, \bibinfo {author} {\bibfnamefont
  {Chris~L.}\ \bibnamefont {{Fryer}}}, \bibinfo {author} {\bibfnamefont
  {Nicole~M.}\ \bibnamefont {{Lloyd-Ronning}}}, and\ \bibinfo {author}
  {\bibfnamefont {Stephan}\ \bibnamefont {{Rosswog}}}} (\bibinfo {year}
  {2019}),\ \bibfield  {title} {\enquote {\bibinfo {title} {{A luminosity
  distribution for kilonovae based on short gamma-ray burst afterglows}},}\
  }\href {https://doi.org/10.1093/mnras/stz891} {\bibfield  {journal} {\bibinfo
   {journal} {Mon. Not. Roy. Astron. Soc.}\ }\textbf {\bibinfo {volume}
  {486}},\ \bibinfo {pages} {672--690}}\BibitemShut {NoStop}%
\bibitem [{\citenamefont {{Asplund}}\ \emph {et~al.}(2009)\citenamefont
  {{Asplund}}, \citenamefont {{Grevesse}}, \citenamefont {{Sauval}},\ and\
  \citenamefont {{Scott}}}]{Asplund.Grevesse.ea:2009}%
  \BibitemOpen
  \bibfield  {author} {\bibinfo {author} {\bibnamefont {{Asplund}},
  \bibfnamefont {M}}, \bibinfo {author} {\bibfnamefont {N.}~\bibnamefont
  {{Grevesse}}}, \bibinfo {author} {\bibfnamefont {A.~J.}\ \bibnamefont
  {{Sauval}}}, and\ \bibinfo {author} {\bibfnamefont {P.}~\bibnamefont
  {{Scott}}}} (\bibinfo {year} {2009}),\ \bibfield  {title} {\enquote {\bibinfo
  {title} {{The Chemical Composition of the Sun}},}\ }\href
  {https://doi.org/10.1146/annurev.astro.46.060407.145222} {\bibfield
  {journal} {\bibinfo  {journal} {Annu. Rev. Astron. Astrophys.}\ }\textbf
  {\bibinfo {volume} {47}},\ \bibinfo {pages} {481--522}}\BibitemShut {NoStop}%
\bibitem [{\citenamefont {{Atanasov}}\ \emph {et~al.}(2015)\citenamefont
  {{Atanasov}} \emph {et~al.}}]{Atanasov2015}%
  \BibitemOpen
  \bibfield  {author} {\bibinfo {author} {\bibnamefont {{Atanasov}},
  \bibfnamefont {D}},  \emph {et~al.}} (\bibinfo {year} {2015}),\ \bibfield
  {title} {\enquote {\bibinfo {title} {{Precision Mass Measurements of
  $^{Cd}$-131$_{129}$ and Their Impact on Stellar Nucleosynthesis via the Rapid
  Neutron Capture Process}},}\ }\href
  {https://doi.org/10.1103/PhysRevLett.115.232501} {\bibfield  {journal}
  {\bibinfo  {journal} {Phys. Rev. Lett.}\ }\textbf {\bibinfo {volume} {115}},\
  \bibinfo {eid} {232501}}\BibitemShut {NoStop}%
\bibitem [{\citenamefont {Audi}\ \emph {et~al.}(2012)\citenamefont {Audi},
  \citenamefont {Kondev}, \citenamefont {Wang}, \citenamefont {Pfeiffer},
  \citenamefont {Sun}, \citenamefont {Blachot},\ and\ \citenamefont
  {MacCormick}}]{Audi.Kondev.ea:2012}%
  \BibitemOpen
  \bibfield  {author} {\bibinfo {author} {\bibnamefont {Audi}, \bibfnamefont
  {G}}, \bibinfo {author} {\bibfnamefont {F.~G.}\ \bibnamefont {Kondev}},
  \bibinfo {author} {\bibfnamefont {M.}~\bibnamefont {Wang}}, \bibinfo {author}
  {\bibfnamefont {B.}~\bibnamefont {Pfeiffer}}, \bibinfo {author}
  {\bibfnamefont {X.}~\bibnamefont {Sun}}, \bibinfo {author} {\bibfnamefont
  {J.}~\bibnamefont {Blachot}}, and\ \bibinfo {author} {\bibfnamefont
  {M.}~\bibnamefont {MacCormick}}} (\bibinfo {year} {2012}),\ \bibfield
  {title} {\enquote {\bibinfo {title} {{The Nubase2012 evaluation of nuclear
  properties}},}\ }\href {https://doi.org/10.1088/1674-1137/36/12/001}
  {\bibfield  {journal} {\bibinfo  {journal} {Chinese Phys. C}\ }\textbf
  {\bibinfo {volume} {36}},\ \bibinfo {pages} {1157}}\BibitemShut {NoStop}%
\bibitem [{\citenamefont {Audi}\ \emph {et~al.}(2003)\citenamefont {Audi},
  \citenamefont {Wapstra},\ and\ \citenamefont
  {Thibault}}]{Audi.Wapstra.Thibault:2003}%
  \BibitemOpen
  \bibfield  {author} {\bibinfo {author} {\bibnamefont {Audi}, \bibfnamefont
  {G}}, \bibinfo {author} {\bibfnamefont {A.H.}\ \bibnamefont {Wapstra}}, and\
  \bibinfo {author} {\bibfnamefont {C.}~\bibnamefont {Thibault}}} (\bibinfo
  {year} {2003}),\ \bibfield  {title} {\enquote {\bibinfo {title} {{The Ame2003
  atomic mass evaluation: (II). Tables, graphs and references}},}\ }\href
  {https://doi.org/10.1016/j.nuclphysa.2003.11.003} {\bibfield  {journal}
  {\bibinfo  {journal} {Nucl. Phys. A}\ }\textbf {\bibinfo {volume} {729}},\
  \bibinfo {pages} {337--676}}\BibitemShut {NoStop}%
\bibitem [{\citenamefont {{Audouze}}\ and\ \citenamefont
  {{Tinsley}}(1976)}]{Audouze:1976}%
  \BibitemOpen
  \bibfield  {author} {\bibinfo {author} {\bibnamefont {{Audouze}},
  \bibfnamefont {J}}, and\ \bibinfo {author} {\bibfnamefont {B.~M.}\
  \bibnamefont {{Tinsley}}}} (\bibinfo {year} {1976}),\ \bibfield  {title}
  {\enquote {\bibinfo {title} {{Chemical evolution of galaxies}},}\ }\href
  {https://doi.org/10.1146/annurev.aa.14.090176.000355} {\bibfield  {journal}
  {\bibinfo  {journal} {Annu. Rev. Astron. Astrophys.}\ }\textbf {\bibinfo
  {volume} {14}},\ \bibinfo {pages} {43--79}}\BibitemShut {NoStop}%
\bibitem [{\citenamefont {{Avrigeanu}}\ and\ \citenamefont
  {{Avrigeanu}}(2016)}]{Avrigeanu2016}%
  \BibitemOpen
  \bibfield  {author} {\bibinfo {author} {\bibnamefont {{Avrigeanu}},
  \bibfnamefont {M}}, and\ \bibinfo {author} {\bibfnamefont {V.}~\bibnamefont
  {{Avrigeanu}}}} (\bibinfo {year} {2016}),\ \bibfield  {title} {\enquote
  {\bibinfo {title} {{On deuteron interactions within surrogate reactions and
  nuclear level density studies}},}\ }in\ \href
  {https://doi.org/10.1088/1742-6596/724/1/012003} {\emph {\bibinfo {booktitle}
  {Journal of Physics Conference Series}}},\ Vol.\ \bibinfo {volume} {724},\
  p.\ \bibinfo {pages} {012003}\BibitemShut {NoStop}%
\bibitem [{\citenamefont {{Avrigeanu}}\ and\ \citenamefont
  {{Avrigeanu}}(2015)}]{Avrigeanu.Avrigeanu:2015}%
  \BibitemOpen
  \bibfield  {author} {\bibinfo {author} {\bibnamefont {{Avrigeanu}},
  \bibfnamefont {V}}, and\ \bibinfo {author} {\bibfnamefont {M.}~\bibnamefont
  {{Avrigeanu}}}} (\bibinfo {year} {2015}),\ \bibfield  {title} {\enquote
  {\bibinfo {title} {{Consistent optical potential for incident and emitted
  low-energy {$\alpha$} particles}},}\ }\href
  {https://doi.org/10.1103/PhysRevC.91.064611} {\bibfield  {journal} {\bibinfo
  {journal} {Phys. Rev. C}\ }\textbf {\bibinfo {volume} {91}},\ \bibinfo {eid}
  {064611}}\BibitemShut {NoStop}%
\bibitem [{\citenamefont {{Avrigeanu}}\ \emph {et~al.}(2014)\citenamefont
  {{Avrigeanu}}, \citenamefont {{Avrigeanu}},\ and\ \citenamefont {{M{\v a}n{\v
  a}ilescu}}}]{Avrigeanu.Avrigeanu.Manailescu:2014}%
  \BibitemOpen
  \bibfield  {author} {\bibinfo {author} {\bibnamefont {{Avrigeanu}},
  \bibfnamefont {V}}, \bibinfo {author} {\bibfnamefont {M.}~\bibnamefont
  {{Avrigeanu}}}, and\ \bibinfo {author} {\bibfnamefont {C.}~\bibnamefont
  {{M{\v a}n{\v a}ilescu}}}} (\bibinfo {year} {2014}),\ \bibfield  {title}
  {\enquote {\bibinfo {title} {{Further explorations of the $\alpha$-particle
  optical model potential at low energies for the mass range $A\approx
  45$--209}},}\ }\href {https://doi.org/10.1103/PhysRevC.90.044612} {\bibfield
  {journal} {\bibinfo  {journal} {Phys. Rev. C}\ }\textbf {\bibinfo {volume}
  {90}},\ \bibinfo {eid} {044612}}\BibitemShut {NoStop}%
\bibitem [{\citenamefont {{Baade}}\ and\ \citenamefont
  {{Zwicky}}(1934)}]{baadezwicky34}%
  \BibitemOpen
  \bibfield  {author} {\bibinfo {author} {\bibnamefont {{Baade}}, \bibfnamefont
  {W}}, and\ \bibinfo {author} {\bibfnamefont {F.}~\bibnamefont {{Zwicky}}}}
  (\bibinfo {year} {1934}),\ \bibfield  {title} {\enquote {\bibinfo {title}
  {{Remarks on Super-Novae and Cosmic Rays}},}\ }\href
  {https://doi.org/10.1103/PhysRev.46.76.2} {\bibfield  {journal} {\bibinfo
  {journal} {Phys. Rev.}\ }\textbf {\bibinfo {volume} {46}},\ \bibinfo {pages}
  {76--77}}\BibitemShut {NoStop}%
\bibitem [{\citenamefont {{Baiotti}}\ and\ \citenamefont
  {{Rezzolla}}(2017)}]{Baiotti.Rezzolla:2017}%
  \BibitemOpen
  \bibfield  {author} {\bibinfo {author} {\bibnamefont {{Baiotti}},
  \bibfnamefont {Luca}}, and\ \bibinfo {author} {\bibfnamefont {Luciano}\
  \bibnamefont {{Rezzolla}}}} (\bibinfo {year} {2017}),\ \bibfield  {title}
  {\enquote {\bibinfo {title} {{Binary neutron star mergers: a review of
  Einstein{\textquoteright}s richest laboratory}},}\ }\href
  {https://doi.org/10.1088/1361-6633/aa67bb} {\bibfield  {journal} {\bibinfo
  {journal} {Rep. Prog. Phys.}\ }\textbf {\bibinfo {volume} {80}},\ \bibinfo
  {eid} {096901}}\BibitemShut {NoStop}%
\bibitem [{\citenamefont {Baldo}\ and\ \citenamefont
  {Burgio}(2016)}]{Baldo.Burgio:2016}%
  \BibitemOpen
  \bibfield  {author} {\bibinfo {author} {\bibnamefont {Baldo}, \bibfnamefont
  {M}}, and\ \bibinfo {author} {\bibfnamefont {G.F.}\ \bibnamefont {Burgio}}}
  (\bibinfo {year} {2016}),\ \bibfield  {title} {\enquote {\bibinfo {title}
  {{The nuclear symmetry energy}},}\ }\href
  {https://doi.org/10.1016/j.ppnp.2016.06.006} {\bibfield  {journal} {\bibinfo
  {journal} {Prog. Part. Nucl. Phys.}\ }\textbf {\bibinfo {volume} {91}},\
  \bibinfo {pages} {203--258}}\BibitemShut {NoStop}%
\bibitem [{\citenamefont {{Banerjee}}\ \emph {et~al.}(2011)\citenamefont
  {{Banerjee}}, \citenamefont {{Haxton}},\ and\ \citenamefont
  {{Qian}}}]{Banerjee.Haxton.Qian:2011}%
  \BibitemOpen
  \bibfield  {author} {\bibinfo {author} {\bibnamefont {{Banerjee}},
  \bibfnamefont {P}}, \bibinfo {author} {\bibfnamefont {W.~C.}\ \bibnamefont
  {{Haxton}}}, and\ \bibinfo {author} {\bibfnamefont {Y.-Z.}\ \bibnamefont
  {{Qian}}}} (\bibinfo {year} {2011}),\ \bibfield  {title} {\enquote {\bibinfo
  {title} {{Long, Cold, Early r Process? Neutrino-Induced Nucleosynthesis in He
  Shells Revisited}},}\ }\href {https://doi.org/10.1103/PhysRevLett.106.201104}
  {\bibfield  {journal} {\bibinfo  {journal} {Phys. Rev. Lett.}\ }\textbf
  {\bibinfo {volume} {106}},\ \bibinfo {eid} {201104}}\BibitemShut {NoStop}%
\bibitem [{\citenamefont {{Banerjee}}\ \emph {et~al.}(2016)\citenamefont
  {{Banerjee}}, \citenamefont {{Qian}}, \citenamefont {{Heger}},\ and\
  \citenamefont {{Haxton}}}]{Banerjee.Qian.ea:2016}%
  \BibitemOpen
  \bibfield  {author} {\bibinfo {author} {\bibnamefont {{Banerjee}},
  \bibfnamefont {P}}, \bibinfo {author} {\bibfnamefont {Y.-Z.}\ \bibnamefont
  {{Qian}}}, \bibinfo {author} {\bibfnamefont {A.}~\bibnamefont {{Heger}}},
  and\ \bibinfo {author} {\bibfnamefont {W.}~\bibnamefont {{Haxton}}}}
  (\bibinfo {year} {2016}),\ \bibfield  {title} {\enquote {\bibinfo {title}
  {{Neutrino-Induced Nucleosynthesis in Helium Shells of Early Core-Collapse
  Supernovae}},}\ }in\ \href {https://doi.org/10.1051/epjconf/201610906001}
  {\emph {\bibinfo {booktitle} {European Physical Journal Web of
  Conferences}}},\ Vol.\ \bibinfo {volume} {109},\ p.\ \bibinfo {pages}
  {06001}\BibitemShut {NoStop}%
\bibitem [{\citenamefont {{Barbieri}}\ \emph {et~al.}(2020)\citenamefont
  {{Barbieri}}, \citenamefont {{Salafia}}, \citenamefont {{Colpi}},
  \citenamefont {{Ghirlanda}},\ and\ \citenamefont
  {{Perego}}}]{Barbieri.Salafia.ea:2020}%
  \BibitemOpen
  \bibfield  {author} {\bibinfo {author} {\bibnamefont {{Barbieri}},
  \bibfnamefont {C}}, \bibinfo {author} {\bibfnamefont {O.~S.}\ \bibnamefont
  {{Salafia}}}, \bibinfo {author} {\bibfnamefont {M.}~\bibnamefont {{Colpi}}},
  \bibinfo {author} {\bibfnamefont {G.}~\bibnamefont {{Ghirlanda}}}, and\
  \bibinfo {author} {\bibfnamefont {A.}~\bibnamefont {{Perego}}}} (\bibinfo
  {year} {2020}),\ \bibfield  {title} {\enquote {\bibinfo {title}
  {{Distinguishing the nature of ``ambiguous'' merging systems hosting a
  neutron star: GW190425 in low-latency}},}\ }\href@noop {} {\bibfield
  {journal} {\bibinfo  {journal} {arXiv e-prints}\ }}\Eprint
  {https://arxiv.org/abs/2002.09395} {arXiv:2002.09395 [astro-ph.HE]}
  \BibitemShut {NoStop}%
\bibitem [{\citenamefont {{Barbieri}}\ \emph {et~al.}(2019)\citenamefont
  {{Barbieri}}, \citenamefont {{Salafia}}, \citenamefont {{Colpi}},
  \citenamefont {{Ghirlanda}}, \citenamefont {{Perego}},\ and\ \citenamefont
  {{Colombo}}}]{Barbieri.Salafia.ea:2019}%
  \BibitemOpen
  \bibfield  {author} {\bibinfo {author} {\bibnamefont {{Barbieri}},
  \bibfnamefont {C}}, \bibinfo {author} {\bibfnamefont {O.~S.}\ \bibnamefont
  {{Salafia}}}, \bibinfo {author} {\bibfnamefont {M.}~\bibnamefont {{Colpi}}},
  \bibinfo {author} {\bibfnamefont {G.}~\bibnamefont {{Ghirlanda}}}, \bibinfo
  {author} {\bibfnamefont {A.}~\bibnamefont {{Perego}}}, and\ \bibinfo {author}
  {\bibfnamefont {A.}~\bibnamefont {{Colombo}}}} (\bibinfo {year} {2019}),\
  \bibfield  {title} {\enquote {\bibinfo {title} {{Filling the Mass Gap: How
  Kilonova Observations Can Unveil the Nature of the Compact Object Merging
  with the Neutron Star}},}\ }\href {https://doi.org/10.3847/2041-8213/ab5c1e}
  {\bibfield  {journal} {\bibinfo  {journal} {Astrophys. J. Lett.}\ }\textbf
  {\bibinfo {volume} {887}},\ \bibinfo {eid} {L35}}\BibitemShut {NoStop}%
\bibitem [{\citenamefont {{Barklem}}\ \emph {et~al.}(2005)\citenamefont
  {{Barklem}}, \citenamefont {{Christlieb}}, \citenamefont {{Beers}},
  \citenamefont {{Hill}}, \citenamefont {{Bessell}}, \citenamefont
  {{Holmberg}}, \citenamefont {{Marsteller}}, \citenamefont {{Rossi}},
  \citenamefont {{Zickgraf}},\ and\ \citenamefont
  {{Reimers}}}]{Barklem.Christlieb.ea:2005}%
  \BibitemOpen
  \bibfield  {author} {\bibinfo {author} {\bibnamefont {{Barklem}},
  \bibfnamefont {P~S}}, \bibinfo {author} {\bibfnamefont {N.}~\bibnamefont
  {{Christlieb}}}, \bibinfo {author} {\bibfnamefont {T.~C.}\ \bibnamefont
  {{Beers}}}, \bibinfo {author} {\bibfnamefont {V.}~\bibnamefont {{Hill}}},
  \bibinfo {author} {\bibfnamefont {M.~S.}\ \bibnamefont {{Bessell}}}, \bibinfo
  {author} {\bibfnamefont {J.}~\bibnamefont {{Holmberg}}}, \bibinfo {author}
  {\bibfnamefont {B.}~\bibnamefont {{Marsteller}}}, \bibinfo {author}
  {\bibfnamefont {S.}~\bibnamefont {{Rossi}}}, \bibinfo {author} {\bibfnamefont
  {F.-J.}\ \bibnamefont {{Zickgraf}}}, and\ \bibinfo {author} {\bibfnamefont
  {D.}~\bibnamefont {{Reimers}}}} (\bibinfo {year} {2005}),\ \bibfield  {title}
  {\enquote {\bibinfo {title} {{The Hamburg/ESO R-process enhanced star survey
  (HERES). II. Spectroscopic analysis of the survey sample}},}\ }\href
  {https://doi.org/10.1051/0004-6361:20052967} {\bibfield  {journal} {\bibinfo
  {journal} {Astron. \& Astrophys.}\ }\textbf {\bibinfo {volume} {439}},\
  \bibinfo {pages} {129--151}}\BibitemShut {NoStop}%
\bibitem [{\citenamefont {{Barnes}}\ \emph {et~al.}(2016)\citenamefont
  {{Barnes}}, \citenamefont {{Kasen}}, \citenamefont {{Wu}},\ and\
  \citenamefont {{Mart{\'{\i}}nez-Pinedo}}}]{barnes16}%
  \BibitemOpen
  \bibfield  {author} {\bibinfo {author} {\bibnamefont {{Barnes}},
  \bibfnamefont {J}}, \bibinfo {author} {\bibfnamefont {D.}~\bibnamefont
  {{Kasen}}}, \bibinfo {author} {\bibfnamefont {M.-R.}\ \bibnamefont {{Wu}}},
  and\ \bibinfo {author} {\bibfnamefont {G.}~\bibnamefont
  {{Mart{\'{\i}}nez-Pinedo}}}} (\bibinfo {year} {2016}),\ \bibfield  {title}
  {\enquote {\bibinfo {title} {{Radioactivity and Thermalization in the Ejecta
  of Compact Object Mergers and Their Impact on Kilonova Light Curves}},}\
  }\href {https://doi.org/10.3847/0004-637X/829/2/110} {\bibfield  {journal}
  {\bibinfo  {journal} {Astrophys. J.}\ }\textbf {\bibinfo {volume} {829}},\
  \bibinfo {eid} {110}}\BibitemShut {NoStop}%
\bibitem [{\citenamefont {Barnes}\ and\ \citenamefont
  {Kasen}(2013)}]{Barnes.Kasen:2013}%
  \BibitemOpen
  \bibfield  {author} {\bibinfo {author} {\bibnamefont {Barnes}, \bibfnamefont
  {Jennifer}}, and\ \bibinfo {author} {\bibfnamefont {Daniel}\ \bibnamefont
  {Kasen}}} (\bibinfo {year} {2013}),\ \bibfield  {title} {\enquote {\bibinfo
  {title} {{Effect of a High Opacity on the Light Curves of Radioactively
  Powered Transients from Compact Object Mergers}},}\ }\href
  {https://doi.org/10.1088/0004-637X/775/1/18} {\bibfield  {journal} {\bibinfo
  {journal} {Astrophys. J.}\ }\textbf {\bibinfo {volume} {775}},\ \bibinfo
  {pages} {18}}\BibitemShut {NoStop}%
\bibitem [{\citenamefont {{Bartos}}\ \emph {et~al.}(2013)\citenamefont
  {{Bartos}}, \citenamefont {{Brady}},\ and\ \citenamefont
  {{M{\'a}rka}}}]{Bartos.Brady.Marka:2013}%
  \BibitemOpen
  \bibfield  {author} {\bibinfo {author} {\bibnamefont {{Bartos}},
  \bibfnamefont {I}}, \bibinfo {author} {\bibfnamefont {P.}~\bibnamefont
  {{Brady}}}, and\ \bibinfo {author} {\bibfnamefont {S.}~\bibnamefont
  {{M{\'a}rka}}}} (\bibinfo {year} {2013}),\ \bibfield  {title} {\enquote
  {\bibinfo {title} {{How gravitational-wave observations can shape the
  gamma-ray burst paradigm}},}\ }\href
  {https://doi.org/10.1088/0264-9381/30/12/123001} {\bibfield  {journal}
  {\bibinfo  {journal} {Class. Quantum Gravity}\ }\textbf {\bibinfo {volume}
  {30}},\ \bibinfo {eid} {123001}}\BibitemShut {NoStop}%
\bibitem [{\citenamefont {{Battistini}}\ and\ \citenamefont
  {{Bensby}}(2016)}]{battistini16}%
  \BibitemOpen
  \bibfield  {author} {\bibinfo {author} {\bibnamefont {{Battistini}},
  \bibfnamefont {C}}, and\ \bibinfo {author} {\bibfnamefont {T.}~\bibnamefont
  {{Bensby}}}} (\bibinfo {year} {2016}),\ \bibfield  {title} {\enquote
  {\bibinfo {title} {{The origin and evolution of r- and s-process elements in
  the Milky Way stellar disk}},}\ }\href
  {https://doi.org/10.1051/0004-6361/201527385} {\bibfield  {journal} {\bibinfo
   {journal} {Astron. \& Astrophys.}\ }\textbf {\bibinfo {volume} {586}},\
  \bibinfo {eid} {A49}}\BibitemShut {NoStop}%
\bibitem [{\citenamefont {{Bauge}}\ \emph {et~al.}(2001)\citenamefont
  {{Bauge}}, \citenamefont {{Delaroche}},\ and\ \citenamefont
  {{Girod}}}]{Bauge.Delaroche.Girod:2001}%
  \BibitemOpen
  \bibfield  {author} {\bibinfo {author} {\bibnamefont {{Bauge}}, \bibfnamefont
  {E}}, \bibinfo {author} {\bibfnamefont {J.~P.}\ \bibnamefont {{Delaroche}}},
  and\ \bibinfo {author} {\bibfnamefont {M.}~\bibnamefont {{Girod}}}} (\bibinfo
  {year} {2001}),\ \bibfield  {title} {\enquote {\bibinfo {title}
  {{Lane-consistent, semimicroscopic nucleon-nucleus optical model}},}\ }\href
  {https://doi.org/10.1103/PhysRevC.63.024607} {\bibfield  {journal} {\bibinfo
  {journal} {Phys. Rev. C}\ }\textbf {\bibinfo {volume} {63}},\ \bibinfo {eid}
  {024607}}\BibitemShut {NoStop}%
\bibitem [{\citenamefont {{Baumgarte}}\ \emph {et~al.}(2000)\citenamefont
  {{Baumgarte}}, \citenamefont {{Shapiro}},\ and\ \citenamefont
  {{Shibata}}}]{Baumgarte.Shapiro.Shibata:2000}%
  \BibitemOpen
  \bibfield  {author} {\bibinfo {author} {\bibnamefont {{Baumgarte}},
  \bibfnamefont {T~W}}, \bibinfo {author} {\bibfnamefont {S.~L.}\ \bibnamefont
  {{Shapiro}}}, and\ \bibinfo {author} {\bibfnamefont {M.}~\bibnamefont
  {{Shibata}}}} (\bibinfo {year} {2000}),\ \bibfield  {title} {\enquote
  {\bibinfo {title} {{On the Maximum Mass of Differentially Rotating Neutron
  Stars}},}\ }\href {https://doi.org/10.1086/312425} {\bibfield  {journal}
  {\bibinfo  {journal} {Astrophys. J. Lett.}\ }\textbf {\bibinfo {volume}
  {528}},\ \bibinfo {pages} {L29--L32}}\BibitemShut {NoStop}%
\bibitem [{\citenamefont {{Bauswein}}\ \emph {et~al.}(2013)\citenamefont
  {{Bauswein}}, \citenamefont {{Baumgarte}},\ and\ \citenamefont
  {{Janka}}}]{Bauswein.Baumgarte.Janka:2013}%
  \BibitemOpen
  \bibfield  {author} {\bibinfo {author} {\bibnamefont {{Bauswein}},
  \bibfnamefont {A}}, \bibinfo {author} {\bibfnamefont {T.~W.}\ \bibnamefont
  {{Baumgarte}}}, and\ \bibinfo {author} {\bibfnamefont {H.-T.}\ \bibnamefont
  {{Janka}}}} (\bibinfo {year} {2013}),\ \bibfield  {title} {\enquote {\bibinfo
  {title} {{Prompt Merger Collapse and the Maximum Mass of Neutron Stars}},}\
  }\href {https://doi.org/10.1103/PhysRevLett.111.131101} {\bibfield  {journal}
  {\bibinfo  {journal} {Phys. Rev. Lett.}\ }\textbf {\bibinfo {volume} {111}},\
  \bibinfo {eid} {131101}}\BibitemShut {NoStop}%
\bibitem [{\citenamefont {Bauswein}\ \emph {et~al.}(2013)\citenamefont
  {Bauswein}, \citenamefont {Goriely},\ and\ \citenamefont
  {Janka}}]{Bauswein.Goriely.Janka:2013}%
  \BibitemOpen
  \bibfield  {author} {\bibinfo {author} {\bibnamefont {Bauswein},
  \bibfnamefont {A}}, \bibinfo {author} {\bibfnamefont {S.}~\bibnamefont
  {Goriely}}, and\ \bibinfo {author} {\bibfnamefont {H.-T.}\ \bibnamefont
  {Janka}}} (\bibinfo {year} {2013}),\ \bibfield  {title} {\enquote {\bibinfo
  {title} {Systematics of dynamical mass ejection, nucleosynthesis, and
  radioactively powered electromagnetic signals from neutron-star mergers},}\
  }\href {https://doi.org/10.1088/0004-637X/773/1/78} {\bibfield  {journal}
  {\bibinfo  {journal} {Astrophys. J.}\ }\textbf {\bibinfo {volume} {773}},\
  \bibinfo {pages} {78}}\BibitemShut {NoStop}%
\bibitem [{\citenamefont {Bauswein}\ and\ \citenamefont
  {Janka}(2012)}]{Bauswein.Janka:2012}%
  \BibitemOpen
  \bibfield  {author} {\bibinfo {author} {\bibnamefont {Bauswein},
  \bibfnamefont {A}}, and\ \bibinfo {author} {\bibfnamefont {H.-T.}\
  \bibnamefont {Janka}}} (\bibinfo {year} {2012}),\ \bibfield  {title}
  {\enquote {\bibinfo {title} {Measuring neutron-star properties via
  gravitational waves from neutron-star mergers},}\ }\href
  {https://doi.org/10.1103/PhysRevLett.108.011101} {\bibfield  {journal}
  {\bibinfo  {journal} {Phys. Rev. Lett.}\ }\textbf {\bibinfo {volume} {108}},\
  \bibinfo {pages} {011101}}\BibitemShut {NoStop}%
\bibitem [{\citenamefont {{Bauswein}}\ and\ \citenamefont
  {{Stergioulas}}(2017)}]{Bauswein.Stergioulas:2017}%
  \BibitemOpen
  \bibfield  {author} {\bibinfo {author} {\bibnamefont {{Bauswein}},
  \bibfnamefont {A}}, and\ \bibinfo {author} {\bibfnamefont {N.}~\bibnamefont
  {{Stergioulas}}}} (\bibinfo {year} {2017}),\ \bibfield  {title} {\enquote
  {\bibinfo {title} {{Semi-analytic derivation of the threshold mass for prompt
  collapse in binary neutron-star mergers}},}\ }\href
  {https://doi.org/10.1093/mnras/stx1983} {\bibfield  {journal} {\bibinfo
  {journal} {Mon. Not. Roy. Astron. Soc.}\ }\textbf {\bibinfo {volume} {471}},\
  \bibinfo {pages} {4956--4965}}\BibitemShut {NoStop}%
\bibitem [{\citenamefont {{Bauswein}}\ and\ \citenamefont
  {{Stergioulas}}(2019)}]{Bauswein.Stergioulas:2019}%
  \BibitemOpen
  \bibfield  {author} {\bibinfo {author} {\bibnamefont {{Bauswein}},
  \bibfnamefont {A}}, and\ \bibinfo {author} {\bibfnamefont {N.}~\bibnamefont
  {{Stergioulas}}}} (\bibinfo {year} {2019}),\ \bibfield  {title} {\enquote
  {\bibinfo {title} {{Spectral classification of gravitational-wave emission
  and equation of state constraints in binary neutron star mergers}},}\ }\href
  {https://doi.org/10.1088/1361-6471/ab2b90} {\bibfield  {journal} {\bibinfo
  {journal} {J. Phys. G: Nucl. Part. Phys.}\ }\textbf {\bibinfo {volume}
  {46}},\ \bibinfo {eid} {113002}}\BibitemShut {NoStop}%
\bibitem [{\citenamefont {{Bauswein}}\ \emph {et~al.}(2016)\citenamefont
  {{Bauswein}}, \citenamefont {{Stergioulas}},\ and\ \citenamefont
  {{Janka}}}]{bauswein16}%
  \BibitemOpen
  \bibfield  {author} {\bibinfo {author} {\bibnamefont {{Bauswein}},
  \bibfnamefont {A}}, \bibinfo {author} {\bibfnamefont {N.}~\bibnamefont
  {{Stergioulas}}}, and\ \bibinfo {author} {\bibfnamefont {H.-T.}\ \bibnamefont
  {{Janka}}}} (\bibinfo {year} {2016}),\ \bibfield  {title} {\enquote {\bibinfo
  {title} {{Exploring properties of high-density matter through remnants of
  neutron-star mergers}},}\ }\href {https://doi.org/10.1140/epja/i2016-16056-7}
  {\bibfield  {journal} {\bibinfo  {journal} {Eur. Phys. J. A}\ }\textbf
  {\bibinfo {volume} {52}},\ \bibinfo {eid} {56}}\BibitemShut {NoStop}%
\bibitem [{\citenamefont {{Bauswein}}\ \emph {et~al.}(2017)\citenamefont
  {{Bauswein}}, \citenamefont {{Just}}, \citenamefont {{Janka}},\ and\
  \citenamefont {{Stergioulas}}}]{Bauswein.Just.ea:2017}%
  \BibitemOpen
  \bibfield  {author} {\bibinfo {author} {\bibnamefont {{Bauswein}},
  \bibfnamefont {Andreas}}, \bibinfo {author} {\bibfnamefont {Oliver}\
  \bibnamefont {{Just}}}, \bibinfo {author} {\bibfnamefont {Hans-Thomas}\
  \bibnamefont {{Janka}}}, and\ \bibinfo {author} {\bibfnamefont {Nikolaos}\
  \bibnamefont {{Stergioulas}}}} (\bibinfo {year} {2017}),\ \bibfield  {title}
  {\enquote {\bibinfo {title} {{Neutron-star Radius Constraints from GW170817
  and Future Detections}},}\ }\href {https://doi.org/10.3847/2041-8213/aa9994}
  {\bibfield  {journal} {\bibinfo  {journal} {Astrophys. J. Lett.}\ }\textbf
  {\bibinfo {volume} {850}},\ \bibinfo {eid} {L34}}\BibitemShut {NoStop}%
\bibitem [{\citenamefont {{Beard}}\ \emph {et~al.}(2014)\citenamefont
  {{Beard}}, \citenamefont {{Uberseder}}, \citenamefont {{Crowter}},\ and\
  \citenamefont {{Wiescher}}}]{Beard2014}%
  \BibitemOpen
  \bibfield  {author} {\bibinfo {author} {\bibnamefont {{Beard}}, \bibfnamefont
  {M}}, \bibinfo {author} {\bibfnamefont {E.}~\bibnamefont {{Uberseder}}},
  \bibinfo {author} {\bibfnamefont {R.}~\bibnamefont {{Crowter}}}, and\
  \bibinfo {author} {\bibfnamefont {M.}~\bibnamefont {{Wiescher}}}} (\bibinfo
  {year} {2014}),\ \bibfield  {title} {\enquote {\bibinfo {title} {{Comparison
  of statistical model calculations for stable isotope neutron capture}},}\
  }\href {https://doi.org/10.1103/PhysRevC.90.034619} {\bibfield  {journal}
  {\bibinfo  {journal} {Phys. Rev. C}\ }\textbf {\bibinfo {volume} {90}},\
  \bibinfo {eid} {034619}}\BibitemShut {NoStop}%
\bibitem [{\citenamefont {{Beers}}\ and\ \citenamefont
  {{Christlieb}}(2005)}]{beers05}%
  \BibitemOpen
  \bibfield  {author} {\bibinfo {author} {\bibnamefont {{Beers}}, \bibfnamefont
  {T~C}}, and\ \bibinfo {author} {\bibfnamefont {N.}~\bibnamefont
  {{Christlieb}}}} (\bibinfo {year} {2005}),\ \bibfield  {title} {\enquote
  {\bibinfo {title} {{The Discovery and Analysis of Very Metal-Poor Stars in
  the Galaxy}},}\ }\href
  {https://doi.org/10.1146/annurev.astro.42.053102.134057} {\bibfield
  {journal} {\bibinfo  {journal} {Annu. Rev. Astron. Astrophys.}\ }\textbf
  {\bibinfo {volume} {43}},\ \bibinfo {pages} {531--580}}\BibitemShut {NoStop}%
\bibitem [{\citenamefont {{Belczynski}}\ \emph {et~al.}(2010)\citenamefont
  {{Belczynski}}, \citenamefont {{Dominik}}, \citenamefont {{Bulik}},
  \citenamefont {{O'Shaughnessy}}, \citenamefont {{Fryer}},\ and\ \citenamefont
  {{Holz}}}]{Belczynski.Dominik.ea:2010}%
  \BibitemOpen
  \bibfield  {author} {\bibinfo {author} {\bibnamefont {{Belczynski}},
  \bibfnamefont {K}}, \bibinfo {author} {\bibfnamefont {M.}~\bibnamefont
  {{Dominik}}}, \bibinfo {author} {\bibfnamefont {T.}~\bibnamefont {{Bulik}}},
  \bibinfo {author} {\bibfnamefont {R.}~\bibnamefont {{O'Shaughnessy}}},
  \bibinfo {author} {\bibfnamefont {C.}~\bibnamefont {{Fryer}}}, and\ \bibinfo
  {author} {\bibfnamefont {D.~E.}\ \bibnamefont {{Holz}}}} (\bibinfo {year}
  {2010}),\ \bibfield  {title} {\enquote {\bibinfo {title} {{The Effect of
  Metallicity on the Detection Prospects for Gravitational Waves}},}\ }\href
  {https://doi.org/10.1088/2041-8205/715/2/L138} {\bibfield  {journal}
  {\bibinfo  {journal} {Astrophys. J. Lett.}\ }\textbf {\bibinfo {volume}
  {715}},\ \bibinfo {pages} {L138--L141}}\BibitemShut {NoStop}%
\bibitem [{\citenamefont {{Beloborodov}}(2003)}]{Beloborodov:2003}%
  \BibitemOpen
  \bibfield  {author} {\bibinfo {author} {\bibnamefont {{Beloborodov}},
  \bibfnamefont {A~M}}} (\bibinfo {year} {2003}),\ \bibfield  {title} {\enquote
  {\bibinfo {title} {{Nuclear Composition of Gamma-Ray Burst Fireballs}},}\
  }\href {https://doi.org/10.1086/374217} {\bibfield  {journal} {\bibinfo
  {journal} {Astrophys. J.}\ }\textbf {\bibinfo {volume} {588}},\ \bibinfo
  {pages} {931--944}}\BibitemShut {NoStop}%
\bibitem [{\citenamefont {{Beloborodov}}(2008)}]{Beloborodov:2008}%
  \BibitemOpen
  \bibfield  {author} {\bibinfo {author} {\bibnamefont {{Beloborodov}},
  \bibfnamefont {A~M}}} (\bibinfo {year} {2008}),\ \bibfield  {title} {\enquote
  {\bibinfo {title} {{Hyper-accreting black holes}},}\ }in\ \href
  {https://doi.org/10.1063/1.3002509} {\emph {\bibinfo {booktitle} {American
  Institute of Physics Conference Series}}},\ Vol.\ \bibinfo {volume} {1054},\
  \bibinfo {editor} {edited by\ \bibinfo {editor} {\bibfnamefont
  {M.}~\bibnamefont {{Axelsson}}}},\ pp.\ \bibinfo {pages} {51--70},\ \Eprint
  {https://arxiv.org/abs/0810.2690} {0810.2690} \BibitemShut {NoStop}%
\bibitem [{\citenamefont {{Bengtsson}}\ and\ \citenamefont
  {{Howard}}(1975)}]{Bengtsson.Howard:1975}%
  \BibitemOpen
  \bibfield  {author} {\bibinfo {author} {\bibnamefont {{Bengtsson}},
  \bibfnamefont {R}}, and\ \bibinfo {author} {\bibfnamefont {W.~M.}\
  \bibnamefont {{Howard}}}} (\bibinfo {year} {1975}),\ \bibfield  {title}
  {\enquote {\bibinfo {title} {{Mass asymmetric fission and the termination of
  the astrophysical r-process}},}\ }\href
  {https://doi.org/10.1016/0370-2693(75)90600-0} {\bibfield  {journal}
  {\bibinfo  {journal} {Phys. Lett. B}\ }\textbf {\bibinfo {volume} {55}},\
  \bibinfo {pages} {281--285}}\BibitemShut {NoStop}%
\bibitem [{\citenamefont {{Beniamini}}\ and\ \citenamefont
  {{Hotokezaka}}(2020)}]{Beniamini.Hotokezaka:2020}%
  \BibitemOpen
  \bibfield  {author} {\bibinfo {author} {\bibnamefont {{Beniamini}},
  \bibfnamefont {Paz}}, and\ \bibinfo {author} {\bibfnamefont {Kenta}\
  \bibnamefont {{Hotokezaka}}}} (\bibinfo {year} {2020}),\ \bibfield  {title}
  {\enquote {\bibinfo {title} {{Turbulent mixing of r-process elements in the
  Milky Way}},}\ }\href {https://doi.org/10.1093/mnras/staa1690} {\bibfield
  {journal} {\bibinfo  {journal} {Mon. Not. Roy. Astron. Soc.}\ }\textbf
  {\bibinfo {volume} {496}}~(\bibinfo {number} {2}),\ \bibinfo {pages}
  {1891--1901}}\BibitemShut {NoStop}%
\bibitem [{\citenamefont {{Beniamini}}\ \emph {et~al.}(2016)\citenamefont
  {{Beniamini}}, \citenamefont {{Hotokezaka}},\ and\ \citenamefont
  {{Piran}}}]{Beniamini.Hotokezaka.Piran:2016}%
  \BibitemOpen
  \bibfield  {author} {\bibinfo {author} {\bibnamefont {{Beniamini}},
  \bibfnamefont {Paz}}, \bibinfo {author} {\bibfnamefont {Kenta}\ \bibnamefont
  {{Hotokezaka}}}, and\ \bibinfo {author} {\bibfnamefont {Tsvi}\ \bibnamefont
  {{Piran}}}} (\bibinfo {year} {2016}),\ \bibfield  {title} {\enquote {\bibinfo
  {title} {{r-process Production Sites as Inferred from Eu Abundances in Dwarf
  Galaxies}},}\ }\href {https://doi.org/10.3847/0004-637X/832/2/149} {\bibfield
   {journal} {\bibinfo  {journal} {Astrophys. J.}\ }\textbf {\bibinfo {volume}
  {832}},\ \bibinfo {eid} {149}}\BibitemShut {NoStop}%
\bibitem [{\citenamefont {{Beniamini}}\ \emph {et~al.}(2019)\citenamefont
  {{Beniamini}}, \citenamefont {{Hotokezaka}}, \citenamefont {{van der
  Horst}},\ and\ \citenamefont
  {{Kouveliotou}}}]{Beniamini.Hotokezaka.Horst.ea:2019}%
  \BibitemOpen
  \bibfield  {author} {\bibinfo {author} {\bibnamefont {{Beniamini}},
  \bibfnamefont {Paz}}, \bibinfo {author} {\bibfnamefont {Kenta}\ \bibnamefont
  {{Hotokezaka}}}, \bibinfo {author} {\bibfnamefont {Alexander}\ \bibnamefont
  {{van der Horst}}}, and\ \bibinfo {author} {\bibfnamefont {Chryssa}\
  \bibnamefont {{Kouveliotou}}}} (\bibinfo {year} {2019}),\ \bibfield  {title}
  {\enquote {\bibinfo {title} {{Formation rates and evolution histories of
  magnetars}},}\ }\href {https://doi.org/10.1093/mnras/stz1391} {\bibfield
  {journal} {\bibinfo  {journal} {Mon. Not. Roy. Astron. Soc.}\ }\textbf
  {\bibinfo {volume} {487}},\ \bibinfo {pages} {1426--1438}}\BibitemShut
  {NoStop}%
\bibitem [{\citenamefont {{Beniamini}}\ and\ \citenamefont
  {{Piran}}(2019)}]{Beniamini.Piran:2019}%
  \BibitemOpen
  \bibfield  {author} {\bibinfo {author} {\bibnamefont {{Beniamini}},
  \bibfnamefont {Paz}}, and\ \bibinfo {author} {\bibfnamefont {Tsvi}\
  \bibnamefont {{Piran}}}} (\bibinfo {year} {2019}),\ \bibfield  {title}
  {\enquote {\bibinfo {title} {{The Gravitational waves merger time
  distribution of binary neutron star systems}},}\ }\href
  {https://doi.org/10.1093/mnras/stz1589} {\bibfield  {journal} {\bibinfo
  {journal} {Mon. Not. Roy. Astron. Soc.}\ }\textbf {\bibinfo {volume} {487}},\
  \bibinfo {pages} {4847--4854}}\BibitemShut {NoStop}%
\bibitem [{\citenamefont {{Bensby}}\ \emph {et~al.}(2014)\citenamefont
  {{Bensby}}, \citenamefont {{Feltzing}},\ and\ \citenamefont
  {{Oey}}}]{bensby14}%
  \BibitemOpen
  \bibfield  {author} {\bibinfo {author} {\bibnamefont {{Bensby}},
  \bibfnamefont {T}}, \bibinfo {author} {\bibfnamefont {S.}~\bibnamefont
  {{Feltzing}}}, and\ \bibinfo {author} {\bibfnamefont {M.~S.}\ \bibnamefont
  {{Oey}}}} (\bibinfo {year} {2014}),\ \bibfield  {title} {\enquote {\bibinfo
  {title} {{Exploring the Milky Way stellar disk. A detailed elemental
  abundance study of 714 F and G dwarf stars in the solar neighbourhood}},}\
  }\href {https://doi.org/10.1051/0004-6361/201322631} {\bibfield  {journal}
  {\bibinfo  {journal} {Astron. \& Astrophys.}\ }\textbf {\bibinfo {volume}
  {562}},\ \bibinfo {eid} {A71}}\BibitemShut {NoStop}%
\bibitem [{\citenamefont {{Bersten}}\ and\ \citenamefont
  {{Mazzali}}(2017)}]{Bersten.Mazzali:2017}%
  \BibitemOpen
  \bibfield  {author} {\bibinfo {author} {\bibnamefont {{Bersten}},
  \bibfnamefont {M~C}}, and\ \bibinfo {author} {\bibfnamefont {P.~A.}\
  \bibnamefont {{Mazzali}}}} (\bibinfo {year} {2017}),\ \enquote {\bibinfo
  {title} {{Light Curves of Type I Supernovae}},}\ in\ \href
  {https://doi.org/10.1007/978-3-319-21846-5_25} {\emph {\bibinfo {booktitle}
  {Handbook of Supernovae}}},\ \bibinfo {editor} {edited by\ \bibinfo {editor}
  {\bibfnamefont {A.~W.}\ \bibnamefont {{Alsabti}}}\ and\ \bibinfo {editor}
  {\bibfnamefont {P.}~\bibnamefont {{Murdin}}}}\ (\bibinfo  {publisher}
  {Springer International Publishing},\ \bibinfo {address} {Cham})\BibitemShut
  {NoStop}%
\bibitem [{\citenamefont {{Bethe}}(1990)}]{Bethe:1990}%
  \BibitemOpen
  \bibfield  {author} {\bibinfo {author} {\bibnamefont {{Bethe}}, \bibfnamefont
  {H~A}}} (\bibinfo {year} {1990}),\ \bibfield  {title} {\enquote {\bibinfo
  {title} {{Supernova mechanisms}},}\ }\href
  {https://doi.org/10.1103/RevModPhys.62.801} {\bibfield  {journal} {\bibinfo
  {journal} {Rev. Mod. Phys.}\ }\textbf {\bibinfo {volume} {62}},\ \bibinfo
  {pages} {801--866}}\BibitemShut {NoStop}%
\bibitem [{\citenamefont {{Bisterzo}}\ \emph {et~al.}(2015)\citenamefont
  {{Bisterzo}}, \citenamefont {{Gallino}}, \citenamefont {{K{\"a}ppeler}},
  \citenamefont {{Wiescher}}, \citenamefont {{Imbriani}}, \citenamefont
  {{Straniero}}, \citenamefont {{Cristallo}}, \citenamefont {{G{\"o}rres}},\
  and\ \citenamefont {{deBoer}}}]{bisterzo15}%
  \BibitemOpen
  \bibfield  {author} {\bibinfo {author} {\bibnamefont {{Bisterzo}},
  \bibfnamefont {S}}, \bibinfo {author} {\bibfnamefont {R.}~\bibnamefont
  {{Gallino}}}, \bibinfo {author} {\bibfnamefont {F.}~\bibnamefont
  {{K{\"a}ppeler}}}, \bibinfo {author} {\bibfnamefont {M.}~\bibnamefont
  {{Wiescher}}}, \bibinfo {author} {\bibfnamefont {G.}~\bibnamefont
  {{Imbriani}}}, \bibinfo {author} {\bibfnamefont {O.}~\bibnamefont
  {{Straniero}}}, \bibinfo {author} {\bibfnamefont {S.}~\bibnamefont
  {{Cristallo}}}, \bibinfo {author} {\bibfnamefont {J.}~\bibnamefont
  {{G{\"o}rres}}}, and\ \bibinfo {author} {\bibfnamefont {R.~J.}\ \bibnamefont
  {{deBoer}}}} (\bibinfo {year} {2015}),\ \bibfield  {title} {\enquote
  {\bibinfo {title} {{The branchings of the main s-process: their sensitivity
  to {$\alpha$}-induced reactions on $^{13}$C and $^{22}$Ne and to the
  uncertainties of the nuclear network}},}\ }\href
  {https://doi.org/10.1093/mnras/stv271} {\bibfield  {journal} {\bibinfo
  {journal} {Mon. Not. Roy. Astron. Soc.}\ }\textbf {\bibinfo {volume} {449}},\
  \bibinfo {pages} {506--527}}\BibitemShut {NoStop}%
\bibitem [{\citenamefont {{Bisterzo}}\ \emph {et~al.}(2017)\citenamefont
  {{Bisterzo}}, \citenamefont {{Travaglio}}, \citenamefont {{Wiescher}},
  \citenamefont {{K{\"a}ppeler}},\ and\ \citenamefont
  {{Gallino}}}]{bisterzo17}%
  \BibitemOpen
  \bibfield  {author} {\bibinfo {author} {\bibnamefont {{Bisterzo}},
  \bibfnamefont {S}}, \bibinfo {author} {\bibfnamefont {C.}~\bibnamefont
  {{Travaglio}}}, \bibinfo {author} {\bibfnamefont {M.}~\bibnamefont
  {{Wiescher}}}, \bibinfo {author} {\bibfnamefont {F.}~\bibnamefont
  {{K{\"a}ppeler}}}, and\ \bibinfo {author} {\bibfnamefont {R.}~\bibnamefont
  {{Gallino}}}} (\bibinfo {year} {2017}),\ \bibfield  {title} {\enquote
  {\bibinfo {title} {{Galactic Chemical Evolution: The Impact of the
  $^{13}$C-pocket Structure on the s-process Distribution}},}\ }\href
  {https://doi.org/10.3847/1538-4357/835/1/97} {\bibfield  {journal} {\bibinfo
  {journal} {Astrophys. J.}\ }\textbf {\bibinfo {volume} {835}},\ \bibinfo
  {eid} {97}}\BibitemShut {NoStop}%
\bibitem [{\citenamefont {{Blake}}\ and\ \citenamefont
  {{Schramm}}(1976)}]{Blake.Schramm:1976}%
  \BibitemOpen
  \bibfield  {author} {\bibinfo {author} {\bibnamefont {{Blake}}, \bibfnamefont
  {J~B}}, and\ \bibinfo {author} {\bibfnamefont {D.~N.}\ \bibnamefont
  {{Schramm}}}} (\bibinfo {year} {1976}),\ \bibfield  {title} {\enquote
  {\bibinfo {title} {{A Possible Alternative to the R-Process}},}\ }\href
  {https://doi.org/10.1086/154782} {\bibfield  {journal} {\bibinfo  {journal}
  {Astrophys. J.}\ }\textbf {\bibinfo {volume} {209}},\ \bibinfo {pages}
  {846--849}}\BibitemShut {NoStop}%
\bibitem [{\citenamefont {{Blaum}}\ \emph {et~al.}(2013)\citenamefont
  {{Blaum}}, \citenamefont {{Dilling}},\ and\ \citenamefont
  {{N{\"o}rtersh{\"a}user}}}]{Blaum2013}%
  \BibitemOpen
  \bibfield  {author} {\bibinfo {author} {\bibnamefont {{Blaum}}, \bibfnamefont
  {K}}, \bibinfo {author} {\bibfnamefont {J.}~\bibnamefont {{Dilling}}}, and\
  \bibinfo {author} {\bibfnamefont {W.}~\bibnamefont
  {{N{\"o}rtersh{\"a}user}}}} (\bibinfo {year} {2013}),\ \bibfield  {title}
  {\enquote {\bibinfo {title} {{Precision atomic physics techniques for nuclear
  physics with radioactive beams}},}\ }\href
  {https://doi.org/10.1088/0031-8949/2013/T152/014017} {\bibfield  {journal}
  {\bibinfo  {journal} {Phys. Scr.}\ }\textbf {\bibinfo {volume} {T152}},\
  \bibinfo {eid} {014017}}\BibitemShut {NoStop}%
\bibitem [{\citenamefont {Blaum}(2006)}]{Blaum:2006}%
  \BibitemOpen
  \bibfield  {author} {\bibinfo {author} {\bibnamefont {Blaum}, \bibfnamefont
  {Klaus}}} (\bibinfo {year} {2006}),\ \bibfield  {title} {\enquote {\bibinfo
  {title} {High-accuracy mass spectrometry with stored ions},}\ }\href
  {https://doi.org/10.1016/j.physrep.2005.10.011} {\bibfield  {journal}
  {\bibinfo  {journal} {Phys. Rep.}\ }\textbf {\bibinfo {volume} {425}},\
  \bibinfo {pages} {1--78}}\BibitemShut {NoStop}%
\bibitem [{\citenamefont {{Bohle}}\ \emph {et~al.}(1984)\citenamefont
  {{Bohle}}, \citenamefont {{Richter}}, \citenamefont {{Steffen}},
  \citenamefont {{Dieperink}}, \citenamefont {{Lo Iudice}}, \citenamefont
  {{Palumbo}},\ and\ \citenamefont {{Scholten}}}]{Bohle.Richter.ea:1984}%
  \BibitemOpen
  \bibfield  {author} {\bibinfo {author} {\bibnamefont {{Bohle}}, \bibfnamefont
  {D}}, \bibinfo {author} {\bibfnamefont {A.}~\bibnamefont {{Richter}}},
  \bibinfo {author} {\bibfnamefont {W.}~\bibnamefont {{Steffen}}}, \bibinfo
  {author} {\bibfnamefont {A.~E.~L.}\ \bibnamefont {{Dieperink}}}, \bibinfo
  {author} {\bibfnamefont {N.}~\bibnamefont {{Lo Iudice}}}, \bibinfo {author}
  {\bibfnamefont {F.}~\bibnamefont {{Palumbo}}}, and\ \bibinfo {author}
  {\bibfnamefont {O.}~\bibnamefont {{Scholten}}}} (\bibinfo {year} {1984}),\
  \bibfield  {title} {\enquote {\bibinfo {title} {New magnetic dipole
  excitation mode studied in the heavy deformed nucleus $^{156}$gd by inelastic
  electron scattering},}\ }\href {https://doi.org/10.1016/0370-2693(84)91099-2}
  {\bibfield  {journal} {\bibinfo  {journal} {Phys. Lett. B}\ }\textbf
  {\bibinfo {volume} {137}},\ \bibinfo {pages} {27--31}}\BibitemShut {NoStop}%
\bibitem [{\citenamefont {Bollig}\ \emph {et~al.}(2017)\citenamefont {Bollig},
  \citenamefont {Janka}, \citenamefont {Lohs}, \citenamefont
  {Mart\'{\i}nez-Pinedo}, \citenamefont {Horowitz},\ and\ \citenamefont
  {Melson}}]{Bollig.Janka.ea:2017}%
  \BibitemOpen
  \bibfield  {author} {\bibinfo {author} {\bibnamefont {Bollig}, \bibfnamefont
  {R}}, \bibinfo {author} {\bibfnamefont {H.-T.}\ \bibnamefont {Janka}},
  \bibinfo {author} {\bibfnamefont {A.}~\bibnamefont {Lohs}}, \bibinfo {author}
  {\bibfnamefont {G.}~\bibnamefont {Mart\'{\i}nez-Pinedo}}, \bibinfo {author}
  {\bibfnamefont {C.~J.}\ \bibnamefont {Horowitz}}, and\ \bibinfo {author}
  {\bibfnamefont {T.}~\bibnamefont {Melson}}} (\bibinfo {year} {2017}),\
  \bibfield  {title} {\enquote {\bibinfo {title} {Muon creation in supernova
  matter facilitates neutrino-driven explosions},}\ }\href
  {https://doi.org/10.1103/PhysRevLett.119.242702} {\bibfield  {journal}
  {\bibinfo  {journal} {Phys. Rev. Lett.}\ }\textbf {\bibinfo {volume} {119}},\
  \bibinfo {pages} {242702}}\BibitemShut {NoStop}%
\bibitem [{\citenamefont {{Bonetti}}\ \emph {et~al.}(2018)\citenamefont
  {{Bonetti}}, \citenamefont {{Perego}}, \citenamefont {{Capelo}},
  \citenamefont {{Dotti}},\ and\ \citenamefont
  {{Miller}}}]{Bonetti.Perego.ea:2018}%
  \BibitemOpen
  \bibfield  {author} {\bibinfo {author} {\bibnamefont {{Bonetti}},
  \bibfnamefont {M}}, \bibinfo {author} {\bibfnamefont {A.}~\bibnamefont
  {{Perego}}}, \bibinfo {author} {\bibfnamefont {P.~R.}\ \bibnamefont
  {{Capelo}}}, \bibinfo {author} {\bibfnamefont {M.}~\bibnamefont {{Dotti}}},
  and\ \bibinfo {author} {\bibfnamefont {M.~C.}\ \bibnamefont {{Miller}}}}
  (\bibinfo {year} {2018}),\ \bibfield  {title} {\enquote {\bibinfo {title}
  {{r-Process Nucleosynthesis in the Early Universe Through Fast Mergers of
  Compact Binaries in Triple Systems}},}\ }\href
  {https://doi.org/10.1017/pasa.2018.11} {\bibfield  {journal} {\bibinfo
  {journal} {Publ. Astron. Soc. Austr.}\ }\textbf {\bibinfo {volume} {35}},\
  \bibinfo {eid} {e017}}\BibitemShut {NoStop}%
\bibitem [{\citenamefont {{Bonetti}}\ \emph {et~al.}(2019)\citenamefont
  {{Bonetti}}, \citenamefont {{Perego}}, \citenamefont {{Dotti}},\ and\
  \citenamefont {{Cescutti}}}]{Bonetti.Perego.Dotti.ea:2019}%
  \BibitemOpen
  \bibfield  {author} {\bibinfo {author} {\bibnamefont {{Bonetti}},
  \bibfnamefont {Matteo}}, \bibinfo {author} {\bibfnamefont {Albino}\
  \bibnamefont {{Perego}}}, \bibinfo {author} {\bibfnamefont {Massimo}\
  \bibnamefont {{Dotti}}}, and\ \bibinfo {author} {\bibfnamefont {Gabriele}\
  \bibnamefont {{Cescutti}}}} (\bibinfo {year} {2019}),\ \bibfield  {title}
  {\enquote {\bibinfo {title} {{Neutron star binary orbits in their host
  potential: effect on early r-process enrichment}},}\ }\href
  {https://doi.org/10.1093/mnras/stz2554} {\bibfield  {journal} {\bibinfo
  {journal} {Mon. Not. Roy. Astron. Soc.}\ }\textbf {\bibinfo {volume} {490}},\
  \bibinfo {pages} {296--311}}\BibitemShut {NoStop}%
\bibitem [{\citenamefont {Borzov}\ and\ \citenamefont
  {Goriely}(2000)}]{Borzov.Goriely:2000}%
  \BibitemOpen
  \bibfield  {author} {\bibinfo {author} {\bibnamefont {Borzov}, \bibfnamefont
  {I}}, and\ \bibinfo {author} {\bibfnamefont {S.}~\bibnamefont {Goriely}}}
  (\bibinfo {year} {2000}),\ \bibfield  {title} {\enquote {\bibinfo {title}
  {Weak interaction rates of neutron-rich nuclei and the r-process
  nucleosynthesis},}\ }\href {https://doi.org/10.1103/PhysRevC.62.035501}
  {\bibfield  {journal} {\bibinfo  {journal} {Phys. Rev. C}\ }\textbf {\bibinfo
  {volume} {62}},\ \bibinfo {pages} {035501}}\BibitemShut {NoStop}%
\bibitem [{\citenamefont {Borzov}(2003)}]{Borzov:2003}%
  \BibitemOpen
  \bibfield  {author} {\bibinfo {author} {\bibnamefont {Borzov}, \bibfnamefont
  {I~N}}} (\bibinfo {year} {2003}),\ \bibfield  {title} {\enquote {\bibinfo
  {title} {{Gamow-Teller and first-forbidden decays near the r-process paths at
  $N=50$, 82, and 126}},}\ }\href {https://doi.org/10.1103/PhysRevC.67.025802}
  {\bibfield  {journal} {\bibinfo  {journal} {Phys. Rev. C}\ }\textbf {\bibinfo
  {volume} {67}},\ \bibinfo {pages} {025802}}\BibitemShut {NoStop}%
\bibitem [{\citenamefont {Borzov}\ \emph {et~al.}(2008)\citenamefont {Borzov},
  \citenamefont {Cuenca-Garc{\'i}a}, \citenamefont {Langanke}, \citenamefont
  {Mart{\'i}nez-Pinedo},\ and\ \citenamefont
  {Montes}}]{Borzov.Cuenca-Garcia.ea:2008}%
  \BibitemOpen
  \bibfield  {author} {\bibinfo {author} {\bibnamefont {Borzov}, \bibfnamefont
  {I~N}}, \bibinfo {author} {\bibfnamefont {J.~J.}\ \bibnamefont
  {Cuenca-Garc{\'i}a}}, \bibinfo {author} {\bibfnamefont {K.}~\bibnamefont
  {Langanke}}, \bibinfo {author} {\bibfnamefont {G.}~\bibnamefont
  {Mart{\'i}nez-Pinedo}}, and\ \bibinfo {author} {\bibfnamefont
  {F.}~\bibnamefont {Montes}}} (\bibinfo {year} {2008}),\ \bibfield  {title}
  {\enquote {\bibinfo {title} {{Beta-decay of $Z<50$ nuclei near the $N=82$
  closed neutron shell}},}\ }\href
  {https://doi.org/10.1016/j.nuclphysa.2008.09.010} {\bibfield  {journal}
  {\bibinfo  {journal} {Nucl. Phys. A}\ }\textbf {\bibinfo {volume} {814}},\
  \bibinfo {pages} {159--173}}\BibitemShut {NoStop}%
\bibitem [{\citenamefont {Bovard}\ \emph {et~al.}(2017)\citenamefont {Bovard},
  \citenamefont {Martin}, \citenamefont {Guercilena}, \citenamefont {Arcones},
  \citenamefont {Rezzolla},\ and\ \citenamefont
  {Korobkin}}]{Bovard.Martin.ea:2017}%
  \BibitemOpen
  \bibfield  {author} {\bibinfo {author} {\bibnamefont {Bovard}, \bibfnamefont
  {Luke}}, \bibinfo {author} {\bibfnamefont {Dirk}\ \bibnamefont {Martin}},
  \bibinfo {author} {\bibfnamefont {Federico}\ \bibnamefont {Guercilena}},
  \bibinfo {author} {\bibfnamefont {Almudena}\ \bibnamefont {Arcones}},
  \bibinfo {author} {\bibfnamefont {Luciano}\ \bibnamefont {Rezzolla}}, and\
  \bibinfo {author} {\bibfnamefont {Oleg}\ \bibnamefont {Korobkin}}} (\bibinfo
  {year} {2017}),\ \bibfield  {title} {\enquote {\bibinfo {title} {$r$-process
  nucleosynthesis from matter ejected in binary neutron star mergers},}\ }\href
  {https://doi.org/10.1103/PhysRevD.96.124005} {\bibfield  {journal} {\bibinfo
  {journal} {Phys. Rev. D}\ }\textbf {\bibinfo {volume} {96}},\ \bibinfo
  {pages} {124005}}\BibitemShut {NoStop}%
\bibitem [{\citenamefont {{Boyd}}\ \emph {et~al.}(2012)\citenamefont {{Boyd}},
  \citenamefont {{Famiano}}, \citenamefont {{Meyer}}, \citenamefont
  {{Motizuki}}, \citenamefont {{Kajino}},\ and\ \citenamefont
  {{Roederer}}}]{Boyd.Famiano.ea:2012}%
  \BibitemOpen
  \bibfield  {author} {\bibinfo {author} {\bibnamefont {{Boyd}}, \bibfnamefont
  {R~N}}, \bibinfo {author} {\bibfnamefont {M.~A.}\ \bibnamefont {{Famiano}}},
  \bibinfo {author} {\bibfnamefont {B.~S.}\ \bibnamefont {{Meyer}}}, \bibinfo
  {author} {\bibfnamefont {Y.}~\bibnamefont {{Motizuki}}}, \bibinfo {author}
  {\bibfnamefont {T.}~\bibnamefont {{Kajino}}}, and\ \bibinfo {author}
  {\bibfnamefont {I.~U.}\ \bibnamefont {{Roederer}}}} (\bibinfo {year}
  {2012}),\ \bibfield  {title} {\enquote {\bibinfo {title} {{The r-process in
  Metal-poor Stars and Black Hole Formation}},}\ }\href
  {https://doi.org/10.1088/2041-8205/744/1/L14} {\bibfield  {journal} {\bibinfo
   {journal} {Astrophys. J.}\ }\textbf {\bibinfo {volume} {744}},\ \bibinfo
  {eid} {L14}}\BibitemShut {NoStop}%
\bibitem [{\citenamefont {{Brauer}}\ \emph {et~al.}(2019)\citenamefont
  {{Brauer}}, \citenamefont {{Ji}}, \citenamefont {{Frebel}}, \citenamefont
  {{Dooley}}, \citenamefont {{G{\'o}mez}},\ and\ \citenamefont
  {{O'Shea}}}]{Brauer.Ji.ea:2019}%
  \BibitemOpen
  \bibfield  {author} {\bibinfo {author} {\bibnamefont {{Brauer}},
  \bibfnamefont {Kaley}}, \bibinfo {author} {\bibfnamefont {Alexander~P.}\
  \bibnamefont {{Ji}}}, \bibinfo {author} {\bibfnamefont {Anna}\ \bibnamefont
  {{Frebel}}}, \bibinfo {author} {\bibfnamefont {Gregory~A.}\ \bibnamefont
  {{Dooley}}}, \bibinfo {author} {\bibfnamefont {Facundo~A.}\ \bibnamefont
  {{G{\'o}mez}}}, and\ \bibinfo {author} {\bibfnamefont {Brian~W.}\
  \bibnamefont {{O'Shea}}}} (\bibinfo {year} {2019}),\ \bibfield  {title}
  {\enquote {\bibinfo {title} {{The Origin of r-process Enhanced Metal-poor
  Halo Stars In Now-destroyed Ultra-faint Dwarf Galaxies}},}\ }\href
  {https://doi.org/10.3847/1538-4357/aafafb} {\bibfield  {journal} {\bibinfo
  {journal} {Astrophys. J.}\ }\textbf {\bibinfo {volume} {871}},\ \bibinfo
  {eid} {247}}\BibitemShut {NoStop}%
\bibitem [{\citenamefont {{Brault}}(1976)}]{brault76}%
  \BibitemOpen
  \bibfield  {author} {\bibinfo {author} {\bibnamefont {{Brault}},
  \bibfnamefont {J~W}}} (\bibinfo {year} {1976}),\ \bibfield  {title} {\enquote
  {\bibinfo {title} {{Rapid-scan high-resolution Fourier spectrometer for the
  visible.}}}\ }\href@noop {} {\bibfield  {journal} {\bibinfo  {journal}
  {Journal of Optics Soc. America}\ }\textbf {\bibinfo {volume} {66}},\
  \bibinfo {pages} {1081}}\BibitemShut {NoStop}%
\bibitem [{\citenamefont {{Bravo}}\ and\ \citenamefont
  {{Garc{\'\i}a-Senz}}(1999)}]{Bravo.Garcia:1999}%
  \BibitemOpen
  \bibfield  {author} {\bibinfo {author} {\bibnamefont {{Bravo}}, \bibfnamefont
  {E}}, and\ \bibinfo {author} {\bibfnamefont {D.}~\bibnamefont
  {{Garc{\'\i}a-Senz}}}} (\bibinfo {year} {1999}),\ \bibfield  {title}
  {\enquote {\bibinfo {title} {{Coulomb corrections to the equation of state of
  nuclear statistical equilibrium matter: implications for SNIa nucleosynthesis
  and the accretion-induced collapse of white dwarfs}},}\ }\href
  {https://doi.org/10.1046/j.1365-8711.1999.02694.x} {\bibfield  {journal}
  {\bibinfo  {journal} {Mon. Not. Roy. Astron. Soc.}\ }\textbf {\bibinfo
  {volume} {307}},\ \bibinfo {pages} {984--992}}\BibitemShut {NoStop}%
\bibitem [{\citenamefont {Brege}\ \emph {et~al.}(2018)\citenamefont {Brege},
  \citenamefont {Duez}, \citenamefont {Foucart}, \citenamefont {Deaton},
  \citenamefont {Caro}, \citenamefont {Hemberger}, \citenamefont {Kidder},
  \citenamefont {O'Connor}, \citenamefont {Pfeiffer},\ and\ \citenamefont
  {Scheel}}]{Brege.Duez.ea:2018}%
  \BibitemOpen
  \bibfield  {author} {\bibinfo {author} {\bibnamefont {Brege}, \bibfnamefont
  {Wyatt}}, \bibinfo {author} {\bibfnamefont {Matthew~D.}\ \bibnamefont
  {Duez}}, \bibinfo {author} {\bibfnamefont {Francois}\ \bibnamefont
  {Foucart}}, \bibinfo {author} {\bibfnamefont {M.~Brett}\ \bibnamefont
  {Deaton}}, \bibinfo {author} {\bibfnamefont {Jesus}\ \bibnamefont {Caro}},
  \bibinfo {author} {\bibfnamefont {Daniel~A.}\ \bibnamefont {Hemberger}},
  \bibinfo {author} {\bibfnamefont {Lawrence~E.}\ \bibnamefont {Kidder}},
  \bibinfo {author} {\bibfnamefont {Evan}\ \bibnamefont {O'Connor}}, \bibinfo
  {author} {\bibfnamefont {Harald~P.}\ \bibnamefont {Pfeiffer}}, and\ \bibinfo
  {author} {\bibfnamefont {Mark~A.}\ \bibnamefont {Scheel}}} (\bibinfo {year}
  {2018}),\ \bibfield  {title} {\enquote {\bibinfo {title} {Black hole-neutron
  star mergers using a survey of finite-temperature equations of state},}\
  }\href {https://doi.org/10.1103/PhysRevD.98.063009} {\bibfield  {journal}
  {\bibinfo  {journal} {Phys. Rev. D}\ }\textbf {\bibinfo {volume} {98}},\
  \bibinfo {pages} {063009}}\BibitemShut {NoStop}%
\bibitem [{\citenamefont {Brink}(1955)}]{Brink:1955}%
  \BibitemOpen
  \bibfield  {author} {\bibinfo {author} {\bibnamefont {Brink}, \bibfnamefont
  {D~M}}} (\bibinfo {year} {1955}),\ \emph {\bibinfo {title} {Some aspects of
  the interaction of light with matter}},\ \href@noop {} {Ph.D. thesis}\
  (\bibinfo  {school} {Oxford University})\BibitemShut {NoStop}%
\bibitem [{\citenamefont {{Brink}}(1957)}]{Brink:1957}%
  \BibitemOpen
  \bibfield  {author} {\bibinfo {author} {\bibnamefont {{Brink}}, \bibfnamefont
  {D~M}}} (\bibinfo {year} {1957}),\ \bibfield  {title} {\enquote {\bibinfo
  {title} {{Individual particle and collective aspects of the nuclear
  photoeffect}},}\ }\href {https://doi.org/10.1016/0029-5582(87)90021-6}
  {\bibfield  {journal} {\bibinfo  {journal} {Nucl. Phys. A}\ }\textbf
  {\bibinfo {volume} {4}},\ \bibinfo {pages} {215--220}}\BibitemShut {NoStop}%
\bibitem [{\citenamefont {{Brown}}\ and\ \citenamefont
  {{Larsen}}(2014)}]{Brown.Larsen:2014}%
  \BibitemOpen
  \bibfield  {author} {\bibinfo {author} {\bibnamefont {{Brown}}, \bibfnamefont
  {B~A}}, and\ \bibinfo {author} {\bibfnamefont {A.~C.}\ \bibnamefont
  {{Larsen}}}} (\bibinfo {year} {2014}),\ \bibfield  {title} {\enquote
  {\bibinfo {title} {{Large Low-Energy $M1$ Strength for $^{56,57}$Fe within
  the Nuclear Shell Model}},}\ }\href
  {https://doi.org/10.1103/PhysRevLett.113.252502} {\bibfield  {journal}
  {\bibinfo  {journal} {Phys. Rev. Lett.}\ }\textbf {\bibinfo {volume} {113}},\
  \bibinfo {eid} {252502}}\BibitemShut {NoStop}%
\bibitem [{\citenamefont {{Bruske}}\ \emph {et~al.}(1981)\citenamefont
  {{Bruske}}, \citenamefont {{Burkard}}, \citenamefont {{H{\"u}ller}},
  \citenamefont {{Kirchner}}, \citenamefont {{Klepper}},\ and\ \citenamefont
  {{Roeckl}}}]{Roeckl}%
  \BibitemOpen
  \bibfield  {author} {\bibinfo {author} {\bibnamefont {{Bruske}},
  \bibfnamefont {C}}, \bibinfo {author} {\bibfnamefont {K.~H.}\ \bibnamefont
  {{Burkard}}}, \bibinfo {author} {\bibfnamefont {W.}~\bibnamefont
  {{H{\"u}ller}}}, \bibinfo {author} {\bibfnamefont {R.}~\bibnamefont
  {{Kirchner}}}, \bibinfo {author} {\bibfnamefont {O.}~\bibnamefont
  {{Klepper}}}, and\ \bibinfo {author} {\bibfnamefont {E.}~\bibnamefont
  {{Roeckl}}}} (\bibinfo {year} {1981}),\ \bibfield  {title} {\enquote
  {\bibinfo {title} {{Status report on the GSI on-line mass separator
  facility}},}\ }\href {https://doi.org/10.1016/0029-554X(81)90888-0}
  {\bibfield  {journal} {\bibinfo  {journal} {Nucl. Instruments Methods Phys.
  Res.}\ }\textbf {\bibinfo {volume} {186}},\ \bibinfo {pages}
  {61--69}}\BibitemShut {NoStop}%
\bibitem [{\citenamefont {{Bugli}}\ \emph {et~al.}(2020)\citenamefont
  {{Bugli}}, \citenamefont {{Guilet}}, \citenamefont {{Obergaulinger}},
  \citenamefont {{Cerd{\'a}-Dur{\'a}n}},\ and\ \citenamefont
  {{Aloy}}}]{Bugli.ea:2020}%
  \BibitemOpen
  \bibfield  {author} {\bibinfo {author} {\bibnamefont {{Bugli}}, \bibfnamefont
  {M}}, \bibinfo {author} {\bibfnamefont {J.}~\bibnamefont {{Guilet}}},
  \bibinfo {author} {\bibfnamefont {M.}~\bibnamefont {{Obergaulinger}}},
  \bibinfo {author} {\bibfnamefont {P.}~\bibnamefont {{Cerd{\'a}-Dur{\'a}n}}},
  and\ \bibinfo {author} {\bibfnamefont {M.~A.}\ \bibnamefont {{Aloy}}}}
  (\bibinfo {year} {2020}),\ \bibfield  {title} {\enquote {\bibinfo {title}
  {{The impact of non-dipolar magnetic fields in core-collapse supernovae}},}\
  }\href {https://doi.org/10.1093/mnras/stz3483} {\bibfield  {journal}
  {\bibinfo  {journal} {Mon. Not. Roy. Astron. Soc.}\ }\textbf {\bibinfo
  {volume} {492}},\ \bibinfo {pages} {58--71}}\BibitemShut {NoStop}%
\bibitem [{\citenamefont {{Burbidge}}\ \emph {et~al.}(1957)\citenamefont
  {{Burbidge}}, \citenamefont {{Burbidge}}, \citenamefont {{Fowler}},\ and\
  \citenamefont {{Hoyle}}}]{Burbidge.Burbidge.ea:1957}%
  \BibitemOpen
  \bibfield  {author} {\bibinfo {author} {\bibnamefont {{Burbidge}},
  \bibfnamefont {E~M}}, \bibinfo {author} {\bibfnamefont {G.~R.}\ \bibnamefont
  {{Burbidge}}}, \bibinfo {author} {\bibfnamefont {W.~A.}\ \bibnamefont
  {{Fowler}}}, and\ \bibinfo {author} {\bibfnamefont {F.}~\bibnamefont
  {{Hoyle}}}} (\bibinfo {year} {1957}),\ \bibfield  {title} {\enquote {\bibinfo
  {title} {{Synthesis of the Elements in Stars}},}\ }\href
  {https://doi.org/10.1103/RevModPhys.29.547} {\bibfield  {journal} {\bibinfo
  {journal} {Rev. Mod. Phys.}\ }\textbf {\bibinfo {volume} {29}},\ \bibinfo
  {pages} {547--650}}\BibitemShut {NoStop}%
\bibitem [{\citenamefont {{Burbidge}}\ \emph {et~al.}(1956)\citenamefont
  {{Burbidge}}, \citenamefont {{Hoyle}}, \citenamefont {{Burbidge}},
  \citenamefont {{Christy}},\ and\ \citenamefont
  {{Fowler}}}]{Burbidge.Hoyle.ea:1956}%
  \BibitemOpen
  \bibfield  {author} {\bibinfo {author} {\bibnamefont {{Burbidge}},
  \bibfnamefont {G~R}}, \bibinfo {author} {\bibfnamefont {F.}~\bibnamefont
  {{Hoyle}}}, \bibinfo {author} {\bibfnamefont {E.~M.}\ \bibnamefont
  {{Burbidge}}}, \bibinfo {author} {\bibfnamefont {R.~F.}\ \bibnamefont
  {{Christy}}}, and\ \bibinfo {author} {\bibfnamefont {W.~A.}\ \bibnamefont
  {{Fowler}}}} (\bibinfo {year} {1956}),\ \bibfield  {title} {\enquote
  {\bibinfo {title} {{Californium-254 and Supernovae}},}\ }\href
  {https://doi.org/10.1103/PhysRev.103.1145} {\bibfield  {journal} {\bibinfo
  {journal} {Phys. Rev.}\ }\textbf {\bibinfo {volume} {103}},\ \bibinfo {pages}
  {1145--1149}}\BibitemShut {NoStop}%
\bibitem [{\citenamefont {{Burris}}\ \emph {et~al.}(2000)\citenamefont
  {{Burris}}, \citenamefont {{Pilachowski}}, \citenamefont {{Armandroff}},
  \citenamefont {{Sneden}}, \citenamefont {{Cowan}},\ and\ \citenamefont
  {{Roe}}}]{Burris.Pilachowski.ea:2000}%
  \BibitemOpen
  \bibfield  {author} {\bibinfo {author} {\bibnamefont {{Burris}},
  \bibfnamefont {D~L}}, \bibinfo {author} {\bibfnamefont {C.~A.}\ \bibnamefont
  {{Pilachowski}}}, \bibinfo {author} {\bibfnamefont {T.~E.}\ \bibnamefont
  {{Armandroff}}}, \bibinfo {author} {\bibfnamefont {C.}~\bibnamefont
  {{Sneden}}}, \bibinfo {author} {\bibfnamefont {J.~J.}\ \bibnamefont
  {{Cowan}}}, and\ \bibinfo {author} {\bibfnamefont {H.}~\bibnamefont {{Roe}}}}
  (\bibinfo {year} {2000}),\ \bibfield  {title} {\enquote {\bibinfo {title}
  {{Neutron-Capture Elements in the Early Galaxy: Insights from a Large Sample
  of Metal-poor Giants}},}\ }\href {https://doi.org/10.1086/317172} {\bibfield
  {journal} {\bibinfo  {journal} {Astrophys. J.}\ }\textbf {\bibinfo {volume}
  {544}},\ \bibinfo {pages} {302--319}}\BibitemShut {NoStop}%
\bibitem [{\citenamefont {{Burrows}}\ \emph {et~al.}(2018)\citenamefont
  {{Burrows}}, \citenamefont {{Vartanyan}}, \citenamefont {{Dolence}},
  \citenamefont {{Skinner}},\ and\ \citenamefont
  {{Radice}}}]{Burrows.Vartanyan.ea:2018}%
  \BibitemOpen
  \bibfield  {author} {\bibinfo {author} {\bibnamefont {{Burrows}},
  \bibfnamefont {A}}, \bibinfo {author} {\bibfnamefont {D.}~\bibnamefont
  {{Vartanyan}}}, \bibinfo {author} {\bibfnamefont {J.~C.}\ \bibnamefont
  {{Dolence}}}, \bibinfo {author} {\bibfnamefont {M.~A.}\ \bibnamefont
  {{Skinner}}}, and\ \bibinfo {author} {\bibfnamefont {D.}~\bibnamefont
  {{Radice}}}} (\bibinfo {year} {2018}),\ \bibfield  {title} {\enquote
  {\bibinfo {title} {{Crucial Physical Dependencies of the Core-Collapse
  Supernova Mechanism}},}\ }\href {https://doi.org/10.1007/s11214-017-0450-9}
  {\bibfield  {journal} {\bibinfo  {journal} {Space Sci. Rev.}\ }\textbf
  {\bibinfo {volume} {214}},\ \bibinfo {eid} {33}}\BibitemShut {NoStop}%
\bibitem [{\citenamefont {Burrows}(2013)}]{Burrows:2013}%
  \BibitemOpen
  \bibfield  {author} {\bibinfo {author} {\bibnamefont {Burrows}, \bibfnamefont
  {Adam}}} (\bibinfo {year} {2013}),\ \bibfield  {title} {\enquote {\bibinfo
  {title} {Colloquium: Perspectives on core-collapse supernova theory},}\
  }\href {https://doi.org/10.1103/RevModPhys.85.245} {\bibfield  {journal}
  {\bibinfo  {journal} {Rev. Mod. Phys.}\ }\textbf {\bibinfo {volume} {85}},\
  \bibinfo {pages} {245--261}}\BibitemShut {NoStop}%
\bibitem [{\citenamefont {{Burrows}}\ \emph {et~al.}(2020)\citenamefont
  {{Burrows}}, \citenamefont {{Radice}}, \citenamefont {{Vartanyan}},
  \citenamefont {{Nagakura}}, \citenamefont {{Skinner}},\ and\ \citenamefont
  {{Dolence}}}]{Burrows.Radice.ea:2020}%
  \BibitemOpen
  \bibfield  {author} {\bibinfo {author} {\bibnamefont {{Burrows}},
  \bibfnamefont {Adam}}, \bibinfo {author} {\bibfnamefont {David}\ \bibnamefont
  {{Radice}}}, \bibinfo {author} {\bibfnamefont {David}\ \bibnamefont
  {{Vartanyan}}}, \bibinfo {author} {\bibfnamefont {Hiroki}\ \bibnamefont
  {{Nagakura}}}, \bibinfo {author} {\bibfnamefont {M.~Aaron}\ \bibnamefont
  {{Skinner}}}, and\ \bibinfo {author} {\bibfnamefont {Joshua~C.}\ \bibnamefont
  {{Dolence}}}} (\bibinfo {year} {2020}),\ \bibfield  {title} {\enquote
  {\bibinfo {title} {{The overarching framework of core-collapse supernova
  explosions as revealed by 3D FORNAX simulations}},}\ }\href
  {https://doi.org/10.1093/mnras/stz3223} {\bibfield  {journal} {\bibinfo
  {journal} {Mon. Not. Roy. Astron. Soc.}\ }\textbf {\bibinfo {volume} {491}},\
  \bibinfo {pages} {2715--2735}}\BibitemShut {NoStop}%
\bibitem [{\citenamefont {{Busso}}\ \emph {et~al.}(1999)\citenamefont
  {{Busso}}, \citenamefont {{Gallino}},\ and\ \citenamefont
  {{Wasserburg}}}]{busso99}%
  \BibitemOpen
  \bibfield  {author} {\bibinfo {author} {\bibnamefont {{Busso}}, \bibfnamefont
  {M}}, \bibinfo {author} {\bibfnamefont {R.}~\bibnamefont {{Gallino}}}, and\
  \bibinfo {author} {\bibfnamefont {G.~J.}\ \bibnamefont {{Wasserburg}}}}
  (\bibinfo {year} {1999}),\ \bibfield  {title} {\enquote {\bibinfo {title}
  {{Nucleosynthesis in Asymptotic Giant Branch Stars: Relevance for Galactic
  Enrichment and Solar System Formation}},}\ }\href
  {https://doi.org/10.1146/annurev.astro.37.1.239} {\bibfield  {journal}
  {\bibinfo  {journal} {Annu. Rev. Astron. Astrophys.}\ }\textbf {\bibinfo
  {volume} {37}},\ \bibinfo {pages} {239--309}}\BibitemShut {NoStop}%
\bibitem [{\citenamefont {{Caballero}}\ \emph {et~al.}(2014)\citenamefont
  {{Caballero}}, \citenamefont {{Arcones}}, \citenamefont {{Borzov}},
  \citenamefont {{Langanke}},\ and\ \citenamefont
  {{Martinez-Pinedo}}}]{Caballero.Arcones.ea:2014}%
  \BibitemOpen
  \bibfield  {author} {\bibinfo {author} {\bibnamefont {{Caballero}},
  \bibfnamefont {O~L}}, \bibinfo {author} {\bibfnamefont {A.}~\bibnamefont
  {{Arcones}}}, \bibinfo {author} {\bibfnamefont {I.~N.}\ \bibnamefont
  {{Borzov}}}, \bibinfo {author} {\bibfnamefont {K.}~\bibnamefont
  {{Langanke}}}, and\ \bibinfo {author} {\bibfnamefont {G.}~\bibnamefont
  {{Martinez-Pinedo}}}} (\bibinfo {year} {2014}),\ \bibfield  {title} {\enquote
  {\bibinfo {title} {{Local and global effects of beta decays on r-process}},}\
  }\href@noop {} {\bibfield  {journal} {\bibinfo  {journal} {ArXiv e-prints}\
  }}\Eprint {https://arxiv.org/abs/1405.0210} {arXiv:1405.0210 [nucl-th]}
  \BibitemShut {NoStop}%
\bibitem [{\citenamefont {{Caballero-Folch}}\ \emph {et~al.}(2016)\citenamefont
  {{Caballero-Folch}} \emph {et~al.}}]{Caballero2016}%
  \BibitemOpen
  \bibfield  {author} {\bibinfo {author} {\bibnamefont {{Caballero-Folch}},
  \bibfnamefont {R}},  \emph {et~al.}} (\bibinfo {year} {2016}),\ \bibfield
  {title} {\enquote {\bibinfo {title} {{First Measurement of Several
  $\beta$-Delayed Neutron Emitting Isotopes Beyond $N=126$}},}\ }\href
  {https://doi.org/10.1103/PhysRevLett.117.012501} {\bibfield  {journal}
  {\bibinfo  {journal} {Phys. Rev. Lett.}\ }\textbf {\bibinfo {volume} {117}},\
  \bibinfo {eid} {012501}}\BibitemShut {NoStop}%
\bibitem [{\citenamefont {{Cabez{\'o}n}}\ \emph {et~al.}(2018)\citenamefont
  {{Cabez{\'o}n}}, \citenamefont {{Pan}}, \citenamefont {{Liebend{\"o}rfer}},
  \citenamefont {{Kuroda}}, \citenamefont {{Ebinger}}, \citenamefont
  {{Heinimann}}, \citenamefont {{Perego}},\ and\ \citenamefont
  {{Thielemann}}}]{Cabezon.Pan.ea:2018}%
  \BibitemOpen
  \bibfield  {author} {\bibinfo {author} {\bibnamefont {{Cabez{\'o}n}},
  \bibfnamefont {Rub{\'e}n~M}}, \bibinfo {author} {\bibfnamefont {Kuo-Chuan}\
  \bibnamefont {{Pan}}}, \bibinfo {author} {\bibfnamefont {Matthias}\
  \bibnamefont {{Liebend{\"o}rfer}}}, \bibinfo {author} {\bibfnamefont
  {Takami}\ \bibnamefont {{Kuroda}}}, \bibinfo {author} {\bibfnamefont {Kevin}\
  \bibnamefont {{Ebinger}}}, \bibinfo {author} {\bibfnamefont {Oliver}\
  \bibnamefont {{Heinimann}}}, \bibinfo {author} {\bibfnamefont {Albino}\
  \bibnamefont {{Perego}}}, and\ \bibinfo {author} {\bibfnamefont
  {Friedrich-Karl}\ \bibnamefont {{Thielemann}}}} (\bibinfo {year} {2018}),\
  \bibfield  {title} {\enquote {\bibinfo {title} {{Core-collapse supernovae in
  the hall of mirrors. A three-dimensional code-comparison project}},}\ }\href
  {https://doi.org/10.1051/0004-6361/201833705} {\bibfield  {journal} {\bibinfo
   {journal} {Astron. \& Astrophys.}\ }\textbf {\bibinfo {volume} {619}},\
  \bibinfo {eid} {A118}}\BibitemShut {NoStop}%
\bibitem [{\citenamefont {Calvi{\~n}o}\ \emph {et~al.}(2014)\citenamefont
  {Calvi{\~n}o} \emph {et~al.}}]{Calvino2014}%
  \BibitemOpen
  \bibfield  {author} {\bibinfo {author} {\bibnamefont {Calvi{\~n}o},
  \bibfnamefont {F}},  \emph {et~al.}} (\bibinfo {year} {2014}),\ \href
  {https://edms.cern.ch/document/1865760} {\emph {\bibinfo {title} {{TDR
  BEta-deLayEd Neutron detector}}}},\ \bibinfo {type} {Technical Design Report
  No.}\ \bibinfo {number} {1865760}\ (\bibinfo  {institution}
  {FAIR})\BibitemShut {NoStop}%
\bibitem [{\citenamefont {{Camelio}}\ \emph {et~al.}(2018)\citenamefont
  {{Camelio}}, \citenamefont {{Dietrich}},\ and\ \citenamefont
  {{Rosswog}}}]{Camelio18}%
  \BibitemOpen
  \bibfield  {author} {\bibinfo {author} {\bibnamefont {{Camelio}},
  \bibfnamefont {Giovanni}}, \bibinfo {author} {\bibfnamefont {Tim}\
  \bibnamefont {{Dietrich}}}, and\ \bibinfo {author} {\bibfnamefont {Stephan}\
  \bibnamefont {{Rosswog}}}} (\bibinfo {year} {2018}),\ \bibfield  {title}
  {\enquote {\bibinfo {title} {{Disc formation in the collapse of supramassive
  neutron stars}},}\ }\href {https://doi.org/10.1093/mnras/sty2181} {\bibfield
  {journal} {\bibinfo  {journal} {Mon. Not. Roy. Astron. Soc.}\ }\textbf
  {\bibinfo {volume} {480}},\ \bibinfo {pages} {5272--5285}}\BibitemShut
  {NoStop}%
\bibitem [{\citenamefont {Cameron}(1957)}]{Cameron:1957}%
  \BibitemOpen
  \bibfield  {author} {\bibinfo {author} {\bibnamefont {Cameron}, \bibfnamefont
  {A~G~W}}} (\bibinfo {year} {1957}),\ \href@noop {} {\emph {\bibinfo {title}
  {Stellar Evolution, Nuclear Astrophysics, and Nucleogenesis}}},\ \bibinfo
  {type} {Report}\ \bibinfo {number} {CRL-41}\ (\bibinfo  {institution} {Chalk
  River})\BibitemShut {NoStop}%
\bibitem [{\citenamefont {{Cameron}}(1959)}]{cameron59}%
  \BibitemOpen
  \bibfield  {author} {\bibinfo {author} {\bibnamefont {{Cameron}},
  \bibfnamefont {A~G~W}}} (\bibinfo {year} {1959}),\ \bibfield  {title}
  {\enquote {\bibinfo {title} {{A Revised Table of Abundances of the
  Elements.}}}\ }\href {https://doi.org/10.1086/146667} {\bibfield  {journal}
  {\bibinfo  {journal} {Astrophys. J.}\ }\textbf {\bibinfo {volume} {129}},\
  \bibinfo {pages} {676}}\BibitemShut {NoStop}%
\bibitem [{\citenamefont {{Cameron}}(2003)}]{cameron03}%
  \BibitemOpen
  \bibfield  {author} {\bibinfo {author} {\bibnamefont {{Cameron}},
  \bibfnamefont {A~G~W}}} (\bibinfo {year} {2003}),\ \bibfield  {title}
  {\enquote {\bibinfo {title} {{Some Nucleosynthesis Effects Associated with
  r-Process Jets}},}\ }\href {https://doi.org/10.1086/368110} {\bibfield
  {journal} {\bibinfo  {journal} {Astrophys. J.}\ }\textbf {\bibinfo {volume}
  {587}},\ \bibinfo {pages} {327--340}}\BibitemShut {NoStop}%
\bibitem [{\citenamefont {{Cameron}}\ \emph {et~al.}(1983)\citenamefont
  {{Cameron}}, \citenamefont {{Cowan}},\ and\ \citenamefont
  {{Truran}}}]{Cameron.Cowan.Truran:1983}%
  \BibitemOpen
  \bibfield  {author} {\bibinfo {author} {\bibnamefont {{Cameron}},
  \bibfnamefont {A~G~W}}, \bibinfo {author} {\bibfnamefont {J.~J.}\
  \bibnamefont {{Cowan}}}, and\ \bibinfo {author} {\bibfnamefont {J.~W.}\
  \bibnamefont {{Truran}}}} (\bibinfo {year} {1983}),\ \bibfield  {title}
  {\enquote {\bibinfo {title} {{The waiting point approximation in R-process
  calculations}},}\ }\href {https://doi.org/10.1007/BF00656112} {\bibfield
  {journal} {\bibinfo  {journal} {Astrophys. Space Sci.}\ }\textbf {\bibinfo
  {volume} {91}},\ \bibinfo {pages} {235--243}}\BibitemShut {NoStop}%
\bibitem [{\citenamefont {{Capano}}\ \emph {et~al.}(2020)\citenamefont
  {{Capano}}, \citenamefont {{Tews}}, \citenamefont {{Brown}}, \citenamefont
  {{Margalit}}, \citenamefont {{De}}, \citenamefont {{Kumar}}, \citenamefont
  {{Brown}}, \citenamefont {{Krishnan}},\ and\ \citenamefont
  {{Reddy}}}]{Capano.Tews.ea:2020}%
  \BibitemOpen
  \bibfield  {author} {\bibinfo {author} {\bibnamefont {{Capano}},
  \bibfnamefont {Collin~D}}, \bibinfo {author} {\bibfnamefont {Ingo}\
  \bibnamefont {{Tews}}}, \bibinfo {author} {\bibfnamefont {Stephanie~M.}\
  \bibnamefont {{Brown}}}, \bibinfo {author} {\bibfnamefont {Ben}\ \bibnamefont
  {{Margalit}}}, \bibinfo {author} {\bibfnamefont {Soumi}\ \bibnamefont
  {{De}}}, \bibinfo {author} {\bibfnamefont {Sumit}\ \bibnamefont {{Kumar}}},
  \bibinfo {author} {\bibfnamefont {Duncan~A.}\ \bibnamefont {{Brown}}},
  \bibinfo {author} {\bibfnamefont {Badri}\ \bibnamefont {{Krishnan}}}, and\
  \bibinfo {author} {\bibfnamefont {Sanjay}\ \bibnamefont {{Reddy}}}} (\bibinfo
  {year} {2020}),\ \bibfield  {title} {\enquote {\bibinfo {title} {{Stringent
  constraints on neutron-star radii from multimessenger observations and
  nuclear theory}},}\ }\href {https://doi.org/10.1038/s41550-020-1014-6}
  {\bibfield  {journal} {\bibinfo  {journal} {Nature Astronomy}\ }\textbf
  {\bibinfo {volume} {4}},\ \bibinfo {pages} {625--632}}\BibitemShut {NoStop}%
\bibitem [{\citenamefont {Caurier}\ \emph {et~al.}(1999)\citenamefont
  {Caurier}, \citenamefont {Langanke}, \citenamefont {Mart{\'{\i}}nez-Pinedo},\
  and\ \citenamefont {Nowacki}}]{Caurier.Langanke.ea:1999}%
  \BibitemOpen
  \bibfield  {author} {\bibinfo {author} {\bibnamefont {Caurier}, \bibfnamefont
  {E}}, \bibinfo {author} {\bibfnamefont {K.}~\bibnamefont {Langanke}},
  \bibinfo {author} {\bibfnamefont {G.}~\bibnamefont {Mart{\'{\i}}nez-Pinedo}},
  and\ \bibinfo {author} {\bibfnamefont {F.}~\bibnamefont {Nowacki}}} (\bibinfo
  {year} {1999}),\ \bibfield  {title} {\enquote {\bibinfo {title} {{Shell model
  calculations of stellar weak interaction rates: I. Gamow-Teller distributions
  and spectra of nuclei in the mass range $A=45$-65}},}\ }\href
  {https://doi.org/10.1016/S0375-9474(99)00240-7} {\bibfield  {journal}
  {\bibinfo  {journal} {Nucl. Phys. A}\ }\textbf {\bibinfo {volume} {653}},\
  \bibinfo {pages} {439--452}}\BibitemShut {NoStop}%
\bibitem [{\citenamefont {Caurier}\ \emph {et~al.}(2005)\citenamefont
  {Caurier}, \citenamefont {Mart\'{\i}nez-Pinedo}, \citenamefont {Nowacki},
  \citenamefont {Poves},\ and\ \citenamefont
  {Zuker}}]{Caurier.Martinez-Pinedo.ea:2005}%
  \BibitemOpen
  \bibfield  {author} {\bibinfo {author} {\bibnamefont {Caurier}, \bibfnamefont
  {E}}, \bibinfo {author} {\bibfnamefont {G.}~\bibnamefont
  {Mart\'{\i}nez-Pinedo}}, \bibinfo {author} {\bibfnamefont {Fr\'ederic}\
  \bibnamefont {Nowacki}}, \bibinfo {author} {\bibfnamefont {A.}~\bibnamefont
  {Poves}}, and\ \bibinfo {author} {\bibfnamefont {A.~P.}\ \bibnamefont
  {Zuker}}} (\bibinfo {year} {2005}),\ \bibfield  {title} {\enquote {\bibinfo
  {title} {{The shell Model as unified view of nuclear structure}},}\ }\href
  {https://doi.org/10.1103/RevModPhys.77.427} {\bibfield  {journal} {\bibinfo
  {journal} {Rev. Mod. Phys.}\ }\textbf {\bibinfo {volume} {77}},\ \bibinfo
  {pages} {427--488}}\BibitemShut {NoStop}%
\bibitem [{\citenamefont {{Cayrel}}\ \emph {et~al.}(2004)\citenamefont
  {{Cayrel}}, \citenamefont {{Depagne}}, \citenamefont {{Spite}}, \citenamefont
  {{Hill}}, \citenamefont {{Spite}}, \citenamefont {{Fran{\c c}ois}},
  \citenamefont {{Plez}}, \citenamefont {{Beers}}, \citenamefont {{Primas}},
  \citenamefont {{Andersen}}, \citenamefont {{Barbuy}}, \citenamefont
  {{Bonifacio}}, \citenamefont {{Molaro}},\ and\ \citenamefont
  {{Nordstr{\"o}m}}}]{cayrel04b}%
  \BibitemOpen
  \bibfield  {author} {\bibinfo {author} {\bibnamefont {{Cayrel}},
  \bibfnamefont {R}}, \bibinfo {author} {\bibfnamefont {E.}~\bibnamefont
  {{Depagne}}}, \bibinfo {author} {\bibfnamefont {M.}~\bibnamefont {{Spite}}},
  \bibinfo {author} {\bibfnamefont {V.}~\bibnamefont {{Hill}}}, \bibinfo
  {author} {\bibfnamefont {F.}~\bibnamefont {{Spite}}}, \bibinfo {author}
  {\bibfnamefont {P.}~\bibnamefont {{Fran{\c c}ois}}}, \bibinfo {author}
  {\bibfnamefont {B.}~\bibnamefont {{Plez}}}, \bibinfo {author} {\bibfnamefont
  {T.}~\bibnamefont {{Beers}}}, \bibinfo {author} {\bibfnamefont
  {F.}~\bibnamefont {{Primas}}}, \bibinfo {author} {\bibfnamefont
  {J.}~\bibnamefont {{Andersen}}}, \bibinfo {author} {\bibfnamefont
  {B.}~\bibnamefont {{Barbuy}}}, \bibinfo {author} {\bibfnamefont
  {P.}~\bibnamefont {{Bonifacio}}}, \bibinfo {author} {\bibfnamefont
  {P.}~\bibnamefont {{Molaro}}}, and\ \bibinfo {author} {\bibfnamefont
  {B.}~\bibnamefont {{Nordstr{\"o}m}}}} (\bibinfo {year} {2004}),\ \bibfield
  {title} {\enquote {\bibinfo {title} {{First stars V - Abundance patterns from
  C to Zn and supernova yields in the early Galaxy}},}\ }\href
  {https://doi.org/10.1051/0004-6361:20034074} {\bibfield  {journal} {\bibinfo
  {journal} {Astron. \& Astrophys.}\ }\textbf {\bibinfo {volume} {416}},\
  \bibinfo {pages} {1117--1138}}\BibitemShut {NoStop}%
\bibitem [{\citenamefont {{Cayrel}}\ \emph {et~al.}(2001)\citenamefont
  {{Cayrel}}, \citenamefont {{Hill}}, \citenamefont {{Beers}}, \citenamefont
  {{Barbuy}}, \citenamefont {{Spite}}, \citenamefont {{Spite}}, \citenamefont
  {{Plez}}, \citenamefont {{Andersen}}, \citenamefont {{Bonifacio}},
  \citenamefont {{Fran{\c c}ois}}, \citenamefont {{Molaro}}, \citenamefont
  {{Nordstr{\"o}m}},\ and\ \citenamefont {{Primas}}}]{cayrel01}%
  \BibitemOpen
  \bibfield  {author} {\bibinfo {author} {\bibnamefont {{Cayrel}},
  \bibfnamefont {R}}, \bibinfo {author} {\bibfnamefont {V.}~\bibnamefont
  {{Hill}}}, \bibinfo {author} {\bibfnamefont {T.~C.}\ \bibnamefont {{Beers}}},
  \bibinfo {author} {\bibfnamefont {B.}~\bibnamefont {{Barbuy}}}, \bibinfo
  {author} {\bibfnamefont {M.}~\bibnamefont {{Spite}}}, \bibinfo {author}
  {\bibfnamefont {F.}~\bibnamefont {{Spite}}}, \bibinfo {author} {\bibfnamefont
  {B.}~\bibnamefont {{Plez}}}, \bibinfo {author} {\bibfnamefont
  {J.}~\bibnamefont {{Andersen}}}, \bibinfo {author} {\bibfnamefont
  {P.}~\bibnamefont {{Bonifacio}}}, \bibinfo {author} {\bibfnamefont
  {P.}~\bibnamefont {{Fran{\c c}ois}}}, \bibinfo {author} {\bibfnamefont
  {P.}~\bibnamefont {{Molaro}}}, \bibinfo {author} {\bibfnamefont
  {B.}~\bibnamefont {{Nordstr{\"o}m}}}, and\ \bibinfo {author} {\bibfnamefont
  {F.}~\bibnamefont {{Primas}}}} (\bibinfo {year} {2001}),\ \bibfield  {title}
  {\enquote {\bibinfo {title} {{Measurement of stellar age from uranium
  decay}},}\ }\href {https://doi.org/10.1038/35055507} {\bibfield  {journal}
  {\bibinfo  {journal} {Nature}\ }\textbf {\bibinfo {volume} {409}},\ \bibinfo
  {pages} {691--692}}\BibitemShut {NoStop}%
\bibitem [{\citenamefont {{Cescutti}}\ \emph {et~al.}(2015)\citenamefont
  {{Cescutti}}, \citenamefont {{Romano}}, \citenamefont {{Matteucci}},
  \citenamefont {{Chiappini}},\ and\ \citenamefont {{Hirschi}}}]{cescutti15}%
  \BibitemOpen
  \bibfield  {author} {\bibinfo {author} {\bibnamefont {{Cescutti}},
  \bibfnamefont {G}}, \bibinfo {author} {\bibfnamefont {D.}~\bibnamefont
  {{Romano}}}, \bibinfo {author} {\bibfnamefont {F.}~\bibnamefont
  {{Matteucci}}}, \bibinfo {author} {\bibfnamefont {C.}~\bibnamefont
  {{Chiappini}}}, and\ \bibinfo {author} {\bibfnamefont {R.}~\bibnamefont
  {{Hirschi}}}} (\bibinfo {year} {2015}),\ \bibfield  {title} {\enquote
  {\bibinfo {title} {{The role of neutron star mergers in the chemical
  evolution of the Galactic halo}},}\ }\href
  {https://doi.org/10.1051/0004-6361/201525698} {\bibfield  {journal} {\bibinfo
   {journal} {Astron. \& Astrophys.}\ }\textbf {\bibinfo {volume} {577}},\
  \bibinfo {eid} {A139}}\BibitemShut {NoStop}%
\bibitem [{\citenamefont {Chawla}\ \emph {et~al.}(2010)\citenamefont {Chawla},
  \citenamefont {Anderson}, \citenamefont {Besselman}, \citenamefont {Lehner},
  \citenamefont {Liebling}, \citenamefont {Motl},\ and\ \citenamefont
  {Neilsen}}]{Chawla.Anderson.ea:2010}%
  \BibitemOpen
  \bibfield  {author} {\bibinfo {author} {\bibnamefont {Chawla}, \bibfnamefont
  {Sarvnipun}}, \bibinfo {author} {\bibfnamefont {Matthew}\ \bibnamefont
  {Anderson}}, \bibinfo {author} {\bibfnamefont {Michael}\ \bibnamefont
  {Besselman}}, \bibinfo {author} {\bibfnamefont {Luis}\ \bibnamefont
  {Lehner}}, \bibinfo {author} {\bibfnamefont {Steven~L.}\ \bibnamefont
  {Liebling}}, \bibinfo {author} {\bibfnamefont {Patrick~M.}\ \bibnamefont
  {Motl}}, and\ \bibinfo {author} {\bibfnamefont {David}\ \bibnamefont
  {Neilsen}}} (\bibinfo {year} {2010}),\ \bibfield  {title} {\enquote {\bibinfo
  {title} {Mergers of magnetized neutron stars with spinning black holes:
  Disruption, accretion, and fallback},}\ }\href
  {https://doi.org/10.1103/PhysRevLett.105.111101} {\bibfield  {journal}
  {\bibinfo  {journal} {Phys. Rev. Lett.}\ }\textbf {\bibinfo {volume} {105}},\
  \bibinfo {pages} {111101}}\BibitemShut {NoStop}%
\bibitem [{\citenamefont {{Chen}}\ \emph {et~al.}(1995)\citenamefont {{Chen}},
  \citenamefont {{Dobaczewski}}, \citenamefont {{Kratz}}, \citenamefont
  {{Langanke}}, \citenamefont {{Pfeiffer}}, \citenamefont {{Thielemann}},\ and\
  \citenamefont {{Vogel}}}]{Chen.Dobaczewski.ea:1995}%
  \BibitemOpen
  \bibfield  {author} {\bibinfo {author} {\bibnamefont {{Chen}}, \bibfnamefont
  {B}}, \bibinfo {author} {\bibfnamefont {J.}~\bibnamefont {{Dobaczewski}}},
  \bibinfo {author} {\bibfnamefont {K.-L.}\ \bibnamefont {{Kratz}}}, \bibinfo
  {author} {\bibfnamefont {K.}~\bibnamefont {{Langanke}}}, \bibinfo {author}
  {\bibfnamefont {B.}~\bibnamefont {{Pfeiffer}}}, \bibinfo {author}
  {\bibfnamefont {F.-K.}\ \bibnamefont {{Thielemann}}}, and\ \bibinfo {author}
  {\bibfnamefont {P.}~\bibnamefont {{Vogel}}}} (\bibinfo {year} {1995}),\
  \bibfield  {title} {\enquote {\bibinfo {title} {{Influence of shell-quenching
  far from stability on the astrophysical r-process}},}\ }\href
  {https://doi.org/10.1016/0370-2693(95)00737-6} {\bibfield  {journal}
  {\bibinfo  {journal} {Phys. Lett. B}\ }\textbf {\bibinfo {volume} {355}},\
  \bibinfo {pages} {37--44}}\BibitemShut {NoStop}%
\bibitem [{\citenamefont {{Chiappini}}\ \emph {et~al.}(2000)\citenamefont
  {{Chiappini}}, \citenamefont {{Matteucci}},\ and\ \citenamefont
  {{Padoan}}}]{Chiappini:2000}%
  \BibitemOpen
  \bibfield  {author} {\bibinfo {author} {\bibnamefont {{Chiappini}},
  \bibfnamefont {C}}, \bibinfo {author} {\bibfnamefont {F.}~\bibnamefont
  {{Matteucci}}}, and\ \bibinfo {author} {\bibfnamefont {P.}~\bibnamefont
  {{Padoan}}}} (\bibinfo {year} {2000}),\ \bibfield  {title} {\enquote
  {\bibinfo {title} {{The Chemical Evolution of the Galaxy with Variable
  Initial Mass Functions}},}\ }\href {https://doi.org/10.1086/308185}
  {\bibfield  {journal} {\bibinfo  {journal} {Astrophys. J.}\ }\textbf
  {\bibinfo {volume} {528}},\ \bibinfo {pages} {711--722}}\BibitemShut
  {NoStop}%
\bibitem [{\citenamefont {{Chiappini}}\ \emph {et~al.}(2001)\citenamefont
  {{Chiappini}}, \citenamefont {{Matteucci}},\ and\ \citenamefont
  {{Romano}}}]{Chiappini:2001a}%
  \BibitemOpen
  \bibfield  {author} {\bibinfo {author} {\bibnamefont {{Chiappini}},
  \bibfnamefont {C}}, \bibinfo {author} {\bibfnamefont {F.}~\bibnamefont
  {{Matteucci}}}, and\ \bibinfo {author} {\bibfnamefont {D.}~\bibnamefont
  {{Romano}}}} (\bibinfo {year} {2001}),\ \bibfield  {title} {\enquote
  {\bibinfo {title} {{Abundance Gradients and the Formation of the Milky
  Way}},}\ }\href {https://doi.org/10.1086/321427} {\bibfield  {journal}
  {\bibinfo  {journal} {Astrophys. J.}\ }\textbf {\bibinfo {volume} {554}},\
  \bibinfo {pages} {1044--1058}}\BibitemShut {NoStop}%
\bibitem [{\citenamefont {{Chornock}}\ \emph {et~al.}(2017)\citenamefont
  {{Chornock}} \emph {et~al.}}]{chornock17}%
  \BibitemOpen
  \bibfield  {author} {\bibinfo {author} {\bibnamefont {{Chornock}},
  \bibfnamefont {R}},  \emph {et~al.}} (\bibinfo {year} {2017}),\ \bibfield
  {title} {\enquote {\bibinfo {title} {{The Electromagnetic Counterpart of the
  Binary Neutron Star Merger LIGO/Virgo GW170817. IV. Detection of
  Near-infrared Signatures of r-process Nucleosynthesis with Gemini-South}},}\
  }\href {https://doi.org/10.3847/2041-8213/aa905c} {\bibfield  {journal}
  {\bibinfo  {journal} {Astrophys. J. Lett.}\ }\textbf {\bibinfo {volume}
  {848}},\ \bibinfo {eid} {L19}}\BibitemShut {NoStop}%
\bibitem [{\citenamefont {{Christie}}\ \emph {et~al.}(2019)\citenamefont
  {{Christie}}, \citenamefont {{Lalakos}}, \citenamefont {{Tchekhovskoy}},
  \citenamefont {{Fern{\'a}ndez}}, \citenamefont {{Foucart}}, \citenamefont
  {{Quataert}},\ and\ \citenamefont {{Kasen}}}]{Christie.Lalakos.ea:2019}%
  \BibitemOpen
  \bibfield  {author} {\bibinfo {author} {\bibnamefont {{Christie}},
  \bibfnamefont {I~M}}, \bibinfo {author} {\bibfnamefont {A.}~\bibnamefont
  {{Lalakos}}}, \bibinfo {author} {\bibfnamefont {A.}~\bibnamefont
  {{Tchekhovskoy}}}, \bibinfo {author} {\bibfnamefont {R.}~\bibnamefont
  {{Fern{\'a}ndez}}}, \bibinfo {author} {\bibfnamefont {F.}~\bibnamefont
  {{Foucart}}}, \bibinfo {author} {\bibfnamefont {E.}~\bibnamefont
  {{Quataert}}}, and\ \bibinfo {author} {\bibfnamefont {D.}~\bibnamefont
  {{Kasen}}}} (\bibinfo {year} {2019}),\ \bibfield  {title} {\enquote {\bibinfo
  {title} {{The role of magnetic field geometry in the evolution of neutron
  star merger accretion discs}},}\ }\href
  {https://doi.org/10.1093/mnras/stz2552} {\bibfield  {journal} {\bibinfo
  {journal} {Mon. Not. Roy. Astron. Soc.}\ }\textbf {\bibinfo {volume} {490}},\
  \bibinfo {pages} {4811--4825}}\BibitemShut {NoStop}%
\bibitem [{\citenamefont {{Ciolfi}}\ \emph {et~al.}(2017)\citenamefont
  {{Ciolfi}}, \citenamefont {{Kastaun}}, \citenamefont {{Giacomazzo}},
  \citenamefont {{Endrizzi}}, \citenamefont {{Siegel}},\ and\ \citenamefont
  {{Perna}}}]{Ciolfi.Kastaun.ea:2017}%
  \BibitemOpen
  \bibfield  {author} {\bibinfo {author} {\bibnamefont {{Ciolfi}},
  \bibfnamefont {Riccardo}}, \bibinfo {author} {\bibfnamefont {Wolfgang}\
  \bibnamefont {{Kastaun}}}, \bibinfo {author} {\bibfnamefont {Bruno}\
  \bibnamefont {{Giacomazzo}}}, \bibinfo {author} {\bibfnamefont {Andrea}\
  \bibnamefont {{Endrizzi}}}, \bibinfo {author} {\bibfnamefont {Daniel~M.}\
  \bibnamefont {{Siegel}}}, and\ \bibinfo {author} {\bibfnamefont {Rosalba}\
  \bibnamefont {{Perna}}}} (\bibinfo {year} {2017}),\ \bibfield  {title}
  {\enquote {\bibinfo {title} {{General relativistic magnetohydrodynamic
  simulations of binary neutron star mergers forming a long-lived neutron
  star}},}\ }\href {https://doi.org/10.1103/PhysRevD.95.063016} {\bibfield
  {journal} {\bibinfo  {journal} {Phys. Rev. D}\ }\textbf {\bibinfo {volume}
  {95}},\ \bibinfo {eid} {063016}}\BibitemShut {NoStop}%
\bibitem [{\citenamefont {{Cizewski}}\ \emph {et~al.}(2018)\citenamefont
  {{Cizewski}}, \citenamefont {{Ratkiewicz}}, \citenamefont {{Escher}},
  \citenamefont {{Lepailleur}}, \citenamefont {{Pain}},\ and\ \citenamefont
  {{Potel}}}]{Cizewski2018}%
  \BibitemOpen
  \bibfield  {author} {\bibinfo {author} {\bibnamefont {{Cizewski}},
  \bibfnamefont {J~A}}, \bibinfo {author} {\bibfnamefont {A.}~\bibnamefont
  {{Ratkiewicz}}}, \bibinfo {author} {\bibfnamefont {J.~E.}\ \bibnamefont
  {{Escher}}}, \bibinfo {author} {\bibfnamefont {A.}~\bibnamefont
  {{Lepailleur}}}, \bibinfo {author} {\bibfnamefont {S.~D.}\ \bibnamefont
  {{Pain}}}, and\ \bibinfo {author} {\bibfnamefont {G.}~\bibnamefont
  {{Potel}}}} (\bibinfo {year} {2018}),\ \bibfield  {title} {\enquote {\bibinfo
  {title} {{Informing neutron capture nucleosynthesis on short-lived nuclei
  with (d,p) reactions}},}\ }in\ \href
  {https://doi.org/10.1051/epjconf/201716501013} {\emph {\bibinfo {booktitle}
  {European Physical Journal Web of Conferences}}},\ Vol.\ \bibinfo {volume}
  {165},\ p.\ \bibinfo {pages} {01013}\BibitemShut {NoStop}%
\bibitem [{\citenamefont {{Clayton}}(1968)}]{Clayton:1968}%
  \BibitemOpen
  \bibfield  {author} {\bibinfo {author} {\bibnamefont {{Clayton}},
  \bibfnamefont {D~D}}} (\bibinfo {year} {1968}),\ \href@noop {} {\emph
  {\bibinfo {title} {{Principles of stellar evolution and nucleosynthesis}}}}\
  (\bibinfo  {publisher} {New York: McGraw-Hill, 1968})\BibitemShut {NoStop}%
\bibitem [{\citenamefont {{Cohen}}\ \emph {et~al.}(2004)\citenamefont
  {{Cohen}}, \citenamefont {{Christlieb}}, \citenamefont {{McWilliam}},
  \citenamefont {{Shectman}}, \citenamefont {{Thompson}}, \citenamefont
  {{Wasserburg}}, \citenamefont {{Ivans}}, \citenamefont {{Dehn}},
  \citenamefont {{Karlsson}},\ and\ \citenamefont {{Melendez}}}]{cohen04}%
  \BibitemOpen
  \bibfield  {author} {\bibinfo {author} {\bibnamefont {{Cohen}}, \bibfnamefont
  {J~G}}, \bibinfo {author} {\bibfnamefont {N.}~\bibnamefont {{Christlieb}}},
  \bibinfo {author} {\bibfnamefont {A.}~\bibnamefont {{McWilliam}}}, \bibinfo
  {author} {\bibfnamefont {S.}~\bibnamefont {{Shectman}}}, \bibinfo {author}
  {\bibfnamefont {I.}~\bibnamefont {{Thompson}}}, \bibinfo {author}
  {\bibfnamefont {G.~J.}\ \bibnamefont {{Wasserburg}}}, \bibinfo {author}
  {\bibfnamefont {I.}~\bibnamefont {{Ivans}}}, \bibinfo {author} {\bibfnamefont
  {M.}~\bibnamefont {{Dehn}}}, \bibinfo {author} {\bibfnamefont
  {T.}~\bibnamefont {{Karlsson}}}, and\ \bibinfo {author} {\bibfnamefont
  {J.}~\bibnamefont {{Melendez}}}} (\bibinfo {year} {2004}),\ \bibfield
  {title} {\enquote {\bibinfo {title} {{Abundances In Very Metal-Poor Dwarf
  Stars}},}\ }\href {https://doi.org/10.1086/422576} {\bibfield  {journal}
  {\bibinfo  {journal} {Astrophys. J.}\ }\textbf {\bibinfo {volume} {612}},\
  \bibinfo {pages} {1107--1135}}\BibitemShut {NoStop}%
\bibitem [{\citenamefont {Cole}\ \emph {et~al.}(2012)\citenamefont {Cole},
  \citenamefont {Anderson}, \citenamefont {Zegers}, \citenamefont {Austin},
  \citenamefont {Brown}, \citenamefont {Valdez}, \citenamefont {Gupta},
  \citenamefont {Hitt},\ and\ \citenamefont {Fawwaz}}]{Cole.Anderson.ea:2012}%
  \BibitemOpen
  \bibfield  {author} {\bibinfo {author} {\bibnamefont {Cole}, \bibfnamefont
  {A~L}}, \bibinfo {author} {\bibfnamefont {T.~S.}\ \bibnamefont {Anderson}},
  \bibinfo {author} {\bibfnamefont {R.~G.~T.}\ \bibnamefont {Zegers}}, \bibinfo
  {author} {\bibfnamefont {Sam~M.}\ \bibnamefont {Austin}}, \bibinfo {author}
  {\bibfnamefont {B.~A.}\ \bibnamefont {Brown}}, \bibinfo {author}
  {\bibfnamefont {L.}~\bibnamefont {Valdez}}, \bibinfo {author} {\bibfnamefont
  {S.}~\bibnamefont {Gupta}}, \bibinfo {author} {\bibfnamefont {G.~W.}\
  \bibnamefont {Hitt}}, and\ \bibinfo {author} {\bibfnamefont {O.}~\bibnamefont
  {Fawwaz}}} (\bibinfo {year} {2012}),\ \bibfield  {title} {\enquote {\bibinfo
  {title} {Gamow-teller strengths and electron-capture rates for $pf$-shell
  nuclei of relevance for late stellar evolution},}\ }\href
  {https://doi.org/10.1103/PhysRevC.86.015809} {\bibfield  {journal} {\bibinfo
  {journal} {Phys. Rev. C}\ }\textbf {\bibinfo {volume} {86}},\ \bibinfo
  {pages} {015809}}\BibitemShut {NoStop}%
\bibitem [{\citenamefont {{Corlies}}\ \emph {et~al.}(2018)\citenamefont
  {{Corlies}}, \citenamefont {{Johnston}},\ and\ \citenamefont
  {{Wise}}}]{Corlies.Johnston.ea:2018}%
  \BibitemOpen
  \bibfield  {author} {\bibinfo {author} {\bibnamefont {{Corlies}},
  \bibfnamefont {Lauren}}, \bibinfo {author} {\bibfnamefont {Kathryn~V.}\
  \bibnamefont {{Johnston}}}, and\ \bibinfo {author} {\bibfnamefont {John~H.}\
  \bibnamefont {{Wise}}}} (\bibinfo {year} {2018}),\ \bibfield  {title}
  {\enquote {\bibinfo {title} {{Exploring simulated early star formation in the
  context of the ultrafaint dwarf galaxies}},}\ }\href
  {https://doi.org/10.1093/mnras/sty064} {\bibfield  {journal} {\bibinfo
  {journal} {Mon. Not. Roy. Astron. Soc.}\ }\textbf {\bibinfo {volume} {475}},\
  \bibinfo {pages} {4868--4880}}\BibitemShut {NoStop}%
\bibitem [{\citenamefont {{C{\^o}t{\'e}}}\ \emph {et~al.}(2017)\citenamefont
  {{C{\^o}t{\'e}}}, \citenamefont {{Belczynski}}, \citenamefont {{Fryer}},
  \citenamefont {{Ritter}}, \citenamefont {{Paul}}, \citenamefont
  {{Wehmeyer}},\ and\ \citenamefont {{O'Shea}}}]{Cote.Belczynski.ea:2017}%
  \BibitemOpen
  \bibfield  {author} {\bibinfo {author} {\bibnamefont {{C{\^o}t{\'e}}},
  \bibfnamefont {B}}, \bibinfo {author} {\bibfnamefont {K.}~\bibnamefont
  {{Belczynski}}}, \bibinfo {author} {\bibfnamefont {C.~L.}\ \bibnamefont
  {{Fryer}}}, \bibinfo {author} {\bibfnamefont {C.}~\bibnamefont {{Ritter}}},
  \bibinfo {author} {\bibfnamefont {A.}~\bibnamefont {{Paul}}}, \bibinfo
  {author} {\bibfnamefont {B.}~\bibnamefont {{Wehmeyer}}}, and\ \bibinfo
  {author} {\bibfnamefont {B.~W.}\ \bibnamefont {{O'Shea}}}} (\bibinfo {year}
  {2017}),\ \bibfield  {title} {\enquote {\bibinfo {title} {{Advanced LIGO
  Constraints on Neutron Star Mergers and r-process Sites}},}\ }\href
  {https://doi.org/10.3847/1538-4357/aa5c8d} {\bibfield  {journal} {\bibinfo
  {journal} {Astrophys. J.}\ }\textbf {\bibinfo {volume} {836}},\ \bibinfo
  {eid} {230}}\BibitemShut {NoStop}%
\bibitem [{\citenamefont {{C{\^o}t{\'e}}}\ \emph {et~al.}(2018)\citenamefont
  {{C{\^o}t{\'e}}}, \citenamefont {{Fryer}}, \citenamefont {{Belczynski}},
  \citenamefont {{Korobkin}}, \citenamefont {{Chru{\'s}li{\'n}ska}},
  \citenamefont {{Vassh}}, \citenamefont {{Mumpower}}, \citenamefont
  {{Lippuner}}, \citenamefont {{Sprouse}}, \citenamefont {{Surman}},\ and\
  \citenamefont {{Wollaeger}}}]{Cote.Fryer.ea:2018}%
  \BibitemOpen
  \bibfield  {author} {\bibinfo {author} {\bibnamefont {{C{\^o}t{\'e}}},
  \bibfnamefont {B}}, \bibinfo {author} {\bibfnamefont {C.~L.}\ \bibnamefont
  {{Fryer}}}, \bibinfo {author} {\bibfnamefont {K.}~\bibnamefont
  {{Belczynski}}}, \bibinfo {author} {\bibfnamefont {O.}~\bibnamefont
  {{Korobkin}}}, \bibinfo {author} {\bibfnamefont {M.}~\bibnamefont
  {{Chru{\'s}li{\'n}ska}}}, \bibinfo {author} {\bibfnamefont {N.}~\bibnamefont
  {{Vassh}}}, \bibinfo {author} {\bibfnamefont {M.~R.}\ \bibnamefont
  {{Mumpower}}}, \bibinfo {author} {\bibfnamefont {J.}~\bibnamefont
  {{Lippuner}}}, \bibinfo {author} {\bibfnamefont {T.~M.}\ \bibnamefont
  {{Sprouse}}}, \bibinfo {author} {\bibfnamefont {R.}~\bibnamefont {{Surman}}},
  and\ \bibinfo {author} {\bibfnamefont {R.}~\bibnamefont {{Wollaeger}}}}
  (\bibinfo {year} {2018}),\ \bibfield  {title} {\enquote {\bibinfo {title}
  {{The Origin of r-process Elements in the Milky Way}},}\ }\href
  {https://doi.org/10.3847/1538-4357/aaad67} {\bibfield  {journal} {\bibinfo
  {journal} {Astrophys. J.}\ }\textbf {\bibinfo {volume} {855}},\ \bibinfo
  {eid} {99}}\BibitemShut {NoStop}%
\bibitem [{\citenamefont {{C{\^o}t{\'e}}}\ \emph
  {et~al.}(2019{\natexlab{a}})\citenamefont {{C{\^o}t{\'e}}}, \citenamefont
  {{Eichler}}, \citenamefont {{Arcones}}, \citenamefont {{Hansen}},
  \citenamefont {{Simonetti}}, \citenamefont {{Frebel}}, \citenamefont
  {{Fryer}}, \citenamefont {{Pignatari}}, \citenamefont {{Reichert}},
  \citenamefont {{Belczynski}},\ and\ \citenamefont
  {{Matteucci}}}]{Cote.Eichler.ea:2019}%
  \BibitemOpen
  \bibfield  {author} {\bibinfo {author} {\bibnamefont {{C{\^o}t{\'e}}},
  \bibfnamefont {Benoit}}, \bibinfo {author} {\bibfnamefont {Marius}\
  \bibnamefont {{Eichler}}}, \bibinfo {author} {\bibfnamefont {Almudena}\
  \bibnamefont {{Arcones}}}, \bibinfo {author} {\bibfnamefont {Camilla~J.}\
  \bibnamefont {{Hansen}}}, \bibinfo {author} {\bibfnamefont {Paolo}\
  \bibnamefont {{Simonetti}}}, \bibinfo {author} {\bibfnamefont {Anna}\
  \bibnamefont {{Frebel}}}, \bibinfo {author} {\bibfnamefont {Chris~L.}\
  \bibnamefont {{Fryer}}}, \bibinfo {author} {\bibfnamefont {Marco}\
  \bibnamefont {{Pignatari}}}, \bibinfo {author} {\bibfnamefont {Moritz}\
  \bibnamefont {{Reichert}}}, \bibinfo {author} {\bibfnamefont {Krzysztof}\
  \bibnamefont {{Belczynski}}}, and\ \bibinfo {author} {\bibfnamefont
  {Francesca}\ \bibnamefont {{Matteucci}}}} (\bibinfo {year}
  {2019}{\natexlab{a}}),\ \bibfield  {title} {\enquote {\bibinfo {title}
  {{Neutron Star Mergers Might Not Be the Only Source of r-process Elements in
  the Milky Way}},}\ }\href {https://doi.org/10.3847/1538-4357/ab10db}
  {\bibfield  {journal} {\bibinfo  {journal} {Astrophys. J.}\ }\textbf
  {\bibinfo {volume} {875}},\ \bibinfo {eid} {106}}\BibitemShut {NoStop}%
\bibitem [{\citenamefont {{C{\^o}t{\'e}}}\ \emph {et~al.}(2020)\citenamefont
  {{C{\^o}t{\'e}}}, \citenamefont {{Eichler}}, \citenamefont {{Yag{\"u}e}},
  \citenamefont {{Vassh}}, \citenamefont {{Mumpower}}, \citenamefont
  {{Vil{\'a}gos}}, \citenamefont {{So{\'o}s}}, \citenamefont {{Arcones}},
  \citenamefont {{Sprouse}}, \citenamefont {{Surman}}, \citenamefont
  {{Pignatari}}, \citenamefont {{Wehmeyer}}, \citenamefont {{Rauscher}},\ and\
  \citenamefont {{Lugaro}}}]{Cote:2020}%
  \BibitemOpen
  \bibfield  {author} {\bibinfo {author} {\bibnamefont {{C{\^o}t{\'e}}},
  \bibfnamefont {Benoit}}, \bibinfo {author} {\bibfnamefont {Marius}\
  \bibnamefont {{Eichler}}}, \bibinfo {author} {\bibfnamefont {Andr{\'e}s}\
  \bibnamefont {{Yag{\"u}e}}}, \bibinfo {author} {\bibfnamefont {Nicole}\
  \bibnamefont {{Vassh}}}, \bibinfo {author} {\bibfnamefont {Matthew~R.}\
  \bibnamefont {{Mumpower}}}, \bibinfo {author} {\bibfnamefont {Blanka}\
  \bibnamefont {{Vil{\'a}gos}}}, \bibinfo {author} {\bibfnamefont
  {Benj{\'a}min}\ \bibnamefont {{So{\'o}s}}}, \bibinfo {author} {\bibfnamefont
  {Almudena}\ \bibnamefont {{Arcones}}}, \bibinfo {author} {\bibfnamefont
  {Trevor~M.}\ \bibnamefont {{Sprouse}}}, \bibinfo {author} {\bibfnamefont
  {Rebecca}\ \bibnamefont {{Surman}}}, \bibinfo {author} {\bibfnamefont
  {Marco}\ \bibnamefont {{Pignatari}}}, \bibinfo {author} {\bibfnamefont
  {Benjamin}\ \bibnamefont {{Wehmeyer}}}, \bibinfo {author} {\bibfnamefont
  {Thomas}\ \bibnamefont {{Rauscher}}}, and\ \bibinfo {author} {\bibfnamefont
  {Maria}\ \bibnamefont {{Lugaro}}}} (\bibinfo {year} {2020}),\ \bibfield
  {title} {\enquote {\bibinfo {title} {{Constraining the Rapid Neutron-Capture
  Process with Meteoritic I-129 and Cm-247}},}\ }\href@noop {} {\bibfield
  {journal} {\bibinfo  {journal} {arXiv e-prints}\ }}\Eprint
  {https://arxiv.org/abs/2006.04833} {arXiv:2006.04833 [astro-ph.SR]}
  \BibitemShut {NoStop}%
\bibitem [{\citenamefont {{C{\^o}t{\'e}}}\ \emph
  {et~al.}(2019{\natexlab{b}})\citenamefont {{C{\^o}t{\'e}}}, \citenamefont
  {{Yag{\"u}e}}, \citenamefont {{Vil{\'a}gos}},\ and\ \citenamefont
  {{Lugaro}}}]{Cote.Yague.ea:2019}%
  \BibitemOpen
  \bibfield  {author} {\bibinfo {author} {\bibnamefont {{C{\^o}t{\'e}}},
  \bibfnamefont {Benoit}}, \bibinfo {author} {\bibfnamefont {Andr{\'e}s}\
  \bibnamefont {{Yag{\"u}e}}}, \bibinfo {author} {\bibfnamefont {Blanka}\
  \bibnamefont {{Vil{\'a}gos}}}, and\ \bibinfo {author} {\bibfnamefont {Maria}\
  \bibnamefont {{Lugaro}}}} (\bibinfo {year} {2019}{\natexlab{b}}),\ \bibfield
  {title} {\enquote {\bibinfo {title} {{Stochastic Chemical Evolution of
  Radioactive Isotopes with a Monte Carlo Approach}},}\ }\href
  {https://doi.org/10.3847/1538-4357/ab5a88} {\bibfield  {journal} {\bibinfo
  {journal} {Astrophys. J.}\ }\textbf {\bibinfo {volume} {887}},\ \bibinfo
  {eid} {213}}\BibitemShut {NoStop}%
\bibitem [{\citenamefont {{Coughlin}}\ \emph {et~al.}(2019)\citenamefont
  {{Coughlin}}, \citenamefont {{Dietrich}}, \citenamefont {{Margalit}},\ and\
  \citenamefont {{Metzger}}}]{Coughlin.Dietrich.ea:2019}%
  \BibitemOpen
  \bibfield  {author} {\bibinfo {author} {\bibnamefont {{Coughlin}},
  \bibfnamefont {Michael~W}}, \bibinfo {author} {\bibfnamefont {Tim}\
  \bibnamefont {{Dietrich}}}, \bibinfo {author} {\bibfnamefont {Ben}\
  \bibnamefont {{Margalit}}}, and\ \bibinfo {author} {\bibfnamefont {Brian~D.}\
  \bibnamefont {{Metzger}}}} (\bibinfo {year} {2019}),\ \bibfield  {title}
  {\enquote {\bibinfo {title} {{Multimessenger Bayesian parameter inference of
  a binary neutron star merger}},}\ }\href
  {https://doi.org/10.1093/mnrasl/slz133} {\bibfield  {journal} {\bibinfo
  {journal} {Mon. Not. Roy. Astron. Soc.}\ }\textbf {\bibinfo {volume} {489}},\
  \bibinfo {pages} {L91--L96}}\BibitemShut {NoStop}%
\bibitem [{\citenamefont {{Coulter}}\ \emph {et~al.}(2017)\citenamefont
  {{Coulter}}, \citenamefont {{Foley}}, \citenamefont {{Kilpatrick}},
  \citenamefont {{Drout}}, \citenamefont {{Piro}}, \citenamefont {{Shappee}},
  \citenamefont {{Siebert}}, \citenamefont {{Simon}}, \citenamefont {{Ulloa}},
  \citenamefont {{Kasen}}, \citenamefont {{Madore}}, \citenamefont
  {{Murguia-Berthier}}, \citenamefont {{Pan}}, \citenamefont {{Prochaska}},
  \citenamefont {{Ramirez-Ruiz}}, \citenamefont {{Rest}},\ and\ \citenamefont
  {{Rojas-Bravo}}}]{coulter17}%
  \BibitemOpen
  \bibfield  {author} {\bibinfo {author} {\bibnamefont {{Coulter}},
  \bibfnamefont {D~A}}, \bibinfo {author} {\bibfnamefont {R.~J.}\ \bibnamefont
  {{Foley}}}, \bibinfo {author} {\bibfnamefont {C.~D.}\ \bibnamefont
  {{Kilpatrick}}}, \bibinfo {author} {\bibfnamefont {M.~R.}\ \bibnamefont
  {{Drout}}}, \bibinfo {author} {\bibfnamefont {A.~L.}\ \bibnamefont {{Piro}}},
  \bibinfo {author} {\bibfnamefont {B.~J.}\ \bibnamefont {{Shappee}}}, \bibinfo
  {author} {\bibfnamefont {M.~R.}\ \bibnamefont {{Siebert}}}, \bibinfo {author}
  {\bibfnamefont {J.~D.}\ \bibnamefont {{Simon}}}, \bibinfo {author}
  {\bibfnamefont {N.}~\bibnamefont {{Ulloa}}}, \bibinfo {author} {\bibfnamefont
  {D.}~\bibnamefont {{Kasen}}}, \bibinfo {author} {\bibfnamefont {B.~F.}\
  \bibnamefont {{Madore}}}, \bibinfo {author} {\bibfnamefont {A.}~\bibnamefont
  {{Murguia-Berthier}}}, \bibinfo {author} {\bibfnamefont {Y.-C.}\ \bibnamefont
  {{Pan}}}, \bibinfo {author} {\bibfnamefont {J.~X.}\ \bibnamefont
  {{Prochaska}}}, \bibinfo {author} {\bibfnamefont {E.}~\bibnamefont
  {{Ramirez-Ruiz}}}, \bibinfo {author} {\bibfnamefont {A.}~\bibnamefont
  {{Rest}}}, and\ \bibinfo {author} {\bibfnamefont {C.}~\bibnamefont
  {{Rojas-Bravo}}}} (\bibinfo {year} {2017}),\ \bibfield  {title} {\enquote
  {\bibinfo {title} {{Swope Supernova Survey 2017a (SSS17a), the optical
  counterpart to a gravitational wave source}},}\ }\href
  {https://doi.org/10.1126/science.aap9811} {\bibfield  {journal} {\bibinfo
  {journal} {Science}\ }\textbf {\bibinfo {volume} {358}},\ \bibinfo {pages}
  {1556--1558}}\BibitemShut {NoStop}%
\bibitem [{\citenamefont {{Cowan}}\ \emph {et~al.}(1980)\citenamefont
  {{Cowan}}, \citenamefont {{Cameron}},\ and\ \citenamefont
  {{Truran}}}]{Cowan.Cameron.Truran:1980}%
  \BibitemOpen
  \bibfield  {author} {\bibinfo {author} {\bibnamefont {{Cowan}}, \bibfnamefont
  {J~J}}, \bibinfo {author} {\bibfnamefont {A.~G.~W.}\ \bibnamefont
  {{Cameron}}}, and\ \bibinfo {author} {\bibfnamefont {J.~W.}\ \bibnamefont
  {{Truran}}}} (\bibinfo {year} {1980}),\ \bibfield  {title} {\enquote
  {\bibinfo {title} {{Seed abundances for r-processing in the helium shells of
  supernovae}},}\ }\href {https://doi.org/10.1086/158424} {\bibfield  {journal}
  {\bibinfo  {journal} {Astrophys. J.}\ }\textbf {\bibinfo {volume} {241}},\
  \bibinfo {pages} {1090--1093}}\BibitemShut {NoStop}%
\bibitem [{\citenamefont {{Cowan}}\ \emph {et~al.}(1983)\citenamefont
  {{Cowan}}, \citenamefont {{Cameron}},\ and\ \citenamefont
  {{Truran}}}]{Cowan.Cameron.Truran:1983}%
  \BibitemOpen
  \bibfield  {author} {\bibinfo {author} {\bibnamefont {{Cowan}}, \bibfnamefont
  {J~J}}, \bibinfo {author} {\bibfnamefont {A.~G.~W.}\ \bibnamefont
  {{Cameron}}}, and\ \bibinfo {author} {\bibfnamefont {J.~W.}\ \bibnamefont
  {{Truran}}}} (\bibinfo {year} {1983}),\ \bibfield  {title} {\enquote
  {\bibinfo {title} {{Explosive helium burning in supernovae - A source of
  r-process elements}},}\ }\href {https://doi.org/10.1086/160687} {\bibfield
  {journal} {\bibinfo  {journal} {Astrophys. J.}\ }\textbf {\bibinfo {volume}
  {265}},\ \bibinfo {pages} {429--442}}\BibitemShut {NoStop}%
\bibitem [{\citenamefont {{Cowan}}\ \emph {et~al.}(1985)\citenamefont
  {{Cowan}}, \citenamefont {{Cameron}},\ and\ \citenamefont
  {{Truran}}}]{Cowan.Cameron.Truran:1985}%
  \BibitemOpen
  \bibfield  {author} {\bibinfo {author} {\bibnamefont {{Cowan}}, \bibfnamefont
  {J~J}}, \bibinfo {author} {\bibfnamefont {A.~G.~W.}\ \bibnamefont
  {{Cameron}}}, and\ \bibinfo {author} {\bibfnamefont {J.~W.}\ \bibnamefont
  {{Truran}}}} (\bibinfo {year} {1985}),\ \bibfield  {title} {\enquote
  {\bibinfo {title} {{R-process nucleosynthesis in dynamic helium-burning
  environments}},}\ }\href {https://doi.org/10.1086/163335} {\bibfield
  {journal} {\bibinfo  {journal} {Astrophys. J.}\ }\textbf {\bibinfo {volume}
  {294}},\ \bibinfo {pages} {656--662}}\BibitemShut {NoStop}%
\bibitem [{\citenamefont {{Cowan}}\ \emph {et~al.}(1999)\citenamefont
  {{Cowan}}, \citenamefont {{Pfeiffer}}, \citenamefont {{Kratz}}, \citenamefont
  {{Thielemann}}, \citenamefont {{Sneden}}, \citenamefont {{Burles}},
  \citenamefont {{Tytler}},\ and\ \citenamefont
  {{Beers}}}]{Cowan.Pfeiffer.ea:1999}%
  \BibitemOpen
  \bibfield  {author} {\bibinfo {author} {\bibnamefont {{Cowan}}, \bibfnamefont
  {J~J}}, \bibinfo {author} {\bibfnamefont {B.}~\bibnamefont {{Pfeiffer}}},
  \bibinfo {author} {\bibfnamefont {K.-L.}\ \bibnamefont {{Kratz}}}, \bibinfo
  {author} {\bibfnamefont {F.-K.}\ \bibnamefont {{Thielemann}}}, \bibinfo
  {author} {\bibfnamefont {C.}~\bibnamefont {{Sneden}}}, \bibinfo {author}
  {\bibfnamefont {S.}~\bibnamefont {{Burles}}}, \bibinfo {author}
  {\bibfnamefont {D.}~\bibnamefont {{Tytler}}}, and\ \bibinfo {author}
  {\bibfnamefont {T.~C.}\ \bibnamefont {{Beers}}}} (\bibinfo {year} {1999}),\
  \bibfield  {title} {\enquote {\bibinfo {title} {{R-Process Abundances and
  Chronometers in Metal-poor Stars}},}\ }\href {https://doi.org/10.1086/307512}
  {\bibfield  {journal} {\bibinfo  {journal} {Astrophys. J.}\ }\textbf
  {\bibinfo {volume} {521}},\ \bibinfo {pages} {194--205}}\BibitemShut
  {NoStop}%
\bibitem [{\citenamefont {{Cowan}}\ and\ \citenamefont
  {{Rose}}(1977)}]{cowan77}%
  \BibitemOpen
  \bibfield  {author} {\bibinfo {author} {\bibnamefont {{Cowan}}, \bibfnamefont
  {J~J}}, and\ \bibinfo {author} {\bibfnamefont {W.~K.}\ \bibnamefont
  {{Rose}}}} (\bibinfo {year} {1977}),\ \bibfield  {title} {\enquote {\bibinfo
  {title} {{Production of C-14 and neutrons in red giants}},}\ }\href
  {https://doi.org/10.1086/155030} {\bibfield  {journal} {\bibinfo  {journal}
  {Astrophys. J.}\ }\textbf {\bibinfo {volume} {212}},\ \bibinfo {pages}
  {149--158}}\BibitemShut {NoStop}%
\bibitem [{\citenamefont {{Cowan}}\ \emph {et~al.}(2005)\citenamefont
  {{Cowan}}, \citenamefont {{Sneden}}, \citenamefont {{Beers}}, \citenamefont
  {{Lawler}}, \citenamefont {{Simmerer}}, \citenamefont {{Truran}},
  \citenamefont {{Primas}}, \citenamefont {{Collier}},\ and\ \citenamefont
  {{Burles}}}]{Cowan.Sneden.ea:2005}%
  \BibitemOpen
  \bibfield  {author} {\bibinfo {author} {\bibnamefont {{Cowan}}, \bibfnamefont
  {J~J}}, \bibinfo {author} {\bibfnamefont {C.}~\bibnamefont {{Sneden}}},
  \bibinfo {author} {\bibfnamefont {T.~C.}\ \bibnamefont {{Beers}}}, \bibinfo
  {author} {\bibfnamefont {J.~E.}\ \bibnamefont {{Lawler}}}, \bibinfo {author}
  {\bibfnamefont {J.}~\bibnamefont {{Simmerer}}}, \bibinfo {author}
  {\bibfnamefont {J.~W.}\ \bibnamefont {{Truran}}}, \bibinfo {author}
  {\bibfnamefont {F.}~\bibnamefont {{Primas}}}, \bibinfo {author}
  {\bibfnamefont {J.}~\bibnamefont {{Collier}}}, and\ \bibinfo {author}
  {\bibfnamefont {S.}~\bibnamefont {{Burles}}}} (\bibinfo {year} {2005}),\
  \bibfield  {title} {\enquote {\bibinfo {title} {{Hubble Space Telescope
  Observations of Heavy Elements in Metal-Poor Galactic Halo Stars}},}\ }\href
  {https://doi.org/10.1086/429952} {\bibfield  {journal} {\bibinfo  {journal}
  {Astrophys. J.}\ }\textbf {\bibinfo {volume} {627}},\ \bibinfo {pages}
  {238--250}}\BibitemShut {NoStop}%
\bibitem [{\citenamefont {{Cowan}}\ \emph {et~al.}(2002)\citenamefont
  {{Cowan}}, \citenamefont {{Sneden}}, \citenamefont {{Burles}}, \citenamefont
  {{Ivans}}, \citenamefont {{Beers}}, \citenamefont {{Truran}}, \citenamefont
  {{Lawler}}, \citenamefont {{Primas}}, \citenamefont {{Fuller}}, \citenamefont
  {{Pfeiffer}},\ and\ \citenamefont {{Kratz}}}]{Cowan.Sneden.ea:2002}%
  \BibitemOpen
  \bibfield  {author} {\bibinfo {author} {\bibnamefont {{Cowan}}, \bibfnamefont
  {J~J}}, \bibinfo {author} {\bibfnamefont {C.}~\bibnamefont {{Sneden}}},
  \bibinfo {author} {\bibfnamefont {S.}~\bibnamefont {{Burles}}}, \bibinfo
  {author} {\bibfnamefont {I.~I.}\ \bibnamefont {{Ivans}}}, \bibinfo {author}
  {\bibfnamefont {T.~C.}\ \bibnamefont {{Beers}}}, \bibinfo {author}
  {\bibfnamefont {J.~W.}\ \bibnamefont {{Truran}}}, \bibinfo {author}
  {\bibfnamefont {J.~E.}\ \bibnamefont {{Lawler}}}, \bibinfo {author}
  {\bibfnamefont {F.}~\bibnamefont {{Primas}}}, \bibinfo {author}
  {\bibfnamefont {G.~M.}\ \bibnamefont {{Fuller}}}, \bibinfo {author}
  {\bibfnamefont {B.}~\bibnamefont {{Pfeiffer}}}, and\ \bibinfo {author}
  {\bibfnamefont {K.-L.}\ \bibnamefont {{Kratz}}}} (\bibinfo {year} {2002}),\
  \bibfield  {title} {\enquote {\bibinfo {title} {{The Chemical Composition and
  Age of the Metal-poor Halo Star BD +17$^{\circ}$3248}},}\ }\href
  {https://doi.org/10.1086/340347} {\bibfield  {journal} {\bibinfo  {journal}
  {Astrophys. J.}\ }\textbf {\bibinfo {volume} {572}},\ \bibinfo {pages}
  {861--879}}\BibitemShut {NoStop}%
\bibitem [{\citenamefont {Cowan}\ \emph {et~al.}(1991)\citenamefont {Cowan},
  \citenamefont {Thielemann},\ and\ \citenamefont
  {Truran}}]{Cowan.Thielemann.Truran:1991}%
  \BibitemOpen
  \bibfield  {author} {\bibinfo {author} {\bibnamefont {Cowan}, \bibfnamefont
  {J~J}}, \bibinfo {author} {\bibfnamefont {F.-K.}\ \bibnamefont {Thielemann}},
  and\ \bibinfo {author} {\bibfnamefont {J.~W.}\ \bibnamefont {Truran}}}
  (\bibinfo {year} {1991}),\ \bibfield  {title} {\enquote {\bibinfo {title}
  {{The r-process and nucleochronology}},}\ }\href
  {https://doi.org/10.1016/0370-1573(91)90070-3} {\bibfield  {journal}
  {\bibinfo  {journal} {Phys. Rep.}\ }\textbf {\bibinfo {volume} {208}},\
  \bibinfo {pages} {267--394}}\BibitemShut {NoStop}%
\bibitem [{\citenamefont {{Cowperthwaite}}\ \emph {et~al.}(2017)\citenamefont
  {{Cowperthwaite}} \emph {et~al.}}]{cowperthwaite17}%
  \BibitemOpen
  \bibfield  {author} {\bibinfo {author} {\bibnamefont {{Cowperthwaite}},
  \bibfnamefont {P~S}},  \emph {et~al.}} (\bibinfo {year} {2017}),\ \bibfield
  {title} {\enquote {\bibinfo {title} {{The Electromagnetic Counterpart of the
  Binary Neutron Star Merger LIGO/Virgo GW170817. II. UV, Optical, and
  Near-infrared Light Curves and Comparison to Kilonova Models}},}\ }\href
  {https://doi.org/10.3847/2041-8213/aa8fc7} {\bibfield  {journal} {\bibinfo
  {journal} {Astrophys. J. Lett.}\ }\textbf {\bibinfo {volume} {848}},\
  \bibinfo {eid} {L17}}\BibitemShut {NoStop}%
\bibitem [{\citenamefont {{Crease}}\ and\ \citenamefont
  {{Seidel}}(2000)}]{Crease2000}%
  \BibitemOpen
  \bibfield  {author} {\bibinfo {author} {\bibnamefont {{Crease}},
  \bibfnamefont {R~P}}, and\ \bibinfo {author} {\bibfnamefont {R.~W.}\
  \bibnamefont {{Seidel}}}} (\bibinfo {year} {2000}),\ \bibfield  {title}
  {\enquote {\bibinfo {title} {{Making Physics: A Biography of Brookhaven
  National Laboratory, 1946 1972}},}\ }\href {https://doi.org/10.1063/1.882943}
  {\bibfield  {journal} {\bibinfo  {journal} {Physics Today}\ }\textbf
  {\bibinfo {volume} {53}},\ \bibinfo {pages} {55}}\BibitemShut {NoStop}%
\bibitem [{\citenamefont {{Curtis}}\ \emph {et~al.}(2019)\citenamefont
  {{Curtis}}, \citenamefont {{Ebinger}}, \citenamefont {{Fr{\"o}hlich}},
  \citenamefont {{Hempel}}, \citenamefont {{Perego}}, \citenamefont
  {{Liebend{\"o}rfer}},\ and\ \citenamefont {{Thielemann}}}]{Curtis.ea:2019}%
  \BibitemOpen
  \bibfield  {author} {\bibinfo {author} {\bibnamefont {{Curtis}},
  \bibfnamefont {Sanjana}}, \bibinfo {author} {\bibfnamefont {Kevin}\
  \bibnamefont {{Ebinger}}}, \bibinfo {author} {\bibfnamefont {Carla}\
  \bibnamefont {{Fr{\"o}hlich}}}, \bibinfo {author} {\bibfnamefont {Matthias}\
  \bibnamefont {{Hempel}}}, \bibinfo {author} {\bibfnamefont {Albino}\
  \bibnamefont {{Perego}}}, \bibinfo {author} {\bibfnamefont {Matthias}\
  \bibnamefont {{Liebend{\"o}rfer}}}, and\ \bibinfo {author} {\bibfnamefont
  {Friedrich-Karl}\ \bibnamefont {{Thielemann}}}} (\bibinfo {year} {2019}),\
  \bibfield  {title} {\enquote {\bibinfo {title} {{PUSHing Core-collapse
  Supernovae to Explosions in Spherical Symmetry. III. Nucleosynthesis
  Yields}},}\ }\href {https://doi.org/10.3847/1538-4357/aae7d2} {\bibfield
  {journal} {\bibinfo  {journal} {Astrophys. J.}\ }\textbf {\bibinfo {volume}
  {870}},\ \bibinfo {eid} {2}}\BibitemShut {NoStop}%
\bibitem [{\citenamefont {{Cyburt}}\ \emph {et~al.}(2016)\citenamefont
  {{Cyburt}}, \citenamefont {{Fields}}, \citenamefont {{Olive}},\ and\
  \citenamefont {{Yeh}}}]{Cyburt.Olive.ea:2016}%
  \BibitemOpen
  \bibfield  {author} {\bibinfo {author} {\bibnamefont {{Cyburt}},
  \bibfnamefont {R~H}}, \bibinfo {author} {\bibfnamefont {B.~D.}\ \bibnamefont
  {{Fields}}}, \bibinfo {author} {\bibfnamefont {K.~A.}\ \bibnamefont
  {{Olive}}}, and\ \bibinfo {author} {\bibfnamefont {T.-H.}\ \bibnamefont
  {{Yeh}}}} (\bibinfo {year} {2016}),\ \bibfield  {title} {\enquote {\bibinfo
  {title} {{Big bang nucleosynthesis: Present status}},}\ }\href
  {https://doi.org/10.1103/RevModPhys.88.015004} {\bibfield  {journal}
  {\bibinfo  {journal} {Rev. Mod. Phys.}\ }\textbf {\bibinfo {volume} {88}},\
  \bibinfo {pages} {015004}}\BibitemShut {NoStop}%
\bibitem [{\citenamefont {{Davies}}\ \emph {et~al.}(1994)\citenamefont
  {{Davies}}, \citenamefont {{Benz}}, \citenamefont {{Piran}},\ and\
  \citenamefont {{Thielemann}}}]{davies94}%
  \BibitemOpen
  \bibfield  {author} {\bibinfo {author} {\bibnamefont {{Davies}},
  \bibfnamefont {M~B}}, \bibinfo {author} {\bibfnamefont {W.}~\bibnamefont
  {{Benz}}}, \bibinfo {author} {\bibfnamefont {T.}~\bibnamefont {{Piran}}},
  and\ \bibinfo {author} {\bibfnamefont {F.~K.}\ \bibnamefont {{Thielemann}}}}
  (\bibinfo {year} {1994}),\ \bibfield  {title} {\enquote {\bibinfo {title}
  {{Merging neutron stars. 1. Initial results for coalescence of noncorotating
  systems}},}\ }\href {https://doi.org/10.1086/174525} {\bibfield  {journal}
  {\bibinfo  {journal} {Astrophys. J.}\ }\textbf {\bibinfo {volume} {431}},\
  \bibinfo {pages} {742--753}}\BibitemShut {NoStop}%
\bibitem [{\citenamefont {{Davis}}\ and\ \citenamefont
  {{McKeegan}}(2014)}]{davis14}%
  \BibitemOpen
  \bibfield  {author} {\bibinfo {author} {\bibnamefont {{Davis}}, \bibfnamefont
  {A~M}}, and\ \bibinfo {author} {\bibfnamefont {K.~D.}\ \bibnamefont
  {{McKeegan}}}} (\bibinfo {year} {2014}),\ \enquote {\bibinfo {title}
  {{Short-Lived Radionuclides and Early Solar System Chronology}},}\ in\
  \href@noop {} {\emph {\bibinfo {booktitle} {Meteorites and Cosmochemical
  Processes}}},\ \bibinfo {editor} {edited by\ \bibinfo {editor} {\bibfnamefont
  {A.~M.}\ \bibnamefont {{Davis}}}},\ pp.\ \bibinfo {pages}
  {361--395}\BibitemShut {NoStop}%
\bibitem [{\citenamefont {De}\ \emph {et~al.}(2018)\citenamefont {De},
  \citenamefont {Finstad}, \citenamefont {Lattimer}, \citenamefont {Brown},
  \citenamefont {Berger},\ and\ \citenamefont {Biwer}}]{De.Finstad.ea:2018}%
  \BibitemOpen
  \bibfield  {author} {\bibinfo {author} {\bibnamefont {De}, \bibfnamefont
  {Soumi}}, \bibinfo {author} {\bibfnamefont {Daniel}\ \bibnamefont {Finstad}},
  \bibinfo {author} {\bibfnamefont {James~M.}\ \bibnamefont {Lattimer}},
  \bibinfo {author} {\bibfnamefont {Duncan~A.}\ \bibnamefont {Brown}}, \bibinfo
  {author} {\bibfnamefont {Edo}\ \bibnamefont {Berger}}, and\ \bibinfo {author}
  {\bibfnamefont {Christopher~M.}\ \bibnamefont {Biwer}}} (\bibinfo {year}
  {2018}),\ \bibfield  {title} {\enquote {\bibinfo {title} {Tidal
  deformabilities and radii of neutron stars from the observation of
  gw170817},}\ }\href {https://doi.org/10.1103/PhysRevLett.121.091102}
  {\bibfield  {journal} {\bibinfo  {journal} {Phys. Rev. Lett.}\ }\textbf
  {\bibinfo {volume} {121}},\ \bibinfo {pages} {091102}}\BibitemShut {NoStop}%
\bibitem [{\citenamefont {{De Donder}}\ and\ \citenamefont
  {{Vanbeveren}}(2004)}]{Dedonder.Vanbeveren:2004}%
  \BibitemOpen
  \bibfield  {author} {\bibinfo {author} {\bibnamefont {{De Donder}},
  \bibfnamefont {E}}, and\ \bibinfo {author} {\bibfnamefont {D.}~\bibnamefont
  {{Vanbeveren}}}} (\bibinfo {year} {2004}),\ \bibfield  {title} {\enquote
  {\bibinfo {title} {{The influence of neutron star mergers on the galactic
  chemical enrichment of r-process elements}},}\ }\href
  {https://doi.org/10.1016/S1384-1076(03)00070-8} {\bibfield  {journal}
  {\bibinfo  {journal} {New Astron.}\ }\textbf {\bibinfo {volume} {9}},\
  \bibinfo {pages} {1--16}}\BibitemShut {NoStop}%
\bibitem [{\citenamefont {{Demetriou}}\ \emph {et~al.}(2002)\citenamefont
  {{Demetriou}}, \citenamefont {{Grama}},\ and\ \citenamefont
  {{Goriely}}}]{Demetriou.Grama.Goriely:2002}%
  \BibitemOpen
  \bibfield  {author} {\bibinfo {author} {\bibnamefont {{Demetriou}},
  \bibfnamefont {P}}, \bibinfo {author} {\bibfnamefont {C.}~\bibnamefont
  {{Grama}}}, and\ \bibinfo {author} {\bibfnamefont {S.}~\bibnamefont
  {{Goriely}}}} (\bibinfo {year} {2002}),\ \bibfield  {title} {\enquote
  {\bibinfo {title} {{Improved global $\alpha$-optical model potentials at low
  energies}},}\ }\href {https://doi.org/10.1016/S0375-9474(02)00756-X}
  {\bibfield  {journal} {\bibinfo  {journal} {Nucl. Phys. A}\ }\textbf
  {\bibinfo {volume} {707}},\ \bibinfo {pages} {253--276}}\BibitemShut
  {NoStop}%
\bibitem [{\citenamefont {{Demetriou}}\ \emph {et~al.}(2003)\citenamefont
  {{Demetriou}}, \citenamefont {{Grama}},\ and\ \citenamefont
  {{Goriely}}}]{Demetriou.Grama.Goriely:2003}%
  \BibitemOpen
  \bibfield  {author} {\bibinfo {author} {\bibnamefont {{Demetriou}},
  \bibfnamefont {P}}, \bibinfo {author} {\bibfnamefont {C.}~\bibnamefont
  {{Grama}}}, and\ \bibinfo {author} {\bibfnamefont {S.}~\bibnamefont
  {{Goriely}}}} (\bibinfo {year} {2003}),\ \bibfield  {title} {\enquote
  {\bibinfo {title} {{Improved global {$\alpha$}-optical model potential at low
  energies}},}\ }\href {https://doi.org/10.1016/S0375-9474(03)00827-3}
  {\bibfield  {journal} {\bibinfo  {journal} {Nucl. Phys. A}\ }\textbf
  {\bibinfo {volume} {718}},\ \bibinfo {pages} {510--512}}\BibitemShut
  {NoStop}%
\bibitem [{\citenamefont {Demorest}\ \emph {et~al.}(2010)\citenamefont
  {Demorest}, \citenamefont {Pennucci}, \citenamefont {Ransom}, \citenamefont
  {Roberts},\ and\ \citenamefont {Hessels}}]{Demorest.Pennucci.ea:2010}%
  \BibitemOpen
  \bibfield  {author} {\bibinfo {author} {\bibnamefont {Demorest},
  \bibfnamefont {P~B}}, \bibinfo {author} {\bibfnamefont {T}~\bibnamefont
  {Pennucci}}, \bibinfo {author} {\bibfnamefont {S~M}\ \bibnamefont {Ransom}},
  \bibinfo {author} {\bibfnamefont {M~S~E}\ \bibnamefont {Roberts}}, and\
  \bibinfo {author} {\bibfnamefont {J~W~T}\ \bibnamefont {Hessels}}} (\bibinfo
  {year} {2010}),\ \bibfield  {title} {\enquote {\bibinfo {title} {{A
  two-solar-mass neutron star measured using Shapiro delay}},}\ }\href
  {https://doi.org/10.1038/nature09466} {\bibfield  {journal} {\bibinfo
  {journal} {Nature}\ }\textbf {\bibinfo {volume} {467}},\ \bibinfo {pages}
  {1081--1083}}\BibitemShut {NoStop}%
\bibitem [{\citenamefont {{Den Hartog}}\ \emph {et~al.}(2002)\citenamefont
  {{Den Hartog}}, \citenamefont {{Wickliffe}},\ and\ \citenamefont
  {{Lawler}}}]{denhartog02}%
  \BibitemOpen
  \bibfield  {author} {\bibinfo {author} {\bibnamefont {{Den Hartog}},
  \bibfnamefont {E~A}}, \bibinfo {author} {\bibfnamefont {M.~E.}\ \bibnamefont
  {{Wickliffe}}}, and\ \bibinfo {author} {\bibfnamefont {J.~E.}\ \bibnamefont
  {{Lawler}}}} (\bibinfo {year} {2002}),\ \bibfield  {title} {\enquote
  {\bibinfo {title} {{Radiative Lifetimes of Eu I, II, and III and Transition
  Probabilities of Eu I}},}\ }\href {https://doi.org/10.1086/340039} {\bibfield
   {journal} {\bibinfo  {journal} {Astrophys. J. Suppl.}\ }\textbf {\bibinfo
  {volume} {141}},\ \bibinfo {pages} {255--265}}\BibitemShut {NoStop}%
\bibitem [{\citenamefont {{Denissenkov}}\ \emph {et~al.}(2017)\citenamefont
  {{Denissenkov}}, \citenamefont {{Herwig}}, \citenamefont {{Battino}},
  \citenamefont {{Ritter}}, \citenamefont {{Pignatari}}, \citenamefont
  {{Jones}},\ and\ \citenamefont {{Paxton}}}]{denissenkov17}%
  \BibitemOpen
  \bibfield  {author} {\bibinfo {author} {\bibnamefont {{Denissenkov}},
  \bibfnamefont {P~A}}, \bibinfo {author} {\bibfnamefont {F.}~\bibnamefont
  {{Herwig}}}, \bibinfo {author} {\bibfnamefont {U.}~\bibnamefont {{Battino}}},
  \bibinfo {author} {\bibfnamefont {C.}~\bibnamefont {{Ritter}}}, \bibinfo
  {author} {\bibfnamefont {M.}~\bibnamefont {{Pignatari}}}, \bibinfo {author}
  {\bibfnamefont {S.}~\bibnamefont {{Jones}}}, and\ \bibinfo {author}
  {\bibfnamefont {B.}~\bibnamefont {{Paxton}}}} (\bibinfo {year} {2017}),\
  \bibfield  {title} {\enquote {\bibinfo {title} {{i-process Nucleosynthesis
  and Mass Retention Efficiency in He-shell Flash Evolution of Rapidly
  Accreting White Dwarfs}},}\ }\href
  {https://doi.org/10.3847/2041-8213/834/2/L10} {\bibfield  {journal} {\bibinfo
   {journal} {Astrophys. J.}\ }\textbf {\bibinfo {volume} {834}},\ \bibinfo
  {eid} {L10}}\BibitemShut {NoStop}%
\bibitem [{\citenamefont {{Desai}}\ \emph {et~al.}(2019)\citenamefont
  {{Desai}}, \citenamefont {{Metzger}},\ and\ \citenamefont
  {{Foucart}}}]{Desai.Metzger.Foucart:2019}%
  \BibitemOpen
  \bibfield  {author} {\bibinfo {author} {\bibnamefont {{Desai}}, \bibfnamefont
  {D}}, \bibinfo {author} {\bibfnamefont {B.~D.}\ \bibnamefont {{Metzger}}},
  and\ \bibinfo {author} {\bibfnamefont {F.}~\bibnamefont {{Foucart}}}}
  (\bibinfo {year} {2019}),\ \bibfield  {title} {\enquote {\bibinfo {title}
  {{Imprints of r-process heating on fall-back accretion: distinguishing black
  hole-neutron star from double neutron star mergers}},}\ }\href
  {https://doi.org/10.1093/mnras/stz644} {\bibfield  {journal} {\bibinfo
  {journal} {Mon. Not. Roy. Astron. Soc.}\ }\textbf {\bibinfo {volume} {485}},\
  \bibinfo {pages} {4404--4412}}\BibitemShut {NoStop}%
\bibitem [{\citenamefont {{Dessart}}\ \emph {et~al.}(2009)\citenamefont
  {{Dessart}}, \citenamefont {{Ott}}, \citenamefont {{Burrows}}, \citenamefont
  {{Rosswog}},\ and\ \citenamefont {{Livne}}}]{Dessart.Ott.ea:2009}%
  \BibitemOpen
  \bibfield  {author} {\bibinfo {author} {\bibnamefont {{Dessart}},
  \bibfnamefont {L}}, \bibinfo {author} {\bibfnamefont {C.~D.}\ \bibnamefont
  {{Ott}}}, \bibinfo {author} {\bibfnamefont {A.}~\bibnamefont {{Burrows}}},
  \bibinfo {author} {\bibfnamefont {S.}~\bibnamefont {{Rosswog}}}, and\
  \bibinfo {author} {\bibfnamefont {E.}~\bibnamefont {{Livne}}}} (\bibinfo
  {year} {2009}),\ \bibfield  {title} {\enquote {\bibinfo {title} {{Neutrino
  Signatures and the Neutrino-Driven Wind in Binary Neutron Star Mergers}},}\
  }\href {https://doi.org/10.1088/0004-637X/690/2/1681} {\bibfield  {journal}
  {\bibinfo  {journal} {Astrophys. J.}\ }\textbf {\bibinfo {volume} {690}},\
  \bibinfo {pages} {1681--1705}}\BibitemShut {NoStop}%
\bibitem [{\citenamefont {{Diehl}}\ and\ \citenamefont
  {{Timmes}}(1998)}]{Diehl.Timmes:1998}%
  \BibitemOpen
  \bibfield  {author} {\bibinfo {author} {\bibnamefont {{Diehl}}, \bibfnamefont
  {R}}, and\ \bibinfo {author} {\bibfnamefont {F.~X.}\ \bibnamefont
  {{Timmes}}}} (\bibinfo {year} {1998}),\ \bibfield  {title} {\enquote
  {\bibinfo {title} {Gamma-ray line emission from radioactive isotopes in stars
  and galaxies},}\ }\href {https://doi.org/10.1086/316169} {\bibfield
  {journal} {\bibinfo  {journal} {Publ. Astron. Soc. Pacific}\ }\textbf
  {\bibinfo {volume} {110}},\ \bibinfo {pages} {637--659}}\BibitemShut
  {NoStop}%
\bibitem [{\citenamefont {{Dillmann}}\ \emph {et~al.}(2003)\citenamefont
  {{Dillmann}}, \citenamefont {{Kratz}}, \citenamefont {{W{\"o}hr}},
  \citenamefont {{Arndt}}, \citenamefont {{Brown}}, \citenamefont {{Hoff}},
  \citenamefont {{Hjorth-Jensen}}, \citenamefont {{K{\"o}ster}}, \citenamefont
  {{Ostrowski}}, \citenamefont {{Pfeiffer}}, \citenamefont {{Seweryniak}},
  \citenamefont {{Shergur}},\ and\ \citenamefont {{Walters}}}]{Dillmann2003}%
  \BibitemOpen
  \bibfield  {author} {\bibinfo {author} {\bibnamefont {{Dillmann}},
  \bibfnamefont {I}}, \bibinfo {author} {\bibfnamefont {K.-L.}\ \bibnamefont
  {{Kratz}}}, \bibinfo {author} {\bibfnamefont {A.}~\bibnamefont {{W{\"o}hr}}},
  \bibinfo {author} {\bibfnamefont {O.}~\bibnamefont {{Arndt}}}, \bibinfo
  {author} {\bibfnamefont {B.~A.}\ \bibnamefont {{Brown}}}, \bibinfo {author}
  {\bibfnamefont {P.}~\bibnamefont {{Hoff}}}, \bibinfo {author} {\bibfnamefont
  {M.}~\bibnamefont {{Hjorth-Jensen}}}, \bibinfo {author} {\bibfnamefont
  {U.}~\bibnamefont {{K{\"o}ster}}}, \bibinfo {author} {\bibfnamefont {A.~N.}\
  \bibnamefont {{Ostrowski}}}, \bibinfo {author} {\bibfnamefont
  {B.}~\bibnamefont {{Pfeiffer}}}, \bibinfo {author} {\bibfnamefont
  {D.}~\bibnamefont {{Seweryniak}}}, \bibinfo {author} {\bibfnamefont
  {J.}~\bibnamefont {{Shergur}}}, and\ \bibinfo {author} {\bibfnamefont
  {W.~B.}\ \bibnamefont {{Walters}}}} (\bibinfo {year} {2003}),\ \bibfield
  {title} {\enquote {\bibinfo {title} {{N=82 Shell Quenching of the Classical
  r-Process ``Waiting-Point'' Nucleus $^{130}$Cd}},}\ }\href
  {https://doi.org/10.1103/PhysRevLett.91.162503} {\bibfield  {journal}
  {\bibinfo  {journal} {Phys. Rev. Lett.}\ }\textbf {\bibinfo {volume} {91}},\
  \bibinfo {eid} {162503}}\BibitemShut {NoStop}%
\bibitem [{\citenamefont {{Drout}}\ \emph {et~al.}(2017)\citenamefont {{Drout}}
  \emph {et~al.}}]{drout17}%
  \BibitemOpen
  \bibfield  {author} {\bibinfo {author} {\bibnamefont {{Drout}}, \bibfnamefont
  {M~R}},  \emph {et~al.}} (\bibinfo {year} {2017}),\ \bibfield  {title}
  {\enquote {\bibinfo {title} {{Light curves of the neutron star merger
  GW170817/SSS17a: Implications for r-process nucleosynthesis}},}\ }\href
  {https://doi.org/10.1126/science.aaq0049} {\bibfield  {journal} {\bibinfo
  {journal} {Science}\ }\textbf {\bibinfo {volume} {358}},\ \bibinfo {pages}
  {1570--1574}}\BibitemShut {NoStop}%
\bibitem [{\citenamefont {Duflo}\ and\ \citenamefont
  {Zuker}(1995)}]{Duflo.Zuker:1995}%
  \BibitemOpen
  \bibfield  {author} {\bibinfo {author} {\bibnamefont {Duflo}, \bibfnamefont
  {J}}, and\ \bibinfo {author} {\bibfnamefont {A.~P.}\ \bibnamefont {Zuker}}}
  (\bibinfo {year} {1995}),\ \bibfield  {title} {\enquote {\bibinfo {title}
  {{Microscopic mass formulas}},}\ }\href
  {https://doi.org/10.1103/PhysRevC.52.R23} {\bibfield  {journal} {\bibinfo
  {journal} {Phys. Rev. C}\ }\textbf {\bibinfo {volume} {52}},\ \bibinfo
  {pages} {R23--R27}}\BibitemShut {NoStop}%
\bibitem [{\citenamefont {{Duncan}}\ \emph {et~al.}(1986)\citenamefont
  {{Duncan}}, \citenamefont {{Shapiro}},\ and\ \citenamefont
  {{Wasserman}}}]{Duncan.Shapiro.Wasserman:1986}%
  \BibitemOpen
  \bibfield  {author} {\bibinfo {author} {\bibnamefont {{Duncan}},
  \bibfnamefont {R~C}}, \bibinfo {author} {\bibfnamefont {S.~L.}\ \bibnamefont
  {{Shapiro}}}, and\ \bibinfo {author} {\bibfnamefont {I.}~\bibnamefont
  {{Wasserman}}}} (\bibinfo {year} {1986}),\ \bibfield  {title} {\enquote
  {\bibinfo {title} {{Neutrino-driven winds from young, hot neutron stars}},}\
  }\href {https://doi.org/10.1086/164587} {\bibfield  {journal} {\bibinfo
  {journal} {Astrophys. J.}\ }\textbf {\bibinfo {volume} {309}},\ \bibinfo
  {pages} {141--160}}\BibitemShut {NoStop}%
\bibitem [{\citenamefont {{Duncan}}\ and\ \citenamefont
  {{Thompson}}(1992)}]{Duncan.Thompson:1992}%
  \BibitemOpen
  \bibfield  {author} {\bibinfo {author} {\bibnamefont {{Duncan}},
  \bibfnamefont {R~C}}, and\ \bibinfo {author} {\bibfnamefont {C.}~\bibnamefont
  {{Thompson}}}} (\bibinfo {year} {1992}),\ \bibfield  {title} {\enquote
  {\bibinfo {title} {{Formation of very strongly magnetized neutron stars -
  Implications for gamma-ray bursts}},}\ }\href
  {https://doi.org/10.1086/186413} {\bibfield  {journal} {\bibinfo  {journal}
  {Astrophys. J. Lett.}\ }\textbf {\bibinfo {volume} {392}},\ \bibinfo {pages}
  {L9--L13}}\BibitemShut {NoStop}%
\bibitem [{\citenamefont {{Dunlop}}\ \emph {et~al.}(2019)\citenamefont
  {{Dunlop}} \emph {et~al.}}]{Dunlop.Svensson.ea:2019}%
  \BibitemOpen
  \bibfield  {author} {\bibinfo {author} {\bibnamefont {{Dunlop}},
  \bibfnamefont {R}},  \emph {et~al.}} (\bibinfo {year} {2019}),\ \bibfield
  {title} {\enquote {\bibinfo {title} {{{\ensuremath{\beta}} decay and
  {\ensuremath{\beta}}-delayed neutron decay of the $N =82$ nucleus
  $^{131}_{49}$In$_{82}$}},}\ }\href
  {https://doi.org/10.1103/PhysRevC.99.045805} {\bibfield  {journal} {\bibinfo
  {journal} {Phys. Rev. C}\ }\textbf {\bibinfo {volume} {99}},\ \bibinfo {eid}
  {045805}}\BibitemShut {NoStop}%
\bibitem [{\citenamefont {{Duquette}}\ \emph {et~al.}(1981)\citenamefont
  {{Duquette}}, \citenamefont {{Salih}},\ and\ \citenamefont
  {{Lawler}}}]{duquette81}%
  \BibitemOpen
  \bibfield  {author} {\bibinfo {author} {\bibnamefont {{Duquette}},
  \bibfnamefont {D~W}}, \bibinfo {author} {\bibfnamefont {S.}~\bibnamefont
  {{Salih}}}, and\ \bibinfo {author} {\bibfnamefont {J.~E.}\ \bibnamefont
  {{Lawler}}}} (\bibinfo {year} {1981}),\ \bibfield  {title} {\enquote
  {\bibinfo {title} {{Radiative lifetimes in MoI using a novel atomic beam
  source}},}\ }\href {https://doi.org/10.1016/0375-9601(81)90826-4} {\bibfield
  {journal} {\bibinfo  {journal} {Phys. Lett. A}\ }\textbf {\bibinfo {volume}
  {83}},\ \bibinfo {pages} {214--216}}\BibitemShut {NoStop}%
\bibitem [{\citenamefont {{Ebinger}}\ \emph {et~al.}(2019)\citenamefont
  {{Ebinger}}, \citenamefont {{Curtis}}, \citenamefont {{Fr{\"o}hlich}},
  \citenamefont {{Hempel}}, \citenamefont {{Perego}}, \citenamefont
  {{Liebend{\"o}rfer}},\ and\ \citenamefont
  {{Thielemann}}}]{Ebinger.Curtis.ea:2019}%
  \BibitemOpen
  \bibfield  {author} {\bibinfo {author} {\bibnamefont {{Ebinger}},
  \bibfnamefont {Kevin}}, \bibinfo {author} {\bibfnamefont {Sanjana}\
  \bibnamefont {{Curtis}}}, \bibinfo {author} {\bibfnamefont {Carla}\
  \bibnamefont {{Fr{\"o}hlich}}}, \bibinfo {author} {\bibfnamefont {Matthias}\
  \bibnamefont {{Hempel}}}, \bibinfo {author} {\bibfnamefont {Albino}\
  \bibnamefont {{Perego}}}, \bibinfo {author} {\bibfnamefont {Matthias}\
  \bibnamefont {{Liebend{\"o}rfer}}}, and\ \bibinfo {author} {\bibfnamefont
  {Friedrich-Karl}\ \bibnamefont {{Thielemann}}}} (\bibinfo {year} {2019}),\
  \bibfield  {title} {\enquote {\bibinfo {title} {{PUSHing Core-collapse
  Supernovae to Explosions in Spherical Symmetry. II. Explodability and Remnant
  Properties}},}\ }\href {https://doi.org/10.3847/1538-4357/aae7c9} {\bibfield
  {journal} {\bibinfo  {journal} {Astrophys. J.}\ }\textbf {\bibinfo {volume}
  {870}},\ \bibinfo {eid} {1}}\BibitemShut {NoStop}%
\bibitem [{\citenamefont {{Ebinger}}\ \emph {et~al.}(2020)\citenamefont
  {{Ebinger}}, \citenamefont {{Curtis}}, \citenamefont {{Ghosh}}, \citenamefont
  {{Fr{\"o}hlich}}, \citenamefont {{Hempel}}, \citenamefont {{Perego}},
  \citenamefont {{Liebend{\"o}rfer}},\ and\ \citenamefont
  {{Thielemann}}}]{Ebinger.Curtis.ea:2020}%
  \BibitemOpen
  \bibfield  {author} {\bibinfo {author} {\bibnamefont {{Ebinger}},
  \bibfnamefont {Kevin}}, \bibinfo {author} {\bibfnamefont {Sanjana}\
  \bibnamefont {{Curtis}}}, \bibinfo {author} {\bibfnamefont {Somdutta}\
  \bibnamefont {{Ghosh}}}, \bibinfo {author} {\bibfnamefont {Carla}\
  \bibnamefont {{Fr{\"o}hlich}}}, \bibinfo {author} {\bibfnamefont {Matthias}\
  \bibnamefont {{Hempel}}}, \bibinfo {author} {\bibfnamefont {Albino}\
  \bibnamefont {{Perego}}}, \bibinfo {author} {\bibfnamefont {Matthias}\
  \bibnamefont {{Liebend{\"o}rfer}}}, and\ \bibinfo {author} {\bibfnamefont
  {Friedrich-Karl}\ \bibnamefont {{Thielemann}}}} (\bibinfo {year} {2020}),\
  \bibfield  {title} {\enquote {\bibinfo {title} {{PUSHing Core-collapse
  Supernovae to Explosions in Spherical Symmetry. IV. Explodability, Remnant
  Properties, and Nucleosynthesis Yields of Low-metallicity Stars}},}\ }\href
  {https://doi.org/10.3847/1538-4357/ab5dcb} {\bibfield  {journal} {\bibinfo
  {journal} {Astrophys. J.}\ }\textbf {\bibinfo {volume} {888}},\ \bibinfo
  {eid} {91}}\BibitemShut {NoStop}%
\bibitem [{\citenamefont {{Eichler}}\ \emph {et~al.}(1989)\citenamefont
  {{Eichler}}, \citenamefont {{Livio}}, \citenamefont {{Piran}},\ and\
  \citenamefont {{Schramm}}}]{Eichler.Livio.ea:1989}%
  \BibitemOpen
  \bibfield  {author} {\bibinfo {author} {\bibnamefont {{Eichler}},
  \bibfnamefont {D}}, \bibinfo {author} {\bibfnamefont {M.}~\bibnamefont
  {{Livio}}}, \bibinfo {author} {\bibfnamefont {T.}~\bibnamefont {{Piran}}},
  and\ \bibinfo {author} {\bibfnamefont {D.~N.}\ \bibnamefont {{Schramm}}}}
  (\bibinfo {year} {1989}),\ \bibfield  {title} {\enquote {\bibinfo {title}
  {{Nucleosynthesis, neutrino bursts and gamma-rays from coalescing neutron
  stars}},}\ }\href {https://doi.org/10.1038/340126a0} {\bibfield  {journal}
  {\bibinfo  {journal} {Nature}\ }\textbf {\bibinfo {volume} {340}},\ \bibinfo
  {pages} {126--128}}\BibitemShut {NoStop}%
\bibitem [{\citenamefont {{Eichler}}\ \emph {et~al.}(2015)\citenamefont
  {{Eichler}}, \citenamefont {{Arcones}}, \citenamefont {{Kelic}},
  \citenamefont {{Korobkin}}, \citenamefont {{Langanke}}, \citenamefont
  {{Marketin}}, \citenamefont {{Mart{\'i}nez-Pinedo}}, \citenamefont {{Panov}},
  \citenamefont {{Rauscher}}, \citenamefont {{Rosswog}}, \citenamefont
  {{Winteler}}, \citenamefont {{Zinner}},\ and\ \citenamefont
  {{Thielemann}}}]{Eichler.Arcones.ea:2015}%
  \BibitemOpen
  \bibfield  {author} {\bibinfo {author} {\bibnamefont {{Eichler}},
  \bibfnamefont {M}}, \bibinfo {author} {\bibfnamefont {A.}~\bibnamefont
  {{Arcones}}}, \bibinfo {author} {\bibfnamefont {A.}~\bibnamefont {{Kelic}}},
  \bibinfo {author} {\bibfnamefont {O.}~\bibnamefont {{Korobkin}}}, \bibinfo
  {author} {\bibfnamefont {K.}~\bibnamefont {{Langanke}}}, \bibinfo {author}
  {\bibfnamefont {T.}~\bibnamefont {{Marketin}}}, \bibinfo {author}
  {\bibfnamefont {G.}~\bibnamefont {{Mart{\'i}nez-Pinedo}}}, \bibinfo {author}
  {\bibfnamefont {I.}~\bibnamefont {{Panov}}}, \bibinfo {author} {\bibfnamefont
  {T.}~\bibnamefont {{Rauscher}}}, \bibinfo {author} {\bibfnamefont
  {S.}~\bibnamefont {{Rosswog}}}, \bibinfo {author} {\bibfnamefont
  {C.}~\bibnamefont {{Winteler}}}, \bibinfo {author} {\bibfnamefont {N.~T.}\
  \bibnamefont {{Zinner}}}, and\ \bibinfo {author} {\bibfnamefont {F.-K.}\
  \bibnamefont {{Thielemann}}}} (\bibinfo {year} {2015}),\ \bibfield  {title}
  {\enquote {\bibinfo {title} {The role of fission in neutron star mergers and
  its impact on the r-process peaks},}\ }\href
  {https://doi.org/10.1088/0004-637X/808/1/30} {\bibfield  {journal} {\bibinfo
  {journal} {Astrophys. J.}\ }\textbf {\bibinfo {volume} {808}},\ \bibinfo
  {eid} {30}}\BibitemShut {NoStop}%
\bibitem [{\citenamefont {{Eichler}}\ \emph {et~al.}(2018)\citenamefont
  {{Eichler}}, \citenamefont {{Nakamura}}, \citenamefont {{Takiwaki}},
  \citenamefont {{Kuroda}}, \citenamefont {{Kotake}}, \citenamefont {{Hempel}},
  \citenamefont {{Cabez{\'o}n}}, \citenamefont {{Liebend{\"o}rfer}},\ and\
  \citenamefont {{Thielemann}}}]{Eichler.Nakamura.ea:2018}%
  \BibitemOpen
  \bibfield  {author} {\bibinfo {author} {\bibnamefont {{Eichler}},
  \bibfnamefont {M}}, \bibinfo {author} {\bibfnamefont {K.}~\bibnamefont
  {{Nakamura}}}, \bibinfo {author} {\bibfnamefont {T.}~\bibnamefont
  {{Takiwaki}}}, \bibinfo {author} {\bibfnamefont {T.}~\bibnamefont
  {{Kuroda}}}, \bibinfo {author} {\bibfnamefont {K.}~\bibnamefont {{Kotake}}},
  \bibinfo {author} {\bibfnamefont {M.}~\bibnamefont {{Hempel}}}, \bibinfo
  {author} {\bibfnamefont {R.}~\bibnamefont {{Cabez{\'o}n}}}, \bibinfo {author}
  {\bibfnamefont {M.}~\bibnamefont {{Liebend{\"o}rfer}}}, and\ \bibinfo
  {author} {\bibfnamefont {F.-K.}\ \bibnamefont {{Thielemann}}}} (\bibinfo
  {year} {2018}),\ \bibfield  {title} {\enquote {\bibinfo {title}
  {{Nucleosynthesis in 2D core-collapse supernovae of 11.2 and 17.0 M$_{\odot}$
  progenitors: implications for Mo and Ru production}},}\ }\href
  {https://doi.org/10.1088/1361-6471/aa8891} {\bibfield  {journal} {\bibinfo
  {journal} {J. Phys. G: Nucl. Part. Phys.}\ }\textbf {\bibinfo {volume}
  {45}},\ \bibinfo {pages} {014001}}\BibitemShut {NoStop}%
\bibitem [{\citenamefont {{Eichler}}\ \emph {et~al.}(2019)\citenamefont
  {{Eichler}}, \citenamefont {{Sayar}}, \citenamefont {{Arcones}},\ and\
  \citenamefont {{Rauscher}}}]{Eichler.Sayar.ea:2019}%
  \BibitemOpen
  \bibfield  {author} {\bibinfo {author} {\bibnamefont {{Eichler}},
  \bibfnamefont {M}}, \bibinfo {author} {\bibfnamefont {W.}~\bibnamefont
  {{Sayar}}}, \bibinfo {author} {\bibfnamefont {A.}~\bibnamefont {{Arcones}}},
  and\ \bibinfo {author} {\bibfnamefont {T.}~\bibnamefont {{Rauscher}}}}
  (\bibinfo {year} {2019}),\ \bibfield  {title} {\enquote {\bibinfo {title}
  {{Probing the Production of Actinides under Different r-process
  Conditions}},}\ }\href {https://doi.org/10.3847/1538-4357/ab24cf} {\bibfield
  {journal} {\bibinfo  {journal} {Astrophys. J.}\ }\textbf {\bibinfo {volume}
  {879}},\ \bibinfo {eid} {47}}\BibitemShut {NoStop}%
\bibitem [{\citenamefont {{Ejnisman}}\ \emph {et~al.}(1998)\citenamefont
  {{Ejnisman}}, \citenamefont {{Goldman}}, \citenamefont {{Krane}},
  \citenamefont {{Mohr}}, \citenamefont {{Nakazawa}}, \citenamefont {{Norman}},
  \citenamefont {{Rauscher}},\ and\ \citenamefont
  {{Reel}}}]{Ejnisman.Goldman.ea:1998}%
  \BibitemOpen
  \bibfield  {author} {\bibinfo {author} {\bibnamefont {{Ejnisman}},
  \bibfnamefont {R}}, \bibinfo {author} {\bibfnamefont {I.~D.}\ \bibnamefont
  {{Goldman}}}, \bibinfo {author} {\bibfnamefont {K.~S.}\ \bibnamefont
  {{Krane}}}, \bibinfo {author} {\bibfnamefont {P.}~\bibnamefont {{Mohr}}},
  \bibinfo {author} {\bibfnamefont {Y.}~\bibnamefont {{Nakazawa}}}, \bibinfo
  {author} {\bibfnamefont {E.~B.}\ \bibnamefont {{Norman}}}, \bibinfo {author}
  {\bibfnamefont {T.}~\bibnamefont {{Rauscher}}}, and\ \bibinfo {author}
  {\bibfnamefont {J.}~\bibnamefont {{Reel}}}} (\bibinfo {year} {1998}),\
  \bibfield  {title} {\enquote {\bibinfo {title} {{Neutron capture cross
  section of $^{44}$Ti}},}\ }\href {https://doi.org/10.1103/PhysRevC.58.2531}
  {\bibfield  {journal} {\bibinfo  {journal} {Phys. Rev. C}\ }\textbf {\bibinfo
  {volume} {58}},\ \bibinfo {pages} {2531--2537}}\BibitemShut {NoStop}%
\bibitem [{\citenamefont {{Eliseev}}\ \emph {et~al.}(2013)\citenamefont
  {{Eliseev}}, \citenamefont {{Blaum}}, \citenamefont {{Block}}, \citenamefont
  {{Droese}}, \citenamefont {{Goncharov}}, \citenamefont {{Minaya Ramirez}},
  \citenamefont {{Nesterenko}}, \citenamefont {{Novikov}},\ and\ \citenamefont
  {{Schweikhard}}}]{Eliseev2013}%
  \BibitemOpen
  \bibfield  {author} {\bibinfo {author} {\bibnamefont {{Eliseev}},
  \bibfnamefont {S}}, \bibinfo {author} {\bibfnamefont {K.}~\bibnamefont
  {{Blaum}}}, \bibinfo {author} {\bibfnamefont {M.}~\bibnamefont {{Block}}},
  \bibinfo {author} {\bibfnamefont {C.}~\bibnamefont {{Droese}}}, \bibinfo
  {author} {\bibfnamefont {M.}~\bibnamefont {{Goncharov}}}, \bibinfo {author}
  {\bibfnamefont {E.}~\bibnamefont {{Minaya Ramirez}}}, \bibinfo {author}
  {\bibfnamefont {D.~A.}\ \bibnamefont {{Nesterenko}}}, \bibinfo {author}
  {\bibfnamefont {Y.~N.}\ \bibnamefont {{Novikov}}}, and\ \bibinfo {author}
  {\bibfnamefont {L.}~\bibnamefont {{Schweikhard}}}} (\bibinfo {year} {2013}),\
  \bibfield  {title} {\enquote {\bibinfo {title} {{Phase-Imaging
  Ion-Cyclotron-Resonance Measurements for Short-Lived Nuclides}},}\ }\href
  {https://doi.org/10.1103/PhysRevLett.110.082501} {\bibfield  {journal}
  {\bibinfo  {journal} {Phys. Rev. Lett.}\ }\textbf {\bibinfo {volume} {110}},\
  \bibinfo {eid} {082501}}\BibitemShut {NoStop}%
\bibitem [{\citenamefont {{Emerick}}\ \emph {et~al.}(2018)\citenamefont
  {{Emerick}}, \citenamefont {{Bryan}}, \citenamefont {{Mac Low}},
  \citenamefont {{C{\^o}t{\'e}}}, \citenamefont {{Johnston}},\ and\
  \citenamefont {{O'Shea}}}]{Emerick.Bryan.ea:2018}%
  \BibitemOpen
  \bibfield  {author} {\bibinfo {author} {\bibnamefont {{Emerick}},
  \bibfnamefont {Andrew}}, \bibinfo {author} {\bibfnamefont {Greg~L.}\
  \bibnamefont {{Bryan}}}, \bibinfo {author} {\bibfnamefont {Mordecai-Mark}\
  \bibnamefont {{Mac Low}}}, \bibinfo {author} {\bibfnamefont {Benoit}\
  \bibnamefont {{C{\^o}t{\'e}}}}, \bibinfo {author} {\bibfnamefont
  {Kathryn~V.}\ \bibnamefont {{Johnston}}}, and\ \bibinfo {author}
  {\bibfnamefont {Brian~W.}\ \bibnamefont {{O'Shea}}}} (\bibinfo {year}
  {2018}),\ \bibfield  {title} {\enquote {\bibinfo {title} {{Metal Mixing and
  Ejection in Dwarf Galaxies Are Dependent on Nucleosynthetic Source}},}\
  }\href {https://doi.org/10.3847/1538-4357/aaec7d} {\bibfield  {journal}
  {\bibinfo  {journal} {Astrophys. J.}\ }\textbf {\bibinfo {volume} {869}},\
  \bibinfo {eid} {94}}\BibitemShut {NoStop}%
\bibitem [{\citenamefont {Engel}\ \emph {et~al.}(1999)\citenamefont {Engel},
  \citenamefont {Bender}, \citenamefont {Dobaczewski}, \citenamefont
  {Nazarewicz},\ and\ \citenamefont {Surman}}]{Engel.Bender.ea:1999}%
  \BibitemOpen
  \bibfield  {author} {\bibinfo {author} {\bibnamefont {Engel}, \bibfnamefont
  {J}}, \bibinfo {author} {\bibfnamefont {M.}~\bibnamefont {Bender}}, \bibinfo
  {author} {\bibfnamefont {J.}~\bibnamefont {Dobaczewski}}, \bibinfo {author}
  {\bibfnamefont {W.}~\bibnamefont {Nazarewicz}}, and\ \bibinfo {author}
  {\bibfnamefont {R.}~\bibnamefont {Surman}}} (\bibinfo {year} {1999}),\
  \bibfield  {title} {\enquote {\bibinfo {title} {Beta decay of r-process
  waiting-point nuclei in a self-consistent approach},}\ }\href
  {https://doi.org/10.1103/PhysRevC.60.014302} {\bibfield  {journal} {\bibinfo
  {journal} {Phys. Rev. C}\ }\textbf {\bibinfo {volume} {60}},\ \bibinfo
  {pages} {014302}}\BibitemShut {NoStop}%
\bibitem [{\citenamefont {Epstein}\ \emph {et~al.}(1988)\citenamefont
  {Epstein}, \citenamefont {Colgate},\ and\ \citenamefont
  {Haxton}}]{Epstein.Colgate.Haxton:1988}%
  \BibitemOpen
  \bibfield  {author} {\bibinfo {author} {\bibnamefont {Epstein}, \bibfnamefont
  {Richard~I}}, \bibinfo {author} {\bibfnamefont {Stirling~A.}\ \bibnamefont
  {Colgate}}, and\ \bibinfo {author} {\bibfnamefont {Wick~C.}\ \bibnamefont
  {Haxton}}} (\bibinfo {year} {1988}),\ \bibfield  {title} {\enquote {\bibinfo
  {title} {{Neutrino-Induced $r$-Process Nucleosynthesis}},}\ }\href
  {https://doi.org/10.1103/PhysRevLett.61.2038} {\bibfield  {journal} {\bibinfo
   {journal} {Phys. Rev. Lett.}\ }\textbf {\bibinfo {volume} {61}},\ \bibinfo
  {pages} {2038--2041}}\BibitemShut {NoStop}%
\bibitem [{\citenamefont {Erler}\ \emph {et~al.}(2012)\citenamefont {Erler},
  \citenamefont {Langanke}, \citenamefont {Loens}, \citenamefont
  {Mart\'{\i}nez-Pinedo},\ and\ \citenamefont
  {Reinhard}}]{Erler.Langanke.ea:2012}%
  \BibitemOpen
  \bibfield  {author} {\bibinfo {author} {\bibnamefont {Erler}, \bibfnamefont
  {J}}, \bibinfo {author} {\bibfnamefont {K.}~\bibnamefont {Langanke}},
  \bibinfo {author} {\bibfnamefont {H.}~\bibnamefont {Loens}}, \bibinfo
  {author} {\bibfnamefont {Gabriel}\ \bibnamefont {Mart\'{\i}nez-Pinedo}}, and\
  \bibinfo {author} {\bibfnamefont {P.-G.}\ \bibnamefont {Reinhard}}} (\bibinfo
  {year} {2012}),\ \bibfield  {title} {\enquote {\bibinfo {title} {{Fission
  properties for r-process nuclei}},}\ }\href
  {https://doi.org/10.1103/PhysRevC.85.025802} {\bibfield  {journal} {\bibinfo
  {journal} {Phys. Rev. C}\ }\textbf {\bibinfo {volume} {85}},\ \bibinfo
  {pages} {025802}}\BibitemShut {NoStop}%
\bibitem [{\citenamefont {{Ertl}}\ \emph {et~al.}(2016)\citenamefont {{Ertl}},
  \citenamefont {{Janka}}, \citenamefont {{Woosley}}, \citenamefont
  {{Sukhbold}},\ and\ \citenamefont {{Ugliano}}}]{Ertl.Janka.ea:2016}%
  \BibitemOpen
  \bibfield  {author} {\bibinfo {author} {\bibnamefont {{Ertl}}, \bibfnamefont
  {T}}, \bibinfo {author} {\bibfnamefont {H.-T.}\ \bibnamefont {{Janka}}},
  \bibinfo {author} {\bibfnamefont {S.~E.}\ \bibnamefont {{Woosley}}}, \bibinfo
  {author} {\bibfnamefont {T.}~\bibnamefont {{Sukhbold}}}, and\ \bibinfo
  {author} {\bibfnamefont {M.}~\bibnamefont {{Ugliano}}}} (\bibinfo {year}
  {2016}),\ \bibfield  {title} {\enquote {\bibinfo {title} {{A Two-parameter
  Criterion for Classifying the Explodability of Massive Stars by the
  Neutrino-driven Mechanism}},}\ }\href
  {https://doi.org/10.3847/0004-637X/818/2/124} {\bibfield  {journal} {\bibinfo
   {journal} {Astrophys. J.}\ }\textbf {\bibinfo {volume} {818}},\ \bibinfo
  {eid} {124}}\BibitemShut {NoStop}%
\bibitem [{\citenamefont {{Ertl}}\ \emph {et~al.}(2020)\citenamefont {{Ertl}},
  \citenamefont {{Woosley}}, \citenamefont {{Sukhbold}},\ and\ \citenamefont
  {{Janka}}}]{Ertl.Woosley.ea:2020}%
  \BibitemOpen
  \bibfield  {author} {\bibinfo {author} {\bibnamefont {{Ertl}}, \bibfnamefont
  {T}}, \bibinfo {author} {\bibfnamefont {S.~E.}\ \bibnamefont {{Woosley}}},
  \bibinfo {author} {\bibfnamefont {Tuguldur}\ \bibnamefont {{Sukhbold}}}, and\
  \bibinfo {author} {\bibfnamefont {H.~T.}\ \bibnamefont {{Janka}}}} (\bibinfo
  {year} {2020}),\ \bibfield  {title} {\enquote {\bibinfo {title} {{The
  Explosion of Helium Stars Evolved with Mass Loss}},}\ }\href
  {https://doi.org/10.3847/1538-4357/ab6458} {\bibfield  {journal} {\bibinfo
  {journal} {Astrophys. J.}\ }\textbf {\bibinfo {volume} {890}},\ \bibinfo
  {eid} {51}}\BibitemShut {NoStop}%
\bibitem [{\citenamefont {{Escher}}\ \emph {et~al.}(2012)\citenamefont
  {{Escher}}, \citenamefont {{Burke}}, \citenamefont {{Dietrich}},
  \citenamefont {{Scielzo}}, \citenamefont {{Thompson}},\ and\ \citenamefont
  {{Younes}}}]{Escher2012}%
  \BibitemOpen
  \bibfield  {author} {\bibinfo {author} {\bibnamefont {{Escher}},
  \bibfnamefont {J~E}}, \bibinfo {author} {\bibfnamefont {J.~T.}\ \bibnamefont
  {{Burke}}}, \bibinfo {author} {\bibfnamefont {F.~S.}\ \bibnamefont
  {{Dietrich}}}, \bibinfo {author} {\bibfnamefont {N.~D.}\ \bibnamefont
  {{Scielzo}}}, \bibinfo {author} {\bibfnamefont {I.~J.}\ \bibnamefont
  {{Thompson}}}, and\ \bibinfo {author} {\bibfnamefont {W.}~\bibnamefont
  {{Younes}}}} (\bibinfo {year} {2012}),\ \bibfield  {title} {\enquote
  {\bibinfo {title} {{Compound-nuclear reaction cross sections from surrogate
  measurements}},}\ }\href {https://doi.org/10.1103/RevModPhys.84.353}
  {\bibfield  {journal} {\bibinfo  {journal} {Rev. Mod. Phys.}\ }\textbf
  {\bibinfo {volume} {84}},\ \bibinfo {pages} {353--397}}\BibitemShut {NoStop}%
\bibitem [{\citenamefont {{Escher}}\ and\ \citenamefont
  {{Dietrich}}(2006)}]{Escher2006}%
  \BibitemOpen
  \bibfield  {author} {\bibinfo {author} {\bibnamefont {{Escher}},
  \bibfnamefont {J~E}}, and\ \bibinfo {author} {\bibfnamefont {F.~S.}\
  \bibnamefont {{Dietrich}}}} (\bibinfo {year} {2006}),\ \bibfield  {title}
  {\enquote {\bibinfo {title} {{Determining (n,f) cross sections for actinide
  nuclei indirectly: Examination of the surrogate ratio method}},}\ }\href
  {https://doi.org/10.1103/PhysRevC.74.054601} {\bibfield  {journal} {\bibinfo
  {journal} {Phys. Rev. C}\ }\textbf {\bibinfo {volume} {74}},\ \bibinfo {eid}
  {054601}}\BibitemShut {NoStop}%
\bibitem [{\citenamefont {{Evans}}\ \emph {et~al.}(2017)\citenamefont {{Evans}}
  \emph {et~al.}}]{evans17}%
  \BibitemOpen
  \bibfield  {author} {\bibinfo {author} {\bibnamefont {{Evans}}, \bibfnamefont
  {P~A}},  \emph {et~al.}} (\bibinfo {year} {2017}),\ \bibfield  {title}
  {\enquote {\bibinfo {title} {{Swift and NuSTAR observations of GW170817:
  Detection of a blue kilonova}},}\ }\href
  {https://doi.org/10.1126/science.aap9580} {\bibfield  {journal} {\bibinfo
  {journal} {Science}\ }\textbf {\bibinfo {volume} {358}},\ \bibinfo {pages}
  {1565--1570}}\BibitemShut {NoStop}%
\bibitem [{\citenamefont {Ezzeddine}\ \emph {et~al.}(2020)\citenamefont
  {Ezzeddine}, \citenamefont {Rasmussen}, \citenamefont {Frebel}, \citenamefont
  {Chiti}, \citenamefont {Hinojisa}, \citenamefont {Placco}, \citenamefont
  {Ji}, \citenamefont {Beers}, \citenamefont {Hansen}, \citenamefont
  {Roederer}, \citenamefont {Sakari},\ and\ \citenamefont
  {Melendez}}]{Ezzeddine_2020}%
  \BibitemOpen
  \bibfield  {author} {\bibinfo {author} {\bibnamefont {Ezzeddine},
  \bibfnamefont {Rana}}, \bibinfo {author} {\bibfnamefont {Kaitlin}\
  \bibnamefont {Rasmussen}}, \bibinfo {author} {\bibfnamefont {Anna}\
  \bibnamefont {Frebel}}, \bibinfo {author} {\bibfnamefont {Anirudh}\
  \bibnamefont {Chiti}}, \bibinfo {author} {\bibfnamefont {Karina}\
  \bibnamefont {Hinojisa}}, \bibinfo {author} {\bibfnamefont {Vinicius~M.}\
  \bibnamefont {Placco}}, \bibinfo {author} {\bibfnamefont {Alexander~P.}\
  \bibnamefont {Ji}}, \bibinfo {author} {\bibfnamefont {Timothy~C.}\
  \bibnamefont {Beers}}, \bibinfo {author} {\bibfnamefont {Terese~T.}\
  \bibnamefont {Hansen}}, \bibinfo {author} {\bibfnamefont {Ian~U.}\
  \bibnamefont {Roederer}}, \bibinfo {author} {\bibfnamefont {Charli~M.}\
  \bibnamefont {Sakari}}, and\ \bibinfo {author} {\bibfnamefont {Jorge}\
  \bibnamefont {Melendez}}} (\bibinfo {year} {2020}),\ \bibfield  {title}
  {\enquote {\bibinfo {title} {The r-process alliance: First magellan/{MIKE}
  release from the southern search for r-process-enhanced stars},}\ }\href
  {https://doi.org/10.3847/1538-4357/ab9d1a} {\bibfield  {journal} {\bibinfo
  {journal} {Astrophys. J.}\ }\textbf {\bibinfo {volume} {898}},\ \bibinfo
  {pages} {150}}\BibitemShut {NoStop}%
\bibitem [{\citenamefont {{Faber}}\ \emph {et~al.}(2006)\citenamefont
  {{Faber}}, \citenamefont {{Baumgarte}}, \citenamefont {{Shapiro}},
  \citenamefont {{Taniguchi}},\ and\ \citenamefont
  {{Rasio}}}]{Faber.Baumgarte.ea:2006}%
  \BibitemOpen
  \bibfield  {author} {\bibinfo {author} {\bibnamefont {{Faber}}, \bibfnamefont
  {Joshua~A}}, \bibinfo {author} {\bibfnamefont {Thomas~W.}\ \bibnamefont
  {{Baumgarte}}}, \bibinfo {author} {\bibfnamefont {Stuart~L.}\ \bibnamefont
  {{Shapiro}}}, \bibinfo {author} {\bibfnamefont {Keisuke}\ \bibnamefont
  {{Taniguchi}}}, and\ \bibinfo {author} {\bibfnamefont {Frederic~A.}\
  \bibnamefont {{Rasio}}}} (\bibinfo {year} {2006}),\ \bibfield  {title}
  {\enquote {\bibinfo {title} {{Black Hole-Neutron Star Binary Merger
  Calculations: GRB Progenitors and the Stability of Mass Transfer}},}\ }in\
  \href {https://doi.org/10.1063/1.2399634} {\emph {\bibinfo {booktitle}
  {Albert Einstein Century International Conference}}},\ \bibinfo {series}
  {American Institute of Physics Conference Series}, Vol.\ \bibinfo {volume}
  {861},\ \bibinfo {editor} {edited by\ \bibinfo {editor} {\bibfnamefont
  {Jean-Michel}\ \bibnamefont {{Alimi}}}\ and\ \bibinfo {editor} {\bibfnamefont
  {Andr{\'e}}\ \bibnamefont {{F{\"u}zfa}}}},\ pp.\ \bibinfo {pages}
  {622--629}\BibitemShut {NoStop}%
\bibitem [{\citenamefont {{Farouqi}}\ \emph {et~al.}(2010)\citenamefont
  {{Farouqi}}, \citenamefont {{Kratz}}, \citenamefont {{Pfeiffer}},
  \citenamefont {{Rauscher}}, \citenamefont {{Thielemann}},\ and\ \citenamefont
  {{Truran}}}]{Farouqi.Kratz.ea:2010}%
  \BibitemOpen
  \bibfield  {author} {\bibinfo {author} {\bibnamefont {{Farouqi}},
  \bibfnamefont {K}}, \bibinfo {author} {\bibfnamefont {K.-L.}\ \bibnamefont
  {{Kratz}}}, \bibinfo {author} {\bibfnamefont {B.}~\bibnamefont {{Pfeiffer}}},
  \bibinfo {author} {\bibfnamefont {T.}~\bibnamefont {{Rauscher}}}, \bibinfo
  {author} {\bibfnamefont {{F.-K.}}\ \bibnamefont {{Thielemann}}}, and\
  \bibinfo {author} {\bibfnamefont {J.~W.}\ \bibnamefont {{Truran}}}} (\bibinfo
  {year} {2010}),\ \bibfield  {title} {\enquote {\bibinfo {title}
  {{Charged-particle and Neutron-capture Processes in the High-entropy Wind of
  Core-collapse Supernovae}},}\ }\href
  {https://doi.org/10.1088/0004-637X/712/2/1359} {\bibfield  {journal}
  {\bibinfo  {journal} {Astrophys. J.}\ }\textbf {\bibinfo {volume} {712}},\
  \bibinfo {pages} {1359--1377}}\BibitemShut {NoStop}%
\bibitem [{\citenamefont {Fattoyev}\ \emph {et~al.}(2018)\citenamefont
  {Fattoyev}, \citenamefont {Piekarewicz},\ and\ \citenamefont
  {Horowitz}}]{Fattoyev.Piekarewicz.Horowitz:2018}%
  \BibitemOpen
  \bibfield  {author} {\bibinfo {author} {\bibnamefont {Fattoyev},
  \bibfnamefont {F~J}}, \bibinfo {author} {\bibfnamefont {J.}~\bibnamefont
  {Piekarewicz}}, and\ \bibinfo {author} {\bibfnamefont {C.~J.}\ \bibnamefont
  {Horowitz}}} (\bibinfo {year} {2018}),\ \bibfield  {title} {\enquote
  {\bibinfo {title} {Neutron skins and neutron stars in the multimessenger
  era},}\ }\href {https://doi.org/10.1103/PhysRevLett.120.172702} {\bibfield
  {journal} {\bibinfo  {journal} {Phys. Rev. Lett.}\ }\textbf {\bibinfo
  {volume} {120}},\ \bibinfo {pages} {172702}}\BibitemShut {NoStop}%
\bibitem [{\citenamefont {{Fern{\'a}ndez}}\ \emph {et~al.}(2015)\citenamefont
  {{Fern{\'a}ndez}}, \citenamefont {{Kasen}}, \citenamefont {{Metzger}},\ and\
  \citenamefont {{Quataert}}}]{fernandez15}%
  \BibitemOpen
  \bibfield  {author} {\bibinfo {author} {\bibnamefont {{Fern{\'a}ndez}},
  \bibfnamefont {R}}, \bibinfo {author} {\bibfnamefont {D.}~\bibnamefont
  {{Kasen}}}, \bibinfo {author} {\bibfnamefont {B.~D.}\ \bibnamefont
  {{Metzger}}}, and\ \bibinfo {author} {\bibfnamefont {E.}~\bibnamefont
  {{Quataert}}}} (\bibinfo {year} {2015}),\ \bibfield  {title} {\enquote
  {\bibinfo {title} {{Outflows from accretion discs formed in neutron star
  mergers: effect of black hole spin}},}\ }\href
  {https://doi.org/10.1093/mnras/stu2112} {\bibfield  {journal} {\bibinfo
  {journal} {Mon. Not. Roy. Astron. Soc.}\ }\textbf {\bibinfo {volume} {446}},\
  \bibinfo {pages} {750--758}}\BibitemShut {NoStop}%
\bibitem [{\citenamefont {{Fern{\'a}ndez}}\ and\ \citenamefont
  {{Metzger}}(2016)}]{fernandez16}%
  \BibitemOpen
  \bibfield  {author} {\bibinfo {author} {\bibnamefont {{Fern{\'a}ndez}},
  \bibfnamefont {R}}, and\ \bibinfo {author} {\bibfnamefont {B.~D.}\
  \bibnamefont {{Metzger}}}} (\bibinfo {year} {2016}),\ \bibfield  {title}
  {\enquote {\bibinfo {title} {{Electromagnetic Signatures of Neutron Star
  Mergers in the Advanced LIGO Era}},}\ }\href
  {https://doi.org/10.1146/annurev-nucl-102115-044819} {\bibfield  {journal}
  {\bibinfo  {journal} {Annu. Rev. Nucl. Part. Sci.}\ }\textbf {\bibinfo
  {volume} {66}},\ \bibinfo {pages} {23--45}}\BibitemShut {NoStop}%
\bibitem [{\citenamefont {Fern{\'a}ndez}\ and\ \citenamefont
  {Metzger}(2013)}]{Fernandez.Metzger:2013}%
  \BibitemOpen
  \bibfield  {author} {\bibinfo {author} {\bibnamefont {Fern{\'a}ndez},
  \bibfnamefont {R}}, and\ \bibinfo {author} {\bibfnamefont {Brian~D.}\
  \bibnamefont {Metzger}}} (\bibinfo {year} {2013}),\ \bibfield  {title}
  {\enquote {\bibinfo {title} {{Delayed outflows from black hole accretion tori
  following neutron star binary coalescence}},}\ }\href
  {https://doi.org/10.1093/mnras/stt1312} {\bibfield  {journal} {\bibinfo
  {journal} {Mon. Not. Roy. Astron. Soc.}\ }\textbf {\bibinfo {volume} {435}},\
  \bibinfo {pages} {502--517}},\ \Eprint {https://arxiv.org/abs/1304.6720}
  {1304.6720} \BibitemShut {NoStop}%
\bibitem [{\citenamefont {{Fern{\'a}ndez}}\ \emph {et~al.}(2019)\citenamefont
  {{Fern{\'a}ndez}}, \citenamefont {{Tchekhovskoy}}, \citenamefont
  {{Quataert}}, \citenamefont {{Foucart}},\ and\ \citenamefont
  {{Kasen}}}]{Fernandez.Tchekhovskoy.ea:2019}%
  \BibitemOpen
  \bibfield  {author} {\bibinfo {author} {\bibnamefont {{Fern{\'a}ndez}},
  \bibfnamefont {Rodrigo}}, \bibinfo {author} {\bibfnamefont {Alexander}\
  \bibnamefont {{Tchekhovskoy}}}, \bibinfo {author} {\bibfnamefont {Eliot}\
  \bibnamefont {{Quataert}}}, \bibinfo {author} {\bibfnamefont {Francois}\
  \bibnamefont {{Foucart}}}, and\ \bibinfo {author} {\bibfnamefont {Daniel}\
  \bibnamefont {{Kasen}}}} (\bibinfo {year} {2019}),\ \bibfield  {title}
  {\enquote {\bibinfo {title} {{Long-term GRMHD simulations of neutron star
  merger accretion discs: implications for electromagnetic counterparts}},}\
  }\href {https://doi.org/10.1093/mnras/sty2932} {\bibfield  {journal}
  {\bibinfo  {journal} {Mon. Not. Roy. Astron. Soc.}\ }\textbf {\bibinfo
  {volume} {482}},\ \bibinfo {pages} {3373--3393}}\BibitemShut {NoStop}%
\bibitem [{\citenamefont {{Fields}}\ \emph {et~al.}(2019)\citenamefont
  {{Fields}} \emph {et~al.}}]{Fields.ea:2019}%
  \BibitemOpen
  \bibfield  {author} {\bibinfo {author} {\bibnamefont {{Fields}},
  \bibfnamefont {Brian}},  \emph {et~al.}} (\bibinfo {year} {2019}),\ \bibfield
   {title} {\enquote {\bibinfo {title} {{Near-Earth Supernova Explosions:
  Evidence, Implications, and Opportunities}},}\ }\href@noop {} {\bibfield
  {journal} {\bibinfo  {journal} {Bull. Am. Astron. Soc.}\ }\textbf {\bibinfo
  {volume} {51}},\ \bibinfo {eid} {410}},\ \Eprint
  {https://arxiv.org/abs/1903.04589} {arXiv:1903.04589 [astro-ph.SR]}
  \BibitemShut {NoStop}%
\bibitem [{\citenamefont {Fischer}\ \emph {et~al.}(2011)\citenamefont
  {Fischer}, \citenamefont {Sagert}, \citenamefont {Pagliara}, \citenamefont
  {Hempel}, \citenamefont {Schaffner-Bielich}, \citenamefont {Rauscher},
  \citenamefont {Thielemann}, \citenamefont {K{\"a}ppeli}, \citenamefont
  {Mart{\'i}nez-Pinedo},\ and\ \citenamefont
  {Liebend{\"o}rfer}}]{Fischer.Sagert.ea:2011}%
  \BibitemOpen
  \bibfield  {author} {\bibinfo {author} {\bibnamefont {Fischer}, \bibfnamefont
  {T}}, \bibinfo {author} {\bibfnamefont {I.}~\bibnamefont {Sagert}}, \bibinfo
  {author} {\bibfnamefont {G.}~\bibnamefont {Pagliara}}, \bibinfo {author}
  {\bibfnamefont {M.}~\bibnamefont {Hempel}}, \bibinfo {author} {\bibfnamefont
  {J.}~\bibnamefont {Schaffner-Bielich}}, \bibinfo {author} {\bibfnamefont
  {T.}~\bibnamefont {Rauscher}}, \bibinfo {author} {\bibfnamefont {F.-K.}\
  \bibnamefont {Thielemann}}, \bibinfo {author} {\bibfnamefont
  {R.}~\bibnamefont {K{\"a}ppeli}}, \bibinfo {author} {\bibfnamefont
  {G.}~\bibnamefont {Mart{\'i}nez-Pinedo}}, and\ \bibinfo {author}
  {\bibfnamefont {M.}~\bibnamefont {Liebend{\"o}rfer}}} (\bibinfo {year}
  {2011}),\ \bibfield  {title} {\enquote {\bibinfo {title} {Core-collapse
  supernova explosions triggered by a quark-hadron phase transition during the
  early post-bounce phase},}\ }\href
  {https://doi.org/10.1088/0067-0049/194/2/39} {\bibfield  {journal} {\bibinfo
  {journal} {Astrophys. J. Suppl.}\ }\textbf {\bibinfo {volume} {194}},\
  \bibinfo {pages} {39}}\BibitemShut {NoStop}%
\bibitem [{\citenamefont {{Fischer}}\ \emph {et~al.}(2010)\citenamefont
  {{Fischer}}, \citenamefont {{Whitehouse}}, \citenamefont {{Mezzacappa}},
  \citenamefont {{Thielemann}},\ and\ \citenamefont
  {{Liebend{\"o}rfer}}}]{fischer10}%
  \BibitemOpen
  \bibfield  {author} {\bibinfo {author} {\bibnamefont {{Fischer}},
  \bibfnamefont {T}}, \bibinfo {author} {\bibfnamefont {S.~C.}\ \bibnamefont
  {{Whitehouse}}}, \bibinfo {author} {\bibfnamefont {A.}~\bibnamefont
  {{Mezzacappa}}}, \bibinfo {author} {\bibfnamefont {F.-K.}\ \bibnamefont
  {{Thielemann}}}, and\ \bibinfo {author} {\bibfnamefont {M.}~\bibnamefont
  {{Liebend{\"o}rfer}}}} (\bibinfo {year} {2010}),\ \bibfield  {title}
  {\enquote {\bibinfo {title} {{Protoneutron star evolution and the
  neutrino-driven wind in general relativistic neutrino radiation hydrodynamics
  simulations}},}\ }\href {https://doi.org/10.1051/0004-6361/200913106}
  {\bibfield  {journal} {\bibinfo  {journal} {Astron. \& Astrophys.}\ }\textbf
  {\bibinfo {volume} {517}},\ \bibinfo {eid} {A80}}\BibitemShut {NoStop}%
\bibitem [{\citenamefont {{Fischer}}\ \emph {et~al.}(2018)\citenamefont
  {{Fischer}}, \citenamefont {{Bastian}}, \citenamefont {{Wu}}, \citenamefont
  {{Baklanov}}, \citenamefont {{Sorokina}}, \citenamefont {{Blinnikov}},
  \citenamefont {{Typel}}, \citenamefont {{Kl{\"a}hn}},\ and\ \citenamefont
  {{Blaschke}}}]{Fischer.Bastian.ea:2018}%
  \BibitemOpen
  \bibfield  {author} {\bibinfo {author} {\bibnamefont {{Fischer}},
  \bibfnamefont {Tobias}}, \bibinfo {author} {\bibfnamefont {Niels-Uwe~F.}\
  \bibnamefont {{Bastian}}}, \bibinfo {author} {\bibfnamefont {Meng-Ru}\
  \bibnamefont {{Wu}}}, \bibinfo {author} {\bibfnamefont {Petr}\ \bibnamefont
  {{Baklanov}}}, \bibinfo {author} {\bibfnamefont {Elena}\ \bibnamefont
  {{Sorokina}}}, \bibinfo {author} {\bibfnamefont {Sergei}\ \bibnamefont
  {{Blinnikov}}}, \bibinfo {author} {\bibfnamefont {Stefan}\ \bibnamefont
  {{Typel}}}, \bibinfo {author} {\bibfnamefont {Thomas}\ \bibnamefont
  {{Kl{\"a}hn}}}, and\ \bibinfo {author} {\bibfnamefont {David~B.}\
  \bibnamefont {{Blaschke}}}} (\bibinfo {year} {2018}),\ \bibfield  {title}
  {\enquote {\bibinfo {title} {{Quark deconfinement as a supernova explosion
  engine for massive blue supergiant stars}},}\ }\href
  {https://doi.org/10.1038/s41550-018-0583-0} {\bibfield  {journal} {\bibinfo
  {journal} {Nature Astronomy}\ }\textbf {\bibinfo {volume} {2}},\ \bibinfo
  {pages} {980--986}}\BibitemShut {NoStop}%
\bibitem [{\citenamefont {{Fischer}}\ \emph
  {et~al.}(2020{\natexlab{a}})\citenamefont {{Fischer}}, \citenamefont {{Guo}},
  \citenamefont {{Dzhioev}}, \citenamefont {{Mart{\'\i}nez-Pinedo}},
  \citenamefont {{Wu}}, \citenamefont {{Lohs}},\ and\ \citenamefont
  {{Qian}}}]{Fischer.Guo.ea:2020}%
  \BibitemOpen
  \bibfield  {author} {\bibinfo {author} {\bibnamefont {{Fischer}},
  \bibfnamefont {Tobias}}, \bibinfo {author} {\bibfnamefont {Gang}\
  \bibnamefont {{Guo}}}, \bibinfo {author} {\bibfnamefont {Alan~A.}\
  \bibnamefont {{Dzhioev}}}, \bibinfo {author} {\bibfnamefont {Gabriel}\
  \bibnamefont {{Mart{\'\i}nez-Pinedo}}}, \bibinfo {author} {\bibfnamefont
  {Meng-Ru}\ \bibnamefont {{Wu}}}, \bibinfo {author} {\bibfnamefont {Andreas}\
  \bibnamefont {{Lohs}}}, and\ \bibinfo {author} {\bibfnamefont {Yong-Zhong}\
  \bibnamefont {{Qian}}}} (\bibinfo {year} {2020}{\natexlab{a}}),\ \bibfield
  {title} {\enquote {\bibinfo {title} {{Neutrino signal from proto-neutron star
  evolution: Effects of opacities from charged-current-neutrino interactions
  and inverse neutron decay}},}\ }\href
  {https://doi.org/10.1103/PhysRevC.101.025804} {\bibfield  {journal} {\bibinfo
   {journal} {Phys. Rev. C}\ }\textbf {\bibinfo {volume} {101}},\ \bibinfo
  {eid} {025804}}\BibitemShut {NoStop}%
\bibitem [{\citenamefont {{Fischer}}\ \emph
  {et~al.}(2020{\natexlab{b}})\citenamefont {{Fischer}}, \citenamefont {{Wu}},
  \citenamefont {{Wehmeyer}}, \citenamefont {{Bastian}}, \citenamefont
  {{Mart{\'\i}nez-Pinedo}},\ and\ \citenamefont
  {{Thielemann}}}]{Fischer.Wu.ea:2020}%
  \BibitemOpen
  \bibfield  {author} {\bibinfo {author} {\bibnamefont {{Fischer}},
  \bibfnamefont {Tobias}}, \bibinfo {author} {\bibfnamefont {Meng-Ru}\
  \bibnamefont {{Wu}}}, \bibinfo {author} {\bibfnamefont {Benjamin}\
  \bibnamefont {{Wehmeyer}}}, \bibinfo {author} {\bibfnamefont {Niels-Uwe~F.}\
  \bibnamefont {{Bastian}}}, \bibinfo {author} {\bibfnamefont {Gabriel}\
  \bibnamefont {{Mart{\'\i}nez-Pinedo}}}, and\ \bibinfo {author} {\bibfnamefont
  {Friedrich-Karl}\ \bibnamefont {{Thielemann}}}} (\bibinfo {year}
  {2020}{\natexlab{b}}),\ \bibfield  {title} {\enquote {\bibinfo {title}
  {{Core-collapse supernova explosions driven by the hadron-quark phase
  transition as rare $r$ process site}},}\ }\href
  {https://doi.org/10.3847/1538-4357/ab86b0} {\bibfield  {journal} {\bibinfo
  {journal} {Astrophys. J.}\ }\textbf {\bibinfo {volume} {894}},\ \bibinfo
  {eid} {9}}\BibitemShut {NoStop}%
\bibitem [{\citenamefont {Fogelberg}\ \emph {et~al.}(2004)\citenamefont
  {Fogelberg}, \citenamefont {Gausemel}, \citenamefont {Mezilev}, \citenamefont
  {Hoff}, \citenamefont {Mach}, \citenamefont {Sanchez-Vega}, \citenamefont
  {Lindroth}, \citenamefont {Ramstr\"om}, \citenamefont {Genevey},
  \citenamefont {Pinston},\ and\ \citenamefont
  {Rejmund}}]{Fogelberg.Gausemel.ea:2004}%
  \BibitemOpen
  \bibfield  {author} {\bibinfo {author} {\bibnamefont {Fogelberg},
  \bibfnamefont {B}}, \bibinfo {author} {\bibfnamefont {H.}~\bibnamefont
  {Gausemel}}, \bibinfo {author} {\bibfnamefont {K.~A.}\ \bibnamefont
  {Mezilev}}, \bibinfo {author} {\bibfnamefont {P.}~\bibnamefont {Hoff}},
  \bibinfo {author} {\bibfnamefont {H.}~\bibnamefont {Mach}}, \bibinfo {author}
  {\bibfnamefont {M.}~\bibnamefont {Sanchez-Vega}}, \bibinfo {author}
  {\bibfnamefont {A.}~\bibnamefont {Lindroth}}, \bibinfo {author}
  {\bibfnamefont {E.}~\bibnamefont {Ramstr\"om}}, \bibinfo {author}
  {\bibfnamefont {J.}~\bibnamefont {Genevey}}, \bibinfo {author} {\bibfnamefont
  {J.~A.}\ \bibnamefont {Pinston}}, and\ \bibinfo {author} {\bibfnamefont
  {M.}~\bibnamefont {Rejmund}}} (\bibinfo {year} {2004}),\ \bibfield  {title}
  {\enquote {\bibinfo {title} {Decays of $^{131}\mathrm{In}$,
  $^{131}\mathrm{Sn}$, and the position of the ${h}_{11/2}$ neutron hole
  state},}\ }\href {https://doi.org/10.1103/PhysRevC.70.034312} {\bibfield
  {journal} {\bibinfo  {journal} {Phys. Rev. C}\ }\textbf {\bibinfo {volume}
  {70}},\ \bibinfo {pages} {034312}}\BibitemShut {NoStop}%
\bibitem [{\citenamefont {{Foglizzo}}\ \emph {et~al.}(2015)\citenamefont
  {{Foglizzo}}, \citenamefont {{Kazeroni}}, \citenamefont {{Guilet}},
  \citenamefont {{Masset}}, \citenamefont {{Gonz{\'a}lez}}, \citenamefont
  {{Krueger}}, \citenamefont {{Novak}}, \citenamefont {{Oertel}}, \citenamefont
  {{Margueron}}, \citenamefont {{Faure}}, \citenamefont {{Martin}},
  \citenamefont {{Blottiau}}, \citenamefont {{Peres}},\ and\ \citenamefont
  {{Durand}}}]{Foglizzo.Kazeroni.ea:2015}%
  \BibitemOpen
  \bibfield  {author} {\bibinfo {author} {\bibnamefont {{Foglizzo}},
  \bibfnamefont {T}}, \bibinfo {author} {\bibfnamefont {R.}~\bibnamefont
  {{Kazeroni}}}, \bibinfo {author} {\bibfnamefont {J.}~\bibnamefont
  {{Guilet}}}, \bibinfo {author} {\bibfnamefont {F.}~\bibnamefont {{Masset}}},
  \bibinfo {author} {\bibfnamefont {M.}~\bibnamefont {{Gonz{\'a}lez}}},
  \bibinfo {author} {\bibfnamefont {B.~K.}\ \bibnamefont {{Krueger}}}, \bibinfo
  {author} {\bibfnamefont {J.}~\bibnamefont {{Novak}}}, \bibinfo {author}
  {\bibfnamefont {M.}~\bibnamefont {{Oertel}}}, \bibinfo {author}
  {\bibfnamefont {J.}~\bibnamefont {{Margueron}}}, \bibinfo {author}
  {\bibfnamefont {J.}~\bibnamefont {{Faure}}}, \bibinfo {author} {\bibfnamefont
  {N.}~\bibnamefont {{Martin}}}, \bibinfo {author} {\bibfnamefont
  {P.}~\bibnamefont {{Blottiau}}}, \bibinfo {author} {\bibfnamefont
  {B.}~\bibnamefont {{Peres}}}, and\ \bibinfo {author} {\bibfnamefont
  {G.}~\bibnamefont {{Durand}}}} (\bibinfo {year} {2015}),\ \bibfield  {title}
  {\enquote {\bibinfo {title} {The explosion mechanism of core-collapse
  supernovae: Progress in supernova theory and experiments},}\ }\href
  {https://doi.org/10.1017/pasa.2015.9} {\bibfield  {journal} {\bibinfo
  {journal} {Publ. Astron. Soc. Austr.}\ }\textbf {\bibinfo {volume} {32}},\
  \bibinfo {pages} {e009}}\BibitemShut {NoStop}%
\bibitem [{\citenamefont {{Foley}}\ \emph {et~al.}(2020)\citenamefont
  {{Foley}}, \citenamefont {{Coulter}}, \citenamefont {{Kilpatrick}},
  \citenamefont {{Piro}}, \citenamefont {{Ramirez-Ruiz}},\ and\ \citenamefont
  {{Schwab}}}]{Foley.Coulter.ea:2020}%
  \BibitemOpen
  \bibfield  {author} {\bibinfo {author} {\bibnamefont {{Foley}}, \bibfnamefont
  {Ryan~J}}, \bibinfo {author} {\bibfnamefont {David~A.}\ \bibnamefont
  {{Coulter}}}, \bibinfo {author} {\bibfnamefont {Charles~D.}\ \bibnamefont
  {{Kilpatrick}}}, \bibinfo {author} {\bibfnamefont {Anthony~L.}\ \bibnamefont
  {{Piro}}}, \bibinfo {author} {\bibfnamefont {Enrico}\ \bibnamefont
  {{Ramirez-Ruiz}}}, and\ \bibinfo {author} {\bibfnamefont {Josiah}\
  \bibnamefont {{Schwab}}}} (\bibinfo {year} {2020}),\ \bibfield  {title}
  {\enquote {\bibinfo {title} {{Updated Parameter Estimates for GW190425 Using
  Astrophysical Arguments and Implications for the Electromagnetic
  Counterpart}},}\ }\href {https://doi.org/10.1093/mnras/staa725} {\bibfield
  {journal} {\bibinfo  {journal} {Mon. Not. Roy. Astron. Soc.}\ }\textbf
  {\bibinfo {volume} {494}},\ \bibinfo {pages} {190--198}}\BibitemShut
  {NoStop}%
\bibitem [{\citenamefont {{Fontes}}\ \emph {et~al.}(2020)\citenamefont
  {{Fontes}}, \citenamefont {{Fryer}}, \citenamefont {{Hungerford}},
  \citenamefont {{Wollaeger}},\ and\ \citenamefont
  {{Korobkin}}}]{Fontes.Fryer.ea:2020}%
  \BibitemOpen
  \bibfield  {author} {\bibinfo {author} {\bibnamefont {{Fontes}},
  \bibfnamefont {C~J}}, \bibinfo {author} {\bibfnamefont {C.~L.}\ \bibnamefont
  {{Fryer}}}, \bibinfo {author} {\bibfnamefont {A.~L.}\ \bibnamefont
  {{Hungerford}}}, \bibinfo {author} {\bibfnamefont {R.~T.}\ \bibnamefont
  {{Wollaeger}}}, and\ \bibinfo {author} {\bibfnamefont {O.}~\bibnamefont
  {{Korobkin}}}} (\bibinfo {year} {2020}),\ \bibfield  {title} {\enquote
  {\bibinfo {title} {{A line-binned treatment of opacities for the spectra and
  light curves from neutron star mergers}},}\ }\href
  {https://doi.org/10.1093/mnras/staa485} {\bibfield  {journal} {\bibinfo
  {journal} {Mon. Not. Roy. Astron. Soc.}\ }\textbf {\bibinfo {volume} {493}},\
  \bibinfo {pages} {4143--4171}}\BibitemShut {NoStop}%
\bibitem [{\citenamefont {{Fontes}}\ \emph {et~al.}(2017)\citenamefont
  {{Fontes}}, \citenamefont {{Fryer}}, \citenamefont {{Hungerford}},
  \citenamefont {{Wollaeger}}, \citenamefont {{Rosswog}},\ and\ \citenamefont
  {{Berger}}}]{Fontes.Fryer.ea:2017}%
  \BibitemOpen
  \bibfield  {author} {\bibinfo {author} {\bibnamefont {{Fontes}},
  \bibfnamefont {Christopher~J}}, \bibinfo {author} {\bibfnamefont {Chris~L.}\
  \bibnamefont {{Fryer}}}, \bibinfo {author} {\bibfnamefont {Aimee~L.}\
  \bibnamefont {{Hungerford}}}, \bibinfo {author} {\bibfnamefont {Ryan~T.}\
  \bibnamefont {{Wollaeger}}}, \bibinfo {author} {\bibfnamefont {Stephan}\
  \bibnamefont {{Rosswog}}}, and\ \bibinfo {author} {\bibfnamefont {Edo}\
  \bibnamefont {{Berger}}}} (\bibinfo {year} {2017}),\ \bibfield  {title}
  {\enquote {\bibinfo {title} {{A line-smeared treatment of opacities for the
  spectra and light curves from macronovae}},}\ }\href@noop {} {\bibfield
  {journal} {\bibinfo  {journal} {arXiv e-prints}\ }}\Eprint
  {https://arxiv.org/abs/1702.02990} {arXiv:1702.02990 [astro-ph.HE]}
  \BibitemShut {NoStop}%
\bibitem [{\citenamefont {Fontes}\ \emph {et~al.}(2015)\citenamefont {Fontes},
  \citenamefont {Fryer}, \citenamefont {Hungerford}, \citenamefont {Hakel},
  \citenamefont {Colgan}, \citenamefont {Kilcrease},\ and\ \citenamefont
  {Sherrill}}]{Fontes.Fryer.ea:2015}%
  \BibitemOpen
  \bibfield  {author} {\bibinfo {author} {\bibnamefont {Fontes}, \bibfnamefont
  {CJ}}, \bibinfo {author} {\bibfnamefont {C.L.}\ \bibnamefont {Fryer}},
  \bibinfo {author} {\bibfnamefont {A.L.}\ \bibnamefont {Hungerford}}, \bibinfo
  {author} {\bibfnamefont {P.}~\bibnamefont {Hakel}}, \bibinfo {author}
  {\bibfnamefont {J.}~\bibnamefont {Colgan}}, \bibinfo {author} {\bibfnamefont
  {D.P.}\ \bibnamefont {Kilcrease}}, and\ \bibinfo {author} {\bibfnamefont
  {M.E.}\ \bibnamefont {Sherrill}}} (\bibinfo {year} {2015}),\ \bibfield
  {title} {\enquote {\bibinfo {title} {{Relativistic opacities for
  astrophysical applications}},}\ }\href
  {https://doi.org/10.1016/j.hedp.2015.06.002} {\bibfield  {journal} {\bibinfo
  {journal} {High Energy Density Phys.}\ }\textbf {\bibinfo {volume} {16}},\
  \bibinfo {pages} {53--59}}\BibitemShut {NoStop}%
\bibitem [{\citenamefont {{Forss{\'e}n}}\ \emph {et~al.}(2007)\citenamefont
  {{Forss{\'e}n}}, \citenamefont {{Dietrich}}, \citenamefont {{Escher}},
  \citenamefont {{Hoffman}},\ and\ \citenamefont {{Kelley}}}]{Forssen2007}%
  \BibitemOpen
  \bibfield  {author} {\bibinfo {author} {\bibnamefont {{Forss{\'e}n}},
  \bibfnamefont {C}}, \bibinfo {author} {\bibfnamefont {F.~S.}\ \bibnamefont
  {{Dietrich}}}, \bibinfo {author} {\bibfnamefont {J.}~\bibnamefont
  {{Escher}}}, \bibinfo {author} {\bibfnamefont {R.~D.}\ \bibnamefont
  {{Hoffman}}}, and\ \bibinfo {author} {\bibfnamefont {K.}~\bibnamefont
  {{Kelley}}}} (\bibinfo {year} {2007}),\ \bibfield  {title} {\enquote
  {\bibinfo {title} {{Determining neutron capture cross sections via the
  surrogate reaction technique}},}\ }\href
  {https://doi.org/10.1103/PhysRevC.75.055807} {\bibfield  {journal} {\bibinfo
  {journal} {Phys. Rev. C}\ }\textbf {\bibinfo {volume} {75}},\ \bibinfo {eid}
  {055807}}\BibitemShut {NoStop}%
\bibitem [{\citenamefont {{Foucart}}\ \emph {et~al.}(2013)\citenamefont
  {{Foucart}}, \citenamefont {{Deaton}}, \citenamefont {{Duez}}, \citenamefont
  {{Kidder}}, \citenamefont {{MacDonald}}, \citenamefont {{Ott}}, \citenamefont
  {{Pfeiffer}}, \citenamefont {{Scheel}}, \citenamefont {{Szilagyi}},\ and\
  \citenamefont {{Teukolsky}}}]{Foucart.Deaton.ea:2013}%
  \BibitemOpen
  \bibfield  {author} {\bibinfo {author} {\bibnamefont {{Foucart}},
  \bibfnamefont {F}}, \bibinfo {author} {\bibfnamefont {M.~B.}\ \bibnamefont
  {{Deaton}}}, \bibinfo {author} {\bibfnamefont {M.~D.}\ \bibnamefont
  {{Duez}}}, \bibinfo {author} {\bibfnamefont {L.~E.}\ \bibnamefont
  {{Kidder}}}, \bibinfo {author} {\bibfnamefont {I.}~\bibnamefont
  {{MacDonald}}}, \bibinfo {author} {\bibfnamefont {C.~D.}\ \bibnamefont
  {{Ott}}}, \bibinfo {author} {\bibfnamefont {H.~P.}\ \bibnamefont
  {{Pfeiffer}}}, \bibinfo {author} {\bibfnamefont {M.~A.}\ \bibnamefont
  {{Scheel}}}, \bibinfo {author} {\bibfnamefont {B.}~\bibnamefont
  {{Szilagyi}}}, and\ \bibinfo {author} {\bibfnamefont {S.~A.}\ \bibnamefont
  {{Teukolsky}}}} (\bibinfo {year} {2013}),\ \bibfield  {title} {\enquote
  {\bibinfo {title} {{Black-hole-neutron-star mergers at realistic mass ratios:
  Equation of state and spin orientation effects}},}\ }\href
  {https://doi.org/10.1103/PhysRevD.87.084006} {\bibfield  {journal} {\bibinfo
  {journal} {Phys. Rev. D}\ }\textbf {\bibinfo {volume} {87}},\ \bibinfo {eid}
  {084006}}\BibitemShut {NoStop}%
\bibitem [{\citenamefont {{Foucart}}\ \emph {et~al.}(2014)\citenamefont
  {{Foucart}}, \citenamefont {{Deaton}}, \citenamefont {{Duez}}, \citenamefont
  {{O'Connor}}, \citenamefont {{Ott}}, \citenamefont {{Haas}}, \citenamefont
  {{Kidder}}, \citenamefont {{Pfeiffer}}, \citenamefont {{Scheel}},\ and\
  \citenamefont {{Szilagyi}}}]{Foucart.Deaton.ea:2014}%
  \BibitemOpen
  \bibfield  {author} {\bibinfo {author} {\bibnamefont {{Foucart}},
  \bibfnamefont {F}}, \bibinfo {author} {\bibfnamefont {M.~B.}\ \bibnamefont
  {{Deaton}}}, \bibinfo {author} {\bibfnamefont {M.~D.}\ \bibnamefont
  {{Duez}}}, \bibinfo {author} {\bibfnamefont {E.}~\bibnamefont {{O'Connor}}},
  \bibinfo {author} {\bibfnamefont {C.~D.}\ \bibnamefont {{Ott}}}, \bibinfo
  {author} {\bibfnamefont {R.}~\bibnamefont {{Haas}}}, \bibinfo {author}
  {\bibfnamefont {L.~E.}\ \bibnamefont {{Kidder}}}, \bibinfo {author}
  {\bibfnamefont {H.~P.}\ \bibnamefont {{Pfeiffer}}}, \bibinfo {author}
  {\bibfnamefont {M.~A.}\ \bibnamefont {{Scheel}}}, and\ \bibinfo {author}
  {\bibfnamefont {B.}~\bibnamefont {{Szilagyi}}}} (\bibinfo {year} {2014}),\
  \bibfield  {title} {\enquote {\bibinfo {title} {{Neutron star-black hole
  mergers with a nuclear equation of state and neutrino cooling: Dependence in
  the binary parameters}},}\ }\href
  {https://doi.org/10.1103/PhysRevD.90.024026} {\bibfield  {journal} {\bibinfo
  {journal} {Phys. Rev. D}\ }\textbf {\bibinfo {volume} {90}},\ \bibinfo {eid}
  {024026}}\BibitemShut {NoStop}%
\bibitem [{\citenamefont {{Foucart}}\ \emph {et~al.}(2018)\citenamefont
  {{Foucart}}, \citenamefont {{Duez}}, \citenamefont {{Kidder}}, \citenamefont
  {{Nguyen}}, \citenamefont {{Pfeiffer}},\ and\ \citenamefont
  {{Scheel}}}]{Foucart.Duez.ea:2018}%
  \BibitemOpen
  \bibfield  {author} {\bibinfo {author} {\bibnamefont {{Foucart}},
  \bibfnamefont {F}}, \bibinfo {author} {\bibfnamefont {M.~D.}\ \bibnamefont
  {{Duez}}}, \bibinfo {author} {\bibfnamefont {L.~E.}\ \bibnamefont
  {{Kidder}}}, \bibinfo {author} {\bibfnamefont {R.}~\bibnamefont {{Nguyen}}},
  \bibinfo {author} {\bibfnamefont {H.~P.}\ \bibnamefont {{Pfeiffer}}}, and\
  \bibinfo {author} {\bibfnamefont {M.~A.}\ \bibnamefont {{Scheel}}}} (\bibinfo
  {year} {2018}),\ \bibfield  {title} {\enquote {\bibinfo {title} {{Evaluating
  radiation transport errors in merger simulations using a Monte Carlo
  algorithm}},}\ }\href {https://doi.org/10.1103/PhysRevD.98.063007} {\bibfield
   {journal} {\bibinfo  {journal} {Phys. Rev. D}\ }\textbf {\bibinfo {volume}
  {98}},\ \bibinfo {eid} {063007}}\BibitemShut {NoStop}%
\bibitem [{\citenamefont {{Foucart}}\ \emph {et~al.}(2015)\citenamefont
  {{Foucart}}, \citenamefont {{O'Connor}}, \citenamefont {{Roberts}},
  \citenamefont {{Duez}}, \citenamefont {{Haas}}, \citenamefont {{Kidder}},
  \citenamefont {{Ott}}, \citenamefont {{Pfeiffer}}, \citenamefont {{Scheel}},\
  and\ \citenamefont {{Szilagyi}}}]{foucart15}%
  \BibitemOpen
  \bibfield  {author} {\bibinfo {author} {\bibnamefont {{Foucart}},
  \bibfnamefont {F}}, \bibinfo {author} {\bibfnamefont {E.}~\bibnamefont
  {{O'Connor}}}, \bibinfo {author} {\bibfnamefont {L.}~\bibnamefont
  {{Roberts}}}, \bibinfo {author} {\bibfnamefont {M.~D.}\ \bibnamefont
  {{Duez}}}, \bibinfo {author} {\bibfnamefont {R.}~\bibnamefont {{Haas}}},
  \bibinfo {author} {\bibfnamefont {L.~E.}\ \bibnamefont {{Kidder}}}, \bibinfo
  {author} {\bibfnamefont {C.~D.}\ \bibnamefont {{Ott}}}, \bibinfo {author}
  {\bibfnamefont {H.~P.}\ \bibnamefont {{Pfeiffer}}}, \bibinfo {author}
  {\bibfnamefont {M.~A.}\ \bibnamefont {{Scheel}}}, and\ \bibinfo {author}
  {\bibfnamefont {B.}~\bibnamefont {{Szilagyi}}}} (\bibinfo {year} {2015}),\
  \bibfield  {title} {\enquote {\bibinfo {title} {{Post-merger evolution of a
  neutron star-black hole binary with neutrino transport}},}\ }\href
  {https://doi.org/10.1103/PhysRevD.91.124021} {\bibfield  {journal} {\bibinfo
  {journal} {Phys. Rev. D}\ }\textbf {\bibinfo {volume} {91}},\ \bibinfo {eid}
  {124021}}\BibitemShut {NoStop}%
\bibitem [{\citenamefont {{Foucart}}(2012)}]{Foucart:2012}%
  \BibitemOpen
  \bibfield  {author} {\bibinfo {author} {\bibnamefont {{Foucart}},
  \bibfnamefont {Francois}}} (\bibinfo {year} {2012}),\ \bibfield  {title}
  {\enquote {\bibinfo {title} {{Black-hole-neutron-star mergers: Disk mass
  predictions}},}\ }\href {https://doi.org/10.1103/PhysRevD.86.124007}
  {\bibfield  {journal} {\bibinfo  {journal} {Phys. Rev. D}\ }\textbf {\bibinfo
  {volume} {86}},\ \bibinfo {eid} {124007}}\BibitemShut {NoStop}%
\bibitem [{\citenamefont {Foucart}\ \emph {et~al.}(2016)\citenamefont
  {Foucart}, \citenamefont {O'Connor}, \citenamefont {Roberts}, \citenamefont
  {Kidder}, \citenamefont {Pfeiffer},\ and\ \citenamefont
  {Scheel}}]{Foucart.Oconnor.ea:2016}%
  \BibitemOpen
  \bibfield  {author} {\bibinfo {author} {\bibnamefont {Foucart}, \bibfnamefont
  {Francois}}, \bibinfo {author} {\bibfnamefont {Evan}\ \bibnamefont
  {O'Connor}}, \bibinfo {author} {\bibfnamefont {Luke}\ \bibnamefont
  {Roberts}}, \bibinfo {author} {\bibfnamefont {Lawrence~E.}\ \bibnamefont
  {Kidder}}, \bibinfo {author} {\bibfnamefont {Harald~P.}\ \bibnamefont
  {Pfeiffer}}, and\ \bibinfo {author} {\bibfnamefont {Mark~A.}\ \bibnamefont
  {Scheel}}} (\bibinfo {year} {2016}),\ \bibfield  {title} {\enquote {\bibinfo
  {title} {Impact of an improved neutrino energy estimate on outflows in
  neutron star merger simulations},}\ }\href
  {https://doi.org/10.1103/PhysRevD.94.123016} {\bibfield  {journal} {\bibinfo
  {journal} {Phys. Rev. D}\ }\textbf {\bibinfo {volume} {94}},\ \bibinfo
  {pages} {123016}}\BibitemShut {NoStop}%
\bibitem [{\citenamefont {{Fowler}}\ and\ \citenamefont
  {{Hoyle}}(1960)}]{Fowler.Hoyle:1960}%
  \BibitemOpen
  \bibfield  {author} {\bibinfo {author} {\bibnamefont {{Fowler}},
  \bibfnamefont {W~A}}, and\ \bibinfo {author} {\bibfnamefont {F.}~\bibnamefont
  {{Hoyle}}}} (\bibinfo {year} {1960}),\ \bibfield  {title} {\enquote {\bibinfo
  {title} {{Nuclear cosmochronology}},}\ }\href
  {https://doi.org/10.1016/0003-4916(60)90025-7} {\bibfield  {journal}
  {\bibinfo  {journal} {Annals of Physics}\ }\textbf {\bibinfo {volume} {10}},\
  \bibinfo {pages} {280--302}}\BibitemShut {NoStop}%
\bibitem [{\citenamefont {{Fran{\c c}ois}}\ \emph {et~al.}(2007)\citenamefont
  {{Fran{\c c}ois}}, \citenamefont {{Depagne}}, \citenamefont {{Hill}},
  \citenamefont {{Spite}}, \citenamefont {{Spite}}, \citenamefont {{Plez}},
  \citenamefont {{Beers}}, \citenamefont {{Andersen}}, \citenamefont {{James}},
  \citenamefont {{Barbuy}}, \citenamefont {{Cayrel}}, \citenamefont
  {{Bonifacio}}, \citenamefont {{Molaro}}, \citenamefont {{Nordstr{\"o}m}},\
  and\ \citenamefont {{Primas}}}]{Francois.Depagne.ea:2007}%
  \BibitemOpen
  \bibfield  {author} {\bibinfo {author} {\bibnamefont {{Fran{\c c}ois}},
  \bibfnamefont {P}}, \bibinfo {author} {\bibfnamefont {E.}~\bibnamefont
  {{Depagne}}}, \bibinfo {author} {\bibfnamefont {V.}~\bibnamefont {{Hill}}},
  \bibinfo {author} {\bibfnamefont {M.}~\bibnamefont {{Spite}}}, \bibinfo
  {author} {\bibfnamefont {F.}~\bibnamefont {{Spite}}}, \bibinfo {author}
  {\bibfnamefont {B.}~\bibnamefont {{Plez}}}, \bibinfo {author} {\bibfnamefont
  {T.~C.}\ \bibnamefont {{Beers}}}, \bibinfo {author} {\bibfnamefont
  {J.}~\bibnamefont {{Andersen}}}, \bibinfo {author} {\bibfnamefont
  {G.}~\bibnamefont {{James}}}, \bibinfo {author} {\bibfnamefont
  {B.}~\bibnamefont {{Barbuy}}}, \bibinfo {author} {\bibfnamefont
  {R.}~\bibnamefont {{Cayrel}}}, \bibinfo {author} {\bibfnamefont
  {P.}~\bibnamefont {{Bonifacio}}}, \bibinfo {author} {\bibfnamefont
  {P.}~\bibnamefont {{Molaro}}}, \bibinfo {author} {\bibfnamefont
  {B.}~\bibnamefont {{Nordstr{\"o}m}}}, and\ \bibinfo {author} {\bibfnamefont
  {F.}~\bibnamefont {{Primas}}}} (\bibinfo {year} {2007}),\ \bibfield  {title}
  {\enquote {\bibinfo {title} {{First stars. VIII. Enrichment of the
  neutron-capture elements in the early Galaxy}},}\ }\href
  {https://doi.org/10.1051/0004-6361:20077706} {\bibfield  {journal} {\bibinfo
  {journal} {Astron. \& Astrophys.}\ }\textbf {\bibinfo {volume} {476}},\
  \bibinfo {pages} {935--950}}\BibitemShut {NoStop}%
\bibitem [{\citenamefont {{Frebel}}\ \emph {et~al.}(2007)\citenamefont
  {{Frebel}}, \citenamefont {{Christlieb}}, \citenamefont {{Norris}},
  \citenamefont {{Thom}}, \citenamefont {{Beers}},\ and\ \citenamefont
  {{Rhee}}}]{frebel07}%
  \BibitemOpen
  \bibfield  {author} {\bibinfo {author} {\bibnamefont {{Frebel}},
  \bibfnamefont {A}}, \bibinfo {author} {\bibfnamefont {N.}~\bibnamefont
  {{Christlieb}}}, \bibinfo {author} {\bibfnamefont {J.~E.}\ \bibnamefont
  {{Norris}}}, \bibinfo {author} {\bibfnamefont {C.}~\bibnamefont {{Thom}}},
  \bibinfo {author} {\bibfnamefont {T.~C.}\ \bibnamefont {{Beers}}}, and\
  \bibinfo {author} {\bibfnamefont {J.}~\bibnamefont {{Rhee}}}} (\bibinfo
  {year} {2007}),\ \bibfield  {title} {\enquote {\bibinfo {title} {{Discovery
  of HE 1523-0901, a Strongly r-Process-enhanced Metal-poor Star with Detected
  Uranium}},}\ }\href {https://doi.org/10.1086/518122} {\bibfield  {journal}
  {\bibinfo  {journal} {Astrophys. J. Lett.}\ }\textbf {\bibinfo {volume}
  {660}},\ \bibinfo {pages} {L117--L120}}\BibitemShut {NoStop}%
\bibitem [{\citenamefont {{Frebel}}\ and\ \citenamefont
  {{Kratz}}(2009)}]{Frebel.Kratz:2009}%
  \BibitemOpen
  \bibfield  {author} {\bibinfo {author} {\bibnamefont {{Frebel}},
  \bibfnamefont {A}}, and\ \bibinfo {author} {\bibfnamefont {K.-L.}\
  \bibnamefont {{Kratz}}}} (\bibinfo {year} {2009}),\ \bibfield  {title}
  {\enquote {\bibinfo {title} {{Stellar age dating with thorium, uranium and
  lead}},}\ }in\ \href {https://doi.org/10.1017/S1743921309032104} {\emph
  {\bibinfo {booktitle} {The Ages of Stars}}},\ \bibinfo {series} {IAU
  Symposium}, Vol.\ \bibinfo {volume} {258},\ \bibinfo {editor} {edited by\
  \bibinfo {editor} {\bibfnamefont {E.~E.}\ \bibnamefont {{Mamajek}}}, \bibinfo
  {editor} {\bibfnamefont {D.~R.}\ \bibnamefont {{Soderblom}}}, \ and\ \bibinfo
  {editor} {\bibfnamefont {R.~F.~G.}\ \bibnamefont {{Wyse}}}},\ pp.\ \bibinfo
  {pages} {449--456}\BibitemShut {NoStop}%
\bibitem [{\citenamefont {{Frebel}}\ and\ \citenamefont
  {{Norris}}(2015)}]{Frebel.Norris:2015}%
  \BibitemOpen
  \bibfield  {author} {\bibinfo {author} {\bibnamefont {{Frebel}},
  \bibfnamefont {A}}, and\ \bibinfo {author} {\bibfnamefont {J.~E.}\
  \bibnamefont {{Norris}}}} (\bibinfo {year} {2015}),\ \bibfield  {title}
  {\enquote {\bibinfo {title} {{Near-Field Cosmology with Extremely Metal-Poor
  Stars}},}\ }\href {https://doi.org/10.1146/annurev-astro-082214-122423}
  {\bibfield  {journal} {\bibinfo  {journal} {Annu. Rev. Astron. Astrophys.}\
  }\textbf {\bibinfo {volume} {53}},\ \bibinfo {pages} {631--688}}\BibitemShut
  {NoStop}%
\bibitem [{\citenamefont {{Freiburghaus}}\ \emph
  {et~al.}(1999{\natexlab{a}})\citenamefont {{Freiburghaus}}, \citenamefont
  {{Rembges}}, \citenamefont {{Rauscher}}, \citenamefont {{Kolbe}},
  \citenamefont {{Thielemann}}, \citenamefont {{Kratz}}, \citenamefont
  {{Pfeiffer}},\ and\ \citenamefont {{Cowan}}}]{Freiburghaus.Rembges.ea:1999}%
  \BibitemOpen
  \bibfield  {author} {\bibinfo {author} {\bibnamefont {{Freiburghaus}},
  \bibfnamefont {C}}, \bibinfo {author} {\bibfnamefont {J.-F.}\ \bibnamefont
  {{Rembges}}}, \bibinfo {author} {\bibfnamefont {T.}~\bibnamefont
  {{Rauscher}}}, \bibinfo {author} {\bibfnamefont {E.}~\bibnamefont {{Kolbe}}},
  \bibinfo {author} {\bibfnamefont {F.-K.}\ \bibnamefont {{Thielemann}}},
  \bibinfo {author} {\bibfnamefont {K.-L.}\ \bibnamefont {{Kratz}}}, \bibinfo
  {author} {\bibfnamefont {B.}~\bibnamefont {{Pfeiffer}}}, and\ \bibinfo
  {author} {\bibfnamefont {J.~J.}\ \bibnamefont {{Cowan}}}} (\bibinfo {year}
  {1999}{\natexlab{a}}),\ \bibfield  {title} {\enquote {\bibinfo {title} {{The
  Astrophysical r-Process: A Comparison of Calculations following Adiabatic
  Expansion with Classical Calculations Based on Neutron Densities and
  Temperatures}},}\ }\href {https://doi.org/10.1086/307072} {\bibfield
  {journal} {\bibinfo  {journal} {Astrophys. J.}\ }\textbf {\bibinfo {volume}
  {516}},\ \bibinfo {pages} {381--398}}\BibitemShut {NoStop}%
\bibitem [{\citenamefont {{Freiburghaus}}\ \emph
  {et~al.}(1999{\natexlab{b}})\citenamefont {{Freiburghaus}}, \citenamefont
  {{Rosswog}},\ and\ \citenamefont
  {{Thielemann}}}]{Freiburghaus.Rosswog.Thielemann:1999}%
  \BibitemOpen
  \bibfield  {author} {\bibinfo {author} {\bibnamefont {{Freiburghaus}},
  \bibfnamefont {C}}, \bibinfo {author} {\bibfnamefont {S.}~\bibnamefont
  {{Rosswog}}}, and\ \bibinfo {author} {\bibfnamefont {F.-K.}\ \bibnamefont
  {{Thielemann}}}} (\bibinfo {year} {1999}{\natexlab{b}}),\ \bibfield  {title}
  {\enquote {\bibinfo {title} {{R-Process in Neutron Star Mergers}},}\ }\href
  {https://doi.org/10.1086/312343} {\bibfield  {journal} {\bibinfo  {journal}
  {Astrophys. J.}\ }\textbf {\bibinfo {volume} {525}},\ \bibinfo {pages}
  {L121--L124}}\BibitemShut {NoStop}%
\bibitem [{\citenamefont {{Frensel}}\ \emph {et~al.}(2017)\citenamefont
  {{Frensel}}, \citenamefont {{Wu}}, \citenamefont {{Volpe}},\ and\
  \citenamefont {{Perego}}}]{frensel17}%
  \BibitemOpen
  \bibfield  {author} {\bibinfo {author} {\bibnamefont {{Frensel}},
  \bibfnamefont {M}}, \bibinfo {author} {\bibfnamefont {M.-R.}\ \bibnamefont
  {{Wu}}}, \bibinfo {author} {\bibfnamefont {C.}~\bibnamefont {{Volpe}}}, and\
  \bibinfo {author} {\bibfnamefont {A.}~\bibnamefont {{Perego}}}} (\bibinfo
  {year} {2017}),\ \bibfield  {title} {\enquote {\bibinfo {title} {{Neutrino
  flavor evolution in binary neutron star merger remnants}},}\ }\href
  {https://doi.org/10.1103/PhysRevD.95.023011} {\bibfield  {journal} {\bibinfo
  {journal} {Phys. Rev. D}\ }\textbf {\bibinfo {volume} {95}},\ \bibinfo {eid}
  {023011}}\BibitemShut {NoStop}%
\bibitem [{\citenamefont {{Frischknecht}}\ \emph {et~al.}(2016)\citenamefont
  {{Frischknecht}}, \citenamefont {{Hirschi}}, \citenamefont {{Pignatari}},
  \citenamefont {{Maeder}}, \citenamefont {{Meynet}}, \citenamefont
  {{Chiappini}}, \citenamefont {{Thielemann}}, \citenamefont {{Rauscher}},
  \citenamefont {{Georgy}},\ and\ \citenamefont
  {{Ekstr{\"o}m}}}]{frischknecht16}%
  \BibitemOpen
  \bibfield  {author} {\bibinfo {author} {\bibnamefont {{Frischknecht}},
  \bibfnamefont {U}}, \bibinfo {author} {\bibfnamefont {R.}~\bibnamefont
  {{Hirschi}}}, \bibinfo {author} {\bibfnamefont {M.}~\bibnamefont
  {{Pignatari}}}, \bibinfo {author} {\bibfnamefont {A.}~\bibnamefont
  {{Maeder}}}, \bibinfo {author} {\bibfnamefont {G.}~\bibnamefont {{Meynet}}},
  \bibinfo {author} {\bibfnamefont {C.}~\bibnamefont {{Chiappini}}}, \bibinfo
  {author} {\bibfnamefont {F.-K.}\ \bibnamefont {{Thielemann}}}, \bibinfo
  {author} {\bibfnamefont {T.}~\bibnamefont {{Rauscher}}}, \bibinfo {author}
  {\bibfnamefont {C.}~\bibnamefont {{Georgy}}}, and\ \bibinfo {author}
  {\bibfnamefont {S.}~\bibnamefont {{Ekstr{\"o}m}}}} (\bibinfo {year} {2016}),\
  \bibfield  {title} {\enquote {\bibinfo {title} {{s-process production in
  rotating massive stars at solar and low metallicities}},}\ }\href
  {https://doi.org/10.1093/mnras/stv2723} {\bibfield  {journal} {\bibinfo
  {journal} {Mon. Not. Roy. Astron. Soc.}\ }\textbf {\bibinfo {volume} {456}},\
  \bibinfo {pages} {1803--1825}}\BibitemShut {NoStop}%
\bibitem [{\citenamefont {Fr{\"o}hlich}\ \emph
  {et~al.}(2006{\natexlab{a}})\citenamefont {Fr{\"o}hlich}, \citenamefont
  {Hauser}, \citenamefont {Liebend{\"o}rfer}, \citenamefont
  {Mart{\'i}nez-Pinedo}, \citenamefont {Thielemann}, \citenamefont {Bravo},
  \citenamefont {Zinner}, \citenamefont {Hix}, \citenamefont {Langanke},
  \citenamefont {Mezzacappa},\ and\ \citenamefont
  {Nomoto}}]{Froehlich.Hauser.ea:2006}%
  \BibitemOpen
  \bibfield  {author} {\bibinfo {author} {\bibnamefont {Fr{\"o}hlich},
  \bibfnamefont {C}}, \bibinfo {author} {\bibfnamefont {P.}~\bibnamefont
  {Hauser}}, \bibinfo {author} {\bibfnamefont {M.}~\bibnamefont
  {Liebend{\"o}rfer}}, \bibinfo {author} {\bibfnamefont {G.}~\bibnamefont
  {Mart{\'i}nez-Pinedo}}, \bibinfo {author} {\bibfnamefont {F.-K.}\
  \bibnamefont {Thielemann}}, \bibinfo {author} {\bibfnamefont
  {E.}~\bibnamefont {Bravo}}, \bibinfo {author} {\bibfnamefont {N.~T.}\
  \bibnamefont {Zinner}}, \bibinfo {author} {\bibfnamefont {W.~R.}\
  \bibnamefont {Hix}}, \bibinfo {author} {\bibfnamefont {K.}~\bibnamefont
  {Langanke}}, \bibinfo {author} {\bibfnamefont {A.}~\bibnamefont
  {Mezzacappa}}, and\ \bibinfo {author} {\bibfnamefont {K.}~\bibnamefont
  {Nomoto}}} (\bibinfo {year} {2006}{\natexlab{a}}),\ \bibfield  {title}
  {\enquote {\bibinfo {title} {{Composition of the Innermost Supernova
  Ejecta}},}\ }\href {https://doi.org/10.1086/498224} {\bibfield  {journal}
  {\bibinfo  {journal} {Astrophys. J.}\ }\textbf {\bibinfo {volume} {637}},\
  \bibinfo {pages} {415--426}}\BibitemShut {NoStop}%
\bibitem [{\citenamefont {Fr{\"o}hlich}\ \emph
  {et~al.}(2006{\natexlab{b}})\citenamefont {Fr{\"o}hlich}, \citenamefont
  {Mart{\'i}nez-Pinedo}, \citenamefont {Liebend{\"o}rfer}, \citenamefont
  {Thielemann}, \citenamefont {Bravo}, \citenamefont {Hix}, \citenamefont
  {Langanke},\ and\ \citenamefont
  {Zinner}}]{Froehlich.Martinez-Pinedo.ea:2006}%
  \BibitemOpen
  \bibfield  {author} {\bibinfo {author} {\bibnamefont {Fr{\"o}hlich},
  \bibfnamefont {C}}, \bibinfo {author} {\bibfnamefont {G.}~\bibnamefont
  {Mart{\'i}nez-Pinedo}}, \bibinfo {author} {\bibfnamefont {M.}~\bibnamefont
  {Liebend{\"o}rfer}}, \bibinfo {author} {\bibfnamefont {F.-K.}\ \bibnamefont
  {Thielemann}}, \bibinfo {author} {\bibfnamefont {E.}~\bibnamefont {Bravo}},
  \bibinfo {author} {\bibfnamefont {W.~R.}\ \bibnamefont {Hix}}, \bibinfo
  {author} {\bibfnamefont {K.}~\bibnamefont {Langanke}}, and\ \bibinfo {author}
  {\bibfnamefont {N.~T.}\ \bibnamefont {Zinner}}} (\bibinfo {year}
  {2006}{\natexlab{b}}),\ \bibfield  {title} {\enquote {\bibinfo {title}
  {{Neutrino-induced nucleosynthesis of $A>64$ nuclei: The $\nu p$-process}},}\
  }\href {https://doi.org/10.1103/PhysRevLett.96.142502} {\bibfield  {journal}
  {\bibinfo  {journal} {Phys. Rev. Lett.}\ }\textbf {\bibinfo {volume} {96}},\
  \bibinfo {eid} {142502}}\BibitemShut {NoStop}%
\bibitem [{\citenamefont {{Fryer}}\ and\ \citenamefont
  {{Kalogera}}(1997)}]{fryer.kalogera97}%
  \BibitemOpen
  \bibfield  {author} {\bibinfo {author} {\bibnamefont {{Fryer}}, \bibfnamefont
  {C}}, and\ \bibinfo {author} {\bibfnamefont {V.}~\bibnamefont {{Kalogera}}}}
  (\bibinfo {year} {1997}),\ \bibfield  {title} {\enquote {\bibinfo {title}
  {{Double Neutron Star Systems and Natal Neutron Star Kicks}},}\ }\href
  {https://doi.org/10.1086/304772} {\bibfield  {journal} {\bibinfo  {journal}
  {Astrophys. J.}\ }\textbf {\bibinfo {volume} {489}},\ \bibinfo {pages}
  {244--253}}\BibitemShut {NoStop}%
\bibitem [{\citenamefont {{Fryer}}\ \emph {et~al.}(2015)\citenamefont
  {{Fryer}}, \citenamefont {{Belczynski}}, \citenamefont {{Ramirez-Ruiz}},
  \citenamefont {{Rosswog}}, \citenamefont {{Shen}},\ and\ \citenamefont
  {{Steiner}}}]{fryer15}%
  \BibitemOpen
  \bibfield  {author} {\bibinfo {author} {\bibnamefont {{Fryer}}, \bibfnamefont
  {C~L}}, \bibinfo {author} {\bibfnamefont {K.}~\bibnamefont {{Belczynski}}},
  \bibinfo {author} {\bibfnamefont {E.}~\bibnamefont {{Ramirez-Ruiz}}},
  \bibinfo {author} {\bibfnamefont {S.}~\bibnamefont {{Rosswog}}}, \bibinfo
  {author} {\bibfnamefont {G.}~\bibnamefont {{Shen}}}, and\ \bibinfo {author}
  {\bibfnamefont {A.~W.}\ \bibnamefont {{Steiner}}}} (\bibinfo {year} {2015}),\
  \bibfield  {title} {\enquote {\bibinfo {title} {{The Fate of the Compact
  Remnant in Neutron Star Mergers}},}\ }\href
  {https://doi.org/10.1088/0004-637X/812/1/24} {\bibfield  {journal} {\bibinfo
  {journal} {Astrophys. J.}\ }\textbf {\bibinfo {volume} {812}},\ \bibinfo
  {eid} {24}}\BibitemShut {NoStop}%
\bibitem [{\citenamefont {{Fujibayashi}}\ \emph {et~al.}(2018)\citenamefont
  {{Fujibayashi}}, \citenamefont {{Kiuchi}}, \citenamefont {{Nishimura}},
  \citenamefont {{Sekiguchi}},\ and\ \citenamefont
  {{Shibata}}}]{Fujibayashi.Kiuchi.ea:2018}%
  \BibitemOpen
  \bibfield  {author} {\bibinfo {author} {\bibnamefont {{Fujibayashi}},
  \bibfnamefont {S}}, \bibinfo {author} {\bibfnamefont {K.}~\bibnamefont
  {{Kiuchi}}}, \bibinfo {author} {\bibfnamefont {N.}~\bibnamefont
  {{Nishimura}}}, \bibinfo {author} {\bibfnamefont {Y.}~\bibnamefont
  {{Sekiguchi}}}, and\ \bibinfo {author} {\bibfnamefont {M.}~\bibnamefont
  {{Shibata}}}} (\bibinfo {year} {2018}),\ \bibfield  {title} {\enquote
  {\bibinfo {title} {{Mass Ejection from the Remnant of a Binary Neutron Star
  Merger: Viscous-radiation Hydrodynamics Study}},}\ }\href
  {https://doi.org/10.3847/1538-4357/aabafd} {\bibfield  {journal} {\bibinfo
  {journal} {Astrophys. J.}\ }\textbf {\bibinfo {volume} {860}},\ \bibinfo
  {eid} {64}}\BibitemShut {NoStop}%
\bibitem [{\citenamefont {{Fujibayashi}}\ \emph {et~al.}(2017)\citenamefont
  {{Fujibayashi}}, \citenamefont {{Sekiguchi}}, \citenamefont {{Kiuchi}},\ and\
  \citenamefont {{Shibata}}}]{Fujibayashi.Sekiguchi.ea:2017}%
  \BibitemOpen
  \bibfield  {author} {\bibinfo {author} {\bibnamefont {{Fujibayashi}},
  \bibfnamefont {S}}, \bibinfo {author} {\bibfnamefont {Y.}~\bibnamefont
  {{Sekiguchi}}}, \bibinfo {author} {\bibfnamefont {K.}~\bibnamefont
  {{Kiuchi}}}, and\ \bibinfo {author} {\bibfnamefont {M.}~\bibnamefont
  {{Shibata}}}} (\bibinfo {year} {2017}),\ \bibfield  {title} {\enquote
  {\bibinfo {title} {{Properties of Neutrino-driven Ejecta from the Remnant of
  a Binary Neutron Star Merger: Pure Radiation Hydrodynamics Case}},}\ }\href
  {https://doi.org/10.3847/1538-4357/aa8039} {\bibfield  {journal} {\bibinfo
  {journal} {Astrophys. J.}\ }\textbf {\bibinfo {volume} {846}},\ \bibinfo
  {eid} {114}}\BibitemShut {NoStop}%
\bibitem [{\citenamefont {{Fujibayashi}}\ \emph {et~al.}(2020)\citenamefont
  {{Fujibayashi}}, \citenamefont {{Shibata}}, \citenamefont {{Wanajo}},
  \citenamefont {{Kiuchi}}, \citenamefont {{Kyutoku}},\ and\ \citenamefont
  {{Sekiguchi}}}]{Fujibayashi.Shibata.ea:2020}%
  \BibitemOpen
  \bibfield  {author} {\bibinfo {author} {\bibnamefont {{Fujibayashi}},
  \bibfnamefont {Sho}}, \bibinfo {author} {\bibfnamefont {Masaru}\ \bibnamefont
  {{Shibata}}}, \bibinfo {author} {\bibfnamefont {Shinya}\ \bibnamefont
  {{Wanajo}}}, \bibinfo {author} {\bibfnamefont {Kenta}\ \bibnamefont
  {{Kiuchi}}}, \bibinfo {author} {\bibfnamefont {Koutarou}\ \bibnamefont
  {{Kyutoku}}}, and\ \bibinfo {author} {\bibfnamefont {Yuichiro}\ \bibnamefont
  {{Sekiguchi}}}} (\bibinfo {year} {2020}),\ \bibfield  {title} {\enquote
  {\bibinfo {title} {{Mass ejection from disks surrounding a low-mass black
  hole: Viscous neutrino-radiation hydrodynamics simulation in full general
  relativity}},}\ }\href {https://doi.org/10.1103/PhysRevD.101.083029}
  {\bibfield  {journal} {\bibinfo  {journal} {Phys. Rev. D}\ }\textbf {\bibinfo
  {volume} {101}}~(\bibinfo {number} {8}),\ \bibinfo {eid}
  {083029}}\BibitemShut {NoStop}%
\bibitem [{\citenamefont {{Fujimoto}}\ \emph {et~al.}(2004)\citenamefont
  {{Fujimoto}}, \citenamefont {{Hashimoto}}, \citenamefont {{Arai}},\ and\
  \citenamefont {{Matsuba}}}]{Fujimoto.Hashimoto.ea:2004}%
  \BibitemOpen
  \bibfield  {author} {\bibinfo {author} {\bibnamefont {{Fujimoto}},
  \bibfnamefont {S-i}}, \bibinfo {author} {\bibfnamefont {M.-a.}\ \bibnamefont
  {{Hashimoto}}}, \bibinfo {author} {\bibfnamefont {K.}~\bibnamefont {{Arai}}},
  and\ \bibinfo {author} {\bibfnamefont {R.}~\bibnamefont {{Matsuba}}}}
  (\bibinfo {year} {2004}),\ \bibfield  {title} {\enquote {\bibinfo {title}
  {{Nucleosynthesis inside an Accretion Disk and Disk Winds Related to
  Gamma-Ray Bursts}},}\ }\href {https://doi.org/10.1086/423778} {\bibfield
  {journal} {\bibinfo  {journal} {Astrophys. J.}\ }\textbf {\bibinfo {volume}
  {614}},\ \bibinfo {pages} {847--857}}\BibitemShut {NoStop}%
\bibitem [{\citenamefont {{Fujimoto}}\ \emph {et~al.}(2003)\citenamefont
  {{Fujimoto}}, \citenamefont {{Hashimoto}}, \citenamefont {{Koike}},
  \citenamefont {{Arai}},\ and\ \citenamefont
  {{Matsuba}}}]{Fujimoto.Hashimoto.ea:2003}%
  \BibitemOpen
  \bibfield  {author} {\bibinfo {author} {\bibnamefont {{Fujimoto}},
  \bibfnamefont {S-i}}, \bibinfo {author} {\bibfnamefont {M.-a.}\ \bibnamefont
  {{Hashimoto}}}, \bibinfo {author} {\bibfnamefont {O.}~\bibnamefont
  {{Koike}}}, \bibinfo {author} {\bibfnamefont {K.}~\bibnamefont {{Arai}}},
  and\ \bibinfo {author} {\bibfnamefont {R.}~\bibnamefont {{Matsuba}}}}
  (\bibinfo {year} {2003}),\ \bibfield  {title} {\enquote {\bibinfo {title}
  {{p-Process Nucleosynthesis inside Supernova-driven Supercritical Accretion
  Disks}},}\ }\href {https://doi.org/10.1086/345982} {\bibfield  {journal}
  {\bibinfo  {journal} {Astrophys. J.}\ }\textbf {\bibinfo {volume} {585}},\
  \bibinfo {pages} {418--428}}\BibitemShut {NoStop}%
\bibitem [{\citenamefont {{Fujimoto}}\ \emph {et~al.}(2007)\citenamefont
  {{Fujimoto}}, \citenamefont {{Hashimoto}}, \citenamefont {{Kotake}},\ and\
  \citenamefont {{Yamada}}}]{fujimoto07}%
  \BibitemOpen
  \bibfield  {author} {\bibinfo {author} {\bibnamefont {{Fujimoto}},
  \bibfnamefont {S-i}}, \bibinfo {author} {\bibfnamefont {M.-a.}\ \bibnamefont
  {{Hashimoto}}}, \bibinfo {author} {\bibfnamefont {K.}~\bibnamefont
  {{Kotake}}}, and\ \bibinfo {author} {\bibfnamefont {S.}~\bibnamefont
  {{Yamada}}}} (\bibinfo {year} {2007}),\ \bibfield  {title} {\enquote
  {\bibinfo {title} {{Heavy-Element Nucleosynthesis in a Collapsar}},}\ }\href
  {https://doi.org/10.1086/509908} {\bibfield  {journal} {\bibinfo  {journal}
  {Astrophys. J.}\ }\textbf {\bibinfo {volume} {656}},\ \bibinfo {pages}
  {382--392}}\BibitemShut {NoStop}%
\bibitem [{\citenamefont {{Fujimoto}}\ \emph {et~al.}(2006)\citenamefont
  {{Fujimoto}}, \citenamefont {{Kotake}}, \citenamefont {{Yamada}},
  \citenamefont {{Hashimoto}},\ and\ \citenamefont {{Sato}}}]{fujimoto06}%
  \BibitemOpen
  \bibfield  {author} {\bibinfo {author} {\bibnamefont {{Fujimoto}},
  \bibfnamefont {S-i}}, \bibinfo {author} {\bibfnamefont {K.}~\bibnamefont
  {{Kotake}}}, \bibinfo {author} {\bibfnamefont {S.}~\bibnamefont {{Yamada}}},
  \bibinfo {author} {\bibfnamefont {M.-a.}\ \bibnamefont {{Hashimoto}}}, and\
  \bibinfo {author} {\bibfnamefont {K.}~\bibnamefont {{Sato}}}} (\bibinfo
  {year} {2006}),\ \bibfield  {title} {\enquote {\bibinfo {title}
  {{Magnetohydrodynamic Simulations of a Rotating Massive Star Collapsing to a
  Black Hole}},}\ }\href {https://doi.org/10.1086/503624} {\bibfield  {journal}
  {\bibinfo  {journal} {Astrophys. J.}\ }\textbf {\bibinfo {volume} {644}},\
  \bibinfo {pages} {1040--1055}}\BibitemShut {NoStop}%
\bibitem [{\citenamefont {{Fujimoto}}\ \emph {et~al.}(2008)\citenamefont
  {{Fujimoto}}, \citenamefont {{Nishimura}},\ and\ \citenamefont
  {{Hashimoto}}}]{Fujimoto.Nishimura.Hashimoto:2008}%
  \BibitemOpen
  \bibfield  {author} {\bibinfo {author} {\bibnamefont {{Fujimoto}},
  \bibfnamefont {S-i}}, \bibinfo {author} {\bibfnamefont {N.}~\bibnamefont
  {{Nishimura}}}, and\ \bibinfo {author} {\bibfnamefont {M.-a.}\ \bibnamefont
  {{Hashimoto}}}} (\bibinfo {year} {2008}),\ \bibfield  {title} {\enquote
  {\bibinfo {title} {{Nucleosynthesis in Magnetically Driven Jets from
  Collapsars}},}\ }\href {https://doi.org/10.1086/529416} {\bibfield  {journal}
  {\bibinfo  {journal} {Astrophys. J.}\ }\textbf {\bibinfo {volume} {680}},\
  \bibinfo {pages} {1350--1358}}\BibitemShut {NoStop}%
\bibitem [{\citenamefont {{Fulbright}}(2000)}]{fulbright00}%
  \BibitemOpen
  \bibfield  {author} {\bibinfo {author} {\bibnamefont {{Fulbright}},
  \bibfnamefont {J~P}}} (\bibinfo {year} {2000}),\ \bibfield  {title} {\enquote
  {\bibinfo {title} {{Abundances and Kinematics of Field Halo and Disk Stars.
  I. Observational Data and Abundance Analysis}},}\ }\href
  {https://doi.org/10.1086/301548} {\bibfield  {journal} {\bibinfo  {journal}
  {Astron. J.}\ }\textbf {\bibinfo {volume} {120}},\ \bibinfo {pages}
  {1841--1852}}\BibitemShut {NoStop}%
\bibitem [{\citenamefont {Fuller}\ \emph {et~al.}(1980)\citenamefont {Fuller},
  \citenamefont {Fowler},\ and\ \citenamefont
  {Newman}}]{Fuller.Fowler.Newman:1980}%
  \BibitemOpen
  \bibfield  {author} {\bibinfo {author} {\bibnamefont {Fuller}, \bibfnamefont
  {G~M}}, \bibinfo {author} {\bibfnamefont {W.~A.}\ \bibnamefont {Fowler}},
  and\ \bibinfo {author} {\bibfnamefont {M.~J.}\ \bibnamefont {Newman}}}
  (\bibinfo {year} {1980}),\ \bibfield  {title} {\enquote {\bibinfo {title}
  {Stellar weak-interaction rates for sd-shell nuclei. i - nuclear matrix
  element systematics with application to al-26 and selected nuclei of
  importance to the supernova problem},}\ }\href
  {https://doi.org/10.1086/190657} {\bibfield  {journal} {\bibinfo  {journal}
  {Astrophys. J. Suppl.}\ }\textbf {\bibinfo {volume} {42}},\ \bibinfo {pages}
  {447--473}}\BibitemShut {NoStop}%
\bibitem [{\citenamefont {Fuller}\ \emph {et~al.}(2017)\citenamefont {Fuller},
  \citenamefont {Kusenko},\ and\ \citenamefont
  {Takhistov}}]{Fuller.Kusenko.Takhistov:2017}%
  \BibitemOpen
  \bibfield  {author} {\bibinfo {author} {\bibnamefont {Fuller}, \bibfnamefont
  {George~M}}, \bibinfo {author} {\bibfnamefont {Alexander}\ \bibnamefont
  {Kusenko}}, and\ \bibinfo {author} {\bibfnamefont {Volodymyr}\ \bibnamefont
  {Takhistov}}} (\bibinfo {year} {2017}),\ \bibfield  {title} {\enquote
  {\bibinfo {title} {{Primordial Black Holes and $r$-Process
  Nucleosynthesis}},}\ }\href {https://doi.org/10.1103/PhysRevLett.119.061101}
  {\bibfield  {journal} {\bibinfo  {journal} {Phys. Rev. Lett.}\ }\textbf
  {\bibinfo {volume} {119}},\ \bibinfo {pages} {061101}}\BibitemShut {NoStop}%
\bibitem [{\citenamefont {{Galama}}\ \emph {et~al.}(1998)\citenamefont
  {{Galama}} \emph {et~al.}}]{Galama.ea:1998}%
  \BibitemOpen
  \bibfield  {author} {\bibinfo {author} {\bibnamefont {{Galama}},
  \bibfnamefont {T~J}},  \emph {et~al.}} (\bibinfo {year} {1998}),\ \bibfield
  {title} {\enquote {\bibinfo {title} {{An unusual supernova in the error box
  of the {\ensuremath{\gamma}}-ray burst of 25 April 1998}},}\ }\href
  {https://doi.org/10.1038/27150} {\bibfield  {journal} {\bibinfo  {journal}
  {Nature}\ }\textbf {\bibinfo {volume} {395}},\ \bibinfo {pages}
  {670--672}}\BibitemShut {NoStop}%
\bibitem [{\citenamefont {{Geissel}}\ \emph {et~al.}(1992)\citenamefont
  {{Geissel}} \emph {et~al.}}]{Geissel}%
  \BibitemOpen
  \bibfield  {author} {\bibinfo {author} {\bibnamefont {{Geissel}},
  \bibfnamefont {H}},  \emph {et~al.}} (\bibinfo {year} {1992}),\ \bibfield
  {title} {\enquote {\bibinfo {title} {{The GSI projectile fragment separator
  (FRS): a versatile magnetic system for relativistic heavy ions}},}\ }\href
  {https://doi.org/10.1016/0168-583X(92)95944-M} {\bibfield  {journal}
  {\bibinfo  {journal} {Nucl. Instruments Methods Phys. Res. Sect. B Beam
  Interact. with Mater. Atoms}\ }\textbf {\bibinfo {volume} {70}},\ \bibinfo
  {pages} {286--297}}\BibitemShut {NoStop}%
\bibitem [{\citenamefont {{Giacomazzo}}\ \emph {et~al.}(2009)\citenamefont
  {{Giacomazzo}}, \citenamefont {{Rezzolla}},\ and\ \citenamefont
  {{Baiotti}}}]{Giacomazzo.Rezzolla.Baiotti:2009}%
  \BibitemOpen
  \bibfield  {author} {\bibinfo {author} {\bibnamefont {{Giacomazzo}},
  \bibfnamefont {B}}, \bibinfo {author} {\bibfnamefont {L.}~\bibnamefont
  {{Rezzolla}}}, and\ \bibinfo {author} {\bibfnamefont {L.}~\bibnamefont
  {{Baiotti}}}} (\bibinfo {year} {2009}),\ \bibfield  {title} {\enquote
  {\bibinfo {title} {{Can magnetic fields be detected during the inspiral of
  binary neutron stars?}}}\ }\href
  {https://doi.org/10.1111/j.1745-3933.2009.00745.x} {\bibfield  {journal}
  {\bibinfo  {journal} {Mon. Not. Roy. Astron. Soc.}\ }\textbf {\bibinfo
  {volume} {399}},\ \bibinfo {pages} {L164--L168}}\BibitemShut {NoStop}%
\bibitem [{\citenamefont {{Giacomazzo}}\ \emph {et~al.}(2015)\citenamefont
  {{Giacomazzo}}, \citenamefont {{Zrake}}, \citenamefont {{Duffell}},
  \citenamefont {{MacFadyen}},\ and\ \citenamefont
  {{Perna}}}]{Giacomazzo.Zrake.ea:2015}%
  \BibitemOpen
  \bibfield  {author} {\bibinfo {author} {\bibnamefont {{Giacomazzo}},
  \bibfnamefont {B}}, \bibinfo {author} {\bibfnamefont {J.}~\bibnamefont
  {{Zrake}}}, \bibinfo {author} {\bibfnamefont {P.~C.}\ \bibnamefont
  {{Duffell}}}, \bibinfo {author} {\bibfnamefont {A.~I.}\ \bibnamefont
  {{MacFadyen}}}, and\ \bibinfo {author} {\bibfnamefont {R.}~\bibnamefont
  {{Perna}}}} (\bibinfo {year} {2015}),\ \bibfield  {title} {\enquote {\bibinfo
  {title} {Producing magnetar magnetic fields in the merger of binary neutron
  stars},}\ }\href {https://doi.org/10.1088/0004-637X/809/1/39} {\bibfield
  {journal} {\bibinfo  {journal} {Astrophys. J.}\ }\textbf {\bibinfo {volume}
  {809}},\ \bibinfo {eid} {39}}\BibitemShut {NoStop}%
\bibitem [{\citenamefont {{Gilroy}}\ \emph {et~al.}(1988)\citenamefont
  {{Gilroy}}, \citenamefont {{Sneden}}, \citenamefont {{Pilachowski}},\ and\
  \citenamefont {{Cowan}}}]{gilroy88}%
  \BibitemOpen
  \bibfield  {author} {\bibinfo {author} {\bibnamefont {{Gilroy}},
  \bibfnamefont {K~K}}, \bibinfo {author} {\bibfnamefont {C.}~\bibnamefont
  {{Sneden}}}, \bibinfo {author} {\bibfnamefont {C.~A.}\ \bibnamefont
  {{Pilachowski}}}, and\ \bibinfo {author} {\bibfnamefont {J.~J.}\ \bibnamefont
  {{Cowan}}}} (\bibinfo {year} {1988}),\ \bibfield  {title} {\enquote {\bibinfo
  {title} {{Abundances of neutron capture elements in Population II stars}},}\
  }\href {https://doi.org/10.1086/166191} {\bibfield  {journal} {\bibinfo
  {journal} {Astrophys. J.}\ }\textbf {\bibinfo {volume} {327}},\ \bibinfo
  {pages} {298--320}}\BibitemShut {NoStop}%
\bibitem [{\citenamefont {Giuliani}\ \emph {et~al.}(2019)\citenamefont
  {Giuliani}, \citenamefont {Matheson}, \citenamefont {Nazarewicz},
  \citenamefont {Olsen}, \citenamefont {Reinhard}, \citenamefont {Sadhukhan},
  \citenamefont {Schuetrumpf}, \citenamefont {Schunck},\ and\ \citenamefont
  {Schwerdtfeger}}]{Giuliani.Matheson.ea:2019}%
  \BibitemOpen
  \bibfield  {author} {\bibinfo {author} {\bibnamefont {Giuliani},
  \bibfnamefont {S~A}}, \bibinfo {author} {\bibfnamefont {Z.}~\bibnamefont
  {Matheson}}, \bibinfo {author} {\bibfnamefont {W.}~\bibnamefont
  {Nazarewicz}}, \bibinfo {author} {\bibfnamefont {E.}~\bibnamefont {Olsen}},
  \bibinfo {author} {\bibfnamefont {P.-G.}\ \bibnamefont {Reinhard}}, \bibinfo
  {author} {\bibfnamefont {J.}~\bibnamefont {Sadhukhan}}, \bibinfo {author}
  {\bibfnamefont {B.}~\bibnamefont {Schuetrumpf}}, \bibinfo {author}
  {\bibfnamefont {N.}~\bibnamefont {Schunck}}, and\ \bibinfo {author}
  {\bibfnamefont {P.}~\bibnamefont {Schwerdtfeger}}} (\bibinfo {year} {2019}),\
  \bibfield  {title} {\enquote {\bibinfo {title} {Colloquium: Superheavy
  elements: Oganesson and beyond},}\ }\href
  {https://doi.org/10.1103/RevModPhys.91.011001} {\bibfield  {journal}
  {\bibinfo  {journal} {Rev. Mod. Phys.}\ }\textbf {\bibinfo {volume} {91}},\
  \bibinfo {eid} {011001}}\BibitemShut {NoStop}%
\bibitem [{\citenamefont {{Giuliani}}\ \emph {et~al.}(2018)\citenamefont
  {{Giuliani}}, \citenamefont {{Mart{\'\i}nez-Pinedo}},\ and\ \citenamefont
  {{Robledo}}}]{Giuliani.Martinez-Pinedo.Robledo:2018}%
  \BibitemOpen
  \bibfield  {author} {\bibinfo {author} {\bibnamefont {{Giuliani}},
  \bibfnamefont {Samuel~A}}, \bibinfo {author} {\bibfnamefont {Gabriel}\
  \bibnamefont {{Mart{\'\i}nez-Pinedo}}}, and\ \bibinfo {author} {\bibfnamefont
  {Luis~M.}\ \bibnamefont {{Robledo}}}} (\bibinfo {year} {2018}),\ \bibfield
  {title} {\enquote {\bibinfo {title} {{Fission properties of superheavy nuclei
  for r-process calculations}},}\ }\href
  {https://doi.org/10.1103/PhysRevC.97.034323} {\bibfield  {journal} {\bibinfo
  {journal} {Phys. Rev. C}\ }\textbf {\bibinfo {volume} {97}},\ \bibinfo {eid}
  {034323}}\BibitemShut {NoStop}%
\bibitem [{\citenamefont {{Giuliani}}\ \emph {et~al.}(2019)\citenamefont
  {{Giuliani}}, \citenamefont {{Mart{\'\i}nez-Pinedo}}, \citenamefont {{Wu}},\
  and\ \citenamefont {{Robledo}}}]{Giuliani.Martinez-Pinedo.ea:2019}%
  \BibitemOpen
  \bibfield  {author} {\bibinfo {author} {\bibnamefont {{Giuliani}},
  \bibfnamefont {Samuel~A}}, \bibinfo {author} {\bibfnamefont {Gabriel}\
  \bibnamefont {{Mart{\'\i}nez-Pinedo}}}, \bibinfo {author} {\bibfnamefont
  {Meng-Ru}\ \bibnamefont {{Wu}}}, and\ \bibinfo {author} {\bibfnamefont
  {Luis~M.}\ \bibnamefont {{Robledo}}}} (\bibinfo {year} {2019}),\ \bibfield
  {title} {\enquote {\bibinfo {title} {{Fission and the r-process
  nucleosynthesis of translead nuclei}},}\ }\href@noop {} {\bibfield  {journal}
  {\bibinfo  {journal} {Phys. Rev. C}\ }}\bibinfo {note} {{in press}},\ \Eprint
  {https://arxiv.org/abs/1904.03733} {arXiv:1904.03733 [nucl-th]} \BibitemShut
  {NoStop}%
\bibitem [{\citenamefont {Giuliani}\ \emph {et~al.}(2014)\citenamefont
  {Giuliani}, \citenamefont {Robledo},\ and\ \citenamefont
  {Rodr{\'\i}guez-Guzm\'an}}]{Giuliani.Robledo.Rodriguez-Guzman:2014}%
  \BibitemOpen
  \bibfield  {author} {\bibinfo {author} {\bibnamefont {Giuliani},
  \bibfnamefont {Samuel~A}}, \bibinfo {author} {\bibfnamefont {Luis~M.}\
  \bibnamefont {Robledo}}, and\ \bibinfo {author} {\bibfnamefont
  {R.}~\bibnamefont {Rodr{\'\i}guez-Guzm\'an}}} (\bibinfo {year} {2014}),\
  \bibfield  {title} {\enquote {\bibinfo {title} {Dynamic versus static fission
  paths with realistic interactions},}\ }\href
  {https://doi.org/10.1103/PhysRevC.90.054311} {\bibfield  {journal} {\bibinfo
  {journal} {Phys. Rev. C}\ }\textbf {\bibinfo {volume} {90}},\ \bibinfo
  {pages} {054311}}\BibitemShut {NoStop}%
\bibitem [{\citenamefont {Giunti}\ \emph {et~al.}(2012)\citenamefont {Giunti},
  \citenamefont {Laveder}, \citenamefont {Li}, \citenamefont {Liu},\ and\
  \citenamefont {Long}}]{Giunti.Laveder.ea:2012}%
  \BibitemOpen
  \bibfield  {author} {\bibinfo {author} {\bibnamefont {Giunti}, \bibfnamefont
  {C}}, \bibinfo {author} {\bibfnamefont {M.}~\bibnamefont {Laveder}}, \bibinfo
  {author} {\bibfnamefont {Y.~F.}\ \bibnamefont {Li}}, \bibinfo {author}
  {\bibfnamefont {Q.~Y.}\ \bibnamefont {Liu}}, and\ \bibinfo {author}
  {\bibfnamefont {H.~W.}\ \bibnamefont {Long}}} (\bibinfo {year} {2012}),\
  \bibfield  {title} {\enquote {\bibinfo {title} {Update of short-baseline
  electron neutrino and antineutrino disappearance},}\ }\href
  {https://doi.org/10.1103/PhysRevD.86.113014} {\bibfield  {journal} {\bibinfo
  {journal} {Phys. Rev. D}\ }\textbf {\bibinfo {volume} {86}},\ \bibinfo
  {pages} {113014}}\BibitemShut {NoStop}%
\bibitem [{\citenamefont {{Godzieba}}\ \emph {et~al.}(2020)\citenamefont
  {{Godzieba}}, \citenamefont {{Radice}},\ and\ \citenamefont
  {{Bernuzzi}}}]{Godzieba.Radice.Bernuzzi:2020}%
  \BibitemOpen
  \bibfield  {author} {\bibinfo {author} {\bibnamefont {{Godzieba}},
  \bibfnamefont {Daniel~A}}, \bibinfo {author} {\bibfnamefont {David}\
  \bibnamefont {{Radice}}}, and\ \bibinfo {author} {\bibfnamefont {Sebastiano}\
  \bibnamefont {{Bernuzzi}}}} (\bibinfo {year} {2020}),\ \bibfield  {title}
  {\enquote {\bibinfo {title} {{On the maximum mass of neutron stars and
  GW190814}},}\ }\href@noop {} {\bibfield  {journal} {\bibinfo  {journal}
  {arXiv e-prints}\ }}\Eprint {https://arxiv.org/abs/2007.10999}
  {arXiv:2007.10999 [astro-ph.HE]} \BibitemShut {NoStop}%
\bibitem [{\citenamefont {{Goldstein}}\ and\ \citenamefont
  {{Kasen}}(2018)}]{Goldstein.Kasen:2018}%
  \BibitemOpen
  \bibfield  {author} {\bibinfo {author} {\bibnamefont {{Goldstein}},
  \bibfnamefont {D~A}}, and\ \bibinfo {author} {\bibfnamefont {D.}~\bibnamefont
  {{Kasen}}}} (\bibinfo {year} {2018}),\ \bibfield  {title} {\enquote {\bibinfo
  {title} {{Evidence for Sub-Chandrasekhar Mass Type Ia Supernovae from an
  Extensive Survey of Radiative Transfer Models}},}\ }\href
  {https://doi.org/10.3847/2041-8213/aaa409} {\bibfield  {journal} {\bibinfo
  {journal} {Astrophys. J.}\ }\textbf {\bibinfo {volume} {852}},\ \bibinfo
  {eid} {L33}}\BibitemShut {NoStop}%
\bibitem [{\citenamefont {{G{\'o}mez-Hornillos}}\ \emph
  {et~al.}(2011)\citenamefont {{G{\'o}mez-Hornillos}} \emph
  {et~al.}}]{Gomez-Hornillos2011}%
  \BibitemOpen
  \bibfield  {author} {\bibinfo {author} {\bibnamefont {{G{\'o}mez-Hornillos}},
  \bibfnamefont {M~B}},  \emph {et~al.}} (\bibinfo {year} {2011}),\ \bibfield
  {title} {\enquote {\bibinfo {title} {{First Measurements with the BEta
  deLayEd Neutron Detector (BELEN-20) at JYFLTRAP}},}\ }in\ \href
  {https://doi.org/10.1088/1742-6596/312/5/052008} {\emph {\bibinfo {booktitle}
  {Journal of Physics Conference Series}}},\ Vol.\ \bibinfo {volume} {312},\
  p.\ \bibinfo {pages} {052008}\BibitemShut {NoStop}%
\bibitem [{\citenamefont {{Gompertz}}\ \emph {et~al.}(2018)\citenamefont
  {{Gompertz}} \emph {et~al.}}]{Gompertz.Others:2018}%
  \BibitemOpen
  \bibfield  {author} {\bibinfo {author} {\bibnamefont {{Gompertz}},
  \bibfnamefont {B~P}},  \emph {et~al.}} (\bibinfo {year} {2018}),\ \bibfield
  {title} {\enquote {\bibinfo {title} {{The Diversity of Kilonova Emission in
  Short Gamma-Ray Bursts}},}\ }\href {https://doi.org/10.3847/1538-4357/aac206}
  {\bibfield  {journal} {\bibinfo  {journal} {Astrophys. J.}\ }\textbf
  {\bibinfo {volume} {860}},\ \bibinfo {eid} {62}}\BibitemShut {NoStop}%
\bibitem [{\citenamefont {{Goriely}}(1997)}]{Goriely:1997}%
  \BibitemOpen
  \bibfield  {author} {\bibinfo {author} {\bibnamefont {{Goriely}},
  \bibfnamefont {S}}} (\bibinfo {year} {1997}),\ \bibfield  {title} {\enquote
  {\bibinfo {title} {Direct neutron captures and the r-process
  nucleosynthesis},}\ }\href@noop {} {\bibfield  {journal} {\bibinfo  {journal}
  {Astron. \& Astrophys.}\ }\textbf {\bibinfo {volume} {325}},\ \bibinfo
  {pages} {414--424}}\BibitemShut {NoStop}%
\bibitem [{\citenamefont {Goriely}(1998)}]{Goriely:1998}%
  \BibitemOpen
  \bibfield  {author} {\bibinfo {author} {\bibnamefont {Goriely}, \bibfnamefont
  {S}}} (\bibinfo {year} {1998}),\ \bibfield  {title} {\enquote {\bibinfo
  {title} {Radiative neutron captures by neutron-rich nuclei and the r-process
  nucleosynthesis},}\ }\href {https://doi.org/10.1016/S0370-2693(98)00907-1}
  {\bibfield  {journal} {\bibinfo  {journal} {Phys. Lett. B}\ }\textbf
  {\bibinfo {volume} {436}},\ \bibinfo {pages} {10--18}}\BibitemShut {NoStop}%
\bibitem [{\citenamefont {Goriely}(1999)}]{Goriely:1999}%
  \BibitemOpen
  \bibfield  {author} {\bibinfo {author} {\bibnamefont {Goriely}, \bibfnamefont
  {S}}} (\bibinfo {year} {1999}),\ \bibfield  {title} {\enquote {\bibinfo
  {title} {{Uncertainties in the solar system r-abundance distribution}},}\
  }\href@noop {} {\bibfield  {journal} {\bibinfo  {journal} {Astron. \&
  Astrophys.}\ }\textbf {\bibinfo {volume} {342}},\ \bibinfo {pages}
  {881--891}}\BibitemShut {NoStop}%
\bibitem [{\citenamefont {Goriely}(2015)}]{Goriely:2015}%
  \BibitemOpen
  \bibfield  {author} {\bibinfo {author} {\bibnamefont {Goriely}, \bibfnamefont
  {S}}} (\bibinfo {year} {2015}),\ \bibfield  {title} {\enquote {\bibinfo
  {title} {The fundamental role of fission during r-process nucleosynthesis in
  neutron star mergers},}\ }\href {https://doi.org/10.1140/epja/i2015-15022-3}
  {\bibfield  {journal} {\bibinfo  {journal} {Eur. Phys. J. A}\ }\textbf
  {\bibinfo {volume} {51}},\ \bibinfo {pages} {22}}\BibitemShut {NoStop}%
\bibitem [{\citenamefont {{Goriely}}\ and\ \citenamefont
  {{Arnould}}(2001)}]{Goriely.Arnould:2001}%
  \BibitemOpen
  \bibfield  {author} {\bibinfo {author} {\bibnamefont {{Goriely}},
  \bibfnamefont {S}}, and\ \bibinfo {author} {\bibfnamefont {M.}~\bibnamefont
  {{Arnould}}}} (\bibinfo {year} {2001}),\ \bibfield  {title} {\enquote
  {\bibinfo {title} {{Actinides: How well do we know their stellar
  production?}}}\ }\href {https://doi.org/10.1051/0004-6361:20011368}
  {\bibfield  {journal} {\bibinfo  {journal} {Astron. \& Astrophys.}\ }\textbf
  {\bibinfo {volume} {379}},\ \bibinfo {pages} {1113--1122}}\BibitemShut
  {NoStop}%
\bibitem [{\citenamefont {{Goriely}}\ \emph {et~al.}(2011)\citenamefont
  {{Goriely}}, \citenamefont {{Bauswein}},\ and\ \citenamefont
  {{Janka}}}]{Goriely.Bauswein.Janka:2011}%
  \BibitemOpen
  \bibfield  {author} {\bibinfo {author} {\bibnamefont {{Goriely}},
  \bibfnamefont {S}}, \bibinfo {author} {\bibfnamefont {A.}~\bibnamefont
  {{Bauswein}}}, and\ \bibinfo {author} {\bibfnamefont {H.-T.}\ \bibnamefont
  {{Janka}}}} (\bibinfo {year} {2011}),\ \bibfield  {title} {\enquote {\bibinfo
  {title} {{r-process Nucleosynthesis in Dynamically Ejected Matter of Neutron
  Star Mergers}},}\ }\href {https://doi.org/10.1088/2041-8205/738/2/L32}
  {\bibfield  {journal} {\bibinfo  {journal} {Astrophys. J.}\ }\textbf
  {\bibinfo {volume} {738}},\ \bibinfo {eid} {L32}}\BibitemShut {NoStop}%
\bibitem [{\citenamefont {{Goriely}}\ \emph {et~al.}(2014)\citenamefont
  {{Goriely}}, \citenamefont {{Bauswein}}, \citenamefont {{Janka}},
  \citenamefont {{Sida}}, \citenamefont {{Lema{\^\i}tre}}, \citenamefont
  {{Panebianco}}, \citenamefont {{Dubray}},\ and\ \citenamefont
  {{Hilaire}}}]{Goriely.Bauswein.ea:2014}%
  \BibitemOpen
  \bibfield  {author} {\bibinfo {author} {\bibnamefont {{Goriely}},
  \bibfnamefont {S}}, \bibinfo {author} {\bibfnamefont {A.}~\bibnamefont
  {{Bauswein}}}, \bibinfo {author} {\bibfnamefont {H.~T.}\ \bibnamefont
  {{Janka}}}, \bibinfo {author} {\bibfnamefont {J.~L.}\ \bibnamefont {{Sida}}},
  \bibinfo {author} {\bibfnamefont {J.~F.}\ \bibnamefont {{Lema{\^\i}tre}}},
  \bibinfo {author} {\bibfnamefont {S.}~\bibnamefont {{Panebianco}}}, \bibinfo
  {author} {\bibfnamefont {N.}~\bibnamefont {{Dubray}}}, and\ \bibinfo {author}
  {\bibfnamefont {S.}~\bibnamefont {{Hilaire}}}} (\bibinfo {year} {2014}),\
  \bibfield  {title} {\enquote {\bibinfo {title} {{The r-process
  nucleosynthesis during the decompression of neutron star crust material}},}\
  }in\ \href {https://doi.org/10.1063/1.4874095} {\emph {\bibinfo {booktitle}
  {American Institute of Physics Conference Series}}},\ Vol.\ \bibinfo {volume}
  {1594},\ \bibinfo {editor} {edited by\ \bibinfo {editor} {\bibfnamefont
  {Sunchan}\ \bibnamefont {{Jeong}}}, \bibinfo {editor} {\bibfnamefont
  {Nobuaki}\ \bibnamefont {{Imai}}}, \bibinfo {editor} {\bibfnamefont
  {Hiroari}\ \bibnamefont {{Miyatake}}}, \ and\ \bibinfo {editor}
  {\bibfnamefont {Toshitaka}\ \bibnamefont {{Kajino}}}},\ pp.\ \bibinfo {pages}
  {357--364}\BibitemShut {NoStop}%
\bibitem [{\citenamefont {{Goriely}}\ \emph {et~al.}(2015)\citenamefont
  {{Goriely}}, \citenamefont {{Bauswein}}, \citenamefont {{Just}},
  \citenamefont {{Pllumbi}},\ and\ \citenamefont
  {{Janka}}}]{Goriely.Bauswein.ea:2015}%
  \BibitemOpen
  \bibfield  {author} {\bibinfo {author} {\bibnamefont {{Goriely}},
  \bibfnamefont {S}}, \bibinfo {author} {\bibfnamefont {A.}~\bibnamefont
  {{Bauswein}}}, \bibinfo {author} {\bibfnamefont {O.}~\bibnamefont {{Just}}},
  \bibinfo {author} {\bibfnamefont {E.}~\bibnamefont {{Pllumbi}}}, and\
  \bibinfo {author} {\bibfnamefont {H.-T.}\ \bibnamefont {{Janka}}}} (\bibinfo
  {year} {2015}),\ \bibfield  {title} {\enquote {\bibinfo {title} {Impact of
  weak interactions of free nucleons on the r-process in dynamical ejecta from
  neutron star mergers},}\ }\href {https://doi.org/10.1093/mnras/stv1526}
  {\bibfield  {journal} {\bibinfo  {journal} {Mon. Not. Roy. Astron. Soc.}\
  }\textbf {\bibinfo {volume} {452}},\ \bibinfo {pages}
  {3894--3904}}\BibitemShut {NoStop}%
\bibitem [{\citenamefont {{Goriely}}\ \emph {et~al.}(2010)\citenamefont
  {{Goriely}}, \citenamefont {{Chamel}},\ and\ \citenamefont
  {{Pearson}}}]{Goriely.Chamel.Pearson:2010}%
  \BibitemOpen
  \bibfield  {author} {\bibinfo {author} {\bibnamefont {{Goriely}},
  \bibfnamefont {S}}, \bibinfo {author} {\bibfnamefont {N.}~\bibnamefont
  {{Chamel}}}, and\ \bibinfo {author} {\bibfnamefont {J.~M.}\ \bibnamefont
  {{Pearson}}}} (\bibinfo {year} {2010}),\ \bibfield  {title} {\enquote
  {\bibinfo {title} {{Further explorations of Skyrme-Hartree-Fock-Bogoliubov
  mass formulas. XII. Stiffness and stability of neutron-star matter}},}\
  }\href {https://doi.org/10.1103/PhysRevC.82.035804} {\bibfield  {journal}
  {\bibinfo  {journal} {Phys. Rev. C}\ }\textbf {\bibinfo {volume} {82}},\
  \bibinfo {eid} {035804}}\BibitemShut {NoStop}%
\bibitem [{\citenamefont {{Goriely}}\ \emph {et~al.}(2016)\citenamefont
  {{Goriely}}, \citenamefont {{Chamel}},\ and\ \citenamefont
  {{Pearson}}}]{Goriely.Chamel.Pearson:2016}%
  \BibitemOpen
  \bibfield  {author} {\bibinfo {author} {\bibnamefont {{Goriely}},
  \bibfnamefont {S}}, \bibinfo {author} {\bibfnamefont {N.}~\bibnamefont
  {{Chamel}}}, and\ \bibinfo {author} {\bibfnamefont {J.~M.}\ \bibnamefont
  {{Pearson}}}} (\bibinfo {year} {2016}),\ \bibfield  {title} {\enquote
  {\bibinfo {title} {{Further explorations of Skyrme-Hartree-Fock-Bogoliubov
  mass formulas. XVI. Inclusion of self-energy effects in pairing}},}\ }\href
  {https://doi.org/10.1103/PhysRevC.93.034337} {\bibfield  {journal} {\bibinfo
  {journal} {Phys. Rev. C}\ }\textbf {\bibinfo {volume} {93}},\ \bibinfo {eid}
  {034337}}\BibitemShut {NoStop}%
\bibitem [{\citenamefont {{Goriely}}\ and\ \citenamefont
  {{Clerbaux}}(1999)}]{Goriely.Clerbaux:1999}%
  \BibitemOpen
  \bibfield  {author} {\bibinfo {author} {\bibnamefont {{Goriely}},
  \bibfnamefont {S}}, and\ \bibinfo {author} {\bibfnamefont {B.}~\bibnamefont
  {{Clerbaux}}}} (\bibinfo {year} {1999}),\ \bibfield  {title} {\enquote
  {\bibinfo {title} {{Uncertainties in the Th cosmochronometry}},}\ }\href@noop
  {} {\bibfield  {journal} {\bibinfo  {journal} {Astron. \& Astrophys.}\
  }\textbf {\bibinfo {volume} {346}},\ \bibinfo {pages} {798--804}}\BibitemShut
  {NoStop}%
\bibitem [{\citenamefont {{Goriely}}\ and\ \citenamefont
  {{Delaroche}}(2007)}]{Goriely.Delaroche:2007}%
  \BibitemOpen
  \bibfield  {author} {\bibinfo {author} {\bibnamefont {{Goriely}},
  \bibfnamefont {S}}, and\ \bibinfo {author} {\bibfnamefont {J.-P.}\
  \bibnamefont {{Delaroche}}}} (\bibinfo {year} {2007}),\ \bibfield  {title}
  {\enquote {\bibinfo {title} {{The isovector imaginary neutron potential: A
  key ingredient for the r-process nucleosynthesis}},}\ }\href
  {https://doi.org/10.1016/j.physletb.2007.07.046} {\bibfield  {journal}
  {\bibinfo  {journal} {Phys. Lett. B}\ }\textbf {\bibinfo {volume} {653}},\
  \bibinfo {pages} {178--183}}\BibitemShut {NoStop}%
\bibitem [{\citenamefont {{Goriely}}\ \emph {et~al.}(2012)\citenamefont
  {{Goriely}}, \citenamefont {{Hilaire}},\ and\ \citenamefont
  {{Girod}}}]{Goriely.Hilaire.Girod:2012}%
  \BibitemOpen
  \bibfield  {author} {\bibinfo {author} {\bibnamefont {{Goriely}},
  \bibfnamefont {S}}, \bibinfo {author} {\bibfnamefont {S.}~\bibnamefont
  {{Hilaire}}}, and\ \bibinfo {author} {\bibfnamefont {M.}~\bibnamefont
  {{Girod}}}} (\bibinfo {year} {2012}),\ \bibfield  {title} {\enquote {\bibinfo
  {title} {{Latest development of the combinatorial model of nuclear level
  densities}},}\ }in\ \href {https://doi.org/10.1088/1742-6596/337/1/012027}
  {\emph {\bibinfo {booktitle} {Journal of Physics Conference Series}}},\ Vol.\
  \bibinfo {volume} {337},\ p.\ \bibinfo {pages} {012027}\BibitemShut {NoStop}%
\bibitem [{\citenamefont {{Goriely}}\ \emph {et~al.}(2008)\citenamefont
  {{Goriely}}, \citenamefont {{Hilaire}},\ and\ \citenamefont
  {{Koning}}}]{Goriely.Hilaire.Koning:2008}%
  \BibitemOpen
  \bibfield  {author} {\bibinfo {author} {\bibnamefont {{Goriely}},
  \bibfnamefont {S}}, \bibinfo {author} {\bibfnamefont {S.}~\bibnamefont
  {{Hilaire}}}, and\ \bibinfo {author} {\bibfnamefont {A.~J.}\ \bibnamefont
  {{Koning}}}} (\bibinfo {year} {2008}),\ \bibfield  {title} {\enquote
  {\bibinfo {title} {{Improved microscopic nuclear level densities within the
  Hartree-Fock-Bogoliubov plus combinatorial method}},}\ }\href
  {https://doi.org/10.1103/PhysRevC.78.064307} {\bibfield  {journal} {\bibinfo
  {journal} {Phys. Rev. C}\ }\textbf {\bibinfo {volume} {78}},\ \bibinfo {eid}
  {064307}}\BibitemShut {NoStop}%
\bibitem [{\citenamefont {Goriely}\ \emph {et~al.}(2009)\citenamefont
  {Goriely}, \citenamefont {Hilaire}, \citenamefont {Koning}, \citenamefont
  {Sin},\ and\ \citenamefont {Capote}}]{Goriely.Hilaire.ea:2009}%
  \BibitemOpen
  \bibfield  {author} {\bibinfo {author} {\bibnamefont {Goriely}, \bibfnamefont
  {S}}, \bibinfo {author} {\bibfnamefont {S.}~\bibnamefont {Hilaire}}, \bibinfo
  {author} {\bibfnamefont {A.~J.}\ \bibnamefont {Koning}}, \bibinfo {author}
  {\bibfnamefont {M.}~\bibnamefont {Sin}}, and\ \bibinfo {author}
  {\bibfnamefont {R.}~\bibnamefont {Capote}}} (\bibinfo {year} {2009}),\
  \bibfield  {title} {\enquote {\bibinfo {title} {Towards a prediction of
  fission cross sections on the basis of microscopic nuclear inputs},}\ }\href
  {https://doi.org/10.1103/PhysRevC.79.024612} {\bibfield  {journal} {\bibinfo
  {journal} {Phys. Rev. C}\ }\textbf {\bibinfo {volume} {79}},\ \bibinfo
  {pages} {024612}}\BibitemShut {NoStop}%
\bibitem [{\citenamefont {Goriely}\ \emph {et~al.}(2018)\citenamefont
  {Goriely}, \citenamefont {Hilaire}, \citenamefont {P\'eru},\ and\
  \citenamefont {Sieja}}]{Goriely.Hilaire.ea:2018}%
  \BibitemOpen
  \bibfield  {author} {\bibinfo {author} {\bibnamefont {Goriely}, \bibfnamefont
  {S}}, \bibinfo {author} {\bibfnamefont {S.}~\bibnamefont {Hilaire}}, \bibinfo
  {author} {\bibfnamefont {S.}~\bibnamefont {P\'eru}}, and\ \bibinfo {author}
  {\bibfnamefont {K.}~\bibnamefont {Sieja}}} (\bibinfo {year} {2018}),\
  \bibfield  {title} {\enquote {\bibinfo {title} {Gogny-hfb+qrpa dipole
  strength function and its application to radiative nucleon capture cross
  section},}\ }\href {https://doi.org/10.1103/PhysRevC.98.014327} {\bibfield
  {journal} {\bibinfo  {journal} {Phys. Rev. C}\ }\textbf {\bibinfo {volume}
  {98}},\ \bibinfo {pages} {014327}}\BibitemShut {NoStop}%
\bibitem [{\citenamefont {{Goriely}}\ and\ \citenamefont
  {{Janka}}(2016)}]{Goriely.Janka:2016}%
  \BibitemOpen
  \bibfield  {author} {\bibinfo {author} {\bibnamefont {{Goriely}},
  \bibfnamefont {S}}, and\ \bibinfo {author} {\bibfnamefont {H.-T.}\
  \bibnamefont {{Janka}}}} (\bibinfo {year} {2016}),\ \bibfield  {title}
  {\enquote {\bibinfo {title} {{Solar r-process-constrained actinide production
  in neutrino-driven winds of supernovae}},}\ }\href
  {https://doi.org/10.1093/mnras/stw946} {\bibfield  {journal} {\bibinfo
  {journal} {Mon. Not. Roy. Astron. Soc.}\ }\textbf {\bibinfo {volume} {459}},\
  \bibinfo {pages} {4174--4182}}\BibitemShut {NoStop}%
\bibitem [{\citenamefont {Goriely}\ and\ \citenamefont
  {Khan}(2002)}]{Goriely.Khan:2002}%
  \BibitemOpen
  \bibfield  {author} {\bibinfo {author} {\bibnamefont {Goriely}, \bibfnamefont
  {S}}, and\ \bibinfo {author} {\bibfnamefont {E.}~\bibnamefont {Khan}}}
  (\bibinfo {year} {2002}),\ \bibfield  {title} {\enquote {\bibinfo {title}
  {{Large-scale QRPA calculation of E1-strength and its impact on the neutron
  capture cross section}},}\ }\href
  {https://doi.org/10.1016/S0375-9474(02)00860-6} {\bibfield  {journal}
  {\bibinfo  {journal} {Nucl. Phys. A}\ }\textbf {\bibinfo {volume} {706}},\
  \bibinfo {pages} {217--232}}\BibitemShut {NoStop}%
\bibitem [{\citenamefont {{Goriely}}\ \emph {et~al.}(2004)\citenamefont
  {{Goriely}}, \citenamefont {{Khan}},\ and\ \citenamefont
  {{Samyn}}}]{Goriely.Khan.Samyn:2004}%
  \BibitemOpen
  \bibfield  {author} {\bibinfo {author} {\bibnamefont {{Goriely}},
  \bibfnamefont {S}}, \bibinfo {author} {\bibfnamefont {E.}~\bibnamefont
  {{Khan}}}, and\ \bibinfo {author} {\bibfnamefont {M.}~\bibnamefont
  {{Samyn}}}} (\bibinfo {year} {2004}),\ \bibfield  {title} {\enquote {\bibinfo
  {title} {{Microscopic HFB + QRPA predictions of dipole strength for
  astrophysics applications}},}\ }\href
  {https://doi.org/10.1016/j.nuclphysa.2004.04.105} {\bibfield  {journal}
  {\bibinfo  {journal} {Nucl. Phys. A}\ }\textbf {\bibinfo {volume} {739}},\
  \bibinfo {pages} {331--352}}\BibitemShut {NoStop}%
\bibitem [{\citenamefont {Goriely}\ and\ \citenamefont
  {Mart{\'i}nez-Pinedo}(2015)}]{Goriely.Martinez-Pinedo:2015}%
  \BibitemOpen
  \bibfield  {author} {\bibinfo {author} {\bibnamefont {Goriely}, \bibfnamefont
  {S}}, and\ \bibinfo {author} {\bibfnamefont {G.}~\bibnamefont
  {Mart{\'i}nez-Pinedo}}} (\bibinfo {year} {2015}),\ \bibfield  {title}
  {\enquote {\bibinfo {title} {The production of transuranium elements by the
  r-process nucleosynthesis},}\ }\href
  {https://doi.org/10.1016/j.nuclphysa.2015.07.020} {\bibfield  {journal}
  {\bibinfo  {journal} {Nucl. Phys. A}\ }\textbf {\bibinfo {volume} {944}},\
  \bibinfo {pages} {158--176}}\BibitemShut {NoStop}%
\bibitem [{\citenamefont {Goriely}\ \emph {et~al.}(2013)\citenamefont
  {Goriely}, \citenamefont {Sida}, \citenamefont {Lema{\^\i}tre}, \citenamefont
  {Panebianco}, \citenamefont {Dubray}, \citenamefont {Hilaire}, \citenamefont
  {Bauswein},\ and\ \citenamefont {Janka}}]{Goriely.Sida.ea:2013}%
  \BibitemOpen
  \bibfield  {author} {\bibinfo {author} {\bibnamefont {Goriely}, \bibfnamefont
  {S}}, \bibinfo {author} {\bibfnamefont {J.-L.}\ \bibnamefont {Sida}},
  \bibinfo {author} {\bibfnamefont {J.-F.}\ \bibnamefont {Lema{\^\i}tre}},
  \bibinfo {author} {\bibfnamefont {S.}~\bibnamefont {Panebianco}}, \bibinfo
  {author} {\bibfnamefont {N.}~\bibnamefont {Dubray}}, \bibinfo {author}
  {\bibfnamefont {S.}~\bibnamefont {Hilaire}}, \bibinfo {author} {\bibfnamefont
  {A.}~\bibnamefont {Bauswein}}, and\ \bibinfo {author} {\bibfnamefont {H.-T.}\
  \bibnamefont {Janka}}} (\bibinfo {year} {2013}),\ \bibfield  {title}
  {\enquote {\bibinfo {title} {New fission fragment distributions and
  $r$-process origin of the rare-earth elements},}\ }\href
  {https://doi.org/10.1103/PhysRevLett.111.242502} {\bibfield  {journal}
  {\bibinfo  {journal} {Phys. Rev. Lett.}\ }\textbf {\bibinfo {volume} {111}},\
  \bibinfo {pages} {242502}}\BibitemShut {NoStop}%
\bibitem [{\citenamefont {G{\"o}rres}\ \emph {et~al.}(1995)\citenamefont
  {G{\"o}rres}, \citenamefont {Wiescher},\ and\ \citenamefont
  {Thielemann}}]{Goerres.Wiescher.Thielemann:1995}%
  \BibitemOpen
  \bibfield  {author} {\bibinfo {author} {\bibnamefont {G{\"o}rres},
  \bibfnamefont {J}}, \bibinfo {author} {\bibfnamefont {M.}~\bibnamefont
  {Wiescher}}, and\ \bibinfo {author} {\bibfnamefont {F.-K.}\ \bibnamefont
  {Thielemann}}} (\bibinfo {year} {1995}),\ \bibfield  {title} {\enquote
  {\bibinfo {title} {{Bridging the waiting points: The role of two-proton
  capture reactions in the rp process}},}\ }\href
  {https://doi.org/10.1103/PhysRevC.51.392} {\bibfield  {journal} {\bibinfo
  {journal} {Phys. Rev. C}\ }\textbf {\bibinfo {volume} {51}},\ \bibinfo
  {pages} {392--400}}\BibitemShut {NoStop}%
\bibitem [{\citenamefont {{Gottlieb}}\ \emph {et~al.}(2018)\citenamefont
  {{Gottlieb}}, \citenamefont {{Nakar}},\ and\ \citenamefont
  {{Piran}}}]{Gottlieb.Nakar.Piran:2018}%
  \BibitemOpen
  \bibfield  {author} {\bibinfo {author} {\bibnamefont {{Gottlieb}},
  \bibfnamefont {Ore}}, \bibinfo {author} {\bibfnamefont {Ehud}\ \bibnamefont
  {{Nakar}}}, and\ \bibinfo {author} {\bibfnamefont {Tsvi}\ \bibnamefont
  {{Piran}}}} (\bibinfo {year} {2018}),\ \bibfield  {title} {\enquote {\bibinfo
  {title} {{The cocoon emission - an electromagnetic counterpart to
  gravitational waves from neutron star mergers}},}\ }\href
  {https://doi.org/10.1093/mnras/stx2357} {\bibfield  {journal} {\bibinfo
  {journal} {Mon. Not. Roy. Astron. Soc.}\ }\textbf {\bibinfo {volume} {473}},\
  \bibinfo {pages} {576--584}}\BibitemShut {NoStop}%
\bibitem [{\citenamefont {{Grawe}}\ \emph {et~al.}(2007)\citenamefont
  {{Grawe}}, \citenamefont {{Langanke}},\ and\ \citenamefont
  {{Mart{\'{\i}}nez-Pinedo}}}]{Grawe.Langanke.Martinez-Pinedo:2007}%
  \BibitemOpen
  \bibfield  {author} {\bibinfo {author} {\bibnamefont {{Grawe}}, \bibfnamefont
  {H}}, \bibinfo {author} {\bibfnamefont {K.}~\bibnamefont {{Langanke}}}, and\
  \bibinfo {author} {\bibfnamefont {G.}~\bibnamefont
  {{Mart{\'{\i}}nez-Pinedo}}}} (\bibinfo {year} {2007}),\ \bibfield  {title}
  {\enquote {\bibinfo {title} {{Nuclear structure and astrophysics}},}\ }\href
  {https://doi.org/10.1088/0034-4885/70/9/R02} {\bibfield  {journal} {\bibinfo
  {journal} {Rep. Prog. Phys.}\ }\textbf {\bibinfo {volume} {70}},\ \bibinfo
  {pages} {1525--1582}}\BibitemShut {NoStop}%
\bibitem [{\citenamefont {{Grebenev}}\ \emph {et~al.}(2012)\citenamefont
  {{Grebenev}}, \citenamefont {{Lutovinov}}, \citenamefont {{Tsygankov}},\ and\
  \citenamefont {{Winkler}}}]{Grebenev.Lutovinov.ea:2012}%
  \BibitemOpen
  \bibfield  {author} {\bibinfo {author} {\bibnamefont {{Grebenev}},
  \bibfnamefont {S~A}}, \bibinfo {author} {\bibfnamefont {A.~A.}\ \bibnamefont
  {{Lutovinov}}}, \bibinfo {author} {\bibfnamefont {S.~S.}\ \bibnamefont
  {{Tsygankov}}}, and\ \bibinfo {author} {\bibfnamefont {C.}~\bibnamefont
  {{Winkler}}}} (\bibinfo {year} {2012}),\ \bibfield  {title} {\enquote
  {\bibinfo {title} {{Hard-X-ray emission lines from the decay of $^{44}$Ti in
  the remnant of supernova 1987A}},}\ }\href
  {https://doi.org/10.1038/nature11473} {\bibfield  {journal} {\bibinfo
  {journal} {Nature}\ }\textbf {\bibinfo {volume} {490}},\ \bibinfo {pages}
  {373--375}}\BibitemShut {NoStop}%
\bibitem [{\citenamefont {{Greiner}}\ \emph {et~al.}(2015)\citenamefont
  {{Greiner}} \emph {et~al.}}]{greiner15}%
  \BibitemOpen
  \bibfield  {author} {\bibinfo {author} {\bibnamefont {{Greiner}},
  \bibfnamefont {J}},  \emph {et~al.}} (\bibinfo {year} {2015}),\ \bibfield
  {title} {\enquote {\bibinfo {title} {{A very luminous magnetar-powered
  supernova associated with an ultra-long {$\gamma$}-ray burst}},}\ }\href
  {https://doi.org/10.1038/nature14579} {\bibfield  {journal} {\bibinfo
  {journal} {Nature}\ }\textbf {\bibinfo {volume} {523}},\ \bibinfo {pages}
  {189--192}}\BibitemShut {NoStop}%
\bibitem [{\citenamefont {{Grichener}}\ and\ \citenamefont
  {{Soker}}(2019)}]{Grichener.Soker:2019}%
  \BibitemOpen
  \bibfield  {author} {\bibinfo {author} {\bibnamefont {{Grichener}},
  \bibfnamefont {Aldana}}, and\ \bibinfo {author} {\bibfnamefont {Noam}\
  \bibnamefont {{Soker}}}} (\bibinfo {year} {2019}),\ \bibfield  {title}
  {\enquote {\bibinfo {title} {{The Common Envelope Jet Supernova (CEJSN)
  r-process Scenario}},}\ }\href {https://doi.org/10.3847/1538-4357/ab1d5d}
  {\bibfield  {journal} {\bibinfo  {journal} {Astrophys. J.}\ }\textbf
  {\bibinfo {volume} {878}},\ \bibinfo {eid} {24}}\BibitemShut {NoStop}%
\bibitem [{\citenamefont {{Griffin}}\ \emph {et~al.}(1982)\citenamefont
  {{Griffin}}, \citenamefont {{Gustafsson}}, \citenamefont {{Vieira}},\ and\
  \citenamefont {{Griffin}}}]{griffin82}%
  \BibitemOpen
  \bibfield  {author} {\bibinfo {author} {\bibnamefont {{Griffin}},
  \bibfnamefont {R}}, \bibinfo {author} {\bibfnamefont {B.}~\bibnamefont
  {{Gustafsson}}}, \bibinfo {author} {\bibfnamefont {T.}~\bibnamefont
  {{Vieira}}}, and\ \bibinfo {author} {\bibfnamefont {R.}~\bibnamefont
  {{Griffin}}}} (\bibinfo {year} {1982}),\ \bibfield  {title} {\enquote
  {\bibinfo {title} {{HD 115444 - A barium star of extreme Population II}},}\
  }\href {https://doi.org/10.1093/mnras/198.3.637} {\bibfield  {journal}
  {\bibinfo  {journal} {Mon. Not. Roy. Astron. Soc.}\ }\textbf {\bibinfo
  {volume} {198}},\ \bibinfo {pages} {637--658}}\BibitemShut {NoStop}%
\bibitem [{\citenamefont {{Grisoni}}\ \emph {et~al.}(2020)\citenamefont
  {{Grisoni}}, \citenamefont {{Cescutti}}, \citenamefont {{Matteucci}},
  \citenamefont {{Forsberg}}, \citenamefont {{J{\"o}nsson}},\ and\
  \citenamefont {{Ryde}}}]{Grisoni.Cescutti.ea:2020}%
  \BibitemOpen
  \bibfield  {author} {\bibinfo {author} {\bibnamefont {{Grisoni}},
  \bibfnamefont {V}}, \bibinfo {author} {\bibfnamefont {G.}~\bibnamefont
  {{Cescutti}}}, \bibinfo {author} {\bibfnamefont {F.}~\bibnamefont
  {{Matteucci}}}, \bibinfo {author} {\bibfnamefont {R.}~\bibnamefont
  {{Forsberg}}}, \bibinfo {author} {\bibfnamefont {H.}~\bibnamefont
  {{J{\"o}nsson}}}, and\ \bibinfo {author} {\bibfnamefont {N.}~\bibnamefont
  {{Ryde}}}} (\bibinfo {year} {2020}),\ \bibfield  {title} {\enquote {\bibinfo
  {title} {{Modelling the chemical evolution of Zr, La, Ce, and Eu in the
  Galactic discs and bulge}},}\ }\href {https://doi.org/10.1093/mnras/staa051}
  {\bibfield  {journal} {\bibinfo  {journal} {Mon. Not. Roy. Astron. Soc.}\
  }\textbf {\bibinfo {volume} {492}},\ \bibinfo {pages}
  {2828--2834}}\BibitemShut {NoStop}%
\bibitem [{\citenamefont {{Grossjean}}\ and\ \citenamefont
  {{Feldmeier}}(1985)}]{Grossjean.Feldmeier:1985}%
  \BibitemOpen
  \bibfield  {author} {\bibinfo {author} {\bibnamefont {{Grossjean}},
  \bibfnamefont {M~K}}, and\ \bibinfo {author} {\bibfnamefont {H.}~\bibnamefont
  {{Feldmeier}}}} (\bibinfo {year} {1985}),\ \bibfield  {title} {\enquote
  {\bibinfo {title} {{Level density of a Fermi gas with pairing
  interactions}},}\ }\href {https://doi.org/10.1016/0375-9474(85)90294-5}
  {\bibfield  {journal} {\bibinfo  {journal} {Nucl. Phys. A}\ }\textbf
  {\bibinfo {volume} {444}},\ \bibinfo {pages} {113--132}}\BibitemShut
  {NoStop}%
\bibitem [{\citenamefont {{Grossman}}\ \emph {et~al.}(2014)\citenamefont
  {{Grossman}}, \citenamefont {{Korobkin}}, \citenamefont {{Rosswog}},\ and\
  \citenamefont {{Piran}}}]{grossman14}%
  \BibitemOpen
  \bibfield  {author} {\bibinfo {author} {\bibnamefont {{Grossman}},
  \bibfnamefont {D}}, \bibinfo {author} {\bibfnamefont {O.}~\bibnamefont
  {{Korobkin}}}, \bibinfo {author} {\bibfnamefont {S.}~\bibnamefont
  {{Rosswog}}}, and\ \bibinfo {author} {\bibfnamefont {T.}~\bibnamefont
  {{Piran}}}} (\bibinfo {year} {2014}),\ \bibfield  {title} {\enquote {\bibinfo
  {title} {{The long-term evolution of neutron star merger remnants - II.
  Radioactively powered transients}},}\ }\href
  {https://doi.org/10.1093/mnras/stt2503} {\bibfield  {journal} {\bibinfo
  {journal} {Mon. Not. Roy. Astron. Soc.}\ }\textbf {\bibinfo {volume} {439}},\
  \bibinfo {pages} {757--770}}\BibitemShut {NoStop}%
\bibitem [{\citenamefont {{Gudin}}\ \emph {et~al.}(2020)\citenamefont
  {{Gudin}}, \citenamefont {{Shank}}, \citenamefont {{Beers}}, \citenamefont
  {{Yuan}}, \citenamefont {{Limberg}}, \citenamefont {{Roederer}},
  \citenamefont {{Placco}}, \citenamefont {{Holmbeck}}, \citenamefont
  {{Dietz}}, \citenamefont {{Rasmussen}}, \citenamefont {{Hansen}},
  \citenamefont {{Sakari}}, \citenamefont {{Ezzeddine}},\ and\ \citenamefont
  {{Frebel}}}]{Gudin.Shank.ea:2020}%
  \BibitemOpen
  \bibfield  {author} {\bibinfo {author} {\bibnamefont {{Gudin}}, \bibfnamefont
  {Dmitrii}}, \bibinfo {author} {\bibfnamefont {Derek}\ \bibnamefont
  {{Shank}}}, \bibinfo {author} {\bibfnamefont {Timothy~C.}\ \bibnamefont
  {{Beers}}}, \bibinfo {author} {\bibfnamefont {Zhen}\ \bibnamefont {{Yuan}}},
  \bibinfo {author} {\bibfnamefont {Guilherme}\ \bibnamefont {{Limberg}}},
  \bibinfo {author} {\bibfnamefont {Ian~U.}\ \bibnamefont {{Roederer}}},
  \bibinfo {author} {\bibfnamefont {Vinicius}\ \bibnamefont {{Placco}}},
  \bibinfo {author} {\bibfnamefont {Erika~M.}\ \bibnamefont {{Holmbeck}}},
  \bibinfo {author} {\bibfnamefont {Sarah}\ \bibnamefont {{Dietz}}}, \bibinfo
  {author} {\bibfnamefont {Kaitlin~C.}\ \bibnamefont {{Rasmussen}}}, \bibinfo
  {author} {\bibfnamefont {Terese~T.}\ \bibnamefont {{Hansen}}}, \bibinfo
  {author} {\bibfnamefont {Charli~M.}\ \bibnamefont {{Sakari}}}, \bibinfo
  {author} {\bibfnamefont {Rana}\ \bibnamefont {{Ezzeddine}}}, and\ \bibinfo
  {author} {\bibfnamefont {Anna}\ \bibnamefont {{Frebel}}}} (\bibinfo {year}
  {2020}),\ \bibfield  {title} {\enquote {\bibinfo {title} {{The R-Process
  Alliance: Chemo-Dynamically Tagged Groups of Halo $r$-Process-Enhanced Stars
  Reveal a Shared Chemical-Evolution History}},}\ }\href@noop {} {\bibfield
  {journal} {\bibinfo  {journal} {arXiv e-prints}\ }}\Eprint
  {https://arxiv.org/abs/2012.13808} {arXiv:2012.13808 [astro-ph.GA]}
  \BibitemShut {NoStop}%
\bibitem [{\citenamefont {{Guerrero}}\ \emph {et~al.}(2017)\citenamefont
  {{Guerrero}}, \citenamefont {{Domingo-Pardo}}, \citenamefont
  {{K{\"a}ppeler}}, \citenamefont {{Lerendegui-Marco}}, \citenamefont
  {{Palomo}}, \citenamefont {{Quesada}},\ and\ \citenamefont
  {{Reifarth}}}]{Guerrero2017}%
  \BibitemOpen
  \bibfield  {author} {\bibinfo {author} {\bibnamefont {{Guerrero}},
  \bibfnamefont {C}}, \bibinfo {author} {\bibfnamefont {C.}~\bibnamefont
  {{Domingo-Pardo}}}, \bibinfo {author} {\bibfnamefont {F.}~\bibnamefont
  {{K{\"a}ppeler}}}, \bibinfo {author} {\bibfnamefont {J.}~\bibnamefont
  {{Lerendegui-Marco}}}, \bibinfo {author} {\bibfnamefont {F.~R.}\ \bibnamefont
  {{Palomo}}}, \bibinfo {author} {\bibfnamefont {J.~M.}\ \bibnamefont
  {{Quesada}}}, and\ \bibinfo {author} {\bibfnamefont {R.}~\bibnamefont
  {{Reifarth}}}} (\bibinfo {year} {2017}),\ \bibfield  {title} {\enquote
  {\bibinfo {title} {{Prospects for direct neutron capture measurements on
  s-process branching point isotopes}},}\ }\href
  {https://doi.org/10.1140/epja/i2017-12261-2} {\bibfield  {journal} {\bibinfo
  {journal} {Eur. Phys. J. A}\ }\textbf {\bibinfo {volume} {53}},\ \bibinfo
  {eid} {87}}\BibitemShut {NoStop}%
\bibitem [{\citenamefont {{Guilet}}\ \emph {et~al.}(2017)\citenamefont
  {{Guilet}}, \citenamefont {{Bauswein}}, \citenamefont {{Just}},\ and\
  \citenamefont {{Janka}}}]{Guilet.Bauswein.ea:2017}%
  \BibitemOpen
  \bibfield  {author} {\bibinfo {author} {\bibnamefont {{Guilet}},
  \bibfnamefont {J}}, \bibinfo {author} {\bibfnamefont {A.}~\bibnamefont
  {{Bauswein}}}, \bibinfo {author} {\bibfnamefont {O.}~\bibnamefont {{Just}}},
  and\ \bibinfo {author} {\bibfnamefont {H.-T.}\ \bibnamefont {{Janka}}}}
  (\bibinfo {year} {2017}),\ \bibfield  {title} {\enquote {\bibinfo {title}
  {Magnetorotational instability in neutron star mergers: impact of
  neutrinos},}\ }\href {https://doi.org/10.1093/mnras/stx1739} {\bibfield
  {journal} {\bibinfo  {journal} {Mon. Not. Roy. Astron. Soc.}\ }\textbf
  {\bibinfo {volume} {471}},\ \bibinfo {pages} {1879--1887}}\BibitemShut
  {NoStop}%
\bibitem [{\citenamefont {{Guttormsen}}\ \emph {et~al.}(1987)\citenamefont
  {{Guttormsen}}, \citenamefont {{Rams{\o}y}},\ and\ \citenamefont
  {{Rekstad}}}]{Guttormsen1987}%
  \BibitemOpen
  \bibfield  {author} {\bibinfo {author} {\bibnamefont {{Guttormsen}},
  \bibfnamefont {M}}, \bibinfo {author} {\bibfnamefont {T.}~\bibnamefont
  {{Rams{\o}y}}}, and\ \bibinfo {author} {\bibfnamefont {J.}~\bibnamefont
  {{Rekstad}}}} (\bibinfo {year} {1987}),\ \bibfield  {title} {\enquote
  {\bibinfo {title} {{The first generation of {$\gamma$}-rays from hot
  nuclei}},}\ }\href {https://doi.org/10.1016/0168-9002(87)91221-6} {\bibfield
  {journal} {\bibinfo  {journal} {Nucl. Instruments Methods Phys. Res. Sect. A
  Accel. Spectrometers, Detect. Assoc. Equip.}\ }\textbf {\bibinfo {volume}
  {255}},\ \bibinfo {pages} {518--523}}\BibitemShut {NoStop}%
\bibitem [{\citenamefont {{Guttormsen}}\ \emph {et~al.}(2005)\citenamefont
  {{Guttormsen}} \emph {et~al.}}]{Guttormsen.Chankova.ea:2005}%
  \BibitemOpen
  \bibfield  {author} {\bibinfo {author} {\bibnamefont {{Guttormsen}},
  \bibfnamefont {M}},  \emph {et~al.}} (\bibinfo {year} {2005}),\ \bibfield
  {title} {\enquote {\bibinfo {title} {{Radiative strength functions in
  $^{93-98}$Mo}},}\ }\href {https://doi.org/10.1103/PhysRevC.71.044307}
  {\bibfield  {journal} {\bibinfo  {journal} {Phys. Rev. C}\ }\textbf {\bibinfo
  {volume} {71}},\ \bibinfo {eid} {044307}}\BibitemShut {NoStop}%
\bibitem [{\citenamefont {Haensel}\ \emph {et~al.}(2007)\citenamefont
  {Haensel}, \citenamefont {Potekhin},\ and\ \citenamefont
  {Yakovlev}}]{Haensel.Potekhin.Yakovlev:2007}%
  \BibitemOpen
  \bibfield  {author} {\bibinfo {author} {\bibnamefont {Haensel}, \bibfnamefont
  {P}}, \bibinfo {author} {\bibfnamefont {A.~Y.}\ \bibnamefont {Potekhin}},
  and\ \bibinfo {author} {\bibfnamefont {D.~G.}\ \bibnamefont {Yakovlev}}}
  (\bibinfo {year} {2007}),\ \href {https://doi.org/10.1007/978-0-387-47301-7}
  {\emph {\bibinfo {title} {{Neutron Stars 1: Equation of State and
  Structure}}}},\ \bibinfo {series} {Astrophysics and Space Science Library},
  Vol.\ \bibinfo {volume} {326}\ (\bibinfo  {publisher} {Springer},\ \bibinfo
  {address} {New York})\BibitemShut {NoStop}%
\bibitem [{\citenamefont {{Halevi}}\ and\ \citenamefont
  {{M{\"o}sta}}(2018)}]{halevi18}%
  \BibitemOpen
  \bibfield  {author} {\bibinfo {author} {\bibnamefont {{Halevi}},
  \bibfnamefont {G}}, and\ \bibinfo {author} {\bibfnamefont {P.}~\bibnamefont
  {{M{\"o}sta}}}} (\bibinfo {year} {2018}),\ \bibfield  {title} {\enquote
  {\bibinfo {title} {{r-Process Nucleosynthesis from Three-dimensional
  Jet-driven Core-Collapse Supernovae with Magnetic Misalignments}},}\ }\href
  {https://doi.org/10.1093/mnras/sty797} {\bibfield  {journal} {\bibinfo
  {journal} {Mon. Not. Roy. Astron. Soc.}\ }\textbf {\bibinfo {volume} {477}},\
  \bibinfo {pages} {2366--2375}}\BibitemShut {NoStop}%
\bibitem [{\citenamefont {{Hamers}}\ and\ \citenamefont
  {{Thompson}}(2019)}]{Hamers.Thompson:2019}%
  \BibitemOpen
  \bibfield  {author} {\bibinfo {author} {\bibnamefont {{Hamers}},
  \bibfnamefont {Adrian~S}}, and\ \bibinfo {author} {\bibfnamefont {Todd~A.}\
  \bibnamefont {{Thompson}}}} (\bibinfo {year} {2019}),\ \bibfield  {title}
  {\enquote {\bibinfo {title} {{Double Neutron Star Mergers from Hierarchical
  Triple-star Systems}},}\ }\href {https://doi.org/10.3847/1538-4357/ab3b06}
  {\bibfield  {journal} {\bibinfo  {journal} {Astrophys. J.}\ }\textbf
  {\bibinfo {volume} {883}},\ \bibinfo {eid} {23}}\BibitemShut {NoStop}%
\bibitem [{\citenamefont {{Hannaford}}\ and\ \citenamefont
  {{Lowe}}(1981)}]{hannaford81}%
  \BibitemOpen
  \bibfield  {author} {\bibinfo {author} {\bibnamefont {{Hannaford}},
  \bibfnamefont {P}}, and\ \bibinfo {author} {\bibfnamefont {R.~M.}\
  \bibnamefont {{Lowe}}}} (\bibinfo {year} {1981}),\ \bibfield  {title}
  {\enquote {\bibinfo {title} {{Determination of atomic lifetimes using pulsed
  laser excitation of sputtered metal vapours}},}\ }\href
  {https://doi.org/10.1088/0022-3700/14/1/002} {\bibfield  {journal} {\bibinfo
  {journal} {Journal of Physics B Atomic Molecular Physics}\ }\textbf {\bibinfo
  {volume} {14}},\ \bibinfo {pages} {L5--L9}}\BibitemShut {NoStop}%
\bibitem [{\citenamefont {{Hansen}}\ and\ \citenamefont
  {{Primas}}(2011)}]{Hansen.Primas:2011}%
  \BibitemOpen
  \bibfield  {author} {\bibinfo {author} {\bibnamefont {{Hansen}},
  \bibfnamefont {C~J}}, and\ \bibinfo {author} {\bibfnamefont {F.}~\bibnamefont
  {{Primas}}}} (\bibinfo {year} {2011}),\ \bibfield  {title} {\enquote
  {\bibinfo {title} {{The origin of palladium and silver}},}\ }\href
  {https://doi.org/10.1051/0004-6361/201015743} {\bibfield  {journal} {\bibinfo
   {journal} {Astron. \& Astrophys.}\ }\textbf {\bibinfo {volume} {525}},\
  \bibinfo {eid} {L5}}\BibitemShut {NoStop}%
\bibitem [{\citenamefont {{Hansen}}\ \emph {et~al.}(2012)\citenamefont
  {{Hansen}}, \citenamefont {{Primas}}, \citenamefont {{Hartman}},
  \citenamefont {{Kratz}}, \citenamefont {{Wanajo}}, \citenamefont
  {{Leibundgut}}, \citenamefont {{Farouqi}}, \citenamefont {{Hallmann}},
  \citenamefont {{Christlieb}},\ and\ \citenamefont
  {{Nilsson}}}]{Hansen.Primas.ea:2012}%
  \BibitemOpen
  \bibfield  {author} {\bibinfo {author} {\bibnamefont {{Hansen}},
  \bibfnamefont {C~J}}, \bibinfo {author} {\bibfnamefont {F.}~\bibnamefont
  {{Primas}}}, \bibinfo {author} {\bibfnamefont {H.}~\bibnamefont {{Hartman}}},
  \bibinfo {author} {\bibfnamefont {K.-L.}\ \bibnamefont {{Kratz}}}, \bibinfo
  {author} {\bibfnamefont {S.}~\bibnamefont {{Wanajo}}}, \bibinfo {author}
  {\bibfnamefont {B.}~\bibnamefont {{Leibundgut}}}, \bibinfo {author}
  {\bibfnamefont {K.}~\bibnamefont {{Farouqi}}}, \bibinfo {author}
  {\bibfnamefont {O.}~\bibnamefont {{Hallmann}}}, \bibinfo {author}
  {\bibfnamefont {N.}~\bibnamefont {{Christlieb}}}, and\ \bibinfo {author}
  {\bibfnamefont {H.}~\bibnamefont {{Nilsson}}}} (\bibinfo {year} {2012}),\
  \bibfield  {title} {\enquote {\bibinfo {title} {{Silver and palladium help
  unveil the nature of a second r-process}},}\ }\href
  {https://doi.org/10.1051/0004-6361/201118643} {\bibfield  {journal} {\bibinfo
   {journal} {Astron. \& Astrophys.}\ }\textbf {\bibinfo {volume} {545}},\
  \bibinfo {eid} {A31}}\BibitemShut {NoStop}%
\bibitem [{\citenamefont {{Hansen}}\ \emph {et~al.}(2017)\citenamefont
  {{Hansen}} \emph {et~al.}}]{Hansen.Simon.ea:2017}%
  \BibitemOpen
  \bibfield  {author} {\bibinfo {author} {\bibnamefont {{Hansen}},
  \bibfnamefont {T~T}},  \emph {et~al.} (\bibinfo {collaboration} {DES
  Collaboration})} (\bibinfo {year} {2017}),\ \bibfield  {title} {\enquote
  {\bibinfo {title} {{An r-process Enhanced Star in the Dwarf Galaxy Tucana
  III}},}\ }\href {https://doi.org/10.3847/1538-4357/aa634a} {\bibfield
  {journal} {\bibinfo  {journal} {Astrophys. J.}\ }\textbf {\bibinfo {volume}
  {838}},\ \bibinfo {eid} {44}}\BibitemShut {NoStop}%
\bibitem [{\citenamefont {{Hansen}}\ \emph {et~al.}(2018)\citenamefont
  {{Hansen}}, \citenamefont {{Holmbeck}}, \citenamefont {{Beers}},
  \citenamefont {{Placco}}, \citenamefont {{Roederer}}, \citenamefont
  {{Frebel}}, \citenamefont {{Sakari}}, \citenamefont {{Simon}},\ and\
  \citenamefont {{Thompson}}}]{Hansen.Holmbeck.ea:2018}%
  \BibitemOpen
  \bibfield  {author} {\bibinfo {author} {\bibnamefont {{Hansen}},
  \bibfnamefont {Terese~T}}, \bibinfo {author} {\bibfnamefont {Erika~M.}\
  \bibnamefont {{Holmbeck}}}, \bibinfo {author} {\bibfnamefont {Timothy~C.}\
  \bibnamefont {{Beers}}}, \bibinfo {author} {\bibfnamefont {Vinicius~M.}\
  \bibnamefont {{Placco}}}, \bibinfo {author} {\bibfnamefont {Ian~U.}\
  \bibnamefont {{Roederer}}}, \bibinfo {author} {\bibfnamefont {Anna}\
  \bibnamefont {{Frebel}}}, \bibinfo {author} {\bibfnamefont {Charli~M.}\
  \bibnamefont {{Sakari}}}, \bibinfo {author} {\bibfnamefont {Joshua~D.}\
  \bibnamefont {{Simon}}}, and\ \bibinfo {author} {\bibfnamefont {Ian~B.}\
  \bibnamefont {{Thompson}}}} (\bibinfo {year} {2018}),\ \bibfield  {title}
  {\enquote {\bibinfo {title} {{The R-process Alliance: First Release from the
  Southern Search for R-process-enhanced Stars in the Galactic Halo}},}\ }\href
  {https://doi.org/10.3847/1538-4357/aabacc} {\bibfield  {journal} {\bibinfo
  {journal} {Astrophys. J.}\ }\textbf {\bibinfo {volume} {858}},\ \bibinfo
  {eid} {92}}\BibitemShut {NoStop}%
\bibitem [{\citenamefont {{Harikae}}\ \emph {et~al.}(2009)\citenamefont
  {{Harikae}}, \citenamefont {{Takiwaki}},\ and\ \citenamefont
  {{Kotake}}}]{harikae09}%
  \BibitemOpen
  \bibfield  {author} {\bibinfo {author} {\bibnamefont {{Harikae}},
  \bibfnamefont {Seiji}}, \bibinfo {author} {\bibfnamefont {Tomoya}\
  \bibnamefont {{Takiwaki}}}, and\ \bibinfo {author} {\bibfnamefont {Kei}\
  \bibnamefont {{Kotake}}}} (\bibinfo {year} {2009}),\ \bibfield  {title}
  {\enquote {\bibinfo {title} {{Long-Term Evolution of Slowly Rotating
  Collapsar in Special Relativistic Magnetohydrodynamics}},}\ }\href
  {https://doi.org/10.1088/0004-637X/704/1/354} {\bibfield  {journal} {\bibinfo
   {journal} {Astrophys. J.}\ }\textbf {\bibinfo {volume} {704}},\ \bibinfo
  {pages} {354--371}}\BibitemShut {NoStop}%
\bibitem [{\citenamefont {{Hartmann}}\ \emph {et~al.}(1985)\citenamefont
  {{Hartmann}}, \citenamefont {{Woosley}},\ and\ \citenamefont {{El
  Eid}}}]{Hartmann.Woosley.El:1985}%
  \BibitemOpen
  \bibfield  {author} {\bibinfo {author} {\bibnamefont {{Hartmann}},
  \bibfnamefont {D}}, \bibinfo {author} {\bibfnamefont {S.~E.}\ \bibnamefont
  {{Woosley}}}, and\ \bibinfo {author} {\bibfnamefont {M.~F.}\ \bibnamefont
  {{El Eid}}}} (\bibinfo {year} {1985}),\ \bibfield  {title} {\enquote
  {\bibinfo {title} {{Nucleosynthesis in neutron-rich supernova ejecta}},}\
  }\href {https://doi.org/10.1086/163580} {\bibfield  {journal} {\bibinfo
  {journal} {Astrophys. J.}\ }\textbf {\bibinfo {volume} {297}},\ \bibinfo
  {pages} {837--845}}\BibitemShut {NoStop}%
\bibitem [{\citenamefont {{Hatarik}}\ \emph {et~al.}(2010)\citenamefont
  {{Hatarik}} \emph {et~al.}}]{Hatarik2010}%
  \BibitemOpen
  \bibfield  {author} {\bibinfo {author} {\bibnamefont {{Hatarik}},
  \bibfnamefont {R}},  \emph {et~al.}} (\bibinfo {year} {2010}),\ \bibfield
  {title} {\enquote {\bibinfo {title} {{Benchmarking a surrogate reaction for
  neutron capture}},}\ }\href {https://doi.org/10.1103/PhysRevC.81.011602}
  {\bibfield  {journal} {\bibinfo  {journal} {Phys. Rev. C}\ }\textbf {\bibinfo
  {volume} {81}},\ \bibinfo {eid} {011602}}\BibitemShut {NoStop}%
\bibitem [{\citenamefont {{Hausmann}}\ \emph {et~al.}(2000)\citenamefont
  {{Hausmann}} \emph {et~al.}}]{Hausmann2000}%
  \BibitemOpen
  \bibfield  {author} {\bibinfo {author} {\bibnamefont {{Hausmann}},
  \bibfnamefont {M}},  \emph {et~al.}} (\bibinfo {year} {2000}),\ \bibfield
  {title} {\enquote {\bibinfo {title} {{First isochronous mass spectrometry at
  the experimental storage ring ESR}},}\ }\href
  {https://doi.org/10.1016/S0168-9002(99)01192-4} {\bibfield  {journal}
  {\bibinfo  {journal} {Nucl. Instruments Methods Phys. Res. Sect. A Accel.
  Spectrometers, Detect. Assoc. Equip.}\ }\textbf {\bibinfo {volume} {446}},\
  \bibinfo {pages} {569--580}}\BibitemShut {NoStop}%
\bibitem [{\citenamefont {{Hausmann}}\ \emph {et~al.}(2001)\citenamefont
  {{Hausmann}} \emph {et~al.}}]{Hausmann2001}%
  \BibitemOpen
  \bibfield  {author} {\bibinfo {author} {\bibnamefont {{Hausmann}},
  \bibfnamefont {M}},  \emph {et~al.}} (\bibinfo {year} {2001}),\ \bibfield
  {title} {\enquote {\bibinfo {title} {{Isochronous Mass Measurements of Hot
  Exotic Nuclei}},}\ }\href {https://doi.org/10.1023/A:1011911720453}
  {\bibfield  {journal} {\bibinfo  {journal} {Hyperfine Interactions}\ }\textbf
  {\bibinfo {volume} {132}},\ \bibinfo {pages} {289--295}}\BibitemShut
  {NoStop}%
\bibitem [{\citenamefont {Haxton}\ \emph {et~al.}(1997)\citenamefont {Haxton},
  \citenamefont {Langanke}, \citenamefont {Qian},\ and\ \citenamefont
  {Vogel}}]{Haxton.Langanke.ea:1997}%
  \BibitemOpen
  \bibfield  {author} {\bibinfo {author} {\bibnamefont {Haxton}, \bibfnamefont
  {W~C}}, \bibinfo {author} {\bibfnamefont {K.}~\bibnamefont {Langanke}},
  \bibinfo {author} {\bibfnamefont {Y.-Z.}\ \bibnamefont {Qian}}, and\ \bibinfo
  {author} {\bibfnamefont {P.}~\bibnamefont {Vogel}}} (\bibinfo {year}
  {1997}),\ \bibfield  {title} {\enquote {\bibinfo {title} {Neutrino-induced
  nucleosynthesis and the site of the $\mathit{r}$ process},}\ }\href
  {https://doi.org/10.1103/PhysRevLett.78.2694} {\bibfield  {journal} {\bibinfo
   {journal} {Phys. Rev. Lett.}\ }\textbf {\bibinfo {volume} {78}},\ \bibinfo
  {pages} {2694--2697}}\BibitemShut {NoStop}%
\bibitem [{\citenamefont {{Hayek}}\ \emph {et~al.}(2009)\citenamefont
  {{Hayek}}, \citenamefont {{Wiesendahl}}, \citenamefont {{Christlieb}},
  \citenamefont {{Eriksson}}, \citenamefont {{Korn}}, \citenamefont
  {{Barklem}}, \citenamefont {{Hill}}, \citenamefont {{Beers}}, \citenamefont
  {{Farouqi}}, \citenamefont {{Pfeiffer}},\ and\ \citenamefont
  {{Kratz}}}]{Hayek.Wiesendahl.ea:2009}%
  \BibitemOpen
  \bibfield  {author} {\bibinfo {author} {\bibnamefont {{Hayek}}, \bibfnamefont
  {W}}, \bibinfo {author} {\bibfnamefont {U.}~\bibnamefont {{Wiesendahl}}},
  \bibinfo {author} {\bibfnamefont {N.}~\bibnamefont {{Christlieb}}}, \bibinfo
  {author} {\bibfnamefont {K.}~\bibnamefont {{Eriksson}}}, \bibinfo {author}
  {\bibfnamefont {A.~J.}\ \bibnamefont {{Korn}}}, \bibinfo {author}
  {\bibfnamefont {P.~S.}\ \bibnamefont {{Barklem}}}, \bibinfo {author}
  {\bibfnamefont {V.}~\bibnamefont {{Hill}}}, \bibinfo {author} {\bibfnamefont
  {T.~C.}\ \bibnamefont {{Beers}}}, \bibinfo {author} {\bibfnamefont
  {K.}~\bibnamefont {{Farouqi}}}, \bibinfo {author} {\bibfnamefont
  {B.}~\bibnamefont {{Pfeiffer}}}, and\ \bibinfo {author} {\bibfnamefont
  {K.-L.}\ \bibnamefont {{Kratz}}}} (\bibinfo {year} {2009}),\ \bibfield
  {title} {\enquote {\bibinfo {title} {{The Hamburg/ESO R-process enhanced star
  survey (HERES). IV. Detailed abundance analysis and age dating of the
  strongly r-process enhanced stars CS 29491-069 and HE 1219-0312}},}\ }\href
  {https://doi.org/10.1051/0004-6361/200811121} {\bibfield  {journal} {\bibinfo
   {journal} {Astron. \& Astrophys.}\ }\textbf {\bibinfo {volume} {504}},\
  \bibinfo {pages} {511--524}}\BibitemShut {NoStop}%
\bibitem [{\citenamefont {{Haynes}}\ and\ \citenamefont
  {{Kobayashi}}(2019)}]{Haynes.Kobayashi:2019}%
  \BibitemOpen
  \bibfield  {author} {\bibinfo {author} {\bibnamefont {{Haynes}},
  \bibfnamefont {Christopher~J}}, and\ \bibinfo {author} {\bibfnamefont
  {Chiaki}\ \bibnamefont {{Kobayashi}}}} (\bibinfo {year} {2019}),\ \bibfield
  {title} {\enquote {\bibinfo {title} {{Galactic simulations of r-process
  elemental abundances}},}\ }\href {https://doi.org/10.1093/mnras/sty3389}
  {\bibfield  {journal} {\bibinfo  {journal} {Mon. Not. Roy. Astron. Soc.}\
  }\textbf {\bibinfo {volume} {483}},\ \bibinfo {pages}
  {5123--5134}}\BibitemShut {NoStop}%
\bibitem [{\citenamefont {{Hebeler}}\ \emph {et~al.}(2015)\citenamefont
  {{Hebeler}}, \citenamefont {{Holt}}, \citenamefont {{Men{\'e}ndez}},\ and\
  \citenamefont {{Schwenk}}}]{Hebeler.Holt.ea:2015}%
  \BibitemOpen
  \bibfield  {author} {\bibinfo {author} {\bibnamefont {{Hebeler}},
  \bibfnamefont {K}}, \bibinfo {author} {\bibfnamefont {J.~D.}\ \bibnamefont
  {{Holt}}}, \bibinfo {author} {\bibfnamefont {J.}~\bibnamefont
  {{Men{\'e}ndez}}}, and\ \bibinfo {author} {\bibfnamefont {A.}~\bibnamefont
  {{Schwenk}}}} (\bibinfo {year} {2015}),\ \bibfield  {title} {\enquote
  {\bibinfo {title} {{Nuclear Forces and Their Impact on Neutron-Rich Nuclei
  and Neutron-Rich Matter}},}\ }\href
  {https://doi.org/10.1146/annurev-nucl-102313-025446} {\bibfield  {journal}
  {\bibinfo  {journal} {Annu. Rev. Nucl. Part. Sci.}\ }\textbf {\bibinfo
  {volume} {65}},\ \bibinfo {pages} {457--484}}\BibitemShut {NoStop}%
\bibitem [{\citenamefont {{Hebeler}}\ \emph {et~al.}(2013)\citenamefont
  {{Hebeler}}, \citenamefont {{Lattimer}}, \citenamefont {{Pethick}},\ and\
  \citenamefont {{Schwenk}}}]{Hebeler.Lattimer.ea:2013}%
  \BibitemOpen
  \bibfield  {author} {\bibinfo {author} {\bibnamefont {{Hebeler}},
  \bibfnamefont {K}}, \bibinfo {author} {\bibfnamefont {J.~M.}\ \bibnamefont
  {{Lattimer}}}, \bibinfo {author} {\bibfnamefont {C.~J.}\ \bibnamefont
  {{Pethick}}}, and\ \bibinfo {author} {\bibfnamefont {A.}~\bibnamefont
  {{Schwenk}}}} (\bibinfo {year} {2013}),\ \bibfield  {title} {\enquote
  {\bibinfo {title} {{Equation of State and Neutron Star Properties Constrained
  by Nuclear Physics and Observation}},}\ }\href
  {https://doi.org/10.1088/0004-637X/773/1/11} {\bibfield  {journal} {\bibinfo
  {journal} {Astrophys. J.}\ }\textbf {\bibinfo {volume} {773}},\ \bibinfo
  {eid} {11}}\BibitemShut {NoStop}%
\bibitem [{\citenamefont {{Heger}}\ \emph {et~al.}(2003)\citenamefont
  {{Heger}}, \citenamefont {{Fryer}}, \citenamefont {{Woosley}}, \citenamefont
  {{Langer}},\ and\ \citenamefont {{Hartmann}}}]{Heger.Fryer.ea:2003}%
  \BibitemOpen
  \bibfield  {author} {\bibinfo {author} {\bibnamefont {{Heger}}, \bibfnamefont
  {A}}, \bibinfo {author} {\bibfnamefont {C.~L.}\ \bibnamefont {{Fryer}}},
  \bibinfo {author} {\bibfnamefont {S.~E.}\ \bibnamefont {{Woosley}}}, \bibinfo
  {author} {\bibfnamefont {N.}~\bibnamefont {{Langer}}}, and\ \bibinfo {author}
  {\bibfnamefont {D.~H.}\ \bibnamefont {{Hartmann}}}} (\bibinfo {year}
  {2003}),\ \bibfield  {title} {\enquote {\bibinfo {title} {{How Massive Single
  Stars End Their Life}},}\ }\href {https://doi.org/10.1086/375341} {\bibfield
  {journal} {\bibinfo  {journal} {Astrophys. J.}\ }\textbf {\bibinfo {volume}
  {591}},\ \bibinfo {pages} {288--300}}\BibitemShut {NoStop}%
\bibitem [{\citenamefont {Heger}\ \emph {et~al.}(2005)\citenamefont {Heger},
  \citenamefont {Kolbe}, \citenamefont {Haxton}, \citenamefont {Langanke},
  \citenamefont {Mart{\'\i}nez-Pinedo},\ and\ \citenamefont
  {Woosley}}]{Heger.Kolbe.ea:2005}%
  \BibitemOpen
  \bibfield  {author} {\bibinfo {author} {\bibnamefont {Heger}, \bibfnamefont
  {A}}, \bibinfo {author} {\bibfnamefont {E.}~\bibnamefont {Kolbe}}, \bibinfo
  {author} {\bibfnamefont {W.C.}\ \bibnamefont {Haxton}}, \bibinfo {author}
  {\bibfnamefont {K.}~\bibnamefont {Langanke}}, \bibinfo {author}
  {\bibfnamefont {G.}~\bibnamefont {Mart{\'\i}nez-Pinedo}}, and\ \bibinfo
  {author} {\bibfnamefont {S.~E.}\ \bibnamefont {Woosley}}} (\bibinfo {year}
  {2005}),\ \bibfield  {title} {\enquote {\bibinfo {title} {Neutrino
  nucleosynthesis},}\ }\href {https://doi.org/10.1016/j.physletb.2004.12.017}
  {\bibfield  {journal} {\bibinfo  {journal} {Phys. Lett. B}\ }\textbf
  {\bibinfo {volume} {606}},\ \bibinfo {pages} {258--264}}\BibitemShut
  {NoStop}%
\bibitem [{\citenamefont {{Hewish}}\ and\ \citenamefont
  {{Okoye}}(1965)}]{hewish65}%
  \BibitemOpen
  \bibfield  {author} {\bibinfo {author} {\bibnamefont {{Hewish}},
  \bibfnamefont {A}}, and\ \bibinfo {author} {\bibfnamefont {S.~E.}\
  \bibnamefont {{Okoye}}}} (\bibinfo {year} {1965}),\ \bibfield  {title}
  {\enquote {\bibinfo {title} {{Evidence for an Unusual Source of High Radio
  Brightness Temperature in the Crab Nebula}},}\ }\href
  {https://doi.org/10.1038/207059a0} {\bibfield  {journal} {\bibinfo  {journal}
  {Nature}\ }\textbf {\bibinfo {volume} {207}},\ \bibinfo {pages}
  {59--60}}\BibitemShut {NoStop}%
\bibitem [{\citenamefont {{Hilaire}}\ \emph {et~al.}(2010)\citenamefont
  {{Hilaire}}, \citenamefont {{Goriely}}, \citenamefont {{Girod}},
  \citenamefont {{Koning}}, \citenamefont {{Capote}},\ and\ \citenamefont
  {{Sin}}}]{Hilaire.Goriely.ea:2010}%
  \BibitemOpen
  \bibfield  {author} {\bibinfo {author} {\bibnamefont {{Hilaire}},
  \bibfnamefont {S}}, \bibinfo {author} {\bibfnamefont {S.}~\bibnamefont
  {{Goriely}}}, \bibinfo {author} {\bibfnamefont {M.}~\bibnamefont {{Girod}}},
  \bibinfo {author} {\bibfnamefont {A.~J.}\ \bibnamefont {{Koning}}}, \bibinfo
  {author} {\bibfnamefont {R.}~\bibnamefont {{Capote}}}, and\ \bibinfo {author}
  {\bibfnamefont {M.}~\bibnamefont {{Sin}}}} (\bibinfo {year} {2010}),\
  \bibfield  {title} {\enquote {\bibinfo {title} {{Combinatorial level
  densities for practical applications}},}\ }in\ \href
  {https://doi.org/10.1051/epjconf/20100204005} {\emph {\bibinfo {booktitle}
  {European Physical Journal Web of Conferences}}},\ Vol.~\bibinfo {volume}
  {2},\ p.\ \bibinfo {pages} {04005}\BibitemShut {NoStop}%
\bibitem [{\citenamefont {{Hill}}\ \emph {et~al.}(2017)\citenamefont {{Hill}},
  \citenamefont {{Christlieb}}, \citenamefont {{Beers}}, \citenamefont
  {{Barklem}}, \citenamefont {{Kratz}}, \citenamefont {{Nordstr{\"o}m}},
  \citenamefont {{Pfeiffer}},\ and\ \citenamefont
  {{Farouqi}}}]{Hill.Christlieb.ea:2017}%
  \BibitemOpen
  \bibfield  {author} {\bibinfo {author} {\bibnamefont {{Hill}}, \bibfnamefont
  {V}}, \bibinfo {author} {\bibfnamefont {N.}~\bibnamefont {{Christlieb}}},
  \bibinfo {author} {\bibfnamefont {T.~C.}\ \bibnamefont {{Beers}}}, \bibinfo
  {author} {\bibfnamefont {P.~S.}\ \bibnamefont {{Barklem}}}, \bibinfo {author}
  {\bibfnamefont {K.-L.}\ \bibnamefont {{Kratz}}}, \bibinfo {author}
  {\bibfnamefont {B.}~\bibnamefont {{Nordstr{\"o}m}}}, \bibinfo {author}
  {\bibfnamefont {B.}~\bibnamefont {{Pfeiffer}}}, and\ \bibinfo {author}
  {\bibfnamefont {K.}~\bibnamefont {{Farouqi}}}} (\bibinfo {year} {2017}),\
  \bibfield  {title} {\enquote {\bibinfo {title} {{The Hamburg/ESO R-process
  Enhanced Star survey (HERES). XI. The highly r-process-enhanced star CS
  29497-004}},}\ }\href {https://doi.org/10.1051/0004-6361/201629092}
  {\bibfield  {journal} {\bibinfo  {journal} {Astron. \& Astrophys.}\ }\textbf
  {\bibinfo {volume} {607}},\ \bibinfo {eid} {A91}}\BibitemShut {NoStop}%
\bibitem [{\citenamefont {{Hill}}\ \emph {et~al.}(2002)\citenamefont {{Hill}},
  \citenamefont {{Plez}}, \citenamefont {{Cayrel}}, \citenamefont {{Beers}},
  \citenamefont {{Nordstr{\"o}m}}, \citenamefont {{Andersen}}, \citenamefont
  {{Spite}}, \citenamefont {{Spite}}, \citenamefont {{Barbuy}}, \citenamefont
  {{Bonifacio}}, \citenamefont {{Depagne}}, \citenamefont {{Fran{\c c}ois}},\
  and\ \citenamefont {{Primas}}}]{hill02}%
  \BibitemOpen
  \bibfield  {author} {\bibinfo {author} {\bibnamefont {{Hill}}, \bibfnamefont
  {V}}, \bibinfo {author} {\bibfnamefont {B.}~\bibnamefont {{Plez}}}, \bibinfo
  {author} {\bibfnamefont {R.}~\bibnamefont {{Cayrel}}}, \bibinfo {author}
  {\bibfnamefont {T.~C.}\ \bibnamefont {{Beers}}}, \bibinfo {author}
  {\bibfnamefont {B.}~\bibnamefont {{Nordstr{\"o}m}}}, \bibinfo {author}
  {\bibfnamefont {J.}~\bibnamefont {{Andersen}}}, \bibinfo {author}
  {\bibfnamefont {M.}~\bibnamefont {{Spite}}}, \bibinfo {author} {\bibfnamefont
  {F.}~\bibnamefont {{Spite}}}, \bibinfo {author} {\bibfnamefont
  {B.}~\bibnamefont {{Barbuy}}}, \bibinfo {author} {\bibfnamefont
  {P.}~\bibnamefont {{Bonifacio}}}, \bibinfo {author} {\bibfnamefont
  {E.}~\bibnamefont {{Depagne}}}, \bibinfo {author} {\bibfnamefont
  {P.}~\bibnamefont {{Fran{\c c}ois}}}, and\ \bibinfo {author} {\bibfnamefont
  {F.}~\bibnamefont {{Primas}}}} (\bibinfo {year} {2002}),\ \bibfield  {title}
  {\enquote {\bibinfo {title} {{First stars. I. The extreme r-element rich,
  iron-poor halo giant CS 31082-001. Implications for the r-process site(s) and
  radioactive cosmochronology}},}\ }\href
  {https://doi.org/10.1051/0004-6361:20020434} {\bibfield  {journal} {\bibinfo
  {journal} {Astron. \& Astrophys.}\ }\textbf {\bibinfo {volume} {387}},\
  \bibinfo {pages} {560--579}}\BibitemShut {NoStop}%
\bibitem [{\citenamefont {{Hill}}\ \emph {et~al.}(2019)\citenamefont {{Hill}},
  \citenamefont {{Sk{\'u}lad{\'o}ttir}}, \citenamefont {{Tolstoy}},
  \citenamefont {{Venn}}, \citenamefont {{Shetrone}}, \citenamefont
  {{Jablonka}}, \citenamefont {{Primas}}, \citenamefont {{Battaglia}},
  \citenamefont {{de Boer}}, \citenamefont {{Fran{\c{c}}ois}}, \citenamefont
  {{Helmi}}, \citenamefont {{Kaufer}}, \citenamefont {{Letarte}}, \citenamefont
  {{Starkenburg}},\ and\ \citenamefont {{Spite}}}]{Hill.Skuladottir.ea:2019}%
  \BibitemOpen
  \bibfield  {author} {\bibinfo {author} {\bibnamefont {{Hill}}, \bibfnamefont
  {V}}, \bibinfo {author} {\bibfnamefont {{\'A}.}~\bibnamefont
  {{Sk{\'u}lad{\'o}ttir}}}, \bibinfo {author} {\bibfnamefont {E.}~\bibnamefont
  {{Tolstoy}}}, \bibinfo {author} {\bibfnamefont {K.~A.}\ \bibnamefont
  {{Venn}}}, \bibinfo {author} {\bibfnamefont {M.~D.}\ \bibnamefont
  {{Shetrone}}}, \bibinfo {author} {\bibfnamefont {P.}~\bibnamefont
  {{Jablonka}}}, \bibinfo {author} {\bibfnamefont {F.}~\bibnamefont
  {{Primas}}}, \bibinfo {author} {\bibfnamefont {G.}~\bibnamefont
  {{Battaglia}}}, \bibinfo {author} {\bibfnamefont {T.~J.~L.}\ \bibnamefont
  {{de Boer}}}, \bibinfo {author} {\bibfnamefont {P.}~\bibnamefont
  {{Fran{\c{c}}ois}}}, \bibinfo {author} {\bibfnamefont {A.}~\bibnamefont
  {{Helmi}}}, \bibinfo {author} {\bibfnamefont {A.}~\bibnamefont {{Kaufer}}},
  \bibinfo {author} {\bibfnamefont {B.}~\bibnamefont {{Letarte}}}, \bibinfo
  {author} {\bibfnamefont {E.}~\bibnamefont {{Starkenburg}}}, and\ \bibinfo
  {author} {\bibfnamefont {M.}~\bibnamefont {{Spite}}}} (\bibinfo {year}
  {2019}),\ \bibfield  {title} {\enquote {\bibinfo {title} {{VLT/FLAMES
  high-resolution chemical abundances in Sculptor: a textbook dwarf spheroidal
  galaxy}},}\ }\href {https://doi.org/10.1051/0004-6361/201833950} {\bibfield
  {journal} {\bibinfo  {journal} {Astron. \& Astrophys.}\ }\textbf {\bibinfo
  {volume} {626}},\ \bibinfo {eid} {A15}}\BibitemShut {NoStop}%
\bibitem [{\citenamefont {{Hillebrandt}}(1978)}]{Hillebrandt:1978}%
  \BibitemOpen
  \bibfield  {author} {\bibinfo {author} {\bibnamefont {{Hillebrandt}},
  \bibfnamefont {W}}} (\bibinfo {year} {1978}),\ \bibfield  {title} {\enquote
  {\bibinfo {title} {{The rapid neutron-capture process and the synthesis of
  heavy and neutron-rich elements}},}\ }\href
  {https://doi.org/10.1007/BF00186236} {\bibfield  {journal} {\bibinfo
  {journal} {Space Sci. Rev.}\ }\textbf {\bibinfo {volume} {21}},\ \bibinfo
  {pages} {639--702}}\BibitemShut {NoStop}%
\bibitem [{\citenamefont {{Hillebrandt}}\ \emph {et~al.}(2013)\citenamefont
  {{Hillebrandt}}, \citenamefont {{Kromer}}, \citenamefont {{R{\"o}pke}},\ and\
  \citenamefont {{Ruiter}}}]{hillebrandt13}%
  \BibitemOpen
  \bibfield  {author} {\bibinfo {author} {\bibnamefont {{Hillebrandt}},
  \bibfnamefont {W}}, \bibinfo {author} {\bibfnamefont {M.}~\bibnamefont
  {{Kromer}}}, \bibinfo {author} {\bibfnamefont {F.~K.}\ \bibnamefont
  {{R{\"o}pke}}}, and\ \bibinfo {author} {\bibfnamefont {A.~J.}\ \bibnamefont
  {{Ruiter}}}} (\bibinfo {year} {2013}),\ \bibfield  {title} {\enquote
  {\bibinfo {title} {{Towards an understanding of Type Ia supernovae from a
  synthesis of theory and observations}},}\ }\href
  {https://doi.org/10.1007/s11467-013-0303-2} {\bibfield  {journal} {\bibinfo
  {journal} {Frontiers of Physics}\ }\textbf {\bibinfo {volume} {8}},\ \bibinfo
  {pages} {116--143}}\BibitemShut {NoStop}%
\bibitem [{\citenamefont {{Hillebrandt}}\ \emph {et~al.}(1976)\citenamefont
  {{Hillebrandt}}, \citenamefont {{Takahashi}},\ and\ \citenamefont
  {{Kodama}}}]{Hillebrandt.Kodama.Takahashi:1976}%
  \BibitemOpen
  \bibfield  {author} {\bibinfo {author} {\bibnamefont {{Hillebrandt}},
  \bibfnamefont {W}}, \bibinfo {author} {\bibfnamefont {K.}~\bibnamefont
  {{Takahashi}}}, and\ \bibinfo {author} {\bibfnamefont {T.}~\bibnamefont
  {{Kodama}}}} (\bibinfo {year} {1976}),\ \bibfield  {title} {\enquote
  {\bibinfo {title} {{R-process Nucleosynthesis: A Dynamical Model}},}\
  }\href@noop {} {\bibfield  {journal} {\bibinfo  {journal} {Astron. \&
  Astrophys.}\ }\textbf {\bibinfo {volume} {52}},\ \bibinfo {pages}
  {63--68}}\BibitemShut {NoStop}%
\bibitem [{\citenamefont {{Hillebrandt}}\ \emph {et~al.}(1981)\citenamefont
  {{Hillebrandt}}, \citenamefont {{Thielemann}}, \citenamefont {{Klapdor}},\
  and\ \citenamefont {{Oda}}}]{Hillebrandt.ea:1981}%
  \BibitemOpen
  \bibfield  {author} {\bibinfo {author} {\bibnamefont {{Hillebrandt}},
  \bibfnamefont {W}}, \bibinfo {author} {\bibfnamefont {F.~K.}\ \bibnamefont
  {{Thielemann}}}, \bibinfo {author} {\bibfnamefont {H.~V.}\ \bibnamefont
  {{Klapdor}}}, and\ \bibinfo {author} {\bibfnamefont {T.}~\bibnamefont
  {{Oda}}}} (\bibinfo {year} {1981}),\ \bibfield  {title} {\enquote {\bibinfo
  {title} {{The r-process during explosive helium burning in supernovae}},}\
  }\href@noop {} {\bibfield  {journal} {\bibinfo  {journal} {Astron. \&
  Astrophys.}\ }\textbf {\bibinfo {volume} {99}},\ \bibinfo {pages}
  {195--198}}\BibitemShut {NoStop}%
\bibitem [{\citenamefont {{Hirai}}\ \emph {et~al.}(2015)\citenamefont
  {{Hirai}}, \citenamefont {{Ishimaru}}, \citenamefont {{Saitoh}},
  \citenamefont {{Fujii}}, \citenamefont {{Hidaka}},\ and\ \citenamefont
  {{Kajino}}}]{hirai15}%
  \BibitemOpen
  \bibfield  {author} {\bibinfo {author} {\bibnamefont {{Hirai}}, \bibfnamefont
  {Y}}, \bibinfo {author} {\bibfnamefont {Y.}~\bibnamefont {{Ishimaru}}},
  \bibinfo {author} {\bibfnamefont {T.~R.}\ \bibnamefont {{Saitoh}}}, \bibinfo
  {author} {\bibfnamefont {M.~S.}\ \bibnamefont {{Fujii}}}, \bibinfo {author}
  {\bibfnamefont {J.}~\bibnamefont {{Hidaka}}}, and\ \bibinfo {author}
  {\bibfnamefont {T.}~\bibnamefont {{Kajino}}}} (\bibinfo {year} {2015}),\
  \bibfield  {title} {\enquote {\bibinfo {title} {{Enrichment of r-process
  Elements in Dwarf Spheroidal Galaxies in Chemo-dynamical Evolution Model}},}\
  }\href {https://doi.org/10.1088/0004-637X/814/1/41} {\bibfield  {journal}
  {\bibinfo  {journal} {Astrophys. J.}\ }\textbf {\bibinfo {volume} {814}},\
  \bibinfo {eid} {41}}\BibitemShut {NoStop}%
\bibitem [{\citenamefont {{Hirai}}\ \emph {et~al.}(2017)\citenamefont
  {{Hirai}}, \citenamefont {{Ishimaru}}, \citenamefont {{Saitoh}},
  \citenamefont {{Fujii}}, \citenamefont {{Hidaka}},\ and\ \citenamefont
  {{Kajino}}}]{hirai17}%
  \BibitemOpen
  \bibfield  {author} {\bibinfo {author} {\bibnamefont {{Hirai}}, \bibfnamefont
  {Y}}, \bibinfo {author} {\bibfnamefont {Y.}~\bibnamefont {{Ishimaru}}},
  \bibinfo {author} {\bibfnamefont {T.~R.}\ \bibnamefont {{Saitoh}}}, \bibinfo
  {author} {\bibfnamefont {M.~S.}\ \bibnamefont {{Fujii}}}, \bibinfo {author}
  {\bibfnamefont {J.}~\bibnamefont {{Hidaka}}}, and\ \bibinfo {author}
  {\bibfnamefont {T.}~\bibnamefont {{Kajino}}}} (\bibinfo {year} {2017}),\
  \bibfield  {title} {\enquote {\bibinfo {title} {{Early chemo-dynamical
  evolution of dwarf galaxies deduced from enrichment of r-process
  elements}},}\ }\href {https://doi.org/10.1093/mnras/stw3342} {\bibfield
  {journal} {\bibinfo  {journal} {Mon. Not. Roy. Astron. Soc.}\ }\textbf
  {\bibinfo {volume} {466}},\ \bibinfo {pages} {2474--2487}}\BibitemShut
  {NoStop}%
\bibitem [{\citenamefont {{Hix}}\ \emph {et~al.}(2016)\citenamefont {{Hix}},
  \citenamefont {{Lentz}}, \citenamefont {{Bruenn}}, \citenamefont
  {{Mezzacappa}}, \citenamefont {{Messer}}, \citenamefont {{Endeve}},
  \citenamefont {{Blondin}}, \citenamefont {{Harris}}, \citenamefont
  {{Marronetti}},\ and\ \citenamefont {{Yakunin}}}]{Hix.Lentz.ea:2016}%
  \BibitemOpen
  \bibfield  {author} {\bibinfo {author} {\bibnamefont {{Hix}}, \bibfnamefont
  {W~R}}, \bibinfo {author} {\bibfnamefont {E.~J.}\ \bibnamefont {{Lentz}}},
  \bibinfo {author} {\bibfnamefont {S.~W.}\ \bibnamefont {{Bruenn}}}, \bibinfo
  {author} {\bibfnamefont {A.}~\bibnamefont {{Mezzacappa}}}, \bibinfo {author}
  {\bibfnamefont {O.~E.~B.}\ \bibnamefont {{Messer}}}, \bibinfo {author}
  {\bibfnamefont {E.}~\bibnamefont {{Endeve}}}, \bibinfo {author}
  {\bibfnamefont {J.~M.}\ \bibnamefont {{Blondin}}}, \bibinfo {author}
  {\bibfnamefont {J.~A.}\ \bibnamefont {{Harris}}}, \bibinfo {author}
  {\bibfnamefont {P.}~\bibnamefont {{Marronetti}}}, and\ \bibinfo {author}
  {\bibfnamefont {K.~N.}\ \bibnamefont {{Yakunin}}}} (\bibinfo {year} {2016}),\
  \bibfield  {title} {\enquote {\bibinfo {title} {{The Multi-dimensional
  Character of Core-collapse Supernovae}},}\ }\href
  {https://doi.org/10.5506/APhysPolB.47.645} {\bibfield  {journal} {\bibinfo
  {journal} {Acta Phys. Pol. B}\ }\textbf {\bibinfo {volume} {47}},\ \bibinfo
  {pages} {645}}\BibitemShut {NoStop}%
\bibitem [{\citenamefont {Hix}\ and\ \citenamefont
  {Meyer}(2006)}]{Hix.Meyer:2006}%
  \BibitemOpen
  \bibfield  {author} {\bibinfo {author} {\bibnamefont {Hix}, \bibfnamefont
  {W~R}}, and\ \bibinfo {author} {\bibfnamefont {B.~S.}\ \bibnamefont {Meyer}}}
  (\bibinfo {year} {2006}),\ \bibfield  {title} {\enquote {\bibinfo {title}
  {{Thermonuclear kinetics in astrophysics}},}\ }\href
  {https://doi.org/10.1016/j.nuclphysa.2004.10.009} {\bibfield  {journal}
  {\bibinfo  {journal} {Nucl. Phys. A}\ }\textbf {\bibinfo {volume} {777}},\
  \bibinfo {pages} {188--207}}\BibitemShut {NoStop}%
\bibitem [{\citenamefont {{Hix}}\ and\ \citenamefont
  {{Thielemann}}(1999)}]{Hix.Thielemann:1999a}%
  \BibitemOpen
  \bibfield  {author} {\bibinfo {author} {\bibnamefont {{Hix}}, \bibfnamefont
  {W~R}}, and\ \bibinfo {author} {\bibfnamefont {F.-K.}\ \bibnamefont
  {{Thielemann}}}} (\bibinfo {year} {1999}),\ \bibfield  {title} {\enquote
  {\bibinfo {title} {{Silicon Burning. II. Quasi-Equilibrium and Explosive
  Burning}},}\ }\href {https://doi.org/10.1086/306692} {\bibfield  {journal}
  {\bibinfo  {journal} {Astrophys. J.}\ }\textbf {\bibinfo {volume} {511}},\
  \bibinfo {pages} {862--875}}\BibitemShut {NoStop}%
\bibitem [{\citenamefont {{Hix}}\ \emph {et~al.}(2007)\citenamefont {{Hix}},
  \citenamefont {{Parete-Koon}}, \citenamefont {{Freiburghaus}},\ and\
  \citenamefont {{Thielemann}}}]{Hix.ea:2007}%
  \BibitemOpen
  \bibfield  {author} {\bibinfo {author} {\bibnamefont {{Hix}}, \bibfnamefont
  {W~Raphael}}, \bibinfo {author} {\bibfnamefont {Suzanne~T.}\ \bibnamefont
  {{Parete-Koon}}}, \bibinfo {author} {\bibfnamefont {Christian}\ \bibnamefont
  {{Freiburghaus}}}, and\ \bibinfo {author} {\bibfnamefont {Friedrich-Karl}\
  \bibnamefont {{Thielemann}}}} (\bibinfo {year} {2007}),\ \bibfield  {title}
  {\enquote {\bibinfo {title} {{The QSE-Reduced Nuclear Reaction Network for
  Silicon Burning}},}\ }\href {https://doi.org/10.1086/520672} {\bibfield
  {journal} {\bibinfo  {journal} {Astrophys. J.}\ }\textbf {\bibinfo {volume}
  {667}},\ \bibinfo {pages} {476--488}}\BibitemShut {NoStop}%
\bibitem [{\citenamefont {Hix}\ and\ \citenamefont
  {Thielemann}(1999)}]{Hix.Thielemann:1999b}%
  \BibitemOpen
  \bibfield  {author} {\bibinfo {author} {\bibnamefont {Hix}, \bibfnamefont
  {W~Raphael}}, and\ \bibinfo {author} {\bibfnamefont {Friedrich-Karl}\
  \bibnamefont {Thielemann}}} (\bibinfo {year} {1999}),\ \bibfield  {title}
  {\enquote {\bibinfo {title} {Computational methods for nucleosynthesis and
  nuclear energy generation},}\ }\href
  {https://doi.org/10.1016/S0377-0427(99)00163-6} {\bibfield  {journal}
  {\bibinfo  {journal} {J. Comput. Appl. Math.}\ }\textbf {\bibinfo {volume}
  {109}},\ \bibinfo {pages} {321--351}}\BibitemShut {NoStop}%
\bibitem [{\citenamefont {{Hoffman}}\ \emph {et~al.}(1997)\citenamefont
  {{Hoffman}}, \citenamefont {{Woosley}},\ and\ \citenamefont
  {{Qian}}}]{hoffman.woosley.qian:1997}%
  \BibitemOpen
  \bibfield  {author} {\bibinfo {author} {\bibnamefont {{Hoffman}},
  \bibfnamefont {R~D}}, \bibinfo {author} {\bibfnamefont {S.~E.}\ \bibnamefont
  {{Woosley}}}, and\ \bibinfo {author} {\bibfnamefont {Y.-Z.}\ \bibnamefont
  {{Qian}}}} (\bibinfo {year} {1997}),\ \bibfield  {title} {\enquote {\bibinfo
  {title} {{Nucleosynthesis in Neutrino-driven Winds. II. Implications for
  Heavy Element Synthesis}},}\ }\href {https://doi.org/10.1086/304181}
  {\bibfield  {journal} {\bibinfo  {journal} {Astrophys. J.}\ }\textbf
  {\bibinfo {volume} {482}},\ \bibinfo {pages} {951--962}}\BibitemShut
  {NoStop}%
\bibitem [{\citenamefont {{Holmbeck}}\ \emph {et~al.}(2018)\citenamefont
  {{Holmbeck}}, \citenamefont {{Beers}}, \citenamefont {{Roederer}},
  \citenamefont {{Placco}}, \citenamefont {{Hansen}}, \citenamefont {{Sakari}},
  \citenamefont {{Sneden}}, \citenamefont {{Liu}}, \citenamefont {{Lee}},
  \citenamefont {{Cowan}},\ and\ \citenamefont
  {{Frebel}}}]{Holmbeck.Beers.ea:2018}%
  \BibitemOpen
  \bibfield  {author} {\bibinfo {author} {\bibnamefont {{Holmbeck}},
  \bibfnamefont {E~M}}, \bibinfo {author} {\bibfnamefont {T.~C.}\ \bibnamefont
  {{Beers}}}, \bibinfo {author} {\bibfnamefont {I.~U.}\ \bibnamefont
  {{Roederer}}}, \bibinfo {author} {\bibfnamefont {V.~M.}\ \bibnamefont
  {{Placco}}}, \bibinfo {author} {\bibfnamefont {T.~T.}\ \bibnamefont
  {{Hansen}}}, \bibinfo {author} {\bibfnamefont {C.~M.}\ \bibnamefont
  {{Sakari}}}, \bibinfo {author} {\bibfnamefont {C.}~\bibnamefont {{Sneden}}},
  \bibinfo {author} {\bibfnamefont {C.}~\bibnamefont {{Liu}}}, \bibinfo
  {author} {\bibfnamefont {Y.~S.}\ \bibnamefont {{Lee}}}, \bibinfo {author}
  {\bibfnamefont {J.~J.}\ \bibnamefont {{Cowan}}}, and\ \bibinfo {author}
  {\bibfnamefont {A.}~\bibnamefont {{Frebel}}}} (\bibinfo {year} {2018}),\
  \bibfield  {title} {\enquote {\bibinfo {title} {{The R-Process Alliance:
  2MASS J09544277+5246414, the Most Actinide-enhanced R-II Star Known}},}\
  }\href {https://doi.org/10.3847/2041-8213/aac722} {\bibfield  {journal}
  {\bibinfo  {journal} {Astrophys. J.}\ }\textbf {\bibinfo {volume} {859}},\
  \bibinfo {eid} {L24}}\BibitemShut {NoStop}%
\bibitem [{\citenamefont {{Holmbeck}}\ \emph
  {et~al.}(2019{\natexlab{a}})\citenamefont {{Holmbeck}}, \citenamefont
  {{Frebel}}, \citenamefont {{McLaughlin}}, \citenamefont {{Mumpower}},
  \citenamefont {{Sprouse}},\ and\ \citenamefont {{Surman}}}]{Holmbeck19b}%
  \BibitemOpen
  \bibfield  {author} {\bibinfo {author} {\bibnamefont {{Holmbeck}},
  \bibfnamefont {Erika~M}}, \bibinfo {author} {\bibfnamefont {Anna}\
  \bibnamefont {{Frebel}}}, \bibinfo {author} {\bibfnamefont {G.~C.}\
  \bibnamefont {{McLaughlin}}}, \bibinfo {author} {\bibfnamefont {Matthew~R.}\
  \bibnamefont {{Mumpower}}}, \bibinfo {author} {\bibfnamefont {Trevor~M.}\
  \bibnamefont {{Sprouse}}}, and\ \bibinfo {author} {\bibfnamefont {Rebecca}\
  \bibnamefont {{Surman}}}} (\bibinfo {year} {2019}{\natexlab{a}}),\ \bibfield
  {title} {\enquote {\bibinfo {title} {{Actinide-rich and Actinide-poor
  r-process-enhanced Metal-poor Stars Do Not Require Separate r-process
  Progenitors}},}\ }\href {https://doi.org/10.3847/1538-4357/ab2a01} {\bibfield
   {journal} {\bibinfo  {journal} {Astrophys. J.}\ }\textbf {\bibinfo {volume}
  {881}},\ \bibinfo {eid} {5}}\BibitemShut {NoStop}%
\bibitem [{\citenamefont {{Holmbeck}}\ \emph
  {et~al.}(2019{\natexlab{b}})\citenamefont {{Holmbeck}}, \citenamefont
  {{Sprouse}}, \citenamefont {{Mumpower}}, \citenamefont {{Vassh}},
  \citenamefont {{Surman}}, \citenamefont {{Beers}},\ and\ \citenamefont
  {{Kawano}}}]{Holmbeck19a}%
  \BibitemOpen
  \bibfield  {author} {\bibinfo {author} {\bibnamefont {{Holmbeck}},
  \bibfnamefont {Erika~M}}, \bibinfo {author} {\bibfnamefont {Trevor~M.}\
  \bibnamefont {{Sprouse}}}, \bibinfo {author} {\bibfnamefont {Matthew~R.}\
  \bibnamefont {{Mumpower}}}, \bibinfo {author} {\bibfnamefont {Nicole}\
  \bibnamefont {{Vassh}}}, \bibinfo {author} {\bibfnamefont {Rebecca}\
  \bibnamefont {{Surman}}}, \bibinfo {author} {\bibfnamefont {Timothy~C.}\
  \bibnamefont {{Beers}}}, and\ \bibinfo {author} {\bibfnamefont {Toshihiko}\
  \bibnamefont {{Kawano}}}} (\bibinfo {year} {2019}{\natexlab{b}}),\ \bibfield
  {title} {\enquote {\bibinfo {title} {{Actinide Production in the Neutron-rich
  Ejecta of a Neutron Star Merger}},}\ }\href
  {https://doi.org/10.3847/1538-4357/aaefef} {\bibfield  {journal} {\bibinfo
  {journal} {Astrophys. J.}\ }\textbf {\bibinfo {volume} {870}},\ \bibinfo
  {eid} {23}}\BibitemShut {NoStop}%
\bibitem [{\citenamefont {{Honda}}\ \emph {et~al.}(2007)\citenamefont
  {{Honda}}, \citenamefont {{Aoki}}, \citenamefont {{Ishimaru}},\ and\
  \citenamefont {{Wanajo}}}]{Honda.Aoki.ea:2007}%
  \BibitemOpen
  \bibfield  {author} {\bibinfo {author} {\bibnamefont {{Honda}}, \bibfnamefont
  {S}}, \bibinfo {author} {\bibfnamefont {W.}~\bibnamefont {{Aoki}}}, \bibinfo
  {author} {\bibfnamefont {Y.}~\bibnamefont {{Ishimaru}}}, and\ \bibinfo
  {author} {\bibfnamefont {S.}~\bibnamefont {{Wanajo}}}} (\bibinfo {year}
  {2007}),\ \bibfield  {title} {\enquote {\bibinfo {title} {{Neutron-Capture
  Elements in the Very Metal-poor Star HD 88609: Another Star with Excesses of
  Light Neutron-Capture Elements}},}\ }\href {https://doi.org/10.1086/520034}
  {\bibfield  {journal} {\bibinfo  {journal} {Astrophys. J.}\ }\textbf
  {\bibinfo {volume} {666}},\ \bibinfo {pages} {1189--1197}}\BibitemShut
  {NoStop}%
\bibitem [{\citenamefont {{Honda}}\ \emph {et~al.}(2006)\citenamefont
  {{Honda}}, \citenamefont {{Aoki}}, \citenamefont {{Ishimaru}}, \citenamefont
  {{Wanajo}},\ and\ \citenamefont {{Ryan}}}]{honda06}%
  \BibitemOpen
  \bibfield  {author} {\bibinfo {author} {\bibnamefont {{Honda}}, \bibfnamefont
  {S}}, \bibinfo {author} {\bibfnamefont {W.}~\bibnamefont {{Aoki}}}, \bibinfo
  {author} {\bibfnamefont {Y.}~\bibnamefont {{Ishimaru}}}, \bibinfo {author}
  {\bibfnamefont {S.}~\bibnamefont {{Wanajo}}}, and\ \bibinfo {author}
  {\bibfnamefont {S.~G.}\ \bibnamefont {{Ryan}}}} (\bibinfo {year} {2006}),\
  \bibfield  {title} {\enquote {\bibinfo {title} {{Neutron-Capture Elements in
  the Very Metal Poor Star HD 122563}},}\ }\href
  {https://doi.org/10.1086/503195} {\bibfield  {journal} {\bibinfo  {journal}
  {Astrophys. J.}\ }\textbf {\bibinfo {volume} {643}},\ \bibinfo {pages}
  {1180--1189}}\BibitemShut {NoStop}%
\bibitem [{\citenamefont {{Honda}}\ \emph {et~al.}(2004)\citenamefont
  {{Honda}}, \citenamefont {{Aoki}}, \citenamefont {{Kajino}}, \citenamefont
  {{Ando}}, \citenamefont {{Beers}}, \citenamefont {{Izumiura}}, \citenamefont
  {{Sadakane}},\ and\ \citenamefont {{Takada-Hidai}}}]{honda04}%
  \BibitemOpen
  \bibfield  {author} {\bibinfo {author} {\bibnamefont {{Honda}}, \bibfnamefont
  {S}}, \bibinfo {author} {\bibfnamefont {W.}~\bibnamefont {{Aoki}}}, \bibinfo
  {author} {\bibfnamefont {T.}~\bibnamefont {{Kajino}}}, \bibinfo {author}
  {\bibfnamefont {H.}~\bibnamefont {{Ando}}}, \bibinfo {author} {\bibfnamefont
  {T.~C.}\ \bibnamefont {{Beers}}}, \bibinfo {author} {\bibfnamefont
  {H.}~\bibnamefont {{Izumiura}}}, \bibinfo {author} {\bibfnamefont
  {K.}~\bibnamefont {{Sadakane}}}, and\ \bibinfo {author} {\bibfnamefont
  {M.}~\bibnamefont {{Takada-Hidai}}}} (\bibinfo {year} {2004}),\ \bibfield
  {title} {\enquote {\bibinfo {title} {{Spectroscopic Studies of Extremely
  Metal-Poor Stars with the Subaru High Dispersion Spectrograph. II. The
  r-Process Elements, Including Thorium}},}\ }\href
  {https://doi.org/10.1086/383406} {\bibfield  {journal} {\bibinfo  {journal}
  {Astrophys. J.}\ }\textbf {\bibinfo {volume} {607}},\ \bibinfo {pages}
  {474--498}}\BibitemShut {NoStop}%
\bibitem [{\citenamefont {Horowitz}(2002)}]{Horowitz:2002}%
  \BibitemOpen
  \bibfield  {author} {\bibinfo {author} {\bibnamefont {Horowitz},
  \bibfnamefont {C~J}}} (\bibinfo {year} {2002}),\ \bibfield  {title} {\enquote
  {\bibinfo {title} {{Weak magnetism for antineutrinos in supernovae}},}\
  }\href {https://doi.org/10.1103/PhysRevD.65.043001} {\bibfield  {journal}
  {\bibinfo  {journal} {Phys. Rev. D}\ }\textbf {\bibinfo {volume} {65}},\
  \bibinfo {pages} {043001}}\BibitemShut {NoStop}%
\bibitem [{\citenamefont {Horowitz}\ \emph {et~al.}(2012)\citenamefont
  {Horowitz}, \citenamefont {Shen}, \citenamefont {O'Connor},\ and\
  \citenamefont {Ott}}]{Horowitz.Shen.ea:2012}%
  \BibitemOpen
  \bibfield  {author} {\bibinfo {author} {\bibnamefont {Horowitz},
  \bibfnamefont {C~J}}, \bibinfo {author} {\bibfnamefont {G.}~\bibnamefont
  {Shen}}, \bibinfo {author} {\bibfnamefont {Evan}\ \bibnamefont {O'Connor}},
  and\ \bibinfo {author} {\bibfnamefont {Christian~D.}\ \bibnamefont {Ott}}}
  (\bibinfo {year} {2012}),\ \bibfield  {title} {\enquote {\bibinfo {title}
  {Charged-current neutrino interactions in core-collapse supernovae in a
  virial expansion},}\ }\href {https://doi.org/10.1103/PhysRevC.86.065806}
  {\bibfield  {journal} {\bibinfo  {journal} {Phys. Rev. C}\ }\textbf {\bibinfo
  {volume} {86}},\ \bibinfo {pages} {065806}}\BibitemShut {NoStop}%
\bibitem [{\citenamefont {{Horowitz}}\ \emph {et~al.}(2019)\citenamefont
  {{Horowitz}} \emph {et~al.}}]{Horowitz.ea:2019}%
  \BibitemOpen
  \bibfield  {author} {\bibinfo {author} {\bibnamefont {{Horowitz}},
  \bibfnamefont {C~J}},  \emph {et~al.}} (\bibinfo {year} {2019}),\ \bibfield
  {title} {\enquote {\bibinfo {title} {{r-process nucleosynthesis: connecting
  rare-isotope beam facilities with the cosmos}},}\ }\href
  {https://doi.org/10.1088/1361-6471/ab0849} {\bibfield  {journal} {\bibinfo
  {journal} {J. Phys. G: Nucl. Part. Phys.}\ }\textbf {\bibinfo {volume}
  {46}},\ \bibinfo {pages} {083001}}\BibitemShut {NoStop}%
\bibitem [{\citenamefont {{Hosmer}}\ \emph {et~al.}(2010)\citenamefont
  {{Hosmer}} \emph {et~al.}}]{CoCu}%
  \BibitemOpen
  \bibfield  {author} {\bibinfo {author} {\bibnamefont {{Hosmer}},
  \bibfnamefont {P}},  \emph {et~al.}} (\bibinfo {year} {2010}),\ \bibfield
  {title} {\enquote {\bibinfo {title} {{Half-lives and branchings for
  {$\beta$}-delayed neutron emission for neutron-rich Co-Cu isotopes in the
  r-process}},}\ }\href {https://doi.org/10.1103/PhysRevC.82.025806} {\bibfield
   {journal} {\bibinfo  {journal} {Phys. Rev. C}\ }\textbf {\bibinfo {volume}
  {82}},\ \bibinfo {eid} {025806}}\BibitemShut {NoStop}%
\bibitem [{\citenamefont {Hosmer}\ \emph {et~al.}(2005)\citenamefont {Hosmer}
  \emph {et~al.}}]{Hosmer.Schatz.ea:2005}%
  \BibitemOpen
  \bibfield  {author} {\bibinfo {author} {\bibnamefont {Hosmer}, \bibfnamefont
  {P~T}},  \emph {et~al.}} (\bibinfo {year} {2005}),\ \bibfield  {title}
  {\enquote {\bibinfo {title} {{Half-Life of the Doubly Magic r-Process Nucleus
  $^{78}$Ni}},}\ }\href {https://doi.org/10.1103/PhysRevLett.94.112501}
  {\bibfield  {journal} {\bibinfo  {journal} {Phys. Rev. Lett.}\ }\textbf
  {\bibinfo {volume} {94}},\ \bibinfo {pages} {112501}}\BibitemShut {NoStop}%
\bibitem [{\citenamefont {{Hosseinzadeh}}\ \emph {et~al.}(2019)\citenamefont
  {{Hosseinzadeh}} \emph {et~al.}}]{Hosseinzadeh.Cowperthwaite.ea:2019}%
  \BibitemOpen
  \bibfield  {author} {\bibinfo {author} {\bibnamefont {{Hosseinzadeh}},
  \bibfnamefont {G}},  \emph {et~al.}} (\bibinfo {year} {2019}),\ \bibfield
  {title} {\enquote {\bibinfo {title} {{Follow-up of the Neutron Star Bearing
  Gravitational-wave Candidate Events S190425z and S190426c with MMT and
  SOAR}},}\ }\href {https://doi.org/10.3847/2041-8213/ab271c} {\bibfield
  {journal} {\bibinfo  {journal} {Astrophys. J. Lett.}\ }\textbf {\bibinfo
  {volume} {880}},\ \bibinfo {eid} {L4}}\BibitemShut {NoStop}%
\bibitem [{\citenamefont {{Hotokezaka}}\ \emph {et~al.}(2018)\citenamefont
  {{Hotokezaka}}, \citenamefont {{Beniamini}},\ and\ \citenamefont
  {{Piran}}}]{hotokezaka18}%
  \BibitemOpen
  \bibfield  {author} {\bibinfo {author} {\bibnamefont {{Hotokezaka}},
  \bibfnamefont {K}}, \bibinfo {author} {\bibfnamefont {P.}~\bibnamefont
  {{Beniamini}}}, and\ \bibinfo {author} {\bibfnamefont {T.}~\bibnamefont
  {{Piran}}}} (\bibinfo {year} {2018}),\ \bibfield  {title} {\enquote {\bibinfo
  {title} {{Neutron star mergers as sites of r-process nucleosynthesis and
  short gamma-ray bursts}},}\ }\href
  {https://doi.org/10.1142/S0218271818420051} {\bibfield  {journal} {\bibinfo
  {journal} {International Journal of Modern Physics D}\ }\textbf {\bibinfo
  {volume} {27}},\ \bibinfo {eid} {1842005}}\BibitemShut {NoStop}%
\bibitem [{\citenamefont {{Hotokezaka}}\ \emph {et~al.}(2013)\citenamefont
  {{Hotokezaka}}, \citenamefont {{Kyutoku}},\ and\ \citenamefont
  {{Shibata}}}]{hotokezaka13}%
  \BibitemOpen
  \bibfield  {author} {\bibinfo {author} {\bibnamefont {{Hotokezaka}},
  \bibfnamefont {K}}, \bibinfo {author} {\bibfnamefont {K.}~\bibnamefont
  {{Kyutoku}}}, and\ \bibinfo {author} {\bibfnamefont {M.}~\bibnamefont
  {{Shibata}}}} (\bibinfo {year} {2013}),\ \bibfield  {title} {\enquote
  {\bibinfo {title} {{Exploring tidal effects of coalescing binary neutron
  stars in numerical relativity}},}\ }\href
  {https://doi.org/10.1103/PhysRevD.87.044001} {\bibfield  {journal} {\bibinfo
  {journal} {Phys. Rev. D}\ }\textbf {\bibinfo {volume} {87}},\ \bibinfo {eid}
  {044001}}\BibitemShut {NoStop}%
\bibitem [{\citenamefont {{Hotokezaka}}\ \emph {et~al.}(2015)\citenamefont
  {{Hotokezaka}}, \citenamefont {{Piran}},\ and\ \citenamefont
  {{Paul}}}]{hotokezaka15}%
  \BibitemOpen
  \bibfield  {author} {\bibinfo {author} {\bibnamefont {{Hotokezaka}},
  \bibfnamefont {K}}, \bibinfo {author} {\bibfnamefont {T.}~\bibnamefont
  {{Piran}}}, and\ \bibinfo {author} {\bibfnamefont {M.}~\bibnamefont
  {{Paul}}}} (\bibinfo {year} {2015}),\ \bibfield  {title} {\enquote {\bibinfo
  {title} {{Short-lived $^{244}$Pu points to compact binary mergers as sites
  for heavy r-process nucleosynthesis}},}\ }\href
  {https://doi.org/10.1038/nphys3574} {\bibfield  {journal} {\bibinfo
  {journal} {Nature Physics}\ }\textbf {\bibinfo {volume} {11}},\ \bibinfo
  {pages} {1042}}\BibitemShut {NoStop}%
\bibitem [{\citenamefont {{Hotokezaka}}\ \emph {et~al.}(2016)\citenamefont
  {{Hotokezaka}}, \citenamefont {{Wanajo}}, \citenamefont {{Tanaka}},
  \citenamefont {{Bamba}}, \citenamefont {{Terada}},\ and\ \citenamefont
  {{Piran}}}]{Hotokezaka.Wanajo.ea:2016}%
  \BibitemOpen
  \bibfield  {author} {\bibinfo {author} {\bibnamefont {{Hotokezaka}},
  \bibfnamefont {K}}, \bibinfo {author} {\bibfnamefont {S.}~\bibnamefont
  {{Wanajo}}}, \bibinfo {author} {\bibfnamefont {M.}~\bibnamefont {{Tanaka}}},
  \bibinfo {author} {\bibfnamefont {A.}~\bibnamefont {{Bamba}}}, \bibinfo
  {author} {\bibfnamefont {Y.}~\bibnamefont {{Terada}}}, and\ \bibinfo {author}
  {\bibfnamefont {T.}~\bibnamefont {{Piran}}}} (\bibinfo {year} {2016}),\
  \bibfield  {title} {\enquote {\bibinfo {title} {{Radioactive decay products
  in neutron star merger ejecta: heating efficiency and {$\gamma$}-ray
  emission}},}\ }\href {https://doi.org/10.1093/mnras/stw404} {\bibfield
  {journal} {\bibinfo  {journal} {Mon. Not. Roy. Astron. Soc.}\ }\textbf
  {\bibinfo {volume} {459}},\ \bibinfo {pages} {35--43}}\BibitemShut {NoStop}%
\bibitem [{\citenamefont {Hotokezaka}\ \emph
  {et~al.}(2013{\natexlab{a}})\citenamefont {Hotokezaka}, \citenamefont
  {Kiuchi}, \citenamefont {Kyutoku}, \citenamefont {Muranushi}, \citenamefont
  {Sekiguchi}, \citenamefont {Shibata},\ and\ \citenamefont
  {Taniguchi}}]{Hotokezaka.Kiuchi.ea:2013b}%
  \BibitemOpen
  \bibfield  {author} {\bibinfo {author} {\bibnamefont {Hotokezaka},
  \bibfnamefont {Kenta}}, \bibinfo {author} {\bibfnamefont {Kenta}\
  \bibnamefont {Kiuchi}}, \bibinfo {author} {\bibfnamefont {Koutarou}\
  \bibnamefont {Kyutoku}}, \bibinfo {author} {\bibfnamefont {Takayuki}\
  \bibnamefont {Muranushi}}, \bibinfo {author} {\bibfnamefont {Yu-ichiro}\
  \bibnamefont {Sekiguchi}}, \bibinfo {author} {\bibfnamefont {Masaru}\
  \bibnamefont {Shibata}}, and\ \bibinfo {author} {\bibfnamefont {Keisuke}\
  \bibnamefont {Taniguchi}}} (\bibinfo {year} {2013}{\natexlab{a}}),\ \bibfield
   {title} {\enquote {\bibinfo {title} {Remnant massive neutron stars of binary
  neutron star mergers: Evolution process and gravitational waveform},}\ }\href
  {https://doi.org/10.1103/PhysRevD.88.044026} {\bibfield  {journal} {\bibinfo
  {journal} {Phys. Rev. D}\ }\textbf {\bibinfo {volume} {88}},\ \bibinfo
  {pages} {044026}}\BibitemShut {NoStop}%
\bibitem [{\citenamefont {Hotokezaka}\ \emph
  {et~al.}(2013{\natexlab{b}})\citenamefont {Hotokezaka}, \citenamefont
  {Kiuchi}, \citenamefont {Kyutoku}, \citenamefont {Okawa}, \citenamefont
  {Sekiguchi}, \citenamefont {Shibata},\ and\ \citenamefont
  {Taniguchi}}]{Hotokezaka.Kiuchi.ea:2013a}%
  \BibitemOpen
  \bibfield  {author} {\bibinfo {author} {\bibnamefont {Hotokezaka},
  \bibfnamefont {Kenta}}, \bibinfo {author} {\bibfnamefont {Kenta}\
  \bibnamefont {Kiuchi}}, \bibinfo {author} {\bibfnamefont {Koutarou}\
  \bibnamefont {Kyutoku}}, \bibinfo {author} {\bibfnamefont {Hirotada}\
  \bibnamefont {Okawa}}, \bibinfo {author} {\bibfnamefont {Yu-ichiro}\
  \bibnamefont {Sekiguchi}}, \bibinfo {author} {\bibfnamefont {Masaru}\
  \bibnamefont {Shibata}}, and\ \bibinfo {author} {\bibfnamefont {Keisuke}\
  \bibnamefont {Taniguchi}}} (\bibinfo {year} {2013}{\natexlab{b}}),\ \bibfield
   {title} {\enquote {\bibinfo {title} {Mass ejection from the merger of binary
  neutron stars},}\ }\href {https://doi.org/10.1103/PhysRevD.87.024001}
  {\bibfield  {journal} {\bibinfo  {journal} {Phys. Rev. D}\ }\textbf {\bibinfo
  {volume} {87}},\ \bibinfo {pages} {024001}}\BibitemShut {NoStop}%
\bibitem [{\citenamefont {Hotokezaka}\ \emph {et~al.}(2011)\citenamefont
  {Hotokezaka}, \citenamefont {Kyutoku}, \citenamefont {Okawa}, \citenamefont
  {Shibata},\ and\ \citenamefont {Kiuchi}}]{Hotokezaka.Kyutoku.ea:2011}%
  \BibitemOpen
  \bibfield  {author} {\bibinfo {author} {\bibnamefont {Hotokezaka},
  \bibfnamefont {Kenta}}, \bibinfo {author} {\bibfnamefont {Koutarou}\
  \bibnamefont {Kyutoku}}, \bibinfo {author} {\bibfnamefont {Hirotada}\
  \bibnamefont {Okawa}}, \bibinfo {author} {\bibfnamefont {Masaru}\
  \bibnamefont {Shibata}}, and\ \bibinfo {author} {\bibfnamefont {Kenta}\
  \bibnamefont {Kiuchi}}} (\bibinfo {year} {2011}),\ \bibfield  {title}
  {\enquote {\bibinfo {title} {Binary neutron star mergers: Dependence on the
  nuclear equation of state},}\ }\href
  {https://doi.org/10.1103/PhysRevD.83.124008} {\bibfield  {journal} {\bibinfo
  {journal} {Phys. Rev. D}\ }\textbf {\bibinfo {volume} {83}},\ \bibinfo
  {pages} {124008}}\BibitemShut {NoStop}%
\bibitem [{\citenamefont {{Hotokezaka}}\ and\ \citenamefont
  {{Nakar}}(2020)}]{Hotokezaka.Nakar:2020}%
  \BibitemOpen
  \bibfield  {author} {\bibinfo {author} {\bibnamefont {{Hotokezaka}},
  \bibfnamefont {Kenta}}, and\ \bibinfo {author} {\bibfnamefont {Ehud}\
  \bibnamefont {{Nakar}}}} (\bibinfo {year} {2020}),\ \bibfield  {title}
  {\enquote {\bibinfo {title} {{Radioactive Heating Rate of r-process Elements
  and Macronova Light Curve}},}\ }\href
  {https://doi.org/10.3847/1538-4357/ab6a98} {\bibfield  {journal} {\bibinfo
  {journal} {Astrophys. J.}\ }\textbf {\bibinfo {volume} {891}},\ \bibinfo
  {eid} {152}}\BibitemShut {NoStop}%
\bibitem [{\citenamefont {Hotokezaka}\ \emph {et~al.}(2017)\citenamefont
  {Hotokezaka}, \citenamefont {Sari},\ and\ \citenamefont
  {Piran}}]{Hotokezaka.Sari.Piran:2017}%
  \BibitemOpen
  \bibfield  {author} {\bibinfo {author} {\bibnamefont {Hotokezaka},
  \bibfnamefont {Kenta}}, \bibinfo {author} {\bibfnamefont {Re'em}\
  \bibnamefont {Sari}}, and\ \bibinfo {author} {\bibfnamefont {Tsvi}\
  \bibnamefont {Piran}}} (\bibinfo {year} {2017}),\ \bibfield  {title}
  {\enquote {\bibinfo {title} {{Analytic heating rate of neutron star merger
  ejecta derived from Fermi's theory of beta decay}},}\ }\href
  {https://doi.org/10.1093/mnras/stx411} {\bibfield  {journal} {\bibinfo
  {journal} {Mon. Not. Roy. Astron. Soc.}\ }\textbf {\bibinfo {volume} {468}},\
  \bibinfo {pages} {91--96}}\BibitemShut {NoStop}%
\bibitem [{\citenamefont {{Howard}}\ \emph {et~al.}(1972)\citenamefont
  {{Howard}}, \citenamefont {{Arnett}}, \citenamefont {{Clayton}},\ and\
  \citenamefont {{Woosley}}}]{Howard.Arnett.ea:1972}%
  \BibitemOpen
  \bibfield  {author} {\bibinfo {author} {\bibnamefont {{Howard}},
  \bibfnamefont {W~M}}, \bibinfo {author} {\bibfnamefont {W.~D.}\ \bibnamefont
  {{Arnett}}}, \bibinfo {author} {\bibfnamefont {D.~D.}\ \bibnamefont
  {{Clayton}}}, and\ \bibinfo {author} {\bibfnamefont {S.~E.}\ \bibnamefont
  {{Woosley}}}} (\bibinfo {year} {1972}),\ \bibfield  {title} {\enquote
  {\bibinfo {title} {{Nucleosynthesis of Rare Nuclei from Seed Nuclei in
  Explosive Carbon Burning}},}\ }\href {https://doi.org/10.1086/151549}
  {\bibfield  {journal} {\bibinfo  {journal} {Astrophys. J.}\ }\textbf
  {\bibinfo {volume} {175}},\ \bibinfo {pages} {201}}\BibitemShut {NoStop}%
\bibitem [{\citenamefont {{Howard}}\ and\ \citenamefont
  {{M{\"o}ller}}(1980)}]{Howard.Moeller:1980}%
  \BibitemOpen
  \bibfield  {author} {\bibinfo {author} {\bibnamefont {{Howard}},
  \bibfnamefont {W~M}}, and\ \bibinfo {author} {\bibfnamefont {P.}~\bibnamefont
  {{M{\"o}ller}}}} (\bibinfo {year} {1980}),\ \bibfield  {title} {\enquote
  {\bibinfo {title} {{Calculated Fission Barriers, Ground-State Masses, and
  Particle Separation Energies for Nuclei with $76 \le Z \le 100$ and $140 \le
  N \le 184$}},}\ }\href {https://doi.org/10.1016/0092-640X(80)90005-4}
  {\bibfield  {journal} {\bibinfo  {journal} {At. Data Nucl. Data Tables}\
  }\textbf {\bibinfo {volume} {25}},\ \bibinfo {pages} {219}}\BibitemShut
  {NoStop}%
\bibitem [{\citenamefont {{H{\"u}depohl}}\ \emph {et~al.}(2010)\citenamefont
  {{H{\"u}depohl}}, \citenamefont {{M{\"u}ller}}, \citenamefont {{Janka}},
  \citenamefont {{Marek}},\ and\ \citenamefont {{Raffelt}}}]{huedepohl10}%
  \BibitemOpen
  \bibfield  {author} {\bibinfo {author} {\bibnamefont {{H{\"u}depohl}},
  \bibfnamefont {L}}, \bibinfo {author} {\bibfnamefont {B.}~\bibnamefont
  {{M{\"u}ller}}}, \bibinfo {author} {\bibfnamefont {H.-T.}\ \bibnamefont
  {{Janka}}}, \bibinfo {author} {\bibfnamefont {A.}~\bibnamefont {{Marek}}},
  and\ \bibinfo {author} {\bibfnamefont {G.~G.}\ \bibnamefont {{Raffelt}}}}
  (\bibinfo {year} {2010}),\ \bibfield  {title} {\enquote {\bibinfo {title}
  {{Neutrino Signal of Electron-Capture Supernovae from Core Collapse to
  Cooling}},}\ }\href {https://doi.org/10.1103/PhysRevLett.104.251101}
  {\bibfield  {journal} {\bibinfo  {journal} {Phys. Rev. Lett.}\ }\textbf
  {\bibinfo {volume} {104}},\ \bibinfo {eid} {251101}}\BibitemShut {NoStop}%
\bibitem [{\citenamefont {{Hulse}}\ and\ \citenamefont
  {{Taylor}}(1975)}]{hulsetaylor75}%
  \BibitemOpen
  \bibfield  {author} {\bibinfo {author} {\bibnamefont {{Hulse}}, \bibfnamefont
  {R~A}}, and\ \bibinfo {author} {\bibfnamefont {J.~H.}\ \bibnamefont
  {{Taylor}}}} (\bibinfo {year} {1975}),\ \bibfield  {title} {\enquote
  {\bibinfo {title} {{Discovery of a pulsar in a binary system}},}\ }\href
  {https://doi.org/10.1086/181708} {\bibfield  {journal} {\bibinfo  {journal}
  {Astrophys. J.}\ }\textbf {\bibinfo {volume} {195}},\ \bibinfo {pages}
  {L51--L53}}\BibitemShut {NoStop}%
\bibitem [{\citenamefont {{Iliadis}}(2007)}]{Iliadis:2007}%
  \BibitemOpen
  \bibfield  {author} {\bibinfo {author} {\bibnamefont {{Iliadis}},
  \bibfnamefont {C}}} (\bibinfo {year} {2007}),\ \href
  {https://doi.org/10.1002/9783527692668} {\emph {\bibinfo {title} {{Nuclear
  Physics of Stars}}}}\ (\bibinfo  {publisher} {Wiley-VCH Verlag},\ \bibinfo
  {address} {Wenheim, Germany})\BibitemShut {NoStop}%
\bibitem [{\citenamefont {{Ishii}}\ \emph {et~al.}(2018)\citenamefont
  {{Ishii}}, \citenamefont {{Shigeyama}},\ and\ \citenamefont
  {{Tanaka}}}]{Ishii.Shigeyama.Tanaka:2018}%
  \BibitemOpen
  \bibfield  {author} {\bibinfo {author} {\bibnamefont {{Ishii}}, \bibfnamefont
  {Ayako}}, \bibinfo {author} {\bibfnamefont {Toshikazu}\ \bibnamefont
  {{Shigeyama}}}, and\ \bibinfo {author} {\bibfnamefont {Masaomi}\ \bibnamefont
  {{Tanaka}}}} (\bibinfo {year} {2018}),\ \bibfield  {title} {\enquote
  {\bibinfo {title} {{Free Neutron Ejection from Shock Breakout in Binary
  Neutron Star Mergers}},}\ }\href {https://doi.org/10.3847/1538-4357/aac385}
  {\bibfield  {journal} {\bibinfo  {journal} {Astrophys. J.}\ }\textbf
  {\bibinfo {volume} {861}},\ \bibinfo {eid} {25}}\BibitemShut {NoStop}%
\bibitem [{\citenamefont {{Ishimaru}}\ and\ \citenamefont
  {{Wanajo}}(1999)}]{Ishimaru.Wanajo:1999}%
  \BibitemOpen
  \bibfield  {author} {\bibinfo {author} {\bibnamefont {{Ishimaru}},
  \bibfnamefont {Y}}, and\ \bibinfo {author} {\bibfnamefont {S.}~\bibnamefont
  {{Wanajo}}}} (\bibinfo {year} {1999}),\ \bibfield  {title} {\enquote
  {\bibinfo {title} {{Enrichment of the R-Process Element Europium in the
  Galactic Halo}},}\ }\href {https://doi.org/10.1086/311829} {\bibfield
  {journal} {\bibinfo  {journal} {Astrophys. J.}\ }\textbf {\bibinfo {volume}
  {511}},\ \bibinfo {pages} {L33--L36}}\BibitemShut {NoStop}%
\bibitem [{\citenamefont {{Ishimaru}}\ \emph {et~al.}(2015)\citenamefont
  {{Ishimaru}}, \citenamefont {{Wanajo}},\ and\ \citenamefont
  {{Prantzos}}}]{ishimaru15}%
  \BibitemOpen
  \bibfield  {author} {\bibinfo {author} {\bibnamefont {{Ishimaru}},
  \bibfnamefont {Y}}, \bibinfo {author} {\bibfnamefont {S.}~\bibnamefont
  {{Wanajo}}}, and\ \bibinfo {author} {\bibfnamefont {N.}~\bibnamefont
  {{Prantzos}}}} (\bibinfo {year} {2015}),\ \bibfield  {title} {\enquote
  {\bibinfo {title} {{Neutron Star Mergers as the Origin of r-process Elements
  in the Galactic Halo Based on the Sub-halo Clustering Scenario}},}\ }\href
  {https://doi.org/10.1088/2041-8205/804/2/L35} {\bibfield  {journal} {\bibinfo
   {journal} {Astrophys. J.}\ }\textbf {\bibinfo {volume} {804}},\ \bibinfo
  {eid} {L35}}\BibitemShut {NoStop}%
\bibitem [{\citenamefont {Ivans}\ \emph {et~al.}(2006)\citenamefont {Ivans},
  \citenamefont {Simmerer}, \citenamefont {Sneden}, \citenamefont {Lawler},
  \citenamefont {Cowan}, \citenamefont {Gallino},\ and\ \citenamefont
  {Bisterzo}}]{Ivans.Simmerer.ea:2006}%
  \BibitemOpen
  \bibfield  {author} {\bibinfo {author} {\bibnamefont {Ivans}, \bibfnamefont
  {I~I}}, \bibinfo {author} {\bibfnamefont {J.}~\bibnamefont {Simmerer}},
  \bibinfo {author} {\bibfnamefont {C.}~\bibnamefont {Sneden}}, \bibinfo
  {author} {\bibfnamefont {J.~E.}\ \bibnamefont {Lawler}}, \bibinfo {author}
  {\bibfnamefont {J.~J.}\ \bibnamefont {Cowan}}, \bibinfo {author}
  {\bibfnamefont {R.}~\bibnamefont {Gallino}}, and\ \bibinfo {author}
  {\bibfnamefont {S.}~\bibnamefont {Bisterzo}}} (\bibinfo {year} {2006}),\
  \bibfield  {title} {\enquote {\bibinfo {title} {{Near-Ultraviolet
  Observations of HD 221170: New Insights into the Nature of r-Process-rich
  Stars}},}\ }\href {https://doi.org/10.1086/504069} {\bibfield  {journal}
  {\bibinfo  {journal} {Astrophys. J.}\ }\textbf {\bibinfo {volume} {645}},\
  \bibinfo {pages} {613--633}}\BibitemShut {NoStop}%
\bibitem [{\citenamefont {{Iwamoto}}\ \emph {et~al.}(1998)\citenamefont
  {{Iwamoto}} \emph {et~al.}}]{Iwamoto.ea:1998}%
  \BibitemOpen
  \bibfield  {author} {\bibinfo {author} {\bibnamefont {{Iwamoto}},
  \bibfnamefont {K}},  \emph {et~al.}} (\bibinfo {year} {1998}),\ \bibfield
  {title} {\enquote {\bibinfo {title} {{A hypernova model for the supernova
  associated with the {\ensuremath{\gamma}}-ray burst of 25 April 1998}},}\
  }\href {https://doi.org/10.1038/27155} {\bibfield  {journal} {\bibinfo
  {journal} {Nature}\ }\textbf {\bibinfo {volume} {395}},\ \bibinfo {pages}
  {672--674}}\BibitemShut {NoStop}%
\bibitem [{\citenamefont {{Janiuk}}(2014)}]{Janiuk:2014}%
  \BibitemOpen
  \bibfield  {author} {\bibinfo {author} {\bibnamefont {{Janiuk}},
  \bibfnamefont {A}}} (\bibinfo {year} {2014}),\ \bibfield  {title} {\enquote
  {\bibinfo {title} {{Nucleosynthesis of elements in gamma-ray burst
  engines}},}\ }\href {https://doi.org/10.1051/0004-6361/201423822} {\bibfield
  {journal} {\bibinfo  {journal} {Astron. \& Astrophys.}\ }\textbf {\bibinfo
  {volume} {568}},\ \bibinfo {eid} {A105}}\BibitemShut {NoStop}%
\bibitem [{\citenamefont {{Janiuk}}(2019{\natexlab{a}})}]{Janiuk:2019}%
  \BibitemOpen
  \bibfield  {author} {\bibinfo {author} {\bibnamefont {{Janiuk}},
  \bibfnamefont {A}}} (\bibinfo {year} {2019}{\natexlab{a}}),\ \bibfield
  {title} {\enquote {\bibinfo {title} {{Nucleosynthesis from dynamical outflows
  of long Gamma-Ray Burst Accretion Disks}},}\ }\href@noop {} {\bibinfo
  {journal} {private communication}\ }\BibitemShut {NoStop}%
\bibitem [{\citenamefont {{Janiuk}}(2017)}]{Janiuk:2017}%
  \BibitemOpen
\bibfield  {journal} {  }\bibfield  {author} {\bibinfo {author} {\bibnamefont
  {{Janiuk}}, \bibfnamefont {Agnieszka}}} (\bibinfo {year} {2017}),\ \bibfield
  {title} {\enquote {\bibinfo {title} {{Microphysics in the Gamma-Ray Burst
  Central Engine}},}\ }\href {https://doi.org/10.3847/1538-4357/aa5f16}
  {\bibfield  {journal} {\bibinfo  {journal} {Astrophys. J.}\ }\textbf
  {\bibinfo {volume} {837}},\ \bibinfo {eid} {39}}\BibitemShut {NoStop}%
\bibitem [{\citenamefont {{Janiuk}}(2019{\natexlab{b}})}]{JaniuksGRB:2019}%
  \BibitemOpen
  \bibfield  {author} {\bibinfo {author} {\bibnamefont {{Janiuk}},
  \bibfnamefont {Agnieszka}}} (\bibinfo {year} {2019}{\natexlab{b}}),\
  \bibfield  {title} {\enquote {\bibinfo {title} {{The r-process
  Nucleosynthesis in the Outflows from Short GRB Accretion Disks}},}\ }\href
  {https://doi.org/10.3847/1538-4357/ab3349} {\bibfield  {journal} {\bibinfo
  {journal} {Astrophys. J.}\ }\textbf {\bibinfo {volume} {882}},\ \bibinfo
  {eid} {163}}\BibitemShut {NoStop}%
\bibitem [{\citenamefont {{Janiuk}}\ and\ \citenamefont
  {{Sapountzis}}(2018)}]{Janiuk.Sapountzis:2018}%
  \BibitemOpen
  \bibfield  {author} {\bibinfo {author} {\bibnamefont {{Janiuk}},
  \bibfnamefont {Agnieszka}}, and\ \bibinfo {author} {\bibfnamefont {Kostas}\
  \bibnamefont {{Sapountzis}}}} (\bibinfo {year} {2018}),\ \bibfield  {title}
  {\enquote {\bibinfo {title} {{Gamma Ray Bursts. Progenitors, accretion in the
  central engine, jet acceleration mechanisms}},}\ }\href@noop {} {\bibfield
  {journal} {\bibinfo  {journal} {arXiv e-prints}\ }}\Eprint
  {https://arxiv.org/abs/1803.07873} {arXiv:1803.07873 [astro-ph.HE]}
  \BibitemShut {NoStop}%
\bibitem [{\citenamefont {{Janiuk}}\ \emph {et~al.}(2018)\citenamefont
  {{Janiuk}}, \citenamefont {{Sukova}},\ and\ \citenamefont
  {{Palit}}}]{Janiuk.Sukova.Palit:2018}%
  \BibitemOpen
  \bibfield  {author} {\bibinfo {author} {\bibnamefont {{Janiuk}},
  \bibfnamefont {Agnieszka}}, \bibinfo {author} {\bibfnamefont {Petra}\
  \bibnamefont {{Sukova}}}, and\ \bibinfo {author} {\bibfnamefont {Ishika}\
  \bibnamefont {{Palit}}}} (\bibinfo {year} {2018}),\ \bibfield  {title}
  {\enquote {\bibinfo {title} {{Accretion in a Dynamical Spacetime and the
  Spinning Up of the Black Hole in the Gamma-Ray Burst Central Engine}},}\
  }\href {https://doi.org/10.3847/1538-4357/aae83f} {\bibfield  {journal}
  {\bibinfo  {journal} {Astrophys. J.}\ }\textbf {\bibinfo {volume} {868}},\
  \bibinfo {eid} {68}}\BibitemShut {NoStop}%
\bibitem [{\citenamefont {{Janka}}(2017{\natexlab{a}})}]{Janka:2017a}%
  \BibitemOpen
  \bibfield  {author} {\bibinfo {author} {\bibnamefont {{Janka}}, \bibfnamefont
  {H-T}}} (\bibinfo {year} {2017}{\natexlab{a}}),\ \enquote {\bibinfo {title}
  {{Neutrino-Driven Explosions}},}\ in\ \href
  {https://doi.org/10.1007/978-3-319-21846-5_109} {\emph {\bibinfo {booktitle}
  {Handbook of Supernovae}}},\ \bibinfo {editor} {edited by\ \bibinfo {editor}
  {\bibfnamefont {A.~W.}\ \bibnamefont {{Alsabti}}}\ and\ \bibinfo {editor}
  {\bibfnamefont {P.}~\bibnamefont {{Murdin}}}}\ (\bibinfo  {publisher}
  {Springer International Publishing},\ \bibinfo {address} {Cham})\BibitemShut
  {NoStop}%
\bibitem [{\citenamefont {{Janka}}(2017{\natexlab{b}})}]{Janka:2016}%
  \BibitemOpen
  \bibfield  {author} {\bibinfo {author} {\bibnamefont {{Janka}}, \bibfnamefont
  {H-T}}} (\bibinfo {year} {2017}{\natexlab{b}}),\ \enquote {\bibinfo {title}
  {{Neutrino Emission from Supernovae}},}\ in\ \href
  {https://doi.org/10.1007/978-3-319-20794-0_4-1} {\emph {\bibinfo {booktitle}
  {Handbook of Supernovae}}},\ \bibinfo {editor} {edited by\ \bibinfo {editor}
  {\bibfnamefont {A.~W.}\ \bibnamefont {{Alsabti}}}\ and\ \bibinfo {editor}
  {\bibfnamefont {P.}~\bibnamefont {{Murdin}}}}\ (\bibinfo  {publisher}
  {Springer International Publishing},\ \bibinfo {address} {Cham})\BibitemShut
  {NoStop}%
\bibitem [{\citenamefont {{Janka}}\ \emph {et~al.}(2008)\citenamefont
  {{Janka}}, \citenamefont {{M{\"u}ller}}, \citenamefont {{Kitaura}},\ and\
  \citenamefont {{Buras}}}]{Janka.Mueller.ea:2008}%
  \BibitemOpen
  \bibfield  {author} {\bibinfo {author} {\bibnamefont {{Janka}}, \bibfnamefont
  {{H-T}}}, \bibinfo {author} {\bibfnamefont {B.}~\bibnamefont {{M{\"u}ller}}},
  \bibinfo {author} {\bibfnamefont {F.~S.}\ \bibnamefont {{Kitaura}}}, and\
  \bibinfo {author} {\bibfnamefont {R.}~\bibnamefont {{Buras}}}} (\bibinfo
  {year} {2008}),\ \bibfield  {title} {\enquote {\bibinfo {title} {{Dynamics of
  shock propagation and nucleosynthesis conditions in O-Ne-Mg core
  supernovae}},}\ }\href {https://doi.org/10.1051/0004-6361:20079334}
  {\bibfield  {journal} {\bibinfo  {journal} {Astron. \& Astrophys.}\ }\textbf
  {\bibinfo {volume} {485}},\ \bibinfo {pages} {199--208}},\ \Eprint
  {https://arxiv.org/abs/0712.4237} {0712.4237} \BibitemShut {NoStop}%
\bibitem [{\citenamefont {Janka}\ \emph {et~al.}(2016)\citenamefont {Janka},
  \citenamefont {Melson},\ and\ \citenamefont
  {Summa}}]{Janka.Melson.Summa:2016}%
  \BibitemOpen
  \bibfield  {author} {\bibinfo {author} {\bibnamefont {Janka}, \bibfnamefont
  {Hans-Thomas}}, \bibinfo {author} {\bibfnamefont {Tobias}\ \bibnamefont
  {Melson}}, and\ \bibinfo {author} {\bibfnamefont {Alexander}\ \bibnamefont
  {Summa}}} (\bibinfo {year} {2016}),\ \bibfield  {title} {\enquote {\bibinfo
  {title} {{Physics of Core-Collapse Supernovae in Three Dimensions: A Sneak
  Preview}},}\ }\href {https://doi.org/10.1146/annurev-nucl-102115-044747}
  {\bibfield  {journal} {\bibinfo  {journal} {Annu. Rev. Nucl. Part. Sci.}\
  }\textbf {\bibinfo {volume} {66}},\ \bibinfo {pages} {341--375}}\BibitemShut
  {NoStop}%
\bibitem [{\citenamefont {{Jeukenne}}\ \emph {et~al.}(1977)\citenamefont
  {{Jeukenne}}, \citenamefont {{Lejeune}},\ and\ \citenamefont
  {{Mahaux}}}]{Jeukenne.Lejeune.Mahaux:1977}%
  \BibitemOpen
  \bibfield  {author} {\bibinfo {author} {\bibnamefont {{Jeukenne}},
  \bibfnamefont {J-P}}, \bibinfo {author} {\bibfnamefont {A.}~\bibnamefont
  {{Lejeune}}}, and\ \bibinfo {author} {\bibfnamefont {C.}~\bibnamefont
  {{Mahaux}}}} (\bibinfo {year} {1977}),\ \bibfield  {title} {\enquote
  {\bibinfo {title} {{Optical-model potential in finite nuclei from Reid's hard
  core interaction}},}\ }\href {https://doi.org/10.1103/PhysRevC.16.80}
  {\bibfield  {journal} {\bibinfo  {journal} {Phys. Rev. C}\ }\textbf {\bibinfo
  {volume} {16}},\ \bibinfo {pages} {80--96}}\BibitemShut {NoStop}%
\bibitem [{\citenamefont {{Ji}}\ \emph {et~al.}(2016)\citenamefont {{Ji}},
  \citenamefont {{Frebel}}, \citenamefont {{Simon}},\ and\ \citenamefont
  {{Chiti}}}]{ji16}%
  \BibitemOpen
  \bibfield  {author} {\bibinfo {author} {\bibnamefont {{Ji}}, \bibfnamefont
  {A~P}}, \bibinfo {author} {\bibfnamefont {A.}~\bibnamefont {{Frebel}}},
  \bibinfo {author} {\bibfnamefont {J.~D.}\ \bibnamefont {{Simon}}}, and\
  \bibinfo {author} {\bibfnamefont {A.}~\bibnamefont {{Chiti}}}} (\bibinfo
  {year} {2016}),\ \bibfield  {title} {\enquote {\bibinfo {title} {{Complete
  Element Abundances of Nine Stars in the r-process Galaxy Reticulum II}},}\
  }\href {https://doi.org/10.3847/0004-637X/830/2/93} {\bibfield  {journal}
  {\bibinfo  {journal} {Astrophys. J.}\ }\textbf {\bibinfo {volume} {830}},\
  \bibinfo {eid} {93}}\BibitemShut {NoStop}%
\bibitem [{\citenamefont {{Ji}}\ \emph
  {et~al.}(2019{\natexlab{a}})\citenamefont {{Ji}}, \citenamefont {{Drout}},\
  and\ \citenamefont {{Hansen}}}]{Ji.Drout.Hansen:2019}%
  \BibitemOpen
  \bibfield  {author} {\bibinfo {author} {\bibnamefont {{Ji}}, \bibfnamefont
  {Alexander~P}}, \bibinfo {author} {\bibfnamefont {Maria~R.}\ \bibnamefont
  {{Drout}}}, and\ \bibinfo {author} {\bibfnamefont {Terese~T.}\ \bibnamefont
  {{Hansen}}}} (\bibinfo {year} {2019}{\natexlab{a}}),\ \bibfield  {title}
  {\enquote {\bibinfo {title} {{The Lanthanide Fraction Distribution in
  Metal-poor Stars: A Test of Neutron Star Mergers as the Dominant r-process
  Site}},}\ }\href {https://doi.org/10.3847/1538-4357/ab3291} {\bibfield
  {journal} {\bibinfo  {journal} {Astrophys. J.}\ }\textbf {\bibinfo {volume}
  {882}},\ \bibinfo {eid} {40}}\BibitemShut {NoStop}%
\bibitem [{\citenamefont {{Ji}}\ and\ \citenamefont
  {{Frebel}}(2018)}]{Ji.Frebel:2018}%
  \BibitemOpen
  \bibfield  {author} {\bibinfo {author} {\bibnamefont {{Ji}}, \bibfnamefont
  {Alexander~P}}, and\ \bibinfo {author} {\bibfnamefont {Anna}\ \bibnamefont
  {{Frebel}}}} (\bibinfo {year} {2018}),\ \bibfield  {title} {\enquote
  {\bibinfo {title} {{From Actinides to Zinc: Using the Full Abundance Pattern
  of the Brightest Star in Reticulum II to Distinguish between Different
  r-process Sites}},}\ }\href {https://doi.org/10.3847/1538-4357/aab14a}
  {\bibfield  {journal} {\bibinfo  {journal} {Astrophys. J.}\ }\textbf
  {\bibinfo {volume} {856}},\ \bibinfo {eid} {138}}\BibitemShut {NoStop}%
\bibitem [{\citenamefont {{Ji}}\ \emph
  {et~al.}(2019{\natexlab{b}})\citenamefont {{Ji}}, \citenamefont {{Simon}},
  \citenamefont {{Frebel}}, \citenamefont {{Venn}},\ and\ \citenamefont
  {{Hansen}}}]{Ji.Simon.ea:2019}%
  \BibitemOpen
  \bibfield  {author} {\bibinfo {author} {\bibnamefont {{Ji}}, \bibfnamefont
  {Alexander~P}}, \bibinfo {author} {\bibfnamefont {Joshua~D.}\ \bibnamefont
  {{Simon}}}, \bibinfo {author} {\bibfnamefont {Anna}\ \bibnamefont
  {{Frebel}}}, \bibinfo {author} {\bibfnamefont {Kim~A.}\ \bibnamefont
  {{Venn}}}, and\ \bibinfo {author} {\bibfnamefont {Terese~T.}\ \bibnamefont
  {{Hansen}}}} (\bibinfo {year} {2019}{\natexlab{b}}),\ \bibfield  {title}
  {\enquote {\bibinfo {title} {{Chemical Abundances in the Ultra-faint Dwarf
  Galaxies Grus I and Triangulum II: Neutron-capture Elements as a Defining
  Feature of the Faintest Dwarfs}},}\ }\href
  {https://doi.org/10.3847/1538-4357/aaf3bb} {\bibfield  {journal} {\bibinfo
  {journal} {Astrophys. J.}\ }\textbf {\bibinfo {volume} {870}},\ \bibinfo
  {eid} {83}}\BibitemShut {NoStop}%
\bibitem [{\citenamefont {{Jin}}\ \emph {et~al.}(2020)\citenamefont {{Jin}},
  \citenamefont {{Covino}}, \citenamefont {{Liao}}, \citenamefont {{Li}},
  \citenamefont {{D'Avanzo}}, \citenamefont {{Fan}},\ and\ \citenamefont
  {{Wei}}}]{Jin.Covino.ea:2020}%
  \BibitemOpen
  \bibfield  {author} {\bibinfo {author} {\bibnamefont {{Jin}}, \bibfnamefont
  {Zhi-Ping}}, \bibinfo {author} {\bibfnamefont {Stefano}\ \bibnamefont
  {{Covino}}}, \bibinfo {author} {\bibfnamefont {Neng-Hui}\ \bibnamefont
  {{Liao}}}, \bibinfo {author} {\bibfnamefont {Xiang}\ \bibnamefont {{Li}}},
  \bibinfo {author} {\bibfnamefont {Paolo}\ \bibnamefont {{D'Avanzo}}},
  \bibinfo {author} {\bibfnamefont {Yi-Zhong}\ \bibnamefont {{Fan}}}, and\
  \bibinfo {author} {\bibfnamefont {Da-Ming}\ \bibnamefont {{Wei}}}} (\bibinfo
  {year} {2020}),\ \bibfield  {title} {\enquote {\bibinfo {title} {{A kilonova
  associated with GRB 070809}},}\ }\href
  {https://doi.org/10.1038/s41550-019-0892-y} {\bibfield  {journal} {\bibinfo
  {journal} {Nature Astronomy}\ }\textbf {\bibinfo {volume} {4}},\ \bibinfo
  {pages} {77--82}}\BibitemShut {NoStop}%
\bibitem [{\citenamefont {{Jin}}\ \emph {et~al.}(2016)\citenamefont {{Jin}},
  \citenamefont {{Hotokezaka}}, \citenamefont {{Li}}, \citenamefont {{Tanaka}},
  \citenamefont {{D'Avanzo}}, \citenamefont {{Fan}}, \citenamefont {{Covino}},
  \citenamefont {{Wei}},\ and\ \citenamefont
  {{Piran}}}]{Jin.Hotokezaka.ea:2016}%
  \BibitemOpen
  \bibfield  {author} {\bibinfo {author} {\bibnamefont {{Jin}}, \bibfnamefont
  {Zhi-Ping}}, \bibinfo {author} {\bibfnamefont {Kenta}\ \bibnamefont
  {{Hotokezaka}}}, \bibinfo {author} {\bibfnamefont {Xiang}\ \bibnamefont
  {{Li}}}, \bibinfo {author} {\bibfnamefont {Masaomi}\ \bibnamefont
  {{Tanaka}}}, \bibinfo {author} {\bibfnamefont {Paolo}\ \bibnamefont
  {{D'Avanzo}}}, \bibinfo {author} {\bibfnamefont {Yi-Zhong}\ \bibnamefont
  {{Fan}}}, \bibinfo {author} {\bibfnamefont {Stefano}\ \bibnamefont
  {{Covino}}}, \bibinfo {author} {\bibfnamefont {Da-Ming}\ \bibnamefont
  {{Wei}}}, and\ \bibinfo {author} {\bibfnamefont {Tsvi}\ \bibnamefont
  {{Piran}}}} (\bibinfo {year} {2016}),\ \bibfield  {title} {\enquote {\bibinfo
  {title} {{The Macronova in GRB 050709 and the GRB-macronova connection}},}\
  }\href {https://doi.org/10.1038/ncomms12898} {\bibfield  {journal} {\bibinfo
  {journal} {Nature Communications}\ }\textbf {\bibinfo {volume} {7}},\
  \bibinfo {eid} {12898}}\BibitemShut {NoStop}%
\bibitem [{\citenamefont {{Jin}}\ \emph {et~al.}(2015)\citenamefont {{Jin}},
  \citenamefont {{Li}}, \citenamefont {{Cano}}, \citenamefont {{Covino}},
  \citenamefont {{Fan}},\ and\ \citenamefont {{Wei}}}]{Jin.Li.ea:2015}%
  \BibitemOpen
  \bibfield  {author} {\bibinfo {author} {\bibnamefont {{Jin}}, \bibfnamefont
  {Zhi-Ping}}, \bibinfo {author} {\bibfnamefont {Xiang}\ \bibnamefont {{Li}}},
  \bibinfo {author} {\bibfnamefont {Zach}\ \bibnamefont {{Cano}}}, \bibinfo
  {author} {\bibfnamefont {Stefano}\ \bibnamefont {{Covino}}}, \bibinfo
  {author} {\bibfnamefont {Yi-Zhong}\ \bibnamefont {{Fan}}}, and\ \bibinfo
  {author} {\bibfnamefont {Da-Ming}\ \bibnamefont {{Wei}}}} (\bibinfo {year}
  {2015}),\ \bibfield  {title} {\enquote {\bibinfo {title} {{The Light Curve of
  the Macronova Associated with the Long-Short Burst GRB 060614}},}\ }\href
  {https://doi.org/10.1088/2041-8205/811/2/L22} {\bibfield  {journal} {\bibinfo
   {journal} {Astrophys. J. Lett.}\ }\textbf {\bibinfo {volume} {811}},\
  \bibinfo {eid} {L22}}\BibitemShut {NoStop}%
\bibitem [{\citenamefont {{Johnson}}\ \emph {et~al.}(1992)\citenamefont
  {{Johnson}}, \citenamefont {{Koonin}}, \citenamefont {{Lang}},\ and\
  \citenamefont {{Ormand}}}]{Johnson.Koonin.ea:1992}%
  \BibitemOpen
  \bibfield  {author} {\bibinfo {author} {\bibnamefont {{Johnson}},
  \bibfnamefont {C~W}}, \bibinfo {author} {\bibfnamefont {S.~E.}\ \bibnamefont
  {{Koonin}}}, \bibinfo {author} {\bibfnamefont {G.~H.}\ \bibnamefont
  {{Lang}}}, and\ \bibinfo {author} {\bibfnamefont {W.~E.}\ \bibnamefont
  {{Ormand}}}} (\bibinfo {year} {1992}),\ \bibfield  {title} {\enquote
  {\bibinfo {title} {{Monte Carlo methods for the nuclear shell model}},}\
  }\href {https://doi.org/10.1103/PhysRevLett.69.3157} {\bibfield  {journal}
  {\bibinfo  {journal} {Phys. Rev. Lett.}\ }\textbf {\bibinfo {volume} {69}},\
  \bibinfo {pages} {3157--3160}}\BibitemShut {NoStop}%
\bibitem [{\citenamefont {{Jones}}\ \emph {et~al.}(2010)\citenamefont {{Jones}}
  \emph {et~al.}}]{nature09048}%
  \BibitemOpen
  \bibfield  {author} {\bibinfo {author} {\bibnamefont {{Jones}}, \bibfnamefont
  {K~L}},  \emph {et~al.}} (\bibinfo {year} {2010}),\ \bibfield  {title}
  {\enquote {\bibinfo {title} {{The magic nature of $^{132}$Sn explored through
  the single-particle states of $^{133}$Sn}},}\ }\href
  {https://doi.org/10.1038/nature09048} {\bibfield  {journal} {\bibinfo
  {journal} {Nature}\ }\textbf {\bibinfo {volume} {465}},\ \bibinfo {pages}
  {454--457}}\BibitemShut {NoStop}%
\bibitem [{\citenamefont {{Jones}}\ \emph {et~al.}(2014)\citenamefont
  {{Jones}}, \citenamefont {{Hirschi}},\ and\ \citenamefont
  {{Nomoto}}}]{jones14}%
  \BibitemOpen
  \bibfield  {author} {\bibinfo {author} {\bibnamefont {{Jones}}, \bibfnamefont
  {S}}, \bibinfo {author} {\bibfnamefont {R.}~\bibnamefont {{Hirschi}}}, and\
  \bibinfo {author} {\bibfnamefont {K.}~\bibnamefont {{Nomoto}}}} (\bibinfo
  {year} {2014}),\ \bibfield  {title} {\enquote {\bibinfo {title} {{The Final
  Fate of Stars that Ignite Neon and Oxygen Off-center: Electron Capture or
  Iron Core-collapse Supernova?}}}\ }\href
  {https://doi.org/10.1088/0004-637X/797/2/83} {\bibfield  {journal} {\bibinfo
  {journal} {Astrophys. J.}\ }\textbf {\bibinfo {volume} {797}},\ \bibinfo
  {eid} {83}}\BibitemShut {NoStop}%
\bibitem [{\citenamefont {{Jones}}\ \emph
  {et~al.}(2016{\natexlab{a}})\citenamefont {{Jones}}, \citenamefont
  {{Ritter}}, \citenamefont {{Herwig}}, \citenamefont {{Fryer}}, \citenamefont
  {{Pignatari}}, \citenamefont {{Bertolli}},\ and\ \citenamefont
  {{Paxton}}}]{Jones.Ritter.ea:2016}%
  \BibitemOpen
  \bibfield  {author} {\bibinfo {author} {\bibnamefont {{Jones}}, \bibfnamefont
  {S}}, \bibinfo {author} {\bibfnamefont {C.}~\bibnamefont {{Ritter}}},
  \bibinfo {author} {\bibfnamefont {F.}~\bibnamefont {{Herwig}}}, \bibinfo
  {author} {\bibfnamefont {C.}~\bibnamefont {{Fryer}}}, \bibinfo {author}
  {\bibfnamefont {M.}~\bibnamefont {{Pignatari}}}, \bibinfo {author}
  {\bibfnamefont {M.~G.}\ \bibnamefont {{Bertolli}}}, and\ \bibinfo {author}
  {\bibfnamefont {B.}~\bibnamefont {{Paxton}}}} (\bibinfo {year}
  {2016}{\natexlab{a}}),\ \bibfield  {title} {\enquote {\bibinfo {title} {{H
  ingestion into He-burning convection zones in super-AGB stellar models as a
  potential site for intermediate neutron-density nucleosynthesis}},}\ }\href
  {https://doi.org/10.1093/mnras/stv2488} {\bibfield  {journal} {\bibinfo
  {journal} {Mon. Not. Roy. Astron. Soc.}\ }\textbf {\bibinfo {volume} {455}},\
  \bibinfo {pages} {3848--3863}}\BibitemShut {NoStop}%
\bibitem [{\citenamefont {{Jones}}\ \emph
  {et~al.}(2016{\natexlab{b}})\citenamefont {{Jones}}, \citenamefont
  {{R{\"o}pke}}, \citenamefont {{Pakmor}}, \citenamefont {{Seitenzahl}},
  \citenamefont {{Ohlmann}},\ and\ \citenamefont
  {{Edelmann}}}]{Jones.Roepke.ea:2016}%
  \BibitemOpen
  \bibfield  {author} {\bibinfo {author} {\bibnamefont {{Jones}}, \bibfnamefont
  {S}}, \bibinfo {author} {\bibfnamefont {F.~K.}\ \bibnamefont {{R{\"o}pke}}},
  \bibinfo {author} {\bibfnamefont {R.}~\bibnamefont {{Pakmor}}}, \bibinfo
  {author} {\bibfnamefont {I.~R.}\ \bibnamefont {{Seitenzahl}}}, \bibinfo
  {author} {\bibfnamefont {S.~T.}\ \bibnamefont {{Ohlmann}}}, and\ \bibinfo
  {author} {\bibfnamefont {P.~V.~F.}\ \bibnamefont {{Edelmann}}}} (\bibinfo
  {year} {2016}{\natexlab{b}}),\ \bibfield  {title} {\enquote {\bibinfo {title}
  {{Do electron-capture supernovae make neutron stars?. First multidimensional
  hydrodynamic simulations of the oxygen deflagration}},}\ }\href
  {https://doi.org/10.1051/0004-6361/201628321} {\bibfield  {journal} {\bibinfo
   {journal} {Astron. \& Astrophys.}\ }\textbf {\bibinfo {volume} {593}},\
  \bibinfo {eid} {A72}}\BibitemShut {NoStop}%
\bibitem [{\citenamefont {{Junghans}}\ \emph {et~al.}(1998)\citenamefont
  {{Junghans}}, \citenamefont {{de Jong}}, \citenamefont {{Clerc}},
  \citenamefont {{Ignatyuk}}, \citenamefont {{Kudyaev}},\ and\ \citenamefont
  {{Schmidt}}}]{Junghans.deJong.ea:1998}%
  \BibitemOpen
  \bibfield  {author} {\bibinfo {author} {\bibnamefont {{Junghans}},
  \bibfnamefont {A~R}}, \bibinfo {author} {\bibfnamefont {M.}~\bibnamefont {{de
  Jong}}}, \bibinfo {author} {\bibfnamefont {H.-G.}\ \bibnamefont {{Clerc}}},
  \bibinfo {author} {\bibfnamefont {A.~V.}\ \bibnamefont {{Ignatyuk}}},
  \bibinfo {author} {\bibfnamefont {G.~A.}\ \bibnamefont {{Kudyaev}}}, and\
  \bibinfo {author} {\bibfnamefont {K.-H.}\ \bibnamefont {{Schmidt}}}}
  (\bibinfo {year} {1998}),\ \bibfield  {title} {\enquote {\bibinfo {title}
  {{Projectile-fragment yields as a probe for the collective enhancement in the
  nuclear level density}},}\ }\href
  {https://doi.org/10.1016/S0375-9474(98)00658-7} {\bibfield  {journal}
  {\bibinfo  {journal} {Nucl. Phys. A}\ }\textbf {\bibinfo {volume} {629}},\
  \bibinfo {pages} {635--655}}\BibitemShut {NoStop}%
\bibitem [{\citenamefont {Juodagalvis}\ \emph {et~al.}(2010)\citenamefont
  {Juodagalvis}, \citenamefont {Langanke}, \citenamefont {Hix}, \citenamefont
  {Mart{\'{i}}nez-Pinedo},\ and\ \citenamefont
  {Sampaio}}]{Juodagalvis.Langanke.ea:2010}%
  \BibitemOpen
  \bibfield  {author} {\bibinfo {author} {\bibnamefont {Juodagalvis},
  \bibfnamefont {A}}, \bibinfo {author} {\bibfnamefont {Karlheinz}\
  \bibnamefont {Langanke}}, \bibinfo {author} {\bibfnamefont {W.R.}\
  \bibnamefont {Hix}}, \bibinfo {author} {\bibfnamefont {Gabriel}\ \bibnamefont
  {Mart{\'{i}}nez-Pinedo}}, and\ \bibinfo {author} {\bibfnamefont {J.M.}\
  \bibnamefont {Sampaio}}} (\bibinfo {year} {2010}),\ \bibfield  {title}
  {\enquote {\bibinfo {title} {{Improved estimate of electron capture rates on
  nuclei during stellar core collapse}},}\ }\href
  {https://doi.org/10.1016/j.nuclphysa.2010.09.012} {\bibfield  {journal}
  {\bibinfo  {journal} {Nucl. Phys. A}\ }\textbf {\bibinfo {volume} {848}},\
  \bibinfo {pages} {454--478}}\BibitemShut {NoStop}%
\bibitem [{\citenamefont {{Just}}\ \emph
  {et~al.}(2015{\natexlab{a}})\citenamefont {{Just}}, \citenamefont
  {{Bauswein}}, \citenamefont {{Pulpillo}}, \citenamefont {{Goriely}},\ and\
  \citenamefont {{Janka}}}]{Just.Bauswein.ea:2015}%
  \BibitemOpen
  \bibfield  {author} {\bibinfo {author} {\bibnamefont {{Just}}, \bibfnamefont
  {O}}, \bibinfo {author} {\bibfnamefont {A.}~\bibnamefont {{Bauswein}}},
  \bibinfo {author} {\bibfnamefont {R.~A.}\ \bibnamefont {{Pulpillo}}},
  \bibinfo {author} {\bibfnamefont {S.}~\bibnamefont {{Goriely}}}, and\
  \bibinfo {author} {\bibfnamefont {H.-T.}\ \bibnamefont {{Janka}}}} (\bibinfo
  {year} {2015}{\natexlab{a}}),\ \bibfield  {title} {\enquote {\bibinfo {title}
  {{Comprehensive nucleosynthesis analysis for ejecta of compact binary
  mergers}},}\ }\href {https://doi.org/10.1093/mnras/stv009} {\bibfield
  {journal} {\bibinfo  {journal} {Mon. Not. Roy. Astron. Soc.}\ }\textbf
  {\bibinfo {volume} {448}},\ \bibinfo {pages} {541--567}}\BibitemShut
  {NoStop}%
\bibitem [{\citenamefont {{Just}}\ \emph
  {et~al.}(2015{\natexlab{b}})\citenamefont {{Just}}, \citenamefont
  {{Obergaulinger}},\ and\ \citenamefont
  {{Janka}}}]{Just.Obergaulinger.Janka:2015}%
  \BibitemOpen
  \bibfield  {author} {\bibinfo {author} {\bibnamefont {{Just}}, \bibfnamefont
  {O}}, \bibinfo {author} {\bibfnamefont {M.}~\bibnamefont {{Obergaulinger}}},
  and\ \bibinfo {author} {\bibfnamefont {H.~T.}\ \bibnamefont {{Janka}}}}
  (\bibinfo {year} {2015}{\natexlab{b}}),\ \bibfield  {title} {\enquote
  {\bibinfo {title} {{A new multidimensional, energy-dependent two-moment
  transport code for neutrino-hydrodynamics}},}\ }\href
  {https://doi.org/10.1093/mnras/stv1892} {\bibfield  {journal} {\bibinfo
  {journal} {Mon. Not. Roy. Astron. Soc.}\ }\textbf {\bibinfo {volume} {453}},\
  \bibinfo {pages} {3386--3413}}\BibitemShut {NoStop}%
\bibitem [{\citenamefont {{Just}}\ \emph {et~al.}(2016)\citenamefont {{Just}},
  \citenamefont {{Obergaulinger}}, \citenamefont {{Janka}}, \citenamefont
  {{Bauswein}},\ and\ \citenamefont {{Schwarz}}}]{just16}%
  \BibitemOpen
  \bibfield  {author} {\bibinfo {author} {\bibnamefont {{Just}}, \bibfnamefont
  {O}}, \bibinfo {author} {\bibfnamefont {M.}~\bibnamefont {{Obergaulinger}}},
  \bibinfo {author} {\bibfnamefont {H.-T.}\ \bibnamefont {{Janka}}}, \bibinfo
  {author} {\bibfnamefont {A.}~\bibnamefont {{Bauswein}}}, and\ \bibinfo
  {author} {\bibfnamefont {N.}~\bibnamefont {{Schwarz}}}} (\bibinfo {year}
  {2016}),\ \bibfield  {title} {\enquote {\bibinfo {title} {{Neutron-star
  Merger Ejecta as Obstacles to Neutrino-powered Jets of Gamma-Ray Bursts}},}\
  }\href {https://doi.org/10.3847/2041-8205/816/2/L30} {\bibfield  {journal}
  {\bibinfo  {journal} {Astrophys. J.}\ }\textbf {\bibinfo {volume} {816}},\
  \bibinfo {eid} {L30}}\BibitemShut {NoStop}%
\bibitem [{\citenamefont {{Kajino}}\ \emph {et~al.}(2019)\citenamefont
  {{Kajino}}, \citenamefont {{Aoki}}, \citenamefont {{Balantekin}},
  \citenamefont {{Diehl}}, \citenamefont {{Famiano}},\ and\ \citenamefont
  {{Mathews}}}]{Kajino.Aoki.ea:2019}%
  \BibitemOpen
  \bibfield  {author} {\bibinfo {author} {\bibnamefont {{Kajino}},
  \bibfnamefont {T}}, \bibinfo {author} {\bibfnamefont {W.}~\bibnamefont
  {{Aoki}}}, \bibinfo {author} {\bibfnamefont {A.~B.}\ \bibnamefont
  {{Balantekin}}}, \bibinfo {author} {\bibfnamefont {R.}~\bibnamefont
  {{Diehl}}}, \bibinfo {author} {\bibfnamefont {M.~A.}\ \bibnamefont
  {{Famiano}}}, and\ \bibinfo {author} {\bibfnamefont {G.~J.}\ \bibnamefont
  {{Mathews}}}} (\bibinfo {year} {2019}),\ \bibfield  {title} {\enquote
  {\bibinfo {title} {{Current status of r-process nucleosynthesis}},}\ }\href
  {https://doi.org/10.1016/j.ppnp.2019.02.008} {\bibfield  {journal} {\bibinfo
  {journal} {Prog. Part. Nucl. Phys.}\ }\textbf {\bibinfo {volume} {107}},\
  \bibinfo {pages} {109--166}}\BibitemShut {NoStop}%
\bibitem [{\citenamefont {{Kajino}}\ and\ \citenamefont
  {{Mathews}}(2017)}]{Kajino.Mathews:2017}%
  \BibitemOpen
  \bibfield  {author} {\bibinfo {author} {\bibnamefont {{Kajino}},
  \bibfnamefont {T}}, and\ \bibinfo {author} {\bibfnamefont {G.~J.}\
  \bibnamefont {{Mathews}}}} (\bibinfo {year} {2017}),\ \bibfield  {title}
  {\enquote {\bibinfo {title} {{Impact of new data for neutron-rich heavy
  nuclei on theoretical models for r-process nucleosynthesis}},}\ }\href
  {https://doi.org/10.1088/1361-6633/aa6a25} {\bibfield  {journal} {\bibinfo
  {journal} {Rep. Prog. Phys.}\ }\textbf {\bibinfo {volume} {80}},\ \bibinfo
  {eid} {084901}}\BibitemShut {NoStop}%
\bibitem [{\citenamefont {{Kalogera}}\ and\ \citenamefont
  {{Fryer}}(1999)}]{Kalogera.Fryer:1999}%
  \BibitemOpen
  \bibfield  {author} {\bibinfo {author} {\bibnamefont {{Kalogera}},
  \bibfnamefont {V}}, and\ \bibinfo {author} {\bibfnamefont {C.~L.}\
  \bibnamefont {{Fryer}}}} (\bibinfo {year} {1999}),\ \bibfield  {title}
  {\enquote {\bibinfo {title} {{Formation of the observed double neutron star
  systems}},}\ }\href {https://doi.org/10.1080/10556799908203009} {\bibfield
  {journal} {\bibinfo  {journal} {Astronomical and Astrophysical Transactions}\
  }\textbf {\bibinfo {volume} {18}},\ \bibinfo {pages} {515--520}}\BibitemShut
  {NoStop}%
\bibitem [{\citenamefont {{Kaplan}}\ \emph {et~al.}(2014)\citenamefont
  {{Kaplan}}, \citenamefont {{Ott}}, \citenamefont {{O'Connor}}, \citenamefont
  {{Kiuchi}}, \citenamefont {{Roberts}},\ and\ \citenamefont
  {{Duez}}}]{Kaplan.Ott.ea:2014}%
  \BibitemOpen
  \bibfield  {author} {\bibinfo {author} {\bibnamefont {{Kaplan}},
  \bibfnamefont {J~D}}, \bibinfo {author} {\bibfnamefont {C.~D.}\ \bibnamefont
  {{Ott}}}, \bibinfo {author} {\bibfnamefont {E.~P.}\ \bibnamefont
  {{O'Connor}}}, \bibinfo {author} {\bibfnamefont {K.}~\bibnamefont
  {{Kiuchi}}}, \bibinfo {author} {\bibfnamefont {L.}~\bibnamefont {{Roberts}}},
  and\ \bibinfo {author} {\bibfnamefont {M.}~\bibnamefont {{Duez}}}} (\bibinfo
  {year} {2014}),\ \bibfield  {title} {\enquote {\bibinfo {title} {{The
  Influence of Thermal Pressure on Equilibrium Models of Hypermassive Neutron
  Star Merger Remnants}},}\ }\href {https://doi.org/10.1088/0004-637X/790/1/19}
  {\bibfield  {journal} {\bibinfo  {journal} {Astrophys. J.}\ }\textbf
  {\bibinfo {volume} {790}},\ \bibinfo {eid} {19}}\BibitemShut {NoStop}%
\bibitem [{\citenamefont {K{\"a}ppeler}(1999)}]{Kaeppeler:1999}%
  \BibitemOpen
  \bibfield  {author} {\bibinfo {author} {\bibnamefont {K{\"a}ppeler},
  \bibfnamefont {F}}} (\bibinfo {year} {1999}),\ \bibfield  {title} {\enquote
  {\bibinfo {title} {{The Origin of the Heavy Elements: The s Process}},}\
  }\href {https://doi.org/10.1016/S0146-6410(99)00098-8} {\bibfield  {journal}
  {\bibinfo  {journal} {Prog. Part. Nucl. Phys.}\ }\textbf {\bibinfo {volume}
  {43}},\ \bibinfo {pages} {419--483}}\BibitemShut {NoStop}%
\bibitem [{\citenamefont {{K{\"a}ppeler}}\ \emph {et~al.}(2011)\citenamefont
  {{K{\"a}ppeler}}, \citenamefont {{Gallino}}, \citenamefont {{Bisterzo}},\
  and\ \citenamefont {{Aoki}}}]{Kaeppeler.Gallino.ea:2011}%
  \BibitemOpen
  \bibfield  {author} {\bibinfo {author} {\bibnamefont {{K{\"a}ppeler}},
  \bibfnamefont {F}}, \bibinfo {author} {\bibfnamefont {R.}~\bibnamefont
  {{Gallino}}}, \bibinfo {author} {\bibfnamefont {S.}~\bibnamefont
  {{Bisterzo}}}, and\ \bibinfo {author} {\bibfnamefont {W.}~\bibnamefont
  {{Aoki}}}} (\bibinfo {year} {2011}),\ \bibfield  {title} {\enquote {\bibinfo
  {title} {{The s process: Nuclear physics, stellar models, and
  observations}},}\ }\href {https://doi.org/10.1103/RevModPhys.83.157}
  {\bibfield  {journal} {\bibinfo  {journal} {Rev. Mod. Phys.}\ }\textbf
  {\bibinfo {volume} {83}},\ \bibinfo {pages} {157--194}}\BibitemShut {NoStop}%
\bibitem [{\citenamefont {{Karakas}}\ and\ \citenamefont
  {{Lattanzio}}(2014)}]{karakas14}%
  \BibitemOpen
  \bibfield  {author} {\bibinfo {author} {\bibnamefont {{Karakas}},
  \bibfnamefont {A~I}}, and\ \bibinfo {author} {\bibfnamefont {J.~C.}\
  \bibnamefont {{Lattanzio}}}} (\bibinfo {year} {2014}),\ \bibfield  {title}
  {\enquote {\bibinfo {title} {{The Dawes Review 2: Nucleosynthesis and Stellar
  Yields of Low- and Intermediate-Mass Single Stars}},}\ }\href
  {https://doi.org/10.1017/pasa.2014.21} {\bibfield  {journal} {\bibinfo
  {journal} {Publ. Astron. Soc. Austr.}\ }\textbf {\bibinfo {volume} {31}},\
  \bibinfo {eid} {e030}}\BibitemShut {NoStop}%
\bibitem [{\citenamefont {{Kasen}}\ \emph {et~al.}(2013)\citenamefont
  {{Kasen}}, \citenamefont {{Badnell}},\ and\ \citenamefont
  {{Barnes}}}]{kasen13}%
  \BibitemOpen
  \bibfield  {author} {\bibinfo {author} {\bibnamefont {{Kasen}}, \bibfnamefont
  {D}}, \bibinfo {author} {\bibfnamefont {N.~R.}\ \bibnamefont {{Badnell}}},
  and\ \bibinfo {author} {\bibfnamefont {J.}~\bibnamefont {{Barnes}}}}
  (\bibinfo {year} {2013}),\ \bibfield  {title} {\enquote {\bibinfo {title}
  {{Opacities and Spectra of the r-process Ejecta from Neutron Star
  Mergers}},}\ }\href {https://doi.org/10.1088/0004-637X/774/1/25} {\bibfield
  {journal} {\bibinfo  {journal} {Astrophys. J.}\ }\textbf {\bibinfo {volume}
  {774}},\ \bibinfo {eid} {25}}\BibitemShut {NoStop}%
\bibitem [{\citenamefont {{Kasen}}\ and\ \citenamefont
  {{Bildsten}}(2010)}]{Kasen.Bildsten:2010}%
  \BibitemOpen
  \bibfield  {author} {\bibinfo {author} {\bibnamefont {{Kasen}}, \bibfnamefont
  {D}}, and\ \bibinfo {author} {\bibfnamefont {L.}~\bibnamefont {{Bildsten}}}}
  (\bibinfo {year} {2010}),\ \bibfield  {title} {\enquote {\bibinfo {title}
  {{Supernova Light Curves Powered by Young Magnetars}},}\ }\href
  {https://doi.org/10.1088/0004-637X/717/1/245} {\bibfield  {journal} {\bibinfo
   {journal} {Astrophys. J.}\ }\textbf {\bibinfo {volume} {717}},\ \bibinfo
  {pages} {245--249}}\BibitemShut {NoStop}%
\bibitem [{\citenamefont {{Kasen}}\ and\ \citenamefont
  {{Barnes}}(2019)}]{Kasen.Barnes:2019}%
  \BibitemOpen
  \bibfield  {author} {\bibinfo {author} {\bibnamefont {{Kasen}}, \bibfnamefont
  {Daniel}}, and\ \bibinfo {author} {\bibfnamefont {Jennifer}\ \bibnamefont
  {{Barnes}}}} (\bibinfo {year} {2019}),\ \bibfield  {title} {\enquote
  {\bibinfo {title} {{Radioactive Heating and Late Time Kilonova Light
  Curves}},}\ }\href {https://doi.org/10.3847/1538-4357/ab06c2} {\bibfield
  {journal} {\bibinfo  {journal} {Astrophys. J.}\ }\textbf {\bibinfo {volume}
  {876}},\ \bibinfo {eid} {128}}\BibitemShut {NoStop}%
\bibitem [{\citenamefont {Kasen}\ \emph {et~al.}(2017)\citenamefont {Kasen},
  \citenamefont {Metzger}, \citenamefont {Barnes}, \citenamefont {Quataert},\
  and\ \citenamefont {Ramirez-Ruiz}}]{Kasen.Metzger.ea:2017}%
  \BibitemOpen
  \bibfield  {author} {\bibinfo {author} {\bibnamefont {Kasen}, \bibfnamefont
  {Daniel}}, \bibinfo {author} {\bibfnamefont {Brian}\ \bibnamefont {Metzger}},
  \bibinfo {author} {\bibfnamefont {Jennifer}\ \bibnamefont {Barnes}}, \bibinfo
  {author} {\bibfnamefont {Eliot}\ \bibnamefont {Quataert}}, and\ \bibinfo
  {author} {\bibfnamefont {Enrico}\ \bibnamefont {Ramirez-Ruiz}}} (\bibinfo
  {year} {2017}),\ \bibfield  {title} {\enquote {\bibinfo {title} {{Origin of
  the heavy elements in binary neutron-star mergers from a gravitational-wave
  event}},}\ }\href {https://doi.org/10.1038/nature24453} {\bibfield  {journal}
  {\bibinfo  {journal} {Nature}\ }\textbf {\bibinfo {volume} {551}},\ \bibinfo
  {pages} {80--84}}\BibitemShut {NoStop}%
\bibitem [{\citenamefont {{Kasliwal}}\ \emph {et~al.}(2017)\citenamefont
  {{Kasliwal}} \emph {et~al.}}]{kasliwal17}%
  \BibitemOpen
  \bibfield  {author} {\bibinfo {author} {\bibnamefont {{Kasliwal}},
  \bibfnamefont {M~M}},  \emph {et~al.}} (\bibinfo {year} {2017}),\ \bibfield
  {title} {\enquote {\bibinfo {title} {{Illuminating gravitational waves: A
  concordant picture of photons from a neutron star merger}},}\ }\href
  {https://doi.org/10.1126/science.aap9455} {\bibfield  {journal} {\bibinfo
  {journal} {Science}\ }\textbf {\bibinfo {volume} {358}},\ \bibinfo {pages}
  {1559--1565}}\BibitemShut {NoStop}%
\bibitem [{\citenamefont {{Kasliwal}}\ \emph {et~al.}(2019)\citenamefont
  {{Kasliwal}}, \citenamefont {{Kasen}}, \citenamefont {{Lau}}, \citenamefont
  {{Perley}}, \citenamefont {{Rosswog}}, \citenamefont {{Ofek}}, \citenamefont
  {{Hotokezaka}}, \citenamefont {{Chary}}, \citenamefont {{Sollerman}},
  \citenamefont {{Goobar}},\ and\ \citenamefont
  {{Kaplan}}}]{Kasliwal.Kasen.ea:2019}%
  \BibitemOpen
  \bibfield  {author} {\bibinfo {author} {\bibnamefont {{Kasliwal}},
  \bibfnamefont {Mansi~M}}, \bibinfo {author} {\bibfnamefont {Daniel}\
  \bibnamefont {{Kasen}}}, \bibinfo {author} {\bibfnamefont {Ryan~M.}\
  \bibnamefont {{Lau}}}, \bibinfo {author} {\bibfnamefont {Daniel~A.}\
  \bibnamefont {{Perley}}}, \bibinfo {author} {\bibfnamefont {Stephan}\
  \bibnamefont {{Rosswog}}}, \bibinfo {author} {\bibfnamefont {Eran~O.}\
  \bibnamefont {{Ofek}}}, \bibinfo {author} {\bibfnamefont {Kenta}\
  \bibnamefont {{Hotokezaka}}}, \bibinfo {author} {\bibfnamefont {Ranga-Ram}\
  \bibnamefont {{Chary}}}, \bibinfo {author} {\bibfnamefont {Jesper}\
  \bibnamefont {{Sollerman}}}, \bibinfo {author} {\bibfnamefont {Ariel}\
  \bibnamefont {{Goobar}}}, and\ \bibinfo {author} {\bibfnamefont {David~L.}\
  \bibnamefont {{Kaplan}}}} (\bibinfo {year} {2019}),\ \bibfield  {title}
  {\enquote {\bibinfo {title} {{Spitzer Mid-Infrared Detections of Neutron Star
  Merger GW170817 Suggests Synthesis of the Heaviest Elements}},}\ }\href
  {https://doi.org/10.1093/mnrasl/slz007} {\bibinfo  {journal} {Mon. Not. Roy.
  Astron. Soc. Lett.}\ ,\ \bibinfo {pages} {L14}}\BibitemShut {NoStop}%
\bibitem [{\citenamefont {{Kaspi}}\ and\ \citenamefont
  {{Beloborodov}}(2017)}]{Kaspi.Belobodorov:2017}%
  \BibitemOpen
\bibfield  {journal} {  }\bibfield  {author} {\bibinfo {author} {\bibnamefont
  {{Kaspi}}, \bibfnamefont {V~M}}, and\ \bibinfo {author} {\bibfnamefont
  {A.~M.}\ \bibnamefont {{Beloborodov}}}} (\bibinfo {year} {2017}),\ \bibfield
  {title} {\enquote {\bibinfo {title} {{Magnetars}},}\ }\href
  {https://doi.org/10.1146/annurev-astro-081915-023329} {\bibfield  {journal}
  {\bibinfo  {journal} {Annu. Rev. Astron. Astrophys.}\ }\textbf {\bibinfo
  {volume} {55}},\ \bibinfo {pages} {261--301}}\BibitemShut {NoStop}%
\bibitem [{\citenamefont {Kastaun}\ \emph {et~al.}(2016)\citenamefont
  {Kastaun}, \citenamefont {Ciolfi},\ and\ \citenamefont
  {Giacomazzo}}]{Kastaun.Ciolfi.Giacomazzo:2016}%
  \BibitemOpen
  \bibfield  {author} {\bibinfo {author} {\bibnamefont {Kastaun}, \bibfnamefont
  {W}}, \bibinfo {author} {\bibfnamefont {R.}~\bibnamefont {Ciolfi}}, and\
  \bibinfo {author} {\bibfnamefont {B.}~\bibnamefont {Giacomazzo}}} (\bibinfo
  {year} {2016}),\ \bibfield  {title} {\enquote {\bibinfo {title} {Structure of
  stable binary neutron star merger remnants: A case study},}\ }\href
  {https://doi.org/10.1103/PhysRevD.94.044060} {\bibfield  {journal} {\bibinfo
  {journal} {Phys. Rev. D}\ }\textbf {\bibinfo {volume} {94}},\ \bibinfo
  {pages} {044060}}\BibitemShut {NoStop}%
\bibitem [{\citenamefont {{Kawaguchi}}\ \emph {et~al.}(2016)\citenamefont
  {{Kawaguchi}}, \citenamefont {{Kyutoku}}, \citenamefont {{Shibata}},\ and\
  \citenamefont {{Tanaka}}}]{Kawaguchi.Kyutoku.ea:2016}%
  \BibitemOpen
  \bibfield  {author} {\bibinfo {author} {\bibnamefont {{Kawaguchi}},
  \bibfnamefont {Kyohei}}, \bibinfo {author} {\bibfnamefont {Koutarou}\
  \bibnamefont {{Kyutoku}}}, \bibinfo {author} {\bibfnamefont {Masaru}\
  \bibnamefont {{Shibata}}}, and\ \bibinfo {author} {\bibfnamefont {Masaomi}\
  \bibnamefont {{Tanaka}}}} (\bibinfo {year} {2016}),\ \bibfield  {title}
  {\enquote {\bibinfo {title} {{Models of Kilonova/Macronova Emission from
  Black Hole-Neutron Star Mergers}},}\ }\href
  {https://doi.org/10.3847/0004-637X/825/1/52} {\bibfield  {journal} {\bibinfo
  {journal} {Astrophys. J.}\ }\textbf {\bibinfo {volume} {825}},\ \bibinfo
  {eid} {52}}\BibitemShut {NoStop}%
\bibitem [{\citenamefont {{Kawaguchi}}\ \emph {et~al.}(2018)\citenamefont
  {{Kawaguchi}}, \citenamefont {{Shibata}},\ and\ \citenamefont
  {{Tanaka}}}]{Kawaguchi.Shibata.Tanaka:2018}%
  \BibitemOpen
  \bibfield  {author} {\bibinfo {author} {\bibnamefont {{Kawaguchi}},
  \bibfnamefont {Kyohei}}, \bibinfo {author} {\bibfnamefont {Masaru}\
  \bibnamefont {{Shibata}}}, and\ \bibinfo {author} {\bibfnamefont {Masaomi}\
  \bibnamefont {{Tanaka}}}} (\bibinfo {year} {2018}),\ \bibfield  {title}
  {\enquote {\bibinfo {title} {{Radiative Transfer Simulation for the Optical
  and Near-infrared Electromagnetic Counterparts to GW170817}},}\ }\href
  {https://doi.org/10.3847/2041-8213/aade02} {\bibfield  {journal} {\bibinfo
  {journal} {Astrophys. J.}\ }\textbf {\bibinfo {volume} {865}},\ \bibinfo
  {eid} {L21}}\BibitemShut {NoStop}%
\bibitem [{\citenamefont {{Kawaguchi}}\ \emph {et~al.}(2020)\citenamefont
  {{Kawaguchi}}, \citenamefont {{Shibata}},\ and\ \citenamefont
  {{Tanaka}}}]{Kawaguchi.Shibata.Tanaka:2020}%
  \BibitemOpen
  \bibfield  {author} {\bibinfo {author} {\bibnamefont {{Kawaguchi}},
  \bibfnamefont {Kyohei}}, \bibinfo {author} {\bibfnamefont {Masaru}\
  \bibnamefont {{Shibata}}}, and\ \bibinfo {author} {\bibfnamefont {Masaomi}\
  \bibnamefont {{Tanaka}}}} (\bibinfo {year} {2020}),\ \bibfield  {title}
  {\enquote {\bibinfo {title} {{Diversity of Kilonova Light Curves}},}\ }\href
  {https://doi.org/10.3847/1538-4357/ab61f6} {\bibfield  {journal} {\bibinfo
  {journal} {Astrophys. J.}\ }\textbf {\bibinfo {volume} {889}},\ \bibinfo
  {eid} {171}}\BibitemShut {NoStop}%
\bibitem [{\citenamefont {{Kelic}}\ \emph {et~al.}(2008)\citenamefont
  {{Kelic}}, \citenamefont {{Ricciardi}},\ and\ \citenamefont
  {{Schmidt}}}]{kelic08}%
  \BibitemOpen
  \bibfield  {author} {\bibinfo {author} {\bibnamefont {{Kelic}}, \bibfnamefont
  {A}}, \bibinfo {author} {\bibfnamefont {M.~V.}\ \bibnamefont {{Ricciardi}}},
  and\ \bibinfo {author} {\bibfnamefont {K.-H.}\ \bibnamefont {{Schmidt}}}}
  (\bibinfo {year} {2008}),\ \bibfield  {title} {\enquote {\bibinfo {title}
  {{New Insight Into the Fission Process from Experiments with Relativistic
  Heavy-Ion Beams}},}\ }in\ \href {https://doi.org/10.1142/9789812837530\_0016}
  {\emph {\bibinfo {booktitle} {Dynamical Aspects of Nuclear Fission}}},\
  \bibinfo {editor} {edited by\ \bibinfo {editor} {\bibfnamefont
  {J.}~\bibnamefont {{Kliman}}}, \bibinfo {editor} {\bibfnamefont {M.~G.}\
  \bibnamefont {{Itkis}}}, \ and\ \bibinfo {editor} {\bibfnamefont {{\v
  S}.}~\bibnamefont {{Gmuca}}}},\ pp.\ \bibinfo {pages} {203--215}\BibitemShut
  {NoStop}%
\bibitem [{\citenamefont {{Kelic}}\ \emph {et~al.}(2009)\citenamefont
  {{Kelic}}, \citenamefont {{Ricciardi}},\ and\ \citenamefont
  {{Schmidt}}}]{Kelic.Ricciardi.Schmidt:2009}%
  \BibitemOpen
  \bibfield  {author} {\bibinfo {author} {\bibnamefont {{Kelic}}, \bibfnamefont
  {A}}, \bibinfo {author} {\bibfnamefont {M.~V.}\ \bibnamefont {{Ricciardi}}},
  and\ \bibinfo {author} {\bibfnamefont {K.-H.}\ \bibnamefont {{Schmidt}}}}
  (\bibinfo {year} {2009}),\ \bibfield  {title} {\enquote {\bibinfo {title}
  {{ABLA07 - towards a complete description of the decay channels of a nuclear
  system from spontaneous fission to multifragmentation}},}\ }\href@noop {}
  {\bibfield  {journal} {\bibinfo  {journal} {arXiv e-prints}\ }}\Eprint
  {https://arxiv.org/abs/0906.4193} {arXiv:0906.4193 [nucl-th]} \BibitemShut
  {NoStop}%
\bibitem [{\citenamefont {{Kirby}}\ \emph {et~al.}(2013)\citenamefont
  {{Kirby}}, \citenamefont {{Cohen}}, \citenamefont {{Guhathakurta}},
  \citenamefont {{Cheng}}, \citenamefont {{Bullock}},\ and\ \citenamefont
  {{Gallazzi}}}]{Kirby.ea:2013}%
  \BibitemOpen
  \bibfield  {author} {\bibinfo {author} {\bibnamefont {{Kirby}}, \bibfnamefont
  {E~N}}, \bibinfo {author} {\bibfnamefont {J.~G.}\ \bibnamefont {{Cohen}}},
  \bibinfo {author} {\bibfnamefont {P.}~\bibnamefont {{Guhathakurta}}},
  \bibinfo {author} {\bibfnamefont {L.}~\bibnamefont {{Cheng}}}, \bibinfo
  {author} {\bibfnamefont {J.~S.}\ \bibnamefont {{Bullock}}}, and\ \bibinfo
  {author} {\bibfnamefont {A.}~\bibnamefont {{Gallazzi}}}} (\bibinfo {year}
  {2013}),\ \bibfield  {title} {\enquote {\bibinfo {title} {{The Universal
  Stellar Mass-Stellar Metallicity Relation for Dwarf Galaxies}},}\ }\href
  {https://doi.org/10.1088/0004-637X/779/2/102} {\bibfield  {journal} {\bibinfo
   {journal} {Astrophys. J.}\ }\textbf {\bibinfo {volume} {779}},\ \bibinfo
  {eid} {102}}\BibitemShut {NoStop}%
\bibitem [{\citenamefont {{Kirsebom}}\ \emph
  {et~al.}(2019{\natexlab{a}})\citenamefont {{Kirsebom}} \emph
  {et~al.}}]{Kirsebom.Jones.ea:2019}%
  \BibitemOpen
  \bibfield  {author} {\bibinfo {author} {\bibnamefont {{Kirsebom}},
  \bibfnamefont {O~S}},  \emph {et~al.}} (\bibinfo {year}
  {2019}{\natexlab{a}}),\ \bibfield  {title} {\enquote {\bibinfo {title}
  {{Discovery of an Exceptionally Strong {\ensuremath{\beta}} -Decay Transition
  of $^{20}$F and Implications for the Fate of Intermediate-Mass Stars}},}\
  }\href {https://doi.org/10.1103/PhysRevLett.123.262701} {\bibfield  {journal}
  {\bibinfo  {journal} {Phys. Rev. Lett.}\ }\textbf {\bibinfo {volume} {123}},\
  \bibinfo {eid} {262701}}\BibitemShut {NoStop}%
\bibitem [{\citenamefont {{Kirsebom}}\ \emph
  {et~al.}(2019{\natexlab{b}})\citenamefont {{Kirsebom}} \emph
  {et~al.}}]{Kirsebom.Hukkanen.ea:2019}%
  \BibitemOpen
  \bibfield  {author} {\bibinfo {author} {\bibnamefont {{Kirsebom}},
  \bibfnamefont {O~S}},  \emph {et~al.}} (\bibinfo {year}
  {2019}{\natexlab{b}}),\ \bibfield  {title} {\enquote {\bibinfo {title}
  {{Measurement of the $2^{+} \rightarrow{} 0^{+}$ ground-state transition in
  the {\ensuremath{\beta}} decay of $^{20}$F}},}\ }\href
  {https://doi.org/10.1103/PhysRevC.100.065805} {\bibfield  {journal} {\bibinfo
   {journal} {Phys. Rev. C}\ }\textbf {\bibinfo {volume} {100}},\ \bibinfo
  {eid} {065805}}\BibitemShut {NoStop}%
\bibitem [{\citenamefont {{Kirson}}(2008)}]{Kirson:2008}%
  \BibitemOpen
  \bibfield  {author} {\bibinfo {author} {\bibnamefont {{Kirson}},
  \bibfnamefont {M~W}}} (\bibinfo {year} {2008}),\ \bibfield  {title} {\enquote
  {\bibinfo {title} {{Mutual influence of terms in a semi-empirical mass
  formula}},}\ }\href {https://doi.org/10.1016/j.nuclphysa.2007.10.011}
  {\bibfield  {journal} {\bibinfo  {journal} {Nucl. Phys. A}\ }\textbf
  {\bibinfo {volume} {798}},\ \bibinfo {pages} {29--60}}\BibitemShut {NoStop}%
\bibitem [{\citenamefont {{Kiss}}\ \emph {et~al.}(2009)\citenamefont {{Kiss}},
  \citenamefont {{Mohr}}, \citenamefont {{F{\"u}l{\"o}p}}, \citenamefont
  {{Galaviz}}, \citenamefont {{Gy{\"u}rky}}, \citenamefont {{Elekes}},
  \citenamefont {{Somorjai}}, \citenamefont {{Kretschmer}}, \citenamefont
  {{Sonnabend}}, \citenamefont {{Zilges}},\ and\ \citenamefont
  {{Avrigeanu}}}]{Kiss.Mohr.ea:2009}%
  \BibitemOpen
  \bibfield  {author} {\bibinfo {author} {\bibnamefont {{Kiss}}, \bibfnamefont
  {G~G}}, \bibinfo {author} {\bibfnamefont {P.}~\bibnamefont {{Mohr}}},
  \bibinfo {author} {\bibfnamefont {Z.}~\bibnamefont {{F{\"u}l{\"o}p}}},
  \bibinfo {author} {\bibfnamefont {D.}~\bibnamefont {{Galaviz}}}, \bibinfo
  {author} {\bibfnamefont {G.}~\bibnamefont {{Gy{\"u}rky}}}, \bibinfo {author}
  {\bibfnamefont {Z.}~\bibnamefont {{Elekes}}}, \bibinfo {author}
  {\bibfnamefont {E.}~\bibnamefont {{Somorjai}}}, \bibinfo {author}
  {\bibfnamefont {A.}~\bibnamefont {{Kretschmer}}}, \bibinfo {author}
  {\bibfnamefont {K.}~\bibnamefont {{Sonnabend}}}, \bibinfo {author}
  {\bibfnamefont {A.}~\bibnamefont {{Zilges}}}, and\ \bibinfo {author}
  {\bibfnamefont {M.}~\bibnamefont {{Avrigeanu}}}} (\bibinfo {year} {2009}),\
  \bibfield  {title} {\enquote {\bibinfo {title} {{High precision
  $^{89}$Y$(\alpha,\alpha){}^{89}$Y scattering at low energies}},}\ }\href
  {https://doi.org/10.1103/PhysRevC.80.045807} {\bibfield  {journal} {\bibinfo
  {journal} {Phys. Rev. C}\ }\textbf {\bibinfo {volume} {80}},\ \bibinfo {eid}
  {045807}}\BibitemShut {NoStop}%
\bibitem [{\citenamefont {{Kitaura}}\ \emph {et~al.}(2006)\citenamefont
  {{Kitaura}}, \citenamefont {{Janka}},\ and\ \citenamefont
  {{Hillebrandt}}}]{Kitaura.Janka.Hillebrandt:2006}%
  \BibitemOpen
  \bibfield  {author} {\bibinfo {author} {\bibnamefont {{Kitaura}},
  \bibfnamefont {F~S}}, \bibinfo {author} {\bibfnamefont {H.-T.}\ \bibnamefont
  {{Janka}}}, and\ \bibinfo {author} {\bibfnamefont {W.}~\bibnamefont
  {{Hillebrandt}}}} (\bibinfo {year} {2006}),\ \bibfield  {title} {\enquote
  {\bibinfo {title} {{Explosions of O-Ne-Mg cores, the Crab supernova, and
  subluminous type II-P supernovae}},}\ }\href
  {https://doi.org/10.1051/0004-6361:20054703} {\bibfield  {journal} {\bibinfo
  {journal} {Astron. \& Astrophys.}\ }\textbf {\bibinfo {volume} {450}},\
  \bibinfo {pages} {345--350}}\BibitemShut {NoStop}%
\bibitem [{\citenamefont {{Kiuchi}}\ \emph {et~al.}(2015)\citenamefont
  {{Kiuchi}}, \citenamefont {{Cerd{\'a}-Dur{\'a}n}}, \citenamefont {{Kyutoku}},
  \citenamefont {{Sekiguchi}},\ and\ \citenamefont {{Shibata}}}]{kiuchi15}%
  \BibitemOpen
  \bibfield  {author} {\bibinfo {author} {\bibnamefont {{Kiuchi}},
  \bibfnamefont {K}}, \bibinfo {author} {\bibfnamefont {P.}~\bibnamefont
  {{Cerd{\'a}-Dur{\'a}n}}}, \bibinfo {author} {\bibfnamefont {K.}~\bibnamefont
  {{Kyutoku}}}, \bibinfo {author} {\bibfnamefont {Y.}~\bibnamefont
  {{Sekiguchi}}}, and\ \bibinfo {author} {\bibfnamefont {M.}~\bibnamefont
  {{Shibata}}}} (\bibinfo {year} {2015}),\ \bibfield  {title} {\enquote
  {\bibinfo {title} {{Efficient magnetic-field amplification due to the
  Kelvin-Helmholtz instability in binary neutron star mergers}},}\ }\href
  {https://doi.org/10.1103/PhysRevD.92.124034} {\bibfield  {journal} {\bibinfo
  {journal} {Phys. Rev. D}\ }\textbf {\bibinfo {volume} {92}},\ \bibinfo {eid}
  {124034}}\BibitemShut {NoStop}%
\bibitem [{\citenamefont {{Kiuchi}}\ \emph {et~al.}(2012)\citenamefont
  {{Kiuchi}}, \citenamefont {{Kyutoku}},\ and\ \citenamefont
  {{Shibata}}}]{Kiuchi.Kyutoku.ea:2012}%
  \BibitemOpen
  \bibfield  {author} {\bibinfo {author} {\bibnamefont {{Kiuchi}},
  \bibfnamefont {Kenta}}, \bibinfo {author} {\bibfnamefont {Koutarou}\
  \bibnamefont {{Kyutoku}}}, and\ \bibinfo {author} {\bibfnamefont {Masaru}\
  \bibnamefont {{Shibata}}}} (\bibinfo {year} {2012}),\ \bibfield  {title}
  {\enquote {\bibinfo {title} {{Three-dimensional evolution of differentially
  rotating magnetized neutron stars}},}\ }\href
  {https://doi.org/10.1103/PhysRevD.86.064008} {\bibfield  {journal} {\bibinfo
  {journal} {Phys. Rev. D}\ }\textbf {\bibinfo {volume} {86}},\ \bibinfo {eid}
  {064008}}\BibitemShut {NoStop}%
\bibitem [{\citenamefont {Kiuchi}\ \emph {et~al.}(2009)\citenamefont {Kiuchi},
  \citenamefont {Sekiguchi}, \citenamefont {Shibata},\ and\ \citenamefont
  {Taniguchi}}]{Kiuchi.Sekiguchi.ea:2009}%
  \BibitemOpen
  \bibfield  {author} {\bibinfo {author} {\bibnamefont {Kiuchi}, \bibfnamefont
  {Kenta}}, \bibinfo {author} {\bibfnamefont {Yuichiro}\ \bibnamefont
  {Sekiguchi}}, \bibinfo {author} {\bibfnamefont {Masaru}\ \bibnamefont
  {Shibata}}, and\ \bibinfo {author} {\bibfnamefont {Keisuke}\ \bibnamefont
  {Taniguchi}}} (\bibinfo {year} {2009}),\ \bibfield  {title} {\enquote
  {\bibinfo {title} {Long-term general relativistic simulation of binary
  neutron stars collapsing to a black hole},}\ }\href
  {https://doi.org/10.1103/PhysRevD.80.064037} {\bibfield  {journal} {\bibinfo
  {journal} {Phys. Rev. D}\ }\textbf {\bibinfo {volume} {80}},\ \bibinfo
  {pages} {064037}}\BibitemShut {NoStop}%
\bibitem [{\citenamefont {{Kizivat}}\ \emph {et~al.}(2010)\citenamefont
  {{Kizivat}}, \citenamefont {{Mart{\'{\i}}nez-Pinedo}}, \citenamefont
  {{Langanke}}, \citenamefont {{Surman}},\ and\ \citenamefont
  {{McLaughlin}}}]{Kizivat.Martinez-Pinedo.ea:2010}%
  \BibitemOpen
  \bibfield  {author} {\bibinfo {author} {\bibnamefont {{Kizivat}},
  \bibfnamefont {L-T}}, \bibinfo {author} {\bibfnamefont {G.}~\bibnamefont
  {{Mart{\'{\i}}nez-Pinedo}}}, \bibinfo {author} {\bibfnamefont
  {K.}~\bibnamefont {{Langanke}}}, \bibinfo {author} {\bibfnamefont
  {R.}~\bibnamefont {{Surman}}}, and\ \bibinfo {author} {\bibfnamefont {G.~C.}\
  \bibnamefont {{McLaughlin}}}} (\bibinfo {year} {2010}),\ \bibfield  {title}
  {\enquote {\bibinfo {title} {{{$\gamma$}-ray bursts black hole accretion
  disks as a site for the {$\nu$}p process}},}\ }\href
  {https://doi.org/10.1103/PhysRevC.81.025802} {\bibfield  {journal} {\bibinfo
  {journal} {Phys. Rev. C}\ }\textbf {\bibinfo {volume} {81}},\ \bibinfo {eid}
  {025802}}\BibitemShut {NoStop}%
\bibitem [{\citenamefont {{Klapdor}}\ \emph {et~al.}(1981)\citenamefont
  {{Klapdor}}, \citenamefont {{Oda}}, \citenamefont {{Metzinger}},
  \citenamefont {{Hillebrandt}},\ and\ \citenamefont
  {{Thielemann}}}]{Klapdor.ea:1981}%
  \BibitemOpen
  \bibfield  {author} {\bibinfo {author} {\bibnamefont {{Klapdor}},
  \bibfnamefont {H~V}}, \bibinfo {author} {\bibfnamefont {T.}~\bibnamefont
  {{Oda}}}, \bibinfo {author} {\bibfnamefont {J.}~\bibnamefont {{Metzinger}}},
  \bibinfo {author} {\bibfnamefont {W.}~\bibnamefont {{Hillebrandt}}}, and\
  \bibinfo {author} {\bibfnamefont {F.~K.}\ \bibnamefont {{Thielemann}}}}
  (\bibinfo {year} {1981}),\ \bibfield  {title} {\enquote {\bibinfo {title}
  {{The beta strength function and the astrophysical site of the r-process}},}\
  }\href {https://doi.org/10.1007/BF01443939} {\bibfield  {journal} {\bibinfo
  {journal} {Zeitschrift f{\"u}r Phys. A Hadron. Nucl.}\ }\textbf {\bibinfo
  {volume} {299}},\ \bibinfo {pages} {213--229}}\BibitemShut {NoStop}%
\bibitem [{\citenamefont {{Knie}}\ \emph {et~al.}(2004)\citenamefont {{Knie}},
  \citenamefont {{Korschinek}}, \citenamefont {{Faestermann}}, \citenamefont
  {{Dorfi}}, \citenamefont {{Rugel}},\ and\ \citenamefont
  {{Wallner}}}]{knie04}%
  \BibitemOpen
  \bibfield  {author} {\bibinfo {author} {\bibnamefont {{Knie}}, \bibfnamefont
  {K}}, \bibinfo {author} {\bibfnamefont {G.}~\bibnamefont {{Korschinek}}},
  \bibinfo {author} {\bibfnamefont {T.}~\bibnamefont {{Faestermann}}}, \bibinfo
  {author} {\bibfnamefont {E.~A.}\ \bibnamefont {{Dorfi}}}, \bibinfo {author}
  {\bibfnamefont {G.}~\bibnamefont {{Rugel}}}, and\ \bibinfo {author}
  {\bibfnamefont {A.}~\bibnamefont {{Wallner}}}} (\bibinfo {year} {2004}),\
  \bibfield  {title} {\enquote {\bibinfo {title} {{$^{60}$Fe Anomaly in a
  Deep-Sea Manganese Crust and Implications for a Nearby Supernova Source}},}\
  }\href {https://doi.org/10.1103/PhysRevLett.93.171103} {\bibfield  {journal}
  {\bibinfo  {journal} {Phys. Rev. Lett.}\ }\textbf {\bibinfo {volume} {93}},\
  \bibinfo {eid} {171103}}\BibitemShut {NoStop}%
\bibitem [{\citenamefont {Kn{\"o}bel}\ \emph {et~al.}(2016)\citenamefont
  {Kn{\"o}bel} \emph {et~al.}}]{Knoebel.Diwisch.ea:2016}%
  \BibitemOpen
  \bibfield  {author} {\bibinfo {author} {\bibnamefont {Kn{\"o}bel},
  \bibfnamefont {R}},  \emph {et~al.}} (\bibinfo {year} {2016}),\ \bibfield
  {title} {\enquote {\bibinfo {title} {{First direct mass measurements of
  stored neutron-rich 129,130,131Cd isotopes with FRS-ESR}},}\ }\href
  {https://doi.org/10.1016/j.physletb.2016.01.039} {\bibfield  {journal}
  {\bibinfo  {journal} {Phys. Lett. B}\ }\textbf {\bibinfo {volume} {754}},\
  \bibinfo {pages} {288--293}}\BibitemShut {NoStop}%
\bibitem [{\citenamefont {{Kobayashi}}(2016)}]{Kobayashi:2016}%
  \BibitemOpen
  \bibfield  {author} {\bibinfo {author} {\bibnamefont {{Kobayashi}},
  \bibfnamefont {C}}} (\bibinfo {year} {2016}),\ \bibfield  {title} {\enquote
  {\bibinfo {title} {{Inhomogeneous chemical enrichment in the Galactic
  Halo}},}\ }in\ \href {https://doi.org/10.1017/S1743921315009783} {\emph
  {\bibinfo {booktitle} {The General Assembly of Galaxy Halos: Structure,
  Origin and Evolution}}},\ \bibinfo {series} {IAU Symposium}, Vol.\ \bibinfo
  {volume} {317},\ \bibinfo {editor} {edited by\ \bibinfo {editor}
  {\bibfnamefont {A.}~\bibnamefont {{Bragaglia}}}, \bibinfo {editor}
  {\bibfnamefont {M.}~\bibnamefont {{Arnaboldi}}}, \bibinfo {editor}
  {\bibfnamefont {M.}~\bibnamefont {{Rejkuba}}}, \ and\ \bibinfo {editor}
  {\bibfnamefont {D.}~\bibnamefont {{Romano}}}},\ pp.\ \bibinfo {pages}
  {57--63}\BibitemShut {NoStop}%
\bibitem [{\citenamefont {{Kobayashi}}\ \emph {et~al.}(2006)\citenamefont
  {{Kobayashi}}, \citenamefont {{Umeda}}, \citenamefont {{Nomoto}},
  \citenamefont {{Tominaga}},\ and\ \citenamefont
  {{Ohkubo}}}]{Kobayashi.Umeda.Nomoto.ea:2006}%
  \BibitemOpen
  \bibfield  {author} {\bibinfo {author} {\bibnamefont {{Kobayashi}},
  \bibfnamefont {C}}, \bibinfo {author} {\bibfnamefont {H.}~\bibnamefont
  {{Umeda}}}, \bibinfo {author} {\bibfnamefont {K.}~\bibnamefont {{Nomoto}}},
  \bibinfo {author} {\bibfnamefont {N.}~\bibnamefont {{Tominaga}}}, and\
  \bibinfo {author} {\bibfnamefont {T.}~\bibnamefont {{Ohkubo}}}} (\bibinfo
  {year} {2006}),\ \bibfield  {title} {\enquote {\bibinfo {title} {{Galactic
  Chemical Evolution: Carbon through Zinc}},}\ }\href
  {https://doi.org/10.1086/508914} {\bibfield  {journal} {\bibinfo  {journal}
  {Astrophys. J.}\ }\textbf {\bibinfo {volume} {653}},\ \bibinfo {pages}
  {1145--1171}}\BibitemShut {NoStop}%
\bibitem [{\citenamefont {Kodama}\ and\ \citenamefont
  {Takahashi}(1975)}]{Kodama.Takahashi:1975}%
  \BibitemOpen
  \bibfield  {author} {\bibinfo {author} {\bibnamefont {Kodama}, \bibfnamefont
  {Takeshi}}, and\ \bibinfo {author} {\bibfnamefont {Kohji}\ \bibnamefont
  {Takahashi}}} (\bibinfo {year} {1975}),\ \bibfield  {title} {\enquote
  {\bibinfo {title} {R-process nucleosynthesis and nuclei far from the region
  of beta-stability},}\ }\href {https://doi.org/10.1016/0375-9474(75)90381-4}
  {\bibfield  {journal} {\bibinfo  {journal} {Nucl. Phys. A}\ }\textbf
  {\bibinfo {volume} {239}},\ \bibinfo {pages} {489--510}}\BibitemShut
  {NoStop}%
\bibitem [{\citenamefont {Kolbe}\ \emph {et~al.}(2003)\citenamefont {Kolbe},
  \citenamefont {Langanke}, \citenamefont {Mart{\'\i}nez-Pinedo},\ and\
  \citenamefont {Vogel}}]{Kolbe.Langanke.ea:2003}%
  \BibitemOpen
  \bibfield  {author} {\bibinfo {author} {\bibnamefont {Kolbe}, \bibfnamefont
  {E}}, \bibinfo {author} {\bibfnamefont {K.}~\bibnamefont {Langanke}},
  \bibinfo {author} {\bibfnamefont {G.}~\bibnamefont {Mart{\'\i}nez-Pinedo}},
  and\ \bibinfo {author} {\bibfnamefont {P.}~\bibnamefont {Vogel}}} (\bibinfo
  {year} {2003}),\ \bibfield  {title} {\enquote {\bibinfo {title} {{Neutrino
  nucleus reactions and nuclear structure}},}\ }\href
  {https://doi.org/10.1088/0954-3899/29/11/010} {\bibfield  {journal} {\bibinfo
   {journal} {J. Phys. G: Nucl. Part. Phys.}\ }\textbf {\bibinfo {volume}
  {29}},\ \bibinfo {pages} {2569--2596}}\BibitemShut {NoStop}%
\bibitem [{\citenamefont {{Komiya}}\ and\ \citenamefont
  {{Shigeyama}}(2016)}]{Komiya.Shigeyama:2016}%
  \BibitemOpen
  \bibfield  {author} {\bibinfo {author} {\bibnamefont {{Komiya}},
  \bibfnamefont {Yutaka}}, and\ \bibinfo {author} {\bibfnamefont {Toshikazu}\
  \bibnamefont {{Shigeyama}}}} (\bibinfo {year} {2016}),\ \bibfield  {title}
  {\enquote {\bibinfo {title} {{Contribution of Neutron Star Mergers to the
  r-Process Chemical Evolution in the Hierarchical Galaxy Formation}},}\ }\href
  {https://doi.org/10.3847/0004-637X/830/2/76} {\bibfield  {journal} {\bibinfo
  {journal} {Astrophys. J.}\ }\textbf {\bibinfo {volume} {830}},\ \bibinfo
  {eid} {76}}\BibitemShut {NoStop}%
\bibitem [{\citenamefont {{Koning}}\ and\ \citenamefont
  {{Delaroche}}(2003)}]{Koning.Delaroche:2003}%
  \BibitemOpen
  \bibfield  {author} {\bibinfo {author} {\bibnamefont {{Koning}},
  \bibfnamefont {A~J}}, and\ \bibinfo {author} {\bibfnamefont {J.~P.}\
  \bibnamefont {{Delaroche}}}} (\bibinfo {year} {2003}),\ \bibfield  {title}
  {\enquote {\bibinfo {title} {{Local and global nucleon optical models from 1
  keV to 200 MeV}},}\ }\href {https://doi.org/10.1016/S0375-9474(02)01321-0}
  {\bibfield  {journal} {\bibinfo  {journal} {Nucl. Phys. A}\ }\textbf
  {\bibinfo {volume} {713}},\ \bibinfo {pages} {231--310}}\BibitemShut
  {NoStop}%
\bibitem [{\citenamefont {{Koning}}\ \emph {et~al.}(2008)\citenamefont
  {{Koning}}, \citenamefont {{Hilaire}},\ and\ \citenamefont
  {{Goriely}}}]{Koning.Hilaire.Goriely:2008}%
  \BibitemOpen
  \bibfield  {author} {\bibinfo {author} {\bibnamefont {{Koning}},
  \bibfnamefont {A~J}}, \bibinfo {author} {\bibfnamefont {S.}~\bibnamefont
  {{Hilaire}}}, and\ \bibinfo {author} {\bibfnamefont {S.}~\bibnamefont
  {{Goriely}}}} (\bibinfo {year} {2008}),\ \bibfield  {title} {\enquote
  {\bibinfo {title} {{Global and local level density models}},}\ }\href
  {https://doi.org/10.1016/j.nuclphysa.2008.06.005} {\bibfield  {journal}
  {\bibinfo  {journal} {Nucl. Phys. A}\ }\textbf {\bibinfo {volume} {810}},\
  \bibinfo {pages} {13--76}}\BibitemShut {NoStop}%
\bibitem [{\citenamefont {{Koonin}}\ \emph {et~al.}(1997)\citenamefont
  {{Koonin}}, \citenamefont {{Dean}},\ and\ \citenamefont
  {{Langanke}}}]{Koonin.Dean.Langanke:1997}%
  \BibitemOpen
  \bibfield  {author} {\bibinfo {author} {\bibnamefont {{Koonin}},
  \bibfnamefont {S~E}}, \bibinfo {author} {\bibfnamefont {D.~J.}\ \bibnamefont
  {{Dean}}}, and\ \bibinfo {author} {\bibfnamefont {K.}~\bibnamefont
  {{Langanke}}}} (\bibinfo {year} {1997}),\ \bibfield  {title} {\enquote
  {\bibinfo {title} {{Shell model Monte Carlo methods}},}\ }\href
  {https://doi.org/10.1016/S0370-1573(96)00017-8} {\bibfield  {journal}
  {\bibinfo  {journal} {Phys. Rep.}\ }\textbf {\bibinfo {volume} {278}},\
  \bibinfo {pages} {1--77}}\BibitemShut {NoStop}%
\bibitem [{\citenamefont {{Korobkin}}\ \emph {et~al.}(2012)\citenamefont
  {{Korobkin}}, \citenamefont {{Rosswog}}, \citenamefont {{Arcones}},\ and\
  \citenamefont {{Winteler}}}]{Korobkin.Rosswog.ea:2012}%
  \BibitemOpen
  \bibfield  {author} {\bibinfo {author} {\bibnamefont {{Korobkin}},
  \bibfnamefont {O}}, \bibinfo {author} {\bibfnamefont {S.}~\bibnamefont
  {{Rosswog}}}, \bibinfo {author} {\bibfnamefont {A.}~\bibnamefont
  {{Arcones}}}, and\ \bibinfo {author} {\bibfnamefont {C.}~\bibnamefont
  {{Winteler}}}} (\bibinfo {year} {2012}),\ \bibfield  {title} {\enquote
  {\bibinfo {title} {{On the astrophysical robustness of the neutron star
  merger r-process}},}\ }\href
  {https://doi.org/10.1111/j.1365-2966.2012.21859.x} {\bibfield  {journal}
  {\bibinfo  {journal} {Mon. Not. Roy. Astron. Soc.}\ }\textbf {\bibinfo
  {volume} {426}},\ \bibinfo {pages} {1940--1949}}\BibitemShut {NoStop}%
\bibitem [{\citenamefont {{Korobkin}}\ \emph
  {et~al.}(2020{\natexlab{a}})\citenamefont {{Korobkin}}, \citenamefont
  {{Hungerford}}, \citenamefont {{Fryer}}, \citenamefont {{Mumpower}},
  \citenamefont {{Misch}}, \citenamefont {{Sprouse}}, \citenamefont
  {{Lippuner}}, \citenamefont {{Surman}}, \citenamefont {{Couture}},
  \citenamefont {{Bloser}}, \citenamefont {{Shirazi}}, \citenamefont {{Even}},
  \citenamefont {{Vestrand}},\ and\ \citenamefont
  {{Miller}}}]{Korobkin.Hungerford.ea:2020}%
  \BibitemOpen
  \bibfield  {author} {\bibinfo {author} {\bibnamefont {{Korobkin}},
  \bibfnamefont {Oleg}}, \bibinfo {author} {\bibfnamefont {Aimee~M.}\
  \bibnamefont {{Hungerford}}}, \bibinfo {author} {\bibfnamefont
  {Christopher~L.}\ \bibnamefont {{Fryer}}}, \bibinfo {author} {\bibfnamefont
  {Matthew~R.}\ \bibnamefont {{Mumpower}}}, \bibinfo {author} {\bibfnamefont
  {G.~Wendell}\ \bibnamefont {{Misch}}}, \bibinfo {author} {\bibfnamefont
  {Trevor~M.}\ \bibnamefont {{Sprouse}}}, \bibinfo {author} {\bibfnamefont
  {Jonas}\ \bibnamefont {{Lippuner}}}, \bibinfo {author} {\bibfnamefont
  {Rebecca}\ \bibnamefont {{Surman}}}, \bibinfo {author} {\bibfnamefont
  {Aaron~J.}\ \bibnamefont {{Couture}}}, \bibinfo {author} {\bibfnamefont
  {Peter~F.}\ \bibnamefont {{Bloser}}}, \bibinfo {author} {\bibfnamefont
  {Farzane}\ \bibnamefont {{Shirazi}}}, \bibinfo {author} {\bibfnamefont
  {Wesley~P.}\ \bibnamefont {{Even}}}, \bibinfo {author} {\bibfnamefont
  {W.~Thomas}\ \bibnamefont {{Vestrand}}}, and\ \bibinfo {author}
  {\bibfnamefont {Richard~S.}\ \bibnamefont {{Miller}}}} (\bibinfo {year}
  {2020}{\natexlab{a}}),\ \bibfield  {title} {\enquote {\bibinfo {title}
  {{Gamma Rays from Kilonova: A Potential Probe of r-process
  Nucleosynthesis}},}\ }\href {https://doi.org/10.3847/1538-4357/ab64d8}
  {\bibfield  {journal} {\bibinfo  {journal} {Astrophys. J.}\ }\textbf
  {\bibinfo {volume} {889}},\ \bibinfo {eid} {168}}\BibitemShut {NoStop}%
\bibitem [{\citenamefont {{Korobkin}}\ \emph
  {et~al.}(2020{\natexlab{b}})\citenamefont {{Korobkin}}, \citenamefont
  {{Wollaeger}}, \citenamefont {{Fryer}}, \citenamefont {{Hungerford}},
  \citenamefont {{Rosswog}}, \citenamefont {{Fontes}}, \citenamefont
  {{Mumpower}}, \citenamefont {{Chase}}, \citenamefont {{Even}}, \citenamefont
  {{Miller}}, \citenamefont {{Misch}},\ and\ \citenamefont
  {{Lippuner}}}]{Korobkin.Wollaeger.ea:2020}%
  \BibitemOpen
  \bibfield  {author} {\bibinfo {author} {\bibnamefont {{Korobkin}},
  \bibfnamefont {Oleg}}, \bibinfo {author} {\bibfnamefont {Ryan}\ \bibnamefont
  {{Wollaeger}}}, \bibinfo {author} {\bibfnamefont {Christopher}\ \bibnamefont
  {{Fryer}}}, \bibinfo {author} {\bibfnamefont {Aimee~L.}\ \bibnamefont
  {{Hungerford}}}, \bibinfo {author} {\bibfnamefont {Stephan}\ \bibnamefont
  {{Rosswog}}}, \bibinfo {author} {\bibfnamefont {Christopher}\ \bibnamefont
  {{Fontes}}}, \bibinfo {author} {\bibfnamefont {Matthew}\ \bibnamefont
  {{Mumpower}}}, \bibinfo {author} {\bibfnamefont {Eve}\ \bibnamefont
  {{Chase}}}, \bibinfo {author} {\bibfnamefont {Wesley}\ \bibnamefont
  {{Even}}}, \bibinfo {author} {\bibfnamefont {Jonah~M.}\ \bibnamefont
  {{Miller}}}, \bibinfo {author} {\bibfnamefont {G.~Wendell}\ \bibnamefont
  {{Misch}}}, and\ \bibinfo {author} {\bibfnamefont {Jonas}\ \bibnamefont
  {{Lippuner}}}} (\bibinfo {year} {2020}{\natexlab{b}}),\ \bibfield  {title}
  {\enquote {\bibinfo {title} {{Axisymmetric Radiative Transfer Models of
  Kilonovae}},}\ }\href@noop {} {\bibfield  {journal} {\bibinfo  {journal}
  {arXiv e-prints}\ }}\Eprint {https://arxiv.org/abs/2004.00102}
  {arXiv:2004.00102 [astro-ph.HE]} \BibitemShut {NoStop}%
\bibitem [{\citenamefont {{Kotake}}\ \emph {et~al.}(2012)\citenamefont
  {{Kotake}}, \citenamefont {{Sumiyoshi}}, \citenamefont {{Yamada}},
  \citenamefont {{Takiwaki}}, \citenamefont {{Kuroda}}, \citenamefont
  {{Suwa}},\ and\ \citenamefont {{Nagakura}}}]{Kotake.Sumiyoshi.ea:2012}%
  \BibitemOpen
  \bibfield  {author} {\bibinfo {author} {\bibnamefont {{Kotake}},
  \bibfnamefont {K}}, \bibinfo {author} {\bibfnamefont {K.}~\bibnamefont
  {{Sumiyoshi}}}, \bibinfo {author} {\bibfnamefont {S.}~\bibnamefont
  {{Yamada}}}, \bibinfo {author} {\bibfnamefont {T.}~\bibnamefont
  {{Takiwaki}}}, \bibinfo {author} {\bibfnamefont {T.}~\bibnamefont
  {{Kuroda}}}, \bibinfo {author} {\bibfnamefont {Y.}~\bibnamefont {{Suwa}}},
  and\ \bibinfo {author} {\bibfnamefont {H.}~\bibnamefont {{Nagakura}}}}
  (\bibinfo {year} {2012}),\ \bibfield  {title} {\enquote {\bibinfo {title}
  {{Core-collapse supernovae as supercomputing science: A status report toward
  six-dimensional simulations with exact Boltzmann neutrino transport in full
  general relativity}},}\ }\href {https://doi.org/10.1093/ptep/pts009}
  {\bibfield  {journal} {\bibinfo  {journal} {Prog. Theor. Exp. Phys.}\
  }\textbf {\bibinfo {volume} {2012}},\ \bibinfo {pages} {01A301}}\BibitemShut
  {NoStop}%
\bibitem [{\citenamefont {Koura}\ \emph {et~al.}(2005)\citenamefont {Koura},
  \citenamefont {Tachibana}, \citenamefont {Uno},\ and\ \citenamefont
  {Yamada}}]{Koura.Tachibana.ea:2005}%
  \BibitemOpen
  \bibfield  {author} {\bibinfo {author} {\bibnamefont {Koura}, \bibfnamefont
  {H}}, \bibinfo {author} {\bibfnamefont {T.}~\bibnamefont {Tachibana}},
  \bibinfo {author} {\bibfnamefont {M.}~\bibnamefont {Uno}}, and\ \bibinfo
  {author} {\bibfnamefont {M.}~\bibnamefont {Yamada}}} (\bibinfo {year}
  {2005}),\ \bibfield  {title} {\enquote {\bibinfo {title} {{Nuclidic Mass
  Formula on a Spherical Basis with an Improved Even-Odd Term}},}\ }\href
  {https://doi.org/10.1143/PTP.113.305} {\bibfield  {journal} {\bibinfo
  {journal} {Prog. Theor. Phys.}\ }\textbf {\bibinfo {volume} {113}},\ \bibinfo
  {pages} {305--325}}\BibitemShut {NoStop}%
\bibitem [{\citenamefont {{Kozub}}\ \emph {et~al.}(2012)\citenamefont {{Kozub}}
  \emph {et~al.}}]{Kozub2012}%
  \BibitemOpen
  \bibfield  {author} {\bibinfo {author} {\bibnamefont {{Kozub}}, \bibfnamefont
  {R~L}},  \emph {et~al.}} (\bibinfo {year} {2012}),\ \bibfield  {title}
  {\enquote {\bibinfo {title} {{Neutron Single Particle Structure in Sn131 and
  Direct Neutron Capture Cross Sections}},}\ }\href
  {https://doi.org/10.1103/PhysRevLett.109.172501} {\bibfield  {journal}
  {\bibinfo  {journal} {Phys. Rev. Lett.}\ }\textbf {\bibinfo {volume} {109}},\
  \bibinfo {eid} {172501}}\BibitemShut {NoStop}%
\bibitem [{\citenamefont {{Kramer}}(2009)}]{Kramer:2009}%
  \BibitemOpen
  \bibfield  {author} {\bibinfo {author} {\bibnamefont {{Kramer}},
  \bibfnamefont {M}}} (\bibinfo {year} {2009}),\ \bibfield  {title} {\enquote
  {\bibinfo {title} {{Pulsars {\&} Magnetars}},}\ }in\ \href
  {https://doi.org/10.1017/S1743921309031159} {\emph {\bibinfo {booktitle}
  {Cosmic Magnetic Fields: From Planets, to Stars and Galaxies}}},\ \bibinfo
  {series} {IAU Symposium}, Vol.\ \bibinfo {volume} {259},\ \bibinfo {editor}
  {edited by\ \bibinfo {editor} {\bibfnamefont {K.~G.}\ \bibnamefont
  {{Strassmeier}}}, \bibinfo {editor} {\bibfnamefont {A.~G.}\ \bibnamefont
  {{Kosovichev}}}, \ and\ \bibinfo {editor} {\bibfnamefont {J.~E.}\
  \bibnamefont {{Beckman}}}},\ pp.\ \bibinfo {pages} {485--492}\BibitemShut
  {NoStop}%
\bibitem [{\citenamefont {Kratz}(1984)}]{Kratz:1984}%
  \BibitemOpen
  \bibfield  {author} {\bibinfo {author} {\bibnamefont {Kratz}, \bibfnamefont
  {K-L}}} (\bibinfo {year} {1984}),\ \bibfield  {title} {\enquote {\bibinfo
  {title} {$\beta$-strength function phenomena of exotic nuclei: A critical
  examination of the significance of nuclear model predictions},}\ }\href
  {https://doi.org/10.1016/0375-9474(84)90407-X} {\bibfield  {journal}
  {\bibinfo  {journal} {Nucl. Phys. A}\ }\textbf {\bibinfo {volume} {417}},\
  \bibinfo {pages} {447--476}}\BibitemShut {NoStop}%
\bibitem [{\citenamefont {{Kratz}}(2001)}]{Kratz}%
  \BibitemOpen
  \bibfield  {author} {\bibinfo {author} {\bibnamefont {{Kratz}}, \bibfnamefont
  {K-L}}} (\bibinfo {year} {2001}),\ \bibfield  {title} {\enquote {\bibinfo
  {title} {{Measurements of r-process nuclei}},}\ }\href
  {https://doi.org/10.1016/S0375-9474(01)00719-9} {\bibfield  {journal}
  {\bibinfo  {journal} {Nucl. Phys. A}\ }\textbf {\bibinfo {volume} {688}},\
  \bibinfo {pages} {308--317}}\BibitemShut {NoStop}%
\bibitem [{\citenamefont {{Kratz}}\ \emph {et~al.}(1993)\citenamefont
  {{Kratz}}, \citenamefont {{Bitouzet}}, \citenamefont {{Thielemann}},
  \citenamefont {{Moeller}},\ and\ \citenamefont
  {{Pfeiffer}}}]{Kratz.Bitouzet.ea:1993}%
  \BibitemOpen
  \bibfield  {author} {\bibinfo {author} {\bibnamefont {{Kratz}}, \bibfnamefont
  {K-L}}, \bibinfo {author} {\bibfnamefont {J.}~\bibnamefont {{Bitouzet}}},
  \bibinfo {author} {\bibfnamefont {F.}~\bibnamefont {{Thielemann}}}, \bibinfo
  {author} {\bibfnamefont {P.}~\bibnamefont {{Moeller}}}, and\ \bibinfo
  {author} {\bibfnamefont {B.}~\bibnamefont {{Pfeiffer}}}} (\bibinfo {year}
  {1993}),\ \bibfield  {title} {\enquote {\bibinfo {title} {Isotopic r-process
  abundances and nuclear structure far from stability - implications for the
  r-process mechanism},}\ }\href {https://doi.org/10.1086/172196} {\bibfield
  {journal} {\bibinfo  {journal} {Astrophys. J.}\ }\textbf {\bibinfo {volume}
  {403}},\ \bibinfo {pages} {216--238}}\BibitemShut {NoStop}%
\bibitem [{\citenamefont {{Kratz}}\ \emph {et~al.}(2014)\citenamefont
  {{Kratz}}, \citenamefont {{Farouqi}},\ and\ \citenamefont
  {{M{\"o}ller}}}]{kratz14}%
  \BibitemOpen
  \bibfield  {author} {\bibinfo {author} {\bibnamefont {{Kratz}}, \bibfnamefont
  {K-L}}, \bibinfo {author} {\bibfnamefont {K.}~\bibnamefont {{Farouqi}}}, and\
  \bibinfo {author} {\bibfnamefont {P.}~\bibnamefont {{M{\"o}ller}}}} (\bibinfo
  {year} {2014}),\ \bibfield  {title} {\enquote {\bibinfo {title} {{A
  High-entropy-wind r-process Study Based on Nuclear-structure Quantities from
  the New Finite-range Droplet Model Frdm(2012)}},}\ }\href
  {https://doi.org/10.1088/0004-637X/792/1/6} {\bibfield  {journal} {\bibinfo
  {journal} {Astrophys. J.}\ }\textbf {\bibinfo {volume} {792}},\ \bibinfo
  {eid} {6}}\BibitemShut {NoStop}%
\bibitem [{\citenamefont {{Kratz}}\ \emph {et~al.}(1986)\citenamefont
  {{Kratz}}, \citenamefont {{Gabelmann}}, \citenamefont {{Hillebrandt}},
  \citenamefont {{Pfeiffer}}, \citenamefont {{Schl{\"o}sser}},\ and\
  \citenamefont {{Thielemann}}}]{Kratz.Gabelmann.ea:1986}%
  \BibitemOpen
  \bibfield  {author} {\bibinfo {author} {\bibnamefont {{Kratz}}, \bibfnamefont
  {K-L}}, \bibinfo {author} {\bibfnamefont {H.}~\bibnamefont {{Gabelmann}}},
  \bibinfo {author} {\bibfnamefont {W.}~\bibnamefont {{Hillebrandt}}}, \bibinfo
  {author} {\bibfnamefont {B.}~\bibnamefont {{Pfeiffer}}}, \bibinfo {author}
  {\bibfnamefont {K.}~\bibnamefont {{Schl{\"o}sser}}}, and\ \bibinfo {author}
  {\bibfnamefont {F.~K.}\ \bibnamefont {{Thielemann}}}} (\bibinfo {year}
  {1986}),\ \bibfield  {title} {\enquote {\bibinfo {title} {{The beta-decay
  half-life of\{$_{48}$/$^{130}$\}Cd$_{82}$ and its importance for
  astrophysical r-process scenarios}},}\ }\href
  {https://doi.org/10.1007/BF01290055} {\bibfield  {journal} {\bibinfo
  {journal} {Zeitschrift fur Physik A Hadron. Nucl.}\ }\textbf {\bibinfo
  {volume} {325}},\ \bibinfo {pages} {489--490}}\BibitemShut {NoStop}%
\bibitem [{\citenamefont {{Kratz}}\ and\ \citenamefont
  {{Herrmann}}(1973)}]{Kratz.Herrmann:1973}%
  \BibitemOpen
  \bibfield  {author} {\bibinfo {author} {\bibnamefont {{Kratz}}, \bibfnamefont
  {K-L}}, and\ \bibinfo {author} {\bibfnamefont {G.}~\bibnamefont
  {{Herrmann}}}} (\bibinfo {year} {1973}),\ \bibfield  {title} {\enquote
  {\bibinfo {title} {{Systematics of neutron emission probabilities from
  delayed neutron precursors}},}\ }\href {https://doi.org/10.1007/BF01391992}
  {\bibfield  {journal} {\bibinfo  {journal} {Zeitschrift f{\"u}r Phys.}\
  }\textbf {\bibinfo {volume} {263}},\ \bibinfo {pages} {435--442}}\BibitemShut
  {NoStop}%
\bibitem [{\citenamefont {Kratz}\ \emph {et~al.}(2000)\citenamefont {Kratz},
  \citenamefont {M{\"o}ller}, \citenamefont {Pfeiffer},\ and\ \citenamefont
  {Walters}}]{Kratz.Moeller.ea:2000}%
  \BibitemOpen
  \bibfield  {author} {\bibinfo {author} {\bibnamefont {Kratz}, \bibfnamefont
  {K-L}}, \bibinfo {author} {\bibfnamefont {P.}~\bibnamefont {M{\"o}ller}},
  \bibinfo {author} {\bibfnamefont {B.}~\bibnamefont {Pfeiffer}}, and\ \bibinfo
  {author} {\bibfnamefont {W.~B.}\ \bibnamefont {Walters}}} (\bibinfo {year}
  {2000}),\ in\ \href@noop {} {\emph {\bibinfo {booktitle} {Capture Gamma-Rays
  and Related Studies}}},\ Vol.\ \bibinfo {volume} {529},\ \bibinfo {editor}
  {edited by\ \bibinfo {editor} {\bibfnamefont {S.}~\bibnamefont {Wender}}}\
  (\bibinfo {organization} {AIP Conf. Proc.})\ p.\ \bibinfo {pages}
  {295}\BibitemShut {NoStop}%
\bibitem [{\citenamefont {{Kratz}}\ \emph {et~al.}(2004)\citenamefont
  {{Kratz}}, \citenamefont {{Pfeiffer}}, \citenamefont {{Cowan}},\ and\
  \citenamefont {{Sneden}}}]{Kratz2004}%
  \BibitemOpen
  \bibfield  {author} {\bibinfo {author} {\bibnamefont {{Kratz}}, \bibfnamefont
  {K-L}}, \bibinfo {author} {\bibfnamefont {B.}~\bibnamefont {{Pfeiffer}}},
  \bibinfo {author} {\bibfnamefont {J.~J.}\ \bibnamefont {{Cowan}}}, and\
  \bibinfo {author} {\bibfnamefont {C.}~\bibnamefont {{Sneden}}}} (\bibinfo
  {year} {2004}),\ \bibfield  {title} {\enquote {\bibinfo {title} {{r-process
  chronometers}},}\ }\href {https://doi.org/10.1016/j.newar.2003.11.014}
  {\bibfield  {journal} {\bibinfo  {journal} {New Astron. Rev.}\ }\textbf
  {\bibinfo {volume} {48}},\ \bibinfo {pages} {105--108}}\BibitemShut {NoStop}%
\bibitem [{\citenamefont {{Kratz}}\ \emph {et~al.}(1979)\citenamefont
  {{Kratz}}, \citenamefont {{Rudolph}}, \citenamefont {{Ohm}}, \citenamefont
  {{Franz}}, \citenamefont {{Zendel}}, \citenamefont {{Herrmann}},
  \citenamefont {{Prussin}}, \citenamefont {{Nuh}}, \citenamefont
  {{Shihab-Eldin}}, \citenamefont {{Slaughter}}, \citenamefont {{Halverson}},\
  and\ \citenamefont {{Klapdor}}}]{Ohm}%
  \BibitemOpen
  \bibfield  {author} {\bibinfo {author} {\bibnamefont {{Kratz}}, \bibfnamefont
  {K-L}}, \bibinfo {author} {\bibfnamefont {W.}~\bibnamefont {{Rudolph}}},
  \bibinfo {author} {\bibfnamefont {H.}~\bibnamefont {{Ohm}}}, \bibinfo
  {author} {\bibfnamefont {H.}~\bibnamefont {{Franz}}}, \bibinfo {author}
  {\bibfnamefont {M.}~\bibnamefont {{Zendel}}}, \bibinfo {author}
  {\bibfnamefont {G.}~\bibnamefont {{Herrmann}}}, \bibinfo {author}
  {\bibfnamefont {S.~G.}\ \bibnamefont {{Prussin}}}, \bibinfo {author}
  {\bibfnamefont {F.~M.}\ \bibnamefont {{Nuh}}}, \bibinfo {author}
  {\bibfnamefont {A.~A.}\ \bibnamefont {{Shihab-Eldin}}}, \bibinfo {author}
  {\bibfnamefont {D.~R.}\ \bibnamefont {{Slaughter}}}, \bibinfo {author}
  {\bibfnamefont {W.}~\bibnamefont {{Halverson}}}, and\ \bibinfo {author}
  {\bibfnamefont {H.~V.}\ \bibnamefont {{Klapdor}}}} (\bibinfo {year} {1979}),\
  \bibfield  {title} {\enquote {\bibinfo {title} {{Investigation of beta
  strength functions by neutron and gamma-ray spectroscopy . (I). The decay of
  $^{87}$Br, $^{137}$I, $^{85}$As and $^{135}$Sb}},}\ }\href
  {https://doi.org/10.1016/0375-9474(79)90486-X} {\bibfield  {journal}
  {\bibinfo  {journal} {Nucl. Phys. A}\ }\textbf {\bibinfo {volume} {317}},\
  \bibinfo {pages} {335--362}}\BibitemShut {NoStop}%
\bibitem [{\citenamefont {{Kratz}}\ \emph {et~al.}(1982)\citenamefont
  {{Kratz}}, \citenamefont {{Schr{\"o}der}}, \citenamefont {{Ohm}},
  \citenamefont {{Zendel}}, \citenamefont {{Gabelmann}}, \citenamefont
  {{Ziegert}}, \citenamefont {{Peuser}}, \citenamefont {{Jung}}, \citenamefont
  {{Pfeiffer}}, \citenamefont {{W{\"u}nsch}}, \citenamefont {{Wollnik}},
  \citenamefont {{Ristori}},\ and\ \citenamefont {{Cran{\c
  c}on}}}]{Kratz.Schroeder.Ohm.ea:1982}%
  \BibitemOpen
  \bibfield  {author} {\bibinfo {author} {\bibnamefont {{Kratz}}, \bibfnamefont
  {K-L}}, \bibinfo {author} {\bibfnamefont {A.}~\bibnamefont {{Schr{\"o}der}}},
  \bibinfo {author} {\bibfnamefont {H.}~\bibnamefont {{Ohm}}}, \bibinfo
  {author} {\bibfnamefont {M.}~\bibnamefont {{Zendel}}}, \bibinfo {author}
  {\bibfnamefont {H.}~\bibnamefont {{Gabelmann}}}, \bibinfo {author}
  {\bibfnamefont {W.}~\bibnamefont {{Ziegert}}}, \bibinfo {author}
  {\bibfnamefont {P.}~\bibnamefont {{Peuser}}}, \bibinfo {author}
  {\bibfnamefont {G.}~\bibnamefont {{Jung}}}, \bibinfo {author} {\bibfnamefont
  {B.}~\bibnamefont {{Pfeiffer}}}, \bibinfo {author} {\bibfnamefont {K.~D.}\
  \bibnamefont {{W{\"u}nsch}}}, \bibinfo {author} {\bibfnamefont
  {H.}~\bibnamefont {{Wollnik}}}, \bibinfo {author} {\bibfnamefont
  {C.}~\bibnamefont {{Ristori}}}, and\ \bibinfo {author} {\bibfnamefont
  {J.}~\bibnamefont {{Cran{\c c}on}}}} (\bibinfo {year} {1982}),\ \bibfield
  {title} {\enquote {\bibinfo {title} {{Beta-delayed neutron emission from$^{93
  100}$Rb to excited states in the residual Sr isotopes}},}\ }\href
  {https://doi.org/10.1007/BF01415126} {\bibfield  {journal} {\bibinfo
  {journal} {Zeitschrift f{\"u}r Phys. A Hadron. Nucl.}\ }\textbf {\bibinfo
  {volume} {306}},\ \bibinfo {pages} {239--257}}\BibitemShut {NoStop}%
\bibitem [{\citenamefont {{Kratz}}\ \emph {et~al.}(1988)\citenamefont
  {{Kratz}}, \citenamefont {{Thielemann}}, \citenamefont {{Hillebrandt}},
  \citenamefont {{M{\"o}ller}}, \citenamefont {{Harms}}, \citenamefont
  {{Wohr}},\ and\ \citenamefont {{Truran}}}]{Kratz.Thielemann.ea:1988}%
  \BibitemOpen
  \bibfield  {author} {\bibinfo {author} {\bibnamefont {{Kratz}}, \bibfnamefont
  {K-L}}, \bibinfo {author} {\bibfnamefont {F.-K.}\ \bibnamefont
  {{Thielemann}}}, \bibinfo {author} {\bibfnamefont {W.}~\bibnamefont
  {{Hillebrandt}}}, \bibinfo {author} {\bibfnamefont {P.}~\bibnamefont
  {{M{\"o}ller}}}, \bibinfo {author} {\bibfnamefont {V.}~\bibnamefont
  {{Harms}}}, \bibinfo {author} {\bibfnamefont {A.}~\bibnamefont {{Wohr}}},
  and\ \bibinfo {author} {\bibfnamefont {J.~W.}\ \bibnamefont {{Truran}}}}
  (\bibinfo {year} {1988}),\ \bibfield  {title} {\enquote {\bibinfo {title}
  {Constraints on r-process conditions from beta-decay properties far off
  stability and r-abundances},}\ }\href
  {https://doi.org/10.1088/0305-4616/14/S/034} {\bibfield  {journal} {\bibinfo
  {journal} {J. Phys. G: Nucl. Part. Phys.}\ }\textbf {\bibinfo {volume}
  {14}},\ \bibinfo {pages} {S331--S342}}\BibitemShut {NoStop}%
\bibitem [{\citenamefont {{Kratz}}\ \emph {et~al.}(1983)\citenamefont
  {{Kratz}}, \citenamefont {{Ziegert}}, \citenamefont {{Hillebrandt}},\ and\
  \citenamefont {{Thielemann}}}]{Kratz1983}%
  \BibitemOpen
  \bibfield  {author} {\bibinfo {author} {\bibnamefont {{Kratz}}, \bibfnamefont
  {K-L}}, \bibinfo {author} {\bibfnamefont {W.}~\bibnamefont {{Ziegert}}},
  \bibinfo {author} {\bibfnamefont {W.}~\bibnamefont {{Hillebrandt}}}, and\
  \bibinfo {author} {\bibfnamefont {F.-K.}\ \bibnamefont {{Thielemann}}}}
  (\bibinfo {year} {1983}),\ \bibfield  {title} {\enquote {\bibinfo {title}
  {{Determination of stellar neutron-capture rates for radioactive nuclei with
  the aid of beta-delayed neutron emission}},}\ }\href@noop {} {\bibfield
  {journal} {\bibinfo  {journal} {Astron. \& Astrophys.}\ }\textbf {\bibinfo
  {volume} {125}},\ \bibinfo {pages} {381--387}}\BibitemShut {NoStop}%
\bibitem [{\citenamefont {{Kratz}}\ \emph {et~al.}(2000)\citenamefont
  {{Kratz}}, \citenamefont {{Pfeiffer}}, \citenamefont {{Thielemann}},\ and\
  \citenamefont {{Walters}}}]{Kratz.Pfeiffer.ea:2000}%
  \BibitemOpen
  \bibfield  {author} {\bibinfo {author} {\bibnamefont {{Kratz}}, \bibfnamefont
  {Karl-Ludwig}}, \bibinfo {author} {\bibfnamefont {Bernd}\ \bibnamefont
  {{Pfeiffer}}}, \bibinfo {author} {\bibfnamefont {Friedrich-Karl}\
  \bibnamefont {{Thielemann}}}, and\ \bibinfo {author} {\bibfnamefont
  {William~B.}\ \bibnamefont {{Walters}}}} (\bibinfo {year} {2000}),\ \bibfield
   {title} {\enquote {\bibinfo {title} {{Nuclear structure studies at ISOLDE
  and their impact on the astrophysical r-process}},}\ }\href
  {https://doi.org/10.1023/A:1012694723985} {\bibfield  {journal} {\bibinfo
  {journal} {Hyperfine Interactions}\ }\textbf {\bibinfo {volume} {129}},\
  \bibinfo {pages} {185--221}}\BibitemShut {NoStop}%
\bibitem [{\citenamefont {{Kugler}}\ \emph {et~al.}(1992)\citenamefont
  {{Kugler}}, \citenamefont {{Fiander}}, \citenamefont {{Johnson}},
  \citenamefont {{Haas}}, \citenamefont {{Przewloka}}, \citenamefont {{Ravn}},
  \citenamefont {{Simon}}, \citenamefont {{Zimmer}},\ and\ \citenamefont
  {{Isolde Collaboration}}}]{Isolde}%
  \BibitemOpen
  \bibfield  {author} {\bibinfo {author} {\bibnamefont {{Kugler}},
  \bibfnamefont {E}}, \bibinfo {author} {\bibfnamefont {D.}~\bibnamefont
  {{Fiander}}}, \bibinfo {author} {\bibfnamefont {B.}~\bibnamefont
  {{Johnson}}}, \bibinfo {author} {\bibfnamefont {H.}~\bibnamefont {{Haas}}},
  \bibinfo {author} {\bibfnamefont {A.}~\bibnamefont {{Przewloka}}}, \bibinfo
  {author} {\bibfnamefont {H.~L.}\ \bibnamefont {{Ravn}}}, \bibinfo {author}
  {\bibfnamefont {D.~J.}\ \bibnamefont {{Simon}}}, \bibinfo {author}
  {\bibfnamefont {K.}~\bibnamefont {{Zimmer}}}, and\ \bibinfo {author}
  {\bibnamefont {{Isolde Collaboration}}}} (\bibinfo {year} {1992}),\ \bibfield
   {title} {\enquote {\bibinfo {title} {{The new CERN-ISOLDE on-line
  mass-separator facility at the PS-Booster}},}\ }\href
  {https://doi.org/10.1016/0168-583X(92)95907-9} {\bibfield  {journal}
  {\bibinfo  {journal} {Nucl. Instruments Methods Phys. Res. Sect. B Beam
  Interact. with Mater. Atoms}\ }\textbf {\bibinfo {volume} {70}},\ \bibinfo
  {pages} {41--49}}\BibitemShut {NoStop}%
\bibitem [{\citenamefont {{Kulkarni}}(2005)}]{Kulkarni:2005}%
  \BibitemOpen
  \bibfield  {author} {\bibinfo {author} {\bibnamefont {{Kulkarni}},
  \bibfnamefont {S~R}}} (\bibinfo {year} {2005}),\ \bibfield  {title} {\enquote
  {\bibinfo {title} {{Modeling Supernova-like Explosions Associated with
  Gamma-ray Bursts with Short Durations}},}\ }\href@noop {} {\bibfield
  {journal} {\bibinfo  {journal} {arXiv e-prints}\ }}\Eprint
  {https://arxiv.org/abs/astro-ph/0510256} {arXiv:astro-ph/0510256 [astro-ph]}
  \BibitemShut {NoStop}%
\bibitem [{\citenamefont {{Kurcewicz}}\ \emph {et~al.}(2012)\citenamefont
  {{Kurcewicz}} \emph {et~al.}}]{Kurcewicz}%
  \BibitemOpen
  \bibfield  {author} {\bibinfo {author} {\bibnamefont {{Kurcewicz}},
  \bibfnamefont {J}},  \emph {et~al.}} (\bibinfo {year} {2012}),\ \bibfield
  {title} {\enquote {\bibinfo {title} {{Discovery and cross-section measurement
  of neutron-rich isotopes in the element range from neodymium to platinum with
  the FRS}},}\ }\href {https://doi.org/10.1016/j.physletb.2012.09.021}
  {\bibfield  {journal} {\bibinfo  {journal} {Phys. Lett. B}\ }\textbf
  {\bibinfo {volume} {717}},\ \bibinfo {pages} {371--375}}\BibitemShut
  {NoStop}%
\bibitem [{\citenamefont {{Kuroda}}\ \emph {et~al.}(2018)\citenamefont
  {{Kuroda}}, \citenamefont {{Kotake}}, \citenamefont {{Takiwaki}},\ and\
  \citenamefont {{Thielemann}}}]{Kuroda.ea:2018}%
  \BibitemOpen
  \bibfield  {author} {\bibinfo {author} {\bibnamefont {{Kuroda}},
  \bibfnamefont {T}}, \bibinfo {author} {\bibfnamefont {K.}~\bibnamefont
  {{Kotake}}}, \bibinfo {author} {\bibfnamefont {T.}~\bibnamefont
  {{Takiwaki}}}, and\ \bibinfo {author} {\bibfnamefont {F.-K.}\ \bibnamefont
  {{Thielemann}}}} (\bibinfo {year} {2018}),\ \bibfield  {title} {\enquote
  {\bibinfo {title} {{A full general relativistic neutrino
  radiation-hydrodynamics simulation of a collapsing very massive star and the
  formation of a black hole}},}\ }\href {https://doi.org/10.1093/mnrasl/sly059}
  {\bibfield  {journal} {\bibinfo  {journal} {Mon. Not. Roy. Astron. Soc.}\
  }\textbf {\bibinfo {volume} {477}},\ \bibinfo {pages} {L80--L84}}\BibitemShut
  {NoStop}%
\bibitem [{\citenamefont {{Kyutoku}}\ \emph {et~al.}(2015)\citenamefont
  {{Kyutoku}}, \citenamefont {{Ioka}}, \citenamefont {{Okawa}}, \citenamefont
  {{Shibata}},\ and\ \citenamefont {{Taniguchi}}}]{Kyutoku.Ioka.ea:2015}%
  \BibitemOpen
  \bibfield  {author} {\bibinfo {author} {\bibnamefont {{Kyutoku}},
  \bibfnamefont {K}}, \bibinfo {author} {\bibfnamefont {K.}~\bibnamefont
  {{Ioka}}}, \bibinfo {author} {\bibfnamefont {H.}~\bibnamefont {{Okawa}}},
  \bibinfo {author} {\bibfnamefont {M.}~\bibnamefont {{Shibata}}}, and\
  \bibinfo {author} {\bibfnamefont {K.}~\bibnamefont {{Taniguchi}}}} (\bibinfo
  {year} {2015}),\ \bibfield  {title} {\enquote {\bibinfo {title} {{Dynamical
  mass ejection from black hole-neutron star binaries}},}\ }\href
  {https://doi.org/10.1103/PhysRevD.92.044028} {\bibfield  {journal} {\bibinfo
  {journal} {Phys. Rev. D}\ }\textbf {\bibinfo {volume} {92}},\ \bibinfo {eid}
  {044028}}\BibitemShut {NoStop}%
\bibitem [{\citenamefont {{Kyutoku}}\ \emph {et~al.}(2013)\citenamefont
  {{Kyutoku}}, \citenamefont {{Ioka}},\ and\ \citenamefont
  {{Shibata}}}]{kyutoku13}%
  \BibitemOpen
  \bibfield  {author} {\bibinfo {author} {\bibnamefont {{Kyutoku}},
  \bibfnamefont {K}}, \bibinfo {author} {\bibfnamefont {K.}~\bibnamefont
  {{Ioka}}}, and\ \bibinfo {author} {\bibfnamefont {M.}~\bibnamefont
  {{Shibata}}}} (\bibinfo {year} {2013}),\ \bibfield  {title} {\enquote
  {\bibinfo {title} {{Anisotropic mass ejection from black hole-neutron star
  binaries: Diversity of electromagnetic counterparts}},}\ }\href
  {https://doi.org/10.1103/PhysRevD.88.041503} {\bibfield  {journal} {\bibinfo
  {journal} {Phys. Rev. D}\ }\textbf {\bibinfo {volume} {88}},\ \bibinfo {eid}
  {041503}}\BibitemShut {NoStop}%
\bibitem [{\citenamefont {{Kyutoku}}\ \emph {et~al.}(2018)\citenamefont
  {{Kyutoku}}, \citenamefont {{Kiuchi}}, \citenamefont {{Sekiguchi}},
  \citenamefont {{Shibata}},\ and\ \citenamefont
  {{Taniguchi}}}]{Kyutoku.Kiuchi.ea:2018}%
  \BibitemOpen
  \bibfield  {author} {\bibinfo {author} {\bibnamefont {{Kyutoku}},
  \bibfnamefont {K}}, \bibinfo {author} {\bibfnamefont {K.}~\bibnamefont
  {{Kiuchi}}}, \bibinfo {author} {\bibfnamefont {Y.}~\bibnamefont
  {{Sekiguchi}}}, \bibinfo {author} {\bibfnamefont {M.}~\bibnamefont
  {{Shibata}}}, and\ \bibinfo {author} {\bibfnamefont {K.}~\bibnamefont
  {{Taniguchi}}}} (\bibinfo {year} {2018}),\ \bibfield  {title} {\enquote
  {\bibinfo {title} {{Neutrino transport in black hole-neutron star binaries:
  Neutrino emission and dynamical mass ejection}},}\ }\href
  {https://doi.org/10.1103/PhysRevD.97.023009} {\bibfield  {journal} {\bibinfo
  {journal} {Phys. Rev. D}\ }\textbf {\bibinfo {volume} {97}},\ \bibinfo {eid}
  {023009}}\BibitemShut {NoStop}%
\bibitem [{\citenamefont {{Kyutoku}}\ \emph {et~al.}(2020)\citenamefont
  {{Kyutoku}}, \citenamefont {{Fujibayashi}}, \citenamefont {{Hayashi}},
  \citenamefont {{Kawaguchi}}, \citenamefont {{Kiuchi}}, \citenamefont
  {{Shibata}},\ and\ \citenamefont {{Tanaka}}}]{Kyutoku.Fujibayashi.ea:2020}%
  \BibitemOpen
  \bibfield  {author} {\bibinfo {author} {\bibnamefont {{Kyutoku}},
  \bibfnamefont {Koutarou}}, \bibinfo {author} {\bibfnamefont {Sho}\
  \bibnamefont {{Fujibayashi}}}, \bibinfo {author} {\bibfnamefont {Kota}\
  \bibnamefont {{Hayashi}}}, \bibinfo {author} {\bibfnamefont {Kyohei}\
  \bibnamefont {{Kawaguchi}}}, \bibinfo {author} {\bibfnamefont {Kenta}\
  \bibnamefont {{Kiuchi}}}, \bibinfo {author} {\bibfnamefont {Masaru}\
  \bibnamefont {{Shibata}}}, and\ \bibinfo {author} {\bibfnamefont {Masaomi}\
  \bibnamefont {{Tanaka}}}} (\bibinfo {year} {2020}),\ \bibfield  {title}
  {\enquote {\bibinfo {title} {{On the Possibility of GW190425 Being a Black
  Hole-Neutron Star Binary Merger}},}\ }\href
  {https://doi.org/10.3847/2041-8213/ab6e70} {\bibfield  {journal} {\bibinfo
  {journal} {Astrophys. J. Lett.}\ }\textbf {\bibinfo {volume} {890}},\
  \bibinfo {eid} {L4}}\BibitemShut {NoStop}%
\bibitem [{\citenamefont {Landau}(1932)}]{Landau:1932}%
  \BibitemOpen
  \bibfield  {author} {\bibinfo {author} {\bibnamefont {Landau}, \bibfnamefont
  {L~D}}} (\bibinfo {year} {1932}),\ \bibfield  {title} {\enquote {\bibinfo
  {title} {{On the theory of star}},}\ }\href@noop {} {\bibfield  {journal}
  {\bibinfo  {journal} {Phys. Z. Sowjet Union}\ }\textbf {\bibinfo {volume}
  {1}},\ \bibinfo {pages} {285--288}}\BibitemShut {NoStop}%
\bibitem [{\citenamefont {{Langanke}}(1998)}]{Langanke:1998}%
  \BibitemOpen
  \bibfield  {author} {\bibinfo {author} {\bibnamefont {{Langanke}},
  \bibfnamefont {K}}} (\bibinfo {year} {1998}),\ \bibfield  {title} {\enquote
  {\bibinfo {title} {{Shell model Monte Carlo level densities for nuclei with
  A\~{}50}},}\ }\href {https://doi.org/10.1016/S0370-2693(98)01047-8}
  {\bibfield  {journal} {\bibinfo  {journal} {Phys. Lett. B}\ }\textbf
  {\bibinfo {volume} {438}},\ \bibinfo {pages} {235--241}}\BibitemShut
  {NoStop}%
\bibitem [{\citenamefont {Langanke}(2006)}]{Langanke:2006}%
  \BibitemOpen
  \bibfield  {author} {\bibinfo {author} {\bibnamefont {Langanke},
  \bibfnamefont {K}}} (\bibinfo {year} {2006}),\ \bibfield  {title} {\enquote
  {\bibinfo {title} {{Shell model Monte Carlo studies of pairing correlations
  and level densities in medium-mass nuclei}},}\ }\href
  {https://doi.org/10.1016/j.nuclphysa.2006.08.005} {\bibfield  {journal}
  {\bibinfo  {journal} {Nucl. Phys. A}\ }\textbf {\bibinfo {volume} {778}},\
  \bibinfo {pages} {233--246}}\BibitemShut {NoStop}%
\bibitem [{\citenamefont {Langanke}\ and\ \citenamefont
  {Kolbe}(2001)}]{Langanke.Kolbe:2001}%
  \BibitemOpen
  \bibfield  {author} {\bibinfo {author} {\bibnamefont {Langanke},
  \bibfnamefont {K}}, and\ \bibinfo {author} {\bibfnamefont {E.}~\bibnamefont
  {Kolbe}}} (\bibinfo {year} {2001}),\ \bibfield  {title} {\enquote {\bibinfo
  {title} {Neutrino-induced charged-current reaction rates for r-process
  nuclei},}\ }\href {https://doi.org/10.1006/adnd.2001.0872} {\bibfield
  {journal} {\bibinfo  {journal} {At. Data Nucl. Data Tables}\ }\textbf
  {\bibinfo {volume} {79}},\ \bibinfo {pages} {293--315}}\BibitemShut {NoStop}%
\bibitem [{\citenamefont {Langanke}\ and\ \citenamefont
  {Kolbe}(2002)}]{Langanke.Kolbe:2002}%
  \BibitemOpen
  \bibfield  {author} {\bibinfo {author} {\bibnamefont {Langanke},
  \bibfnamefont {K}}, and\ \bibinfo {author} {\bibfnamefont {E.}~\bibnamefont
  {Kolbe}}} (\bibinfo {year} {2002}),\ \bibfield  {title} {\enquote {\bibinfo
  {title} {Neutrino-induced neutral-current reaction rates for r-process
  nuclei},}\ }\href {https://doi.org/10.1006/adnd.2002.0883} {\bibfield
  {journal} {\bibinfo  {journal} {At. Data Nucl. Data Tables}\ }\textbf
  {\bibinfo {volume} {82}},\ \bibinfo {pages} {191--209}}\BibitemShut {NoStop}%
\bibitem [{\citenamefont {Langanke}\ and\ \citenamefont
  {Mart{\'\i}nez-Pinedo}(2000)}]{Langanke.Martinez-Pinedo:2000}%
  \BibitemOpen
  \bibfield  {author} {\bibinfo {author} {\bibnamefont {Langanke},
  \bibfnamefont {K}}, and\ \bibinfo {author} {\bibfnamefont {G.}~\bibnamefont
  {Mart{\'\i}nez-Pinedo}}} (\bibinfo {year} {2000}),\ \bibfield  {title}
  {\enquote {\bibinfo {title} {{Shell-model calculations of stellar weak
  interaction rates: II. Weak rates for nuclei in the mass range $A=45$--65 in
  supernovae environment}},}\ }\href
  {https://doi.org/10.1016/S0375-9474(00)00131-7} {\bibfield  {journal}
  {\bibinfo  {journal} {Nucl. Phys. A}\ }\textbf {\bibinfo {volume} {673}},\
  \bibinfo {pages} {481--508}}\BibitemShut {NoStop}%
\bibitem [{\citenamefont {{Langanke}}\ and\ \citenamefont
  {{Mart{\'\i}nez-Pinedo}}(2001)}]{Langanke.Martinez-Pinedo:2001}%
  \BibitemOpen
  \bibfield  {author} {\bibinfo {author} {\bibnamefont {{Langanke}},
  \bibfnamefont {K}}, and\ \bibinfo {author} {\bibfnamefont {G.}~\bibnamefont
  {{Mart{\'\i}nez-Pinedo}}}} (\bibinfo {year} {2001}),\ \bibfield  {title}
  {\enquote {\bibinfo {title} {{Rate Tables for the Weak Processes of
  $pf$-shell Nuclei in Stellar Environments}},}\ }\href
  {https://doi.org/10.1006/adnd.2001.0865} {\bibfield  {journal} {\bibinfo
  {journal} {At. Data Nucl. Data Tables}\ }\textbf {\bibinfo {volume} {79}},\
  \bibinfo {pages} {1--46}}\BibitemShut {NoStop}%
\bibitem [{\citenamefont {Langanke}\ and\ \citenamefont
  {Mart{\'\i}nez-Pinedo}(2003)}]{Langanke.Martinez-Pinedo:2003}%
  \BibitemOpen
  \bibfield  {author} {\bibinfo {author} {\bibnamefont {Langanke},
  \bibfnamefont {K}}, and\ \bibinfo {author} {\bibfnamefont {G.}~\bibnamefont
  {Mart{\'\i}nez-Pinedo}}} (\bibinfo {year} {2003}),\ \bibfield  {title}
  {\enquote {\bibinfo {title} {{Nuclear weak-interaction processes in
  stars}},}\ }\href {https://doi.org/10.1103/RevModPhys.75.819} {\bibfield
  {journal} {\bibinfo  {journal} {Rev. Mod. Phys.}\ }\textbf {\bibinfo {volume}
  {75}},\ \bibinfo {pages} {819--862}}\BibitemShut {NoStop}%
\bibitem [{\citenamefont {{Larsen}}\ \emph {et~al.}(2006)\citenamefont
  {{Larsen}}, \citenamefont {{Chankova}}, \citenamefont {{Guttormsen}},
  \citenamefont {{Ingebretsen}}, \citenamefont {{Messelt}}, \citenamefont
  {{Rekstad}}, \citenamefont {{Siem}}, \citenamefont {{Syed}}, \citenamefont
  {{{\O}deg{\aa}rd}}, \citenamefont {{L{\"o}nnroth}}, \citenamefont
  {{Schiller}},\ and\ \citenamefont {{Voinov}}}]{Larsen.Chankova.ea:2006}%
  \BibitemOpen
  \bibfield  {author} {\bibinfo {author} {\bibnamefont {{Larsen}},
  \bibfnamefont {A~C}}, \bibinfo {author} {\bibfnamefont {R.}~\bibnamefont
  {{Chankova}}}, \bibinfo {author} {\bibfnamefont {M.}~\bibnamefont
  {{Guttormsen}}}, \bibinfo {author} {\bibfnamefont {F.}~\bibnamefont
  {{Ingebretsen}}}, \bibinfo {author} {\bibfnamefont {S.}~\bibnamefont
  {{Messelt}}}, \bibinfo {author} {\bibfnamefont {J.}~\bibnamefont
  {{Rekstad}}}, \bibinfo {author} {\bibfnamefont {S.}~\bibnamefont {{Siem}}},
  \bibinfo {author} {\bibfnamefont {N.~U.~H.}\ \bibnamefont {{Syed}}}, \bibinfo
  {author} {\bibfnamefont {S.~W.}\ \bibnamefont {{{\O}deg{\aa}rd}}}, \bibinfo
  {author} {\bibfnamefont {T.}~\bibnamefont {{L{\"o}nnroth}}}, \bibinfo
  {author} {\bibfnamefont {A.}~\bibnamefont {{Schiller}}}, and\ \bibinfo
  {author} {\bibfnamefont {A.}~\bibnamefont {{Voinov}}}} (\bibinfo {year}
  {2006}),\ \bibfield  {title} {\enquote {\bibinfo {title} {{Microcanonical
  entropies and radiative strength functions of $^{50,51}$V}},}\ }\href
  {https://doi.org/10.1103/PhysRevC.73.064301} {\bibfield  {journal} {\bibinfo
  {journal} {Phys. Rev. C}\ }\textbf {\bibinfo {volume} {73}},\ \bibinfo {eid}
  {064301}}\BibitemShut {NoStop}%
\bibitem [{\citenamefont {{Larsen}}\ and\ \citenamefont
  {{Goriely}}(2010)}]{Larsen.Goriely:2010}%
  \BibitemOpen
  \bibfield  {author} {\bibinfo {author} {\bibnamefont {{Larsen}},
  \bibfnamefont {A~C}}, and\ \bibinfo {author} {\bibfnamefont {S.}~\bibnamefont
  {{Goriely}}}} (\bibinfo {year} {2010}),\ \bibfield  {title} {\enquote
  {\bibinfo {title} {{Impact of a low-energy enhancement in the {$\gamma$}-ray
  strength function on the neutron-capture cross section}},}\ }\href
  {https://doi.org/10.1103/PhysRevC.82.014318} {\bibfield  {journal} {\bibinfo
  {journal} {Phys. Rev. C}\ }\textbf {\bibinfo {volume} {82}},\ \bibinfo {eid}
  {014318}}\BibitemShut {NoStop}%
\bibitem [{\citenamefont {{Larsen}}\ \emph {et~al.}(2007)\citenamefont
  {{Larsen}}, \citenamefont {{Guttormsen}}, \citenamefont {{Chankova}},
  \citenamefont {{Ingebretsen}}, \citenamefont {{L{\"o}nnroth}}, \citenamefont
  {{Messelt}}, \citenamefont {{Rekstad}}, \citenamefont {{Schiller}},
  \citenamefont {{Siem}}, \citenamefont {{Syed}},\ and\ \citenamefont
  {{Voinov}}}]{Larsen.Guttormsen.ea:2007}%
  \BibitemOpen
  \bibfield  {author} {\bibinfo {author} {\bibnamefont {{Larsen}},
  \bibfnamefont {A~C}}, \bibinfo {author} {\bibfnamefont {M.}~\bibnamefont
  {{Guttormsen}}}, \bibinfo {author} {\bibfnamefont {R.}~\bibnamefont
  {{Chankova}}}, \bibinfo {author} {\bibfnamefont {F.}~\bibnamefont
  {{Ingebretsen}}}, \bibinfo {author} {\bibfnamefont {T.}~\bibnamefont
  {{L{\"o}nnroth}}}, \bibinfo {author} {\bibfnamefont {S.}~\bibnamefont
  {{Messelt}}}, \bibinfo {author} {\bibfnamefont {J.}~\bibnamefont
  {{Rekstad}}}, \bibinfo {author} {\bibfnamefont {A.}~\bibnamefont
  {{Schiller}}}, \bibinfo {author} {\bibfnamefont {S.}~\bibnamefont {{Siem}}},
  \bibinfo {author} {\bibfnamefont {N.~U.~H.}\ \bibnamefont {{Syed}}}, and\
  \bibinfo {author} {\bibfnamefont {A.}~\bibnamefont {{Voinov}}}} (\bibinfo
  {year} {2007}),\ \bibfield  {title} {\enquote {\bibinfo {title} {{Nuclear
  level densities and {$\gamma$}-ray strength functions in Sc44,45}},}\ }\href
  {https://doi.org/10.1103/PhysRevC.76.044303} {\bibfield  {journal} {\bibinfo
  {journal} {Phys. Rev. C}\ }\textbf {\bibinfo {volume} {76}},\ \bibinfo {eid}
  {044303}}\BibitemShut {NoStop}%
\bibitem [{\citenamefont {{Larsen}}\ \emph {et~al.}(2019)\citenamefont
  {{Larsen}}, \citenamefont {{Spyrou}}, \citenamefont {{Liddick}},\ and\
  \citenamefont {{Guttormsen}}}]{Larsen.Spyrou.ea:2019}%
  \BibitemOpen
  \bibfield  {author} {\bibinfo {author} {\bibnamefont {{Larsen}},
  \bibfnamefont {A~C}}, \bibinfo {author} {\bibfnamefont {A.}~\bibnamefont
  {{Spyrou}}}, \bibinfo {author} {\bibfnamefont {S.~N.}\ \bibnamefont
  {{Liddick}}}, and\ \bibinfo {author} {\bibfnamefont {M.}~\bibnamefont
  {{Guttormsen}}}} (\bibinfo {year} {2019}),\ \bibfield  {title} {\enquote
  {\bibinfo {title} {{Novel techniques for constraining neutron-capture rates
  relevant for r-process heavy-element nucleosynthesis}},}\ }\href
  {https://doi.org/10.1016/j.ppnp.2019.04.002} {\bibfield  {journal} {\bibinfo
  {journal} {Prog. Part. Nucl. Phys.}\ }\textbf {\bibinfo {volume} {107}},\
  \bibinfo {pages} {69--108}}\BibitemShut {NoStop}%
\bibitem [{\citenamefont {{Larsen}}\ \emph {et~al.}(2015)\citenamefont
  {{Larsen}} \emph {et~al.}}]{Larsen.Goriely.ea:2015}%
  \BibitemOpen
  \bibfield  {author} {\bibinfo {author} {\bibnamefont {{Larsen}},
  \bibfnamefont {A~C}},  \emph {et~al.}} (\bibinfo {year} {2015}),\ \bibfield
  {title} {\enquote {\bibinfo {title} {{Upbend and M1 Scissors Mode in
  Neutron-rich Nuclei -- Consequences for r-process (n,gamma ) Reaction
  Rates}},}\ }\href {https://doi.org/10.5506/APhysPolB.46.509} {\bibfield
  {journal} {\bibinfo  {journal} {Acta Physica Polonica B}\ }\textbf {\bibinfo
  {volume} {46}},\ \bibinfo {pages} {509}}\BibitemShut {NoStop}%
\bibitem [{\citenamefont {{Larson}}\ \emph {et~al.}(2013)\citenamefont
  {{Larson}}, \citenamefont {{Liddick}}, \citenamefont {{Bennett}},
  \citenamefont {{Bowe}}, \citenamefont {{Chemey}}, \citenamefont {{Prokop}},
  \citenamefont {{Simon}}, \citenamefont {{Spyrou}}, \citenamefont {{Suchyta}},
  \citenamefont {{Quinn}}, \citenamefont {{Tabor}}, \citenamefont {{Tai}},
  \citenamefont {{Tripathi}},\ and\ \citenamefont {{VonMoss}}}]{Larson2013}%
  \BibitemOpen
  \bibfield  {author} {\bibinfo {author} {\bibnamefont {{Larson}},
  \bibfnamefont {N}}, \bibinfo {author} {\bibfnamefont {S.~N.}\ \bibnamefont
  {{Liddick}}}, \bibinfo {author} {\bibfnamefont {M.}~\bibnamefont
  {{Bennett}}}, \bibinfo {author} {\bibfnamefont {A.}~\bibnamefont {{Bowe}}},
  \bibinfo {author} {\bibfnamefont {A.}~\bibnamefont {{Chemey}}}, \bibinfo
  {author} {\bibfnamefont {C.}~\bibnamefont {{Prokop}}}, \bibinfo {author}
  {\bibfnamefont {A.}~\bibnamefont {{Simon}}}, \bibinfo {author} {\bibfnamefont
  {A.}~\bibnamefont {{Spyrou}}}, \bibinfo {author} {\bibfnamefont
  {S.}~\bibnamefont {{Suchyta}}}, \bibinfo {author} {\bibfnamefont {S.~J.}\
  \bibnamefont {{Quinn}}}, \bibinfo {author} {\bibfnamefont {S.~L.}\
  \bibnamefont {{Tabor}}}, \bibinfo {author} {\bibfnamefont {P.~L.}\
  \bibnamefont {{Tai}}}, \bibinfo {author} {\bibfnamefont {V.}~\bibnamefont
  {{Tripathi}}}, and\ \bibinfo {author} {\bibfnamefont {J.~M.}\ \bibnamefont
  {{VonMoss}}}} (\bibinfo {year} {2013}),\ \bibfield  {title} {\enquote
  {\bibinfo {title} {{High efficiency beta-decay spectroscopy using a planar
  germanium double-sided strip detector}},}\ }\href
  {https://doi.org/10.1016/j.nima.2013.06.027} {\bibfield  {journal} {\bibinfo
  {journal} {Nucl. Instruments Methods Phys. Res. Sect. A Accel. Spectrometers,
  Detect. Assoc. Equip.}\ }\textbf {\bibinfo {volume} {727}},\ \bibinfo {pages}
  {59--64}}\BibitemShut {NoStop}%
\bibitem [{\citenamefont {{Lattimer}}(2012)}]{Lattimer:2012}%
  \BibitemOpen
  \bibfield  {author} {\bibinfo {author} {\bibnamefont {{Lattimer}},
  \bibfnamefont {J~M}}} (\bibinfo {year} {2012}),\ \bibfield  {title} {\enquote
  {\bibinfo {title} {{The Nuclear Equation of State and Neutron Star
  Masses}},}\ }\href {https://doi.org/10.1146/annurev-nucl-102711-095018}
  {\bibfield  {journal} {\bibinfo  {journal} {Annu. Rev. Nucl. Part. Sci.}\
  }\textbf {\bibinfo {volume} {62}},\ \bibinfo {pages} {485--515}}\BibitemShut
  {NoStop}%
\bibitem [{\citenamefont {{Lattimer}}\ and\ \citenamefont
  {{Schramm}}(1974)}]{Lattimer.Schramm:1974}%
  \BibitemOpen
  \bibfield  {author} {\bibinfo {author} {\bibnamefont {{Lattimer}},
  \bibfnamefont {J~M}}, and\ \bibinfo {author} {\bibfnamefont {D.~N.}\
  \bibnamefont {{Schramm}}}} (\bibinfo {year} {1974}),\ \bibfield  {title}
  {\enquote {\bibinfo {title} {{Black-hole-neutron-star collisions}},}\ }\href
  {https://doi.org/10.1086/181612} {\bibfield  {journal} {\bibinfo  {journal}
  {Astrophys. J.}\ }\textbf {\bibinfo {volume} {192}},\ \bibinfo {pages}
  {L145--L147}}\BibitemShut {NoStop}%
\bibitem [{\citenamefont {{Lattimer}}\ and\ \citenamefont
  {{Schramm}}(1976)}]{lattimer76}%
  \BibitemOpen
  \bibfield  {author} {\bibinfo {author} {\bibnamefont {{Lattimer}},
  \bibfnamefont {J~M}}, and\ \bibinfo {author} {\bibfnamefont {D.~N.}\
  \bibnamefont {{Schramm}}}} (\bibinfo {year} {1976}),\ \bibfield  {title}
  {\enquote {\bibinfo {title} {{The tidal disruption of neutron stars by black
  holes in close binaries}},}\ }\href {https://doi.org/10.1086/154860}
  {\bibfield  {journal} {\bibinfo  {journal} {Astrophys. J.}\ }\textbf
  {\bibinfo {volume} {210}},\ \bibinfo {pages} {549--567}}\BibitemShut
  {NoStop}%
\bibitem [{\citenamefont {{Lattimer}}(2019)}]{Lattimer:2019b}%
  \BibitemOpen
  \bibfield  {author} {\bibinfo {author} {\bibnamefont {{Lattimer}},
  \bibfnamefont {James~M}}} (\bibinfo {year} {2019}),\ \bibfield  {title}
  {\enquote {\bibinfo {title} {{The Properties of a Black Hole-Neutron Star
  Merger Candidate}},}\ }\href@noop {} {\bibfield  {journal} {\bibinfo
  {journal} {arXiv e-prints}\ }}\Eprint {https://arxiv.org/abs/1908.03622}
  {arXiv:1908.03622 [astro-ph.HE]} \BibitemShut {NoStop}%
\bibitem [{\citenamefont {{Lawler}}\ \emph {et~al.}(2006)\citenamefont
  {{Lawler}}, \citenamefont {{Den Hartog}}, \citenamefont {{Sneden}},\ and\
  \citenamefont {{Cowan}}}]{lawler06}%
  \BibitemOpen
  \bibfield  {author} {\bibinfo {author} {\bibnamefont {{Lawler}},
  \bibfnamefont {J~E}}, \bibinfo {author} {\bibfnamefont {E.~A.}\ \bibnamefont
  {{Den Hartog}}}, \bibinfo {author} {\bibfnamefont {C.}~\bibnamefont
  {{Sneden}}}, and\ \bibinfo {author} {\bibfnamefont {J.~J.}\ \bibnamefont
  {{Cowan}}}} (\bibinfo {year} {2006}),\ \bibfield  {title} {\enquote {\bibinfo
  {title} {{Improved Laboratory Transition Probabilities for Sm II and
  Application to the Samarium Abundances of the Sun and Three r-Process-rich,
  Metal-poor Stars}},}\ }\href {https://doi.org/10.1086/498213} {\bibfield
  {journal} {\bibinfo  {journal} {Astrophys. J. Suppl.}\ }\textbf {\bibinfo
  {volume} {162}},\ \bibinfo {pages} {227--260}}\BibitemShut {NoStop}%
\bibitem [{\citenamefont {{Lawler}}\ \emph {et~al.}(2008)\citenamefont
  {{Lawler}}, \citenamefont {{Hartog}}, \citenamefont {{Sneden}},\ and\
  \citenamefont {{Cowan}}}]{lawler08}%
  \BibitemOpen
  \bibfield  {author} {\bibinfo {author} {\bibnamefont {{Lawler}},
  \bibfnamefont {J~E}}, \bibinfo {author} {\bibfnamefont {E.~A.}\ \bibnamefont
  {{Hartog}}}, \bibinfo {author} {\bibfnamefont {C.}~\bibnamefont {{Sneden}}},
  and\ \bibinfo {author} {\bibfnamefont {J.~J.}\ \bibnamefont {{Cowan}}}}
  (\bibinfo {year} {2008}),\ \bibfield  {title} {\enquote {\bibinfo {title}
  {{Comparison of Sm II transition probabilities}},}\ }\href
  {https://doi.org/10.1139/P08-020} {\bibfield  {journal} {\bibinfo  {journal}
  {Canadian Journal of Physics}\ }\textbf {\bibinfo {volume} {86}},\ \bibinfo
  {pages} {1033--1038}}\BibitemShut {NoStop}%
\bibitem [{\citenamefont {{Lawler}}\ \emph {et~al.}(2009)\citenamefont
  {{Lawler}}, \citenamefont {{Sneden}}, \citenamefont {{Cowan}}, \citenamefont
  {{Ivans}},\ and\ \citenamefont {{Den Hartog}}}]{lawler09}%
  \BibitemOpen
  \bibfield  {author} {\bibinfo {author} {\bibnamefont {{Lawler}},
  \bibfnamefont {J~E}}, \bibinfo {author} {\bibfnamefont {C.}~\bibnamefont
  {{Sneden}}}, \bibinfo {author} {\bibfnamefont {J.~J.}\ \bibnamefont
  {{Cowan}}}, \bibinfo {author} {\bibfnamefont {I.~I.}\ \bibnamefont
  {{Ivans}}}, and\ \bibinfo {author} {\bibfnamefont {E.~A.}\ \bibnamefont {{Den
  Hartog}}}} (\bibinfo {year} {2009}),\ \bibfield  {title} {\enquote {\bibinfo
  {title} {{Improved Laboratory Transition Probabilities for Ce II, Application
  to the Cerium Abundances of the Sun and Five r-Process-Rich, Metal-Poor
  Stars, and Rare Earth Lab Data Summary}},}\ }\href
  {https://doi.org/10.1088/0067-0049/182/1/51} {\bibfield  {journal} {\bibinfo
  {journal} {Astrophys. J. Suppl.}\ }\textbf {\bibinfo {volume} {182}},\
  \bibinfo {pages} {51--79}}\BibitemShut {NoStop}%
\bibitem [{\citenamefont {Lee}\ and\ \citenamefont
  {Ramirez-Ruiz}(2007)}]{Lee.Ramirez-Ruiz:2007}%
  \BibitemOpen
  \bibfield  {author} {\bibinfo {author} {\bibnamefont {Lee}, \bibfnamefont
  {William~H}}, and\ \bibinfo {author} {\bibfnamefont {Enrico}\ \bibnamefont
  {Ramirez-Ruiz}}} (\bibinfo {year} {2007}),\ \bibfield  {title} {\enquote
  {\bibinfo {title} {{The progenitors of short gamma-ray bursts}},}\ }\href
  {https://doi.org/10.1088/1367-2630/9/1/017} {\bibfield  {journal} {\bibinfo
  {journal} {New J. Phys.}\ }\textbf {\bibinfo {volume} {9}},\ \bibinfo {pages}
  {17--17}}\BibitemShut {NoStop}%
\bibitem [{\citenamefont {{Lehner}}\ \emph {et~al.}(2016)\citenamefont
  {{Lehner}}, \citenamefont {{Liebling}}, \citenamefont {{Palenzuela}},
  \citenamefont {{Caballero}}, \citenamefont {{O'Connor}}, \citenamefont
  {{Anderson}},\ and\ \citenamefont {{Neilsen}}}]{Lehner.Liebling.ea:2016}%
  \BibitemOpen
  \bibfield  {author} {\bibinfo {author} {\bibnamefont {{Lehner}},
  \bibfnamefont {L}}, \bibinfo {author} {\bibfnamefont {S.~L.}\ \bibnamefont
  {{Liebling}}}, \bibinfo {author} {\bibfnamefont {C.}~\bibnamefont
  {{Palenzuela}}}, \bibinfo {author} {\bibfnamefont {O.~L.}\ \bibnamefont
  {{Caballero}}}, \bibinfo {author} {\bibfnamefont {E.}~\bibnamefont
  {{O'Connor}}}, \bibinfo {author} {\bibfnamefont {M.}~\bibnamefont
  {{Anderson}}}, and\ \bibinfo {author} {\bibfnamefont {D.}~\bibnamefont
  {{Neilsen}}}} (\bibinfo {year} {2016}),\ \bibfield  {title} {\enquote
  {\bibinfo {title} {{Unequal mass binary neutron star mergers and
  multimessenger signals}},}\ }\href
  {https://doi.org/10.1088/0264-9381/33/18/184002} {\bibfield  {journal}
  {\bibinfo  {journal} {Class. Quantum Gravity}\ }\textbf {\bibinfo {volume}
  {33}},\ \bibinfo {eid} {184002}}\BibitemShut {NoStop}%
\bibitem [{\citenamefont {{Leist}}\ \emph {et~al.}(1985)\citenamefont
  {{Leist}}, \citenamefont {{Ziegert}}, \citenamefont {{Wiescher}},
  \citenamefont {{Kratz}},\ and\ \citenamefont {{Thielemann}}}]{Leist1985}%
  \BibitemOpen
  \bibfield  {author} {\bibinfo {author} {\bibnamefont {{Leist}}, \bibfnamefont
  {B}}, \bibinfo {author} {\bibfnamefont {W.}~\bibnamefont {{Ziegert}}},
  \bibinfo {author} {\bibfnamefont {M.}~\bibnamefont {{Wiescher}}}, \bibinfo
  {author} {\bibfnamefont {K.-L.}\ \bibnamefont {{Kratz}}}, and\ \bibinfo
  {author} {\bibfnamefont {F.-K.}\ \bibnamefont {{Thielemann}}}} (\bibinfo
  {year} {1985}),\ \bibfield  {title} {\enquote {\bibinfo {title} {{Neutron
  capture cross sections for neutron-rich isotopes}},}\ }\href
  {https://doi.org/10.1007/BF01412093} {\bibfield  {journal} {\bibinfo
  {journal} {Zeitschrift f{\"u}r Phys. A Hadron. Nucl.}\ }\textbf {\bibinfo
  {volume} {322}},\ \bibinfo {pages} {531--532}}\BibitemShut {NoStop}%
\bibitem [{\citenamefont {{Lhersonneau}}\ \emph
  {et~al.}(1995{\natexlab{a}})\citenamefont {{Lhersonneau}}, \citenamefont
  {{Gabelmann}}, \citenamefont {{Pfeiffer}},\ and\ \citenamefont
  {{Kratz}}}]{Lhersonneau.Gabelmann.ea:1995}%
  \BibitemOpen
  \bibfield  {author} {\bibinfo {author} {\bibnamefont {{Lhersonneau}},
  \bibfnamefont {G}}, \bibinfo {author} {\bibfnamefont {H.}~\bibnamefont
  {{Gabelmann}}}, \bibinfo {author} {\bibfnamefont {B.}~\bibnamefont
  {{Pfeiffer}}}, and\ \bibinfo {author} {\bibfnamefont {K.-L.}\ \bibnamefont
  {{Kratz}}}} (\bibinfo {year} {1995}{\natexlab{a}}),\ \bibfield  {title}
  {\enquote {\bibinfo {title} {{Structure of the highly deformed
  nucleus$^{101}$Sr$_{63}$ and evidence for identical K=3/2 bands}},}\ }\href
  {https://doi.org/10.1007/BF01289502} {\bibfield  {journal} {\bibinfo
  {journal} {Zeitschrift f{\"u}r Phys. A Hadron. Nucl.}\ }\textbf {\bibinfo
  {volume} {352}},\ \bibinfo {pages} {293--301}}\BibitemShut {NoStop}%
\bibitem [{\citenamefont {{Lhersonneau}}\ \emph
  {et~al.}(1995{\natexlab{b}})\citenamefont {{Lhersonneau}}, \citenamefont
  {{Pfeiffer}}, \citenamefont {{Huhta}}, \citenamefont {{W{\"o}hr}},
  \citenamefont {{Kl{\"o}ckl}}, \citenamefont {{Kratz}},\ and\ \citenamefont
  {{{\"A}yst{\"o}}}}]{Lhersonneau.Pfeiffer.ea:1995}%
  \BibitemOpen
  \bibfield  {author} {\bibinfo {author} {\bibnamefont {{Lhersonneau}},
  \bibfnamefont {G}}, \bibinfo {author} {\bibfnamefont {B.}~\bibnamefont
  {{Pfeiffer}}}, \bibinfo {author} {\bibfnamefont {M.}~\bibnamefont {{Huhta}}},
  \bibinfo {author} {\bibfnamefont {A.}~\bibnamefont {{W{\"o}hr}}}, \bibinfo
  {author} {\bibfnamefont {I.}~\bibnamefont {{Kl{\"o}ckl}}}, \bibinfo {author}
  {\bibfnamefont {K.~L.}\ \bibnamefont {{Kratz}}}, and\ \bibinfo {author}
  {\bibfnamefont {J.}~\bibnamefont {{{\"A}yst{\"o}}}}} (\bibinfo {year}
  {1995}{\natexlab{b}}),\ \bibfield  {title} {\enquote {\bibinfo {title}
  {{First evidence for the 2$^{+}$ level in the very neutron-rich
  nucleus$^{102}$Sr}},}\ }\href {https://doi.org/10.1007/BF01291137} {\bibfield
   {journal} {\bibinfo  {journal} {Zeitschrift f{\"u}r Phys. A Hadron. Nucl.}\
  }\textbf {\bibinfo {volume} {351}},\ \bibinfo {pages} {357--358}}\BibitemShut
  {NoStop}%
\bibitem [{\citenamefont {{Li}}\ \emph {et~al.}(2015)\citenamefont {{Li}},
  \citenamefont {{Aoki}}, \citenamefont {{Honda}}, \citenamefont {{Zhao}},
  \citenamefont {{Christlieb}},\ and\ \citenamefont
  {{Suda}}}]{Li.Aoki.ea:2015}%
  \BibitemOpen
  \bibfield  {author} {\bibinfo {author} {\bibnamefont {{Li}}, \bibfnamefont
  {Hai-Ning}}, \bibinfo {author} {\bibfnamefont {Wako}\ \bibnamefont {{Aoki}}},
  \bibinfo {author} {\bibfnamefont {Satoshi}\ \bibnamefont {{Honda}}}, \bibinfo
  {author} {\bibfnamefont {Gang}\ \bibnamefont {{Zhao}}}, \bibinfo {author}
  {\bibfnamefont {Norbert}\ \bibnamefont {{Christlieb}}}, and\ \bibinfo
  {author} {\bibfnamefont {Takuma}\ \bibnamefont {{Suda}}}} (\bibinfo {year}
  {2015}),\ \bibfield  {title} {\enquote {\bibinfo {title} {{Discovery of a
  strongly r-process enhanced extremely metal-poor star LAMOST
  J110901.22+075441.8}},}\ }\href {https://doi.org/10.1088/1674-4527/15/8/011}
  {\bibfield  {journal} {\bibinfo  {journal} {Res. Astron. Astrophys.}\
  }\textbf {\bibinfo {volume} {15}},\ \bibinfo {eid} {1264}}\BibitemShut
  {NoStop}%
\bibitem [{\citenamefont {{Li}}\ and\ \citenamefont
  {{Paczy{\'n}ski}}(1998)}]{lipacz97}%
  \BibitemOpen
  \bibfield  {author} {\bibinfo {author} {\bibnamefont {{Li}}, \bibfnamefont
  {L-X}}, and\ \bibinfo {author} {\bibfnamefont {B.}~\bibnamefont
  {{Paczy{\'n}ski}}}} (\bibinfo {year} {1998}),\ \bibfield  {title} {\enquote
  {\bibinfo {title} {{Transient Events from Neutron Star Mergers}},}\ }\href
  {https://doi.org/10.1086/311680} {\bibfield  {journal} {\bibinfo  {journal}
  {Astrophys. J.}\ }\textbf {\bibinfo {volume} {507}},\ \bibinfo {pages}
  {L59--L62}}\BibitemShut {NoStop}%
\bibitem [{\citenamefont {{Li}}(2019)}]{Li:2019}%
  \BibitemOpen
  \bibfield  {author} {\bibinfo {author} {\bibnamefont {{Li}}, \bibfnamefont
  {Li-Xin}}} (\bibinfo {year} {2019}),\ \bibfield  {title} {\enquote {\bibinfo
  {title} {{Line Expansion Opacity in Relativistically Expanding Media}},}\
  }\href {https://doi.org/10.3847/1538-4357/ab5306} {\bibfield  {journal}
  {\bibinfo  {journal} {Astrophys. J.}\ }\textbf {\bibinfo {volume} {887}},\
  \bibinfo {eid} {60}}\BibitemShut {NoStop}%
\bibitem [{\citenamefont {{Liddick}}\ \emph {et~al.}(2016)\citenamefont
  {{Liddick}} \emph {et~al.}}]{Liddick2016}%
  \BibitemOpen
  \bibfield  {author} {\bibinfo {author} {\bibnamefont {{Liddick}},
  \bibfnamefont {S~N}},  \emph {et~al.}} (\bibinfo {year} {2016}),\ \bibfield
  {title} {\enquote {\bibinfo {title} {{Experimental Neutron Capture Rate
  Constraint Far from Stability}},}\ }\href
  {https://doi.org/10.1103/PhysRevLett.116.242502} {\bibfield  {journal}
  {\bibinfo  {journal} {Phys. Rev. Lett.}\ }\textbf {\bibinfo {volume} {116}},\
  \bibinfo {eid} {242502}}\BibitemShut {NoStop}%
\bibitem [{\citenamefont {{Liebend{\"o}rfer}}\ \emph
  {et~al.}(2005)\citenamefont {{Liebend{\"o}rfer}}, \citenamefont {{Rampp}},
  \citenamefont {{Janka}},\ and\ \citenamefont
  {{Mezzacappa}}}]{Liebendoerfer.Rampp.ea:2005}%
  \BibitemOpen
  \bibfield  {author} {\bibinfo {author} {\bibnamefont {{Liebend{\"o}rfer}},
  \bibfnamefont {M}}, \bibinfo {author} {\bibfnamefont {M.}~\bibnamefont
  {{Rampp}}}, \bibinfo {author} {\bibfnamefont {H.-T.}\ \bibnamefont
  {{Janka}}}, and\ \bibinfo {author} {\bibfnamefont {A.}~\bibnamefont
  {{Mezzacappa}}}} (\bibinfo {year} {2005}),\ \bibfield  {title} {\enquote
  {\bibinfo {title} {{Supernova Simulations with Boltzmann Neutrino Transport:
  A Comparison of Methods}},}\ }\href {https://doi.org/10.1086/427203}
  {\bibfield  {journal} {\bibinfo  {journal} {Astrophys. J.}\ }\textbf
  {\bibinfo {volume} {620}},\ \bibinfo {pages} {840--860}}\BibitemShut
  {NoStop}%
\bibitem [{\citenamefont {{Liebend{\"o}rfer}}\ \emph
  {et~al.}(2009)\citenamefont {{Liebend{\"o}rfer}}, \citenamefont
  {{Whitehouse}},\ and\ \citenamefont
  {{Fischer}}}]{Liebendoerfer.Whitehouse.ea:2009}%
  \BibitemOpen
  \bibfield  {author} {\bibinfo {author} {\bibnamefont {{Liebend{\"o}rfer}},
  \bibfnamefont {M}}, \bibinfo {author} {\bibfnamefont {S.~C.}\ \bibnamefont
  {{Whitehouse}}}, and\ \bibinfo {author} {\bibfnamefont {T.}~\bibnamefont
  {{Fischer}}}} (\bibinfo {year} {2009}),\ \bibfield  {title} {\enquote
  {\bibinfo {title} {{The Isotropic Diffusion Source Approximation for
  Supernova Neutrino Transport}},}\ }\href
  {https://doi.org/10.1088/0004-637X/698/2/1174} {\bibfield  {journal}
  {\bibinfo  {journal} {Astrophys. J.}\ }\textbf {\bibinfo {volume} {698}},\
  \bibinfo {pages} {1174--1190}}\BibitemShut {NoStop}%
\bibitem [{\citenamefont {{Limongi}}\ and\ \citenamefont
  {{Chieffi}}(2018)}]{limongi18}%
  \BibitemOpen
  \bibfield  {author} {\bibinfo {author} {\bibnamefont {{Limongi}},
  \bibfnamefont {M}}, and\ \bibinfo {author} {\bibfnamefont {A.}~\bibnamefont
  {{Chieffi}}}} (\bibinfo {year} {2018}),\ \bibfield  {title} {\enquote
  {\bibinfo {title} {{Presupernova Evolution and Explosive Nucleosynthesis of
  Rotating Massive Stars in the Metallicity Range $-3 \le [\text{Fe/H}]\le
  0$}},}\ }\href {https://doi.org/10.3847/1538-4365/aacb24} {\bibfield
  {journal} {\bibinfo  {journal} {Astrophys. J. Suppl.}\ }\textbf {\bibinfo
  {volume} {237}},\ \bibinfo {eid} {13}}\BibitemShut {NoStop}%
\bibitem [{\citenamefont {{Lippuner}}\ \emph {et~al.}(2017)\citenamefont
  {{Lippuner}}, \citenamefont {{Fern{\'a}ndez}}, \citenamefont {{Roberts}},
  \citenamefont {{Foucart}}, \citenamefont {{Kasen}}, \citenamefont
  {{Metzger}},\ and\ \citenamefont
  {{Ott}}}]{Lippuner.Fernandez.Roberts.ea:2017}%
  \BibitemOpen
  \bibfield  {author} {\bibinfo {author} {\bibnamefont {{Lippuner}},
  \bibfnamefont {J}}, \bibinfo {author} {\bibfnamefont {R.}~\bibnamefont
  {{Fern{\'a}ndez}}}, \bibinfo {author} {\bibfnamefont {L.~F.}\ \bibnamefont
  {{Roberts}}}, \bibinfo {author} {\bibfnamefont {F.}~\bibnamefont
  {{Foucart}}}, \bibinfo {author} {\bibfnamefont {D.}~\bibnamefont {{Kasen}}},
  \bibinfo {author} {\bibfnamefont {B.~D.}\ \bibnamefont {{Metzger}}}, and\
  \bibinfo {author} {\bibfnamefont {C.~D.}\ \bibnamefont {{Ott}}}} (\bibinfo
  {year} {2017}),\ \bibfield  {title} {\enquote {\bibinfo {title} {{Signatures
  of hypermassive neutron star lifetimes on r-process nucleosynthesis in the
  disc ejecta from neutron star mergers}},}\ }\href
  {https://doi.org/10.1093/mnras/stx1987} {\bibfield  {journal} {\bibinfo
  {journal} {Mon. Not. Roy. Astron. Soc.}\ }\textbf {\bibinfo {volume} {472}},\
  \bibinfo {pages} {904--918}}\BibitemShut {NoStop}%
\bibitem [{\citenamefont {{Lippuner}}\ and\ \citenamefont
  {{Roberts}}(2017)}]{Lippuner.Roberts:2017}%
  \BibitemOpen
  \bibfield  {author} {\bibinfo {author} {\bibnamefont {{Lippuner}},
  \bibfnamefont {J}}, and\ \bibinfo {author} {\bibfnamefont {L.~F.}\
  \bibnamefont {{Roberts}}}} (\bibinfo {year} {2017}),\ \bibfield  {title}
  {\enquote {\bibinfo {title} {{SkyNet: A Modular Nuclear Reaction Network
  Library}},}\ }\href {https://doi.org/10.3847/1538-4365/aa94cb} {\bibfield
  {journal} {\bibinfo  {journal} {Astrophys. J. Suppl.}\ }\textbf {\bibinfo
  {volume} {233}},\ \bibinfo {eid} {18}}\BibitemShut {NoStop}%
\bibitem [{\citenamefont {{Lippuner}}\ and\ \citenamefont
  {{Roberts}}(2015)}]{Lippuner.Roberts:2015}%
  \BibitemOpen
  \bibfield  {author} {\bibinfo {author} {\bibnamefont {{Lippuner}},
  \bibfnamefont {Jonas}}, and\ \bibinfo {author} {\bibfnamefont {Luke~F.}\
  \bibnamefont {{Roberts}}}} (\bibinfo {year} {2015}),\ \bibfield  {title}
  {\enquote {\bibinfo {title} {{r-process Lanthanide Production and Heating
  Rates in Kilonovae}},}\ }\href {https://doi.org/10.1088/0004-637X/815/2/82}
  {\bibfield  {journal} {\bibinfo  {journal} {Astrophys. J.}\ }\textbf
  {\bibinfo {volume} {815}},\ \bibinfo {eid} {82}}\BibitemShut {NoStop}%
\bibitem [{\citenamefont {{Litvinov}}\ \emph {et~al.}(2004)\citenamefont
  {{Litvinov}} \emph {et~al.}}]{Litvinov2004}%
  \BibitemOpen
  \bibfield  {author} {\bibinfo {author} {\bibnamefont {{Litvinov}},
  \bibfnamefont {Y~A}},  \emph {et~al.}} (\bibinfo {year} {2004}),\ \bibfield
  {title} {\enquote {\bibinfo {title} {{Precision experiments with
  time-resolved Schottky mass spectrometry}},}\ }\href
  {https://doi.org/10.1016/j.nuclphysa.2004.01.089} {\bibfield  {journal}
  {\bibinfo  {journal} {Nucl. Phys. A}\ }\textbf {\bibinfo {volume} {734}},\
  \bibinfo {pages} {473--476}}\BibitemShut {NoStop}%
\bibitem [{\citenamefont {{Litvinova}}\ and\ \citenamefont
  {{Belov}}(2013)}]{Litvinova.Belov:2013}%
  \BibitemOpen
  \bibfield  {author} {\bibinfo {author} {\bibnamefont {{Litvinova}},
  \bibfnamefont {E}}, and\ \bibinfo {author} {\bibfnamefont {N.}~\bibnamefont
  {{Belov}}}} (\bibinfo {year} {2013}),\ \bibfield  {title} {\enquote {\bibinfo
  {title} {{Low-energy limit of the radiative dipole strength in nuclei}},}\
  }\href {https://doi.org/10.1103/PhysRevC.88.031302} {\bibfield  {journal}
  {\bibinfo  {journal} {Phys. Rev. C}\ }\textbf {\bibinfo {volume} {88}},\
  \bibinfo {eid} {031302}}\BibitemShut {NoStop}%
\bibitem [{\citenamefont {Litvinova}\ \emph {et~al.}(2009)\citenamefont
  {Litvinova}, \citenamefont {Loens}, \citenamefont {Langanke}, \citenamefont
  {{Mart{\'i}nez-Pinedo}}, \citenamefont {Rauscher}, \citenamefont {Ring},
  \citenamefont {Thielemann},\ and\ \citenamefont
  {Tselyaev}}]{Litvinova.Loens.ea:2009}%
  \BibitemOpen
  \bibfield  {author} {\bibinfo {author} {\bibnamefont {Litvinova},
  \bibfnamefont {E}}, \bibinfo {author} {\bibfnamefont {{H.P.}}\ \bibnamefont
  {Loens}}, \bibinfo {author} {\bibfnamefont {K.}~\bibnamefont {Langanke}},
  \bibinfo {author} {\bibfnamefont {G.}~\bibnamefont {{Mart{\'i}nez-Pinedo}}},
  \bibinfo {author} {\bibfnamefont {T.}~\bibnamefont {Rauscher}}, \bibinfo
  {author} {\bibfnamefont {P.}~\bibnamefont {Ring}}, \bibinfo {author}
  {\bibfnamefont {{F.-K.}}\ \bibnamefont {Thielemann}}, and\ \bibinfo {author}
  {\bibfnamefont {V.}~\bibnamefont {Tselyaev}}} (\bibinfo {year} {2009}),\
  \bibfield  {title} {\enquote {\bibinfo {title} {Low-lying dipole response in
  the relativistic quasiparticle time blocking approximation and its influence
  on neutron capture cross sections},}\ }\href
  {https://doi.org/10.1016/j.nuclphysa.2009.03.009} {\bibfield  {journal}
  {\bibinfo  {journal} {Nucl. Phys. A}\ }\textbf {\bibinfo {volume} {823}},\
  \bibinfo {pages} {26--37}}\BibitemShut {NoStop}%
\bibitem [{\citenamefont {Liu}\ \emph {et~al.}(2011)\citenamefont {Liu},
  \citenamefont {Wang}, \citenamefont {Deng},\ and\ \citenamefont
  {Wu}}]{Liu.Wang.ea:2011}%
  \BibitemOpen
  \bibfield  {author} {\bibinfo {author} {\bibnamefont {Liu}, \bibfnamefont
  {Min}}, \bibinfo {author} {\bibfnamefont {Ning}\ \bibnamefont {Wang}},
  \bibinfo {author} {\bibfnamefont {Yangge}\ \bibnamefont {Deng}}, and\
  \bibinfo {author} {\bibfnamefont {Xizhen}\ \bibnamefont {Wu}}} (\bibinfo
  {year} {2011}),\ \bibfield  {title} {\enquote {\bibinfo {title} {{Further
  improvements on a global nuclear mass model}},}\ }\href
  {https://doi.org/10.1103/PhysRevC.84.014333} {\bibfield  {journal} {\bibinfo
  {journal} {Phys. Rev. C}\ }\textbf {\bibinfo {volume} {84}},\ \bibinfo
  {pages} {014333}}\BibitemShut {NoStop}%
\bibitem [{\citenamefont {Liu}\ \emph {et~al.}(2008)\citenamefont {Liu},
  \citenamefont {Shapiro}, \citenamefont {Etienne},\ and\ \citenamefont
  {Taniguchi}}]{Liu.Shapiro.ea:2008}%
  \BibitemOpen
  \bibfield  {author} {\bibinfo {author} {\bibnamefont {Liu}, \bibfnamefont
  {Yuk~Tung}}, \bibinfo {author} {\bibfnamefont {Stuart~L.}\ \bibnamefont
  {Shapiro}}, \bibinfo {author} {\bibfnamefont {Zachariah~B.}\ \bibnamefont
  {Etienne}}, and\ \bibinfo {author} {\bibfnamefont {Keisuke}\ \bibnamefont
  {Taniguchi}}} (\bibinfo {year} {2008}),\ \bibfield  {title} {\enquote
  {\bibinfo {title} {General relativistic simulations of magnetized binary
  neutron star mergers},}\ }\href {https://doi.org/10.1103/PhysRevD.78.024012}
  {\bibfield  {journal} {\bibinfo  {journal} {Phys. Rev. D}\ }\textbf {\bibinfo
  {volume} {78}},\ \bibinfo {pages} {024012}}\BibitemShut {NoStop}%
\bibitem [{\citenamefont {{Livio}}\ and\ \citenamefont
  {{Mazzali}}(2018)}]{Livio.Mazzali:2018}%
  \BibitemOpen
  \bibfield  {author} {\bibinfo {author} {\bibnamefont {{Livio}}, \bibfnamefont
  {M}}, and\ \bibinfo {author} {\bibfnamefont {P.}~\bibnamefont {{Mazzali}}}}
  (\bibinfo {year} {2018}),\ \bibfield  {title} {\enquote {\bibinfo {title}
  {{On the progenitors of Type Ia supernovae}},}\ }\href
  {https://doi.org/10.1016/j.physrep.2018.02.002} {\bibfield  {journal}
  {\bibinfo  {journal} {Phys. Rep.}\ }\textbf {\bibinfo {volume} {736}},\
  \bibinfo {pages} {1--23}}\BibitemShut {NoStop}%
\bibitem [{\citenamefont {{Lodders}}\ \emph {et~al.}(2009)\citenamefont
  {{Lodders}}, \citenamefont {{Palme}},\ and\ \citenamefont
  {{Gail}}}]{lodders09}%
  \BibitemOpen
  \bibfield  {author} {\bibinfo {author} {\bibnamefont {{Lodders}},
  \bibfnamefont {K}}, \bibinfo {author} {\bibfnamefont {H.}~\bibnamefont
  {{Palme}}}, and\ \bibinfo {author} {\bibfnamefont {H.-P.}\ \bibnamefont
  {{Gail}}}} (\bibinfo {year} {2009}),\ \bibfield  {title} {\enquote {\bibinfo
  {title} {{Abundances of the Elements in the Solar System}},}\ }\href
  {https://doi.org/10.1007/978-3-540-88055-4_34} {\bibfield  {journal}
  {\bibinfo  {journal} {Landolt B\"ornstein}\ }\textbf {\bibinfo {volume}
  {4B}},\ \bibinfo {pages} {712}}\BibitemShut {NoStop}%
\bibitem [{\citenamefont {{Loens}}\ \emph {et~al.}(2012)\citenamefont
  {{Loens}}, \citenamefont {{Langanke}}, \citenamefont
  {{Mart{\'{\i}}nez-Pinedo}},\ and\ \citenamefont
  {{Sieja}}}]{Loens.Langanke.ea:2012}%
  \BibitemOpen
  \bibfield  {author} {\bibinfo {author} {\bibnamefont {{Loens}}, \bibfnamefont
  {H~P}}, \bibinfo {author} {\bibfnamefont {K.}~\bibnamefont {{Langanke}}},
  \bibinfo {author} {\bibfnamefont {G.}~\bibnamefont
  {{Mart{\'{\i}}nez-Pinedo}}}, and\ \bibinfo {author} {\bibfnamefont
  {K.}~\bibnamefont {{Sieja}}}} (\bibinfo {year} {2012}),\ \bibfield  {title}
  {\enquote {\bibinfo {title} {{M1 strength functions from large-scale
  shell-model calculations and their effect on astrophysical neutron capture
  cross-sections}},}\ }\href {https://doi.org/10.1140/epja/i2012-12034-5}
  {\bibfield  {journal} {\bibinfo  {journal} {Eur. Phys. J. A}\ }\textbf
  {\bibinfo {volume} {48}},\ \bibinfo {eid} {34}}\BibitemShut {NoStop}%
\bibitem [{\citenamefont {Loens}\ \emph {et~al.}(2008)\citenamefont {Loens},
  \citenamefont {Langanke}, \citenamefont {Mart{\'i}nez-Pinedo}, \citenamefont
  {Rauscher},\ and\ \citenamefont {Thielemann}}]{Loens.Langanke.ea:2008}%
  \BibitemOpen
  \bibfield  {author} {\bibinfo {author} {\bibnamefont {Loens}, \bibfnamefont
  {HP}}, \bibinfo {author} {\bibfnamefont {K.}~\bibnamefont {Langanke}},
  \bibinfo {author} {\bibfnamefont {G.}~\bibnamefont {Mart{\'i}nez-Pinedo}},
  \bibinfo {author} {\bibfnamefont {T.}~\bibnamefont {Rauscher}}, and\ \bibinfo
  {author} {\bibfnamefont {F.-K.}\ \bibnamefont {Thielemann}}} (\bibinfo {year}
  {2008}),\ \bibfield  {title} {\enquote {\bibinfo {title} {Complete inclusion
  of parity-dependent level densities in the statistical description of
  astrophysical reaction rates},}\ }\href
  {https://doi.org/10.1016/j.physletb.2008.07.073} {\bibfield  {journal}
  {\bibinfo  {journal} {Phys. Lett. B}\ }\textbf {\bibinfo {volume} {666}},\
  \bibinfo {pages} {395--399}}\BibitemShut {NoStop}%
\bibitem [{\citenamefont {Lorusso}\ \emph {et~al.}(2015)\citenamefont {Lorusso}
  \emph {et~al.}}]{Lorusso.Nishimura.ea:2015}%
  \BibitemOpen
  \bibfield  {author} {\bibinfo {author} {\bibnamefont {Lorusso}, \bibfnamefont
  {G}},  \emph {et~al.}} (\bibinfo {year} {2015}),\ \bibfield  {title}
  {\enquote {\bibinfo {title} {{$\beta$-Decay Half-Lives of 110 Neutron-Rich
  Nuclei across the N=82 Shell Gap: Implications for the Mechanism and
  Universality of the Astrophysical r-Process}},}\ }\href
  {https://doi.org/10.1103/PhysRevLett.114.192501} {\bibfield  {journal}
  {\bibinfo  {journal} {Phys. Rev. Lett.}\ }\textbf {\bibinfo {volume} {114}},\
  \bibinfo {pages} {192501}}\BibitemShut {NoStop}%
\bibitem [{\citenamefont {{Ludwig}}\ \emph {et~al.}(2010)\citenamefont
  {{Ludwig}}, \citenamefont {{Caffau}}, \citenamefont {{Steffen}},
  \citenamefont {{Bonifacio}},\ and\ \citenamefont
  {{Sbordone}}}]{Ludwig.Caffau.ea:2010}%
  \BibitemOpen
  \bibfield  {author} {\bibinfo {author} {\bibnamefont {{Ludwig}},
  \bibfnamefont {H~G}}, \bibinfo {author} {\bibfnamefont {E.}~\bibnamefont
  {{Caffau}}}, \bibinfo {author} {\bibfnamefont {M.}~\bibnamefont {{Steffen}}},
  \bibinfo {author} {\bibfnamefont {P.}~\bibnamefont {{Bonifacio}}}, and\
  \bibinfo {author} {\bibfnamefont {L.}~\bibnamefont {{Sbordone}}}} (\bibinfo
  {year} {2010}),\ \bibfield  {title} {\enquote {\bibinfo {title} {{Accuracy of
  spectroscopy-based radioactive dating of stars}},}\ }\href
  {https://doi.org/10.1051/0004-6361/200810780} {\bibfield  {journal} {\bibinfo
   {journal} {Astron. \& Astrophys.}\ }\textbf {\bibinfo {volume} {509}},\
  \bibinfo {eid} {A84}}\BibitemShut {NoStop}%
\bibitem [{\citenamefont {{Ludwig}}\ \emph {et~al.}(2016)\citenamefont
  {{Ludwig}} \emph {et~al.}}]{ludwig16}%
  \BibitemOpen
  \bibfield  {author} {\bibinfo {author} {\bibnamefont {{Ludwig}},
  \bibfnamefont {P}},  \emph {et~al.}} (\bibinfo {year} {2016}),\ \bibfield
  {title} {\enquote {\bibinfo {title} {{Time-resolved 2-million-year-old
  supernova activity discovered in Earth's microfossil record}},}\ }\href
  {https://doi.org/10.1073/pnas.1601040113} {\bibfield  {journal} {\bibinfo
  {journal} {Proceedings of the National Academy of Science}\ }\textbf
  {\bibinfo {volume} {113}},\ \bibinfo {pages} {9232--9237}}\BibitemShut
  {NoStop}%
\bibitem [{\citenamefont {{Lugaro}}\ \emph {et~al.}(2018)\citenamefont
  {{Lugaro}}, \citenamefont {{Ott}},\ and\ \citenamefont
  {{Kereszturi}}}]{Lugaro.Ott.ea:2018}%
  \BibitemOpen
  \bibfield  {author} {\bibinfo {author} {\bibnamefont {{Lugaro}},
  \bibfnamefont {M}}, \bibinfo {author} {\bibfnamefont {U.}~\bibnamefont
  {{Ott}}}, and\ \bibinfo {author} {\bibfnamefont {{\'A}.}~\bibnamefont
  {{Kereszturi}}}} (\bibinfo {year} {2018}),\ \bibfield  {title} {\enquote
  {\bibinfo {title} {{Radioactive nuclei from cosmochronology to
  habitability}},}\ }\href {https://doi.org/10.1016/j.ppnp.2018.05.002}
  {\bibfield  {journal} {\bibinfo  {journal} {Prog. Part. Nucl. Phys.}\
  }\textbf {\bibinfo {volume} {102}},\ \bibinfo {pages} {1--47}}\BibitemShut
  {NoStop}%
\bibitem [{\citenamefont {{MacFadyen}}\ and\ \citenamefont
  {{Woosley}}(1999)}]{macfadyen99}%
  \BibitemOpen
  \bibfield  {author} {\bibinfo {author} {\bibnamefont {{MacFadyen}},
  \bibfnamefont {A~I}}, and\ \bibinfo {author} {\bibfnamefont {S.~E.}\
  \bibnamefont {{Woosley}}}} (\bibinfo {year} {1999}),\ \bibfield  {title}
  {\enquote {\bibinfo {title} {{Collapsars: Gamma-Ray Bursts and Explosions in
  ``Failed Supernovae''}},}\ }\href {https://doi.org/10.1086/307790} {\bibfield
   {journal} {\bibinfo  {journal} {Astrophys. J.}\ }\textbf {\bibinfo {volume}
  {524}},\ \bibinfo {pages} {262--289}}\BibitemShut {NoStop}%
\bibitem [{\citenamefont {{MacFadyen}}\ \emph {et~al.}(2001)\citenamefont
  {{MacFadyen}}, \citenamefont {{Woosley}},\ and\ \citenamefont
  {{Heger}}}]{macfadyen01}%
  \BibitemOpen
  \bibfield  {author} {\bibinfo {author} {\bibnamefont {{MacFadyen}},
  \bibfnamefont {A~I}}, \bibinfo {author} {\bibfnamefont {S.~E.}\ \bibnamefont
  {{Woosley}}}, and\ \bibinfo {author} {\bibfnamefont {A.}~\bibnamefont
  {{Heger}}}} (\bibinfo {year} {2001}),\ \bibfield  {title} {\enquote {\bibinfo
  {title} {{Supernovae, Jets, and Collapsars}},}\ }\href
  {https://doi.org/10.1086/319698} {\bibfield  {journal} {\bibinfo  {journal}
  {Astrophys. J.}\ }\textbf {\bibinfo {volume} {550}},\ \bibinfo {pages}
  {410--425}}\BibitemShut {NoStop}%
\bibitem [{\citenamefont {{Macias}}\ and\ \citenamefont
  {{Ramirez-Ruiz}}(2019)}]{Macias.Ramirez:2019}%
  \BibitemOpen
  \bibfield  {author} {\bibinfo {author} {\bibnamefont {{Macias}},
  \bibfnamefont {Phillip}}, and\ \bibinfo {author} {\bibfnamefont {Enrico}\
  \bibnamefont {{Ramirez-Ruiz}}}} (\bibinfo {year} {2019}),\ \bibfield  {title}
  {\enquote {\bibinfo {title} {{Constraining Collapsar r-process Models through
  Stellar Abundances}},}\ }\href {https://doi.org/10.3847/2041-8213/ab2049}
  {\bibfield  {journal} {\bibinfo  {journal} {Astrophys. J. Lett.}\ }\textbf
  {\bibinfo {volume} {877}},\ \bibinfo {eid} {L24}}\BibitemShut {NoStop}%
\bibitem [{\citenamefont {{Malkus}}\ \emph {et~al.}(2012)\citenamefont
  {{Malkus}}, \citenamefont {{Kneller}}, \citenamefont {{McLaughlin}},\ and\
  \citenamefont {{Surman}}}]{malkus12}%
  \BibitemOpen
  \bibfield  {author} {\bibinfo {author} {\bibnamefont {{Malkus}},
  \bibfnamefont {A}}, \bibinfo {author} {\bibfnamefont {J.~P.}\ \bibnamefont
  {{Kneller}}}, \bibinfo {author} {\bibfnamefont {G.~C.}\ \bibnamefont
  {{McLaughlin}}}, and\ \bibinfo {author} {\bibfnamefont {R.}~\bibnamefont
  {{Surman}}}} (\bibinfo {year} {2012}),\ \bibfield  {title} {\enquote
  {\bibinfo {title} {{Neutrino oscillations above black hole accretion disks:
  Disks with electron-flavor emission}},}\ }\href
  {https://doi.org/10.1103/PhysRevD.86.085015} {\bibfield  {journal} {\bibinfo
  {journal} {Phys. Rev. D}\ }\textbf {\bibinfo {volume} {86}},\ \bibinfo {eid}
  {085015}}\BibitemShut {NoStop}%
\bibitem [{\citenamefont {{Malkus}}\ \emph {et~al.}(2016)\citenamefont
  {{Malkus}}, \citenamefont {{McLaughlin}},\ and\ \citenamefont
  {{Surman}}}]{malkus16}%
  \BibitemOpen
  \bibfield  {author} {\bibinfo {author} {\bibnamefont {{Malkus}},
  \bibfnamefont {A}}, \bibinfo {author} {\bibfnamefont {G.~C.}\ \bibnamefont
  {{McLaughlin}}}, and\ \bibinfo {author} {\bibfnamefont {R.}~\bibnamefont
  {{Surman}}}} (\bibinfo {year} {2016}),\ \bibfield  {title} {\enquote
  {\bibinfo {title} {{Symmetric and standard matter neutrino resonances above
  merging compact objects}},}\ }\href
  {https://doi.org/10.1103/PhysRevD.93.045021} {\bibfield  {journal} {\bibinfo
  {journal} {Phys. Rev. D}\ }\textbf {\bibinfo {volume} {93}},\ \bibinfo {eid}
  {045021}}\BibitemShut {NoStop}%
\bibitem [{\citenamefont {Mamdouh}\ \emph {et~al.}(2001)\citenamefont
  {Mamdouh}, \citenamefont {Pearson}, \citenamefont {Rayet},\ and\
  \citenamefont {Tondeur}}]{Mamdouh.Pearson.ea:2001}%
  \BibitemOpen
  \bibfield  {author} {\bibinfo {author} {\bibnamefont {Mamdouh}, \bibfnamefont
  {A}}, \bibinfo {author} {\bibfnamefont {J.M.}\ \bibnamefont {Pearson}},
  \bibinfo {author} {\bibfnamefont {M.}~\bibnamefont {Rayet}}, and\ \bibinfo
  {author} {\bibfnamefont {F.}~\bibnamefont {Tondeur}}} (\bibinfo {year}
  {2001}),\ \bibfield  {title} {\enquote {\bibinfo {title} {Fission barriers of
  neutron-rich and superheavy nuclei calculated with the etfsi method},}\
  }\href@noop {} {\bibfield  {journal} {\bibinfo  {journal} {Nucl. Phys. A}\
  }\textbf {\bibinfo {volume} {679}},\ \bibinfo {pages} {337--358}}\BibitemShut
  {NoStop}%
\bibitem [{\citenamefont {{Manning}}\ \emph {et~al.}(2019)\citenamefont
  {{Manning}} \emph {et~al.}}]{Manning.Arbanas.ea:2019}%
  \BibitemOpen
  \bibfield  {author} {\bibinfo {author} {\bibnamefont {{Manning}},
  \bibfnamefont {B}},  \emph {et~al.}} (\bibinfo {year} {2019}),\ \bibfield
  {title} {\enquote {\bibinfo {title} {{Informing direct neutron capture on tin
  isotopes near the N =82 shell closure}},}\ }\href
  {https://doi.org/10.1103/PhysRevC.99.041302} {\bibfield  {journal} {\bibinfo
  {journal} {Phys. Rev. C}\ }\textbf {\bibinfo {volume} {99}},\ \bibinfo {eid}
  {041302}}\BibitemShut {NoStop}%
\bibitem [{\citenamefont {{Maoz}}\ \emph {et~al.}(2014)\citenamefont {{Maoz}},
  \citenamefont {{Mannucci}},\ and\ \citenamefont {{Nelemans}}}]{maoz14}%
  \BibitemOpen
  \bibfield  {author} {\bibinfo {author} {\bibnamefont {{Maoz}}, \bibfnamefont
  {D}}, \bibinfo {author} {\bibfnamefont {F.}~\bibnamefont {{Mannucci}}}, and\
  \bibinfo {author} {\bibfnamefont {G.}~\bibnamefont {{Nelemans}}}} (\bibinfo
  {year} {2014}),\ \bibfield  {title} {\enquote {\bibinfo {title}
  {{Observational Clues to the Progenitors of Type Ia Supernovae}},}\ }\href
  {https://doi.org/10.1146/annurev-astro-082812-141031} {\bibfield  {journal}
  {\bibinfo  {journal} {Annu. Rev. Astron. Astrophys.}\ }\textbf {\bibinfo
  {volume} {52}},\ \bibinfo {pages} {107--170}}\BibitemShut {NoStop}%
\bibitem [{\citenamefont {{Marek}}\ \emph {et~al.}(2006)\citenamefont
  {{Marek}}, \citenamefont {{Dimmelmeier}}, \citenamefont {{Janka}},
  \citenamefont {{M{\"u}ller}},\ and\ \citenamefont {{Buras}}}]{Marek.ea:2006}%
  \BibitemOpen
  \bibfield  {author} {\bibinfo {author} {\bibnamefont {{Marek}}, \bibfnamefont
  {A}}, \bibinfo {author} {\bibfnamefont {H.}~\bibnamefont {{Dimmelmeier}}},
  \bibinfo {author} {\bibfnamefont {H.~Th.}\ \bibnamefont {{Janka}}}, \bibinfo
  {author} {\bibfnamefont {E.}~\bibnamefont {{M{\"u}ller}}}, and\ \bibinfo
  {author} {\bibfnamefont {R.}~\bibnamefont {{Buras}}}} (\bibinfo {year}
  {2006}),\ \bibfield  {title} {\enquote {\bibinfo {title} {{Exploring the
  relativistic regime with Newtonian hydrodynamics: an improved effective
  gravitational potential for supernova simulations}},}\ }\href
  {https://doi.org/10.1051/0004-6361:20052840} {\bibfield  {journal} {\bibinfo
  {journal} {Astron. \& Astrophys.}\ }\textbf {\bibinfo {volume} {445}},\
  \bibinfo {pages} {273--289}}\BibitemShut {NoStop}%
\bibitem [{\citenamefont {Margalit}\ and\ \citenamefont
  {Metzger}(2017)}]{Margalit.Metzger:2017}%
  \BibitemOpen
  \bibfield  {author} {\bibinfo {author} {\bibnamefont {Margalit},
  \bibfnamefont {Ben}}, and\ \bibinfo {author} {\bibfnamefont {Brian}\
  \bibnamefont {Metzger}}} (\bibinfo {year} {2017}),\ \bibfield  {title}
  {\enquote {\bibinfo {title} {{Constraining the Maximum Mass of Neutron Stars
  From Multi-Messenger Observations of GW170817}},}\ }\href
  {https://doi.org/10.3847/2041-8213/aa991c} {\bibfield  {journal} {\bibinfo
  {journal} {Astrophys. J. Lett.}\ }\textbf {\bibinfo {volume} {850}},\
  \bibinfo {pages} {L19}}\BibitemShut {NoStop}%
\bibitem [{\citenamefont {{Margalit}}\ and\ \citenamefont
  {{Metzger}}(2019)}]{Margalit.Metzger:2019}%
  \BibitemOpen
  \bibfield  {author} {\bibinfo {author} {\bibnamefont {{Margalit}},
  \bibfnamefont {Ben}}, and\ \bibinfo {author} {\bibfnamefont {Brian~D.}\
  \bibnamefont {{Metzger}}}} (\bibinfo {year} {2019}),\ \bibfield  {title}
  {\enquote {\bibinfo {title} {{The Multi-messenger Matrix: The Future of
  Neutron Star Merger Constraints on the Nuclear Equation of State}},}\ }\href
  {https://doi.org/10.3847/2041-8213/ab2ae2} {\bibfield  {journal} {\bibinfo
  {journal} {Astrophys. J. Lett.}\ }\textbf {\bibinfo {volume} {880}},\
  \bibinfo {eid} {L15}}\BibitemShut {NoStop}%
\bibitem [{\citenamefont {{Marketin}}\ \emph
  {et~al.}(2016{\natexlab{a}})\citenamefont {{Marketin}}, \citenamefont
  {{Huther}},\ and\ \citenamefont
  {{Mart{\'{\i}}nez-Pinedo}}}]{Marketin.Huther.Martinez-Pinedo:2016}%
  \BibitemOpen
  \bibfield  {author} {\bibinfo {author} {\bibnamefont {{Marketin}},
  \bibfnamefont {T}}, \bibinfo {author} {\bibfnamefont {L.}~\bibnamefont
  {{Huther}}}, and\ \bibinfo {author} {\bibfnamefont {G.}~\bibnamefont
  {{Mart{\'{\i}}nez-Pinedo}}}} (\bibinfo {year} {2016}{\natexlab{a}}),\
  \bibfield  {title} {\enquote {\bibinfo {title} {{Large-scale evaluation of
  {$\beta$} -decay rates of r -process nuclei with the inclusion of
  first-forbidden transitions}},}\ }\href
  {https://doi.org/10.1103/PhysRevC.93.025805} {\bibfield  {journal} {\bibinfo
  {journal} {Phys. Rev. C}\ }\textbf {\bibinfo {volume} {93}},\ \bibinfo {eid}
  {025805}}\BibitemShut {NoStop}%
\bibitem [{\citenamefont {{Marketin}}\ \emph
  {et~al.}(2016{\natexlab{b}})\citenamefont {{Marketin}}, \citenamefont
  {{Huther}}, \citenamefont {{Petkovi{\'c}}}, \citenamefont {{Paar}},\ and\
  \citenamefont {{Mart{\'{\i}}nez-Pinedo}}}]{Marketin.Huther.ea:2016}%
  \BibitemOpen
  \bibfield  {author} {\bibinfo {author} {\bibnamefont {{Marketin}},
  \bibfnamefont {T}}, \bibinfo {author} {\bibfnamefont {L.}~\bibnamefont
  {{Huther}}}, \bibinfo {author} {\bibfnamefont {J.}~\bibnamefont
  {{Petkovi{\'c}}}}, \bibinfo {author} {\bibfnamefont {N.}~\bibnamefont
  {{Paar}}}, and\ \bibinfo {author} {\bibfnamefont {G.}~\bibnamefont
  {{Mart{\'{\i}}nez-Pinedo}}}} (\bibinfo {year} {2016}{\natexlab{b}}),\
  \bibfield  {title} {\enquote {\bibinfo {title} {{Beta decay rates of
  neutron-rich nuclei}},}\ }in\ \href {https://doi.org/10.1063/1.4953298}
  {\emph {\bibinfo {booktitle} {American Institute of Physics Conference
  Series}}},\ Vol.\ \bibinfo {volume} {1743},\ p.\ \bibinfo {pages}
  {040006}\BibitemShut {NoStop}%
\bibitem [{\citenamefont {Marketin}\ \emph {et~al.}(2007)\citenamefont
  {Marketin}, \citenamefont {Vretenar},\ and\ \citenamefont
  {Ring}}]{Marketin.Vretenar.Ring:2007}%
  \BibitemOpen
  \bibfield  {author} {\bibinfo {author} {\bibnamefont {Marketin},
  \bibfnamefont {T}}, \bibinfo {author} {\bibfnamefont {D.}~\bibnamefont
  {Vretenar}}, and\ \bibinfo {author} {\bibfnamefont {P.}~\bibnamefont {Ring}}}
  (\bibinfo {year} {2007}),\ \bibfield  {title} {\enquote {\bibinfo {title}
  {Calculation of beta-decay rates in a relativistic model with
  momentum-dependent self-energies},}\ }\href
  {https://doi.org/10.1103/PhysRevC.75.024304} {\bibfield  {journal} {\bibinfo
  {journal} {Phys. Rev. C}\ }\textbf {\bibinfo {volume} {75}},\ \bibinfo
  {pages} {024304}}\BibitemShut {NoStop}%
\bibitem [{\citenamefont {{Marshall}}\ \emph {et~al.}(2019)\citenamefont
  {{Marshall}} \emph {et~al.}}]{Marshall.ea:2019}%
  \BibitemOpen
  \bibfield  {author} {\bibinfo {author} {\bibnamefont {{Marshall}},
  \bibfnamefont {J~L}},  \emph {et~al.} (\bibinfo {collaboration} {DES
  Collaboration})} (\bibinfo {year} {2019}),\ \bibfield  {title} {\enquote
  {\bibinfo {title} {{Chemical Abundance Analysis of Tucana III, the Second
  r-process Enhanced Ultra-faint Dwarf Galaxy}},}\ }\href
  {https://doi.org/10.3847/1538-4357/ab3653} {\bibfield  {journal} {\bibinfo
  {journal} {Astrophys. J.}\ }\textbf {\bibinfo {volume} {882}},\ \bibinfo
  {eid} {177}}\BibitemShut {NoStop}%
\bibitem [{\citenamefont {Martin}\ \emph {et~al.}(2016)\citenamefont {Martin},
  \citenamefont {Arcones}, \citenamefont {Nazarewicz},\ and\ \citenamefont
  {Olsen}}]{Martin.Arcones.ea:2016}%
  \BibitemOpen
  \bibfield  {author} {\bibinfo {author} {\bibnamefont {Martin}, \bibfnamefont
  {D}}, \bibinfo {author} {\bibfnamefont {A.}~\bibnamefont {Arcones}}, \bibinfo
  {author} {\bibfnamefont {W.}~\bibnamefont {Nazarewicz}}, and\ \bibinfo
  {author} {\bibfnamefont {E.}~\bibnamefont {Olsen}}} (\bibinfo {year}
  {2016}),\ \bibfield  {title} {\enquote {\bibinfo {title} {Impact of nuclear
  mass uncertainties on the $r$ process},}\ }\href
  {https://doi.org/10.1103/PhysRevLett.116.121101} {\bibfield  {journal}
  {\bibinfo  {journal} {Phys. Rev. Lett.}\ }\textbf {\bibinfo {volume} {116}},\
  \bibinfo {pages} {121101}}\BibitemShut {NoStop}%
\bibitem [{\citenamefont {{Martin}}\ \emph {et~al.}(2015)\citenamefont
  {{Martin}}, \citenamefont {{Perego}}, \citenamefont {{Arcones}},
  \citenamefont {{Thielemann}}, \citenamefont {{Korobkin}},\ and\ \citenamefont
  {{Rosswog}}}]{martin15}%
  \BibitemOpen
  \bibfield  {author} {\bibinfo {author} {\bibnamefont {{Martin}},
  \bibfnamefont {D}}, \bibinfo {author} {\bibfnamefont {A.}~\bibnamefont
  {{Perego}}}, \bibinfo {author} {\bibfnamefont {A.}~\bibnamefont {{Arcones}}},
  \bibinfo {author} {\bibfnamefont {F.-K.}\ \bibnamefont {{Thielemann}}},
  \bibinfo {author} {\bibfnamefont {O.}~\bibnamefont {{Korobkin}}}, and\
  \bibinfo {author} {\bibfnamefont {S.}~\bibnamefont {{Rosswog}}}} (\bibinfo
  {year} {2015}),\ \bibfield  {title} {\enquote {\bibinfo {title}
  {{Neutrino-driven Winds in the Aftermath of a Neutron Star Merger:
  Nucleosynthesis and Electromagnetic Transients}},}\ }\href
  {https://doi.org/10.1088/0004-637X/813/1/2} {\bibfield  {journal} {\bibinfo
  {journal} {Astrophys. J.}\ }\textbf {\bibinfo {volume} {813}},\ \bibinfo
  {eid} {2}}\BibitemShut {NoStop}%
\bibitem [{\citenamefont {{Martin}}\ \emph {et~al.}(2018)\citenamefont
  {{Martin}}, \citenamefont {{Perego}}, \citenamefont {{Kastaun}},\ and\
  \citenamefont {{Arcones}}}]{Martin.Perego.ea:2018}%
  \BibitemOpen
  \bibfield  {author} {\bibinfo {author} {\bibnamefont {{Martin}},
  \bibfnamefont {D}}, \bibinfo {author} {\bibfnamefont {A.}~\bibnamefont
  {{Perego}}}, \bibinfo {author} {\bibfnamefont {W.}~\bibnamefont {{Kastaun}}},
  and\ \bibinfo {author} {\bibfnamefont {A.}~\bibnamefont {{Arcones}}}}
  (\bibinfo {year} {2018}),\ \bibfield  {title} {\enquote {\bibinfo {title}
  {{The role of weak interactions in dynamic ejecta from binary neutron star
  mergers}},}\ }\href {https://doi.org/10.1088/1361-6382/aa9f5a} {\bibfield
  {journal} {\bibinfo  {journal} {Class. Quantum Gravity}\ }\textbf {\bibinfo
  {volume} {35}},\ \bibinfo {eid} {034001}}\BibitemShut {NoStop}%
\bibitem [{\citenamefont {Martin}\ \emph {et~al.}(1971)\citenamefont {Martin},
  \citenamefont {{Zalubas}},\ and\ \citenamefont {{Hagan}}}]{martin71}%
  \BibitemOpen
  \bibfield  {author} {\bibinfo {author} {\bibnamefont {Martin}, \bibfnamefont
  {W~C}}, \bibinfo {author} {\bibfnamefont {R.}~\bibnamefont {{Zalubas}}}, and\
  \bibinfo {author} {\bibfnamefont {L.}~\bibnamefont {{Hagan}}}} (\bibinfo
  {year} {1971}),\ \href@noop {} {\emph {\bibinfo {title} {{Atomic Energy
  Levels - The Rare Earth Elements, Nat. Ref. Data Ser. 60}}}}\ (\bibinfo
  {publisher} {National Bureau of Standards},\ \bibinfo {address} {Washington:
  US GPO})\BibitemShut {NoStop}%
\bibitem [{\citenamefont {Mart{\'{i}}nez-Pinedo}\ \emph
  {et~al.}(2014)\citenamefont {Mart{\'{i}}nez-Pinedo}, \citenamefont
  {Fischer},\ and\ \citenamefont
  {Huther}}]{Martinez-Pinedo.Fischer.Huther:2014}%
  \BibitemOpen
  \bibfield  {author} {\bibinfo {author} {\bibnamefont {Mart{\'{i}}nez-Pinedo},
  \bibfnamefont {G}}, \bibinfo {author} {\bibfnamefont {T}~\bibnamefont
  {Fischer}}, and\ \bibinfo {author} {\bibfnamefont {L}~\bibnamefont {Huther}}}
  (\bibinfo {year} {2014}),\ \bibfield  {title} {\enquote {\bibinfo {title}
  {{Supernova neutrinos and nucleosynthesis}},}\ }\href
  {https://doi.org/10.1088/0954-3899/41/4/044008} {\bibfield  {journal}
  {\bibinfo  {journal} {J. Phys. G: Nucl. Part. Phys.}\ }\textbf {\bibinfo
  {volume} {41}},\ \bibinfo {pages} {044008}}\BibitemShut {NoStop}%
\bibitem [{\citenamefont {Mart{\'\i}nez-Pinedo}\ \emph
  {et~al.}(2012)\citenamefont {Mart{\'\i}nez-Pinedo}, \citenamefont {Fischer},
  \citenamefont {Lohs},\ and\ \citenamefont
  {Huther}}]{Martinez-Pinedo.Fischer.ea:2012}%
  \BibitemOpen
  \bibfield  {author} {\bibinfo {author} {\bibnamefont {Mart{\'\i}nez-Pinedo},
  \bibfnamefont {G}}, \bibinfo {author} {\bibfnamefont {T.}~\bibnamefont
  {Fischer}}, \bibinfo {author} {\bibfnamefont {A.}~\bibnamefont {Lohs}}, and\
  \bibinfo {author} {\bibfnamefont {L.}~\bibnamefont {Huther}}} (\bibinfo
  {year} {2012}),\ \bibfield  {title} {\enquote {\bibinfo {title}
  {Charged-current weak interaction processes in hot and dense matter and its
  impact on the spectra of neutrinos emitted from protoneutron star cooling},}\
  }\href {https://doi.org/10.1103/PhysRevLett.109.251104} {\bibfield  {journal}
  {\bibinfo  {journal} {Phys. Rev. Lett.}\ }\textbf {\bibinfo {volume} {109}},\
  \bibinfo {pages} {251104}}\BibitemShut {NoStop}%
\bibitem [{\citenamefont {Mart{\'i}nez-Pinedo}\ \emph
  {et~al.}(2014)\citenamefont {Mart{\'i}nez-Pinedo}, \citenamefont {Lam},
  \citenamefont {Langanke}, \citenamefont {Zegers},\ and\ \citenamefont
  {Sullivan}}]{Martinez-Pinedo.Lam.ea:2014}%
  \BibitemOpen
  \bibfield  {author} {\bibinfo {author} {\bibnamefont {Mart{\'i}nez-Pinedo},
  \bibfnamefont {G}}, \bibinfo {author} {\bibfnamefont {Y.~H.}\ \bibnamefont
  {Lam}}, \bibinfo {author} {\bibfnamefont {K.}~\bibnamefont {Langanke}},
  \bibinfo {author} {\bibfnamefont {R.~G.~T.}\ \bibnamefont {Zegers}}, and\
  \bibinfo {author} {\bibfnamefont {C.}~\bibnamefont {Sullivan}}} (\bibinfo
  {year} {2014}),\ \bibfield  {title} {\enquote {\bibinfo {title}
  {{Astrophysical weak-interaction rates for selected $A=20$ and $A=24$
  nuclei}},}\ }\href {https://doi.org/10.1103/PhysRevC.89.045806} {\bibfield
  {journal} {\bibinfo  {journal} {Phys. Rev. C}\ }\textbf {\bibinfo {volume}
  {89}},\ \bibinfo {pages} {045806}}\BibitemShut {NoStop}%
\bibitem [{\citenamefont {Mart{\'\i}nez-Pinedo}\ \emph
  {et~al.}(2007)\citenamefont {Mart{\'\i}nez-Pinedo}, \citenamefont {Mocelj},
  \citenamefont {Zinner}, \citenamefont {Keli{\'c}}, \citenamefont {Langankea},
  \citenamefont {Panov}, \citenamefont {Pfeiffer}, \citenamefont {Rauscher},
  \citenamefont {Schmidt},\ and\ \citenamefont
  {Thielemann}}]{Martinez-Pinedo.Mocelj.ea:2007}%
  \BibitemOpen
  \bibfield  {author} {\bibinfo {author} {\bibnamefont {Mart{\'\i}nez-Pinedo},
  \bibfnamefont {G}}, \bibinfo {author} {\bibfnamefont {D.}~\bibnamefont
  {Mocelj}}, \bibinfo {author} {\bibfnamefont {N.T.}\ \bibnamefont {Zinner}},
  \bibinfo {author} {\bibfnamefont {A.}~\bibnamefont {Keli{\'c}}}, \bibinfo
  {author} {\bibfnamefont {K.}~\bibnamefont {Langankea}}, \bibinfo {author}
  {\bibfnamefont {I.}~\bibnamefont {Panov}}, \bibinfo {author} {\bibfnamefont
  {B.}~\bibnamefont {Pfeiffer}}, \bibinfo {author} {\bibfnamefont
  {T.}~\bibnamefont {Rauscher}}, \bibinfo {author} {\bibfnamefont {K.-H.}\
  \bibnamefont {Schmidt}}, and\ \bibinfo {author} {\bibfnamefont {F.-K.}\
  \bibnamefont {Thielemann}}} (\bibinfo {year} {2007}),\ \bibfield  {title}
  {\enquote {\bibinfo {title} {{The role of fission in the r-process}},}\
  }\href {https://doi.org/10.1016/j.ppnp.2007.01.018} {\bibfield  {journal}
  {\bibinfo  {journal} {Prog. Part. Nucl. Phys.}\ }\textbf {\bibinfo {volume}
  {59}},\ \bibinfo {pages} {199--205}}\BibitemShut {NoStop}%
\bibitem [{\citenamefont {Mart{\'{i}}nez-Pinedo}\ \emph
  {et~al.}(2017)\citenamefont {Mart{\'{i}}nez-Pinedo}, \citenamefont {Fischer},
  \citenamefont {Langanke}, \citenamefont {Lohs}, \citenamefont {Sieverding},\
  and\ \citenamefont {Wu}}]{Martinez-Pinedo.Fischer.ea:2016}%
  \BibitemOpen
  \bibfield  {author} {\bibinfo {author} {\bibnamefont {Mart{\'{i}}nez-Pinedo},
  \bibfnamefont {Gabriel}}, \bibinfo {author} {\bibfnamefont {Tobias}\
  \bibnamefont {Fischer}}, \bibinfo {author} {\bibfnamefont {Karlheinz}\
  \bibnamefont {Langanke}}, \bibinfo {author} {\bibfnamefont {Andreas}\
  \bibnamefont {Lohs}}, \bibinfo {author} {\bibfnamefont {Andre}\ \bibnamefont
  {Sieverding}}, and\ \bibinfo {author} {\bibfnamefont {Meng-Ru}\ \bibnamefont
  {Wu}}} (\bibinfo {year} {2017}),\ \enquote {\bibinfo {title} {{Neutrinos and
  Their Impact on Core-Collapse Supernova Nucleosynthesis}},}\ in\ \href
  {https://doi.org/10.1007/978-3-319-20794-0_78-1} {\emph {\bibinfo {booktitle}
  {Handbook of Supernovae}}},\ \bibinfo {editor} {edited by\ \bibinfo {editor}
  {\bibfnamefont {A.~W.}\ \bibnamefont {{Alsabti}}}\ and\ \bibinfo {editor}
  {\bibfnamefont {P.}~\bibnamefont {{Murdin}}}}\ (\bibinfo  {publisher}
  {Springer International Publishing},\ \bibinfo {address} {Cham})\BibitemShut
  {NoStop}%
\bibitem [{\citenamefont {{Mashonkina}}\ \emph {et~al.}(2014)\citenamefont
  {{Mashonkina}}, \citenamefont {{Christlieb}},\ and\ \citenamefont
  {{Eriksson}}}]{Mashonkina:2014}%
  \BibitemOpen
  \bibfield  {author} {\bibinfo {author} {\bibnamefont {{Mashonkina}},
  \bibfnamefont {L}}, \bibinfo {author} {\bibfnamefont {N.}~\bibnamefont
  {{Christlieb}}}, and\ \bibinfo {author} {\bibfnamefont {K.}~\bibnamefont
  {{Eriksson}}}} (\bibinfo {year} {2014}),\ \bibfield  {title} {\enquote
  {\bibinfo {title} {{The Hamburg/ESO R-process Enhanced Star survey (HERES).
  X. HE 2252-4225, one more r-process enhanced and actinide-boost halo
  star}},}\ }\href {https://doi.org/10.1051/0004-6361/201424017} {\bibfield
  {journal} {\bibinfo  {journal} {Astron. \& Astrophys.}\ }\textbf {\bibinfo
  {volume} {569}},\ \bibinfo {eid} {A43}}\BibitemShut {NoStop}%
\bibitem [{\citenamefont {{Mathews}}\ \emph {et~al.}(1983)\citenamefont
  {{Mathews}}, \citenamefont {{Mengoni}}, \citenamefont {{Thielemann}},\ and\
  \citenamefont {{Fowler}}}]{Mathews1983}%
  \BibitemOpen
  \bibfield  {author} {\bibinfo {author} {\bibnamefont {{Mathews}},
  \bibfnamefont {G~J}}, \bibinfo {author} {\bibfnamefont {A.}~\bibnamefont
  {{Mengoni}}}, \bibinfo {author} {\bibfnamefont {F.-K.}\ \bibnamefont
  {{Thielemann}}}, and\ \bibinfo {author} {\bibfnamefont {W.~A.}\ \bibnamefont
  {{Fowler}}}} (\bibinfo {year} {1983}),\ \bibfield  {title} {\enquote
  {\bibinfo {title} {{Neutron capture rates in the r-process - The role of
  direct radiative capture}},}\ }\href {https://doi.org/10.1086/161164}
  {\bibfield  {journal} {\bibinfo  {journal} {Astrophys. J.}\ }\textbf
  {\bibinfo {volume} {270}},\ \bibinfo {pages} {740--745}}\BibitemShut
  {NoStop}%
\bibitem [{\citenamefont {{Matteucci}}\ and\ \citenamefont
  {{Chiappini}}(2001)}]{Matteucci.Chiappini:2001}%
  \BibitemOpen
  \bibfield  {author} {\bibinfo {author} {\bibnamefont {{Matteucci}},
  \bibfnamefont {F}}, and\ \bibinfo {author} {\bibfnamefont {C.}~\bibnamefont
  {{Chiappini}}}} (\bibinfo {year} {2001}),\ \bibfield  {title} {\enquote
  {\bibinfo {title} {{The evolution of the oxygen abundance in the Galaxy}},}\
  }\href {https://doi.org/10.1016/S1387-6473(01)00128-2} {\bibfield  {journal}
  {\bibinfo  {journal} {New Astron. Rev.}\ }\textbf {\bibinfo {volume} {45}},\
  \bibinfo {pages} {567--570}}\BibitemShut {NoStop}%
\bibitem [{\citenamefont {{Matteucci}}\ and\ \citenamefont
  {{Greggio}}(1986)}]{matteucci86}%
  \BibitemOpen
  \bibfield  {author} {\bibinfo {author} {\bibnamefont {{Matteucci}},
  \bibfnamefont {F}}, and\ \bibinfo {author} {\bibfnamefont {L.}~\bibnamefont
  {{Greggio}}}} (\bibinfo {year} {1986}),\ \bibfield  {title} {\enquote
  {\bibinfo {title} {{Relative roles of type I and II supernovae in the
  chemical enrichment of the interstellar gas}},}\ }\href@noop {} {\bibfield
  {journal} {\bibinfo  {journal} {Astron. \& Astrophys.}\ }\textbf {\bibinfo
  {volume} {154}},\ \bibinfo {pages} {279--287}}\BibitemShut {NoStop}%
\bibitem [{\citenamefont {{Matteucci}}\ \emph {et~al.}(2014)\citenamefont
  {{Matteucci}}, \citenamefont {{Romano}}, \citenamefont {{Arcones}},
  \citenamefont {{Korobkin}},\ and\ \citenamefont
  {{Rosswog}}}]{Matteucci.Romano.Arcones.ea:2014}%
  \BibitemOpen
  \bibfield  {author} {\bibinfo {author} {\bibnamefont {{Matteucci}},
  \bibfnamefont {F}}, \bibinfo {author} {\bibfnamefont {D.}~\bibnamefont
  {{Romano}}}, \bibinfo {author} {\bibfnamefont {A.}~\bibnamefont {{Arcones}}},
  \bibinfo {author} {\bibfnamefont {O.}~\bibnamefont {{Korobkin}}}, and\
  \bibinfo {author} {\bibfnamefont {S.}~\bibnamefont {{Rosswog}}}} (\bibinfo
  {year} {2014}),\ \bibfield  {title} {\enquote {\bibinfo {title} {{Europium
  production: neutron star mergers versus core-collapse supernovae}},}\ }\href
  {https://doi.org/10.1093/mnras/stt2350} {\bibfield  {journal} {\bibinfo
  {journal} {Mon. Not. Roy. Astron. Soc.}\ }\textbf {\bibinfo {volume} {438}},\
  \bibinfo {pages} {2177--2185}}\BibitemShut {NoStop}%
\bibitem [{\citenamefont {Matteucci}(2012)}]{Matteucci:2012}%
  \BibitemOpen
  \bibfield  {author} {\bibinfo {author} {\bibnamefont {Matteucci},
  \bibfnamefont {Francesca}}} (\bibinfo {year} {2012}),\ \href
  {https://doi.org/10.1007/978-3-642-22491-1} {\emph {\bibinfo {title}
  {{Chemical Evolution of Galaxies}}}},\ Astronomy and Astrophysics Library\
  (\bibinfo  {publisher} {Springer-Verlag},\ \bibinfo {address} {Berlin
  Heidelberg})\BibitemShut {NoStop}%
\bibitem [{\citenamefont {{McConnell}}\ and\ \citenamefont
  {{Talbert}}(1975)}]{TristanI}%
  \BibitemOpen
  \bibfield  {author} {\bibinfo {author} {\bibnamefont {{McConnell}},
  \bibfnamefont {J~R}}, and\ \bibinfo {author} {\bibfnamefont {W.~L.}\
  \bibnamefont {{Talbert}}}} (\bibinfo {year} {1975}),\ \bibfield  {title}
  {\enquote {\bibinfo {title} {{The Tristan on-line isotope separator
  facility}},}\ }\href {https://doi.org/10.1016/0029-554X(75)90672-2}
  {\bibfield  {journal} {\bibinfo  {journal} {Nuclear Instruments and Methods}\
  }\textbf {\bibinfo {volume} {128}},\ \bibinfo {pages} {227--243}}\BibitemShut
  {NoStop}%
\bibitem [{\citenamefont {{McFadden}}\ and\ \citenamefont
  {{Satchler}}(1966)}]{McFadden.Satchler:1966}%
  \BibitemOpen
  \bibfield  {author} {\bibinfo {author} {\bibnamefont {{McFadden}},
  \bibfnamefont {L}}, and\ \bibinfo {author} {\bibfnamefont {G.~R.}\
  \bibnamefont {{Satchler}}}} (\bibinfo {year} {1966}),\ \bibfield  {title}
  {\enquote {\bibinfo {title} {{Optical-model analysis of the scattering of
  24.7 MeV alpha particles}},}\ }\href
  {https://doi.org/10.1016/0029-5582(66)90441-X} {\bibfield  {journal}
  {\bibinfo  {journal} {Nucl. Phys. A}\ }\textbf {\bibinfo {volume} {84}},\
  \bibinfo {pages} {177--200}}\BibitemShut {NoStop}%
\bibitem [{\citenamefont {{McKinney}}\ \emph {et~al.}(2013)\citenamefont
  {{McKinney}}, \citenamefont {{Tchekhovskoy}},\ and\ \citenamefont
  {{Blandford}}}]{mckinney13}%
  \BibitemOpen
  \bibfield  {author} {\bibinfo {author} {\bibnamefont {{McKinney}},
  \bibfnamefont {J~C}}, \bibinfo {author} {\bibfnamefont {A.}~\bibnamefont
  {{Tchekhovskoy}}}, and\ \bibinfo {author} {\bibfnamefont {R.~D.}\
  \bibnamefont {{Blandford}}}} (\bibinfo {year} {2013}),\ \bibfield  {title}
  {\enquote {\bibinfo {title} {{Alignment of Magnetized Accretion Disks and
  Relativistic Jets with Spinning Black Holes}},}\ }\href
  {https://doi.org/10.1126/science.1230811} {\bibfield  {journal} {\bibinfo
  {journal} {Science}\ }\textbf {\bibinfo {volume} {339}},\ \bibinfo {pages}
  {49}}\BibitemShut {NoStop}%
\bibitem [{\citenamefont {McLaughlin}\ \emph {et~al.}(1999)\citenamefont
  {McLaughlin}, \citenamefont {Fetter}, \citenamefont {Balantekin},\ and\
  \citenamefont {Fuller}}]{Mclaughlin.Fetter.ea:1999}%
  \BibitemOpen
  \bibfield  {author} {\bibinfo {author} {\bibnamefont {McLaughlin},
  \bibfnamefont {G~C}}, \bibinfo {author} {\bibfnamefont {J.~M.}\ \bibnamefont
  {Fetter}}, \bibinfo {author} {\bibfnamefont {A.~B.}\ \bibnamefont
  {Balantekin}}, and\ \bibinfo {author} {\bibfnamefont {G.~M.}\ \bibnamefont
  {Fuller}}} (\bibinfo {year} {1999}),\ \bibfield  {title} {\enquote {\bibinfo
  {title} {Active-sterile neutrino transformation solution for r-process
  nucleosynthesis},}\ }\href {https://doi.org/10.1103/PhysRevC.59.2873}
  {\bibfield  {journal} {\bibinfo  {journal} {Phys. Rev. C}\ }\textbf {\bibinfo
  {volume} {59}},\ \bibinfo {pages} {2873--2887}}\BibitemShut {NoStop}%
\bibitem [{\citenamefont {{McLaughlin}}\ and\ \citenamefont
  {{Surman}}(2005)}]{Mclaughlin.Surman:2005}%
  \BibitemOpen
  \bibfield  {author} {\bibinfo {author} {\bibnamefont {{McLaughlin}},
  \bibfnamefont {G~C}}, and\ \bibinfo {author} {\bibfnamefont {R.}~\bibnamefont
  {{Surman}}}} (\bibinfo {year} {2005}),\ \bibfield  {title} {\enquote
  {\bibinfo {title} {{Prospects for obtaining an r process from Gamma Ray Burst
  Disk Winds}},}\ }\href {https://doi.org/10.1016/j.nuclphysa.2005.05.036}
  {\bibfield  {journal} {\bibinfo  {journal} {Nucl. Phys. A}\ }\textbf
  {\bibinfo {volume} {758}},\ \bibinfo {pages} {189--196}}\BibitemShut
  {NoStop}%
\bibitem [{\citenamefont {{Meisel}}\ and\ \citenamefont
  {{George}}(2013)}]{Meisel.George:2013}%
  \BibitemOpen
  \bibfield  {author} {\bibinfo {author} {\bibnamefont {{Meisel}},
  \bibfnamefont {Z}}, and\ \bibinfo {author} {\bibfnamefont {S.}~\bibnamefont
  {{George}}}} (\bibinfo {year} {2013}),\ \bibfield  {title} {\enquote
  {\bibinfo {title} {{Time-of-flight mass spectrometry of very exotic
  systems}},}\ }\href {https://doi.org/10.1016/j.ijms.2013.03.022} {\bibfield
  {journal} {\bibinfo  {journal} {Int. J. Mass Spectrom.}\ }\textbf {\bibinfo
  {volume} {349-350}},\ \bibinfo {pages} {145--150}}\BibitemShut {NoStop}%
\bibitem [{\citenamefont {Mendoza-Temis}(2014)}]{Mendoza-Temis:2014}%
  \BibitemOpen
  \bibfield  {author} {\bibinfo {author} {\bibnamefont {Mendoza-Temis},
  \bibfnamefont {Joel~Jes{\'u}s}}} (\bibinfo {year} {2014}),\ \emph {\bibinfo
  {title} {Nuclear masses and their impact in r-process nucleosynthesis}},\
  \href@noop {} {Ph.D. thesis}\ (\bibinfo  {school} {Technische Universit{\"a}t
  Darmstadt})\BibitemShut {NoStop}%
\bibitem [{\citenamefont {Mendoza-Temis}\ \emph {et~al.}(2015)\citenamefont
  {Mendoza-Temis}, \citenamefont {Wu}, \citenamefont {Langanke}, \citenamefont
  {Mart\'{\i}nez-Pinedo}, \citenamefont {Bauswein},\ and\ \citenamefont
  {Janka}}]{Mendoza-Temis.Wu.ea:2015}%
  \BibitemOpen
  \bibfield  {author} {\bibinfo {author} {\bibnamefont {Mendoza-Temis},
  \bibfnamefont {Joel~Jes\'us}}, \bibinfo {author} {\bibfnamefont {Meng-Ru}\
  \bibnamefont {Wu}}, \bibinfo {author} {\bibfnamefont {Karlheinz}\
  \bibnamefont {Langanke}}, \bibinfo {author} {\bibfnamefont {Gabriel}\
  \bibnamefont {Mart\'{\i}nez-Pinedo}}, \bibinfo {author} {\bibfnamefont
  {Andreas}\ \bibnamefont {Bauswein}}, and\ \bibinfo {author} {\bibfnamefont
  {Hans-Thomas}\ \bibnamefont {Janka}}} (\bibinfo {year} {2015}),\ \bibfield
  {title} {\enquote {\bibinfo {title} {Nuclear robustness of the $r$ process in
  neutron-star mergers},}\ }\href {https://doi.org/10.1103/PhysRevC.92.055805}
  {\bibfield  {journal} {\bibinfo  {journal} {Phys. Rev. C}\ }\textbf {\bibinfo
  {volume} {92}},\ \bibinfo {pages} {055805}}\BibitemShut {NoStop}%
\bibitem [{\citenamefont {{Mennekens}}\ and\ \citenamefont
  {{Vanbeveren}}(2014)}]{mennekens14}%
  \BibitemOpen
  \bibfield  {author} {\bibinfo {author} {\bibnamefont {{Mennekens}},
  \bibfnamefont {N}}, and\ \bibinfo {author} {\bibfnamefont {D.}~\bibnamefont
  {{Vanbeveren}}}} (\bibinfo {year} {2014}),\ \bibfield  {title} {\enquote
  {\bibinfo {title} {{Massive double compact object mergers: gravitational wave
  sources and r-process element production sites}},}\ }\href
  {https://doi.org/10.1051/0004-6361/201322198} {\bibfield  {journal} {\bibinfo
   {journal} {Astron. \& Astrophys.}\ }\textbf {\bibinfo {volume} {564}},\
  \bibinfo {eid} {A134}}\BibitemShut {NoStop}%
\bibitem [{\citenamefont {{Mennekens}}\ and\ \citenamefont
  {{Vanbeveren}}(2016)}]{mennekens16}%
  \BibitemOpen
  \bibfield  {author} {\bibinfo {author} {\bibnamefont {{Mennekens}},
  \bibfnamefont {N}}, and\ \bibinfo {author} {\bibfnamefont {D.}~\bibnamefont
  {{Vanbeveren}}}} (\bibinfo {year} {2016}),\ \bibfield  {title} {\enquote
  {\bibinfo {title} {{The delay time distribution of massive double compact
  star mergers}},}\ }\href {https://doi.org/10.1051/0004-6361/201628193}
  {\bibfield  {journal} {\bibinfo  {journal} {Astron. \& Astrophys.}\ }\textbf
  {\bibinfo {volume} {589}},\ \bibinfo {eid} {A64}}\BibitemShut {NoStop}%
\bibitem [{\citenamefont {Mention}\ \emph {et~al.}(2011)\citenamefont
  {Mention}, \citenamefont {Fechner}, \citenamefont {Lasserre}, \citenamefont
  {Mueller}, \citenamefont {Lhuillier}, \citenamefont {Cribier},\ and\
  \citenamefont {Letourneau}}]{Mention.Fechner.ea:2011}%
  \BibitemOpen
  \bibfield  {author} {\bibinfo {author} {\bibnamefont {Mention}, \bibfnamefont
  {G}}, \bibinfo {author} {\bibfnamefont {M.}~\bibnamefont {Fechner}}, \bibinfo
  {author} {\bibfnamefont {Th.}\ \bibnamefont {Lasserre}}, \bibinfo {author}
  {\bibfnamefont {Th.~A.}\ \bibnamefont {Mueller}}, \bibinfo {author}
  {\bibfnamefont {D.}~\bibnamefont {Lhuillier}}, \bibinfo {author}
  {\bibfnamefont {M.}~\bibnamefont {Cribier}}, and\ \bibinfo {author}
  {\bibfnamefont {A.}~\bibnamefont {Letourneau}}} (\bibinfo {year} {2011}),\
  \bibfield  {title} {\enquote {\bibinfo {title} {Reactor antineutrino
  anomaly},}\ }\href {https://doi.org/10.1103/PhysRevD.83.073006} {\bibfield
  {journal} {\bibinfo  {journal} {Phys. Rev. D}\ }\textbf {\bibinfo {volume}
  {83}},\ \bibinfo {pages} {073006}}\BibitemShut {NoStop}%
\bibitem [{\citenamefont {{Merrill}}(1952)}]{merrill52}%
  \BibitemOpen
  \bibfield  {author} {\bibinfo {author} {\bibnamefont {{Merrill}},
  \bibfnamefont {P~W}}} (\bibinfo {year} {1952}),\ \bibfield  {title} {\enquote
  {\bibinfo {title} {{Spectroscopic Observations of Stars of Class}},}\ }\href
  {https://doi.org/10.1086/145589} {\bibfield  {journal} {\bibinfo  {journal}
  {Astrophys. J.}\ }\textbf {\bibinfo {volume} {116}},\ \bibinfo {pages}
  {21}}\BibitemShut {NoStop}%
\bibitem [{\citenamefont {{Metzger}}(2017{\natexlab{a}})}]{metzger17}%
  \BibitemOpen
  \bibfield  {author} {\bibinfo {author} {\bibnamefont {{Metzger}},
  \bibfnamefont {B~D}}} (\bibinfo {year} {2017}{\natexlab{a}}),\ \bibfield
  {title} {\enquote {\bibinfo {title} {{Kilonovae}},}\ }\href
  {https://doi.org/10.1007/s41114-017-0006-z} {\bibfield  {journal} {\bibinfo
  {journal} {Living Reviews in Relativity}\ }\textbf {\bibinfo {volume} {20}},\
  \bibinfo {eid} {3}}\BibitemShut {NoStop}%
\bibitem [{\citenamefont {{Metzger}}(2017{\natexlab{b}})}]{Metzger:2017}%
  \BibitemOpen
  \bibfield  {author} {\bibinfo {author} {\bibnamefont {{Metzger}},
  \bibfnamefont {B~D}}} (\bibinfo {year} {2017}{\natexlab{b}}),\ \bibfield
  {title} {\enquote {\bibinfo {title} {{Welcome to the Multi-Messenger Era!
  Lessons from a Neutron Star Merger and the Landscape Ahead}},}\ }\href@noop
  {} {\bibfield  {journal} {\bibinfo  {journal} {ArXiv e-prints}\ }}\Eprint
  {https://arxiv.org/abs/1710.05931} {arXiv:1710.05931 [astro-ph.HE]}
  \BibitemShut {NoStop}%
\bibitem [{\citenamefont {{Metzger}}\ \emph
  {et~al.}(2010{\natexlab{a}})\citenamefont {{Metzger}}, \citenamefont
  {{Arcones}}, \citenamefont {{Quataert}},\ and\ \citenamefont
  {{Mart{\'{\i}}nez-Pinedo}}}]{Metzger.Arcones.ea:2010}%
  \BibitemOpen
  \bibfield  {author} {\bibinfo {author} {\bibnamefont {{Metzger}},
  \bibfnamefont {B~D}}, \bibinfo {author} {\bibfnamefont {A.}~\bibnamefont
  {{Arcones}}}, \bibinfo {author} {\bibfnamefont {E.}~\bibnamefont
  {{Quataert}}}, and\ \bibinfo {author} {\bibfnamefont {G.}~\bibnamefont
  {{Mart{\'{\i}}nez-Pinedo}}}} (\bibinfo {year} {2010}{\natexlab{a}}),\
  \bibfield  {title} {\enquote {\bibinfo {title} {{The effects of r-process
  heating on fallback accretion in compact object mergers}},}\ }\href
  {https://doi.org/10.1111/j.1365-2966.2009.16107.x} {\bibfield  {journal}
  {\bibinfo  {journal} {Mon. Not. Roy. Astron. Soc.}\ }\textbf {\bibinfo
  {volume} {402}},\ \bibinfo {pages} {2771--2777}}\BibitemShut {NoStop}%
\bibitem [{\citenamefont {{Metzger}}\ \emph {et~al.}(2015)\citenamefont
  {{Metzger}}, \citenamefont {{Bauswein}}, \citenamefont {{Goriely}},\ and\
  \citenamefont {{Kasen}}}]{Metzger.Bauswein.ea:2015}%
  \BibitemOpen
  \bibfield  {author} {\bibinfo {author} {\bibnamefont {{Metzger}},
  \bibfnamefont {B~D}}, \bibinfo {author} {\bibfnamefont {A.}~\bibnamefont
  {{Bauswein}}}, \bibinfo {author} {\bibfnamefont {S.}~\bibnamefont
  {{Goriely}}}, and\ \bibinfo {author} {\bibfnamefont {D.}~\bibnamefont
  {{Kasen}}}} (\bibinfo {year} {2015}),\ \bibfield  {title} {\enquote {\bibinfo
  {title} {{Neutron-powered precursors of kilonovae}},}\ }\href
  {https://doi.org/10.1093/mnras/stu2225} {\bibfield  {journal} {\bibinfo
  {journal} {Mon. Not. Roy. Astron. Soc.}\ }\textbf {\bibinfo {volume} {446}},\
  \bibinfo {pages} {1115--1120}}\BibitemShut {NoStop}%
\bibitem [{\citenamefont {{Metzger}}\ and\ \citenamefont
  {{Berger}}(2012)}]{Metzger.Berger:2012}%
  \BibitemOpen
  \bibfield  {author} {\bibinfo {author} {\bibnamefont {{Metzger}},
  \bibfnamefont {B~D}}, and\ \bibinfo {author} {\bibfnamefont {E.}~\bibnamefont
  {{Berger}}}} (\bibinfo {year} {2012}),\ \bibfield  {title} {\enquote
  {\bibinfo {title} {{What is the Most Promising Electromagnetic Counterpart of
  a Neutron Star Binary Merger?}}}\ }\href
  {https://doi.org/10.1088/0004-637X/746/1/48} {\bibfield  {journal} {\bibinfo
  {journal} {Astrophys. J.}\ }\textbf {\bibinfo {volume} {746}},\ \bibinfo
  {eid} {48}}\BibitemShut {NoStop}%
\bibitem [{\citenamefont {{Metzger}}\ and\ \citenamefont
  {{Fern{\'a}ndez}}(2014)}]{metzger14}%
  \BibitemOpen
  \bibfield  {author} {\bibinfo {author} {\bibnamefont {{Metzger}},
  \bibfnamefont {B~D}}, and\ \bibinfo {author} {\bibfnamefont {R.}~\bibnamefont
  {{Fern{\'a}ndez}}}} (\bibinfo {year} {2014}),\ \bibfield  {title} {\enquote
  {\bibinfo {title} {{Red or blue? A potential kilonova imprint of the delay
  until black hole formation following a neutron star merger}},}\ }\href
  {https://doi.org/10.1093/mnras/stu802} {\bibfield  {journal} {\bibinfo
  {journal} {Mon. Not. Roy. Astron. Soc.}\ }\textbf {\bibinfo {volume} {441}},\
  \bibinfo {pages} {3444--3453}}\BibitemShut {NoStop}%
\bibitem [{\citenamefont {{Metzger}}\ \emph
  {et~al.}(2010{\natexlab{b}})\citenamefont {{Metzger}}, \citenamefont
  {{Mart{\'{\i}}nez-Pinedo}}, \citenamefont {{Darbha}}, \citenamefont
  {{Quataert}}, \citenamefont {{Arcones}}, \citenamefont {{Kasen}},
  \citenamefont {{Thomas}}, \citenamefont {{Nugent}}, \citenamefont {{Panov}},\
  and\ \citenamefont {{Zinner}}}]{Metzger.Martinez.ea:2010}%
  \BibitemOpen
  \bibfield  {author} {\bibinfo {author} {\bibnamefont {{Metzger}},
  \bibfnamefont {B~D}}, \bibinfo {author} {\bibfnamefont {G.}~\bibnamefont
  {{Mart{\'{\i}}nez-Pinedo}}}, \bibinfo {author} {\bibfnamefont
  {S.}~\bibnamefont {{Darbha}}}, \bibinfo {author} {\bibfnamefont
  {E.}~\bibnamefont {{Quataert}}}, \bibinfo {author} {\bibfnamefont
  {A.}~\bibnamefont {{Arcones}}}, \bibinfo {author} {\bibfnamefont
  {D.}~\bibnamefont {{Kasen}}}, \bibinfo {author} {\bibfnamefont
  {R.}~\bibnamefont {{Thomas}}}, \bibinfo {author} {\bibfnamefont
  {P.}~\bibnamefont {{Nugent}}}, \bibinfo {author} {\bibfnamefont {I.~V.}\
  \bibnamefont {{Panov}}}, and\ \bibinfo {author} {\bibfnamefont {N.~T.}\
  \bibnamefont {{Zinner}}}} (\bibinfo {year} {2010}{\natexlab{b}}),\ \bibfield
  {title} {\enquote {\bibinfo {title} {{Electromagnetic counterparts of compact
  object mergers powered by the radioactive decay of r-process nuclei}},}\
  }\href {https://doi.org/10.1111/j.1365-2966.2010.16864.x} {\bibfield
  {journal} {\bibinfo  {journal} {Mon. Not. Roy. Astron. Soc.}\ }\textbf
  {\bibinfo {volume} {406}},\ \bibinfo {pages} {2650--2662}}\BibitemShut
  {NoStop}%
\bibitem [{\citenamefont {{Metzger}}\ \emph {et~al.}(2009)\citenamefont
  {{Metzger}}, \citenamefont {{Piro}},\ and\ \citenamefont
  {{Quataert}}}]{Metzger.Piro.Quataert:2009}%
  \BibitemOpen
  \bibfield  {author} {\bibinfo {author} {\bibnamefont {{Metzger}},
  \bibfnamefont {B~D}}, \bibinfo {author} {\bibfnamefont {A.~L.}\ \bibnamefont
  {{Piro}}}, and\ \bibinfo {author} {\bibfnamefont {E.}~\bibnamefont
  {{Quataert}}}} (\bibinfo {year} {2009}),\ \bibfield  {title} {\enquote
  {\bibinfo {title} {{Neutron-rich freeze-out in viscously spreading accretion
  discs formed from compact object mergers}},}\ }\href
  {https://doi.org/10.1111/j.1365-2966.2008.14380.x} {\bibfield  {journal}
  {\bibinfo  {journal} {Mon. Not. Roy. Astron. Soc.}\ }\textbf {\bibinfo
  {volume} {396}},\ \bibinfo {pages} {304--314}}\BibitemShut {NoStop}%
\bibitem [{\citenamefont {{Metzger}}\ \emph {et~al.}(2008)\citenamefont
  {{Metzger}}, \citenamefont {{Thompson}},\ and\ \citenamefont
  {{Quataert}}}]{Metzger.Thompson.Quataert:2008}%
  \BibitemOpen
  \bibfield  {author} {\bibinfo {author} {\bibnamefont {{Metzger}},
  \bibfnamefont {B~D}}, \bibinfo {author} {\bibfnamefont {T.~A.}\ \bibnamefont
  {{Thompson}}}, and\ \bibinfo {author} {\bibfnamefont {E.}~\bibnamefont
  {{Quataert}}}} (\bibinfo {year} {2008}),\ \bibfield  {title} {\enquote
  {\bibinfo {title} {{On the Conditions for Neutron-rich Gamma-Ray Burst
  Outflows}},}\ }\href {https://doi.org/10.1086/526418} {\bibfield  {journal}
  {\bibinfo  {journal} {Astrophys. J.}\ }\textbf {\bibinfo {volume} {676}},\
  \bibinfo {pages} {1130--1150}}\BibitemShut {NoStop}%
\bibitem [{\citenamefont {{Metzger}}\ \emph {et~al.}(2018)\citenamefont
  {{Metzger}}, \citenamefont {{Thompson}},\ and\ \citenamefont
  {{Quataert}}}]{Metzger.Thompson.Quataert:2018}%
  \BibitemOpen
  \bibfield  {author} {\bibinfo {author} {\bibnamefont {{Metzger}},
  \bibfnamefont {B~D}}, \bibinfo {author} {\bibfnamefont {T.~A.}\ \bibnamefont
  {{Thompson}}}, and\ \bibinfo {author} {\bibfnamefont {E.}~\bibnamefont
  {{Quataert}}}} (\bibinfo {year} {2018}),\ \bibfield  {title} {\enquote
  {\bibinfo {title} {{A Magnetar Origin for the Kilonova Ejecta in
  GW170817}},}\ }\href {https://doi.org/10.3847/1538-4357/aab095} {\bibfield
  {journal} {\bibinfo  {journal} {Astrophys. J.}\ }\textbf {\bibinfo {volume}
  {856}},\ \bibinfo {eid} {101}}\BibitemShut {NoStop}%
\bibitem [{\citenamefont {{Meyer}}(1993)}]{Meyer:1993}%
  \BibitemOpen
  \bibfield  {author} {\bibinfo {author} {\bibnamefont {{Meyer}}, \bibfnamefont
  {B~S}}} (\bibinfo {year} {1993}),\ \bibfield  {title} {\enquote {\bibinfo
  {title} {{Entropy and nucleosynthesis}},}\ }\href
  {https://doi.org/10.1016/0370-1573(93)90071-K} {\bibfield  {journal}
  {\bibinfo  {journal} {Phys. Rep.}\ }\textbf {\bibinfo {volume} {227}},\
  \bibinfo {pages} {257--267}}\BibitemShut {NoStop}%
\bibitem [{\citenamefont {{Meyer}}\ \emph {et~al.}(1992)\citenamefont
  {{Meyer}}, \citenamefont {{Mathews}}, \citenamefont {{Howard}}, \citenamefont
  {{Woosley}},\ and\ \citenamefont {{Hoffman}}}]{meyer92}%
  \BibitemOpen
  \bibfield  {author} {\bibinfo {author} {\bibnamefont {{Meyer}}, \bibfnamefont
  {B~S}}, \bibinfo {author} {\bibfnamefont {G.~J.}\ \bibnamefont {{Mathews}}},
  \bibinfo {author} {\bibfnamefont {W.~M.}\ \bibnamefont {{Howard}}}, \bibinfo
  {author} {\bibfnamefont {S.~E.}\ \bibnamefont {{Woosley}}}, and\ \bibinfo
  {author} {\bibfnamefont {R.~D.}\ \bibnamefont {{Hoffman}}}} (\bibinfo {year}
  {1992}),\ \bibfield  {title} {\enquote {\bibinfo {title} {{R-process
  nucleosynthesis in the high-entropy supernova bubble}},}\ }\href
  {https://doi.org/10.1086/171957} {\bibfield  {journal} {\bibinfo  {journal}
  {Astrophys. J.}\ }\textbf {\bibinfo {volume} {399}},\ \bibinfo {pages}
  {656--664}}\BibitemShut {NoStop}%
\bibitem [{\citenamefont {{Meyer}}\ \emph {et~al.}(1998)\citenamefont
  {{Meyer}}, \citenamefont {{McLaughlin}},\ and\ \citenamefont
  {{Fuller}}}]{Meyer.Mclaughlin.Fuller:1998}%
  \BibitemOpen
  \bibfield  {author} {\bibinfo {author} {\bibnamefont {{Meyer}}, \bibfnamefont
  {B~S}}, \bibinfo {author} {\bibfnamefont {G.~C.}\ \bibnamefont
  {{McLaughlin}}}, and\ \bibinfo {author} {\bibfnamefont {G.~M.}\ \bibnamefont
  {{Fuller}}}} (\bibinfo {year} {1998}),\ \bibfield  {title} {\enquote
  {\bibinfo {title} {{Neutrino capture and r-process nucleosynthesis}},}\
  }\href {https://doi.org/10.1103/PhysRevC.58.3696} {\bibfield  {journal}
  {\bibinfo  {journal} {Phys. Rev. C}\ }\textbf {\bibinfo {volume} {58}},\
  \bibinfo {pages} {3696--3710}}\BibitemShut {NoStop}%
\bibitem [{\citenamefont {{Meyer}}\ and\ \citenamefont
  {{Schramm}}(1988)}]{meyer88}%
  \BibitemOpen
  \bibfield  {author} {\bibinfo {author} {\bibnamefont {{Meyer}}, \bibfnamefont
  {B~S}}, and\ \bibinfo {author} {\bibfnamefont {D.~N.}\ \bibnamefont
  {{Schramm}}}} (\bibinfo {year} {1988}),\ \bibfield  {title} {\enquote
  {\bibinfo {title} {{Nucleosynthesis from the Decompression of Initially Cold
  Neutron Star Matter}},}\ }in\ \href@noop {} {\emph {\bibinfo {booktitle}
  {Origin and Distribution of the Elements}}},\ \bibinfo {editor} {edited by\
  \bibinfo {editor} {\bibfnamefont {G.~J.}\ \bibnamefont {{Mathews}}}},\ p.\
  \bibinfo {pages} {610}\BibitemShut {NoStop}%
\bibitem [{\citenamefont {{Minchev}}\ \emph {et~al.}(2014)\citenamefont
  {{Minchev}}, \citenamefont {{Chiappini}},\ and\ \citenamefont
  {{Martig}}}]{Minchev.ea:2014}%
  \BibitemOpen
  \bibfield  {author} {\bibinfo {author} {\bibnamefont {{Minchev}},
  \bibfnamefont {I}}, \bibinfo {author} {\bibfnamefont {C.}~\bibnamefont
  {{Chiappini}}}, and\ \bibinfo {author} {\bibfnamefont {M.}~\bibnamefont
  {{Martig}}}} (\bibinfo {year} {2014}),\ \bibfield  {title} {\enquote
  {\bibinfo {title} {{Chemodynamical evolution of the Milky Way disk. II.
  Variations with Galactic radius and height above the disk plane}},}\ }\href
  {https://doi.org/10.1051/0004-6361/201423487} {\bibfield  {journal} {\bibinfo
   {journal} {Astron. \& Astrophys.}\ }\textbf {\bibinfo {volume} {572}},\
  \bibinfo {eid} {A92}}\BibitemShut {NoStop}%
\bibitem [{\citenamefont {{Mirizzi}}(2015)}]{Mirizzi:2015}%
  \BibitemOpen
  \bibfield  {author} {\bibinfo {author} {\bibnamefont {{Mirizzi}},
  \bibfnamefont {A}}} (\bibinfo {year} {2015}),\ \bibfield  {title} {\enquote
  {\bibinfo {title} {{Breaking the symmetries in self-induced flavor
  conversions of neutrino beams from a ring}},}\ }\href
  {https://doi.org/10.1103/PhysRevD.92.105020} {\bibfield  {journal} {\bibinfo
  {journal} {Phys. Rev. D}\ }\textbf {\bibinfo {volume} {92}},\ \bibinfo {eid}
  {105020}}\BibitemShut {NoStop}%
\bibitem [{\citenamefont {{Mirizzi}}\ \emph {et~al.}(2015)\citenamefont
  {{Mirizzi}}, \citenamefont {{Mangano}},\ and\ \citenamefont
  {{Saviano}}}]{mirizzi15}%
  \BibitemOpen
  \bibfield  {author} {\bibinfo {author} {\bibnamefont {{Mirizzi}},
  \bibfnamefont {A}}, \bibinfo {author} {\bibfnamefont {G.}~\bibnamefont
  {{Mangano}}}, and\ \bibinfo {author} {\bibfnamefont {N.}~\bibnamefont
  {{Saviano}}}} (\bibinfo {year} {2015}),\ \bibfield  {title} {\enquote
  {\bibinfo {title} {{Self-induced flavor instabilities of a dense neutrino
  stream in a two-dimensional model}},}\ }\href
  {https://doi.org/10.1103/PhysRevD.92.021702} {\bibfield  {journal} {\bibinfo
  {journal} {Phys. Rev. D}\ }\textbf {\bibinfo {volume} {92}},\ \bibinfo {eid}
  {021702}}\BibitemShut {NoStop}%
\bibitem [{\citenamefont {{Mirizzi}}\ \emph {et~al.}(2016)\citenamefont
  {{Mirizzi}}, \citenamefont {{Tamborra}}, \citenamefont {{Janka}},
  \citenamefont {{Saviano}}, \citenamefont {{Scholberg}}, \citenamefont
  {{Bollig}}, \citenamefont {{H{\"u}depohl}},\ and\ \citenamefont
  {{Chakraborty}}}]{mirizzi15b}%
  \BibitemOpen
  \bibfield  {author} {\bibinfo {author} {\bibnamefont {{Mirizzi}},
  \bibfnamefont {A}}, \bibinfo {author} {\bibfnamefont {I.}~\bibnamefont
  {{Tamborra}}}, \bibinfo {author} {\bibfnamefont {H.-T.}\ \bibnamefont
  {{Janka}}}, \bibinfo {author} {\bibfnamefont {N.}~\bibnamefont {{Saviano}}},
  \bibinfo {author} {\bibfnamefont {K.}~\bibnamefont {{Scholberg}}}, \bibinfo
  {author} {\bibfnamefont {R.}~\bibnamefont {{Bollig}}}, \bibinfo {author}
  {\bibfnamefont {L.}~\bibnamefont {{H{\"u}depohl}}}, and\ \bibinfo {author}
  {\bibfnamefont {S.}~\bibnamefont {{Chakraborty}}}} (\bibinfo {year} {2016}),\
  \bibfield  {title} {\enquote {\bibinfo {title} {{Supernova neutrinos:
  production, oscillations and detection}},}\ }\href
  {https://doi.org/10.1393/ncr/i2016-10120-8} {\bibfield  {journal} {\bibinfo
  {journal} {Nuovo Cimento Rivista Serie}\ }\textbf {\bibinfo {volume} {39}},\
  \bibinfo {pages} {1--112}}\BibitemShut {NoStop}%
\bibitem [{\citenamefont {{Mishenina}}\ \emph
  {et~al.}(2019{\natexlab{a}})\citenamefont {{Mishenina}}, \citenamefont
  {{Pignatari}}, \citenamefont {{Gorbaneva}}, \citenamefont {{Bisterzo}},
  \citenamefont {{Travaglio}}, \citenamefont {{Thielemann}},\ and\
  \citenamefont {{Soubiran}}}]{Mishenina.Pignatari.ea:2019a}%
  \BibitemOpen
  \bibfield  {author} {\bibinfo {author} {\bibnamefont {{Mishenina}},
  \bibfnamefont {T}}, \bibinfo {author} {\bibfnamefont {M.}~\bibnamefont
  {{Pignatari}}}, \bibinfo {author} {\bibfnamefont {T.}~\bibnamefont
  {{Gorbaneva}}}, \bibinfo {author} {\bibfnamefont {S.}~\bibnamefont
  {{Bisterzo}}}, \bibinfo {author} {\bibfnamefont {C.}~\bibnamefont
  {{Travaglio}}}, \bibinfo {author} {\bibfnamefont {F.~K.}\ \bibnamefont
  {{Thielemann}}}, and\ \bibinfo {author} {\bibfnamefont {C.}~\bibnamefont
  {{Soubiran}}}} (\bibinfo {year} {2019}{\natexlab{a}}),\ \bibfield  {title}
  {\enquote {\bibinfo {title} {{Enrichment of the Galactic disc with neutron
  capture elements: Sr}},}\ }\href {https://doi.org/10.1093/mnras/stz178}
  {\bibfield  {journal} {\bibinfo  {journal} {Mon. Not. Roy. Astron. Soc.}\
  }\textbf {\bibinfo {volume} {484}},\ \bibinfo {pages}
  {3846--3864}}\BibitemShut {NoStop}%
\bibitem [{\citenamefont {{Mishenina}}\ \emph
  {et~al.}(2019{\natexlab{b}})\citenamefont {{Mishenina}}, \citenamefont
  {{Pignatari}}, \citenamefont {{Gorbaneva}}, \citenamefont {{Travaglio}},
  \citenamefont {{C{\^o}t{\'e}}}, \citenamefont {{Thielemann}},\ and\
  \citenamefont {{Soubiran}}}]{Mishenina.Pignatari.ea:2019}%
  \BibitemOpen
  \bibfield  {author} {\bibinfo {author} {\bibnamefont {{Mishenina}},
  \bibfnamefont {T}}, \bibinfo {author} {\bibfnamefont {M.}~\bibnamefont
  {{Pignatari}}}, \bibinfo {author} {\bibfnamefont {T.}~\bibnamefont
  {{Gorbaneva}}}, \bibinfo {author} {\bibfnamefont {C.}~\bibnamefont
  {{Travaglio}}}, \bibinfo {author} {\bibfnamefont {B.}~\bibnamefont
  {{C{\^o}t{\'e}}}}, \bibinfo {author} {\bibfnamefont {F.~K.}\ \bibnamefont
  {{Thielemann}}}, and\ \bibinfo {author} {\bibfnamefont {C.}~\bibnamefont
  {{Soubiran}}}} (\bibinfo {year} {2019}{\natexlab{b}}),\ \bibfield  {title}
  {\enquote {\bibinfo {title} {{Enrichment of the Galactic disc with
  neutron-capture elements: Mo and Ru}},}\ }\href
  {https://doi.org/10.1093/mnras/stz2202} {\bibfield  {journal} {\bibinfo
  {journal} {Mon. Not. Roy. Astron. Soc.}\ }\textbf {\bibinfo {volume} {489}},\
  \bibinfo {pages} {1697--1708}}\BibitemShut {NoStop}%
\bibitem [{\citenamefont {Mocelj}\ \emph {et~al.}(2007)\citenamefont {Mocelj},
  \citenamefont {Rauscher}, \citenamefont {Mart{\'i}nez-Pinedo}, \citenamefont
  {Langanke}, \citenamefont {Pacearescu}, \citenamefont {Faessler},
  \citenamefont {Thielemann},\ and\ \citenamefont
  {Alhassid}}]{Mocelj.Rauscher.ea:2007}%
  \BibitemOpen
  \bibfield  {author} {\bibinfo {author} {\bibnamefont {Mocelj}, \bibfnamefont
  {D}}, \bibinfo {author} {\bibfnamefont {T.}~\bibnamefont {Rauscher}},
  \bibinfo {author} {\bibfnamefont {G.}~\bibnamefont {Mart{\'i}nez-Pinedo}},
  \bibinfo {author} {\bibfnamefont {K.}~\bibnamefont {Langanke}}, \bibinfo
  {author} {\bibfnamefont {L.}~\bibnamefont {Pacearescu}}, \bibinfo {author}
  {\bibfnamefont {A.}~\bibnamefont {Faessler}}, \bibinfo {author}
  {\bibfnamefont {F.-K.}\ \bibnamefont {Thielemann}}, and\ \bibinfo {author}
  {\bibfnamefont {Y.}~\bibnamefont {Alhassid}}} (\bibinfo {year} {2007}),\
  \bibfield  {title} {\enquote {\bibinfo {title} {Large-scale prediction of the
  parity distribution in the nuclear level density and application to
  astrophysical reaction rates},}\ }\href
  {https://doi.org/10.1103/PhysRevC.75.045805} {\bibfield  {journal} {\bibinfo
  {journal} {Phys. Rev. C}\ }\textbf {\bibinfo {volume} {75}},\ \bibinfo
  {pages} {045805}}\BibitemShut {NoStop}%
\bibitem [{\citenamefont {Mohr}\ \emph {et~al.}(2020)\citenamefont {Mohr},
  \citenamefont {F\"ul\"op}, \citenamefont {Gy\"urky}, \citenamefont {Kiss},\
  and\ \citenamefont {Sz\"ucs}}]{Mohr.Fueloep.ea:2020}%
  \BibitemOpen
  \bibfield  {author} {\bibinfo {author} {\bibnamefont {Mohr}, \bibfnamefont
  {P}}, \bibinfo {author} {\bibfnamefont {Zs.}\ \bibnamefont {F\"ul\"op}},
  \bibinfo {author} {\bibfnamefont {Gy.}\ \bibnamefont {Gy\"urky}}, \bibinfo
  {author} {\bibfnamefont {G.~G.}\ \bibnamefont {Kiss}}, and\ \bibinfo {author}
  {\bibfnamefont {T.}~\bibnamefont {Sz\"ucs}}} (\bibinfo {year} {2020}),\
  \bibfield  {title} {\enquote {\bibinfo {title} {Successful prediction of
  total $\ensuremath{\alpha}$-induced reaction cross sections at
  astrophysically relevant sub-coulomb energies using a novel approach},}\
  }\href {https://doi.org/10.1103/PhysRevLett.124.252701} {\bibfield  {journal}
  {\bibinfo  {journal} {Phys. Rev. Lett.}\ }\textbf {\bibinfo {volume} {124}},\
  \bibinfo {pages} {252701}}\BibitemShut {NoStop}%
\bibitem [{\citenamefont {{Mohr}}\ \emph {et~al.}(2017)\citenamefont {{Mohr}},
  \citenamefont {{Gy{\"u}rky}},\ and\ \citenamefont
  {{F{\"u}l{\"o}p}}}]{Mohr.Gyurku.Fulop:2017}%
  \BibitemOpen
  \bibfield  {author} {\bibinfo {author} {\bibnamefont {{Mohr}}, \bibfnamefont
  {P}}, \bibinfo {author} {\bibfnamefont {G.}~\bibnamefont {{Gy{\"u}rky}}},
  and\ \bibinfo {author} {\bibfnamefont {Z.}~\bibnamefont {{F{\"u}l{\"o}p}}}}
  (\bibinfo {year} {2017}),\ \bibfield  {title} {\enquote {\bibinfo {title}
  {{Statistical model analysis of {$\alpha$} -induced reaction cross sections
  of $^{64}$Zn at low energies}},}\ }\href
  {https://doi.org/10.1103/PhysRevC.95.015807} {\bibfield  {journal} {\bibinfo
  {journal} {Phys. Rev. C}\ }\textbf {\bibinfo {volume} {95}},\ \bibinfo {eid}
  {015807}}\BibitemShut {NoStop}%
\bibitem [{\citenamefont {{Mohr}}\ \emph {et~al.}(2013)\citenamefont {{Mohr}},
  \citenamefont {{Kiss}}, \citenamefont {{F{\"u}l{\"o}p}}, \citenamefont
  {{Galaviz}}, \citenamefont {{Gy{\"u}rky}},\ and\ \citenamefont
  {{Somorjai}}}]{Mohr.Kiss.ea:2013}%
  \BibitemOpen
  \bibfield  {author} {\bibinfo {author} {\bibnamefont {{Mohr}}, \bibfnamefont
  {P}}, \bibinfo {author} {\bibfnamefont {G.~G.}\ \bibnamefont {{Kiss}}},
  \bibinfo {author} {\bibfnamefont {Z.}~\bibnamefont {{F{\"u}l{\"o}p}}},
  \bibinfo {author} {\bibfnamefont {D.}~\bibnamefont {{Galaviz}}}, \bibinfo
  {author} {\bibfnamefont {G.}~\bibnamefont {{Gy{\"u}rky}}}, and\ \bibinfo
  {author} {\bibfnamefont {E.}~\bibnamefont {{Somorjai}}}} (\bibinfo {year}
  {2013}),\ \bibfield  {title} {\enquote {\bibinfo {title} {{Elastic alpha
  scattering experiments and the alpha-nucleus optical potential at low
  energies}},}\ }\href {https://doi.org/10.1016/j.adt.2012.10.003} {\bibfield
  {journal} {\bibinfo  {journal} {At. Data Nucl. Data Tables}\ }\textbf
  {\bibinfo {volume} {99}},\ \bibinfo {pages} {651--679}}\BibitemShut {NoStop}%
\bibitem [{\citenamefont {{M{\"o}ller}}\ \emph
  {et~al.}(2012{\natexlab{a}})\citenamefont {{M{\"o}ller}}, \citenamefont
  {{Myers}}, \citenamefont {{Sagawa}},\ and\ \citenamefont
  {{Yoshida}}}]{moeller12}%
  \BibitemOpen
  \bibfield  {author} {\bibinfo {author} {\bibnamefont {{M{\"o}ller}},
  \bibfnamefont {P}}, \bibinfo {author} {\bibfnamefont {W.~D.}\ \bibnamefont
  {{Myers}}}, \bibinfo {author} {\bibfnamefont {H.}~\bibnamefont {{Sagawa}}},
  and\ \bibinfo {author} {\bibfnamefont {S.}~\bibnamefont {{Yoshida}}}}
  (\bibinfo {year} {2012}{\natexlab{a}}),\ \bibfield  {title} {\enquote
  {\bibinfo {title} {{New Finite-Range Droplet Mass Model and Equation-of-State
  Parameters}},}\ }\href {https://doi.org/10.1103/PhysRevLett.108.052501}
  {\bibfield  {journal} {\bibinfo  {journal} {Phys. Rev. Lett.}\ }\textbf
  {\bibinfo {volume} {108}},\ \bibinfo {eid} {052501}}\BibitemShut {NoStop}%
\bibitem [{\citenamefont {M{\"o}ller}\ \emph {et~al.}(1997)\citenamefont
  {M{\"o}ller}, \citenamefont {Nix},\ and\ \citenamefont
  {Kratz}}]{Moeller.Nix.Kratz:1997}%
  \BibitemOpen
  \bibfield  {author} {\bibinfo {author} {\bibnamefont {M{\"o}ller},
  \bibfnamefont {P}}, \bibinfo {author} {\bibfnamefont {J.~R.}\ \bibnamefont
  {Nix}}, and\ \bibinfo {author} {\bibfnamefont {K.-L.}\ \bibnamefont {Kratz}}}
  (\bibinfo {year} {1997}),\ \bibfield  {title} {\enquote {\bibinfo {title}
  {{Nuclear Properties for Astrophysical and Radioactive-Ion Beam
  Applications}},}\ }\href {https://doi.org/10.1006/adnd.1997.0746} {\bibfield
  {journal} {\bibinfo  {journal} {At. Data Nucl. Data Tables}\ }\textbf
  {\bibinfo {volume} {66}},\ \bibinfo {pages} {131--343}}\BibitemShut {NoStop}%
\bibitem [{\citenamefont {{M{\"o}ller}}\ \emph {et~al.}(1995)\citenamefont
  {{M{\"o}ller}}, \citenamefont {{Nix}}, \citenamefont {{Myers}},\ and\
  \citenamefont {{Swiatecki}}}]{moeller95}%
  \BibitemOpen
  \bibfield  {author} {\bibinfo {author} {\bibnamefont {{M{\"o}ller}},
  \bibfnamefont {P}}, \bibinfo {author} {\bibfnamefont {J.~R.}\ \bibnamefont
  {{Nix}}}, \bibinfo {author} {\bibfnamefont {W.~D.}\ \bibnamefont {{Myers}}},
  and\ \bibinfo {author} {\bibfnamefont {W.~J.}\ \bibnamefont {{Swiatecki}}}}
  (\bibinfo {year} {1995}),\ \bibfield  {title} {\enquote {\bibinfo {title}
  {{Nuclear Ground-State Masses and Deformations}},}\ }\href
  {https://doi.org/10.1006/adnd.1995.1002} {\bibfield  {journal} {\bibinfo
  {journal} {At. Data Nucl. Data Tables}\ }\textbf {\bibinfo {volume} {59}},\
  \bibinfo {pages} {185}}\BibitemShut {NoStop}%
\bibitem [{\citenamefont {M\"oller}\ \emph {et~al.}(2003)\citenamefont
  {M\"oller}, \citenamefont {Pfeiffer},\ and\ \citenamefont
  {Kratz}}]{Moeller.Pfeiffer.Kratz:2003}%
  \BibitemOpen
  \bibfield  {author} {\bibinfo {author} {\bibnamefont {M\"oller},
  \bibfnamefont {P}}, \bibinfo {author} {\bibfnamefont {B.}~\bibnamefont
  {Pfeiffer}}, and\ \bibinfo {author} {\bibfnamefont {K.-L.}\ \bibnamefont
  {Kratz}}} (\bibinfo {year} {2003}),\ \bibfield  {title} {\enquote {\bibinfo
  {title} {New calculations of gross beta-decay properties for astrophysical
  applications: Speeding-up the classical r process},}\ }\href
  {https://doi.org/10.1103/PhysRevC.67.055802} {\bibfield  {journal} {\bibinfo
  {journal} {Phys. Rev. C}\ }\textbf {\bibinfo {volume} {67}},\ \bibinfo {eid}
  {055802}}\BibitemShut {NoStop}%
\bibitem [{\citenamefont {{M{\"o}ller}}\ \emph
  {et~al.}(2012{\natexlab{b}})\citenamefont {{M{\"o}ller}}, \citenamefont
  {{Sierk}}, \citenamefont {{Bengtsson}}, \citenamefont {{Sagawa}},\ and\
  \citenamefont {{Ichikawa}}}]{Moeller.Sierk.ea:2012}%
  \BibitemOpen
  \bibfield  {author} {\bibinfo {author} {\bibnamefont {{M{\"o}ller}},
  \bibfnamefont {P}}, \bibinfo {author} {\bibfnamefont {A.~J.}\ \bibnamefont
  {{Sierk}}}, \bibinfo {author} {\bibfnamefont {R.}~\bibnamefont
  {{Bengtsson}}}, \bibinfo {author} {\bibfnamefont {H.}~\bibnamefont
  {{Sagawa}}}, and\ \bibinfo {author} {\bibfnamefont {T.}~\bibnamefont
  {{Ichikawa}}}} (\bibinfo {year} {2012}{\natexlab{b}}),\ \bibfield  {title}
  {\enquote {\bibinfo {title} {{Nuclear shape isomers}},}\ }\href
  {https://doi.org/10.1016/j.adt.2010.09.002} {\bibfield  {journal} {\bibinfo
  {journal} {At. Data Nucl. Data Tables}\ }\textbf {\bibinfo {volume} {98}},\
  \bibinfo {pages} {149--300}}\BibitemShut {NoStop}%
\bibitem [{\citenamefont {{M{\"o}ller}}\ \emph {et~al.}(2015)\citenamefont
  {{M{\"o}ller}}, \citenamefont {{Sierk}}, \citenamefont {{Ichikawa}},
  \citenamefont {{Iwamoto}},\ and\ \citenamefont
  {{Mumpower}}}]{Moeller.Sierk.ea:2015}%
  \BibitemOpen
  \bibfield  {author} {\bibinfo {author} {\bibnamefont {{M{\"o}ller}},
  \bibfnamefont {P}}, \bibinfo {author} {\bibfnamefont {A.~J.}\ \bibnamefont
  {{Sierk}}}, \bibinfo {author} {\bibfnamefont {T.}~\bibnamefont {{Ichikawa}}},
  \bibinfo {author} {\bibfnamefont {A.}~\bibnamefont {{Iwamoto}}}, and\
  \bibinfo {author} {\bibfnamefont {M.}~\bibnamefont {{Mumpower}}}} (\bibinfo
  {year} {2015}),\ \bibfield  {title} {\enquote {\bibinfo {title} {{Fission
  barriers at the end of the chart of the nuclides}},}\ }\href
  {https://doi.org/10.1103/PhysRevC.91.024310} {\bibfield  {journal} {\bibinfo
  {journal} {Phys. Rev. C}\ }\textbf {\bibinfo {volume} {91}},\ \bibinfo {eid}
  {024310}}\BibitemShut {NoStop}%
\bibitem [{\citenamefont {{M{\"o}ller}}\ \emph {et~al.}(2016)\citenamefont
  {{M{\"o}ller}}, \citenamefont {{Sierk}}, \citenamefont {{Ichikawa}},\ and\
  \citenamefont {{Sagawa}}}]{moeller15}%
  \BibitemOpen
  \bibfield  {author} {\bibinfo {author} {\bibnamefont {{M{\"o}ller}},
  \bibfnamefont {P}}, \bibinfo {author} {\bibfnamefont {A.~J.}\ \bibnamefont
  {{Sierk}}}, \bibinfo {author} {\bibfnamefont {T.}~\bibnamefont {{Ichikawa}}},
  and\ \bibinfo {author} {\bibfnamefont {H.}~\bibnamefont {{Sagawa}}}}
  (\bibinfo {year} {2016}),\ \bibfield  {title} {\enquote {\bibinfo {title}
  {{Nuclear ground-state masses and deformations: FRDM(2012)}},}\ }\href
  {https://doi.org/10.1016/j.adt.2015.10.002} {\bibfield  {journal} {\bibinfo
  {journal} {At. Data Nucl. Data Tables}\ }\textbf {\bibinfo {volume} {109}},\
  \bibinfo {pages} {1--204}}\BibitemShut {NoStop}%
\bibitem [{\citenamefont {{Montes}}\ \emph {et~al.}(2006)\citenamefont
  {{Montes}} \emph {et~al.}}]{Montes}%
  \BibitemOpen
  \bibfield  {author} {\bibinfo {author} {\bibnamefont {{Montes}},
  \bibfnamefont {F}},  \emph {et~al.}} (\bibinfo {year} {2006}),\ \bibfield
  {title} {\enquote {\bibinfo {title} {{{$\beta$}-decay half-lives and
  {$\beta$}-delayed neutron emission probabilities for neutron rich nuclei
  close to the N=82r-process path}},}\ }\href
  {https://doi.org/10.1103/PhysRevC.73.035801} {\bibfield  {journal} {\bibinfo
  {journal} {Phys. Rev. C}\ }\textbf {\bibinfo {volume} {73}},\ \bibinfo {eid}
  {035801}}\BibitemShut {NoStop}%
\bibitem [{\citenamefont {{Montes}}\ \emph {et~al.}(2016)\citenamefont
  {{Montes}}, \citenamefont {{Ramirez-Ruiz}}, \citenamefont {{Naiman}},
  \citenamefont {{Shen}},\ and\ \citenamefont
  {{Lee}}}]{Montes.Ramirez.ea:2016}%
  \BibitemOpen
  \bibfield  {author} {\bibinfo {author} {\bibnamefont {{Montes}},
  \bibfnamefont {G}}, \bibinfo {author} {\bibfnamefont {E.}~\bibnamefont
  {{Ramirez-Ruiz}}}, \bibinfo {author} {\bibfnamefont {J.}~\bibnamefont
  {{Naiman}}}, \bibinfo {author} {\bibfnamefont {S.}~\bibnamefont {{Shen}}},
  and\ \bibinfo {author} {\bibfnamefont {W.~H.}\ \bibnamefont {{Lee}}}}
  (\bibinfo {year} {2016}),\ \bibfield  {title} {\enquote {\bibinfo {title}
  {{Transport and Mixing of r-process Elements in Neutron Star Binary Merger
  Blast Waves}},}\ }\href {https://doi.org/10.3847/0004-637X/830/1/12}
  {\bibfield  {journal} {\bibinfo  {journal} {Astrophys. J.}\ }\textbf
  {\bibinfo {volume} {830}},\ \bibinfo {eid} {12}}\BibitemShut {NoStop}%
\bibitem [{\citenamefont {Mooley}\ \emph {et~al.}(2018)\citenamefont {Mooley}
  \emph {et~al.}}]{Mooley.Nakar.ea:2018}%
  \BibitemOpen
  \bibfield  {author} {\bibinfo {author} {\bibnamefont {Mooley}, \bibfnamefont
  {K~P}},  \emph {et~al.}} (\bibinfo {year} {2018}),\ \bibfield  {title}
  {\enquote {\bibinfo {title} {{A mildly relativistic wide-angle outflow in the
  neutron-star merger event GW170817}},}\ }\href
  {https://doi.org/10.1038/nature25452} {\bibfield  {journal} {\bibinfo
  {journal} {Nature}\ }\textbf {\bibinfo {volume} {554}},\ \bibinfo {pages}
  {207--210}}\BibitemShut {NoStop}%
\bibitem [{\citenamefont {{Moore}}(1971)}]{moore71}%
  \BibitemOpen
  \bibfield  {author} {\bibinfo {author} {\bibnamefont {{Moore}}, \bibfnamefont
  {C~E}}} (\bibinfo {year} {1971}),\ \href@noop {} {\emph {\bibinfo {title}
  {{Atomic Energy Levels, Nat. Ref. Data Ser. 60}}}}\ (\bibinfo  {publisher}
  {National Bureau of Standards},\ \bibinfo {address} {Washington: US
  GPO})\BibitemShut {NoStop}%
\bibitem [{\citenamefont {{Moriya}}\ \emph {et~al.}(2014)\citenamefont
  {{Moriya}}, \citenamefont {{Tominaga}}, \citenamefont {{Langer}},
  \citenamefont {{Nomoto}}, \citenamefont {{Blinnikov}},\ and\ \citenamefont
  {{Sorokina}}}]{Moriya.Tominaga.ea:2014}%
  \BibitemOpen
  \bibfield  {author} {\bibinfo {author} {\bibnamefont {{Moriya}},
  \bibfnamefont {Takashi~J}}, \bibinfo {author} {\bibfnamefont {Nozomu}\
  \bibnamefont {{Tominaga}}}, \bibinfo {author} {\bibfnamefont {Norbert}\
  \bibnamefont {{Langer}}}, \bibinfo {author} {\bibfnamefont {Ken'ichi}\
  \bibnamefont {{Nomoto}}}, \bibinfo {author} {\bibfnamefont {Sergei~I.}\
  \bibnamefont {{Blinnikov}}}, and\ \bibinfo {author} {\bibfnamefont
  {Elena~I.}\ \bibnamefont {{Sorokina}}}} (\bibinfo {year} {2014}),\ \bibfield
  {title} {\enquote {\bibinfo {title} {{Electron-capture supernovae exploding
  within their progenitor wind}},}\ }\href
  {https://doi.org/10.1051/0004-6361/201424264} {\bibfield  {journal} {\bibinfo
   {journal} {Astron. \& Astrophys.}\ }\textbf {\bibinfo {volume} {569}},\
  \bibinfo {eid} {A57}}\BibitemShut {NoStop}%
\bibitem [{\citenamefont {Most}\ \emph {et~al.}(2018)\citenamefont {Most},
  \citenamefont {Weih}, \citenamefont {Rezzolla},\ and\ \citenamefont
  {Schaffner-Bielich}}]{Most.Weih.ea:2018}%
  \BibitemOpen
  \bibfield  {author} {\bibinfo {author} {\bibnamefont {Most}, \bibfnamefont
  {Elias~R}}, \bibinfo {author} {\bibfnamefont {Lukas~R.}\ \bibnamefont
  {Weih}}, \bibinfo {author} {\bibfnamefont {Luciano}\ \bibnamefont
  {Rezzolla}}, and\ \bibinfo {author} {\bibfnamefont {J\"urgen}\ \bibnamefont
  {Schaffner-Bielich}}} (\bibinfo {year} {2018}),\ \bibfield  {title} {\enquote
  {\bibinfo {title} {{New Constraints on Radii and Tidal Deformabilities of
  Neutron Stars from GW170817}},}\ }\href
  {https://doi.org/10.1103/PhysRevLett.120.261103} {\bibfield  {journal}
  {\bibinfo  {journal} {Phys. Rev. Lett.}\ }\textbf {\bibinfo {volume} {120}},\
  \bibinfo {pages} {261103}}\BibitemShut {NoStop}%
\bibitem [{\citenamefont {{M{\"o}sta}}\ \emph {et~al.}(2015)\citenamefont
  {{M{\"o}sta}}, \citenamefont {{Ott}}, \citenamefont {{Radice}}, \citenamefont
  {{Roberts}}, \citenamefont {{Schnetter}},\ and\ \citenamefont
  {{Haas}}}]{Moesta.ea:2015}%
  \BibitemOpen
  \bibfield  {author} {\bibinfo {author} {\bibnamefont {{M{\"o}sta}},
  \bibfnamefont {P}}, \bibinfo {author} {\bibfnamefont {C.~D.}\ \bibnamefont
  {{Ott}}}, \bibinfo {author} {\bibfnamefont {D.}~\bibnamefont {{Radice}}},
  \bibinfo {author} {\bibfnamefont {L.~F.}\ \bibnamefont {{Roberts}}}, \bibinfo
  {author} {\bibfnamefont {E.}~\bibnamefont {{Schnetter}}}, and\ \bibinfo
  {author} {\bibfnamefont {R.}~\bibnamefont {{Haas}}}} (\bibinfo {year}
  {2015}),\ \bibfield  {title} {\enquote {\bibinfo {title} {{A large-scale
  dynamo and magnetoturbulence in rapidly rotating core-collapse
  supernovae}},}\ }\href {https://doi.org/10.1038/nature15755} {\bibfield
  {journal} {\bibinfo  {journal} {Nature}\ }\textbf {\bibinfo {volume} {528}},\
  \bibinfo {pages} {376--379}}\BibitemShut {NoStop}%
\bibitem [{\citenamefont {{M{\"o}sta}}\ \emph {et~al.}(2014)\citenamefont
  {{M{\"o}sta}}, \citenamefont {{Richers}}, \citenamefont {{Ott}},
  \citenamefont {{Haas}}, \citenamefont {{Piro}}, \citenamefont {{Boydstun}},
  \citenamefont {{Abdikamalov}}, \citenamefont {{Reisswig}},\ and\
  \citenamefont {{Schnetter}}}]{Moesta.Richers.ea:2014}%
  \BibitemOpen
  \bibfield  {author} {\bibinfo {author} {\bibnamefont {{M{\"o}sta}},
  \bibfnamefont {P}}, \bibinfo {author} {\bibfnamefont {S.}~\bibnamefont
  {{Richers}}}, \bibinfo {author} {\bibfnamefont {C.~D.}\ \bibnamefont
  {{Ott}}}, \bibinfo {author} {\bibfnamefont {R.}~\bibnamefont {{Haas}}},
  \bibinfo {author} {\bibfnamefont {A.~L.}\ \bibnamefont {{Piro}}}, \bibinfo
  {author} {\bibfnamefont {K.}~\bibnamefont {{Boydstun}}}, \bibinfo {author}
  {\bibfnamefont {E.}~\bibnamefont {{Abdikamalov}}}, \bibinfo {author}
  {\bibfnamefont {C.}~\bibnamefont {{Reisswig}}}, and\ \bibinfo {author}
  {\bibfnamefont {E.}~\bibnamefont {{Schnetter}}}} (\bibinfo {year} {2014}),\
  \bibfield  {title} {\enquote {\bibinfo {title} {Magnetorotational
  core-collapse supernovae in three dimensions},}\ }\href
  {https://doi.org/10.1088/2041-8205/785/2/L29} {\bibfield  {journal} {\bibinfo
   {journal} {Astrophys. J.}\ }\textbf {\bibinfo {volume} {785}},\ \bibinfo
  {eid} {L29}}\BibitemShut {NoStop}%
\bibitem [{\citenamefont {{M{\"o}sta}}\ \emph {et~al.}(2018)\citenamefont
  {{M{\"o}sta}}, \citenamefont {{Roberts}}, \citenamefont {{Halevi}},
  \citenamefont {{Ott}}, \citenamefont {{Lippuner}}, \citenamefont {{Haas}},\
  and\ \citenamefont {{Schnetter}}}]{moesta18}%
  \BibitemOpen
  \bibfield  {author} {\bibinfo {author} {\bibnamefont {{M{\"o}sta}},
  \bibfnamefont {P}}, \bibinfo {author} {\bibfnamefont {L.~F.}\ \bibnamefont
  {{Roberts}}}, \bibinfo {author} {\bibfnamefont {G.}~\bibnamefont {{Halevi}}},
  \bibinfo {author} {\bibfnamefont {C.~D.}\ \bibnamefont {{Ott}}}, \bibinfo
  {author} {\bibfnamefont {J.}~\bibnamefont {{Lippuner}}}, \bibinfo {author}
  {\bibfnamefont {R.}~\bibnamefont {{Haas}}}, and\ \bibinfo {author}
  {\bibfnamefont {E.}~\bibnamefont {{Schnetter}}}} (\bibinfo {year} {2018}),\
  \bibfield  {title} {\enquote {\bibinfo {title} {{r-process Nucleosynthesis
  from Three-dimensional Magnetorotational Core-collapse Supernovae}},}\ }\href
  {https://doi.org/10.3847/1538-4357/aad6ec} {\bibfield  {journal} {\bibinfo
  {journal} {Astrophys. J.}\ }\textbf {\bibinfo {volume} {864}},\ \bibinfo
  {eid} {171}}\BibitemShut {NoStop}%
\bibitem [{\citenamefont {{Mueller}}(1986)}]{Mueller:1986}%
  \BibitemOpen
  \bibfield  {author} {\bibinfo {author} {\bibnamefont {{Mueller}},
  \bibfnamefont {E}}} (\bibinfo {year} {1986}),\ \bibfield  {title} {\enquote
  {\bibinfo {title} {{Nuclear-reaction networks and stellar evolution codes -
  The coupling of composition changes and energy release in explosive nuclear
  burning}},}\ }\href@noop {} {\bibfield  {journal} {\bibinfo  {journal}
  {Astron. \& Astrophys.}\ }\textbf {\bibinfo {volume} {162}},\ \bibinfo
  {pages} {103--108}}\BibitemShut {NoStop}%
\bibitem [{\citenamefont {{M{\"u}ller}}(2016)}]{Mueller:2016}%
  \BibitemOpen
  \bibfield  {author} {\bibinfo {author} {\bibnamefont {{M{\"u}ller}},
  \bibfnamefont {B}}} (\bibinfo {year} {2016}),\ \bibfield  {title} {\enquote
  {\bibinfo {title} {{The Status of Multi-Dimensional Core-Collapse Supernova
  Models}},}\ }\href {https://doi.org/10.1017/pasa.2016.40} {\bibfield
  {journal} {\bibinfo  {journal} {Publ. Astron. Soc. Austr.}\ }\textbf
  {\bibinfo {volume} {33}},\ \bibinfo {pages} {e048}}\BibitemShut {NoStop}%
\bibitem [{\citenamefont {{M{\"u}ller}}\ \emph {et~al.}(2018)\citenamefont
  {{M{\"u}ller}}, \citenamefont {{Gay}}, \citenamefont {{Heger}}, \citenamefont
  {{Tauris}},\ and\ \citenamefont {{Sim}}}]{Mueller18}%
  \BibitemOpen
  \bibfield  {author} {\bibinfo {author} {\bibnamefont {{M{\"u}ller}},
  \bibfnamefont {B}}, \bibinfo {author} {\bibfnamefont {D.~W.}\ \bibnamefont
  {{Gay}}}, \bibinfo {author} {\bibfnamefont {A.}~\bibnamefont {{Heger}}},
  \bibinfo {author} {\bibfnamefont {T.~M.}\ \bibnamefont {{Tauris}}}, and\
  \bibinfo {author} {\bibfnamefont {S.~A.}\ \bibnamefont {{Sim}}}} (\bibinfo
  {year} {2018}),\ \bibfield  {title} {\enquote {\bibinfo {title}
  {{Multidimensional simulations of ultrastripped supernovae to shock
  breakout}},}\ }\href {https://doi.org/10.1093/mnras/sty1683} {\bibfield
  {journal} {\bibinfo  {journal} {Mon. Not. Roy. Astron. Soc.}\ }\textbf
  {\bibinfo {volume} {479}},\ \bibinfo {pages} {3675--3689}}\BibitemShut
  {NoStop}%
\bibitem [{\citenamefont {{M{\"u}ller}}\ \emph {et~al.}(2019)\citenamefont
  {{M{\"u}ller}}, \citenamefont {{Tauris}}, \citenamefont {{Heger}},
  \citenamefont {{Banerjee}}, \citenamefont {{Qian}}, \citenamefont {{Powell}},
  \citenamefont {{Chan}}, \citenamefont {{Gay}},\ and\ \citenamefont
  {{Langer}}}]{MuellerB19}%
  \BibitemOpen
  \bibfield  {author} {\bibinfo {author} {\bibnamefont {{M{\"u}ller}},
  \bibfnamefont {Bernhard}}, \bibinfo {author} {\bibfnamefont {Thomas~M.}\
  \bibnamefont {{Tauris}}}, \bibinfo {author} {\bibfnamefont {Alexander}\
  \bibnamefont {{Heger}}}, \bibinfo {author} {\bibfnamefont {Projjwal}\
  \bibnamefont {{Banerjee}}}, \bibinfo {author} {\bibfnamefont {Yong-Zhong}\
  \bibnamefont {{Qian}}}, \bibinfo {author} {\bibfnamefont {Jade}\ \bibnamefont
  {{Powell}}}, \bibinfo {author} {\bibfnamefont {Conrad}\ \bibnamefont
  {{Chan}}}, \bibinfo {author} {\bibfnamefont {Daniel~W.}\ \bibnamefont
  {{Gay}}}, and\ \bibinfo {author} {\bibfnamefont {Norbert}\ \bibnamefont
  {{Langer}}}} (\bibinfo {year} {2019}),\ \bibfield  {title} {\enquote
  {\bibinfo {title} {{Three-dimensional simulations of neutrino-driven
  core-collapse supernovae from low-mass single and binary star
  progenitors}},}\ }\href {https://doi.org/10.1093/mnras/stz216} {\bibfield
  {journal} {\bibinfo  {journal} {Mon. Not. Roy. Astron. Soc.}\ }\textbf
  {\bibinfo {volume} {484}},\ \bibinfo {pages} {3307--3324}}\BibitemShut
  {NoStop}%
\bibitem [{\citenamefont {{Mumpower}}\ \emph {et~al.}(2020)\citenamefont
  {{Mumpower}}, \citenamefont {{Jaffke}}, \citenamefont {{Verriere}},\ and\
  \citenamefont {{Randrup}}}]{Mumpower.Jaffke.ea:2019}%
  \BibitemOpen
  \bibfield  {author} {\bibinfo {author} {\bibnamefont {{Mumpower}},
  \bibfnamefont {M~R}}, \bibinfo {author} {\bibfnamefont {P.}~\bibnamefont
  {{Jaffke}}}, \bibinfo {author} {\bibfnamefont {M.}~\bibnamefont
  {{Verriere}}}, and\ \bibinfo {author} {\bibfnamefont {J.}~\bibnamefont
  {{Randrup}}}} (\bibinfo {year} {2020}),\ \bibfield  {title} {\enquote
  {\bibinfo {title} {{Primary fission fragment mass yields across the chart of
  nuclides}},}\ }\href {https://doi.org/10.1103/PhysRevC.101.054607} {\bibfield
   {journal} {\bibinfo  {journal} {Phys. Rev. C}\ }\textbf {\bibinfo {volume}
  {101}}~(\bibinfo {number} {5}),\ \bibinfo {eid} {054607}}\BibitemShut
  {NoStop}%
\bibitem [{\citenamefont {{Mumpower}}\ \emph {et~al.}(2018)\citenamefont
  {{Mumpower}}, \citenamefont {{Kawano}}, \citenamefont {{Sprouse}},
  \citenamefont {{Vassh}}, \citenamefont {{Holmbeck}}, \citenamefont
  {{Surman}},\ and\ \citenamefont {{M{\"o}ller}}}]{Mumpower.Kawano.ea:2018}%
  \BibitemOpen
  \bibfield  {author} {\bibinfo {author} {\bibnamefont {{Mumpower}},
  \bibfnamefont {M~R}}, \bibinfo {author} {\bibfnamefont {T.}~\bibnamefont
  {{Kawano}}}, \bibinfo {author} {\bibfnamefont {T.~M.}\ \bibnamefont
  {{Sprouse}}}, \bibinfo {author} {\bibfnamefont {N.}~\bibnamefont {{Vassh}}},
  \bibinfo {author} {\bibfnamefont {E.~M.}\ \bibnamefont {{Holmbeck}}},
  \bibinfo {author} {\bibfnamefont {R.}~\bibnamefont {{Surman}}}, and\ \bibinfo
  {author} {\bibfnamefont {P.}~\bibnamefont {{M{\"o}ller}}}} (\bibinfo {year}
  {2018}),\ \bibfield  {title} {\enquote {\bibinfo {title}
  {{{\ensuremath{\beta}}-delayed Fission in r-process Nucleosynthesis}},}\
  }\href {https://doi.org/10.3847/1538-4357/aaeaca} {\bibfield  {journal}
  {\bibinfo  {journal} {Astrophys. J.}\ }\textbf {\bibinfo {volume} {869}},\
  \bibinfo {eid} {14}}\BibitemShut {NoStop}%
\bibitem [{\citenamefont {{Mumpower}}\ \emph
  {et~al.}(2017{\natexlab{a}})\citenamefont {{Mumpower}}, \citenamefont
  {{Kawano}}, \citenamefont {{Ullmann}}, \citenamefont {{Krti{\v c}ka}},\ and\
  \citenamefont {{Sprouse}}}]{Mumpower.Kawano.ea:2017}%
  \BibitemOpen
  \bibfield  {author} {\bibinfo {author} {\bibnamefont {{Mumpower}},
  \bibfnamefont {M~R}}, \bibinfo {author} {\bibfnamefont {T.}~\bibnamefont
  {{Kawano}}}, \bibinfo {author} {\bibfnamefont {J.~L.}\ \bibnamefont
  {{Ullmann}}}, \bibinfo {author} {\bibfnamefont {M.}~\bibnamefont {{Krti{\v
  c}ka}}}, and\ \bibinfo {author} {\bibfnamefont {T.~M.}\ \bibnamefont
  {{Sprouse}}}} (\bibinfo {year} {2017}{\natexlab{a}}),\ \bibfield  {title}
  {\enquote {\bibinfo {title} {{Estimation of M1 scissors mode strength for
  deformed nuclei in the medium- to heavy-mass region by statistical
  Hauser-Feshbach model calculations}},}\ }\href
  {https://doi.org/10.1103/PhysRevC.96.024612} {\bibfield  {journal} {\bibinfo
  {journal} {Phys. Rev. C}\ }\textbf {\bibinfo {volume} {96}},\ \bibinfo {eid}
  {024612}}\BibitemShut {NoStop}%
\bibitem [{\citenamefont {{Mumpower}}\ \emph
  {et~al.}(2017{\natexlab{b}})\citenamefont {{Mumpower}}, \citenamefont
  {{McLaughlin}}, \citenamefont {{Surman}},\ and\ \citenamefont
  {{Steiner}}}]{Mumpower2017}%
  \BibitemOpen
  \bibfield  {author} {\bibinfo {author} {\bibnamefont {{Mumpower}},
  \bibfnamefont {M~R}}, \bibinfo {author} {\bibfnamefont {G.~C.}\ \bibnamefont
  {{McLaughlin}}}, \bibinfo {author} {\bibfnamefont {R.}~\bibnamefont
  {{Surman}}}, and\ \bibinfo {author} {\bibfnamefont {A.~W.}\ \bibnamefont
  {{Steiner}}}} (\bibinfo {year} {2017}{\natexlab{b}}),\ \bibfield  {title}
  {\enquote {\bibinfo {title} {{Reverse engineering nuclear properties from
  rare earth abundances in the r process}},}\ }\href
  {https://doi.org/10.1088/1361-6471/44/3/034003} {\bibfield  {journal}
  {\bibinfo  {journal} {J. Phys. G: Nucl. Part. Phys.}\ }\textbf {\bibinfo
  {volume} {44}},\ \bibinfo {eid} {034003}}\BibitemShut {NoStop}%
\bibitem [{\citenamefont {{Mumpower}}\ \emph {et~al.}(2015)\citenamefont
  {{Mumpower}}, \citenamefont {{Surman}}, \citenamefont {{Fang}}, \citenamefont
  {{Beard}}, \citenamefont {{M{\"o}ller}}, \citenamefont {{Kawano}},\ and\
  \citenamefont {{Aprahamian}}}]{Mumpower2015}%
  \BibitemOpen
  \bibfield  {author} {\bibinfo {author} {\bibnamefont {{Mumpower}},
  \bibfnamefont {M~R}}, \bibinfo {author} {\bibfnamefont {R.}~\bibnamefont
  {{Surman}}}, \bibinfo {author} {\bibfnamefont {D.-L.}\ \bibnamefont
  {{Fang}}}, \bibinfo {author} {\bibfnamefont {M.}~\bibnamefont {{Beard}}},
  \bibinfo {author} {\bibfnamefont {P.}~\bibnamefont {{M{\"o}ller}}}, \bibinfo
  {author} {\bibfnamefont {T.}~\bibnamefont {{Kawano}}}, and\ \bibinfo {author}
  {\bibfnamefont {A.}~\bibnamefont {{Aprahamian}}}} (\bibinfo {year} {2015}),\
  \bibfield  {title} {\enquote {\bibinfo {title} {{Impact of individual nuclear
  masses on r-process abundances}},}\ }\href
  {https://doi.org/10.1103/PhysRevC.92.035807} {\bibfield  {journal} {\bibinfo
  {journal} {Phys. Rev. C}\ }\textbf {\bibinfo {volume} {92}},\ \bibinfo {eid}
  {035807}}\BibitemShut {NoStop}%
\bibitem [{\citenamefont {{Mumpower}}\ \emph {et~al.}(2016)\citenamefont
  {{Mumpower}}, \citenamefont {{Surman}}, \citenamefont {{McLaughlin}},\ and\
  \citenamefont {{Aprahamian}}}]{Mumpower.Surman.ea:2016}%
  \BibitemOpen
  \bibfield  {author} {\bibinfo {author} {\bibnamefont {{Mumpower}},
  \bibfnamefont {M~R}}, \bibinfo {author} {\bibfnamefont {R.}~\bibnamefont
  {{Surman}}}, \bibinfo {author} {\bibfnamefont {G.~C.}\ \bibnamefont
  {{McLaughlin}}}, and\ \bibinfo {author} {\bibfnamefont {A.}~\bibnamefont
  {{Aprahamian}}}} (\bibinfo {year} {2016}),\ \bibfield  {title} {\enquote
  {\bibinfo {title} {{The impact of individual nuclear properties on r-process
  nucleosynthesis}},}\ }\href {https://doi.org/10.1016/j.ppnp.2015.09.001}
  {\bibfield  {journal} {\bibinfo  {journal} {Prog. Part. Nucl. Phys.}\
  }\textbf {\bibinfo {volume} {86}},\ \bibinfo {pages} {86--126}}\BibitemShut
  {NoStop}%
\bibitem [{\citenamefont {{Munson}}\ \emph {et~al.}(2018)\citenamefont
  {{Munson}} \emph {et~al.}}]{Munson.Siegl.ea:2018}%
  \BibitemOpen
  \bibfield  {author} {\bibinfo {author} {\bibnamefont {{Munson}},
  \bibfnamefont {J~M}},  \emph {et~al.}} (\bibinfo {year} {2018}),\ \bibfield
  {title} {\enquote {\bibinfo {title} {{Recoil-ion detection efficiency for
  complex {\ensuremath{\beta}} decays studied using the Beta-decay Paul
  Trap}},}\ }\href {https://doi.org/10.1016/j.nima.2018.04.063} {\bibfield
  {journal} {\bibinfo  {journal} {Nuclear Instruments and Methods in Physics
  Research A}\ }\textbf {\bibinfo {volume} {898}},\ \bibinfo {pages}
  {60--66}}\BibitemShut {NoStop}%
\bibitem [{\citenamefont {{M{\"u}nzel}}\ \emph {et~al.}(1981)\citenamefont
  {{M{\"u}nzel}}, \citenamefont {{Wollnik}}, \citenamefont {{Pfeiffer}},\ and\
  \citenamefont {{Jung}}}]{OstisII}%
  \BibitemOpen
  \bibfield  {author} {\bibinfo {author} {\bibnamefont {{M{\"u}nzel}},
  \bibfnamefont {J}}, \bibinfo {author} {\bibfnamefont {H.}~\bibnamefont
  {{Wollnik}}}, \bibinfo {author} {\bibfnamefont {B.}~\bibnamefont
  {{Pfeiffer}}}, and\ \bibinfo {author} {\bibfnamefont {G.}~\bibnamefont
  {{Jung}}}} (\bibinfo {year} {1981}),\ \bibfield  {title} {\enquote {\bibinfo
  {title} {{A high-temperature ion source for the on-line separator OSTIS}},}\
  }\href {https://doi.org/10.1016/0029-554X(81)90924-1} {\bibfield  {journal}
  {\bibinfo  {journal} {Nucl. Instruments Methods Phys. Res.}\ }\textbf
  {\bibinfo {volume} {186}},\ \bibinfo {pages} {343--347}}\BibitemShut
  {NoStop}%
\bibitem [{\citenamefont {{Mustafa}}\ \emph {et~al.}(1992)\citenamefont
  {{Mustafa}}, \citenamefont {{Blann}}, \citenamefont {{Ignatyuk}},\ and\
  \citenamefont {{Grimes}}}]{Mustafa.Blann.ea:1992}%
  \BibitemOpen
  \bibfield  {author} {\bibinfo {author} {\bibnamefont {{Mustafa}},
  \bibfnamefont {M~G}}, \bibinfo {author} {\bibfnamefont {M.}~\bibnamefont
  {{Blann}}}, \bibinfo {author} {\bibfnamefont {A.~V.}\ \bibnamefont
  {{Ignatyuk}}}, and\ \bibinfo {author} {\bibfnamefont {S.~M.}\ \bibnamefont
  {{Grimes}}}} (\bibinfo {year} {1992}),\ \bibfield  {title} {\enquote
  {\bibinfo {title} {{Nuclear level densities at high excitations}},}\ }\href
  {https://doi.org/10.1103/PhysRevC.45.1078} {\bibfield  {journal} {\bibinfo
  {journal} {Phys. Rev. C}\ }\textbf {\bibinfo {volume} {45}},\ \bibinfo
  {pages} {1078--1083}}\BibitemShut {NoStop}%
\bibitem [{\citenamefont {{Mustonen}}\ and\ \citenamefont
  {{Engel}}(2016)}]{Mustonen.Engel:2016}%
  \BibitemOpen
  \bibfield  {author} {\bibinfo {author} {\bibnamefont {{Mustonen}},
  \bibfnamefont {M~T}}, and\ \bibinfo {author} {\bibfnamefont {J.}~\bibnamefont
  {{Engel}}}} (\bibinfo {year} {2016}),\ \bibfield  {title} {\enquote {\bibinfo
  {title} {{Global description of {$\beta$}$^{-}$ decay in even-even nuclei
  with the axially-deformed Skyrme finite-amplitude method}},}\ }\href
  {https://doi.org/10.1103/PhysRevC.93.014304} {\bibfield  {journal} {\bibinfo
  {journal} {Phys. Rev. C}\ }\textbf {\bibinfo {volume} {93}},\ \bibinfo {eid}
  {014304}}\BibitemShut {NoStop}%
\bibitem [{\citenamefont {Myers}\ and\ \citenamefont
  {{\'S}wia\c{t}ecki}(1999)}]{Myers.Swiatecki:1999}%
  \BibitemOpen
  \bibfield  {author} {\bibinfo {author} {\bibnamefont {Myers}, \bibfnamefont
  {W~D}}, and\ \bibinfo {author} {\bibfnamefont {W.~J.}\ \bibnamefont
  {{\'S}wia\c{t}ecki}}} (\bibinfo {year} {1999}),\ \bibfield  {title} {\enquote
  {\bibinfo {title} {{Thomas-Fermi fission barriers}},}\ }\href
  {https://doi.org/10.1103/PhysRevC.60.014606} {\bibfield  {journal} {\bibinfo
  {journal} {Phys. Rev. C}\ }\textbf {\bibinfo {volume} {60}},\ \bibinfo {eid}
  {014606}}\BibitemShut {NoStop}%
\bibitem [{\citenamefont {{Nadyozhin}}\ and\ \citenamefont
  {{Panov}}(2007)}]{Nadyozhin.Panov:2007}%
  \BibitemOpen
  \bibfield  {author} {\bibinfo {author} {\bibnamefont {{Nadyozhin}},
  \bibfnamefont {D~K}}, and\ \bibinfo {author} {\bibfnamefont {I.~V.}\
  \bibnamefont {{Panov}}}} (\bibinfo {year} {2007}),\ \bibfield  {title}
  {\enquote {\bibinfo {title} {{Weak r-process component as a result of the
  neutrino interaction with the helium shell of a supernova}},}\ }\href
  {https://doi.org/10.1134/S1063773707060035} {\bibfield  {journal} {\bibinfo
  {journal} {Astronomy Letters}\ }\textbf {\bibinfo {volume} {33}},\ \bibinfo
  {pages} {385--389}}\BibitemShut {NoStop}%
\bibitem [{\citenamefont {{Nagataki}}(2011)}]{nagataki11}%
  \BibitemOpen
  \bibfield  {author} {\bibinfo {author} {\bibnamefont {{Nagataki}},
  \bibfnamefont {S}}} (\bibinfo {year} {2011}),\ \bibfield  {title} {\enquote
  {\bibinfo {title} {{Grb-Sn Connection:. Central Engine of Long GRBs and
  Explosive Nucleosynthesis}},}\ }\href
  {https://doi.org/10.1142/S0218271811020032} {\bibfield  {journal} {\bibinfo
  {journal} {International Journal of Modern Physics D}\ }\textbf {\bibinfo
  {volume} {20}},\ \bibinfo {pages} {1975--1978}}\BibitemShut {NoStop}%
\bibitem [{\citenamefont {{Nagataki}}\ \emph {et~al.}(2007)\citenamefont
  {{Nagataki}}, \citenamefont {{Takahashi}}, \citenamefont {{Mizuta}},\ and\
  \citenamefont {{Takiwaki}}}]{nagataki07}%
  \BibitemOpen
  \bibfield  {author} {\bibinfo {author} {\bibnamefont {{Nagataki}},
  \bibfnamefont {S}}, \bibinfo {author} {\bibfnamefont {R.}~\bibnamefont
  {{Takahashi}}}, \bibinfo {author} {\bibfnamefont {A.}~\bibnamefont
  {{Mizuta}}}, and\ \bibinfo {author} {\bibfnamefont {T.}~\bibnamefont
  {{Takiwaki}}}} (\bibinfo {year} {2007}),\ \bibfield  {title} {\enquote
  {\bibinfo {title} {{Numerical Study of Gamma-Ray Burst Jet Formation in
  Collapsars}},}\ }\href {https://doi.org/10.1086/512057} {\bibfield  {journal}
  {\bibinfo  {journal} {Astrophys. J.}\ }\textbf {\bibinfo {volume} {659}},\
  \bibinfo {pages} {512--529}}\BibitemShut {NoStop}%
\bibitem [{\citenamefont {{Nakada}}\ and\ \citenamefont
  {{Alhassid}}(1997)}]{Nakada.Alhassid:1997}%
  \BibitemOpen
  \bibfield  {author} {\bibinfo {author} {\bibnamefont {{Nakada}},
  \bibfnamefont {H}}, and\ \bibinfo {author} {\bibfnamefont {Y.}~\bibnamefont
  {{Alhassid}}}} (\bibinfo {year} {1997}),\ \bibfield  {title} {\enquote
  {\bibinfo {title} {{Total and Parity-Projected Level Densities of Iron-Region
  Nuclei in the Auxiliary Fields Monte Carlo Shell Model}},}\ }\href
  {https://doi.org/10.1103/PhysRevLett.79.2939} {\bibfield  {journal} {\bibinfo
   {journal} {Phys. Rev. Lett.}\ }\textbf {\bibinfo {volume} {79}},\ \bibinfo
  {pages} {2939--2942}}\BibitemShut {NoStop}%
\bibitem [{\citenamefont {{Nakamura}}\ \emph {et~al.}(2015)\citenamefont
  {{Nakamura}}, \citenamefont {{Kajino}}, \citenamefont {{Mathews}},
  \citenamefont {{Sato}},\ and\ \citenamefont
  {{Harikae}}}]{Nakamura.Kajino.ea:2015}%
  \BibitemOpen
  \bibfield  {author} {\bibinfo {author} {\bibnamefont {{Nakamura}},
  \bibfnamefont {K}}, \bibinfo {author} {\bibfnamefont {T.}~\bibnamefont
  {{Kajino}}}, \bibinfo {author} {\bibfnamefont {G.~J.}\ \bibnamefont
  {{Mathews}}}, \bibinfo {author} {\bibfnamefont {S.}~\bibnamefont {{Sato}}},
  and\ \bibinfo {author} {\bibfnamefont {S.}~\bibnamefont {{Harikae}}}}
  (\bibinfo {year} {2015}),\ \bibfield  {title} {\enquote {\bibinfo {title}
  {{r-process nucleosynthesis in the MHD+neutrino-heated collapsar jet}},}\
  }\href {https://doi.org/10.1051/0004-6361/201526110} {\bibfield  {journal}
  {\bibinfo  {journal} {Astron. \& Astrophys.}\ }\textbf {\bibinfo {volume}
  {582}},\ \bibinfo {eid} {A34}}\BibitemShut {NoStop}%
\bibitem [{\citenamefont {{Nakamura}}\ \emph {et~al.}(2001)\citenamefont
  {{Nakamura}}, \citenamefont {{Umeda}}, \citenamefont {{Iwamoto}},
  \citenamefont {{Nomoto}}, \citenamefont {{Hashimoto}}, \citenamefont
  {{Hix}},\ and\ \citenamefont {{Thielemann}}}]{Nakamura.Umeda.ea:2001}%
  \BibitemOpen
  \bibfield  {author} {\bibinfo {author} {\bibnamefont {{Nakamura}},
  \bibfnamefont {T}}, \bibinfo {author} {\bibfnamefont {H.}~\bibnamefont
  {{Umeda}}}, \bibinfo {author} {\bibfnamefont {K.}~\bibnamefont {{Iwamoto}}},
  \bibinfo {author} {\bibfnamefont {K.}~\bibnamefont {{Nomoto}}}, \bibinfo
  {author} {\bibfnamefont {M.-a.}\ \bibnamefont {{Hashimoto}}}, \bibinfo
  {author} {\bibfnamefont {W.~R.}\ \bibnamefont {{Hix}}}, and\ \bibinfo
  {author} {\bibfnamefont {F.-K.}\ \bibnamefont {{Thielemann}}}} (\bibinfo
  {year} {2001}),\ \bibfield  {title} {\enquote {\bibinfo {title} {{Explosive
  Nucleosynthesis in Hypernovae}},}\ }\href {https://doi.org/10.1086/321495}
  {\bibfield  {journal} {\bibinfo  {journal} {Astrophys. J.}\ }\textbf
  {\bibinfo {volume} {555}},\ \bibinfo {pages} {880--899}}\BibitemShut
  {NoStop}%
\bibitem [{\citenamefont {Nakar}(2007)}]{Nakar:2007}%
  \BibitemOpen
  \bibfield  {author} {\bibinfo {author} {\bibnamefont {Nakar}, \bibfnamefont
  {Ehud}}} (\bibinfo {year} {2007}),\ \bibfield  {title} {\enquote {\bibinfo
  {title} {{Short-hard gamma-ray bursts}},}\ }\href
  {https://doi.org/10.1016/j.physrep.2007.02.005} {\bibfield  {journal}
  {\bibinfo  {journal} {Phys. Rep.}\ }\textbf {\bibinfo {volume} {442}},\
  \bibinfo {pages} {166--236}}\BibitemShut {NoStop}%
\bibitem [{\citenamefont {{National Research Council}}(2003)}]{NAP10079}%
  \BibitemOpen
  \bibfield  {author} {\bibinfo {author} {\bibnamefont {{National Research
  Council}},}} (\bibinfo {year} {2003}),\ \href
  {https://doi.org/10.17226/10079} {\emph {\bibinfo {title} {Connecting Quarks
  with the Cosmos: Eleven Science Questions for the New Century}}}\ (\bibinfo
  {publisher} {The National Academies Press},\ \bibinfo {address} {Washington,
  DC})\BibitemShut {NoStop}%
\bibitem [{\citenamefont {Neufcourt}\ \emph {et~al.}(2018)\citenamefont
  {Neufcourt}, \citenamefont {Cao}, \citenamefont {Nazarewicz},\ and\
  \citenamefont {Viens}}]{Neufcourt.Cao.ea:2018}%
  \BibitemOpen
  \bibfield  {author} {\bibinfo {author} {\bibnamefont {Neufcourt},
  \bibfnamefont {L\'eo}}, \bibinfo {author} {\bibfnamefont {Yuchen}\
  \bibnamefont {Cao}}, \bibinfo {author} {\bibfnamefont {Witold}\ \bibnamefont
  {Nazarewicz}}, and\ \bibinfo {author} {\bibfnamefont {Frederi}\ \bibnamefont
  {Viens}}} (\bibinfo {year} {2018}),\ \bibfield  {title} {\enquote {\bibinfo
  {title} {Bayesian approach to model-based extrapolation of nuclear
  observables},}\ }\href {https://doi.org/10.1103/PhysRevC.98.034318}
  {\bibfield  {journal} {\bibinfo  {journal} {Phys. Rev. C}\ }\textbf {\bibinfo
  {volume} {98}},\ \bibinfo {pages} {034318}}\BibitemShut {NoStop}%
\bibitem [{\citenamefont {{Ney}}\ \emph {et~al.}(2020)\citenamefont {{Ney}},
  \citenamefont {{Engel}},\ and\ \citenamefont
  {{Schunck}}}]{Ney.Engel.Schunck:2020}%
  \BibitemOpen
  \bibfield  {author} {\bibinfo {author} {\bibnamefont {{Ney}}, \bibfnamefont
  {E~M}}, \bibinfo {author} {\bibfnamefont {J.}~\bibnamefont {{Engel}}}, and\
  \bibinfo {author} {\bibfnamefont {N.}~\bibnamefont {{Schunck}}}} (\bibinfo
  {year} {2020}),\ \bibfield  {title} {\enquote {\bibinfo {title} {{Global
  Description of Beta Decay with the Axially-Deformed Skyrme Finite Amplitude
  Method: Extension to Odd-Mass and Odd-Odd Nuclei}},}\ }\href@noop {}
  {\bibfield  {journal} {\bibinfo  {journal} {arXiv e-prints}\ }}\Eprint
  {https://arxiv.org/abs/2005.12883} {arXiv:2005.12883 [nucl-th]} \BibitemShut
  {NoStop}%
\bibitem [{\citenamefont {{Nicholl}}\ \emph
  {et~al.}(2017{\natexlab{a}})\citenamefont {{Nicholl}}, \citenamefont
  {{Guillochon}},\ and\ \citenamefont {{Berger}}}]{Nicholl.ea:2017}%
  \BibitemOpen
  \bibfield  {author} {\bibinfo {author} {\bibnamefont {{Nicholl}},
  \bibfnamefont {M}}, \bibinfo {author} {\bibfnamefont {J.}~\bibnamefont
  {{Guillochon}}}, and\ \bibinfo {author} {\bibfnamefont {E.}~\bibnamefont
  {{Berger}}}} (\bibinfo {year} {2017}{\natexlab{a}}),\ \bibfield  {title}
  {\enquote {\bibinfo {title} {{The Magnetar Model for Type I Superluminous
  Supernovae. I. Bayesian Analysis of the Full Multicolor Light-curve Sample
  with MOSFiT}},}\ }\href {https://doi.org/10.3847/1538-4357/aa9334} {\bibfield
   {journal} {\bibinfo  {journal} {Astrophys. J.}\ }\textbf {\bibinfo {volume}
  {850}},\ \bibinfo {eid} {55}}\BibitemShut {NoStop}%
\bibitem [{\citenamefont {{Nicholl}}\ \emph
  {et~al.}(2017{\natexlab{b}})\citenamefont {{Nicholl}} \emph
  {et~al.}}]{nicholl17}%
  \BibitemOpen
  \bibfield  {author} {\bibinfo {author} {\bibnamefont {{Nicholl}},
  \bibfnamefont {M}},  \emph {et~al.}} (\bibinfo {year} {2017}{\natexlab{b}}),\
  \bibfield  {title} {\enquote {\bibinfo {title} {{The Electromagnetic
  Counterpart of the Binary Neutron Star Merger LIGO/Virgo GW170817. III.
  Optical and UV Spectra of a Blue Kilonova from Fast Polar Ejecta}},}\ }\href
  {https://doi.org/10.3847/2041-8213/aa9029} {\bibfield  {journal} {\bibinfo
  {journal} {Astrophys. J. Lett.}\ }\textbf {\bibinfo {volume} {848}},\
  \bibinfo {eid} {L18}}\BibitemShut {NoStop}%
\bibitem [{\citenamefont {{Nishimura}}\ \emph {et~al.}(2018)\citenamefont
  {{Nishimura}}, \citenamefont {{Rauscher}}, \citenamefont {{Hirschi}},
  \citenamefont {{Murphy}}, \citenamefont {{Cescutti}},\ and\ \citenamefont
  {{Travaglio}}}]{Nishimura.Rauscher.ea:2018}%
  \BibitemOpen
  \bibfield  {author} {\bibinfo {author} {\bibnamefont {{Nishimura}},
  \bibfnamefont {N}}, \bibinfo {author} {\bibfnamefont {T.}~\bibnamefont
  {{Rauscher}}}, \bibinfo {author} {\bibfnamefont {R.}~\bibnamefont
  {{Hirschi}}}, \bibinfo {author} {\bibfnamefont {A.~S.~J.}\ \bibnamefont
  {{Murphy}}}, \bibinfo {author} {\bibfnamefont {G.}~\bibnamefont
  {{Cescutti}}}, and\ \bibinfo {author} {\bibfnamefont {C.}~\bibnamefont
  {{Travaglio}}}} (\bibinfo {year} {2018}),\ \bibfield  {title} {\enquote
  {\bibinfo {title} {{Uncertainties in the production of p nuclides in
  thermonuclear supernovae determined by Monte Carlo variations}},}\ }\href
  {https://doi.org/10.1093/mnras/stx3033} {\bibfield  {journal} {\bibinfo
  {journal} {Mon. Not. Roy. Astron. Soc.}\ }\textbf {\bibinfo {volume} {474}},\
  \bibinfo {pages} {3133--3139}}\BibitemShut {NoStop}%
\bibitem [{\citenamefont {{Nishimura}}\ \emph {et~al.}(2017)\citenamefont
  {{Nishimura}}, \citenamefont {{Sawai}}, \citenamefont {{Takiwaki}},
  \citenamefont {{Yamada}},\ and\ \citenamefont
  {{Thielemann}}}]{Nishimura.Sawai.ea:2017}%
  \BibitemOpen
  \bibfield  {author} {\bibinfo {author} {\bibnamefont {{Nishimura}},
  \bibfnamefont {N}}, \bibinfo {author} {\bibfnamefont {H.}~\bibnamefont
  {{Sawai}}}, \bibinfo {author} {\bibfnamefont {T.}~\bibnamefont {{Takiwaki}}},
  \bibinfo {author} {\bibfnamefont {S.}~\bibnamefont {{Yamada}}}, and\ \bibinfo
  {author} {\bibfnamefont {F.-K.}\ \bibnamefont {{Thielemann}}}} (\bibinfo
  {year} {2017}),\ \bibfield  {title} {\enquote {\bibinfo {title} {{The
  Intermediate r-process in Core-collapse Supernovae Driven by the
  Magneto-rotational Instability}},}\ }\href
  {https://doi.org/10.3847/2041-8213/aa5dee} {\bibfield  {journal} {\bibinfo
  {journal} {Astrophys. J.}\ }\textbf {\bibinfo {volume} {836}},\ \bibinfo
  {eid} {L21}}\BibitemShut {NoStop}%
\bibitem [{\citenamefont {{Nishimura}}\ \emph {et~al.}(2015)\citenamefont
  {{Nishimura}}, \citenamefont {{Takiwaki}},\ and\ \citenamefont
  {{Thielemann}}}]{Nishimura.Takiwaki.Thielemann:2015}%
  \BibitemOpen
  \bibfield  {author} {\bibinfo {author} {\bibnamefont {{Nishimura}},
  \bibfnamefont {N}}, \bibinfo {author} {\bibfnamefont {T.}~\bibnamefont
  {{Takiwaki}}}, and\ \bibinfo {author} {\bibfnamefont {F.-K.}\ \bibnamefont
  {{Thielemann}}}} (\bibinfo {year} {2015}),\ \bibfield  {title} {\enquote
  {\bibinfo {title} {{The r-process Nucleosynthesis in the Various Jet-like
  Explosions of Magnetorotational Core-collapse Supernovae}},}\ }\href
  {https://doi.org/10.1088/0004-637X/810/2/109} {\bibfield  {journal} {\bibinfo
   {journal} {Astrophys. J.}\ }\textbf {\bibinfo {volume} {810}},\ \bibinfo
  {eid} {109}}\BibitemShut {NoStop}%
\bibitem [{\citenamefont {{Nishimura}}\ \emph {et~al.}(2006)\citenamefont
  {{Nishimura}}, \citenamefont {{Kotake}}, \citenamefont {{Hashimoto}},
  \citenamefont {{Yamada}}, \citenamefont {{Nishimura}}, \citenamefont
  {{Fujimoto}},\ and\ \citenamefont {{Sato}}}]{nishimura06}%
  \BibitemOpen
  \bibfield  {author} {\bibinfo {author} {\bibnamefont {{Nishimura}},
  \bibfnamefont {S}}, \bibinfo {author} {\bibfnamefont {K.}~\bibnamefont
  {{Kotake}}}, \bibinfo {author} {\bibfnamefont {M.-a.}\ \bibnamefont
  {{Hashimoto}}}, \bibinfo {author} {\bibfnamefont {S.}~\bibnamefont
  {{Yamada}}}, \bibinfo {author} {\bibfnamefont {N.}~\bibnamefont
  {{Nishimura}}}, \bibinfo {author} {\bibfnamefont {S.}~\bibnamefont
  {{Fujimoto}}}, and\ \bibinfo {author} {\bibfnamefont {K.}~\bibnamefont
  {{Sato}}}} (\bibinfo {year} {2006}),\ \bibfield  {title} {\enquote {\bibinfo
  {title} {{r-Process Nucleosynthesis in Magnetohydrodynamic Jet Explosions of
  Core-Collapse Supernovae}},}\ }\href {https://doi.org/10.1086/500786}
  {\bibfield  {journal} {\bibinfo  {journal} {Astrophys. J.}\ }\textbf
  {\bibinfo {volume} {642}},\ \bibinfo {pages} {410--419}}\BibitemShut
  {NoStop}%
\bibitem [{\citenamefont {{Nolden}}\ \emph {et~al.}(2011)\citenamefont
  {{Nolden}}, \citenamefont {{H{\"u}lsmann}}, \citenamefont {{Litvinov}},
  \citenamefont {{Moritz}}, \citenamefont {{Peschke}}, \citenamefont {{Petri}},
  \citenamefont {{Sanjari}}, \citenamefont {{Steck}}, \citenamefont {{Weick}},
  \citenamefont {{Wu}}, \citenamefont {{Zang}}, \citenamefont {{Zhang}},\ and\
  \citenamefont {{Zhao}}}]{Nolden2011}%
  \BibitemOpen
  \bibfield  {author} {\bibinfo {author} {\bibnamefont {{Nolden}},
  \bibfnamefont {F}}, \bibinfo {author} {\bibfnamefont {P.}~\bibnamefont
  {{H{\"u}lsmann}}}, \bibinfo {author} {\bibfnamefont {Y.~A.}\ \bibnamefont
  {{Litvinov}}}, \bibinfo {author} {\bibfnamefont {P.}~\bibnamefont
  {{Moritz}}}, \bibinfo {author} {\bibfnamefont {C.}~\bibnamefont {{Peschke}}},
  \bibinfo {author} {\bibfnamefont {P.}~\bibnamefont {{Petri}}}, \bibinfo
  {author} {\bibfnamefont {M.~S.}\ \bibnamefont {{Sanjari}}}, \bibinfo {author}
  {\bibfnamefont {M.}~\bibnamefont {{Steck}}}, \bibinfo {author} {\bibfnamefont
  {H.}~\bibnamefont {{Weick}}}, \bibinfo {author} {\bibfnamefont {J.~X.}\
  \bibnamefont {{Wu}}}, \bibinfo {author} {\bibfnamefont {Y.~D.}\ \bibnamefont
  {{Zang}}}, \bibinfo {author} {\bibfnamefont {S.~H.}\ \bibnamefont {{Zhang}}},
  and\ \bibinfo {author} {\bibfnamefont {T.~C.}\ \bibnamefont {{Zhao}}}}
  (\bibinfo {year} {2011}),\ \bibfield  {title} {\enquote {\bibinfo {title} {{A
  fast and sensitive resonant Schottky pick-up for heavy ion storage rings}},}\
  }\href {https://doi.org/10.1016/j.nima.2011.06.058} {\bibfield  {journal}
  {\bibinfo  {journal} {Nucl. Instruments Methods Phys. Res. Sect. A Accel.
  Spectrometers, Detect. Assoc. Equip.}\ }\textbf {\bibinfo {volume} {659}},\
  \bibinfo {pages} {69--77}}\BibitemShut {NoStop}%
\bibitem [{\citenamefont {{Nomoto}}(2017)}]{nomoto17}%
  \BibitemOpen
  \bibfield  {author} {\bibinfo {author} {\bibnamefont {{Nomoto}},
  \bibfnamefont {K}}} (\bibinfo {year} {2017}),\ \enquote {\bibinfo {title}
  {{Nucleosynthesis in Hypernovae Associated with Gamma-Ray Bursts}},}\ in\
  \href {https://doi.org/10.1007/978-3-319-21846-5_86} {\emph {\bibinfo
  {booktitle} {Handbook of Supernovae}}},\ \bibinfo {editor} {edited by\
  \bibinfo {editor} {\bibfnamefont {A.~W.}\ \bibnamefont {{Alsabti}}}\ and\
  \bibinfo {editor} {\bibfnamefont {P.}~\bibnamefont {{Murdin}}}}\ (\bibinfo
  {publisher} {Springer International Publishing})\ p.\ \bibinfo {pages}
  {1931}\BibitemShut {NoStop}%
\bibitem [{\citenamefont {{Nomoto}}\ \emph {et~al.}(2013)\citenamefont
  {{Nomoto}}, \citenamefont {{Kobayashi}},\ and\ \citenamefont
  {{Tominaga}}}]{nomoto13}%
  \BibitemOpen
  \bibfield  {author} {\bibinfo {author} {\bibnamefont {{Nomoto}},
  \bibfnamefont {K}}, \bibinfo {author} {\bibfnamefont {C.}~\bibnamefont
  {{Kobayashi}}}, and\ \bibinfo {author} {\bibfnamefont {N.}~\bibnamefont
  {{Tominaga}}}} (\bibinfo {year} {2013}),\ \bibfield  {title} {\enquote
  {\bibinfo {title} {{Nucleosynthesis in Stars and the Chemical Enrichment of
  Galaxies}},}\ }\href {https://doi.org/10.1146/annurev-astro-082812-140956}
  {\bibfield  {journal} {\bibinfo  {journal} {Annu. Rev. Astron. Astrophys.}\
  }\textbf {\bibinfo {volume} {51}},\ \bibinfo {pages} {457--509}}\BibitemShut
  {NoStop}%
\bibitem [{\citenamefont {{Nomoto}}\ \emph {et~al.}(1985)\citenamefont
  {{Nomoto}}, \citenamefont {{Thielemann}},\ and\ \citenamefont
  {{Miyaji}}}]{nomoto85}%
  \BibitemOpen
  \bibfield  {author} {\bibinfo {author} {\bibnamefont {{Nomoto}},
  \bibfnamefont {K}}, \bibinfo {author} {\bibfnamefont {F.-K.}\ \bibnamefont
  {{Thielemann}}}, and\ \bibinfo {author} {\bibfnamefont {S.}~\bibnamefont
  {{Miyaji}}}} (\bibinfo {year} {1985}),\ \bibfield  {title} {\enquote
  {\bibinfo {title} {{The triple alpha reaction at low temperatures in
  accreting white dwarfs and neutron stars}},}\ }\href@noop {} {\bibfield
  {journal} {\bibinfo  {journal} {Astron. \& Astrophys.}\ }\textbf {\bibinfo
  {volume} {149}},\ \bibinfo {pages} {239--245}}\BibitemShut {NoStop}%
\bibitem [{\citenamefont {{Nomoto}}\ \emph {et~al.}(2006)\citenamefont
  {{Nomoto}}, \citenamefont {{Tominaga}}, \citenamefont {{Umeda}},
  \citenamefont {{Kobayashi}},\ and\ \citenamefont
  {{Maeda}}}]{Nomoto.Tominaga.ea:2006}%
  \BibitemOpen
  \bibfield  {author} {\bibinfo {author} {\bibnamefont {{Nomoto}},
  \bibfnamefont {K}}, \bibinfo {author} {\bibfnamefont {N.}~\bibnamefont
  {{Tominaga}}}, \bibinfo {author} {\bibfnamefont {H.}~\bibnamefont {{Umeda}}},
  \bibinfo {author} {\bibfnamefont {C.}~\bibnamefont {{Kobayashi}}}, and\
  \bibinfo {author} {\bibfnamefont {K.}~\bibnamefont {{Maeda}}}} (\bibinfo
  {year} {2006}),\ \bibfield  {title} {\enquote {\bibinfo {title}
  {{Nucleosynthesis yields of core-collapse supernovae and hypernovae, and
  galactic chemical evolution}},}\ }\href
  {https://doi.org/10.1016/j.nuclphysa.2006.05.008} {\bibfield  {journal}
  {\bibinfo  {journal} {Nucl. Phys. A}\ }\textbf {\bibinfo {volume} {777}},\
  \bibinfo {pages} {424--458}}\BibitemShut {NoStop}%
\bibitem [{\citenamefont {Nomoto}\ and\ \citenamefont
  {Leung}(2017{\natexlab{a}})}]{Nomoto.Leung:2017b}%
  \BibitemOpen
  \bibfield  {author} {\bibinfo {author} {\bibnamefont {Nomoto}, \bibfnamefont
  {Ke'nichi}}, and\ \bibinfo {author} {\bibfnamefont {Shing-Chi}\ \bibnamefont
  {Leung}}} (\bibinfo {year} {2017}{\natexlab{a}}),\ \enquote {\bibinfo {title}
  {{Electron Capture Supernovae from Super Asymptotic Giant Branch Stars}},}\
  in\ \href {https://doi.org/10.1007/978-3-319-20794-0_118-1} {\emph {\bibinfo
  {booktitle} {Handbook of Supernovae}}},\ \bibinfo {editor} {edited by\
  \bibinfo {editor} {\bibfnamefont {A.~W.}\ \bibnamefont {{Alsabti}}}\ and\
  \bibinfo {editor} {\bibfnamefont {P.}~\bibnamefont {{Murdin}}}}\ (\bibinfo
  {publisher} {Springer International Publishing},\ \bibinfo {address}
  {Cham})\BibitemShut {NoStop}%
\bibitem [{\citenamefont {Nomoto}\ and\ \citenamefont
  {Leung}(2017{\natexlab{b}})}]{Nomoto.Leung:2017a}%
  \BibitemOpen
  \bibfield  {author} {\bibinfo {author} {\bibnamefont {Nomoto}, \bibfnamefont
  {Ken'ichi}}, and\ \bibinfo {author} {\bibfnamefont {Shing-Chi}\ \bibnamefont
  {Leung}}} (\bibinfo {year} {2017}{\natexlab{b}}),\ \enquote {\bibinfo {title}
  {{Thermonuclear Explosions of Chandrasekhar Mass White Dwarfs}},}\ in\ \href
  {https://doi.org/10.1007/978-3-319-20794-0_62-1} {\emph {\bibinfo {booktitle}
  {Handbook of Supernovae}}},\ \bibinfo {editor} {edited by\ \bibinfo {editor}
  {\bibfnamefont {Athem~W.}\ \bibnamefont {Alsabti}}\ and\ \bibinfo {editor}
  {\bibfnamefont {Paul}\ \bibnamefont {Murdin}}}\ (\bibinfo  {publisher}
  {Springer International Publishing},\ \bibinfo {address} {Cham})\BibitemShut
  {NoStop}%
\bibitem [{\citenamefont {{Nomoto}}\ \emph {et~al.}(2010)\citenamefont
  {{Nomoto}}, \citenamefont {{Tanaka}}, \citenamefont {{Tominaga}},\ and\
  \citenamefont {{Maeda}}}]{Nomoto.Tanaka.ea:2010}%
  \BibitemOpen
  \bibfield  {author} {\bibinfo {author} {\bibnamefont {{Nomoto}},
  \bibfnamefont {Ken'ichi}}, \bibinfo {author} {\bibfnamefont {Masaomi}\
  \bibnamefont {{Tanaka}}}, \bibinfo {author} {\bibfnamefont {Nozomu}\
  \bibnamefont {{Tominaga}}}, and\ \bibinfo {author} {\bibfnamefont {Keiichi}\
  \bibnamefont {{Maeda}}}} (\bibinfo {year} {2010}),\ \bibfield  {title}
  {\enquote {\bibinfo {title} {{Hypernovae, gamma-ray bursts, and first
  stars}},}\ }\href {https://doi.org/10.1016/j.newar.2010.09.022} {\bibfield
  {journal} {\bibinfo  {journal} {New Astron. Rev.}\ }\textbf {\bibinfo
  {volume} {54}},\ \bibinfo {pages} {191--200}}\BibitemShut {NoStop}%
\bibitem [{\citenamefont {Nunokawa}\ \emph {et~al.}(1997)\citenamefont
  {Nunokawa}, \citenamefont {Peltoniemi}, \citenamefont {Rossi},\ and\
  \citenamefont {Valle}}]{Nunokawa.Peltoniemi.ea:1997}%
  \BibitemOpen
  \bibfield  {author} {\bibinfo {author} {\bibnamefont {Nunokawa},
  \bibfnamefont {H}}, \bibinfo {author} {\bibfnamefont {J.~T.}\ \bibnamefont
  {Peltoniemi}}, \bibinfo {author} {\bibfnamefont {A.}~\bibnamefont {Rossi}},
  and\ \bibinfo {author} {\bibfnamefont {J.~W.~F.}\ \bibnamefont {Valle}}}
  (\bibinfo {year} {1997}),\ \bibfield  {title} {\enquote {\bibinfo {title}
  {Supernova bounds on resonant active-sterile neutrino conversions},}\ }\href
  {https://doi.org/10.1103/PhysRevD.56.1704} {\bibfield  {journal} {\bibinfo
  {journal} {Phys. Rev. D}\ }\textbf {\bibinfo {volume} {56}},\ \bibinfo
  {pages} {1704--1713}}\BibitemShut {NoStop}%
\bibitem [{\citenamefont {{Obergaulinger}}\ and\ \citenamefont
  {{Aloy}}(2020)}]{Obergaulinger.Aloy:2020}%
  \BibitemOpen
  \bibfield  {author} {\bibinfo {author} {\bibnamefont {{Obergaulinger}},
  \bibfnamefont {M}}, and\ \bibinfo {author} {\bibfnamefont {M.~{\'A}.}\
  \bibnamefont {{Aloy}}}} (\bibinfo {year} {2020}),\ \bibfield  {title}
  {\enquote {\bibinfo {title} {{Magnetorotational core collapse of possible GRB
  progenitors - I. Explosion mechanisms}},}\ }\href
  {https://doi.org/10.1093/mnras/staa096} {\bibfield  {journal} {\bibinfo
  {journal} {Mon. Not. Roy. Astron. Soc.}\ }\textbf {\bibinfo {volume} {492}},\
  \bibinfo {pages} {4613--4634}}\BibitemShut {NoStop}%
\bibitem [{\citenamefont {{Obergaulinger}}\ \emph {et~al.}(2010)\citenamefont
  {{Obergaulinger}}, \citenamefont {{Aloy}},\ and\ \citenamefont
  {{M{\"u}ller}}}]{Obergaulinger.Aloy.Mueller:2010}%
  \BibitemOpen
  \bibfield  {author} {\bibinfo {author} {\bibnamefont {{Obergaulinger}},
  \bibfnamefont {M}}, \bibinfo {author} {\bibfnamefont {M.~A.}\ \bibnamefont
  {{Aloy}}}, and\ \bibinfo {author} {\bibfnamefont {E.}~\bibnamefont
  {{M{\"u}ller}}}} (\bibinfo {year} {2010}),\ \bibfield  {title} {\enquote
  {\bibinfo {title} {{Local simulations of the magnetized Kelvin-Helmholtz
  instability in neutron-star mergers}},}\ }\href
  {https://doi.org/10.1051/0004-6361/200913386} {\bibfield  {journal} {\bibinfo
   {journal} {Astron. \& Astrophys.}\ }\textbf {\bibinfo {volume} {515}},\
  \bibinfo {pages} {A30}}\BibitemShut {NoStop}%
\bibitem [{\citenamefont {{Obergaulinger}}\ \emph {et~al.}(2018)\citenamefont
  {{Obergaulinger}}, \citenamefont {{Just}},\ and\ \citenamefont
  {{Aloy}}}]{Obergaulinger.Just.Aloy:2018}%
  \BibitemOpen
  \bibfield  {author} {\bibinfo {author} {\bibnamefont {{Obergaulinger}},
  \bibfnamefont {M}}, \bibinfo {author} {\bibfnamefont {O.}~\bibnamefont
  {{Just}}}, and\ \bibinfo {author} {\bibfnamefont {M.~A.}\ \bibnamefont
  {{Aloy}}}} (\bibinfo {year} {2018}),\ \bibfield  {title} {\enquote {\bibinfo
  {title} {{Core collapse with magnetic fields and rotation}},}\ }\href
  {https://doi.org/10.1088/1361-6471/aac982} {\bibfield  {journal} {\bibinfo
  {journal} {J. Phys. G: Nucl. Part. Phys.}\ }\textbf {\bibinfo {volume}
  {45}},\ \bibinfo {pages} {084001}}\BibitemShut {NoStop}%
\bibitem [{\citenamefont {{Oechslin}}\ \emph {et~al.}(2007)\citenamefont
  {{Oechslin}}, \citenamefont {{Janka}},\ and\ \citenamefont
  {{Marek}}}]{oechslin07}%
  \BibitemOpen
  \bibfield  {author} {\bibinfo {author} {\bibnamefont {{Oechslin}},
  \bibfnamefont {R}}, \bibinfo {author} {\bibfnamefont {H.-T.}\ \bibnamefont
  {{Janka}}}, and\ \bibinfo {author} {\bibfnamefont {A.}~\bibnamefont
  {{Marek}}}} (\bibinfo {year} {2007}),\ \bibfield  {title} {\enquote {\bibinfo
  {title} {{Relativistic neutron star merger simulations with non-zero
  temperature equations of state. I. Variation of binary parameters and
  equation of state}},}\ }\href {https://doi.org/10.1051/0004-6361:20066682}
  {\bibfield  {journal} {\bibinfo  {journal} {Astron. \& Astrophys.}\ }\textbf
  {\bibinfo {volume} {467}},\ \bibinfo {pages} {395--409}}\BibitemShut
  {NoStop}%
\bibitem [{\citenamefont {{Oechslin}}\ \emph {et~al.}(2002)\citenamefont
  {{Oechslin}}, \citenamefont {{Rosswog}},\ and\ \citenamefont
  {{Thielemann}}}]{oechslin02}%
  \BibitemOpen
  \bibfield  {author} {\bibinfo {author} {\bibnamefont {{Oechslin}},
  \bibfnamefont {R}}, \bibinfo {author} {\bibfnamefont {S.}~\bibnamefont
  {{Rosswog}}}, and\ \bibinfo {author} {\bibfnamefont {F.-K.}\ \bibnamefont
  {{Thielemann}}}} (\bibinfo {year} {2002}),\ \bibfield  {title} {\enquote
  {\bibinfo {title} {{Conformally flat smoothed particle hydrodynamics
  application to neutron star mergers}},}\ }\href
  {https://doi.org/10.1103/PhysRevD.65.103005} {\bibfield  {journal} {\bibinfo
  {journal} {Phys. Rev. D}\ }\textbf {\bibinfo {volume} {65}},\ \bibinfo {eid}
  {103005}}\BibitemShut {NoStop}%
\bibitem [{\citenamefont {{Oechslin}}\ \emph {et~al.}(2004)\citenamefont
  {{Oechslin}}, \citenamefont {{Ury{\= u}}}, \citenamefont {{Poghosyan}},\ and\
  \citenamefont {{Thielemann}}}]{oechslin04}%
  \BibitemOpen
  \bibfield  {author} {\bibinfo {author} {\bibnamefont {{Oechslin}},
  \bibfnamefont {R}}, \bibinfo {author} {\bibfnamefont {K.}~\bibnamefont
  {{Ury{\= u}}}}, \bibinfo {author} {\bibfnamefont {G.}~\bibnamefont
  {{Poghosyan}}}, and\ \bibinfo {author} {\bibfnamefont {F.~K.}\ \bibnamefont
  {{Thielemann}}}} (\bibinfo {year} {2004}),\ \bibfield  {title} {\enquote
  {\bibinfo {title} {{The influence of quark matter at high densities on binary
  neutron star mergers}},}\ }\href
  {https://doi.org/10.1111/j.1365-2966.2004.07621.x} {\bibfield  {journal}
  {\bibinfo  {journal} {Mon. Not. Roy. Astron. Soc.}\ }\textbf {\bibinfo
  {volume} {349}},\ \bibinfo {pages} {1469--1480}}\BibitemShut {NoStop}%
\bibitem [{\citenamefont {{Oertel}}\ \emph {et~al.}(2017)\citenamefont
  {{Oertel}}, \citenamefont {{Hempel}}, \citenamefont {{Kl{\"a}hn}},\ and\
  \citenamefont {{Typel}}}]{Oertel.Hempel.ea:2017}%
  \BibitemOpen
  \bibfield  {author} {\bibinfo {author} {\bibnamefont {{Oertel}},
  \bibfnamefont {M}}, \bibinfo {author} {\bibfnamefont {M.}~\bibnamefont
  {{Hempel}}}, \bibinfo {author} {\bibfnamefont {T.}~\bibnamefont
  {{Kl{\"a}hn}}}, and\ \bibinfo {author} {\bibfnamefont {S.}~\bibnamefont
  {{Typel}}}} (\bibinfo {year} {2017}),\ \bibfield  {title} {\enquote {\bibinfo
  {title} {{Equations of state for supernovae and compact stars}},}\ }\href
  {https://doi.org/10.1103/RevModPhys.89.015007} {\bibfield  {journal}
  {\bibinfo  {journal} {Rev. Mod. Phys.}\ }\textbf {\bibinfo {volume} {89}},\
  \bibinfo {eid} {015007}}\BibitemShut {NoStop}%
\bibitem [{\citenamefont {{Ojima}}\ \emph {et~al.}(2018)\citenamefont
  {{Ojima}}, \citenamefont {{Ishimaru}}, \citenamefont {{Wanajo}},
  \citenamefont {{Prantzos}},\ and\ \citenamefont {{Fran{\c c}ois}}}]{ojima18}%
  \BibitemOpen
  \bibfield  {author} {\bibinfo {author} {\bibnamefont {{Ojima}}, \bibfnamefont
  {T}}, \bibinfo {author} {\bibfnamefont {Y.}~\bibnamefont {{Ishimaru}}},
  \bibinfo {author} {\bibfnamefont {S.}~\bibnamefont {{Wanajo}}}, \bibinfo
  {author} {\bibfnamefont {N.}~\bibnamefont {{Prantzos}}}, and\ \bibinfo
  {author} {\bibfnamefont {P.}~\bibnamefont {{Fran{\c c}ois}}}} (\bibinfo
  {year} {2018}),\ \bibfield  {title} {\enquote {\bibinfo {title} {{Stochastic
  Chemical Evolution of Galactic Subhalos and the Origin of r-process
  Elements}},}\ }\href {https://doi.org/10.3847/1538-4357/aada11} {\bibfield
  {journal} {\bibinfo  {journal} {Astrophys. J.}\ }\textbf {\bibinfo {volume}
  {865}},\ \bibinfo {eid} {87}}\BibitemShut {NoStop}%
\bibitem [{\citenamefont {{Ono}}\ \emph {et~al.}(2012)\citenamefont {{Ono}},
  \citenamefont {{Hashimoto}}, \citenamefont {{Fujimoto}}, \citenamefont
  {{Kotake}},\ and\ \citenamefont {{Yamada}}}]{Ono.Hashimoto.ea:2012}%
  \BibitemOpen
  \bibfield  {author} {\bibinfo {author} {\bibnamefont {{Ono}}, \bibfnamefont
  {M}}, \bibinfo {author} {\bibfnamefont {M.}~\bibnamefont {{Hashimoto}}},
  \bibinfo {author} {\bibfnamefont {S.}~\bibnamefont {{Fujimoto}}}, \bibinfo
  {author} {\bibfnamefont {K.}~\bibnamefont {{Kotake}}}, and\ \bibinfo {author}
  {\bibfnamefont {S.}~\bibnamefont {{Yamada}}}} (\bibinfo {year} {2012}),\
  \bibfield  {title} {\enquote {\bibinfo {title} {{Explosive Nucleosynthesis in
  Magnetohydrodynamical Jets from Collapsars. II --- Heavy-Element
  Nucleosynthesis of s, p, r-Processes}},}\ }\href
  {https://doi.org/10.1143/PTP.128.741} {\bibfield  {journal} {\bibinfo
  {journal} {Prog. Theor. Phys.}\ }\textbf {\bibinfo {volume} {128}},\ \bibinfo
  {pages} {741--765}}\BibitemShut {NoStop}%
\bibitem [{\citenamefont {Orford}\ \emph {et~al.}(2018)\citenamefont {Orford}
  \emph {et~al.}}]{Orford.Vassh.ea:2018}%
  \BibitemOpen
  \bibfield  {author} {\bibinfo {author} {\bibnamefont {Orford}, \bibfnamefont
  {R}},  \emph {et~al.}} (\bibinfo {year} {2018}),\ \bibfield  {title}
  {\enquote {\bibinfo {title} {Precision mass measurements of neutron-rich
  neodymium and samarium isotopes and their role in understanding rare-earth
  peak formation},}\ }\href {https://doi.org/10.1103/PhysRevLett.120.262702}
  {\bibfield  {journal} {\bibinfo  {journal} {Phys. Rev. Lett.}\ }\textbf
  {\bibinfo {volume} {120}},\ \bibinfo {pages} {262702}}\BibitemShut {NoStop}%
\bibitem [{\citenamefont {{Ormand}}(1997)}]{Ormand:1997}%
  \BibitemOpen
  \bibfield  {author} {\bibinfo {author} {\bibnamefont {{Ormand}},
  \bibfnamefont {W~E}}} (\bibinfo {year} {1997}),\ \bibfield  {title} {\enquote
  {\bibinfo {title} {{Estimating the nuclear level density with the Monte Carlo
  shell model}},}\ }\href {https://doi.org/10.1103/PhysRevC.56.R1678}
  {\bibfield  {journal} {\bibinfo  {journal} {Phys. Rev. C}\ }\textbf {\bibinfo
  {volume} {56}},\ \bibinfo {pages} {R1678--R1682}}\BibitemShut {NoStop}%
\bibitem [{\citenamefont {{Otsuki}}\ \emph {et~al.}(2010)\citenamefont
  {{Otsuki}}, \citenamefont {{Burrows}}, \citenamefont {{Martinez-Pinedo}},
  \citenamefont {{Typel}}, \citenamefont {{Langanke}},\ and\ \citenamefont
  {{Matos}}}]{Otsuki.Typel.Martinez-Pinedo:2010}%
  \BibitemOpen
  \bibfield  {author} {\bibinfo {author} {\bibnamefont {{Otsuki}},
  \bibfnamefont {K}}, \bibinfo {author} {\bibfnamefont {A.}~\bibnamefont
  {{Burrows}}}, \bibinfo {author} {\bibfnamefont {G.}~\bibnamefont
  {{Martinez-Pinedo}}}, \bibinfo {author} {\bibfnamefont {S.}~\bibnamefont
  {{Typel}}}, \bibinfo {author} {\bibfnamefont {K.}~\bibnamefont {{Langanke}}},
  and\ \bibinfo {author} {\bibfnamefont {M.}~\bibnamefont {{Matos}}}} (\bibinfo
  {year} {2010}),\ \bibfield  {title} {\enquote {\bibinfo {title} {{r-process
  in Type II supernovae and the role of direct capture}},}\ }in\ \href
  {https://doi.org/10.1063/1.3455940} {\emph {\bibinfo {booktitle} {American
  Institute of Physics Conference Series}}},\ Vol.\ \bibinfo {volume} {1238},\
  \bibinfo {editor} {edited by\ \bibinfo {editor} {\bibfnamefont
  {H.}~\bibnamefont {{Susa}}}, \bibinfo {editor} {\bibfnamefont
  {M.}~\bibnamefont {{Arnould}}}, \bibinfo {editor} {\bibfnamefont
  {S.}~\bibnamefont {{Gales}}}, \bibinfo {editor} {\bibfnamefont
  {T.}~\bibnamefont {{Motobayashi}}}, \bibinfo {editor} {\bibfnamefont
  {C.}~\bibnamefont {{Scheidenberger}}}, \ and\ \bibinfo {editor}
  {\bibfnamefont {H.}~\bibnamefont {{Utsunomiya}}}},\ pp.\ \bibinfo {pages}
  {240--242}\BibitemShut {NoStop}%
\bibitem [{\citenamefont {Otsuki}\ \emph {et~al.}(2000)\citenamefont {Otsuki},
  \citenamefont {Tagoshi}, \citenamefont {Kajino},\ and\ \citenamefont
  {Wanajo}}]{Otsuki.Tagoshi.ea:2000}%
  \BibitemOpen
  \bibfield  {author} {\bibinfo {author} {\bibnamefont {Otsuki}, \bibfnamefont
  {K}}, \bibinfo {author} {\bibfnamefont {H.}~\bibnamefont {Tagoshi}}, \bibinfo
  {author} {\bibfnamefont {T.}~\bibnamefont {Kajino}}, and\ \bibinfo {author}
  {\bibfnamefont {S.}~\bibnamefont {Wanajo}}} (\bibinfo {year} {2000}),\
  \bibfield  {title} {\enquote {\bibinfo {title} {General relativistic effects
  on neutrino-driven winds from young, hot neutron stars and r-process
  nucleosynthesis},}\ }\href {https://doi.org/10.1086/308632} {\bibfield
  {journal} {\bibinfo  {journal} {Astrophys. J.}\ }\textbf {\bibinfo {volume}
  {533}},\ \bibinfo {pages} {424--439}}\BibitemShut {NoStop}%
\bibitem [{\citenamefont {{\"O}zel}\ and\ \citenamefont
  {Freire}(2016)}]{Oezel.Freire:2016}%
  \BibitemOpen
  \bibfield  {author} {\bibinfo {author} {\bibnamefont {{\"O}zel},
  \bibfnamefont {Feryal}}, and\ \bibinfo {author} {\bibfnamefont {Paulo}\
  \bibnamefont {Freire}}} (\bibinfo {year} {2016}),\ \bibfield  {title}
  {\enquote {\bibinfo {title} {Masses, radii, and the equation of state of
  neutron stars},}\ }\href
  {https://doi.org/10.1146/annurev-astro-081915-023322} {\bibfield  {journal}
  {\bibinfo  {journal} {Annu. Rev. Astron. Astrophys.}\ }\textbf {\bibinfo
  {volume} {54}},\ \bibinfo {pages} {401--440}}\BibitemShut {NoStop}%
\bibitem [{\citenamefont {{{\"O}zen}}\ \emph {et~al.}(2015)\citenamefont
  {{{\"O}zen}}, \citenamefont {{Alhassid}},\ and\ \citenamefont
  {{Nakada}}}]{Ozen.Alhassid.Nakada:2015}%
  \BibitemOpen
  \bibfield  {author} {\bibinfo {author} {\bibnamefont {{{\"O}zen}},
  \bibfnamefont {C}}, \bibinfo {author} {\bibfnamefont {Y.}~\bibnamefont
  {{Alhassid}}}, and\ \bibinfo {author} {\bibfnamefont {H.}~\bibnamefont
  {{Nakada}}}} (\bibinfo {year} {2015}),\ \bibfield  {title} {\enquote
  {\bibinfo {title} {{Nuclear state densities of odd-mass heavy nuclei in the
  shell model Monte Carlo approach}},}\ }\href
  {https://doi.org/10.1103/PhysRevC.91.034329} {\bibfield  {journal} {\bibinfo
  {journal} {Phys. Rev. C}\ }\textbf {\bibinfo {volume} {91}},\ \bibinfo {eid}
  {034329}}\BibitemShut {NoStop}%
\bibitem [{\citenamefont {{Pagel}}(2009)}]{Pagel:2009}%
  \BibitemOpen
  \bibfield  {author} {\bibinfo {author} {\bibnamefont {{Pagel}}, \bibfnamefont
  {B~E~J}}} (\bibinfo {year} {2009}),\ \href@noop {} {\emph {\bibinfo {title}
  {{Nucleosynthesis and Chemical Evolution of Galaxies}}}}\ (\bibinfo
  {publisher} {Cambridge University Press},\ \bibinfo {address} {Cambridge,
  UK})\BibitemShut {NoStop}%
\bibitem [{\citenamefont {{Pain}}\ \emph {et~al.}(2007)\citenamefont {{Pain}},
  \citenamefont {{Cizewski}}, \citenamefont {{Hatarik}}, \citenamefont
  {{Jones}}, \citenamefont {{Thomas}}, \citenamefont {{Bardayan}},
  \citenamefont {{Blackmon}}, \citenamefont {{Nesaraja}}, \citenamefont
  {{Smith}}, \citenamefont {{Kozub}},\ and\ \citenamefont
  {{Johnson}}}]{Pain2007}%
  \BibitemOpen
  \bibfield  {author} {\bibinfo {author} {\bibnamefont {{Pain}}, \bibfnamefont
  {S~D}}, \bibinfo {author} {\bibfnamefont {J.~A.}\ \bibnamefont {{Cizewski}}},
  \bibinfo {author} {\bibfnamefont {R.}~\bibnamefont {{Hatarik}}}, \bibinfo
  {author} {\bibfnamefont {K.~L.}\ \bibnamefont {{Jones}}}, \bibinfo {author}
  {\bibfnamefont {J.~S.}\ \bibnamefont {{Thomas}}}, \bibinfo {author}
  {\bibfnamefont {D.~W.}\ \bibnamefont {{Bardayan}}}, \bibinfo {author}
  {\bibfnamefont {J.~C.}\ \bibnamefont {{Blackmon}}}, \bibinfo {author}
  {\bibfnamefont {C.~D.}\ \bibnamefont {{Nesaraja}}}, \bibinfo {author}
  {\bibfnamefont {M.~S.}\ \bibnamefont {{Smith}}}, \bibinfo {author}
  {\bibfnamefont {R.~L.}\ \bibnamefont {{Kozub}}}, and\ \bibinfo {author}
  {\bibfnamefont {M.~S.}\ \bibnamefont {{Johnson}}}} (\bibinfo {year} {2007}),\
  \bibfield  {title} {\enquote {\bibinfo {title} {{Development of a high
  solid-angle silicon detector array for measurement of transfer reactions in
  inverse kinematics}},}\ }\href {https://doi.org/10.1016/j.nimb.2007.04.289}
  {\bibfield  {journal} {\bibinfo  {journal} {Nucl. Instruments Methods Phys.
  Res. Sect. B Beam Interact. with Mater. Atoms}\ }\textbf {\bibinfo {volume}
  {261}},\ \bibinfo {pages} {1122--1125}}\BibitemShut {NoStop}%
\bibitem [{\citenamefont {{Pakhomov}}\ \emph {et~al.}(2019)\citenamefont
  {{Pakhomov}}, \citenamefont {{Mashonkina}}, \citenamefont {{Sitnova}},\ and\
  \citenamefont {{Jablonka}}}]{Pakhomov.Mashonkina.ea:2019}%
  \BibitemOpen
  \bibfield  {author} {\bibinfo {author} {\bibnamefont {{Pakhomov}},
  \bibfnamefont {Yu~V}}, \bibinfo {author} {\bibfnamefont {L.~I.}\ \bibnamefont
  {{Mashonkina}}}, \bibinfo {author} {\bibfnamefont {T.~M.}\ \bibnamefont
  {{Sitnova}}}, and\ \bibinfo {author} {\bibfnamefont {P.}~\bibnamefont
  {{Jablonka}}}} (\bibinfo {year} {2019}),\ \bibfield  {title} {\enquote
  {\bibinfo {title} {{Contribution of Type Ia Supernovae to the Chemical
  Enrichment of the Ultra-Faint Dwarf Galaxy Bo{\"o}tes I}},}\ }\href
  {https://doi.org/10.1134/S1063773719050050} {\bibfield  {journal} {\bibinfo
  {journal} {Astronomy Letters}\ }\textbf {\bibinfo {volume} {45}},\ \bibinfo
  {pages} {259--275}}\BibitemShut {NoStop}%
\bibitem [{\citenamefont {Palenzuela}\ \emph {et~al.}(2015)\citenamefont
  {Palenzuela}, \citenamefont {Liebling}, \citenamefont {Neilsen},
  \citenamefont {Lehner}, \citenamefont {Caballero}, \citenamefont {O'Connor},\
  and\ \citenamefont {Anderson}}]{Palenzuela.Liebling.ea:2015}%
  \BibitemOpen
  \bibfield  {author} {\bibinfo {author} {\bibnamefont {Palenzuela},
  \bibfnamefont {Carlos}}, \bibinfo {author} {\bibfnamefont {Steven~L.}\
  \bibnamefont {Liebling}}, \bibinfo {author} {\bibfnamefont {David}\
  \bibnamefont {Neilsen}}, \bibinfo {author} {\bibfnamefont {Luis}\
  \bibnamefont {Lehner}}, \bibinfo {author} {\bibfnamefont {O.~L.}\
  \bibnamefont {Caballero}}, \bibinfo {author} {\bibfnamefont {Evan}\
  \bibnamefont {O'Connor}}, and\ \bibinfo {author} {\bibfnamefont {Matthew}\
  \bibnamefont {Anderson}}} (\bibinfo {year} {2015}),\ \bibfield  {title}
  {\enquote {\bibinfo {title} {Effects of the microphysical equation of state
  in the mergers of magnetized neutron stars with neutrino cooling},}\ }\href
  {https://doi.org/10.1103/PhysRevD.92.044045} {\bibfield  {journal} {\bibinfo
  {journal} {Phys. Rev. D}\ }\textbf {\bibinfo {volume} {92}},\ \bibinfo
  {pages} {044045}}\BibitemShut {NoStop}%
\bibitem [{\citenamefont {{Pan}}\ \emph {et~al.}(2018)\citenamefont {{Pan}},
  \citenamefont {{Liebend{\"o}rfer}}, \citenamefont {{Couch}},\ and\
  \citenamefont {{Thielemann}}}]{Pan.ea:2018}%
  \BibitemOpen
  \bibfield  {author} {\bibinfo {author} {\bibnamefont {{Pan}}, \bibfnamefont
  {K-C}}, \bibinfo {author} {\bibfnamefont {M.}~\bibnamefont
  {{Liebend{\"o}rfer}}}, \bibinfo {author} {\bibfnamefont {S.~M.}\ \bibnamefont
  {{Couch}}}, and\ \bibinfo {author} {\bibfnamefont {F.-K.}\ \bibnamefont
  {{Thielemann}}}} (\bibinfo {year} {2018}),\ \bibfield  {title} {\enquote
  {\bibinfo {title} {{Equation of State Dependent Dynamics and Multi-messenger
  Signals from Stellar-mass Black Hole Formation}},}\ }\href
  {https://doi.org/10.3847/1538-4357/aab71d} {\bibfield  {journal} {\bibinfo
  {journal} {Astrophys. J.}\ }\textbf {\bibinfo {volume} {857}},\ \bibinfo
  {eid} {13}}\BibitemShut {NoStop}%
\bibitem [{\citenamefont {{Pan}}\ \emph {et~al.}(2019)\citenamefont {{Pan}},
  \citenamefont {{Mattes}}, \citenamefont {{O'Connor}}, \citenamefont
  {{Couch}}, \citenamefont {{Perego}},\ and\ \citenamefont
  {{Arcones}}}]{Pan.Mattes.ea:2019}%
  \BibitemOpen
  \bibfield  {author} {\bibinfo {author} {\bibnamefont {{Pan}}, \bibfnamefont
  {Kuo-Chuan}}, \bibinfo {author} {\bibfnamefont {Carlos}\ \bibnamefont
  {{Mattes}}}, \bibinfo {author} {\bibfnamefont {Evan~P.}\ \bibnamefont
  {{O'Connor}}}, \bibinfo {author} {\bibfnamefont {Sean~M.}\ \bibnamefont
  {{Couch}}}, \bibinfo {author} {\bibfnamefont {Albino}\ \bibnamefont
  {{Perego}}}, and\ \bibinfo {author} {\bibfnamefont {Almudena}\ \bibnamefont
  {{Arcones}}}} (\bibinfo {year} {2019}),\ \bibfield  {title} {\enquote
  {\bibinfo {title} {{The impact of different neutrino transport methods on
  multidimensional core-collapse supernova simulations}},}\ }\href
  {https://doi.org/10.1088/1361-6471/aaed51} {\bibfield  {journal} {\bibinfo
  {journal} {J. Phys. G: Nucl. Part. Phys.}\ }\textbf {\bibinfo {volume}
  {46}},\ \bibinfo {pages} {014001}}\BibitemShut {NoStop}%
\bibitem [{\citenamefont {{Panov}}\ \emph {et~al.}(2005)\citenamefont
  {{Panov}}, \citenamefont {{Kolbe}}, \citenamefont {{Pfeiffer}}, \citenamefont
  {{Rauscher}}, \citenamefont {{Kratz}},\ and\ \citenamefont
  {{Thielemann}}}]{Panov.Kolbe.ea:2005}%
  \BibitemOpen
  \bibfield  {author} {\bibinfo {author} {\bibnamefont {{Panov}}, \bibfnamefont
  {I~V}}, \bibinfo {author} {\bibfnamefont {E.}~\bibnamefont {{Kolbe}}},
  \bibinfo {author} {\bibfnamefont {B.}~\bibnamefont {{Pfeiffer}}}, \bibinfo
  {author} {\bibfnamefont {T.}~\bibnamefont {{Rauscher}}}, \bibinfo {author}
  {\bibfnamefont {K.-L.}\ \bibnamefont {{Kratz}}}, and\ \bibinfo {author}
  {\bibfnamefont {F.-K.}\ \bibnamefont {{Thielemann}}}} (\bibinfo {year}
  {2005}),\ \bibfield  {title} {\enquote {\bibinfo {title} {{Calculations of
  fission rates for r-process nucleosynthesis}},}\ }\href
  {https://doi.org/10.1016/j.nuclphysa.2004.09.115} {\bibfield  {journal}
  {\bibinfo  {journal} {Nucl. Phys. A}\ }\textbf {\bibinfo {volume} {747}},\
  \bibinfo {pages} {633--654}}\BibitemShut {NoStop}%
\bibitem [{\citenamefont {{Panov}}\ \emph {et~al.}(2013)\citenamefont
  {{Panov}}, \citenamefont {{Korneev}}, \citenamefont {{Lutostansky}},\ and\
  \citenamefont {{Thielemann}}}]{panov13}%
  \BibitemOpen
  \bibfield  {author} {\bibinfo {author} {\bibnamefont {{Panov}}, \bibfnamefont
  {I~V}}, \bibinfo {author} {\bibfnamefont {I.~Y.}\ \bibnamefont {{Korneev}}},
  \bibinfo {author} {\bibfnamefont {Y.~S.}\ \bibnamefont {{Lutostansky}}}, and\
  \bibinfo {author} {\bibfnamefont {F.-K.}\ \bibnamefont {{Thielemann}}}}
  (\bibinfo {year} {2013}),\ \bibfield  {title} {\enquote {\bibinfo {title}
  {{Probabilities of delayed processes for nuclei involved in the
  r-process}},}\ }\href {https://doi.org/10.1134/S1063778813010080} {\bibfield
  {journal} {\bibinfo  {journal} {Phys. Atom. Nuclei}\ }\textbf {\bibinfo
  {volume} {76}},\ \bibinfo {pages} {88--101}}\BibitemShut {NoStop}%
\bibitem [{\citenamefont {{Panov}}\ \emph {et~al.}(2010)\citenamefont
  {{Panov}}, \citenamefont {{Korneev}}, \citenamefont {{Rauscher}},
  \citenamefont {{Mart{\'{\i}}nez-Pinedo}}, \citenamefont {{Keli{\'c}-Heil}},
  \citenamefont {{Zinner}},\ and\ \citenamefont
  {{Thielemann}}}]{Panov.Korneev.ea:2010}%
  \BibitemOpen
  \bibfield  {author} {\bibinfo {author} {\bibnamefont {{Panov}}, \bibfnamefont
  {I~V}}, \bibinfo {author} {\bibfnamefont {I.~Y.}\ \bibnamefont {{Korneev}}},
  \bibinfo {author} {\bibfnamefont {T.}~\bibnamefont {{Rauscher}}}, \bibinfo
  {author} {\bibfnamefont {G.}~\bibnamefont {{Mart{\'{\i}}nez-Pinedo}}},
  \bibinfo {author} {\bibfnamefont {A.}~\bibnamefont {{Keli{\'c}-Heil}}},
  \bibinfo {author} {\bibfnamefont {N.~T.}\ \bibnamefont {{Zinner}}}, and\
  \bibinfo {author} {\bibfnamefont {{F.-K.}}\ \bibnamefont {{Thielemann}}}}
  (\bibinfo {year} {2010}),\ \bibfield  {title} {\enquote {\bibinfo {title}
  {Neutron-induced astrophysical reaction rates for translead nuclei},}\ }\href
  {https://doi.org/10.1051/0004-6361/200911967} {\bibfield  {journal} {\bibinfo
   {journal} {Astron. \& Astrophys.}\ }\textbf {\bibinfo {volume} {513}},\
  \bibinfo {pages} {A61}}\BibitemShut {NoStop}%
\bibitem [{\citenamefont {{Panov}}\ \emph {et~al.}(2008)\citenamefont
  {{Panov}}, \citenamefont {{Korneev}},\ and\ \citenamefont
  {{Thielemann}}}]{panov08}%
  \BibitemOpen
  \bibfield  {author} {\bibinfo {author} {\bibnamefont {{Panov}}, \bibfnamefont
  {I~V}}, \bibinfo {author} {\bibfnamefont {I.~Y.}\ \bibnamefont {{Korneev}}},
  and\ \bibinfo {author} {\bibfnamefont {F.-K.}\ \bibnamefont {{Thielemann}}}}
  (\bibinfo {year} {2008}),\ \bibfield  {title} {\enquote {\bibinfo {title}
  {{The r-Process in the region of transuranium elements and the contribution
  of fission products to the nucleosynthesis of nuclei with $A \le 130$}},}\
  }\href {https://doi.org/10.1007/s11443-008-3006-1} {\bibfield  {journal}
  {\bibinfo  {journal} {Astronomy Letters}\ }\textbf {\bibinfo {volume} {34}},\
  \bibinfo {pages} {189--197}}\BibitemShut {NoStop}%
\bibitem [{\citenamefont {{Panov}}\ \emph {et~al.}(2017)\citenamefont
  {{Panov}}, \citenamefont {{Lutostansky}}, \citenamefont {{Eichler}},\ and\
  \citenamefont {{Thielemann}}}]{Panov2017}%
  \BibitemOpen
  \bibfield  {author} {\bibinfo {author} {\bibnamefont {{Panov}}, \bibfnamefont
  {I~V}}, \bibinfo {author} {\bibfnamefont {Y.~S.}\ \bibnamefont
  {{Lutostansky}}}, \bibinfo {author} {\bibfnamefont {M.}~\bibnamefont
  {{Eichler}}}, and\ \bibinfo {author} {\bibfnamefont {F.-K.}\ \bibnamefont
  {{Thielemann}}}} (\bibinfo {year} {2017}),\ \bibfield  {title} {\enquote
  {\bibinfo {title} {{Determination of the Galaxy age by the method of
  uranium-thorium-plutonium isotopic ratios}},}\ }\href
  {https://doi.org/10.1134/S1063778817040202} {\bibfield  {journal} {\bibinfo
  {journal} {Phys. Atom. Nuclei}\ }\textbf {\bibinfo {volume} {80}},\ \bibinfo
  {pages} {657--665}}\BibitemShut {NoStop}%
\bibitem [{\citenamefont {{Panov}}\ \emph {et~al.}(2016)\citenamefont
  {{Panov}}, \citenamefont {{Lutostansky}},\ and\ \citenamefont
  {{Thielemann}}}]{panov16}%
  \BibitemOpen
  \bibfield  {author} {\bibinfo {author} {\bibnamefont {{Panov}}, \bibfnamefont
  {I~V}}, \bibinfo {author} {\bibfnamefont {Y.~S.}\ \bibnamefont
  {{Lutostansky}}}, and\ \bibinfo {author} {\bibfnamefont {F.-K.}\ \bibnamefont
  {{Thielemann}}}} (\bibinfo {year} {2016}),\ \bibfield  {title} {\enquote
  {\bibinfo {title} {{Beta-decay half-lives for the r-process nuclei}},}\
  }\href {https://doi.org/10.1016/j.nuclphysa.2015.12.001} {\bibfield
  {journal} {\bibinfo  {journal} {Nucl. Phys. A}\ }\textbf {\bibinfo {volume}
  {947}},\ \bibinfo {pages} {1--11}}\BibitemShut {NoStop}%
\bibitem [{\citenamefont {{Panov}}\ and\ \citenamefont
  {{Thielemann}}(2004)}]{panov04}%
  \BibitemOpen
  \bibfield  {author} {\bibinfo {author} {\bibnamefont {{Panov}}, \bibfnamefont
  {I~V}}, and\ \bibinfo {author} {\bibfnamefont {F.-K.}\ \bibnamefont
  {{Thielemann}}}} (\bibinfo {year} {2004}),\ \bibfield  {title} {\enquote
  {\bibinfo {title} {{Fission and the r-Process: Competition between
  Neutron-Induced and Beta-Delayed Fission}},}\ }\href
  {https://doi.org/10.1134/1.1795953} {\bibfield  {journal} {\bibinfo
  {journal} {Astronomy Letters}\ }\textbf {\bibinfo {volume} {30}},\ \bibinfo
  {pages} {647--655}}\BibitemShut {NoStop}%
\bibitem [{\citenamefont {{Papenfort}}\ \emph {et~al.}(2018)\citenamefont
  {{Papenfort}}, \citenamefont {{Gold}},\ and\ \citenamefont
  {{Rezzolla}}}]{Papenfort.Gold.Rezzolla:2018}%
  \BibitemOpen
  \bibfield  {author} {\bibinfo {author} {\bibnamefont {{Papenfort}},
  \bibfnamefont {L~Jens}}, \bibinfo {author} {\bibfnamefont {Roman}\
  \bibnamefont {{Gold}}}, and\ \bibinfo {author} {\bibfnamefont {Luciano}\
  \bibnamefont {{Rezzolla}}}} (\bibinfo {year} {2018}),\ \bibfield  {title}
  {\enquote {\bibinfo {title} {{Dynamical ejecta and nucleosynthetic yields
  from eccentric binary neutron-star mergers}},}\ }\href
  {https://doi.org/10.1103/PhysRevD.98.104028} {\bibfield  {journal} {\bibinfo
  {journal} {Phys. Rev. D}\ }\textbf {\bibinfo {volume} {98}},\ \bibinfo {eid}
  {104028}}\BibitemShut {NoStop}%
\bibitem [{\citenamefont {Pardo}\ \emph {et~al.}(2016)\citenamefont {Pardo},
  \citenamefont {Savard},\ and\ \citenamefont
  {Janssens}}]{Pardo.Savard.Janssens:2016}%
  \BibitemOpen
  \bibfield  {author} {\bibinfo {author} {\bibnamefont {Pardo}, \bibfnamefont
  {Richard~C}}, \bibinfo {author} {\bibfnamefont {Guy}\ \bibnamefont {Savard}},
  and\ \bibinfo {author} {\bibfnamefont {Robert V.~F.}\ \bibnamefont
  {Janssens}}} (\bibinfo {year} {2016}),\ \bibfield  {title} {\enquote
  {\bibinfo {title} {{ATLAS with CARIBU: A Laboratory Portrait}},}\ }\href@noop
  {} {\bibfield  {journal} {\bibinfo  {journal} {Nuclear Physics News}\
  }\textbf {\bibinfo {volume} {26}},\ \bibinfo {pages} {5--11}}\BibitemShut
  {NoStop}%
\bibitem [{\citenamefont {{Patat}}\ \emph {et~al.}(1998)\citenamefont
  {{Patat}}, \citenamefont {{Piemonte}}, \citenamefont {{Lidman}},
  \citenamefont {{Augusteijn}}, \citenamefont {{Hainaut}}, \citenamefont
  {{Boehnhardt}}, \citenamefont {{Leibundgut}},\ and\ \citenamefont
  {{Brewer}}}]{Patat.Piemonte.ea:1998}%
  \BibitemOpen
  \bibfield  {author} {\bibinfo {author} {\bibnamefont {{Patat}}, \bibfnamefont
  {F}}, \bibinfo {author} {\bibfnamefont {A.}~\bibnamefont {{Piemonte}}},
  \bibinfo {author} {\bibfnamefont {C.}~\bibnamefont {{Lidman}}}, \bibinfo
  {author} {\bibfnamefont {T.}~\bibnamefont {{Augusteijn}}}, \bibinfo {author}
  {\bibfnamefont {O.~R.}\ \bibnamefont {{Hainaut}}}, \bibinfo {author}
  {\bibfnamefont {H.}~\bibnamefont {{Boehnhardt}}}, \bibinfo {author}
  {\bibfnamefont {B.}~\bibnamefont {{Leibundgut}}}, and\ \bibinfo {author}
  {\bibfnamefont {J.}~\bibnamefont {{Brewer}}}} (\bibinfo {year} {1998}),\
  \bibfield  {title} {\enquote {\bibinfo {title} {{Supernova 1998bw in ESO
  184-G82}},}\ }\href@noop {} {\bibfield  {journal} {\bibinfo  {journal}
  {IAU~Circ.}\ }\textbf {\bibinfo {volume} {7017}},\ \bibinfo {pages}
  {2}}\BibitemShut {NoStop}%
\bibitem [{\citenamefont {{Perego}}\ \emph
  {et~al.}(2017{\natexlab{a}})\citenamefont {{Perego}}, \citenamefont
  {{Radice}},\ and\ \citenamefont {{Bernuzzi}}}]{Perego.Radice.Bernuzzi:2017}%
  \BibitemOpen
  \bibfield  {author} {\bibinfo {author} {\bibnamefont {{Perego}},
  \bibfnamefont {A}}, \bibinfo {author} {\bibfnamefont {D.}~\bibnamefont
  {{Radice}}}, and\ \bibinfo {author} {\bibfnamefont {S.}~\bibnamefont
  {{Bernuzzi}}}} (\bibinfo {year} {2017}{\natexlab{a}}),\ \bibfield  {title}
  {\enquote {\bibinfo {title} {{AT 2017gfo: An Anisotropic and Three-component
  Kilonova Counterpart of GW170817}},}\ }\href
  {https://doi.org/10.3847/2041-8213/aa9ab9} {\bibfield  {journal} {\bibinfo
  {journal} {Astrophys. J.}\ }\textbf {\bibinfo {volume} {850}},\ \bibinfo
  {eid} {L37}}\BibitemShut {NoStop}%
\bibitem [{\citenamefont {{Perego}}\ \emph {et~al.}(2014)\citenamefont
  {{Perego}}, \citenamefont {{Rosswog}}, \citenamefont {{Cabez{\'o}n}},
  \citenamefont {{Korobkin}}, \citenamefont {{K{\"a}ppeli}}, \citenamefont
  {{Arcones}},\ and\ \citenamefont {{Liebend{\"o}rfer}}}]{perego14}%
  \BibitemOpen
  \bibfield  {author} {\bibinfo {author} {\bibnamefont {{Perego}},
  \bibfnamefont {A}}, \bibinfo {author} {\bibfnamefont {S.}~\bibnamefont
  {{Rosswog}}}, \bibinfo {author} {\bibfnamefont {R.~M.}\ \bibnamefont
  {{Cabez{\'o}n}}}, \bibinfo {author} {\bibfnamefont {O.}~\bibnamefont
  {{Korobkin}}}, \bibinfo {author} {\bibfnamefont {R.}~\bibnamefont
  {{K{\"a}ppeli}}}, \bibinfo {author} {\bibfnamefont {A.}~\bibnamefont
  {{Arcones}}}, and\ \bibinfo {author} {\bibfnamefont {M.}~\bibnamefont
  {{Liebend{\"o}rfer}}}} (\bibinfo {year} {2014}),\ \bibfield  {title}
  {\enquote {\bibinfo {title} {{Neutrino-driven winds from neutron star merger
  remnants}},}\ }\href {https://doi.org/10.1093/mnras/stu1352} {\bibfield
  {journal} {\bibinfo  {journal} {Mon. Not. Roy. Astron. Soc.}\ }\textbf
  {\bibinfo {volume} {443}},\ \bibinfo {pages} {3134--3156}}\BibitemShut
  {NoStop}%
\bibitem [{\citenamefont {{Perego}}\ \emph
  {et~al.}(2017{\natexlab{b}})\citenamefont {{Perego}}, \citenamefont
  {{Yasin}},\ and\ \citenamefont {{Arcones}}}]{Perego.Yasin.Arcones:2017}%
  \BibitemOpen
  \bibfield  {author} {\bibinfo {author} {\bibnamefont {{Perego}},
  \bibfnamefont {A}}, \bibinfo {author} {\bibfnamefont {H.}~\bibnamefont
  {{Yasin}}}, and\ \bibinfo {author} {\bibfnamefont {A.}~\bibnamefont
  {{Arcones}}}} (\bibinfo {year} {2017}{\natexlab{b}}),\ \bibfield  {title}
  {\enquote {\bibinfo {title} {{Neutrino pair annihilation above merger
  remnants: implications of a long-lived massive neutron star}},}\ }\href
  {https://doi.org/10.1088/1361-6471/aa7bdc} {\bibfield  {journal} {\bibinfo
  {journal} {J. Phys. G: Nucl. Part. Phys.}\ }\textbf {\bibinfo {volume}
  {44}},\ \bibinfo {pages} {084007}}\BibitemShut {NoStop}%
\bibitem [{\citenamefont {{Perego}}\ \emph {et~al.}(2019)\citenamefont
  {{Perego}}, \citenamefont {{Bernuzzi}},\ and\ \citenamefont
  {{Radice}}}]{Perego.Bernuzzi.Radice:2019}%
  \BibitemOpen
  \bibfield  {author} {\bibinfo {author} {\bibnamefont {{Perego}},
  \bibfnamefont {Albino}}, \bibinfo {author} {\bibfnamefont {Sebastiano}\
  \bibnamefont {{Bernuzzi}}}, and\ \bibinfo {author} {\bibfnamefont {David}\
  \bibnamefont {{Radice}}}} (\bibinfo {year} {2019}),\ \bibfield  {title}
  {\enquote {\bibinfo {title} {{Thermodynamics conditions of matter in neutron
  star mergers}},}\ }\href {https://doi.org/10.1140/epja/i2019-12810-7}
  {\bibfield  {journal} {\bibinfo  {journal} {Eur. Phys. J. A}\ }\textbf
  {\bibinfo {volume} {55}},\ \bibinfo {eid} {124}}\BibitemShut {NoStop}%
\bibitem [{\citenamefont {{Pereira}}\ \emph {et~al.}(2009)\citenamefont
  {{Pereira}} \emph {et~al.}}]{Pereira}%
  \BibitemOpen
  \bibfield  {author} {\bibinfo {author} {\bibnamefont {{Pereira}},
  \bibfnamefont {J}},  \emph {et~al.}} (\bibinfo {year} {2009}),\ \bibfield
  {title} {\enquote {\bibinfo {title} {{{$\beta$}-decay half-lives and
  {$\beta$}-delayed neutron emission probabilities of nuclei in the region
  $A\lesssim 110$, relevant for the r process}},}\ }\href
  {https://doi.org/10.1103/PhysRevC.79.035806} {\bibfield  {journal} {\bibinfo
  {journal} {Phys. Rev. C}\ }\textbf {\bibinfo {volume} {79}},\ \bibinfo {eid}
  {035806}}\BibitemShut {NoStop}%
\bibitem [{\citenamefont {{Pereira}}\ \emph {et~al.}(2010)\citenamefont
  {{Pereira}} \emph {et~al.}}]{Nero}%
  \BibitemOpen
  \bibfield  {author} {\bibinfo {author} {\bibnamefont {{Pereira}},
  \bibfnamefont {J}},  \emph {et~al.}} (\bibinfo {year} {2010}),\ \bibfield
  {title} {\enquote {\bibinfo {title} {{The neutron long counter NERO for
  studies of {$\beta$}-delayed neutron emission in the r-process}},}\ }\href
  {https://doi.org/10.1016/j.nima.2010.02.262} {\bibfield  {journal} {\bibinfo
  {journal} {Nucl. Instruments Methods Phys. Res. Sect. A Accel. Spectrometers,
  Detect. Assoc. Equip.}\ }\textbf {\bibinfo {volume} {618}},\ \bibinfo {pages}
  {275--283}}\BibitemShut {NoStop}%
\bibitem [{\citenamefont {{Petermann}}\ \emph {et~al.}(2012)\citenamefont
  {{Petermann}}, \citenamefont {{Langanke}}, \citenamefont
  {{Mart{\'{\i}}nez-Pinedo}}, \citenamefont {{Panov}}, \citenamefont
  {{Reinhard}},\ and\ \citenamefont
  {{Thielemann}}}]{Petermann.Langanke.ea:2012}%
  \BibitemOpen
  \bibfield  {author} {\bibinfo {author} {\bibnamefont {{Petermann}},
  \bibfnamefont {I}}, \bibinfo {author} {\bibfnamefont {K.}~\bibnamefont
  {{Langanke}}}, \bibinfo {author} {\bibfnamefont {G.}~\bibnamefont
  {{Mart{\'{\i}}nez-Pinedo}}}, \bibinfo {author} {\bibfnamefont {I.~V.}\
  \bibnamefont {{Panov}}}, \bibinfo {author} {\bibfnamefont {P.-G.}\
  \bibnamefont {{Reinhard}}}, and\ \bibinfo {author} {\bibfnamefont {F.-K.}\
  \bibnamefont {{Thielemann}}}} (\bibinfo {year} {2012}),\ \bibfield  {title}
  {\enquote {\bibinfo {title} {{Have superheavy elements been produced in
  nature?}}}\ }\href {https://doi.org/10.1140/epja/i2012-12122-6} {\bibfield
  {journal} {\bibinfo  {journal} {Eur. Phys. J. A}\ }\textbf {\bibinfo {volume}
  {48}},\ \bibinfo {pages} {122}}\BibitemShut {NoStop}%
\bibitem [{\citenamefont {{Pfeiffer}}\ \emph {et~al.}(1997)\citenamefont
  {{Pfeiffer}}, \citenamefont {{Kratz}},\ and\ \citenamefont
  {{Thielemann}}}]{Pfeiffer.Kratz.Thielemann:1997}%
  \BibitemOpen
  \bibfield  {author} {\bibinfo {author} {\bibnamefont {{Pfeiffer}},
  \bibfnamefont {B}}, \bibinfo {author} {\bibfnamefont {K.~L.}\ \bibnamefont
  {{Kratz}}}, and\ \bibinfo {author} {\bibfnamefont {F.~K.}\ \bibnamefont
  {{Thielemann}}}} (\bibinfo {year} {1997}),\ \bibfield  {title} {\enquote
  {\bibinfo {title} {{Analysis of the solar-system r-process abundance pattern
  with the new ETFSI-Q mass formula}},}\ }\href
  {https://doi.org/10.1007/s002180050237} {\bibfield  {journal} {\bibinfo
  {journal} {Zeitschrift f{\"u}r Phys. A Hadron. Nucl.}\ }\textbf {\bibinfo
  {volume} {357}},\ \bibinfo {pages} {235--238}}\BibitemShut {NoStop}%
\bibitem [{\citenamefont {Pfeiffer}\ \emph {et~al.}(2001)\citenamefont
  {Pfeiffer}, \citenamefont {Kratz}, \citenamefont {Thielemann},\ and\
  \citenamefont {Walters}}]{Pfeiffer.Kratz.ea:2001}%
  \BibitemOpen
  \bibfield  {author} {\bibinfo {author} {\bibnamefont {Pfeiffer},
  \bibfnamefont {B}}, \bibinfo {author} {\bibfnamefont {K.-L.}\ \bibnamefont
  {Kratz}}, \bibinfo {author} {\bibfnamefont {F.-K.}\ \bibnamefont
  {Thielemann}}, and\ \bibinfo {author} {\bibfnamefont {W.~B.}\ \bibnamefont
  {Walters}}} (\bibinfo {year} {2001}),\ \bibfield  {title} {\enquote {\bibinfo
  {title} {Nuclear structure studies for the astrophysical r-process},}\ }\href
  {https://doi.org/10.1016/S0375-9474(01)01141-1} {\bibfield  {journal}
  {\bibinfo  {journal} {Nucl. Phys. A}\ }\textbf {\bibinfo {volume} {693}},\
  \bibinfo {pages} {282--324}}\BibitemShut {NoStop}%
\bibitem [{\citenamefont {{Pian}}\ \emph {et~al.}(2017)\citenamefont {{Pian}}
  \emph {et~al.}}]{Pian.Others:2017}%
  \BibitemOpen
  \bibfield  {author} {\bibinfo {author} {\bibnamefont {{Pian}}, \bibfnamefont
  {E}},  \emph {et~al.}} (\bibinfo {year} {2017}),\ \bibfield  {title}
  {\enquote {\bibinfo {title} {{Spectroscopic identification of r-process
  nucleosynthesis in a double neutron-star merger}},}\ }\href
  {https://doi.org/10.1038/nature24298} {\bibfield  {journal} {\bibinfo
  {journal} {Nature}\ }\textbf {\bibinfo {volume} {551}},\ \bibinfo {pages}
  {67--70}}\BibitemShut {NoStop}%
\bibitem [{\citenamefont {{Pinto}}\ and\ \citenamefont
  {{Eastman}}(2000{\natexlab{a}})}]{Pinto.Eastman:2000a}%
  \BibitemOpen
  \bibfield  {author} {\bibinfo {author} {\bibnamefont {{Pinto}}, \bibfnamefont
  {P~A}}, and\ \bibinfo {author} {\bibfnamefont {R.~G.}\ \bibnamefont
  {{Eastman}}}} (\bibinfo {year} {2000}{\natexlab{a}}),\ \bibfield  {title}
  {\enquote {\bibinfo {title} {{The Physics of Type IA Supernova Light Curves.
  I. Analytic Results and Time Dependence}},}\ }\href
  {https://doi.org/10.1086/308376} {\bibfield  {journal} {\bibinfo  {journal}
  {Astrophys. J.}\ }\textbf {\bibinfo {volume} {530}},\ \bibinfo {pages}
  {744--756}}\BibitemShut {NoStop}%
\bibitem [{\citenamefont {{Pinto}}\ and\ \citenamefont
  {{Eastman}}(2000{\natexlab{b}})}]{Pinto.Eastman:2000b}%
  \BibitemOpen
  \bibfield  {author} {\bibinfo {author} {\bibnamefont {{Pinto}}, \bibfnamefont
  {P~A}}, and\ \bibinfo {author} {\bibfnamefont {R.~G.}\ \bibnamefont
  {{Eastman}}}} (\bibinfo {year} {2000}{\natexlab{b}}),\ \bibfield  {title}
  {\enquote {\bibinfo {title} {{The Physics of Type IA Supernova Light Curves.
  II. Opacity and Diffusion}},}\ }\href {https://doi.org/10.1086/308380}
  {\bibfield  {journal} {\bibinfo  {journal} {Astrophys. J.}\ }\textbf
  {\bibinfo {volume} {530}},\ \bibinfo {pages} {757--776}}\BibitemShut
  {NoStop}%
\bibitem [{\citenamefont {{Piran}}(2005)}]{Piran:2004}%
  \BibitemOpen
  \bibfield  {author} {\bibinfo {author} {\bibnamefont {{Piran}}, \bibfnamefont
  {T}}} (\bibinfo {year} {2005}),\ \bibfield  {title} {\enquote {\bibinfo
  {title} {{The physics of gamma-ray bursts}},}\ }\href
  {https://doi.org/10.1103/RevModPhys.76.1143} {\bibfield  {journal} {\bibinfo
  {journal} {Rev. Mod. Phys.}\ }\textbf {\bibinfo {volume} {76}},\ \bibinfo
  {pages} {1143--1210}}\BibitemShut {NoStop}%
\bibitem [{\citenamefont {{Piran}}\ \emph {et~al.}(2013)\citenamefont
  {{Piran}}, \citenamefont {{Nakar}},\ and\ \citenamefont
  {{Rosswog}}}]{Piran.Nakar.Rosswog:2013}%
  \BibitemOpen
  \bibfield  {author} {\bibinfo {author} {\bibnamefont {{Piran}}, \bibfnamefont
  {T}}, \bibinfo {author} {\bibfnamefont {E.}~\bibnamefont {{Nakar}}}, and\
  \bibinfo {author} {\bibfnamefont {S.}~\bibnamefont {{Rosswog}}}} (\bibinfo
  {year} {2013}),\ \bibfield  {title} {\enquote {\bibinfo {title} {{The
  electromagnetic signals of compact binary mergers}},}\ }\href
  {https://doi.org/10.1093/mnras/stt037} {\bibfield  {journal} {\bibinfo
  {journal} {Mon. Not. Roy. Astron. Soc.}\ }\textbf {\bibinfo {volume} {430}},\
  \bibinfo {pages} {2121--2136}}\BibitemShut {NoStop}%
\bibitem [{\citenamefont {{Pitrou}}\ \emph {et~al.}(2018)\citenamefont
  {{Pitrou}}, \citenamefont {{Coc}}, \citenamefont {{Uzan}},\ and\
  \citenamefont {{Vangioni}}}]{Pitrou.Coc.ea:2018}%
  \BibitemOpen
  \bibfield  {author} {\bibinfo {author} {\bibnamefont {{Pitrou}},
  \bibfnamefont {C}}, \bibinfo {author} {\bibfnamefont {A.}~\bibnamefont
  {{Coc}}}, \bibinfo {author} {\bibfnamefont {J.-P.}\ \bibnamefont {{Uzan}}},
  and\ \bibinfo {author} {\bibfnamefont {E.}~\bibnamefont {{Vangioni}}}}
  (\bibinfo {year} {2018}),\ \bibfield  {title} {\enquote {\bibinfo {title}
  {{Precision big bang nucleosynthesis with improved Helium-4 predictions}},}\
  }\href {https://doi.org/10.1016/j.physrep.2018.04.005} {\bibfield  {journal}
  {\bibinfo  {journal} {Phys. Rep.}\ }\textbf {\bibinfo {volume} {754}},\
  \bibinfo {pages} {1--66}}\BibitemShut {NoStop}%
\bibitem [{\citenamefont {{Placco}}\ \emph {et~al.}(2017)\citenamefont
  {{Placco}}, \citenamefont {{Holmbeck}}, \citenamefont {{Frebel}},
  \citenamefont {{Beers}}, \citenamefont {{Surman}}, \citenamefont {{Ji}},
  \citenamefont {{Ezzeddine}}, \citenamefont {{Points}}, \citenamefont
  {{Kaleida}}, \citenamefont {{Hansen}}, \citenamefont {{Sakari}},\ and\
  \citenamefont {{Casey}}}]{Placco.Holmbeck.ea:2017}%
  \BibitemOpen
  \bibfield  {author} {\bibinfo {author} {\bibnamefont {{Placco}},
  \bibfnamefont {V~M}}, \bibinfo {author} {\bibfnamefont {E.~M.}\ \bibnamefont
  {{Holmbeck}}}, \bibinfo {author} {\bibfnamefont {A.}~\bibnamefont
  {{Frebel}}}, \bibinfo {author} {\bibfnamefont {T.~C.}\ \bibnamefont
  {{Beers}}}, \bibinfo {author} {\bibfnamefont {R.~A.}\ \bibnamefont
  {{Surman}}}, \bibinfo {author} {\bibfnamefont {A.~P.}\ \bibnamefont {{Ji}}},
  \bibinfo {author} {\bibfnamefont {R.}~\bibnamefont {{Ezzeddine}}}, \bibinfo
  {author} {\bibfnamefont {S.~D.}\ \bibnamefont {{Points}}}, \bibinfo {author}
  {\bibfnamefont {C.~C.}\ \bibnamefont {{Kaleida}}}, \bibinfo {author}
  {\bibfnamefont {T.~T.}\ \bibnamefont {{Hansen}}}, \bibinfo {author}
  {\bibfnamefont {C.~M.}\ \bibnamefont {{Sakari}}}, and\ \bibinfo {author}
  {\bibfnamefont {A.~R.}\ \bibnamefont {{Casey}}}} (\bibinfo {year} {2017}),\
  \bibfield  {title} {\enquote {\bibinfo {title} {{RAVE J203843.2-002333: The
  First Highly R-process-enhanced Star Identified in the RAVE Survey}},}\
  }\href {https://doi.org/10.3847/1538-4357/aa78ef} {\bibfield  {journal}
  {\bibinfo  {journal} {Astrophys. J.}\ }\textbf {\bibinfo {volume} {844}},\
  \bibinfo {eid} {18}}\BibitemShut {NoStop}%
\bibitem [{\citenamefont {{Pllumbi}}\ \emph {et~al.}(2015)\citenamefont
  {{Pllumbi}}, \citenamefont {{Tamborra}}, \citenamefont {{Wanajo}},
  \citenamefont {{Janka}},\ and\ \citenamefont
  {{H{\"u}depohl}}}]{Pllumbi.Tamborra.ea:2015}%
  \BibitemOpen
  \bibfield  {author} {\bibinfo {author} {\bibnamefont {{Pllumbi}},
  \bibfnamefont {Else}}, \bibinfo {author} {\bibfnamefont {Irene}\ \bibnamefont
  {{Tamborra}}}, \bibinfo {author} {\bibfnamefont {Shinya}\ \bibnamefont
  {{Wanajo}}}, \bibinfo {author} {\bibfnamefont {Hans-Thomas}\ \bibnamefont
  {{Janka}}}, and\ \bibinfo {author} {\bibfnamefont {Lorenz}\ \bibnamefont
  {{H{\"u}depohl}}}} (\bibinfo {year} {2015}),\ \bibfield  {title} {\enquote
  {\bibinfo {title} {{Impact of Neutrino Flavor Oscillations on the
  Neutrino-driven Wind Nucleosynthesis of an Electron-capture Supernova}},}\
  }\href {https://doi.org/10.1088/0004-637X/808/2/188} {\bibfield  {journal}
  {\bibinfo  {journal} {Astrophys. J.}\ }\textbf {\bibinfo {volume} {808}},\
  \bibinfo {eid} {188}}\BibitemShut {NoStop}%
\bibitem [{\citenamefont {{Potel}}\ \emph {et~al.}(2015)\citenamefont
  {{Potel}}, \citenamefont {{Nunes}},\ and\ \citenamefont
  {{Thompson}}}]{Potel2015}%
  \BibitemOpen
  \bibfield  {author} {\bibinfo {author} {\bibnamefont {{Potel}}, \bibfnamefont
  {G}}, \bibinfo {author} {\bibfnamefont {F.~M.}\ \bibnamefont {{Nunes}}}, and\
  \bibinfo {author} {\bibfnamefont {I.~J.}\ \bibnamefont {{Thompson}}}}
  (\bibinfo {year} {2015}),\ \bibfield  {title} {\enquote {\bibinfo {title}
  {{Establishing a theory for deuteron-induced surrogate reactions}},}\ }\href
  {https://doi.org/10.1103/PhysRevC.92.034611} {\bibfield  {journal} {\bibinfo
  {journal} {Phys. Rev. C}\ }\textbf {\bibinfo {volume} {92}},\ \bibinfo {eid}
  {034611}}\BibitemShut {NoStop}%
\bibitem [{\citenamefont {{Prantzos}}(2012)}]{Prantzos:2012}%
  \BibitemOpen
  \bibfield  {author} {\bibinfo {author} {\bibnamefont {{Prantzos}},
  \bibfnamefont {N}}} (\bibinfo {year} {2012}),\ \bibfield  {title} {\enquote
  {\bibinfo {title} {{Production and evolution of Li, Be, and B isotopes in the
  Galaxy}},}\ }\href {https://doi.org/10.1051/0004-6361/201219043} {\bibfield
  {journal} {\bibinfo  {journal} {Astron. \& Astrophys.}\ }\textbf {\bibinfo
  {volume} {542}},\ \bibinfo {eid} {A67}}\BibitemShut {NoStop}%
\bibitem [{\citenamefont {{Price}}\ and\ \citenamefont
  {{Rosswog}}(2006)}]{Price.Rosswog:2006}%
  \BibitemOpen
  \bibfield  {author} {\bibinfo {author} {\bibnamefont {{Price}}, \bibfnamefont
  {D~J}}, and\ \bibinfo {author} {\bibfnamefont {S.}~\bibnamefont {{Rosswog}}}}
  (\bibinfo {year} {2006}),\ \bibfield  {title} {\enquote {\bibinfo {title}
  {{Producing Ultrastrong Magnetic Fields in Neutron Star Mergers}},}\ }\href
  {https://doi.org/10.1126/science.1125201} {\bibfield  {journal} {\bibinfo
  {journal} {Science}\ }\textbf {\bibinfo {volume} {312}},\ \bibinfo {pages}
  {719--722}}\BibitemShut {NoStop}%
\bibitem [{\citenamefont {Pruet}\ \emph {et~al.}(2006)\citenamefont {Pruet},
  \citenamefont {Hoffman}, \citenamefont {Woosley}, \citenamefont {Janka},\
  and\ \citenamefont {Buras}}]{Pruet.Hoffman.ea:2006}%
  \BibitemOpen
  \bibfield  {author} {\bibinfo {author} {\bibnamefont {Pruet}, \bibfnamefont
  {J}}, \bibinfo {author} {\bibfnamefont {R.~D.}\ \bibnamefont {Hoffman}},
  \bibinfo {author} {\bibfnamefont {S.~E.}\ \bibnamefont {Woosley}}, \bibinfo
  {author} {\bibfnamefont {H.-T.}\ \bibnamefont {Janka}}, and\ \bibinfo
  {author} {\bibfnamefont {R.}~\bibnamefont {Buras}}} (\bibinfo {year}
  {2006}),\ \bibfield  {title} {\enquote {\bibinfo {title} {{Nucleosynthesis in
  Early Supernova Winds II: The Role of Neutrinos}},}\ }\href
  {https://doi.org/10.1086/503891} {\bibfield  {journal} {\bibinfo  {journal}
  {Astrophys. J.}\ }\textbf {\bibinfo {volume} {644}},\ \bibinfo {pages}
  {1028--1039}}\BibitemShut {NoStop}%
\bibitem [{\citenamefont {{Pruet}}\ \emph {et~al.}(2004)\citenamefont
  {{Pruet}}, \citenamefont {{Thompson}},\ and\ \citenamefont
  {{Hoffman}}}]{Pruet.Thompson.Hoffman:2004}%
  \BibitemOpen
  \bibfield  {author} {\bibinfo {author} {\bibnamefont {{Pruet}}, \bibfnamefont
  {J}}, \bibinfo {author} {\bibfnamefont {T.~A.}\ \bibnamefont {{Thompson}}},
  and\ \bibinfo {author} {\bibfnamefont {R.~D.}\ \bibnamefont {{Hoffman}}}}
  (\bibinfo {year} {2004}),\ \bibfield  {title} {\enquote {\bibinfo {title}
  {{Nucleosynthesis in Outflows from the Inner Regions of Collapsars}},}\
  }\href {https://doi.org/10.1086/382036} {\bibfield  {journal} {\bibinfo
  {journal} {Astrophys. J.}\ }\textbf {\bibinfo {volume} {606}},\ \bibinfo
  {pages} {1006--1018}}\BibitemShut {NoStop}%
\bibitem [{\citenamefont {{Pruet}}\ \emph {et~al.}(2003)\citenamefont
  {{Pruet}}, \citenamefont {{Woosley}},\ and\ \citenamefont
  {{Hoffman}}}]{Pruet.Woosley.Hoffman:2003}%
  \BibitemOpen
  \bibfield  {author} {\bibinfo {author} {\bibnamefont {{Pruet}}, \bibfnamefont
  {J}}, \bibinfo {author} {\bibfnamefont {S.~E.}\ \bibnamefont {{Woosley}}},
  and\ \bibinfo {author} {\bibfnamefont {R.~D.}\ \bibnamefont {{Hoffman}}}}
  (\bibinfo {year} {2003}),\ \bibfield  {title} {\enquote {\bibinfo {title}
  {{Nucleosynthesis in Gamma-Ray Burst Accretion Disks}},}\ }\href
  {https://doi.org/10.1086/367957} {\bibfield  {journal} {\bibinfo  {journal}
  {Astrophys. J.}\ }\textbf {\bibinfo {volume} {586}},\ \bibinfo {pages}
  {1254--1261}}\BibitemShut {NoStop}%
\bibitem [{\citenamefont {Qian}\ \emph {et~al.}(1997)\citenamefont {Qian},
  \citenamefont {Haxton}, \citenamefont {Langanke},\ and\ \citenamefont
  {Vogel}}]{Qian.Haxton.ea:1997}%
  \BibitemOpen
  \bibfield  {author} {\bibinfo {author} {\bibnamefont {Qian}, \bibfnamefont
  {Y-Z}}, \bibinfo {author} {\bibfnamefont {W.~C.}\ \bibnamefont {Haxton}},
  \bibinfo {author} {\bibfnamefont {K.}~\bibnamefont {Langanke}}, and\ \bibinfo
  {author} {\bibfnamefont {P.}~\bibnamefont {Vogel}}} (\bibinfo {year}
  {1997}),\ \bibfield  {title} {\enquote {\bibinfo {title} {{Neutrino-induced
  neutron spallation and supernova $r$-process nucleosynthesis}},}\ }\href
  {https://doi.org/10.1103/PhysRevC.55.1532} {\bibfield  {journal} {\bibinfo
  {journal} {Phys. Rev. C}\ }\textbf {\bibinfo {volume} {55}},\ \bibinfo
  {pages} {1532--1544}}\BibitemShut {NoStop}%
\bibitem [{\citenamefont {{Qian}}\ \emph {et~al.}(1998)\citenamefont {{Qian}},
  \citenamefont {{Vogel}},\ and\ \citenamefont
  {{Wasserburg}}}]{Qian.Vogel.Wasserburg:1998}%
  \BibitemOpen
  \bibfield  {author} {\bibinfo {author} {\bibnamefont {{Qian}}, \bibfnamefont
  {Y-Z}}, \bibinfo {author} {\bibfnamefont {P.}~\bibnamefont {{Vogel}}}, and\
  \bibinfo {author} {\bibfnamefont {G.~J.}\ \bibnamefont {{Wasserburg}}}}
  (\bibinfo {year} {1998}),\ \bibfield  {title} {\enquote {\bibinfo {title}
  {{Supernovae as the Site of the r-Process: Implications for Gamma-Ray
  Astronomy}},}\ }\href {https://doi.org/10.1086/306285} {\bibfield  {journal}
  {\bibinfo  {journal} {Astrophys. J.}\ }\textbf {\bibinfo {volume} {506}},\
  \bibinfo {pages} {868--873}}\BibitemShut {NoStop}%
\bibitem [{\citenamefont {{Qian}}\ \emph {et~al.}(1999)\citenamefont {{Qian}},
  \citenamefont {{Vogel}},\ and\ \citenamefont
  {{Wasserburg}}}]{Qian.Vogel.Wasserburg:1999}%
  \BibitemOpen
  \bibfield  {author} {\bibinfo {author} {\bibnamefont {{Qian}}, \bibfnamefont
  {Y-Z}}, \bibinfo {author} {\bibfnamefont {P.}~\bibnamefont {{Vogel}}}, and\
  \bibinfo {author} {\bibfnamefont {G.~J.}\ \bibnamefont {{Wasserburg}}}}
  (\bibinfo {year} {1999}),\ \bibfield  {title} {\enquote {\bibinfo {title}
  {{Probing r-Process Production of Nuclei Beyond $^{209}$Bi with Gamma
  Rays}},}\ }\href {https://doi.org/10.1086/307805} {\bibfield  {journal}
  {\bibinfo  {journal} {Astrophys. J.}\ }\textbf {\bibinfo {volume} {524}},\
  \bibinfo {pages} {213--219}}\BibitemShut {NoStop}%
\bibitem [{\citenamefont {{Qian}}\ and\ \citenamefont
  {{Wasserburg}}(2007)}]{Qian.Wasserburg:2007}%
  \BibitemOpen
  \bibfield  {author} {\bibinfo {author} {\bibnamefont {{Qian}}, \bibfnamefont
  {Y-Z}}, and\ \bibinfo {author} {\bibfnamefont {G.~J.}\ \bibnamefont
  {{Wasserburg}}}} (\bibinfo {year} {2007}),\ \bibfield  {title} {\enquote
  {\bibinfo {title} {{Where, oh where has the r-process gone?}}}\ }\href
  {https://doi.org/10.1016/j.physrep.2007.02.006} {\bibfield  {journal}
  {\bibinfo  {journal} {Phys. Rep.}\ }\textbf {\bibinfo {volume} {442}},\
  \bibinfo {pages} {237--268}}\BibitemShut {NoStop}%
\bibitem [{\citenamefont {{Qian}}\ and\ \citenamefont
  {{Woosley}}(1996)}]{Qian.Woosley:1996}%
  \BibitemOpen
  \bibfield  {author} {\bibinfo {author} {\bibnamefont {{Qian}}, \bibfnamefont
  {Y-Z}}, and\ \bibinfo {author} {\bibfnamefont {S.~E.}\ \bibnamefont
  {{Woosley}}}} (\bibinfo {year} {1996}),\ \bibfield  {title} {\enquote
  {\bibinfo {title} {{Nucleosynthesis in Neutrino-driven Winds. I. The Physical
  Conditions}},}\ }\href {https://doi.org/10.1086/177973} {\bibfield  {journal}
  {\bibinfo  {journal} {Astrophys. J.}\ }\textbf {\bibinfo {volume} {471}},\
  \bibinfo {pages} {331--351}}\BibitemShut {NoStop}%
\bibitem [{\citenamefont {Qian}(2014)}]{Qian:2014}%
  \BibitemOpen
  \bibfield  {author} {\bibinfo {author} {\bibnamefont {Qian}, \bibfnamefont
  {Yong-Zhong}}} (\bibinfo {year} {2014}),\ \bibfield  {title} {\enquote
  {\bibinfo {title} {Diverse, massive-star-associated sources for elements
  heavier than fe and the roles of neutrinos},}\ }\href
  {https://doi.org/10.1088/0954-3899/41/4/044002} {\bibfield  {journal}
  {\bibinfo  {journal} {J. Phys. G: Nucl. Part. Phys.}\ }\textbf {\bibinfo
  {volume} {41}},\ \bibinfo {pages} {044002}}\BibitemShut {NoStop}%
\bibitem [{\citenamefont {{Quinn}}\ \emph {et~al.}(2012)\citenamefont {{Quinn}}
  \emph {et~al.}}]{Quinn}%
  \BibitemOpen
  \bibfield  {author} {\bibinfo {author} {\bibnamefont {{Quinn}}, \bibfnamefont
  {M}},  \emph {et~al.}} (\bibinfo {year} {2012}),\ \bibfield  {title}
  {\enquote {\bibinfo {title} {{{$\beta$} decay of nuclei around $^{90}$Se:
  Search for signatures of a N=56 subshell closure relevant to the r
  process}},}\ }\href {https://doi.org/10.1103/PhysRevC.85.035807} {\bibfield
  {journal} {\bibinfo  {journal} {Phys. Rev. C}\ }\textbf {\bibinfo {volume}
  {85}},\ \bibinfo {eid} {035807}}\BibitemShut {NoStop}%
\bibitem [{\citenamefont {{Radice}}\ \emph {et~al.}(2016)\citenamefont
  {{Radice}}, \citenamefont {{Galeazzi}}, \citenamefont {{Lippuner}},
  \citenamefont {{Roberts}}, \citenamefont {{Ott}},\ and\ \citenamefont
  {{Rezzolla}}}]{Radice.Galeazzi.Lippuner.ea:2016}%
  \BibitemOpen
  \bibfield  {author} {\bibinfo {author} {\bibnamefont {{Radice}},
  \bibfnamefont {D}}, \bibinfo {author} {\bibfnamefont {F.}~\bibnamefont
  {{Galeazzi}}}, \bibinfo {author} {\bibfnamefont {J.}~\bibnamefont
  {{Lippuner}}}, \bibinfo {author} {\bibfnamefont {L.~F.}\ \bibnamefont
  {{Roberts}}}, \bibinfo {author} {\bibfnamefont {C.~D.}\ \bibnamefont
  {{Ott}}}, and\ \bibinfo {author} {\bibfnamefont {L.}~\bibnamefont
  {{Rezzolla}}}} (\bibinfo {year} {2016}),\ \bibfield  {title} {\enquote
  {\bibinfo {title} {{Dynamical mass ejection from binary neutron star
  mergers}},}\ }\href {https://doi.org/10.1093/mnras/stw1227} {\bibfield
  {journal} {\bibinfo  {journal} {Mon. Not. Roy. Astron. Soc.}\ }\textbf
  {\bibinfo {volume} {460}},\ \bibinfo {pages} {3255--3271}}\BibitemShut
  {NoStop}%
\bibitem [{\citenamefont {{Radice}}\ \emph
  {et~al.}(2018{\natexlab{a}})\citenamefont {{Radice}}, \citenamefont
  {{Perego}}, \citenamefont {{Hotokezaka}}, \citenamefont {{Bernuzzi}},
  \citenamefont {{Fromm}},\ and\ \citenamefont
  {{Roberts}}}]{Radice.Perego.ea:2018b}%
  \BibitemOpen
  \bibfield  {author} {\bibinfo {author} {\bibnamefont {{Radice}},
  \bibfnamefont {D}}, \bibinfo {author} {\bibfnamefont {A.}~\bibnamefont
  {{Perego}}}, \bibinfo {author} {\bibfnamefont {K.}~\bibnamefont
  {{Hotokezaka}}}, \bibinfo {author} {\bibfnamefont {S.}~\bibnamefont
  {{Bernuzzi}}}, \bibinfo {author} {\bibfnamefont {S.~A.}\ \bibnamefont
  {{Fromm}}}, and\ \bibinfo {author} {\bibfnamefont {L.~F.}\ \bibnamefont
  {{Roberts}}}} (\bibinfo {year} {2018}{\natexlab{a}}),\ \bibfield  {title}
  {\enquote {\bibinfo {title} {{Viscous-dynamical Ejecta from Binary Neutron
  Star Mergers}},}\ }\href {https://doi.org/10.3847/2041-8213/aaf053}
  {\bibfield  {journal} {\bibinfo  {journal} {Astrophys. J. Lett.}\ }\textbf
  {\bibinfo {volume} {869}},\ \bibinfo {eid} {L35}}\BibitemShut {NoStop}%
\bibitem [{\citenamefont {{Radice}}\ \emph
  {et~al.}(2018{\natexlab{b}})\citenamefont {{Radice}}, \citenamefont
  {{Perego}}, \citenamefont {{Hotokezaka}}, \citenamefont {{Fromm}},
  \citenamefont {{Bernuzzi}},\ and\ \citenamefont
  {{Roberts}}}]{Radice.Perego.ea:2018a}%
  \BibitemOpen
  \bibfield  {author} {\bibinfo {author} {\bibnamefont {{Radice}},
  \bibfnamefont {D}}, \bibinfo {author} {\bibfnamefont {A.}~\bibnamefont
  {{Perego}}}, \bibinfo {author} {\bibfnamefont {K.}~\bibnamefont
  {{Hotokezaka}}}, \bibinfo {author} {\bibfnamefont {S.~A.}\ \bibnamefont
  {{Fromm}}}, \bibinfo {author} {\bibfnamefont {S.}~\bibnamefont {{Bernuzzi}}},
  and\ \bibinfo {author} {\bibfnamefont {L.~F.}\ \bibnamefont {{Roberts}}}}
  (\bibinfo {year} {2018}{\natexlab{b}}),\ \bibfield  {title} {\enquote
  {\bibinfo {title} {{Binary Neutron Star Mergers: Mass Ejection,
  Electromagnetic Counterparts, and Nucleosynthesis}},}\ }\href
  {https://doi.org/10.3847/1538-4357/aaf054} {\bibfield  {journal} {\bibinfo
  {journal} {Astrophys. J.}\ }\textbf {\bibinfo {volume} {869}},\ \bibinfo
  {eid} {130}}\BibitemShut {NoStop}%
\bibitem [{\citenamefont {{Radice}}\ \emph {et~al.}(2020)\citenamefont
  {{Radice}}, \citenamefont {{Bernuzzi}},\ and\ \citenamefont
  {{Perego}}}]{Radice.Bernuzzi.Perego:2020}%
  \BibitemOpen
  \bibfield  {author} {\bibinfo {author} {\bibnamefont {{Radice}},
  \bibfnamefont {David}}, \bibinfo {author} {\bibfnamefont {Sebastiano}\
  \bibnamefont {{Bernuzzi}}}, and\ \bibinfo {author} {\bibfnamefont {Albino}\
  \bibnamefont {{Perego}}}} (\bibinfo {year} {2020}),\ \bibfield  {title}
  {\enquote {\bibinfo {title} {{The Dynamics of Binary Neutron Star Mergers and
  of GW170817}},}\ }\href {https://doi.org/10.1146/annurev-nucl-013120-114541}
  {\bibfield  {journal} {\bibinfo  {journal} {Annu. Rev. Nucl. Part. Sci.}\
  }\textbf {\bibinfo {volume} {70}},\ \bibinfo {pages} {95--119}}\BibitemShut
  {NoStop}%
\bibitem [{\citenamefont {{Radon}}\ \emph {et~al.}(2000)\citenamefont {{Radon}}
  \emph {et~al.}}]{Radon2000}%
  \BibitemOpen
  \bibfield  {author} {\bibinfo {author} {\bibnamefont {{Radon}}, \bibfnamefont
  {T}},  \emph {et~al.}} (\bibinfo {year} {2000}),\ \bibfield  {title}
  {\enquote {\bibinfo {title} {{Schottky mass measurements of stored and cooled
  neutron-deficient projectile fragments in the element range of $57\le Z \le
  84$}},}\ }\href {https://doi.org/10.1016/S0375-9474(00)00304-3} {\bibfield
  {journal} {\bibinfo  {journal} {Nucl. Phys. A}\ }\textbf {\bibinfo {volume}
  {677}},\ \bibinfo {pages} {75--99}}\BibitemShut {NoStop}%
\bibitem [{\citenamefont {{Rad{\v{z}}i{\={u}}t{\.{e}}}}\ \emph
  {et~al.}(2020)\citenamefont {{Rad{\v{z}}i{\={u}}t{\.{e}}}}, \citenamefont
  {{Gaigalas}}, \citenamefont {{Kato}}, \citenamefont {{Rynkun}},\ and\
  \citenamefont {{Tanaka}}}]{Radziute.Gaigalas.ea:2020}%
  \BibitemOpen
  \bibfield  {author} {\bibinfo {author} {\bibnamefont
  {{Rad{\v{z}}i{\={u}}t{\.{e}}}}, \bibfnamefont {Laima}}, \bibinfo {author}
  {\bibfnamefont {Gediminas}\ \bibnamefont {{Gaigalas}}}, \bibinfo {author}
  {\bibfnamefont {Daiji}\ \bibnamefont {{Kato}}}, \bibinfo {author}
  {\bibfnamefont {Pavel}\ \bibnamefont {{Rynkun}}}, and\ \bibinfo {author}
  {\bibfnamefont {Masaomi}\ \bibnamefont {{Tanaka}}}} (\bibinfo {year}
  {2020}),\ \bibfield  {title} {\enquote {\bibinfo {title} {{Extended
  calculations of energy levels and transition rates for singly ionized
  lanthanide elements I: Pr - Gd}},}\ }\href
  {https://doi.org/10.3847/1538-4365/ab8312} {\bibfield  {journal} {\bibinfo
  {journal} {Astrophys. J. Suppl.}\ }\textbf {\bibinfo {volume} {248}},\
  \bibinfo {eid} {17}}\BibitemShut {NoStop}%
\bibitem [{\citenamefont {{Raman}}\ \emph {et~al.}(1983)\citenamefont
  {{Raman}}, \citenamefont {{Fogelberg}}, \citenamefont {{Harvey}},
  \citenamefont {{Macklin}}, \citenamefont {{Stelson}}, \citenamefont
  {{Schr{\"o}der}},\ and\ \citenamefont {{Kratz}}}]{Raman1983}%
  \BibitemOpen
  \bibfield  {author} {\bibinfo {author} {\bibnamefont {{Raman}}, \bibfnamefont
  {S}}, \bibinfo {author} {\bibfnamefont {B.}~\bibnamefont {{Fogelberg}}},
  \bibinfo {author} {\bibfnamefont {J.~A.}\ \bibnamefont {{Harvey}}}, \bibinfo
  {author} {\bibfnamefont {R.~L.}\ \bibnamefont {{Macklin}}}, \bibinfo {author}
  {\bibfnamefont {P.~H.}\ \bibnamefont {{Stelson}}}, \bibinfo {author}
  {\bibfnamefont {A.}~\bibnamefont {{Schr{\"o}der}}}, and\ \bibinfo {author}
  {\bibfnamefont {K.-L.}\ \bibnamefont {{Kratz}}}} (\bibinfo {year} {1983}),\
  \bibfield  {title} {\enquote {\bibinfo {title} {{Overlapping beta decay and
  resonance neutron spectroscopy of levels in $^{87}$Kr}},}\ }\href
  {https://doi.org/10.1103/PhysRevC.28.602} {\bibfield  {journal} {\bibinfo
  {journal} {Phys. Rev. C}\ }\textbf {\bibinfo {volume} {28}},\ \bibinfo
  {pages} {602--622}}\BibitemShut {NoStop}%
\bibitem [{\citenamefont {{Ramirez-Ruiz}}\ \emph {et~al.}(2015)\citenamefont
  {{Ramirez-Ruiz}}, \citenamefont {{Trenti}}, \citenamefont {{MacLeod}},
  \citenamefont {{Roberts}}, \citenamefont {{Lee}},\ and\ \citenamefont
  {{Saladino-Rosas}}}]{Ramirez-Ruiz.Trenti.ea:2015}%
  \BibitemOpen
  \bibfield  {author} {\bibinfo {author} {\bibnamefont {{Ramirez-Ruiz}},
  \bibfnamefont {E}}, \bibinfo {author} {\bibfnamefont {M.}~\bibnamefont
  {{Trenti}}}, \bibinfo {author} {\bibfnamefont {M.}~\bibnamefont {{MacLeod}}},
  \bibinfo {author} {\bibfnamefont {L.~F.}\ \bibnamefont {{Roberts}}}, \bibinfo
  {author} {\bibfnamefont {W.~H.}\ \bibnamefont {{Lee}}}, and\ \bibinfo
  {author} {\bibfnamefont {M.~I.}\ \bibnamefont {{Saladino-Rosas}}}} (\bibinfo
  {year} {2015}),\ \bibfield  {title} {\enquote {\bibinfo {title} {{Compact
  Stellar Binary Assembly in the First Nuclear Star Clusters and r-process
  Synthesis in the Early Universe}},}\ }\href
  {https://doi.org/10.1088/2041-8205/802/2/L22} {\bibfield  {journal} {\bibinfo
   {journal} {Astrophys. J.}\ }\textbf {\bibinfo {volume} {802}},\ \bibinfo
  {eid} {L22}}\BibitemShut {NoStop}%
\bibitem [{\citenamefont {{Rauscher}}(2011)}]{Rauscher:2011}%
  \BibitemOpen
  \bibfield  {author} {\bibinfo {author} {\bibnamefont {{Rauscher}},
  \bibfnamefont {T}}} (\bibinfo {year} {2011}),\ \bibfield  {title} {\enquote
  {\bibinfo {title} {{The Path to Improved Reaction Rates for Astrophysics}},}\
  }\href {https://doi.org/10.1142/S021830131101840X} {\bibfield  {journal}
  {\bibinfo  {journal} {Int. J. Mod. Phys. E}\ }\textbf {\bibinfo {volume}
  {20}},\ \bibinfo {pages} {1071--1169}}\BibitemShut {NoStop}%
\bibitem [{\citenamefont {{Rauscher}}(2013)}]{Rauscher:2013}%
  \BibitemOpen
  \bibfield  {author} {\bibinfo {author} {\bibnamefont {{Rauscher}},
  \bibfnamefont {T}}} (\bibinfo {year} {2013}),\ \bibfield  {title} {\enquote
  {\bibinfo {title} {{Suppression of excited-state contributions to stellar
  reaction rates}},}\ }\href {https://doi.org/10.1103/PhysRevC.88.035803}
  {\bibfield  {journal} {\bibinfo  {journal} {Phys. Rev. C}\ }\textbf {\bibinfo
  {volume} {88}},\ \bibinfo {eid} {035803}}\BibitemShut {NoStop}%
\bibitem [{\citenamefont {{Rauscher}}\ \emph {et~al.}(1998)\citenamefont
  {{Rauscher}}, \citenamefont {{Bieber}}, \citenamefont {{Oberhummer}},
  \citenamefont {{Kratz}}, \citenamefont {{Dobaczewski}}, \citenamefont
  {{M{\"o}ller}},\ and\ \citenamefont {{Sharma}}}]{Rauscher.Bieber.ea:1998}%
  \BibitemOpen
  \bibfield  {author} {\bibinfo {author} {\bibnamefont {{Rauscher}},
  \bibfnamefont {T}}, \bibinfo {author} {\bibfnamefont {R.}~\bibnamefont
  {{Bieber}}}, \bibinfo {author} {\bibfnamefont {H.}~\bibnamefont
  {{Oberhummer}}}, \bibinfo {author} {\bibfnamefont {K.~L.}\ \bibnamefont
  {{Kratz}}}, \bibinfo {author} {\bibfnamefont {J.}~\bibnamefont
  {{Dobaczewski}}}, \bibinfo {author} {\bibfnamefont {P.}~\bibnamefont
  {{M{\"o}ller}}}, and\ \bibinfo {author} {\bibfnamefont {M.~M.}\ \bibnamefont
  {{Sharma}}}} (\bibinfo {year} {1998}),\ \bibfield  {title} {\enquote
  {\bibinfo {title} {{Dependence of direct neutron capture on nuclear-structure
  models}},}\ }\href {https://doi.org/10.1103/PhysRevC.57.2031} {\bibfield
  {journal} {\bibinfo  {journal} {Phys. Rev. C}\ }\textbf {\bibinfo {volume}
  {57}},\ \bibinfo {pages} {2031--2039}}\BibitemShut {NoStop}%
\bibitem [{\citenamefont {{Rauscher}}\ \emph {et~al.}(2013)\citenamefont
  {{Rauscher}}, \citenamefont {{Dauphas}}, \citenamefont {{Dillmann}},
  \citenamefont {{Fr{\"o}hlich}}, \citenamefont {{F{\"u}l{\"o}p}},\ and\
  \citenamefont {{Gy{\"u}rky}}}]{rauscher13}%
  \BibitemOpen
  \bibfield  {author} {\bibinfo {author} {\bibnamefont {{Rauscher}},
  \bibfnamefont {T}}, \bibinfo {author} {\bibfnamefont {N.}~\bibnamefont
  {{Dauphas}}}, \bibinfo {author} {\bibfnamefont {I.}~\bibnamefont
  {{Dillmann}}}, \bibinfo {author} {\bibfnamefont {C.}~\bibnamefont
  {{Fr{\"o}hlich}}}, \bibinfo {author} {\bibfnamefont {Z.}~\bibnamefont
  {{F{\"u}l{\"o}p}}}, and\ \bibinfo {author} {\bibfnamefont {G.}~\bibnamefont
  {{Gy{\"u}rky}}}} (\bibinfo {year} {2013}),\ \bibfield  {title} {\enquote
  {\bibinfo {title} {{Constraining the astrophysical origin of the p-nuclei
  through nuclear physics and meteoritic data}},}\ }\href
  {https://doi.org/10.1088/0034-4885/76/6/066201} {\bibfield  {journal}
  {\bibinfo  {journal} {Rep. Prog. Phys.}\ }\textbf {\bibinfo {volume} {76}},\
  \bibinfo {eid} {066201}}\BibitemShut {NoStop}%
\bibitem [{\citenamefont {{Rauscher}}\ and\ \citenamefont
  {{Thielemann}}(2000)}]{Rauscher.Thielemann:2000}%
  \BibitemOpen
  \bibfield  {author} {\bibinfo {author} {\bibnamefont {{Rauscher}},
  \bibfnamefont {T}}, and\ \bibinfo {author} {\bibfnamefont {F.-K.}\
  \bibnamefont {{Thielemann}}}} (\bibinfo {year} {2000}),\ \bibfield  {title}
  {\enquote {\bibinfo {title} {{Astrophysical Reaction Rates From Statistical
  Model Calculations}},}\ }\href {https://doi.org/10.1006/adnd.2000.0834}
  {\bibfield  {journal} {\bibinfo  {journal} {At. Data Nucl. Data Tables}\
  }\textbf {\bibinfo {volume} {75}},\ \bibinfo {pages} {1--351}}\BibitemShut
  {NoStop}%
\bibitem [{\citenamefont {{Rauscher}}\ and\ \citenamefont
  {{Thielemann}}(2001)}]{Rauscher.Thielemann:2001}%
  \BibitemOpen
  \bibfield  {author} {\bibinfo {author} {\bibnamefont {{Rauscher}},
  \bibfnamefont {T}}, and\ \bibinfo {author} {\bibfnamefont {F.-K.}\
  \bibnamefont {{Thielemann}}}} (\bibinfo {year} {2001}),\ \bibfield  {title}
  {\enquote {\bibinfo {title} {{Tables of Nuclear Cross Sections and Reaction
  Rates: An Addendum to the Paper ``ASTROPHYSICAL Reaction Rates from
  Statistical Model Calculations''}},}\ }\href
  {https://doi.org/10.1006/adnd.2001.0863} {\bibfield  {journal} {\bibinfo
  {journal} {At. Data Nucl. Data Tables}\ }\textbf {\bibinfo {volume} {79}},\
  \bibinfo {pages} {47--64}}\BibitemShut {NoStop}%
\bibitem [{\citenamefont {{Rauscher}}\ \emph {et~al.}(1997)\citenamefont
  {{Rauscher}}, \citenamefont {{Thielemann}},\ and\ \citenamefont
  {{Kratz}}}]{Rauscher.Thielemann.Kratz:1997}%
  \BibitemOpen
  \bibfield  {author} {\bibinfo {author} {\bibnamefont {{Rauscher}},
  \bibfnamefont {T}}, \bibinfo {author} {\bibfnamefont {F.-K.}\ \bibnamefont
  {{Thielemann}}}, and\ \bibinfo {author} {\bibfnamefont {K.-L.}\ \bibnamefont
  {{Kratz}}}} (\bibinfo {year} {1997}),\ \bibfield  {title} {\enquote {\bibinfo
  {title} {{Nuclear level density and the determination of thermonuclear rates
  for astrophysics}},}\ }\href {https://doi.org/10.1103/PhysRevC.56.1613}
  {\bibfield  {journal} {\bibinfo  {journal} {Phys. Rev. C}\ }\textbf {\bibinfo
  {volume} {56}},\ \bibinfo {pages} {1613--1625}}\BibitemShut {NoStop}%
\bibitem [{\citenamefont {{Rauscher}}(2008)}]{Rauscher:2008}%
  \BibitemOpen
  \bibfield  {author} {\bibinfo {author} {\bibnamefont {{Rauscher}},
  \bibfnamefont {Thomas}}} (\bibinfo {year} {2008}),\ \bibfield  {title}
  {\enquote {\bibinfo {title} {{Astrophysical relevance of
  {\ensuremath{\gamma}} transition energies}},}\ }\href
  {https://doi.org/10.1103/PhysRevC.78.032801} {\bibfield  {journal} {\bibinfo
  {journal} {Phys. Rev. C}\ }\textbf {\bibinfo {volume} {78}},\ \bibinfo {eid}
  {032801}}\BibitemShut {NoStop}%
\bibitem [{\citenamefont {{Ravn}}\ \emph {et~al.}(1975)\citenamefont {{Ravn}},
  \citenamefont {{Sundell}},\ and\ \citenamefont {{Westgaard}}}]{Ravn}%
  \BibitemOpen
  \bibfield  {author} {\bibinfo {author} {\bibnamefont {{Ravn}}, \bibfnamefont
  {H~L}}, \bibinfo {author} {\bibfnamefont {S.}~\bibnamefont {{Sundell}}}, and\
  \bibinfo {author} {\bibfnamefont {L.}~\bibnamefont {{Westgaard}}}} (\bibinfo
  {year} {1975}),\ \bibfield  {title} {\enquote {\bibinfo {title} {{Target
  Techniques for the isolde on-line isotope separator}},}\ }\href
  {https://doi.org/10.1016/0029-554X(75)90089-0} {\bibfield  {journal}
  {\bibinfo  {journal} {Nuclear Instruments and Methods}\ }\textbf {\bibinfo
  {volume} {123}},\ \bibinfo {pages} {131--144}}\BibitemShut {NoStop}%
\bibitem [{\citenamefont {{Recchi}}\ \emph {et~al.}(2009)\citenamefont
  {{Recchi}}, \citenamefont {{Calura}},\ and\ \citenamefont
  {{Kroupa}}}]{Recchi:2009}%
  \BibitemOpen
  \bibfield  {author} {\bibinfo {author} {\bibnamefont {{Recchi}},
  \bibfnamefont {S}}, \bibinfo {author} {\bibfnamefont {F.}~\bibnamefont
  {{Calura}}}, and\ \bibinfo {author} {\bibfnamefont {P.}~\bibnamefont
  {{Kroupa}}}} (\bibinfo {year} {2009}),\ \bibfield  {title} {\enquote
  {\bibinfo {title} {{The chemical evolution of galaxies within the IGIMF
  theory: the [{$\alpha$}/Fe] ratios and downsizing}},}\ }\href
  {https://doi.org/10.1051/0004-6361/200811472} {\bibfield  {journal} {\bibinfo
   {journal} {Astron. \& Astrophys.}\ }\textbf {\bibinfo {volume} {499}},\
  \bibinfo {pages} {711--722}},\ \Eprint {https://arxiv.org/abs/0903.2395}
  {0903.2395} \BibitemShut {NoStop}%
\bibitem [{\citenamefont {{Recchi}}\ \emph {et~al.}(2001)\citenamefont
  {{Recchi}}, \citenamefont {{Matteucci}},\ and\ \citenamefont
  {{D'Ercole}}}]{Recchi:2001}%
  \BibitemOpen
  \bibfield  {author} {\bibinfo {author} {\bibnamefont {{Recchi}},
  \bibfnamefont {S}}, \bibinfo {author} {\bibfnamefont {F.}~\bibnamefont
  {{Matteucci}}}, and\ \bibinfo {author} {\bibfnamefont {A.}~\bibnamefont
  {{D'Ercole}}}} (\bibinfo {year} {2001}),\ \bibfield  {title} {\enquote
  {\bibinfo {title} {{Dynamical and chemical evolution of gas-rich dwarf
  galaxies}},}\ }\href {https://doi.org/10.1046/j.1365-8711.2001.04189.x}
  {\bibfield  {journal} {\bibinfo  {journal} {Mon. Not. Roy. Astron. Soc.}\
  }\textbf {\bibinfo {volume} {322}},\ \bibinfo {pages} {800--820}}\BibitemShut
  {NoStop}%
\bibitem [{\citenamefont {{Reddy}}\ \emph {et~al.}(2006)\citenamefont
  {{Reddy}}, \citenamefont {{Lambert}},\ and\ \citenamefont {{Allende
  Prieto}}}]{reddy06}%
  \BibitemOpen
  \bibfield  {author} {\bibinfo {author} {\bibnamefont {{Reddy}}, \bibfnamefont
  {B~E}}, \bibinfo {author} {\bibfnamefont {D.~L.}\ \bibnamefont {{Lambert}}},
  and\ \bibinfo {author} {\bibfnamefont {C.}~\bibnamefont {{Allende Prieto}}}}
  (\bibinfo {year} {2006}),\ \bibfield  {title} {\enquote {\bibinfo {title}
  {{Elemental abundance survey of the Galactic thick disc}},}\ }\href
  {https://doi.org/10.1111/j.1365-2966.2006.10148.x} {\bibfield  {journal}
  {\bibinfo  {journal} {Mon. Not. Roy. Astron. Soc.}\ }\textbf {\bibinfo
  {volume} {367}},\ \bibinfo {pages} {1329--1366}}\BibitemShut {NoStop}%
\bibitem [{\citenamefont {{Reddy}}\ \emph {et~al.}(2003)\citenamefont
  {{Reddy}}, \citenamefont {{Tomkin}}, \citenamefont {{Lambert}},\ and\
  \citenamefont {{Allende Prieto}}}]{reddy03}%
  \BibitemOpen
  \bibfield  {author} {\bibinfo {author} {\bibnamefont {{Reddy}}, \bibfnamefont
  {B~E}}, \bibinfo {author} {\bibfnamefont {J.}~\bibnamefont {{Tomkin}}},
  \bibinfo {author} {\bibfnamefont {D.~L.}\ \bibnamefont {{Lambert}}}, and\
  \bibinfo {author} {\bibfnamefont {C.}~\bibnamefont {{Allende Prieto}}}}
  (\bibinfo {year} {2003}),\ \bibfield  {title} {\enquote {\bibinfo {title}
  {{The chemical compositions of Galactic disc F and G dwarfs}},}\ }\href
  {https://doi.org/10.1046/j.1365-8711.2003.06305.x} {\bibfield  {journal}
  {\bibinfo  {journal} {Mon. Not. Roy. Astron. Soc.}\ }\textbf {\bibinfo
  {volume} {340}},\ \bibinfo {pages} {304--340}}\BibitemShut {NoStop}%
\bibitem [{\citenamefont {{Rehse}}\ \emph {et~al.}(2006)\citenamefont
  {{Rehse}}, \citenamefont {{Li}}, \citenamefont {{Scholl}}, \citenamefont
  {{Sharikova}}, \citenamefont {{Chatelain}}, \citenamefont {{Holt}},\ and\
  \citenamefont {{Rosner}}}]{rehse06}%
  \BibitemOpen
  \bibfield  {author} {\bibinfo {author} {\bibnamefont {{Rehse}}, \bibfnamefont
  {S~J}}, \bibinfo {author} {\bibfnamefont {R.}~\bibnamefont {{Li}}}, \bibinfo
  {author} {\bibfnamefont {T.~J.}\ \bibnamefont {{Scholl}}}, \bibinfo {author}
  {\bibfnamefont {A.}~\bibnamefont {{Sharikova}}}, \bibinfo {author}
  {\bibfnamefont {R.}~\bibnamefont {{Chatelain}}}, \bibinfo {author}
  {\bibfnamefont {R.~A.}\ \bibnamefont {{Holt}}}, and\ \bibinfo {author}
  {\bibfnamefont {S.~D.}\ \bibnamefont {{Rosner}}}} (\bibinfo {year} {2006}),\
  \bibfield  {title} {\enquote {\bibinfo {title} {{Fast-ion-beam
  laser-induced-fluorescence measurements of spontaneous-emission branching
  ratios and oscillator strengths in SmII}},}\ }\href
  {https://doi.org/10.1139/P06-063} {\bibfield  {journal} {\bibinfo  {journal}
  {Canadian Journal of Physics}\ }\textbf {\bibinfo {volume} {84}},\ \bibinfo
  {pages} {723--771}}\BibitemShut {NoStop}%
\bibitem [{\citenamefont {{Reichert}}\ \emph {et~al.}(2020)\citenamefont
  {{Reichert}}, \citenamefont {{Obergaulinger}}, \citenamefont {{Eichler}},
  \citenamefont {{Aloy}},\ and\ \citenamefont
  {{Arcones}}}]{Reichert.Obergaulinger.ea:2020}%
  \BibitemOpen
  \bibfield  {author} {\bibinfo {author} {\bibnamefont {{Reichert}},
  \bibfnamefont {Moritz}}, \bibinfo {author} {\bibfnamefont {Martin}\
  \bibnamefont {{Obergaulinger}}}, \bibinfo {author} {\bibfnamefont {Marius}\
  \bibnamefont {{Eichler}}}, \bibinfo {author} {\bibfnamefont
  {Miguel-{\'A}ngel}\ \bibnamefont {{Aloy}}}, and\ \bibinfo {author}
  {\bibfnamefont {Almudena}\ \bibnamefont {{Arcones}}}} (\bibinfo {year}
  {2020}),\ \bibfield  {title} {\enquote {\bibinfo {title} {{Nucleosynthesis in
  magneto-rotational supernovae}},}\ }\href@noop {} {\bibfield  {journal}
  {\bibinfo  {journal} {arXiv e-prints}\ }}\Eprint
  {https://arxiv.org/abs/2010.02227} {arXiv:2010.02227 [astro-ph.HE]}
  \BibitemShut {NoStop}%
\bibitem [{\citenamefont {{Reifarth}}\ \emph {et~al.}(2017)\citenamefont
  {{Reifarth}}, \citenamefont {{G{\"o}bel}}, \citenamefont {{Heftrich}},
  \citenamefont {{Weigand}}, \citenamefont {{Jurado}}, \citenamefont
  {{K{\"a}ppeler}},\ and\ \citenamefont {{Litvinov}}}]{Reifarth2017}%
  \BibitemOpen
  \bibfield  {author} {\bibinfo {author} {\bibnamefont {{Reifarth}},
  \bibfnamefont {R}}, \bibinfo {author} {\bibfnamefont {K.}~\bibnamefont
  {{G{\"o}bel}}}, \bibinfo {author} {\bibfnamefont {T.}~\bibnamefont
  {{Heftrich}}}, \bibinfo {author} {\bibfnamefont {M.}~\bibnamefont
  {{Weigand}}}, \bibinfo {author} {\bibfnamefont {B.}~\bibnamefont {{Jurado}}},
  \bibinfo {author} {\bibfnamefont {F.}~\bibnamefont {{K{\"a}ppeler}}}, and\
  \bibinfo {author} {\bibfnamefont {Y.~A.}\ \bibnamefont {{Litvinov}}}}
  (\bibinfo {year} {2017}),\ \bibfield  {title} {\enquote {\bibinfo {title}
  {{Spallation-based neutron target for direct studies of neutron-induced
  reactions in inverse kinematics}},}\ }\href
  {https://doi.org/10.1103/PhysRevAccelBeams.20.044701} {\bibfield  {journal}
  {\bibinfo  {journal} {Phys. Rev. Accel. Beams}\ }\textbf {\bibinfo {volume}
  {20}},\ \bibinfo {eid} {044701}}\BibitemShut {NoStop}%
\bibitem [{\citenamefont {{Reifarth}}\ \emph {et~al.}(2014)\citenamefont
  {{Reifarth}}, \citenamefont {{Lederer}},\ and\ \citenamefont
  {{K{\"a}ppeler}}}]{Reifarth.Lederer.Kaeppeler:2014}%
  \BibitemOpen
  \bibfield  {author} {\bibinfo {author} {\bibnamefont {{Reifarth}},
  \bibfnamefont {R}}, \bibinfo {author} {\bibfnamefont {C.}~\bibnamefont
  {{Lederer}}}, and\ \bibinfo {author} {\bibfnamefont {F.}~\bibnamefont
  {{K{\"a}ppeler}}}} (\bibinfo {year} {2014}),\ \bibfield  {title} {\enquote
  {\bibinfo {title} {{Neutron reactions in astrophysics}},}\ }\href
  {https://doi.org/10.1088/0954-3899/41/5/053101} {\bibfield  {journal}
  {\bibinfo  {journal} {J. Phys. G: Nucl. Part. Phys.}\ }\textbf {\bibinfo
  {volume} {41}},\ \bibinfo {eid} {053101}}\BibitemShut {NoStop}%
\bibitem [{\citenamefont {{Reifarth}}\ and\ \citenamefont
  {{Litvinov}}(2014)}]{Reifarth2014}%
  \BibitemOpen
  \bibfield  {author} {\bibinfo {author} {\bibnamefont {{Reifarth}},
  \bibfnamefont {R}}, and\ \bibinfo {author} {\bibfnamefont {Y.~A.}\
  \bibnamefont {{Litvinov}}}} (\bibinfo {year} {2014}),\ \bibfield  {title}
  {\enquote {\bibinfo {title} {{Measurements of neutron-induced reactions in
  inverse kinematics}},}\ }\href
  {https://doi.org/10.1103/PhysRevSTAB.17.014701} {\bibfield  {journal}
  {\bibinfo  {journal} {Phys. Rev. Accel. Beams}\ }\textbf {\bibinfo {volume}
  {17}},\ \bibinfo {eid} {014701}}\BibitemShut {NoStop}%
\bibitem [{\citenamefont {Reiter}\ \emph {et~al.}(2020)\citenamefont {Reiter}
  \emph {et~al.}}]{Reiter.Ayet.ea:2020}%
  \BibitemOpen
  \bibfield  {author} {\bibinfo {author} {\bibnamefont {Reiter}, \bibfnamefont
  {M~P}},  \emph {et~al.}} (\bibinfo {year} {2020}),\ \bibfield  {title}
  {\enquote {\bibinfo {title} {Mass measurements of neutron-rich gallium
  isotopes refine production of nuclei of the first $r$-process abundance peak
  in neutron-star merger calculations},}\ }\href
  {https://doi.org/10.1103/PhysRevC.101.025803} {\bibfield  {journal} {\bibinfo
   {journal} {Phys. Rev. C}\ }\textbf {\bibinfo {volume} {101}},\ \bibinfo
  {pages} {025803}}\BibitemShut {NoStop}%
\bibitem [{\citenamefont {{Rembges}}\ \emph {et~al.}(1997)\citenamefont
  {{Rembges}}, \citenamefont {{Freiburghaus}}, \citenamefont {{Rauscher}},
  \citenamefont {{Thielemann}}, \citenamefont {{Schatz}},\ and\ \citenamefont
  {{Wiescher}}}]{rembges97}%
  \BibitemOpen
  \bibfield  {author} {\bibinfo {author} {\bibnamefont {{Rembges}},
  \bibfnamefont {F}}, \bibinfo {author} {\bibfnamefont {C.}~\bibnamefont
  {{Freiburghaus}}}, \bibinfo {author} {\bibfnamefont {T.}~\bibnamefont
  {{Rauscher}}}, \bibinfo {author} {\bibfnamefont {F.-K.}\ \bibnamefont
  {{Thielemann}}}, \bibinfo {author} {\bibfnamefont {H.}~\bibnamefont
  {{Schatz}}}, and\ \bibinfo {author} {\bibfnamefont {M.}~\bibnamefont
  {{Wiescher}}}} (\bibinfo {year} {1997}),\ \bibfield  {title} {\enquote
  {\bibinfo {title} {{An Approximation for the rp-Process}},}\ }\href
  {https://doi.org/10.1086/304300} {\bibfield  {journal} {\bibinfo  {journal}
  {Astrophys. J.}\ }\textbf {\bibinfo {volume} {484}},\ \bibinfo {pages}
  {412--423}}\BibitemShut {NoStop}%
\bibitem [{\citenamefont {{Renaud}}\ \emph {et~al.}(2006)\citenamefont
  {{Renaud}}, \citenamefont {{Vink}}, \citenamefont {{Decourchelle}},
  \citenamefont {{Lebrun}}, \citenamefont {{den Hartog}}, \citenamefont
  {{Terrier}}, \citenamefont {{Couvreur}}, \citenamefont {{Kn{\"o}dlseder}},
  \citenamefont {{Martin}}, \citenamefont {{Prantzos}}, \citenamefont
  {{Bykov}},\ and\ \citenamefont {{Bloemen}}}]{Renaud.Vink.ea:2006}%
  \BibitemOpen
  \bibfield  {author} {\bibinfo {author} {\bibnamefont {{Renaud}},
  \bibfnamefont {M}}, \bibinfo {author} {\bibfnamefont {J.}~\bibnamefont
  {{Vink}}}, \bibinfo {author} {\bibfnamefont {A.}~\bibnamefont
  {{Decourchelle}}}, \bibinfo {author} {\bibfnamefont {F.}~\bibnamefont
  {{Lebrun}}}, \bibinfo {author} {\bibfnamefont {P.~R.}\ \bibnamefont {{den
  Hartog}}}, \bibinfo {author} {\bibfnamefont {R.}~\bibnamefont {{Terrier}}},
  \bibinfo {author} {\bibfnamefont {C.}~\bibnamefont {{Couvreur}}}, \bibinfo
  {author} {\bibfnamefont {J.}~\bibnamefont {{Kn{\"o}dlseder}}}, \bibinfo
  {author} {\bibfnamefont {P.}~\bibnamefont {{Martin}}}, \bibinfo {author}
  {\bibfnamefont {N.}~\bibnamefont {{Prantzos}}}, \bibinfo {author}
  {\bibfnamefont {A.~M.}\ \bibnamefont {{Bykov}}}, and\ \bibinfo {author}
  {\bibfnamefont {H.}~\bibnamefont {{Bloemen}}}} (\bibinfo {year} {2006}),\
  \bibfield  {title} {\enquote {\bibinfo {title} {{The Signature of $^{44}$Ti
  in Cassiopeia A Revealed by IBIS/ISGRI on INTEGRAL}},}\ }\href
  {https://doi.org/10.1086/507300} {\bibfield  {journal} {\bibinfo  {journal}
  {Astrophys. J. Lett.}\ }\textbf {\bibinfo {volume} {647}},\ \bibinfo {pages}
  {L41--L44}}\BibitemShut {NoStop}%
\bibitem [{\citenamefont {{Rezzolla}}\ \emph {et~al.}(2018)\citenamefont
  {{Rezzolla}}, \citenamefont {{Most}},\ and\ \citenamefont
  {{Weih}}}]{Rezzolla.Most.Weih:2018}%
  \BibitemOpen
  \bibfield  {author} {\bibinfo {author} {\bibnamefont {{Rezzolla}},
  \bibfnamefont {Luciano}}, \bibinfo {author} {\bibfnamefont {Elias~R.}\
  \bibnamefont {{Most}}}, and\ \bibinfo {author} {\bibfnamefont {Lukas~R.}\
  \bibnamefont {{Weih}}}} (\bibinfo {year} {2018}),\ \bibfield  {title}
  {\enquote {\bibinfo {title} {{Using Gravitational-wave Observations and
  Quasi-universal Relations to Constrain the Maximum Mass of Neutron Stars}},}\
  }\href {https://doi.org/10.3847/2041-8213/aaa401} {\bibfield  {journal}
  {\bibinfo  {journal} {Astrophys. J.}\ }\textbf {\bibinfo {volume} {852}},\
  \bibinfo {eid} {L25}}\BibitemShut {NoStop}%
\bibitem [{\citenamefont {{Richers}}\ \emph {et~al.}(2017)\citenamefont
  {{Richers}}, \citenamefont {{Nagakura}}, \citenamefont {{Ott}}, \citenamefont
  {{Dolence}}, \citenamefont {{Sumiyoshi}},\ and\ \citenamefont
  {{Yamada}}}]{Richers.Nagakura.ea:2017}%
  \BibitemOpen
  \bibfield  {author} {\bibinfo {author} {\bibnamefont {{Richers}},
  \bibfnamefont {S}}, \bibinfo {author} {\bibfnamefont {H.}~\bibnamefont
  {{Nagakura}}}, \bibinfo {author} {\bibfnamefont {C.~D.}\ \bibnamefont
  {{Ott}}}, \bibinfo {author} {\bibfnamefont {J.}~\bibnamefont {{Dolence}}},
  \bibinfo {author} {\bibfnamefont {K.}~\bibnamefont {{Sumiyoshi}}}, and\
  \bibinfo {author} {\bibfnamefont {S.}~\bibnamefont {{Yamada}}}} (\bibinfo
  {year} {2017}),\ \bibfield  {title} {\enquote {\bibinfo {title} {{A Detailed
  Comparison of Multidimensional Boltzmann Neutrino Transport Methods in
  Core-collapse Supernovae}},}\ }\href
  {https://doi.org/10.3847/1538-4357/aa8bb2} {\bibfield  {journal} {\bibinfo
  {journal} {Astrophys. J.}\ }\textbf {\bibinfo {volume} {847}},\ \bibinfo
  {eid} {133}}\BibitemShut {NoStop}%
\bibitem [{\citenamefont {{Ripley}}\ \emph {et~al.}(2014)\citenamefont
  {{Ripley}}, \citenamefont {{Metzger}}, \citenamefont {{Arcones}},\ and\
  \citenamefont {{Mart{\'{\i}}nez-Pinedo}}}]{Ripley.Metzger.ea:2014}%
  \BibitemOpen
  \bibfield  {author} {\bibinfo {author} {\bibnamefont {{Ripley}},
  \bibfnamefont {J~L}}, \bibinfo {author} {\bibfnamefont {B.~D.}\ \bibnamefont
  {{Metzger}}}, \bibinfo {author} {\bibfnamefont {A.}~\bibnamefont
  {{Arcones}}}, and\ \bibinfo {author} {\bibfnamefont {G.}~\bibnamefont
  {{Mart{\'{\i}}nez-Pinedo}}}} (\bibinfo {year} {2014}),\ \bibfield  {title}
  {\enquote {\bibinfo {title} {{X-ray decay lines from heavy nuclei in
  supernova remnants as a probe of the r-process origin and the birth periods
  of magnetars}},}\ }\href {https://doi.org/10.1093/mnras/stt2434} {\bibfield
  {journal} {\bibinfo  {journal} {Mon. Not. Roy. Astron. Soc.}\ }\textbf
  {\bibinfo {volume} {438}},\ \bibinfo {pages} {3243--3254}}\BibitemShut
  {NoStop}%
\bibitem [{\citenamefont {{Roberts}}(2012)}]{Roberts:2012}%
  \BibitemOpen
  \bibfield  {author} {\bibinfo {author} {\bibnamefont {{Roberts}},
  \bibfnamefont {L~F}}} (\bibinfo {year} {2012}),\ \bibfield  {title} {\enquote
  {\bibinfo {title} {{A New Code for Proto-neutron Star Evolution}},}\ }\href
  {https://doi.org/10.1088/0004-637X/755/2/126} {\bibfield  {journal} {\bibinfo
   {journal} {Astrophys. J.}\ }\textbf {\bibinfo {volume} {755}},\ \bibinfo
  {eid} {126}}\BibitemShut {NoStop}%
\bibitem [{\citenamefont {{Roberts}}\ \emph {et~al.}(2017)\citenamefont
  {{Roberts}}, \citenamefont {{Lippuner}}, \citenamefont {{Duez}},
  \citenamefont {{Faber}}, \citenamefont {{Foucart}}, \citenamefont
  {{Lombardi}}, \citenamefont {{Ning}}, \citenamefont {{Ott}},\ and\
  \citenamefont {{Ponce}}}]{roberts17}%
  \BibitemOpen
  \bibfield  {author} {\bibinfo {author} {\bibnamefont {{Roberts}},
  \bibfnamefont {L~F}}, \bibinfo {author} {\bibfnamefont {J.}~\bibnamefont
  {{Lippuner}}}, \bibinfo {author} {\bibfnamefont {M.~D.}\ \bibnamefont
  {{Duez}}}, \bibinfo {author} {\bibfnamefont {J.~A.}\ \bibnamefont {{Faber}}},
  \bibinfo {author} {\bibfnamefont {F.}~\bibnamefont {{Foucart}}}, \bibinfo
  {author} {\bibfnamefont {J.~C.}\ \bibnamefont {{Lombardi}}, \bibfnamefont
  {Jr.}}, \bibinfo {author} {\bibfnamefont {S.}~\bibnamefont {{Ning}}},
  \bibinfo {author} {\bibfnamefont {C.~D.}\ \bibnamefont {{Ott}}}, and\
  \bibinfo {author} {\bibfnamefont {M.}~\bibnamefont {{Ponce}}}} (\bibinfo
  {year} {2017}),\ \bibfield  {title} {\enquote {\bibinfo {title} {{The
  influence of neutrinos on r-process nucleosynthesis in the ejecta of black
  hole-neutron star mergers}},}\ }\href {https://doi.org/10.1093/mnras/stw2622}
  {\bibfield  {journal} {\bibinfo  {journal} {Mon. Not. Roy. Astron. Soc.}\
  }\textbf {\bibinfo {volume} {464}},\ \bibinfo {pages}
  {3907--3919}}\BibitemShut {NoStop}%
\bibitem [{\citenamefont {{Roberts}}\ \emph {et~al.}(2012)\citenamefont
  {{Roberts}}, \citenamefont {{Reddy}},\ and\ \citenamefont
  {{Shen}}}]{Roberts.Reddy.Shen:2012}%
  \BibitemOpen
  \bibfield  {author} {\bibinfo {author} {\bibnamefont {{Roberts}},
  \bibfnamefont {L~F}}, \bibinfo {author} {\bibfnamefont {S.}~\bibnamefont
  {{Reddy}}}, and\ \bibinfo {author} {\bibfnamefont {G.}~\bibnamefont
  {{Shen}}}} (\bibinfo {year} {2012}),\ \bibfield  {title} {\enquote {\bibinfo
  {title} {{Medium modification of the charged-current neutrino opacity and its
  implications}},}\ }\href {https://doi.org/10.1103/PhysRevC.86.065803}
  {\bibfield  {journal} {\bibinfo  {journal} {Phys. Rev. C}\ }\textbf {\bibinfo
  {volume} {86}},\ \bibinfo {eid} {065803}}\BibitemShut {NoStop}%
\bibitem [{\citenamefont {Roberts}\ \emph {et~al.}(2012)\citenamefont
  {Roberts}, \citenamefont {Shen}, \citenamefont {Cirigliano}, \citenamefont
  {Pons}, \citenamefont {Reddy},\ and\ \citenamefont
  {Woosley}}]{Roberts.Shen.ea:2012}%
  \BibitemOpen
  \bibfield  {author} {\bibinfo {author} {\bibnamefont {Roberts}, \bibfnamefont
  {L~F}}, \bibinfo {author} {\bibfnamefont {G.}~\bibnamefont {Shen}}, \bibinfo
  {author} {\bibfnamefont {V.}~\bibnamefont {Cirigliano}}, \bibinfo {author}
  {\bibfnamefont {J.~A.}\ \bibnamefont {Pons}}, \bibinfo {author}
  {\bibfnamefont {S.}~\bibnamefont {Reddy}}, and\ \bibinfo {author}
  {\bibfnamefont {S.~E.}\ \bibnamefont {Woosley}}} (\bibinfo {year} {2012}),\
  \bibfield  {title} {\enquote {\bibinfo {title} {Protoneutron star cooling
  with convection: The effect of the symmetry energy},}\ }\href
  {https://doi.org/10.1103/PhysRevLett.108.061103} {\bibfield  {journal}
  {\bibinfo  {journal} {Phys. Rev. Lett.}\ }\textbf {\bibinfo {volume} {108}},\
  \bibinfo {pages} {061103}}\BibitemShut {NoStop}%
\bibitem [{\citenamefont {{Roberts}}\ \emph {et~al.}(2010)\citenamefont
  {{Roberts}}, \citenamefont {{Woosley}},\ and\ \citenamefont
  {{Hoffman}}}]{Roberts.Woosley.Hoffman:2010}%
  \BibitemOpen
  \bibfield  {author} {\bibinfo {author} {\bibnamefont {{Roberts}},
  \bibfnamefont {L~F}}, \bibinfo {author} {\bibfnamefont {S.~E.}\ \bibnamefont
  {{Woosley}}}, and\ \bibinfo {author} {\bibfnamefont {R.~D.}\ \bibnamefont
  {{Hoffman}}}} (\bibinfo {year} {2010}),\ \bibfield  {title} {\enquote
  {\bibinfo {title} {{Integrated Nucleosynthesis in Neutrino-driven Winds}},}\
  }\href {https://doi.org/10.1088/0004-637X/722/1/954} {\bibfield  {journal}
  {\bibinfo  {journal} {Astrophys. J.}\ }\textbf {\bibinfo {volume} {722}},\
  \bibinfo {pages} {954--967}}\BibitemShut {NoStop}%
\bibitem [{\citenamefont {Roberts}\ \emph {et~al.}(2011)\citenamefont
  {Roberts}, \citenamefont {Kasen}, \citenamefont {Lee},\ and\ \citenamefont
  {Ramirez-Ruiz}}]{Roberts.Kasen.ea:2011}%
  \BibitemOpen
  \bibfield  {author} {\bibinfo {author} {\bibnamefont {Roberts}, \bibfnamefont
  {Luke~F}}, \bibinfo {author} {\bibfnamefont {Dan}\ \bibnamefont {Kasen}},
  \bibinfo {author} {\bibfnamefont {William~H.}\ \bibnamefont {Lee}}, and\
  \bibinfo {author} {\bibfnamefont {Enrico}\ \bibnamefont {Ramirez-Ruiz}}}
  (\bibinfo {year} {2011}),\ \bibfield  {title} {\enquote {\bibinfo {title}
  {{Electromagnetic Transients Powered by Nuclear Decay in the Tidal Tails of
  Coalescing Compact Binaries}},}\ }\href
  {https://doi.org/10.1088/2041-8205/736/1/L21} {\bibfield  {journal} {\bibinfo
   {journal} {Astrophys. J.}\ }\textbf {\bibinfo {volume} {736}},\ \bibinfo
  {pages} {L21}}\BibitemShut {NoStop}%
\bibitem [{\citenamefont {Roberts}\ and\ \citenamefont
  {Reddy}(2017)}]{Roberts.Reddy:2017}%
  \BibitemOpen
  \bibfield  {author} {\bibinfo {author} {\bibnamefont {Roberts}, \bibfnamefont
  {Luke~F}}, and\ \bibinfo {author} {\bibfnamefont {Sanjay}\ \bibnamefont
  {Reddy}}} (\bibinfo {year} {2017}),\ \bibfield  {title} {\enquote {\bibinfo
  {title} {Charged current neutrino interactions in hot and dense matter},}\
  }\href {https://doi.org/10.1103/PhysRevC.95.045807} {\bibfield  {journal}
  {\bibinfo  {journal} {Phys. Rev. C}\ }\textbf {\bibinfo {volume} {95}},\
  \bibinfo {pages} {045807}}\BibitemShut {NoStop}%
\bibitem [{\citenamefont {Rodr\'{i}guez}\ \emph {et~al.}(2015)\citenamefont
  {Rodr\'{i}guez}, \citenamefont {Arzhanov},\ and\ \citenamefont
  {Mart\'{i}nez-Pinedo}}]{Rodriguez.Arzhanov.Martinez-Pinedo:2015}%
  \BibitemOpen
  \bibfield  {author} {\bibinfo {author} {\bibnamefont {Rodr\'{i}guez},
  \bibfnamefont {Tom\'as~R}}, \bibinfo {author} {\bibfnamefont {Alexander}\
  \bibnamefont {Arzhanov}}, and\ \bibinfo {author} {\bibfnamefont {Gabriel}\
  \bibnamefont {Mart\'{i}nez-Pinedo}}} (\bibinfo {year} {2015}),\ \bibfield
  {title} {\enquote {\bibinfo {title} {Toward global beyond-mean-field
  calculations of nuclear masses and low-energy spectra},}\ }\href
  {https://doi.org/10.1103/PhysRevC.91.044315} {\bibfield  {journal} {\bibinfo
  {journal} {Phys. Rev. C}\ }\textbf {\bibinfo {volume} {91}},\ \bibinfo
  {pages} {044315}}\BibitemShut {NoStop}%
\bibitem [{\citenamefont {{Roederer}}(2013)}]{Roederer:2013}%
  \BibitemOpen
  \bibfield  {author} {\bibinfo {author} {\bibnamefont {{Roederer}},
  \bibfnamefont {I~U}}} (\bibinfo {year} {2013}),\ \bibfield  {title} {\enquote
  {\bibinfo {title} {{Are There Any Stars Lacking Neutron-capture Elements?
  Evidence from Strontium and Barium}},}\ }\href
  {https://doi.org/10.1088/0004-6256/145/1/26} {\bibfield  {journal} {\bibinfo
  {journal} {Astron. J.}\ }\textbf {\bibinfo {volume} {145}},\ \bibinfo {eid}
  {26}}\BibitemShut {NoStop}%
\bibitem [{\citenamefont {{Roederer}}(2017)}]{Roederer:2017}%
  \BibitemOpen
  \bibfield  {author} {\bibinfo {author} {\bibnamefont {{Roederer}},
  \bibfnamefont {I~U}}} (\bibinfo {year} {2017}),\ \bibfield  {title} {\enquote
  {\bibinfo {title} {{The Origin of the Heaviest Metals in Most Ultra-faint
  Dwarf Galaxies}},}\ }\href {https://doi.org/10.3847/1538-4357/835/1/23}
  {\bibfield  {journal} {\bibinfo  {journal} {Astrophys. J.}\ }\textbf
  {\bibinfo {volume} {835}},\ \bibinfo {eid} {23}}\BibitemShut {NoStop}%
\bibitem [{\citenamefont {{Roederer}}\ \emph
  {et~al.}(2010{\natexlab{a}})\citenamefont {{Roederer}}, \citenamefont
  {{Cowan}}, \citenamefont {{Karakas}}, \citenamefont {{Kratz}}, \citenamefont
  {{Lugaro}}, \citenamefont {{Simmerer}}, \citenamefont {{Farouqi}},\ and\
  \citenamefont {{Sneden}}}]{Roederer.Cowan.ea:2010}%
  \BibitemOpen
  \bibfield  {author} {\bibinfo {author} {\bibnamefont {{Roederer}},
  \bibfnamefont {I~U}}, \bibinfo {author} {\bibfnamefont {J.~J.}\ \bibnamefont
  {{Cowan}}}, \bibinfo {author} {\bibfnamefont {A.~I.}\ \bibnamefont
  {{Karakas}}}, \bibinfo {author} {\bibfnamefont {K.-L.}\ \bibnamefont
  {{Kratz}}}, \bibinfo {author} {\bibfnamefont {M.}~\bibnamefont {{Lugaro}}},
  \bibinfo {author} {\bibfnamefont {J.}~\bibnamefont {{Simmerer}}}, \bibinfo
  {author} {\bibfnamefont {K.}~\bibnamefont {{Farouqi}}}, and\ \bibinfo
  {author} {\bibfnamefont {C.}~\bibnamefont {{Sneden}}}} (\bibinfo {year}
  {2010}{\natexlab{a}}),\ \bibfield  {title} {\enquote {\bibinfo {title} {{The
  Ubiquity of the Rapid Neutron-capture Process}},}\ }\href
  {https://doi.org/10.1088/0004-637X/724/2/975} {\bibfield  {journal} {\bibinfo
   {journal} {Astrophys. J.}\ }\textbf {\bibinfo {volume} {724}},\ \bibinfo
  {pages} {975--993}}\BibitemShut {NoStop}%
\bibitem [{\citenamefont {{Roederer}}\ \emph {et~al.}(2016)\citenamefont
  {{Roederer}}, \citenamefont {{Karakas}}, \citenamefont {{Pignatari}},\ and\
  \citenamefont {{Herwig}}}]{roederer16}%
  \BibitemOpen
  \bibfield  {author} {\bibinfo {author} {\bibnamefont {{Roederer}},
  \bibfnamefont {I~U}}, \bibinfo {author} {\bibfnamefont {A.~I.}\ \bibnamefont
  {{Karakas}}}, \bibinfo {author} {\bibfnamefont {M.}~\bibnamefont
  {{Pignatari}}}, and\ \bibinfo {author} {\bibfnamefont {F.}~\bibnamefont
  {{Herwig}}}} (\bibinfo {year} {2016}),\ \bibfield  {title} {\enquote
  {\bibinfo {title} {{The Diverse Origins of Neutron-capture Elements in the
  Metal-poor Star HD 94028: Possible Detection of Products of i-Process
  Nucleosynthesis}},}\ }\href {https://doi.org/10.3847/0004-637X/821/1/37}
  {\bibfield  {journal} {\bibinfo  {journal} {Astrophys. J.}\ }\textbf
  {\bibinfo {volume} {821}},\ \bibinfo {eid} {37}}\BibitemShut {NoStop}%
\bibitem [{\citenamefont {{Roederer}}\ \emph {et~al.}(2009)\citenamefont
  {{Roederer}}, \citenamefont {{Kratz}}, \citenamefont {{Frebel}},
  \citenamefont {{Christlieb}}, \citenamefont {{Pfeiffer}}, \citenamefont
  {{Cowan}},\ and\ \citenamefont {{Sneden}}}]{Roederer.Kratz.ea:2009}%
  \BibitemOpen
  \bibfield  {author} {\bibinfo {author} {\bibnamefont {{Roederer}},
  \bibfnamefont {I~U}}, \bibinfo {author} {\bibfnamefont {K.-L.}\ \bibnamefont
  {{Kratz}}}, \bibinfo {author} {\bibfnamefont {A.}~\bibnamefont {{Frebel}}},
  \bibinfo {author} {\bibfnamefont {N.}~\bibnamefont {{Christlieb}}}, \bibinfo
  {author} {\bibfnamefont {B.}~\bibnamefont {{Pfeiffer}}}, \bibinfo {author}
  {\bibfnamefont {J.~J.}\ \bibnamefont {{Cowan}}}, and\ \bibinfo {author}
  {\bibfnamefont {C.}~\bibnamefont {{Sneden}}}} (\bibinfo {year} {2009}),\
  \bibfield  {title} {\enquote {\bibinfo {title} {{The End of Nucleosynthesis:
  Production of Lead and Thorium in the Early Galaxy}},}\ }\href
  {https://doi.org/10.1088/0004-637X/698/2/1963} {\bibfield  {journal}
  {\bibinfo  {journal} {Astrophys. J.}\ }\textbf {\bibinfo {volume} {698}},\
  \bibinfo {pages} {1963--1980}}\BibitemShut {NoStop}%
\bibitem [{\citenamefont {{Roederer}}\ and\ \citenamefont
  {{Lawler}}(2012)}]{roederer12a}%
  \BibitemOpen
  \bibfield  {author} {\bibinfo {author} {\bibnamefont {{Roederer}},
  \bibfnamefont {I~U}}, and\ \bibinfo {author} {\bibfnamefont {J.~E.}\
  \bibnamefont {{Lawler}}}} (\bibinfo {year} {2012}),\ \bibfield  {title}
  {\enquote {\bibinfo {title} {{Detection of Elements at All Three r-process
  Peaks in the Metal-poor Star HD 160617}},}\ }\href
  {https://doi.org/10.1088/0004-637X/750/1/76} {\bibfield  {journal} {\bibinfo
  {journal} {Astrophys. J.}\ }\textbf {\bibinfo {volume} {750}},\ \bibinfo
  {eid} {76}}\BibitemShut {NoStop}%
\bibitem [{\citenamefont {{Roederer}}\ \emph {et~al.}(2012)\citenamefont
  {{Roederer}}, \citenamefont {{Lawler}}, \citenamefont {{Cowan}},
  \citenamefont {{Beers}}, \citenamefont {{Frebel}}, \citenamefont {{Ivans}},
  \citenamefont {{Schatz}}, \citenamefont {{Sobeck}},\ and\ \citenamefont
  {{Sneden}}}]{roederer12b}%
  \BibitemOpen
  \bibfield  {author} {\bibinfo {author} {\bibnamefont {{Roederer}},
  \bibfnamefont {I~U}}, \bibinfo {author} {\bibfnamefont {J.~E.}\ \bibnamefont
  {{Lawler}}}, \bibinfo {author} {\bibfnamefont {J.~J.}\ \bibnamefont
  {{Cowan}}}, \bibinfo {author} {\bibfnamefont {T.~C.}\ \bibnamefont
  {{Beers}}}, \bibinfo {author} {\bibfnamefont {A.}~\bibnamefont {{Frebel}}},
  \bibinfo {author} {\bibfnamefont {I.~I.}\ \bibnamefont {{Ivans}}}, \bibinfo
  {author} {\bibfnamefont {H.}~\bibnamefont {{Schatz}}}, \bibinfo {author}
  {\bibfnamefont {J.~S.}\ \bibnamefont {{Sobeck}}}, and\ \bibinfo {author}
  {\bibfnamefont {C.}~\bibnamefont {{Sneden}}}} (\bibinfo {year} {2012}),\
  \bibfield  {title} {\enquote {\bibinfo {title} {{Detection of the Second
  r-process Peak Element Tellurium in Metal-poor Stars}},}\ }\href
  {https://doi.org/10.1088/2041-8205/747/1/L8} {\bibfield  {journal} {\bibinfo
  {journal} {Astrophys. J. Suppl.}\ }\textbf {\bibinfo {volume} {747}},\
  \bibinfo {eid} {L8}}\BibitemShut {NoStop}%
\bibitem [{\citenamefont {{Roederer}}\ \emph
  {et~al.}(2014{\natexlab{a}})\citenamefont {{Roederer}}, \citenamefont
  {{Schatz}}, \citenamefont {{Lawler}}, \citenamefont {{Beers}}, \citenamefont
  {{Cowan}}, \citenamefont {{Frebel}}, \citenamefont {{Ivans}}, \citenamefont
  {{Sneden}},\ and\ \citenamefont {{Sobeck}}}]{roederer14}%
  \BibitemOpen
  \bibfield  {author} {\bibinfo {author} {\bibnamefont {{Roederer}},
  \bibfnamefont {I~U}}, \bibinfo {author} {\bibfnamefont {H.}~\bibnamefont
  {{Schatz}}}, \bibinfo {author} {\bibfnamefont {J.~E.}\ \bibnamefont
  {{Lawler}}}, \bibinfo {author} {\bibfnamefont {T.~C.}\ \bibnamefont
  {{Beers}}}, \bibinfo {author} {\bibfnamefont {J.~J.}\ \bibnamefont
  {{Cowan}}}, \bibinfo {author} {\bibfnamefont {A.}~\bibnamefont {{Frebel}}},
  \bibinfo {author} {\bibfnamefont {I.~I.}\ \bibnamefont {{Ivans}}}, \bibinfo
  {author} {\bibfnamefont {C.}~\bibnamefont {{Sneden}}}, and\ \bibinfo {author}
  {\bibfnamefont {J.~S.}\ \bibnamefont {{Sobeck}}}} (\bibinfo {year}
  {2014}{\natexlab{a}}),\ \bibfield  {title} {\enquote {\bibinfo {title} {{New
  Detections of Arsenic, Selenium, and Other Heavy Elements in Two Metal-poor
  Stars}},}\ }\href {https://doi.org/10.1088/0004-637X/791/1/32} {\bibfield
  {journal} {\bibinfo  {journal} {Astrophys. J.}\ }\textbf {\bibinfo {volume}
  {791}},\ \bibinfo {eid} {32}}\BibitemShut {NoStop}%
\bibitem [{\citenamefont {{Roederer}}\ \emph
  {et~al.}(2010{\natexlab{b}})\citenamefont {{Roederer}}, \citenamefont
  {{Sneden}}, \citenamefont {{Lawler}},\ and\ \citenamefont
  {{Cowan}}}]{Roederer.Sneden.ea:2010}%
  \BibitemOpen
  \bibfield  {author} {\bibinfo {author} {\bibnamefont {{Roederer}},
  \bibfnamefont {I~U}}, \bibinfo {author} {\bibfnamefont {C.}~\bibnamefont
  {{Sneden}}}, \bibinfo {author} {\bibfnamefont {J.~E.}\ \bibnamefont
  {{Lawler}}}, and\ \bibinfo {author} {\bibfnamefont {J.~J.}\ \bibnamefont
  {{Cowan}}}} (\bibinfo {year} {2010}{\natexlab{b}}),\ \bibfield  {title}
  {\enquote {\bibinfo {title} {{New Abundance Determinations of Cadmium,
  Lutetium, and Osmium in the r-process Enriched Star BD +17 3248}},}\ }\href
  {https://doi.org/10.1088/2041-8205/714/1/L123} {\bibfield  {journal}
  {\bibinfo  {journal} {Astrophys. J.}\ }\textbf {\bibinfo {volume} {714}},\
  \bibinfo {pages} {L123--L127}}\BibitemShut {NoStop}%
\bibitem [{\citenamefont {{Roederer}}\ \emph {et~al.}(2018)\citenamefont
  {{Roederer}}, \citenamefont {{Hattori}},\ and\ \citenamefont
  {{Valluri}}}]{Roederer.Hattori.Valluri:2018}%
  \BibitemOpen
  \bibfield  {author} {\bibinfo {author} {\bibnamefont {{Roederer}},
  \bibfnamefont {Ian~U}}, \bibinfo {author} {\bibfnamefont {Kohei}\
  \bibnamefont {{Hattori}}}, and\ \bibinfo {author} {\bibfnamefont {Monica}\
  \bibnamefont {{Valluri}}}} (\bibinfo {year} {2018}),\ \bibfield  {title}
  {\enquote {\bibinfo {title} {{Kinematics of Highly r-process-enhanced Field
  Stars: Evidence for an Accretion Origin and Detection of Several Groups from
  Disrupted Satellites}},}\ }\href {https://doi.org/10.3847/1538-3881/aadd9c}
  {\bibfield  {journal} {\bibinfo  {journal} {Astron. J.}\ }\textbf {\bibinfo
  {volume} {156}},\ \bibinfo {eid} {179}}\BibitemShut {NoStop}%
\bibitem [{\citenamefont {{Roederer}}\ \emph
  {et~al.}(2014{\natexlab{b}})\citenamefont {{Roederer}}, \citenamefont
  {{Preston}}, \citenamefont {{Thompson}}, \citenamefont {{Shectman}},
  \citenamefont {{Sneden}}, \citenamefont {{Burley}},\ and\ \citenamefont
  {{Kelson}}}]{Roederer.Preston.ea:2014}%
  \BibitemOpen
  \bibfield  {author} {\bibinfo {author} {\bibnamefont {{Roederer}},
  \bibfnamefont {Ian~U}}, \bibinfo {author} {\bibfnamefont {George~W.}\
  \bibnamefont {{Preston}}}, \bibinfo {author} {\bibfnamefont {Ian~B.}\
  \bibnamefont {{Thompson}}}, \bibinfo {author} {\bibfnamefont {Stephen~A.}\
  \bibnamefont {{Shectman}}}, \bibinfo {author} {\bibfnamefont {Christopher}\
  \bibnamefont {{Sneden}}}, \bibinfo {author} {\bibfnamefont {Gregory~S.}\
  \bibnamefont {{Burley}}}, and\ \bibinfo {author} {\bibfnamefont {Daniel~D.}\
  \bibnamefont {{Kelson}}}} (\bibinfo {year} {2014}{\natexlab{b}}),\ \bibfield
  {title} {\enquote {\bibinfo {title} {{A Search for Stars of Very Low Metal
  Abundance. VI. Detailed Abundances of 313 Metal-poor Stars}},}\ }\href
  {https://doi.org/10.1088/0004-6256/147/6/136} {\bibfield  {journal} {\bibinfo
   {journal} {Astron. J.}\ }\textbf {\bibinfo {volume} {147}},\ \bibinfo {eid}
  {136}}\BibitemShut {NoStop}%
\bibitem [{\citenamefont {Rolfs}\ and\ \citenamefont
  {Rodney}(1988)}]{rolfs.rodney:1988}%
  \BibitemOpen
  \bibfield  {author} {\bibinfo {author} {\bibnamefont {Rolfs}, \bibfnamefont
  {CE}}, and\ \bibinfo {author} {\bibfnamefont {W.S.}\ \bibnamefont {Rodney}}}
  (\bibinfo {year} {1988}),\ \href@noop {} {\emph {\bibinfo {title} {{Cauldrons
  in the Cosmos: Nuclear Astrophysics}}}}\ (\bibinfo  {publisher} {University
  of Chicago Press},\ \bibinfo {address} {Chicago})\BibitemShut {NoStop}%
\bibitem [{\citenamefont {{Rosswog}}(2005)}]{rosswog05}%
  \BibitemOpen
  \bibfield  {author} {\bibinfo {author} {\bibnamefont {{Rosswog}},
  \bibfnamefont {S}}} (\bibinfo {year} {2005}),\ \bibfield  {title} {\enquote
  {\bibinfo {title} {{Mergers of Neutron Star-Black Hole Binaries with Small
  Mass Ratios: Nucleosynthesis, Gamma-Ray Bursts, and Electromagnetic
  Transients}},}\ }\href {https://doi.org/10.1086/497062} {\bibfield  {journal}
  {\bibinfo  {journal} {Astrophys. J.}\ }\textbf {\bibinfo {volume} {634}},\
  \bibinfo {pages} {1202--1213}}\BibitemShut {NoStop}%
\bibitem [{\citenamefont {{Rosswog}}(2007)}]{Rosswog:2007}%
  \BibitemOpen
  \bibfield  {author} {\bibinfo {author} {\bibnamefont {{Rosswog}},
  \bibfnamefont {S}}} (\bibinfo {year} {2007}),\ \bibfield  {title} {\enquote
  {\bibinfo {title} {{Fallback accretion in the aftermath of a compact binary
  merger}},}\ }\href {https://doi.org/10.1111/j.1745-3933.2007.00284.x}
  {\bibfield  {journal} {\bibinfo  {journal} {Mon. Not. Roy. Astron. Soc.}\
  }\textbf {\bibinfo {volume} {376}},\ \bibinfo {pages} {L48--L51}}\BibitemShut
  {NoStop}%
\bibitem [{\citenamefont {{Rosswog}}\ \emph {et~al.}(2000)\citenamefont
  {{Rosswog}}, \citenamefont {{Davies}}, \citenamefont {{Thielemann}},\ and\
  \citenamefont {{Piran}}}]{Rosswog.Davies.ea:2000}%
  \BibitemOpen
  \bibfield  {author} {\bibinfo {author} {\bibnamefont {{Rosswog}},
  \bibfnamefont {S}}, \bibinfo {author} {\bibfnamefont {M.~B.}\ \bibnamefont
  {{Davies}}}, \bibinfo {author} {\bibfnamefont {F.-K.}\ \bibnamefont
  {{Thielemann}}}, and\ \bibinfo {author} {\bibfnamefont {T.}~\bibnamefont
  {{Piran}}}} (\bibinfo {year} {2000}),\ \bibfield  {title} {\enquote {\bibinfo
  {title} {{Merging neutron stars: asymmetric systems}},}\ }\href@noop {}
  {\bibfield  {journal} {\bibinfo  {journal} {Astron. \& Astrophys.}\ }\textbf
  {\bibinfo {volume} {360}},\ \bibinfo {pages} {171--184}}\BibitemShut
  {NoStop}%
\bibitem [{\citenamefont {{Rosswog}}\ \emph {et~al.}(2017)\citenamefont
  {{Rosswog}}, \citenamefont {{Feindt}}, \citenamefont {{Korobkin}},
  \citenamefont {{Wu}}, \citenamefont {{Sollerman}}, \citenamefont {{Goobar}},\
  and\ \citenamefont {{Martinez-Pinedo}}}]{Rosswog.Feindt.ea:2017}%
  \BibitemOpen
  \bibfield  {author} {\bibinfo {author} {\bibnamefont {{Rosswog}},
  \bibfnamefont {S}}, \bibinfo {author} {\bibfnamefont {U.}~\bibnamefont
  {{Feindt}}}, \bibinfo {author} {\bibfnamefont {O.}~\bibnamefont
  {{Korobkin}}}, \bibinfo {author} {\bibfnamefont {M.-R.}\ \bibnamefont
  {{Wu}}}, \bibinfo {author} {\bibfnamefont {J.}~\bibnamefont {{Sollerman}}},
  \bibinfo {author} {\bibfnamefont {A.}~\bibnamefont {{Goobar}}}, and\ \bibinfo
  {author} {\bibfnamefont {G.}~\bibnamefont {{Martinez-Pinedo}}}} (\bibinfo
  {year} {2017}),\ \bibfield  {title} {\enquote {\bibinfo {title}
  {{Detectability of compact binary merger macronovae}},}\ }\href
  {https://doi.org/10.1088/1361-6382/aa68a9} {\bibfield  {journal} {\bibinfo
  {journal} {Class. Quantum Gravity}\ }\textbf {\bibinfo {volume} {34}},\
  \bibinfo {eid} {104001}}\BibitemShut {NoStop}%
\bibitem [{\citenamefont {{Rosswog}}\ \emph {et~al.}(2014)\citenamefont
  {{Rosswog}}, \citenamefont {{Korobkin}}, \citenamefont {{Arcones}},
  \citenamefont {{Thielemann}},\ and\ \citenamefont
  {{Piran}}}]{Rosswog.Korobkin.ea:2014}%
  \BibitemOpen
  \bibfield  {author} {\bibinfo {author} {\bibnamefont {{Rosswog}},
  \bibfnamefont {S}}, \bibinfo {author} {\bibfnamefont {O.}~\bibnamefont
  {{Korobkin}}}, \bibinfo {author} {\bibfnamefont {A.}~\bibnamefont
  {{Arcones}}}, \bibinfo {author} {\bibfnamefont {F.-K.}\ \bibnamefont
  {{Thielemann}}}, and\ \bibinfo {author} {\bibfnamefont {T.}~\bibnamefont
  {{Piran}}}} (\bibinfo {year} {2014}),\ \bibfield  {title} {\enquote {\bibinfo
  {title} {{The long-term evolution of neutron star merger remnants - I. The
  impact of r-process nucleosynthesis}},}\ }\href
  {https://doi.org/10.1093/mnras/stt2502} {\bibfield  {journal} {\bibinfo
  {journal} {Mon. Not. Roy. Astron. Soc.}\ }\textbf {\bibinfo {volume} {439}},\
  \bibinfo {pages} {744--756}}\BibitemShut {NoStop}%
\bibitem [{\citenamefont {{Rosswog}}\ \emph {et~al.}(1999)\citenamefont
  {{Rosswog}}, \citenamefont {{Liebend{\"o}rfer}}, \citenamefont
  {{Thielemann}}, \citenamefont {{Davies}}, \citenamefont {{Benz}},\ and\
  \citenamefont {{Piran}}}]{rosswog99}%
  \BibitemOpen
  \bibfield  {author} {\bibinfo {author} {\bibnamefont {{Rosswog}},
  \bibfnamefont {S}}, \bibinfo {author} {\bibfnamefont {M.}~\bibnamefont
  {{Liebend{\"o}rfer}}}, \bibinfo {author} {\bibfnamefont {F.-K.}\ \bibnamefont
  {{Thielemann}}}, \bibinfo {author} {\bibfnamefont {M.~B.}\ \bibnamefont
  {{Davies}}}, \bibinfo {author} {\bibfnamefont {W.}~\bibnamefont {{Benz}}},
  and\ \bibinfo {author} {\bibfnamefont {T.}~\bibnamefont {{Piran}}}} (\bibinfo
  {year} {1999}),\ \bibfield  {title} {\enquote {\bibinfo {title} {{Mass
  ejection in neutron star mergers}},}\ }\href@noop {} {\bibfield  {journal}
  {\bibinfo  {journal} {Astron. \& Astrophys.}\ }\textbf {\bibinfo {volume}
  {341}},\ \bibinfo {pages} {499--526}}\BibitemShut {NoStop}%
\bibitem [{\citenamefont {{Rosswog}}\ \emph {et~al.}(2018)\citenamefont
  {{Rosswog}}, \citenamefont {{Sollerman}}, \citenamefont {{Feindt}},
  \citenamefont {{Goobar}}, \citenamefont {{Korobkin}}, \citenamefont
  {{Wollaeger}}, \citenamefont {{Fremling}},\ and\ \citenamefont
  {{Kasliwal}}}]{rosswog18}%
  \BibitemOpen
  \bibfield  {author} {\bibinfo {author} {\bibnamefont {{Rosswog}},
  \bibfnamefont {S}}, \bibinfo {author} {\bibfnamefont {J.}~\bibnamefont
  {{Sollerman}}}, \bibinfo {author} {\bibfnamefont {U.}~\bibnamefont
  {{Feindt}}}, \bibinfo {author} {\bibfnamefont {A.}~\bibnamefont {{Goobar}}},
  \bibinfo {author} {\bibfnamefont {O.}~\bibnamefont {{Korobkin}}}, \bibinfo
  {author} {\bibfnamefont {R.}~\bibnamefont {{Wollaeger}}}, \bibinfo {author}
  {\bibfnamefont {C.}~\bibnamefont {{Fremling}}}, and\ \bibinfo {author}
  {\bibfnamefont {M.~M.}\ \bibnamefont {{Kasliwal}}}} (\bibinfo {year}
  {2018}),\ \bibfield  {title} {\enquote {\bibinfo {title} {{The first direct
  double neutron star merger detection: Implications for cosmic
  nucleosynthesis}},}\ }\href {https://doi.org/10.1051/0004-6361/201732117}
  {\bibfield  {journal} {\bibinfo  {journal} {Astron. \& Astrophys.}\ }\textbf
  {\bibinfo {volume} {615}},\ \bibinfo {eid} {A132}}\BibitemShut {NoStop}%
\bibitem [{\citenamefont {Rosswog}(2015)}]{Rosswog:2015}%
  \BibitemOpen
  \bibfield  {author} {\bibinfo {author} {\bibnamefont {Rosswog}, \bibfnamefont
  {Stephan}}} (\bibinfo {year} {2015}),\ \bibfield  {title} {\enquote {\bibinfo
  {title} {The multi-messenger picture of compact binary mergers},}\ }\href
  {https://doi.org/10.1142/S0218271815300128} {\bibfield  {journal} {\bibinfo
  {journal} {Int. J. Mod. Phys. D}\ }\textbf {\bibinfo {volume} {24}},\
  \bibinfo {pages} {1530012}}\BibitemShut {NoStop}%
\bibitem [{\citenamefont {Rrapaj}\ \emph {et~al.}(2015)\citenamefont {Rrapaj},
  \citenamefont {Holt}, \citenamefont {Bartl}, \citenamefont {Reddy},\ and\
  \citenamefont {Schwenk}}]{Rrapaj.Holt.ea:2015}%
  \BibitemOpen
  \bibfield  {author} {\bibinfo {author} {\bibnamefont {Rrapaj}, \bibfnamefont
  {Ermal}}, \bibinfo {author} {\bibfnamefont {J.~W.}\ \bibnamefont {Holt}},
  \bibinfo {author} {\bibfnamefont {Alexander}\ \bibnamefont {Bartl}}, \bibinfo
  {author} {\bibfnamefont {Sanjay}\ \bibnamefont {Reddy}}, and\ \bibinfo
  {author} {\bibfnamefont {A.}~\bibnamefont {Schwenk}}} (\bibinfo {year}
  {2015}),\ \bibfield  {title} {\enquote {\bibinfo {title} {Charged-current
  reactions in the supernova neutrino-sphere},}\ }\href
  {https://doi.org/10.1103/PhysRevC.91.035806} {\bibfield  {journal} {\bibinfo
  {journal} {Phys. Rev. C}\ }\textbf {\bibinfo {volume} {91}},\ \bibinfo
  {pages} {035806}}\BibitemShut {NoStop}%
\bibitem [{\citenamefont {{Ruffert}}\ and\ \citenamefont
  {{Janka}}(2001)}]{Ruffert.Janka:2001}%
  \BibitemOpen
  \bibfield  {author} {\bibinfo {author} {\bibnamefont {{Ruffert}},
  \bibfnamefont {M}}, and\ \bibinfo {author} {\bibfnamefont {H.-T.}\
  \bibnamefont {{Janka}}}} (\bibinfo {year} {2001}),\ \bibfield  {title}
  {\enquote {\bibinfo {title} {{Coalescing neutron stars - A step towards
  physical models. III. Improved numerics and different neutron star masses and
  spins}},}\ }\href {https://doi.org/10.1051/0004-6361:20011453} {\bibfield
  {journal} {\bibinfo  {journal} {Astron. \& Astrophys.}\ }\textbf {\bibinfo
  {volume} {380}},\ \bibinfo {pages} {544--577}}\BibitemShut {NoStop}%
\bibitem [{\citenamefont {{Ruffert}}\ \emph {et~al.}(1996)\citenamefont
  {{Ruffert}}, \citenamefont {{Janka}},\ and\ \citenamefont
  {{Schaefer}}}]{ruffert96}%
  \BibitemOpen
  \bibfield  {author} {\bibinfo {author} {\bibnamefont {{Ruffert}},
  \bibfnamefont {M}}, \bibinfo {author} {\bibfnamefont {H.-T.}\ \bibnamefont
  {{Janka}}}, and\ \bibinfo {author} {\bibfnamefont {G.}~\bibnamefont
  {{Schaefer}}}} (\bibinfo {year} {1996}),\ \bibfield  {title} {\enquote
  {\bibinfo {title} {{Coalescing neutron stars - a step towards physical
  models. I. Hydrodynamic evolution and gravitational-wave emission.}}}\
  }\href@noop {} {\bibfield  {journal} {\bibinfo  {journal} {Astron. \&
  Astrophys.}\ }\textbf {\bibinfo {volume} {311}},\ \bibinfo {pages}
  {532--566}}\BibitemShut {NoStop}%
\bibitem [{\citenamefont {{Ruffert}}\ \emph {et~al.}(1997)\citenamefont
  {{Ruffert}}, \citenamefont {{Janka}}, \citenamefont {{Takahashi}},\ and\
  \citenamefont {{Schaefer}}}]{Ruffert.Janka.ea:1997}%
  \BibitemOpen
  \bibfield  {author} {\bibinfo {author} {\bibnamefont {{Ruffert}},
  \bibfnamefont {M}}, \bibinfo {author} {\bibfnamefont {H.-T.}\ \bibnamefont
  {{Janka}}}, \bibinfo {author} {\bibfnamefont {K.}~\bibnamefont
  {{Takahashi}}}, and\ \bibinfo {author} {\bibfnamefont {G.}~\bibnamefont
  {{Schaefer}}}} (\bibinfo {year} {1997}),\ \bibfield  {title} {\enquote
  {\bibinfo {title} {{Coalescing neutron stars - a step towards physical
  models. II. Neutrino emission, neutron tori, and gamma-ray bursts}},}\
  }\href@noop {} {\bibfield  {journal} {\bibinfo  {journal} {Astron. \&
  Astrophys.}\ }\textbf {\bibinfo {volume} {319}},\ \bibinfo {pages}
  {122--153}}\BibitemShut {NoStop}%
\bibitem [{\citenamefont {Ruiz}\ \emph {et~al.}(2018)\citenamefont {Ruiz},
  \citenamefont {Shapiro},\ and\ \citenamefont
  {Tsokaros}}]{Ruiz.Shapiro.Tsokaros:2018}%
  \BibitemOpen
  \bibfield  {author} {\bibinfo {author} {\bibnamefont {Ruiz}, \bibfnamefont
  {Milton}}, \bibinfo {author} {\bibfnamefont {Stuart~L.}\ \bibnamefont
  {Shapiro}}, and\ \bibinfo {author} {\bibfnamefont {Antonios}\ \bibnamefont
  {Tsokaros}}} (\bibinfo {year} {2018}),\ \bibfield  {title} {\enquote
  {\bibinfo {title} {Gw170817, general relativistic magnetohydrodynamic
  simulations, and the neutron star maximum mass},}\ }\href
  {https://doi.org/10.1103/PhysRevD.97.021501} {\bibfield  {journal} {\bibinfo
  {journal} {Phys. Rev. D}\ }\textbf {\bibinfo {volume} {97}},\ \bibinfo
  {pages} {021501}}\BibitemShut {NoStop}%
\bibitem [{\citenamefont {{Sadhukhan}}\ \emph {et~al.}(2016)\citenamefont
  {{Sadhukhan}}, \citenamefont {{Nazarewicz}},\ and\ \citenamefont
  {{Schunck}}}]{Sadhukhan.Nazarewicz.Schunck:2016}%
  \BibitemOpen
  \bibfield  {author} {\bibinfo {author} {\bibnamefont {{Sadhukhan}},
  \bibfnamefont {J}}, \bibinfo {author} {\bibfnamefont {W.}~\bibnamefont
  {{Nazarewicz}}}, and\ \bibinfo {author} {\bibfnamefont {N.}~\bibnamefont
  {{Schunck}}}} (\bibinfo {year} {2016}),\ \bibfield  {title} {\enquote
  {\bibinfo {title} {{Microscopic modeling of mass and charge distributions in
  the spontaneous fission of $^{240}$Pu}},}\ }\href
  {https://doi.org/10.1103/PhysRevC.93.011304} {\bibfield  {journal} {\bibinfo
  {journal} {Phys. Rev. C}\ }\textbf {\bibinfo {volume} {93}},\ \bibinfo {eid}
  {011304}}\BibitemShut {NoStop}%
\bibitem [{\citenamefont {{Sadhukhan}}\ \emph {et~al.}(2020)\citenamefont
  {{Sadhukhan}}, \citenamefont {{Giuliani}}, \citenamefont {{Matheson}},\ and\
  \citenamefont {{Nazarewicz}}}]{Sadhukhan.Giuliani.ea:2020}%
  \BibitemOpen
  \bibfield  {author} {\bibinfo {author} {\bibnamefont {{Sadhukhan}},
  \bibfnamefont {Jhilam}}, \bibinfo {author} {\bibfnamefont {Samuel~A.}\
  \bibnamefont {{Giuliani}}}, \bibinfo {author} {\bibfnamefont {Zachary}\
  \bibnamefont {{Matheson}}}, and\ \bibinfo {author} {\bibfnamefont {Witold}\
  \bibnamefont {{Nazarewicz}}}} (\bibinfo {year} {2020}),\ \bibfield  {title}
  {\enquote {\bibinfo {title} {{Efficient method for estimation of fission
  fragment yields of r-process nuclei}},}\ }\href
  {https://doi.org/10.1103/PhysRevC.101.065803} {\bibfield  {journal} {\bibinfo
   {journal} {Phys. Rev. C}\ }\textbf {\bibinfo {volume} {101}},\ \bibinfo
  {eid} {065803}}\BibitemShut {NoStop}%
\bibitem [{\citenamefont {Sadhukhan}\ \emph {et~al.}(2013)\citenamefont
  {Sadhukhan}, \citenamefont {Mazurek}, \citenamefont {Baran}, \citenamefont
  {Dobaczewski}, \citenamefont {Nazarewicz},\ and\ \citenamefont
  {Sheikh}}]{Sadhukhan.Mazurek.ea:2013}%
  \BibitemOpen
  \bibfield  {author} {\bibinfo {author} {\bibnamefont {Sadhukhan},
  \bibfnamefont {Jhilam}}, \bibinfo {author} {\bibfnamefont {K.}~\bibnamefont
  {Mazurek}}, \bibinfo {author} {\bibfnamefont {A.}~\bibnamefont {Baran}},
  \bibinfo {author} {\bibfnamefont {J.}~\bibnamefont {Dobaczewski}}, \bibinfo
  {author} {\bibfnamefont {W.}~\bibnamefont {Nazarewicz}}, and\ \bibinfo
  {author} {\bibfnamefont {J.~A.}\ \bibnamefont {Sheikh}}} (\bibinfo {year}
  {2013}),\ \bibfield  {title} {\enquote {\bibinfo {title} {Spontaneous fission
  lifetimes from the minimization of self-consistent collective action},}\
  }\href {https://doi.org/10.1103/PhysRevC.88.064314} {\bibfield  {journal}
  {\bibinfo  {journal} {Phys. Rev. C}\ }\textbf {\bibinfo {volume} {88}},\
  \bibinfo {pages} {064314}}\BibitemShut {NoStop}%
\bibitem [{\citenamefont {{Safarzadeh}}\ \emph {et~al.}(2019)\citenamefont
  {{Safarzadeh}}, \citenamefont {{Ramirez-Ruiz}}, \citenamefont {{Andrews}},
  \citenamefont {{Macias}}, \citenamefont {{Fragos}},\ and\ \citenamefont
  {{Scannapieco}}}]{Safarzadeh.Ramirez-Ruiz.ea:2019}%
  \BibitemOpen
  \bibfield  {author} {\bibinfo {author} {\bibnamefont {{Safarzadeh}},
  \bibfnamefont {Mohammadtaher}}, \bibinfo {author} {\bibfnamefont {Enrico}\
  \bibnamefont {{Ramirez-Ruiz}}}, \bibinfo {author} {\bibfnamefont {Jeff.~J.}\
  \bibnamefont {{Andrews}}}, \bibinfo {author} {\bibfnamefont {Phillip}\
  \bibnamefont {{Macias}}}, \bibinfo {author} {\bibfnamefont {Tassos}\
  \bibnamefont {{Fragos}}}, and\ \bibinfo {author} {\bibfnamefont {Evan}\
  \bibnamefont {{Scannapieco}}}} (\bibinfo {year} {2019}),\ \bibfield  {title}
  {\enquote {\bibinfo {title} {{r-process Enrichment of the Ultra-faint Dwarf
  Galaxies by Fast-merging Double-neutron Stars}},}\ }\href
  {https://doi.org/10.3847/1538-4357/aafe0e} {\bibfield  {journal} {\bibinfo
  {journal} {Astrophys. J.}\ }\textbf {\bibinfo {volume} {872}},\ \bibinfo
  {eid} {105}}\BibitemShut {NoStop}%
\bibitem [{\citenamefont {{Sagert}}\ \emph {et~al.}(2009)\citenamefont
  {{Sagert}}, \citenamefont {{Fischer}}, \citenamefont {{Hempel}},
  \citenamefont {{Pagliara}}, \citenamefont {{Schaffner-Bielich}},
  \citenamefont {{Mezzacappa}}, \citenamefont {{Thielemann}},\ and\
  \citenamefont {{Liebend{\"o}rfer}}}]{sagert09}%
  \BibitemOpen
  \bibfield  {author} {\bibinfo {author} {\bibnamefont {{Sagert}},
  \bibfnamefont {I}}, \bibinfo {author} {\bibfnamefont {T.}~\bibnamefont
  {{Fischer}}}, \bibinfo {author} {\bibfnamefont {M.}~\bibnamefont {{Hempel}}},
  \bibinfo {author} {\bibfnamefont {G.}~\bibnamefont {{Pagliara}}}, \bibinfo
  {author} {\bibfnamefont {J.}~\bibnamefont {{Schaffner-Bielich}}}, \bibinfo
  {author} {\bibfnamefont {A.}~\bibnamefont {{Mezzacappa}}}, \bibinfo {author}
  {\bibfnamefont {F.-K.}\ \bibnamefont {{Thielemann}}}, and\ \bibinfo {author}
  {\bibfnamefont {M.}~\bibnamefont {{Liebend{\"o}rfer}}}} (\bibinfo {year}
  {2009}),\ \bibfield  {title} {\enquote {\bibinfo {title} {{Signals of the QCD
  Phase Transition in Core-Collapse Supernovae}},}\ }\href
  {https://doi.org/10.1103/PhysRevLett.102.081101} {\bibfield  {journal}
  {\bibinfo  {journal} {Phys. Rev. Lett.}\ }\textbf {\bibinfo {volume} {102}},\
  \bibinfo {pages} {081101}}\BibitemShut {NoStop}%
\bibitem [{\citenamefont {{Sakari}}\ \emph {et~al.}(2018)\citenamefont
  {{Sakari}} \emph {et~al.}}]{Sakari.Placco.ea:2018}%
  \BibitemOpen
  \bibfield  {author} {\bibinfo {author} {\bibnamefont {{Sakari}},
  \bibfnamefont {Charli~M}},  \emph {et~al.}} (\bibinfo {year} {2018}),\
  \bibfield  {title} {\enquote {\bibinfo {title} {{The R-Process Alliance:
  First Release from the Northern Search for r-process-enhanced Metal-poor
  Stars in the Galactic Halo}},}\ }\href
  {https://doi.org/10.3847/1538-4357/aae9df} {\bibfield  {journal} {\bibinfo
  {journal} {Astrophys. J.}\ }\textbf {\bibinfo {volume} {868}},\ \bibinfo
  {eid} {110}}\BibitemShut {NoStop}%
\bibitem [{\citenamefont {{Salih}}\ and\ \citenamefont
  {{Lawler}}(1983)}]{salih83}%
  \BibitemOpen
  \bibfield  {author} {\bibinfo {author} {\bibnamefont {{Salih}}, \bibfnamefont
  {S}}, and\ \bibinfo {author} {\bibfnamefont {J.~E.}\ \bibnamefont
  {{Lawler}}}} (\bibinfo {year} {1983}),\ \bibfield  {title} {\enquote
  {\bibinfo {title} {{Pulsed ion source for laser spectroscopy: Application to
  Nb II}},}\ }\href {https://doi.org/10.1103/PhysRevA.28.3653} {\bibfield
  {journal} {\bibinfo  {journal} {Phys. Rev. A}\ }\textbf {\bibinfo {volume}
  {28}},\ \bibinfo {pages} {3653--3655}}\BibitemShut {NoStop}%
\bibitem [{\citenamefont {{Sato}}(1974)}]{Sato:1974}%
  \BibitemOpen
  \bibfield  {author} {\bibinfo {author} {\bibnamefont {{Sato}}, \bibfnamefont
  {K}}} (\bibinfo {year} {1974}),\ \bibfield  {title} {\enquote {\bibinfo
  {title} {{Formation of Elements in Neutron Rich Ejected Matter of Supernovae.
  II ---Dynamical r-Process Stage---}},}\ }\href
  {https://doi.org/10.1143/PTP.51.726} {\bibfield  {journal} {\bibinfo
  {journal} {Progress of Theoretical Physics}\ }\textbf {\bibinfo {volume}
  {51}},\ \bibinfo {pages} {726--744}}\BibitemShut {NoStop}%
\bibitem [{\citenamefont {{Sch{\"a}fer}}\ \emph {et~al.}(1966)\citenamefont
  {{Sch{\"a}fer}}, \citenamefont {{Schmidt}},\ and\ \citenamefont
  {{Volze}}}]{schafer66}%
  \BibitemOpen
  \bibfield  {author} {\bibinfo {author} {\bibnamefont {{Sch{\"a}fer}},
  \bibfnamefont {F~P}}, \bibinfo {author} {\bibfnamefont {W.}~\bibnamefont
  {{Schmidt}}}, and\ \bibinfo {author} {\bibfnamefont {J.}~\bibnamefont
  {{Volze}}}} (\bibinfo {year} {1966}),\ \bibfield  {title} {\enquote {\bibinfo
  {title} {{Organic Dye Solution Laser}},}\ }\href
  {https://doi.org/10.1063/1.1754762} {\bibfield  {journal} {\bibinfo
  {journal} {Appl. Phys. Lett.}\ }\textbf {\bibinfo {volume} {9}},\ \bibinfo
  {pages} {306--309}}\BibitemShut {NoStop}%
\bibitem [{\citenamefont {{Schatz}}\ \emph {et~al.}(2002)\citenamefont
  {{Schatz}}, \citenamefont {{Toenjes}}, \citenamefont {{Pfeiffer}},
  \citenamefont {{Beers}}, \citenamefont {{Cowan}}, \citenamefont {{Hill}},\
  and\ \citenamefont {{Kratz}}}]{Schatz.ea:2002}%
  \BibitemOpen
  \bibfield  {author} {\bibinfo {author} {\bibnamefont {{Schatz}},
  \bibfnamefont {H}}, \bibinfo {author} {\bibfnamefont {R.}~\bibnamefont
  {{Toenjes}}}, \bibinfo {author} {\bibfnamefont {B.}~\bibnamefont
  {{Pfeiffer}}}, \bibinfo {author} {\bibfnamefont {T.~C.}\ \bibnamefont
  {{Beers}}}, \bibinfo {author} {\bibfnamefont {J.~J.}\ \bibnamefont
  {{Cowan}}}, \bibinfo {author} {\bibfnamefont {V.}~\bibnamefont {{Hill}}},
  and\ \bibinfo {author} {\bibfnamefont {K.-L.}\ \bibnamefont {{Kratz}}}}
  (\bibinfo {year} {2002}),\ \bibfield  {title} {\enquote {\bibinfo {title}
  {{Thorium and Uranium Chronometers Applied to CS 31082-001}},}\ }\href
  {https://doi.org/10.1086/342939} {\bibfield  {journal} {\bibinfo  {journal}
  {Astrophys. J.}\ }\textbf {\bibinfo {volume} {579}},\ \bibinfo {pages}
  {626--638}}\BibitemShut {NoStop}%
\bibitem [{\citenamefont {{Schiller}}\ \emph {et~al.}(2000)\citenamefont
  {{Schiller}}, \citenamefont {{Bergholt}}, \citenamefont {{Guttormsen}},
  \citenamefont {{Melby}}, \citenamefont {{Rekstad}},\ and\ \citenamefont
  {{Siem}}}]{Schiller.Bergholt.ea:2000}%
  \BibitemOpen
  \bibfield  {author} {\bibinfo {author} {\bibnamefont {{Schiller}},
  \bibfnamefont {A}}, \bibinfo {author} {\bibfnamefont {L.}~\bibnamefont
  {{Bergholt}}}, \bibinfo {author} {\bibfnamefont {M.}~\bibnamefont
  {{Guttormsen}}}, \bibinfo {author} {\bibfnamefont {E.}~\bibnamefont
  {{Melby}}}, \bibinfo {author} {\bibfnamefont {J.}~\bibnamefont {{Rekstad}}},
  and\ \bibinfo {author} {\bibfnamefont {S.}~\bibnamefont {{Siem}}}} (\bibinfo
  {year} {2000}),\ \bibfield  {title} {\enquote {\bibinfo {title} {{Extraction
  of level density and $\gamma$ strength function from primary $\gamma$
  spectra}},}\ }\href {https://doi.org/10.1016/S0168-9002(99)01187-0}
  {\bibfield  {journal} {\bibinfo  {journal} {Nucl. Instruments Methods Phys.
  Res. Sect. A Accel. Spectrometers, Detect. Assoc. Equip.}\ }\textbf {\bibinfo
  {volume} {447}},\ \bibinfo {pages} {498--511}}\BibitemShut {NoStop}%
\bibitem [{\citenamefont {Schmidt}\ \emph {et~al.}(2016)\citenamefont
  {Schmidt}, \citenamefont {Jurado}, \citenamefont {Amouroux},\ and\
  \citenamefont {Schmitt}}]{Schmidt.Jurado.ea:2016}%
  \BibitemOpen
  \bibfield  {author} {\bibinfo {author} {\bibnamefont {Schmidt}, \bibfnamefont
  {K-H}}, \bibinfo {author} {\bibfnamefont {B.}~\bibnamefont {Jurado}},
  \bibinfo {author} {\bibfnamefont {C.}~\bibnamefont {Amouroux}}, and\ \bibinfo
  {author} {\bibfnamefont {C.}~\bibnamefont {Schmitt}}} (\bibinfo {year}
  {2016}),\ \bibfield  {title} {\enquote {\bibinfo {title} {{General
  Description of Fission Observables: GEF Model Code}},}\ }\href
  {https://doi.org/10.1016/j.nds.2015.12.009} {\bibfield  {journal} {\bibinfo
  {journal} {Nucl. Data Sheets}\ }\textbf {\bibinfo {volume} {131}},\ \bibinfo
  {pages} {107--221}}\BibitemShut {NoStop}%
\bibitem [{\citenamefont {Schmidt}\ and\ \citenamefont
  {Jurado}(2018)}]{Schmidt.Jurado:2018}%
  \BibitemOpen
  \bibfield  {author} {\bibinfo {author} {\bibnamefont {Schmidt}, \bibfnamefont
  {Karl-Heinz}}, and\ \bibinfo {author} {\bibfnamefont {Beatriz}\ \bibnamefont
  {Jurado}}} (\bibinfo {year} {2018}),\ \bibfield  {title} {\enquote {\bibinfo
  {title} {Review on the progress in nuclear fission--experimental methods and
  theoretical descriptions},}\ }\href
  {https://doi.org/10.1088/1361-6633/aacfa7} {\bibfield  {journal} {\bibinfo
  {journal} {Rep. Prog. Phys.}\ }\textbf {\bibinfo {volume} {81}},\ \bibinfo
  {pages} {106301}}\BibitemShut {NoStop}%
\bibitem [{\citenamefont {Schmitt}\ \emph {et~al.}(2018)\citenamefont
  {Schmitt}, \citenamefont {Schmidt},\ and\ \citenamefont
  {Jurado}}]{Schmitt.Schmidt.Jurado:2018}%
  \BibitemOpen
  \bibfield  {author} {\bibinfo {author} {\bibnamefont {Schmitt}, \bibfnamefont
  {C}}, \bibinfo {author} {\bibfnamefont {K.-H.}\ \bibnamefont {Schmidt}}, and\
  \bibinfo {author} {\bibfnamefont {B.}~\bibnamefont {Jurado}}} (\bibinfo
  {year} {2018}),\ \bibfield  {title} {\enquote {\bibinfo {title} {{Benchmark
  of the GEF code for fission-fragment yields over an enlarged range in
  fissioning nucleus mass, excitation energy, and angular momentum}},}\ }\href
  {https://doi.org/10.1103/PhysRevC.98.044605} {\bibfield  {journal} {\bibinfo
  {journal} {Phys. Rev. C}\ }\textbf {\bibinfo {volume} {98}},\ \bibinfo
  {pages} {044605}}\BibitemShut {NoStop}%
\bibitem [{\citenamefont {{Sch{\"o}nrich}}\ and\ \citenamefont
  {{Weinberg}}(2019)}]{Schonrich:2019}%
  \BibitemOpen
  \bibfield  {author} {\bibinfo {author} {\bibnamefont {{Sch{\"o}nrich}},
  \bibfnamefont {Ralph~A}}, and\ \bibinfo {author} {\bibfnamefont {David~H.}\
  \bibnamefont {{Weinberg}}}} (\bibinfo {year} {2019}),\ \bibfield  {title}
  {\enquote {\bibinfo {title} {{The chemical evolution of r-process elements
  from neutron star mergers: the role of a 2-phase interstellar medium}},}\
  }\href {https://doi.org/10.1093/mnras/stz1126} {\bibfield  {journal}
  {\bibinfo  {journal} {Mon. Not. Roy. Astron. Soc.}\ }\textbf {\bibinfo
  {volume} {487}},\ \bibinfo {pages} {580--594}}\BibitemShut {NoStop}%
\bibitem [{\citenamefont {{Schramm}}(1973)}]{Schramm:1973}%
  \BibitemOpen
  \bibfield  {author} {\bibinfo {author} {\bibnamefont {{Schramm}},
  \bibfnamefont {D~N}}} (\bibinfo {year} {1973}),\ \bibfield  {title} {\enquote
  {\bibinfo {title} {{Explosive r-PROCESS Nucleosynthesis}},}\ }\href
  {https://doi.org/10.1086/152416} {\bibfield  {journal} {\bibinfo  {journal}
  {Astrophys. J.}\ }\textbf {\bibinfo {volume} {185}},\ \bibinfo {pages}
  {293--302}}\BibitemShut {NoStop}%
\bibitem [{\citenamefont {{Schramm}}\ and\ \citenamefont
  {{Wasserburg}}(1970)}]{Schramm.Wasserburg:1970}%
  \BibitemOpen
  \bibfield  {author} {\bibinfo {author} {\bibnamefont {{Schramm}},
  \bibfnamefont {D~N}}, and\ \bibinfo {author} {\bibfnamefont {G.~J.}\
  \bibnamefont {{Wasserburg}}}} (\bibinfo {year} {1970}),\ \bibfield  {title}
  {\enquote {\bibinfo {title} {{Nucleochronologies and the Mean Age of the
  Elements}},}\ }\href {https://doi.org/10.1086/150634} {\bibfield  {journal}
  {\bibinfo  {journal} {Astrophys. J.}\ }\textbf {\bibinfo {volume} {162}},\
  \bibinfo {pages} {57}}\BibitemShut {NoStop}%
\bibitem [{\citenamefont {{Schwengner}}\ \emph {et~al.}(2017)\citenamefont
  {{Schwengner}}, \citenamefont {{Frauendorf}},\ and\ \citenamefont
  {{Brown}}}]{Schwengner.Frauendorf.Brown:2017}%
  \BibitemOpen
  \bibfield  {author} {\bibinfo {author} {\bibnamefont {{Schwengner}},
  \bibfnamefont {R}}, \bibinfo {author} {\bibfnamefont {S.}~\bibnamefont
  {{Frauendorf}}}, and\ \bibinfo {author} {\bibfnamefont {B.~A.}\ \bibnamefont
  {{Brown}}}} (\bibinfo {year} {2017}),\ \bibfield  {title} {\enquote {\bibinfo
  {title} {{Low-Energy Magnetic Dipole Radiation in Open-Shell Nuclei}},}\
  }\href {https://doi.org/10.1103/PhysRevLett.118.092502} {\bibfield  {journal}
  {\bibinfo  {journal} {Phys. Rev. Lett.}\ }\textbf {\bibinfo {volume} {118}},\
  \bibinfo {eid} {092502}}\BibitemShut {NoStop}%
\bibitem [{\citenamefont {{Schwengner}}\ \emph {et~al.}(2013)\citenamefont
  {{Schwengner}}, \citenamefont {{Frauendorf}},\ and\ \citenamefont
  {{Larsen}}}]{Schwengner.Frauendorf.Larsen:2013}%
  \BibitemOpen
  \bibfield  {author} {\bibinfo {author} {\bibnamefont {{Schwengner}},
  \bibfnamefont {R}}, \bibinfo {author} {\bibfnamefont {S.}~\bibnamefont
  {{Frauendorf}}}, and\ \bibinfo {author} {\bibfnamefont {A.~C.}\ \bibnamefont
  {{Larsen}}}} (\bibinfo {year} {2013}),\ \bibfield  {title} {\enquote
  {\bibinfo {title} {{Low-Energy Enhancement of Magnetic Dipole Radiation}},}\
  }\href {https://doi.org/10.1103/PhysRevLett.111.232504} {\bibfield  {journal}
  {\bibinfo  {journal} {Phys. Rev. Lett.}\ }\textbf {\bibinfo {volume} {111}},\
  \bibinfo {eid} {232504}}\BibitemShut {NoStop}%
\bibitem [{\citenamefont {{Seeger}}\ \emph {et~al.}(1965)\citenamefont
  {{Seeger}}, \citenamefont {{Fowler}},\ and\ \citenamefont
  {{Clayton}}}]{seeger.fowler.clayton:1965}%
  \BibitemOpen
  \bibfield  {author} {\bibinfo {author} {\bibnamefont {{Seeger}},
  \bibfnamefont {P~A}}, \bibinfo {author} {\bibfnamefont {W.~A.}\ \bibnamefont
  {{Fowler}}}, and\ \bibinfo {author} {\bibfnamefont {D.~D.}\ \bibnamefont
  {{Clayton}}}} (\bibinfo {year} {1965}),\ \bibfield  {title} {\enquote
  {\bibinfo {title} {{Nucleosynthesis of Heavy Elements by Neutron Capture.}}}\
  }\href {https://doi.org/10.1086/190111} {\bibfield  {journal} {\bibinfo
  {journal} {Astrophys. J. Suppl.}\ }\textbf {\bibinfo {volume} {11}},\
  \bibinfo {pages} {121--166}}\BibitemShut {NoStop}%
\bibitem [{\citenamefont {{Seitenzahl}}\ \emph {et~al.}(2019)\citenamefont
  {{Seitenzahl}}, \citenamefont {{Ghavamian}}, \citenamefont {{Laming}},\ and\
  \citenamefont {{Vogt}}}]{Seitenzahl.Chavamian.ea:2019}%
  \BibitemOpen
  \bibfield  {author} {\bibinfo {author} {\bibnamefont {{Seitenzahl}},
  \bibfnamefont {I~R}}, \bibinfo {author} {\bibfnamefont {P.}~\bibnamefont
  {{Ghavamian}}}, \bibinfo {author} {\bibfnamefont {J.~M.}\ \bibnamefont
  {{Laming}}}, and\ \bibinfo {author} {\bibfnamefont {F.~P.~A.}\ \bibnamefont
  {{Vogt}}}} (\bibinfo {year} {2019}),\ \bibfield  {title} {\enquote {\bibinfo
  {title} {{Optical Tomography of Chemical Elements Synthesized in Type Ia
  Supernovae}},}\ }\href {https://doi.org/10.1103/PhysRevLett.123.041101}
  {\bibfield  {journal} {\bibinfo  {journal} {Phys. Rev. Lett.}\ }\textbf
  {\bibinfo {volume} {123}},\ \bibinfo {eid} {041101}}\BibitemShut {NoStop}%
\bibitem [{\citenamefont {{Seitenzahl}}\ \emph {et~al.}(2014)\citenamefont
  {{Seitenzahl}}, \citenamefont {{Timmes}},\ and\ \citenamefont
  {{Magkotsios}}}]{Seitenzahl.Timmes.Magkotsios:2014}%
  \BibitemOpen
  \bibfield  {author} {\bibinfo {author} {\bibnamefont {{Seitenzahl}},
  \bibfnamefont {I~R}}, \bibinfo {author} {\bibfnamefont {F.~X.}\ \bibnamefont
  {{Timmes}}}, and\ \bibinfo {author} {\bibfnamefont {G.}~\bibnamefont
  {{Magkotsios}}}} (\bibinfo {year} {2014}),\ \bibfield  {title} {\enquote
  {\bibinfo {title} {{The Light Curve of SN 1987A Revisited: Constraining
  Production Masses of Radioactive Nuclides}},}\ }\href
  {https://doi.org/10.1088/0004-637X/792/1/10} {\bibfield  {journal} {\bibinfo
  {journal} {Astrophys. J.}\ }\textbf {\bibinfo {volume} {792}},\ \bibinfo
  {eid} {10}}\BibitemShut {NoStop}%
\bibitem [{\citenamefont {{Seitenzahl}}\ and\ \citenamefont
  {{Townsley}}(2017)}]{seitenzahl17}%
  \BibitemOpen
  \bibfield  {author} {\bibinfo {author} {\bibnamefont {{Seitenzahl}},
  \bibfnamefont {IR}}, and\ \bibinfo {author} {\bibfnamefont {D.M.}\
  \bibnamefont {{Townsley}}}} (\bibinfo {year} {2017}),\ \enquote {\bibinfo
  {title} {{Nucleosynthesis in Thermonuclear Supernovae}},}\ in\ \href
  {https://doi.org/10.1007/978-3-319-20794-0_87-1} {\emph {\bibinfo {booktitle}
  {Handbook of Supernovae}}},\ \bibinfo {editor} {edited by\ \bibinfo {editor}
  {\bibfnamefont {A.~W.}\ \bibnamefont {{Alsabti}}}\ and\ \bibinfo {editor}
  {\bibfnamefont {P.}~\bibnamefont {{Murdin}}}}\ (\bibinfo  {publisher}
  {Springer International Publishing},\ \bibinfo {address} {Cham})\BibitemShut
  {NoStop}%
\bibitem [{\citenamefont {{Sekiguchi}}\ \emph {et~al.}(2015)\citenamefont
  {{Sekiguchi}}, \citenamefont {{Kiuchi}}, \citenamefont {{Kyutoku}},\ and\
  \citenamefont {{Shibata}}}]{sekiguchi15}%
  \BibitemOpen
  \bibfield  {author} {\bibinfo {author} {\bibnamefont {{Sekiguchi}},
  \bibfnamefont {Y}}, \bibinfo {author} {\bibfnamefont {K.}~\bibnamefont
  {{Kiuchi}}}, \bibinfo {author} {\bibfnamefont {K.}~\bibnamefont {{Kyutoku}}},
  and\ \bibinfo {author} {\bibfnamefont {M.}~\bibnamefont {{Shibata}}}}
  (\bibinfo {year} {2015}),\ \bibfield  {title} {\enquote {\bibinfo {title}
  {{Dynamical mass ejection from binary neutron star mergers:
  Radiation-hydrodynamics study in general relativity}},}\ }\href
  {https://doi.org/10.1103/PhysRevD.91.064059} {\bibfield  {journal} {\bibinfo
  {journal} {Phys. Rev. D}\ }\textbf {\bibinfo {volume} {91}},\ \bibinfo {eid}
  {064059}}\BibitemShut {NoStop}%
\bibitem [{\citenamefont {{Sekiguchi}}\ \emph {et~al.}(2016)\citenamefont
  {{Sekiguchi}}, \citenamefont {{Kiuchi}}, \citenamefont {{Kyutoku}},
  \citenamefont {{Shibata}},\ and\ \citenamefont {{Taniguchi}}}]{sekiguchi16}%
  \BibitemOpen
  \bibfield  {author} {\bibinfo {author} {\bibnamefont {{Sekiguchi}},
  \bibfnamefont {Y}}, \bibinfo {author} {\bibfnamefont {K.}~\bibnamefont
  {{Kiuchi}}}, \bibinfo {author} {\bibfnamefont {K.}~\bibnamefont {{Kyutoku}}},
  \bibinfo {author} {\bibfnamefont {M.}~\bibnamefont {{Shibata}}}, and\
  \bibinfo {author} {\bibfnamefont {K.}~\bibnamefont {{Taniguchi}}}} (\bibinfo
  {year} {2016}),\ \bibfield  {title} {\enquote {\bibinfo {title} {{Dynamical
  mass ejection from the merger of asymmetric binary neutron stars:
  Radiation-hydrodynamics study in general relativity}},}\ }\href
  {https://doi.org/10.1103/PhysRevD.93.124046} {\bibfield  {journal} {\bibinfo
  {journal} {Phys. Rev. D}\ }\textbf {\bibinfo {volume} {93}},\ \bibinfo {eid}
  {124046}}\BibitemShut {NoStop}%
\bibitem [{\citenamefont {{Sekiguchi}}\ and\ \citenamefont
  {{Shibata}}(2011)}]{sekiguchi11}%
  \BibitemOpen
  \bibfield  {author} {\bibinfo {author} {\bibnamefont {{Sekiguchi}},
  \bibfnamefont {Y}}, and\ \bibinfo {author} {\bibfnamefont {M.}~\bibnamefont
  {{Shibata}}}} (\bibinfo {year} {2011}),\ \bibfield  {title} {\enquote
  {\bibinfo {title} {{Formation of Black Hole and Accretion Disk in a Massive
  High-entropy Stellar Core Collapse}},}\ }\href
  {https://doi.org/10.1088/0004-637X/737/1/6} {\bibfield  {journal} {\bibinfo
  {journal} {Astrophys. J.}\ }\textbf {\bibinfo {volume} {737}},\ \bibinfo
  {eid} {6}}\BibitemShut {NoStop}%
\bibitem [{\citenamefont {Sekiguchi}\ \emph {et~al.}(2011)\citenamefont
  {Sekiguchi}, \citenamefont {Kiuchi}, \citenamefont {Kyutoku},\ and\
  \citenamefont {Shibata}}]{Sekiguchi.Kiuchi.ea:2011}%
  \BibitemOpen
  \bibfield  {author} {\bibinfo {author} {\bibnamefont {Sekiguchi},
  \bibfnamefont {Yuichiro}}, \bibinfo {author} {\bibfnamefont {Kenta}\
  \bibnamefont {Kiuchi}}, \bibinfo {author} {\bibfnamefont {Koutarou}\
  \bibnamefont {Kyutoku}}, and\ \bibinfo {author} {\bibfnamefont {Masaru}\
  \bibnamefont {Shibata}}} (\bibinfo {year} {2011}),\ \bibfield  {title}
  {\enquote {\bibinfo {title} {Gravitational waves and neutrino emission from
  the merger of binary neutron stars},}\ }\href
  {https://doi.org/10.1103/PhysRevLett.107.051102} {\bibfield  {journal}
  {\bibinfo  {journal} {Phys. Rev. Lett.}\ }\textbf {\bibinfo {volume} {107}},\
  \bibinfo {pages} {051102}}\BibitemShut {NoStop}%
\bibitem [{\citenamefont {Shafer}\ \emph {et~al.}(2016)\citenamefont {Shafer},
  \citenamefont {Engel}, \citenamefont {Fr\"ohlich}, \citenamefont
  {McLaughlin}, \citenamefont {Mumpower},\ and\ \citenamefont
  {Surman}}]{Shafer.Engel.ea:2016}%
  \BibitemOpen
  \bibfield  {author} {\bibinfo {author} {\bibnamefont {Shafer}, \bibfnamefont
  {T}}, \bibinfo {author} {\bibfnamefont {J.}~\bibnamefont {Engel}}, \bibinfo
  {author} {\bibfnamefont {C.}~\bibnamefont {Fr\"ohlich}}, \bibinfo {author}
  {\bibfnamefont {G.~C.}\ \bibnamefont {McLaughlin}}, \bibinfo {author}
  {\bibfnamefont {M.}~\bibnamefont {Mumpower}}, and\ \bibinfo {author}
  {\bibfnamefont {R.}~\bibnamefont {Surman}}} (\bibinfo {year} {2016}),\
  \bibfield  {title} {\enquote {\bibinfo {title} {$\ensuremath{\beta}$ decay of
  deformed $r$-process nuclei near $a=80$ and $a=160$, including odd-$a$ and
  odd-odd nuclei, with the skyrme finite-amplitude method},}\ }\href
  {https://doi.org/10.1103/PhysRevC.94.055802} {\bibfield  {journal} {\bibinfo
  {journal} {Phys. Rev. C}\ }\textbf {\bibinfo {volume} {94}},\ \bibinfo
  {pages} {055802}}\BibitemShut {NoStop}%
\bibitem [{\citenamefont {{Shen}}\ \emph {et~al.}(2015)\citenamefont {{Shen}},
  \citenamefont {{Cooke}}, \citenamefont {{Ramirez-Ruiz}}, \citenamefont
  {{Madau}}, \citenamefont {{Mayer}},\ and\ \citenamefont {{Guedes}}}]{shen15}%
  \BibitemOpen
  \bibfield  {author} {\bibinfo {author} {\bibnamefont {{Shen}}, \bibfnamefont
  {S}}, \bibinfo {author} {\bibfnamefont {R.~J.}\ \bibnamefont {{Cooke}}},
  \bibinfo {author} {\bibfnamefont {E.}~\bibnamefont {{Ramirez-Ruiz}}},
  \bibinfo {author} {\bibfnamefont {P.}~\bibnamefont {{Madau}}}, \bibinfo
  {author} {\bibfnamefont {L.}~\bibnamefont {{Mayer}}}, and\ \bibinfo {author}
  {\bibfnamefont {J.}~\bibnamefont {{Guedes}}}} (\bibinfo {year} {2015}),\
  \bibfield  {title} {\enquote {\bibinfo {title} {{The History of R-Process
  Enrichment in the Milky Way}},}\ }\href
  {https://doi.org/10.1088/0004-637X/807/2/115} {\bibfield  {journal} {\bibinfo
   {journal} {Astrophys. J.}\ }\textbf {\bibinfo {volume} {807}},\ \bibinfo
  {eid} {115}}\BibitemShut {NoStop}%
\bibitem [{\citenamefont {{Shetrone}}\ \emph {et~al.}(2003)\citenamefont
  {{Shetrone}}, \citenamefont {{Venn}}, \citenamefont {{Tolstoy}},
  \citenamefont {{Primas}}, \citenamefont {{Hill}},\ and\ \citenamefont
  {{Kaufer}}}]{Shetrone.Venn.ea:2003}%
  \BibitemOpen
  \bibfield  {author} {\bibinfo {author} {\bibnamefont {{Shetrone}},
  \bibfnamefont {M}}, \bibinfo {author} {\bibfnamefont {K.~A.}\ \bibnamefont
  {{Venn}}}, \bibinfo {author} {\bibfnamefont {E.}~\bibnamefont {{Tolstoy}}},
  \bibinfo {author} {\bibfnamefont {F.}~\bibnamefont {{Primas}}}, \bibinfo
  {author} {\bibfnamefont {V.}~\bibnamefont {{Hill}}}, and\ \bibinfo {author}
  {\bibfnamefont {A.}~\bibnamefont {{Kaufer}}}} (\bibinfo {year} {2003}),\
  \bibfield  {title} {\enquote {\bibinfo {title} {{VLT/UVES Abundances in Four
  Nearby Dwarf Spheroidal Galaxies. I. Nucleosynthesis and Abundance
  Ratios}},}\ }\href {https://doi.org/10.1086/345966} {\bibfield  {journal}
  {\bibinfo  {journal} {Astron. J.}\ }\textbf {\bibinfo {volume} {125}},\
  \bibinfo {pages} {684--706}}\BibitemShut {NoStop}%
\bibitem [{\citenamefont {{Shetrone}}\ \emph {et~al.}(1998)\citenamefont
  {{Shetrone}}, \citenamefont {{Bolte}},\ and\ \citenamefont
  {{Stetson}}}]{Shetrone.Bolte.Stetson:1998}%
  \BibitemOpen
  \bibfield  {author} {\bibinfo {author} {\bibnamefont {{Shetrone}},
  \bibfnamefont {M~D}}, \bibinfo {author} {\bibfnamefont {M.}~\bibnamefont
  {{Bolte}}}, and\ \bibinfo {author} {\bibfnamefont {P.~B.}\ \bibnamefont
  {{Stetson}}}} (\bibinfo {year} {1998}),\ \bibfield  {title} {\enquote
  {\bibinfo {title} {{Keck HIRES Abundances in the Dwarf Spheroidal Galaxy
  Draco}},}\ }\href {https://doi.org/10.1086/300341} {\bibfield  {journal}
  {\bibinfo  {journal} {Astron. J.}\ }\textbf {\bibinfo {volume} {115}},\
  \bibinfo {pages} {1888--1893}}\BibitemShut {NoStop}%
\bibitem [{\citenamefont {{Shibagaki}}\ \emph {et~al.}(2016)\citenamefont
  {{Shibagaki}}, \citenamefont {{Kajino}}, \citenamefont {{Mathews}},
  \citenamefont {{Chiba}}, \citenamefont {{Nishimura}},\ and\ \citenamefont
  {{Lorusso}}}]{Shibagaki.Kajino.ea:2016}%
  \BibitemOpen
  \bibfield  {author} {\bibinfo {author} {\bibnamefont {{Shibagaki}},
  \bibfnamefont {S}}, \bibinfo {author} {\bibfnamefont {T.}~\bibnamefont
  {{Kajino}}}, \bibinfo {author} {\bibfnamefont {G.~J.}\ \bibnamefont
  {{Mathews}}}, \bibinfo {author} {\bibfnamefont {S.}~\bibnamefont {{Chiba}}},
  \bibinfo {author} {\bibfnamefont {S.}~\bibnamefont {{Nishimura}}}, and\
  \bibinfo {author} {\bibfnamefont {G.}~\bibnamefont {{Lorusso}}}} (\bibinfo
  {year} {2016}),\ \bibfield  {title} {\enquote {\bibinfo {title} {{Relative
  Contributions of the Weak, Main, and Fission-recycling r-process}},}\ }\href
  {https://doi.org/10.3847/0004-637X/816/2/79} {\bibfield  {journal} {\bibinfo
  {journal} {Astrophys. J.}\ }\textbf {\bibinfo {volume} {816}},\ \bibinfo
  {eid} {79}}\BibitemShut {NoStop}%
\bibitem [{\citenamefont {{Shibata}}\ and\ \citenamefont
  {{Taniguchi}}(2011)}]{Shibata.Taniguchi:2011}%
  \BibitemOpen
  \bibfield  {author} {\bibinfo {author} {\bibnamefont {{Shibata}},
  \bibfnamefont {M}}, and\ \bibinfo {author} {\bibfnamefont {K.}~\bibnamefont
  {{Taniguchi}}}} (\bibinfo {year} {2011}),\ \bibfield  {title} {\enquote
  {\bibinfo {title} {{Coalescence of Black Hole-Neutron Star Binaries}},}\
  }\href {https://doi.org/10.12942/lrr-2011-6} {\bibfield  {journal} {\bibinfo
  {journal} {Living Reviews in Relativity}\ }\textbf {\bibinfo {volume} {14}},\
  \bibinfo {eid} {6}}\BibitemShut {NoStop}%
\bibitem [{\citenamefont {Shibata}(2018)}]{Shibata:2018}%
  \BibitemOpen
  \bibfield  {author} {\bibinfo {author} {\bibnamefont {Shibata}, \bibfnamefont
  {Masaru}}} (\bibinfo {year} {2018}),\ \href
  {https://indico.gsi.de/event/7091/contribution/1/material/slides/0.pdf}
  {\enquote {\bibinfo {title} {{Mass ejection from neutron star mergers}},}\
  }\bibinfo {note} {Talk at the EMMI Rapid Reaction Task Force: The physics of
  neutron star mergers at GSI/FAIR}\BibitemShut {NoStop}%
\bibitem [{\citenamefont {Shibata}\ \emph {et~al.}(2017)\citenamefont
  {Shibata}, \citenamefont {Fujibayashi}, \citenamefont {Hotokezaka},
  \citenamefont {Kiuchi}, \citenamefont {Kyutoku}, \citenamefont {Sekiguchi},\
  and\ \citenamefont {Tanaka}}]{Shibata.Fujibayashi.ea:2017}%
  \BibitemOpen
  \bibfield  {author} {\bibinfo {author} {\bibnamefont {Shibata}, \bibfnamefont
  {Masaru}}, \bibinfo {author} {\bibfnamefont {Sho}\ \bibnamefont
  {Fujibayashi}}, \bibinfo {author} {\bibfnamefont {Kenta}\ \bibnamefont
  {Hotokezaka}}, \bibinfo {author} {\bibfnamefont {Kenta}\ \bibnamefont
  {Kiuchi}}, \bibinfo {author} {\bibfnamefont {Koutarou}\ \bibnamefont
  {Kyutoku}}, \bibinfo {author} {\bibfnamefont {Yuichiro}\ \bibnamefont
  {Sekiguchi}}, and\ \bibinfo {author} {\bibfnamefont {Masaomi}\ \bibnamefont
  {Tanaka}}} (\bibinfo {year} {2017}),\ \bibfield  {title} {\enquote {\bibinfo
  {title} {{Modeling GW170817 based on numerical relativity and its
  implications}},}\ }\href {https://doi.org/10.1103/PhysRevD.96.123012}
  {\bibfield  {journal} {\bibinfo  {journal} {Phys. Rev. D}\ }\textbf {\bibinfo
  {volume} {96}},\ \bibinfo {pages} {123012}}\BibitemShut {NoStop}%
\bibitem [{\citenamefont {{Shibata}}\ and\ \citenamefont
  {{Hotokezaka}}(2019)}]{Shibata.Hotokezaka:2019}%
  \BibitemOpen
  \bibfield  {author} {\bibinfo {author} {\bibnamefont {{Shibata}},
  \bibfnamefont {Masaru}}, and\ \bibinfo {author} {\bibfnamefont {Kenta}\
  \bibnamefont {{Hotokezaka}}}} (\bibinfo {year} {2019}),\ \bibfield  {title}
  {\enquote {\bibinfo {title} {{Merger and Mass Ejection of Neutron Star
  Binaries}},}\ }\href {https://doi.org/10.1146/annurev-nucl-101918-023625}
  {\bibfield  {journal} {\bibinfo  {journal} {Annu. Rev. Nucl. Part. Sci.}\
  }\textbf {\bibinfo {volume} {69}},\ \bibinfo {pages} {41--64}}\BibitemShut
  {NoStop}%
\bibitem [{\citenamefont {Shibata}\ and\ \citenamefont
  {Taniguchi}(2006)}]{Shibata.Taniguchi:2006}%
  \BibitemOpen
  \bibfield  {author} {\bibinfo {author} {\bibnamefont {Shibata}, \bibfnamefont
  {Masaru}}, and\ \bibinfo {author} {\bibfnamefont {Keisuke}\ \bibnamefont
  {Taniguchi}}} (\bibinfo {year} {2006}),\ \bibfield  {title} {\enquote
  {\bibinfo {title} {Merger of binary neutron stars to a black hole: Disk mass,
  short gamma-ray bursts, and quasinormal mode ringing},}\ }\href
  {https://doi.org/10.1103/PhysRevD.73.064027} {\bibfield  {journal} {\bibinfo
  {journal} {Phys. Rev. D}\ }\textbf {\bibinfo {volume} {73}},\ \bibinfo
  {pages} {064027}}\BibitemShut {NoStop}%
\bibitem [{\citenamefont {Shibata}\ \emph {et~al.}(2005)\citenamefont
  {Shibata}, \citenamefont {Taniguchi},\ and\ \citenamefont
  {Ury{\=u}}}]{Shibata.Taniguchi.Uryu:2005}%
  \BibitemOpen
  \bibfield  {author} {\bibinfo {author} {\bibnamefont {Shibata}, \bibfnamefont
  {Masaru}}, \bibinfo {author} {\bibfnamefont {Keisuke}\ \bibnamefont
  {Taniguchi}}, and\ \bibinfo {author} {\bibfnamefont {K{\=o}ji}\ \bibnamefont
  {Ury{\=u}}}} (\bibinfo {year} {2005}),\ \bibfield  {title} {\enquote
  {\bibinfo {title} {Merger of binary neutron stars with realistic equations of
  state in full general relativity},}\ }\href
  {https://doi.org/10.1103/PhysRevD.71.084021} {\bibfield  {journal} {\bibinfo
  {journal} {Phys. Rev. D}\ }\textbf {\bibinfo {volume} {71}},\ \bibinfo
  {pages} {084021}}\BibitemShut {NoStop}%
\bibitem [{\citenamefont {Shibata}\ and\ \citenamefont
  {Ury{\=u}}(2000)}]{Shibata.Uryu:2000}%
  \BibitemOpen
  \bibfield  {author} {\bibinfo {author} {\bibnamefont {Shibata}, \bibfnamefont
  {Masaru}}, and\ \bibinfo {author} {\bibfnamefont {K{\=o}ji}\ \bibnamefont
  {Ury{\=u}}}} (\bibinfo {year} {2000}),\ \bibfield  {title} {\enquote
  {\bibinfo {title} {{Simulation of merging binary neutron stars in full
  general relativity: $\ensuremath{\Gamma}=2$ case}},}\ }\href
  {https://doi.org/10.1103/PhysRevD.61.064001} {\bibfield  {journal} {\bibinfo
  {journal} {Phys. Rev. D}\ }\textbf {\bibinfo {volume} {61}},\ \bibinfo
  {pages} {064001}}\BibitemShut {NoStop}%
\bibitem [{\citenamefont {Shibata}\ and\ \citenamefont
  {Ury{\=u}}(2006)}]{Shibata.Uryu:2006}%
  \BibitemOpen
  \bibfield  {author} {\bibinfo {author} {\bibnamefont {Shibata}, \bibfnamefont
  {Masaru}}, and\ \bibinfo {author} {\bibfnamefont {K{\=o}ji}\ \bibnamefont
  {Ury{\=u}}}} (\bibinfo {year} {2006}),\ \bibfield  {title} {\enquote
  {\bibinfo {title} {{Merger of black hole-neutron star binaries: Nonspinning
  black hole case}},}\ }\href {https://doi.org/10.1103/PhysRevD.74.121503}
  {\bibfield  {journal} {\bibinfo  {journal} {Phys. Rev. D}\ }\textbf {\bibinfo
  {volume} {74}},\ \bibinfo {pages} {121503}}\BibitemShut {NoStop}%
\bibitem [{\citenamefont {{Siegel}}\ and\ \citenamefont
  {{Metzger}}(2018)}]{Siegel.Metzger:2018}%
  \BibitemOpen
  \bibfield  {author} {\bibinfo {author} {\bibnamefont {{Siegel}},
  \bibfnamefont {D~M}}, and\ \bibinfo {author} {\bibfnamefont {B.~D.}\
  \bibnamefont {{Metzger}}}} (\bibinfo {year} {2018}),\ \bibfield  {title}
  {\enquote {\bibinfo {title} {{Three-dimensional GRMHD Simulations of
  Neutrino-cooled Accretion Disks from Neutron Star Mergers}},}\ }\href
  {https://doi.org/10.3847/1538-4357/aabaec} {\bibfield  {journal} {\bibinfo
  {journal} {Astrophys. J.}\ }\textbf {\bibinfo {volume} {858}},\ \bibinfo
  {eid} {52}}\BibitemShut {NoStop}%
\bibitem [{\citenamefont {{Siegel}}(2019)}]{Siegel:2019}%
  \BibitemOpen
  \bibfield  {author} {\bibinfo {author} {\bibnamefont {{Siegel}},
  \bibfnamefont {Daniel~M}}} (\bibinfo {year} {2019}),\ \bibfield  {title}
  {\enquote {\bibinfo {title} {{GW170817 --the first observed neutron star
  merger and its kilonova: Implications for the astrophysical site of the
  r-process}},}\ }\href {https://doi.org/10.1140/epja/i2019-12888-9} {\bibfield
   {journal} {\bibinfo  {journal} {Eur. Phys. J. A}\ }\textbf {\bibinfo
  {volume} {55}},\ \bibinfo {eid} {203}}\BibitemShut {NoStop}%
\bibitem [{\citenamefont {{Siegel}}\ \emph {et~al.}(2019)\citenamefont
  {{Siegel}}, \citenamefont {{Barnes}},\ and\ \citenamefont
  {{Metzger}}}]{Siegel.Barnes.Metzger:2019}%
  \BibitemOpen
  \bibfield  {author} {\bibinfo {author} {\bibnamefont {{Siegel}},
  \bibfnamefont {Daniel~M}}, \bibinfo {author} {\bibfnamefont {Jennifer}\
  \bibnamefont {{Barnes}}}, and\ \bibinfo {author} {\bibfnamefont {Brian~D.}\
  \bibnamefont {{Metzger}}}} (\bibinfo {year} {2019}),\ \bibfield  {title}
  {\enquote {\bibinfo {title} {{Collapsars as a major source of r-process
  elements}},}\ }\href {https://doi.org/10.1038/s41586-019-1136-0} {\bibfield
  {journal} {\bibinfo  {journal} {Nature}\ }\textbf {\bibinfo {volume} {569}},\
  \bibinfo {pages} {241--244}}\BibitemShut {NoStop}%
\bibitem [{\citenamefont {{Siegel}}\ \emph {et~al.}(2014)\citenamefont
  {{Siegel}}, \citenamefont {{Ciolfi}},\ and\ \citenamefont
  {{Rezzolla}}}]{Siegel.Ciolfi.ea:2012}%
  \BibitemOpen
  \bibfield  {author} {\bibinfo {author} {\bibnamefont {{Siegel}},
  \bibfnamefont {Daniel~M}}, \bibinfo {author} {\bibfnamefont {Riccardo}\
  \bibnamefont {{Ciolfi}}}, and\ \bibinfo {author} {\bibfnamefont {Luciano}\
  \bibnamefont {{Rezzolla}}}} (\bibinfo {year} {2014}),\ \bibfield  {title}
  {\enquote {\bibinfo {title} {{Magnetically Driven Winds from Differentially
  Rotating Neutron Stars and X-Ray Afterglows of Short Gamma-Ray Bursts}},}\
  }\href {https://doi.org/10.1088/2041-8205/785/1/L6} {\bibfield  {journal}
  {\bibinfo  {journal} {Astrophys. J. Lett.}\ }\textbf {\bibinfo {volume}
  {785}},\ \bibinfo {eid} {L6}}\BibitemShut {NoStop}%
\bibitem [{\citenamefont {Siegel}\ and\ \citenamefont
  {Metzger}(2017)}]{Siegel.Metzger:2017}%
  \BibitemOpen
  \bibfield  {author} {\bibinfo {author} {\bibnamefont {Siegel}, \bibfnamefont
  {Daniel~M}}, and\ \bibinfo {author} {\bibfnamefont {Brian~D.}\ \bibnamefont
  {Metzger}}} (\bibinfo {year} {2017}),\ \bibfield  {title} {\enquote {\bibinfo
  {title} {{Three-Dimensional General-Relativistic Magnetohydrodynamic
  Simulations of Remnant Accretion Disks from Neutron Star Mergers: Outflows
  and $r$-Process Nucleosynthesis}},}\ }\href
  {https://doi.org/10.1103/PhysRevLett.119.231102} {\bibfield  {journal}
  {\bibinfo  {journal} {Phys. Rev. Lett.}\ }\textbf {\bibinfo {volume} {119}},\
  \bibinfo {pages} {231102}}\BibitemShut {NoStop}%
\bibitem [{\citenamefont {{Siegl}}\ \emph {et~al.}(2018)\citenamefont {{Siegl}}
  \emph {et~al.}}]{Siegl.Scielzo.ea:2018}%
  \BibitemOpen
  \bibfield  {author} {\bibinfo {author} {\bibnamefont {{Siegl}}, \bibfnamefont
  {K}},  \emph {et~al.}} (\bibinfo {year} {2018}),\ \bibfield  {title}
  {\enquote {\bibinfo {title} {{Recoil ions from the {\ensuremath{\beta}} decay
  of $^{134}$Sb confined in a Paul trap}},}\ }\href
  {https://doi.org/10.1103/PhysRevC.97.035504} {\bibfield  {journal} {\bibinfo
  {journal} {Phys. Rev. C}\ }\textbf {\bibinfo {volume} {97}}~(\bibinfo
  {number} {3}),\ \bibinfo {eid} {035504}}\BibitemShut {NoStop}%
\bibitem [{\citenamefont {{Sieja}}(2017)}]{Sieja:2017}%
  \BibitemOpen
  \bibfield  {author} {\bibinfo {author} {\bibnamefont {{Sieja}}, \bibfnamefont
  {K}}} (\bibinfo {year} {2017}),\ \bibfield  {title} {\enquote {\bibinfo
  {title} {{Electric and Magnetic Dipole Strength at Low Energy}},}\ }\href
  {https://doi.org/10.1103/PhysRevLett.119.052502} {\bibfield  {journal}
  {\bibinfo  {journal} {Phys. Rev. Lett.}\ }\textbf {\bibinfo {volume} {119}},\
  \bibinfo {eid} {052502}}\BibitemShut {NoStop}%
\bibitem [{\citenamefont {Sieja}(2018)}]{Sieja:2018}%
  \BibitemOpen
  \bibfield  {author} {\bibinfo {author} {\bibnamefont {Sieja}, \bibfnamefont
  {K}}} (\bibinfo {year} {2018}),\ \bibfield  {title} {\enquote {\bibinfo
  {title} {{Shell-model study of the $M1$ dipole strength at low energy in the
  $A > 100$ nuclei}},}\ }\href {https://doi.org/10.1103/PhysRevC.98.064312}
  {\bibfield  {journal} {\bibinfo  {journal} {Phys. Rev. C}\ }\textbf {\bibinfo
  {volume} {98}},\ \bibinfo {pages} {064312}}\BibitemShut {NoStop}%
\bibitem [{\citenamefont {{Sieverding}}\ \emph {et~al.}(2019)\citenamefont
  {{Sieverding}}, \citenamefont {{Langanke}}, \citenamefont
  {{Mart{\'i}nez-Pinedo}}, \citenamefont {{Bollig}}, \citenamefont {{Janka}},\
  and\ \citenamefont {{Heger}}}]{Sieverding.Martinez-Pinedo.ea:2019}%
  \BibitemOpen
  \bibfield  {author} {\bibinfo {author} {\bibnamefont {{Sieverding}},
  \bibfnamefont {A}}, \bibinfo {author} {\bibfnamefont {K.}~\bibnamefont
  {{Langanke}}}, \bibinfo {author} {\bibfnamefont {G.}~\bibnamefont
  {{Mart{\'i}nez-Pinedo}}}, \bibinfo {author} {\bibfnamefont {R.}~\bibnamefont
  {{Bollig}}}, \bibinfo {author} {\bibfnamefont {H.~T.}\ \bibnamefont
  {{Janka}}}, and\ \bibinfo {author} {\bibfnamefont {A.}~\bibnamefont
  {{Heger}}}} (\bibinfo {year} {2019}),\ \bibfield  {title} {\enquote {\bibinfo
  {title} {{The {\ensuremath{\nu}}-process with Fully Time-dependent Supernova
  Neutrino Emission Spectra}},}\ }\href
  {https://doi.org/10.3847/1538-4357/ab17e2} {\bibfield  {journal} {\bibinfo
  {journal} {Astrophys. J.}\ }\textbf {\bibinfo {volume} {876}},\ \bibinfo
  {eid} {151}}\BibitemShut {NoStop}%
\bibitem [{\citenamefont {{Simmerer}}\ \emph {et~al.}(2004)\citenamefont
  {{Simmerer}}, \citenamefont {{Sneden}}, \citenamefont {{Cowan}},
  \citenamefont {{Collier}}, \citenamefont {{Woolf}},\ and\ \citenamefont
  {{Lawler}}}]{Simmerer.Sneden.ea:2004}%
  \BibitemOpen
  \bibfield  {author} {\bibinfo {author} {\bibnamefont {{Simmerer}},
  \bibfnamefont {J}}, \bibinfo {author} {\bibfnamefont {C.}~\bibnamefont
  {{Sneden}}}, \bibinfo {author} {\bibfnamefont {J.~J.}\ \bibnamefont
  {{Cowan}}}, \bibinfo {author} {\bibfnamefont {J.}~\bibnamefont {{Collier}}},
  \bibinfo {author} {\bibfnamefont {V.~M.}\ \bibnamefont {{Woolf}}}, and\
  \bibinfo {author} {\bibfnamefont {J.~E.}\ \bibnamefont {{Lawler}}}} (\bibinfo
  {year} {2004}),\ \bibfield  {title} {\enquote {\bibinfo {title} {{The Rise of
  the s-Process in the Galaxy}},}\ }\href {https://doi.org/10.1086/424504}
  {\bibfield  {journal} {\bibinfo  {journal} {Astrophys. J.}\ }\textbf
  {\bibinfo {volume} {617}},\ \bibinfo {pages} {1091--1114}}\BibitemShut
  {NoStop}%
\bibitem [{\citenamefont {{Simon}}(2019)}]{Simon:2019}%
  \BibitemOpen
  \bibfield  {author} {\bibinfo {author} {\bibnamefont {{Simon}}, \bibfnamefont
  {Joshua~D}}} (\bibinfo {year} {2019}),\ \bibfield  {title} {\enquote
  {\bibinfo {title} {{The Faintest Dwarf Galaxies}},}\ }\href
  {https://doi.org/10.1146/annurev-astro-091918-104453} {\bibfield  {journal}
  {\bibinfo  {journal} {Annu. Rev. Astron. Astrophys.}\ }\textbf {\bibinfo
  {volume} {57}},\ \bibinfo {pages} {375--415}}\BibitemShut {NoStop}%
\bibitem [{\citenamefont {{Simonetti}}\ \emph {et~al.}(2019)\citenamefont
  {{Simonetti}}, \citenamefont {{Matteucci}}, \citenamefont {{Greggio}},\ and\
  \citenamefont {{Cescutti}}}]{Simonetti.Matteucci.ea:2019}%
  \BibitemOpen
  \bibfield  {author} {\bibinfo {author} {\bibnamefont {{Simonetti}},
  \bibfnamefont {Paolo}}, \bibinfo {author} {\bibfnamefont {Francesca}\
  \bibnamefont {{Matteucci}}}, \bibinfo {author} {\bibfnamefont {Laura}\
  \bibnamefont {{Greggio}}}, and\ \bibinfo {author} {\bibfnamefont {Gabriele}\
  \bibnamefont {{Cescutti}}}} (\bibinfo {year} {2019}),\ \bibfield  {title}
  {\enquote {\bibinfo {title} {{A new delay time distribution for merging
  neutron stars tested against Galactic and cosmic data}},}\ }\href
  {https://doi.org/10.1093/mnras/stz991} {\bibfield  {journal} {\bibinfo
  {journal} {Mon. Not. Roy. Astron. Soc.}\ }\textbf {\bibinfo {volume} {486}},\
  \bibinfo {pages} {2896--2909}}\BibitemShut {NoStop}%
\bibitem [{\citenamefont {{Siqueira Mello}}\ \emph {et~al.}(2013)\citenamefont
  {{Siqueira Mello}}, \citenamefont {{Spite}}, \citenamefont {{Barbuy}},
  \citenamefont {{Spite}}, \citenamefont {{Caffau}}, \citenamefont {{Hill}},
  \citenamefont {{Wanajo}}, \citenamefont {{Primas}}, \citenamefont {{Plez}},
  \citenamefont {{Cayrel}}, \citenamefont {{Andersen}}, \citenamefont
  {{Nordstr{\"o}m}}, \citenamefont {{Sneden}}, \citenamefont {{Beers}},
  \citenamefont {{Bonifacio}}, \citenamefont {{Fran{\c c}ois}},\ and\
  \citenamefont {{Molaro}}}]{siqueira13}%
  \BibitemOpen
  \bibfield  {author} {\bibinfo {author} {\bibnamefont {{Siqueira Mello}},
  \bibfnamefont {C}}, \bibinfo {author} {\bibfnamefont {M.}~\bibnamefont
  {{Spite}}}, \bibinfo {author} {\bibfnamefont {B.}~\bibnamefont {{Barbuy}}},
  \bibinfo {author} {\bibfnamefont {F.}~\bibnamefont {{Spite}}}, \bibinfo
  {author} {\bibfnamefont {E.}~\bibnamefont {{Caffau}}}, \bibinfo {author}
  {\bibfnamefont {V.}~\bibnamefont {{Hill}}}, \bibinfo {author} {\bibfnamefont
  {S.}~\bibnamefont {{Wanajo}}}, \bibinfo {author} {\bibfnamefont
  {F.}~\bibnamefont {{Primas}}}, \bibinfo {author} {\bibfnamefont
  {B.}~\bibnamefont {{Plez}}}, \bibinfo {author} {\bibfnamefont
  {R.}~\bibnamefont {{Cayrel}}}, \bibinfo {author} {\bibfnamefont
  {J.}~\bibnamefont {{Andersen}}}, \bibinfo {author} {\bibfnamefont
  {B.}~\bibnamefont {{Nordstr{\"o}m}}}, \bibinfo {author} {\bibfnamefont
  {C.}~\bibnamefont {{Sneden}}}, \bibinfo {author} {\bibfnamefont {T.~C.}\
  \bibnamefont {{Beers}}}, \bibinfo {author} {\bibfnamefont {P.}~\bibnamefont
  {{Bonifacio}}}, \bibinfo {author} {\bibfnamefont {P.}~\bibnamefont {{Fran{\c
  c}ois}}}, and\ \bibinfo {author} {\bibfnamefont {P.}~\bibnamefont
  {{Molaro}}}} (\bibinfo {year} {2013}),\ \bibfield  {title} {\enquote
  {\bibinfo {title} {{First stars. XVI. HST/STIS abundances of heavy elements
  in the uranium-rich metal-poor star CS 31082-001}},}\ }\href
  {https://doi.org/10.1051/0004-6361/201219949} {\bibfield  {journal} {\bibinfo
   {journal} {Astron. \& Astrophys.}\ }\textbf {\bibinfo {volume} {550}},\
  \bibinfo {eid} {A122}}\BibitemShut {NoStop}%
\bibitem [{\citenamefont {{Sk{\'u}lad{\'o}ttir}}\ \emph
  {et~al.}(2019)\citenamefont {{Sk{\'u}lad{\'o}ttir}}, \citenamefont
  {{Hansen}}, \citenamefont {{Salvadori}},\ and\ \citenamefont
  {{Choplin}}}]{Skuladottir.Hansen.ea:2019}%
  \BibitemOpen
  \bibfield  {author} {\bibinfo {author} {\bibnamefont {{Sk{\'u}lad{\'o}ttir}},
  \bibfnamefont {{\'A}}}, \bibinfo {author} {\bibfnamefont {C.~J.}\
  \bibnamefont {{Hansen}}}, \bibinfo {author} {\bibfnamefont {S.}~\bibnamefont
  {{Salvadori}}}, and\ \bibinfo {author} {\bibfnamefont {A.}~\bibnamefont
  {{Choplin}}}} (\bibinfo {year} {2019}),\ \bibfield  {title} {\enquote
  {\bibinfo {title} {{Neutron-capture elements in dwarf galaxies. I. Chemical
  clocks and the short timescale of the r-process}},}\ }\href
  {https://doi.org/10.1051/0004-6361/201936125} {\bibfield  {journal} {\bibinfo
   {journal} {Astron. \& Astrophys.}\ }\textbf {\bibinfo {volume} {631}},\
  \bibinfo {eid} {A171}}\BibitemShut {NoStop}%
\bibitem [{\citenamefont {{Sk{\'u}lad{\'o}ttir}}\ and\ \citenamefont
  {{Salvadori}}(2020)}]{Skuladottir.Salvadori:2020}%
  \BibitemOpen
  \bibfield  {author} {\bibinfo {author} {\bibnamefont {{Sk{\'u}lad{\'o}ttir}},
  \bibfnamefont {{\'A}}}, and\ \bibinfo {author} {\bibfnamefont
  {S.}~\bibnamefont {{Salvadori}}}} (\bibinfo {year} {2020}),\ \bibfield
  {title} {\enquote {\bibinfo {title} {{Evidence for {\ensuremath{\gtrsim}}4
  Gyr timescales of neutron star mergers from Galactic archaeology}},}\ }\href
  {https://doi.org/10.1051/0004-6361/201937293} {\bibfield  {journal} {\bibinfo
   {journal} {Astron. \& Astrophys.}\ }\textbf {\bibinfo {volume} {634}},\
  \bibinfo {eid} {L2}}\BibitemShut {NoStop}%
\bibitem [{\citenamefont {{Smartt}}\ \emph {et~al.}(2017)\citenamefont
  {{Smartt}} \emph {et~al.}}]{smartt17}%
  \BibitemOpen
  \bibfield  {author} {\bibinfo {author} {\bibnamefont {{Smartt}},
  \bibfnamefont {S~J}},  \emph {et~al.}} (\bibinfo {year} {2017}),\ \bibfield
  {title} {\enquote {\bibinfo {title} {{A kilonova as the electromagnetic
  counterpart to a gravitational-wave source}},}\ }\href
  {https://doi.org/10.1038/nature24303} {\bibfield  {journal} {\bibinfo
  {journal} {Nature}\ }\textbf {\bibinfo {volume} {551}},\ \bibinfo {pages}
  {75--79}}\BibitemShut {NoStop}%
\bibitem [{\citenamefont {{Sneden}}\ \emph {et~al.}(2008)\citenamefont
  {{Sneden}}, \citenamefont {{Cowan}},\ and\ \citenamefont
  {{Gallino}}}]{sneden08}%
  \BibitemOpen
  \bibfield  {author} {\bibinfo {author} {\bibnamefont {{Sneden}},
  \bibfnamefont {C}}, \bibinfo {author} {\bibfnamefont {J.~J.}\ \bibnamefont
  {{Cowan}}}, and\ \bibinfo {author} {\bibfnamefont {R.}~\bibnamefont
  {{Gallino}}}} (\bibinfo {year} {2008}),\ \bibfield  {title} {\enquote
  {\bibinfo {title} {{Neutron-Capture Elements in the Early Galaxy}},}\ }\href
  {https://doi.org/10.1146/annurev.astro.46.060407.145207} {\bibfield
  {journal} {\bibinfo  {journal} {Annu. Rev. Astron. Astrophys.}\ }\textbf
  {\bibinfo {volume} {46}},\ \bibinfo {pages} {241--288}}\BibitemShut {NoStop}%
\bibitem [{\citenamefont {{Sneden}}\ \emph {et~al.}(2003)\citenamefont
  {{Sneden}}, \citenamefont {{Cowan}}, \citenamefont {{Lawler}}, \citenamefont
  {{Ivans}}, \citenamefont {{Burles}}, \citenamefont {{Beers}}, \citenamefont
  {{Primas}}, \citenamefont {{Hill}}, \citenamefont {{Truran}}, \citenamefont
  {{Fuller}}, \citenamefont {{Pfeiffer}},\ and\ \citenamefont
  {{Kratz}}}]{sneden03}%
  \BibitemOpen
  \bibfield  {author} {\bibinfo {author} {\bibnamefont {{Sneden}},
  \bibfnamefont {C}}, \bibinfo {author} {\bibfnamefont {J.~J.}\ \bibnamefont
  {{Cowan}}}, \bibinfo {author} {\bibfnamefont {J.~E.}\ \bibnamefont
  {{Lawler}}}, \bibinfo {author} {\bibfnamefont {I.~I.}\ \bibnamefont
  {{Ivans}}}, \bibinfo {author} {\bibfnamefont {S.}~\bibnamefont {{Burles}}},
  \bibinfo {author} {\bibfnamefont {T.~C.}\ \bibnamefont {{Beers}}}, \bibinfo
  {author} {\bibfnamefont {F.}~\bibnamefont {{Primas}}}, \bibinfo {author}
  {\bibfnamefont {V.}~\bibnamefont {{Hill}}}, \bibinfo {author} {\bibfnamefont
  {J.~W.}\ \bibnamefont {{Truran}}}, \bibinfo {author} {\bibfnamefont {G.~M.}\
  \bibnamefont {{Fuller}}}, \bibinfo {author} {\bibfnamefont {B.}~\bibnamefont
  {{Pfeiffer}}}, and\ \bibinfo {author} {\bibfnamefont {K.-L.}\ \bibnamefont
  {{Kratz}}}} (\bibinfo {year} {2003}),\ \bibfield  {title} {\enquote {\bibinfo
  {title} {{The Extremely Metal-poor, Neutron Capture-rich Star CS 22892-052: A
  Comprehensive Abundance Analysis}},}\ }\href {https://doi.org/10.1086/375491}
  {\bibfield  {journal} {\bibinfo  {journal} {Astrophys. J.}\ }\textbf
  {\bibinfo {volume} {591}},\ \bibinfo {pages} {936--953}}\BibitemShut
  {NoStop}%
\bibitem [{\citenamefont {{Sneden}}\ \emph {et~al.}(2009)\citenamefont
  {{Sneden}}, \citenamefont {{Lawler}}, \citenamefont {{Cowan}}, \citenamefont
  {{Ivans}},\ and\ \citenamefont {{Den Hartog}}}]{sneden09}%
  \BibitemOpen
  \bibfield  {author} {\bibinfo {author} {\bibnamefont {{Sneden}},
  \bibfnamefont {C}}, \bibinfo {author} {\bibfnamefont {J.~E.}\ \bibnamefont
  {{Lawler}}}, \bibinfo {author} {\bibfnamefont {J.~J.}\ \bibnamefont
  {{Cowan}}}, \bibinfo {author} {\bibfnamefont {I.~I.}\ \bibnamefont
  {{Ivans}}}, and\ \bibinfo {author} {\bibfnamefont {E.~A.}\ \bibnamefont {{Den
  Hartog}}}} (\bibinfo {year} {2009}),\ \bibfield  {title} {\enquote {\bibinfo
  {title} {{New Rare Earth Element Abundance Distributions for the Sun and Five
  r-Process-Rich Very Metal-Poor Stars}},}\ }\href
  {https://doi.org/10.1088/0067-0049/182/1/80} {\bibfield  {journal} {\bibinfo
  {journal} {Astrophys. J. Suppl.}\ }\textbf {\bibinfo {volume} {182}},\
  \bibinfo {pages} {80--96}}\BibitemShut {NoStop}%
\bibitem [{\citenamefont {{Sneden}}\ \emph {et~al.}(1996)\citenamefont
  {{Sneden}}, \citenamefont {{McWilliam}}, \citenamefont {{Preston}},
  \citenamefont {{Cowan}}, \citenamefont {{Burris}},\ and\ \citenamefont
  {{Armosky}}}]{sneden96}%
  \BibitemOpen
  \bibfield  {author} {\bibinfo {author} {\bibnamefont {{Sneden}},
  \bibfnamefont {C}}, \bibinfo {author} {\bibfnamefont {A.}~\bibnamefont
  {{McWilliam}}}, \bibinfo {author} {\bibfnamefont {G.~W.}\ \bibnamefont
  {{Preston}}}, \bibinfo {author} {\bibfnamefont {J.~J.}\ \bibnamefont
  {{Cowan}}}, \bibinfo {author} {\bibfnamefont {D.~L.}\ \bibnamefont
  {{Burris}}}, and\ \bibinfo {author} {\bibfnamefont {B.~J.}\ \bibnamefont
  {{Armosky}}}} (\bibinfo {year} {1996}),\ \bibfield  {title} {\enquote
  {\bibinfo {title} {{The Ultra--Metal-poor, Neutron-Capture--rich Giant Star
  CS 22892-052}},}\ }\href {https://doi.org/10.1086/177656} {\bibfield
  {journal} {\bibinfo  {journal} {Astrophys. J.}\ }\textbf {\bibinfo {volume}
  {467}},\ \bibinfo {pages} {819}}\BibitemShut {NoStop}%
\bibitem [{\citenamefont {{Sneden}}\ and\ \citenamefont
  {{Parthasarathy}}(1983)}]{sneden83}%
  \BibitemOpen
  \bibfield  {author} {\bibinfo {author} {\bibnamefont {{Sneden}},
  \bibfnamefont {C}}, and\ \bibinfo {author} {\bibfnamefont {M.}~\bibnamefont
  {{Parthasarathy}}}} (\bibinfo {year} {1983}),\ \bibfield  {title} {\enquote
  {\bibinfo {title} {{The r- and s-process nuclei in the early history of the
  galaxy - HD 122563}},}\ }\href {https://doi.org/10.1086/160913} {\bibfield
  {journal} {\bibinfo  {journal} {Astrophys. J.}\ }\textbf {\bibinfo {volume}
  {267}},\ \bibinfo {pages} {757--778}}\BibitemShut {NoStop}%
\bibitem [{\citenamefont {{Sneden}}\ \emph {et~al.}(1994)\citenamefont
  {{Sneden}}, \citenamefont {{Preston}}, \citenamefont {{McWilliam}},\ and\
  \citenamefont {{Searle}}}]{sneden94}%
  \BibitemOpen
  \bibfield  {author} {\bibinfo {author} {\bibnamefont {{Sneden}},
  \bibfnamefont {C}}, \bibinfo {author} {\bibfnamefont {G.~W.}\ \bibnamefont
  {{Preston}}}, \bibinfo {author} {\bibfnamefont {A.}~\bibnamefont
  {{McWilliam}}}, and\ \bibinfo {author} {\bibfnamefont {L.}~\bibnamefont
  {{Searle}}}} (\bibinfo {year} {1994}),\ \bibfield  {title} {\enquote
  {\bibinfo {title} {{Ultrametal-poor halo stars: The remarkable spectrum of CS
  22892-052}},}\ }\href {https://doi.org/10.1086/187464} {\bibfield  {journal}
  {\bibinfo  {journal} {Astrophys. J. Lett.}\ }\textbf {\bibinfo {volume}
  {431}},\ \bibinfo {pages} {L27--L30}}\BibitemShut {NoStop}%
\bibitem [{\citenamefont {{Sneden}}\ and\ \citenamefont
  {{Cowan}}(2003)}]{Sneden.Cowan:2003}%
  \BibitemOpen
  \bibfield  {author} {\bibinfo {author} {\bibnamefont {{Sneden}},
  \bibfnamefont {Christopher}}, and\ \bibinfo {author} {\bibfnamefont
  {John~J.}\ \bibnamefont {{Cowan}}}} (\bibinfo {year} {2003}),\ \bibfield
  {title} {\enquote {\bibinfo {title} {{Genesis of the Heaviest Elements in the
  Milky Way Galaxy}},}\ }\href {https://doi.org/10.1126/science.1077506}
  {\bibfield  {journal} {\bibinfo  {journal} {Science}\ }\textbf {\bibinfo
  {volume} {299}},\ \bibinfo {pages} {70--75}}\BibitemShut {NoStop}%
\bibitem [{\citenamefont {{Soares-Santos}}\ \emph {et~al.}(2017)\citenamefont
  {{Soares-Santos}} \emph {et~al.}}]{soaressantos17}%
  \BibitemOpen
  \bibfield  {author} {\bibinfo {author} {\bibnamefont {{Soares-Santos}},
  \bibfnamefont {M}},  \emph {et~al.}} (\bibinfo {year} {2017}),\ \bibfield
  {title} {\enquote {\bibinfo {title} {{The Electromagnetic Counterpart of the
  Binary Neutron Star Merger LIGO/Virgo GW170817. I. Discovery of the Optical
  Counterpart Using the Dark Energy Camera}},}\ }\href
  {https://doi.org/10.3847/2041-8213/aa9059} {\bibfield  {journal} {\bibinfo
  {journal} {Astrophys. J. Lett.}\ }\textbf {\bibinfo {volume} {848}},\
  \bibinfo {eid} {L16}}\BibitemShut {NoStop}%
\bibitem [{\citenamefont {{S{\"o}derstr{\"o}m}}\ \emph
  {et~al.}(2013)\citenamefont {{S{\"o}derstr{\"o}m}} \emph
  {et~al.}}]{Soderstrom2013}%
  \BibitemOpen
  \bibfield  {author} {\bibinfo {author} {\bibnamefont {{S{\"o}derstr{\"o}m}},
  \bibfnamefont {P-A}},  \emph {et~al.}} (\bibinfo {year} {2013}),\ \bibfield
  {title} {\enquote {\bibinfo {title} {{Installation and commissioning of
  EURICA - Euroball-RIKEN Cluster Array}},}\ }\href
  {https://doi.org/10.1016/j.nimb.2013.03.018} {\bibfield  {journal} {\bibinfo
  {journal} {Nucl. Instruments Methods Phys. Res. Sect. B Beam Interact. with
  Mater. Atoms}\ }\textbf {\bibinfo {volume} {317}},\ \bibinfo {pages}
  {649--652}}\BibitemShut {NoStop}%
\bibitem [{\citenamefont {{S{\o}rensen}}\ \emph {et~al.}(2017)\citenamefont
  {{S{\o}rensen}}, \citenamefont {{Svensmark}},\ and\ \citenamefont {{Gr{\aa}e
  J{\o}rgensen}}}]{Soerensen.Svensmark.Grae:2017}%
  \BibitemOpen
  \bibfield  {author} {\bibinfo {author} {\bibnamefont {{S{\o}rensen}},
  \bibfnamefont {M}}, \bibinfo {author} {\bibfnamefont {H.}~\bibnamefont
  {{Svensmark}}}, and\ \bibinfo {author} {\bibfnamefont {U.}~\bibnamefont
  {{Gr{\aa}e J{\o}rgensen}}}} (\bibinfo {year} {2017}),\ \bibfield  {title}
  {\enquote {\bibinfo {title} {{Near-Earth supernova activity during the past
  35 Myr}},}\ }\href@noop {} {\bibfield  {journal} {\bibinfo  {journal} {ArXiv
  e-prints}\ }}\Eprint {https://arxiv.org/abs/1708.08248} {arXiv:1708.08248
  [astro-ph.SR]} \BibitemShut {NoStop}%
\bibitem [{\citenamefont {{Sorokin}}\ and\ \citenamefont
  {{Lankard}}(1966)}]{sorokin66}%
  \BibitemOpen
  \bibfield  {author} {\bibinfo {author} {\bibnamefont {{Sorokin}},
  \bibfnamefont {P~P}}, and\ \bibinfo {author} {\bibfnamefont {J.~R.}\
  \bibnamefont {{Lankard}}}} (\bibinfo {year} {1966}),\ \bibfield  {title}
  {\enquote {\bibinfo {title} {{Stimulated Emission Observed from an Organic
  Dye Chloro-aluminum Phthalocynine}},}\ }\href
  {https://doi.org/10.1147/rd.102.0162} {\bibfield  {journal} {\bibinfo
  {journal} {IBM J. Res. Dev.}\ }\textbf {\bibinfo {volume} {10}},\ \bibinfo
  {pages} {162--163}}\BibitemShut {NoStop}%
\bibitem [{\citenamefont {{Spitoni}}\ \emph {et~al.}(2009)\citenamefont
  {{Spitoni}}, \citenamefont {{Matteucci}}, \citenamefont {{Recchi}},
  \citenamefont {{Cescutti}},\ and\ \citenamefont {{Pipino}}}]{Spitoni:2009}%
  \BibitemOpen
  \bibfield  {author} {\bibinfo {author} {\bibnamefont {{Spitoni}},
  \bibfnamefont {E}}, \bibinfo {author} {\bibfnamefont {F.}~\bibnamefont
  {{Matteucci}}}, \bibinfo {author} {\bibfnamefont {S.}~\bibnamefont
  {{Recchi}}}, \bibinfo {author} {\bibfnamefont {G.}~\bibnamefont
  {{Cescutti}}}, and\ \bibinfo {author} {\bibfnamefont {A.}~\bibnamefont
  {{Pipino}}}} (\bibinfo {year} {2009}),\ \bibfield  {title} {\enquote
  {\bibinfo {title} {{Effects of galactic fountains and delayed mixing in the
  chemical evolution of the Milky Way}},}\ }\href
  {https://doi.org/10.1051/0004-6361/200911768} {\bibfield  {journal} {\bibinfo
   {journal} {Astron. \& Astrophys.}\ }\textbf {\bibinfo {volume} {504}},\
  \bibinfo {pages} {87--96}}\BibitemShut {NoStop}%
\bibitem [{\citenamefont {{Spyrou}}\ \emph {et~al.}(2014)\citenamefont
  {{Spyrou}} \emph {et~al.}}]{Spyrou2014}%
  \BibitemOpen
  \bibfield  {author} {\bibinfo {author} {\bibnamefont {{Spyrou}},
  \bibfnamefont {A}},  \emph {et~al.}} (\bibinfo {year} {2014}),\ \bibfield
  {title} {\enquote {\bibinfo {title} {{Novel technique for Constraining
  r-Process $(n,\gamma)$ Reaction Rates}},}\ }\href
  {https://doi.org/10.1103/PhysRevLett.113.232502} {\bibfield  {journal}
  {\bibinfo  {journal} {Phys. Rev. Lett.}\ }\textbf {\bibinfo {volume} {113}},\
  \bibinfo {eid} {232502}}\BibitemShut {NoStop}%
\bibitem [{\citenamefont {{Spyrou}}\ \emph {et~al.}(2017)\citenamefont
  {{Spyrou}} \emph {et~al.}}]{Spyrou2017}%
  \BibitemOpen
  \bibfield  {author} {\bibinfo {author} {\bibnamefont {{Spyrou}},
  \bibfnamefont {A}},  \emph {et~al.}} (\bibinfo {year} {2017}),\ \bibfield
  {title} {\enquote {\bibinfo {title} {{Neutron-capture rates for explosive
  nucleosynthesis: the case of $^{68}$Ni$(n,\gamma)^{69}$Ni}},}\ }\href
  {https://doi.org/10.1088/1361-6471/aa5ae7} {\bibfield  {journal} {\bibinfo
  {journal} {J. Phys. G: Nucl. Part. Phys.}\ }\textbf {\bibinfo {volume}
  {44}},\ \bibinfo {eid} {044002}}\BibitemShut {NoStop}%
\bibitem [{\citenamefont {{Steinberg}}\ and\ \citenamefont
  {{Wilkins}}(1978)}]{Steinberg.Wilkins:1978}%
  \BibitemOpen
  \bibfield  {author} {\bibinfo {author} {\bibnamefont {{Steinberg}},
  \bibfnamefont {E~P}}, and\ \bibinfo {author} {\bibfnamefont {B.~D.}\
  \bibnamefont {{Wilkins}}}} (\bibinfo {year} {1978}),\ \bibfield  {title}
  {\enquote {\bibinfo {title} {{Implications of fission mass distributions for
  the astrophysical r-process}},}\ }\href {https://doi.org/10.1086/156334}
  {\bibfield  {journal} {\bibinfo  {journal} {Astrophys. J.}\ }\textbf
  {\bibinfo {volume} {223}},\ \bibinfo {pages} {1000--1014}}\BibitemShut
  {NoStop}%
\bibitem [{\citenamefont {{Suda}}\ \emph {et~al.}(2008)\citenamefont {{Suda}},
  \citenamefont {{Katsuta}}, \citenamefont {{Yamada}}, \citenamefont {{Suwa}},
  \citenamefont {{Ishizuka}}, \citenamefont {{Komiya}}, \citenamefont
  {{Sorai}}, \citenamefont {{Aikawa}},\ and\ \citenamefont
  {{Fujimoto}}}]{Sagadatabase}%
  \BibitemOpen
  \bibfield  {author} {\bibinfo {author} {\bibnamefont {{Suda}}, \bibfnamefont
  {Takuma}}, \bibinfo {author} {\bibfnamefont {Yutaka}\ \bibnamefont
  {{Katsuta}}}, \bibinfo {author} {\bibfnamefont {Shimako}\ \bibnamefont
  {{Yamada}}}, \bibinfo {author} {\bibfnamefont {Tamon}\ \bibnamefont
  {{Suwa}}}, \bibinfo {author} {\bibfnamefont {Chikako}\ \bibnamefont
  {{Ishizuka}}}, \bibinfo {author} {\bibfnamefont {Yutaka}\ \bibnamefont
  {{Komiya}}}, \bibinfo {author} {\bibfnamefont {Kazuo}\ \bibnamefont
  {{Sorai}}}, \bibinfo {author} {\bibfnamefont {Masayuki}\ \bibnamefont
  {{Aikawa}}}, and\ \bibinfo {author} {\bibfnamefont {Masayuki~Y.}\
  \bibnamefont {{Fujimoto}}}} (\bibinfo {year} {2008}),\ \bibfield  {title}
  {\enquote {\bibinfo {title} {{Stellar Abundances for the Galactic Archeology
  (SAGA) Database --- Compilation of the Characteristics of Known Extremely
  Metal-Poor Stars}},}\ }\href {https://doi.org/10.1093/pasj/60.5.1159}
  {\bibfield  {journal} {\bibinfo  {journal} {Publ. Astron. Soc. Japan}\
  }\textbf {\bibinfo {volume} {60}},\ \bibinfo {pages} {1159}}\BibitemShut
  {NoStop}%
\bibitem [{\citenamefont {{Sukhbold}}\ \emph {et~al.}(2016)\citenamefont
  {{Sukhbold}}, \citenamefont {{Ertl}}, \citenamefont {{Woosley}},
  \citenamefont {{Brown}},\ and\ \citenamefont
  {{Janka}}}]{Sukhbold.Ertl.ea:2016}%
  \BibitemOpen
  \bibfield  {author} {\bibinfo {author} {\bibnamefont {{Sukhbold}},
  \bibfnamefont {T}}, \bibinfo {author} {\bibfnamefont {T.}~\bibnamefont
  {{Ertl}}}, \bibinfo {author} {\bibfnamefont {S.~E.}\ \bibnamefont
  {{Woosley}}}, \bibinfo {author} {\bibfnamefont {J.~M.}\ \bibnamefont
  {{Brown}}}, and\ \bibinfo {author} {\bibfnamefont {H.-T.}\ \bibnamefont
  {{Janka}}}} (\bibinfo {year} {2016}),\ \bibfield  {title} {\enquote {\bibinfo
  {title} {{Core-collapse Supernovae from 9 to 120 Solar Masses Based on
  Neutrino-powered Explosions}},}\ }\href
  {https://doi.org/10.3847/0004-637X/821/1/38} {\bibfield  {journal} {\bibinfo
  {journal} {Astrophys. J.}\ }\textbf {\bibinfo {volume} {821}},\ \bibinfo
  {eid} {38}}\BibitemShut {NoStop}%
\bibitem [{\citenamefont {{Sumiyoshi}}\ \emph {et~al.}(2001)\citenamefont
  {{Sumiyoshi}}, \citenamefont {{Terasawa}}, \citenamefont {{Mathews}},
  \citenamefont {{Kajino}}, \citenamefont {{Yamada}},\ and\ \citenamefont
  {{Suzuki}}}]{sumiyoshi01}%
  \BibitemOpen
  \bibfield  {author} {\bibinfo {author} {\bibnamefont {{Sumiyoshi}},
  \bibfnamefont {K}}, \bibinfo {author} {\bibfnamefont {M.}~\bibnamefont
  {{Terasawa}}}, \bibinfo {author} {\bibfnamefont {G.~J.}\ \bibnamefont
  {{Mathews}}}, \bibinfo {author} {\bibfnamefont {T.}~\bibnamefont {{Kajino}}},
  \bibinfo {author} {\bibfnamefont {S.}~\bibnamefont {{Yamada}}}, and\ \bibinfo
  {author} {\bibfnamefont {H.}~\bibnamefont {{Suzuki}}}} (\bibinfo {year}
  {2001}),\ \bibfield  {title} {\enquote {\bibinfo {title} {{r-Process in
  Prompt Supernova Explosions Revisited}},}\ }\href
  {https://doi.org/10.1086/323524} {\bibfield  {journal} {\bibinfo  {journal}
  {Astrophys. J.}\ }\textbf {\bibinfo {volume} {562}},\ \bibinfo {pages}
  {880--886}}\BibitemShut {NoStop}%
\bibitem [{\citenamefont {{Sun}}\ \emph {et~al.}(2008)\citenamefont {{Sun}}
  \emph {et~al.}}]{Sun2008}%
  \BibitemOpen
  \bibfield  {author} {\bibinfo {author} {\bibnamefont {{Sun}}, \bibfnamefont
  {B}},  \emph {et~al.}} (\bibinfo {year} {2008}),\ \bibfield  {title}
  {\enquote {\bibinfo {title} {{Nuclear structure studies of short-lived
  neutron-rich nuclei with the novel large-scale isochronous mass spectrometry
  at the FRS-ESR facility}},}\ }\href
  {https://doi.org/10.1016/j.nuclphysa.2008.08.013} {\bibfield  {journal}
  {\bibinfo  {journal} {Nucl. Phys. A}\ }\textbf {\bibinfo {volume} {812}},\
  \bibinfo {pages} {1--12}}\BibitemShut {NoStop}%
\bibitem [{\citenamefont {{Sun}}\ and\ \citenamefont
  {{Meng}}(2008)}]{Sun.Meng:2008}%
  \BibitemOpen
  \bibfield  {author} {\bibinfo {author} {\bibnamefont {{Sun}}, \bibfnamefont
  {B-H}}, and\ \bibinfo {author} {\bibfnamefont {J.}~\bibnamefont {{Meng}}}}
  (\bibinfo {year} {2008}),\ \bibfield  {title} {\enquote {\bibinfo {title}
  {{Challenge on the Astrophysical R-Process Calculation with Nuclear Mass
  Models}},}\ }\href {https://doi.org/10.1088/0256-307X/25/7/027} {\bibfield
  {journal} {\bibinfo  {journal} {Chinese Phys. Lett.}\ }\textbf {\bibinfo
  {volume} {25}},\ \bibinfo {pages} {2429--2431}}\BibitemShut {NoStop}%
\bibitem [{\citenamefont {{Surman}}\ \emph {et~al.}(2014)\citenamefont
  {{Surman}}, \citenamefont {{Caballero}}, \citenamefont {{McLaughlin}},
  \citenamefont {{Just}},\ and\ \citenamefont {{Janka}}}]{surman14}%
  \BibitemOpen
  \bibfield  {author} {\bibinfo {author} {\bibnamefont {{Surman}},
  \bibfnamefont {R}}, \bibinfo {author} {\bibfnamefont {O.~L.}\ \bibnamefont
  {{Caballero}}}, \bibinfo {author} {\bibfnamefont {G.~C.}\ \bibnamefont
  {{McLaughlin}}}, \bibinfo {author} {\bibfnamefont {O.}~\bibnamefont
  {{Just}}}, and\ \bibinfo {author} {\bibfnamefont {H.-T.}\ \bibnamefont
  {{Janka}}}} (\bibinfo {year} {2014}),\ \bibfield  {title} {\enquote {\bibinfo
  {title} {{Production of $^{56}$Ni in black hole-neutron star merger accretion
  disc outflows}},}\ }\href {https://doi.org/10.1088/0954-3899/41/4/044006}
  {\bibfield  {journal} {\bibinfo  {journal} {J. Phys. G: Nucl. Part. Phys.}\
  }\textbf {\bibinfo {volume} {41}},\ \bibinfo {eid} {044006}}\BibitemShut
  {NoStop}%
\bibitem [{\citenamefont {Surman}\ \emph {et~al.}(1997)\citenamefont {Surman},
  \citenamefont {Engel}, \citenamefont {Bennett},\ and\ \citenamefont
  {Meyer}}]{Surman.Engel.ea:1997}%
  \BibitemOpen
  \bibfield  {author} {\bibinfo {author} {\bibnamefont {Surman}, \bibfnamefont
  {R}}, \bibinfo {author} {\bibfnamefont {J.}~\bibnamefont {Engel}}, \bibinfo
  {author} {\bibfnamefont {J.~R.}\ \bibnamefont {Bennett}}, and\ \bibinfo
  {author} {\bibfnamefont {B.~S.}\ \bibnamefont {Meyer}}} (\bibinfo {year}
  {1997}),\ \bibfield  {title} {\enquote {\bibinfo {title} {Source of the
  rare-earth element peak in r-process nucleosynthesis},}\ }\href
  {https://doi.org/10.1103/PhysRevLett.79.1809} {\bibfield  {journal} {\bibinfo
   {journal} {Phys. Rev. Lett.}\ }\textbf {\bibinfo {volume} {79}},\ \bibinfo
  {pages} {1809--1812}}\BibitemShut {NoStop}%
\bibitem [{\citenamefont {{Surman}}\ \emph {et~al.}(2006)\citenamefont
  {{Surman}}, \citenamefont {{McLaughlin}},\ and\ \citenamefont
  {{Hix}}}]{surman06}%
  \BibitemOpen
  \bibfield  {author} {\bibinfo {author} {\bibnamefont {{Surman}},
  \bibfnamefont {R}}, \bibinfo {author} {\bibfnamefont {G.~C.}\ \bibnamefont
  {{McLaughlin}}}, and\ \bibinfo {author} {\bibfnamefont {W.~R.}\ \bibnamefont
  {{Hix}}}} (\bibinfo {year} {2006}),\ \bibfield  {title} {\enquote {\bibinfo
  {title} {{Nucleosynthesis in the Outflow from Gamma-Ray Burst Accretion
  Disks}},}\ }\href {https://doi.org/10.1086/501116} {\bibfield  {journal}
  {\bibinfo  {journal} {Astrophys. J.}\ }\textbf {\bibinfo {volume} {643}},\
  \bibinfo {pages} {1057--1064}}\BibitemShut {NoStop}%
\bibitem [{\citenamefont {{Surman}}\ \emph {et~al.}(2008)\citenamefont
  {{Surman}}, \citenamefont {{McLaughlin}}, \citenamefont {{Ruffert}},
  \citenamefont {{Janka}},\ and\ \citenamefont {{Hix}}}]{surman08}%
  \BibitemOpen
  \bibfield  {author} {\bibinfo {author} {\bibnamefont {{Surman}},
  \bibfnamefont {R}}, \bibinfo {author} {\bibfnamefont {G.~C.}\ \bibnamefont
  {{McLaughlin}}}, \bibinfo {author} {\bibfnamefont {M.}~\bibnamefont
  {{Ruffert}}}, \bibinfo {author} {\bibfnamefont {H.-T.}\ \bibnamefont
  {{Janka}}}, and\ \bibinfo {author} {\bibfnamefont {W.~R.}\ \bibnamefont
  {{Hix}}}} (\bibinfo {year} {2008}),\ \bibfield  {title} {\enquote {\bibinfo
  {title} {{r-Process Nucleosynthesis in Hot Accretion Disk Flows from Black
  Hole-Neutron Star Mergers}},}\ }\href {https://doi.org/10.1086/589507}
  {\bibfield  {journal} {\bibinfo  {journal} {Astrophys. J.}\ }\textbf
  {\bibinfo {volume} {679}},\ \bibinfo {eid} {L117}}\BibitemShut {NoStop}%
\bibitem [{\citenamefont {{Suzuki}}\ and\ \citenamefont
  {{Kajino}}(2013)}]{Suzuki.Kajino:2013}%
  \BibitemOpen
  \bibfield  {author} {\bibinfo {author} {\bibnamefont {{Suzuki}},
  \bibfnamefont {T}}, and\ \bibinfo {author} {\bibfnamefont {T.}~\bibnamefont
  {{Kajino}}}} (\bibinfo {year} {2013}),\ \bibfield  {title} {\enquote
  {\bibinfo {title} {{Element synthesis in the supernova environment and
  neutrino oscillations}},}\ }\href
  {https://doi.org/10.1088/0954-3899/40/8/083101} {\bibfield  {journal}
  {\bibinfo  {journal} {J. Phys. G: Nucl. Part. Phys.}\ }\textbf {\bibinfo
  {volume} {40}},\ \bibinfo {eid} {083101}}\BibitemShut {NoStop}%
\bibitem [{\citenamefont {Suzuki}\ \emph {et~al.}(2012)\citenamefont {Suzuki},
  \citenamefont {Yoshida}, \citenamefont {Kajino},\ and\ \citenamefont
  {Otsuka}}]{Suzuki.Yoshida.ea:2012}%
  \BibitemOpen
  \bibfield  {author} {\bibinfo {author} {\bibnamefont {Suzuki}, \bibfnamefont
  {Toshio}}, \bibinfo {author} {\bibfnamefont {Takashi}\ \bibnamefont
  {Yoshida}}, \bibinfo {author} {\bibfnamefont {Toshitaka}\ \bibnamefont
  {Kajino}}, and\ \bibinfo {author} {\bibfnamefont {Takaharu}\ \bibnamefont
  {Otsuka}}} (\bibinfo {year} {2012}),\ \bibfield  {title} {\enquote {\bibinfo
  {title} {{$\beta$ decays of isotones with neutron magic number of $N=126$ and
  r-process nucleosynthesis}},}\ }\href
  {https://doi.org/10.1103/PhysRevC.85.015802} {\bibfield  {journal} {\bibinfo
  {journal} {Phys. Rev. C}\ }\textbf {\bibinfo {volume} {85}},\ \bibinfo
  {pages} {015802}}\BibitemShut {NoStop}%
\bibitem [{\citenamefont {{Svanberg}}\ \emph {et~al.}(1994)\citenamefont
  {{Svanberg}}, \citenamefont {{Larsson}}, \citenamefont {{Persson}},\ and\
  \citenamefont {{Wahlstr{\"o}m}}}]{svanberg94}%
  \BibitemOpen
  \bibfield  {author} {\bibinfo {author} {\bibnamefont {{Svanberg}},
  \bibfnamefont {S}}, \bibinfo {author} {\bibfnamefont {J.}~\bibnamefont
  {{Larsson}}}, \bibinfo {author} {\bibfnamefont {A.}~\bibnamefont
  {{Persson}}}, and\ \bibinfo {author} {\bibfnamefont {C.-G.}\ \bibnamefont
  {{Wahlstr{\"o}m}}}} (\bibinfo {year} {1994}),\ \bibfield  {title} {\enquote
  {\bibinfo {title} {{Lund high-power laser facility - systems and first
  results}},}\ }\href {https://doi.org/10.1088/0031-8949/49/2/009} {\bibfield
  {journal} {\bibinfo  {journal} {Phys. Scr.}\ }\textbf {\bibinfo {volume}
  {49}},\ \bibinfo {pages} {187--197}}\BibitemShut {NoStop}%
\bibitem [{\citenamefont {{Symbalisty}}\ and\ \citenamefont
  {{Schramm}}(1982)}]{symbalisty82}%
  \BibitemOpen
  \bibfield  {author} {\bibinfo {author} {\bibnamefont {{Symbalisty}},
  \bibfnamefont {E}}, and\ \bibinfo {author} {\bibfnamefont {D.~N.}\
  \bibnamefont {{Schramm}}}} (\bibinfo {year} {1982}),\ \bibfield  {title}
  {\enquote {\bibinfo {title} {{Neutron star collisions and the r-process}},}\
  }\href@noop {} {\bibfield  {journal} {\bibinfo  {journal} {Astrophys.~Lett.}\
  }\textbf {\bibinfo {volume} {22}},\ \bibinfo {pages} {143--145}}\BibitemShut
  {NoStop}%
\bibitem [{\citenamefont {{Symbalisty}}\ \emph {et~al.}(1985)\citenamefont
  {{Symbalisty}}, \citenamefont {{Schramm}},\ and\ \citenamefont
  {{Wilson}}}]{symbalisty85}%
  \BibitemOpen
  \bibfield  {author} {\bibinfo {author} {\bibnamefont {{Symbalisty}},
  \bibfnamefont {E~M~D}}, \bibinfo {author} {\bibfnamefont {D.~N.}\
  \bibnamefont {{Schramm}}}, and\ \bibinfo {author} {\bibfnamefont {J.~R.}\
  \bibnamefont {{Wilson}}}} (\bibinfo {year} {1985}),\ \bibfield  {title}
  {\enquote {\bibinfo {title} {{An expanding vortex site for the r-process in
  rotating stellar collapse}},}\ }\href {https://doi.org/10.1086/184448}
  {\bibfield  {journal} {\bibinfo  {journal} {Astrophys. J.}\ }\textbf
  {\bibinfo {volume} {291}},\ \bibinfo {pages} {L11--L14}}\BibitemShut
  {NoStop}%
\bibitem [{\citenamefont {Sz{\"u}cs}\ \emph {et~al.}(2020)\citenamefont
  {Sz{\"u}cs}, \citenamefont {Mohr}, \citenamefont {Gy{\"u}rky}, \citenamefont
  {Hal{\'a}sz}, \citenamefont {Husz{\"a}nk}, \citenamefont {Kiss},
  \citenamefont {Szegedi}, \citenamefont {T{\"o}r{\"o}k},\ and\ \citenamefont
  {F{\"u}l{\"o}p}}]{Szuecs.Mohr.ea:2020}%
  \BibitemOpen
  \bibfield  {author} {\bibinfo {author} {\bibnamefont {Sz{\"u}cs},
  \bibfnamefont {T}}, \bibinfo {author} {\bibfnamefont {P}~\bibnamefont
  {Mohr}}, \bibinfo {author} {\bibfnamefont {Gy.}\ \bibnamefont {Gy{\"u}rky}},
  \bibinfo {author} {\bibfnamefont {Z.}~\bibnamefont {Hal{\'a}sz}}, \bibinfo
  {author} {\bibfnamefont {R}~\bibnamefont {Husz{\"a}nk}}, \bibinfo {author}
  {\bibfnamefont {G.~G.}\ \bibnamefont {Kiss}}, \bibinfo {author}
  {\bibfnamefont {T.~N.}\ \bibnamefont {Szegedi}}, \bibinfo {author}
  {\bibfnamefont {{Zs.}}\ \bibnamefont {T{\"o}r{\"o}k}}, and\ \bibinfo {author}
  {\bibfnamefont {{Zs.}}\ \bibnamefont {F{\"u}l{\"o}p}}} (\bibinfo {year}
  {2020}),\ \bibfield  {title} {\enquote {\bibinfo {title} {Activation
  measurement of $\alpha$-induced cross sections for $^{197}$au: analysis in
  the statistical model and beyond},}\ }in\ \href
  {https://doi.org/10.1088/1742-6596/1668/1/012042} {\emph {\bibinfo
  {booktitle} {Journal of Physics Conference Series}}},\ \bibinfo {series}
  {Journal of Physics Conference Series}, Vol.\ \bibinfo {volume} {1668},\ p.\
  \bibinfo {pages} {012042}\BibitemShut {NoStop}%
\bibitem [{\citenamefont {Tachibana}\ \emph {et~al.}(1990)\citenamefont
  {Tachibana}, \citenamefont {Yamada},\ and\ \citenamefont
  {Yoshida}}]{Tachibana.Yamada.Yoshida:1990}%
  \BibitemOpen
  \bibfield  {author} {\bibinfo {author} {\bibnamefont {Tachibana},
  \bibfnamefont {T}}, \bibinfo {author} {\bibfnamefont {M.}~\bibnamefont
  {Yamada}}, and\ \bibinfo {author} {\bibfnamefont {Y.}~\bibnamefont
  {Yoshida}}} (\bibinfo {year} {1990}),\ \bibfield  {title} {\enquote {\bibinfo
  {title} {{Improvement of the Gross Theory of -Decay. II: One-Particle
  Strength Function}},}\ }\href {https://doi.org/10.1143/ptp/84.4.641}
  {\bibfield  {journal} {\bibinfo  {journal} {Prog. Theor. Phys.}\ }\textbf
  {\bibinfo {volume} {84}},\ \bibinfo {pages} {641--657}}\BibitemShut {NoStop}%
\bibitem [{\citenamefont {Takahashi}\ \emph {et~al.}(1994)\citenamefont
  {Takahashi}, \citenamefont {Witti},\ and\ \citenamefont
  {Janka}}]{Takahashi.Witti.Janka:1994}%
  \BibitemOpen
  \bibfield  {author} {\bibinfo {author} {\bibnamefont {Takahashi},
  \bibfnamefont {K}}, \bibinfo {author} {\bibfnamefont {J.}~\bibnamefont
  {Witti}}, and\ \bibinfo {author} {\bibfnamefont {H.-Th.}\ \bibnamefont
  {Janka}}} (\bibinfo {year} {1994}),\ \bibfield  {title} {\enquote {\bibinfo
  {title} {{Nucleosynthesis in neutrino-driven winds from protoneutron stars
  II. The r-process}},}\ }\href@noop {} {\bibfield  {journal} {\bibinfo
  {journal} {Astron. \& Astrophys.}\ }\textbf {\bibinfo {volume} {286}},\
  \bibinfo {pages} {857--869}}\BibitemShut {NoStop}%
\bibitem [{\citenamefont {{Takiwaki}}\ and\ \citenamefont
  {{Kotake}}(2011)}]{takiwaki11}%
  \BibitemOpen
  \bibfield  {author} {\bibinfo {author} {\bibnamefont {{Takiwaki}},
  \bibfnamefont {T}}, and\ \bibinfo {author} {\bibfnamefont {K.}~\bibnamefont
  {{Kotake}}}} (\bibinfo {year} {2011}),\ \bibfield  {title} {\enquote
  {\bibinfo {title} {{Gravitational Wave Signatures of Magnetohydrodynamically
  Driven Core-collapse Supernova Explosions}},}\ }\href
  {https://doi.org/10.1088/0004-637X/743/1/30} {\bibfield  {journal} {\bibinfo
  {journal} {Astrophys. J.}\ }\textbf {\bibinfo {volume} {743}},\ \bibinfo
  {eid} {30}}\BibitemShut {NoStop}%
\bibitem [{\citenamefont {{Takiwaki}}\ \emph {et~al.}(2009)\citenamefont
  {{Takiwaki}}, \citenamefont {{Kotake}},\ and\ \citenamefont
  {{Sato}}}]{takiwaki09}%
  \BibitemOpen
  \bibfield  {author} {\bibinfo {author} {\bibnamefont {{Takiwaki}},
  \bibfnamefont {T}}, \bibinfo {author} {\bibfnamefont {K.}~\bibnamefont
  {{Kotake}}}, and\ \bibinfo {author} {\bibfnamefont {K.}~\bibnamefont
  {{Sato}}}} (\bibinfo {year} {2009}),\ \bibfield  {title} {\enquote {\bibinfo
  {title} {{Special Relativistic Simulations of Magnetically Dominated Jets in
  Collapsing Massive Stars}},}\ }\href
  {https://doi.org/10.1088/0004-637X/691/2/1360} {\bibfield  {journal}
  {\bibinfo  {journal} {Astrophys. J.}\ }\textbf {\bibinfo {volume} {691}},\
  \bibinfo {pages} {1360--1379}}\BibitemShut {NoStop}%
\bibitem [{\citenamefont {{Talbert}}\ \emph {et~al.}(1979)\citenamefont
  {{Talbert}}, \citenamefont {{Wohn}}, \citenamefont {{Hill}}, \citenamefont
  {{Landin}}, \citenamefont {{Cullison}},\ and\ \citenamefont
  {{Gill}}}]{TristanII}%
  \BibitemOpen
  \bibfield  {author} {\bibinfo {author} {\bibnamefont {{Talbert}},
  \bibfnamefont {W~L}}, \bibinfo {author} {\bibfnamefont {F.~K.}\ \bibnamefont
  {{Wohn}}}, \bibinfo {author} {\bibfnamefont {J.~C.}\ \bibnamefont {{Hill}}},
  \bibinfo {author} {\bibfnamefont {A.~R.}\ \bibnamefont {{Landin}}}, \bibinfo
  {author} {\bibfnamefont {M.~A.}\ \bibnamefont {{Cullison}}}, and\ \bibinfo
  {author} {\bibfnamefont {R.~L.}\ \bibnamefont {{Gill}}}} (\bibinfo {year}
  {1979}),\ \bibfield  {title} {\enquote {\bibinfo {title} {{Tristan II:
  Extension of the Tristan facility to the study of non-gaseous activities}},}\
  }\href {https://doi.org/10.1016/0029-554X(79)90416-6} {\bibfield  {journal}
  {\bibinfo  {journal} {Nuclear Instruments and Methods}\ }\textbf {\bibinfo
  {volume} {161}},\ \bibinfo {pages} {431--437}}\BibitemShut {NoStop}%
\bibitem [{\citenamefont {Tamborra}\ \emph {et~al.}(2012)\citenamefont
  {Tamborra}, \citenamefont {M{\"u}ller}, \citenamefont {H{\"u}depohl},
  \citenamefont {Janka},\ and\ \citenamefont
  {Raffelt}}]{Tamborra.Mueller.ea:2012}%
  \BibitemOpen
  \bibfield  {author} {\bibinfo {author} {\bibnamefont {Tamborra},
  \bibfnamefont {Irene}}, \bibinfo {author} {\bibfnamefont {Bernhard}\
  \bibnamefont {M{\"u}ller}}, \bibinfo {author} {\bibfnamefont {Lorenz}\
  \bibnamefont {H{\"u}depohl}}, \bibinfo {author} {\bibfnamefont {Hans-Thomas}\
  \bibnamefont {Janka}}, and\ \bibinfo {author} {\bibfnamefont {Georg}\
  \bibnamefont {Raffelt}}} (\bibinfo {year} {2012}),\ \bibfield  {title}
  {\enquote {\bibinfo {title} {High-resolution supernova neutrino spectra
  represented by a simple fit},}\ }\href
  {https://doi.org/10.1103/PhysRevD.86.125031} {\bibfield  {journal} {\bibinfo
  {journal} {Phys. Rev. D}\ }\textbf {\bibinfo {volume} {86}},\ \bibinfo
  {pages} {125031}}\BibitemShut {NoStop}%
\bibitem [{\citenamefont {{Tanaka}}(2016)}]{Tanaka:2016}%
  \BibitemOpen
  \bibfield  {author} {\bibinfo {author} {\bibnamefont {{Tanaka}},
  \bibfnamefont {M}}} (\bibinfo {year} {2016}),\ \bibfield  {title} {\enquote
  {\bibinfo {title} {{Kilonova/Macronova Emission from Compact Binary
  Mergers}},}\ }\href {https://doi.org/10.1155/2016/6341974} {\bibfield
  {journal} {\bibinfo  {journal} {Adv. Astron.}\ }\textbf {\bibinfo {volume}
  {2016}},\ \bibinfo {eid} {634197}}\BibitemShut {NoStop}%
\bibitem [{\citenamefont {{Tanaka}}\ \emph {et~al.}(2018)\citenamefont
  {{Tanaka}}, \citenamefont {{Kato}}, \citenamefont {{Gaigalas}}, \citenamefont
  {{Rynkun}}, \citenamefont {{Rad{\v z}i{\= u}t{\.e}}}, \citenamefont
  {{Wanajo}}, \citenamefont {{Sekiguchi}}, \citenamefont {{Nakamura}},
  \citenamefont {{Tanuma}}, \citenamefont {{Murakami}},\ and\ \citenamefont
  {{Sakaue}}}]{Tanaka.Kato.ea:2018}%
  \BibitemOpen
  \bibfield  {author} {\bibinfo {author} {\bibnamefont {{Tanaka}},
  \bibfnamefont {M}}, \bibinfo {author} {\bibfnamefont {D.}~\bibnamefont
  {{Kato}}}, \bibinfo {author} {\bibfnamefont {G.}~\bibnamefont {{Gaigalas}}},
  \bibinfo {author} {\bibfnamefont {P.}~\bibnamefont {{Rynkun}}}, \bibinfo
  {author} {\bibfnamefont {L.}~\bibnamefont {{Rad{\v z}i{\= u}t{\.e}}}},
  \bibinfo {author} {\bibfnamefont {S.}~\bibnamefont {{Wanajo}}}, \bibinfo
  {author} {\bibfnamefont {Y.}~\bibnamefont {{Sekiguchi}}}, \bibinfo {author}
  {\bibfnamefont {N.}~\bibnamefont {{Nakamura}}}, \bibinfo {author}
  {\bibfnamefont {H.}~\bibnamefont {{Tanuma}}}, \bibinfo {author}
  {\bibfnamefont {I.}~\bibnamefont {{Murakami}}}, and\ \bibinfo {author}
  {\bibfnamefont {H.~A.}\ \bibnamefont {{Sakaue}}}} (\bibinfo {year} {2018}),\
  \bibfield  {title} {\enquote {\bibinfo {title} {{Properties of Kilonovae from
  Dynamical and Post-merger Ejecta of Neutron Star Mergers}},}\ }\href
  {https://doi.org/10.3847/1538-4357/aaa0cb} {\bibfield  {journal} {\bibinfo
  {journal} {Astrophys. J.}\ }\textbf {\bibinfo {volume} {852}},\ \bibinfo
  {eid} {109}}\BibitemShut {NoStop}%
\bibitem [{\citenamefont {{Tanaka}}\ \emph {et~al.}(2017)\citenamefont
  {{Tanaka}} \emph {et~al.}}]{Tanaka.Utsumi.ea:2017}%
  \BibitemOpen
  \bibfield  {author} {\bibinfo {author} {\bibnamefont {{Tanaka}},
  \bibfnamefont {M}},  \emph {et~al.}} (\bibinfo {year} {2017}),\ \bibfield
  {title} {\enquote {\bibinfo {title} {Kilonova from post-merger ejecta as an
  optical and near-infrared counterpart of gw170817},}\ }\href
  {https://doi.org/10.1093/pasj/psx121} {\bibfield  {journal} {\bibinfo
  {journal} {Publ. Astron. Soc. Japan}\ }\textbf {\bibinfo {volume} {69}},\
  \bibinfo {eid} {102}}\BibitemShut {NoStop}%
\bibitem [{\citenamefont {{Tanaka}}\ and\ \citenamefont
  {{Hotokezaka}}(2013)}]{Tanaka.Hotokezaka:2013}%
  \BibitemOpen
  \bibfield  {author} {\bibinfo {author} {\bibnamefont {{Tanaka}},
  \bibfnamefont {Masaomi}}, and\ \bibinfo {author} {\bibfnamefont {Kenta}\
  \bibnamefont {{Hotokezaka}}}} (\bibinfo {year} {2013}),\ \bibfield  {title}
  {\enquote {\bibinfo {title} {{Radiative Transfer Simulations of Neutron Star
  Merger Ejecta}},}\ }\href {https://doi.org/10.1088/0004-637X/775/2/113}
  {\bibfield  {journal} {\bibinfo  {journal} {Astrophys. J.}\ }\textbf
  {\bibinfo {volume} {775}},\ \bibinfo {eid} {113}}\BibitemShut {NoStop}%
\bibitem [{\citenamefont {{Tanaka}}\ \emph {et~al.}(2020)\citenamefont
  {{Tanaka}}, \citenamefont {{Kato}}, \citenamefont {{Gaigalas}},\ and\
  \citenamefont {{Kawaguchi}}}]{Tanaka.Kato.ea:2020}%
  \BibitemOpen
  \bibfield  {author} {\bibinfo {author} {\bibnamefont {{Tanaka}},
  \bibfnamefont {Masaomi}}, \bibinfo {author} {\bibfnamefont {Daiji}\
  \bibnamefont {{Kato}}}, \bibinfo {author} {\bibfnamefont {Gediminas}\
  \bibnamefont {{Gaigalas}}}, and\ \bibinfo {author} {\bibfnamefont {Kyohei}\
  \bibnamefont {{Kawaguchi}}}} (\bibinfo {year} {2020}),\ \bibfield  {title}
  {\enquote {\bibinfo {title} {{Systematic opacity calculations for
  kilonovae}},}\ }\href {https://doi.org/10.1093/mnras/staa1576} {\bibfield
  {journal} {\bibinfo  {journal} {Mon. Not. Roy. Astron. Soc.}\ }\textbf
  {\bibinfo {volume} {496}}~(\bibinfo {number} {2}),\ \bibinfo {pages}
  {1369--1392}}\BibitemShut {NoStop}%
\bibitem [{\citenamefont {Tang}\ \emph {et~al.}(2020)\citenamefont {Tang} \emph
  {et~al.}}]{Tang.Key.ea:2020}%
  \BibitemOpen
  \bibfield  {author} {\bibinfo {author} {\bibnamefont {Tang}, \bibfnamefont
  {T~L}},  \emph {et~al.}} (\bibinfo {year} {2020}),\ \bibfield  {title}
  {\enquote {\bibinfo {title} {{First Exploration of Neutron Shell Structure
  below Lead and beyond $N=126$}},}\ }\href
  {https://doi.org/10.1103/PhysRevLett.124.062502} {\bibfield  {journal}
  {\bibinfo  {journal} {Phys. Rev. Lett.}\ }\textbf {\bibinfo {volume} {124}},\
  \bibinfo {pages} {062502}}\BibitemShut {NoStop}%
\bibitem [{\citenamefont {{Tanvir}}\ \emph {et~al.}(2013)\citenamefont
  {{Tanvir}}, \citenamefont {{Levan}}, \citenamefont {{Fruchter}},
  \citenamefont {{Hjorth}}, \citenamefont {{Hounsell}}, \citenamefont
  {{Wiersema}},\ and\ \citenamefont {{Tunnicliffe}}}]{tanvir13}%
  \BibitemOpen
  \bibfield  {author} {\bibinfo {author} {\bibnamefont {{Tanvir}},
  \bibfnamefont {N~R}}, \bibinfo {author} {\bibfnamefont {A.~J.}\ \bibnamefont
  {{Levan}}}, \bibinfo {author} {\bibfnamefont {A.~S.}\ \bibnamefont
  {{Fruchter}}}, \bibinfo {author} {\bibfnamefont {J.}~\bibnamefont
  {{Hjorth}}}, \bibinfo {author} {\bibfnamefont {R.~A.}\ \bibnamefont
  {{Hounsell}}}, \bibinfo {author} {\bibfnamefont {K.}~\bibnamefont
  {{Wiersema}}}, and\ \bibinfo {author} {\bibfnamefont {R.~L.}\ \bibnamefont
  {{Tunnicliffe}}}} (\bibinfo {year} {2013}),\ \bibfield  {title} {\enquote
  {\bibinfo {title} {{A `kilonova' associated with the short-duration
  {$\gamma$}-ray burst GRB 130603B}},}\ }\href
  {https://doi.org/10.1038/nature12505} {\bibfield  {journal} {\bibinfo
  {journal} {Nature}\ }\textbf {\bibinfo {volume} {500}},\ \bibinfo {pages}
  {547--549}}\BibitemShut {NoStop}%
\bibitem [{\citenamefont {{Tanvir}}\ \emph {et~al.}(2017)\citenamefont
  {{Tanvir}} \emph {et~al.}}]{Tanvir.Others:2017}%
  \BibitemOpen
  \bibfield  {author} {\bibinfo {author} {\bibnamefont {{Tanvir}},
  \bibfnamefont {N~R}},  \emph {et~al.}} (\bibinfo {year} {2017}),\ \bibfield
  {title} {\enquote {\bibinfo {title} {{The Emergence of a Lanthanide-rich
  Kilonova Following the Merger of Two Neutron Stars}},}\ }\href
  {https://doi.org/10.3847/2041-8213/aa90b6} {\bibfield  {journal} {\bibinfo
  {journal} {Astrophys. J. Lett.}\ }\textbf {\bibinfo {volume} {848}},\
  \bibinfo {eid} {L27}}\BibitemShut {NoStop}%
\bibitem [{\citenamefont {{Tarumi}}\ \emph {et~al.}(2020)\citenamefont
  {{Tarumi}}, \citenamefont {{Yoshida}},\ and\ \citenamefont
  {{Inoue}}}]{Tarumi.Yoshida.ea:2020}%
  \BibitemOpen
  \bibfield  {author} {\bibinfo {author} {\bibnamefont {{Tarumi}},
  \bibfnamefont {Yuta}}, \bibinfo {author} {\bibfnamefont {Naoki}\ \bibnamefont
  {{Yoshida}}}, and\ \bibinfo {author} {\bibfnamefont {Shigeki}\ \bibnamefont
  {{Inoue}}}} (\bibinfo {year} {2020}),\ \bibfield  {title} {\enquote {\bibinfo
  {title} {{R-process enrichment in ultrafaint dwarf galaxies}},}\ }\href
  {https://doi.org/10.1093/mnras/staa720} {\bibfield  {journal} {\bibinfo
  {journal} {Mon. Not. Roy. Astron. Soc.}\ }\textbf {\bibinfo {volume} {494}},\
  \bibinfo {pages} {120--128}}\BibitemShut {NoStop}%
\bibitem [{\citenamefont {{Terasawa}}\ \emph {et~al.}(2004)\citenamefont
  {{Terasawa}}, \citenamefont {{Langanke}}, \citenamefont {{Kajino}},
  \citenamefont {{Mathews}},\ and\ \citenamefont
  {{Kolbe}}}]{Terasawa.Langanke.ea:2004}%
  \BibitemOpen
  \bibfield  {author} {\bibinfo {author} {\bibnamefont {{Terasawa}},
  \bibfnamefont {M}}, \bibinfo {author} {\bibfnamefont {K.}~\bibnamefont
  {{Langanke}}}, \bibinfo {author} {\bibfnamefont {T.}~\bibnamefont
  {{Kajino}}}, \bibinfo {author} {\bibfnamefont {G.~J.}\ \bibnamefont
  {{Mathews}}}, and\ \bibinfo {author} {\bibfnamefont {E.}~\bibnamefont
  {{Kolbe}}}} (\bibinfo {year} {2004}),\ \bibfield  {title} {\enquote {\bibinfo
  {title} {{Neutrino Effects before, during, and after the Freezeout of the
  r-Process}},}\ }\href {https://doi.org/10.1086/386359} {\bibfield  {journal}
  {\bibinfo  {journal} {Astrophys. J.}\ }\textbf {\bibinfo {volume} {608}},\
  \bibinfo {pages} {470--479}}\BibitemShut {NoStop}%
\bibitem [{\citenamefont {{Terasawa}}\ \emph {et~al.}(2001)\citenamefont
  {{Terasawa}}, \citenamefont {{Sumiyoshi}}, \citenamefont {{Kajino}},
  \citenamefont {{Mathews}},\ and\ \citenamefont
  {{Tanihata}}}]{Terasawa.Sumiyoshi.ea:2001b}%
  \BibitemOpen
  \bibfield  {author} {\bibinfo {author} {\bibnamefont {{Terasawa}},
  \bibfnamefont {M}}, \bibinfo {author} {\bibfnamefont {K.}~\bibnamefont
  {{Sumiyoshi}}}, \bibinfo {author} {\bibfnamefont {T.}~\bibnamefont
  {{Kajino}}}, \bibinfo {author} {\bibfnamefont {G.~J.}\ \bibnamefont
  {{Mathews}}}, and\ \bibinfo {author} {\bibfnamefont {I.}~\bibnamefont
  {{Tanihata}}}} (\bibinfo {year} {2001}),\ \bibfield  {title} {\enquote
  {\bibinfo {title} {New nuclear reaction flow during r-process nucleosynthesis
  in supernovae: Critical role of light, neutron-rich nuclei},}\ }\href
  {https://doi.org/10.1086/323526} {\bibfield  {journal} {\bibinfo  {journal}
  {Astrophys. J.}\ }\textbf {\bibinfo {volume} {562}},\ \bibinfo {pages}
  {470--479}}\BibitemShut {NoStop}%
\bibitem [{\citenamefont {Tews}\ \emph {et~al.}(2018)\citenamefont {Tews},
  \citenamefont {Margueron},\ and\ \citenamefont
  {Reddy}}]{Tews.Margueron.Reddy:2018}%
  \BibitemOpen
  \bibfield  {author} {\bibinfo {author} {\bibnamefont {Tews}, \bibfnamefont
  {I}}, \bibinfo {author} {\bibfnamefont {J.}~\bibnamefont {Margueron}}, and\
  \bibinfo {author} {\bibfnamefont {S.}~\bibnamefont {Reddy}}} (\bibinfo {year}
  {2018}),\ \bibfield  {title} {\enquote {\bibinfo {title} {Critical
  examination of constraints on the equation of state of dense matter obtained
  from gw170817},}\ }\href {https://doi.org/10.1103/PhysRevC.98.045804}
  {\bibfield  {journal} {\bibinfo  {journal} {Phys. Rev. C}\ }\textbf {\bibinfo
  {volume} {98}},\ \bibinfo {pages} {045804}}\BibitemShut {NoStop}%
\bibitem [{\citenamefont {{Tews}}\ \emph {et~al.}(2019)\citenamefont {{Tews}},
  \citenamefont {{Margueron}},\ and\ \citenamefont
  {{Reddy}}}]{Tews.Margueron.Reddy:2019}%
  \BibitemOpen
  \bibfield  {author} {\bibinfo {author} {\bibnamefont {{Tews}}, \bibfnamefont
  {I}}, \bibinfo {author} {\bibfnamefont {J.}~\bibnamefont {{Margueron}}}, and\
  \bibinfo {author} {\bibfnamefont {S.}~\bibnamefont {{Reddy}}}} (\bibinfo
  {year} {2019}),\ \bibfield  {title} {\enquote {\bibinfo {title} {{Confronting
  gravitational-wave observations with modern nuclear physics constraints}},}\
  }\href {https://doi.org/10.1140/epja/i2019-12774-6} {\bibfield  {journal}
  {\bibinfo  {journal} {Eur. Phys. J. A}\ }\textbf {\bibinfo {volume} {55}},\
  \bibinfo {eid} {97}}\BibitemShut {NoStop}%
\bibitem [{\citenamefont {{Thielemann}}\ \emph {et~al.}(1979)\citenamefont
  {{Thielemann}}, \citenamefont {{Arnould}},\ and\ \citenamefont
  {{Hillebrandt}}}]{Thielemann.Arnould.Hillebrandt:1979}%
  \BibitemOpen
  \bibfield  {author} {\bibinfo {author} {\bibnamefont {{Thielemann}},
  \bibfnamefont {F-K}}, \bibinfo {author} {\bibfnamefont {M.}~\bibnamefont
  {{Arnould}}}, and\ \bibinfo {author} {\bibfnamefont {W.}~\bibnamefont
  {{Hillebrandt}}}} (\bibinfo {year} {1979}),\ \bibfield  {title} {\enquote
  {\bibinfo {title} {{Meteoritic anomalies and explosive neutron processing of
  helium-burning shells}},}\ }\href@noop {} {\bibfield  {journal} {\bibinfo
  {journal} {Astron. \& Astrophys.}\ }\textbf {\bibinfo {volume} {74}},\
  \bibinfo {pages} {175--185}}\BibitemShut {NoStop}%
\bibitem [{\citenamefont {{Thielemann}}\ \emph {et~al.}(1989)\citenamefont
  {{Thielemann}}, \citenamefont {{Cameron}},\ and\ \citenamefont
  {{Cowan}}}]{Thielemann.Cameron.Cowan:1989}%
  \BibitemOpen
  \bibfield  {author} {\bibinfo {author} {\bibnamefont {{Thielemann}},
  \bibfnamefont {F-K}}, \bibinfo {author} {\bibfnamefont {A.~G.~W.}\
  \bibnamefont {{Cameron}}}, and\ \bibinfo {author} {\bibfnamefont {J.~J.}\
  \bibnamefont {{Cowan}}}} (\bibinfo {year} {1989}),\ \bibfield  {title}
  {\enquote {\bibinfo {title} {{Fission in the Astrophysical r-Process}},}\
  }in\ \href {http://www.ans.org/store/item-690091/} {\emph {\bibinfo
  {booktitle} {50 Years with Nuclear Fission}}},\ \bibinfo {editor} {edited by\
  \bibinfo {editor} {\bibfnamefont {J.}~\bibnamefont {{Behrens}}}\ and\
  \bibinfo {editor} {\bibfnamefont {A.D.}\ \bibnamefont {{Carlson}}}},\ p.\
  \bibinfo {pages} {592}\BibitemShut {NoStop}%
\bibitem [{\citenamefont {{Thielemann}}\ \emph {et~al.}(2018)\citenamefont
  {{Thielemann}}, \citenamefont {{Diehl}}, \citenamefont {{Heger}},
  \citenamefont {{Hirschi}},\ and\ \citenamefont
  {{Liebend{\"o}rfer}}}]{thielemann18}%
  \BibitemOpen
  \bibfield  {author} {\bibinfo {author} {\bibnamefont {{Thielemann}},
  \bibfnamefont {F-K}}, \bibinfo {author} {\bibfnamefont {R.}~\bibnamefont
  {{Diehl}}}, \bibinfo {author} {\bibfnamefont {A.}~\bibnamefont {{Heger}}},
  \bibinfo {author} {\bibfnamefont {R.}~\bibnamefont {{Hirschi}}}, and\
  \bibinfo {author} {\bibfnamefont {M.}~\bibnamefont {{Liebend{\"o}rfer}}}}
  (\bibinfo {year} {2018}),\ \enquote {\bibinfo {title} {{Massive Stars and
  Their Supernovae}},}\ in\ \href {https://doi.org/10.1007/978-3-319-91929-4_4}
  {\emph {\bibinfo {booktitle} {Astrophysics with Radioactive Isotopes}}},\
  \bibinfo {series} {Astrophysics and Space Science Library}, Vol.\ \bibinfo
  {volume} {453},\ \bibinfo {editor} {edited by\ \bibinfo {editor}
  {\bibfnamefont {R.}~\bibnamefont {{Diehl}}}, \bibinfo {editor} {\bibfnamefont
  {D.~H.}\ \bibnamefont {{Hartmann}}}, \ and\ \bibinfo {editor} {\bibfnamefont
  {N.}~\bibnamefont {{Prantzos}}}}\ (\bibinfo  {publisher} {Springer},\
  \bibinfo {address} {Cham})\ pp.\ \bibinfo {pages} {173--286}\BibitemShut
  {NoStop}%
\bibitem [{\citenamefont {{Thielemann}}\ \emph
  {et~al.}(2017{\natexlab{a}})\citenamefont {{Thielemann}}, \citenamefont
  {{Eichler}}, \citenamefont {{Panov}}, \citenamefont {{Pignatari}},\ and\
  \citenamefont {{Wehmeyer}}}]{thielemann17a}%
  \BibitemOpen
  \bibfield  {author} {\bibinfo {author} {\bibnamefont {{Thielemann}},
  \bibfnamefont {F-K}}, \bibinfo {author} {\bibfnamefont {M.}~\bibnamefont
  {{Eichler}}}, \bibinfo {author} {\bibfnamefont {I.~V.}\ \bibnamefont
  {{Panov}}}, \bibinfo {author} {\bibfnamefont {M.}~\bibnamefont
  {{Pignatari}}}, and\ \bibinfo {author} {\bibfnamefont {B.}~\bibnamefont
  {{Wehmeyer}}}} (\bibinfo {year} {2017}{\natexlab{a}}),\ \enquote {\bibinfo
  {title} {{Making the Heaviest Elements in a Rare Class of Supernovae}},}\ in\
  \href {https://doi.org/10.1007/978-3-319-20794-0_81-1} {\emph {\bibinfo
  {booktitle} {Handbook of Supernovae}}},\ \bibinfo {editor} {edited by\
  \bibinfo {editor} {\bibfnamefont {A.~W.}\ \bibnamefont {{Alsabti}}}\ and\
  \bibinfo {editor} {\bibfnamefont {P.}~\bibnamefont {{Murdin}}}}\ (\bibinfo
  {publisher} {Springer International Publishing},\ \bibinfo {address}
  {Cham})\BibitemShut {NoStop}%
\bibitem [{\citenamefont {{Thielemann}}\ \emph
  {et~al.}(2017{\natexlab{b}})\citenamefont {{Thielemann}}, \citenamefont
  {{Eichler}}, \citenamefont {{Panov}},\ and\ \citenamefont
  {{Wehmeyer}}}]{Thielemann.Eichler.ea:2017}%
  \BibitemOpen
  \bibfield  {author} {\bibinfo {author} {\bibnamefont {{Thielemann}},
  \bibfnamefont {F-K}}, \bibinfo {author} {\bibfnamefont {M.}~\bibnamefont
  {{Eichler}}}, \bibinfo {author} {\bibfnamefont {I.~V.}\ \bibnamefont
  {{Panov}}}, and\ \bibinfo {author} {\bibfnamefont {B.}~\bibnamefont
  {{Wehmeyer}}}} (\bibinfo {year} {2017}{\natexlab{b}}),\ \bibfield  {title}
  {\enquote {\bibinfo {title} {{Neutron Star Mergers and Nucleosynthesis of
  Heavy Elements}},}\ }\href
  {https://doi.org/10.1146/annurev-nucl-101916-123246} {\bibfield  {journal}
  {\bibinfo  {journal} {Annu. Rev. Nucl. Part. Sci.}\ }\textbf {\bibinfo
  {volume} {67}},\ \bibinfo {pages} {253--274}}\BibitemShut {NoStop}%
\bibitem [{\citenamefont {Thielemann}\ \emph
  {et~al.}(2011{\natexlab{a}})\citenamefont {Thielemann}, \citenamefont
  {Hirschi}, \citenamefont {Liebend{\"o}rfer},\ and\ \citenamefont
  {Diehl}}]{thielemann11}%
  \BibitemOpen
  \bibfield  {author} {\bibinfo {author} {\bibnamefont {Thielemann},
  \bibfnamefont {F-K}}, \bibinfo {author} {\bibfnamefont {R.}~\bibnamefont
  {Hirschi}}, \bibinfo {author} {\bibfnamefont {M.}~\bibnamefont
  {Liebend{\"o}rfer}}, and\ \bibinfo {author} {\bibfnamefont {R.}~\bibnamefont
  {Diehl}}} (\bibinfo {year} {2011}{\natexlab{a}}),\ \enquote {\bibinfo {title}
  {{Massive Stars and Their Supernovae}},}\ in\ \href
  {https://doi.org/10.1007/978-3-642-12698-7_4} {\emph {\bibinfo {booktitle}
  {Astronomy with Radioactivities}}},\ \bibinfo {series} {Lecture Notes in
  Physics}, Vol.\ \bibinfo {volume} {812},\ \bibinfo {editor} {edited by\
  \bibinfo {editor} {\bibfnamefont {Roland}\ \bibnamefont {Diehl}}, \bibinfo
  {editor} {\bibfnamefont {Dieter~H.}\ \bibnamefont {Hartmann}}, \ and\
  \bibinfo {editor} {\bibfnamefont {Nikos}\ \bibnamefont {Prantzos}}}\
  (\bibinfo  {publisher} {Springer},\ \bibinfo {address} {Berlin, Heidelberg})\
  pp.\ \bibinfo {pages} {153--231}\BibitemShut {NoStop}%
\bibitem [{\citenamefont {{Thielemann}}\ \emph {et~al.}(1983)\citenamefont
  {{Thielemann}}, \citenamefont {{Metzinger}},\ and\ \citenamefont
  {{Klapdor}}}]{Thielemann.Metzinger.Klapdor:1983}%
  \BibitemOpen
  \bibfield  {author} {\bibinfo {author} {\bibnamefont {{Thielemann}},
  \bibfnamefont {F-K}}, \bibinfo {author} {\bibfnamefont {J.}~\bibnamefont
  {{Metzinger}}}, and\ \bibinfo {author} {\bibfnamefont {H.~V.}\ \bibnamefont
  {{Klapdor}}}} (\bibinfo {year} {1983}),\ \bibfield  {title} {\enquote
  {\bibinfo {title} {{Beta-delayed fission and neutron emission: Consequences
  for the astrophysical r-process and the age of the galaxy}},}\ }\href
  {https://doi.org/10.1007/BF01413833} {\bibfield  {journal} {\bibinfo
  {journal} {Zeitschrift f{\"u}r Phys. A Hadron. Nucl.}\ }\textbf {\bibinfo
  {volume} {309}},\ \bibinfo {pages} {301--317}}\BibitemShut {NoStop}%
\bibitem [{\citenamefont {{Thielemann}}\ \emph {et~al.}(1986)\citenamefont
  {{Thielemann}}, \citenamefont {{Nomoto}},\ and\ \citenamefont
  {{Yokoi}}}]{Thielemann.Nomoto.Yokoi:1986}%
  \BibitemOpen
  \bibfield  {author} {\bibinfo {author} {\bibnamefont {{Thielemann}},
  \bibfnamefont {F~K}}, \bibinfo {author} {\bibfnamefont {K.}~\bibnamefont
  {{Nomoto}}}, and\ \bibinfo {author} {\bibfnamefont {K.}~\bibnamefont
  {{Yokoi}}}} (\bibinfo {year} {1986}),\ \bibfield  {title} {\enquote {\bibinfo
  {title} {{Explosive nucleosynthesis in carbon deflagration models of Type I
  supernovae}},}\ }\href@noop {} {\bibfield  {journal} {\bibinfo  {journal}
  {Astron. \& Astrophys.}\ }\textbf {\bibinfo {volume} {158}}~(\bibinfo
  {number} {1-2}),\ \bibinfo {pages} {17--33}}\BibitemShut {NoStop}%
\bibitem [{\citenamefont {Thielemann}\ \emph
  {et~al.}(2011{\natexlab{b}})\citenamefont {Thielemann} \emph
  {et~al.}}]{Thielemann.Arcones.ea:2011}%
  \BibitemOpen
  \bibfield  {author} {\bibinfo {author} {\bibnamefont {Thielemann},
  \bibfnamefont {F-K}},  \emph {et~al.}} (\bibinfo {year}
  {2011}{\natexlab{b}}),\ \bibfield  {title} {\enquote {\bibinfo {title} {What
  are the astrophysical sites for the -process and the production of heavy
  elements?}}\ }\href {https://doi.org/10.1016/j.ppnp.2011.01.032} {\bibfield
  {journal} {\bibinfo  {journal} {Prog. Part. Nucl. Phys.}\ }\textbf {\bibinfo
  {volume} {66}},\ \bibinfo {pages} {346--353}}\BibitemShut {NoStop}%
\bibitem [{\citenamefont {{Thielemann}}\ \emph {et~al.}(1990)\citenamefont
  {{Thielemann}}, \citenamefont {{Hashimoto}},\ and\ \citenamefont
  {{Nomoto}}}]{Thielemann.Hashimoto.ea:1990}%
  \BibitemOpen
  \bibfield  {author} {\bibinfo {author} {\bibnamefont {{Thielemann}},
  \bibfnamefont {Friedrich-Karl}}, \bibinfo {author} {\bibfnamefont {Masa-Aki}\
  \bibnamefont {{Hashimoto}}}, and\ \bibinfo {author} {\bibfnamefont
  {Ken'ichi}\ \bibnamefont {{Nomoto}}}} (\bibinfo {year} {1990}),\ \bibfield
  {title} {\enquote {\bibinfo {title} {{Explosive Nucleosynthesis in SN 1987A.
  II. Composition, Radioactivities, and the Neutron Star Mass}},}\ }\href
  {https://doi.org/10.1086/168308} {\bibfield  {journal} {\bibinfo  {journal}
  {Astrophys. J.}\ }\textbf {\bibinfo {volume} {349}},\ \bibinfo {pages}
  {222}}\BibitemShut {NoStop}%
\bibitem [{\citenamefont {{Thielemann}}\ \emph {et~al.}(1994)\citenamefont
  {{Thielemann}}, \citenamefont {{Kratz}}, \citenamefont {{Pfeiffer}},
  \citenamefont {{Rauscher}}, \citenamefont {{van Wormer}},\ and\ \citenamefont
  {{Wiescher}}}]{Thielemann.Kratz.ea:1994}%
  \BibitemOpen
  \bibfield  {author} {\bibinfo {author} {\bibnamefont {{Thielemann}},
  \bibfnamefont {Friedrich-Karl}}, \bibinfo {author} {\bibfnamefont
  {Karl-Ludwig}\ \bibnamefont {{Kratz}}}, \bibinfo {author} {\bibfnamefont
  {Bernd}\ \bibnamefont {{Pfeiffer}}}, \bibinfo {author} {\bibfnamefont
  {Thomas}\ \bibnamefont {{Rauscher}}}, \bibinfo {author} {\bibfnamefont
  {Laura}\ \bibnamefont {{van Wormer}}}, and\ \bibinfo {author} {\bibfnamefont
  {Michael~C.}\ \bibnamefont {{Wiescher}}}} (\bibinfo {year} {1994}),\
  \bibfield  {title} {\enquote {\bibinfo {title} {{Astrophysics and nuclei far
  from stabilitys}},}\ }\href {https://doi.org/10.1016/0375-9474(94)90299-2}
  {\bibfield  {journal} {\bibinfo  {journal} {Nucl. Phys. A}\ }\textbf
  {\bibinfo {volume} {570}},\ \bibinfo {pages} {329--343}}\BibitemShut
  {NoStop}%
\bibitem [{\citenamefont {{Thielemann}}\ \emph {et~al.}(1996)\citenamefont
  {{Thielemann}}, \citenamefont {{Nomoto}},\ and\ \citenamefont
  {{Hashimoto}}}]{Thielemann.Nomoto.Hashimoto:1996}%
  \BibitemOpen
  \bibfield  {author} {\bibinfo {author} {\bibnamefont {{Thielemann}},
  \bibfnamefont {Friedrich-Karl}}, \bibinfo {author} {\bibfnamefont {Ken'ichi}\
  \bibnamefont {{Nomoto}}}, and\ \bibinfo {author} {\bibfnamefont {Masa-Aki}\
  \bibnamefont {{Hashimoto}}}} (\bibinfo {year} {1996}),\ \bibfield  {title}
  {\enquote {\bibinfo {title} {{Core-Collapse Supernovae and Their Ejecta}},}\
  }\href {https://doi.org/10.1086/176980} {\bibfield  {journal} {\bibinfo
  {journal} {Astrophys. J.}\ }\textbf {\bibinfo {volume} {460}},\ \bibinfo
  {pages} {408}}\BibitemShut {NoStop}%
\bibitem [{\citenamefont {{Thielemann}}\ and\ \citenamefont
  {{Truran}}(1986)}]{Thielemann.Truran:1986}%
  \BibitemOpen
  \bibfield  {author} {\bibinfo {author} {\bibnamefont {{Thielemann}},
  \bibfnamefont {Friedrich-Karl}}, and\ \bibinfo {author} {\bibfnamefont
  {James~W.}\ \bibnamefont {{Truran}}}} (\bibinfo {year} {1986}),\ \bibfield
  {title} {\enquote {\bibinfo {title} {{General implementation of screening
  effects in thermonuclear reaction rates}},}\ }in\ \href@noop {} {\emph
  {\bibinfo {booktitle} {Advances in Nuclear Astrophysics}}},\ \bibinfo
  {editor} {edited by\ \bibinfo {editor} {\bibfnamefont {Elisabeth}\
  \bibnamefont {{Vangioni-Flam}}}, \bibinfo {editor} {\bibfnamefont {Jean}\
  \bibnamefont {{Audouze}}}, \bibinfo {editor} {\bibfnamefont {Michel}\
  \bibnamefont {{Casse}}}, \bibinfo {editor} {\bibfnamefont {Jean-Pierre}\
  \bibnamefont {{Chieze}}}, \ and\ \bibinfo {editor} {\bibfnamefont
  {J.}~\bibnamefont {{Tran Thanh Van}}}},\ pp.\ \bibinfo {pages}
  {541--551}\BibitemShut {NoStop}%
\bibitem [{\citenamefont {Thompson}\ \emph {et~al.}(2001)\citenamefont
  {Thompson}, \citenamefont {Burrows},\ and\ \citenamefont
  {Meyer}}]{Thompson.Burrows.Meyer:2001}%
  \BibitemOpen
  \bibfield  {author} {\bibinfo {author} {\bibnamefont {Thompson},
  \bibfnamefont {T~A}}, \bibinfo {author} {\bibfnamefont {A.}~\bibnamefont
  {Burrows}}, and\ \bibinfo {author} {\bibfnamefont {B.~S.}\ \bibnamefont
  {Meyer}}} (\bibinfo {year} {2001}),\ \bibfield  {title} {\enquote {\bibinfo
  {title} {{The Physics of Proto-Neutron Star Winds: Implications for r-Process
  Nucleosynthesis}},}\ }\href {https://doi.org/10.1086/323861} {\bibfield
  {journal} {\bibinfo  {journal} {Astrophys. J.}\ }\textbf {\bibinfo {volume}
  {562}},\ \bibinfo {pages} {887--908}}\BibitemShut {NoStop}%
\bibitem [{\citenamefont {{Timmes}}(1999)}]{timmes99}%
  \BibitemOpen
  \bibfield  {author} {\bibinfo {author} {\bibnamefont {{Timmes}},
  \bibfnamefont {F~X}}} (\bibinfo {year} {1999}),\ \bibfield  {title} {\enquote
  {\bibinfo {title} {{Integration of Nuclear Reaction Networks for Stellar
  Hydrodynamics}},}\ }\href {https://doi.org/10.1086/313257} {\bibfield
  {journal} {\bibinfo  {journal} {Astrophys. J. Suppl.}\ }\textbf {\bibinfo
  {volume} {124}},\ \bibinfo {pages} {241--263}}\BibitemShut {NoStop}%
\bibitem [{\citenamefont {Timmes}\ \emph {et~al.}(1995)\citenamefont {Timmes},
  \citenamefont {Woosley},\ and\ \citenamefont
  {Weaver}}]{Timmes.Woosley.Weaver:1995}%
  \BibitemOpen
  \bibfield  {author} {\bibinfo {author} {\bibnamefont {Timmes}, \bibfnamefont
  {F~X}}, \bibinfo {author} {\bibfnamefont {S.~E.}\ \bibnamefont {Woosley}},
  and\ \bibinfo {author} {\bibfnamefont {T.~A.}\ \bibnamefont {Weaver}}}
  (\bibinfo {year} {1995}),\ \bibfield  {title} {\enquote {\bibinfo {title}
  {{Galactic Chemical Evolution: Hydrogen through Zinc}},}\ }\href
  {https://doi.org/10.1086/192172} {\bibfield  {journal} {\bibinfo  {journal}
  {Astrophys. J. Suppl.}\ }\textbf {\bibinfo {volume} {98}},\ \bibinfo {pages}
  {617}}\BibitemShut {NoStop}%
\bibitem [{\citenamefont {{Tinsley}}(1980)}]{Tinsley:1980}%
  \BibitemOpen
  \bibfield  {author} {\bibinfo {author} {\bibnamefont {{Tinsley}},
  \bibfnamefont {B~M}}} (\bibinfo {year} {1980}),\ \bibfield  {title} {\enquote
  {\bibinfo {title} {{Evolution of the Stars and Gas in Galaxies}},}\
  }\href@noop {} {\bibfield  {journal} {\bibinfo  {journal} {Fundamentals of
  Cosmic Physics}\ }\textbf {\bibinfo {volume} {5}},\ \bibinfo {pages}
  {287--388}}\BibitemShut {NoStop}%
\bibitem [{\citenamefont {{Tornyi}}\ \emph {et~al.}(2014)\citenamefont
  {{Tornyi}}, \citenamefont {{Guttormsen}}, \citenamefont {{Eriksen}},
  \citenamefont {{G{\"o}rgen}}, \citenamefont {{Giacoppo}}, \citenamefont
  {{Hagen}}, \citenamefont {{Krasznahorkay}}, \citenamefont {{Larsen}},
  \citenamefont {{Renstr{\o}m}}, \citenamefont {{Rose}}, \citenamefont
  {{Siem}},\ and\ \citenamefont {{Tveten}}}]{Tornyi2014}%
  \BibitemOpen
  \bibfield  {author} {\bibinfo {author} {\bibnamefont {{Tornyi}},
  \bibfnamefont {T~G}}, \bibinfo {author} {\bibfnamefont {M.}~\bibnamefont
  {{Guttormsen}}}, \bibinfo {author} {\bibfnamefont {T.~K.}\ \bibnamefont
  {{Eriksen}}}, \bibinfo {author} {\bibfnamefont {A.}~\bibnamefont
  {{G{\"o}rgen}}}, \bibinfo {author} {\bibfnamefont {F.}~\bibnamefont
  {{Giacoppo}}}, \bibinfo {author} {\bibfnamefont {T.~W.}\ \bibnamefont
  {{Hagen}}}, \bibinfo {author} {\bibfnamefont {A.}~\bibnamefont
  {{Krasznahorkay}}}, \bibinfo {author} {\bibfnamefont {A.~C.}\ \bibnamefont
  {{Larsen}}}, \bibinfo {author} {\bibfnamefont {T.}~\bibnamefont
  {{Renstr{\o}m}}}, \bibinfo {author} {\bibfnamefont {S.~J.}\ \bibnamefont
  {{Rose}}}, \bibinfo {author} {\bibfnamefont {S.}~\bibnamefont {{Siem}}}, and\
  \bibinfo {author} {\bibfnamefont {G.~M.}\ \bibnamefont {{Tveten}}}} (\bibinfo
  {year} {2014}),\ \bibfield  {title} {\enquote {\bibinfo {title} {{Level
  density and {$\gamma$}-ray strength function in the odd-odd Np238
  nucleus}},}\ }\href {https://doi.org/10.1103/PhysRevC.89.044323} {\bibfield
  {journal} {\bibinfo  {journal} {Phys. Rev. C}\ }\textbf {\bibinfo {volume}
  {89}},\ \bibinfo {eid} {044323}}\BibitemShut {NoStop}%
\bibitem [{\citenamefont {{Travaglio}}\ \emph {et~al.}(1999)\citenamefont
  {{Travaglio}}, \citenamefont {{Galli}}, \citenamefont {{Gallino}},
  \citenamefont {{Busso}}, \citenamefont {{Ferrini}},\ and\ \citenamefont
  {{Straniero}}}]{Travaglio.Galli.Gallino.ea:1999}%
  \BibitemOpen
  \bibfield  {author} {\bibinfo {author} {\bibnamefont {{Travaglio}},
  \bibfnamefont {C}}, \bibinfo {author} {\bibfnamefont {D.}~\bibnamefont
  {{Galli}}}, \bibinfo {author} {\bibfnamefont {R.}~\bibnamefont {{Gallino}}},
  \bibinfo {author} {\bibfnamefont {M.}~\bibnamefont {{Busso}}}, \bibinfo
  {author} {\bibfnamefont {F.}~\bibnamefont {{Ferrini}}}, and\ \bibinfo
  {author} {\bibfnamefont {O.}~\bibnamefont {{Straniero}}}} (\bibinfo {year}
  {1999}),\ \bibfield  {title} {\enquote {\bibinfo {title} {{Galactic Chemical
  Evolution of Heavy Elements: From Barium to Europium}},}\ }\href
  {https://doi.org/10.1086/307571} {\bibfield  {journal} {\bibinfo  {journal}
  {Astrophys. J.}\ }\textbf {\bibinfo {volume} {521}},\ \bibinfo {pages}
  {691--702}}\BibitemShut {NoStop}%
\bibitem [{\citenamefont {{Travaglio}}\ \emph {et~al.}(2004)\citenamefont
  {{Travaglio}}, \citenamefont {{Gallino}}, \citenamefont {{Arnone}},
  \citenamefont {{Cowan}}, \citenamefont {{Jordan}},\ and\ \citenamefont
  {{Sneden}}}]{Travaglio.Gallino.ea:2004}%
  \BibitemOpen
  \bibfield  {author} {\bibinfo {author} {\bibnamefont {{Travaglio}},
  \bibfnamefont {C}}, \bibinfo {author} {\bibfnamefont {R.}~\bibnamefont
  {{Gallino}}}, \bibinfo {author} {\bibfnamefont {E.}~\bibnamefont {{Arnone}}},
  \bibinfo {author} {\bibfnamefont {J.}~\bibnamefont {{Cowan}}}, \bibinfo
  {author} {\bibfnamefont {F.}~\bibnamefont {{Jordan}}}, and\ \bibinfo {author}
  {\bibfnamefont {C.}~\bibnamefont {{Sneden}}}} (\bibinfo {year} {2004}),\
  \bibfield  {title} {\enquote {\bibinfo {title} {{Galactic Evolution of Sr, Y,
  And Zr: A Multiplicity of Nucleosynthetic Processes}},}\ }\href
  {https://doi.org/10.1086/380507} {\bibfield  {journal} {\bibinfo  {journal}
  {Astrophys. J.}\ }\textbf {\bibinfo {volume} {601}},\ \bibinfo {pages}
  {864--884}}\BibitemShut {NoStop}%
\bibitem [{\citenamefont {{Travaglio}}\ \emph {et~al.}(2018)\citenamefont
  {{Travaglio}}, \citenamefont {{Rauscher}}, \citenamefont {{Heger}},
  \citenamefont {{Pignatari}},\ and\ \citenamefont
  {{West}}}]{Travaglio.Rauscher.ea:2018}%
  \BibitemOpen
  \bibfield  {author} {\bibinfo {author} {\bibnamefont {{Travaglio}},
  \bibfnamefont {C}}, \bibinfo {author} {\bibfnamefont {T.}~\bibnamefont
  {{Rauscher}}}, \bibinfo {author} {\bibfnamefont {A.}~\bibnamefont {{Heger}}},
  \bibinfo {author} {\bibfnamefont {M.}~\bibnamefont {{Pignatari}}}, and\
  \bibinfo {author} {\bibfnamefont {C.}~\bibnamefont {{West}}}} (\bibinfo
  {year} {2018}),\ \bibfield  {title} {\enquote {\bibinfo {title} {{Role of
  Core-collapse Supernovae in Explaining Solar System Abundances of p
  Nuclides}},}\ }\href {https://doi.org/10.3847/1538-4357/aaa4f7} {\bibfield
  {journal} {\bibinfo  {journal} {Astrophys. J.}\ }\textbf {\bibinfo {volume}
  {854}},\ \bibinfo {eid} {18}}\BibitemShut {NoStop}%
\bibitem [{\citenamefont {{Truran}}\ \emph {et~al.}(1978)\citenamefont
  {{Truran}}, \citenamefont {{Cowan}},\ and\ \citenamefont
  {{Cameron}}}]{Truran.Cowan.Cameron:1978}%
  \BibitemOpen
  \bibfield  {author} {\bibinfo {author} {\bibnamefont {{Truran}},
  \bibfnamefont {J~W}}, \bibinfo {author} {\bibfnamefont {J.~J.}\ \bibnamefont
  {{Cowan}}}, and\ \bibinfo {author} {\bibfnamefont {A.~G.~W.}\ \bibnamefont
  {{Cameron}}}} (\bibinfo {year} {1978}),\ \bibfield  {title} {\enquote
  {\bibinfo {title} {{The helium-driven r-process in supernovae}},}\ }\href
  {https://doi.org/10.1086/182693} {\bibfield  {journal} {\bibinfo  {journal}
  {Astrophys. J.}\ }\textbf {\bibinfo {volume} {222}},\ \bibinfo {pages}
  {L63--L67}}\BibitemShut {NoStop}%
\bibitem [{\citenamefont {{Tsujimoto}}\ and\ \citenamefont
  {{Shigeyama}}(2014)}]{Tsujimoto.Shigeyama:2014}%
  \BibitemOpen
  \bibfield  {author} {\bibinfo {author} {\bibnamefont {{Tsujimoto}},
  \bibfnamefont {T}}, and\ \bibinfo {author} {\bibfnamefont {T.}~\bibnamefont
  {{Shigeyama}}}} (\bibinfo {year} {2014}),\ \bibfield  {title} {\enquote
  {\bibinfo {title} {{Enrichment history of r-process elements shaped by a
  merger of neutron star pairs}},}\ }\href
  {https://doi.org/10.1051/0004-6361/201423751} {\bibfield  {journal} {\bibinfo
   {journal} {Astron. \& Astrophys.}\ }\textbf {\bibinfo {volume} {565}},\
  \bibinfo {eid} {L5}}\BibitemShut {NoStop}%
\bibitem [{\citenamefont {{Tsujimoto}}\ \emph {et~al.}(1997)\citenamefont
  {{Tsujimoto}}, \citenamefont {{Yoshii}}, \citenamefont {{Nomoto}},
  \citenamefont {{Matteucci}}, \citenamefont {{Thielemann}},\ and\
  \citenamefont {{Hashimoto}}}]{Tsujimoto.Yoshii.ea:1997}%
  \BibitemOpen
  \bibfield  {author} {\bibinfo {author} {\bibnamefont {{Tsujimoto}},
  \bibfnamefont {T}}, \bibinfo {author} {\bibfnamefont {Y.}~\bibnamefont
  {{Yoshii}}}, \bibinfo {author} {\bibfnamefont {K.}~\bibnamefont {{Nomoto}}},
  \bibinfo {author} {\bibfnamefont {F.}~\bibnamefont {{Matteucci}}}, \bibinfo
  {author} {\bibfnamefont {F.-K.}\ \bibnamefont {{Thielemann}}}, and\ \bibinfo
  {author} {\bibfnamefont {M.}~\bibnamefont {{Hashimoto}}}} (\bibinfo {year}
  {1997}),\ \bibfield  {title} {\enquote {\bibinfo {title} {{A New Approach to
  Determine the Initial Mass Function in the Solar Neighborhood}},}\ }\href
  {https://doi.org/10.1086/304215} {\bibfield  {journal} {\bibinfo  {journal}
  {Astrophys. J.}\ }\textbf {\bibinfo {volume} {483}},\ \bibinfo {pages}
  {228--234}}\BibitemShut {NoStop}%
\bibitem [{\citenamefont {{Ural}}\ \emph {et~al.}(2015)\citenamefont {{Ural}},
  \citenamefont {{Cescutti}}, \citenamefont {{Koch}}, \citenamefont {{Kleyna}},
  \citenamefont {{Feltzing}},\ and\ \citenamefont
  {{Wilkinson}}}]{Ural.Cescutti.ea:2015}%
  \BibitemOpen
  \bibfield  {author} {\bibinfo {author} {\bibnamefont {{Ural}}, \bibfnamefont
  {U}}, \bibinfo {author} {\bibfnamefont {G.}~\bibnamefont {{Cescutti}}},
  \bibinfo {author} {\bibfnamefont {A.}~\bibnamefont {{Koch}}}, \bibinfo
  {author} {\bibfnamefont {J.}~\bibnamefont {{Kleyna}}}, \bibinfo {author}
  {\bibfnamefont {S.}~\bibnamefont {{Feltzing}}}, and\ \bibinfo {author}
  {\bibfnamefont {M.~I.}\ \bibnamefont {{Wilkinson}}}} (\bibinfo {year}
  {2015}),\ \bibfield  {title} {\enquote {\bibinfo {title} {{An inefficient
  dwarf: chemical abundances and the evolution of the Ursa Minor dwarf
  spheroidal galaxy}},}\ }\href {https://doi.org/10.1093/mnras/stv294}
  {\bibfield  {journal} {\bibinfo  {journal} {Mon. Not. Roy. Astron. Soc.}\
  }\textbf {\bibinfo {volume} {449}},\ \bibinfo {pages} {761--770}}\BibitemShut
  {NoStop}%
\bibitem [{\citenamefont {{van de Voort}}\ \emph {et~al.}(2015)\citenamefont
  {{van de Voort}}, \citenamefont {{Quataert}}, \citenamefont {{Hopkins}},
  \citenamefont {{Kere{\v s}}},\ and\ \citenamefont
  {{Faucher-Gigu{\`e}re}}}]{vandervoort15}%
  \BibitemOpen
  \bibfield  {author} {\bibinfo {author} {\bibnamefont {{van de Voort}},
  \bibfnamefont {F}}, \bibinfo {author} {\bibfnamefont {E.}~\bibnamefont
  {{Quataert}}}, \bibinfo {author} {\bibfnamefont {P.~F.}\ \bibnamefont
  {{Hopkins}}}, \bibinfo {author} {\bibfnamefont {D.}~\bibnamefont {{Kere{\v
  s}}}}, and\ \bibinfo {author} {\bibfnamefont {C.-A.}\ \bibnamefont
  {{Faucher-Gigu{\`e}re}}}} (\bibinfo {year} {2015}),\ \bibfield  {title}
  {\enquote {\bibinfo {title} {{Galactic r-process enrichment by neutron star
  mergers in cosmological simulations of a Milky Way-mass galaxy}},}\ }\href
  {https://doi.org/10.1093/mnras/stu2404} {\bibfield  {journal} {\bibinfo
  {journal} {Mon. Not. Roy. Astron. Soc.}\ }\textbf {\bibinfo {volume} {447}},\
  \bibinfo {pages} {140--148}}\BibitemShut {NoStop}%
\bibitem [{\citenamefont {{van de Voort}}\ \emph {et~al.}(2020)\citenamefont
  {{van de Voort}}, \citenamefont {{Pakmor}}, \citenamefont {{Grand}},
  \citenamefont {{Springel}}, \citenamefont {{G{\'o}mez}},\ and\ \citenamefont
  {{Marinacci}}}]{VandeVoort.ea:2019}%
  \BibitemOpen
  \bibfield  {author} {\bibinfo {author} {\bibnamefont {{van de Voort}},
  \bibfnamefont {Freeke}}, \bibinfo {author} {\bibfnamefont {R{\"u}diger}\
  \bibnamefont {{Pakmor}}}, \bibinfo {author} {\bibfnamefont {Robert J.~J.}\
  \bibnamefont {{Grand}}}, \bibinfo {author} {\bibfnamefont {Volker}\
  \bibnamefont {{Springel}}}, \bibinfo {author} {\bibfnamefont {Facundo~A.}\
  \bibnamefont {{G{\'o}mez}}}, and\ \bibinfo {author} {\bibfnamefont
  {Federico}\ \bibnamefont {{Marinacci}}}} (\bibinfo {year} {2020}),\ \bibfield
   {title} {\enquote {\bibinfo {title} {{Neutron star mergers and rare
  core-collapse supernovae as sources of r-process enrichment in simulated
  galaxies}},}\ }\href {https://doi.org/10.1093/mnras/staa754} {\bibfield
  {journal} {\bibinfo  {journal} {Mon. Not. Roy. Astron. Soc.}\ }\textbf
  {\bibinfo {volume} {494}}~(\bibinfo {number} {4}),\ \bibinfo {pages}
  {4867--4883}}\BibitemShut {NoStop}%
\bibitem [{\citenamefont {{Vangioni}}\ \emph {et~al.}(2016)\citenamefont
  {{Vangioni}}, \citenamefont {{Goriely}}, \citenamefont {{Daigne}},
  \citenamefont {{Fran{\c c}ois}},\ and\ \citenamefont
  {{Belczynski}}}]{Vangioni.ea:2016}%
  \BibitemOpen
  \bibfield  {author} {\bibinfo {author} {\bibnamefont {{Vangioni}},
  \bibfnamefont {E}}, \bibinfo {author} {\bibfnamefont {S.}~\bibnamefont
  {{Goriely}}}, \bibinfo {author} {\bibfnamefont {F.}~\bibnamefont {{Daigne}}},
  \bibinfo {author} {\bibfnamefont {P.}~\bibnamefont {{Fran{\c c}ois}}}, and\
  \bibinfo {author} {\bibfnamefont {K.}~\bibnamefont {{Belczynski}}}} (\bibinfo
  {year} {2016}),\ \bibfield  {title} {\enquote {\bibinfo {title} {{Cosmic
  neutron-star merger rate and gravitational waves constrained by the r-process
  nucleosynthesis}},}\ }\href {https://doi.org/10.1093/mnras/stv2296}
  {\bibfield  {journal} {\bibinfo  {journal} {Mon. Not. Roy. Astron. Soc.}\
  }\textbf {\bibinfo {volume} {455}},\ \bibinfo {pages} {17--34}}\BibitemShut
  {NoStop}%
\bibitem [{\citenamefont {{Vassh}}\ \emph {et~al.}(2019)\citenamefont
  {{Vassh}}, \citenamefont {{Vogt}}, \citenamefont {{Surman}}, \citenamefont
  {{Randrup}}, \citenamefont {{Sprouse}}, \citenamefont {{Mumpower}},
  \citenamefont {{Jaffke}}, \citenamefont {{Shaw}}, \citenamefont {{Holmbeck}},
  \citenamefont {{Zhu}},\ and\ \citenamefont
  {{McLaughlin}}}]{Vassh.Vogt.ea:2019}%
  \BibitemOpen
  \bibfield  {author} {\bibinfo {author} {\bibnamefont {{Vassh}}, \bibfnamefont
  {N}}, \bibinfo {author} {\bibfnamefont {R.}~\bibnamefont {{Vogt}}}, \bibinfo
  {author} {\bibfnamefont {R.}~\bibnamefont {{Surman}}}, \bibinfo {author}
  {\bibfnamefont {J.}~\bibnamefont {{Randrup}}}, \bibinfo {author}
  {\bibfnamefont {T.~M.}\ \bibnamefont {{Sprouse}}}, \bibinfo {author}
  {\bibfnamefont {M.~R.}\ \bibnamefont {{Mumpower}}}, \bibinfo {author}
  {\bibfnamefont {P.}~\bibnamefont {{Jaffke}}}, \bibinfo {author}
  {\bibfnamefont {D.}~\bibnamefont {{Shaw}}}, \bibinfo {author} {\bibfnamefont
  {E.~M.}\ \bibnamefont {{Holmbeck}}}, \bibinfo {author} {\bibfnamefont
  {Y.}~\bibnamefont {{Zhu}}}, and\ \bibinfo {author} {\bibfnamefont {G.~C.}\
  \bibnamefont {{McLaughlin}}}} (\bibinfo {year} {2019}),\ \bibfield  {title}
  {\enquote {\bibinfo {title} {{Using excitation-energy dependent fission
  yields to identify key fissioning nuclei in r-process nucleosynthesis}},}\
  }\href {https://doi.org/10.1088/1361-6471/ab0bea} {\bibfield  {journal}
  {\bibinfo  {journal} {J. Phys. G: Nucl. Part. Phys.}\ }\textbf {\bibinfo
  {volume} {46}},\ \bibinfo {pages} {065202}}\BibitemShut {NoStop}%
\bibitem [{\citenamefont {{Vassh}}\ \emph {et~al.}(2020)\citenamefont
  {{Vassh}}, \citenamefont {{Mumpower}}, \citenamefont {{McLaughlin}},
  \citenamefont {{Sprouse}},\ and\ \citenamefont
  {{Surman}}}]{Vassh.Mumpower.ea:2019}%
  \BibitemOpen
  \bibfield  {author} {\bibinfo {author} {\bibnamefont {{Vassh}}, \bibfnamefont
  {Nicole}}, \bibinfo {author} {\bibfnamefont {Matthew~R.}\ \bibnamefont
  {{Mumpower}}}, \bibinfo {author} {\bibfnamefont {Gail~C.}\ \bibnamefont
  {{McLaughlin}}}, \bibinfo {author} {\bibfnamefont {Trevor~M.}\ \bibnamefont
  {{Sprouse}}}, and\ \bibinfo {author} {\bibfnamefont {Rebecca}\ \bibnamefont
  {{Surman}}}} (\bibinfo {year} {2020}),\ \bibfield  {title} {\enquote
  {\bibinfo {title} {{Co-production of light and heavy $r$-process elements via
  fission deposition}},}\ }\href {https://doi.org/10.3847/1538-4357/ab91a9}
  {\bibfield  {journal} {\bibinfo  {journal} {Astrophys. J.}\ }\textbf
  {\bibinfo {volume} {896}}~(\bibinfo {number} {1}),\ \bibinfo {eid}
  {28}}\BibitemShut {NoStop}%
\bibitem [{\citenamefont {{Venn}}\ \emph {et~al.}(2003)\citenamefont {{Venn}},
  \citenamefont {{Tolstoy}}, \citenamefont {{Kaufer}}, \citenamefont
  {{Skillman}}, \citenamefont {{Clarkson}}, \citenamefont {{Smartt}},
  \citenamefont {{Lennon}},\ and\ \citenamefont
  {{Kudritzki}}}]{Venn.Tolstoy.ea:2003}%
  \BibitemOpen
  \bibfield  {author} {\bibinfo {author} {\bibnamefont {{Venn}}, \bibfnamefont
  {K~A}}, \bibinfo {author} {\bibfnamefont {E.}~\bibnamefont {{Tolstoy}}},
  \bibinfo {author} {\bibfnamefont {A.}~\bibnamefont {{Kaufer}}}, \bibinfo
  {author} {\bibfnamefont {E.~D.}\ \bibnamefont {{Skillman}}}, \bibinfo
  {author} {\bibfnamefont {S.~M.}\ \bibnamefont {{Clarkson}}}, \bibinfo
  {author} {\bibfnamefont {S.~J.}\ \bibnamefont {{Smartt}}}, \bibinfo {author}
  {\bibfnamefont {D.~J.}\ \bibnamefont {{Lennon}}}, and\ \bibinfo {author}
  {\bibfnamefont {R.~P.}\ \bibnamefont {{Kudritzki}}}} (\bibinfo {year}
  {2003}),\ \bibfield  {title} {\enquote {\bibinfo {title} {{The Chemical
  Composition of Two Supergiants in the Dwarf Irregular Galaxy WLM}},}\ }\href
  {https://doi.org/10.1086/377345} {\bibfield  {journal} {\bibinfo  {journal}
  {Astron. J.}\ }\textbf {\bibinfo {volume} {126}},\ \bibinfo {pages}
  {1326--1345}}\BibitemShut {NoStop}%
\bibitem [{\citenamefont {{Vescovi}}\ \emph {et~al.}(2018)\citenamefont
  {{Vescovi}} \emph {et~al.}}]{Vescavi.ea:2018}%
  \BibitemOpen
  \bibfield  {author} {\bibinfo {author} {\bibnamefont {{Vescovi}},
  \bibfnamefont {D}},  \emph {et~al.}} (\bibinfo {year} {2018}),\ \bibfield
  {title} {\enquote {\bibinfo {title} {{On the Origin of Early Solar System
  Radioactivities: Problems with the Asymptotic Giant Branch and Massive Star
  Scenarios}},}\ }\href {https://doi.org/10.3847/1538-4357/aad191} {\bibfield
  {journal} {\bibinfo  {journal} {Astrophys. J.}\ }\textbf {\bibinfo {volume}
  {863}}~(\bibinfo {number} {2}),\ \bibinfo {eid} {115}}\BibitemShut {NoStop}%
\bibitem [{\citenamefont {{Vigna-G{\'o}mez}}\ \emph {et~al.}(2018)\citenamefont
  {{Vigna-G{\'o}mez}}, \citenamefont {{Neijssel}}, \citenamefont {{Stevenson}},
  \citenamefont {{Barrett}}, \citenamefont {{Belczynski}}, \citenamefont
  {{Justham}}, \citenamefont {{de Mink}}, \citenamefont {{M{\"u}ller}},
  \citenamefont {{Podsiadlowski}}, \citenamefont {{Renzo}}, \citenamefont
  {{Sz{\'e}csi}},\ and\ \citenamefont {{Mandel}}}]{Vigna-Gomez.ea:2018}%
  \BibitemOpen
  \bibfield  {author} {\bibinfo {author} {\bibnamefont {{Vigna-G{\'o}mez}},
  \bibfnamefont {Alejandro}}, \bibinfo {author} {\bibfnamefont {Coenraad~J.}\
  \bibnamefont {{Neijssel}}}, \bibinfo {author} {\bibfnamefont {Simon}\
  \bibnamefont {{Stevenson}}}, \bibinfo {author} {\bibfnamefont {Jim~W.}\
  \bibnamefont {{Barrett}}}, \bibinfo {author} {\bibfnamefont {Krzysztof}\
  \bibnamefont {{Belczynski}}}, \bibinfo {author} {\bibfnamefont {Stephen}\
  \bibnamefont {{Justham}}}, \bibinfo {author} {\bibfnamefont {Selma~E.}\
  \bibnamefont {{de Mink}}}, \bibinfo {author} {\bibfnamefont {Bernhard}\
  \bibnamefont {{M{\"u}ller}}}, \bibinfo {author} {\bibfnamefont {Philipp}\
  \bibnamefont {{Podsiadlowski}}}, \bibinfo {author} {\bibfnamefont {Mathieu}\
  \bibnamefont {{Renzo}}}, \bibinfo {author} {\bibfnamefont {Dorottya}\
  \bibnamefont {{Sz{\'e}csi}}}, and\ \bibinfo {author} {\bibfnamefont {Ilya}\
  \bibnamefont {{Mandel}}}} (\bibinfo {year} {2018}),\ \bibfield  {title}
  {\enquote {\bibinfo {title} {{On the formation history of Galactic double
  neutron stars}},}\ }\href {https://doi.org/10.1093/mnras/sty2463} {\bibfield
  {journal} {\bibinfo  {journal} {Mon. Not. Roy. Astron. Soc.}\ }\textbf
  {\bibinfo {volume} {481}},\ \bibinfo {pages} {4009--4029}}\BibitemShut
  {NoStop}%
\bibitem [{\citenamefont {{Vilen}}\ \emph {et~al.}(2018)\citenamefont
  {{Vilen}}, \citenamefont {{Kelly}}, \citenamefont {{Kankainen}},
  \citenamefont {{Brodeur}}, \citenamefont {{Aprahamian}}, \citenamefont
  {{Canete}}, \citenamefont {{Eronen}}, \citenamefont {{Jokinen}},
  \citenamefont {{Kuta}}, \citenamefont {{Moore}}, \citenamefont {{Mumpower}},
  \citenamefont {{Nesterenko}}, \citenamefont {{Penttil{\"a}}}, \citenamefont
  {{Pohjalainen}}, \citenamefont {{Porter}}, \citenamefont {{Rinta-Antila}},
  \citenamefont {{Surman}}, \citenamefont {{Voss}},\ and\ \citenamefont
  {{{\"A}yst{\"o}}}}]{Vilen.Kelly.ea:2018}%
  \BibitemOpen
  \bibfield  {author} {\bibinfo {author} {\bibnamefont {{Vilen}}, \bibfnamefont
  {M}}, \bibinfo {author} {\bibfnamefont {J.~M.}\ \bibnamefont {{Kelly}}},
  \bibinfo {author} {\bibfnamefont {A.}~\bibnamefont {{Kankainen}}}, \bibinfo
  {author} {\bibfnamefont {M.}~\bibnamefont {{Brodeur}}}, \bibinfo {author}
  {\bibfnamefont {A.}~\bibnamefont {{Aprahamian}}}, \bibinfo {author}
  {\bibfnamefont {L.}~\bibnamefont {{Canete}}}, \bibinfo {author}
  {\bibfnamefont {T.}~\bibnamefont {{Eronen}}}, \bibinfo {author}
  {\bibfnamefont {A.}~\bibnamefont {{Jokinen}}}, \bibinfo {author}
  {\bibfnamefont {T.}~\bibnamefont {{Kuta}}}, \bibinfo {author} {\bibfnamefont
  {I.~D.}\ \bibnamefont {{Moore}}}, \bibinfo {author} {\bibfnamefont {M.~R.}\
  \bibnamefont {{Mumpower}}}, \bibinfo {author} {\bibfnamefont {D.~A.}\
  \bibnamefont {{Nesterenko}}}, \bibinfo {author} {\bibfnamefont
  {H.}~\bibnamefont {{Penttil{\"a}}}}, \bibinfo {author} {\bibfnamefont
  {I.}~\bibnamefont {{Pohjalainen}}}, \bibinfo {author} {\bibfnamefont {W.~S.}\
  \bibnamefont {{Porter}}}, \bibinfo {author} {\bibfnamefont {S.}~\bibnamefont
  {{Rinta-Antila}}}, \bibinfo {author} {\bibfnamefont {R.}~\bibnamefont
  {{Surman}}}, \bibinfo {author} {\bibfnamefont {A.}~\bibnamefont {{Voss}}},
  and\ \bibinfo {author} {\bibfnamefont {J.}~\bibnamefont {{{\"A}yst{\"o}}}}}
  (\bibinfo {year} {2018}),\ \bibfield  {title} {\enquote {\bibinfo {title}
  {{Precision Mass Measurements on Neutron-Rich Rare-Earth Isotopes at
  JYFLTRAP: Reduced Neutron Pairing and Implications for r-Process
  Calculations}},}\ }\href {https://doi.org/10.1103/PhysRevLett.120.262701}
  {\bibfield  {journal} {\bibinfo  {journal} {Phys. Rev. Lett.}\ }\textbf
  {\bibinfo {volume} {120}},\ \bibinfo {eid} {262701}}\BibitemShut {NoStop}%
\bibitem [{\citenamefont {{Villar}}\ \emph {et~al.}(2018)\citenamefont
  {{Villar}}, \citenamefont {{Cowperthwaite}}, \citenamefont {{Berger}},
  \citenamefont {{Blanchard}}, \citenamefont {{Gomez}}, \citenamefont
  {{Alexander}}, \citenamefont {{Margutti}}, \citenamefont {{Chornock}},
  \citenamefont {{Eftekhari}}, \citenamefont {{Fazio}}, \citenamefont
  {{Guillochon}}, \citenamefont {{Hora}}, \citenamefont {{Nicholl}},\ and\
  \citenamefont {{Williams}}}]{Villar.Others:2018}%
  \BibitemOpen
  \bibfield  {author} {\bibinfo {author} {\bibnamefont {{Villar}},
  \bibfnamefont {V~A}}, \bibinfo {author} {\bibfnamefont {P.~S.}\ \bibnamefont
  {{Cowperthwaite}}}, \bibinfo {author} {\bibfnamefont {E.}~\bibnamefont
  {{Berger}}}, \bibinfo {author} {\bibfnamefont {P.~K.}\ \bibnamefont
  {{Blanchard}}}, \bibinfo {author} {\bibfnamefont {S.}~\bibnamefont
  {{Gomez}}}, \bibinfo {author} {\bibfnamefont {K.~D.}\ \bibnamefont
  {{Alexander}}}, \bibinfo {author} {\bibfnamefont {R.}~\bibnamefont
  {{Margutti}}}, \bibinfo {author} {\bibfnamefont {R.}~\bibnamefont
  {{Chornock}}}, \bibinfo {author} {\bibfnamefont {T.}~\bibnamefont
  {{Eftekhari}}}, \bibinfo {author} {\bibfnamefont {G.~G.}\ \bibnamefont
  {{Fazio}}}, \bibinfo {author} {\bibfnamefont {J.}~\bibnamefont
  {{Guillochon}}}, \bibinfo {author} {\bibfnamefont {J.~L.}\ \bibnamefont
  {{Hora}}}, \bibinfo {author} {\bibfnamefont {M.}~\bibnamefont {{Nicholl}}},
  and\ \bibinfo {author} {\bibfnamefont {P.~K.~G.}\ \bibnamefont {{Williams}}}}
  (\bibinfo {year} {2018}),\ \bibfield  {title} {\enquote {\bibinfo {title}
  {{Spitzer Space Telescope Infrared Observations of the Binary Neutron Star
  Merger GW170817}},}\ }\href {https://doi.org/10.3847/2041-8213/aad281}
  {\bibfield  {journal} {\bibinfo  {journal} {Astrophys. J. Lett.}\ }\textbf
  {\bibinfo {volume} {862}},\ \bibinfo {eid} {L11}}\BibitemShut {NoStop}%
\bibitem [{\citenamefont {{Villar}}\ \emph {et~al.}(2017)\citenamefont
  {{Villar}}, \citenamefont {{Guillochon}}, \citenamefont {{Berger}},
  \citenamefont {{Metzger}}, \citenamefont {{Cowperthwaite}}, \citenamefont
  {{Nicholl}}, \citenamefont {{Alexander}}, \citenamefont {{Blanchard}},
  \citenamefont {{Chornock}}, \citenamefont {{Eftekhari}}, \citenamefont
  {{Fong}}, \citenamefont {{Margutti}},\ and\ \citenamefont
  {{Williams}}}]{Villar.ea:2017}%
  \BibitemOpen
  \bibfield  {author} {\bibinfo {author} {\bibnamefont {{Villar}},
  \bibfnamefont {V~A}}, \bibinfo {author} {\bibfnamefont {J.}~\bibnamefont
  {{Guillochon}}}, \bibinfo {author} {\bibfnamefont {E.}~\bibnamefont
  {{Berger}}}, \bibinfo {author} {\bibfnamefont {B.~D.}\ \bibnamefont
  {{Metzger}}}, \bibinfo {author} {\bibfnamefont {P.~S.}\ \bibnamefont
  {{Cowperthwaite}}}, \bibinfo {author} {\bibfnamefont {M.}~\bibnamefont
  {{Nicholl}}}, \bibinfo {author} {\bibfnamefont {K.~D.}\ \bibnamefont
  {{Alexander}}}, \bibinfo {author} {\bibfnamefont {P.~K.}\ \bibnamefont
  {{Blanchard}}}, \bibinfo {author} {\bibfnamefont {R.}~\bibnamefont
  {{Chornock}}}, \bibinfo {author} {\bibfnamefont {T.}~\bibnamefont
  {{Eftekhari}}}, \bibinfo {author} {\bibfnamefont {W.}~\bibnamefont {{Fong}}},
  \bibinfo {author} {\bibfnamefont {R.}~\bibnamefont {{Margutti}}}, and\
  \bibinfo {author} {\bibfnamefont {P.~K.~G.}\ \bibnamefont {{Williams}}}}
  (\bibinfo {year} {2017}),\ \bibfield  {title} {\enquote {\bibinfo {title}
  {{The Combined Ultraviolet, Optical, and Near-infrared Light Curves of the
  Kilonova Associated with the Binary Neutron Star Merger GW170817: Unified
  Data Set, Analytic Models, and Physical Implications}},}\ }\href
  {https://doi.org/10.3847/2041-8213/aa9c84} {\bibfield  {journal} {\bibinfo
  {journal} {Astrophys. J. Lett.}\ }\textbf {\bibinfo {volume} {851}},\
  \bibinfo {eid} {L21}}\BibitemShut {NoStop}%
\bibitem [{\citenamefont {{Vink}}\ \emph {et~al.}(2001)\citenamefont {{Vink}},
  \citenamefont {{Laming}}, \citenamefont {{Kaastra}}, \citenamefont
  {{Bleeker}}, \citenamefont {{Bloemen}},\ and\ \citenamefont
  {{Oberlack}}}]{Vink.Laming.ea:2001}%
  \BibitemOpen
  \bibfield  {author} {\bibinfo {author} {\bibnamefont {{Vink}}, \bibfnamefont
  {J}}, \bibinfo {author} {\bibfnamefont {J.~M.}\ \bibnamefont {{Laming}}},
  \bibinfo {author} {\bibfnamefont {J.~S.}\ \bibnamefont {{Kaastra}}}, \bibinfo
  {author} {\bibfnamefont {J.~A.~M.}\ \bibnamefont {{Bleeker}}}, \bibinfo
  {author} {\bibfnamefont {H.}~\bibnamefont {{Bloemen}}}, and\ \bibinfo
  {author} {\bibfnamefont {U.}~\bibnamefont {{Oberlack}}}} (\bibinfo {year}
  {2001}),\ \bibfield  {title} {\enquote {\bibinfo {title} {{Detection of the
  67.9 and 78.4 keV Lines Associated with the Radioactive Decay of $^{44}$Ti in
  Cassiopeia A}},}\ }\href {https://doi.org/10.1086/324172} {\bibfield
  {journal} {\bibinfo  {journal} {Astrophys. J. Lett.}\ }\textbf {\bibinfo
  {volume} {560}},\ \bibinfo {pages} {L79--L82}}\BibitemShut {NoStop}%
\bibitem [{\citenamefont {{Vink}}(2008)}]{Vink:2008}%
  \BibitemOpen
  \bibfield  {author} {\bibinfo {author} {\bibnamefont {{Vink}}, \bibfnamefont
  {Jacco}}} (\bibinfo {year} {2008}),\ \bibfield  {title} {\enquote {\bibinfo
  {title} {{Supernova remnants with magnetars: Clues to magnetar formation}},}\
  }\href {https://doi.org/10.1016/j.asr.2007.06.042} {\bibfield  {journal}
  {\bibinfo  {journal} {Advances in Space Research}\ }\textbf {\bibinfo
  {volume} {41}},\ \bibinfo {pages} {503--511}}\BibitemShut {NoStop}%
\bibitem [{\citenamefont {Vlasenko}\ and\ \citenamefont
  {McLaughlin}(2018)}]{Vlasenko.Mclaughlin:2018}%
  \BibitemOpen
  \bibfield  {author} {\bibinfo {author} {\bibnamefont {Vlasenko},
  \bibfnamefont {Alexey}}, and\ \bibinfo {author} {\bibfnamefont {G.~C.}\
  \bibnamefont {McLaughlin}}} (\bibinfo {year} {2018}),\ \bibfield  {title}
  {\enquote {\bibinfo {title} {{Matter-neutrino resonance in a multiangle
  neutrino bulb model}},}\ }\href {https://doi.org/10.1103/PhysRevD.97.083011}
  {\bibfield  {journal} {\bibinfo  {journal} {Phys. Rev. D}\ }\textbf {\bibinfo
  {volume} {97}},\ \bibinfo {pages} {083011}}\BibitemShut {NoStop}%
\bibitem [{\citenamefont {{Wallner}}\ \emph {et~al.}(2015)\citenamefont
  {{Wallner}}, \citenamefont {{Faestermann}}, \citenamefont {{Feige}},
  \citenamefont {{Feldstein}}, \citenamefont {{Knie}}, \citenamefont
  {{Korschinek}}, \citenamefont {{Kutschera}}, \citenamefont {{Ofan}},
  \citenamefont {{Paul}}, \citenamefont {{Quinto}}, \citenamefont {{Rugel}},\
  and\ \citenamefont {{Steier}}}]{Wallner.Faestermann.ea:2015}%
  \BibitemOpen
  \bibfield  {author} {\bibinfo {author} {\bibnamefont {{Wallner}},
  \bibfnamefont {A}}, \bibinfo {author} {\bibfnamefont {T.}~\bibnamefont
  {{Faestermann}}}, \bibinfo {author} {\bibfnamefont {J.}~\bibnamefont
  {{Feige}}}, \bibinfo {author} {\bibfnamefont {C.}~\bibnamefont
  {{Feldstein}}}, \bibinfo {author} {\bibfnamefont {K.}~\bibnamefont {{Knie}}},
  \bibinfo {author} {\bibfnamefont {G.}~\bibnamefont {{Korschinek}}}, \bibinfo
  {author} {\bibfnamefont {W.}~\bibnamefont {{Kutschera}}}, \bibinfo {author}
  {\bibfnamefont {A.}~\bibnamefont {{Ofan}}}, \bibinfo {author} {\bibfnamefont
  {M.}~\bibnamefont {{Paul}}}, \bibinfo {author} {\bibfnamefont
  {F.}~\bibnamefont {{Quinto}}}, \bibinfo {author} {\bibfnamefont
  {G.}~\bibnamefont {{Rugel}}}, and\ \bibinfo {author} {\bibfnamefont
  {P.}~\bibnamefont {{Steier}}}} (\bibinfo {year} {2015}),\ \bibfield  {title}
  {\enquote {\bibinfo {title} {{Abundance of live $^{244}$Pu in deep-sea
  reservoirs on Earth points to rarity of actinide nucleosynthesis}},}\ }\href
  {https://doi.org/10.1038/ncomms6956} {\bibfield  {journal} {\bibinfo
  {journal} {Nature Communications}\ }\textbf {\bibinfo {volume} {6}},\
  \bibinfo {eid} {5956}}\BibitemShut {NoStop}%
\bibitem [{\citenamefont {{Wallner}}\ \emph {et~al.}(2016)\citenamefont
  {{Wallner}}, \citenamefont {{Feige}}, \citenamefont {{Kinoshita}},
  \citenamefont {{Paul}}, \citenamefont {{Fifield}}, \citenamefont {{Golser}},
  \citenamefont {{Honda}}, \citenamefont {{Linnemann}}, \citenamefont
  {{Matsuzaki}}, \citenamefont {{Merchel}}, \citenamefont {{Rugel}},
  \citenamefont {{Tims}}, \citenamefont {{Steier}}, \citenamefont
  {{Yamagata}},\ and\ \citenamefont {{Winkler}}}]{Wallner:2016}%
  \BibitemOpen
  \bibfield  {author} {\bibinfo {author} {\bibnamefont {{Wallner}},
  \bibfnamefont {A}}, \bibinfo {author} {\bibfnamefont {J.}~\bibnamefont
  {{Feige}}}, \bibinfo {author} {\bibfnamefont {N.}~\bibnamefont
  {{Kinoshita}}}, \bibinfo {author} {\bibfnamefont {M.}~\bibnamefont {{Paul}}},
  \bibinfo {author} {\bibfnamefont {L.~K.}\ \bibnamefont {{Fifield}}}, \bibinfo
  {author} {\bibfnamefont {R.}~\bibnamefont {{Golser}}}, \bibinfo {author}
  {\bibfnamefont {M.}~\bibnamefont {{Honda}}}, \bibinfo {author} {\bibfnamefont
  {U.}~\bibnamefont {{Linnemann}}}, \bibinfo {author} {\bibfnamefont
  {H.}~\bibnamefont {{Matsuzaki}}}, \bibinfo {author} {\bibfnamefont
  {S.}~\bibnamefont {{Merchel}}}, \bibinfo {author} {\bibfnamefont
  {G.}~\bibnamefont {{Rugel}}}, \bibinfo {author} {\bibfnamefont {S.~G.}\
  \bibnamefont {{Tims}}}, \bibinfo {author} {\bibfnamefont {P.}~\bibnamefont
  {{Steier}}}, \bibinfo {author} {\bibfnamefont {T.}~\bibnamefont
  {{Yamagata}}}, and\ \bibinfo {author} {\bibfnamefont {S.~R.}\ \bibnamefont
  {{Winkler}}}} (\bibinfo {year} {2016}),\ \bibfield  {title} {\enquote
  {\bibinfo {title} {{Recent near-Earth supernovae probed by global deposition
  of interstellar radioactive $^{60}$Fe}},}\ }\href
  {https://doi.org/10.1038/nature17196} {\bibfield  {journal} {\bibinfo
  {journal} {Nature}\ }\textbf {\bibinfo {volume} {532}},\ \bibinfo {pages}
  {69--72}}\BibitemShut {NoStop}%
\bibitem [{\citenamefont {{Wallner}}\ \emph {et~al.}(2019)\citenamefont
  {{Wallner}} \emph {et~al.}}]{Wallner.Others:2019}%
  \BibitemOpen
  \bibfield  {author} {\bibinfo {author} {\bibnamefont {{Wallner}},
  \bibfnamefont {A}},  \emph {et~al.}} (\bibinfo {year} {2019}),\ \href@noop {}
  {}\bibinfo {note} {{private communication}}\BibitemShut {NoStop}%
\bibitem [{\citenamefont {{Wanajo}}(2007)}]{Wanajo:2007}%
  \BibitemOpen
  \bibfield  {author} {\bibinfo {author} {\bibnamefont {{Wanajo}},
  \bibfnamefont {S}}} (\bibinfo {year} {2007}),\ \bibfield  {title} {\enquote
  {\bibinfo {title} {{Cold r-Process in Neutrino-driven Winds}},}\ }\href
  {https://doi.org/10.1086/521724} {\bibfield  {journal} {\bibinfo  {journal}
  {Astrophys. J.}\ }\textbf {\bibinfo {volume} {666}},\ \bibinfo {pages}
  {L77--L80}}\BibitemShut {NoStop}%
\bibitem [{\citenamefont {Wanajo}\ and\ \citenamefont
  {Ishimaru}(2006)}]{Wanajo.Ishimaru:2006}%
  \BibitemOpen
  \bibfield  {author} {\bibinfo {author} {\bibnamefont {Wanajo}, \bibfnamefont
  {S}}, and\ \bibinfo {author} {\bibfnamefont {Y.}~\bibnamefont {Ishimaru}}}
  (\bibinfo {year} {2006}),\ \bibfield  {title} {\enquote {\bibinfo {title}
  {{r-process Calculations and Galactic Chemical Evolution}},}\ }\href
  {https://doi.org/10.1016/j.nuclphysa.2005.10.012} {\bibfield  {journal}
  {\bibinfo  {journal} {Nucl. Phys. A}\ }\textbf {\bibinfo {volume} {777}},\
  \bibinfo {pages} {676--699}}\BibitemShut {NoStop}%
\bibitem [{\citenamefont {{Wanajo}}\ and\ \citenamefont
  {{Janka}}(2012)}]{Wanajo.Janka:2012}%
  \BibitemOpen
  \bibfield  {author} {\bibinfo {author} {\bibnamefont {{Wanajo}},
  \bibfnamefont {S}}, and\ \bibinfo {author} {\bibfnamefont {H.-T.}\
  \bibnamefont {{Janka}}}} (\bibinfo {year} {2012}),\ \bibfield  {title}
  {\enquote {\bibinfo {title} {{The r-process in the Neutrino-driven Wind from
  a Black-hole Torus}},}\ }\href {https://doi.org/10.1088/0004-637X/746/2/180}
  {\bibfield  {journal} {\bibinfo  {journal} {Astrophys. J.}\ }\textbf
  {\bibinfo {volume} {746}},\ \bibinfo {eid} {180}}\BibitemShut {NoStop}%
\bibitem [{\citenamefont {{Wanajo}}\ \emph {et~al.}(2013)\citenamefont
  {{Wanajo}}, \citenamefont {{Janka}},\ and\ \citenamefont
  {{M{\"u}ller}}}]{wanajo13}%
  \BibitemOpen
  \bibfield  {author} {\bibinfo {author} {\bibnamefont {{Wanajo}},
  \bibfnamefont {S}}, \bibinfo {author} {\bibfnamefont {H.-T.}\ \bibnamefont
  {{Janka}}}, and\ \bibinfo {author} {\bibfnamefont {B.}~\bibnamefont
  {{M{\"u}ller}}}} (\bibinfo {year} {2013}),\ \bibfield  {title} {\enquote
  {\bibinfo {title} {{Electron-capture Supernovae as Sources of $^{60}$Fe}},}\
  }\href {https://doi.org/10.1088/2041-8205/774/1/L6} {\bibfield  {journal}
  {\bibinfo  {journal} {Astrophys. J.}\ }\textbf {\bibinfo {volume} {774}},\
  \bibinfo {eid} {L6}}\BibitemShut {NoStop}%
\bibitem [{\citenamefont {{Wanajo}}\ \emph {et~al.}(2001)\citenamefont
  {{Wanajo}}, \citenamefont {{Kajino}}, \citenamefont {{Mathews}},\ and\
  \citenamefont {{Otsuki}}}]{Wanajo.Kajino.ea:2001}%
  \BibitemOpen
  \bibfield  {author} {\bibinfo {author} {\bibnamefont {{Wanajo}},
  \bibfnamefont {S}}, \bibinfo {author} {\bibfnamefont {T.}~\bibnamefont
  {{Kajino}}}, \bibinfo {author} {\bibfnamefont {G.~J.}\ \bibnamefont
  {{Mathews}}}, and\ \bibinfo {author} {\bibfnamefont {K.}~\bibnamefont
  {{Otsuki}}}} (\bibinfo {year} {2001}),\ \bibfield  {title} {\enquote
  {\bibinfo {title} {{The r-Process in Neutrino-driven Winds from Nascent,
  ``Compact'' Neutron Stars of Core-Collapse Supernovae}},}\ }\href@noop {}
  {\bibfield  {journal} {\bibinfo  {journal} {Astrophys. J.}\ }\textbf
  {\bibinfo {volume} {554}},\ \bibinfo {pages} {578--586}}\BibitemShut
  {NoStop}%
\bibitem [{\citenamefont {{Wanajo}}\ \emph {et~al.}(2018)\citenamefont
  {{Wanajo}}, \citenamefont {{M{\"u}ller}}, \citenamefont {{Janka}},\ and\
  \citenamefont {{Heger}}}]{Wanajo.Mueller.ea:2018}%
  \BibitemOpen
  \bibfield  {author} {\bibinfo {author} {\bibnamefont {{Wanajo}},
  \bibfnamefont {S}}, \bibinfo {author} {\bibfnamefont {B.}~\bibnamefont
  {{M{\"u}ller}}}, \bibinfo {author} {\bibfnamefont {H.-T.}\ \bibnamefont
  {{Janka}}}, and\ \bibinfo {author} {\bibfnamefont {A.}~\bibnamefont
  {{Heger}}}} (\bibinfo {year} {2018}),\ \bibfield  {title} {\enquote {\bibinfo
  {title} {{Nucleosynthesis in the Innermost Ejecta of Neutrino-driven
  Supernova Explosions in Two Dimensions}},}\ }\href
  {https://doi.org/10.3847/1538-4357/aa9d97} {\bibfield  {journal} {\bibinfo
  {journal} {Astrophys. J.}\ }\textbf {\bibinfo {volume} {852}},\ \bibinfo
  {pages} {40}}\BibitemShut {NoStop}%
\bibitem [{\citenamefont {{Wanajo}}\ \emph {et~al.}(2009)\citenamefont
  {{Wanajo}}, \citenamefont {{Nomoto}}, \citenamefont {{Janka}}, \citenamefont
  {{Kitaura}},\ and\ \citenamefont {{M{\"u}ller}}}]{Wanajo.Nomoto.ea:2009}%
  \BibitemOpen
  \bibfield  {author} {\bibinfo {author} {\bibnamefont {{Wanajo}},
  \bibfnamefont {S}}, \bibinfo {author} {\bibfnamefont {K.}~\bibnamefont
  {{Nomoto}}}, \bibinfo {author} {\bibfnamefont {{H.-T.}}\ \bibnamefont
  {{Janka}}}, \bibinfo {author} {\bibfnamefont {F.~S.}\ \bibnamefont
  {{Kitaura}}}, and\ \bibinfo {author} {\bibfnamefont {B.}~\bibnamefont
  {{M{\"u}ller}}}} (\bibinfo {year} {2009}),\ \bibfield  {title} {\enquote
  {\bibinfo {title} {{Nucleosynthesis in Electron Capture Supernovae of
  Asymptotic Giant Branch Stars}},}\ }\href
  {https://doi.org/10.1088/0004-637X/695/1/208} {\bibfield  {journal} {\bibinfo
   {journal} {Astrophys. J.}\ }\textbf {\bibinfo {volume} {695}},\ \bibinfo
  {pages} {208--220}}\BibitemShut {NoStop}%
\bibitem [{\citenamefont {{Wanajo}}\ \emph {et~al.}(2014)\citenamefont
  {{Wanajo}}, \citenamefont {{Sekiguchi}}, \citenamefont {{Nishimura}},
  \citenamefont {{Kiuchi}}, \citenamefont {{Kyutoku}},\ and\ \citenamefont
  {{Shibata}}}]{Wanajo.Sekiguchi.ea:2014}%
  \BibitemOpen
  \bibfield  {author} {\bibinfo {author} {\bibnamefont {{Wanajo}},
  \bibfnamefont {S}}, \bibinfo {author} {\bibfnamefont {Y.}~\bibnamefont
  {{Sekiguchi}}}, \bibinfo {author} {\bibfnamefont {N.}~\bibnamefont
  {{Nishimura}}}, \bibinfo {author} {\bibfnamefont {K.}~\bibnamefont
  {{Kiuchi}}}, \bibinfo {author} {\bibfnamefont {K.}~\bibnamefont {{Kyutoku}}},
  and\ \bibinfo {author} {\bibfnamefont {M.}~\bibnamefont {{Shibata}}}}
  (\bibinfo {year} {2014}),\ \bibfield  {title} {\enquote {\bibinfo {title}
  {{Production of All the r-process Nuclides in the Dynamical Ejecta of Neutron
  Star Mergers}},}\ }\href {https://doi.org/10.1088/2041-8205/789/2/L39}
  {\bibfield  {journal} {\bibinfo  {journal} {Astrophys. J.}\ }\textbf
  {\bibinfo {volume} {789}},\ \bibinfo {eid} {L39}}\BibitemShut {NoStop}%
\bibitem [{\citenamefont {Wanajo}(2006)}]{Wanajo:2006}%
  \BibitemOpen
  \bibfield  {author} {\bibinfo {author} {\bibnamefont {Wanajo}, \bibfnamefont
  {Shinya}}} (\bibinfo {year} {2006}),\ \bibfield  {title} {\enquote {\bibinfo
  {title} {{The rp-Process in Neutrino-driven Winds}},}\ }\href
  {https://doi.org/10.1086/505483} {\bibfield  {journal} {\bibinfo  {journal}
  {Astrophys. J.}\ }\textbf {\bibinfo {volume} {647}},\ \bibinfo {pages}
  {1323--1340}}\BibitemShut {NoStop}%
\bibitem [{\citenamefont {{Wanajo}}(2018)}]{Wanajo:2018}%
  \BibitemOpen
  \bibfield  {author} {\bibinfo {author} {\bibnamefont {{Wanajo}},
  \bibfnamefont {Shinya}}} (\bibinfo {year} {2018}),\ \bibfield  {title}
  {\enquote {\bibinfo {title} {{Physical Conditions for the r-process. I.
  Radioactive Energy Sources of Kilonovae}},}\ }\href
  {https://doi.org/10.3847/1538-4357/aae0f2} {\bibfield  {journal} {\bibinfo
  {journal} {Astrophys. J.}\ }\textbf {\bibinfo {volume} {868}},\ \bibinfo
  {eid} {65}}\BibitemShut {NoStop}%
\bibitem [{\citenamefont {Wanajo}\ \emph {et~al.}(2011)\citenamefont {Wanajo},
  \citenamefont {Janka},\ and\ \citenamefont
  {M{\"u}ller}}]{Wanajo.Janka.Mueller:2011}%
  \BibitemOpen
  \bibfield  {author} {\bibinfo {author} {\bibnamefont {Wanajo}, \bibfnamefont
  {Shinya}}, \bibinfo {author} {\bibfnamefont {Hans-Thomas}\ \bibnamefont
  {Janka}}, and\ \bibinfo {author} {\bibfnamefont {Bernhard}\ \bibnamefont
  {M{\"u}ller}}} (\bibinfo {year} {2011}),\ \bibfield  {title} {\enquote
  {\bibinfo {title} {Electron-capture supernovae as the origin of elements
  beyond iron},}\ }\href {https://doi.org/10.1088/2041-8205/726/2/L15}
  {\bibfield  {journal} {\bibinfo  {journal} {Astrophys. J.}\ }\textbf
  {\bibinfo {volume} {726}},\ \bibinfo {pages} {L15}}\BibitemShut {NoStop}%
\bibitem [{\citenamefont {{Wanderman}}\ and\ \citenamefont
  {{Piran}}(2015)}]{wanderman15}%
  \BibitemOpen
  \bibfield  {author} {\bibinfo {author} {\bibnamefont {{Wanderman}},
  \bibfnamefont {D}}, and\ \bibinfo {author} {\bibfnamefont {T.}~\bibnamefont
  {{Piran}}}} (\bibinfo {year} {2015}),\ \bibfield  {title} {\enquote {\bibinfo
  {title} {{The rate, luminosity function and time delay of non-Collapsar short
  GRBs}},}\ }\href {https://doi.org/10.1093/mnras/stv123} {\bibfield  {journal}
  {\bibinfo  {journal} {Mon. Not. Roy. Astron. Soc.}\ }\textbf {\bibinfo
  {volume} {448}},\ \bibinfo {pages} {3026--3037}}\BibitemShut {NoStop}%
\bibitem [{\citenamefont {Wang}\ \emph {et~al.}(2012)\citenamefont {Wang},
  \citenamefont {Audi}, \citenamefont {Wapstra}, \citenamefont {Kondev},
  \citenamefont {MacCormick}, \citenamefont {Xu},\ and\ \citenamefont
  {Pfeiffer}}]{Wang.Audi.ea:2012}%
  \BibitemOpen
  \bibfield  {author} {\bibinfo {author} {\bibnamefont {Wang}, \bibfnamefont
  {M}}, \bibinfo {author} {\bibfnamefont {G.}~\bibnamefont {Audi}}, \bibinfo
  {author} {\bibfnamefont {A.H.}\ \bibnamefont {Wapstra}}, \bibinfo {author}
  {\bibfnamefont {F.G.}\ \bibnamefont {Kondev}}, \bibinfo {author}
  {\bibfnamefont {M.}~\bibnamefont {MacCormick}}, \bibinfo {author}
  {\bibfnamefont {X.}~\bibnamefont {Xu}}, and\ \bibinfo {author} {\bibfnamefont
  {B.}~\bibnamefont {Pfeiffer}}} (\bibinfo {year} {2012}),\ \bibfield  {title}
  {\enquote {\bibinfo {title} {{The Ame2012 atomic mass evaluation: (II)
  Tables, graphs and references}},}\ }\href
  {https://doi.org/10.1088/1674-1137/36/12/003} {\bibfield  {journal} {\bibinfo
   {journal} {Chinese Phys. C}\ }\textbf {\bibinfo {volume} {36}},\ \bibinfo
  {pages} {1603--2014}}\BibitemShut {NoStop}%
\bibitem [{\citenamefont {Wang}\ \emph {et~al.}(2010)\citenamefont {Wang},
  \citenamefont {Liu},\ and\ \citenamefont {Wu}}]{Wang.Liu.Wu:2010}%
  \BibitemOpen
  \bibfield  {author} {\bibinfo {author} {\bibnamefont {Wang}, \bibfnamefont
  {Ning}}, \bibinfo {author} {\bibfnamefont {Min}\ \bibnamefont {Liu}}, and\
  \bibinfo {author} {\bibfnamefont {Xizhen}\ \bibnamefont {Wu}}} (\bibinfo
  {year} {2010}),\ \bibfield  {title} {\enquote {\bibinfo {title}
  {{Modification of nuclear mass formula by considering isospin effects}},}\
  }\href {https://doi.org/10.1103/PhysRevC.81.044322} {\bibfield  {journal}
  {\bibinfo  {journal} {Phys. Rev. C}\ }\textbf {\bibinfo {volume} {81}},\
  \bibinfo {pages} {044322}}\BibitemShut {NoStop}%
\bibitem [{\citenamefont {{Watson}}\ \emph {et~al.}(2019)\citenamefont
  {{Watson}}, \citenamefont {{Hansen}}, \citenamefont {{Selsing}},
  \citenamefont {{Koch}}, \citenamefont {{Malesani}}, \citenamefont
  {{Andersen}}, \citenamefont {{Fynbo}}, \citenamefont {{Arcones}},
  \citenamefont {{Bauswein}}, \citenamefont {{Covino}}, \citenamefont
  {{Grado}}, \citenamefont {{Heintz}}, \citenamefont {{Hunt}}, \citenamefont
  {{Kouveliotou}}, \citenamefont {{Leloudas}}, \citenamefont {{Levan}},
  \citenamefont {{Mazzali}},\ and\ \citenamefont
  {{Pian}}}]{Watson.Hansen.ea:2019}%
  \BibitemOpen
  \bibfield  {author} {\bibinfo {author} {\bibnamefont {{Watson}},
  \bibfnamefont {Darach}}, \bibinfo {author} {\bibfnamefont {Camilla~J.}\
  \bibnamefont {{Hansen}}}, \bibinfo {author} {\bibfnamefont {Jonatan}\
  \bibnamefont {{Selsing}}}, \bibinfo {author} {\bibfnamefont {Andreas}\
  \bibnamefont {{Koch}}}, \bibinfo {author} {\bibfnamefont {Daniele~B.}\
  \bibnamefont {{Malesani}}}, \bibinfo {author} {\bibfnamefont {Anja~C.}\
  \bibnamefont {{Andersen}}}, \bibinfo {author} {\bibfnamefont {Johan P.~U.}\
  \bibnamefont {{Fynbo}}}, \bibinfo {author} {\bibfnamefont {Almudena}\
  \bibnamefont {{Arcones}}}, \bibinfo {author} {\bibfnamefont {Andreas}\
  \bibnamefont {{Bauswein}}}, \bibinfo {author} {\bibfnamefont {Stefano}\
  \bibnamefont {{Covino}}}, \bibinfo {author} {\bibfnamefont {Aniello}\
  \bibnamefont {{Grado}}}, \bibinfo {author} {\bibfnamefont {Kasper~E.}\
  \bibnamefont {{Heintz}}}, \bibinfo {author} {\bibfnamefont {Leslie}\
  \bibnamefont {{Hunt}}}, \bibinfo {author} {\bibfnamefont {Chryssa}\
  \bibnamefont {{Kouveliotou}}}, \bibinfo {author} {\bibfnamefont {Giorgos}\
  \bibnamefont {{Leloudas}}}, \bibinfo {author} {\bibfnamefont {Andrew~J.}\
  \bibnamefont {{Levan}}}, \bibinfo {author} {\bibfnamefont {Paolo}\
  \bibnamefont {{Mazzali}}}, and\ \bibinfo {author} {\bibfnamefont {Elena}\
  \bibnamefont {{Pian}}}} (\bibinfo {year} {2019}),\ \bibfield  {title}
  {\enquote {\bibinfo {title} {{Identification of strontium in the merger of
  two neutron stars}},}\ }\href {https://doi.org/10.1038/s41586-019-1676-3}
  {\bibfield  {journal} {\bibinfo  {journal} {Nature}\ }\textbf {\bibinfo
  {volume} {574}},\ \bibinfo {pages} {497--500}}\BibitemShut {NoStop}%
\bibitem [{\citenamefont {{Waxman}}\ \emph {et~al.}(2018)\citenamefont
  {{Waxman}}, \citenamefont {{Ofek}}, \citenamefont {{Kushnir}},\ and\
  \citenamefont {{Gal-Yam}}}]{Waxman.Ofek.ea:2017}%
  \BibitemOpen
  \bibfield  {author} {\bibinfo {author} {\bibnamefont {{Waxman}},
  \bibfnamefont {E}}, \bibinfo {author} {\bibfnamefont {E.~O.}\ \bibnamefont
  {{Ofek}}}, \bibinfo {author} {\bibfnamefont {D.}~\bibnamefont {{Kushnir}}},
  and\ \bibinfo {author} {\bibfnamefont {A.}~\bibnamefont {{Gal-Yam}}}}
  (\bibinfo {year} {2018}),\ \bibfield  {title} {\enquote {\bibinfo {title}
  {{Constraints on the ejecta of the GW170817 neutron star merger from its
  electromagnetic emission}},}\ }\href {https://doi.org/10.1093/mnras/sty2441}
  {\bibfield  {journal} {\bibinfo  {journal} {Mon. Not. Roy. Astron. Soc.}\
  }\textbf {\bibinfo {volume} {481}},\ \bibinfo {pages}
  {3423--3441}}\BibitemShut {NoStop}%
\bibitem [{\citenamefont {{Waxman}}\ \emph {et~al.}(2019)\citenamefont
  {{Waxman}}, \citenamefont {{Ofek}},\ and\ \citenamefont
  {{Kushnir}}}]{Waxman.Ofek.Kushnir:2019}%
  \BibitemOpen
  \bibfield  {author} {\bibinfo {author} {\bibnamefont {{Waxman}},
  \bibfnamefont {Eli}}, \bibinfo {author} {\bibfnamefont {Eran~O.}\
  \bibnamefont {{Ofek}}}, and\ \bibinfo {author} {\bibfnamefont {Doron}\
  \bibnamefont {{Kushnir}}}} (\bibinfo {year} {2019}),\ \bibfield  {title}
  {\enquote {\bibinfo {title} {{Late-time Kilonova Light Curves and
  Implications to GW170817}},}\ }\href
  {https://doi.org/10.3847/1538-4357/ab1f71} {\bibfield  {journal} {\bibinfo
  {journal} {Astrophys. J.}\ }\textbf {\bibinfo {volume} {878}},\ \bibinfo
  {eid} {93}}\BibitemShut {NoStop}%
\bibitem [{\citenamefont {Way}\ and\ \citenamefont
  {Wigner}(1948)}]{Way.Wigner:1948}%
  \BibitemOpen
  \bibfield  {author} {\bibinfo {author} {\bibnamefont {Way}, \bibfnamefont
  {K}}, and\ \bibinfo {author} {\bibfnamefont {E.~P.}\ \bibnamefont {Wigner}}}
  (\bibinfo {year} {1948}),\ \bibfield  {title} {\enquote {\bibinfo {title}
  {{The Rate of Decay of Fission Products}},}\ }\href
  {https://doi.org/10.1103/PhysRev.73.1318} {\bibfield  {journal} {\bibinfo
  {journal} {Phys. Rev.}\ }\textbf {\bibinfo {volume} {73}},\ \bibinfo {pages}
  {1318--1330}}\BibitemShut {NoStop}%
\bibitem [{\citenamefont {{Wehmeyer}}\ \emph {et~al.}(2019)\citenamefont
  {{Wehmeyer}}, \citenamefont {{Fr{\"o}hlich}}, \citenamefont {{C{\^o}t{\'e}}},
  \citenamefont {{Pignatari}},\ and\ \citenamefont
  {{Thielemann}}}]{Wehmeyer.ea:2019}%
  \BibitemOpen
  \bibfield  {author} {\bibinfo {author} {\bibnamefont {{Wehmeyer}},
  \bibfnamefont {B}}, \bibinfo {author} {\bibfnamefont {C.}~\bibnamefont
  {{Fr{\"o}hlich}}}, \bibinfo {author} {\bibfnamefont {B.}~\bibnamefont
  {{C{\^o}t{\'e}}}}, \bibinfo {author} {\bibfnamefont {M.}~\bibnamefont
  {{Pignatari}}}, and\ \bibinfo {author} {\bibfnamefont {F.~K.}\ \bibnamefont
  {{Thielemann}}}} (\bibinfo {year} {2019}),\ \bibfield  {title} {\enquote
  {\bibinfo {title} {{Using failed supernovae to constrain the Galactic
  r-process element production}},}\ }\href
  {https://doi.org/10.1093/mnras/stz1310} {\bibfield  {journal} {\bibinfo
  {journal} {Mon. Not. Roy. Astron. Soc.}\ }\textbf {\bibinfo {volume} {487}},\
  \bibinfo {pages} {1745--1753}}\BibitemShut {NoStop}%
\bibitem [{\citenamefont {{Wehmeyer}}\ \emph {et~al.}(2015)\citenamefont
  {{Wehmeyer}}, \citenamefont {{Pignatari}},\ and\ \citenamefont
  {{Thielemann}}}]{wehmeyer15}%
  \BibitemOpen
  \bibfield  {author} {\bibinfo {author} {\bibnamefont {{Wehmeyer}},
  \bibfnamefont {B}}, \bibinfo {author} {\bibfnamefont {M.}~\bibnamefont
  {{Pignatari}}}, and\ \bibinfo {author} {\bibfnamefont {F.-K.}\ \bibnamefont
  {{Thielemann}}}} (\bibinfo {year} {2015}),\ \bibfield  {title} {\enquote
  {\bibinfo {title} {{Galactic evolution of rapid neutron capture process
  abundances: the inhomogeneous approach}},}\ }\href
  {https://doi.org/10.1093/mnras/stv1352} {\bibfield  {journal} {\bibinfo
  {journal} {Mon. Not. Roy. Astron. Soc.}\ }\textbf {\bibinfo {volume} {452}},\
  \bibinfo {pages} {1970--1981}}\BibitemShut {NoStop}%
\bibitem [{\citenamefont {{Wehmeyer}}\ \emph {et~al.}(2018)\citenamefont
  {{Wehmeyer}} \emph {et~al.}}]{Wehmeyer.Others:2018}%
  \BibitemOpen
  \bibfield  {author} {\bibinfo {author} {\bibnamefont {{Wehmeyer}},
  \bibfnamefont {B}},  \emph {et~al.}} (\bibinfo {year} {2018}),\ \href@noop {}
  {}\bibinfo {note} {{private communication}}\BibitemShut {NoStop}%
\bibitem [{\citenamefont {{Weisberg}}\ and\ \citenamefont
  {{Huang}}(2016)}]{Weisberg.Huang:2016}%
  \BibitemOpen
  \bibfield  {author} {\bibinfo {author} {\bibnamefont {{Weisberg}},
  \bibfnamefont {J~M}}, and\ \bibinfo {author} {\bibfnamefont {Y.}~\bibnamefont
  {{Huang}}}} (\bibinfo {year} {2016}),\ \bibfield  {title} {\enquote {\bibinfo
  {title} {{Relativistic Measurements from Timing the Binary Pulsar PSR
  B1913+16}},}\ }\href {https://doi.org/10.3847/0004-637X/829/1/55} {\bibfield
  {journal} {\bibinfo  {journal} {Astrophys. J.}\ }\textbf {\bibinfo {volume}
  {829}},\ \bibinfo {pages} {55}}\BibitemShut {NoStop}%
\bibitem [{\citenamefont {{Westin}}\ \emph {et~al.}(2000)\citenamefont
  {{Westin}}, \citenamefont {{Sneden}}, \citenamefont {{Gustafsson}},\ and\
  \citenamefont {{Cowan}}}]{Westin.Sneden.ea:2000}%
  \BibitemOpen
  \bibfield  {author} {\bibinfo {author} {\bibnamefont {{Westin}},
  \bibfnamefont {J}}, \bibinfo {author} {\bibfnamefont {C.}~\bibnamefont
  {{Sneden}}}, \bibinfo {author} {\bibfnamefont {B.}~\bibnamefont
  {{Gustafsson}}}, and\ \bibinfo {author} {\bibfnamefont {J.~J.}\ \bibnamefont
  {{Cowan}}}} (\bibinfo {year} {2000}),\ \bibfield  {title} {\enquote {\bibinfo
  {title} {{The r-Process-enriched Low-Metallicity Giant HD 115444}},}\ }\href
  {https://doi.org/10.1086/308407} {\bibfield  {journal} {\bibinfo  {journal}
  {Astrophys. J.}\ }\textbf {\bibinfo {volume} {530}},\ \bibinfo {pages}
  {783--799}}\BibitemShut {NoStop}%
\bibitem [{\citenamefont {{Whaling}}\ \emph {et~al.}(1993)\citenamefont
  {{Whaling}}, \citenamefont {{Carle}},\ and\ \citenamefont
  {{Pitt}}}]{whaling93}%
  \BibitemOpen
  \bibfield  {author} {\bibinfo {author} {\bibnamefont {{Whaling}},
  \bibfnamefont {W}}, \bibinfo {author} {\bibfnamefont {M.~T.}\ \bibnamefont
  {{Carle}}}, and\ \bibinfo {author} {\bibfnamefont {M.~L.}\ \bibnamefont
  {{Pitt}}}} (\bibinfo {year} {1993}),\ \bibfield  {title} {\enquote {\bibinfo
  {title} {{Argon branching ratios for spectrometer response calibration.}}}\
  }\href {https://doi.org/10.1016/0022-4073(93)90124-Z} {\bibfield  {journal}
  {\bibinfo  {journal} {J. Quant Spec. Rad. Trans.}\ }\textbf {\bibinfo
  {volume} {50}},\ \bibinfo {pages} {7--18}}\BibitemShut {NoStop}%
\bibitem [{\citenamefont {{Wheeler}}\ \emph {et~al.}(1998)\citenamefont
  {{Wheeler}}, \citenamefont {{Cowan}},\ and\ \citenamefont
  {{Hillebrandt}}}]{Wheeler.Cowan.Hillebrandt:1998}%
  \BibitemOpen
  \bibfield  {author} {\bibinfo {author} {\bibnamefont {{Wheeler}},
  \bibfnamefont {J~C}}, \bibinfo {author} {\bibfnamefont {J.~J.}\ \bibnamefont
  {{Cowan}}}, and\ \bibinfo {author} {\bibfnamefont {W.}~\bibnamefont
  {{Hillebrandt}}}} (\bibinfo {year} {1998}),\ \bibfield  {title} {\enquote
  {\bibinfo {title} {{The r-Process in Collapsing O/Ne/Mg Cores}},}\ }\href
  {https://doi.org/10.1086/311133} {\bibfield  {journal} {\bibinfo  {journal}
  {Astrophys. J.}\ }\textbf {\bibinfo {volume} {493}},\ \bibinfo {pages}
  {L101--L104}}\BibitemShut {NoStop}%
\bibitem [{\citenamefont {{Wheeler}}\ \emph {et~al.}(1989)\citenamefont
  {{Wheeler}}, \citenamefont {{Sneden}},\ and\ \citenamefont
  {{Truran}}}]{Wheeler.Sneden.Truran:1989}%
  \BibitemOpen
  \bibfield  {author} {\bibinfo {author} {\bibnamefont {{Wheeler}},
  \bibfnamefont {J~C}}, \bibinfo {author} {\bibfnamefont {C.}~\bibnamefont
  {{Sneden}}}, and\ \bibinfo {author} {\bibfnamefont {J.~W.}\ \bibnamefont
  {{Truran}}, \bibfnamefont {Jr.}}} (\bibinfo {year} {1989}),\ \bibfield
  {title} {\enquote {\bibinfo {title} {{Abundance ratios as a function of
  metallicity}},}\ }\href {https://doi.org/10.1146/annurev.aa.27.090189.001431}
  {\bibfield  {journal} {\bibinfo  {journal} {Annu. Rev. Astron. Astrophys.}\
  }\textbf {\bibinfo {volume} {27}},\ \bibinfo {pages} {279--349}}\BibitemShut
  {NoStop}%
\bibitem [{\citenamefont {{Winteler}}\ \emph {et~al.}(2012)\citenamefont
  {{Winteler}}, \citenamefont {{K{\"a}ppeli}}, \citenamefont {{Perego}},
  \citenamefont {{Arcones}}, \citenamefont {{Vasset}}, \citenamefont
  {{Nishimura}}, \citenamefont {{Liebend{\"o}rfer}},\ and\ \citenamefont
  {{Thielemann}}}]{Winteler.Kaeppeli.ea:2012}%
  \BibitemOpen
  \bibfield  {author} {\bibinfo {author} {\bibnamefont {{Winteler}},
  \bibfnamefont {C}}, \bibinfo {author} {\bibfnamefont {R.}~\bibnamefont
  {{K{\"a}ppeli}}}, \bibinfo {author} {\bibfnamefont {A.}~\bibnamefont
  {{Perego}}}, \bibinfo {author} {\bibfnamefont {A.}~\bibnamefont {{Arcones}}},
  \bibinfo {author} {\bibfnamefont {N.}~\bibnamefont {{Vasset}}}, \bibinfo
  {author} {\bibfnamefont {N.}~\bibnamefont {{Nishimura}}}, \bibinfo {author}
  {\bibfnamefont {M.}~\bibnamefont {{Liebend{\"o}rfer}}}, and\ \bibinfo
  {author} {\bibfnamefont {F.-K.}\ \bibnamefont {{Thielemann}}}} (\bibinfo
  {year} {2012}),\ \bibfield  {title} {\enquote {\bibinfo {title}
  {{Magnetorotationally Driven Supernovae as the Origin of Early Galaxy
  r-process Elements?}}}\ }\href {https://doi.org/10.1088/2041-8205/750/1/L22}
  {\bibfield  {journal} {\bibinfo  {journal} {Astrophys. J.}\ }\textbf
  {\bibinfo {volume} {750}},\ \bibinfo {eid} {L22}}\BibitemShut {NoStop}%
\bibitem [{\citenamefont {{Wisshak}}\ \emph {et~al.}(2000)\citenamefont
  {{Wisshak}}, \citenamefont {{Voss}}, \citenamefont {{Arlandini}},
  \citenamefont {{K{\"a}ppeler}},\ and\ \citenamefont
  {{Kazakov}}}]{Wisshak2000}%
  \BibitemOpen
  \bibfield  {author} {\bibinfo {author} {\bibnamefont {{Wisshak}},
  \bibfnamefont {K}}, \bibinfo {author} {\bibfnamefont {F.}~\bibnamefont
  {{Voss}}}, \bibinfo {author} {\bibfnamefont {C.}~\bibnamefont {{Arlandini}}},
  \bibinfo {author} {\bibfnamefont {F.}~\bibnamefont {{K{\"a}ppeler}}}, and\
  \bibinfo {author} {\bibfnamefont {L.}~\bibnamefont {{Kazakov}}}} (\bibinfo
  {year} {2000}),\ \bibfield  {title} {\enquote {\bibinfo {title} {{Stellar
  neutron capture cross sections of the Yb isotopes}},}\ }\href
  {https://doi.org/10.1103/PhysRevC.61.065801} {\bibfield  {journal} {\bibinfo
  {journal} {Phys. Rev. C}\ }\textbf {\bibinfo {volume} {61}},\ \bibinfo {eid}
  {065801}}\BibitemShut {NoStop}%
\bibitem [{\citenamefont {Witti}\ \emph {et~al.}(1994)\citenamefont {Witti},
  \citenamefont {Janka},\ and\ \citenamefont
  {Takahashi}}]{Witti.Janka.Takahashi:1994}%
  \BibitemOpen
  \bibfield  {author} {\bibinfo {author} {\bibnamefont {Witti}, \bibfnamefont
  {J}}, \bibinfo {author} {\bibfnamefont {H.-Th.}\ \bibnamefont {Janka}}, and\
  \bibinfo {author} {\bibfnamefont {K.}~\bibnamefont {Takahashi}}} (\bibinfo
  {year} {1994}),\ \bibfield  {title} {\enquote {\bibinfo {title}
  {{Nucleosynthesis in neutrino-driven winds from protoneutron stars I. The
  $\alpha$-process}},}\ }\href@noop {} {\bibfield  {journal} {\bibinfo
  {journal} {Astron. \& Astrophys.}\ }\textbf {\bibinfo {volume} {286}},\
  \bibinfo {pages} {841--856}}\BibitemShut {NoStop}%
\bibitem [{\citenamefont {{Wojczuk}}\ and\ \citenamefont
  {{Janiuk}}(2018)}]{Wojczuk.Janiuk:2018}%
  \BibitemOpen
  \bibfield  {author} {\bibinfo {author} {\bibnamefont {{Wojczuk}},
  \bibfnamefont {K}}, and\ \bibinfo {author} {\bibfnamefont {A.}~\bibnamefont
  {{Janiuk}}}} (\bibinfo {year} {2018}),\ \bibfield  {title} {\enquote
  {\bibinfo {title} {{Nucleosynthesis in black hole accretion flows feeding
  short GRBs}},}\ }in\ \href@noop {} {\emph {\bibinfo {booktitle} {XXXVIII
  Polish Astronomical Society Meeting}}},\ Vol.~\bibinfo {volume} {7},\
  \bibinfo {editor} {edited by\ \bibinfo {editor} {\bibfnamefont
  {A.}~\bibnamefont {{R{\'o}{\.z}a{\'n}ska}}}},\ pp.\ \bibinfo {pages}
  {337--340}\BibitemShut {NoStop}%
\bibitem [{\citenamefont {{Woosley}}(1993)}]{Woosley:1993}%
  \BibitemOpen
  \bibfield  {author} {\bibinfo {author} {\bibnamefont {{Woosley}},
  \bibfnamefont {S~E}}} (\bibinfo {year} {1993}),\ \bibfield  {title} {\enquote
  {\bibinfo {title} {{Gamma-Ray Bursts from Stellar Mass Accretion Disks around
  Black Holes}},}\ }\href {https://doi.org/10.1086/172359} {\bibfield
  {journal} {\bibinfo  {journal} {Astrophys. J.}\ }\textbf {\bibinfo {volume}
  {405}},\ \bibinfo {pages} {273}}\BibitemShut {NoStop}%
\bibitem [{\citenamefont {{Woosley}}\ \emph {et~al.}(1973)\citenamefont
  {{Woosley}}, \citenamefont {{Arnett}},\ and\ \citenamefont
  {{Clayton}}}]{Woosley.Arnett.Clayton:1973}%
  \BibitemOpen
  \bibfield  {author} {\bibinfo {author} {\bibnamefont {{Woosley}},
  \bibfnamefont {S~E}}, \bibinfo {author} {\bibfnamefont {W.~D.}\ \bibnamefont
  {{Arnett}}}, and\ \bibinfo {author} {\bibfnamefont {D.~D.}\ \bibnamefont
  {{Clayton}}}} (\bibinfo {year} {1973}),\ \bibfield  {title} {\enquote
  {\bibinfo {title} {{The Explosive Burning of Oxygen and Silicon}},}\ }\href
  {https://doi.org/10.1086/190282} {\bibfield  {journal} {\bibinfo  {journal}
  {Astrophys. J. Suppl.}\ }\textbf {\bibinfo {volume} {26}},\ \bibinfo {pages}
  {231}}\BibitemShut {NoStop}%
\bibitem [{\citenamefont {{Woosley}}\ \emph {et~al.}(1999)\citenamefont
  {{Woosley}}, \citenamefont {{Eastman}},\ and\ \citenamefont
  {{Schmidt}}}]{Woosley.Eastman.ea:1999}%
  \BibitemOpen
  \bibfield  {author} {\bibinfo {author} {\bibnamefont {{Woosley}},
  \bibfnamefont {S~E}}, \bibinfo {author} {\bibfnamefont {Ronald~G.}\
  \bibnamefont {{Eastman}}}, and\ \bibinfo {author} {\bibfnamefont {Brian~P.}\
  \bibnamefont {{Schmidt}}}} (\bibinfo {year} {1999}),\ \bibfield  {title}
  {\enquote {\bibinfo {title} {{Gamma-Ray Bursts and Type IC Supernova SN
  1998BW}},}\ }\href {https://doi.org/10.1086/307131} {\bibfield  {journal}
  {\bibinfo  {journal} {Astrophys. J.}\ }\textbf {\bibinfo {volume} {516}},\
  \bibinfo {pages} {788--796}}\BibitemShut {NoStop}%
\bibitem [{\citenamefont {{Woosley}}\ \emph {et~al.}(1990)\citenamefont
  {{Woosley}}, \citenamefont {{Hartmann}}, \citenamefont {{Hoffman}},\ and\
  \citenamefont {{Haxton}}}]{Woosley.Hartmann.ea:1990}%
  \BibitemOpen
  \bibfield  {author} {\bibinfo {author} {\bibnamefont {{Woosley}},
  \bibfnamefont {S~E}}, \bibinfo {author} {\bibfnamefont {D.~H.}\ \bibnamefont
  {{Hartmann}}}, \bibinfo {author} {\bibfnamefont {R.~D.}\ \bibnamefont
  {{Hoffman}}}, and\ \bibinfo {author} {\bibfnamefont {W.~C.}\ \bibnamefont
  {{Haxton}}}} (\bibinfo {year} {1990}),\ \bibfield  {title} {\enquote
  {\bibinfo {title} {{The $\nu$-process}},}\ }\href
  {https://doi.org/10.1086/168839} {\bibfield  {journal} {\bibinfo  {journal}
  {Astrophys. J.}\ }\textbf {\bibinfo {volume} {356}},\ \bibinfo {pages}
  {272--301}}\BibitemShut {NoStop}%
\bibitem [{\citenamefont {{Woosley}}\ and\ \citenamefont
  {{Heger}}(2007)}]{Woosley.Heger:2007}%
  \BibitemOpen
  \bibfield  {author} {\bibinfo {author} {\bibnamefont {{Woosley}},
  \bibfnamefont {S~E}}, and\ \bibinfo {author} {\bibfnamefont {A.}~\bibnamefont
  {{Heger}}}} (\bibinfo {year} {2007}),\ \bibfield  {title} {\enquote {\bibinfo
  {title} {{Nucleosynthesis and remnants in massive stars of solar
  metallicity}},}\ }\href {https://doi.org/10.1016/j.physrep.2007.02.009}
  {\bibfield  {journal} {\bibinfo  {journal} {Phys. Rep.}\ }\textbf {\bibinfo
  {volume} {442}},\ \bibinfo {pages} {269--283}}\BibitemShut {NoStop}%
\bibitem [{\citenamefont {Woosley}\ \emph {et~al.}(2002)\citenamefont
  {Woosley}, \citenamefont {Heger},\ and\ \citenamefont
  {Weaver}}]{Woosley.Heger.Weaver:2002}%
  \BibitemOpen
  \bibfield  {author} {\bibinfo {author} {\bibnamefont {Woosley}, \bibfnamefont
  {S~E}}, \bibinfo {author} {\bibfnamefont {A.}~\bibnamefont {Heger}}, and\
  \bibinfo {author} {\bibfnamefont {T.~A.}\ \bibnamefont {Weaver}}} (\bibinfo
  {year} {2002}),\ \bibfield  {title} {\enquote {\bibinfo {title} {{The
  evolution and explosion of massive stars}},}\ }\href
  {https://doi.org/10.1103/RevModPhys.74.1015} {\bibfield  {journal} {\bibinfo
  {journal} {Rev. Mod. Phys.}\ }\textbf {\bibinfo {volume} {74}},\ \bibinfo
  {pages} {1015--1071}}\BibitemShut {NoStop}%
\bibitem [{\citenamefont {{Woosley}}\ and\ \citenamefont
  {{Hoffman}}(1992)}]{woosley92}%
  \BibitemOpen
  \bibfield  {author} {\bibinfo {author} {\bibnamefont {{Woosley}},
  \bibfnamefont {S~E}}, and\ \bibinfo {author} {\bibfnamefont {R.~D.}\
  \bibnamefont {{Hoffman}}}} (\bibinfo {year} {1992}),\ \bibfield  {title}
  {\enquote {\bibinfo {title} {{The $\alpha$-process and the $r$-process}},}\
  }\href {https://doi.org/10.1086/171644} {\bibfield  {journal} {\bibinfo
  {journal} {Astrophys. J.}\ }\textbf {\bibinfo {volume} {395}},\ \bibinfo
  {pages} {202--239}}\BibitemShut {NoStop}%
\bibitem [{\citenamefont {Woosley}\ \emph {et~al.}(1994)\citenamefont
  {Woosley}, \citenamefont {Wilson}, \citenamefont {Mathews}, \citenamefont
  {Hoffman},\ and\ \citenamefont {Meyer}}]{Woosley.Wilson.ea:1994}%
  \BibitemOpen
  \bibfield  {author} {\bibinfo {author} {\bibnamefont {Woosley}, \bibfnamefont
  {S~E}}, \bibinfo {author} {\bibfnamefont {J.~R.}\ \bibnamefont {Wilson}},
  \bibinfo {author} {\bibfnamefont {G.~J.}\ \bibnamefont {Mathews}}, \bibinfo
  {author} {\bibfnamefont {R.~D.}\ \bibnamefont {Hoffman}}, and\ \bibinfo
  {author} {\bibfnamefont {B.~S.}\ \bibnamefont {Meyer}}} (\bibinfo {year}
  {1994}),\ \bibfield  {title} {\enquote {\bibinfo {title} {{The r-process and
  neutrino-heated supernova ejecta}},}\ }\href {https://doi.org/10.1086/174638}
  {\bibfield  {journal} {\bibinfo  {journal} {Astrophys. J.}\ }\textbf
  {\bibinfo {volume} {433}},\ \bibinfo {pages} {229--246}}\BibitemShut
  {NoStop}%
\bibitem [{\citenamefont {Wu}\ \emph {et~al.}(2017)\citenamefont {Wu} \emph
  {et~al.}}]{Wu.Nishimura.ea:2017}%
  \BibitemOpen
  \bibfield  {author} {\bibinfo {author} {\bibnamefont {Wu}, \bibfnamefont
  {J}},  \emph {et~al.}} (\bibinfo {year} {2017}),\ \bibfield  {title}
  {\enquote {\bibinfo {title} {94 $\beta$-decay half-lives of neutron-rich
  $_{55}\mathrm{Cs}$ to $_{67}\mathrm{Ho}$: Experimental feedback and
  evaluation of the $r$-process rare-earth peak formation},}\ }\href
  {https://doi.org/10.1103/PhysRevLett.118.072701} {\bibfield  {journal}
  {\bibinfo  {journal} {Phys. Rev. Lett.}\ }\textbf {\bibinfo {volume} {118}},\
  \bibinfo {pages} {072701}}\BibitemShut {NoStop}%
\bibitem [{\citenamefont {{Wu}}\ \emph {et~al.}(2016)\citenamefont {{Wu}},
  \citenamefont {{Fern{\'a}ndez}}, \citenamefont {{Mart{\'{\i}}nez-Pinedo}},\
  and\ \citenamefont {{Metzger}}}]{Wu.Fernandez.Martinez.ea:2016}%
  \BibitemOpen
  \bibfield  {author} {\bibinfo {author} {\bibnamefont {{Wu}}, \bibfnamefont
  {M-R}}, \bibinfo {author} {\bibfnamefont {R.}~\bibnamefont
  {{Fern{\'a}ndez}}}, \bibinfo {author} {\bibfnamefont {G.}~\bibnamefont
  {{Mart{\'{\i}}nez-Pinedo}}}, and\ \bibinfo {author} {\bibfnamefont {B.~D.}\
  \bibnamefont {{Metzger}}}} (\bibinfo {year} {2016}),\ \bibfield  {title}
  {\enquote {\bibinfo {title} {{Production of the entire range of r-process
  nuclides by black hole accretion disc outflows from neutron star mergers}},}\
  }\href {https://doi.org/10.1093/mnras/stw2156} {\bibfield  {journal}
  {\bibinfo  {journal} {Mon. Not. Roy. Astron. Soc.}\ }\textbf {\bibinfo
  {volume} {463}},\ \bibinfo {pages} {2323--2334}}\BibitemShut {NoStop}%
\bibitem [{\citenamefont {{Wu}}\ \emph {et~al.}(2014)\citenamefont {{Wu}},
  \citenamefont {{Fischer}}, \citenamefont {{Huther}}, \citenamefont
  {{Mart{\'{\i}}nez-Pinedo}},\ and\ \citenamefont
  {{Qian}}}]{Wu.Fischer.ea:2014}%
  \BibitemOpen
  \bibfield  {author} {\bibinfo {author} {\bibnamefont {{Wu}}, \bibfnamefont
  {M-R}}, \bibinfo {author} {\bibfnamefont {T.}~\bibnamefont {{Fischer}}},
  \bibinfo {author} {\bibfnamefont {L.}~\bibnamefont {{Huther}}}, \bibinfo
  {author} {\bibfnamefont {G.}~\bibnamefont {{Mart{\'{\i}}nez-Pinedo}}}, and\
  \bibinfo {author} {\bibfnamefont {Y.-Z.}\ \bibnamefont {{Qian}}}} (\bibinfo
  {year} {2014}),\ \bibfield  {title} {\enquote {\bibinfo {title} {{Impact of
  active-sterile neutrino mixing on supernova explosion and
  nucleosynthesis}},}\ }\href {https://doi.org/10.1103/PhysRevD.89.061303}
  {\bibfield  {journal} {\bibinfo  {journal} {Phys. Rev. D}\ }\textbf {\bibinfo
  {volume} {89}},\ \bibinfo {pages} {061303(R)}}\BibitemShut {NoStop}%
\bibitem [{\citenamefont {{Wu}}\ \emph {et~al.}(2017)\citenamefont {{Wu}} \emph
  {et~al.}}]{Wu:2017}%
  \BibitemOpen
  \bibfield  {author} {\bibinfo {author} {\bibnamefont {{Wu}}, \bibfnamefont
  {M-R}},  \emph {et~al.}} (\bibinfo {year} {2017}),\ \href@noop {} {}\bibinfo
  {note} {{private communication}}\BibitemShut {NoStop}%
\bibitem [{\citenamefont {{Wu}}\ \emph {et~al.}(2019)\citenamefont {{Wu}},
  \citenamefont {{Banerjee}}, \citenamefont {{Metzger}}, \citenamefont
  {{Mart{\'\i}nez-Pinedo}}, \citenamefont {{Aramaki}}, \citenamefont {{Burns}},
  \citenamefont {{Hailey}}, \citenamefont {{Barnes}},\ and\ \citenamefont
  {{Karagiorgi}}}]{Wu.Banerjee.ea:2019}%
  \BibitemOpen
  \bibfield  {author} {\bibinfo {author} {\bibnamefont {{Wu}}, \bibfnamefont
  {Meng-Ru}}, \bibinfo {author} {\bibfnamefont {Projjwal}\ \bibnamefont
  {{Banerjee}}}, \bibinfo {author} {\bibfnamefont {Brian~D.}\ \bibnamefont
  {{Metzger}}}, \bibinfo {author} {\bibfnamefont {Gabriel}\ \bibnamefont
  {{Mart{\'\i}nez-Pinedo}}}, \bibinfo {author} {\bibfnamefont {Tsuguo}\
  \bibnamefont {{Aramaki}}}, \bibinfo {author} {\bibfnamefont {Eric}\
  \bibnamefont {{Burns}}}, \bibinfo {author} {\bibfnamefont {Charles~J.}\
  \bibnamefont {{Hailey}}}, \bibinfo {author} {\bibfnamefont {Jennifer}\
  \bibnamefont {{Barnes}}}, and\ \bibinfo {author} {\bibfnamefont {Georgia}\
  \bibnamefont {{Karagiorgi}}}} (\bibinfo {year} {2019}),\ \bibfield  {title}
  {\enquote {\bibinfo {title} {{Finding the Remnants of the Milky Way's Last
  Neutron Star Mergers}},}\ }\href {https://doi.org/10.3847/1538-4357/ab2593}
  {\bibfield  {journal} {\bibinfo  {journal} {Astrophys. J.}\ }\textbf
  {\bibinfo {volume} {880}},\ \bibinfo {eid} {23}}\BibitemShut {NoStop}%
\bibitem [{\citenamefont {Wu}\ \emph {et~al.}(2019)\citenamefont {Wu},
  \citenamefont {Barnes}, \citenamefont {Mart\'{\i}nez-Pinedo},\ and\
  \citenamefont {Metzger}}]{Wu.Barnes.ea:2019}%
  \BibitemOpen
  \bibfield  {author} {\bibinfo {author} {\bibnamefont {Wu}, \bibfnamefont
  {Meng-Ru}}, \bibinfo {author} {\bibfnamefont {J.}~\bibnamefont {Barnes}},
  \bibinfo {author} {\bibfnamefont {G.}~\bibnamefont {Mart\'{\i}nez-Pinedo}},
  and\ \bibinfo {author} {\bibfnamefont {B.~D.}\ \bibnamefont {Metzger}}}
  (\bibinfo {year} {2019}),\ \bibfield  {title} {\enquote {\bibinfo {title}
  {{Fingerprints of Heavy-Element Nucleosynthesis in the Late-Time Lightcurves
  of Kilonovae}},}\ }\href {https://doi.org/10.1103/PhysRevLett.122.062701}
  {\bibfield  {journal} {\bibinfo  {journal} {Phys. Rev. Lett.}\ }\textbf
  {\bibinfo {volume} {122}},\ \bibinfo {pages} {062701}}\BibitemShut {NoStop}%
\bibitem [{\citenamefont {Wu}\ \emph {et~al.}(2015)\citenamefont {Wu},
  \citenamefont {Qian}, \citenamefont {Mart\'inez-Pinedo}, \citenamefont
  {Fischer},\ and\ \citenamefont {Huther}}]{Wu.Qian.ea:2015}%
  \BibitemOpen
  \bibfield  {author} {\bibinfo {author} {\bibnamefont {Wu}, \bibfnamefont
  {Meng-Ru}}, \bibinfo {author} {\bibfnamefont {Yong-Zhong}\ \bibnamefont
  {Qian}}, \bibinfo {author} {\bibfnamefont {Gabriel}\ \bibnamefont
  {Mart\'inez-Pinedo}}, \bibinfo {author} {\bibfnamefont {Tobias}\ \bibnamefont
  {Fischer}}, and\ \bibinfo {author} {\bibfnamefont {Lutz}\ \bibnamefont
  {Huther}}} (\bibinfo {year} {2015}),\ \bibfield  {title} {\enquote {\bibinfo
  {title} {Effects of neutrino oscillations on nucleosynthesis and neutrino
  signals for an 18~{M}$_{\odot}$ supernova model},}\ }\href
  {https://doi.org/10.1103/PhysRevD.91.065016} {\bibfield  {journal} {\bibinfo
  {journal} {Phys. Rev. D}\ }\textbf {\bibinfo {volume} {91}},\ \bibinfo {eid}
  {065016}}\BibitemShut {NoStop}%
\bibitem [{\citenamefont {Wu}\ and\ \citenamefont
  {Tamborra}(2017)}]{Wu.Tamborra:2017}%
  \BibitemOpen
  \bibfield  {author} {\bibinfo {author} {\bibnamefont {Wu}, \bibfnamefont
  {Meng-Ru}}, and\ \bibinfo {author} {\bibfnamefont {Irene}\ \bibnamefont
  {Tamborra}}} (\bibinfo {year} {2017}),\ \bibfield  {title} {\enquote
  {\bibinfo {title} {{Fast neutrino conversions: Ubiquitous in compact binary
  merger remnants}},}\ }\href {https://doi.org/10.1103/PhysRevD.95.103007}
  {\bibfield  {journal} {\bibinfo  {journal} {Phys. Rev. D}\ }\textbf {\bibinfo
  {volume} {95}},\ \bibinfo {pages} {103007}}\BibitemShut {NoStop}%
\bibitem [{\citenamefont {Wu}\ \emph {et~al.}(2017)\citenamefont {Wu},
  \citenamefont {Tamborra}, \citenamefont {Just},\ and\ \citenamefont
  {Janka}}]{Wu.Tamborra.ea:2017}%
  \BibitemOpen
  \bibfield  {author} {\bibinfo {author} {\bibnamefont {Wu}, \bibfnamefont
  {Meng-Ru}}, \bibinfo {author} {\bibfnamefont {Irene}\ \bibnamefont
  {Tamborra}}, \bibinfo {author} {\bibfnamefont {Oliver}\ \bibnamefont {Just}},
  and\ \bibinfo {author} {\bibfnamefont {Hans-Thomas}\ \bibnamefont {Janka}}}
  (\bibinfo {year} {2017}),\ \bibfield  {title} {\enquote {\bibinfo {title}
  {Imprints of neutrino-pair flavor conversions on nucleosynthesis in ejecta
  from neutron-star merger remnants},}\ }\href
  {https://doi.org/10.1103/PhysRevD.96.123015} {\bibfield  {journal} {\bibinfo
  {journal} {Phys. Rev. D}\ }\textbf {\bibinfo {volume} {96}},\ \bibinfo
  {pages} {123015}}\BibitemShut {NoStop}%
\bibitem [{\citenamefont {Wu}\ and\ \citenamefont
  {MacFadyen}(2018)}]{Wu.Macfadyen:2018}%
  \BibitemOpen
  \bibfield  {author} {\bibinfo {author} {\bibnamefont {Wu}, \bibfnamefont
  {Yiyang}}, and\ \bibinfo {author} {\bibfnamefont {Andrew}\ \bibnamefont
  {MacFadyen}}} (\bibinfo {year} {2018}),\ \bibfield  {title} {\enquote
  {\bibinfo {title} {{Constraining the Outflow Structure of the Binary Neutron
  Star Merger Event GW170817/GRB170817A with a Markov Chain Monte Carlo
  Analysis}},}\ }\href {https://doi.org/10.3847/1538-4357/aae9de} {\bibfield
  {journal} {\bibinfo  {journal} {Astrophys. J.}\ }\textbf {\bibinfo {volume}
  {869}},\ \bibinfo {pages} {55}}\BibitemShut {NoStop}%
\bibitem [{\citenamefont {{W{\"u}nsch}}(1978)}]{OstisI}%
  \BibitemOpen
  \bibfield  {author} {\bibinfo {author} {\bibnamefont {{W{\"u}nsch}},
  \bibfnamefont {K-D}}} (\bibinfo {year} {1978}),\ \bibfield  {title} {\enquote
  {\bibinfo {title} {{An on-line mass-separator for thermically ionisable
  fission products: OSTIS}},}\ }\href
  {https://doi.org/10.1016/0029-554X(78)90516-5} {\bibfield  {journal}
  {\bibinfo  {journal} {Nuclear Instruments and Methods}\ }\textbf {\bibinfo
  {volume} {155}},\ \bibinfo {pages} {347--351}}\BibitemShut {NoStop}%
\bibitem [{\citenamefont {{Xu}}\ and\ \citenamefont
  {{Goriely}}(2012)}]{Xu.Goriely:2012}%
  \BibitemOpen
  \bibfield  {author} {\bibinfo {author} {\bibnamefont {{Xu}}, \bibfnamefont
  {Y}}, and\ \bibinfo {author} {\bibfnamefont {S.}~\bibnamefont {{Goriely}}}}
  (\bibinfo {year} {2012}),\ \bibfield  {title} {\enquote {\bibinfo {title}
  {{Systematic study of direct neutron capture}},}\ }\href
  {https://doi.org/10.1103/PhysRevC.86.045801} {\bibfield  {journal} {\bibinfo
  {journal} {Phys. Rev. C}\ }\textbf {\bibinfo {volume} {86}},\ \bibinfo {eid}
  {045801}}\BibitemShut {NoStop}%
\bibitem [{\citenamefont {{Xu}}\ \emph {et~al.}(2014)\citenamefont {{Xu}},
  \citenamefont {{Goriely}}, \citenamefont {{Koning}},\ and\ \citenamefont
  {{Hilaire}}}]{Xu.Goriely.ea:2014}%
  \BibitemOpen
  \bibfield  {author} {\bibinfo {author} {\bibnamefont {{Xu}}, \bibfnamefont
  {Y}}, \bibinfo {author} {\bibfnamefont {S.}~\bibnamefont {{Goriely}}},
  \bibinfo {author} {\bibfnamefont {A.~J.}\ \bibnamefont {{Koning}}}, and\
  \bibinfo {author} {\bibfnamefont {S.}~\bibnamefont {{Hilaire}}}} (\bibinfo
  {year} {2014}),\ \bibfield  {title} {\enquote {\bibinfo {title} {{Systematic
  study of neutron capture including the compound, pre-equilibrium, and direct
  mechanisms}},}\ }\href {https://doi.org/10.1103/PhysRevC.90.024604}
  {\bibfield  {journal} {\bibinfo  {journal} {Phys. Rev. C}\ }\textbf {\bibinfo
  {volume} {90}},\ \bibinfo {eid} {024604}}\BibitemShut {NoStop}%
\bibitem [{\citenamefont {{Yakovlev}}\ \emph {et~al.}(2006)\citenamefont
  {{Yakovlev}}, \citenamefont {{Gasques}}, \citenamefont {{Afanasjev}},
  \citenamefont {{Beard}},\ and\ \citenamefont
  {{Wiescher}}}]{Yakovlev.ea:2006}%
  \BibitemOpen
  \bibfield  {author} {\bibinfo {author} {\bibnamefont {{Yakovlev}},
  \bibfnamefont {D~G}}, \bibinfo {author} {\bibfnamefont {L.~R.}\ \bibnamefont
  {{Gasques}}}, \bibinfo {author} {\bibfnamefont {A.~V.}\ \bibnamefont
  {{Afanasjev}}}, \bibinfo {author} {\bibfnamefont {M.}~\bibnamefont
  {{Beard}}}, and\ \bibinfo {author} {\bibfnamefont {M.}~\bibnamefont
  {{Wiescher}}}} (\bibinfo {year} {2006}),\ \bibfield  {title} {\enquote
  {\bibinfo {title} {{Fusion reactions in multicomponent dense matter}},}\
  }\href {https://doi.org/10.1103/PhysRevC.74.035803} {\bibfield  {journal}
  {\bibinfo  {journal} {Phys. Rev. C}\ }\textbf {\bibinfo {volume} {74}},\
  \bibinfo {eid} {035803}}\BibitemShut {NoStop}%
\bibitem [{\citenamefont {{Yan}}\ \emph {et~al.}(2016)\citenamefont {{Yan}},
  \citenamefont {{Blaum}}, \citenamefont {{Litvinov}}, \citenamefont {{Tu}},
  \citenamefont {{Xu}}, \citenamefont {{Zhang}},\ and\ \citenamefont
  {{Zhou}}}]{Yan2016}%
  \BibitemOpen
  \bibfield  {author} {\bibinfo {author} {\bibnamefont {{Yan}}, \bibfnamefont
  {X~L}}, \bibinfo {author} {\bibfnamefont {K.}~\bibnamefont {{Blaum}}},
  \bibinfo {author} {\bibfnamefont {Y.~A.}\ \bibnamefont {{Litvinov}}},
  \bibinfo {author} {\bibfnamefont {X.~L.}\ \bibnamefont {{Tu}}}, \bibinfo
  {author} {\bibfnamefont {H.~S.}\ \bibnamefont {{Xu}}}, \bibinfo {author}
  {\bibfnamefont {Y.~H.}\ \bibnamefont {{Zhang}}}, and\ \bibinfo {author}
  {\bibfnamefont {X.~H.}\ \bibnamefont {{Zhou}}} (\bibinfo {collaboration} {ESR
  and CSRe Collaborations on In-Ring Mass Measurements})} (\bibinfo {year}
  {2016}),\ \bibfield  {title} {\enquote {\bibinfo {title} {{Recent results on
  mass measurements of exotic nuclides in storage rings}},}\ }in\ \href
  {https://doi.org/10.1088/1742-6596/665/1/012053} {\emph {\bibinfo {booktitle}
  {Journal of Physics Conference Series}}},\ Vol.\ \bibinfo {volume} {665},\
  p.\ \bibinfo {pages} {012053}\BibitemShut {NoStop}%
\bibitem [{\citenamefont {{Yang}}\ \emph {et~al.}(2015)\citenamefont {{Yang}},
  \citenamefont {{Jin}}, \citenamefont {{Li}}, \citenamefont {{Covino}},
  \citenamefont {{Zheng}}, \citenamefont {{Hotokezaka}}, \citenamefont {{Fan}},
  \citenamefont {{Piran}},\ and\ \citenamefont {{Wei}}}]{Yang.Jin.ea:2015}%
  \BibitemOpen
  \bibfield  {author} {\bibinfo {author} {\bibnamefont {{Yang}}, \bibfnamefont
  {Bin}}, \bibinfo {author} {\bibfnamefont {Zhi-Ping}\ \bibnamefont {{Jin}}},
  \bibinfo {author} {\bibfnamefont {Xiang}\ \bibnamefont {{Li}}}, \bibinfo
  {author} {\bibfnamefont {Stefano}\ \bibnamefont {{Covino}}}, \bibinfo
  {author} {\bibfnamefont {Xian-Zhong}\ \bibnamefont {{Zheng}}}, \bibinfo
  {author} {\bibfnamefont {Kenta}\ \bibnamefont {{Hotokezaka}}}, \bibinfo
  {author} {\bibfnamefont {Yi-Zhong}\ \bibnamefont {{Fan}}}, \bibinfo {author}
  {\bibfnamefont {Tsvi}\ \bibnamefont {{Piran}}}, and\ \bibinfo {author}
  {\bibfnamefont {Da-Ming}\ \bibnamefont {{Wei}}}} (\bibinfo {year} {2015}),\
  \bibfield  {title} {\enquote {\bibinfo {title} {{A possible macronova in the
  late afterglow of the long-short burst GRB 060614}},}\ }\href
  {https://doi.org/10.1038/ncomms8323} {\bibfield  {journal} {\bibinfo
  {journal} {Nature Communications}\ }\textbf {\bibinfo {volume} {6}},\
  \bibinfo {eid} {7323}}\BibitemShut {NoStop}%
\bibitem [{\citenamefont {{Yee}}\ \emph {et~al.}(2013)\citenamefont {{Yee}}
  \emph {et~al.}}]{Yee2013}%
  \BibitemOpen
  \bibfield  {author} {\bibinfo {author} {\bibnamefont {{Yee}}, \bibfnamefont
  {R~M}},  \emph {et~al.}} (\bibinfo {year} {2013}),\ \bibfield  {title}
  {\enquote {\bibinfo {title} {{{$\beta$}-Delayed Neutron Spectroscopy Using
  Trapped Radioactive Ions}},}\ }\href
  {https://doi.org/10.1103/PhysRevLett.110.092501} {\bibfield  {journal}
  {\bibinfo  {journal} {Phys. Rev. Lett.}\ }\textbf {\bibinfo {volume} {110}},\
  \bibinfo {eid} {092501}}\BibitemShut {NoStop}%
\bibitem [{\citenamefont {Yeh}\ \emph {et~al.}(1983)\citenamefont {Yeh},
  \citenamefont {Clark}, \citenamefont {Scharff-Goldhaber}, \citenamefont
  {Chrien}, \citenamefont {Yuan}, \citenamefont {Shmid}, \citenamefont {Gill},
  \citenamefont {Evans}, \citenamefont {Dautet},\ and\ \citenamefont
  {Lee}}]{Yeh.Clark.ea:1983}%
  \BibitemOpen
  \bibfield  {author} {\bibinfo {author} {\bibnamefont {Yeh}, \bibfnamefont
  {T-R}}, \bibinfo {author} {\bibfnamefont {D.~D.}\ \bibnamefont {Clark}},
  \bibinfo {author} {\bibfnamefont {G.}~\bibnamefont {Scharff-Goldhaber}},
  \bibinfo {author} {\bibfnamefont {R.~E.}\ \bibnamefont {Chrien}}, \bibinfo
  {author} {\bibfnamefont {L.-J.}\ \bibnamefont {Yuan}}, \bibinfo {author}
  {\bibfnamefont {M.}~\bibnamefont {Shmid}}, \bibinfo {author} {\bibfnamefont
  {R.~L.}\ \bibnamefont {Gill}}, \bibinfo {author} {\bibfnamefont {A.~E.}\
  \bibnamefont {Evans}}, \bibinfo {author} {\bibfnamefont {H.}~\bibnamefont
  {Dautet}}, and\ \bibinfo {author} {\bibfnamefont {J.}~\bibnamefont {Lee}}}
  (\bibinfo {year} {1983}),\ \bibfield  {title} {\enquote {\bibinfo {title}
  {High resolution measurements of delayed neutron emission spectra from
  fission products},}\ }in\ \href
  {https://doi.org/10.1007/978-94-009-7099-1_55} {\emph {\bibinfo {booktitle}
  {Nuclear Data for Science and Technology}}},\ \bibinfo {editor} {edited by\
  \bibinfo {editor} {\bibfnamefont {K.~H.}\ \bibnamefont {B{\"o}ckhoff}}}\
  (\bibinfo  {publisher} {Springer Netherlands},\ \bibinfo {address}
  {Dordrecht})\ pp.\ \bibinfo {pages} {261--264}\BibitemShut {NoStop}%
\bibitem [{\citenamefont {{Yong}}\ \emph {et~al.}(2017)\citenamefont {{Yong}},
  \citenamefont {{Norris}}, \citenamefont {{Da Costa}}, \citenamefont
  {{Stanford}}, \citenamefont {{Karakas}}, \citenamefont {{Shingles}},
  \citenamefont {{Hirschi}},\ and\ \citenamefont
  {{Pignatari}}}]{Yong.Morris.ea:2017}%
  \BibitemOpen
  \bibfield  {author} {\bibinfo {author} {\bibnamefont {{Yong}}, \bibfnamefont
  {D}}, \bibinfo {author} {\bibfnamefont {J.~E.}\ \bibnamefont {{Norris}}},
  \bibinfo {author} {\bibfnamefont {G.~S.}\ \bibnamefont {{Da Costa}}},
  \bibinfo {author} {\bibfnamefont {L.~M.}\ \bibnamefont {{Stanford}}},
  \bibinfo {author} {\bibfnamefont {A.~I.}\ \bibnamefont {{Karakas}}}, \bibinfo
  {author} {\bibfnamefont {L.~J.}\ \bibnamefont {{Shingles}}}, \bibinfo
  {author} {\bibfnamefont {R.}~\bibnamefont {{Hirschi}}}, and\ \bibinfo
  {author} {\bibfnamefont {M.}~\bibnamefont {{Pignatari}}}} (\bibinfo {year}
  {2017}),\ \bibfield  {title} {\enquote {\bibinfo {title} {{A Chemical
  Signature from Fast-rotating Low-metallicity Massive Stars: ROA 276 in
  {$\omega$} Centauri}},}\ }\href {https://doi.org/10.3847/1538-4357/aa6250}
  {\bibfield  {journal} {\bibinfo  {journal} {Astrophys. J.}\ }\textbf
  {\bibinfo {volume} {837}},\ \bibinfo {eid} {176}}\BibitemShut {NoStop}%
\bibitem [{\citenamefont {{Yuan}}\ \emph {et~al.}(2020)\citenamefont {{Yuan}},
  \citenamefont {{Myeong}}, \citenamefont {{Beers}}, \citenamefont {{Evans}},
  \citenamefont {{Lee}}, \citenamefont {{Banerjee}}, \citenamefont {{Gudin}},
  \citenamefont {{Hattori}}, \citenamefont {{Li}}, \citenamefont {{Matsuno}},
  \citenamefont {{Placco}}, \citenamefont {{Smith}}, \citenamefont
  {{Whitten}},\ and\ \citenamefont {{Zhao}}}]{Yuan.Myeong.ea:2020}%
  \BibitemOpen
  \bibfield  {author} {\bibinfo {author} {\bibnamefont {{Yuan}}, \bibfnamefont
  {Zhen}}, \bibinfo {author} {\bibfnamefont {G.~C.}\ \bibnamefont {{Myeong}}},
  \bibinfo {author} {\bibfnamefont {Timothy~C.}\ \bibnamefont {{Beers}}},
  \bibinfo {author} {\bibfnamefont {N.~W.}\ \bibnamefont {{Evans}}}, \bibinfo
  {author} {\bibfnamefont {Young~Sun}\ \bibnamefont {{Lee}}}, \bibinfo {author}
  {\bibfnamefont {Projjwal}\ \bibnamefont {{Banerjee}}}, \bibinfo {author}
  {\bibfnamefont {Dmitrii}\ \bibnamefont {{Gudin}}}, \bibinfo {author}
  {\bibfnamefont {Kohei}\ \bibnamefont {{Hattori}}}, \bibinfo {author}
  {\bibfnamefont {Haining}\ \bibnamefont {{Li}}}, \bibinfo {author}
  {\bibfnamefont {Tadafumi}\ \bibnamefont {{Matsuno}}}, \bibinfo {author}
  {\bibfnamefont {Vinicius~M.}\ \bibnamefont {{Placco}}}, \bibinfo {author}
  {\bibfnamefont {M.~C.}\ \bibnamefont {{Smith}}}, \bibinfo {author}
  {\bibfnamefont {Devin~D.}\ \bibnamefont {{Whitten}}}, and\ \bibinfo {author}
  {\bibfnamefont {Gang}\ \bibnamefont {{Zhao}}}} (\bibinfo {year} {2020}),\
  \bibfield  {title} {\enquote {\bibinfo {title} {{Dynamical Relics of the
  Ancient Galactic Halo}},}\ }\href {https://doi.org/10.3847/1538-4357/ab6ef7}
  {\bibfield  {journal} {\bibinfo  {journal} {Astrophys. J.}\ }\textbf
  {\bibinfo {volume} {891}},\ \bibinfo {eid} {39}}\BibitemShut {NoStop}%
\bibitem [{\citenamefont {{Zampieri}}(2017)}]{Zampieri:2017}%
  \BibitemOpen
  \bibfield  {author} {\bibinfo {author} {\bibnamefont {{Zampieri}},
  \bibfnamefont {L}}} (\bibinfo {year} {2017}),\ \enquote {\bibinfo {title}
  {{Light Curves of Type II Supernovae}},}\ in\ \href
  {https://doi.org/10.1007/978-3-319-21846-5_26} {\emph {\bibinfo {booktitle}
  {Handbook of Supernovae}}},\ \bibinfo {editor} {edited by\ \bibinfo {editor}
  {\bibfnamefont {A.~W.}\ \bibnamefont {{Alsabti}}}\ and\ \bibinfo {editor}
  {\bibfnamefont {P.}~\bibnamefont {{Murdin}}}}\ (\bibinfo  {publisher}
  {Springer International Publishing},\ \bibinfo {address} {Cham})\BibitemShut
  {NoStop}%
\bibitem [{\citenamefont {{Zhang}}\ \emph {et~al.}(2016)\citenamefont
  {{Zhang}}, \citenamefont {{Schuetrumpf}},\ and\ \citenamefont
  {{Nazarewicz}}}]{Zhang.Schuetrumpf.Nazarewicz:2016}%
  \BibitemOpen
  \bibfield  {author} {\bibinfo {author} {\bibnamefont {{Zhang}}, \bibfnamefont
  {C~L}}, \bibinfo {author} {\bibfnamefont {B.}~\bibnamefont {{Schuetrumpf}}},
  and\ \bibinfo {author} {\bibfnamefont {W.}~\bibnamefont {{Nazarewicz}}}}
  (\bibinfo {year} {2016}),\ \bibfield  {title} {\enquote {\bibinfo {title}
  {{Nucleon localization and fragment formation in nuclear fission}},}\ }\href
  {https://doi.org/10.1103/PhysRevC.94.064323} {\bibfield  {journal} {\bibinfo
  {journal} {Phys. Rev. C}\ }\textbf {\bibinfo {volume} {94}},\ \bibinfo {eid}
  {064323}}\BibitemShut {NoStop}%
\bibitem [{\citenamefont {Zhi}\ \emph {et~al.}(2013)\citenamefont {Zhi},
  \citenamefont {Caurier}, \citenamefont {Cuenca-Garc{\'\i}a}, \citenamefont
  {Langanke}, \citenamefont {Mart{\'\i}nez-Pinedo},\ and\ \citenamefont
  {Sieja}}]{Zhi.Caurier.ea:2013}%
  \BibitemOpen
  \bibfield  {author} {\bibinfo {author} {\bibnamefont {Zhi}, \bibfnamefont
  {Q}}, \bibinfo {author} {\bibfnamefont {E.}~\bibnamefont {Caurier}}, \bibinfo
  {author} {\bibfnamefont {J.~J.}\ \bibnamefont {Cuenca-Garc{\'\i}a}}, \bibinfo
  {author} {\bibfnamefont {K.}~\bibnamefont {Langanke}}, \bibinfo {author}
  {\bibfnamefont {G.}~\bibnamefont {Mart{\'\i}nez-Pinedo}}, and\ \bibinfo
  {author} {\bibfnamefont {K.}~\bibnamefont {Sieja}}} (\bibinfo {year}
  {2013}),\ \bibfield  {title} {\enquote {\bibinfo {title} {Shell-model
  half-lives including first-forbidden contributions for $r$-process
  waiting-point nuclei},}\ }\href {https://doi.org/10.1103/PhysRevC.87.025803}
  {\bibfield  {journal} {\bibinfo  {journal} {Phys. Rev. C}\ }\textbf {\bibinfo
  {volume} {87}},\ \bibinfo {pages} {025803}}\BibitemShut {NoStop}%
\bibitem [{\citenamefont {{Zhou}}\ \emph {et~al.}(2019)\citenamefont {{Zhou}},
  \citenamefont {{Vink}}, \citenamefont {{Safi-Harb}},\ and\ \citenamefont
  {{Miceli}}}]{Zhou.Vink.ea:2019}%
  \BibitemOpen
  \bibfield  {author} {\bibinfo {author} {\bibnamefont {{Zhou}}, \bibfnamefont
  {Ping}}, \bibinfo {author} {\bibfnamefont {Jacco}\ \bibnamefont {{Vink}}},
  \bibinfo {author} {\bibfnamefont {Samar}\ \bibnamefont {{Safi-Harb}}}, and\
  \bibinfo {author} {\bibfnamefont {Marco}\ \bibnamefont {{Miceli}}}} (\bibinfo
  {year} {2019}),\ \bibfield  {title} {\enquote {\bibinfo {title} {{Spatially
  resolved X-ray study of supernova remnants that host magnetars: Implication
  of their fossil field origin}},}\ }\href
  {https://doi.org/10.1051/0004-6361/201936002} {\bibfield  {journal} {\bibinfo
   {journal} {Astron. \& Astrophys.}\ }\textbf {\bibinfo {volume} {629}},\
  \bibinfo {eid} {A51}}\BibitemShut {NoStop}%
\bibitem [{\citenamefont {{Zhu}}\ \emph {et~al.}(2020)\citenamefont {{Zhu}},
  \citenamefont {{Yang}}, \citenamefont {{Liu}}, \citenamefont {{Huang}},
  \citenamefont {{Zhang}}, \citenamefont {{Li}}, \citenamefont {{Yu}},\ and\
  \citenamefont {{Gao}}}]{Zhu.Yang.ea:2020}%
  \BibitemOpen
  \bibfield  {author} {\bibinfo {author} {\bibnamefont {{Zhu}}, \bibfnamefont
  {Jin-Ping}}, \bibinfo {author} {\bibfnamefont {Yuan-Pei}\ \bibnamefont
  {{Yang}}}, \bibinfo {author} {\bibfnamefont {Liang-Duan}\ \bibnamefont
  {{Liu}}}, \bibinfo {author} {\bibfnamefont {Yan}\ \bibnamefont {{Huang}}},
  \bibinfo {author} {\bibfnamefont {Bing}\ \bibnamefont {{Zhang}}}, \bibinfo
  {author} {\bibfnamefont {Zhuo}\ \bibnamefont {{Li}}}, \bibinfo {author}
  {\bibfnamefont {Yun-Wei}\ \bibnamefont {{Yu}}}, and\ \bibinfo {author}
  {\bibfnamefont {He}~\bibnamefont {{Gao}}}} (\bibinfo {year} {2020}),\
  \bibfield  {title} {\enquote {\bibinfo {title} {{Kilonova Emission From Black
  Hole-Neutron Star Mergers. I. Viewing-Angle-Dependent Lightcurves}},}\ }\href
  {https://doi.org/10.3847/1538-4357/ab93bf} {\bibfield  {journal} {\bibinfo
  {journal} {Astrophys. J.}\ }\textbf {\bibinfo {volume} {897}},\ \bibinfo
  {eid} {20}}\BibitemShut {NoStop}%
\bibitem [{\citenamefont {{Zhu}}\ \emph {et~al.}(2018)\citenamefont {{Zhu}}
  \emph {et~al.}}]{Zhu.Wollaeger.ea:2018}%
  \BibitemOpen
  \bibfield  {author} {\bibinfo {author} {\bibnamefont {{Zhu}}, \bibfnamefont
  {Y}},  \emph {et~al.}} (\bibinfo {year} {2018}),\ \bibfield  {title}
  {\enquote {\bibinfo {title} {{Californium-254 and Kilonova Light Curves}},}\
  }\href {https://doi.org/10.3847/2041-8213/aad5de} {\bibfield  {journal}
  {\bibinfo  {journal} {Astrophys. J.}\ }\textbf {\bibinfo {volume} {863}},\
  \bibinfo {eid} {L23}}\BibitemShut {NoStop}%
\bibitem [{\citenamefont {{Zhu}}\ \emph {et~al.}(2016)\citenamefont {{Zhu}},
  \citenamefont {{Perego}},\ and\ \citenamefont {{McLaughlin}}}]{zhu16}%
  \BibitemOpen
  \bibfield  {author} {\bibinfo {author} {\bibnamefont {{Zhu}}, \bibfnamefont
  {Y~L}}, \bibinfo {author} {\bibfnamefont {A.}~\bibnamefont {{Perego}}}, and\
  \bibinfo {author} {\bibfnamefont {G.~C.}\ \bibnamefont {{McLaughlin}}}}
  (\bibinfo {year} {2016}),\ \bibfield  {title} {\enquote {\bibinfo {title}
  {{Matter-neutrino resonance transitions above a neutron star merger
  remnant}},}\ }\href {https://doi.org/10.1103/PhysRevD.94.105006} {\bibfield
  {journal} {\bibinfo  {journal} {Phys. Rev. D}\ }\textbf {\bibinfo {volume}
  {94}},\ \bibinfo {eid} {105006}}\BibitemShut {NoStop}%
\bibitem [{\citenamefont {{Zrake}}\ and\ \citenamefont
  {{MacFadyen}}(2013)}]{Zrake.Macfadyen:2013}%
  \BibitemOpen
  \bibfield  {author} {\bibinfo {author} {\bibnamefont {{Zrake}}, \bibfnamefont
  {J}}, and\ \bibinfo {author} {\bibfnamefont {A.~I.}\ \bibnamefont
  {{MacFadyen}}}} (\bibinfo {year} {2013}),\ \bibfield  {title} {\enquote
  {\bibinfo {title} {{Magnetic Energy Production by Turbulence in Binary
  Neutron Star Mergers}},}\ }\href
  {https://doi.org/10.1088/2041-8205/769/2/L29} {\bibfield  {journal} {\bibinfo
   {journal} {Astrophys. J.}\ }\textbf {\bibinfo {volume} {769}},\ \bibinfo
  {eid} {L29}}\BibitemShut {NoStop}%
\end{thebibliography}%

\end{document}